\documentclass[11pt]{article}
\usepackage[margin=1in]{geometry}

\usepackage{stackrel}
\usepackage{amsthm}
\usepackage{enumitem}
\usepackage{mathtools}

\newtheorem{theorem}{Theorem}[section]
\newtheorem{lemma}[theorem]{Lemma}
\newtheorem{definition}[theorem]{Definition}
\newtheorem{corollary}[theorem]{Corollary}

\newtheorem{remark}[theorem]{Remark}

\newtheorem{proposition}[theorem]{Proposition}

\usepackage[lined, algoruled, linesnumbered]{algorithm2e}
\usepackage{hyperref,xcolor}
\definecolor{winered}{rgb}{0.5,0,0}
\hypersetup
{
    pdfborder={0 0 0},
    colorlinks=true,
    linkcolor={winered},
    urlcolor={winered},
    filecolor={winered},
    citecolor={winered},
    linktoc=all,
}

\newcommand{\ignore}[1]{}

\title{Computing the $5$-Edge-Connected Components in Linear Time}

\usepackage{wrapfig}

\begin{document}

\author{Evangelos Kosinas\thanks{Department of Computer Science \& Engineering,
University of Ioannina, Greece.
E-mail: \texttt{ekosinas@cs.uoi.gr}. 
The research work was supported by the Hellenic
Foundation for Research and Innovation (HFRI) under the 3rd Call for HFRI PhD Fellowships (Fellowship Number:
6547).}} 

\maketitle

\thispagestyle{empty}

\begin{abstract}
We provide a deterministic algorithm for computing the $5$-edge-connected components of an undirected multigraph in linear time. There were probably good indications that this computation can be performed in linear time, but no such algorithm was actually known prior to this work. Thus, our paper answers a theoretical question, and sheds light on the possibility that a solution may exist for general $k$. Furthermore, although the algorithm that we provide is quite extensive and broken up into several pieces, it can have an almost-linear time implementation with the use of elementary data structures. A key component in our algorithm is an oracle for answering connectivity queries for pairs of vertices in the presence of at most four edge-failures. Specifically, the oracle has size $O(n)$, it can be constructed in linear time, and it answers connectivity queries in the presence of at most four edge-failures in $O(1)$ time, where $n$ denotes the number of vertices of the graph. We note that this is a result of independent interest.

Our paper can be considered as a follow-up of recent work on computing the $4$-edge-connected components in linear time. Specifically, we follow a DFS-based approach in order to compute a collection of $4$-edge cuts, that is rich enough in properties for our purposes. Furthermore, we expand the toolkit of DFS-based concepts, and demonstrate its general usefulness. In particular, our oracle for connectivity queries is also based on them. However, in dealing with the computation of the $5$-edge-connected components, we are faced with unique challenges that do not appear when dealing with lower connectivity. The problem is that the $4$-edge cuts in $3$-edge-connected graphs are entangled in various complicated ways, that make it difficult to organize them in a compact way. Here we provide a novel analysis of those cuts, that reveals the existence of various interesting structures. These can be exploited so that we can disentangle and collect only those cuts that are essential in computing the $5$-edge-connected components. This analysis may provide a clue for a general solution for the $k$-edge-connected components, or other related graph connectivity problems.
\end{abstract}

\clearpage

\pagestyle{empty}

\small

\tableofcontents

\clearpage

\pagestyle{plain}

\normalsize

\setcounter{page}{1}

\section{Introduction}

\subsection{Problem definition}
We assume that the reader is familiar with standard graph terminology, as provided e.g. in \cite{DBLP:books/daglib/0030488} or \cite{DBLP:books/cu/NI2008}. Let $G=(V,E)$ be an undirected multigraph. We say that two vertices $x$ and $y$ of $G$ are $k$-edge-connected if we have to remove at least $k$ edges from $G$ in order to destroy all paths from $x$ to $y$. In general, a set of $k$ edges with the property that its removal from $G$ disconnects at least one pair of vertices, is called a $k$-edge cut of $G$. Equivalently, by Menger's theorem we have that $x$ and $y$ are $k$-edge-connected if there are at least $k$ edge-disjoint paths from $x$ to $y$ (see, e.g., \cite{DBLP:books/cu/NI2008}). We denote this condition at $x\equiv_{k}y$. It is easy to see that $\equiv_k$ is an equivalence relation on $V$. The equivalence classes of $\equiv_k$ are called the $k$-edge-connected components of $G$. 

Determining the $k$-edge-connectivity relation is a fundamental graph connectivity problem. The case $k=1$ coincides with the computation of the connected components, and can be solved easily with a standard graph traversal (like BFS or DFS). For $k=2$, Tarjan~\cite{DBLP:journals/siamcomp/Tarjan72} provided a linear-time algorithm, that essentially finds all the bridges of the graph. The case $k=3$ was initially solved in linear time through the reduction of Galil and Italiano \cite{DBLP:journals/sigact/GalilI91} from the triconnectivity algorithm of Hopcroft and Tarjan \cite{DBLP:journals/siamcomp/HopcroftT73}. Afterwards, more linear-time algorithms were developed for $k=3$, that did not rely on this reduction and were much simpler (see e.g., \cite{DBLP:books/cu/NI2008, DBLP:journals/jda/Tsin09}). Relatively recently, linear-time algorithms for the case $k=4$ were presented \cite{DBLP:conf/esa/GeorgiadisIK21, DBLP:conf/esa/NadaraRSS21}. Although we are not aware of any specific application of the case $k=5$ (or beyond), it is not known if we can compute the $k$-edge-connected components in linear time for $k\geq 5$ (not even with randomized algorithms), and this seems to be an intriguing problem. Thus, considering the case $k=5$ seems to be the natural next step in order to determine whether this computation is possible in linear time for general fixed $k$.

\subsection{Related work}
The best time bounds that we have for computing the $5$-edge-connected components are almost linear, and they are derived from solutions of more general versions of the problem that we consider. Specifically, Dinitz and Nossenson \cite{DBLP:conf/swat/DinitzN00} have provided an algorithm for maintaining the relation of $5$-edge-connectivity in incremental graphs. More precisely, starting from an empty graph, they show how to process a sequence of $n$ insertions of vertices and $m$ insertions of edges in $O(m+n\log^2 n)$ time in total, so that, at any point in this sequence, we can answer $5$-edge-connectivity queries for pairs of vertices in constant time. Since the relation of $5$-edge-connectivity with this algorithm is essentially maintained with the use of a disjoint-set union data structure (DSU), we can use this incremental algorithm in order to report the $5$-edge-connected components of a graph $G$, after we have started from the empty graph and we have inserted from the beginning all the edges of $G$. Thus, we have an $O(m+n\log^2 n)$-time algorithm for computing the $5$-edge-connected components. Since this comes from an incremental algorithm, it is reasonable to expect that this computation can be performed even faster on a static graph. It seems difficult to achieve this from the work of Dinitz and Nossenson for the following reasons. First, this comes from an extended abstract, but we were not able to find the journal version that would contain the full details. And second, this algorithm relies on the $2$-level cactus of the $(\mathit{minimum}+1)$-cuts \cite{DBLP:conf/stoc/DinitzN95}, which is quite involved, and we do not know how to construct it in linear time. Instead, we start anew the analysis of the structure of $4$-cuts in $3$-edge-connected graphs, that enables us to compute enough of them in linear time, so that we can derive the $5$-edge-connected components.

We note that the problem of maintaining the $k$-edge-connectivity relation in dynamic graphs is a problem that has received a lot of attention. First, for the case $k\in\{2,3\}$ there are optimal and almost optimal solutions for incremental graphs, that process a sequence of $n$ vertex insertions and $m$ edge insertions and queries in $O(n+m\alpha(m,n))$ time, where $\alpha$ is an inverse of Ackermann's function (see \cite{DBLP:journals/algorithmica/WestbrookT92, DBLP:journals/siamcomp/GalilI93, DBLP:journals/dm/PoutreLO93, DBLP:journals/siamcomp/Poutre00}). For $k=4$, Dinitz and Westbrook~\cite{DBLP:journals/algorithmica/DinitzW98} presented an algorithm that processes a sequence of $n$ vertex insertions and $m$ edge insertions in $O(m+n\log n)$ time, so that, in the meantime, we can answer $4$-edge-connectivity queries in constant time. Very recently, Jin and Sun \cite{DBLP:conf/focs/0001S21} presented a deterministic algorithm for answering $k$-edge-connectivity queries in a fully dynamic graph in $n^{o(1)}$ worst case update and query time for any positive integer $k = (\log n)^{o(1)}$ for a graph with $n$ vertices. This is a very remarkable result, but it is highly complicated, and it does not seem to provide an algorithm for computing the $k$-edge-connected components in, say, $O(n\cdot n^{o(1)})$ time, because it only computes the answer to the queries in response to them, without maintaining explicitly the $k$-edge-connected components (as do the algorithms e.g. in \cite{DBLP:journals/algorithmica/WestbrookT92, DBLP:journals/siamcomp/GalilI93, DBLP:journals/dm/PoutreLO93, DBLP:journals/algorithmica/DinitzW98}). However, this result, since it applies to fully dynamic graphs, makes it seem reasonable that the $k$-edge-connected components can be computed in (almost) linear time, for general fixed $k$. 

Another route for computing the $k$-edge-connected components is given by {Gomory-Hu trees}~\cite{doi:10.1137/0109047Gomory}. A Gomory-Hu tree of a graph $G$ is a weighted tree on the same vertex set as $G$, with the property that $(1)$ the minimum weight of an edge on the tree-path that connects any two vertices $x$ and $y$ coincides with the edge-connectivity of $x$ and $y$ in $G$, and $(2)$ by taking the connected components of the tree after removing such an edge we get a minimum cut of $G$ that separates $x$ and $y$. Thus, given a Gomory-Hu tree, we can easily compute the $k$-edge-connected components in linear time, for any fixed $k$, by simply removing all edges with weight less than $k$ from the tree, and then gathering the connected components. However, the computation of the Gomory-Hu tree itself is very demanding. The original algorithm of Gomory and Hu can take as much as $\Omega(mn)$ time for a graph with $m$ edges and $n$ vertices. In a recent breakthrough, Abboud et al. \cite{DBLP:conf/focs/AbboudK0PST22} provided a randomized Monte Carlo construction of Gomory-Hu trees that takes $\widetilde{O}(n^2)$\footnote{The $\widetilde{O}$ notation hides polylogarithmic factors.} time in general weighted graphs with $n$ vertices. Furthermore, using the recent $m^{1+o(1)}$-time max-flow algorithm of Chen et al.~\cite{DBLP:conf/focs/ChenKLPGS22}, Abboud et al. \cite{DBLP:conf/focs/AbboudK0PST22} provide a randomized Monte Carlo algorithm that runs in $m^{1+o(1)}$ time in unweighted graphs with $m$ edges. Thus, we can compute the $k$-edge-connected components, for any fixed $k$, with a randomized Monte Carlo algorithm in $m^{1+o(1)}$ time. For our purposes, it seems more fitting to use a \emph{partial} Gomory-Hu tree, introduced by Hariharan et al.~\cite{DBLP:conf/soda/HariharanKP07}. This has the same properties as a general Gomory-Hu tree (i.e., $(1)$ and $(2)$), but it captures the $k$-edge-connectivity relation only up to a bounded $k$. Hariharan et al.~\cite{DBLP:conf/soda/HariharanKP07} showed how to compute a partial Gomory-Hu tree, for edge-connectivity up to a fixed $k$, in expected $O(m+kn\log n)$ time. Thus, we get an algorithm of expected $O(m+kn\log n)$ time for computing the $k$-edge-connected components, for any fixed $k$. 

From this general overview of the history of this subject (which omits several other related advances, such as determining the vertex-connectivity relation~\cite{DBLP:conf/stoc/PettieSY22}, or computing the $k$-edge-connected components in directed graphs~\cite{DBLP:conf/soda/GeorgiadisKPP23}), we can see that determining various notions of edge-connectivity is an area of active interest. However, the precise computation of the $k$-edge-connectivity relation in linear time, for general fixed $k$, is still an elusive open problem, that demands a deeper understanding of the structure of cuts in undirected graphs.

\subsection{Our contribution}
Here we present a deterministic linear-time algorithm for computing the $5$-edge-connected components of an undirected multigraph. This result relies on a novel analysis of the structure of $4$-cuts in $3$-edge-connected graphs. This analysis is crucial in order to guide us to a selection of enough $4$-cuts that can provide the partition of the $5$-edge-connected components. The second half of this work is devoted to the development of a linear-time algorithm that computes a compact representation of all $4$-cuts of a $3$-edge-connected graph. (The precise meaning of this term will be given in the Technical Overview, in the following section.) The state of the art in deterministically computing even a single $4$-cut is the algorithm of Gabow~\cite{DBLP:journals/talg/Gabow16} that runs in $O(m+n\log n)$ time in a graph with $n$ vertices and $m$ edges. Thus, we present the first deterministic algorithm that computes a $4$-cut of a graph, and tests the $5$-edge-connectivity in linear time.

In addition to computing the $5$-edge-connected components, we also provide a linear-time construction of an oracle that can answer in constant time queries of the form ``given two vertices $x$ and $y$, report a $4$-cut that separates $x$ and $y$, or determine that no such $4$-cut exists" (see Corollary~\ref{corollary:partial-gomory-hu}). 
In essence, we provide a data structure that retains the full functionality of a partial Gomory-Hu tree for $5$-edge-connectivity.

An indispensable tool in our analysis is the concept of the \emph{essential} $4$-cuts. These are the $4$-cuts that separate at least one pair of vertices that are $4$-edge-connected. We do not know if this concept (or its generalization) has been used before in the literature, but it is reasonable to care about the essential $4$-cuts when we want to compute the relation of $5$-edge-connectivity. In fact, by retaining only the essential $4$-cuts in the end, we have some convenient properties that enable us to derive efficiently the partition of the $5$-edge-connected components. 

We show how to process a graph in linear time, so that we can check the essentiality of any given $4$-cut in constant time. This relies on an oracle for answering connectivity queries in the presence of at most four edge-failures. Specifically, we show how to preprocess a graph $G$ with $n$ vertices and $m$ edges in $O(m+n)$ time, in order to derive an oracle of size $O(n)$ that can answer in $O(1)$ time queries of the form ``given a set of edges $E'$ with $|E'|\leq 4$ and two vertices $x$ and $y$, are $x$ and $y$ connected in $G\setminus E'$?". We achieve this result by using a DFS-tree of the graph, and by making a creative use of some DFS-based parameters, in order to reconstruct on a high-level the connected components of the graph upon removal of a set of (at most four) edges.

We note that this oracle is a special instance of the general problem of designing an oracle that answers connectivity queries in the presence of edge-failures \cite{DBLP:conf/focs/PatrascuT07}. The currently best bounds for this problem are given by Duan and Pettie in \cite{DBLP:journals/siamcomp/DuanP20}, where they show how to construct an oracle of $O(m\log{\log n})$ (or $O(m)$) size\footnote{The time-bounds for constructing the oracle are not specified.}, so that, given a set $E'$ of at most $d$ edges, one can answer connectivity queries in $G\setminus E'$ in $O(\mathit{min}\{\frac{\log{\log n}}{\log{\log{\log n}}},\frac{\log d}{\log{\log n}}\})$ time, after a $O(d^2\log{\log n})$-time (or $O(d^2\log^{\epsilon} n)$-time, for any $\epsilon>0$) preprocessing. Thus, the oracle that we provide improves on the state of the art in the case where $d$ is a fixed constant, upper bounded by $4$. It is an interesting question whether we can achieve the same bounds for larger fixed $d$. We believe that this is probably the case, but it appears that this is a very challenging combinatorial problem. 

Finally, we note that our algorithm for computing the $5$-edge-connected components, although it is quite extensive and broken up into several pieces, has an almost linear-time implementation with the use of elementary data structures. Specifically, the only sophisticated data structures that we use in order to achieve linear time are the DSU data structure of Gabow and Tarjan \cite{DBLP:conf/stoc/GabowT83}, 
and any linear-time algorithm for answering off-line NCA queries (e.g., \cite{DBLP:journals/siamcomp/HarelT84} or \cite{DBLP:journals/siamcomp/BuchsbaumGKRTW08}). We note that, in particular, the DSU data structure of Gabow and Tarjan utilizes the power of the RAM model of computation. Thus, it is still an open question whether the computation of the $5$-edge-connected components can be performed in linear time without using the power of the RAM model. For practical purposes, however, there are implementations for those data structures that run in almost linear time, with an overhead  of only an inverse of Ackermann's function \cite{DBLP:journals/jacm/Tarjan75}. Thus, in practice, one could use those implementations for our algorithm, in order to achieve almost-linear time.

\subsection{Technical overview}
First, let us fix some terminology that we will use throughout. In this work, all graphs are undirected multigraphs. It is convenient to consider only \emph{edge-minimal} cuts. Thus, whenever we consider a $k$-edge cut $C$ of a connected graph $G$, we assume that $C$ is minimal w.r.t. the property that $G\setminus C$ is disconnected. For simplicity, we call $C$ a \emph{$k$-cut} of $G$. If two vertices $x$ and $y$ belong to different connected components of $G\setminus C$, then we say that $C$ separates $x$ and $y$. Notice that two vertices of $G$ are $5$-edge-connected if and only if there is no $k$-cut that separates them, for any $k\leq 4$. A $k$-cut that separates at least one pair of $k$-edge-connected vertices is called an \emph{essential} $k$-cut.

There is a duality between cuts of $G$ and bipartitions of $V(G)$. ($V(G)$ denotes the vertex set of $G$.) Specifically, if $C$ is a cut of $G$ and $X,Y$ are the connected components of $G\setminus C$, then we have $E_G[X,Y]=C$ (where $E_G[X,Y]=\{(x,y)\in E(G)\mid x\in X \mbox{ and } y\in Y\}$). Thus, we can view $C$ either as a set of edges, or as the bipartition $\{X,Y\}$ of $V(G)$. $X$ and $Y$ are also called the \emph{sides} of $C$. If $X$ is a subset of $V(G)$, then we denote $E_G[X,V(G)\setminus X]$ as $\partial(X)$. If $r$ is a vertex of $G$, and $X$ is the side of a cut $C$ of $G$ that does not contain $r$, then we call $|X|$ the \emph{$r$-size} of $C$.

Let $C$ and $C'$ be two cuts of $G$, with sides $X,Y$ and $X',Y'$, respectively. If at least one of the intersections $X\cap X'$, $X\cap Y'$, $Y\cap X'$ or $Y\cap Y'$ is empty, then we say that $C$ and $C'$ are \emph{parallel}. A collection $\mathcal{C}$ of cuts of $G$ that are pairwise parallel is called a \emph{parallel family} of cuts of $G$. It is a known fact that a parallel family of cuts of a graph with $n$ vertices contains at most $O(n)$ cuts (see, e.g., \cite{DBLP:conf/stoc/DinitzN95}).

If $\mathcal{P}$ is a collection of partitions of a set $V$, then we let $\mathit{atoms}(\mathcal{P})$ denote the partition of $V$ that is given by the mutual refinement of all partitions in $\mathcal{P}$. In other words, $\mathit{atoms}(\mathcal{P})$ is defined by the property that two elements $x$ and $y$ of $V$ belong to two different sets in $\mathit{atoms}(\mathcal{P})$ if and only if there is a partition $P\in\mathcal{P}$ such that $x$ and $y$ belong to different sets in $P$. This terminology is convenient for the following reason. Let $\mathcal{C}_\mathit{kcuts}$ denote the collection of all $k$-cuts of a connected graph $G$, for every $k\geq 1$. Then the partition of the $5$-edge-connected components of $G$ is given by $\mathit{atoms}(\mathcal{C}_\mathit{1cuts}\cup\mathcal{C}_\mathit{2cuts}\cup \mathcal{C}_\mathit{3cuts}\cup \mathcal{C}_\mathit{4cuts})$.

In Section~\ref{section:using-a-dfs-tree} we provide the following results. First, we show that, given a graph $G$ and a vertex $r$ of $G$, there is a linear-time preprocessing of $G$ such that we can report the $r$-size of any $4$-cut of $G$ in $O(1)$ time (see Lemma~\ref{lemma:determine-components}). Second, there is a linear-time preprocessing of $G$ such that, given a $4$-cut $C$ of $G$, we can check if $C$ is an essential $4$-cut in $O(1)$ time (see Proposition~\ref{proposition:essentiality}). And third, given a parallel family $\mathcal{C}$ of $4$-cuts of $G$, we can compute the atoms of $\mathcal{C}$ in linear time (see Proposition~\ref{proposition:algorithm:atoms-of-parallel}). In order to establish Proposition~\ref{proposition:essentiality}, we utilize the oracle that we develop in Section~\ref{section:4eoracle}, for answering connectivity queries in the presence of at most four edge-failures.

\subsubsection{Reduction to $3$-edge-connected graphs}
\label{section:intro-reducing}

We rely on a construction that was described by Dinitz \cite{DBLP:conf/wg/Dinitz92}, that enables us to reduce the computation of the $5$-edge-connected components to $3$-edge-connected graphs. (We note that this was also used by \cite{DBLP:conf/esa/GeorgiadisIK21} and \cite{DBLP:conf/esa/NadaraRSS21} in order to compute the $4$-edge-connected components.) Specifically, \cite{DBLP:conf/wg/Dinitz92} provided the following result. Let $G$ be a graph, and let $S_1,\dots,S_t$ be the $3$-edge-connected components of $G$. Then, we can augment the graphs $G[S_1],\dots,G[S_t]$ with the addition of $O(|V(G)|)$ artificial edges, so that the resulting graphs $G'[S_1],\dots,G'[S_t]$ have the property that $(1)$ $G'[S_i]$ is $3$-edge-connected for every $i\in\{1,\dots,t\}$, $(2)$ for every $k$-edge-connected component $S$ of $G$, for $k\geq 3$, there is an $i\in\{1,\dots,t\}$ such that $G'[S_i]$ contains $S$ as a $k$-edge-connected component, and $(3)$ for every $i\in\{1,\dots,t\}$, and every $k\geq 3$, a $k$-edge-connected component of $G'[S_i]$ is also a $k$-edge-connected component of $G$. We note that the auxiliary graphs $G'[S_1],\dots,G'[S_t]$ can be constructed easily in linear time in total, after computing the $3$-edge-connected components of $G$ (using, e.g., the algorithm from \cite{DBLP:journals/jda/Tsin09}). Thus, in order to compute the $5$-edge-connected components of $G$ in linear time, properties $(1)$, $(2)$ and $(3)$ imply that it is enough to know how to compute in linear time the $5$-edge-connected components of a $3$-edge-connected graph. 

We note that we do not know how to produce auxiliary $4$-edge-connected graphs, with properties like $(1)$, $(2)$ and $(3)$, that can provide the $5$-edge-connected components.
However, even if we knew how to do that, we would still be faced with the problem of computing enough $4$-cuts in order to derive the $5$-edge-connected components. Although the work of Gabow~\cite{DBLP:journals/talg/Gabow16} shows that we can compute the $k$-edge-connected components of a $(k-1)$-edge-connected graph in $O(m+k^2n\log (n/k))$ time, there are no indications that computing the $5$-edge-connected components of a $4$-edge-connected graph in \emph{linear time} is a much easier problem than working directly on $3$-edge-connected graphs. 

\subsubsection{Computing enough $4$-cuts of a $3$-edge-connected graph}
Let $G$ be a $3$-edge-connected graph with $n$ vertices and $m$ edges.
In order to compute the $5$-edge-connected components of $G$, we have to solve simultaneously the following two problems.
First, we have to compute a collection $\mathcal{C}$ of $4$-cuts that are enough in order to provide the $5$-edge-connected components. At the same time, we must be able to efficiently compute the atoms of $\mathcal{C}$ (in order to derive the $5$-edge-connected components). The straightforward way to compute these atoms is to compute the bipartition of the connected components after the removal of every $4$-cut in $\mathcal{C}$, and then return the mutual refinement of all those bipartitions. However, if the number of $4$-cuts in $\mathcal{C}$ is $\Omega(n)$, then this procedure will take $\Omega(nm)$ time in total, which is very far from our linear-time goal. Nevertheless, if $\mathcal{C}$ is a parallel family of $4$-cuts, then the computation of the atoms of $\mathcal{C}$ can be performed in $O(m+n)$ time (see Proposition~\ref{proposition:algorithm:atoms-of-parallel}). Furthermore, in this case we can construct in linear time an oracle of $O(n)$ size that can report in constant time a $4$-cut that separates any given pair of vertices, or determine that no such $4$-cut exists (see Corollary~\ref{corollary:atoms-of-parallel}). Thus, our goal is precisely to compute a parallel family of $4$-cuts that can provide the $5$-edge-connected components. 

It turns out that this is a highly non-trivial task. First of all, even computing a single $4$-cut takes $O(m+n\log n)$ time with the state-of-the-art method (which is Gabow's mincut algorithm~\cite{DBLP:journals/talg/Gabow16}). On the other hand, it would be impractical to compute \emph{all} $4$-cuts of the graph, no matter the algorithm used, since the number of all $4$-cuts in a $3$-edge-connected graph can be as high as $\Omega(n^2)$ even in graphs with $O(n)$ edges. Our approach, instead, is to compute a compact collection of all $4$-cuts that has size $O(n)$. When we say a ``compact collection", we mean that there is a procedure, through which, from this collection of $4$-cuts, we can essentially derive all $4$-cuts. At this point, it is necessary to precisely define our concepts. First, we have the following property of $4$-cuts in $3$-edge-connected graphs.

\begin{lemma}[Implied $4$-cut]
\label{lemma:intro-implied-4-cut}
Let $\{e_1,e_2,e_3,e_4\}$ and $\{e_3,e_4,e_5,e_6\}$ be two distinct $4$-cuts of a $3$-edge-connected graph $G$. Then $\{e_1,e_2,e_5,e_6\}$ is also a $4$-cut of $G$. 
\end{lemma}   
\begin{proof}
See Lemma~\ref{lemma:implied4cut}.
\end{proof}


Then, Lemma~\ref{lemma:intro-implied-4-cut} motivates the following.

\begin{definition}[Implicating sequences of $4$-cuts]
\normalfont
Let $\mathcal{C}$ be a collection of $4$-cuts of a $3$-edge-connected graph $G$. Let $p_1,\dots,p_{k+1}$ be a sequence of pairs of edges, and let $C_1,\dots,C_k$ be a sequence of $4$-cuts from $\mathcal{C}$, such that $C_i=p_i\cup p_{i+1}$ for every $i\in\{1,\dots,k\}$, and $C=p_1\cup p_{k+1}$ is a $4$-cut of $G$. Then we say that $C$ is implied from $\mathcal{C}$ through the pair of edges $p_1$ (or equivalently: through the pair of edges $p_{k+1}$). In this case, we say that $C_1,\dots,C_k$ is an implicating sequence of $\mathcal{C}$. If $\mathcal{C}$ implies every $4$-cut of $G$, then we say that $\mathcal{C}$ is a \emph{complete collection} of $4$-cuts of $G$.
\end{definition} 

One of our main results is the following.

\begin{theorem}
\label{theorem:intro-main}
Let $G$ be a $3$-edge-connected graph with $m$ edges and $n$ vertices. Then, in $O(m+n)$ time, we can compute a complete collection $\mathcal{C}$ of $4$-cuts of $G$ with $|\mathcal{C}|=O(n)$.
\end{theorem}
\begin{proof}
See Theorem~\ref{theorem:main}.
\end{proof}

It is not at all obvious why a complete collection of $4$-cuts with size $O(n)$ should exist. Furthermore, computing such a collection in linear time seems to be a very difficult problem, considering that it is not even known how to compute a single $4$-cut in linear time.
In particular, with Theorem~\ref{theorem:intro-main} we improve on the state of the art in computing a mincut of bounded cardinality as follows.

\begin{corollary}
Let $G$ be any graph. Then, in linear time, we can compute a $k$-cut of $G$, with $k\leq 4$, or determine that $G$ is $5$-edge-connected.
\end{corollary}
\begin{proof}
From previous work~\cite{DBLP:journals/siamcomp/Tarjan72, DBLP:journals/jda/Tsin09, DBLP:conf/esa/GeorgiadisIK21,DBLP:conf/esa/NadaraRSS21}, we know that, in linear time, we can compute a $k$-cut of $G$, with $k\leq 3$, or determine that $G$ is $4$-edge-connected. So let us assume that $G$ is $4$-edge-connected. Then, Theorem~\ref{theorem:intro-main} implies that, in linear time, we can compute a $4$-cut of $G$, or determine that $G$ is $5$-edge-connected. 
\end{proof}

More than half of this paper is devoted to establishing Theorem~\ref{theorem:intro-main}. The high-level idea is to identify the $4$-cuts on a DFS-tree of the graph. We can distinguish various types of $4$-cuts on a DFS-tree, and there is enough structure that enables us to compute a specific selection of them, that implies all $4$-cuts of the graph. For a detailed elaboration on this idea we refer to Section~\ref{section:computing-4-cuts}. We note that the bulk of this work would be significantly reduced if we had a simpler algorithm for computing a complete collection of $4$-cuts with at most linear size. At the moment, we do not know any alternative method to do this. However, even computing a near-linear sized complete collection of $4$-cuts (in near-linear time), would still provide a near-linear time algorithm for computing the $5$-edge-connected components, by following the same analysis. Thus, there is room for simplifying the computation a lot, by relaxing the strictness of the linear complexity. In any case, given a complete collection of $4$-cuts, we are faced with the problem of how to use this package of information in order to derive the $5$-edge-connected components. This is what we discuss next.

\subsubsection{Unpacking the implicating sequences of a complete collection of $4$-cuts}
Given a complete collection $\mathcal{C}$ of $4$-cuts of $G$, the challenge is to unpack as many $4$-cuts as are needed, in order to derive the $5$-edge-connected components. The first thing we do is to implicitly expand all implicating sequences of $\mathcal{C}$, and organize them in collections of pairs of edges that generate, in total, all the $4$-cuts of the graph. The concept of \emph{generating} $4$-cuts is made precise in the following.

\begin{definition}
\normalfont
Let $\mathcal{F}=\{p_1,\dots,p_k\}$ be a collection of pairs of edges of $G$, with $k\geq 2$, such that $p_i\cup p_j$ is a $4$-cut of $G$ for every $i,j\in\{1,\dots,k\}$ with $i\neq j$. Then we say that $\mathcal{F}$ generates the collection of $4$-cuts $\{p_i\cup p_j\mid i,j\in\{1,\dots,k\}, i\neq j\}$.
\end{definition} 

An important intermediate result that we use throughout is the following.

\begin{proposition}
\label{proposition:intro-generate-implied}
Let $\mathcal{C}$ be a collection of $4$-cuts of $G$. Then, in $O(n+|\mathcal{C}|)$ time, we can construct a set $\{\mathcal{F}_1,\dots,\mathcal{F}_k\}$ of collections of pairs of edges, with $|\mathcal{F}_i|\geq 2$ for every $i\in\{1,\dots,k\}$, such that $\mathcal{F}_i$ generates a collection of $4$-cuts implied by $\mathcal{C}$, and every $4$-cut implied by $\mathcal{C}$ is generated by $\mathcal{F}_i$, for some $i\in\{1,\dots,k\}$. The total size of $\{\mathcal{F}_1,\dots,\mathcal{F}_k\}$ is $O(|\mathcal{C}|)$ (i.e., $|\mathcal{F}_1|+\dots+|\mathcal{F}_k|=O(|\mathcal{C}|)$).
\end{proposition}
\begin{proof}
See Proposition~\ref{proposition:cyclic_families_algorithm}.
\end{proof}

Thus, given a complete collection $\mathcal{C}$ of $4$-cuts, the collections of pairs of edges that we get in Proposition~\ref{proposition:intro-generate-implied} constitute an alternative compact representation of the collection of all $4$-cuts of the graph. In order to establish Proposition~\ref{proposition:intro-generate-implied}, we use an algorithm that breaks up every $4$-cut from $\mathcal{C}$ into its three different partitions into pairs of edges, and then greedily reassembles all implicating sequences of $\mathcal{C}$, by constructing maximal collections of pairs of edges that participate in an implicating sequence. Specifically, for every bipartition $\{p,q\}$ of a $4$-cut $C\in\mathcal{C}$ into pairs of edges, we generate two elements $(C,p)$ and $(C,q)$. Then, we consider the elements $(C,p)$ and $(C,q)$ as connected, by introducing an artificial edge that joins them. Notice that $\mathcal{F}=\{p,q\}$ is a collection of pairs of edges that generates a $4$-cut implied by $\mathcal{C}$. Then, we try to expand $\mathcal{F}$ as much as possible, into a collection of pairs of edges that generates $4$-cuts implied by $\mathcal{C}$, by tracing the implicating sequences of $\mathcal{C}$ that use $p$ or $q$. Thus, if e.g. another $4$-cut $C'\in\mathcal{C}$ contains the pair of edges $p$, and is partitioned as $C'=p\cup q'$, then we also consider the element $(C',p)$ connected with $(C',q')$, and the element $(C',p)$ connected with $(C,p)$, so that $\mathcal{F}'=\{p,q,q'\}$ is a collection of pairs of edges that generates $4$-cuts implied by $\mathcal{C}$. The precise method by which we create these collections of pairs of edges is shown in Algorithm~\ref{algorithm:generatefamilies}.
The output of Algorithm~\ref{algorithm:generatefamilies} has some nice properties that we analyze in Sections~\ref{section:generating-collections}, \ref{section:iso-and-quasi-iso} and \ref{section:additional-properties}. 

Let $\mathcal{F}$ be a collection of pairs of edges that is returned by Algorithm~\ref{algorithm:generatefamilies}. Then we distinguish three different cases for $\mathcal{F}$: either $|\mathcal{F}|>3$, or $|\mathcal{F}|=3$, or $|\mathcal{F}|=2$. 
The collections of pairs of edges that have size more than $3$ generate collections of $4$-cuts that have a very convenient structure for computational purposes. These are discussed next.

\subsubsection{Cyclic families of $4$-cuts, and minimal $4$-cuts}

Notice that if we have a collection of $k$ pairs of edges that generates a collection $\mathcal{C}$ of $4$-cuts, then $|\mathcal{C}|=k(k-1)/2$.
Now, the reason that the number of $4$-cuts in $3$-edge-connected graphs with $n$ vertices can be as high as $\Omega(n^2)$ is essentially the existence of some families of $4$-cuts that are captured in the following.\footnote{We note that Definition~\ref{definition:intro-cyclic-families} is similar to the concept of a \emph{circular partition} (given e.g. in \cite{DBLP:books/cu/NI2008} or \cite{DBLP:journals/jal/Fleischer99}), that is used in the construction of the cactus representation of the minimum cuts of a graph. The difference is that here the $4$-cuts are not necessarily mincuts. Thus, some convenient properties like Lemma 5.1 in \cite{DBLP:books/cu/NI2008} or Lemma 2.5 in \cite{DBLP:journals/jal/Fleischer99} fail to hold. However, organizing the $4$-cuts in cyclic families is still very useful for computational purposes. In particular, the cyclic families of $4$-cuts that are produced by the output of Algorithm~\ref{algorithm:generatefamilies} on a complete collection of $4$-cuts have some very convenient properties that we explore in Section~\ref{section:structure-of} (most importantly, see Lemma~\ref{lemma:non-crossing-of-minimal}).}

\begin{definition}[Cyclic family of $4$-cuts]
\label{definition:intro-cyclic-families}
\normalfont
Let $\{p_1,\dots,p_k\}$, with $k\geq 3$, be a collection of pairs of edges of $G$ that generates a collection $\mathcal{C}$ of $4$-cuts of $G$. Suppose that there is a partition $\{X_1,\dots,X_k\}$ of $V(G)$ with the property that $(1)$ $G[X_i]$ is connected for every $i\in\{1,\dots,k\}$, $(2)$ $E[X_i,X_{i+1}]=p_i$ for every $i\in\{1,\dots,k-1\}$, and $(3)$ $E[X_k,X_1]=p_k$. Then $\mathcal{C}$ is called a cyclic family of $4$-cuts. (See Figure~\ref{figure:cyclic-family}.)
\end{definition} 

Now, our claim above is supported by Proposition~\ref{proposition:cyclic_family}, Proposition~\ref{proposition:cyclic_families_algorithm}, and Theorem~\ref{theorem:main}. Specifically, Proposition~\ref{proposition:cyclic_family} basically states that a collection of pairs of edges with more than $3$ pairs of edges generates a cyclic family of $4$-cuts. Theorem~\ref{theorem:main} implies that there is a complete collection $\mathcal{C}$ of $4$-cuts with size $O(n)$, and then Proposition~\ref{proposition:cyclic_families_algorithm} (applied on $\mathcal{C}$) implies that there is a set $\{\mathcal{F}_1,\dots,\mathcal{F}_k\}$ of collections of pairs of edges, with $|\mathcal{F}_1|+\dots+|\mathcal{F}_k|=O(|\mathcal{C}|)=O(n)$, that generate in total all $4$-cuts of $G$. Thus, we have the following combinatorial result.\footnote{It is possible that Corollary~\ref{corollary:intro-number} can also be derived from the $2$-level cactus representation of the $(\mathit{minimum}+1)$-cuts of a graph \cite{DBLP:conf/stoc/DinitzN95}. However, here perhaps it is clearer why the number of $4$-cuts in $3$-edge-connected graphs is bounded by $O(n^2)$, and what are the responsible structures that this number can be as high as $\Omega(n^2)$.}

\begin{corollary}
\label{corollary:intro-number}
The number of $4$-cuts in a $3$-edge-connected graph with $n$ vertices is $O(n^2)$.
\end{corollary}

Now, given a cyclic family of $4$-cuts $\mathcal{C}$ as in Definition~\ref{definition:intro-cyclic-families}, by Lemma~\ref{lemma:edges_between_minimal} 
we have $\partial(X_i)=p_i\cup p_{i-1}$ for every $i\in\{2,\dots,k\}$, and $\partial(X_1)=p_1\cup p_k$, and therefore we have $\{\partial(X_1),\dots,\partial(X_k)\}\subseteq\mathcal{C}$.
The collection of $4$-cuts $\mathcal{M}:=\{\partial(X_1),\dots,\partial(X_k)\}$ is of particular importance, and we call it the collection of the \emph{$\mathcal{C}$-minimal} $4$-cuts. These $4$-cuts are $\mathcal{C}$-``minimal" in the sense that one of their sides (i.e., $X_i$), is a subset of one of the sides of every $4$-cut in $\mathcal{C}$. (Lemma~\ref{lemma:edges_between_minimal} describes the structure of the sides of the $4$-cuts in a cyclic family of $4$-cuts; this can also be inferred from Figure~\ref{figure:cyclic-family}.)
The main reasons that $\mathcal{M}$ is important are the following. First, $\mathcal{M}$ is a parallel family of $4$-cuts. Second, the atoms of $\mathcal{M}$ coincide with the atoms of $\mathcal{C}$. And third, given the collection $\mathcal{F}$ of pairs of edges that generates $\mathcal{C}$, we can compute $\mathcal{M}$ in $O(n+|\mathcal{F}|)$ time. 

The first two points are almost immediate from the definition of minimal $4$-cuts (see also Figure~\ref{figure:cyclic-family}). On the other hand, the computation of the minimal $4$-cuts is not entirely trivial. The problem is that, given the collection $\mathcal{F}$ of pairs of edges that generates a cyclic family of $4$-cuts $\mathcal{C}$, it is not necessary that the pairs of edges in $\mathcal{F}$ are given in the order that is needed in order to form the $\mathcal{C}$-minimal $4$-cuts. Thus, we have to determine the sequence of the pairs of edges in $\mathcal{F}$ that provides the $\mathcal{C}$-minimal $4$-cuts. One way to achieve this can be roughly described as follows. First, we take any vertex $r\in V(G)$, and let us assume w.l.o.g. that $r\in X_1$. Then we pick any pair of edges $p_i$ from $\mathcal{F}$. Then, notice that, among all $4$-cuts of the form $p_i\cup p_j$, for $j\in\{1,\dots,k\}\setminus\{i\}$, we have that either $p_i\cup p_1$ or $p_i\cup p_k$ has the maximum $r$-size. Thus, we can determine one of the two pairs of edges that are incident to $X_1$ (i.e., either $p_1$ or $p_k$), by taking the maximum $r$-size of all $4$-cuts of the form $p_i\cup p_j$, for $j\in\{1,\dots,k\}\setminus\{i\}$. So let suppose that we have determined that $p_1$ is one of the two pairs of edges from $\mathcal{F}$ that is incident to $X_1$. Then, notice that the $4$-cuts $p_1\cup p_2,\dots,p_1\cup p_k$ are sorted in increasing order w.r.t. their $r$-size. Thus, it is sufficient to form all $4$-cuts of the form $\{p_1\cup p_i\mid i\in\{2,\dots,k\}\}$, and then sort them in increasing order w.r.t. their $r$-size. Then we can extract the sequence of pairs of edges $p_1,\dots,p_k$, which is what we need in order to find the $\mathcal{C}$-minimal $4$-cuts (by taking the union of every two consecutive pairs of edges in this sequence, plus $p_1\cup p_k$). This method demands $O(n)$ time in order to perform the sorting of the $4$-cuts of the form $\{p_1\cup p_i\mid i\in\{2,\dots,k\}\}$ (with bucket-sort). Thus, this method in itself is impractical for our purposes, because we may have to compute the minimal $4$-cuts for $\Omega(n)$ collections of pairs of edges. However, given all the collections of pairs of edges beforehand, we can use this method to compute the minimal $4$-cuts of the cyclic families that are generated by those collections with only one bucket-sort (that sorts all the $4$-cuts that we will form and we need to have sorted, in increasing order w.r.t. their $r$-size). Thus, Algorithm~\ref{algorithm:minimal-4cuts} shows how we can compute all $\mathcal{C}_1$-,$\dots$,$\mathcal{C}_t$-minimal $4$-cuts, where $\mathcal{C}_1,\dots,\mathcal{C}_t$ are cyclic families of $4$-cuts that are generated by the collections of pairs of edges $\mathcal{F}_1,\dots,\mathcal{F}_t$, respectively. The running time of this algorithm is $O(n+|\mathcal{F}_1|+\dots+|\mathcal{F}_t|)$, as shown in Proposition~\ref{proposition:algorithm:minimal-4cuts}.

Now let $\mathcal{C}$ be a complete collection of $4$-cuts, and let $\mathcal{F}_1,\dots,\mathcal{F}_t$ be the collections of pairs of edges that are returned by Algorithm~\ref{algorithm:generatefamilies} on input $\mathcal{C}$ and have the property that $|\mathcal{F}_i|>3$, for every $i\in\{1,\dots,t\}$. Then, Proposition~\ref{proposition:cyclic_families_algorithm} implies that $\mathcal{F}_i$ generates a collection $\mathcal{C}_i$ of $4$-cuts implied by $\mathcal{C}$, for every $i\in\{1,\dots,t\}$, and by Proposition~\ref{proposition:cyclic_family} we have that $\mathcal{C}_i$ is a cyclic family of $4$-cuts. Thus, we can apply Algorithm~\ref{algorithm:minimal-4cuts} in order to derive the collection $\mathcal{M}_i$ of the $\mathcal{C}_i$-miminal $4$-cuts, for every $i\in\{1,\dots,t\}$, in $O(n+|\mathcal{F}_1|+\dots+|\mathcal{F}_t|)$ time in total. As noted above, we have that $\mathit{atoms}(\mathcal{M}_i)=\mathit{atoms}(\mathcal{C}_i)$, and $\mathcal{M}_i$ is a parallel family of $4$-cuts, for every $i\in\{1,\dots,t\}$. Thus, we have $\mathit{atoms}(\mathcal{C}_1\cup\dots\cup\mathcal{C}_t)=\mathit{atoms}(\mathcal{M}_1\cup\dots\cup\mathcal{M}_t)$. 
We note that this formula is not very useful for computing $\mathit{atoms}(\mathcal{C}_1\cup\dots\cup\mathcal{C}_t)$, because there is no guarantee that $\mathcal{M}_1\cup\dots\cup\mathcal{M}_t$ is a parallel family of $4$-cuts. (In fact, Figure~\ref{figure:crossing_minimal} provides a counterexample.) 
However, if we keep the subcollection $\mathcal{M}'$ of the essential $4$-cuts in $\mathcal{M}_1\cup\dots\cup\mathcal{M}_t$, then we have that $\mathcal{M}'$ is a parallel family of $4$-cuts, as a consequence of Lemma~\ref{lemma:non-crossing-of-minimal}. Thus, it is sufficient to keep only $\mathcal{M}'$ and compute $\mathit{atoms}(\mathcal{M}')$.

Now let $\mathcal{F}$ be a collection of pairs of edges that is returned by Algorithm~\ref{algorithm:generatefamilies} on input $\mathcal{C}$ and has $|\mathcal{F}|=3$. By Proposition~\ref{proposition:cyclic_families_algorithm} we have that $\mathcal{F}$ generates a collection $\mathcal{C}'$ of $4$-cuts implied by $\mathcal{C}$. 
If $\mathcal{C}'$ is not a cyclic family of $4$-cuts, then we say that $\mathcal{C}'$ is a \emph{degenerate family} of $4$-cuts, and Lemma~\ref{lemma:maximal3family} describes the structure of such a family (i.e., this is given by Figure~\ref{figure:implied4cut}(a), if $\mathcal{F}=\{\{e_1,e_2\},\{e_3,e_4\},\{e_5,e_6\}\}$).
Then, by Corollary~\ref{corollary:degenerate_non-essential} we have that $\mathcal{C}'$, if it is not a cyclic family of $4$-cuts, it has the property that all its $4$-cuts are non-essential. 
Thus, if $\mathcal{C}'$ consists of three non-essential $4$-cuts, then we can discard them. Otherwise, we have that $\mathcal{C}'$ is a cyclic family of $4$-cuts. Since $|\mathcal{F}|=3$, we have that all $4$-cuts in $\mathcal{C}'$ are $\mathcal{C}'$-minimal. Thus, it is sufficient to keep only the essential $4$-cuts from $\mathcal{C}'$. Then, these are all parallel among themselves, and also parallel with the $4$-cuts in $\mathcal{M}'$ (due to Lemma~\ref{lemma:non-crossing-of-minimal}). 

Thus, we have shown how to extract enough $4$-cuts from the collections of pairs of edges that are returned by Algorithm~\ref{algorithm:generatefamilies} and have size at least three. Now it remains to consider the collections of pairs of edges that have size $2$.

\subsubsection{Isolated and quasi-isolated $4$-cuts}
Let $\mathcal{C}$ be a complete collection of $4$-cuts, and let $\mathcal{F}$ be a collection of pairs of edges that is returned by Algorithm~\ref{algorithm:generatefamilies} on input $\mathcal{C}$ and has $|\mathcal{F}|=2$. They, by Proposition~\ref{proposition:cyclic_families_algorithm} we have that $\mathcal{F}$ generates a $4$-cut $C$ implied by $\mathcal{C}$. More precisely, by Lemma~\ref{lemma:2-pair-collection} we have $C\in\mathcal{C}$. Now, if there is another collection of pairs of edges $\mathcal{F}'$ that is returned by Algorithm~\ref{algorithm:generatefamilies} on input $\mathcal{C}$ that also generates $C$ and has $|\mathcal{F}'|>2$, then we have collected enough $4$-cuts in order to capture the separation of $V(G)$ induced by $C$. Otherwise, we have that all collections of pairs of edges that are returned by Algorithm~\ref{algorithm:generatefamilies} on input $\mathcal{C}$ and generate $C$ have size $2$. Then, since $C\in\mathcal{C}$, these collections are the three different partitions of $C$ into pairs of edges. We note that we can determine in linear time what are the $4$-cuts from $\mathcal{C}$ with the property that all three partitions of them into pairs of edges are returned by Algorithm~\ref{algorithm:generatefamilies} on input $\mathcal{C}$. Now, for every such $4$-cut $C$, we distinguish two different cases: either $(1)$ there is no collection $\mathcal{F}$ of pairs of edges with $|\mathcal{F}|>2$ that generates a collection of $4$-cuts that includes $C$, or $(2)$ the contrary of $(1)$ is true. In case $(1)$, we say that $C$ is an \emph{isolated} $4$-cut. In case $(2)$, we say that $C$ is a \emph{quasi-isolated} $4$-cut. (Notice that the concept of quasi-isolated $4$-cuts is relative to a collection $\mathcal{C}$ of $4$-cuts that is given as input to Algorithm~\ref{algorithm:generatefamilies}.)

The distinction between isolated and quasi-isolated $4$-cuts is important, because by Corollary~\ref{corollary:iso-essential-parallel} we have that an essential isolated $4$-cut is parallel with every essential $4$-cut. On the other hand, there are examples where two essential quasi-isolated $4$-cuts may cross (see Figure~\ref{figure:crossing_quasi_isolated}). However, there are two nice things that are very helpful here. First, the quasi-isolated $4$-cuts are basically not needed for our purposes. More precisely, by Lemma~\ref{lemma:quasi-isolated-replaceable} we have that every pair of vertices that are separated by an essential quasi-isolated $4$-cut, are also separated by a $4$-cut that is generated by a collection of pairs of edges with size more than $2$ that is returned by Algorithm~\ref{algorithm:generatefamilies}. 
And second, we have enough information in order to identify the quasi-isolated $4$-cuts, so that we can discard them. Specifically, by Corollary~\ref{corollary:quasi-iso-pair} we have that every essential quasi-isolated $4$-cut shares a pair of edges with an essential $\mathcal{C}'$-minimal $4$-cut, where $\mathcal{C}'$ is a cyclic family of $4$-cuts that is generated by a collection of pairs of edges (with size at least $3$) that is returned by Algorithm~\ref{algorithm:generatefamilies} on input $\mathcal{C}$. This is a property that distinguishes the quasi-isolated from the isolated $4$-cuts, and we can use it in order to identify all the essential isolated $4$-cuts, as shown in Proposition~\ref{proposition:algorithm:isolated-4cuts}.

\subsubsection{The full algorithm}
In summary, these are the steps that we follow in order to compute the $5$-edge-connected components of a $3$-edge-connected graph $G$.

\begin{enumerate}
\item{Compute the partition $\mathcal{P}_4$ of the $4$-edge-connected components of $G$.}
\item{Compute a complete collection $\mathcal{C}$ of $4$-cuts of $G$ with size $O(n)$.}
\item{Compute the collections of pairs of edges $\mathcal{F}_1,\dots,\mathcal{F}_k$ that are returned by Algorithm~\ref{algorithm:generatefamilies} on input $\mathcal{C}$.}
\item{Let $I\subseteq\{1,\dots,k\}$ be the collection of indices such that, for every $i\in I$, either $|\mathcal{F}_i|>3$, or $|\mathcal{F}_i|=3$ and $\mathcal{F}_i$ generates at least one essential $4$-cut. Let $\mathcal{C}_i$ be the cyclic family of $4$-cuts generated by $\mathcal{F}_i$, for every $i\in I$.}
\item{Compute the collection $\mathcal{M}_i$ of the $\mathcal{C}_i$-minimal $4$-cuts, for every $i\in I$.}
\item{Compute the subcollection $\mathcal{M}'$ of the essential $4$-cuts in $\bigcup_{i\in I}{\mathcal{M}_i}$.}
\item{Compute the collection $\mathcal{ISO}$ of the essential isolated $4$-cuts of $G$.}
\item{Let $\mathcal{P}_5$ be the refinement of $\mathit{atoms}(\mathcal{M}')$ with $\mathit{atoms}(\mathcal{ISO})$.}
\item{Return $\mathcal{P}_5$ refined by $\mathcal{P}_4$.}
\end{enumerate}

Step $1$ can be performed in linear time from previous results (see \cite{DBLP:conf/esa/GeorgiadisIK21} or \cite{DBLP:conf/esa/NadaraRSS21}). By Theorem~\ref{theorem:main}, Step $2$ can be performed in linear time. Step $3$ takes $O(n)$ time, according to Proposition~\ref{proposition:cyclic_families_algorithm}. By Proposition~\ref{proposition:cyclic_family}, we know that every $\mathcal{F}_i$ with $|\mathcal{F}_i|>3$ generates a cyclic family of $4$-cuts. By Corollary~\ref{corollary:essential_implies_cyclic} we have that every $\mathcal{F}_i$ with $|\mathcal{F}_i|=3$ that generates at least one essential $4$-cut generates a cyclic family of $4$-cuts. Thus, if we let $I$ be the collection of indices in Step $4$, then we have that $\mathcal{F}_i$ generates a cyclic family of $4$-cuts for every $i\in I$. Then, it makes sense to perform Step $5$, and this can be completed in $O(n)$ time, according to Proposition~\ref{proposition:algorithm:minimal-4cuts}. Then, we can extract the subcollection $\mathcal{M}'$ of the essential $4$-cuts in $\bigcup_{i\in I}{\mathcal{M}_i}$ in $O(n)$ time, after we have performed the preprocessing described in Proposition~\ref{proposition:essentiality}. Thus, Step $6$ takes linear time. The computation in Step $7$ also takes linear time, according to Proposition~\ref{proposition:algorithm:isolated-4cuts}. Since we have that $\mathcal{M}'$ and $\mathcal{ISO}$ are parallel families of $4$-cuts, we can compute $\mathit{atoms}(\mathcal{M}')$ and $\mathit{atoms}(\mathcal{ISO})$ in linear time, according to Proposition~\ref{proposition:algorithm:atoms-of-parallel}. Then, the mutual refinement of those partitions in Step $8$ takes $O(n)$ time, by using bucket-sort. Finally, the refinement in Step $9$ also takes $O(n)$ time with bucket-sort.

Now we will demonstrate the correctness of this procedure. First, notice that the partition of the $5$-edge-connected components is a refinement of the partition returned in Step $9$ (because the latter is a refinement of the partition of the $4$-edge-connected components with the atoms of a specific collection of $4$-cuts). Conversely, let $x$ and $y$ be two vertices of $G$ that are not $5$-edge-connected. Then, there is either a $3$-cut or a $4$-cut that separates $x$ and $y$. If there is a $3$-cut that separates $x$ and $y$, then $x$ and $y$ belong to different $4$-edge-connected components, and therefore they belong to different sets in the partition returned in Step $9$. Otherwise, if $x$ and $y$ are $4$-edge-connected, then there is an essential $4$-cut $C$ that separates them. Since $\mathcal{C}$ is a complete collection of $4$-cuts, Proposition~\ref{proposition:cyclic_families_algorithm} implies that there is an $i\in\{1,\dots,k\}$ such that $C$ is generated by $\mathcal{F}_i$. If $|\mathcal{F}_i|\geq 3$, then, since $C$ is an essential $4$-cut, Proposition~\ref{proposition:cyclic_family} and Corollary~\ref{corollary:essential_implies_cyclic} imply that $\mathcal{F}_i$ generates a cyclic family $\mathcal{C}_i$ of $4$-cuts. Therefore, by Lemma~\ref{lemma:replace-with-minimal} we have that $x$ and $y$ are separated by an essential $\mathcal{C}_i$-minimal $4$-cut. Thus, $x$ and $y$ belong to different sets in $\mathit{atoms}(\mathcal{M}')$. Otherwise, $C$ is either an isolated or a quasi-isolated $4$-cut. If $C$ is isolated, then it belongs to $\mathcal{ISO}$, and therefore $x$ and $y$ belong to different sets in $\mathit{atoms}(\mathcal{ISO})$. Otherwise, by Lemma~\ref{lemma:quasi-isolated-replaceable} we have that $x$ and $y$ are separated by an essential $\mathcal{C}_i$-minimal $4$-cut, for some $i\in I$. Thus, $x$ and $y$ belong to different sets in $\mathit{atoms}(\mathcal{M}')$. In either case, then, we have that $x$ and $y$ belong to different sets in the partition returned by Step $9$. We conclude that this partition coincides with that of the $5$-edge-connected components. 

\subsection{Organization of this paper}
In Section~\ref{section:preliminaries} we introduce some preliminary concepts and notation that we will use throughout. In Section~\ref{section:structure-of} we study the structure of $4$-cuts in $3$-edge-connected graphs. In Section~\ref{section:using-a-dfs-tree} we present some applications of identifying $4$-cuts on a DFS-tree, that we will need in order to establish our main result. In Section~\ref{section:computing-5} we present the algorithm for computing the $5$-edge-connected components of a $3$-edge-connected graph. Section~\ref{section:DFS} introduces various DFS-based concepts, analyzes properties of them, and demonstrates how to compute them efficiently. Section~\ref{section:4eoracle} presents the oracle for answering connectivity queries in the presence of at most four edge-failures. Section~\ref{section:computing-4-cuts} gives an overview of the algorithm for computing a complete collection of $4$-cuts of a $3$-edge-connected graph. There, we provide a DFS-based classification of all $4$-cuts of a $3$-edge-connected graph, and briefly discuss the methods that we employ in order to compute the $4$-cuts of each class. In Sections~\ref{section:type-2}, \ref{section:type-3a} and \ref{section:type-3b} we provide the full details for computing the most demanding classes of $4$-cuts.

\section{Preliminaries}
\label{section:preliminaries}

\subsection{Basic graph terminology}
In this work, all graphs considered are undirected multigraphs (i.e., they may have parallel edges). We use standard graph-theoretic terminology, that can be found e.g. in \cite{DBLP:books/daglib/0030488} or \cite{DBLP:books/cu/NI2008}. Let $G=(V,E)$ be a graph. We let $V(G)$ and $E(G)$ denote the vertex-set and the edge-set of $G$, respectively. (That is, we have $V(G)=V$ and $E(G)=E$.) Since $G$ is a multigraph, it may have multiple edges of the form $(x,y)$, for two vertices $x$ and $y$. Thus, in order to be precise, we should also include an index to specify the edge we are referring to. That is, every edge $e\in E$ should be written as $(x,y,i)$, where $x$ and $y$ are the endpoints of $e$, and $i$ is a unique identifier of $e\in E$. However, we keep our notation simple (i.e., we identify edges just with the tuple of their endpoints), and this will not affect our arguments. 

For every two subsets $X$ and $Y$ of $V$, we let $E_G[X,Y]$ denote the set of the edges of $G$ with one endpoint in $X$ and the other endpoint in $Y$. We may skip the subscript ``$G$" from this notation if the reference graph is clear from the context. Although it is not necessary that $X$ and $Y$ are disjoint, wherever we use this notation in the sequel we have that $X$ and $Y$ are disjoint. We let $\partial_G(X)$ denote $E_G[X,V\setminus X]$. Again, we may skip the subscript ``$G$" when no confusion arises. For a singleton $\{v\}$ that consists of a vertex $v$, we may simply write $\partial(v)$ instead of $\partial(\{v\})$. If $X$ is a set of vertices of $G$, we let $G[X]$ denote the induced subgraph on $X$. This is the graph with vertex set $X$ and edge set $\{(x,y)\in E\mid x\in X \mbox{ and } y\in X\}$. If $C$ is a subset of edges of $G$, we use $G\setminus C$ to denote the graph $(V,E\setminus C)$. If $C$ consists of a single edge $e$, we may use the simplified notation $G\setminus e:=G\setminus\{e\}$.

A \emph{path} $P$ in $G$ is an alternating sequence $x_1,e_1,\dots,x_{k-1},e_{k-1},x_k$, with $k\geq 1$, of vertices and edges of $G$, starting with a vertex $x_1$ and ending with a vertex $x_k$, such that $e_i=(x_i,x_{i+1})$ for every $i\in\{1,\dots,k-1\}$. In this case, we say that $P$ is a path from $x_1$ to $x_k$ in $G$, and these are called the beginning and the end, respectively, of $P$. Furthermore, we say that $P$ passes from the vertices $x_1,\dots,x_k$, and uses the edges $e_1,\dots,e_{k-1}$. If $P$ and $Q$ are two paths such that the end of $P$ coincides with the beginning of $Q$, then $P+Q$ denotes the path that is formed by the concatenation of $P$ and $Q$ (by discarding either the end of $P$ or the beginning of $Q$, in order to keep a single copy of it as the concatenation point). Thus, if $P$ is a path from $x$ to $y$, and $Q$ is a path from $y$ to $z$, then $P+Q$ is a path from $x$ to $z$.
For every two vertices $x$ and $y$ of $G$, we say that $x$ is connected with $y$ if there is a path from $x$ to $y$ in $G$. Notice that this defines an equivalence relation on $V(G)$; its equivalence classes are called the \emph{connected components} of $G$. 
In particular, if $V(G)$ is the only connected component of $G$, then $G$ is called a \emph{connected} graph. Otherwise, it is called disconnected.

Let $G$ be a connected graph. A bipartition $\{X,V\setminus X\}$ of $V(G)$ is called a \emph{cut} of $G$. The corresponding edge-set $C=E[X,V\setminus X]$ has the property that its removal from $G$ increases the number of connected components at least by one. Since $G$ is connected, it is not difficult to verify that $C$ is uniquely determined by $X$ (i.e., $\{X,V\setminus X\}$ is the only bipartition of $V(G)$ whose corresponding edge-set is $C$). Thus, we also call $C$ a cut of $G$. We will be using the term ``cut" to denote interchangeably a bipartition of $G$ and the edge-set that is derived from it. It will be clear from the context whether the term cut refers to a bipartition or its corresponding edge-set. In particular, whenever we speak of the ``edges" of a cut, we consider it as an edge-set. Notice that a set $C$ of edges with the property that $G\setminus C$ is disconnected is not necessarily a cut of $G$. However, this is definitely the case when $C$ is minimal w.r.t. this property (i.e., if no proper subset $C'$ of $C$ has the property that $G\setminus C'$ is disconnected). In this work, we will deal exclusively with \emph{edge-minimal} cuts. An edge-minimal cut $C$ is a set of edges with the property that $G\setminus C$ is disconnected, but $G\setminus C'$ is connected for every proper subset $C'$ of $C$. Notice that $C$ has the property that $G\setminus C$ consists of two connected components $X$ and $V\setminus X$. These are called the \emph{sides} of $C$, and they have the property that $C=E[X,V\setminus X]$. 
From now on, the term ``cut" will always mean ``edge-minimal cut". A cut with $k$ edges is called a $k$-cut. We let $\mathcal{C}_\mathit{kcuts}(G)$ denote the collection of all $k$-cuts of $G$. Since the reference graph will always be clear from the context, we will simply denote this as $\mathcal{C}_\mathit{kcuts}$. Following standard terminology, we refer to the $1$-cuts as \emph{bridges}. 

\subsection{Partitions and atoms}
Let $P$ be a partition of a set $V$. Then we say that $P$ \emph{separates} two elements $x,y\in V$ if and only if $x$ and $y$ belong to different sets from $P$. A \emph{refinement} $P'$ of $P$ is a partition with the property that every set in $P'$ is a subset of a set in $P$. Equivalently, $P'$ is a refinement of $P$ if and only if every two elements separated by $P$ are also separated by $P'$. The common refinement of two partitions $P$ and $Q$ is the unique partition $R$ with the property that two elements are separated by $R$ if and only if they are separated by either $P$ or $Q$. Equivalently, $R$ is given by the collection of the non-empty intersections of the form $X\cap Y$, where $X\in P$ and $Y\in Q$.

Let $\mathcal{P}$ be a collection of partitions of a set $V$. Then, the \emph{atoms} of $\mathcal{P}$, denoted as $\mathit{atoms}(\mathcal{P})$, is the common refinement of all partitions in $\mathcal{P}$. In other words, the partition $\mathit{atoms}(\mathcal{P})$ is defined by the property that two elements $x$ and $y$ of $V$ are separated by $\mathit{atoms}(\mathcal{P})$ if and only if they are separated by a partition in $\mathcal{P}$.
%

We are particularly interested in collections of bipartitions. Two bipartitions $P=\{X,Y\}$ and $Q=\{X',Y'\}$ of $V$ are called \emph{parallel} if at least one of the intersections $X\cap X'$, $X\cap Y'$, $Y\cap X'$, $Y\cap Y'$ is empty. (Notice that, in general, at most one of those intersections may be empty, unless $P=Q$, in which case precisely two of those intersections are empty.) Otherwise, if none of the intersections $X\cap X'$, $X\cap Y'$, $Y\cap X'$, $Y\cap Y'$ is empty, then we say that $P$ and $Q$ \emph{cross}. Notice that a collection of bipartitions of a set with $n$ elements can have size $\Omega(2^n)$. However, the number of partitions in a collection of bipartitions that are pairwise parallel is bounded by $O(n)$ \cite{DBLP:conf/stoc/DinitzN95}. 

This terminology concerning partitions can be naturally applied to (collections of) cuts, since these are defined as bipartitions of the vertex set of a graph. Thus, we may speak of cuts that are parallel, or that cross. Also, if $\mathcal{C}$ is a collection of cuts, then the partition $\mathit{atoms}(\mathcal{C})$ is defined. (Regardless of whether $\mathcal{C}$ was considered as a collection of sets of edges, the expression $\mathit{atoms}(\mathcal{C})$ interprets $\mathcal{C}$ as a collection of bipartitions.) A collection of cuts that are pairwise parallel is called a parallel family of cuts. Thus, a parallel family of cuts of a graph $G$ contains $O(|V(G)|)$ cuts.

\subsection{Edge-connectivity and $k$-edge-connected components}
It is natural to ask how ``well connected" is a graph, and various concepts have been developed to capture this notion. Let $G$ be a connected graph. We say that two vertices $x$ and $y$ of $G$ are \emph{$k$-edge-connected}, if we have to remove at least $k$ edges from $G$ in order to disconnect them. Equivalently, by Menger's theorem we have that $x$ and $y$ are $k$-edge-connected if and only if there are $k$ edge-disjoint paths from $x$ to $y$ (see, e.g., \cite{DBLP:books/cu/NI2008}). The maximum $k$ such that $x$ and $y$ are $k$-edge-connected is called the \emph{edge-connectivity} of $x$ and $y$, denoted as $\lambda(x,y)$. 

It is not difficult to see that the relation of $k$-edge-connectivity defines an equivalence relation on $V(G)$; its equivalence classes are called the \emph{$k$-edge-connected components} of $G$. Equivalently, a $k$-edge-connected component of $G$ is a maximal set of vertices with the property that every pair of vertices $x,y$ in it has $\lambda(x,y)\geq k$ (and so we have to remove at least $k$ edges in order to disconnect them). Notice that, if two vertices $x$ and $y$ are not $k$-edge-connected, then there is a $k'$-cut that separates them, for some $k'<k$. Thus, the collection of the $k$-edge-connected components is given by $\mathit{atoms}(\mathcal{C}_\mathit{1cuts}\cup\dots\cup\mathcal{C}_\mathit{(k-1)cuts})$. If $V(G)$ is the unique $k$-edge-connected component of $G$, then $G$ is called $k$-edge-connected. Equivalently, $G$ is $k$-edge-connected if and only if we have to remove at least $k$ edges in order to disconnect it. 

Our goal in this work is to show how to compute the $5$-edge-connected components of a graph in linear time. There is a linear-time construction that allows us to reduce this computation to a $3$-edge-connected graph \cite{DBLP:conf/wg/Dinitz92}. The general idea is that we first compute the $3$-edge-connected components of the graph, and then we augment the subgraphs that are induced by them with some auxiliary edges, so that the $5$-edge-connected components of the original graph are given by the collection of the $5$-edge-connected components of the auxiliary graphs (for more details, see e.g. \cite{DBLP:conf/esa/GeorgiadisIK21}). 

Let $G$ be a $3$-edge-connected graph. Then, by definition, $G$ is connected, and it has no $1$-cuts or $2$-cuts. Thus, the $5$-edge-connected components of $G$ are given by $\mathit{atoms}(\mathcal{C}_\mathit{3cuts}\cup\mathcal{C}_\mathit{4cuts})$. We know how to compute the collection $\mathcal{C}_\mathit{3cuts}$ in linear time (see \cite{DBLP:conf/esa/GeorgiadisIK21} or \cite{DBLP:conf/esa/NadaraRSS21}). Here there is implicit the fact that the collection of all $3$-cuts of a $3$-edge-connected graph with $n$ vertices has size $O(n)$. This is known for several decades now, and it is a consequence of the structure of $3$-cuts in $3$-edge-connected graphs: i.e., these form a parallel family of bipartitions \cite{DBLP:conf/stoc/DinitzN95}. However, the number of $4$-cuts of a $3$-edge-connected graph with $n$ vertices can be as high as $\Omega(n^2)$, even in graphs with $O(n)$ edges. Thus, the computation of $\mathit{atoms}(\mathcal{C}_\mathit{3cuts}\cup\mathcal{C}_\mathit{4cuts})$ must be performed without explicitly computing all $4$-cuts of $G$.

A key idea in our analysis is to consider only those $4$-cuts that separate at least one pair of vertices that are $4$-edge-connected. We call those $4$-cuts \emph{essential}, because they strictly refine the collection of the $4$-edge-connected components. On the other hand, there are $4$-cuts that separate vertices that are only $3$-edge-connected, but not $4$-edge-connected. These $4$-cuts can be discarded for the purpose of computing the atoms of $\mathcal{C}_\mathit{3cuts}\cup\mathcal{C}_\mathit{4cuts}$, because the separations that are induced by them are captured by the separations according to the $\mathcal{C}_\mathit{3cuts}$ (i.e., the $4$-edge-connected components). 


\subsection{Notation}
Here we introduce some notation that we will use throughout.

For any two positive integers $i$ and $k$ such that $i\in\{1,\dots,k\}$, we define the ``cyclic" addition and subtraction as:\\

$i+_{k}1=
\left\{
\begin{array}{ll}
		i+1  & \mbox{if } i<k\\
		1 & \mbox{if } i=k
\end{array}
\right.$ 

$i-_{k}1=
\left\{
\begin{array}{ll}
		i-1  & \mbox{if } i>1\\
		k & \mbox{if } i=1
\end{array}
\right.$\\

For any two sets $A$ and $B$ such that $A\cap B=\emptyset$, we use the notation $A\sqcup B$ to denote the union $A\cup B$, while emphasizing the fact that $A\cap B=\emptyset$. This notation will be very convenient for our argumentation, because it packs more information in a single symbol. Also, whenever we use the expression $A\subset B$, we imply that $A$ is a \emph{proper} subset of $B$ (and thus $A\neq B$). Otherwise, if $A=B$ is allowed, then we write $A\subseteq B$.

If $L$ is a sorted list of elements and $x$ is an element in $L$, we use $\mathit{next}_L(x)$ and $\mathit{prev}_L(x)$ to denote the successor and the predecessor, respectively, of $x$ in $L$. We use $\bot$ to denote the end-of-list element. A \emph{segment} of $L$ is a sublist of consecutive elements. For convenience, we may handle a list as set (and write, e.g., $x\in L$).  

If $f:X\rightarrow Y$ is a function and $\mathcal{C}$ is a collection of subsets of $Y$, we let $f^{-1}(\mathcal{C})$ denote the collection $\{f^{-1}(C)\mid C\in\mathcal{C}\}$ of subsets of $X$.
 
For every graph that we consider, we assume a total ordering of its edge set (e.g., lexicographic order). If $p=\{e,e'\}$ is a pair of edges of a graph and $e<e'$, then we let $\vec{p}$ denote the ordered pair of edges $(e,e')$.

\section{Properties of $4$-cuts in $3$-edge-connected graphs}
\label{section:structure-of}
Throughout this section we assume that $G$ is a $3$-edge-connected graph. We also assume a total ordering of the edges of $G$ (e.g., lexicographic order w.r.t. their endpoints). This is needed for Algorithm~\ref{algorithm:generatefamilies}, and for analyzing its output. We let $V$ denote $V(G)$. All graph-related elements, such as vertices, edges, cuts, etc., refer to $G$. 

\begin{lemma}
\label{lemma:no-common-three-edges}
Let $C_1$ and $C_2$ be two distinct $4$-cuts of $G$. Then $|C_1\cap C_2|\neq 3$.
\end{lemma}
\begin{proof}
Let us suppose, for the sake of contradiction, that $|C_1\cap C_2|=3$. Since $C_1$ and $C_2$ are $4$-cuts, we have that $G'=G\setminus(C_1\cap C_2)$ is connected. Let $e_1=C_1\setminus(C_1\cap C_2)$ and $e_2=C_2\setminus(C_1\cap C_2)$. Since $C_1$ and $C_2$ are distinct, we have that $e_1\neq e_2$. And since both $C_1$ and $C_2$ are $4$-cuts, we have that both $G'\setminus e_1$ and $G'\setminus e_2$ are disconnected. Thus, $e_1$ and $e_2$ are two distinct bridges of $G'$, and so $V(G')$ can be partitioned into three sets $X$, $Y$, and $Z$, such that $E_{G'}[X,Z]=\{e_1\}$, $E_{G'}[Y,Z]=\{e_2\}$, and $E_{G'}[X,Y]=\emptyset$.
Since $G$ is $3$-edge-connected, we have that $|\partial(X)|\geq 3$, $|\partial(Y)|\geq 3$ and $|\partial(Z)|\geq 3$. This can only be true if either $(1)$ two of the edges from $C_1\cap C_2$ connect $X$ and $Y$, and the other edge connects either $X$ and $Z$, or $Y$ and $Z$, or $(2)$ one edge from $C_1\cap C_2$ connects $X$ and $Y$, one edge from $C_1\cap C_2$ connects $X$ and $Z$, and one edge from $C_1\cap C_2$ connects $Y$ and $Z$. Let us consider case $(1)$ first, and let us assume w.l.o.g. that an edge from $C_1\cap C_2$ connects $X$ and $Z$. But now we have that the two edges from $C_1\cap C_2$ that connect $X$ and $Y$, plus $e_2$, constitute a $3$-cut of $G$ (with sides $Y$ and $V\setminus Y$). This contradicts the fact that $C_2$ is a $4$-cut of $G$. Thus, only case $(2)$ can be true. But then we have that $\partial(X)$ consists of $e_1$ and two edges from $C_1\cap C_2$. Therefore $\partial(X)$ is a proper subset of $C_1$ that disconnects $G$ upon removal, contradicting the fact that $C_1$ is a $4$-cut of $G$.
We conclude that it is impossible to have $|C_1\cap C_2|=3$. 
\end{proof}

\begin{lemma}
\label{lemma:bridge_in_corner}
Let $X$ be a subset of $V(G)$ such that $|\partial(X)|=4$. Then $G[X]$ is connected, and it has at most one bridge. If $G[X]$ has a bridge $e$, then $G[X]\setminus e$ consists of two connected components $Y_1$ and $Y_2$ such that $|E[Y_1,V\setminus X]|=2$ and $|E[Y_2,V\setminus X]|=2$.
\end{lemma}
\begin{proof}
First, let us suppose, for the sake of contradiction, that $G[X]$ is not connected. Then, let $S$ and $S'$ be two distinct connected components of $G[X]$. Then we have that $\partial(S)\subseteq\partial(X)$, $\partial(S')\subseteq\partial(X)$, and $\partial(S)\cap\partial(S')=\emptyset$. This implies that $|\partial(S)|+|\partial(S')|=|\partial(S)\cup\partial(S')|\leq|\partial(X)|=4$. Thus, at least one of $|\partial(S)|$ and $|\partial(S')|$ must be lower than $3$, in contradiction to the fact that $G$ is $3$-edge-connected. This shows that $G[X]$ is connected.

Now, let us suppose, for the sake of contradiction, that $G[X]$ has at least two bridges $e_1$ and $e_2$. Then there is a partition $\{Z_1,Z_2,Z_3\}$ of $X$, such that all of $G[Z_1]$, $G[Z_2]$ and $G[Z_3]$, are connected, and such that $E[Z_1,Z_2]=\{e_1\}$, $E[Z_2,Z_3]=\{e_2\}$ and $E[Z_1,Z_3]=\emptyset$. Let $E_i=E[Z_i,V\setminus X]$, for $i\in\{1,2,3\}$. Then we have $E_1\sqcup E_2\sqcup E_3=\partial(X)$, $\partial(Z_1)=E_1\sqcup\{e_1\}$, $\partial(Z_3)=E_3\sqcup\{e_2\}$, and $\partial(Z_2)=E_2\sqcup\{e_1,e_2\}$. Since $|E_1|+|E_2|+|E_3|=4$, $|\partial(Z_1)|=|E_1|+1$, $|\partial(Z_3)|=|E_3|+1$, and $|\partial(Z_2)|=|E_2|+2$, we infer that at least one of $|\partial(Z_i)|$, for $i\in\{1,2,3\}$, is at most $2$. This contradicts the fact that $G$ is $3$-edge-connected. Thus, $G[X]$ can have at most one bridge.

Now let $e$ be a bridge of $G[X]$. Then $G[X]\setminus e$ consists of two connected components $Y_1$ and $Y_2$, and we have $E[Y_1,Y_2]=\{e\}$. Since $G$ is $3$-edge-connected, we have that both $|\partial(Y_1)|$ and $|\partial(Y_2)|$ must be at least $3$. We have that $\partial(Y_i)=E[Y_i,V\setminus X]\cup\{e\}$, for $i\in\{1,2\}$. Thus, both $E[Y_1,V\setminus X]$ and $E[Y_2,V\setminus X]$ must contain at least $2$ edges. Since $E[Y_1,V\setminus X]\sqcup E[Y_2,V\setminus X]=\partial(X)$, this implies that $|E[Y_1,V\setminus X]|=2$ and $|E[Y_2,V\setminus X]|=2$ (due to $|\partial(X)|=4$).
\end{proof}

\subsection{The structure of crossing $4$-cuts of a $3$-edge-connected graph}
The following lemma is one of the cornerstones of our work. It motivates several concepts that we develop in order to analyze the structure of $4$-cuts in $3$-edge-connected graphs.

\begin{figure}[t!]\centering
\includegraphics[trim={0 14cm 0 0}, clip=true, width=0.8\linewidth]{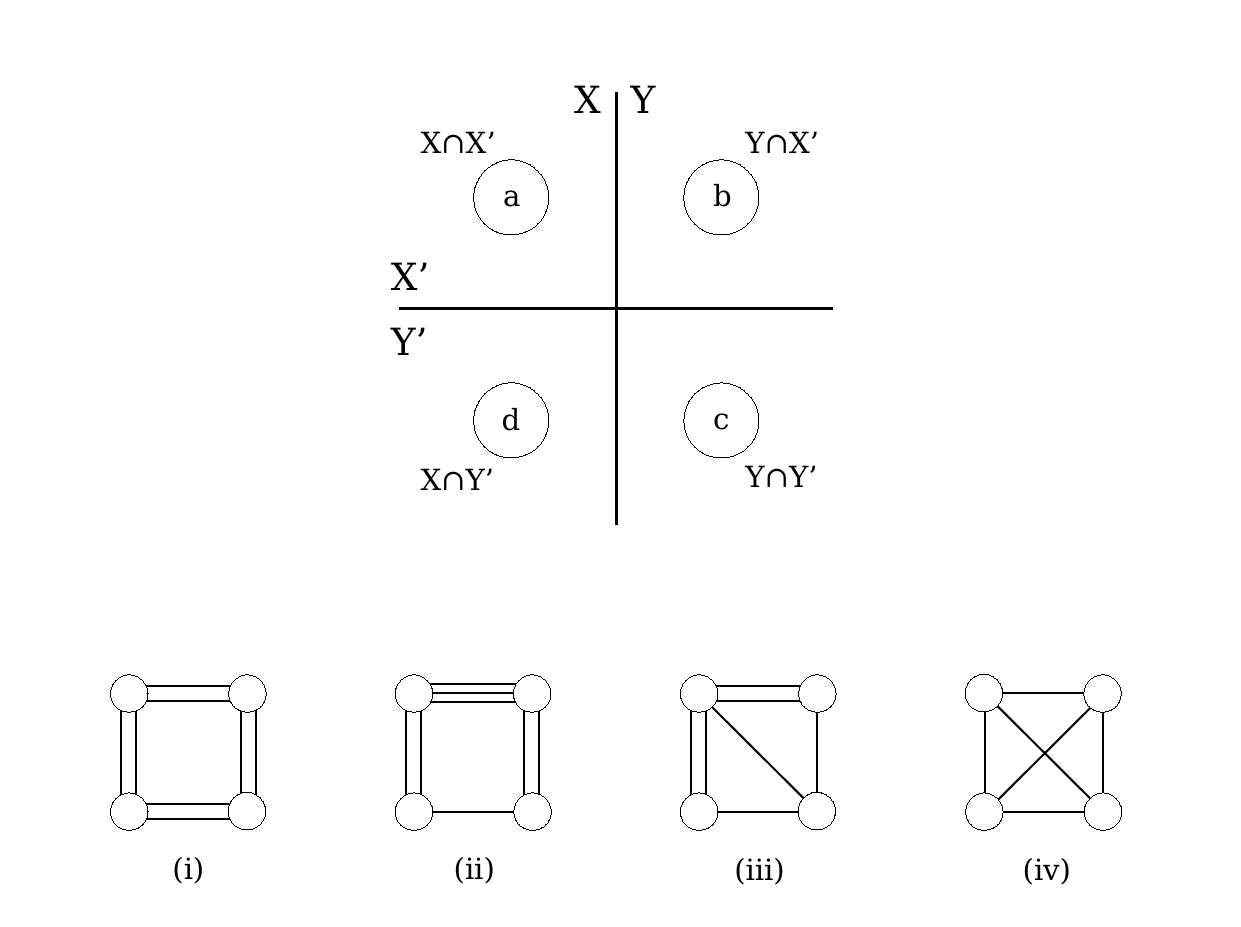}
\caption{\small{All possible crossings of two $4$-cuts $C=\{X,Y\}$ and $C'=\{X',Y'\}$.}}\label{figure:crossing4cuts}
\end{figure}

\begin{lemma}[The Structure of Crossing $4$-cuts]
\label{lemma:crossing4cuts}
Let $C$ and $C'$ be two $4$-cuts of a $3$-edge-connected graph. Then, cases \textbf{(i)} to \textbf{(iv)} in Figure~\ref{figure:crossing4cuts} show all the different ways in which $C$ and $C'$ may cross.
\end{lemma}
\begin{proof}
Let $X$ and $Y$ be the sides of $C$, and let $X'$ and $Y'$ be the sides of $C'$. We let $a=X\cap X'$, $b=Y\cap X'$, $c=Y\cap Y'$, and $d=X\cap Y'$ (see Figure~\ref{figure:crossing4cuts}). We will analyze all the different (non-isomorphic) ways in which $a$, $b$, $c$ and $d$ may be connected with edges. Thus, we have to determine all the different combinations of values for $|E[a,b]|$, $|E[a,c]|$, $|E[a,d]|$, $|E[b,c]|$, $|E[b,d]|$ and $|E[c,d]|$. 
There are two properties that guide us. First, the graph that is formed by $a$, $b$, $c$ and $d$ does not have any $1$-cuts or $2$-cuts, since the original graph is $3$-edge-connected. In particular, we have $|\partial(a)|\geq 3$, $|\partial(b)|\geq 3$, $|\partial(c)|\geq 3$ and $|\partial(d)|\geq 3$. And second, we have $C=E[a,b]\cup E[a,c]\cup E[d,b]\cup E[d,c]$, $C'=E[a,c]\cup E[a,d]\cup E[b,c]\cup E[b,d]$, and $|C|=|C'|=4$. 

Since $|C|=4$, there are at most four edges in $E[a,b]$. Thus, we will consider all possible values for $|E[a,b]|$. We will start by showing that $|E[a,b]|\neq 4$ and $|E[a,b]|\neq 0$. 

First, suppose that $E[a,b]$ consists of four edges. Then, since $|C|=4$, there are no edges in $E[a,c]$, $E[d,b]$ or $E[d,c]$. Thus, since $|\partial(d)|\geq 3$, there are at least three edges in $E[d,a]$. Then, since $|\partial(c)|\geq 3$, and since $E[a,c]=E[d,c]=\emptyset$, there must be at least three edges in $E[c,b]$. But then, since $E[d,a]\cup E[c,b]\subseteq C'$, we have $|E[d,a]|+|E[c,b]|\leq|C'|=4$, which is impossible, because $|E[d,a]|+|E[c,b]|\geq 6$. This shows that there cannot be four edges in $E[a,b]$.

Now suppose that $E[a,b]$ is empty. Then, since $|\partial(a)|\geq 3$, we have $|E[a,c]|+|E[a,d]|\geq 3$. Since $C'=E[a,c]\cup E[a,d]\cup E[b,c]\cup E[b,d]$ and $|C'|=4$, this implies that $|E[b,c]|+|E[b,d]|\leq 1$. But then, since $|E[a,b]|=0$, we have a contradiction to the fact $|\partial(b)|\geq 3$. This shows that $E[a,b]$ must contain at least one edge.

Now suppose that $E[a,b]$ consists of three edges. Then $E[a,c]$ contains at most one edge. Suppose that $E[a,c]$ contains one edge. Then, since $|C|=4$, we have $E[d,b]=E[d,c]=\emptyset$. Then, since $|\partial(d)|\geq 3$, we have at least three edges in $E[d,a]$. Furthermore, since $|\partial(c)|\geq 3$ and $|E[a,c]|=1$ and $E[d,c]=\emptyset$, we have at least two edges in $E[c,b]$. But then, since $E[d,a]\cup E[c,b]\subseteq C'$, we have $|E[d,a]|+|E[c,b]|\leq|C'|=4$, which is impossible, because $|E[d,a]|+|E[c,b]|\geq 5$. This shows that $E[a,c]=\emptyset$. Then, since $|C|=4$, we have that one of $E[d,b]$ and $E[d,c]$ consists of one edge, and the other is empty. Suppose that $E[d,b]$ contains one edge (and therefore $E[d,c]=\emptyset$). Then, since $|\partial(c)|\geq 3$ and $E[a,c]=E[d,c]=\emptyset$, we have that $E[c,b]$ contains at least three edges. Then, since $E[c,b]\cup E[d,b]\cup E[d,a]\subseteq C'$ and $|C'|=4$, we have that $E[d,a]$ must be empty. But then we have $|\partial(d)|=|E[d,b]|=1$, contradicting the fact that $|\partial(d)|\geq 3$. This shows that $E[d,b]$ is empty, and $E[d,c]$ consists of one edge. Then, since $|\partial(d)|\geq 3$, we have that $E[a,d]$ contains at least two edges. Similarly, since $|\partial(c)|\geq 3$, and $E[a,c]=\emptyset$ and $|E[c,d]|=1$, we have that $E[b,c]$ contains at least two edges. Then, since $|C'|=4$ and $E[a,d]\cup E[b,c]\subseteq C'$, we have $|E[a,d]|=2$ and $|E[b,c]|=2$. Thus, we are in case \textbf{(ii)}.

Now suppose that $E[a,b]$ consists of two edges. Then, since $|C|=4$ and $E[a,b]\cup E[a,c]\subseteq C$, we have that $E[a,c]$ contains at most two edges. Let us suppose, for the sake of contradiction, that $|E[a,c]|=2$. Then, since $|C|=4$, we have $C=E[a,b]\cup E[a,c]$. Since $E[d,b]\cup E[d,c]\subseteq C$, this implies that $E[d,b]=E[d,c]=\emptyset$. Then, since $|\partial(d)|\geq 3$, we have that $E[d,a]$ contains at least three edges. But then, since $E[a,c]\cup E[a,d]\subseteq C'$, we have $|E[a,c]|+|E[a,d]|\leq|C'|=4$, which contradicts the fact that $|E[a,c]|+|E[a,d]|\geq 5$. This shows that $E[a,c]$ contains less than two edges. 

Let us assume first that $E[a,c]=1$. Then, since $|C|=4$ and $C=E[a,b]\cup E[a,c]\cup E[d,b]\cup E[d,c]$, we have that one of $E[d,b]$ and $E[d,c]$ consists of one edge, and the other is empty. Let us suppose, for the sake of contradiction, that $E[d,b]$ contains one edge and $E[d,c]$ is empty. Then, since $|C'|=4$ and $C'=E[a,c]\cup E[a,d]\cup E[b,c]\cup E[b,d]$, we have $|E[a,d]|+|E[b,c]|=2$. This implies that it cannot be that both $E[a,d]$ and $E[b,c]$ contain at least two edges. Now, if $E[a,d]$ contains less than two edges, then $|E[d,b]|=1$ and $E[d,c]=\emptyset$ imply that $|\partial(d)|<3$, which is impossible. And if $E[b,c]$ contains less than two edges, then $|E[a,c]|=1$ and $E[d,c]=\emptyset$ imply that $|\partial(c)|<3$, which is also impossible. Thus, we have that $E[d,b]=\emptyset$ and $E[d,c]$ consists of one edge. Then, since $|C'|=4$ and $C'=E[a,c]\cup E[a,d]\cup E[b,c]\cup E[b,d]$, we have $|E[a,d]|+|E[b,c]|=3$. And then, since $|\partial(d)|\geq 3$ and $|\partial(c)|\geq 3$, we have $|E[a,d]|=2$ and $|E[b,c]|=1$. Thus, we are in case \textbf{(iii)}. 

Now let us assume that $E[a,c]=\emptyset$. Then, since $|C|=4$ and $C=E[a,b]\cup E[a,c]\cup E[d,b]\cup E[d,c]$, we have $|E[d,b]|+|E[d,c]|=2$. This implies that either $|E[d,b]|=0$ and $|E[d,c]|=2$, or $|E[d,b]|=1$ and $|E[d,c]|=1$, or $|E[d,b]|=2$ and $|E[d,c]|=0$. Let us suppose first that $|E[d,b]|=0$ and $|E[d,c]|=2$. Then, since $|C'|=4$ and $C'=E[a,c]\cup E[a,d]\cup E[b,c]\cup E[b,d]$, we have $|E[a,d]|+|E[b,c]|=4$ (because $E[a,c]=E[b,d]=\emptyset$). Then, since $|\partial(d)|\geq 3$ and $|\partial(c)|\geq 3$, we have that either $|E[a,d]|=3$ and $|E[b,c]|=1$, or $|E[a,d]|=|E[b,c]|=2$, or $|E[a,d]|=1$ and $|E[b,c]|=3$. The first and the third case correspond to \textbf{(ii)} (by permuting the labels $a$, $b$, $c$ and $d$). The second case is precisely \textbf{(i)}. 
Now let us suppose that $|E[d,b]|=1$ and $|E[d,c]|=1$. Then, since $|C'|=4$ and $C'=E[a,c]\cup E[a,d]\cup E[b,c]\cup E[b,d]$, we have $|E[a,d]|+|E[b,c]|=3$. And then, since $|\partial(d)|\geq 3$ and $|\partial(c)|\geq 3$, we have $|E[a,d]|=1$ and $|E[b,c]|=2$. Thus, we are in case \textbf{(iii)} (by permuting the labels $a$, $b$, $c$ and $d$). 
Finally, let us suppose that $|E[d,b]|=2$ and $|E[d,c]|=0$. Then, since $E[a,c]=E[d,c]=\emptyset$ and $|\partial(c)|\geq 3$, we have that $E[b,c]$ contains at least three edges. But then, since $E[b,c]\cup E[d,b]\subseteq C'$, we have $|E[b,c]|+|E[d,b]|\leq|C'|=4$, which contradicts the fact that $|E[b,c]|+|E[d,b]|\geq 5$. Thus, this case is impossible.

Finally, suppose that $E[a,b]$ consists of one edge. Since we have considered all other cases for $|E[a,b]|$, it is sufficient to assume, due to the symmetry of our situation, that $|E[b,c]|=1$, $|E[c,d]|=1$ and $|E[d,a]|=1$. (Because, if at least one of those values is different than $1$, then we can properly relabel the corners of the square, and revert to one of the previous cases.) It remains to determine the values $|E[a,c]|$ and $|E[d,b]|$. Since $|C|=4$ and $|E[a,b]|=|E[d,c]|=1$, we have that either $|E[a,c]|=2$ and $|E[d,b]|=0$, or $|E[a,c]|=1$ and $|E[d,b]|=1$, or $|E[a,c]|=0$ and $|E[d,b]|=2$. The first and the last case are rejected, because they imply that $|\partial(d)|=2$ and $|\partial(a)|=2$, respectively. Thus, $|E[a,c]|=1$ and $|E[d,b]|=1$ are the only viable options, and thus we are in case \textbf{(iv)}.
\end{proof}

The following is an obvious corollary of Lemma~\ref{lemma:crossing4cuts}.

\begin{figure}[h!]\centering
\includegraphics[trim={0 23cm 0 0}, clip=true, width=0.8\linewidth]{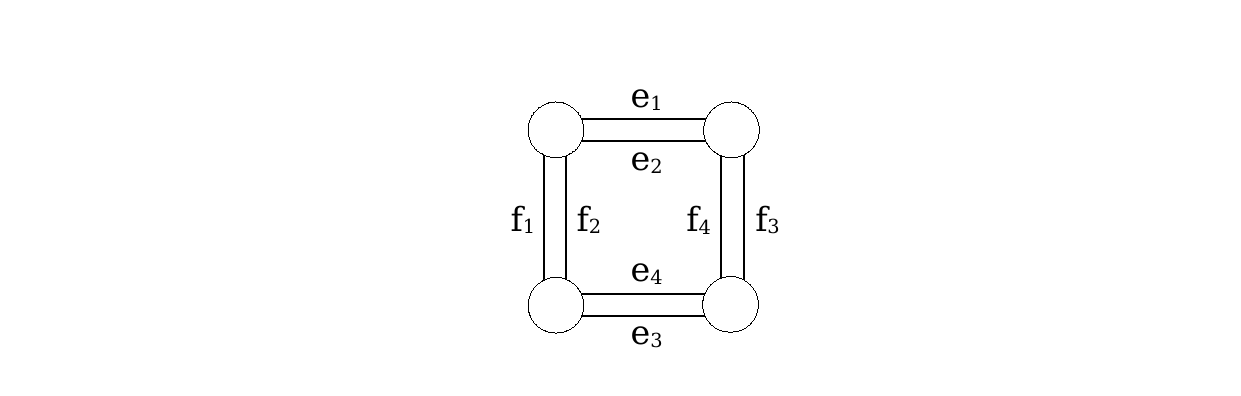}
\caption{\small{The crossing square of two essential $4$-cuts $\{e_1,e_2,e_3,e_4\}$ and $\{f_1,f_2,f_3,f_4\}$.}}\label{figure:crossing-essential}
\end{figure}

\begin{corollary}
\label{corollary:crossing-essential}
Let $C_1=\{e_1,e_2,e_3,e_4\}$ and $C_2=\{f_1,f_2,f_3,f_4\}$ be two essential $4$-cuts that cross. Then, $C_1$ and $C_2$ cross as in Figure~\ref{figure:crossing-essential} (up to permuting the labels of the edges of each cut).
\end{corollary}
\begin{proof}
Lemma~\ref{lemma:crossing4cuts} implies that the possible crossings of $C_1$ and $C_2$ are given by cases \textbf{(i)} to \textbf{(iv)} of Figure~\ref{figure:crossing4cuts}. Notice that only in case \textbf{(i)} we have that both of the $4$-cuts that cross are essential. To see this, observe that in either of cases \textbf{(ii)} to \textbf{(iv)} we have that the lower corners of the square have degree $3$. Therefore, no vertex that is contained in those corners can be $4$-edge-connected with a vertex that is not contained in them. But the union of the lower corners is precisely a side of one of the $4$-cuts $C_1$ and $C_2$. Thus, it cannot be that both $C_1$ and $C_2$ are essential $4$-cuts in those cases.
Therefore, we may assume w.l.o.g. (i.e., by possibly permuting the labels of the edges of each cut) that $C_1$ and $C_2$ cross as in Figure~\ref{figure:crossing-essential}.
\end{proof}

\subsection{Implied $4$-cuts, and cyclic families of $4$-cuts}

The following is one of the implications of Lemma~\ref{lemma:crossing4cuts}.

\begin{figure}[t!]\centering
\includegraphics[trim={0 22cm 0 0}, clip=true, width=0.8\linewidth]{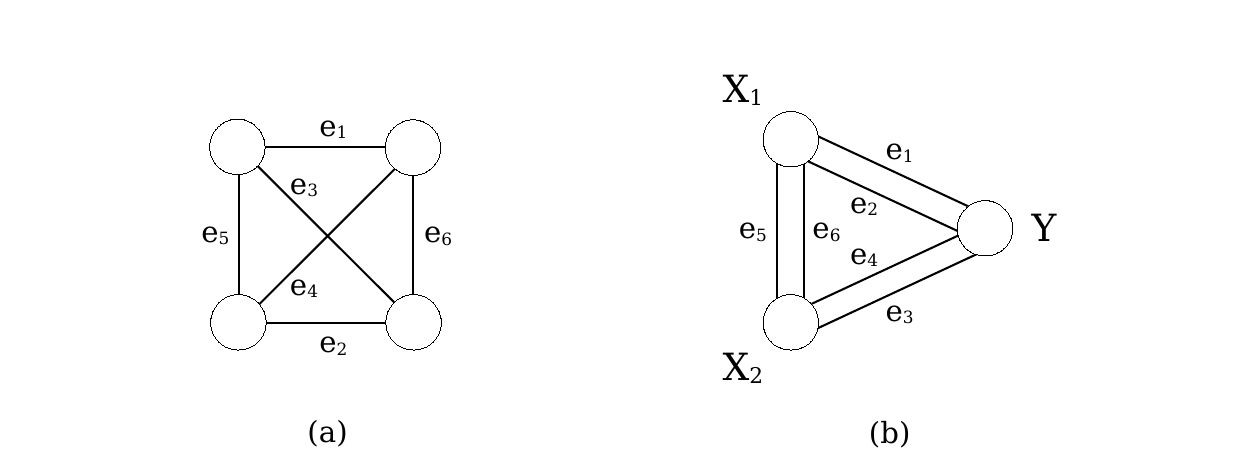}
\caption{\small{The possible arrangements of two distinct $4$-cuts of the form $C_1=\{e_1,e_2,e_3,e_4\}$ and $C_2=\{e_3,e_4,e_5,e_6\}$ (up to swapping the labels $e_3$ and $e_4$). In $(a)$, $C_1$ and $C_2$ cross. In $(b)$, $C_1$ and $C_2$ are parallel. In either case, we have that $\{e_1,e_2,e_5,e_6\}$ is also a $4$-cut.}}\label{figure:implied4cut}
\end{figure}

\begin{lemma}[Implied $4$-cut]
\label{lemma:implied4cut}
Let $C_1=\{e_1,e_2,e_3,e_4\}$ and $C_2=\{e_3,e_4,e_5,e_6\}$ be two distinct $4$-cuts of a graph $G$. Then $C_3=\{e_1,e_2,e_5,e_6\}$ is also a $4$-cut of $G$.
\end{lemma}
\begin{proof}
First, let us assume that $C_1$ and $C_2$ cross. By Lemma~\ref{lemma:crossing4cuts}, we have that Figure~\ref{figure:crossing4cuts} shows all the different ways in which $C_1$ and $C_2$ may cross. Then, since $C_1$ and $C_2$ share a pair of edges, notice that only case \textbf{(iv)} applies here, because this is the only case in which the crossing $4$-cuts share two edges. Thus, $C_1$ and $C_2$ cross as in $(a)$ of Figure~\ref{figure:implied4cut}. Then it is easy to see that Lemma~\ref{lemma:bridge_in_corner} implies that the four corners of this square are connected subgraphs of $G$. Then we observe that $\{e_1,e_2,e_5,e_6\}$ is indeed a $4$-cut of $G$, because its deletion splits the graph into two connected components, but no proper subset of it has this property.

Now suppose that $C_1$ and $C_2$ are parallel. Let $X$ be one of the two sides of $C_1$ and let $X'$ be one of the two sides of $C_2$. Then, since $C_1$ and $C_2$ are parallel and distinct, we may assume w.l.o.g. that $X'\subset X$.
Since $C_1$ and $C_2$ are distinct, we have $|C_1\cap C_2|\neq 4$. By Lemma~\ref{lemma:no-common-three-edges} we have $|C_1\cap C_2|\neq 3$. Thus, $\{e_3,e_4\}\subseteq C_1\cap C_2$ implies that $\{e_3,e_4\}= C_1\cap C_2$. This implies that $\{e_1,e_2\}\cap\{e_5,e_6\}=\emptyset$. Since $C_2$ is a $4$-cut, it is a $4$-element set, and therefore $\{e_3,e_4\}\cap\{e_5,e_6\}=\emptyset$. Thus, we have $C_1\cap\{e_5,e_6\}=\emptyset$. Since $C_1$ is a $4$-cut of $G$, we have that the edges in $C_1$ are the only edges of $G$ that join the sides of $C_1$. Thus, we have that either of $e_5$ and $e_6$ lies entirely within either $G[X]$ or $G[V\setminus X]$. Since $X'\subset X$ and both $e_5$ and $e_6$ have one endpoint in $X'$, we infer that both $e_5$ and $e_6$ lie in $G[X]$.

Let $G'=G\setminus\{e_5,e_6\}$. Since $G[X]$ is connected, we have that $G'[X]$ consists of at most three connected components. Since $X$ is one of the sides of $C_1$, Lemma~\ref{lemma:bridge_in_corner} implies that $G[X]$ contains at most one bridge. Thus, it cannot be the case that $G'[X]$ consists of three connected components (because otherwise $e_5$ and $e_6$ would be two distinct bridges of $G[X]$). 
Let us suppose, for the sake of contradiction, that $G'[X]$ is connected. Let $G''=G'\setminus\{e_3,e_4\}$. Then, since neither of $e_3$ and $e_4$ lies within $G[X]$, we have that $G''[X]$ is connected. Furthermore, since neither of $\{e_3,e_4,e_5,e_6\}$ lies within $G[V\setminus X]$, we have that $G''[V\setminus X]$ is connected. Then, notice that we have $E_{G''}[X,V\setminus X]=\{e_1,e_2\}$. But this implies that $G''=G'\setminus\{e_3,e_4\}=G\setminus C_2$ is connected, in contradiction to the fact that $C_2$ is a $4$-cut of $G$. This shows that $G'[X]$ is disconnected. 
Therefore, we have that $G'[X]$ consists of two connected components $X_1$ and $X_2$ such that $E[X_1,X_2]=\{e_5,e_6\}$. Notice that none of the subgraphs $G[X_1]$, $G[X_2]$ and $G[V\setminus X]$ contains edges from $\{e_1,e_2,e_3,e_4,e_5,e_6\}$. 

Let $Y=V\setminus X$. Now we will determine the edge-sets $E[X_1,Y]$ and $E[X_2,Y]$. Since $X$ is one of the sides of $C_1$, we have $C_1=E[X,Y]$. Thus, since $E[X,Y]=E[X_1,Y]\sqcup E[X_2,Y]$, we have $E[X_1,Y]\sqcup E[X_2,Y]=\{e_1,e_2,e_3,e_4\}$. Notice that $\partial(X_1)=\{e_5,e_6\}\sqcup E[X_1,Y]$ and $\partial(X_2)=\{e_5,e_6\}\sqcup E[X_2,Y]$. Since $G$ is $3$-edge-connected, we have $|\partial(X_1)|\geq 3$ and $|\partial(X_2)|\geq 3$. Thus, we have that both $E[X_1,Y]$ and $E[X_2,Y]$ contain at least one edge from $C_1$. 

Let us suppose, for the sake of contradiction, that one of $E[X_1,Y]$ and $E[X_2,Y]$ contains precisely one edge from $C_1$. Then we may assume w.l.o.g. that $E[X_1,Y]$ contains precisely one edge from $C_1$. First, let us assume that $E[X_1,Y]=\{e_1\}$. Then we have $E[X_2,Y]=\{e_2,e_3,e_4\}$. But then we have that $G\setminus\{e_3,e_4,e_5,e_6\}$ is connected, because, in $G\setminus C_2$, $X_1$ is connected with $V\setminus X$ through $e_1$, and $X_2$ is connected with $V\setminus X$ through $e_2$. This contradicts the fact that $C_2$ is a $4$-cut of $G$. Similarly, if we assume that $E[X_1,Y]=\{e_2\}$, then with the same reasoning we get a contradiction to the fact that $C_2$ is a $4$-cut of $G$. Now let us assume that $E[X_1,Y]=\{e_3\}$. Then we have that $\partial(X_1)=\{e_3,e_5,e_6\}$. This implies that $\{e_3,e_5,e_6\}$ is a $3$-cut of $G$, in contradiction to the fact that $C_2$ is a $4$-cut of $G$. Similarly, if we assume that $E[X_1,Y]=\{e_4\}$, then with the same reasoning we get a contradiction to the fact that $C_2$ is a $4$-cut of $G$. This shows that both $E[X_1,Y]$ and $E[X_2,Y]$ contain at least two edges from $C_1$. Since $E[X_1,Y]\sqcup E[X_2,Y]=C_1 $ and $|C_1|=4$, this implies that $|E[X_1,Y]|=|E[X_2,Y]|=2$. 

Let us suppose, for the sake of contradiction, that one of $E[X_1,Y]$ and $E[X_2,Y]$ is $\{e_1,e_3\}$. Then we may assume w.l.o.g. that $E[X_1,Y]=\{e_1,e_3\}$. Then it is easy to see that $\partial(X_1)=\{e_5,e_6,e_1,e_3\}$ is a $4$-cut of $G$. But then we have $|\partial(X_1)\cap C_2|=3$, contradicting Lemma~\ref{lemma:no-common-three-edges}. This shows that none of $E[X_1,Y]$ and $E[X_2,Y]$ is $\{e_1,e_3\}$. Similarly, we can show that none of $E[X_1,Y]$ and $E[X_2,Y]$ is $\{e_1,e_4\}$. Thus, we have that one of $E[X_1,Y]$ and $E[X_2,Y]$ is $\{e_1,e_2\}$, and the other is $\{e_3,e_4\}$. Let us assume w.l.o.g. that $E[X_1,Y]=\{e_1,e_2\}$ and $E[X_2,Y]=\{e_3,e_4\}$. Then we have a situation as that depicted in $(b)$ of Figure~\ref{figure:implied4cut}. Then we observe that $\{e_1,e_2,e_5,e_6\}$ is indeed a $4$-cut of $G$, since its deletion splits the graph into two connected components, but no proper subset of it has this property.
\end{proof}

%
Lemma~\ref{lemma:implied4cut} motivates the following definition.

\begin{definition}[Implicating Sequence]
\label{definition:implied_4cut}
\normalfont
Let $\mathcal{C}$ be a collection of $4$-cuts of $G$, let $p_1,\dots,p_{k+1}$ be a sequence of pairs of edges, and let $C_1,\dots,C_k$ be a sequence of $4$-cuts from $\mathcal{C}$, such that $C_i=p_i\cup p_{i+1}$ for every $i\in\{1,\dots,k\}$. Then $C_1,\dots,C_k$ is called an implicating sequence of $\mathcal{C}$. Furthermore, if $C=p_1\cup p_{k+1}$ is a $4$-cut of $G$, then we say that $\mathcal{C}$ implies $C$ through the pair of edges $p_1$ (or equivalently: through the pair of edges $p_{k+1}$).
\end{definition}

\begin{remark}
\normalfont
We note the following two facts, which are immediate consequences of Definition~\ref{definition:implied_4cut} that we will use throughout. First, every $4$-cut $C\in\mathcal{C}$ is implied by $\mathcal{C}$, through any pair of edges that is contained in $C$. And second, if $C_1,\dots,C_k$ is an implicating sequence of $\mathcal{C}$, then, for every $i\in\{1,\dots,k-1\}$, either $C_i=C_{i+1}$ or $C_i\cap C_{i+1}=p_{i+1}$ (as a consequence of $C_i=p_i\cup p_{i+1}$, $C_{i+1}=p_{i+1}\cup p_{i+2}$, and Lemma~\ref{lemma:no-common-three-edges}).
\end{remark}

\begin{lemma}
\label{lemma:implicating-sequence}
Let $\mathcal{C}$ be a collection of $4$-cuts of $G$, and let $C_1,\dots,C_k$ be an implicating sequence of $\mathcal{C}$ with $k>1$. Let $p=C_1\setminus C_2$, let $q=C_k\setminus C_{k-1}$, and suppose that $\emptyset\neq p\neq q\neq\emptyset$. Then $p\cup q$ is a $4$-cut implied by $\mathcal{C}$ through $p$. 
\end{lemma}
\begin{proof}
This follows inductively by a repeated application of Lemma~\ref{lemma:implied4cut}. First we consider the case $k=2$. Thus, let $C_1,C_2$ be an implicating sequence of $\mathcal{C}$ such that $C_1\setminus C_2\neq C_2\setminus C_1$. Then we have that $C_1\cap C_2$ is a pair of edges. Thus, Lemma~\ref{lemma:implied4cut} implies that $C'=(C_1\setminus C_2)\cup (C_2\setminus C_1)$ is a $4$-cut. By definition, we have that $C'$ is implied by $\mathcal{C}$ through the pair of edges $C_1\setminus C_2$.

Now let us suppose that the conclusion of the lemma holds for a $k\geq 2$. We will show that it also holds for $k+1$. So let $C_1,\dots,C_{k+1}$ be an implicating sequence of $\mathcal{C}$ such that $\emptyset\neq C_1\setminus C_2\neq C_{k+1}\setminus C_k\neq\emptyset$. Then there is a sequence $p_1,\dots,p_{k+2}$ of pairs of edges such that $C_i=p_i\cup p_{i+1}$, for every $i\in\{1,\dots,k+1\}$. If $C_k=C_{k-1}$, then $C_1,\dots,C_{k-1},C_{k+1}$ is an implicating sequence of $\mathcal{C}$ with $\emptyset\neq C_1\setminus C_2\neq C_{k+1}\setminus C_{k-1}\neq\emptyset$, and the conclusion holds due to the inductive hypothesis. So let us assume that $C_k\neq C_{k-1}$. Then we have $p_1=C_1\setminus C_2$, $p_{k+1}=C_k\setminus C_{k-1}$, $p_{k+2}=C_{k+1}\setminus C_k$, and $p_1\neq p_{k+2}$. 

Our goal is to show that $p_1\cup p_{k+2}$ is a $4$-cut of $G$ (then, by definition, $C_1,\dots,C_{k+1}$ is an implicating sequence of $\mathcal{C}$ that demonstrates that $p_1\cup p_{k+2}$ is implied by $\mathcal{C}$ through $p_1$). Due to the inductive hypothesis, we have that either $p_1=p_{k+1}$, or $p_1\cup p_{k+1}$ is a $4$-cut of $G$. If $p_1=p_{k+1}$, then, since $C_{k+1}=p_{k+1}\cup p_{k+2}$, we have that $p_1\cup p_{k+2}$ is a $4$-cut of $G$. So let us assume that $p_1\cup p_{k+1}$ is a $4$-cut of $G$. Then, since $C_{k+1}=p_{k+1}\cup p_{k+2}$ and $p_1\neq p_{k+2}$, we have that $p_1\cup p_{k+1}$ and $p_{k+1}\cup p_{k+2}$ are two distinct $4$-cuts. Thus, Lemma~\ref{lemma:implied4cut} implies that $p_1\cup p_{k+2}$ is a $4$-cut of $G$. This concludes the proof.
\end{proof}

We extend the terminology of Definition~\ref{definition:implied_4cut} as follows. Let $\mathcal{C}$ and $\mathcal{C}'$ be two collections of $4$-cuts such that every $4$-cut in $\mathcal{C}'$ is implied by $\mathcal{C}$. Then we say that $\mathcal{C}$ implies the collection of $4$-cuts $\mathcal{C}'$.

\begin{figure}[h!]\centering
\includegraphics[trim={0 22cm 0 0}, clip=true, width=0.8\linewidth]{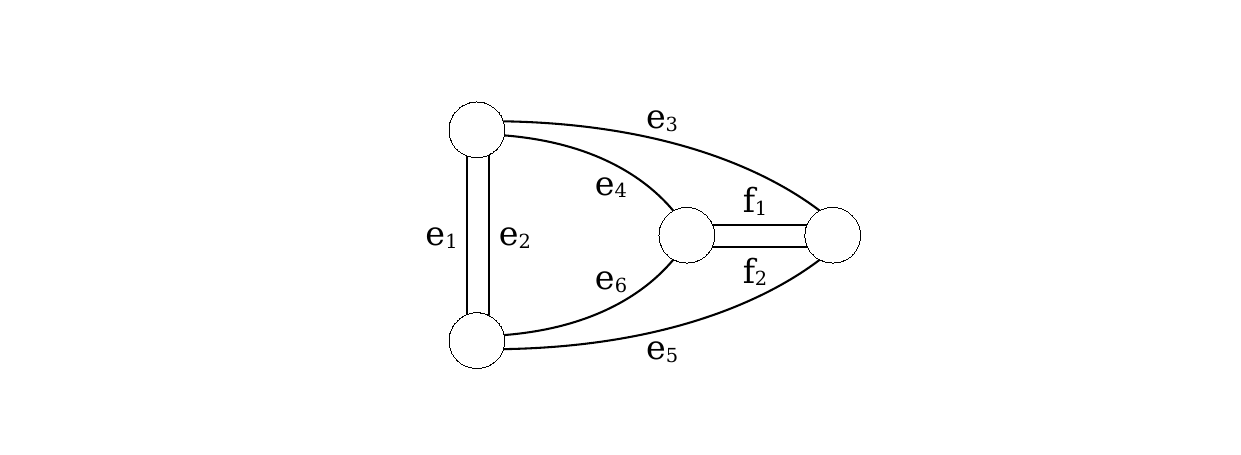}
\caption{\small{Let $\mathcal{C}$ be the collection of $4$-cuts $\{\{e_1,e_2,e_3,e_4\},\{e_3,e_5,f_1,f_2\},\{e_4,e_6,f_1,f_2\}\}$. Then the collection of all $4$-cuts implied by $\mathcal{C}$ is given by $\mathcal{C}'=\mathcal{C}\cup\{\{e_3,e_4,e_5,e_5\}\}$. Now notice that $\mathcal{C}'$ implies the $4$-cut $\{e_1,e_2,e_5,e_6\}$. However, this $4$-cut is not implied by $\mathcal{C}$.}}\label{figure:not-implied}
\end{figure}

\begin{remark}
\normalfont
Despite what its name suggests, we note that the relation of implication between collections of $4$-cuts is not transitive. In other words, if a collection of $4$-cuts $\mathcal{C}$ implies a collection of $4$-cuts $\mathcal{C}'$, and $\mathcal{C}'$ implies a collection of $4$-cuts $\mathcal{C}''$, then it is not necessarily true that $\mathcal{C}$ implies $\mathcal{C}''$. An example for that is given in Figure~\ref{figure:not-implied}. The next lemma provides a sufficient condition under which a kind of transitivity holds.
\end{remark}

\begin{lemma}
\label{lemma:implied_from_union}
Let $\mathcal{C}$ be a collection of $4$-cuts, and let $C$ and $C'$ be two $4$-cuts such that $C\in\mathcal{C}$ and $\mathcal{C}$ implies $C'$ through a pair of edges $p$. Suppose that $\{C,C'\}$ implies a $4$-cut $C''$ through $p$. Then $\mathcal{C}$ implies $C''$ through $p$.
\end{lemma}
\begin{proof}
Let us assume that $C''\notin\{C,C'\}$, because otherwise the lemma follows trivially. Since $C'$ is implied by $\mathcal{C}$ through the pair of edges $p$, we have that $p\subset C'$. Let $q=C'\setminus p$. Then we have $C'=p\cup q$. Since $C''$ is implied by $C$ and $C'$ through $p$, and since $p\subset C'$ and $C''\neq C'$, we have that $q\subset C$ and $p\subset C''$. Furthermore, let $q'=C\setminus q$. Then, $C=q\cup q'$ and $C''=p\cup q'$. Now, if $C'\in\mathcal{C}$, then we obviously have that $C''$ is implied by $\mathcal{C}$ through the pair of edges $p$. Otherwise, since $C'$ is implied by $\mathcal{C}$ through $p$, we have that there is a sequence of $4$-cuts $C_1,\dots,C_k$ in $\mathcal{C}$, and a sequence $p_1,\dots,p_{k+1}$ of pairs of edges, such that $p_1=p$, $p_{k+1}=q$, and $C_i=p_i\cup p_{i+1}$, for every $i\in\{1,\dots,k\}$. Thus, the existence of the sequence of $4$-cuts $C_1,\dots,C_k,C$ in $\mathcal{C}$, demonstrates that $C''$ is implied by $\mathcal{C}$ through the pair of edges $p$.
\end{proof}

A collection of $4$-cuts that implies all $4$-cuts of $G$, is called a \emph{complete} collection of $4$-cuts of $G$.
One of our main contributions in this paper is the following:

\begin{theorem}
\label{theorem:basic}
There is a linear-time algorithm that, given a $3$-edge-connected graph $G$, computes a complete collection of $4$-cuts of $G$, that has size $O(n)$.
\end{theorem}

We prove Theorem~\ref{theorem:basic} in Section~\ref{section:computing-4-cuts}. We note that it seems non-trivial to even establish the existence of a complete collection of $4$-cuts of $G$ that has size $O(n)$. This fact is implied through the analysis of the algorithm that computes $\mathcal{C}$.

\begin{definition}
\label{definition:cyclic_family}
\normalfont
Let $\mathcal{F}=\{p_1,\dots,p_k\}$, with $k\geq 2$, be a collection of pairs of edges with the property that $p_i\cup p_j$ is a $4$-cut of $G$, for every $1\leq i< j\leq k$, and let $\mathcal{C}$ be the collection of all such $4$-cuts. Then we say that $\mathcal{F}$ generates $\mathcal{C}$.
\end{definition}

The following concept is motivated by the structure of crossing $4$-cuts that appears in \textbf{(i)} of Figure~\ref{figure:crossing4cuts}.  

\begin{figure}[h!]\centering
\includegraphics[trim={0 21cm 0 0}, clip=true, width=0.8\linewidth]{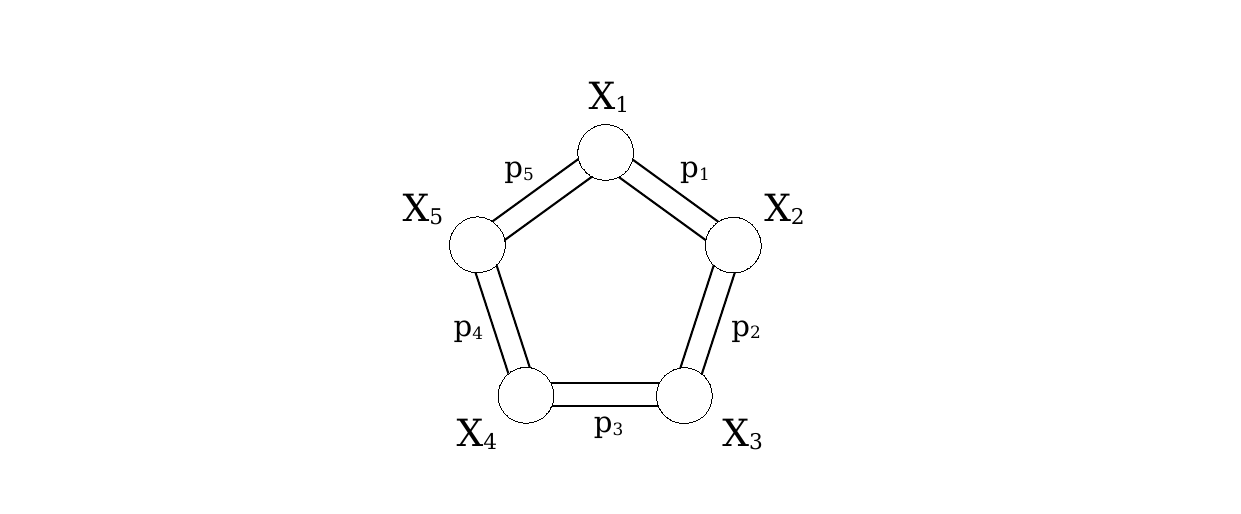}
\caption{\small{A cyclic family of $4$-cuts generated by the collection of pairs of edges $\{p_1,p_2,p_3,p_4,p_5\}$.}}\label{figure:cyclic-family}
\end{figure}

\begin{definition}
\normalfont{(Cyclic families of $4$-cuts.)}
Let $\{p_1,\dots,p_k\}$, with $k\geq 3$, be a collection of pairs of edges of $G$ that generates a collection $\mathcal{C}$ of $4$-cuts of $G$. Suppose that there is a partition $\{X_1,\dots,X_k\}$ of $V(G)$ with the property that $G[X_i]$ is connected for every $i\in\{1,\dots,k\}$, and $E[X_i,X_{i+_{k}1}]=p_i$ for every $i\in\{1,\dots,k\}$. Then $\mathcal{C}$ is called a cyclic family of $4$-cuts. (See Figure~\ref{figure:cyclic-family}.)
\end{definition}

In turns out that a cyclic family of $4$-cuts has a lot of properties that are helpful in order to derive efficiently the $5$-edge-connected components from a complete collection of $4$-cuts. The name ``cyclic" refers to the structure of the graph that is formed by shrinking every $X_i$, $i\in\{1,\dots,k\}$, into a single vertex. This has the structure of a cycle (if we ignore edge multiplicities), as proved in the following lemma. Throughout this work, we may use Lemma~\ref{lemma:edges_between_minimal} without explicit mention. 

\begin{lemma}
\label{lemma:edges_between_minimal}
Let $\mathcal{F}=\{p_1,\dots,p_k\}$, with $k\geq 3$, be a collection of pairs of edges that generates a collection of $4$-cuts of $G$ such that there is a partition $\{X_1,\dots,X_k\}$ of $V(G)$ with the property that $G[X_i]$ is connected for every $i\in\{1,\dots,k\}$, and $E[X_i,X_{i+_{k}1}]=p_i$ for every $i\in\{1,\dots,k\}$. Then $\partial(X_i)=p_i\cup p_{i-_{k}1}$, for every $i\in\{1,\dots,k\}$. Furthermore, for every $i,j\in\{1,\dots,k\}$ with $i\neq j$, the connected components of $G\setminus(p_i\cup p_j)$ are given by $X_{i+_{k}1}\cup X_{i+_{k}2}\cup\dots\cup X_j$ and $X_{j+_{k}1}\cup X_{j+_{k}2}\cup\dots\cup X_i$. (See Figure~\ref{figure:cyclic-family}.)
\end{lemma}
\begin{proof}
Let $i\in\{1,\dots,k\}$. Since $\{X_1,\dots,X_k\}$ is a partition of $V$, we have $\partial(X_i)=(E[X_i,X_1]\cup\dots\cup E[X_i,X_k])\setminus E[X_i,X_i]$. Since $E[X_i,X_{i+_{k}1}]=p_i$ and $E[X_i,X_{i-_{k}1}]=p_{i-_{k}1}$, this implies that $p_i\cup p_{i-_{k}1}\subseteq\partial(X_i)$. If we assume that $E[X_i,X_j]\neq\emptyset$ for some $j\in\{1,\dots,k\}\setminus\{i-_{k}1,i,i+_{k}1\}$, then it is easy to see that $G\setminus(p_i\cup p_{i-_{k}1})$ remains connected, in contradiction to the fact that $\mathcal{F}$ generates a collection of $4$-cuts of $G$. This shows that $\partial(X_i)=p_i\cup p_{i-_{k}1}$.

Now let $i$ and $j$ be two distinct indices in $\{1,\dots,k\}$. If $|i-j|=1$ or $\{i,j\}=\{1,k\}$, then we may assume w.l.o.g. that $j=i-_{k}1$. Then we have shown that $p_i\cup p_j=\partial(X_i)$, and therefore the connected components of $G$ are given by $X_i$ and $(X_1\cup\dots\cup X_k)\setminus X_i$. So let us assume that $|i-j|>1$ and $\{i,j\}\neq\{1,k\}$. Then we have that none of $p_i$ and $p_j$ intersects with any of $E[X_{i+_{k}1},X_{i+_{k}2}],\dots,E[X_{j-_{k}1},X_j]$ or $E[X_{j+_{k}1},X_{j+_{k}2}],\dots,E[X_{i-_{k}1},X_i]$. Furthermore, we have that $E[X_i,X_{i+_{k}1}]\setminus(p_i\cup p_j)=\emptyset$ and $E[X_j,X_{j+_{k}1}]\setminus(p_i\cup p_j)=\emptyset$. Finally, we have that all graphs $G[X_1]\setminus(p_i\cup p_j),\dots,G[X_k]\setminus(p_i\cup p_j)$ remain connected. Thus, we can see that the connected components of $G\setminus(p_i\cup p_j)$ are given by $X_{i+_{k}1}\cup X_{i+_{k}2}\cup\dots\cup X_j$ and $X_{j+_{k}1}\cup X_{j+_{k}2}\cup\dots\cup X_i$.
\end{proof}

In Proposition~\ref{proposition:cyclic_family}, we show that if a collection of pairs of edges $\mathcal{F}$, with $|\mathcal{F}|>3$, generates a collection of $4$-cuts $\mathcal{C}$, then $\mathcal{C}$ is a cyclic family of $4$-cuts.
The next lemma analyzes the case $|\mathcal{F}|=3$, which provides the base step in order to prove inductively Proposition~\ref{proposition:cyclic_family}.

\begin{lemma}
\label{lemma:maximal3family}
Let $\mathcal{F}=\{p_1,p_2,p_3\}$ be a collection of pairs of edges that generates a collection of $4$-cuts of $G$. Then $G'=G\setminus(p_1\cup p_2\cup p_3)$ consists of either three or four connected components. If $G'$ consists of three connected components $X_1,X_2,X_3$, then (by possibly permuting the indices) we have $E[X_1,X_2]=p_1$, $E[X_2,X_3]=p_2$, and $E[X_3,X_1]=p_3$ (i.e., $\mathcal{F}$ generates a cyclic family of $4$-cuts of $G$). If $G'$ consists of four connected components, then $\mathcal{F}$ is maximal w.r.t. the property of generating a collection of $4$-cuts of $G$. Furthermore, if $p_1=\{e_1,e_2\}$, $p_2=\{e_3,e_4\}$ and $p_3=\{e_5,e_6\}$, then the quotient graph of $G$ that is formed by shrinking the connected components of $G'$ into single vertices is shown in $(a)$ of Figure~\ref{figure:implied4cut} (after possibly swapping the labels $e_3$ and $e_4$).
\end{lemma}
\begin{proof}
Let $p_1=\{e_1,e_2\}$, $p_2=\{e_3,e_4\}$ and $p_3=\{e_5,e_6\}$. By assumption we have that $p_1\cup p_2$ is a $4$-cut of $G$, and so let $X,V\setminus X$ be the two connected components of $G\setminus(p_1\cup p_2)$. Then, either $(1)$ $p_3$ is contained entirely within $G[X]$ or $G[V\setminus X]$, or $(2)$ both $G[X]$ and $G[V\setminus X]$ contain an edge from $p_3$. We will show that in case $(1)$ $G'$ consists of three connected components, in case $(2)$ $G'$ consists of four connected components, and in either case the claims of lemma hold true.

Let us consider case $(1)$ first. Then we may assume, w.l.o.g., that $p_3$ lies entirely within $G[X]$. Since $G[X]$ is connected, we have that $G[X]\setminus p_3$ is split into at most three connected components. Now, if $G[X]\setminus p_3$ is connected, then $G\setminus(p_2\cup p_3)$ is also connected (since $E_{G\setminus(p_2\cup p_3)}[X,V\setminus X]=p_1$, and both $(G\setminus(p_2\cup p_3))[X]$ and $(G\setminus(p_2\cup p_3))[V\setminus X]$ remain connected), contradicting the fact that $p_2\cup p_3$ is a $4$-cut of $G$. Thus, $G[X]\setminus p_3$ is split into either two or three connected components. Let us assume, for the sake of contradiction, that $G[X]\setminus p_3$ is split into three connected components. 
This implies that both edges of $p_3$ are bridges of $G[X]$. But since $|\partial(X)|=|p_1\cup p_2|=4$, Lemma~\ref{lemma:bridge_in_corner} implies that $G[X]$ contains at most one bridge, a contradiction.   
Thus, we have that $G[X]\setminus p_3$ consists of two connected components, $C_1$ and $C_2$, and $E[C_1,C_2]=p_3$.

Now we claim that either of $E[C_1,V\setminus X]$ and $E[C_2,V\setminus X]$ contains exactly two edges from $p_1\cup p_2$. To see this, first notice that $\partial(C_1)=E[C_1,C_2]\sqcup E[C_1,V\setminus X]$, $\partial(C_2)=E[C_1,C_2]\sqcup E[C_2,V\setminus X]$, and $E[C_1,V\setminus X]\sqcup E[C_2,V\setminus X]=p_1\cup p_2$. Since $E[C_1,C_2]=p_3$ and $G$ is $3$-edge-connected, this implies that either of $E[C_1,V\setminus X]$ and $E[C_2,V\setminus X]$ contains at least one edge from $p_1\cup p_2$. Let us suppose, for the sake of contradiction, that one of $E[C_1,V\setminus X]$ and $E[C_2,V\setminus X]$ contains exactly one edge from $p_1\cup p_2$. Then, w.l.o.g., we may assume that $E[C_1,V\setminus X]=\{e_1\}$ (the other cases are treated similarly). Then we have $\partial(C_1)=\{e_1,e_5,e_6\}$, and therefore $G\setminus\{e_1,e_5,e_6\}$ is not connected. But this contradicts the fact that $p_1\cup p_3$ is a $4$-cut of $G$. This shows that 
either of $E[C_1,V\setminus X]$ and $E[C_2,V\setminus X]$ contains at least two edges from $p_1\cup p_2$. Then, since $E[C_1,V\setminus X]\sqcup E[C_2,V\setminus X]=p_1\cup p_2$ and $|p_1\cup p_2|=4$, we infer that either of $E[C_1,V\setminus X]$ and $E[C_2,V\setminus X]$ contains exactly two edges from $p_1\cup p_2$.

Now, if $E[C_1,V\setminus X]=p_1$ (and $E[C_2,V\setminus X]=p_2$), or $E[C_1,V\setminus X]=p_2$ (and $E[C_2,V\setminus X]=p_1$), then we basically have the lemma for the case that $G'$ consists of three connected components. But if we assume the contrary, then, w.l.o.g., let $E[C_1,V\setminus X]=\{e_1,e_3\}$ (and $E[C_2,V\setminus X]=\{e_2,e_4\}$). But this means that $p_3\cup p_1$ is not a $4$-cut of $G$ (because $C_1$ remains connected with $V\setminus X$ in $G\setminus(p_3\cup p_1)$ through $e_3$, and $C_2$ remains connected with $V\setminus X$ in $G\setminus(p_3\cup p_1)$ through $e_4$), a contradiction.

Now let us consider case $(2)$. Then we may assume, w.l.o.g., that $e_5$ is contained in $G[X]$ and $e_6$ is contained in $G[V\setminus X]$. Then $G[X]\setminus e_5$ is split into at most two connected components. Let us suppose, for the sake of contradiction, that $G[X]\setminus e_5$ is connected. Then $G[X]\setminus (p_3\cup p_1)$ is also connected (because the only edge from $p_3\cup p_1$ that lies in $G[X]$ is $e_5$), and therefore $p_3\cup p_1$ is not a $4$-cut of $G$ (because the endpoints of $e_5$ remain connected in $G\setminus(p_3\cup p_1)$), a contradiction. Thus, we have that $e_5$ is a bridge of $G[X]$. Similarly, $e_6$ is a bridge of $G[V\setminus X]$.
So let $C_1,C_2$ be the connected components of $G[X]\setminus e_5$, and let $C_3,C_4$ be the connected components of $G[V\setminus X]\setminus e_6$. Then we have $E[C_1,C_2]=e_5$ and $E[C_3,C_4]=e_6$. This shows that $G'$ consists of four connected components, $C_1,C_2,C_3,C_4$. 

Now we have to consider how the vertex sets $C_1,C_2,C_3,C_4$ are interconnected using the edges from $p_1\cup p_2$. First, it is not difficult to see that every one of $C_1,C_2,C_3,C_4$ must have exactly two edges from $p_1\cup p_2$ as boundary edges, because otherwise we violate the fact that $G$ is $3$-edge-connected. Then, we can see that no one of $C_1,C_2,C_3,C_4$ can have either both edges from $p_1$ or both edges from $p_2$ as boundary edges, because otherwise we violate the fact that $p_1\cup p_3$ and $p_2\cup p_3$ are $4$-cuts of $G$ (and therefore no proper subsets of them can destroy the connectivity of $G$ upon removal). Finally, we can see that both $C_1$ and $C_2$ must be connected with both $C_3$ and $C_4$ using edges from $p_1\cup p_2$, because otherwise $p_3$ is a $2$-cut of $G$. Thus, we may assume, w.l.o.g., that $E[C_1,C_3]=e_1$, $E[C_1,C_4]=e_3$, $E[C_2,C_3]=e_4$, and $E[C_2,C_4]=e_2$. Notice that this is precisely the situation depicted in $(a)$ of Figure~\ref{figure:implied4cut}.

It remains to show (still being in case $(2)$), that there is no pair of edges $p_4=\{e_7,e_8\}$, with $p_4\notin\mathcal{F}$, such that $\mathcal{F}\cup\{p_4\}$ generates a collection of $4$-cuts of $G$. So let us assume the contrary. Notice that this implies that $\{e_7,e_8\}\cap (p_1\cup p_2\cup p_3)=\emptyset$, because otherwise $p_4\cup p_i$ is not a $4$-element set, for some $i\in\{1,2,3\}$. This means that either both edges from $p_4$ lie entirely within $G[C_i]$, for some $i\in\{1,2,3,4\}$, or that both $G[C_i]$ and $G[C_j]$ contain edges from $p_4$, for some $i,j\in\{1,2,3,4\}$ with $i\neq j$. Let us consider the first case first, and so let us assume w.l.o.g. that $p_4$ is contained within $G[C_1]$. Then $G[C_1]\setminus p_4$ consists of at most three connected components. We will show that all three cases concerning the number of connected components of $G[C_1]\setminus p_4$ lead to a contradiction. 
If we assume that $G[C_1]\setminus p_4$ is connected, then we contradict the fact that $p_1\cup p_4$ is a $4$-cut of $G$ (because the endpoints of $p_4$ remain connected in $G\setminus(p_1\cup p_4)$). Let us assume that $G[C_1]\setminus p_4$ consists of two connected components $D_1$ and $D_2$. Then we have $E[D_1,D_2]=p_4$. Now, if we consider all the different combinations of the incidence relation of the boundary edges of $C_1$ with $D_1$ and $D_2$, we will see that, in every possible case, we either contradict the fact that $G$ is $3$-edge-connected (e.g., if all the boundary edges of $C_1$ are incident to $D_1$), or the fact that $p_4$ must form a $4$-cut of $G$ will all of $p_1,p_2,p_3$ (so, e.g., if $e_1$ and $e_3$ are incident to $D_1$, and $e_5$ is incident to $D_2$, then we have that $p_4\cup\{e_5\}$ is a $3$-cut of $G$, contradicting the fact that $p_4\cup p_3$ is a $4$-cut of $G$). Finally, let us assume that $G[C_1]\setminus p_4$ consists of three connected components $D_1$, $D_2$ and $D_3$. Then w.l.o.g. we may assume that $E[D_1,D_2]=\{e_7\}$ and $E[D_2,D_3]=\{e_8\}$. But now, since $C_1$ has precisely three boundary edges in $G$, we have that, no matter the incidence relation of those boundary edges to $D_1,D_2,D_3$, we violate the fact that $G$ is $3$-edge-connected. 

Finally, it remains to consider the case that the edges from $p_4$ lie in two different subgraphs $G[C_i],G[C_j]$, for some $i,j\in\{1,2,3,4\}$ with $i\neq j$. Due to the symmetry of the interconnections between $C_1,C_2,C_3,C_4$, we may assume w.l.o.g. that $e_7$ is contained in $G[C_1]$ and $e_8$ is contained in $G[C_2]$. Observe that $G[C_1]\setminus e_7$ and $G[C_2]\setminus e_8$ cannot both be connected, because otherwise $p_1\cup p_4$ (or $p_2\cup p_4$, or $p_3\cup p_4$) is not a $4$-cut of $G$. Thus we may assume w.l.o.g. that $G[C_1]\setminus e_7$ is not connected, and let $D_1,D_2$ be its connected components. Then we have $E[D_1,D_2]=e_7$. But since $C_1$ has only three boundary edges in $G$, this violates the fact that $G$ is $3$-edge-connected (because either $e_7$ is a bridge of $G$, or it forms a $2$-cut with one of the boundary edges of $C_1$ in $G$). Thus we have shown that $\mathcal{F}$ cannot be extended with one more pair of edges $p_4$ such that $\mathcal{F}\cup\{p_4\}$ generates a collection of $4$-cuts of $G$.
\end{proof}

We consider the last case in Lemma~\ref{lemma:maximal3family} to be degenerate, because the pairs of edges $p_1,p_2,p_3$ are so entangled with the components induced by the three $4$-cuts generated by them, that this collection of pairs of edges cannot be extended to a larger collection of pairs of edges that also generates a collection of $4$-cuts. This singularity cannot occur with larger collections of pairs of edges that generate $4$-cuts, because (loosely speaking) the ability of every pair of them to provide a $4$-cut forces them to produce a more organized system of $4$-cuts (see Proposition~\ref{proposition:cyclic_family}).
Formally, if $\mathcal{F}$ is collection of pairs of edges with $|\mathcal{F}|=3$ that generates a collection of $4$-cuts $\mathcal{C}$ such that $G\setminus{\bigcup{\mathcal{F}}}$ consists of four connected components, then $\mathcal{C}$ is a called a \emph{degenerate} family of $4$-cuts.

\begin{corollary}
\label{corollary:degenerate_non-essential}
Let $\mathcal{C}$ be a degenerate family of $4$-cuts. Then $\mathcal{C}$ consists of three non-essential $4$-cuts.
\end{corollary}
\begin{proof}
By definition, we have that $\mathcal{C}$ is generated by a collection of pairs of edges $\mathcal{F}$ with $|\mathcal{F}|=3$ such that $G\setminus{\bigcup{\mathcal{F}}}$ consists of four connected components. Let $\mathcal{F}=\{\{e_1,e_2\},\{e_3,e_4\},\{e_5,e_6\}\}$. Then, by Lemma~\ref{lemma:maximal3family}, we can assume w.l.o.g. that the edges in $\bigcup{\mathcal{F}}$ are arranged as in $(a)$ of Figure~\ref{figure:implied4cut}. Thus, it is easy to see that every pair of vertices that are separated by a $4$-cut from $\mathcal{C}$ can also be separated by a $3$-cut. We conclude that none of the three $4$-cuts in $\mathcal{C}$ is essential.
\end{proof}

\begin{proposition}
\label{proposition:cyclic_family}
Let $\mathcal{C}$ be a collection of $4$-cuts of $G$ that is generated by a collection $\mathcal{F}=\{p_1,\dots,p_k\}$ of $k$ pairs of edges, with $k\geq 4$. Then there is a partition $\{X_1,\dots,X_k\}$ of $V(G)$, such that $G[X_i]$ is connected for every $i\in\{1,\dots,k\}$, and $E[X_i,X_{i+_{k}1}]=p_i$ for every $i\in\{1,\dots,k\}$. In other words, $\mathcal{C}$ is a cyclic family of $4$-cuts of $G$.
\end{proposition}
\begin{proof}
For every integer $k'$ with $3\leq k'\leq k$, we define the proposition $\Pi(k')\equiv$ ``there is a partition $\{X_1,\dots,X_{k'}\}$ of $V(G)$, such that $G[X_i]$ is connected for every $i\in\{1,\dots,k'\}$, and $E[X_i,X_{i+_{k'}1}]=p_i$ for every $i\in\{1,\dots,k'\}$". We will show inductively that $\Pi(k)$ is true.

If we take the subcollection $\{p_1,p_2,p_3\}$ of $\mathcal{F}$, then this generates a collection of $4$-cuts of $G$; but since $k\geq 4$, this collection is not maximal w.r.t. the property of generating a collection of $4$-cuts of $G$. Thus, by Lemma~\ref{lemma:maximal3family} we have that there is a partition $\{X_1,X_2,X_3\}$ of $V$, such that $G[X_i]$ is connected for every $i\in\{1,2,3\}$, and $E[X_i,X_{i+_{3}1}]=p_i$ for every $i\in\{1,2,3\}$. This establishes $\Pi(3)$ (the base step of our induction).

Now let us suppose that $\Pi(k')$ is true, for some $k'$ with $3\leq k'<k$. This means that there exists a partition $\{X_1,\dots,X_{k'}\}$ of $V(G)$ such that $G[X_i]$ is connected for every $i\in\{1,\dots,k'\}$, and $E[X_i,X_{i+_{k'}1}]=p_i$ for every $i\in\{1,\dots,k'\}$. 
We will show that $\Pi(k'+1)$ is also true.

Since $p_{k'+1}$ forms a $4$-cut with every pair of edges in $\{p_1,\dots,p_{k'}\}$, we have that none of the edges in $p_{k'+1}$ lies in $p_1\cup\dots\cup p_{k'}$ (because otherwise $p_{k'+1}\cup p_i$ would not be a $4$-element set, for some $i\in\{1,\dots,k'\}$). Thus, either $p_{k'+1}$ lies entirely within $G[X_i]$, for some $i\in\{1,\dots,k'\}$, or there are $i,j\in\{1,\dots,k'\}$, with $i\neq j$, such that both $G[X_i]$ and $G[X_j]$ contain an edge from $p_{k'+1}$. Let us suppose, for the sake of contradiction, that the second case is true. Let $p_{k'+1}=\{e_1,e_2\}$, and let us assume, w.l.o.g., that $e_1$ is contained in $G[X_i]$ and $e_2$ is contained in $G[X_j]$. Since $k'\geq 3$, we may assume w.l.o.g. that $j\neq i-_{k'}1$. 
Now, since $p_i\cup p_{k'+1}$ is a $4$-cut of $G$ and $e_1$ is the only edge from $p_i\cup p_{k'+1}$ that lies in $G[X_i]$, we have that $e_1$ is a bridge of $G[X_i]$ (because otherwise the endpoints of $e_1$ would remain connected in $G[X_i]\setminus(p_i\cup p_{k'+1})$). So let $C_1,C_2$ be the connected components of $G[X_i]\setminus e_1$. Thus, we have $E[C_1,C_2]=\{e_1\}$. By applying Lemma~\ref{lemma:edges_between_minimal} on the collection of pairs of edges $\{p_1,\dots,p_{k'}\}$, we have that $\partial(X_i)=p_i\cup p_{i-_{k'}1}$. Notice that $\partial(C_1)=E[C_1,C_2]\sqcup E[C_1,V\setminus X_i]$ and $\partial(C_2)=E[C_1,C_2]\sqcup E[C_2,V\setminus X_i]$. Thus, since the graph is $3$-edge-connected and $E[C_1,V\setminus X_i]\sqcup E[C_2,V\setminus X_i]=\partial(X_i)$ and $|\partial(X_i)|=4$, we have that both $E[C_1,V\setminus X_i]$ and $E[C_2,V\setminus X_i]$ must contain precisely two edges from $p_i\cup p_{i-_{k'}1}$. If $E[C_1,V\setminus X_i]=p_i$, then we have that $p_i\cup\{e_1\}$ is a $3$-cut of $G$, contradicting the fact that $p_i\cup\{e_1,e_2\}$ is a $4$-cut of $G$. Similarly, we can reject the case $E[C_1,V\setminus X_i]=p_{i-_{k'}1}$. Thus, we have that both $E[C_1,V\setminus X_i]$ and $E[C_2,V\setminus X_i]$ must intersect with both $p_i$ and $p_{i-_{k'}1}$. But then, since $G[X_{i-_{k'}1}]$ is connected, and $e_1,e_2$ do not lie in $G[X_{i-_{k'}1}]$, we have that $C_1$ and $C_2$ remain connected in $G\setminus (p_i\cup\{e_1,e_2\})$ (because both of them remain connected with $X_{i-_{k'}1}$ in $G\setminus(p_i\cup\{e_1,e_2\})$ through the edges from $p_{i-_{k'}1}$). This implies that the endpoints of $e_1$ remain connected in $G\setminus (p_i\cup\{e_1,e_2\})$, contradicting the fact that $p_i\cup\{e_1,e_2\}$ is a $4$-cut of $G$. Thus, we have shown that $p_{k'+1}$ lies entirely within $G[X_i]$, for some $i\in\{1,\dots,k'\}$. 

Now, since $G[X_i]$ is connected, we have that $G[X_i]\setminus p_{k'+1}$ is split into at most three connected components. It cannot be the case that $G[X_i]\setminus p_{k'+1}$ is connected, because otherwise e.g. $p_1\cup p_{k'+1}$ is not a $4$-cut of $G$ (because the endpoints of the edges from $p_{k'+1}$ would remain connected in $G\setminus(p_1\cup p_{k'+1})$). Now let us suppose, for the sake of contradiction, that $G[X_i]\setminus p_{k'+1}$ is split into three connected components. This implies that both edges from $p_{k'+1}$ are bridges of $G[X_i]$. By applying Lemma~\ref{lemma:edges_between_minimal} on the collection of pairs of edges $\{p_1,\dots,p_{k'}\}$, we have that the boundary of $X_i$ in $G$ contains exactly four edges. Then, Lemma~\ref{lemma:bridge_in_corner} implies that $G[X_i]$ contains at most one bridge -- a contradiction. This shows that $G[X_i]\setminus p_{k'+1}$ consists of two connected components $C_1$ and $C_2$. Then we have $E[C_1,C_2]=p_{k'+1}$. 

It remains to determine the incidence relation between $C_1$ and $C_2$ and the edges from $\partial(X_i)=p_i\cup p_{i-_{k'}1}$. Notice that $\partial(C_1)=E[C_1,C_2]\sqcup E[C_1,V\setminus X_i]$ and $\partial(C_2)=E[C_1,C_2]\sqcup E[C_2,V\setminus X_i]$. Thus, since the graph is $3$-edge-connected, we have that either of $E[C_1,V\setminus X_i]$ and $E[C_2,V\setminus X_i]$ must contain at least one edge from $p_i\cup p_{i-_{k'}1}$. Therefore, since $|\partial(X_i)|=4$, we can see that either of $E[C_1,V\setminus X_i]$ and $E[C_2,V\setminus X_i]$ must contain exactly two edges from $p_i\cup p_{i-_{k'}1}$, because otherwise we contradict the fact that $p_{k'+1}\cup p_i$ and $p_{k'+1}\cup p_{i-_{k'}1}$ are $4$-cuts of $G$ (and therefore no proper subset of them can disconnect $G$ upon removal).

Now, we either have $(1)$ $E[C_1,X_{i+_{k'}1}]=p_i$ (and $E[C_2,X_{i-_{k'}1}]=p_{i-_{k'}1}$), or $(2)$ $E[C_1,X_{i-_{k'}1}]=p_{i-_{k'}1}$ (and $E[C_2,X_{i+_{k'}1}]=p_i$), or $(3)$ both $E[C_1,X_{i+_{k'}1}]$ and $E[C_2,X_{i+_{k'}1}]$ intersect with $p_i$ (and both $E[C_1,X_{i-_{k'}1}]$ and $E[C_2,X_{i-_{k'}1}]$ intersect with $p_{i-_{k'}1}$). Let us suppose, for the sake of contradiction, that the third case is true. But then we have that $p_i\cup p_{k'+1}$ is not a $4$-cut of $G$, since both $C_1$ and $C_2$ remain connected with $X_{i-_{k'}1}$ through the edges from $p_{i-_{k'}1}$. Thus, we may assume, w.l.o.g., that $E[C_1,X_{i+_{k'}1}]=p_i$ and $E[C_2,X_{i-_{k'}1}]=p_{i-_{k'}1}$. Now we observe that, by renaming appropriately the sets $C_1$, $C_2$ and $\{X_1,\dots,X_{k'}\}\setminus\{X_i\}$, we have that there is a partition $\{X_1,\dots,X_{k'+1}\}$ of $V$, such that $G[X_i]$ is connected for every $i\in\{1,\dots,k'+1\}$, and $E[X_i,X_{i+_{k'+1}1}]=p_i$ for every $i\in\{1,\dots,k'+1\}$. Thus, $\Pi(k'+1)$ is also true, and the result follows inductively.
\end{proof}

\begin{corollary}
\label{corollary:essential_implies_cyclic}
Let $\mathcal{C}$ be a collection of $4$-cuts with $|\mathcal{C}|\geq 3$ that is generated by a collection of pairs of edges. Suppose that $\mathcal{C}$ contains an essential $4$-cut. Then $\mathcal{C}$ is a cyclic family of $4$-cuts.
\end{corollary}
\begin{proof}
If $|\mathcal{C}|=3$, then, since $\mathcal{C}$ contains an essential $4$-cut, Corollary~\ref{corollary:degenerate_non-essential} implies that $\mathcal{C}$ cannot be a degenerate family of $4$-cuts. Thus, Lemma~\ref{lemma:maximal3family} implies that $\mathcal{C}$ is a cyclic family of $4$-cuts. Otherwise, if $|\mathcal{C}|>3$, then $\mathcal{C}$ is generated by a collection of pairs of edges that has size at least four, and therefore Proposition~\ref{proposition:cyclic_family} implies that $\mathcal{C}$ is a cyclic family of $4$-cuts.
\end{proof}

A collection $\mathcal{C}$ of $4$-cuts is called trivial if $|\mathcal{C}|=1$.

\begin{lemma}
\label{lemma:uniqueness_of_family}
Let $\mathcal{F}$ be a collection of pairs of edges of $G$ that generates a non-trivial collection $\mathcal{C}$ of $4$-cuts of $G$. Then, $\mathcal{F}$ is unique w.r.t. the property of generating $\mathcal{C}$.
\end{lemma}
\begin{proof}
Let us suppose, for the sake of contradiction, that there are two distinct collections $\mathcal{F}_1$ and $\mathcal{F}_2$ of pairs of edges of $G$ that generate $\mathcal{C}$. Then, since $\mathcal{F}_1$ and $\mathcal{F}_2$ generate the same non-trivial collection of $4$-cuts, we have that $|\mathcal{F}_1|=|\mathcal{F}_2|>2$. Now, since $\mathcal{F}_1$ and $\mathcal{F}_2$ are distinct and $|\mathcal{F}_1|=|\mathcal{F}_2|$, there is a pair of edges $\{e,e'\}\in\mathcal{F}_2\setminus\mathcal{F}_1$. Let $C=\{e_1,e_2,e_3,e_4\}$ be a $4$-cut in $\mathcal{C}$. Since $\mathcal{F}_1$ generates $C$, we may assume w.l.o.g. that $\{\{e_1,e_2\},\{e_3,e_4\}\}\subset\mathcal{F}_1$. Now let us assume, for the sake of contradiction, that $\{e_1,e_2\}\in\mathcal{F}_2$. Since $\{e_1,e_2\}\in\mathcal{F}_1$ and $\{e,e'\}\notin\mathcal{F}_1$, we have $\{e_1,e_2\}\neq\{e,e'\}$. Then, since $\{e,e'\}\in\mathcal{F}_2$, we have that $C'=\{e_1,e_2,e,e'\}$ is a $4$-cut in $\mathcal{C}$. But since $\{e,e'\}\notin\mathcal{F}_1$ and $\mathcal{F}_1$ also generates $C'$, we have that either $\{e_1,e\}\in\mathcal{F}_1$ or $\{e_1,e'\}\in\mathcal{F}_1$. (Notice that none of $e,e'$ can be $e_2$, for otherwise $C'$ would not be a $4$-element set.) But then we have that either $\{e_1,e_2\}\cup\{e_1,e\}$ or $\{e_1,e_2\}\cup\{e_1,e'\}$ is a $4$-cut of $G$, a contradiction. This shows that $\{e_1,e_2\}\notin\mathcal{F}_2$. Thus, since $\mathcal{F}_2$ generates $C$, we may assume w.l.o.g. that $\{\{e_1,e_3\},\{e_2,e_4\}\}\subset\mathcal{F}_2$. 


Now, since $|\mathcal{F}_1|\geq 3$, there is a pair of edges $\{e_5,e_6\}\in\mathcal{F}_1\setminus\{\{e_1,e_2\},\{e_3,e_4\}\}$. 
Then we have that $\mathcal{F}_1$ generates $\{e_1,e_2,e_5,e_6\}$, and therefore $\mathcal{F}_2$ also generates $\{e_1,e_2,e_5,e_6\}$.
Then, since $\{e_1,e_2\}\notin\mathcal{F}_2$, we have that either $\{e_1,e_5\}\in\mathcal{F}_2$ or $\{e_1,e_6\}\in\mathcal{F}_2$. Notice that $e_3\notin\{e_5,e_6\}$, because otherwise we would have that $\{e_3,e_4\}\cup\{e_5,e_6\}$ is not a $4$-element set, contradicting the fact that $\mathcal{F}_1$ generates a collection of $4$-cuts. This implies that $\{e_1,e_3\}\neq\{e_1,e_5\}$ and $\{e_1,e_3\}\neq\{e_1,e_6\}$. But then, since  $\{e_1,e_3\}\in\mathcal{F}_2$, we have that either $\{e_1,e_3\}\cup\{e_1,e_5\}$ or $\{e_1,e_3\}\cup\{e_1,e_6\}$ is a $4$-cut of $G$, which is absurd. We conclude that there is a unique collection of pairs of edges of $G$ that generates $\mathcal{C}$. 

\end{proof}

Let $\mathcal{C}$ be a cyclic family of $4$-cuts. Then, Lemma~\ref{lemma:uniqueness_of_family} implies that there is a unique collection $\mathcal{F}=\{p_1,\dots,p_k\}$ of pair of edges that generates $\mathcal{C}$. Now, by definition, we have that there is a partition $\{X_1,\dots,X_k\}$ of $V(G)$ such that $G[X_i]$ is connected for every $i\in\{1,\dots,k\}$, and (w.l.o.g.) $E[X_i,X_{i+_{k}1}]=p_i$ for every $i\in\{1,\dots,k\}$. Thus, by Lemma~\ref{lemma:edges_between_minimal} we have that the connected components of $G\setminus{\bigcup{\mathcal{F}}}$ are precisely $X_1,\dots,X_k$. Then, we call $X_1,\dots, X_k$ the \emph{corners} of the cyclic family $\mathcal{C}$ (considered either as vertex sets, or as subgraphs of $G$). For every $i\in\{1,\dots,k\}$ we have that $\partial(X_i)=p_i\cup p_{i-_{k}1}$. We call the $4$-cuts $\partial(X_1),\dots,\partial(X_k)$ the \emph{$\mathcal{C}$-minimal $4$-cuts}. 

Let $C$ be a $4$-cut of $G$. If there is no collection $\mathcal{F}$ of pairs of edges with $|\mathcal{F}|>2$ that generates (a collection of $4$-cuts that contains) $C$, then $C$ is called an \emph{isolated} $4$-cut. 

\begin{corollary}
\label{corollary:iso-essential-parallel}
Let $C$ be an essential isolated $4$-cut of $G$. Then, $C$ is parallel with every essential $4$-cut of $G$.
\end{corollary}
\begin{proof}
Let us suppose, for the sake of contradiction, that there is an essential $4$-cut $C'$ such that $C$ and $C'$ cross. Let $C=\{e_1,e_2,e_3,e_4\}$, and let $C'=\{f_1,f_2,f_3,f_4\}$. Then, by Corollary~\ref{corollary:crossing-essential} we have that $C$ and $C'$ must cross as in Figure~\ref{figure:crossing-essential}. By Lemma~\ref{lemma:bridge_in_corner}, we have that the four corners of this figure are connected, and therefore all of them constitute $4$-cuts of $G$. But this implies that $\{\{e_1,e_2\},\{e_3,e_4\},\{f_1,f_2\},\{f_3,f_4\}\}$ is a collection of pairs of edges that generates $C$ -- in contradiction to the fact that $C$ is an isolated $4$-cut. Thus, we conclude that $C$ does not cross with other essential $4$-cuts.
\end{proof}

\subsection{Properties of cyclic families of $4$-cuts}

The following two lemmata demonstrate the importance of minimal $4$-cuts. Both of them are a consequence of the structure of the sides of the minimal $4$-cuts (provided by Lemma~\ref{lemma:edges_between_minimal}).

\begin{lemma}
\label{lemma:minimal-are-parallel}
Let $\mathcal{C}$ be a cyclic family of $4$-cuts, and let $C$ and $C'$ be two $\mathcal{C}$-minimal $4$-cuts. Then $C$ and $C'$ are parallel.
\end{lemma}
\begin{proof}
Since $\mathcal{C}$ is a cyclic family of $4$-cuts, there is a partition $\{X_1,\dots,X_k\}$ of $V(G)$, and a collection of pairs of edges $\{p_1,\dots,p_k\}$, such that $G[X_i]$ is connected for every $i\in\{1,\dots,k\}$, $E[X_i,X_{i+_{k}1}]=p_i$ for every $i\in\{1,\dots,k\}$, and $\mathcal{C}=\{p_i\cup p_j\mid i,j\in\{1,\dots,k\} \mbox{ with } i\neq j\}$. Then, since $C$ and $C'$ are $\mathcal{C}$-minimal $4$-cuts, there are $i,j\in\{1,\dots,k\}$ such that $C=p_i\cup p_{i+_{k}1}$ and $C'=p_j\cup p_{j+_{k}1}$. Then, by Lemma~\ref{lemma:edges_between_minimal} we have that the connected components of $G\setminus C$ are $X_i$ and $V\setminus X_i$, and the connected components of $G\setminus C'$ are $X_j$ and $V\setminus X_j$. Since $\{X_1,\dots,X_k\}$ is a partition of $V(G)$, we have that either $X_i=X_j$ or $X_i\cap X_j=\emptyset$. Thus, the $4$-cuts $C$ and $C'$ are parallel (as an immediate consequence of the definition).
\end{proof}

\begin{lemma}[A non-minimal $4$-cut can be replaced by minimal $4$-cuts]
\label{lemma:replace-with-minimal}
Let $\mathcal{C}$ be a cyclic family of $4$-cuts, and let $C$ be a $4$-cut in $\mathcal{C}$ that separates two vertices $x$ and $y$. Then, there is a $\mathcal{C}$-minimal $4$-cut $C'$ that also separates $x$ and $y$.
\end{lemma}
\begin{proof}
Since $\mathcal{C}$ is a cyclic family of $4$-cuts, there is a partition $\{X_1,\dots,X_k\}$ of $V(G)$, and a collection of pairs of edges $\{p_1,\dots,p_k\}$, such that $G[X_i]$ is connected for every $i\in\{1,\dots,k\}$, $E[X_i,X_{i+_{k}1}]=p_i$ for every $i\in\{1,\dots,k\}$, and $\mathcal{C}=\{p_i\cup p_j\mid i,j\in\{1,\dots,k\} \mbox{ with } i\neq j\}$. Then, there are $i,j\in\{1,\dots,k\}$ with $i\neq j$, such that $C=p_i\cup p_j$. By Lemma~\ref{lemma:edges_between_minimal}, we have that the connected components of $G\setminus C$ are given by $X_{i+_{k}1}\cup X_{i+_{k}2}\cup\dots\cup X_j$ and $X_{j+_{k}1}\cup X_{j+_{k}2}\cup\dots\cup X_i$. Since $x$ and $y$ are separated by $C$, we have that there are $t\in\{i+_{k}1,i+_{k}2,\dots,j\}$ and $t'\in\{j+_{k}1,j+_{k}2,\dots,i\}$, such that $x\in X_t$ and $y\in X_{t'}$. Consider the $4$-cut $C'=p_{t-_{k}1}\cup p_t$. Then, Lemma~\ref{lemma:edges_between_minimal} implies that $C'=\partial(X_t)$, and therefore $C'$ is a $\mathcal{C}$-minimal $4$-cut (by definition). Notice that the connected components of $G\setminus C'$ are given by $X_t$ and $V\setminus X_t$. Thus, $C'$ separates $x$ and $y$. 
\end{proof}

\begin{corollary}
\label{corollary:minimal-4cuts}
Let $\mathcal{C}$ be a cyclic family of $4$-cuts, and let $\mathcal{M}$ be the collection of the $\mathcal{C}$-minimal $4$-cuts. Then $\mathcal{M}$ is a parallel family of $4$-cuts with $\mathit{atoms}(\mathcal{M})=\mathit{atoms}(\mathcal{C})$.
\end{corollary}
\begin{proof}
An immediate consequence of Lemmata~\ref{lemma:minimal-are-parallel} and \ref{lemma:replace-with-minimal}.
\end{proof}

\subsection{Generating the implied $4$-cuts}
\label{section:generating-collections}

Let $\mathcal{C}$ be a collection of $4$-cuts of $G$. Our goal is to construct a linear-space representation of all $4$-cuts implied by $\mathcal{C}$, that will be convenient in order to essentially process all of them simultaneously and derive $\mathit{atoms}(\mathcal{C})$. We will show how to do this in linear time, by constructing a set $\{\mathcal{F}_1,\dots,\mathcal{F}_k\}$ of collections of pairs of edges, each one of which generates a collection of $4$-cuts implied by $\mathcal{C}$, with the property that every $4$-cut implied by $\mathcal{C}$ is generated by at least one of those collections.

Intuitively speaking, the idea is to partition every $4$-cut $C\in\mathcal{C}$ into pairs of pairs of edges in all possible ways, in order to (implicitly) trace all the implicating sequences of $\mathcal{C}$ that use $C$. Thus, if $C=\{e_1,e_2,e_3,e_4\}$, then we produce the three partitions $\{\{e_1,e_2\},\{e_3,e_4\}\}$, $\{\{e_1,e_3\},\{e_2,e_4\}\}$ and $\{\{e_1,e_4\},\{e_2,e_3\}\}$ of $C$ into pairs of edges. Every one of those partitions has the potential to participate in an implicating sequence for a $4$-cut. For example, if there is a $4$-cut $C'=\{e_3,e_4,e_5,e_6\}\in\mathcal{C}$ with $C'\neq C$, then $\{e_1,e_2,e_5,e_6\}$ is a $4$-cut implied by $C$ and $C'$, and in order to derive this implication we have to (conceptually) partition $C$ into $\{\{e_1,e_2\},\{e_3,e_4\}\}$ and $C'$ into $\{\{e_3,e_4\},\{e_5,e_6\}\}$. Thus, $\mathcal{F}=\{\{e_1,e_2\},\{e_3,e_4\},\{e_5,e_6\}\}$ is a collection of pairs of edges that generates a collection of $4$-cuts implied by $\mathcal{C}$. Now we would like to extend this collection as much as possible, by considering the partition of another $4$-cut from $\mathcal{C}$ into pairs of edges that includes one of the pairs of edges in $\mathcal{F}$. This is easy to do if we have broken every $4$-cut from $\mathcal{C}$ into all its possible bipartitions of pairs of edges. 

In order to implement this idea, 
for every $4$-cut $C=\{e_1,e_2,e_3,e_4\}\in\mathcal{C}$, we produce six elements $(C,\{e_1,e_2\})$, $(C,\{e_1,e_3\})$, $(C,\{e_1,e_4\})$, $(C,\{e_2,e_3\})$, $(C,\{e_2,e_4\})$ and $(C,\{e_3,e_4\})$. 
Then we introduce three artificial (undirected) edges $\{(C,\{e_1,e_2\}),(C,\{e_3,e_4\})\}$, $\{(C,\{e_1,e_3\}),(C,\{e_2,e_4\})\}$ and $\{(C,\{e_1,e_4\}),(C,\{e_2,e_3\})\}$. The purpose of those edges is to maintain the information that their endpoints correspond to a specific partition of $C$ into pairs of edges. Now suppose that there are two $4$-cuts $C,C'\in\mathcal{C}$ that participate in an implicating sequence as consecutive $4$-cuts. This means that there are two elements of the form $(C,\{e,e'\})$ and $(C',\{e,e'\})$. Then we would like to have those elements connected with a new artificial edge, in order to maintain the information that the $4$-cuts $C,C'$ intersect in the pair $\{e,e'\}$. However, it would be inefficient to introduce such an edge in all those cases, because we would need $\Omega(|\mathcal{C}|^2)$ time in the worst case scenario. Instead, if $C_1,\dots,C_k$ are all the $4$-cuts from $\mathcal{C}$ that contain the pair of edges $\{e,e'\}$, then we ensure that we have all elements $(C_1,\{e,e'\}),\dots,(C_k,\{e,e'\})$ in a sequence, and then we connect every consecutive pair of elements in this sequence with a new artificial edge. In total, this results in an undirected graph $\mathcal{G}$, that basically represents all the partitions of the $4$-cuts from $\mathcal{C}$ into two pairs of edges, and all intersections of the $4$-cuts from $\mathcal{C}$ in a pair of edges. Then we can efficiently derive this information if we simply compute the connected components of this graph. We can prove that the connected components of $\mathcal{G}$ correspond to collections of pairs of edges that generate collections of $4$-cuts implied by $\mathcal{C}$ (see Proposition~\ref{proposition:cyclic_families_algorithm}). 

The implementation of this idea is shown in Algorithm~\ref{algorithm:generatefamilies}. We use a total ordering of the edges of $G$ (e.g., lexicographic order), so that the order of edges in a pair of edges is fixed. This is needed because, if $C$ and $C'$ are two distinct $4$-cuts that contain a pair of edges $\{e,e'\}$, then we would like to have the elements $(C,\{e,e'\})$ and $(C',\{e,e'\})$ (that are generated internally by the algorithm) in a maximal sequence of elements of this form. Thus, the order of $e$ and $e'$ should be fixed, so that the tuples $(C,\{e,e'\})$ and $(C',\{e,e'\})$ can be recognized as having the same second component. If $p$ is a pair of edges, then we let $\vec{p}$ denote the corresponding ordered pair of edges that respects the total ordering of $E(G)$. Whenever $(e,e')$ denotes an ordered pair of edges, we assume that this order respects the total ordering of $E(G)$. 

The remainder of this section is devoted to an exploration of the properties of the output of Algorithm~\ref{algorithm:generatefamilies}.

\begin{algorithm}[h!]
\caption{\textsf{Return a set of collections of pairs of edges that generate in total all the $4$-cuts that are implied by a collection of $4$-cuts $\mathcal{C}$}}
\label{algorithm:generatefamilies}
\LinesNumbered
\DontPrintSemicolon
\SetKwInOut{Input}{input}
\SetKwInOut{Output}{output}
\Input{a collection $\mathcal{C}$ of $4$-cuts of $G$}
\Output{a set $\mathcal{F}_1,\dots,\mathcal{F}_k$ of collections of pairs of edges that generate collections of $4$-cuts of $G$ that contain in total all the $4$-cuts implied by $\mathcal{C}$}
Let $\mathit{P}\leftarrow\emptyset$, $\mathit{J}\leftarrow\emptyset$\;
\ForEach{$C=\{e_1,e_2,e_3,e_4\}\in\mathcal{C}$}{
  let $p_1\leftarrow \{e_1,e_2\}$, $p_2\leftarrow\{e_3,e_4\}$, $p_3\leftarrow\{e_1,e_3\}$, $p_4\leftarrow\{e_2,e_4\}$, 
    $p_5\leftarrow\{e_1,e_4\}$, $p_6\leftarrow\{e_2,e_3\}$\;
  generate the elements $(C,\vec{p}_1)$, $(C,\vec{p}_2)$, $(C,\vec{p}_3)$, $(C,\vec{p}_4)$, $(C,\vec{p}_5)$, $(C,\vec{p}_6)$\;
  add those elements to $\mathit{P}$\;
  add to $\mathit{J}$ the edges $\{(C,\vec{p}_1),(C,\vec{p}_2)\}$, $\{(C,\vec{p}_3),(C,\vec{p}_4)\}$, $\{(C,\vec{p}_5),(C,\vec{p}_6)\}$\; 
  \label{line:generatefamilies:edges1}
}
sort the elements of $\mathit{P}$ lexicographically w.r.t. their second component\;
\label{line:generatefamilies:sorting}
\ForEach{pair of consecutive elements $(C,p),(C',p)$ of $\mathit{P}$ with the same second component}{
  add to $\mathit{J}$ the edge $\{(C,p),(C',p)\}$\;
  \label{line:generatefamilies:addedges}
}
compute the connected components $S_1,\dots,S_k$ of the graph $\mathcal{G}=(\mathit{P},\mathit{J})$\;
\label{line:generatefamilies:components}
\ForEach{$i\in\{1,\dots,k\}$}{
  $\mathcal{F}_i \leftarrow \{\{e,e'\}\mid\exists (C,(e,e'))\in S_i\}$ \tcp{consider $\mathcal{F}_i$ as a simple set}
  \label{line:generatefamilies:final}
}
\textbf{return} $\mathcal{F}_1,\dots,\mathcal{F}_k$\;
\end{algorithm}

\begin{proposition}
\label{proposition:cyclic_families_algorithm}
Let $\mathcal{C}$ be a collection of $4$-cuts of a graph $G$, and let $\mathcal{F}_1,\dots,\mathcal{F}_k$ be the output of Algorithm~\ref{algorithm:generatefamilies} on input $\mathcal{C}$. Then every $\mathcal{F}_i$ is a collection of pairs of edges that generates a collection of $4$-cuts of $G$ that are implied by $\mathcal{C}$. Conversely, for every $4$-cut $C$ implied by $\mathcal{C}$, there is at least one $i\in\{1,\dots,k\}$ such that $C$ belongs to the collection of $4$-cuts generated by $\mathcal{F}_i$. The running time of Algorithm~\ref{algorithm:generatefamilies} is $O(n+|\mathcal{C}|)$, where $n=|V(G)|$. The output of Algorithm~\ref{algorithm:generatefamilies} has size $O(|\mathcal{C}|)$ (i.e., $O(|\mathcal{F}_1|+\dots+|\mathcal{F}_k|)=O(|\mathcal{C}|)$). 
\end{proposition}
\begin{proof}
Let $i\in\{1,\dots,k\}$. We will show that $\mathcal{F}_i$ generates a collection of $4$-cuts implied by $\mathcal{C}$. Let $\mathcal{G}$ be the graph that is generated internally by the algorithm in Line~\ref{line:generatefamilies:components}, and let $S_i$ be the connected component of $\mathcal{G}$ from which $\mathcal{F}_i$ is derived in Line~\ref{line:generatefamilies:final}. First we will show that $|\mathcal{F}_i|\geq 2$. Let $(C,(e,e'))$ be an element in $S_i$. Then $C$ is a $4$-cut in $\mathcal{C}$, and let $\{e'',e'''\}=C\setminus\{e,e'\}$. Due to the construction of $\mathcal{G}$, we have w.l.o.g., (i.e., by possibly changing the order of the edges), that $(C,(e'',e'''))$ is a vertex of $\mathcal{G}$. Then, there is an edge in $\mathcal{G}$ with endpoints $(C,(e,e'))$  and $(C,(e'',e'''))$ (see Line~\ref{line:generatefamilies:edges1}). This implies that $(C,(e,e'))$ and $(C,(e'',e'''))$ belong to the same connected component of $\mathcal{G}$, and therefore we have $(C,(e'',e'''))\in S_i$. This shows that $\{\{e,e'\},\{e'',e'''\}\}\subseteq\mathcal{F}$. Since $C$ is a $4$-element set, we have $\{e,e'\}\neq\{e'',e'''\}$. This shows that $|\mathcal{F}_i|\geq 2$.


Now let $p$ and $q$ be two distinct pairs of edges that are contained in $\mathcal{F}_i$. Then there are $4$-cuts $C$ and $C'$ in $\mathcal{C}$ such that there are elements $(C,\vec{p})$ and $(C',\vec{q})$ that are contained in $S_i$. 
Then, since $(C,\vec{p})$ and $(C',\vec{q})$ are in the same connected component of $\mathcal{G}$, there is a path from $(C,\vec{p})$ to $(C',\vec{q})$ in $\mathcal{G}$ that passes from distinct vertices. This implies that there is sequence of pairs of edges $p_1,\dots,p_N$ of $G$, with $N\geq 2$, and a sequence $C_1,\dots,C_N$ of $4$-cuts from $\mathcal{C}$, such that $(C_1,\vec{p}_1)=(C,\vec{p})$, $(C_N,\vec{p}_N)=(C',\vec{q})$, and for every $i\in\{1,\dots,N-1\}$ there is an edge of $\mathcal{G}$ with endpoints $(C_i,\vec{p}_i)$ and $(C_{i+1},\vec{p}_{i+1})$. Since the edges of $\mathcal{G}$ are generated in Lines~\ref{line:generatefamilies:edges1} and \ref{line:generatefamilies:addedges}, for every $i\in\{1,\dots,N-1\}$ we have that either $C_i=C_{i+1}$ and $C_i=p_i\cup p_{i+1}$, or $C_i\cap C_{i+1}=p_i=p_{i+1}$ $(*)$. 

Now we define a sequence of indexes $t(1),t(2),\dots,t(N')$, for some $N'\leq N$, as follows. First, we let $t(1)$ be the maximum index $i\geq 1$ such that $p_1=p_2=\dots=p_i$. Now suppose that we have defined $t(i)$, for some $i\geq 1$, and $p_{t(i)}\neq p_N$. Then we let $t(i+1)$ be the maximum index $j>t(i)$ such that $p_{t(i)+1}=p_{t(i)+2}=\dots=p_j$. This construction is terminated when we reach the first $N'$ such that $p_{t(N')}=p_N$.
Notice that $N'\geq 2$, since $p_1\neq p_N$.
By construction, we have $p_{t(i)+1}\neq p_{t(i)}$ and $p_{t(i+1)}=p_{t(i)+1}$, for every $i\in\{1,\dots,N'-1\}$. Thus, for every $i\in\{1,\dots,N'-1\}$, by $(*)$ we have $C_{t(i)}=p_{t(i)}\cup p_{t(i)+1}$, and therefore $C_{t(i)}=p_{t(i)}\cup p_{t(i+1)}$. Thus, $C_{t(1)},\dots,C_{t(N'-1)}$ is an implicating sequence of $\mathcal{C}$. If $N'=2$, then we have $C_{t(1)}=p_{t(1)}\cup p_{t(2)}=p_1\cup p_N=p\cup q$. Thus, $p\cup q\in\mathcal{C}$. Otherwise, Lemma~\ref{lemma:implicating-sequence} implies that $p_{t(1)}\cup p_{t(N')}$ is a $4$-cut implied by $\mathcal{C}$, and therefore $p_1\cup p_N=p\cup q$ is a $4$-cut implied by $\mathcal{C}$. In any case then, we have that $p\cup q$ is a $4$-cut implied by $\mathcal{C}$.


Conversely, let $C$ be a $4$-cut implied by $\mathcal{C}$. This means that there is a sequence $p_1,\dots,p_{k+1}$ of pairs of edges of $G$, and a sequence $C_1,\dots,C_k$ of $4$-cuts in $\mathcal{C}$, with $k\geq 1$, such that $C=p_1\cup p_{k+1}$, and $C_i=p_i\cup p_{i+1}$ for every $i\in\{1,\dots,k\}$. Then, for every $i\in\{1,\dots,k\}$, there is an edge of $\mathcal{G}$ with endpoints $(C_i,\vec{p}_i)$ and $(C_i,\vec{p}_{i+1})$ (see Line~\ref{line:generatefamilies:edges1}). Furthermore, for every $i\in\{1,\dots,k-1\}$, there is path from $(C_i,\vec{p}_{i+1})$ to $(C_{i+1},\vec{p}_{i+1})$ in $\mathcal{G}$ (due to the existence of the edges in Line~\ref{line:generatefamilies:addedges}). Thus, all pairs of the form $(C_i,\vec{p}_i)$, for $i\in\{1,\dots,k\}$, belong to the same connected component $S$ of $\mathcal{G}$. Furthermore, $(C_k,\vec{p}_{k+1})$ also belongs to $S$, due to the existence of the edge with endpoints $(C_k,\vec{p}_k)$ and $(C_k,\vec{p}_{k+1})$ (see Line~\ref{line:generatefamilies:edges1}). Thus, there is a collection of pairs of edges $\mathcal{F}$ that is returned by Algorithm~\ref{algorithm:generatefamilies} on input $\mathcal{C}$ such that $\{p_1,\dots,p_{k+1}\}\subseteq\mathcal{F}$ (see Line~\ref{line:generatefamilies:final}). Then we have that $C=p_1\cup p_{k+1}$ is generated by $\mathcal{F}$.

We can easily see that Algorithm~\ref{algorithm:generatefamilies} runs in $O(n+|\mathcal{C}|)$ time. For every $C\in\mathcal{C}$, we generate six elements of $O(1)$ size, and three edges of $O(1)$ size. Line~\ref{line:generatefamilies:sorting} takes $O(n+|\mathcal{C}|)$ time if implemented with bucket sort, since the components of the tuples that we sort are edges of the graph, and so their endpoints lie in the range $\{1,\dots,n\}$. Line~\ref{line:generatefamilies:addedges} adds $O(|\mathcal{C}|)$ edges of $O(1)$ size. The computation of the connected components in Line~\ref{line:generatefamilies:components} takes $O(|V(\mathcal{G})|+|E(\mathcal{G})|)=O(|\mathcal{C}|)$ time, and Line~\ref{line:generatefamilies:final} takes $O(|V(\mathcal{G})|)=O(|\mathcal{C}|)$ time. Thus, the running time of Algorithm~\ref{algorithm:generatefamilies} is $O(n+|\mathcal{C}|)$. Finally, for every $i\in\{1,\dots,k\}$, let $S_i$ be the connected component of $\mathcal{G}$ from which $\mathcal{F}_i$ is derived (in Line~\ref{line:generatefamilies:final}). Then we have $O(|\mathcal{F}_1|+\dots+|\mathcal{F}_k|)=O(|S_1|+\dots+|S_k|)=O(|V(\mathcal{G})|=O(|\mathcal{C}|)$. The second equality is due to the fact that $S_1,\dots,S_k$ are the connected components of $\mathcal{G}$.
\end{proof}


\begin{lemma}
\label{lemma:returned_is_disjoint}
Let $\mathcal{C}$ be a collection of $4$-cuts of a graph $G$, and let $\mathcal{F}$ and $\mathcal{F}'$ be two distinct collections of pairs of edges that are returned by Algorithm~\ref{algorithm:generatefamilies} on input $\mathcal{C}$. Then $\mathcal{F}\cap\mathcal{F}'=\emptyset$.
\end{lemma}
\begin{proof}
Let $\mathcal{F}$ and $\mathcal{F}'$ be two collections of pairs of edges returned by Algorithm~\ref{algorithm:generatefamilies} on input $\mathcal{C}$ with $\mathcal{F}\neq\mathcal{F}'$. Let $S$ and $S'$ be the connected components of the graph $\mathcal{G}$ generated in Line~\ref{line:generatefamilies:components}, from which $\mathcal{F}$ and $\mathcal{F}'$, respectively, are derived in Line~\ref{line:generatefamilies:final}. Let us assume, for the sake of contradiction, that there is a pair of edges $\{e,e'\}\in\mathcal{F}\cap\mathcal{F}'$. Then, w.l.o.g., (i.e., by possibly changing the order of the edges), there are elements $(C,(e,e'))\in S$ and $(C',(e,e'))\in S'$, such that $C$ and $C'$ are $4$-cuts in $\mathcal{C}$. But then, due to the existence of the edges in Line~\ref{line:generatefamilies:addedges}, we have that $(C,(e,e'))$ is connected with $(C',(e,e'))$ in $\mathcal{G}$. This implies that $S=S'$, which further implies that $\mathcal{F}=\mathcal{F}'$, a contradiction. We conclude that $\mathcal{F}\cap\mathcal{F}'=\emptyset$.
\end{proof}

\begin{lemma}
\label{lemma:not-included-4cut}
Let $\mathcal{C}$ be a collection of $4$-cuts of a graph $G$, and let $C=\{e_1,e_2,e_3,e_4\}$ be a $4$-cut of $G$ that is implied by $\mathcal{C}$ through the pair of edges $\{e_1,e_2\}$. Then, in the output of Algorithm~\ref{algorithm:generatefamilies} on input $\mathcal{C}$, there is a collection $\mathcal{F}$ of pairs of edges such that $\{\{e_1,e_2\},\{e_3,e_4\}\}\subseteq\mathcal{F}$. Furthermore, if $C\notin\mathcal{C}$, then this inclusion is proper (i.e., $|\mathcal{F}|>2$).
\end{lemma}
\begin{proof}
We may assume w.l.o.g. that $e_1<e_2$ and $e_3<e_4$.
Suppose first that $C\in\mathcal{C}$. Let $\mathcal{G}$ be the graph generated by Algorithm~\ref{algorithm:generatefamilies} on input $\mathcal{C}$ in Line~\ref{line:generatefamilies:components}. Then, the elements $(C,(e_1,e_2))$ and $(C,(e_3,e_4))$ are vertices of $\mathcal{G}$ that are connected with an edge (due to Line~\ref{line:generatefamilies:edges1}). Thus, let $S$ be the connected component of $\mathcal{G}$ that contains $(C,(e_1,e_2))$ and $(C,(e_3,e_4))$, and let $\mathcal{F}$ be the collection of pairs of edges that is derived from $S$ in Line~\ref{line:generatefamilies:final}. Then, we have $\{\{e_1,e_2\},\{e_3,e_4\}\}\subseteq\mathcal{F}$.

Now let us suppose that $C\notin\mathcal{C}$. Since $C$ is implied by $\mathcal{C}$ through the pair of edges $\{e_1,e_2\}$, there is a sequence $p_1,\dots,p_{k+1}$ of pairs of edges, and a sequence $C_1,\dots,C_k$ of $4$-cuts from $\mathcal{C}$, such that $p_1=\{e_1,e_2\}$, $p_{k+1}=\{e_3,e_4\}$, and $C_i=p_i\cup p_{i+1}$ for every $i\in\{1,\dots,k\}$. Then, for every $i\in\{1,\dots,k-1\}$, we have that either $C_i=C_{i+1}$ or $C_i\cap C_{i+1}=p_{i+1}$. 
Now let $i\in\{1,\dots,k\}$ be an index. Since $C_i=p_i\cup p_{i+1}$, we have that $(C_i,\vec{p}_i)$ and $(C_i,\vec{p}_{i+1})$ are the endpoints of an edge of $\mathcal{G}$ (see Line~\ref{line:generatefamilies:edges1}). Now let $i<k$. If $C_i=C_{i+1}$, then we have $(C_i,\vec{p}_{i+1})=(C_{i+1},\vec{p}_{i+1})$. Otherwise, we have $C_i\cap C_{i+1}=p_{i+1}$, and therefore the elements $(C_i,\vec{p}_{i+1})$ and $(C_{i+1},\vec{p}_{i+1})$ are vertices in the same connected component of $\mathcal{G}$ (due to the existence of the edges in Line~\ref{line:generatefamilies:addedges}). This shows that the vertices $(C_i,\vec{p}_i)$, for $i\in\{1,\dots,k-1\}$, are in the same connected component $S$ of $\mathcal{G}$. Furthermore, since there is an edge of $\mathcal{G}$ with endpoints $(C_k,\vec{p}_k)$ and $(C_k,\vec{p}_{k+1})$, we have that $(C_k,\vec{p}_{k+1})$ is also in $S$. Now let $\mathcal{F}$ be the collection of pairs of edges that is derived from $S$ in Line~\ref{line:generatefamilies:final}. Then, we have $\{p_1,\dots,p_{k+1}\}\subseteq\mathcal{F}$. Since $C\notin\mathcal{C}$ and $p_1\cup p_2\in\mathcal{C}$, we have $C\neq p_1\cup p_2$. Thus, since $C=p_1\cup p_{k+1}$, we have $p_2\neq p_{k+1}$. Finally, since $p_1\cup p_2$ and $p_1\cup p_{k+1}$ are $4$-cuts, we have $p_1\neq p_2$ and $p_1\neq p_{k+1}$. Thus, we have that $p_1,p_2,p_{k+1}$ are three distinct pairs of edges contained in $\mathcal{F}$.
\end{proof}

\begin{lemma}
\label{lemma:2-pair-collection}
Let $\mathcal{C}$ be a collection of $4$-cuts of $G$, and let $\mathcal{F}$ be a collection of pairs of edges that is returned by Algorithm~\ref{algorithm:generatefamilies} on input $\mathcal{C}$. Suppose that $|\mathcal{F}|=2$. Then, $\bigcup{\mathcal{F}}\in\mathcal{C}$.
\end{lemma}
\begin{proof}
Let $\mathcal{F}=\{p,p'\}$. By Proposition~\ref{proposition:cyclic_families_algorithm}, we have that $p\cup p'$ is a $4$-cut implied by $\mathcal{C}$. Let $\mathcal{G}$ be the graph that is generated internally by the algorithm in Line~\ref{line:generatefamilies:components}, and let $S$ be the connected component of $\mathcal{G}$ from which $\mathcal{F}$ is derived in Line~\ref{line:generatefamilies:final}. Then there are $4$-cuts $C$ and $C'$ in $\mathcal{C}$ such that the tuples $(C,\vec{p})$ and $(C',\vec{p'})$ are in $S$, and we have $p\subset C$ and $p'\subset C'$. If $C=C'$, then we obviously have $p\cup p' \in\mathcal{C}$. So let us assume that $C\neq C'$. 

Since the tuples $(C,\vec{p})$ and $(C',\vec{p'})$ are in the same connected component of $\mathcal{G}$, this means that there is a path $P$ from $(C,\vec{p})$ to $(C',\vec{p'})$ in $\mathcal{G}$ that passes from distinct vertices. Let $(C'',\vec{p''})$ be the second vertex on $P$ (where $C''\in\mathcal{C}$ and $p''$ is a pair of edges in $C''$). Since the edges of $\mathcal{G}$ are generated in Lines~\ref{line:generatefamilies:edges1} and \ref{line:generatefamilies:addedges}, and since $p\neq p'$ and $C\neq C'$, we have that $(1)$ either $C''=C$ and $p''\neq p$, or $(2)$ $C''=C'$ and $p''\neq p'$, or $(3)$ $C''\neq C$ and $p''=p$, or $(4)$ $C''\neq C'$ and $p''=p'$. Notice that, between $(1)$ and $(2)$, we may assume w.l.o.g. $(1)$. Also, between $(3)$ and $(4)$, we may assume w.l.o.g. $(3)$. 

First, let us assume that $(1)$ is true. Due to the existence of $P$, we have that $(C'',\vec{p''})$ lies in $S$. This implies that $p''\in\mathcal{F}$. Therefore, since $p''\neq p$ and $\mathcal{F}=\{p,p'\}$, we have $p''=p'$. Let us suppose, for the sake of contradiction, that $p\cup p'\neq C$. Since $(C,\vec{p})$ and $(C'',\vec{p'})=(C,\vec{p'})$ are vertices of $\mathcal{G}$, we have $p\subset C$ and $p'\subset C$. Then, since $p\neq p'$ and $p\cup p'\neq C$, we have $|C\cap(p\cup p')|=3$. But this contradicts Lemma~\ref{lemma:no-common-three-edges}. This shows that $C=p\cup p'$, and therefore $p\cup p'$ is a $4$-cut in $\mathcal{C}$.

Now let us assume that $(3)$ is true. Due to the existence of $P$, we have that $(C'',\vec{p''})$ lies in $S$. Let $q=C''\setminus p''$. Then there is an edge of $\mathcal{G}$ with endpoints $(C'',\vec{p''})$ and $(C'',\vec{q})$ (see Line~\ref{line:generatefamilies:edges1}). Thus, $(C'',\vec{q})$ also lies in $S$. This implies that $q\in\mathcal{F}$. Thus, since $q\neq p''=p$, we have $q=p'$. Therefore, we have that $C''=p''\cup q=p\cup p'$ is a $4$-cut in $\mathcal{C}$. 
\end{proof}



\subsection{Isolated and quasi-isolated $4$-cuts}
\label{section:iso-and-quasi-iso}
Let $\mathcal{C}$ be a collection of $4$-cuts of $G$, and let $C$ be a $4$-cut implied by $\mathcal{C}$. Then, it may be that there is no collection $\mathcal{F}$ of pairs of edges with $|\mathcal{F}|>2$ that generates a collection of $4$-cuts implied by $\mathcal{C}$ that includes $C$. In this case, we call $C$ a \emph{$\mathcal{C}$-isolated} $4$-cut. Notice that, if $\mathcal{C}$ is a complete collection of $4$-cuts of $G$ and $C$ is a $\mathcal{C}$-isolated $4$-cut, then $C$ is an isolated $4$-cut. 

The following lemma provides a necessary condition that must be satisfied by a $\mathcal{C}$-isolated $4$-cut.

\begin{lemma}
\label{lemma:isolated_necessary}
Let $\mathcal{C}$ be a collection of $4$-cuts of $G$, and let $C=\{e_1,e_2,e_3,e_4\}$ be a $\mathcal{C}$-isolated $4$-cut. Then we have $C\in\mathcal{C}$, and the $2$-element collections of pairs of edges $\{\{e_1,e_2\},\{e_3,e_4\}\}$, $\{\{e_1,e_3\},\{e_2,e_4\}\}$ and $\{\{e_1,e_4\},\{e_2,e_3\}\}$, are part of the output of Algorithm~\ref{algorithm:generatefamilies} on input $\mathcal{C}$.  
\end{lemma}
\begin{proof}
Let us suppose, for the sake of contradiction, that $C\notin\mathcal{C}$. Since $C$ is a $\mathcal{C}$-isolated $4$-cut, we have that $\mathcal{C}$ implies $C$. Thus, we may assume w.l.o.g. that $\mathcal{C}$ implies $C$ through the pair of edges $\{e_1,e_2\}$. Then, Lemma~\ref{lemma:not-included-4cut} implies that there is a collection $\mathcal{F}$ of pairs of edges that is returned by Algorithm~\ref{algorithm:generatefamilies}  on input $\mathcal{C}$ such that $\{\{e_1,e_2\},\{e_3,e_4\}\}\subset\mathcal{F}$. Proposition~\ref{proposition:cyclic_families_algorithm} implies that $\mathcal{F}$ generates a collection $\mathcal{C}'$ of $4$-cuts implied by $\mathcal{C}$. Thus, since $|\mathcal{F}|>2$ and $C\in\mathcal{C}'$, we have a contradiction to the fact that $C$ is $\mathcal{C}$-isolated $4$-cut. This shows that $C\in\mathcal{C}$.

Now, since $C\in\mathcal{C}$, we have that $\mathcal{C}$ trivially implies $C$ through the pair of edges $\{e_1,e_2\}$. Thus, Lemma~\ref{lemma:not-included-4cut} implies that there is a collection $\mathcal{F}$ of pairs of edges that is returned by Algorithm~\ref{algorithm:generatefamilies}  on input $\mathcal{C}$ such that $\{\{e_1,e_2\},\{e_3,e_4\}\}\subseteq\mathcal{F}$. Proposition~\ref{proposition:cyclic_families_algorithm} implies that $\mathcal{F}$ generates a collection $\mathcal{C}'$ of $4$-cuts implied by $\mathcal{C}$. Thus, since $C\in\mathcal{C}'$, we have that $|\mathcal{F}|=2$, because $C$ is $\mathcal{C}$-isolated. Similarly, since $\mathcal{C}$ implies $C$ through the pairs of edges $\{e_1,e_3\}$ and $\{e_1,e_4\}$, we have that the $2$-element collections of pairs of edges $\{\{e_1,e_3\},\{e_2,e_4\}\}$ and $\{\{e_1,e_4\},\{e_2,e_3\}\}$ are part of the output of Algorithm~\ref{algorithm:generatefamilies} on input $\mathcal{C}$.
\end{proof}

We note that the condition provided by Lemma~\ref{lemma:isolated_necessary} is only necessary, but not sufficient. In other words, it may be that there is a $4$-cut $C=\{e_1,e_2,e_3,e_4\}$ such that the collections of pairs of edges $\{\{e_1,e_2\},\{e_3,e_4\}\}$, $\{\{e_1,e_3\},\{e_2,e_4\}\}$ and $\{\{e_1,e_4\},\{e_2,e_3\}\}$, are part of the output of Algorithm~\ref{algorithm:generatefamilies} on input $\mathcal{C}$, but $C$ is not a $\mathcal{C}$-isolated $4$-cut. In this case, we call $C$ a \emph{quasi $\mathcal{C}$-isolated} $4$-cut.

\begin{corollary}
\label{corollary:quasi-isolated_in_C}
Let $\mathcal{C}$ be a collection of $4$-cuts of $G$, and let $C$ be a quasi $\mathcal{C}$-isolated $4$-cut. Then $C\in\mathcal{C}$.
\end{corollary}
\begin{proof}
Let $C=\{e_1,e_2,e_3,e_4\}$. Since $C$ is quasi $\mathcal{C}$-isolated, we have that the three collections of pairs of edges $\{\{e_1,e_2\},\{e_3,e_4\}\}$, $\{\{e_1,e_3\},\{e_2,e_4\}\}$ and $\{\{e_1,e_4\},\{e_2,e_3\}\}$, are part of the output of Algorithm~\ref{algorithm:generatefamilies} on input $\mathcal{C}$. Thus, Lemma~\ref{lemma:2-pair-collection} implies that $C\in\mathcal{C}$. 
\end{proof}

The following lemma, which concerns essential quasi $\mathcal{C}$-isolated $4$-cuts, will be very useful in computing all the essential $\mathcal{C}$-isolated $4$-cuts, because it provides a criterion with which we can distinguish the $\mathcal{C}$-isolated $4$-cuts from the quasi $\mathcal{C}$-isolated $4$-cuts (see Corollary~\ref{corollary:quasi-iso-pair}). 

\begin{figure}[t!]\centering
\includegraphics[trim={0 21cm 0 0}, clip=true, width=0.9\linewidth]{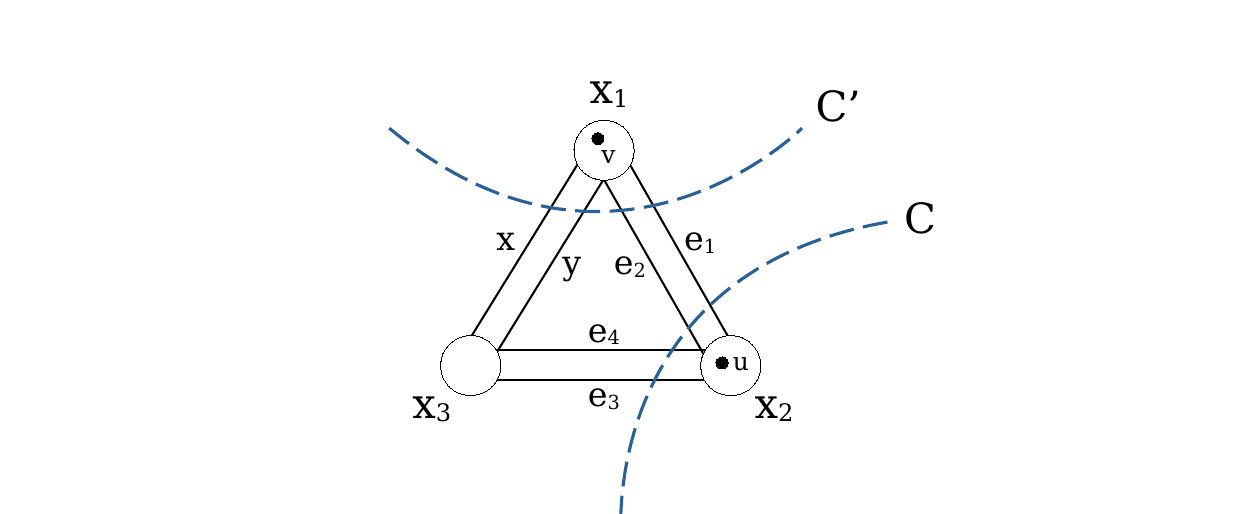}
\caption{\small{A depiction of the situation analyzed in Lemma~\ref{lemma:quasi-iso-pair}.}}\label{figure:quasi_triangle}
\end{figure}

\begin{lemma}[An essential quasi-isolated $4$-cut shares a pair of edges with a minimal $4$-cut]
\label{lemma:quasi-iso-pair}
Let $\mathcal{C}$ be a collection of $4$-cuts of $G$, and let $C$ be an essential quasi $\mathcal{C}$-isolated $4$-cut. Then, there is a pair of edges $p= C\cap C'$, where $C'$ is an essential $\mathcal{C}'$-minimal $4$-cut, where $\mathcal{C}'$ is a cyclic family of $4$-cuts that is generated by a collection $\mathcal{F}'$ of pairs of edges with $|\mathcal{F}'|\geq 3$ that is returned by Algorithm~\ref{algorithm:generatefamilies} on input $\mathcal{C}$.
\end{lemma}
\begin{proof}
Let $C=\{e_1,e_2,e_3,e_4\}$.
Since $C$ is a quasi $\mathcal{C}$-isolated $4$-cut, we have that $C$ is not $\mathcal{C}$-isolated. This means that there is a collection $\mathcal{F}$ of pairs of edges with $|\mathcal{F}|>2$ that generates a collection $\widetilde{\mathcal{C}}$ of $4$-cuts that are implied by $\mathcal{C}$ such that $C\in\widetilde{\mathcal{C}}$. Thus, we may assume w.l.o.g. that $\{\{e_1,e_2\},\{e_3,e_4\},\{x,y\}\}\subseteq\mathcal{F}$, where $\{x,y\}$ is a pair of edges with $\{x,y\}\notin\{\{e_1,e_2\},\{e_3,e_4\}\}$. Since $C$ is an essential $4$-cut, by Corollary~\ref{corollary:essential_implies_cyclic} we have that $\widetilde{\mathcal{C}}$ is a cyclic family of $4$-cuts. This implies that there is a partition $\{X_1,X_2,X_3\}$ of $V(G)$ such that the subgraphs $G[X_1]$, $G[X_2]$ and $G[X_3]$ are connected, and $E[X_1,X_2]=\{e_1,e_2\}$, $E[X_2,X_3]=\{e_3,e_4\}$, $E[X_3,X_1]=\{x,y\}$. (See Figure~\ref{figure:quasi_triangle}.) 
Since $C$ is an essential $4$-cut and the connected components of $G\setminus C$ are $X_2$ and $X_1\cup X_3$, we have that there is a pair $u,v$ of $4$-edge-connected vertices such that $u\in X_2$ and $v\in X_1\cup X_3$. We may assume w.l.o.g. that $v\in X_1$. 

Let $C'=\{e_1,e_2,x,y\}$. Then we have that $C'$ is a $4$-cut implied by $\mathcal{C}$. Let us suppose, for the sake of contradiction, that $\mathcal{C}$ implies $C'$ through the pair of edges $\{e_1,e_2\}$. Then, Lemma~\ref{lemma:not-included-4cut} implies that there is a collection $\mathcal{F}'$ of pairs of edges that is returned by Algorithm~\ref{algorithm:generatefamilies} on input $\mathcal{C}$ such that $\{\{e_1,e_2\},\{x,y\}\}\subseteq\mathcal{F}'$. Since $C$ is a quasi $\mathcal{C}$-isolated $4$-cut, by definition we have that the collection of pairs of edges $\mathcal{F}''=\{\{e_1,e_2\},\{e_3,e_4\}\}$ is returned by Algorithm~\ref{algorithm:generatefamilies} on input $\mathcal{C}$. Since $\{x,y\}\neq\{e_3,e_4\}$, we have that $\mathcal{F}'\neq\mathcal{F}''$. Therefore, Lemma~\ref{lemma:returned_is_disjoint} implies that $\mathcal{F}'\cap\mathcal{F}''=\emptyset$, in contradiction to the fact that $\{e_1,e_2\}\in\mathcal{F}'\cap\mathcal{F}''$. Thus, we have that $\mathcal{C}$ does not imply $C'$ through the pair of edges $\{e_1,e_2\}$. Thus, we may assume w.l.o.g. that $\mathcal{C}$ implies $C'$ through the pair of edges $\{e_1,x\}$. 

Notice that, since $\mathcal{C}$ does not imply $C'$ through the pair of edges $\{e_1,e_2\}\subset C'$, we have that $C'\notin\mathcal{C}$. Thus, since $\mathcal{C}$ implies $C'$ through the pair of edges $\{e_1,x\}$, by Lemma~\ref{lemma:not-included-4cut} we have that there is a collection $\mathcal{F}'$ of pairs of edges that is returned by Algorithm~\ref{algorithm:generatefamilies} on input $\mathcal{C}$ such that $\{\{e_1,x\},\{e_2,y\}\}\subset\mathcal{F}'$. By Proposition~\ref{proposition:cyclic_families_algorithm}, we have that $\mathcal{F}'$ generates a collection $\mathcal{C}'$ of $4$-cuts that are implied by $\mathcal{C}$. We have that the connected components of $G\setminus C'$ are $X_1$ and $X_2\cup X_3$. Since $u\in X_2$ and $v\in X_1$, and $u,v$ are $4$-edge-connected vertices, this implies that $C'$ is an essential $4$-cut. Thus, since $|\mathcal{F}'|\geq 3$, by Corollary~\ref{corollary:essential_implies_cyclic} we have that $\mathcal{C}'$ is a cyclic family of $4$-cuts. We will prove that $X_2\cup X_3$ is a corner of $\mathcal{C}'$, and therefore $C'$ is $\mathcal{C}'$-minimal $4$-cut.

So let us suppose, for the sake of contradiction, that $X_2\cup X_3$ is not a corner of $\mathcal{C}'$. Since $X_2\cup X_3$ is one of the connected components of $G\setminus C'$, this implies that there is a pair of edges $\{z,w\}\in\mathcal{F}'$, such that $\{z,w\}\subset E(G[X_2\cup X_3])$. Then, since $\{e_1,x\}\in\mathcal{F}'$ and $\mathcal{F}'$ generates $4$-cuts of $G$, we have that $C''=\{e_1,x,z,w\}$ is a $4$-cut of $G$. Let $G'=G\setminus\{z,w\}$. Since $\{z,w\}\subset E(G[X_2\cup X_3])$, we have that $G'[X_1]$ is connected. Thus, it cannot be that either $G'[X_2]$ or $G'[X_3]$ is connected, because otherwise the endpoints of $e_1$ or $x$, respectively, remain connected in $G'\setminus\{e_1,x\}$, in contradiction to the fact that $C''$ is a $4$-cut of $G$. Thus, we have that one of $z,w$ is a bridge of $G[X_2]$, and the other is a bridge of $G[X_3]$. Thus, we may assume w.l.o.g. that $z$ is a bridge of $G[X_2]$, and let $Y_1$ and $Y_2$ be the connected components of $G[X_2]\setminus z$. Then we have that $E[Y_1,Y_2]=\{z\}$. Since $|\partial(X_2)|=4$, by Lemma~\ref{lemma:bridge_in_corner} we have that $|E[Y_1,V\setminus X_2]|=2$ and $|E[Y_2,V\setminus X_2]|=2$. Since $E[Y_1,Y_2]=\{z\}$, this implies that $|\partial(Y_1)|=|\partial(Y_2)|=3$. But then we have that $u$ is not $4$-edge-connected with $v$, since either $\partial(Y_1)$ or $\partial(Y_2)$ (depending on whether $Y_1$ or $Y_2$ contains $u$, respectively) is a $3$-cut that separates $u$ from $v$ -- a contradiction. 

Thus, we have shown that $X_2\cup X_3$ is a corner of $\mathcal{C}'$. Therefore, since $X_2\cup X_3$ is one of the connected components of $G\setminus C'$, we have that $C'$ is a $\mathcal{C}'$-minimal $4$-cut. Furthermore, since $v\in X_1$ and $u\in X_2$, we have that $C'$ is an essential $4$-cut. Finally, we have that $\{e_1,e_2\}= C\cap C'$, and $\mathcal{C}'$ is generated by $\mathcal{F}'$, where $\mathcal{F}'$ is one of the collections of pairs of edges that are returned by Algorithm~\ref{algorithm:generatefamilies} on input $\mathcal{C}$. Thus, the proof is complete.
\end{proof}

\begin{corollary}
\label{corollary:quasi-iso-pair}
Let $\mathcal{C}$ be a complete collection of $4$-cuts of $G$, and let $C$ be an essential $4$-cut of $G$. Let $\mathcal{F}_1,\dots,\mathcal{F}_k$ be the collections of pairs of edges that are returned by Algorithm~\ref{algorithm:generatefamilies} on input $\mathcal{C}$, and let $\mathcal{C}_1,\dots,\mathcal{C}_k$ be the collections of $4$-cuts that they generate, respectively. Then, $C$ is a quasi $\mathcal{C}$-isolated $4$-cut if and only if:
\begin{enumerate}[label={(\arabic*)}]
\item{$C\in\mathcal{C}$.}
\item{All three partitions of $C$ into pairs of edges are contained in $\{\mathcal{F}_1,\dots,\mathcal{F}_k\}$.}
\item{There is a pair of edges $p$ in $C$ such that $p=C\cap C'$, where $C'$ is an essential $\mathcal{C}_i$-minimal $4$-cut, for some $i\in\{1,\dots,k\}$.}
\end{enumerate}
\end{corollary}
\begin{proof}
($\Rightarrow$) We have $C\in\mathcal{C}$ by Corollary~\ref{corollary:quasi-isolated_in_C}. $(2)$ is an immediate consequence of the definition of quasi $\mathcal{C}$-isolated $4$-cuts. $(3)$ is ensured by Lemma~\ref{lemma:quasi-iso-pair}, since $C$ is essential.

($\Leftarrow$)
Since $C\cap C'=p$, we have that $p'=C'\setminus C$ and $q=C\setminus C'$ are two distinct pairs of edges. Thus, we have $C=p\cup q$ and $C'=p\cup p'$. Then, by Lemma~\ref{lemma:implied4cut} we have that $C''=q\cup p'$ is a $4$-cut of $G$. Thus, since $\mathcal{C}$ is a complete collection of $4$-cuts of $G$, we have that $\{p,p',q\}$ is a collection of pairs of edges that generates $4$-cuts implied by $\mathcal{C}$, including $C$. This shows that $C$ is not a $\mathcal{C}$-isolated $4$-cut. Thus, since $(2)$ is satisfied, by definition we have that $C$ is a quasi $\mathcal{C}$-isolated $4$-cut.

\end{proof}

\subsection{Some additional properties satisfied by the output of Algorithm~\ref{algorithm:generatefamilies} }
\label{section:additional-properties}

\begin{lemma}
\label{lemma:uniting_pairs1}
Let $\mathcal{C}$ be a collection of $4$-cuts of $G$, and let $\mathcal{F}$ and $\mathcal{F}'$ be two distinct collections of pairs of edges that are returned by Algorithm~\ref{algorithm:generatefamilies} on input $\mathcal{C}$. Let $p$ and $p'$ be two pairs of edges such that $p\in\mathcal{F}$ and $p'\in\mathcal{F}'$. Then, $p\cup p'$ (if it is a $4$-cut of $G$) is not implied by $\mathcal{C}$ through the pair of edges $p$. 
\end{lemma}
\begin{proof}
Let $\mathcal{G}$ be the graph that is generated internally by Algorithm~\ref{algorithm:generatefamilies} on input $\mathcal{C}$ (in Line~\ref{line:generatefamilies:components}). Since $\mathcal{F}$ and $\mathcal{F}'$ are returned by Algorithm~\ref{algorithm:generatefamilies} on input $\mathcal{C}$, there are connected components $S$ and $S'$ of $\mathcal{G}$, such that $\mathcal{F}$ is derived from $S$, and $\mathcal{F}'$ is derived from $S'$ (in Line~\ref{line:generatefamilies:final}). Since $p\in\mathcal{F}$ and $p'\in\mathcal{F}'$, there are $4$-cuts $C$ and $C'$ in $\mathcal{C}$ such that $(C,\vec{p})\in S$ and $(C',\vec{p'})\in S'$. 

Let us assume, for the sake of contradiction, that $C''=p\cup p'$ is a $4$-cut of $G$ that is implied by $\mathcal{C}$ through the pair of edges $p$. (We note that it is not even necessary that $C''$ is a $4$-element set, but our assumption implies that.) This means that there is a sequence $p_1,\dots,p_{k+1}$ of pairs of edges, and a sequence $C_1,\dots,C_k$ of $4$-cuts from $\mathcal{C}$, such that $p_1=p$, $p_{k+1}=p'$, and $C_i=p_i\cup p_{i+1}$ for every $i\in\{1,\dots,k\}$. Then, for every $i\in\{1,\dots,k-1\}$, Lemma~\ref{lemma:no-common-three-edges} implies that either $C_i=C_{i+1}$ or $C_i\cap C_{i+1}=p_{i+1}$. Now let $i\in\{1,\dots,k\}$ be an index. Since $C_i=p_i\cup p_{i+1}$, we have that $(C_i,\vec{p}_i)$ and $(C_i,\vec{p}_{i+1})$ are the endpoints of an edge of $\mathcal{G}$ (see Line~\ref{line:generatefamilies:edges1}). Now let $i<k$. If $C_i=C_{i+1}$, then we have $(C_i,\vec{p}_{i+1})=(C_{i+1},\vec{p}_{i+1})$. Otherwise, we have $C_i\cap C_{i+1}=p_{i+1}$, and therefore the elements $(C_i,\vec{p}_{i+1})$ and $(C_{i+1},\vec{p}_{i+1})$ are vertices in the same connected component of $\mathcal{G}$ (due to the existence of the edges in Line~\ref{line:generatefamilies:addedges}). This shows that the vertices $(C_i,\vec{p}_i)$, for $i\in\{1,\dots,k-1\}$, are in the same connected component $S''$ of $\mathcal{G}$. Furthermore, since $C_k=p_k\cup p_{k+1}$, there is an edge of $\mathcal{G}$ with endpoints $(C_k,\vec{p}_k)$ and $(C_k,\vec{p}_{k+1})$. Thus, we have that $(C_k,\vec{p}_{k+1})$ is also in $S''$. Now let $\mathcal{F}''$ be the collection of pairs of edges that is derived from $S''$ in Line~\ref{line:generatefamilies:final}. Then, we have $\{p_1,\dots,p_{k+1}\}\subseteq\mathcal{F}''$. Since $p_1=p$ and $p_{k+1}=p'$, this implies that $\mathcal{F}''\cap\mathcal{F}\neq\emptyset$ and $\mathcal{F}''\cap\mathcal{F}'\neq\emptyset$. But this contradicts Lemma~\ref{lemma:returned_is_disjoint}. We conclude that $C''$ (if it is a $4$-cut of $G$) is not implied by $\mathcal{C}$ through the pair of edges $p$.
\end{proof}

\begin{figure}[t!]\centering
\includegraphics[trim={0 20cm 0 0}, clip=true, width=0.9\linewidth]{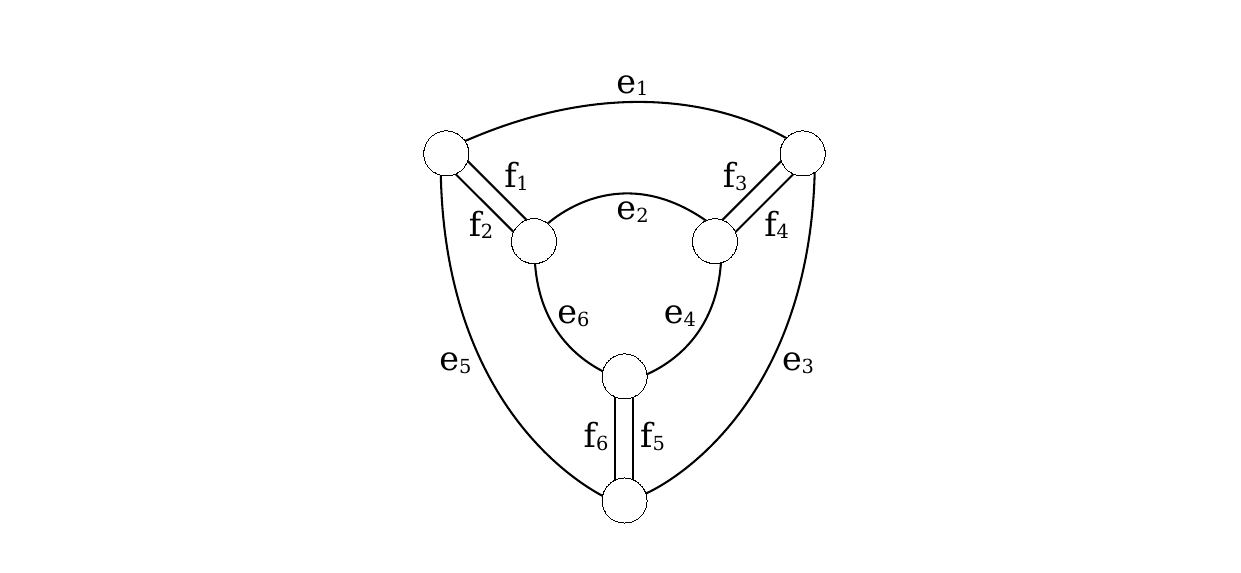}
\caption{\small{This is a $3$-edge-connected graph with $4$-cuts $C_1=\{e_1,e_2,e_3,e_4\}$, $C_2=\{e_1,e_2,e_5,e_5\}$, $C_3=\{e_3,e_4,e_5,e_6\}$, $D_1=\{e_1,e_5,f_1,f_2\}$, $D_2=\{e_2,e_6,f_1,f_2\}$, $E_1=\{e_1,e_3,f_3,f_4\}$, $E_2=\{e_2,e_4,f_3,f_4\}$, $F_1=\{e_3,e_5,f_5,f_6\}$ and $F_2=\{e_4,e_6,f_5,f_6\}$. It is easy to see that $\mathcal{C}=\{D_1,D_2,E_1,E_2,F_1,F_2\}$ is a collection of $4$-cuts that implies all $4$-cuts of this graph. In particular, $\mathcal{C}$ implies the cyclic family of $4$-cuts $\{C_1,C_2,C_3\}$, which is generated by the collection of pairs of edges $\{\{e_1,e_2\},\{e_3,e_4\},\{e_5,e_6\}\}$. However, notice that $C_1$ is not implied by $\mathcal{C}$ through the pair of edges $\{e_1,e_2\}$ (it is only implied by $\mathcal{C}$ through the pair of edges $\{e_1,e_3\}$ or $\{e_2,e_4\}$, with the implicating sequence $E_1=\{e_1,e_3,f_3,f_4\}$, $E_2=\{f_3,f_4,e_2,e_4\}$).}}\label{figure:no_imply_triangle}
\end{figure}
\begin{figure}[t!]\centering
\includegraphics[trim={0 18cm 0 0}, clip=true, width=0.8\linewidth]{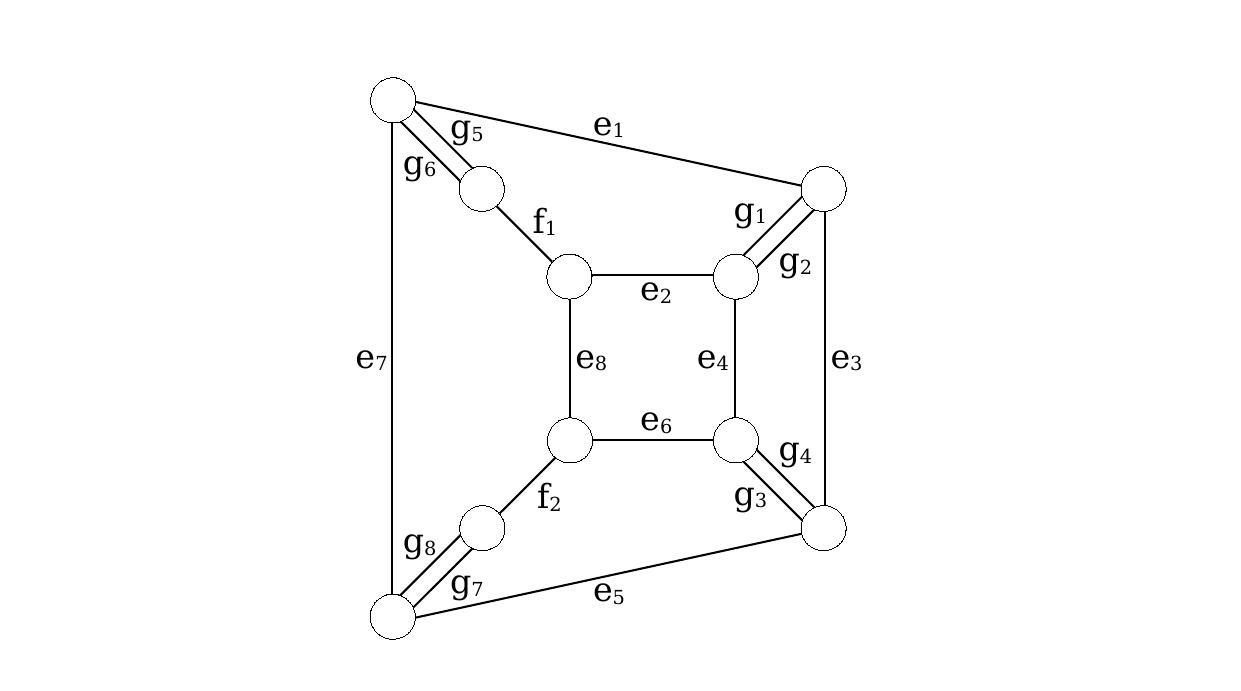}
\caption{\small{This is a $3$-edge-connected graph with $4$-cuts $C_1=\{e_1,e_2,e_3,e_4\}$, $C_2=\{e_1,e_2,e_5,e_6\}$, $C_3=\{e_1,e_2,e_7,e_8\}$, $C_4=\{e_3,e_4,e_5,e_6\}$, $C_5=\{e_3,e_4,e_7,e_8\}$, $C_6=\{e_5,e_6,e_7,e_8\}$, $D_1=\{e_1,e_3,g_1,g_2\}$, $D_2=\{e_2,e_4,g_1,g_2\}$, $E_1=\{e_3,e_5,g_3,g_4\}$, $E_2=\{e_4,e_6,g_3,g_4\}$, $F_1=\{e_1,e_5,f_1,f_2\}$, $F_2=\{e_2,e_6,f_1,f_2\}$, $G_1=\{e_1,e_7,g_5,g_6\}$, $G_2=\{e_2,e_8,g_5,g_6\}$, $H_1=\{e_5,e_7,g_7,g_8\}$ and $H_2=\{e_6,e_8,g_7,g_8\}$. Notice that $C_1$ is implied by $\{D_1,D_2\}$, $C_2$ is implied by $\{F_1,F_2\}$, $C_3$ is implied by $\{G_1,G_2\}$, $C_4$ is implied by $\{E_1,E_2\}$, and $C_6$ is implied by $\{H_1,H_2\}$. Thus, we have that $\mathcal{C}=\{C_5,D_1,D_2,E_1,E_2,F_1,F_2,G_1,G_2,H_1,H_2\}$ is a collection of $4$-cuts that implies all $4$-cuts of this graph. In particular, $\mathcal{C}$ implies the cyclic family of $4$-cuts $\mathcal{C}'=\{C_1,C_2,C_3,C_4,C_5,C_6\}$, which is generated by the collection of pairs of edges $\{\{e_1,e_2\},\{e_3,e_4\},\{e_5,e_6\},\{e_7,e_8\}\}$. However, notice that $C_2=\{e_1,e_2,e_5,e_6\}$ is not implied by $\mathcal{C}$ through the pair of edges $\{e_1,e_2\}$, and the pairs of edges $\{e_1,e_2\}$ and $\{e_5,e_6\}$ have distance $2$ in $\mathcal{C}'$.}}\label{figure:no_imply_square}
\end{figure}
\begin{figure}[t!]\centering
\includegraphics[trim={0 21cm 0 0}, clip=true, width=0.8\linewidth]{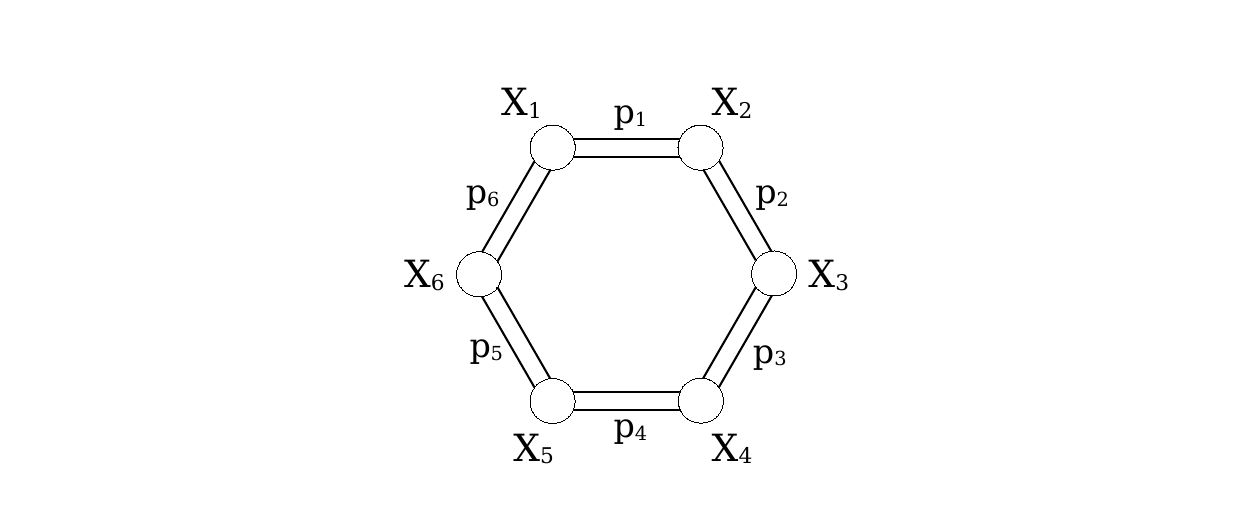}
\caption{\small{A cyclic family of $4$-cuts $\mathcal{C}_6$ generated by the collection of pairs of edges $\{p_1,p_2,p_3,p_4,p_5,p_6\}$. If $i$ and $j$ are two indices in $\{1,\dots,6\}$, then we say that the pairs of edges $p_i$ and $p_j$ have distance $\mathit{min}\{(i-j+6)\mathit{mod}6, (j-i+6)\mathit{mod}6\}$ in $\mathcal{C}_6$. Thus, $\{p_1,p_4\}$, $\{p_2,p_5\}$ and $\{p_3,p_6\}$ are the only pairs of pairs of edges that have distance $3$ in $\mathcal{C}_6$. Notice that these pairs of pairs of edges are antipodal in this figure. The point of Lemma~\ref{lemma:antipodal_in_hexagon} is that if there is a collection $\mathcal{C}$ of $4$-cuts that implies $\mathcal{C}_6$, then the $4$-cuts $p_1\cup p_4$, $p_2\cup p_5$ and $p_3\cup p_6$ are implied by $\mathcal{C}$ through the pairs of edges $p_1$, $p_2$ and $p_3$, respectively. In general, it is not necessary that this is the case for $4$-cuts that are generated by pairs of pairs of edges that have distance less than $3$. This is demonstrated in the previous figures, \ref{figure:no_imply_triangle} and \ref{figure:no_imply_square}.
}}\label{figure:hexagon}
\end{figure}

In order to appreciate the following lemma, we need to discuss a subtle point that concerns the way in which cyclic families of $4$-cuts are implied by collections of $4$-cuts. Suppose that we have a collection $\mathcal{C}$ of $4$-cuts that implies the cyclic family $\mathcal{C}'$ of $4$-cuts that is generated by the collection of pairs of edges $\mathcal{F}=\{p_1,p_2,p_3\}$. Then, $\mathcal{C}'$ consists of the $4$-cuts $\{p_1\cup p_2,p_1\cup p_3,p_2\cup p_3\}$. However, it is not necessary that $\mathcal{C}$ implies, say, $p_1\cup p_2$ through the pair of edges $p_1$. In other words, it is not necessary that $\mathcal{C}$ implies the $4$-cuts in $\mathcal{C}'$ through the pairs of edges from which they are generated. An example for that is given in Figure~\ref{figure:no_imply_triangle}. Moreover, the same is true even if $\mathcal{C'}$ is generated by a collection of four pairs of edges, as we can see in  Figure~\ref{figure:no_imply_square}. However, if $\mathcal{C}'$ is generated by a collection $\mathcal{F}$ of six or more pairs of edges, then there are some pairs of pairs of edges in $\mathcal{F}$ that have ``distance" at least three (there is no need to define precisely this term, but we refer to Figure~\ref{figure:hexagon} for an intuitive understanding of it). The $4$-cuts that are formed by the union of such pairs of edges have the property that they are implied by $\mathcal{C}$ through them. In particular, if $\mathcal{F}$ consists of six pairs of edges, then there are three pairs of pairs of edges in $\mathcal{F}$ that are ``antipodal" (see Figure~\ref{figure:hexagon}). In this case, the following lemma establishes our claim. The intuitive idea behind Lemma~\ref{lemma:antipodal_in_hexagon} is that every $4$-cut that is formed by the union of two pairs of edges that have distance at least three, has the property that the pairs of edges that form it are not entangled with other edges in forming $4$-cuts in a way that would interfere with the straightforward way through which we would expect $\mathcal{C}$ to imply it. 

\begin{lemma}[Antipodal pairs of edges in a hexagonal family of $4$-cuts]
\label{lemma:antipodal_in_hexagon}
Let $\mathcal{C}$ be a collection of $4$-cuts of $G$, and let $\{p_1,\dots,p_6\}$ be a collection of pairs of edges that generates a cyclic family $\mathcal{C}'$ of $4$-cuts that is implied by $\mathcal{C}$. We may assume w.l.o.g. that there is a partition $\{X_1,\dots,X_6\}$ of $V(G)$ such that $G[X_i]$ is connected for every $i\in\{1,\dots,6\}$, and $E[X_i,X_{i+_{6}1}]=p_i$ for every $i\in\{1,\dots,6\}$. (See Figure~\ref{figure:hexagon}). 
Then, $p_1\cup p_4$ is implied by $\mathcal{C}$ through the pair of edges $p_1$. 
\end{lemma}
\begin{proof}
Let us suppose, for the sake of contradiction, that $C=p_1\cup p_4$ is not implied by $\mathcal{C}$ through the pair of edges $p_1$. In particular, this implies that $C\notin\mathcal{C}$. Let $p_1=\{e_1,e_2\}$ and let $p_4=\{e_3,e_4\}$. Since $C$ is implied by $\mathcal{C}$, but not through the pair of edges $p_1$, we may assume w.l.o.g. that $C$ is implied by $\mathcal{C}$ through the pair of edges $\{e_1,e_3\}$. Then, by Lemma~\ref{lemma:not-included-4cut} we have that there is a collection $\mathcal{F}$ of pairs of edges that is returned by Algorithm~\ref{algorithm:generatefamilies} on input $\mathcal{C}$, such that $\{\{e_1,e_3\},\{e_2,e_4\}\}\subset\mathcal{F}$. So let $\{f_1,f_2\}$ be a pair of edges in $\mathcal{F}\setminus\{\{e_1,e_3\},\{e_2,e_4\}\}$. Since $\mathcal{F}$ is returned by Algorithm~\ref{algorithm:generatefamilies} on input $\mathcal{C}$, Proposition~\ref{proposition:cyclic_families_algorithm} implies that $C'=\{e_1,e_3,f_1,f_2\}$ is a $4$-cut implied by $\mathcal{C}$. 

Let $G'=G\setminus\{e_1,e_3\}$. Then the subgraphs $G'[X_1]$, $G'[X_2]$, $G'[X_4]$ and $G'[X_5]$ remain connected, and we have $E_{G'}[X_1,X_2]=\{e_2\}$ and $E_{G'}[X_4,X_5]=\{e_4\}$. Thus, it cannot be that both $(G'\setminus\{f_1,f_2\})[X_1]$ and $(G'\setminus\{f_1,f_2\})[X_2]$ are connected, or that both $(G'\setminus\{f_1,f_2\})[X_4]$ and $(G'\setminus\{f_1,f_2\})[X_5]$ are connected, because otherwise the endpoints of $e_1$ or $e_3$, respectively, would remain connected in $G\setminus\{e_1,e_3,f_1,f_2\}$ -- in contradiction to the fact that $C'$ is a $4$-cut of $G$. Thus, we have that one of $f_1,f_2$ is a bridge of either $G'[X_1]$ or $G'[X_2]$, and the other is a bridge of either $G'[X_4]$ or $G'[X_5]$. Thus, we may assume w.l.o.g. (considering the symmetry of Figure~\ref{figure:hexagon}), 
that $f_1$ is a bridge of $G'[X_1]=G[X_1]$ (and therefore $f_2$ lies in either $G[X_4]$ or $G[X_5]$). Let $Y_1$ and $Y_2$ be the connected components of $G'[X_1]\setminus f_1$.
Since $\partial(X_1)=p_1\cup p_6$, Lemma~\ref{lemma:bridge_in_corner} implies that $|E[Y_1,V\setminus X_1]|=2$ and $|E[Y_2,V\setminus X_1]|=2$. Now there are three possibilities to consider: either $(1)$ $E[Y_1,V\setminus X_1]=p_1$ (and $E[Y_2,V\setminus X_1]=p_6$), or $(2)$ $E[Y_1,V\setminus X_1]=p_6$ (and $E[Y_2,V\setminus X_1]=p_1$), or $(3)$ both $E[Y_1,V\setminus X_1]$ and $E[Y_2,V\setminus X_1]$ intersect with both $p_1$ and $p_6$. 

Let us suppose that $(1)$ is true. Then we have that $E[Y_1,X_2]=\{e_1,e_2\}$. Since neither of $e_1,e_3,f_1,f_2$ lies in $G[X_2]$, we have that $(G\setminus C')[X_2]$ remains connected, and $E_{G\setminus C'}[Y_1,X_2]=\{e_2\}$. Thus, the endpoints of $e_1$ remain connected in $G\setminus C'$, in contradiction to the fact that $C'$ is a $4$-cut of $G$. Thus, case $(1)$ is rejected. With the analogous argument, we can reject case $(2)$. Thus, only case $(3)$ can be true. This implies that neither of $E[Y_1,X_6]$ and $E[Y_2,X_6]$ is empty (because each contains one edge from $p_6$). But then we have that $Y_1$ and $Y_2$ (and therefore the endpoints of $f_1$) remain connected in $G\setminus C'$ (because $(G\setminus C')[X_6]$ is connected, and neither of $E[Y_1,X_6]$ and $E[Y_2,X_6]$ intersects with $C'$), in contradiction to the fact that $C'$ is a $4$-cut of $G$. Thus, we conclude that our initial supposition cannot be true, and therefore $p_1\cup p_4$ is implied by $\mathcal{C}$ through the pair of edges $p_1$. 
\end{proof}

\begin{figure}[h]\centering
\includegraphics[trim={0 18cm 0 0}, clip=true, width=0.9\linewidth]{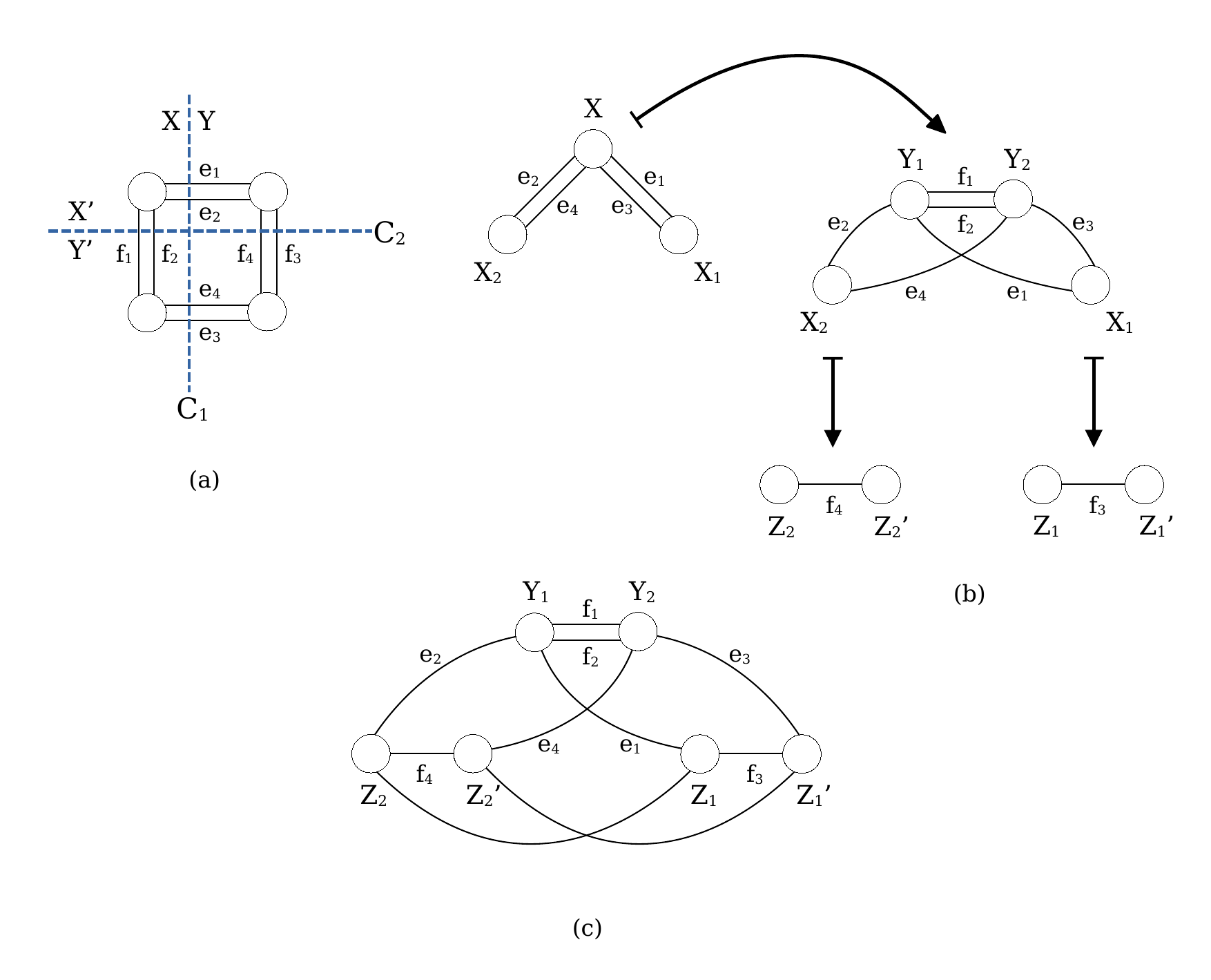}
\caption{\small{Companion figures to Lemma~\ref{lemma:non-crossing-of-minimal}. $(a)$ The square of the crossing $4$-cuts $C_1$ and $C_2$. The sides of $C_1$ are $X$ and $Y$, and the sides of $C_2$ are $X'$ and $Y'$. $(b)$ We assume that $X$ is the corner of $C_1$ in $\mathcal{C}_1$, and $\{\{e_1,e_3\},\{e_2,e_4\}\}\subset\mathcal{F}_1$. $X_1$ and $X_2$ are the neighboring corners of $X$ in $\mathcal{C}_1$. Then, we have w.l.o.g. that $f_3$ is a bridge of $G[X_1]$ and $f_4$ is a bridge of $G[X_2]$. $(c)$ We infer this situation, which contradicts the essentiality of $C_1$.}}\label{figure:crossing_minimal_essential}
\end{figure}

\begin{lemma}[Minimal essential $4$-cuts do not cross]
\label{lemma:non-crossing-of-minimal}
Let $\mathcal{C}$ be a complete collection of $4$-cuts of $G$, and let $\mathcal{F}_1$ and $\mathcal{F}_2$ be two distinct collections of pairs of edges with $|\mathcal{F}_1|>2$ and $|\mathcal{F}_2|>2$ that are returned by Algorithm~\ref{algorithm:generatefamilies} on input $\mathcal{C}$. Let $\mathcal{C}_1$ and $\mathcal{C}_2$ be the collections of $4$-cuts that are generated by $\mathcal{F}_1$ and $\mathcal{F}_2$, respectively. Let $C_1$ be an essential $\mathcal{C}_1$-minimal $4$-cut, and let $C_2$ be an essential $\mathcal{C}_2$-minimal $4$-cut. Then, $C_1$ and $C_2$ are parallel.
\end{lemma}
\begin{proof}
Let $C_1=\{e_1,e_2,e_3,e_4\}$ and let $C_2=\{f_1,f_2,f_3,f_4\}$. Let us suppose, for the sake of contradiction, that $C_1$ and $C_2$ cross. Since $C_1$ and $C_2$ are essential $4$-cuts, by Corollary~\ref{corollary:crossing-essential} we may assume w.l.o.g. that $C_1$ and $C_2$ cross as in Figure~\ref{figure:crossing-essential}. Notice that, by Lemma~\ref{lemma:bridge_in_corner}, we have that the four corners of Figure~\ref{figure:crossing-essential} are connected subgraphs of $G$. Thus, the boundaries of those corners are $4$-cuts of $G$, and therefore these are $4$-cuts implied by $\mathcal{C}$ (because $\mathcal{C}$ implies all $4$-cuts of $G$). Now let $X$ be the connected component of $G\setminus C_1$ that contains $f_1$ and $f_2$, and let $Y$ be the connected component of $G\setminus C_1$ that contains $f_3$ and $f_4$ (see Figure~\ref{figure:crossing_minimal_essential}$(a)$). 
Since $C_1$ is $\mathcal{C}_1$-minimal, we have that either $X$ or $Y$ is a corner of $\mathcal{C}_1$. So let us assume, w.l.o.g., that $X$ is a corner of $\mathcal{C}_1$. Similarly, let $X'$ be the connected component of $G\setminus C_2$ that contains $e_1$ and $e_2$, and let $Y'$ be the connected component of $G\setminus C_2$ that contains $e_3$ and $e_4$. Since $C_2$ is $\mathcal{C}_2$-minimal, we have that either $X'$ or $Y'$ is a corner of $\mathcal{C}_2$. Due to the symmetry of Figure~\ref{figure:crossing_minimal_essential}$(a)$, we may assume w.l.o.g. that $X'$ is a corner of $\mathcal{C}_2$. (I.e., although we have assumed that $X$ is a corner of $\mathcal{C}_1$, there is no loss of generality in assuming that $X'$ is a corner of $\mathcal{C}_2$.)

Now, since $C_1\in\mathcal{C}_1$ and $\mathcal{C}_1$ is generated by $\mathcal{F}_1$, there are three possibilities to consider: either $(1)$ $\{\{e_1,e_2\},\{e_3,e_4\}\}\subset\mathcal{F}_1$, or $(2)$ $\{\{e_1,e_3\},\{e_2,e_4\}\}\subset\mathcal{F}_1$, or $(3)$ $\{\{e_1,e_4\},\{e_2,e_3\}\}\subset\mathcal{F}_1$. Let us consider case $(2)$ first. Since $C_1$ is an essential $4$-cut and $|\mathcal{C}_1|\geq 3$, by Corollary~\ref{corollary:essential_implies_cyclic} we have that $\mathcal{C}_1$ is a cyclic family of $4$-cuts. Thus, we may consider the neighboring corners $X_1$ and $X_2$ of $X$ in $\mathcal{C}_1$ such that $E[X,X_1]=\{e_1,e_3\}$ and $E[X,X_2]=\{e_2,e_4\}$. Let $Y_1$ and $Y_2$ be the connected components of $G[X]\setminus\{f_1,f_2\}$. Then, since we are in the situation depicted in Figure~\ref{figure:crossing_minimal_essential}$(b)$, 
we have that both $E[Y_1,V\setminus X]$ and $E[Y_2,V\setminus X]$ intersect with both $\{e_1,e_3\}$ and $\{e_2,e_4\}$. Thus we may assume, w.l.o.g., that $E[Y_1,X_1]=\{e_1\}$, $E[Y_1,X_2]=\{e_2\}$, $E[Y_2,X_1]=\{e_3\}$ and $E[Y_2,X_2]=\{e_4\}$ (see Figure~\ref{figure:crossing_minimal_essential}$(b)$). 
Then, notice that it cannot be the case that either $(G\setminus C_2)[X_1]$ or $(G\setminus C_2)[X_2]$ is connected, because otherwise the endpoints of $f_1$ and $f_2$ would remain connected in $G\setminus C_2$, in contradiction to the fact that $C_2$ is a $4$-cut of $G$. Thus, we have that either $f_3$ is a bridge of $G[X_1]$ and $f_4$ is a bridge of $G[X_2]$, or reversely. So let us assume, w.l.o.g., that $f_3$ is a bridge of $G[X_1]$ and $f_4$ is a bridge of $G[X_2]$. Let $Z_1$ and $Z_1'$ be the connected components of $G[X_1]\setminus f_3$, and let $Z_2$ and $Z_2'$ be the connected components of $G[X_2]\setminus f_4$. Then we have $E[Z_1,Z_1']=\{f_3\}$ and $E[Z_2,Z_2']=\{f_4\}$.

Now consider the neighboring corner $X_1'$ of $X_1$ in $\mathcal{C}_1$ that is different from $X$. We claim that both $E[Z_1,X_1']$ and $E[Z_1',X_1']$ are non-empty. To see this, suppose the contrary. Then we may assume w.l.o.g. that $E[Z_1,X_1']=\emptyset$. Since the graph is $3$-edge-connected, we have $|\partial(Z_1)|\geq 3$. Thus, since $\partial(Z_1)=E[Z_1,Z_1']\cup E[Z_1,X_1']\cup E[Z_1,X]$, we have that $E[Z_1,X]$ contains at least two edges (because $E[Z_1,Z_1']=\{f_3\}$). Thus, $E[Z_1,X]$ consists of $\{e_1,e_3\}$. But this implies that $\partial(Z_1)=\{e_1,e_3,f_3\}$ is a $3$-cut of $G$, which is impossible (see Figure~\ref{figure:crossing_minimal_essential}$(a)$). 
This shows that both $E[Z_1,X_1']$ and $E[Z_1',X_1']$ are non-empty. Similarly, if we let $X_2'$ denote the  neighboring corner of $X_2$ in $\mathcal{C}_1$ that is different from $X$, then we have that both $E[Z_2,X_2']$ and $E[Z_2',X_2']$ are non-empty.

Then, if $(G\setminus C_2)[X_1']$ is connected, we have that $Z_1$ and $Z_1'$ (and therefore the endpoints of $f_3$) remain connected in $G\setminus C_2$, in contradiction to the fact that $C_2$ is a $4$-cut of $G$. Thus, we have that $(G\setminus C_2)[X_1']$ is disconnected. This implies that an edge from $C_2$ is in $G[X_1']$, and the only candidate is $f_4$. Thus, we have that $X_1'=X_2$ (and $X_2'=X_1$), and so we have a situation like that depicted in Figure~\ref{figure:crossing_minimal_essential}$(c)$. 
But then we have that $C_1=\{e_1,e_2,e_3,e_4\}$ is a non-essential $4$-cut, because the connected components of $G\setminus C_1$ are $Y_1\cup Y_2$ and $Z_1\cup Z_1'\cup Z_2\cup Z_2'$, and there is no pair of vertices in those components that are $4$-edge-connected (because $|\partial(Z_1)|=|\partial(Z_1')|=|\partial(Z_2)|=|\partial(Z_2')|=3$). Thus, case $(2)$ cannot be true. With the analogous argument we can see that case $(3)$ also cannot be true. (To see this, just switch the labels of edges $e_3$ and $e_4$.)
Thus, only case $(1)$ is true. Similarly, if we consider the three possibilities for the collection $\mathcal{F}_2$ -- i.e., either $\{\{f_1,f_2\},\{f_3,f_4\}\}\subset\mathcal{F}_2$, or $\{\{f_1,f_3\},\{f_2,f_4\}\}\subset\mathcal{F}_2$, or $\{\{f_1,f_4\},\{f_2,f_3\}\}\subset\mathcal{F}_2$ --, then we can see that only $\{\{f_1,f_2\},\{f_3,f_4\}\}\subset\mathcal{F}_2$ can be true. 

Now consider a pair of edges $\{e_5,e_6\}\in\mathcal{F}_1\setminus\{\{e_1,e_2\},\{e_3,e_4\}\}$, and a pair of edges $\{f_5,f_6\}\in\mathcal{F}_2\setminus\{\{f_1,f_2\},\{f_3,f_4\}\}$. Then, Proposition~\ref{proposition:cyclic_families_algorithm} implies that $\{e_1,e_2,e_5,e_6\}$ and $\{f_1,f_2,f_5,f_6\}$ are $4$-cuts implied by $\mathcal{C}$. 
Therefore, a repeated application of Lemma~\ref{lemma:implied4cut} implies that $\mathcal{F}_6=\{\{e_1,e_2\},\{e_3,e_4\},\{e_5,e_6\},\{f_1,f_2\},\{f_3,f_4\},\{f_5,f_6\}\}$ is a collection of pairs of edges that generates a collection $\mathcal{C}_6$ of $4$-cuts. Lemma~\ref{lemma:returned_is_disjoint} implies that $\mathcal{F}_1\cap\mathcal{F}_2=\emptyset$, and therefore $|\mathcal{F}_6|=6$. Then, Proposition~\ref{proposition:cyclic_family} implies that $\mathcal{C}_6$ is a cyclic family of $4$-cuts. 

\begin{figure}[t!]\centering
\includegraphics[trim={0 15cm 0 0}, clip=true, width=0.85\linewidth]{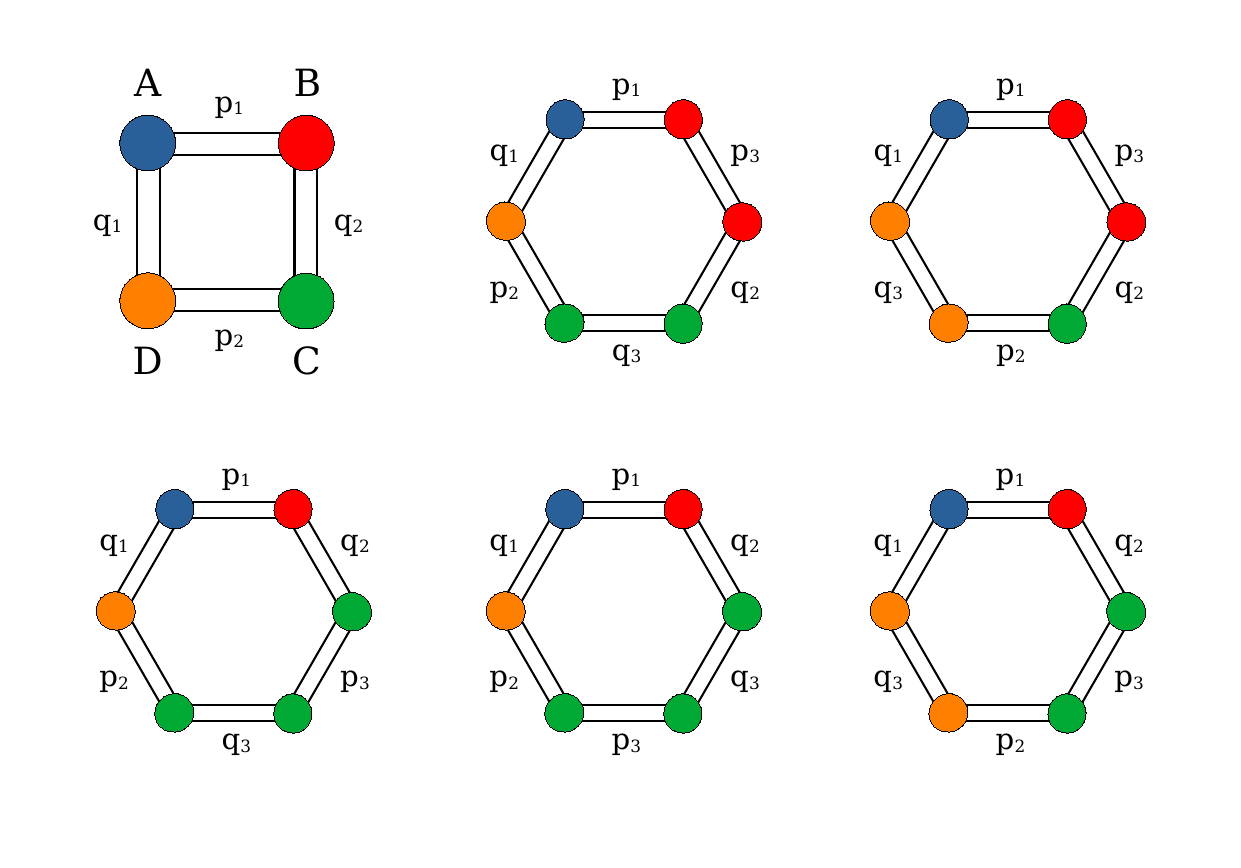}
\caption{\small{$\mathcal{F}=\{p_1,p_2,q_1,q_2\}$ is a collection of pairs of edges that generates a cyclic family of $4$-cuts $\mathcal{C}$ with corners $A$, $B$, $C$ and $D$. We consider all the different ways in which $\mathcal{C}$ can be expanded into a cyclic family of $4$-cuts $\mathcal{C}_6$, where $\mathcal{C}_6$ is generated by the collection of pairs of edges $\mathcal{F}\cup\{p_3,q_3\}$, under the restriction that $p_3\in B\cup C$ and $q_3\in C\cup D$. The colors of the corners of $\mathcal{C}_6$ correspond to the colors of the corners of the square that got expanded. Notice that, in either case, we have that there are $i,j\in\{1,2,3\}$ such that $p_i$ and $q_j$ are antipodal pairs of edges in $\mathcal{C}_6$.}}\label{figure:expanding-square}
\end{figure}

For every $i\in\{1,2,3\}$, let $p_i=\{e_{2i-1},e_{2i}\}$ and let $q_i=\{f_{2i-1},f_{2i}\}$. Now we will demonstrate that there are $i\in\{1,2,3\}$ and $j\in\{1,2,3\}$, such that $p_i$ and $q_j$ are antipodal pairs of edges in $\mathcal{C}_6$. 
Let $A,B,C,D$ be the corners of the square of the crossing $4$-cuts $C_1$ and $C_2$, as shown in Figure~\ref{figure:expanding-square}. 
%
Since $G[X]$ is a corner of $\mathcal{C}_1$, we have that the pair of edges $\{e_5,e_6\}$ lies in $G[Y]=B\cup C$. Then, since the collection of pairs of edges $\{\{e_1,e_2\},\{e_3,e_4\},\{f_1,f_2\},\{f_3,f_4\}\}$ generates a cyclic family of $4$-cuts, and can be extended into the collection $\mathcal{F}_6$, we have that the pair of edges $\{e_5,e_6\}$ lies entirely within $B$ or $C$. Similarly, since $G[X']$ is a corner of $\mathcal{C}_2$, we can infer that the pair of edges $\{f_5,f_6\}$ lies entirely within $C$ or $D$. Thus, all the possible configurations for the pairs of edges in $\mathcal{F}_6$ on the hexagon of the corners of $\mathcal{C}_6$ are shown in Figure~\ref{figure:expanding-square}. There, we can see that, in either case, there are $i\in\{1,2,3\}$ and $j\in\{1,2,3\}$, such that $p_i$ and $q_j$ are antipodal pairs of edges in $\mathcal{F}_6$.

Now let $C=p_i\cup q_j$. Then, by Lemma~\ref{lemma:antipodal_in_hexagon} we have that $C$ is implied by $\mathcal{C}$ through the pair of edges $p_i$. But Lemma~\ref{lemma:uniting_pairs1} implies that $\mathcal{C}$ does not imply $C$ through the pair of edges $p_i$ (because $p_i\in\mathcal{F}_1$ and $q_j\in\mathcal{F}_2$), a contradiction. Thus, our initial assumption cannot be true, and therefore $C_1$ and $C_2$ do not cross.

\end{proof}

\begin{figure}[t!]\centering
\includegraphics[trim={0 21cm 0 0cm}, clip=true, width=0.9\linewidth]{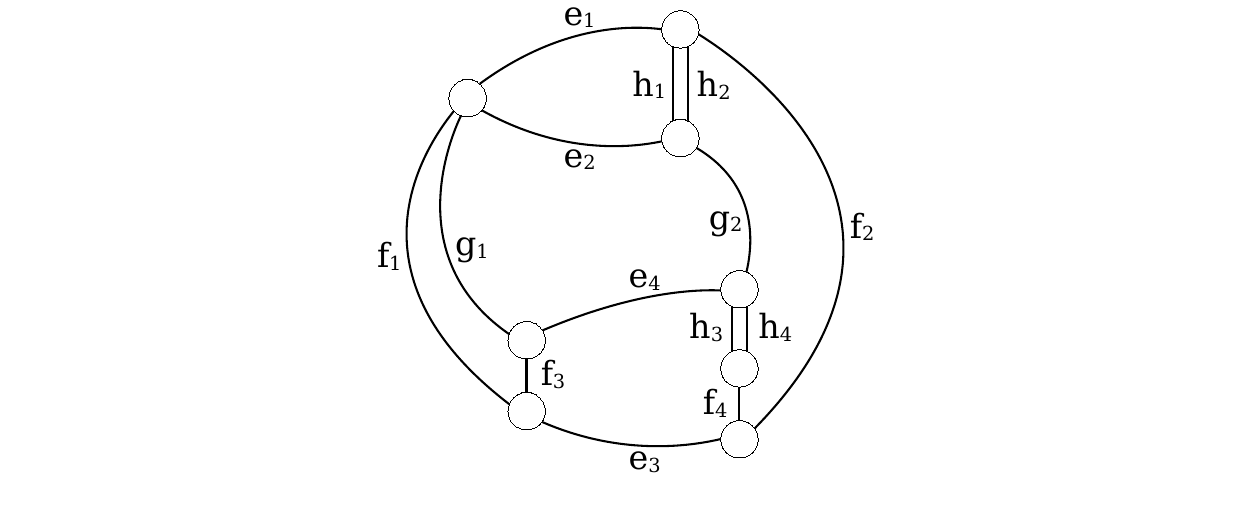}
\caption{{This is a $3$-edge-connected graph with $4$-cuts: $C_1=\{e_1,e_2,e_3,e_4\}$, $C_2=\{e_1,e_2,f_1,g_1\}$, $C_3=\{e_1,e_2,f_2,g_2\}$, $C_4=\{e_3,e_4,f_1,g_1\}$, $C_5=\{e_3,e_4,f_2,g_2\}$, $C_6=\{f_1,f_2,g_1,g_2\}$, $C_7=\{f_3,f_4,g_1,g_2\}$, $C_8=\{f_1,f_2,f_3,f_4\}$, $C_9=\{e_1,f_2,h_1,h_2\}$, $C_{10}=\{e_2,g_2,h_1,h_2\}$, $C_{11}=\{f_2,e_3,h_3,h_4\}$ and $C_{12}=\{g_2,e_4,h_3,h_4\}$. 
We have that $C_3$ is implied by $\{C_9,C_{10}\}$, $C_4$ is implied by $\{C_1,C_2\}$, $C_5$ is implied by $\{C_{11},C_{12}\}$, and $C_6$ is implied by $\{C_7,C_8\}$.
Thus, $\mathcal{C}=\{C_1,C_2,C_7,C_8,C_9,C_{10},C_{11},C_{12}\}$ is a complete collection of $4$-cuts of this graph.
If we apply Algorithm~\ref{algorithm:generatefamilies} on $\mathcal{C}$, we will get as a result the collections of pairs of edges $\mathcal{F}_1=\{\{e_1,e_2\},\{f_1,g_1\},\{e_3,e_4\}\}$, $\mathcal{F}_2=\{\{f_1,f_2\},\{g_1,g_2\},\{f_3,f_4\}\}$, $\mathcal{F}_3=\{\{e_1,f_2\},\{e_2,g_2\},\{h_1,h_2\}\}$, and $\mathcal{F}_4=\{\{f_2,e_3\},\{g_2,e_4\},\{h_3,h_4\}\}$. Notice that $\mathcal{F}_1$ and $\mathcal{F}_2$ generate the cyclic families of $4$-cuts $\mathcal{C}_1=\{C_1,C_2,C_4\}$ and $\mathcal{C}_2=\{C_6,C_7,C_8\}$, respectively. Since $|\mathcal{C}_1|=3$, we have that every $4$-cut in $\mathcal{C}_1$ is $\mathcal{C}_1$-minimal. Similarly, every $4$-cut contained in $\mathcal{C}_2$ is $\mathcal{C}_2$-minimal. Thus, $C_1$ is a $\mathcal{C}_1$-minimal $4$-cut, and $C_8$ is a $\mathcal{C}_2$-minimal $4$-cut. However, we have that $C_1$ and $C_8$ cross. Notice that $C_8$ is not an essential $4$-cut (because one of its sides consists of the endpoints of $e_3$, both of which have degree $3$, and therefore they are not $4$-edge-connected with any other vertex of the graph). On the other hand, $C_1$ is an essential $4$-cut (because e.g. the endpoints of $e_1$ are $4$-edge-connected). Thus, the condition of essentiality of both $4$-cuts in the statement of Lemma~\ref{lemma:non-crossing-of-minimal} cannot be removed.}}
\label{figure:crossing_minimal}
\end{figure}

We note that the condition of essentiality of both $4$-cuts $C_1$ and $C_2$ in Lemma~\ref{lemma:non-crossing-of-minimal} cannot be removed without destroying the inference of the lemma. This is demonstrated in Figure~\ref{figure:crossing_minimal}.

\begin{figure}[t!]\centering
\includegraphics[trim={0 20.5cm 0 0}, clip=true, width=0.8\linewidth]{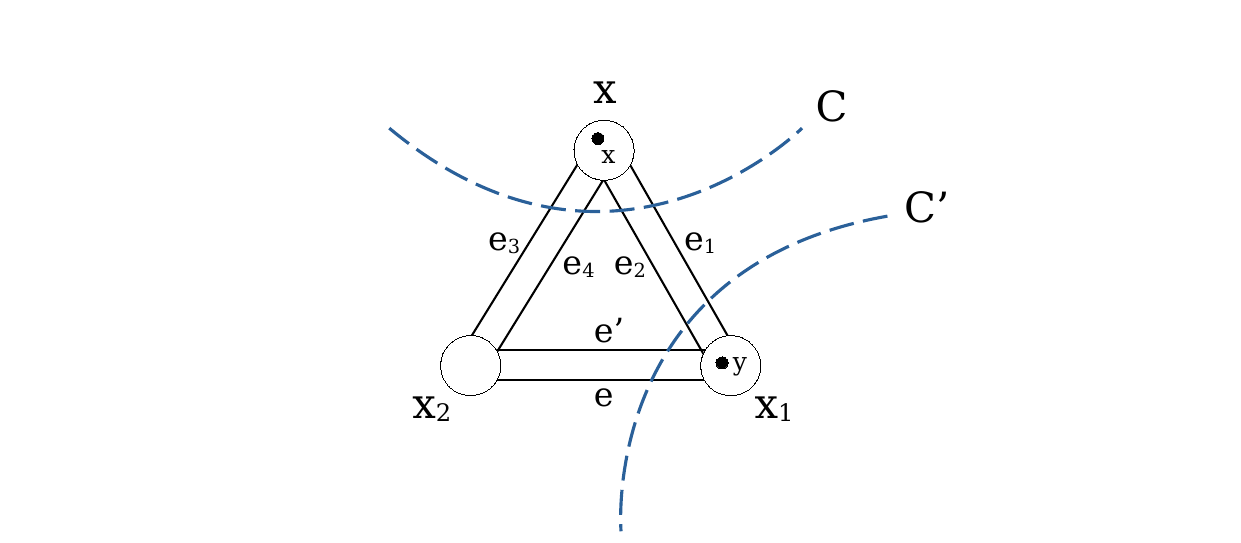}
\caption{\small{A depiction of the situation analyzed in Lemma~\ref{lemma:quasi-isolated-replaceable}.}}\label{figure:essential_quasi_triangle}
\end{figure}

\begin{lemma}[The essential quasi-isolated $4$-cuts are replaceable]
\label{lemma:quasi-isolated-replaceable}
Let $\mathcal{C}$ be a collection of $4$-cuts of $G$, and let $C$ be an essential quasi $\mathcal{C}$-isolated $4$-cut. Let $x$ and $y$ be two vertices that are separated by $C$. Then, there is a collection of pairs of edges $\mathcal{F}$ with $|\mathcal{F}|>2$ that is returned by Algorithm~\ref{algorithm:generatefamilies} on input $\mathcal{C}$, such that $\mathcal{F}$ generates a $4$-cut that separates $x$ and $y$.
\end{lemma}
\begin{proof}
Let $C=\{e_1,e_2,e_3,e_4\}$. Since $C$ is quasi $\mathcal{C}$-isolated, we may assume w.l.o.g. that there is a pair of edges $\{e,e'\}\notin\{\{e_1,e_2\},\{e_3,e_4\}\}$ such that $\mathcal{F}'=\{\{e_1,e_2\},\{e_3,e_4\},\{e,e'\}\}$ generates a collection $\mathcal{C}'$ of $4$-cuts  implied by $\mathcal{C}$. Since $C$ is an essential $4$-cut, by Corollary~\ref{corollary:degenerate_non-essential} we have that $\mathcal{C}'$ cannot be a degenerate family of $4$-cuts. Therefore, by Lemma~\ref{lemma:maximal3family} we have that $\mathcal{C}'$ is a cyclic family of $4$-cuts. Thus, there is a partition $\{X_1,X_2,X\}$ of $V(G)$, such that $E[X,X_1]=\{e_1,e_2\}$, $E[X,X_2]=\{e_3,e_4\}$ and $E[X_1,X_2]=\{e,e'\}$ (see Figure~\ref{figure:essential_quasi_triangle}). 
Notice that the connected components of $G\setminus C$ are $X$ and $X_1\cup X_2$. Thus, since $x,y$ are separated by $C$, we may assume w.l.o.g. that $x\in X$ and $y\in X_1$. 

Let $C'=\{e_1,e_2,e,e'\}$. Since $C'$ is generated by $\mathcal{F}'$, we have that $C'$ is a $4$-cut implied by $\mathcal{C}$. Let us suppose, for the sake of contradiction, that $C'$ is implied by $\mathcal{C}$ through the pair of edges $\{e_1,e_2\}$. Then, Lemma~\ref{lemma:not-included-4cut} implies that there is a collection $\mathcal{F}$ of pairs of edges that is returned by Algorithm~\ref{algorithm:generatefamilies} on input $\mathcal{C}$ such that $\{\{e_1,e_2\},\{e,e'\}\}\subseteq\mathcal{F}$. Since $C$ is a quasi $\mathcal{C}$-isolated $4$-cut, we have that the collection of pairs of edges $\mathcal{F}''=\{\{e_1,e_2\},\{e_3,e_4\}\}$ is returned by Algorithm~\ref{algorithm:generatefamilies} on input $\mathcal{C}$. Since $\{e,e'\}\neq\{e_3,e_4\}$, we have that $\mathcal{F}\neq\mathcal{F}''$. But then, Lemma~\ref{lemma:returned_is_disjoint} implies that $\mathcal{F}\cap\mathcal{F}''=\emptyset$, in contradiction to the fact that $\mathcal{F}\cap\mathcal{F}''=\{e_1,e_2\}$. Thus, we have shown that $C'$ is not implied by $\mathcal{C}$ through the pair of edges $\{e_1,e_2\}$. This further implies that $C'\notin\mathcal{C}$.

Since $C'$ is nonetheless implied by $\mathcal{C}$, we may assume w.l.o.g. that $C'$ is implied by $\mathcal{C}$ through the pair of edges $\{e_1,e\}$. Then, Lemma~\ref{lemma:not-included-4cut} implies that there is a collection $\mathcal{F}$ of pairs of edges with $\{\{e_1,e\},\{e_2,e'\}\}\subset\mathcal{F}$ that is returned by Algorithm~\ref{algorithm:generatefamilies} on input $\mathcal{C}$. Thus we have that $|\mathcal{F}|>2$, and $\mathcal{F}$ generates $C'$. Notice that the $4$-cut $C'=\{e_1,e_2,e,e'\}$ separates $x$ and $y$. Thus, the proof is complete.
\end{proof}

\section{Using a DFS-tree for some problems concerning $4$-cuts}
\label{section:using-a-dfs-tree}

In this section we present some applications of identifying $4$-cuts on a DFS-tree. First, there is a linear-time preprocessing, after which we can report the $r$-size of any $4$-cut in constant time (Lemma~\ref{lemma:determine-components}). Second, there is a linear-time preprocessing, after which we can check the essentiality of any $4$-cut in constant time (Proposition~\ref{proposition:essentiality}). And third, given a parallel family of $4$-cuts $\mathcal{C}$, we can compute in linear time the atoms of $\mathcal{C}$ (Proposition~\ref{proposition:algorithm:atoms-of-parallel}), as well as an oracle that can answer queries of the form ``given two vertices $x$ and $y$, return a $4$-cut from $\mathcal{C}$ that separates $x$ and $y$, or determine that no such $4$-cut exists", in constant time (Corollary~\ref{corollary:atoms-of-parallel}).  

Let $G$ be a $3$-edge-connected graph, and let $r$ be a vertex of $G$. Consider the following problems. Given a $4$-cut $C$ of $G$ (as an edge-set), what is the size of the side of $C$ that does not contain $r$? Also, which endpoints of the edges in $C$ lie in the connected component of $G\setminus C$ that contains $r$? We will show how we can answer those questions in constant time, provided that we have computed a DFS-tree of $G$. Given a $4$-cut $C$ of $G$, the number of vertices of the part of $C$ that does not contain $r$ is called the \emph{$r$-size} of $C$. 

So let $T$ be a DFS-tree of $G$ with start vertex (root) $r$ \cite{DBLP:journals/siamcomp/Tarjan72}. We identify the vertices of $G$ with their order of visit by the DFS. Thus, $r=1$, and the last vertex visited by $G$ is $n$. For a vertex $v\neq r$ of $G$, we let $p(v)$ denote the parent of $v$ on $T$. (Thus, $v$ is a child of $p(v)$.) For every two vertices $u$ and $v$, we let $T[u,v]$ denote the simple tree-path from $u$ to $v$. A vertex $v$ is called an ancestor of $u$, if $v$ lies on the tree-path $T[r,u]$. (Equivalently, $u$ is a descendant of $v$.) The set of all descendants of $v$ is denoted as $T(v)$. (In particular, we have $v\in T(v)$.) The number of descendants of $v$ is denoted as $\mathit{ND}(v)$. In other words, $\mathit{ND}(v)=|T(v)|$. The $\mathit{ND}$ values can be computed easily during the DFS, because they satisfy the recursive formula $\mathit{ND}(v)=\mathit{ND}(c_1)+\dots+\mathit{ND}(c_k)+1$, where $c_1,\dots,c_k$ are the children of $v$. 
We can use the $\mathit{ND}$ values in order to check the ancestry relation in constant time. Specifically, given two vertices $u$ and $v$, we have that $u$ is a descendant of $v$ if and only if $v\leq u\leq v+\mathit{ND}(v)-1$. Equivalently, we have $T(v)=\{v,v+1,\dots,v+\mathit{ND}(v)-1\}$.

\subsection{Computing the $r$-size of $4$-cuts}
\label{section:computing-r-size}
Let $C$ be a $4$-cut of $G$. We will show how to answer each of the questions above in constant time. To do this, we first consider the connected components of $T\setminus C$. These are determined by the tree-edges in $C$. Notice that $C$ must contain at least one tree-edge (because otherwise $G\setminus C$ remains connected through the tree-edges from $T$). We distinguish the following cases.

First, let us consider the case that $C$ contains only one tree-edge $(u,p(u))$. Then the connected components of $T\setminus C$ are $T(u)$ and $T(r)\setminus T(u)$. Thus, the $r$-size of $C$ is $\mathit{ND}(u)$. Furthermore, for each non-tree edge $(x,y)\in C$, we can easily determine in constant time which of $x$ and $y$ lies in $T(u)$, and which lies in $T(r)\setminus T(u)$.

Now let us consider the case that $C$ contains exactly two tree-edges $(u,p(u))$ and $(v,p(v))$. By Lemma~\ref{lemma:type3-4cuts} in Section~\ref{subsection:dfs-properties} we have that one of $u$ and $v$ must be an ancestor of the other. Thus, we may assume w.l.o.g. that $v$ is a proper ancestor of $u$. Then the connected components of $T\setminus C$ are given by $T(u)$, $T(v)\setminus T(u)$, and $T(r)\setminus T(v)$. Thus, the connected components of $G\setminus C$ are given by the union of two of those subtrees, plus the other subtree. Now we are guided by the fact that the endpoints of the edges in $C$ lie in different connected components of $G\setminus C$. Thus, $T(u)\cup (T(r)\setminus T(v))$ lies in a distinct connected component of $G\setminus C$ than $T(v)\setminus T(v)$. (Because $u\in T(u)$ whereas $p(u)\in T(v)\setminus T(v)$, and $v\in T(v)\setminus T(u)$ whereas $p(v)\in T(r)\setminus T(v)$.) Therefore, the connected components of $G\setminus C$ are given by $T(u)\cup (T(r)\setminus T(v))$ and $T(v)\setminus T(u)$. This implies that the $r$-size of $C$ is $\mathit{ND}(v)-\mathit{ND}(u)$. Furthermore, it is easy to determine which endpoints of the edges in $C$ lie in which connected components of $G\setminus C$.

Now let us consider the case that $C$ contains exactly three tree-edges $(u,p(u))$, $(v,p(v))$ and $(w,p(w))$. Then, by Lemma~\ref{lemma:type3-4cuts} in Section~\ref{subsection:dfs-properties} we have that one of $u$, $v$ and $w$ must be an ancestor of the other two. Thus, we may assume w.l.o.g. that $w$ is an ancestor of both $u$ and $v$. Now there are two cases to consider: either $u$ and $v$ are not related as ancestor and descendant, or one of $u$ and $v$ is an ancestor of the other. Let us consider the first case first. Then the connected components of $T\setminus C$ are given by $T(u)$, $T(v)$, $T(w)\setminus(T(u)\cup T(v))$ and $T(r)\setminus T(w)$. Then, in order to determine the connected components of $G\setminus C$, we are guided by the property that the endpoints of the edges in $C$ lie in different connected components of $G\setminus C$. Thus, it is not difficult to see that the connected components of $G\setminus C$ are given by $T(u)\cup T(v)\cup (T(r)\setminus T(w))$ and $T(w)\setminus(T(u)\cup T(v))$. Thus, the $r$-size of $C$ is $\mathit{ND}(w)-\mathit{ND}(u)-\mathit{ND}(v)$. Also, given an edge from $C$, it is easy to determine which endpoints of the edges in $C$ lie in which connected component of $G\setminus C$. Now let us consider the case that one of $u$ and $v$ is an ancestor of the other. We may assume w.l.o.g. that $v$ is an ancestor of $u$. Then we can see as previously that the connected components of $G\setminus C$ are $T(u)\cup (T(w)\setminus T(v))$ and $(T(v)\setminus T(u))\cup(T(r)\setminus T(w))$. Thus, the $r$-size of $C$ is $\mathit{ND}(u)+\mathit{ND}(w)-\mathit{ND}(v)$. Furthermore, it is easy to determine which endpoints of the edges in $C$ lie in the connected component of $G\setminus C$ that contains $r$.

In the case that $C$ consists of four tree-edges we follow the same arguments as previously. In each case, the connected components of $G\setminus C$ are given by unions and differences of five subtrees of $T$. Thus, given any vertex $x$, we can check in constant time whether $x$ belongs to the connected component of $G\setminus C$ that contains $r$. Furthermore, we can easily compute the $r$-size of $C$ as previously.

The results of this section are summarized in the following.

\begin{lemma}
\label{lemma:determine-components}
Let $G$ be a $3$-edge-connected graph, and let $r$ be a vertex of $G$. Then there is a linear-time preprocessing of $G$, such that we can answer queries of the form ``given a $4$-cut $C$ of $G$, determine the $r$-size of $C$" and ``given a $4$-cut $C$ of $G$, determine the endpoints of the edges in $C$ that lie in the connected component of $G\setminus C$ that contains $r$", in constant time.
\end{lemma}

\subsection{Checking the essentiality of $4$-cuts}

Here we provide an oracle for performing essentiality checks for $4$-cuts of a $3$-edge-connected graph $G$. Specifically, after a linear-time preprocessing of $G$, if we are given a $4$-cut $C$ of $G$ (as an edge-set), we can determine in constant time whether $C$ is an essential $4$-cut of $G$. 

\begin{proposition}
\label{proposition:essentiality}
Let $G$ be a $3$-edge-connected graph. We can preprocess $G$ in linear time, so that we can perform essentiality checks for $4$-cuts of $G$ in constant time.
\end{proposition}
\begin{proof}
First, we compute the $4$-edge-connected components of $G$. This can be done in linear time (see \cite{DBLP:conf/esa/GeorgiadisIK21} or \cite{DBLP:conf/esa/NadaraRSS21}). Then, for every $4$-edge-connected component $S$ of $G$, we connect all vertices of $S$ in a path, by introducing artificial edges in $G$. Specifically, if $x_1,\dots,x_k$ are the vertices in $S$, then we introduce the artificial edges $(x_1,x_2),(x_2,x_3),\dots,(x_{k-1},x_k)$ into $G$. This takes $O(n)$ time in total, where $n$ is the number of vertices of $G$. Let $G'$ be the resulting graph. (Thus, $G'$ is given by $G$, plus the artificial edges we have introduced.) Then we perform the linear-time preprocessing described in Proposition~\ref{proposition:4e-fault} on $G'$, so that we can answer connectivity queries for pairs of vertices of $G'$, in the presence of at most four edge-failures, in constant time. 

Now let $C$ be a $4$-cut of $G$. Let $(x,y)$ be any edge in $C$. We claim that $C$ is an essential $4$-cut of $G$ if and only if $x$ and $y$ are connected in $G'\setminus C$. To see this, let us assume first that $C$ is an essential $4$-cut of $G$. This means that there are two vertices $u$ and $v$ that are $4$-edge-connected and lie in different connected components of $G\setminus C$. Then, by construction of $G'$, we have that $u$ and $v$ are connected in $G'$. This implies that the two connected components of $G\setminus C$ are connected in $G'\setminus C$. (To be precise: if $X$ and $Y$ are the connected components of $G\setminus C$, then there is at least one edge between $X$ and $Y$ in $G'\setminus C$.) Thus, $G'\setminus C$ is connected, and therefore $x$ and $y$ are connected in $G'\setminus C$. Conversely, suppose that $x$ and $y$ are connected in $G'\setminus C$. Since $C$ is a $4$-cut of $G$, we have that $x$ and $y$ are disconnected in $G\setminus C$. Let $X$ and $Y$ be the connected components of $G\setminus C$. We may assume w.l.o.g. that $x\in X$ and $y\in Y$. Then, since $x$ and $y$ are connected in $G'\setminus C$, there is a path $P$ from $x$ to $y$ in $G'\setminus C$. Since this path starts from a vertex in $X$ and ends in a vertex in $Y$, it must use an edge $(u,v)$ such that $u\in X$ and $v\in Y$. The edge $(u,v)$ does not exist in $G$ (because otherwise $G\setminus C$ would be connected). Thus, it is one of the artificial edges that we have introduced. This means that $u$ and $v$ belong to the same $4$-edge-connected component of $G$. Thus, $C$ separates a pair of $4$-edge-connected vertices of $G$, and therefore it is an essential $4$-cut of $G$.

Thus, we can determine if $C$ is an essential $4$-cut of $G$, by simply checking whether the endpoints of an edge in $C$ are connected in $G'\setminus C$. Since $C$ is a $4$-element set, we can perform this check in constant time, due to the preprocessing of $G'$ according to Proposition~\ref{proposition:4e-fault}.
\end{proof}

\subsection{Computing the atoms of a parallel family of $4$-cuts}
\label{section:computing-atoms}
Let $\mathcal{C}$ be a parallel family of $4$-cuts. This implies that $\mathcal{C}$ has size $O(n)$ (see, e.g., \cite{DBLP:conf/stoc/DinitzN95}). Our goal is to show how to compute efficiently the collection $\mathit{atoms}(\mathcal{C})$. That is, we want to compute the partition $\mathcal{P}$ with the property that two vertices are separated by a set in $\mathcal{P}$ if and only if they are separated by a $4$-cut in $\mathcal{C}$. Here we follow the idea in \cite{DBLP:conf/esa/GeorgiadisIK21}, that computes the $4$-edge connected components of a $3$-edge-connected graph given its collection of $3$-cuts. \cite{DBLP:conf/esa/GeorgiadisIK21} essentialy provides an algorithm to compute $\mathit{atom}(\mathcal{C}_\mathit{3cuts})$. On a high level, the idea is to break the graph into two components according to every $4$-cut $C\in\mathcal{C}$. Specifically, let $C=\{(x_1,y_1),(x_2,y_2),(x_3,y_3),(x_4,y_4)\}$ be a $4$-cut in $\mathcal{C}$. Then, let $X$ and $Y$ be the two connected components of $G\setminus C$, and assume w.l.o.g. that $\{x_1,x_2,x_3,x_4\}\subseteq X$ and $\{y_1,y_2,y_3,y_4\}\subseteq Y$. Then we attach an auxiliary vertex $y$ to $X$, and the edges $(x_1,y),(x_2,y),(x_3,y),(x_4,y)$. Similarly, we attach an auxiliary vertex $x$ to $Y$, and the edges $(x,y_1),(x,y_2),(x,y_3),(x,y_4)$. The purpose of those auxiliary vertices and edges is to simulate for each part $X$ and $Y$ the existence of the other part, while maintaining the same connections. Let $G'$ denote the resulting graph; we call this the result of splitting $G$ according to $C$. Notice that $V(G')=V(G)\sqcup\{x,y\}$. The non-auxiliary vertices of $G'$ (that is, the vertices in $V(G)$) are called ordinary. Now, if $C$ is the unique $4$-cut in $\mathcal{C}$, then $\mathit{atoms}(\mathcal{C})$ is given by the connected components of $G'$. More precisely, we compute the two connected components $X'$ and $Y'$ of $G'$, and we keep the collection of the ordinary vertices from each component (thus, we get $X$ and $Y$). If there are more $4$-cuts in $\mathcal{C}$, then we keep doing the same process, this time splitting $G'$. Due to the parallelicity of $\mathcal{C}$, we have that a $4$-cut $C'\in\mathcal{C}$ with $C'\neq C$ lies entirely within a connected component of $G'$. And due to the construction of $G'$, we have that (the edge-set corresponding to) $C'$ is a $4$-cut of $G'$. Thus, it makes sense to split $G'$ according to its $4$-cut $C'$. When no more splittings are possible (i.e., when we have used every $4$-cut in $\mathcal{C}$ for a splitting), then we compute the connected components of the final graph, and we collect the subsets of the ordinary vertices from each connected component. Thus, we get $\mathit{atoms}(\mathcal{C})$. 

In order to prove the correctness of the above procedure, we need to formalize the concept of splitting a graph according to a $4$-cut. A key-concept that we will use throughout is the quotient map to a split graph.

\begin{figure}[h!]\centering
\includegraphics[trim={0 33cm 0 0}, clip=true, width=0.9\linewidth]{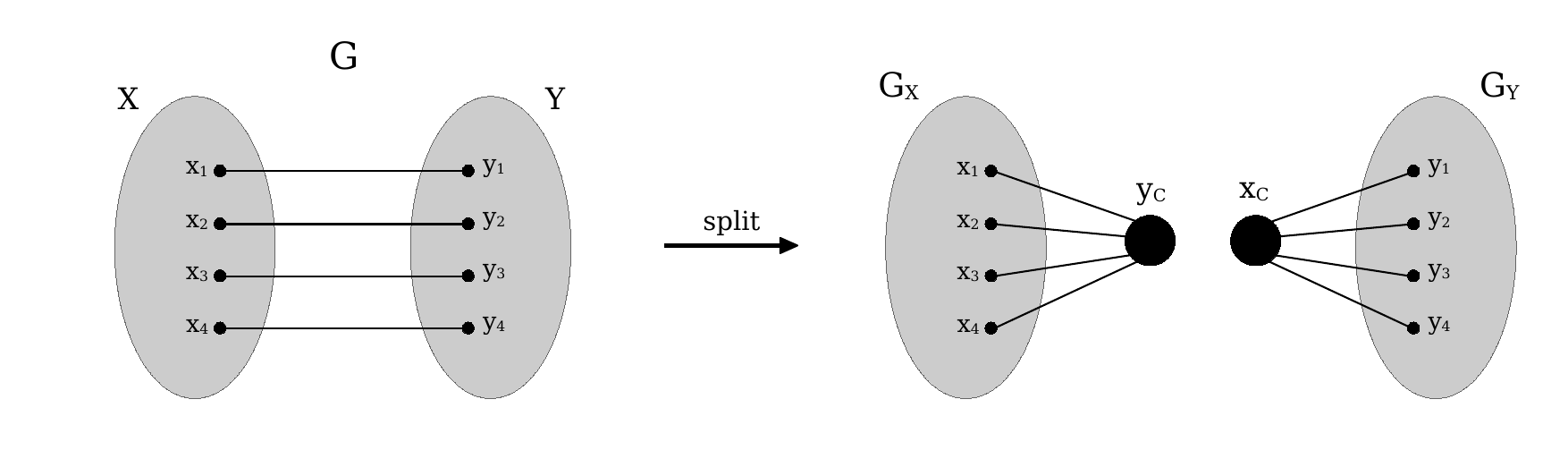}
\caption{\small{Splitting a graph $G$ according to a $4$-cut $C=\{(x_1,y_1),(x_2,y_2),(x_3,y_3),(x_4,y_4)\}$ with sides $X$ and $Y$. We introduce two new auxiliary vertices $x_C$ and $y_C$, that simulate the parts $X$ and $Y$, respectively.}}\label{figure:split_4cut}
\end{figure}

\begin{definition}[Splitting a graph according to a $4$-cut] 
\label{definition:splitting-a-graph}
\normalfont
Let $G$ be a connected graph, let $C=\{(x_1,y_1),(x_2,y_2),(x_3,y_3),(x_4,y_4)\}$ be a $4$-cut of $G$, and let $X$ and $Y$ be the two sides of $C$. We may assume w.l.o.g. that $\{x_1,x_2,x_3,x_4\}\subseteq X$ and $\{y_1,y_2,y_3,y_4\}\subseteq Y$. Then we define the two split graphs $G_X$ and $G_Y$ of $G$ according to $C$ as follows. We introduce two auxiliary vertices $x_C$ and $y_C$ (that simulate the parts $X$ and $Y$, respectively). Then the vertex set of $G_X$ is $X\cup\{y_C\}$, and the edge set of $G_X$ is $E(G[X])\cup\{(x_1,y_C),(x_2,y_C),(x_3,y_C),(x_4,y_C)\}$. Similarly, the vertex set of $G_Y$ is $Y\cup\{x_C\}$, and the edge set of $G_Y$ is $E(G[Y])\cup\{(x_C,y_1),(x_C,y_2),(x_C,y_3),(x_C,y_4)\}$. (See Figure~\ref{figure:split_4cut} for an example.) 

$X$ is called the set of the ordinary vertices of $G_X$, and $Y$ is called the set of the ordinary vertices of $G_Y$. Notice that $G_X$ and $G_Y$ is the quotient graph that is formed from $G$ by shriking $Y$ and $X$, respectively, into a single node. Then we also define the quotient maps $q_X$ and $q_Y$ from $V(G)$ to $V(G_X)$ and $V(G_Y)$, respectively. $q_X$ coincides with the identity map on $X$, and it sends $Y$ onto $y_C$. (In other words, $q_X(v)=v$ for every $v\in X$, and $q_X(v)=y_C$ for every $v\in Y$.) Similarly, $q_Y$ coincides with the identity map on $Y$, and it sends $X$ onto $x_C$. (In other words, $q_Y(v)=v$ for every $v\in Y$, and $q_Y(v)=x_C$ for every $v\in X$.) These maps induce a natural correspondence between edges of $G$ and edges of $G_X$ and $G_Y$. Specifically, for every edge $(u,v)$ of $G$, we let $q_X((u,v)):=(q_X(u),q_X(v))$ and $q_Y((u,v)):=(q_Y(u),q_Y(v))$.\footnote{More precisely, since $G$ is a multigraph, every edge $e=(u,v)$ of $G$ has a unique edge-identifier $i$. Thus, we can consider this edge as a triple $(u,v,i)$. Then $q_X$ maps $(u,v,i)$ into $(q_X(u),q_X(v),i)$, so that, if $q_X(u)\neq q_X(v)$, then $q_X(e)$ is a unique edge of $G_X$ (i.e., it is not the image of any other edge of $G$ through $q_X$). However, in order to keep our notation and our arguments simple, we will drop this consideration, and we will keep considering the edges of $G$ as pairs of vertices.}  
\end{definition}

The following lemma shows that the operation of splitting a graph according to a $4$-cut maintains families of parallel $4$-cuts inside the split graphs.

\begin{lemma}
\label{lemma:split-graph}
Let $C$ be a $4$-cut of $G$, let $X$ and $Y$ be the two sides of $C$, and let $(G_X,q_X)$ and $(G_Y,q_Y)$ be the corresponding split graphs of $G$ according to $C$, together with the respective quotient maps. Then we have the following.
\begin{enumerate}[label=(\alph*)]
\item{Let $C'$ be a $4$-cut of $G$, distinct from $C$, that is parallel with $C$. Then one of $q_X(C')$ and $q_Y(C')$ contains at least one self-loop, and the other is a set of four edges.} 
\item{Let $C'$ be as in $(a)$, and suppose that $q_X(C')$ is a set of four edges (of $G_X$). Then $q_X(C')$ is a $4$-cut of $G_X$. Let $X'$ and $Y'$ be the sides of $q_X(C')$ in $G_X$. Then $q_X^{-1}(X')$ and $q_X^{-1}(Y')$ are the two sides of $C'$ in $G$.}
\item{Let $C_1$ and $C_2$ be two $4$-cuts of $G$, distinct from $C$, that are parallel with $C$ and among themselves. Suppose that $q_X(C_1)$ and $q_X(C_2)$ are $4$-cuts of $G_X$. Then $q_X(C_1)$ and $q_X(C_2)$ are two distinct parallel $4$-cuts of $G_X$.}
\end{enumerate}
\end{lemma}
\begin{proof}
By Definition~\ref{definition:splitting-a-graph}, there are two auxiliary vertices $x_C$ and $y_C$, such that $V(G_X)=X\cup\{y_C\}$, $V(G_Y)=Y\cup\{x_C\}$, $q_X$ coincides with the identity map on $X$ and $q_X(Y)=\{y_C\}$, and $q_Y$ coincides with the identity map on $Y$ and $q_Y(X)=\{x_C\}$.

\textbf{(a)} Let $X'$ and $Y'$ be the two sides of $C'$ in $G$. Then, since $C$ and $C'$ are parallel, we have that one of $X'$ and $Y'$ lies entirely within $X$ or $Y$. Thus, we may assume w.l.o.g. that $X'\subset X$. (We have strict inclusion, because $C'\neq C$.) Let $(x,y)$ be an edge in $C'$. Then we may assume w.l.o.g. that $x\in X'$ and $y\in Y'$. Since $X'\subset X$, we have $q_X(x)=x$. If $y\in X$, then we have $q_X(y)=y$. Otherwise, we have $q_X(y)=y_C$. Thus, in either case we have $q_X(x)\neq q_X(y)$, and therefore $q_X((x,y))$ is an edge of $G_X$. This shows that $q_X(C')$ is a set of four edges of $G_X$. 

On the other hand, let $C'=\{(x_1,y_1),(x_2,y_2),(x_3,y_3),(x_4,y_4)\}$, and let us assume w.l.o.g. that $\{x_1,x_2,x_3,x_4\}\subseteq X'$ and $\{y_1,y_2,y_3,y_4\}\subseteq Y'$. Then, since $X'\subset X$, for every $i\in\{1,2,3,4\}$ we have $q_Y(x_i)=x_C$. Since $E_G[X,Y]=C$ and $C\neq C'$, there must be an $i\in\{1,2,3,4\}$ such that $(x_i,y_i)\notin E_G[X,Y]$. Thus, since $x_i\in X$, we have $y_i\notin Y$, and therefore $y_i\in X$. This implies that $q_Y(y_i)=x_C$, and therefore $q_Y((x_i,y_i))$ is a self-loop.

\textbf{(b)} Let $X'$ and $Y'$ be the two sides of $C'$ in $G$. Then, since $C$ and $C'$ are parallel, we have that one of $X'$ and $Y'$ lies entirely within $X$ or $Y$. Since $q_X(C')$ is a set of four edges of $G_X$, by $(a)$ we have that $q_Y(C')$ contains at least one self-loop. Then, by following the argument of $(a)$, we have that it cannot be that one of $X'$ and $Y'$ lies entirely within $Y$ (because then we would have that $q_Y(C')$ consists of four edges of $G_Y$). Thus, one of $X'$ and $Y'$ lies entirely within $X$. Then, we may assume w.l.o.g. that $X'\subset X$. This implies that $Y\subset Y'$.

Now we will establish a correspondence between paths in $G$ and paths in $G_X$, that satisfies some useful properties. Let $P$ be a path in $G$ with endpoints $x$ and $y$. We define a path $\widetilde{P}$ in $G_X$ as follows. First, suppose that both $x$ and $y$ are in $X$. If $P$ uses edges only from $G[X]$, then $\widetilde{P}=P$. Otherwise, let $v_1,(v_1,v_2),v_2,\dots,(v_{k-1},v_k),v_k$ be a maximal segment of $P$ such that $v_1\in X$, $\{v_2,\dots,v_{k-1}\}\subseteq Y$, and $v_k\in X$. (Notice that, since $P$ starts from $X$ and ends in $X$, there must exist such a maximal segment of $P$, and it has $k\geq 3$.) Then we replace this segment with $v_1,(v_1,y_C),y_C,(y_C,v_k),v_k$. We repeat this process until we arrive at a sequence $\widetilde{P}$ of alternating vertices and edges that does not use vertices from $Y$. Then we can see that $\widetilde{P}$ is a path in $G_X$ from $x$ to $y$. Furthermore, $\widetilde{P}$ has the following properties.
First, every occurrence of a vertex from $X$ in $P$ is maintained in $\widetilde{P}$. Second, every maximal segment of occurrences of vertices from $Y$ in $P$ is replaced by a single occurrence of $y_C$. Third, every occurrence of an edge $(z,w)$ such that not both $z$ and $w$ are in $Y$, is replaced by an occurrence of $(q_X(z),q_X(w))$. And fourth, all the vertices and edges used by $\widetilde{P}$ are essentially given by the previous three properties.

Now suppose that one of $x$ and $y$ lies in $X$, and the other lies in $Y$. Then let us suppose that $x\in X$ and $y\in Y$ (in the other case, we have a similar definition and properties). Then, let $w_1,(w_1,w_2),w_2,\dots,(w_{l-1},w_l),w_l$ be the final segment of $P$ that satisfies $w_1\in X$ and $\{w_2,\dots, w_l\}\subseteq Y$. (Notice that, since $P$ starts from $X$ and ends in $Y$, this is indeed the form of the final part of $P$, and it has $l\geq 2$.) Then we replace this segment with $w_1,(w_1,y_C),y_C$. Then, we perform the substitutions that we described previously for segments of $P$ of the form $v_1,(v_1,v_2),v_2,\dots,(v_{k-1},v_k),v_k$ that are maximal w.r.t. $v_1\in X$, $\{v_2,\dots,v_{k-1}\}\subseteq Y$, and $v_k\in X$. Let $\widetilde{P}$ be the result after we have applied all those substitutions. Then we can see that $\widetilde{P}$ is a path in $G_X$ from $x$ to $y_C$. Furthermore, $\widetilde{P}$ satisfies the four properties that we described previously (that essentially define $\widetilde{P}$).

Now let $Y''=q_X(Y')$. (Thus, we have $Y''=(Y'\cap X)\cup\{y_C\}$.) Notice that $q_X^{-1}(X')=X'$ and $q_X^{-1}(Y'')=Y'$. We will show that $q_X(C')$ is a $4$-cut of $G_X$ with sides $X'$ and $Y''$. 
First, notice that $\{X',Y''\}$ constitutes a partition of $V(G_X)$, and $E_{G_X}[X',Y'']=q_X(C')$. Thus, it is sufficient to show that both $X'$ and $Y''$ induce a connected subgraph of $G_X\setminus q_X(C')$. We will derive this result as a consequence of the correspondence between paths in $G$ and paths in $G_X$.

So let $x$ and $y$ be two vertices in $X'$. Then, since $X'$ is a connected component of $G\setminus C'$, there is a path $P$ from $x$ to $y$ in $G\setminus C'$ that uses vertices only from $X'$. Thus, $\widetilde{P}$ is a path from $x$ to $y$ that avoids the edges from $q_X(C')$ and uses vertices only from $X'$. This shows that $X'$ induces a connected subgraph of $G_X\setminus q_X(C')$.

Now let $x$ and $y$ be two vertices in $Y''$. Let us assume first that none of $x$ and $y$ is $y_C$. (Thus, we have that both $x$ and $y$ are in $X$.) Since $Y'$ is a connected component of $G\setminus C'$, there is a path $P$ from $x$ to $y$ in $G\setminus C'$ that uses vertices only from $Y'$. Then $\widetilde{P}$ is a path from $x$ to $y$ that uses vertices only from $q_X(Y')$ and avoids the edges from $q_X(C')$. Now let us assume that one of $x$ and $y$ is $y_C$, and the other is not. Then we may assume w.l.o.g. that $y=y_C$ and $x\in Y'$. Since $x$ is a vertex in $V(G_X)\setminus\{y_C\}$, notice that $x\in X$. Now let $y_0$ be a vertex in $Y$. Then, since $Y'$ is a connected component of $G\setminus C'$ that contains $Y$, there is a path $P$ from $x$ to $y_0$ in $G\setminus C'$ that uses vertices only from $Y'$. Then $\widetilde{P}$ is a path from $x$ to $y_C$ that uses vertices only from $q(Y')$ and avoids the edges from $q_X(C')$. Thus, in either case we have that $x$ and $y$ are connected in $G_X\setminus q_X(C')$ through a path that uses vertices only from $Y''$. This shows that $Y''$ induces a connected subgraph of $G_X\setminus q_X(C')$.
Thus, we have that $q_X(C')$ is a $4$-cut of $G_X$, with sides $X'$ and $Y''$. Since we have $q_X^{-1}(X')=X'$ and $q_X^{-1}(Y'')=Y'$, this completes the proof. 

\textbf{(c)} Let $X_1$ and $Y_1$ be the sides of $q_X(C_1)$ in $G_X$, and let $X_2$ and $Y_2$ be the sides of $q_X(C_2)$ in $G_X$. Then, by $(b)$ we have that $q_X^{-1}(X_1)$ and $q_X^{-1}(Y_1)$ are the two sides of $C_1$ in $G$, and $q_X^{-1}(X_2)$ and $q_X^{-1}(Y_2)$ are the two sides of $C_2$ in $G$. Thus, since $C_1$ and $C_2$ are distinct, we have $\{q_X^{-1}(X_1),q_X^{-1}(Y_1)\}\neq\{q_X^{-1}(X_2),q_X^{-1}(Y_2)\}$, and therefore $\{X_1,Y_1\}\neq\{X_2,Y_2\}$. This means that $q_X(C_1)$ and $q_X(C_2)$ are distinct $4$-cuts of $G_X$. Since $C_1$ and $C_2$ are parallel $4$-cuts of $G$, at least one of the intersections $q_X^{-1}(X_1)\cap q_X^{-1}(X_2)$, $q_X^{-1}(X_1)\cap q_X^{-1}(Y_2)$, $q_X^{-1}(Y_1)\cap q_X^{-1}(X_2)$, $q_X^{-1}(Y_1)\cap q_X^{-1}(Y_2)$ is empty, and therefore at least one of the inverse images $q_X^{-1}(X_1\cap X_2)$, $q_X^{-1}(X_1\cap Y_2)$, $q_X^{-1}(Y_1\cap X_2)$, $q_X^{-1}(Y_1\cap Y_2)$ is empty. Since $q_X:G\rightarrow G_X$ is a surjective map, this implies that at least one of the intersections $X_1\cap X_2$, $X_1\cap Y_2$, $Y_1\cap X_2$, $Y_1\cap Y_2$ is empty. This means that the $4$-cuts $q_X(C_1)$ and $q_X(C_2)$ are parallel. 
\end{proof}

Lemma~\ref{lemma:split-graph} implies that if we have a parallel family $\mathcal{C}$ of $4$-cuts, then we can successively partition the graph according to all $4$-cuts in $\mathcal{C}$. This is made precise in the following.

\begin{definition}[Splitting a graph according to a parallel family of $4$-cuts] 
\label{definition:split-graphs}
\normalfont
Let $G$ be a connected graph, let $\mathcal{C}$ be a parallel family of $4$-cuts of $G$, and let $C$ be a $4$-cut in $\mathcal{C}$. Let $X$ and $Y$ be the sides of $C$ in $G$, and let $(G_X,q_X)$ and $(G_Y,q_Y)$ be the corresponding split graphs, together with the respective quotient maps. Then, Lemma~\ref{lemma:split-graph} implies that $q_X$ maps some of the $4$-cuts from $\mathcal{C}\setminus\{C\}$ into a collection $\mathcal{C}_X$ of parallel $4$-cuts of $G_X$, and $q_Y$ maps the remaining $4$-cuts from $\mathcal{C}\setminus\{C\}$ into a collection $\mathcal{C}_Y$ of parallel $4$-cuts of $G_Y$. Then we can repeat the same process into $G_X$ and $G_Y$, with the collections of $4$-cuts $\mathcal{C}_X$ and $\mathcal{C}_Y$, respectively. Let $G_1,\dots,G_k$ be the final split graphs that we get, after we have completed this process. (We note that $k=|\mathcal{C}|+1$.) For every $i\in\{1,\dots,k\}$, we denote $V(G_i)\cap V(G)$ as $G_i^{o}$, and we call it the set of the ordinary vertices of $G_i$. 

Every split graph $G_i$, for $i\in\{1,\dots,k\}$, comes together with the respective quotient map $q_i:V(G)\rightarrow V(G_i)$, that is formed by the repeated composition of the quotient maps that we used in order to arrive at $G_i$. 
More precisely, let $G'$ be one of the split graphs in $\{G_1,\dots,G_k\}$. Then there is a sequence $C_1,\dots,C_t$ of $4$-cuts from $\mathcal{C}$, for a $t\geq 1$, and a sequence $G_0',\dots,G_t'$ of graphs, such that $G_0'=G$, $G_t'=G'$, and $G_i'$ is derived from the splitting of $G_{i-1}'$ according to $C_i$, for every $i\in\{1,\dots,t\}$. Then, according to Definition~\ref{definition:splitting-a-graph}, we get a quotient map $q_i':V(G_{i-1}')\rightarrow V(G_i')$ for every $i\in\{1,\dots,t\}$, that corresponds to the splitting of $G_{i-1}'$ into $G_i'$ according to $C_i$. Then the composition $q_t'\circ\dots\circ q_1'$ is the quotient map from $V(G)$ to $V(G')$. 
\end{definition}

Notice that the split graphs that we get in Definition~\ref{definition:split-graphs} from a parallel family $\mathcal{C}$ of $4$-cuts depend on the order in which we use the $4$-cuts from $\mathcal{C}$ in order to perform the splittings. However, this order is irrelevant if we only care about deriving the atoms of $\mathcal{C}$, as shown in the following.

\begin{lemma}
\label{lemma:get-atoms}
Let $G$ be a connected graph, let $\mathcal{C}$ be a parallel family of $4$-cuts of $G$, and let $G_1,\dots,G_k$ be the split graphs that we get from $G$ by splitting it according to the $4$-cuts from $\mathcal{C}$ (in any order). Then $\mathit{atoms}(\mathcal{C})=\{G_1^{o},\dots,G_k^{o}\}$.
\end{lemma}
\begin{proof}
First, we note that when we use a $4$-cut in order to split a graph $G$ into two graphs $G_X$ and $G_Y$, we have that $\{G_X^{o},G_Y^{o}\}$ is a partition of $V(G)$. Thus, since the collection of graphs $\{G_1,\dots,G_k\}$ is formed by repeated splittings of $G$, we have that $\{G_1^{o},\dots,G_k^{o}\}$ is a partition of $V(G)$.
 
Now let $x$ and $y$ be two vertices of $G$ that belong to different sets in $\mathit{atoms}(\mathcal{C})$. Then there is a $4$-cut $C\in\mathcal{C}$ that separates $x$ and $y$. Thus, we may consider the first $4$-cut $C\in\mathcal{C}$ that we used for the splittings and has the property that it separates $x$ and $y$. Then there is a sequence $C_1,\dots,C_t$ of $4$-cuts from $\mathcal{C}$, with $t\geq 0$, that were succesively used in order to split $G$, until we arrived at a split graph $G'$ with the property $x,y\in V(G')$, and it was time to split $G'$ using $C$. (We allow $t=0$, because this corresponds to the case that $C$ is the first $4$-cut from $\mathcal{C}$ that was used in order to split $G$.) Let $G_X$ and $G_Y$ be the two split graphs that we get by splitting $G'$ according to $C$ (more precisely: according to the image of $C$ in $G'$ through the quotient map). Then, since $C$ is a $4$-cut of $G$ that separates $x$ and $y$, as a consequence of Lemma~\ref{lemma:split-graph} we have that (the image of) $C$ separates $x$ and $y$ in $G'$. Thus, w.l.o.g., we may assume that $x$ is a vertex of $G_X$, but not of $G_Y$, and $y$ is a vertex of $G_Y$, but not of $G_X$. Then, we have that the graph in $G_1,\dots,G_k$ that contains $x$ is either $G_X$, or it is derived by further splitting $G_X$. Similarly, the graph in $G_1,\dots,G_k$ that contains $y$ is either $G_Y$, or it is derived by further splitting $G_Y$. This implies that $x$ and $y$ belong to different sets from $\{G_1^{o},\dots,G_k^{o}\}$.

Conversely, let $x$ and $y$ be two vertices that belong to different sets from $\{G_1^{o},\dots,G_k^{o}\}$. Thus, there is a sequence $C_1,\dots,C_t$ of $4$-cuts from $\mathcal{C}$, with $t\geq 1$, that led to the separation of $x$ and $y$ into different split graphs. More precisely, there is a sequence $C_1,\dots,C_t$ of $4$-cuts from $\mathcal{C}$, with $t\geq 1$, and a sequence of graphs $G_0,G_1,\dots,G_t$, such that: $(1)$ $G_0=G$, $(2)$ $G_i$ was derived from the splitting of $G_{i-1}$ according to $C_i$, for every $i\in\{1,\dots,t\}$, and $(3)$ $G_{t-1}$ contains both $x$ and $y$, but $G_t$ contains only one of $x$ and $y$. Due to $(3)$, we may assume w.l.o.g. that $x\in V(G_t)$ and $y\notin V(G_t)$. We will show that $C_t$ separates $x$ and $y$ in $G$. Let $q_i$ be the quotient map from $V(G_{i-1})$ to $V(G_i)$, for every $i\in\{1,\dots,t\}$, and let $q_0$ be the identity map on $V(G)$. Thus, by $(2)$ we have that $q_{i-1}(\dots q_0(C_i)\dots)$ is a $4$-cut of $G_{i-1}$, for every $i\in\{1,\dots,t\}$. Let $C_t'=q_{t-1}(\dots q_0(C_t)\dots)$, and let $X$ be the side of $C_t'$ in $G_{t-1}$ from which $G_t$ is derived. Then, by a repeated application of Lemma~\ref{lemma:split-graph} we get that $X'=q_1^{-1}(\dots q_{t-1}^{-1}(X)\dots)$ is one of the sides of $C_t$ in $G$. Then, since $x\in X$, we have $x\in X'$. (Because the inverses of the quotient mappings maintain the vertices from $V(G)$.) Since $y\notin V(G_t)$, we have $y\notin X$. We claim that $y\notin X'$. To see this, assume the contrary. Then, since all graphs $G_0,\dots,G_{t-1}$ contain $y$, we have that the only vertex $z\in V(G_{t-1})$ that satisfies $q_1^{-1}(\dots q_{t-1}^{-1}(z)\dots)=y$ is $z=y$. But then, since $y\in X'=q_1^{-1}(\dots q_{t-1}^{-1}(X)\dots)$, this implies that $y\in X$, a contradiction. This shows that $y\notin X'$. We conclude that $C_t$ separates $x$ and $y$ in $G$, and therefore $x$ and $y$ belong to different sets in $\mathit{atoms}(\mathcal{C})$.
\end{proof}

Thus, in order to compute the atoms of $\mathcal{C}$, it is sufficient to split $G$ according to the $4$-cuts in $\mathcal{C}$, and then collect the sets of ordinary vertices of the split graphs. In order to implement this idea efficiently, as in \cite{DBLP:conf/esa/GeorgiadisIK21} we have to take care of three things. First, given a $4$-cut $C$ in $\mathcal{C}$, we need to know the distribution of the endpoints of the edges of $C$ in the connected components of $G\setminus C$. We want to achieve this without explicitly computing the connected components of $G\setminus C$. Second, for every $4$-cut that we process, we have to be able to determine the split graph that contains it. And third, since every splitting removes the edges of a $4$-cut and substitutes them with new auxiliary edges, now given a new $4$-cut $C$ for splitting, we must know if some of its edges correspond to auxiliary edges of the split graph (so that we have to remove its auxiliary counterparts, and not the original edges of $C$). We solve these problems by locating the $4$-cuts from $\mathcal{C}$ on a DFS-tree rooted at $r$, and by processing them in increasing order w.r.t. their $r$-size. By locating a $4$-cut $C$ on the DFS-tree we can determine easily in constant time how the endpoints of the edges of $C$ are separated in the connected components of $G\setminus C$, according to Lemma~\ref{lemma:determine-components}. And by processing the $4$-cuts from $\mathcal{C}$ in increasing order w.r.t. their $r$-size, we can be certain that whenever we process a $4$-cut, this essentially lies within the split graph that contains $r$, as shown in the following. 

\begin{lemma}
\label{lemma:split-in-increasing-r-size}
Let $G$ be a connected graph, let $r$ be a vertex of $G$, and let $\mathcal{C}$ be a parallel family of $4$-cuts of $G$. Let $C$ be a $4$-cut of $G$ that is not in $\mathcal{C}$, such that $C$ is parallel with every $4$-cut in $\mathcal{C}$, and the $r$-size of $C$ is at least as great as the maximum $r$-size of all $4$-cuts in $\mathcal{C}$. Suppose that we split $G$ according to the $4$-cuts in $\mathcal{C}$, and let $(G',q')$ be the split graph that contains $r$, together with the respective quotient map. Then $q'(C)$ is a $4$-cut of $G'$.
\end{lemma}
\begin{proof}
Let $(G_1,q_1),\dots,(G_k,q_k)$ be all the split graphs, together with the respective quotient maps, that we get after splitting $G$ according to the $4$-cuts in $\mathcal{C}$. (We note that $q_i$ is a map from $G$ to $G_i$, for every $i\in\{1,\dots,k\}$.) Then it is a consequence of Lemma~\ref{lemma:split-graph} that there is a unique $(G',q')\in\{(G_1,q_1),\dots,(G_k,q_k)\}$  such that $q'(C)$ is a $4$-cut of $G'$ (because, in all other cases, we have that the image of $C$ through a quotient map contains at least one self-loop). By construction of the split graphs, we have that $\{V(G_i)\cap V(G)\mid i\in\{1,\dots,k\}\}$ is a partition of $V(G)$. Thus, there is only one split graph that contains $r$. We will show that $G'$ is the graph that contains $r$. To do this, we will use induction on the number of splittings that we had to perform in order to reach the split graph in which $C$ is mapped as a $4$-cut.

As the base step of our induction, we can consider the case that no splittings took place at all. In this case, we let the ``quotient" map $q'$ be the identity map on $V(G)$. Then, it is obviously true that $q'(C)$ is a $4$-cut of $G$, and $G$ is the ``split" graph that contains $r$. Now let us assume that we have performed $t$ consecutive splittings, for $t\geq 0$, which resulted in a graph $G_0$ with quotient map $q_0$, with the property that $r\in V(G_0)$ and $q_0(C)$ is a $4$-cut of $G_0$. Now suppose that we split $G_0$ once more according to a $4$-cut $C'\in\mathcal{C}$. Thus, we have that $q_0(C')$ is a $4$-cut of $G_0$, and let $X$ and $Y$ be the two sides of $q_0(C')$. Let $(G_X,q_X)$ and $(G_Y,q_Y)$ be the resulting split graphs, together with the respective quotient maps. Then there are two auxiliary vertices $x_{C'}$ and $y_{C'}$, such that $V(G_X)=X\cup\{y_{C'}\}$, $V(G_Y)=Y\cup\{x_{C'}\}$, $q_X(Y)=\{y_{C'}\}$ and $q_Y(X)=\{x_{C'}\}$. We may assume w.l.o.g. that $r\in X$. Thus, we have $q_Y(r)=x_{C'}$. 

Since $q_0(C)$ is a $4$-cut of $G_0$ that is parallel with $q_0(C')$, by Lemma~\ref{lemma:split-graph} we have one of $q_X(q_0(C))$ and $q_Y(q_0(C))$ contains a self-loop, and the other is a set of four edges. So let us suppose, for the sake of contradiction, that $q_X(q_0(C))$ contains a self-loop. Then, Lemma~\ref{lemma:split-graph} implies that $q_Y(q_0(C))$ is a $4$-cut of $G_Y$. Let $X'$ and $Y'$ be the two sides of $q_Y(q_0(C))$ (in $G_Y$). Then, Lemma~\ref{lemma:split-graph} implies that $q_Y^{-1}(X')$ and $q_Y^{-1}(Y')$ are the two sides of $q_0(C)$ (in $G_0$). We may assume w.l.o.g. that $q_Y^{-1}(X')$ is the side of $q_0(C)$ that contains $r$. This implies that $x_{C'}\in X'$, and therefore $q_Y^{-1}(X')$ contains $X$. Since $C$ and $C'$ are distinct $4$-cuts of $G$, Lemma~\ref{lemma:split-graph} implies that $q_0(C)$ and $q_0(C')$ are distinct $4$-cuts of $G_0$. Thus, we cannot have $q_Y^{-1}(X')=X$, and therefore we have that $q_Y^{-1}(X')$ contains $X$ as a proper subset. This implies that $q_Y^{-1}(Y')$ is a proper subset of $Y$. By Lemma~\ref{lemma:split-graph} we have that $q_0^{-1}(X)$ and $q_0^{-1}(Y)$ are the two sides of $C'$ (in $G$). Thus, since $r\in X$, we have that $r\in q_0^{-1}(X)$, and therefore the $r$-size of $C'$ is $|q_0^{-1}(Y)|$. Similarly, by Lemma~\ref{lemma:split-graph} we have that $q_0^{-1}(q_Y^{-1}(X'))$ and $q_0^{-1}(q_Y^{-1}(Y'))$ are the two sides of $C$ (in $G$). Thus, since $r\in q_Y^{-1}(X')$, we have that $r\in q_0^{-1}(q_Y^{-1}(X'))$, and therefore the $r$-size of $C$ is $|q_0^{-1}(q_Y^{-1}(Y'))|$. Since $q_Y^{-1}(Y')$ is a proper subset of $Y$, we have that $q_0^{-1}(q_Y^{-1}(Y'))$ is a proper subset of $q_0^{-1}(Y)$ (because $q_0$ is a surjective function). This implies that $|q_0^{-1}(q_Y^{-1}(Y'))|<|q_0^{-1}(Y)|$, in contradiction to the fact that the $r$-size of $C$ is at least as great as the $r$-size of $C'$. This shows that $q_X(q_0(C))$ does not contain a self-loop. Thus, by Lemma~\ref{lemma:split-graph} we have that $q_X(q_0(C))$ is a $4$-cut of $G_X$. This shows that, after $t+1$ splittings, $C$ is mapped as a $4$-cut to the split graph that contains $r$. Thus, the lemma follows inductively.
\end{proof}


Thus, if we process the $4$-cuts from $\mathcal{C}$ in increasing order w.r.t. their $r$-size, then we can be certain that every $4$-cut $C\in\mathcal{C}$ that we process lies within the split graph $G'$ that contains $r$. Therefore, it is sufficient to maintain only the quotient map to $G'$. We use $v'$ to denote the image of a vertex $v\in V(G)$ to $G'$, and we let $C'$ denote the translation of $C$ within $G'$ through the quotient map. Thus, whenever we process a $4$-cut $C\in\mathcal{C}$, we translate the endpoints of every edge $(x,y)\in C$ to their corresponding vertices $x'$ and $y'$ in $G'$. Then we delete $(x',y')$ from $G'$, and we substitute it with two auxiliary edges $(x',y_C)$ and $(x_C,y')$ (i.e., we create one copy of $(x',y')$ for each of the connected components of $G'\setminus C$). Let $(x_C,y')$ be the copy of $(x',y')$ that is contained in the connected component of $G'\setminus C'$ that contains $r$. Then we update the pointer of $x$ to $G'$ as $x'\leftarrow x_C$.

The procedure for computing $\mathit{atoms}(\mathcal{C})$ is shown in Algorithm~\ref{algorithm:atoms-of-parallel}. The proof of correctness and linear complexity is given in Proposition~\ref{proposition:algorithm:atoms-of-parallel}. As a corollary of this method for computing $\mathit{atoms}(\mathcal{C})$, we can construct in linear time a data structure that we can use in order to answer queries of the form ``given two vertices $x$ and $y$, determine whether $x$ and $y$ are separated by a $4$-cut in $\mathcal{C}$, and, if yes, report a $4$-cut in $\mathcal{C}$ that separates them" in constant time. This is proved in Corollary~\ref{corollary:atoms-of-parallel}. We note that these results are essentially independent of the fact that we consider $4$-cuts, and so they generalize to any parallel family of cuts (of various cardinalities), provided that there is a fixed upped bound on their number of edges.

\begin{algorithm}[h!]
\caption{\textsf{Compute $\mathit{atoms}(\mathcal{C})$ of a parallel family of $4$-cuts $\mathcal{C}$}}
\label{algorithm:atoms-of-parallel}
\LinesNumbered
\DontPrintSemicolon
sort the $4$-cuts in $\mathcal{C}$ in increasing order w.r.t. their $r$-size\;
\label{line:atoms-sort}
\ForEach{vertex $v$}{
\tcp{initialize the pointers of the vertices to the split graph that contains $r$}
  set $v'\leftarrow v$\;
}
let $G'\leftarrow G$ \tcp{we maintain throughout the collection of the split graphs as a single graph $G'$}
\label{line:atoms-initialize}
\ForEach{$C\in\mathcal{C}$}{
\label{line:atoms-for}
  let $C=\{(x_1,y_1),(x_2,y_2),(x_3,y_3),(x_4,y_4)\}$\;
  determine the endpoints of the edges in $C$ that are in the connected component of $G\setminus C$ 
  that contains $r$; let those endpoints be $y_1,y_2,y_3,y_4$\;
  \label{line:atoms-determine}
  remove the edges $(x_1',y_1'),(x_2',y_2'),(x_3',y_3'),(x_4',y_4')$ from $G'$\;
  \label{line:atoms-delete}
  insert two new vertices $x_C$ and $y_C$ to $G'$\;
  insert the edges $(x_1',y_C),(x_2',y_C),(x_3',y_C),(x_4',y_C)$ and
  $(x_C,y_1'),(x_C,y_2'),(x_C,y_3'),(x_C,y_4')$ to $G'$\;
  \label{line:atoms-insert}
  set $x_1'\leftarrow x_C$, $x_2'\leftarrow x_C$, $x_3'\leftarrow x_C$, $x_4'\leftarrow x_C$\;
}
compute the connected components $S_1,\dots,S_k$ of $G'$\;
\label{line:atoms-components}
\ForEach{$i\in\{1,\dots,k\}$}{
\label{line:atoms-for-2}
  let $S_i'$ be the set of the ordinary vertices in $S_i$\;
  \label{line:atoms-ord}
}
\textbf{return} $S_1',\dots,S_k'$\;
\end{algorithm}

\begin{proposition}
\label{proposition:algorithm:atoms-of-parallel}
Algorithm~\ref{algorithm:atoms-of-parallel} correctly computes the atoms of a parallel family of $4$-cuts $\mathcal{C}$. Furthermore, it has a linear-time implementation.
\end{proposition}
\begin{proof}
By Lemma~\ref{lemma:get-atoms}, it is sufficient to split the graph according to the $4$-cuts in $\mathcal{C}$, and then return the sets of ordinary vertices of the split graphs. Since the vertex-sets of the split graphs are pairwise disjoint, we can consider the collection of all of them as a single graph $G'$. Thus, we can equivalently compute the connected components of $G'$, and then return the sets of ordinary vertices of its connected components (see Lines~\ref{line:atoms-components} and \ref{line:atoms-ord}).

Let $r$ be a vertex of $G$. Then Lemma~\ref{lemma:split-in-increasing-r-size} implies that if we use the $4$-cuts from $\mathcal{C}$ in increasing order w.r.t. their $r$-size for the splittings, then, every time we pick a $4$-cut for the splitting, this is mapped as a $4$-cut into the split graph that contains $r$. Thus, it is sufficient to maintain throughout only the quotient map from $V(G)$ to the split graph that contains $r$. We denote this as $v'$, for every vertex $v\in V(G)$. 

Now let $C=\{(x_1,y_1),(x_2,y_2),(x_3,y_3),(x_4,y_4)\}$ be a $4$-cut in $\mathcal{C}$, and let us assume w.l.o.g. that $y_1,y_2,y_3,y_4$ are the endpoints of the edges in $C$ that lie in the connected component of $G\setminus C$ that contains $r$. Then we have that $C'=\{(x_1',y_1'),(x_2',y_2'),(x_3',y_3'),(x_4',y_4')\}$ is a $4$-cut of the split graph that contains $r$, and Lemma~\ref{lemma:split-graph}(b) implies that $y_1',y_2',y_3',y_4'$ are the endpoints of the edges in $C'$ that lie in the same side of $C'$ as $r$. Now, in order to perform the splitting induced by $C$, we introduce two new auxiliary vertices $x_C$ and $y_C$ to $G'$, we delete the edges of $C'$ from $G'$, we introduce the new edges $(x_1',y_C),(x_2',y_C),(x_3',y_C),(x_4',y_C)$ and $(x_C,y_1'),(x_C,y_2'),(x_C,y_3'),(x_C,y_4')$ to $G'$, and we update the quotient map to the split graph that contains $r$ as $x_1'\leftarrow x_C, x_2'\leftarrow x_C, x_3'\leftarrow x_C, x_4'\leftarrow x_C$. (The images of $y_1,y_2,y_3,y_4$ into the split graph that contains $r$ have not changed.) These are precisely the operations that take place during the processing of every $4$-cut $C\in\mathcal{C}$ during the course of the \textbf{for} loop in Line~\ref{line:atoms-for}. (We assume that the \textbf{for} loop in Line~\ref{line:atoms-for} processes the $4$-cuts in $\mathcal{C}$ in increasing order w.r.t. their $r$-size, according to the sorting that took place in Line~\ref{line:atoms-sort}.)

Now it remains to argue about the complexity of Algorithm~\ref{algorithm:atoms-of-parallel}. First, after the linear-time preprocessing described in Lemma~\ref{lemma:determine-components}, we have that the $r$-sizes of the $4$-cuts in $\mathcal{C}$ can be computed in $O(|\mathcal{C}|)$ time in total. Since $\mathcal{C}$ is a parallel family of $4$-cuts of $G$, it contains $O(n)$ $4$-cuts (see, e.g., \cite{DBLP:conf/stoc/DinitzN95}). Then, we can sort the $4$-cuts in $\mathcal{C}$ in increasing order w.r.t. their $r$-size using bucket-sort. Thus, Line~\ref{line:atoms-sort} can be performed in linear time. After the preprocessing described in Lemma~\ref{lemma:determine-components}, we can also compute in constant time, for every $4$-cut $C\in\mathcal{C}$, the endpoints of the edges in $C$ that lie in the same connected component of $G\setminus C$ that contains $r$. Thus, Line~\ref{line:atoms-determine} incurs total cost $O(n)$. In order to perform efficiently the deletions and insertions of edges in Lines~\ref{line:atoms-delete} and \ref{line:atoms-insert}, respectively, we just process them in an off-line manner. Thus, we first collect every $\mathtt{insert\_edge}(x,y)$ query as a triple $(x,y,+)$, and we collect every $\mathtt{delete\_edge}(x,y)$ query as $(x,y,-)$. Furthermore, we also collect every edge $(x,y)$ of the original graph $G$ as a triple $(x,y,+)$; this corresponds to the initialization in Line~\ref{line:atoms-initialize}. Then we sort all those triples lexicographically (giving, e.g., priority to $+$). Since their total number is $O(m+|\mathcal{C}|)=O(m+n)$, we can perform this sorting in $O(m+n)$ time with bucket-sort. Let $L$ be the resulting list of triples. Then, for every maximal segment of $L$ that consists of triples whose first two components coincide -- and so these correspond to insertions and deletions of the same edge $(x,y)$ --, we just determine whether the number of ``pluses" dominates the number of ``minuses" for $(x,y)$. If these values are equal, then we do not include the edge $(x,y)$ in the final graph $G'$. Otherwise, we create as many copies of $(x,y)$ for $G'$, as is the difference between the number of pluses and the number of minuses for $(x,y)$. Thus, we can create $G'$ in linear time in total. Therefore, the computation in Line~\ref{line:atoms-components} takes $O(m+n)$ time. Then, the \textbf{for} loop in Line~\ref{line:atoms-for-2} takes $O(|V(G')|)=O(n)$ time, because we can easily check whether a vertex of $G'$ is auxiliary (by maintaining a bit that signifies it). We conclude that Algorithm~\ref{algorithm:atoms-of-parallel} has a linear-time implementation.
\end{proof}

\begin{corollary}
\label{corollary:atoms-of-parallel}
Let $\mathcal{C}$ be a parallel family of $4$-cuts of $G$. Then we can construct in linear time a data structure of $O(n)$ size that we can use in order to answer queries of the form ``given two vertices $x$ and $y$, determine if $x$ and $y$ are separated by a $4$-cut from $\mathcal{C}$, and, if yes, report a $4$-cut from $\mathcal{C}$ that separates them" in constant time. 
\end{corollary}
\begin{proof}
This is an easy consequence of the method that we use in order to compute the atoms of $\mathcal{C}$. First, by using Algorithm~\ref{algorithm:atoms-of-parallel}, we can compute the split graphs according to $\mathcal{C}$ (w.r.t. an ordering of the $4$-cuts from $\mathcal{C}$ in increasing order w.r.t. their $r$-size). By Proposition~\ref{proposition:algorithm:atoms-of-parallel}, all these split graphs can be computed in linear time in total. Since the split graphs are pairwise vertex-disjoint, we can consider the collection of them as a graph $G'$ (whose connected components are the split graphs). Let $C$ be a $4$-cut from $\mathcal{C}$, and let $x_C$ and $y_C$ be the auxiliary vertices that were introduced due to the splitting according to $C$. Then, we insert an artificial edge $(x_C,y_C)$ to $G'$. We perform this for all $4$-cuts from $\mathcal{C}$, and let $G''$ denote the resulting graph. (Notice that $|V(G'')|=O(n)$ and $|E(G'')|=O(m)$.) Then, for every $4$-cut $C\in\mathcal{C}$, we have that $(x_C,y_C)$ is a bridge of $G''$, whose sides correspond to the connected components of $G\setminus C$ (if we keep the ordinary vertices from each side). Notice that the $2$-edge-connected components of $G''$ are in a bijective correspondence with the split graphs of $G$. Thus, our construction is as follows. First, we compute the tree $T$ of the $2$-edge-connected components of $G''$. Thus, the nodes of $T$ are the split graphs of $G$, and the edges of $T$ are the new artificial edges of the form $(x_C,y_C)$ that we have created, for every $4$-cut $C\in\mathcal{C}$. 
Finally, we perform a DFS on $T$, and we keep the parent pointers $p$, the parent edges, the values $\mathit{ND}$ for all vertices, and a pointer from every parent edge to the $4$-cut from $\mathcal{C}$ that corresponds to it, and reversely. It is easy to see that this whole construction can be completed in linear time and takes $O(n)$ space. 

Now, given two vertices $x$ and $y$, and a query that asks for a $4$-cut from $\mathcal{C}$ that separates $x$ and $y$, we first retrieve the nodes $u$ and $v$ of $T$ that contain $x$ and $y$, respectively. If these coincide, then there is no $4$-cut from $\mathcal{C}$ that separates $x$ and $y$. Otherwise, we first check whether $u$ and $v$ are related as ancestor and descendant. If that is the case, let us assume w.l.o.g. that $u$ is a descendant of $v$. Then, the $4$-cut from $\mathcal{C}$ that corresponds to the parent edge $(u,p(u))$ is a $4$-cut that separates $x$ and $y$. Otherwise, if $u$ and $v$ are not related as ancestor and descendant, we may assume w.l.o.g. that $u$ is not the root of $T$. Then, the $4$-cut from $\mathcal{C}$ that corresponds to the parent edge $(u,p(u))$ is a $4$-cut that separates $x$ and $y$. Thus, we can see that in $O(1)$ time we can determine a $4$-cut from $\mathcal{C}$ that separates $x$ and $y$, or report that no such $4$-cut exists.

\end{proof}

As a concluding remark, we note that Proposition~\ref{proposition:algorithm:atoms-of-parallel} and Corollary~\ref{corollary:atoms-of-parallel} hold for general parallel families of cuts (even of mixed cardinalities), provided that there is a fixed upper bound on their cardinality. This is because, in all the arguments in this section, we did not rely in an essential way on the fact that we consider collections of $4$-cuts. Even Lemma~\ref{lemma:determine-components}, that concerns the computation of the $r$-size, and the determining of the endpoints of the edges of a cut that lie in the same connected component as $r$, holds for general collections of cuts (of bounded cardinality). This is an easy combinatorial problem, that relies on the fact that, when we traverse a tree-path, every time that we cross an edge that participates in a cut, we move to the other side of the cut.

\section{Computing the $5$-edge-connected components}
\label{section:computing-5}

\subsection{Overview}
In this section we present the linear-time algorithm for computing the $5$-edge-connected components of a $3$-edge-connected graph $G$. 
On a high level, the idea is to collect enough $4$-cuts of $G$ so that: $(1)$ the collection of those $4$-cuts is sufficient to provide the partition of $G$ into its $5$-edge-connected components, and $(2)$ there is a linear-time algorithm for computing the partition of $V(G)$ induced by this collection.

Solving $(1)$ and $(2)$ simultaneously is a very complex problem. First of all, computing a collection of $4$-cuts of $G$ that has $O(|V(G)|)$ size and is enough in order to provide the $5$-edge-connected components, is in itself a highly non-trivial task. This is because the number of all $4$-cuts of $G$ can be as high as $\Omega(|V(G)|^2)$. Thus, we must discover a structure in $G$ that allows us to select enough $4$-cuts for our purpose. This seems very demanding, because, to the best of our knowledge, no linear-time algorithm exists even for checking the existence of a $4$-cut. Furthermore, even if we had such a collection of $4$-cuts, we are faced with problem $(2)$, which is basically to compute the atoms induced by this collection. We do not know how to do this in linear time for general collections of $4$-cuts. However, if the collection is a parallel family of $4$-cuts, then Proposition~\ref{proposition:algorithm:atoms-of-parallel} establishes that we can compute its atoms in linear time.

Our solution to both $(1)$ and $(2)$ is as follows. First, we solve $(1)$ indirectly, by computing a complete collection $\mathcal{C}$ of $4$-cuts of $G$ that has  size $O(|V(G)|)$. We note that it is not even clear why such a collection should exist. A large part of this work is devoted to establishing Theorem~\ref{theorem:main}, which guarantees the existence of such a collection, and also that it can be computed in linear time. Now, provided with $\mathcal{C}$, we basically have a compact representation of all $4$-cuts of $G$. However, we cannot expand all the implicating sequences of $\mathcal{C}$ in order to derive all $4$-cuts of $G$, as this could demand $\Omega(|V(G)|^2)$ time. Instead, we apply Algorithm~\ref{algorithm:generatefamilies} on $\mathcal{C}$, which implicitly expands all the implicating sequences of $\mathcal{C}$, and packs them into a set $\{\mathcal{F}_1,\dots,\mathcal{F}_k\}$ of collections of pairs of edges, that have the property that $(i)$ they generate $4$-cuts implied by $\mathcal{C}$, and $(ii)$ every $4$-cut implied by $\mathcal{C}$ is generated by at least one such collection. 

\begin{figure}[t!]\centering
\includegraphics[trim={0 22cm 0 0}, clip=true, width=0.9\linewidth]{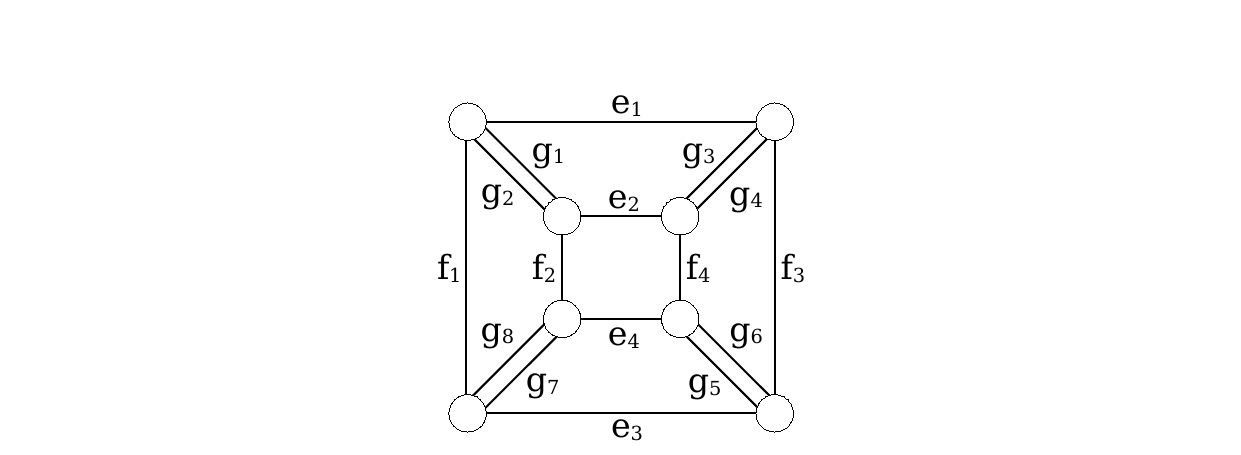}
\caption{\small{This is a $3$-edge-connected graph with $4$-cuts $C_1=\{e_1,e_2,e_3,e_4\}$, $C_2=\{f_1,f_2,f_3,f_4\}$, $D_1=\{e_1,f_1,g_1,g_2\}$, $D_2=\{e_2,f_2,g_1,g_2\}$, $D_3=\{e_1,f_1,e_2,f_2\}$, $E_1=\{e_1,f_3,g_3,g_4\}$, $E_2=\{e_2,f_4,g_3,g_4\}$, $E_3=\{e_1,f_3,e_2,f_4\}$, $F_1=\{e_3,f_3,g_5,g_6\}$, $F_2=\{e_4,f_4,g_5,g_6\}$, $F_3=\{e_3,f_3,e_4,f_4\}$, $G_1=\{e_3,f_1,g_7,g_8\}$, $G_2=\{e_4,f_2,g_7,g_8\}$ and $G_3=\{e_3,f_1,e_4,f_2\}$. 
Thus, it is not difficult to see that $\mathcal{C}=\{C_1,C_2,D_1,D_2,E_1,E_2,F_1,F_2,G_1,G_2\}$ is a complete collection of $4$-cuts of this graph. If we apply Algorithm~\ref{algorithm:generatefamilies} on $\mathcal{C}$, we will get as a result the collections of pairs of edges $\mathcal{F}_1=\{\{e_1,e_2\},\{e_3,e_4\}\}$, $\mathcal{F}_2=\{\{e_1,e_3\},\{e_2,e_4\}\}$, $\mathcal{F}_3=\{\{e_1,e_4\},\{e_2,e_3\}\}$, $\mathcal{F}_4=\{\{f_1,f_2\},\{f_3,f_4\}\}$, $\mathcal{F}_5=\{\{f_1,f_3\},\{f_2,f_4\}\}$, $\mathcal{F}_6=\{\{f_1,f_4\},\{f_2,f_3\}\}$, $\mathcal{F}_7=\{\{e_1,f_1\},\{e_2,f_2\},\{g_1,g_2\}\}$, $\mathcal{F}_8=\{\{e_1,f_3\},\{e_2,f_4\},\{g_3,g_4\}\}$, $\mathcal{F}_9=\{\{e_3,f_3\},\{e_4,f_4\},\{g_5,g_6\}\}$ and $\mathcal{F}_{10}=\{e_3,f_1\},\{e_4,f_2\},\{g_7,g_8\}\}$. Notice that $\{\{e_1,e_2\},\{e_3,e_4\},\{f_1,f_2\},\{f_3,f_4\}\}$ is a collection of pairs of edges that generates a cyclic family of $4$-cuts of this graph, that includes $C_1$ and $C_2$. Thus, $C_1$ and $C_2$ are not isolated $4$-cuts. Therefore, since all three partitions into pairs of edges of $C_1$ and $C_2$ are returned by Algorithm~\ref{algorithm:generatefamilies} on input $\mathcal{C}$, we have that $C_1$ and $C_2$ are quasi $\mathcal{C}$-isolated $4$-cuts. Notice that $C_1$ and $C_2$ are essential and cross.}}\label{figure:crossing_quasi_isolated}
\end{figure}

Now, a tempting idea would be to compute all $\mathcal{C}_i$-minimal $4$-cuts, for every $i\in\{1,\dots,k\}$, where $\mathcal{C}_i$ is the collection of $4$-cuts generated by $\mathcal{F}_i$. This is because the collection of all $\mathcal{C}_i$-minimal $4$-cuts is a parallel family of $4$-cuts, that provides the same atoms as $\mathcal{C}_i$ (Corollary~\ref{corollary:minimal-4cuts}). According to Proposition~\ref{proposition:algorithm:minimal-4cuts}, we can indeed compute the $\mathcal{C}_i$-minimal $4$-cuts, for all $i\in\{1,\dots,k\}$ such that $\mathcal{C}_i$ is a cyclic family of $4$-cuts, in linear time in total. However, we cannot use Algorithm~\ref{algorithm:atoms-of-parallel} in order to compute the atoms provided by the $\mathcal{C}_i$-minimal $4$-cuts, for every $i\in\{1,\dots,k\}$ separately, because this would take $O(k|V(G)|)$ time in total, and $k$ can be as large as $\Omega(|V(G)|)$. 
On the other hand, we cannot compute the atoms provided by all $\mathcal{C}_i$-minimal $4$-cuts, for all $i\in\{1,\dots,k\}$ simultaneously, because there is no guaranteee that this is a parallel family of $4$-cuts. Thus, we have to carefully select enough $\mathcal{C}_i$-minimal $4$-cuts, for every $i\in\{1,\dots,k\}$, so that $(1)$ and $(2)$ are satisfied simultaneously. This is still a challenging task. A first step towards resolving it is to keep only the essential $\mathcal{C}_i$-minimal $4$-cuts, for every $i\in\{1,\dots,k\}$. 
This is enough in order to provide the $5$-edge-connected components.
Furthermore, according to Lemma~\ref{lemma:non-crossing-of-minimal}, this provides a parallel family of $4$-cuts. However, so far we have overlooked the fact that there may be some collections of pairs of edges in $\{\mathcal{F}_1,\dots,\mathcal{F}_k\}$ that have size $2$, and yet can be expanded into larger collections of pairs of edges that generate $4$-cuts implied by $\mathcal{C}$. 
Then, the $4$-cuts that are generated by such collections may cross with other such $4$-cuts.
Furthermore, this can be true even if those $4$-cuts are essential. (See Figure~\ref{figure:crossing_quasi_isolated} for an example.) In this case Lemma~\ref{lemma:quasi-isolated-replaceable} is very useful, because it establishes that we can drop from our consideration those $4$-cuts. Thus, it is sufficient to consider only the essential $\mathcal{C}$-isolated $4$-cuts. By Corollary~\ref{corollary:iso-essential-parallel}, these have the property that they are parallel with every other essential $4$-cut. 

This is the general idea for computing the $5$-edge-connected components in linear time. Our result is summarized in Proposition~\ref{proposition:algorithm:get-5comp}. In order to establish this proposition, we need to provide efficient algorithms for computing the minimal $4$-cuts and the essential isolated $4$-cuts. We perform these tasks in Sections~\ref{section:minimal} and \ref{section:computing-iso}, respectively. In Section~\ref{section:computing-enough-4cuts} we describe the procedure for achieving both $(1)$ and $(2)$, given a complete collection of $4$-cuts.

Throughout this chapter, we assume that $G$ is a $3$-edge-connected graph, and all graph-related elements refer to $G$. 
Also, we assume that we have performed the linear-time preprocessings that are described in Lemma~\ref{lemma:determine-components} and Proposition~\ref{proposition:essentiality}. (These are for reporting the $r$-size and for testing the essentiality of $4$-cuts in constant time.)

\subsection{Computing the minimal $4$-cuts}
\label{section:minimal}
Let $\mathcal{C}$ be a collection of $4$-cuts of $G$, and let $\mathcal{F}$ be a collection of pairs of edges that is returned by Algorithm~\ref{algorithm:generatefamilies} on input $\mathcal{C}$. By Proposition~\ref{proposition:cyclic_families_algorithm}, we have that $\mathcal{F}$ generates a collection $\mathcal{C}'$ of $4$-cuts implied by $\mathcal{C}$. Suppose that $\mathcal{C}'$ is a cyclic family of $4$-cuts. Then we want to find all the $\mathcal{C}'$-minimal $4$-cuts. Let us recall precisely what this means. Let $\mathcal{F}=\{p_1,\dots,p_k\}$, where $k\geq 3$. Then, since $\mathcal{F}$ generates a cyclic family of $4$-cuts, we may assume w.l.o.g. that there is a partition $\{X_1,\dots,X_k\}$ of $V(G)$ such that $G[X_i]$ is connected for every $i\in\{1,\dots,k\}$, and $E[X_i,X_{i+_{k}1}]=p_i$ for every $i\in\{1,\dots,k\}$. 
Then the collection of all $\mathcal{C}'$-minimal $4$-cuts is $\{p_i\cup p_{i+_{k}1}\mid i\in\{1,\dots,k\}\}$. Of course, the problem is that we do not receive the collection $\mathcal{F}$ in such an orderly fashion, and the number of all $4$-cuts generated by $\mathcal{F}$ is $\Theta(k^2)$, which can be as large as $\Omega(n^2)$.

We propose the following method to compute the $\mathcal{C}'$-minimal $4$-cuts. Suppose that we have fixed a vertex $r$, and assume w.l.o.g. that $r\in X_1$. If we knew at least one of the two pairs of edges in $\mathcal{F}$ that are incident to $X_1$ (i.e., either $p_1$ or $p_k$), then it would be easy to find all the $\mathcal{C}'$-minimal $4$-cuts. Specifically, suppose that we know that $p_1$ is incident to the corner of $\mathcal{C}'$ that contains $r$. Then we have that the $4$-cuts $p_1\cup p_2, p_1\cup p_3,\dots, p_1\cup p_k$ are sorted in increasing order w.r.t. their $r$-size. Thus, if we have sorted the collection of $4$-cuts $\{p_1\cup p_i\mid i\in\{2,\dots,k\}\}$ in increasing order w.r.t. the $r$-size, then we can retrieve the sequence $p_1,p_2,p_3,\dots,p_k$, which is enough to provide the $\mathcal{C}'$-minimal $4$-cuts (i.e., by collecting the union of every pair of consecutive elements in this sequence, plus $p_k\cup p_1$).

Thus, the problem is how to identify one of the two pairs of edges in $\mathcal{F}$ that are incident to the corner of $\mathcal{C}'$ that contains $r$. To do this, we start with any pair of edges $p\in\mathcal{F}$. Then, it is easy to see that the $4$-cut $p\cup p_i$ in $\{p\cup p_i\mid i\in\{1,\dots,k\} \mbox{ and } p_i\neq p\}$ with the \emph{maximum} $r$-size has the property that $p_i$ is a pair of edges incident to $X_1$ (i.e., $i$ is either $1$ or $k$).

If we put those two ideas together, then we can see that Algorithm~\ref{algorithm:minimal-4cuts} computes, simultaneously, all the $\mathcal{C}_1$-, $\dots$, $\mathcal{C}_t$-minimal $4$-cuts, for every set of collections of pairs of edges $\mathcal{F}_1,\dots,\mathcal{F}_t$ that generate the cyclic families of $4$-cuts $\mathcal{C}_1,\dots,\mathcal{C}_t$, respectively. The analysis of Algorithm~\ref{algorithm:minimal-4cuts}, as well as its proof of correctness, is given in Proposition~\ref{proposition:algorithm:minimal-4cuts}.

\begin{algorithm}[t!]
\caption{\textsf{Compute all the $\mathcal{C}_1$-, $\dots$ , $\mathcal{C}_t$-minimal $4$-cuts, for a given set of collections of pairs of edges $\mathcal{F}_1,\dots,\mathcal{F}_t$ that generate the cyclic families of $4$-cuts $\mathcal{C}_1,\dots,\mathcal{C}_t$, respectively}}
\label{algorithm:minimal-4cuts}
\LinesNumbered
\DontPrintSemicolon
let $r$ be any fixed vertex\;
\label{line:minimal-root}
let $q_i$ be a \emph{null} pointer to a pair of edges in $\mathcal{F}_i$, for every $i\in\{1,\dots,t\}$\;
\ForEach{$i\in\{1,\dots,t\}$}{
\label{line:minimal-get-q}
  \tcp{find a pair of edges $q_i\in\mathcal{F}_i$ that is incident to the corner of $\mathcal{C}_i$ that contains $r$}
  let $p$ be any pair of edges in $\mathcal{F}_i$\;
  let $\mathit{max}\leftarrow 0$\;
  \ForEach{pair of edges $q$ in $\mathcal{F}_i\setminus\{p\}$}{
    let $\mathit{size}$ be the $r$-size of the $4$-cut $p\cup q$\;
    \If{$\mathit{size}>\mathit{max}$}{
      $q_i\leftarrow q$\;
      $\mathit{size}\leftarrow\mathit{max}$\;
    }
  }
}
initialize $\mathcal{P}\leftarrow\emptyset$\;
\label{line:minimal-P}
\ForEach{$i\in\{1,\dots,t\}$}{
  \ForEach{$p\in\mathcal{F}_i\setminus\{q_i\}$}{
    insert the pair $(q_i\cup p, i)$ into $\mathcal{P}$\;
  }
} 
sort $\mathcal{P}$ in increasing order w.r.t. the $r$-size of the first component of its elements\;
\label{line:minimal-sort}
initialize $\mathcal{M}\leftarrow\emptyset$\; 
\label{line:minimal-M1}
initialize $p_i\leftarrow q_i$, for every $i\in\{1,\dots,t\}$\;
\ForEach{pair $(q_i\cup p, i)\in\mathcal{P}$}{
  insert the $4$-cut $p_i\cup p$ into $\mathcal{M}$\;
  set $p_i\leftarrow p$\;
}
\lForEach{$i\in\{1,\dots,t\}$}{insert the $4$-cut $p_i\cup q_i$ into $\mathcal{M}$}
\label{line:minimal-M2}
\textbf{return} $\mathcal{M}$\;
\end{algorithm}

\begin{proposition}
\label{proposition:algorithm:minimal-4cuts}
Let $\mathcal{F}_1,\dots,\mathcal{F}_t$ be a set of collections of pairs of edges that generate the cyclic families of $4$-cuts $\mathcal{C}_1,\dots,\mathcal{C}_t$. Then, the output of Algorithm~\ref{algorithm:minimal-4cuts} on input $\mathcal{F}_1,\dots,\mathcal{F}_t$ is the collection of all $\mathcal{C}_1$-, $\dots$, $\mathcal{C}_t$-minimal $4$-cuts. The running time of Algorithm~\ref{algorithm:minimal-4cuts} is $O(n+|\mathcal{F}_1|+\dots+|\mathcal{F}_t|)$.
\end{proposition}
\begin{proof}
Let $\mathcal{F}=\{p_1,\dots,p_k\}$ be one of the collections of pairs of edges in $\{\mathcal{F}_1,\dots,\mathcal{F}_t\}$. Since $\mathcal{F}$ generates a cyclic family $\mathcal{C}$ of $4$-cuts, we may assume w.l.o.g. that there is a partition $\{X_1,\dots,X_k\}$ of $V(G)$, such that $G[X_i]$ is connected for every $i\in\{1,\dots,k\}$, and $E[X_i,X_{i+_{k}1}]=p_i$ for every $i\in\{1,\dots,k\}$. Then, 
the $\mathcal{C}$-minimal $4$-cuts are given by $\{p_i\cup p_{i+_{k}1}\mid i\in\{1,\dots,k\}\}$. Let $r$ be any fixed vertex, and let us assume w.l.o.g. that $r\in X_1$. 

Let $p$ be any pair of edges in $\mathcal{F}$. Thus, there is an $i\in\{1,\dots,k\}$ such that $p=p_i$. Then, for every $j\in\{1,\dots,i-1\}$, the two sides of the $4$-cut $p_i\cup p_j$ are given by $X_{j+1}\cup\dots\cup X_i$ and $X_{i+_{k}1}\cup\dots\cup X_j$. Thus, the $r$-size of $p_i\cup p_j$ is given by $|X_{j+1}|+\dots+|X_i|$. Also, for every $j\in\{i+1,\dots,k\}$, the two sides of the $4$-cut $p_i\cup p_j$ are given by $X_{i+1}\cup\dots\cup X_j$ and $X_{j+_{k}1}\cup\dots\cup X_i$. Thus, the $r$-size of $p_i\cup p_j$ is given by $|X_{i+1}|+\dots+|X_j|$. Thus, if $j\in\{1,\dots,i-1\}$, then the $r$-size of $p_i\cup p_j$ is maximized for $j=1$. And if $j\in\{i+1,\dots,k\}$, then the $r$-size of $p_i\cup p_j$ is maximized for $j=k$. This shows that, if we consider the collection of $4$-cuts of the form $\{p_i\cup p_j\mid j\in\{1,\dots,k\}\setminus\{i\}\}$, then the $4$-cut with the maximum $r$-size in this collection is either $p_i\cup p_1$ or $p_i\cup p_k$. In either case, then, we receive one of the two pairs of edges in $\mathcal{F}$ that are incident to the corner of $\mathcal{C}$ that contains $r$. This shows that the \textbf{for} loop in Line~\ref{line:minimal-get-q} correctly computes a pair of edges $q_i\in\mathcal{F}_i$ that is incident to the corner of $\mathcal{C}_i$ that contains $r$, for every $i\in\{1,\dots,t\}$. 

Now consider again the collection of pairs of edges $\mathcal{F}$. Suppose that we have determined that $p_1$ is one of the pairs of edges in $\mathcal{F}$ that is incident to $X_1$. Then, for every $i\in\{2,\dots,k\}$, the two sides of $p_1\cup p_i$ are $X_2\cup\dots\cup X_i$ and $X_{i+_{k}1}\cup\dots\cup X_1$. Thus, the $r$-size of $p_1\cup p_i$ is $|X_2|+\dots+|X_i|$. This shows that the $4$-cuts $p_1\cup p_2,\dots,p_1\cup p_k$ are sorted in increasing order w.r.t. their $r$-size. Notice that, if we knew the sequence of pairs of edges $p_1,\dots,p_k$ then we could collect all $\mathcal{C}$-minimal $4$-cuts, by forming the union of every two consecutive pairs of edges in this sequence, plus $p_k\cup p_1$. Thus, the idea is to collect the $4$-cuts of the form $\{p_1\cup p_i\mid i\in\{2,\dots,k\}\}$, sort them in increasing order w.r.t. their $r$-size, and then gather the sequence $p_2,\dots,p_k$ by taking the difference from $p_1$. (The argument is similar if the pair of edges that was determined to be incident to $X_1$ is $p_k$.)

In order to sort the collection $\{p_1\cup p_i\mid i\in\{2,\dots,k\}\}$, we use bucket-sort. This takes $O(n)$ time. However, we cannot apply this procedure separately for every collection $\mathcal{F}_i$, because otherwise we will need $\Omega(kn)$ time, and $k$ can be as large as $\Omega(n)$. Thus, we have to collect the $4$-cuts from all those collections in a set $\mathcal{P}$, and sort them simultaneously with bucket-sort. 
In order to retrieve the information for every $4$-cut in $\mathcal{P}$ what is the collection of pairs of edges from which it was generated, we index every $4$-cut that we put in $\mathcal{P}$ with the index of the collection of pairs of edges from which it was generated. This is implemented in Lines~\ref{line:minimal-P} to \ref{line:minimal-sort}. Since we perform the sorting with bucket-sort, this takes $O(|\mathcal{F}_1|+\dots+|\mathcal{F}_t|+n)$ time in total. Finally, Lines~\ref{line:minimal-M1} to \ref{line:minimal-M2} implement the idea that we explained above in order to collect all $\mathcal{C}_i$-minimal $4$-cuts, for every $i\in\{1,\dots,t\}$. 

It should be clear that the expression $O(|\mathcal{F}_1|+\dots+|\mathcal{F}_t|+n)$ dominates the total running time. We only note that, given any $4$-cut, we can easily compute its $r$-size in $O(1)$ time, according to the ancestry relation of the tree-edges that are contained in it. This was explained in Section~\ref{section:computing-r-size} (see Lemma~\ref{lemma:determine-components}).
\end{proof}

\subsection{Computing the essential isolated $4$-cuts}
\label{section:computing-iso}
Let $\mathcal{C}$ be a complete collection of $4$-cuts of $G$. We will provide an algorithm that computes all the essential $\mathcal{C}$-isolated $4$-cuts. In other words, we want to compute all the essential $4$-cuts $C\in\mathcal{C}$ that have the property that there is no collection $\mathcal{F}$ of pairs of edges with $|\mathcal{F}|>2$ that generates a collection $\mathcal{C}'$ of $4$-cuts implied by $\mathcal{C}$ such that $C\in\mathcal{C}'$. By Lemma~\ref{lemma:isolated_necessary}, we have that every $\mathcal{C}$-isolated $4$-cut $C$ has the property that the three different partitions of $C$ into pairs of edges are returned by Algorithm~\ref{algorithm:generatefamilies}  on input $\mathcal{C}$ $(*)$. However, property $(*)$ is also satisfied by the \emph{quasi} $\mathcal{C}$-isolated $4$-cuts. Therefore, given that a $4$-cut $C\in\mathcal{C}$ satisfies property $(*)$, we have to be able to determine whether $C$ is $\mathcal{C}$-isolated or quasi $\mathcal{C}$-isolated. 

Since we actually care only about the \emph{essential} $4$-cuts, Lemma~\ref{lemma:quasi-iso-pair} is very useful in this situation. Because Lemma~\ref{lemma:quasi-iso-pair} shows that every essential quasi $\mathcal{C}$-isolated $4$-cut shares a pair of edges with an essential $\mathcal{C}'$-minimal $4$-cut, where $\mathcal{C}'$ is a cyclic family of $4$-cuts that is generated by a collection of pairs of edges that is returned by Algorithm~\ref{algorithm:generatefamilies}  on input $\mathcal{C}$. The reason that this property is very useful, is that the number of minimal $4$-cuts that are extracted from the collections of pairs of edges that are returned by Algorithm~\ref{algorithm:generatefamilies}  is bounded by $O(n)$ (if $|\mathcal{C}|=O(n)$), and therefore the search space for intersections of quasi $\mathcal{C}$-isolated $4$-cuts with other $4$-cuts implied by $\mathcal{C}$ is conveniently small.

Thus, our strategy for computing the essential $\mathcal{C}$-isolated $4$-cuts can be summarized as follows. First, we collect the output $\mathcal{F}_1,\dots,\mathcal{F}_k$ of Algorithm~\ref{algorithm:generatefamilies} on input $\mathcal{C}$. Then, we find all $4$-cuts $C\in\mathcal{C}$ that satisfy property $(*)$: i.e., we collect all $4$-cuts $C\in\mathcal{C}$ such that all three partitions of $C$ into pairs of edges are included in $\mathcal{F}_1,\dots,\mathcal{F}_k$. Then, among those $4$-cuts, we keep only the essential. Let $\widetilde{\mathcal{C}}$ be the resulting collection. Thus, we have that all the essential $\mathcal{C}$-isolated $4$-cuts are contained in $\widetilde{\mathcal{C}}$. However, the problem is that $\widetilde{\mathcal{C}}$ may also contain some quasi $\mathcal{C}$-isolated $4$-cuts, which we have to identify in order to discard. For this purpose, we rely on Corollary~\ref{corollary:quasi-iso-pair} in the following way. We apply Algorithm~\ref{algorithm:minimal-4cuts}, in order to compute the collection $\mathcal{M}$ of all the essential $\mathcal{C}'$-minimal $4$-cuts, for every cyclic family of $4$-cuts $\mathcal{C}'$ for which there exists an $i\in\{1,\dots,k\}$ such that $\mathcal{C'}$ is generated by $\mathcal{F}_i$. Then, for every $4$-cut $C\in\widetilde{\mathcal{C}}$, and every $4$-cut $C'\in\mathcal{M}$, we have to check whether $C$ and $C'$ share a pair of edges.

To perform this efficiently, for every $4$-cut $C\in\mathcal{M}$, and every pair of edges $p\subset C$, we create a pair $(p,*)$. Also, for every $4$-cut $C\in\widetilde{\mathcal{C}}$, and every pair of edges $p\subset C$, we create a pair $(p,C)$. Now, we collect all those pairs $(p,*)$ and $(p,C)$ into a collection $\mathcal{P}$, which we sort in lexicographic order, giving priority to $*$. Then, we simply check, for every pair $(p,C)$ in $\mathcal{P}$, whether it is preceded by a pair of the form $(p,*)$. If that is the case, then we know that $C$ (which is a $4$-cut in $\widetilde{\mathcal{C}}$) shares a pair of edges with a $\mathcal{C}'$-minimal $4$-cut, where $\mathcal{C}'$ is the cyclic family of $4$-cuts that is generated by some $\mathcal{F}_i$, for an $i\in\{1,\dots,k\}$. Thus, Corollary~\ref{corollary:quasi-iso-pair} implies that $C$ is a quasi $\mathcal{C}$-isolated $4$-cut. Conversely, we can prove that, if $C$ is an essential quasi $\mathcal{C}$-isolated $4$-cut, then there is a pair of edges $p\subset C$ such that $(p,C)$ is preceded by $(p,*)$ in $\mathcal{P}$. Thus, we can collect all the $4$-cuts in $\widetilde{\mathcal{C}}$ that are provably not quasi $\mathcal{C}$-isolated: these are precisely the essential $\mathcal{C}$-isolated $4$-cuts.
This procedure is shown in Algorithm~\ref{algorithm:isolated-4cuts}. Its correctness is established in Proposition~\ref{proposition:algorithm:isolated-4cuts}. 

\begin{algorithm}[h!]
\caption{\textsf{Compute all the essential $\mathcal{C}$-isolated $4$-cuts, where $\mathcal{C}$ is a complete collection of $4$-cuts of $G$}}
\label{algorithm:isolated-4cuts}
\LinesNumbered
\DontPrintSemicolon
compute the collections of pairs of edges $\mathcal{F}_1,\dots,\mathcal{F}_k$ that are returned by Algorithm~\ref{algorithm:generatefamilies} on input $\mathcal{C}$\;
\label{line:iso-1}
initialize a counter $\mathit{Count}(C)\leftarrow 0$, for every $C\in\mathcal{C}$\;
\ForEach{$i\in\{1,\dots,k\}$}{
  \If{$|\mathcal{F}_i|=2$}{
    let $C$ be the $4$-cut in $\mathcal{C}$ from which $\mathcal{F}_i$ is derived\;
    \label{line:iso-infer-c}
    set $\mathit{Count}(C)\leftarrow\mathit{Count}(C)+1$\;
    \label{line:iso-count+1}
  }
}
initialize an empty collection $\widetilde{\mathcal{C}}$\;
\ForEach{$C\in\mathcal{C}$}{
\label{line:iso-for-c}
  \If{$\mathit{Count}(C)=3$ \textbf{and} $C$ is an essential $4$-cut of $G$}{
  \label{line:iso-check-ess}
    insert $C$ into $\widetilde{\mathcal{C}}$\;
    \label{line:iso-create-c}
  }
}
compute the collection $\mathcal{M}$ of all the essential $\mathcal{C}_i$-minimal $4$-cuts, for every $i\in\{1,\dots,k\}$ such that $\mathcal{F}_i$ generates a cyclic family $\mathcal{C}_i$ of $4$-cuts\;
\label{line:iso-compute-m}
initialize an empty collection $\mathcal{P}$\;
\ForEach{$C\in\mathcal{M}$}{
  \ForEach{ordered pair of edges $p$ in $C$}{
    insert a pair $(p,*)$ into $\mathcal{P}$\;
  }
}
\ForEach{$C\in\widetilde{\mathcal{C}}$}{
  \ForEach{ordered pair of edges $p$ in $C$}{
    insert a pair $(p,C)$ into $\mathcal{P}$\;
  }
}
sort $\mathcal{P}$ in lexicographic order, giving priority to $*$\;
\label{line:iso-bucket}
initialize an empty collection $\mathcal{Q}$\;
\ForEach{pair $(p,C)\in\mathcal{P}$}{
  \If{the predecessor of $(p,C)$ in $\mathcal{P}$ is $(p,*)$}{
    insert $C$ into $\mathcal{Q}$\;
    \label{line:iso-iso}
  }
}
\textbf{return} $\widetilde{\mathcal{C}}\setminus \mathcal{Q}$\;
\label{line:iso-final}
\end{algorithm}

\begin{proposition}
\label{proposition:algorithm:isolated-4cuts}
Let $\mathcal{C}$ be a complete collection of $4$-cuts of $G$. Then, Algorithm~\ref{algorithm:isolated-4cuts} correctly computes the collection of all the essential $\mathcal{C}$-isolated $4$-cuts. The running time of Algorithm~\ref{algorithm:isolated-4cuts} is $O(n+|\mathcal{C}|)$.
\end{proposition}
\begin{proof}
Let $\mathcal{F}_1,\dots,\mathcal{F}_k$ be the output of Algorithm~\ref{algorithm:generatefamilies}  on input $\mathcal{C}$. By Lemma~\ref{lemma:isolated_necessary}, we have that every $\mathcal{C}$-isolated $4$-cut $C$ has the property that $C\in\mathcal{C}$ and all three partitions of $C$ into pairs of edges are contained in the set $\{\mathcal{F}_1,\dots,\mathcal{F}_k\}$ $(*)$. Thus, the first step is to find all $C\in\mathcal{C}$ that have this property. To do this, we first find all $\mathcal{F}_i$, for $i\in\{1,\dots,k\}$, that satisfy $|\mathcal{F}_i|=2$. If for an $i\in\{1,\dots,k\}$ we have $|\mathcal{F}_i|=2$, then by Lemma~\ref{lemma:2-pair-collection} we have that $C=\bigcup{\mathcal{F}_i}\in\mathcal{C}$. Thus, for the $4$-cut $C$, we increase the counter $\mathit{Count}(C)$ by one (in Line~\ref{line:iso-count+1}), signifying that we have found one more partition of $C$ into pairs of edges within $\{\mathcal{F}_1,\dots,\mathcal{F}_k\}$. Thus, if for a $4$-cut $C\in\mathcal{C}$ we have $\mathit{Count}(C)=3$, then we know that all partitions of $C$ into pairs of edges are contained in $\{\mathcal{F}_1,\dots,\mathcal{F}_k\}$, and so we insert this $4$-cut into $\widetilde{\mathcal{C}}$ if it is essential (Line~\ref{line:iso-create-c}). The purpose of $\widetilde{\mathcal{C}}$ is precisely to contain all the essential $4$-cuts $C\in\mathcal{C}$ that satisfy property $(*)$. 

Let us provide a simple extension to Algorithm~\ref{algorithm:generatefamilies}, in order to maintain the information that $\mathcal{F}_i$ generates precisely the $4$-cut $C$, whenever $|\mathcal{F}_i|=2$, for $i\in\{1,\dots,k\}$. (This is needed in Line~\ref{line:iso-infer-c}.) Let $C=\{e_1,e_2,e_3,e_4\}\in\mathcal{C}$, and let us assume w.l.o.g. that $\mathcal{F}_i=\{\{e_1,e_2\},\{e_3,e_4\}\}$. Let us also assume, w.l.o.g., that $e_1<e_2$ and $e_3<e_4$. Then we have that $\mathcal{F}_i$ is derived (in Line~\ref{line:generatefamilies:final} of Algorithm~\ref{algorithm:generatefamilies}) from a connected component $S$ of the graph $\mathcal{G}$ that is generated internally by Algorithm~\ref{algorithm:generatefamilies} in Line~\ref{line:generatefamilies:components}. Thus, we have that $S$ contains at least two elements $(C',(e_1,e_2))$ and $(C'',(e_3,e_4))$, for some $4$-cuts $C',C''\in\mathcal{C}$. Since $C\in\mathcal{C}$, Algorithm~\ref{algorithm:generatefamilies} also generates the elements $(C,(e_1,e_2))$ and $(C,(e_3,e_4))$. By construction of $\mathcal{G}$, we have that $(C,(e_1,e_2))$ is connected with $(C',(e_1,e_2))$ in $\mathcal{G}$, and $(C,(e_3,e_4))$ is connected with $(C'',(e_3,e_4))$ in $\mathcal{G}$. Let us suppose, for the sake of contradiction, that $C'\neq C$. Then, let $\{x,y\}$ be the pair of edges such that $C'=\{e_1,e_2,x,y\}$, and let us assume w.l.o.g. that $x<y$. Then, we have that Algorithm~\ref{algorithm:generatefamilies} generates the element $(C',(x,y))$, and this is connected with $(C',(e_1,e_2))$ (see Line~\ref{line:generatefamilies:edges1}). Thus, $(C',(x,y))$ is also in $S$, and therefore $\mathcal{F}_i$ must also contain $\{x,y\}$ (see Line~\ref{line:generatefamilies:final}). Since $C'\neq C$, we have that $\{x,y\}\neq\{e_3,e_4\}$. And since $C'$ is a $4$-cut, it is a $4$-element set, and therefore $\{x,y\}\neq\{e_1,e_2\}$. This implies that $|\mathcal{F}_i|\geq 3$, a contradiction. Thus, we have that $C'=C$. Similarly, we have that $C''=C$. Thus, we can see that the only elements of $S$ are $(C,(e_1,e_2))$ and $(C,(e_3,e_4))$. Thus, whenever a connected component $S$ of $\mathcal{G}$ contains precisely two elements of the form $(C,(e_1,e_2))$ and $(C,(e_3,e_4))$, then we can simply associate with the collection of pairs of edges that is derived from $S$ (in Line~\ref{line:generatefamilies:final}) the information that it generates $C$.

Thus, when we reach Line~\ref{line:iso-compute-m}, we can be certain that $\widetilde{\mathcal{C}}$ contains precisely all the essential $4$-cuts $C\in\mathcal{C}$ that satisfy property $(*)$. Now, among all the $4$-cuts in $\widetilde{\mathcal{C}}$, we have to identify and discard those that are quasi $\mathcal{C}$-isolated. Then, by definition, we will be left with the (essential) $\mathcal{C}$-isolated $4$-cuts. 
Let $\mathcal{M}$ be the collection of all the essential $\mathcal{C}_i$-minimal $4$-cuts, for all $i\in\{1,\dots,k\}$ such that $\mathcal{F}_i$ generates a cyclic family $\mathcal{C}_i$.
Now, according to Corollary~\ref{corollary:quasi-iso-pair}, in order to determine whether a $4$-cut in $\widetilde{\mathcal{C}}$ is quasi $\mathcal{C}$-isolated, it is sufficient to check whether it intersects with a $4$-cut in $\mathcal{M}$ in a pair of edges. Thus, the idea is basically to break every $4$-cut in $\mathcal{M}$ into all different combinations of (ordered) pairs of edges, and then check, for every $4$-cut $C\in\widetilde{\mathcal{C}}$, whether $C$ contains any of those pairs. If that is the case, then by Corollary~\ref{corollary:quasi-iso-pair} we have that $C$ is quasi $\mathcal{C}$-isolated; otherwise, by Corollary~\ref{corollary:quasi-iso-pair} we have that $C$ is $\mathcal{C}$-isolated. 

In order to perform this checking efficiently, we collect every pair of edges $p$ that is contained in a $4$-cut in $\mathcal{M}$, and then we form a pair $(p,*)$. We demand that those pairs are ordered (according to any total ordering of the edges). Thus, for every $4$-cut $C$ in $\mathcal{M}$, we have that $C$ generates six pairs of the form $(p,*)$. The symbol $*$ is simply to signify that $p$ is contained in a $4$-cut in $\mathcal{M}$. Now, we basically do the same for every $4$-cut $C$ in $\widetilde{\mathcal{C}}$: for every pair of edges $p$ in $C$, we create a pair $(p,C)$. Notice that here we maintain the information, what is the $4$-cut in $\widetilde{\mathcal{C}}$ from which $(p,C)$ is derived. Now, we collect all those pairs in a collection $\mathcal{P}$, which we sort lexicographically (with bucket-sort), giving priority to $*$. Thus, if there is a $4$-cut $C\in\widetilde{\mathcal{C}}$ that contains a pair of edges $p$ which is shared by a $4$-cut in $\mathcal{M}$, then we have that the element $(p,C)$ is preceded by an element $(p,*)$ in $\mathcal{P}$. Moreover, we have that $(p,*)$ is precisely the predecessor of $(p,C)$ in $\mathcal{P}$. To see this, suppose the contrary. Then, we have that there is a $4$-cut $C'\in\widetilde{\mathcal{C}}$ with $C'\neq C$ that contains the pair of edges $p$. Let $q=C'\setminus p$. Since $C'\in\widetilde{\mathcal{C}}$, we have that the collection of pairs of edges $\{p,q\}$ is contained in $\{\mathcal{F}_1,\dots,\mathcal{F}_k\}$. Let $p'=C\setminus p$. Then, since $C\in\widetilde{\mathcal{C}}$, we have that the collection of pairs of edges $\{p,p'\}$ is contained in $\{\mathcal{F}_1,\dots,\mathcal{F}_k\}$. Since $C'\neq C$, we have that $\{p,q\}\neq\{p,p'\}$. Therefore, Lemma~\ref{lemma:returned_is_disjoint} implies that $\{p,q\}\cap\{p,p'\}=\emptyset$, a contradiction. Thus, we have that there is no $4$-cut $C'\in\widetilde{\mathcal{C}}$ with $C'\neq C$ that contains the pair of edges $p$. Therefore, the predecessor of $(p,C)$ in $\mathcal{P}$ is precisely $(p,*)$. (We note that this is because we consider the \emph{ordered} pairs of edges that are produced by the pairs of edges that are contained in the $4$-cuts. Otherwise, $(p,*)$ could be the predecessor of the predecessor of $(p,C)$.) Thus, we can infer that $C$ shares a pair of edges with an essential $\mathit{C}_i$-minimal $4$-cut, for some $i\in \{1,\dots,k\}$. Thus, in Line~\ref{line:iso-iso} we insert $C$ into the collection $\mathcal{Q}$ (which is to contain all the essential quasi $\mathcal{C}$-isolated $4$-cuts). Conversely, if $C$ does not have this property, then obviously the predecessor of $(p,C)$ in $\mathcal{P}$ cannot have the form $(p,*)$, for any pair of edges $p\subset C$. Thus, by Corollary~\ref{corollary:quasi-iso-pair} we infer that $C$ is not a quasi $\mathcal{C}$-isolated $4$-cut, and therefore, since $C\in\widetilde{\mathcal{C}}$, it must be an essential $\mathcal{C}$-isolated $4$-cut. Thus, when we reach Line~\ref{line:iso-final}, $\mathcal{Q}$ contains precisely all the essential quasi $\mathcal{C}$-isolated $4$-cuts, and therefore we only have to return $\widetilde{\mathcal{C}}\setminus\mathcal{Q}$, because this is now the collection of all the essential $\mathcal{C}$-isolated $4$-cuts.

To conclude the proof of correctness, we have to provide a method to compute the collection $\mathcal{M}$.
To do this, we perform for every $i\in\{1,\dots,k\}$ the following check. First of all, by Proposition~\ref{proposition:cyclic_families_algorithm} we can be certain that $\mathcal{F}_i$ generates a collection of $4$-cuts of $G$. Now, if $|\mathcal{F}_i|=2$, then $i$ is ignored. If $|\mathcal{F}_i|>3$, then Proposition~\ref{proposition:cyclic_family} implies that $\mathcal{F}_i$ generates a cyclic family of $4$-cuts, and so we keep $i$. If $|\mathcal{F}_i|=3$, then we compute the three $4$-cuts $C_1$, $C_2$ and $C_3$, that are generated by $\mathcal{F}_i$. If either of $C_1,C_2,C_3$ is essential, then Corollary~\ref{corollary:essential_implies_cyclic} implies that $\mathcal{F}_i$ generates a cyclic family of $4$-cuts, and so we keep $i$. Otherwise, either $\mathcal{F}_i$ does not generate a cyclic family of $4$-cuts, or, if it does, all of its minimal $4$-cuts are non-essential, and so we can ignore $i$. Now, let $I$ be the collection of all the indices that we have collected. Then, for every $i\in I$, we have that $\mathcal{F}_i$ generates a cyclic family of $4$-cuts. Thus, we can apply Algorithm~\ref{algorithm:minimal-4cuts} on the collection $\{\mathcal{F}_i\mid i\in I\}$, which will produce the collection $\mathcal{M}'$ of all $\mathcal{C}_i$-minimal $4$-cuts, where $\mathcal{C}_i$ is the cyclic family of $4$-cuts generated by $\mathcal{F}_i$, for $i\in I$. Then, we only keep from $\mathcal{M}'$ the essential $4$-cuts, and so we compute the collectiom $\mathcal{M}$.

Now let us provide the time-bounds for the non-trivial steps of Algorithm~\ref{algorithm:isolated-4cuts}. First, by Proposition~\ref{proposition:cyclic_families_algorithm} we have that Line~\ref{line:iso-1} takes $O(n+|\mathcal{C}|)$ time. Furthermore, Proposition~\ref{proposition:cyclic_families_algorithm} ensures that $|\mathcal{F}_1|+\dots+|\mathcal{F}_k|=O(|\mathcal{C}|)$. In order to check the essentiality in Line~\ref{line:iso-check-ess}, we rely on the assumption we have made at the beginning of this section: that is, we have completed the linear-time preprocessing of the graph that is described in Proposition~\ref{proposition:essentiality}, so that we can check the essentiality of any $4$-cut in $O(1)$ time. Thus, the \textbf{for} loop in Line~\ref{line:iso-for-c} takes $O(|\mathcal{C}|)$ time in total. 
By Proposition~\ref{proposition:algorithm:minimal-4cuts}, we can compute the collection $\mathcal{M}'$ of all the $\mathcal{C}_i$-minimal $4$-cuts, for every $i\in I$, in $O(n+\sum_{i\in I}|\mathcal{F}_i|)=O(n+|\mathcal{F}_1|+\dots+|\mathcal{F}_k|)=O(n+|\mathcal{C}|)$ time. Notice that the size of $\mathcal{M}'$ is $O(|\mathcal{F}_1|+\dots+|\mathcal{F}_k|)=O(|\mathcal{C}|)$, and there are $O(|\mathcal{C}|)$ essentiality checks that are involved in the computation of $\mathcal{M}$. Finally, notice that the size of $\mathcal{P}$ is $O(|\mathcal{C}|)$, and therefore Line~\ref{line:iso-bucket} takes $O(n+|\mathcal{C}|)$ time if we perform the sorting with bucket-sort. Also, we can compute the set difference $\widetilde{\mathcal{C}}\setminus\mathcal{Q}$ in Line~\ref{line:iso-final} with bucket-sort. Thus, the total running time of Algorithm~\ref{algorithm:isolated-4cuts} is $O(n+|\mathcal{C}|)$.
\end{proof}

\subsection{Computing enough $4$-cuts in order to derive the $5$-edge-connected components}
\label{section:computing-enough-4cuts}
Let $\mathcal{C}$ be a complete collection of $4$-cuts of $G$. Then, Algorithm~\ref{algorithm:get-4cuts} shows how we can extract a parallel collection $\mathcal{C}'$ of $4$-cuts from $\mathcal{C}$, that contains enough $4$-cuts in order to separate all pairs of vertices $x,y$ with $\lambda(x,y)=4$. The proof of correctness is given in Proposition~\ref{proposition:algorithm:get-4cuts}.

\begin{algorithm}[h!]
\caption{\textsf{Generate a parallel collection of $4$-cuts from a complete collection $\mathcal{C}$ of $4$-cuts, that contains enough $4$-cuts in order to separate all pairs of vertices $x,y$ with $\lambda(x,y)=4$}}
\label{algorithm:get-4cuts}
\LinesNumbered
\DontPrintSemicolon
compute the collections of pairs of edges $\mathcal{F}_1,\dots,\mathcal{F}_k$ that are returned by Algorithm~\ref{algorithm:generatefamilies} on input $\mathcal{C}$\;
\label{line:get-4cuts-f}
compute the collection $\mathcal{M}$ of all the essential $\mathcal{C}_i$-minimal $4$-cuts, where $\mathcal{C}_i$ is the cyclic family of $4$-cuts that is generated by $\mathcal{F}_i$, for some $i\in\{1,\dots,k\}$\;
\label{line:get-4cuts-m}
compute the collection $\mathcal{ISO}$ of the essential $\mathcal{C}$-isolated $4$-cuts\;
\label{line:get-4cuts-iso}
\textbf{return} $\mathcal{M}\cup\mathcal{ISO}$\;
\end{algorithm}

\begin{proposition}
\label{proposition:algorithm:get-4cuts}
Let $\mathcal{C}$ be a complete collection of $4$-cuts of $G$. Then, the output of Algorithm~\ref{algorithm:get-4cuts} on input $\mathcal{C}$ is a parallel family $\mathcal{C}'$ of $4$-cuts such that: for every pair of vertices $x,y$ of $G$ with $\lambda(x,y)=4$, there is a $4$-cut in $\mathcal{C}'$ that separates $x$ and $y$. The running time of Algorithm~\ref{algorithm:get-4cuts} is $O(n+|\mathcal{C}|)$.
\end{proposition}
\begin{proof}
Let $\mathcal{F}_1,\dots,\mathcal{F}_k$ be the collections of pairs of edges that are returned by Algorithm~\ref{algorithm:generatefamilies} on input $\mathcal{C}$. Then, by Proposition~\ref{proposition:cyclic_families_algorithm} we have that $\mathcal{F}_i$ generates a collection of $4$-cuts implied by $\mathcal{C}$, for every $i\in\{1,\dots,k\}$. Thus, let $\mathcal{C}_i$ be the collection of $4$-cuts generated by $\mathcal{F}_i$, for every $i\in\{1,\dots,k\}$. By Proposition~\ref{proposition:cyclic_families_algorithm}, we also have that every $4$-cut implied by $\mathcal{C}$ is contained in $\mathcal{C}_i$, for some $i\in\{1,\dots,k\}$. Thus, since $\mathcal{C}$ is a complete collection of $4$-cuts, we have that every $4$-cut of $G$ is contained in some $\mathcal{C}_i$, for an $i\in\{1,\dots,k\}$. Let $\mathcal{C}'$ be the output of Algorithm~\ref{algorithm:get-4cuts} on input $\mathcal{C}$. Let also $\mathcal{M}$ and $\mathcal{ISO}$ be the collections that are constructed in Lines~\ref{line:get-4cuts-m} and \ref{line:get-4cuts-iso}, respectively, on input $\mathcal{C}$. Notice that every $4$-cut in $\mathcal{ISO}$ is an essential \emph{isolated} $4$-cut, because $\mathcal{C}$ is a complete collection of $4$-cuts (and therefore it implies all the $4$-cuts of $G$).

Now let $x,y$ be a pair of vertices of $G$ with $\lambda(x,y)=4$. This means that there is a $4$-cut $C$ of $G$ that separates $x$ and $y$. By definition, $C$ is an essential $4$-cut. Since $C$ is a $4$-cut of $G$, there is an $i\in\{1,\dots,k\}$ such that $C\in\mathcal{C}_i$. Thus, we can distinguish two cases: either $(1)$ there is an $i\in\{1,\dots,k\}$ such that $C\in\mathcal{C}_i$ and $|\mathcal{F}_i|>2$ , or $(2)$ for every $i\in\{1,\dots,k\}$ such that $C\in\mathcal{C}_i$ we have that $|\mathcal{F}_i|=2$. 
Let us consider case $(1)$ first. Thus, there is an $i\in\{1,\dots,k\}$ such that $C\in\mathcal{C}_i$ and $|\mathcal{F}_i|>2$. Then, since $C$ is an essential $4$-cut in $\mathcal{C}_i$, by Corollary~\ref{corollary:essential_implies_cyclic} we have that $\mathcal{C}_i$ is a cyclic family of $4$-cuts. Then, since $C$ separates $x$ and $y$ and $\lambda(x,y)=4$, by Lemma~\ref{lemma:replace-with-minimal} we have that there is an essential $\mathcal{C}_i$-minimal $4$-cut $C'$ that separates $x$ and $y$. Then, we have that $C'\in\mathcal{M}$, and therefore $C'\in\mathcal{C}'$.

Now let us consider case $(2)$. Thus, we have that, for every $i\in\{1,\dots,k\}$ such that $C\in\mathcal{C}_i$, we have $|\mathcal{F}_i|=2$. Let us consider an $i\in\{1,\dots,k\}$ such that $C\in\mathcal{C}_i$ (we have already shown that such an $i$ exists). Then, we have that $|\mathcal{F}_i|=2$. Thus, Lemma~\ref{lemma:2-pair-collection} implies that $C\in\mathcal{C}$. Then, by Lemma~\ref{lemma:not-included-4cut} we have that every partition $\mathcal{F}'$ of $C$ into pairs of edges is contained in some $\mathcal{F}_j$, for $j\in\{1,\dots,k\}$. Then we have that $C\in\mathcal{C}_j$, and therefore $|\mathcal{F}_j|=2$. This implies that $\mathcal{F}'=\mathcal{F}_j$. This shows that all partitions of $C$ into pairs of edges are contained in the output of Algorithm~\ref{algorithm:generatefamilies} on input $\mathcal{C}$. Thus, we can distinguish two cases: either $(2.1)$ $C$ is a $\mathcal{C}$-isolated $4$-cut, or $(2.2)$ $C$ is a quasi $\mathcal{C}$-isolated $4$-cut. In case $(2.1)$, we have that $C\in\mathcal{ISO}$ (because $C$ is essential). In case $(2.2)$, we can evoke Lemma~\ref{lemma:quasi-isolated-replaceable}: this implies that there is a $t\in\{1,\dots,k\}$ such that $|\mathcal{F}_t|>2$ and $\mathcal{C}_t$ contains a $4$-cut $C'$ that separates $x$ and $y$. By definition, we have that $C'$ is an essential $4$-cut. Thus, Corollary~\ref{corollary:essential_implies_cyclic} implies that $\mathcal{C}_t$ is a cyclic family of $4$-cuts. Therefore, since $C'\in\mathcal{C}_t$ separates $x$ and $y$ (which are $4$-edge-connected), Lemma~\ref{lemma:replace-with-minimal} implies that there is an essential $\mathcal{C}_t$-minimal $4$-cut $C''$ that separates $x$ and $y$. Thus, we have that $C''\in\mathcal{M}$, and therefore $C''\in\mathcal{C}'$. 
Thus, we have shown that, for every pair of vertices $x,y$ of $G$ with $\lambda(x,y)=4$, there is a $4$-cut in $\mathcal{C}'$ that separates $x$ and $y$.

Now we will show that $\mathcal{C}'$ is a parallel collection of $4$-cuts. Let $C,C'$ be two distinct $4$-cuts in $\mathcal{C}'$. If at least one of $C,C'$ is in $\mathcal{ISO}$, then it is an essential isolated $4$-cut, and so Corollary~\ref{corollary:iso-essential-parallel} implies that it is parallel with every other essential $4$-cut. Thus, let us assume that both $C$ and $C'$ are in $\mathcal{M}$. 
Then there are $i,j\in\{1,\dots,k\}$ such that $C$ is a $\mathcal{C}_i$-minimal $4$-cut, and $C'$ is a $\mathcal{C}_j$-minimal $4$-cut. If $i=j$, then Lemma~\ref{lemma:minimal-are-parallel} implies that $C$ and $C'$ are parallel $4$-cuts. Otherwise, since both $C$ and $C'$ are essential $4$-cuts, Lemma~\ref{lemma:non-crossing-of-minimal} implies that $C$ and $C'$ are parallel. Thus, we have shown that $\mathcal{C}'$ is a parallel collection of $4$-cuts.

Finally, let us consider the running time of Algorithm~\ref{algorithm:get-4cuts}. By Proposition~\ref{proposition:cyclic_families_algorithm}, we have that Line~\ref{line:get-4cuts-f} takes time $O(n+|\mathcal{C}|)$, and the output $\mathcal{F}_1,\dots,\mathcal{F}_k$ has size $O(|\mathcal{F}_1|+\dots+|\mathcal{F}_k|)=O(|\mathcal{C}|)$. Then, we can implement Line~\ref{line:get-4cuts-m} with the same idea as Line~\ref{line:iso-compute-m} of Algorithm~\ref{algorithm:isolated-4cuts} (see the proof of Proposition~\ref{proposition:algorithm:isolated-4cuts}). This will take $O(n+|\mathcal{C}|)$ time, provided that we have made the linear-time preprocessing on $G$ that is described in Proposition~\ref{proposition:essentiality}, in order to be able to perform essentiality checks in $O(1)$ worst-case time per $4$-cut. Finally, by Proposition~\ref{proposition:algorithm:isolated-4cuts}, we have that Line~\ref{line:get-4cuts-iso} takes time $O(n+|\mathcal{C}|)$. We conclude that the running time of Algorithm~\ref{algorithm:get-4cuts} is $O(n+|\mathcal{C}|)$. 

\end{proof}

\subsection{The algorithm}
The full algorithm for computing the $5$-edge-connected components of a $3$-edge-connected graph in linear time is shown in Algorithm~\ref{algorithm:get-5comp}. The proof of correctness is given in Proposition~\ref{proposition:algorithm:get-5comp}.
 
\begin{algorithm}[h!]
\caption{\textsf{Compute the $5$-edge-connected components of a $3$-edge-connected graph $G$}}
\label{algorithm:get-5comp}
\LinesNumbered
\DontPrintSemicolon
compute the partition $\mathcal{P}_4$ of the $4$-edge-connected components of $G$\;
\label{line:get-5-4comp}
compute a complete collection $\mathcal{C}$ of $4$-cuts of $G$\;
\label{line:get-5-complete}
compute the output $\mathcal{C}'$ of Algorithm~\ref{algorithm:get-4cuts} on input $\mathcal{C}$\;
\label{line:get-5-c}
compute the partition $\mathcal{P}_5=\mathit{atoms}(\mathcal{C}')$\;
\label{line:get-5-part}
\textbf{return} $\mathcal{P}_4$ refined by $\mathcal{P}_5$\;
\label{line:get-5-refine}
\end{algorithm}

\begin{proposition}
\label{proposition:algorithm:get-5comp}
Algorithm~\ref{algorithm:get-5comp} correctly computes the $5$-edge-connected components of a $3$-edge-connected graph. Furthermore, it has a linear-time implementation.
\end{proposition}
\begin{proof}
Let $\mathcal{P}_4$, $\mathcal{C}$, $\mathcal{C}'$, and $\mathcal{P}_5$, be as defined in Lines~\ref{line:get-5-4comp}, \ref{line:get-5-complete}, \ref{line:get-5-c}, and \ref{line:get-5-part}, respectively.
Let $\mathcal{P}$ be the output of Algorithm~\ref{algorithm:get-5comp}. Notice that $\mathcal{P}$ is a partition of the vertex set of $G$. We will show that, for every pair of vertices $x,y$ of $G$, we have $\lambda(x,y)<5$ if and only if $x$ and $y$ are separated by $\mathcal{P}$. So let $x,y$ be a pair of vertices of $G$ with $\lambda(x,y)<5$. Since $G$ is $3$-edge-connected, we have that either $\lambda(x,y)=3$, or $\lambda(x,y)=4$. If $\lambda(x,y)=3$, then $x$ and $y$ lie in different $4$-edge-connected components of $G$. Thus, $x$ and $y$ are separated by $\mathcal{P}_4$, and therefore they are separated by $\mathcal{P}$, since $\mathcal{P}$ is a refinement of $\mathcal{P}_4$. Now let us assume that $\lambda(x,y)=4$. Then, Proposition~\ref{proposition:algorithm:get-4cuts} implies that there is a $4$-cut $C\in\mathcal{C}'$ that separates $x$ and $y$. Thus, $x$ and $y$ are separated by $\mathcal{P}_5=\mathit{atoms}(\mathcal{C}')$, and therefore they are separated by $\mathcal{P}$, since $\mathcal{P}$ is a refinement of $\mathcal{P}_5$. Thus, for every pair of vertices $x,y$ of $G$ with $\lambda(x,y)<5$, we have that $x$ and $y$ are separated by $\mathcal{P}$. Conversely, every pair of vertices that are separated by $\mathcal{P}$, are separated by either $\mathcal{P}_4$ or $\mathcal{P}_5$, and therefore they are separated by either a $3$-cut or a $4$-cut of $G$, and therefore they are not $5$-edge-connected. This shows that $\mathcal{P}$ is the collection of the $5$-edge-connected components of $G$.

By previous work, we know that Line~\ref{line:get-5-4comp} can be implemented in linear time (see \cite{DBLP:conf/esa/GeorgiadisIK21} or \cite{DBLP:conf/esa/NadaraRSS21}). By Theorem~\ref{theorem:main}, we have that a complete collection $\mathcal{C}$ of $4$-cuts of $G$ with size $O(n)$ can be computed in linear time (in Line~\ref{line:get-5-complete}). Thus, by Proposition~\ref{proposition:algorithm:get-4cuts} we have that the output $\mathcal{C}'$ of Algorithm~\ref{algorithm:get-4cuts} on input $\mathcal{C}$ can be computed in $O(n+|\mathcal{C}|)=O(n)$ time. Furthermore, by Proposition~\ref{proposition:algorithm:get-4cuts} we have that $\mathcal{C}'$ is a parallel family of $4$-cuts. Thus, by Proposition~\ref{proposition:algorithm:atoms-of-parallel} we have that the computation of $\mathit{atoms}(\mathcal{C}')$ can be performed in $O(n)$ time. Finally, the common refinement of $\mathcal{P}_4$ and $\mathcal{P}_5$ can be computed in $O(n)$ time with bucket-sort, since these are partitions of $V(G)$. We conclude that Algorithm~\ref{algorithm:get-5comp} has a linear-time implementation.
\end{proof}

We note that this method for computing the $5$-edge-connected components is also useful for constructing a data structure that has the same functionality as a partial Gomory-Hu tree that retains all connectivities up to $5$ \cite{DBLP:conf/soda/HariharanKP07}. Specifically, we have the following.

\begin{corollary}
\label{corollary:partial-gomory-hu}
Given a $3$-edge-connected graph $G$ with $n$ vertices, there is a linear-time preprocessing of $G$ that constructs a data structure of size $O(n)$, such that, given two $4$-edge-connected vertices $x$ and $y$ of $G$, we can determine in $O(1)$ time a $4$-cut of $G$ that separates $x$ and $y$, or report that no such $4$-cut exists. 
\end{corollary}
\begin{proof}
This is a consequence of the fact that Algorithm~\ref{algorithm:get-5comp} computes a parallel family $\mathcal{C}$ of $4$-cuts of $G$, with the property that every two $4$-edge-connected vertices that are separated by a $4$-cut of $G$, are also separated by a $4$-cut from $\mathcal{C}$. Then, we can apply Corollary~\ref{corollary:atoms-of-parallel} on $\mathcal{C}$.
\end{proof}


\section{Concepts defined on DFS-tree}
\label{section:DFS}

Throughout this chapter we assume that $G$ is a connected graph with $n$ vertices and $m$ edges, and $r$ is a vertex of $G$. In Section~\ref{subsection:basic-dfs} we present the parameters that are defined w.r.t. a DFS-tree of $G$, and we will use throughout in the rest of this work. In Section~\ref{subsection:dfs-properties} we state and prove some simple properties that are satisfied by the DFS parameters. In Sections~\ref{subsection:low}, \ref{subsection:high}, \ref{subsection:leftmost} and \ref{subsection:M}, we show how to compute the $\mathit{low}$ edges, the $\mathit{high}$ edges, the leftmost and the rightmost edges, and the $M$ points, respectively. In Section~\ref{subsection:paths} we prove two lemmata that concern the structure of paths w.r.t. a DFS-tree. In Section~\ref{subsection:back-edge-oracle} we present an oracle for back-edge queries that we will use in our oracle for connectivity queries in the presence of at most four edge-failures in Section~\ref{section:4eoracle}. Finally, we conclude with Section~\ref{subsection:segments} that deals with the computation of the decreasingly ordered segments that consist of vertices that have the same $\mathit{high}$ (or $\mathit{high}_2$) point, and are maximal w.r.t. the property that their elements are related as ancestor and descendant. These segments are involved in the computation of Type-$3\beta ii$ $4$-cuts, in Section~\ref{subsection:type-3b-ii}. 

\subsection{Basic definitions}
\label{subsection:basic-dfs}
Let $T$ be a DFS-tree of $G$ with start vertex $r$ \cite{DBLP:journals/siamcomp/Tarjan72}. We identify the vertices of $G$ with their order of visit by the DFS. (Thus, $r=1$, and the last vertex visited by $G$ is $n$.) For a vertex $v\neq r$ of $G$, we let $p(v)$ denote the parent of $v$ on $T$. (Thus, $v$ is a child of $p(v)$.) We let $T[v,u]$ denote the simple path from $v$ to $u$ on $T$, for any two vertices $v$ and $u$ of $G$. We use $T[v,u)$, $T(v,u]$ or $T(u,v)$, in order to denote the path $T[v,u]$ minus the vertex on the side of the parenthesis. If $v$ lies on the tree-path $T[r,u]$, then we say that $v$ is an ancestor of $u$ (equivalently, $u$ is a descendant of $v$). Notice that if $v$ is an ancestor of $u$, then $v\leq u$. (The converse is not necessarily true.) If $v$ is an ancestor of $u$ such that $v\neq u$, then we say that $v$ is a proper ancestor of $u$ (equivalently, $u$ is a proper descendant of $v$). We extend the ancestry relation to tree-edges. If $(u,p(u))$ and $(v,p(v))$ are two tree-edges, then we say that $(v,p(v))$ is an ancestor of $(u,p(u))$ (or equivalently, $(u,p(u))$ is a descendant of $(v,p(v))$) if and only if $v$ is an ancestor of $u$. We let $T(v)$ denote the set of descendants of a vertex $v$. (Notice that this is a subtree of $T$.) The number of descendants of $v$ is denoted as $\mathit{ND}(v)$ (i.e., $\mathit{ND}(v)=|T(v)|$). We note that $\mathit{ND}(v)$ can be computed easily during the DFS, because it satisfies the recursive formula $\mathit{ND}(v)=\mathit{ND}(c_1)+\dots+\mathit{ND}(c_k)+1$, where $c_1,\dots,c_k$ are the children of $v$.
We use the $\mathit{ND}$ values in order to check the ancestry relation in constant time. Specifically, given two vertices $u$ and $v$, we have that $u$ is a descendant of $v$ if and only if $v\leq u\leq v+\mathit{ND}(v)-1$. Equivalently, we have $T(v)=\{v,v+1,\dots,v+\mathit{ND}(v)-1\}$.

A DFS traversal imposes an organization of the edges of the graph that is very rich in properties. Specifically, every non-tree of $G$ has its endpoints related as ancestor and descendant on $T$ \cite{DBLP:journals/siamcomp/Tarjan72}. Thus, the non-tree edges of $G$ are called ``back-edges". Whenever we let $(x,y)$ denote a back-edge, we always assume that $x$ is the higher endpoint of $(x,y)$ (i.e., $x>y$). Thus, $x$ is the endpoint of $(x,y)$ that is a descendant of $y$. For a vertex $v\neq r$, we say that a back-edge $(x,y)$ leaps over $v$ if $x$ is a descendant of $v$ and $y$ is a proper ancestor of $v$. We let $B(v)$ denote the set of the back-edges that leap over $v$. Recently, the sets of leaping back-edges were used in order to solve various graph-connectivity problems (see \cite{DBLP:conf/isaac/GeorgiadisK20,DBLP:conf/esa/GeorgiadisIK21}). The usefulness of those sets is intimated by the fact that if we delete the tree-edge $(v,p(v))$ from $G$, then the subtree $T(v)$ of $T$ is connected with the rest of the graph through the back-edges in $B(v)$. Thus, e.g., we can test if an edge $(v,p(v))$ is a bridge by checking whether the set $B(v)$ is non-empty. In general, we can extract a lot of useful information from those sets, that can help us solve various connectivity problems. We note that we do not explicitly compute the sets of leaping back-edges, as their total size can be excessively large (i.e., it can be $\Omega(n^2)$ even in graphs with $O(n)$ number of edges). Instead, we compute some parameters that summarize the information that is contained in those sets (e.g., by considering the distribution of the endpoints of the back-edges that are contained in them). Here we will define some parameters that we will use throughout. Others that are more specialized, and probably of a more restricted scope, are developed and analyzed on the spot, in the sections that follow.
 
Let $v\neq r$ be a vertex. We let $\mathit{bcount}(v)$ denote the number of back-edges that leap over $v$ (i.e., $\mathit{bcount}(v)=|B(v)|$). We let $\mathit{SumDesc}(v)$ denote the sum of the higher endpoints of the back-edges in $B(v)$, and we let $\mathit{SumAnc}(v)$ denote the sum of the lower endpoints of the back-edges in $B(v)$. Similarly, we let $\mathit{XorDesc}(v)$ denote the $\mathit{XOR}$ of the higher endpoints of the back-edges in $B(v)$, and we let $\mathit{XorAnc}(v)$ denote the $\mathit{XOR}$ of the lower endpoints of the back-edges in $B(v)$. 
We introduce use values $\mathit{XorDesc}(v)$ and $\mathit{XorAnc}(v)$ because they help us retrieve back-edges from $B(v)$. Specifically, supposing that we know the $\mathit{XOR}$ $X$ of the higher endpoints of the back-edges in $B(v)\setminus\{e\}$, and the $\mathit{XOR}$ $Y$ of the lower endpoints of the back-edges in $B(v)\setminus\{e\}$, where $e$ is a back-edge in $B(v)$, then we can retrieve the endpoints of $e$ with the values $X\oplus\mathit{XorDesc}(v)$ and $Y\oplus\mathit{XorAnc}(v)$. The values $\mathit{SumDesc}$ and $\mathit{SumAnc}$ are used in order to draw inferences for the existence of back-edges.
We note that these parameters satisfy a recursive formula that allows us to compute them in linear time in total, for all vertices. Specifically, let $\mathit{In}(z)$ denote the set of the back-edges with lower endpoint $z$, for every vertex $z$. Also, let $\mathit{Out}(z)$ denote the set of the back-edges with higher endpoint $z$, for every vertex $z$. Then, $\mathit{bcount}(v)=\mathit{bcount}(c_1)+\dots+\mathit{bcount}(c_k)+|\mathit{Out}(v)|-|\mathit{In}(v)|$, where $c_1,\dots,c_k$ are the children of $v$. Similarly, we have $\mathit{SumDesc}(v)=\mathit{SumDesc}(c_1)+\dots+\mathit{SumDesc}(c_k)+\mathit{SumDesc}(\mathit{Out}(v))-\mathit{SumDesc}(\mathit{In}(v))$, where we let $\mathit{SumDesc}(S)$ denote the sum of the higher endpoints of the back-edges in a set $S$ of back-edges. The analogous relations hold for $\mathit{SumAnc}(v)$, $\mathit{XorAnc}(v)$ and $\mathit{XorDesc}(v)$. Thus, we can compute all these parameters with a bottom-up procedure (e.g., during the backtracking of the DFS), in total linear time, for all vertices $v\neq r$.

\subsubsection{$\mathit{low}$ and $\mathit{high}$ edges}
Now we consider parameters that are defined in relation to the lower endpoints of the back-edges in $B(v)$, for a vertex $v\neq r$. First, let $(v,z_1),\dots,(v,z_s)$ be the list of the back-edges with higher endpoint $v$, sorted in increasing order w.r.t. their lower endpoint. (Notice that these back-edges belong to $B(v)$.) Then we let $l_i(v)$ denote $z_i$, for every $i\in\{1,\dots,s\}$. If $i>s$, then we let $l_i(v)=v$. The vertex $l_1(v)$ is of particular importance, and we may denote it simply as $l(v)$. Thus, we can know e.g. if there is a back-edge that stems from $v$, by checking whether $l(v)<v$. 

Now let $(x_1,y_1),\dots,(x_k,y_k)$ be the list of the back-edges in $B(v)$ sorted in increasing order w.r.t. their lower endpoint. We note that such a sorting may not be unique, but we suppose that we have fixed one. Then (w.r.t. this sorting) we call $(x_i,y_i)$ the $\mathit{low}_i$-edge of $v$, for every $i\in\{1,\dots,k\}$. The lower endpoint of the $\mathit{low}_i$-edge of $v$ is called the $\mathit{low}_i$ point of $v$, and we denote it as $\mathit{low}_i(v)$ (i.e., we have $\mathit{low}_i(v)=y_i$). Notice that the definition of the $\mathit{low}_i$ points of $v$ is independent of the sorting of the back-edges in $B(v)$, provided only that this is in increasing order w.r.t. the lower endpoints. If we want to reference the $\mathit{low}_i$-edge of $v$ with its endpoints, then we denote it as $(\mathit{lowD}_i(v),\mathit{low}_i(v))$. Of particular importance is the $\mathit{low}_1$ point of $v$, which we may simply denote as $\mathit{low}(v)$. The $\mathit{low}$ points have been introduced several decades ago, in order to solve various graph problems with a DFS-based approach (see, e.g., \cite{DBLP:journals/siamcomp/Tarjan72}). In Section~\ref{subsection:low} we show how to compute the $\mathit{low}_i$-edges of all vertices, for every $i\in\{1,\dots,k\}$, where $k$ is a fixed integer, in total linear time (see Proposition~\ref{proposition:low}). 

Now let $v$ be a vertex, and let $c_1,\dots,c_t$ be the children of $v$ sorted in increasing order w.r.t. their $\mathit{low}$ point (breaking ties arbitrarily). In other words, we have $\mathit{low}(c_1)\leq\dots\leq\mathit{low}(c_t)$. Then we call $c_i$ the $\mathit{lowi}$ child of $v$. Once we have computed the $\mathit{low}$ points of all vertices, we note that it is easy to construct the lists of the $\mathit{low}$ children of all vertices in $O(n)$ time in total, using bucket-sort.

Now let $v\neq r$ be a vertex, and let $(x_1,y_1),\dots,(x_k,y_k)$ be the list of the back-edges in $B(v)$ sorted in decreasing order w.r.t. their lower endpoint. Again, we note that such a sorting may not be unique, but we suppose that we have fixed one. Then (w.r.t. this sorting) we call $(x_i,y_i)$ the $\mathit{high}_i$-edge of $v$, for every $i\in\{1,\dots,k\}$. The lower endpoint of the $\mathit{high}_i$-edge of $v$ is called the $\mathit{high}_i$ point of $v$, and we denote it as $\mathit{high}_i(v)$ (i.e., we have $\mathit{high}_i(v)=y_i$). Notice that the definition of the $\mathit{high}_i$ points of $v$ is independent of the sorting of the back-edges in $B(v)$, provided only that this is in decreasing order w.r.t. the lower endpoints. If we want to reference the $\mathit{high}_i$-edge of $v$ with its endpoints, then we denote it as $(\mathit{highD}_i(v),\mathit{high}_i(v))$. Of particular importance is the $\mathit{high}_1$ point of $v$, which we may simply denote as $\mathit{high}(v)$. Also, we denote the $\mathit{high}_1$-edge of $v$ as $e_\mathit{high}(v)$. The $\mathit{high}$ points have been introduced relatively recently (as a concept dual to the $\mathit{low}$ points) in order to solve various problems of low connectivity with a DFS-based approach (see, e.g., \cite{DBLP:conf/isaac/GeorgiadisK20,DBLP:conf/esa/GeorgiadisIK21}). 
One of the reasons that the $\mathit{high}$ points are useful is that they let us know whether there exists a back-edge that leaps over a vertex $u$, but not over a specific proper ancestor $v$ of $u$. (Specifically, this is equivalent to $\mathit{high}(u)\geq v$.) Thus, if that is the case, then we know that if we remove both $(u,p(u))$ and $(v,p(v))$ from the graph, then $u$ is connected with $p(u)$ through a path that uses $e_\mathit{high}(u)$. In Section~\ref{subsection:high} we show how to compute the $\mathit{high}_i$-edges of all vertices, for every $i\in\{1,\dots,k\}$, where $k$ is a fixed integer, in total linear time (see Proposition~\ref{proposition:high}).

\subsubsection{Maximum points, leftmost and rightmost edges}
Now we consider concepts that are defined in relation to the higher endpoints of the leaping back-edges. Let $v\neq r$ be a vertex. We let $M(v)$ denote the maximum vertex that is an ancestor of the higher endpoints of the back-edges in $B(v)$. Equivalently, $M(v)$ is the nearest common ancestor of the higher endpoints of the back-edges in $B(v)$. (If $B(v)=\emptyset$, then we let $M(v):=\bot$.) Notice that $M(v)$ (if it exists) is a descendant of $v$. For every vertex $x$, we let $M^{-1}(x)$ denote the list of all vertices $v$ with $M(v)=x$, sorted in decreasing order. Thus, we have that all vertices in $M^{-1}(x)$ have $x$ as a common descendant, and therefore they are related as ancestor and descendant. For every vertex $v\in M^{-1}(x)$, we let $\mathit{nextM}(v)$ and $\mathit{prevM}(v)$ denote the successor and the predecessor, respectively, of $v$ in $M^{-1}(x)$. Equivalently, we have that $\mathit{nextM}(v)$ (resp., $\mathit{prevM}(v)$) is the greatest proper ancestor (resp., the lowest proper descendant) $u$ of $v$ such that $M(u)=M(v)$. We also let $\mathit{lastM}(v)$ denote the lowest vertex in $M^{-1}(M(v))$. 

We extend the concept of the $M$ points in general sets of back-edges. Thus, if $S$ is a set of back-edges, then we let $M(S)$ denote the nearest common ancestor of the higher endpoints of the back-edges in $S$. (Thus, we have $M(v)=M(B(v))$.) We introduce a notation for the $M$ points of some special sets of back-edges. Let $c$ be a descendant of a vertex $v\neq r$, and let $S$ be the set of the back-edges that leap over $v$ and stem from the subtree of $c$ (i.e., $S=\{(x,y)\in B(v)\mid x\mbox{ is a descendant of }c\}$). Then we denote $M(S)$ as $M(v,c)$. Also, let $\widetilde{S}$ be the set of the back-edges in $B(v)$ that stem from proper descendants of $M(v)$ (i.e., $\widetilde{S}=\{(x,y)\in B(v)\mid x\mbox{ is a proper descendant of }M(v)\}$). Then we denote $M(\widetilde{S})$ as $\widetilde{M}(v)$. In Section~\ref{subsection:M} we deal with the computation of the $M$ points. This relies on the computation of the leftmost and the rightmost points, which we define next. 

Let $v\neq r$ be a vertex, and let $(x_1,y_1),\dots,(x_k,y_k)$ be the list of the back-edges in $B(v)$ sorted in increasing order w.r.t. their higher endpoint. We note that such a sorting may not be unique, but we suppose that we have fixed one. Then we call $(x_i,y_i)$ the $i$-th leftmost edge of $v$. We denote $x_i$ as $L_i(v)$, and we call it the $i$-th leftmost point of $v$. Of particular importance is the first leftmost edge of $v$, which we denote as $e_L(v)$. On the other hand, let $(x_1,y_1),\dots,(x_k,y_k)$ be the list of the back-edges in $B(v)$ sorted in decreasing order w.r.t. their higher endpoint. Again, such a sorting may not be unique, but we suppose that we have fixed one. Then we call $(x_i,y_i)$ the $i$-th rightmost edge of $v$. We denote $x_i$ as $R_i(v)$, and we call it the $i$-th rightmost point of $v$. Of particular importance is the first rightmost edge of $v$, which we denote as $e_R(v)$. (We note that the edges $e_L(v)$ and $e_R(v)$ were used in \cite{DBLP:conf/esa/NadaraRSS21}, with different notation.) In Section~\ref{subsection:leftmost} we extend the concepts of the leftmost and the rightmost edges, and provide an efficient method to compute them (see Proposition~\ref{proposition:L-sets}).

\subsection{Properties of the DFS parameters}
\label{subsection:dfs-properties}

\begin{lemma}
\label{lemma:M}
Let $u$ and $v$ be two vertices $\neq r$ such that $v$ is an ancestor of $u$ and $M(v)$ is a descendant of $u$. Then $M(v)$ is a descendant of $M(u)$.
\end{lemma}
\begin{proof}
Let $(x,y)$ be a back-edge in $B(v)$. Then $x$ is a descendant of $M(v)$, and therefore a descendant of $u$. Futhermore, $y$ is a proper ancestor of $v$, and therefore a proper ancestor of $u$. This shows that $(x,y)\in B(u)$, and thus $x$ is a descendant of $M(u)$. Due to the generality of $(x,y)\in B(v)$, this implies that $M(v)$ is a descendant of $M(u)$.
\end{proof}

\begin{lemma}
\label{lemma:same_m_subset_B}
Let $u$ and $v$ be two vertices $\neq r$ such that $v$ is an ancestor of $u$ and $M(v)$ is a descendant of $M(u)$. Then $B(v)\subseteq B(u)$.
\end{lemma}
\begin{proof}
Let $(x,y)$ be a back-edge in $B(v)$. Then $x$ is a descendant of $M(v)$, and therefore a descendant of $M(u)$. Furthermore, $y$ is a proper ancestor of $v$, and therefore a proper ancestor of $u$. This shows that $(x,y)\in B(u)$. Due to the generality of $(x,y)\in B(v)$, we conclude that $B(v)\subseteq B(u)$.
\end{proof}

\begin{lemma}
\label{lemma:same_high}
Let $u$ and $v$ be two vertices $\neq r$ such that $u$ is a descendant of $v$ and $\mathit{high}(u)=\mathit{high}(v)$. Then $B(u)\subseteq B(v)$.
\end{lemma}
\begin{proof}
Let $(x,y)$ be a back-edge in $B(u)$. Then $x$ is a descendant of $u$, and therefore a descendant of $v$. Furthermore, $y$ is an ancestor of $\mathit{high}(u)$, and therefore an ancestor of $\mathit{high}(v)$, and therefore a proper ancestor of $v$. This shows that $(x,y)\in B(v)$. Due to the generality of $(x,y)\in B(u)$, we conclude that $B(u)\subseteq B(v)$.
\end{proof}

\begin{lemma}
\label{lemma:same_m_same_low}
Let $v$ and $v'$ be two vertices such that $M(v)=M(v')$. Then $\mathit{low}(v)=\mathit{low}(v')$.
\end{lemma}
\begin{proof}
Since $M(v)=M(v')$, we have that $v$ and $v'$ are related as ancestor and descendant. Thus, we may assume w.l.o.g. that $v'$ is an ancestor of $v$. Then, Lemma~\ref{lemma:same_m_subset_B} implies that $B(v')\subseteq B(v)$. This implies that $\mathit{low}(v)\leq\mathit{low}(v')$. Now let $(x,y)$ be a back-edge in $B(v)$ such that $y=\mathit{low}(v)$. Then $x$ is a descendant of $v$, and therefore a descendant of $v'$. Furthermore, both $y$ and $v'$ have $v$ as a common descendant, and therefore they are related as ancestor and descendant. Then, $y=\mathit{low}(v)\leq\mathit{low}(v')<v'$ implies that $y$ is a proper ancestor of $v'$. This shows that $(x,y)\in B(v')$, and therefore $y\geq\mathit{low}(v')$. Thus, we conclude that $\mathit{low}(v)=\mathit{low}(v')$.
\end{proof}

\begin{lemma}
\label{lemma:high_and_low}
Let $u$ and $v$ be two vertices $\neq r$ such that $v$ is an ancestor of $u$ and $\mathit{high}(u)=\mathit{high}(v)$. Then $\mathit{low}(v)\leq\mathit{low}(u)$.
\end{lemma}
\begin{proof}
By Lemma~\ref{lemma:same_high} we have that $B(u)\subseteq B(v)$, and thus we get $\mathit{low}(v)\leq\mathit{low}(u)$ as an immediate consequence.  
\end{proof} 

\begin{lemma}
\label{lemma:same_M_dif_B_lower}
Let $u$ and $v$ be two vertices $\neq r$ such that $M(u)=M(v)$, $v$ is a proper ancestor of $u$, and $B(u)\neq B(v)$. Then $\mathit{high}(u)$ is a descendant of $v$. 
\end{lemma}
\begin{proof}
By Lemma~\ref{lemma:same_m_subset_B} we have $B(v)\subseteq B(u)$. Since $u$ is a common descendant of $v$ and $\mathit{high}(u)$, we have that $v$ and $\mathit{high}(u)$ are related as ancestor and descendant. Now let us suppose, for the sake of contradiction, that $\mathit{high}(u)$ is not a descendant of $v$. This implies that $\mathit{high}(u)$ is a proper ancestor of $v$. Now let $(x,y)$ be a back-edge in $B(u)$. Then we have that $x$ is a descendant of $u$, and therefore a descendant of $v$. Furthermore, $y$ is an ancestor of $\mathit{high}(u)$, and therefore a proper ancestor of $v$. This shows that $(x,y)\in B(v)$. Due to the generality of $(x,y)\in B(u)$, this implies that $B(u)\subseteq B(v)$. Thus, since $B(v)\subseteq B(u)$, we have that $B(u)=B(v)$, in contradiction to the assumption $B(u)\neq B(v)$. Thus, we conclude that $\mathit{high}(u)$ is a descendant of $v$. 
\end{proof}

\begin{lemma}
\label{lemma:same_M_same_high}
Let $u$ and $v$ be two vertices $\neq r$. Then the following are equivalent:
\begin{enumerate}[label=(\arabic*)]
\item{$B(u)=B(v)$}
\item{$M(u)=M(v)$ and $\mathit{high}(u)=\mathit{high}(v)$}
\item{$M(u)=M(v)$ and $\mathit{bcount}(u)=\mathit{bcount}(v)$} 
\end{enumerate}
\end{lemma}
\begin{proof}
$(1)$ obviously implies $(2)$ and $(3)$. Conversely, we will show that either of $(2)$ and $(3)$ also implies $(1)$. Let us first assume $(2)$. Let $(x,y)$ be a back-edge in $B(u)$. Then $x$ is a descendant of $M(u)$, and therefore a descendant of $M(v)$. Furthermore, $y$ is an ancestor of $\mathit{high}(u)$, and therefore an ancestor of $\mathit{high}(v)$, and therefore a proper ancestor of $v$. This shows that $(x,y)\in B(v)$. Due to the generality of $(x,y)\in B(u)$, this implies that $B(u)\subseteq B(v)$. Similarly, we can show the reverse inclusion, and therefore we have $B(u)=B(v)$. 

Now let us assume $(3)$. If $B(u)=\emptyset$, then $M(u)=\bot$, and therefore $M(v)=\bot$, and therefore $B(v)=\emptyset$. So let us assume that $B(u)\neq\emptyset$. Then $M(u)$ is defined, and therefore $M(u)=M(v)$ implies that $M(u)$ is a common descendant of $u$ and $v$, and therefore $u$ and $v$ are related as ancestor and descendant. Thus, we may assume w.l.o.g. that $u$ is a descendant of $v$. Then, since $M(u)=M(v)$, Lemma~\ref{lemma:same_m_subset_B} implies that $B(v)\subseteq B(u)$. Therefore, $\mathit{bcount}(u)=\mathit{bcount}(v)$ implies that $B(u)=B(v)$.
\end{proof}

The following proposition provides a criterion that characterizes $3$-edge-connected graphs. This has been established in \cite{DBLP:conf/esa/GeorgiadisIK21}, and we will use it throughout without explicit mention.

\begin{proposition}[\cite{DBLP:conf/esa/GeorgiadisIK21}]
\label{proposition:3-edge-conn}
$G$ is $3$-edge-connected if and only if: for every vertex $v\neq r$ we have $\mathit{bcount}(v)>1$, and for every two distinct vertices $u$ and $v$ such that $r\notin\{u,v\}$ we have $B(u)\neq B(v)$.
\end{proposition}

\begin{lemma}
\label{lemma:lower_than_high}
Let $v$ and $v'$ be two vertices with $M(v)=M(v')$ such that $v'$ is a proper ancestor of $v$. If $G$ is $3$-edge-connected, then $v'$ is an ancestor of $\mathit{high}(v)$.
\end{lemma}
\begin{proof}
We have that $\mathit{high}(v)$ is a proper ancestor of $v$. Thus, since $v'$ and $\mathit{high}(v)$ have $v$ as a common descendant, they are related as ancestor and descendant. Let us suppose, for the sake of contradiction, that $v'$ is not an ancestor of $\mathit{high}(v)$. Then, we have that $v'$ is a proper descendant of $\mathit{high}(v)$. Since $M(v)=M(v')$ and $v'$ is a proper ancestor of $v$, by Lemma~\ref{lemma:same_m_subset_B} we have that $B(v')\subseteq B(v)$. Now let $(x,y)$ be a back-edge in $B(v)$. Then $x$ is a descendant of $v$, and therefore a descendant of $v'$. Furthermore, $y$ is an ancestor of $\mathit{high}(v)$, and therefore a proper ancestor of $v'$. This shows that $(x,y)\in B(v')$. Due to the generality of $(x,y)\in B(v)$, this implies that $B(v)\subseteq B(v')$. But then, since $B(v')\subseteq B(v)$, we have that $B(v')=B(v)$, in contradiction to the fact that the graph is $3$-edge-connected. Thus, we conclude that $v'$ is an ancestor of $\mathit{high}(v)$. 
\end{proof}

\begin{lemma}
\label{lemma:e_L-e_R}
Let $v\neq r$ be a vertex and let $e$ be a back-edge in $B(v)$ such that $M(B(v)\setminus\{e\})\neq M(v)$. Then, either $e=e_L(v)$ or $e=e_R(v)$.
\end{lemma}
\begin{proof}
Let $X$ be the set of the higher endpoints of the back-edges in $B(v)$. Then we have $M(v)=\mathit{nca}(X)$, $L_1(v)=\mathit{min}(X)$ and $R_1(v)=\mathit{max}(X)$. We claim that $M(v)=\mathit{nca}\{L_1(v),R_1(v)\}$. First, we obviously have that $M(v)$ is an ancestor of $\mathit{nca}\{L_1(v),R_1(v)\}$. Conversely, $\mathit{nca}\{L_1(v),R_1(v)\}$ is an ancestor of both $L_1(v)$ and $R_1(v)$. Since $L_1(v)\leq R_1(v)$, this implies that $\mathit{nca}\{L_1(v),R_1(v)\}$ is an ancestor of every vertex $z$ such that $L_1(v)\leq z\leq R_1(v)$. Thus, since $L_1(v)=\mathit{min}(X)$ and $R_1(v)=\mathit{max}(X)$, we have that $\mathit{nca}\{L_1(v),R_1(v)\}$ is an ancestor of every vertex in $X$, and therefore $\mathit{nca}\{L_1(v),R_1(v)\}$ is an ancestor of $M(v)$. Thus, we have $M(v)=\mathit{nca}\{L_1(v),R_1(v)\}$.

Now let $X'$ be the set of the higher endpoints of the back-edges in $B(v)\setminus\{e\}$. Then we have $X'\subseteq X$ and $M(B(v)\setminus\{e\})=\mathit{nca}(X')$. Since $M(B(v)\setminus\{e\})\neq M(v)$ and $M(v)=\mathit{nca}\{L_1(v),R_1(v)\}$, this implies that we cannot have both $L_1(v)$ and $R_1(v)$ in $X'$. Thus, either $L_1(v)\notin X'$ or $R_1(v)\notin X'$. Let us assume that $L_1(v)\notin X'$. This implies that $e$ is the only back-edge in $B(v)$ whose higher endpoint is $L_1(v)$. Thus, by definition we have $e=e_L(v)$.
Similarly, if we have $R_1(v)\notin X'$, then we can infer that $e=e_R(v)$.
\end{proof}

\begin{lemma}
\label{lemma:w-v-u-not-related}
Let $w\neq r$ be a vertex, and let $u$ and $v$ be two descendants of $w$ such that $M(u)=M(w,x)\neq\bot$ and $M(v)=M(w,y)\neq\bot$, where $x$ and $y$ are descendants of different children of $M(w)$. Then $u$ and $v$ are not related as ancestor and descendant.
\end{lemma}
\begin{proof}
Let $c_1$ be the child of $M(w)$ that is an ancestor of $x$, and let $c_2$ be the child of $M(w)$ that is an ancestor of $y$. By assumption we have that $c_1\neq c_2$. Now let us suppose, for the sake of contradiction, that $u$ is not a descendant of $c_1$. Since $M(u)=M(w,x)$, we have that $M(u)$ is a common descendant of $u$ and $x$, and therefore $u$ and $x$ are related as ancestor and descendant. Since $u$ is not a descendant of $c_1$, we cannot have that $u$ is a descendant of $x$. Thus, $u$ is a proper ancestor of $x$. Then, we have that $x$ is a common descendant of $u$ and $c_1$, and therefore $u$ and $c_1$ are related as ancestor and descendant. Thus, we have that $u$ is a proper ancestor of $c_1$. This implies that $u$ is an ancestor of $M(w)$, and therefore an ancestor of $c_2$. Now, since $M(w,y)$ is defined, we have that there is a back-edge $(z,t)\in B(w)$ such that $z$ is a descendant of $y$. Then, $z$ is a descendant of $c_2$, and therefore a descendant of $u$. Furthermore, $t$ is a proper ancestor of $w$, and therefore a proper ancestor of $u$. This shows that $(z,t)\in B(u)$. This implies that $z$ is a descendant of $M(u)$. Since $M(u)=M(w,x)\neq\bot$, we have that there is a back-edge $(z',t')\in B(u)$ such that $z'$ is a descendant of $x$. Since $(z,t)\in B(u)$ and $(z',t')\in B(u)$, we have that $M(u)$ is an ancestor of $\mathit{nca}\{z,z'\}$. But $z$ is a descendant of $c_1$, whereas $z'$ is a descendant of $c_2$. This implies that $\mathit{nca}\{z,z'\}=M(w)$, and therefore $M(u)$ is an ancestor of $M(w)$, contradicting the fact $M(u)=M(w,x)$ (which implies that $M(u)$ is a descendant of $x$, and therefore of $c_1$). Thus, we have that $u$ is a descendant of $c_1$. Similarly, we can show that $v$ is a descendant of $c_2$. Thus, since $u$ and $v$ are descendants of different children of $M(w)$, we have that they cannot be related as ancestor and descendant.
\end{proof}

\begin{lemma}
\label{lemma:no_M(v)}
Let $v\neq r$ be a vertex with $B(v)\neq\emptyset$ such that there is no back-edge of the form $(M(v),z)$ in $B(v)$. Then $\mathit{low}(c_2)<v$, where $c_2$ is the $\mathit{low2}$ child of $M(v)$.
\end{lemma}
\begin{proof}
Let $(x,y)$ be a back-edge in $B(v)$. Then we have that $x$ is a descendant of $M(v)$. Since there is no back-edge of the form $(M(v),z)$ in $B(v)$, we have that $x\neq M(v)$. Thus, $x$ is a proper descendant of $M(v)$. This shows that $M(v)$ has at least one child. Furthermore, it cannot be the case that $M(v)$ has only one child, because otherwise all the back-edges of $B(v)$ would stem from the subtree of this child, and so $M(v)$ would be a descendant of its child, which is absurd. Thus, $M(v)$ has at least two children, and so it makes sense to consider the $\mathit{low2}$ child $c_2$ of $M(v)$.

Let us suppose, for the sake of contradiction, that $\mathit{low}(c_2)\geq v$. This implies that, of all the children of $M(v)$, only the $\mathit{low1}$ child $c_1$ of $M(v)$ may have $\mathit{low}(c_1)<v$. Let $(x,y)$ be a back-edge in $B(v)$. Then we have that $x$ is a descendant of $M(v)$. Since there is no back-edge of the form $(M(v),z)$ in $B(v)$, we have that $x\neq M(v)$, and therefore $x$ is a proper descendant of $M(v)$. Let $c$ be the child of $M(v)$ that is an ancestor of $x$. Since $(x,y)\in B(v)$, we have that $y$ is a proper ancestor of $v$, and therefore a proper ancestor of $M(v)$, and therefore a proper ancestor of $c$. This shows that $(x,y)\in B(c)$. Since $y$ is a proper ancestor of $v$, we have $y<v$. Thus, since $(x,y)\in B(c)$, we have that $\mathit{low}(c)\leq y<v$. This implies that $c=c_1$. Due to the generality of $(x,y)\in B(v)$, this implies that $M(v)$ is a descendant of $c_1$, which is absurd. Thus, our supposition cannot be true, and therefore we have that $\mathit{low}(c_2)<v$.
\end{proof}

\begin{lemma}
\label{lemma:type3-4cuts}
Let $C$ be an edge-minimal cut of $G$. Let $(v_1,p(v_1)),\dots,(v_k,p(v_k))$ be the list of the tree-edges in $C$. Let us assume w.l.o.g. that $v_1$ is the lowest among $v_1,\dots,v_k$. Then $v_1$ is a common ancestor of $\{v_1,\dots,v_k\}$. Furthermore, suppose that $C$ contains a back-edge $e$. Then $e\in B(v_1)\cup\dots\cup B(v_k)$.
\end{lemma}
\begin{proof}
Let us suppose, for the sake of contradiction, that $v_1$ is not a common ancestor of $\{v_1,\dots,v_k\}$. This means that at least one among $\{v_1,\dots,v_k\}$ is not a descendant of $v_1$. Now let $I$ be the collection of all indices in $\{1,\dots,k\}$ such that $v_i$ is a descendant of $v_1$, for every $i\in I$. Thus, we have $I\subset \{1,\dots,k\}$. Now let $C_I=\{(v_i,p(v_i))\mid i\in I\}$, and let $C'$ be the subset of $C$ that consists of all the back-edges in $C$. Then, since $I\subset \{1,\dots,k\}$, we have $C_I\cup C'\subset C$. Thus, since $C$ is an edge-minimal cut of $G$, we have that $G'=G\setminus(C_I\cup C')$ is connected. Thus, there is a path $P$ from $v_1$ to $p(v_1)$ in $G'$. Then, by Lemma~\ref{lemma:back-edge-or-tree-edge} we have that the first occurrence of an edge in $P$ that leads outside of $T(v_1)$ is either $(v_1,p(v_1))$ or a back-edge that leaps over $v_1$. The first case is rejected, since $(v_1,p(v_1))\in C_I$. Thus, the first occurrence of an edge in $P$ that leads outside of $T(v_1)$ is a back-edge $(x,y)$ that leaps over $v_1$. Now consider the part $P'$ of $P$ from $v_1$ up to, and including, $x$. Then we have that $P'$ avoids the tree-edges from $C$ that have the form $(v,p(v))$ where $v$ is a descendant of $v_1$ (since $P$ has this property). Also, $P'$ avoids all the back-edges from $C$ (since $P$ has this property). Furthermore, $P'$ avoids the tree-edges of the form $(v,p(v))$ where $v$ is not a descendant of $v_1$, because it is the initial part of $P$ that lies entirely within $T(v_1)$. This shows that $P'$ is a path in $G\setminus C$. Since $v_1$ is the minimum among $\{v_1,\dots,v_k\}$, we have that no vertex in $\{v_1,\dots,v_k\}$ is a proper ancestor of $v_1$. Thus, the tree-path $T[p(v_1),r]$ remains intact in $G\setminus C$. But then, $P'+(x,y)+T[y,p(v_1)]$ is a path from $v_1$ to $p(v_1)$ in $G\setminus C$, in contradiction to the fact that $C$ is an edge-minimal cut of $G$. This shows that $v_1$ is a common ancestor of $\{v_1,\dots,v_k\}$.

Now let $e$ be a back-edge in $C$. Let us suppose, for the sake of contradiction, that $e\notin B(v_1)\cup\dots\cup B(v_k)$. Let $e=(x,y)$. Then we have that none of $v_1,\dots,v_k$, can be on the tree-path $T[x,y]$. Thus, $T[x,y]$ remains intact in $G\setminus C$, and therefore the endpoints of $e$ remain connected in $G\setminus C$, in contradiction to the fact that $C$ is an edge-minimal cut of $G$. We conclude that $e\in B(v_1)\cup\dots\cup B(v_k)$.
\end{proof}

\subsection{Computing the $\mathit{low}$-edges}
\label{subsection:low}
Let $v\neq r$ be a vertex. The definition of the $\mathit{low}_i$-edges of $v$, for $i=1,2,\dots$, assumes any ordering of the back-edges in $B(v)$ that it is increasing w.r.t. the lower endpoints. For computational purposes (basically, for convenience in our arguments), we will fix such an ordering for sets of back-edges, which we call the $\mathit{low}$ ordering. Let $(x_1,y_1),\dots,(x_t,y_t)$ be a list of back-edges. Then we say that this list is sorted in the $\mathit{low}$ ordering if it is increasing w.r.t. the lower endpoints, and also satisfies $x_i\leq x_{i+1}$, for every $i\in\{1,\dots,t-1\}$ such that $y_i=y_{i+1}$.\footnote{Here we have to be a little more precise. Since we consider multigraphs, we may have several back-edges of the form $(x,y)$ in a set of back-edges. Then we assume a unique integer identifier that is assigned to every edge of the graph, and the ties here are broken according to those identifiers. However, for the sake of simplicity, we will not make explicit use of this information in what follows.}  

Now let $v\neq r$ be a vertex, and let $(x_1,y_1),\dots,(x_t,y_t)$ be the list of the back-edges in $B(v)$ sorted in the $\mathit{low}$ ordering. Then we let $(x_i,y_i)$ be the $\mathit{low}_i$-edge of $v$, for every $i\in\{1,\dots,t\}$. We assume that the $\mathit{low}$ ordering is applied for every set of leaping back-edges, and the $\mathit{low}_i$-edges correspond to this ordering. Then we have the following.

\begin{lemma}
\label{lemma:lowk}
Let $v\neq r$ be a vertex, and let $(v,z_1),\dots,(v,z_s)$ be the list of the back-edges with higher endpoint $v$, sorted in the $\mathit{low}$ ordering. Let $e$ be the $\mathit{low}_k$-edge of $v$, for some $k\geq 1$. Then, either $e\in\{(v,z_1),\dots,(v,z_k)\}$, or there is a child $c$ of $v$ such that $e$ is the $\mathit{low}_{k'}$-edge of $c$, for some $k'\leq k$.
\end{lemma}
\begin{proof}
Let $(x_1,y_1),\dots,(x_t,y_t)$ be the list of the back-edges in $B(v)$ sorted in the $\mathit{low}$ ordering. Then we have $e=(x_k,y_k)$.

First, let us suppose, for the sake of contradiction, that $s>k$ and $e\in\{(v,z_{k+1}),\dots,(v,z_s)\}$. Then, since the back-edges in $(v,z_1),\dots,(v,z_s)$ are sorted in increasing order w.r.t. their lower endpoint, we have that, for every $i\in\{1,\dots,s\}$, there is a $j\in\{i,\dots,t\}$ such that $(v,z_i)=(x_j,y_j)$. Since $e\in\{(v,z_{k+1})),\dots,(v,z_s)\}$, there is an $i\in\{k+1,\dots,s\}$ such that $e=(v,z_i)$. But then we have $e=(x_j,y_j)$ for some $j\in\{k+1,\dots,t\}$, contradicting the fact that $e=(x_k,y_k)$. This shows that either $e\in\{(v,z_1),\dots,(v,z_k)\}$, or $e$ does not belong to the set $\{(v,z_1),\dots,(v,z_s)\}$ at all.

Now let us assume that $e\notin\{(v,z_1),\dots,(v,z_s)\}$. Then, since $e=(x_k,y_k)$, we have $x_k\neq v$, and therefore $x_k$ is a proper descendant of $v$. So let $c$ be the child of $v$ that is an ancestor of $x_k$. Then we have $e\in B(c)$. Let $(x_1',y_1'),\dots,(x_{t'}',y_{t'}')$ be the list of the back-edges in $B(c)$ sorted in the $\mathit{low}$ ordering. Then we have that the $\mathit{low}_{i}$-edge of $c$, for any $i\in\{1,\dots,t'\}$, is $(x_i',y_i')$. Now let us suppose, for the sake of contradiction, that $t'>k$ and $e\in\{(x_{k+1}',y_{k+1}'),\dots,(x_{t'}',y_{t'}')\}$. So let $i$ be the index in $\{k+1,\dots,t'\}$ such that $e=(x_i',y_i')$. Since $e=(x_k,y_k)\in B(v)$ we have that $y_k$ is a proper ancestor of $v$, and therefore $y_k<v$. Thus, since $y_k=y_i'$, we have $y_i'<v$. Since the back-edges in $(x_1',y_1'),\dots,(x_{t'}',y_{t'}')$ are sorted in increasing order w.r.t. their lower endpoint, we have $y_j'\leq y_i'$, and therefore $y_j'<v$, for every $j\in\{1,\dots,i\}$. Now let $j$ be an index in $\{1,\dots,i\}$. Then we have that $x_j'$ is a descendant of $c$, and therefore a descendant of $v$. Since $(x_j',y_j')$ is a back-edge, we have that $x_j'$ is a descendant of $y_j'$. Thus, $x_j'$ is a common descendant of $v$ and $y_j'$, and therefore $v$ and $y_j'$ are related as ancestor and descendant. Since $j\in\{1,\dots,i\}$, we have $y_j'<v$, and therefore $y_j'$ is a proper ancestor of $v$. This shows that all back-edges in $(x_1',y_1'),\dots,(x_i',y_i')$ leap over $v$. Thus, since $(x_1',y_1'),\dots,(x_i',y_i')$ and $(x_1,y_1),\dots,(x_t,y_t)$  are sequences of back-edges sorted in the $\mathit{low}$ ordering and $\{(x_1',y_1'),\dots,(x_i',y_i')\}\subseteq \{(x_1,y_1),\dots,(x_t,y_t)\}$, we have $(x_j',y_j')=(x_{j'},y_{j'})$, where $j'\geq j$, for every $j\in\{1,\dots,i\}$. But since $i\geq k+1$, this implies that $e=(x_i',y_i')=(x_{i'},y_{i'})$ for some $i'\geq k+1$, contradicting the fact that $e=(x_k,y_k)$. This shows that $e\in\{(x_1',y_1'),\dots,(x_k',y_k')\}$.
\end{proof}

Lemma~\ref{lemma:lowk} provides enough information in order to find the $\mathit{low}_i$-edge of $v$, for any $i\geq 1$. Specifically, it is sufficient to search for it in the list of the first $i$ back-edges of the form $(v,z)$ with the lowest lower endpoint, plus the lists of the $\mathit{low}_1$-,$\dots$,$\mathit{low}_i$-edges of all the children of $v$. A procedure that computes the $\mathit{low}_1$-,$\dots$,$\mathit{low}_k$-edges of all vertices, for any fixed $k$, is shown in Algorithm~\ref{algorithm:low}. 
The idea is to process the vertices in a bottom-up fashion (e.g., in decreasing order w.r.t. the DFS numbering). Thus, for every vertex $v$ that we process, we have computed the $\mathit{low}_1$-,$\dots$,$\mathit{low}_k$-edges of its children, and therefore we have to check among those, plus the $k$ back-edges of the form $(v,z)$ with the lowest lower endpoint, in order to get the $\mathit{low}_1$-,$\dots$,$\mathit{low}_k$-edges of $v$. For our purposes in this work, $k$ will be a fixed constant (at most $4$), and thus the dependency of the running time on $k$ does not matter for us. However, we use balanced binary-search trees (BST) in order to get an algorithm with $O(m+nk\log k)$ time.
Our result is summarized in Proposition~\ref{proposition:low}.

\begin{algorithm}[h!]
\caption{\textsf{Compute the $\mathit{low}_i$-edges of all vertices, for all $i\in\{1,\dots,k\}$}}
\label{algorithm:low}
\LinesNumbered
\DontPrintSemicolon
compute the $\mathit{low}$ ordering of the adjacency list of every vertex\;
\label{line:low-sort}
initialize an empty balanced binary-search tree $\mathit{BST}[v]$ that stores back-edges, for every vertex $v$\;
\tcp{$\mathit{BST}[v]$ sorts the edges it stores w.r.t. the $\mathit{low}$ ordering}
\For{$v\leftarrow n$ to $v=2$}{
\label{line:low-for}
  let $L=(v,z_1),\dots,(v,z_s)$ be the list of the back-edges with higher endpoint $v$, sorted in the $\mathit{low}$ ordering\;
  fill $\mathit{BST}[v]$ with the first $k$ (non-$\mathit{null}$) back-edges from $L$\;
  \label{line:low-l(v)}
  let $c_1,\dots,c_t$ be the children of $v$ sorted in increasing order\;
  \For{$c\leftarrow c_1$ to $c=c_t$}{
    \ForEach{$i\in\{1,\dots,k\}$}{
    \label{line:low-for-cback-edges}
      let $(x,y)$ be the $\mathit{low}_i$-edge of $c$\;
          \label{line:low-for-cback-edges-select}
      \lIf{$y\geq v$}{\textbf{continue}}
      \If{$\mathit{BST}[v]$ has less than $k$ entries}{
        insert $(x,y)$ into $\mathit{BST}[v]$\;
        \label{line:low-insert-1}
      } 
      \Else{
        let $(x',y')$ be the $k$-th entry of $\mathit{BST}[v]$\;
        \If{$y<y'$}{
        \label{line:low-condition}
          delete the $k$-th entry of $\mathit{BST}[v]$\;
          \label{line:low-delete}
          insert $(x,y)$ into $\mathit{BST}[v]$\;
          \label{line:low-insert-2}
        }
      }
    }  
  }
  \ForEach{$i\in\{1,\dots,k\}$}{
  \label{line:low-fin}
    let the $\mathit{low}_i$-edge of $v$ be the $i$-th entry in $\mathit{BST}[v]$\; 
  }
}
\end{algorithm}

\begin{proposition}
\label{proposition:low}
Let $k$ be any fixed integer. Algorithm~\ref{algorithm:low} computes the $\mathit{low}_i$-edges of all vertices, for all $i\in\{1,\dots,k\}$. Furthermore, it runs in $O(m+nk\log{k})$ time.
\end{proposition}
\begin{proof}
We will prove correctness inductively, by establishing that, whenever the \textbf{for} loop in Line~\ref{line:low-for} processes a vertex $v$, we have that the $\mathit{low}_i$-edges of the children of $v$ have been correctly computed, for every $i\in\{1,\dots,k\}$. Initially, this is trivially true (because we process the vertices in decreasing DFS order, and so the first vertex that we process is a leaf). So let us suppose that the \textbf{for} loop in Line~\ref{line:low-for} starts processing a vertex $v$ for which we have computed the $\mathit{low}_i$-edges of its children, for every $i\in\{1,\dots,k\}$. It is sufficient to show that, by the time the processing of $v$ is done, we have correctly computed the $\mathit{low}_i$-edge of $v$, for every $i\in\{1,\dots,k\}$. 

Let $(v,z_1),\dots,(v,z_s)$ be the list of the back-edges with higher endpoint $v$, sorted in the $\mathit{low}$ ordering (and thus in increasing order w.r.t. their lower endpoint). Let also $c_1,\dots,c_t$ be the children of $v$ sorted in increasing order. 
In Line~\ref{line:low-l(v)} we fill $\mathit{BST}[v]$ with the first $k$ (non-null) back-edges from $\{(v,z_1),\dots,(v,z_s)\}$. Notice that these are all back-edges that leap over $v$. Then, we may insert more back-edges into $\mathit{BST}[v]$ in Line~\ref{line:low-insert-1} or in Line~\ref{line:low-insert-2}. In either case, the back-edge $(x,y)$ that we insert into $\mathit{BST}[v]$ is a back-edge in $B(c)$, for a child $c$ of $v$, that satisfies $y<v$. Since $(x,y)\in B(c)$, we have that $x$ is a descendant of $c$, and therefore a descendant of $v$. Since $(x,y)$ is a back-edge, we have that $x$ is a descendant of $y$. Thus, $x$ is a common descendant of $v$ and $y$, and therefore $v$ and $y$ are related as ancestor and descendant. Then, $y<v$ implies that $y$ is a proper ancestor of $v$. Thus, we have that $(x,y)$ is a back-edge in $B(v)$. This shows that, when we reach Line~\ref{line:low-fin}, we have that all elements of $\mathit{BST}[v]$ are back-edges that leap over $v$. Furthermore, notice that, when a back-edge is deleted from $\mathit{BST}[v]$ in Line~\ref{line:low-delete}, this is because its lower endpoint is greater than that of the back-edge that is to be inserted. Thus, when we reach Line~\ref{line:low-fin}, we have that the back-edges in $\mathit{BST}[v]$ are those with the $k$-th lowest lower endpoints among the back-edges that we have met during the processing of $v$. 

Let $(x_1,y_1),\dots,(x_N,y_N)$ be the list of the back-edges in $B(v)$ sorted in the $\mathit{low}$ ordering.
Let $i$ be an index in $\{1,\dots,k\}$, and let $(x,y)$ be the $\mathit{low}_i$-edge of $v$. Thus, we have $(x,y)=(x_i,y_i)$. According to Lemma~\ref{lemma:lowk}, we have that $(x,y)$ is either in $\{(v,z_1),\dots,(v,z_i)\}$, or it is the $\mathit{low}_{i'}$-edge of a child $c$ of $v$, for an $i'\in\{1,\dots,i\}$. In the first case, we have that $(x,y)$ was inserted into $\mathit{BST}[v]$ in Line~\ref{line:low-l(v)}. In the second case, due to the inductive hypothesis, we have that $(x,y)$ was processed at some point during the \textbf{for} loop in Line~\ref{line:low-for-cback-edges} (during the processing of a child $c$ of $v$). We will show that, when we reach Line~\ref{line:low-fin}, we have that $(x,y)$ is contained in $\mathit{BST}[v]$. 

First, let us suppose, for the sake of contradiction, that $(x,y)$ was inserted into $\mathit{BST}[v]$ at some point, but later on it was deleted. Then the deletion took place in Line~\ref{line:low-delete}, due to the existence of a back-edge $(z,w)\in B(c)$ that has $w<y$, for a child $c$ of $v$, and $(x,y)$ was the $k$-th entry of $\mathit{BST}[v]$. But then, due to the sorting of $\mathit{BST}[v]$, this implies that there are $k$ back-edges in $B(v)$ that precede $(x,y)$ in the $\mathit{low}$ ordering, contradicting the fact that $(x,y)$ is the $\mathit{low}_i$-edge of $v$, for some $i\leq k$. Thus, it is impossible that $(x,y)$ was inserted into $\mathit{BST}[v]$ at some point, but later on it was deleted. Now let us suppose, for the sake of contradiction, that $(x,y)$ was processed at some point, but it was not inserted into $\mathit{BST}[v]$. This implies that $(x,y)$ was met during the processing of a child $c$ of $v$, Line~\ref{line:low-condition} was reached, but the condition in this line was not satisfied. Thus, the $k$-th entry $(x',y')$ of $\mathit{BST}[v]$ had $y'\leq y$. Since the back-edges in $\mathit{BST}[v]$ are sorted in the $\mathit{low}$ ordering, we cannot have that $(x',y')$ is a predecessor of $(x,y)$ in this ordering, because otherwise we contradict the fact that $(x,y)$ is the $\mathit{low}_i$-edge of $v$, where $i\leq k$. Thus, we have that either $y<y'$, or $y=y'$ and $x<x'$, (or $(x,y)=(x',y')$, but the precedence is given to $(x,y)$). Thus, since $y'\leq y$, we have $y=y'$, and therefore $x\leq x'$. Let $c'$ be the child whose processing led to the insertion of $(x',y')$ in $\mathit{BST}[v]$. Then, since we process the children of $v$ in increasing order, we have that $c'\leq c$. Thus, since $x$ is a descendant of $c$, and $x'$ is a descendant of $c'$, and $x\leq x'$, we have that $c'=c$. But then, since $(x,y)$ precedes $(x',y')$ in the $\mathit{low}$ orderding, we have that $(x,y)$ was met before $(x',y')$ during the \textbf{for} loop in Line~\ref{line:low-for-cback-edges}, due to the selection in Line~\ref{line:low-for-cback-edges-select}, a contradiction. This shows that, when we reach Line~\ref{line:low-fin}, we have that $(x,y)$ is contained in $\mathit{BST}[v]$.   


Now let $k'$ be the maximum index in $\{1,\dots,k\}$ such that the $\mathit{low}_{k'}$-edge of $v$ is not null. (We note that, if $k'<k$, then $|B(v)|=k'$.) Then we have that, when we reach Line~\ref{line:low-fin}, $\mathit{BST}[v]$ is filled with $k'$ back-edges from $B(v)$. Furthermore, we have established that the $\mathit{low}_i$-edge of $v$ is included in $\mathit{BST}[v]$, for every $i\in\{1,\dots,k'\}$. Thus, since the sorting of $\mathit{BST}[v]$ corresponds to the sorting of the list of back-edges in $B(v)$ that provides the $\mathit{low}$-edges, we have that the $i$-th entry of $\mathit{BST}[v]$ is the $\mathit{low}_i$-edge of $v$, for every $i\in\{1,\dots,k'\}$.  

This establishes the correctness of Algorithm~\ref{algorithm:low}. It is not difficult to argue about its complexity. First, the sorting of the adjacency lists in Line~\ref{line:low-sort} can be performed in $O(m)$ time with bucket-sort (since the $\mathit{low}$ ordering is basically a variant of lexicographic order). Now, whenever the \textbf{for} loop in Line~\ref{line:low-for} processes a vertex $v$, we have that $\mathit{BST}[v]$ is filled with at most $k$ edges in Line~\ref{line:low-l(v)}. Then, the \textbf{for} loop in Line~\ref{line:low-for-cback-edges} processes at most $k\cdot \mathit{numChild}(v)$ back-edges in total, where $\mathit{numChild}(v)$ denotes the number of the children of $v$. Every one of those back-edges may be inserted into $\mathit{BST}[v]$ either in Line~\ref{line:low-insert-1} or in Line~\ref{line:low-insert-2}. Furthermore, it may force a deletion from $\mathit{BST}[v]$ in Line~\ref{line:low-delete}. Thus, the total number of $\mathit{BST}$ operations during the processing of $v$ is $O(k(\mathit{numChild}(v)+1))$. Thus, the total number of $\mathit{BST}$ operations during the course of Algorithm~\ref{algorithm:low} is $O(kn)$. Since every one of those operations is performed on a balanced binary-search tree with no more than $k$ entries, we have that it incurs cost $O(\log k)$. Thus, the total cost of $\mathit{BST}$ operations during the course of Algorithm~\ref{algorithm:low} is $O(nk\log k)$. We conclude that Algorithm~\ref{algorithm:low} runs in $O(m+nk\log k)$ time.
\end{proof}

\subsection{Computing the $\mathit{high}$-edges}
\label{subsection:high}

Let $v\neq r$ be a vertex. The definition of the $\mathit{high}_i$-edges of $v$, for $i=1,2,\dots$, assumes any ordering of the back-edges in $B(v)$ that it is decreasing w.r.t. the lower endpoints. For convenience in our arguments, we will fix such an ordering for sets of back-edges, which we call the $\mathit{high}$ ordering. Let $(x_1,y_1),\dots,(x_t,y_t)$ be a list of back-edges sorted in decreasing order w.r.t. the lower endpoints, while also satisfying $x_i\leq x_{i+1}$, for every $i\in\{1,\dots,t-1\}$ such that $y_i=y_{i+1}$. Then we say that this list is sorted in the $\mathit{high}$ ordering. If $(x_1,y_1),\dots,(x_t,y_t)$ is the list of the back-edges in $B(v)$ sorted in the $\mathit{high}$ ordering, then we let $(x_i,y_i)$ be the $\mathit{high}_i$-edge of $v$, for every $i\in\{1,\dots,t\}$. We assume that the $\mathit{high}$ ordering is applied for every set of leaping back-edges, and the $\mathit{high}_i$-edges correspond to this ordering.

Now let $k$ be a fixed positive integer. In order to compute the $\mathit{high}_i$-edges of all vertices, for every $i\in\{1,\dots,k\}$, the idea is to process all the back-edges according to the $\mathit{high}$ ordering. For every vertex $v$, we maintain a variable $\mathit{minIndex}[v]$, that stores the minimum $i$ such that the $\mathit{high}_i$-edge of $v$ is not yet computed. (Initially, we set $\mathit{minIndex}[v]\leftarrow 1$.) Then, for every back-edge $(x,y)$ that we process, we ascend the tree-path $T[x,y)$, and we have that the $\mathit{high}_i$-edge of $v$ is $(x,y)$, for every $v\in T[x,y)$ such that $\mathit{minIndex}[v]=i$. In order to implement this idea efficiently, whenever we ascend the path $T[x,y)$ during the processing of a back-edge $(x,y)$, we have to avoid the vertices for which the $\mathit{high}_i$ edges are computed, for every $i\in\{1,\dots,k\}$. (In other words, we have to avoid the vertices that have $\mathit{minIndex}=k+1$.) We can achieve this with the use of a DSU data structure. Specifically, the DSU data structure maintains sets of vertices that have $\mathit{minIndex}=k+1$, or they are singletons. The sets maintained by the data structure are subtrees of $T$. The operations supported by this data structure are $\mathtt{find}(v)$ and $\mathtt{unite}(u,v)$. The operation $\mathtt{find}(v)$ returns the root of the subtree maintained by the DSU that contains $v$. The operation $\mathtt{unite}(u,v)$ unites the subtree that contains $u$ with the subtree that contains $v$ into a larger subtree. Whenever $\mathtt{unite}(u,v)$ is called, we have $v=p(u)$. We use those operations whenever we ascend the tree-path $T[x,y)$, during the processing of a back-edge $(x,y)$. Thus, whenever we meet a vertex $v$ which has $\mathit{minIndex}[v]=k+1$, we first unite it with its parent $p(v)$, if $p(v)$ also has $\mathit{minIndex}[p(v)]=k+1$, and then we move to $\mathtt{find}(p(v))$. Thus, the next time that we meet $v$, we can jump immediately to the root of the subtree maintained by the DSU that contains $p(v)$. This idea for computing the $\mathit{high}_i$-edges of all vertices, for every $i\in\{1,\dots,k\}$, is shown in Algorithm~\ref{algorithm:high}. The proof of correctness and the complexity of this algorithm is given in Proposition~\ref{proposition:high}. 

\begin{algorithm}[h!]
\caption{\textsf{Compute the $\mathit{high}_i$-edges of all vertices, for all $i\in\{1,\dots,k\}$}}
\label{algorithm:high}
\LinesNumbered
\DontPrintSemicolon
sort the adjacency list of every vertex in increasing order w.r.t. the higher endpoints of its edges\;
\label{line:high-sort}
initialize an array $\mathit{minIndex}$, for every vertex $v\neq r$\;
\ForEach{vertex $v\neq r$}{
  \ForEach{$i\in\{1,\dots,k\}$}{
    let the $\mathit{high}_i$-edge of $v$ be $\bot$\;
  }  
  let $\mathit{minIndex}[v]\leftarrow 1$\;
}
\For{$y\leftarrow n-1$ to $y=1$}{
\label{line:high-for-y}
  \ForEach{back-edge $(x,y)$ in the adjacency list of $y$}{
  \label{line:high-for-back-edge}
    let $v\leftarrow x$\;
    \label{line:high-set-v}
    \While{$v\neq y$}{
    \label{line:high-while}
      \If{$\mathit{minIndex}[v]=k+1$}{
      \label{line:high-if}
        \While{$\mathit{minIndex}[p[v]]=k+1$}{
        \label{line:high-while-2}
          $\mathtt{unite}(v,p(v))$\;
          \label{line:high-unite}
          $v\leftarrow\mathtt{find}(v)$\; 
          \label{line:high-find}
        }
        $v\leftarrow p(v)$\;
        \label{line:high-go-to-p(v)}
      }
      \lIf{$v=y$}{\textbf{break}}
      \label{line:high-break}
      let $i\leftarrow\mathit{minIndex}[v]$\;
      \label{line:high-index-i}
      let the $\mathit{high}_i$-edge of $v$ be $(x,y)$\;
      \label{line:high-assign}
      $\mathit{minIndex}[v]\leftarrow i+1$\;
      \label{line:high-minindex+}
      $v\leftarrow p(v)$\; 
      \label{line:high-next}   
    }
  }
}
\end{algorithm}

\begin{proposition}
\label{proposition:high}
Let $k$ be any fixed positive integer. Then Algorithm~\ref{algorithm:high} computes the $\mathit{high}_i$-edges of all vertices $\neq r$, for every $i\in\{1,\dots,k\}$, in $O(m+kn)$ time in total.
\end{proposition}
\begin{proof}
We will prove correctness inductively, by establishing that whenever we reach the condition of the \textbf{while} loop in Line~\ref{line:high-while}, we have:
\begin{enumerate}[label=(\arabic*)]
\item{For every vertex $z$ such that $\mathit{minIndex}[z]\leq k$, the $\mathit{high}_i$-edge of $z$ has been correctly computed, for every $i<\mathit{minIndex}[z]$.}
\item{For every vertex $z$ such that $\mathit{minIndex}[z]\leq k$, no back-edge that has been processed so far by the \textbf{for} loop in Line~\ref{line:high-for-back-edge} prior to the processing of $(x,y)$ is the $\mathit{high}_i$-edge of $z$, for any $i\geq\mathit{minIndex}[z]$.}
\item{$v$ lies on the tree-path $T[x,y]$.}
\item{Every vertex $v'$ that lies on the tree-path $T[x,y)$ and is a proper descendant of $v$ such that $(x,y)$ is the $\mathit{high}_i$-edge of $v'$ for some $i\in\{1,\dots,k\}$, has its $\mathit{high}_i$-edge correctly computed and $\mathit{minIndex}[v']=i+1$.}
\item{For every vertex $z$ such that $\mathit{minIndex}[z]>k$, we have $\mathit{minIndex}[z]=k+1$. In this case, the $\mathit{high}_i$-edge of $z$ has been correctly computed, for every $i\in\{1,\dots,k\}$. Furthermore, the set that is maintained by the DSU data structure and contains $z$ is a subtree of $T$.}
\item{Every set that is maintained by the DSU data structure and is not a singleton, has the property that all its vertices have $\mathit{minIndex}=k+1$.}
\end{enumerate}

First, let us consider the first time that we reach the condition of the \textbf{while} loop in Line~\ref{line:high-while}. Then, all vertices have $\mathit{minIndex}$ $1$. Thus, properties $(1)$ and $(5)$ are trivially satisfied. Furthermore, since the \textbf{for} loop in Line~\ref{line:high-for-back-edge} has not processed any back-edge prior to $(x,y)$, condition $(2)$ is trivially satisfied. Also, since $v=x$ and $x$ has no proper descendants on the tree-path $T[x,y)$, we have that $(3)$ and $(4)$ are trivially satisfied. Finally, we have not performed any DSU operations yet, and so every set maintained by the DSU data structure is a singleton. Thus, $(6)$ is also satisfied. This establishes the base step of our induction. 

Now suppose that we reach the condition of the \textbf{while} loop in Line~\ref{line:high-while} and our inductive hypothesis is true. First, suppose that $v=y$. This implies that either the computation will stop (in which case there is nothing to show), or the \textbf{for} loop in Line~\ref{line:high-for-back-edge} will start processing a new back-edge $(x',y')$ (we note that $y'$ is not necessarily $y$, because the \textbf{for} loop in Line~\ref{line:high-for-y} may have changed the value of the variable ``$y$"). Then, the invariants $(1)$, $(5)$ and $(6)$ are obviously maintained. Property $(4)$ implies that every vertex $v'$ whose $\mathit{high}_i$-edge is $(x,y)$, for some $i\in\{1,\dots,k\}$, has its $\mathit{high}_i$-edge correctly computed and $\mathit{minIndex}[v']=i+1$. This fact, in conjunction with $(2)$, implies that $(2)$ will also hold true when we reach for the first time the condition of the \textbf{while} loop in Line~\ref{line:high-while} during the processing of $(x',y')$. Finally, we will have $(3)$ (since $v=x'$) and condition $(4)$ will be trivially true (because $x'$ has no proper descendants on the tree-path $T[x',y')$). Thus, the inductive hypothesis will still be true. So let us suppose that, when we reach the condition of the \textbf{while} loop in Line~\ref{line:high-while}, our inductive hypothesis is true and $v\neq y$. Then we will enter the \textbf{while} loop in Line~\ref{line:high-while}. Here we distinguish two cases: either $(i)$ $\mathit{minIndex}[v]\leq k$, or $(ii)$ $\mathit{minIndex}[v]>k$.

Let us consider case $(i)$ first. Then the condition in Line~\ref{line:high-if} is not satisfied, and therefore we go to Line~\ref{line:high-index-i}, and we will set $i\leftarrow\mathit{minIndex}[v]$. By property $(2)$ we have that no back-edge that has been processed so far by the \textbf{for} loop in Line~\ref{line:high-for-back-edge} prior to the processing of $(x,y)$ is the $\mathit{high}_i$-edge of $v$. By property $(3)$ we have that $(x,y)$ is a back-edge in $B(v)$. Thus, since the \textbf{for} loop in Line~\ref{line:high-for-y} processes the vertices ``$y$" in decreasing order, and since the \textbf{for} loop in Line~\ref{line:high-for-back-edge} processes the incoming back-edges to $y$ in increasing order w.r.t. their higher endpoint (due to the sorting in Line~\ref{line:high-sort}), we have that the $\mathit{high}_i$-edge of $v$ is $(x,y)$. Thus, the assignment in Line~\ref{line:high-assign} is correct. Then, in Line~\ref{line:high-minindex+} we let $\mathit{minIndex}[v]\leftarrow i+1$. Let $v'=p(v)$. Then we reach the condition of the \textbf{while} loop in Line~\ref{line:high-while}, and the ``$v$" variable holds the value $v'$. The invariant $(1)$ is maintained, because we have only increased $\mathit{minIndex}[v]$ by one, and we have correctly computed all $\mathit{high}_j$-edges of $v$, for $j<\mathit{minIndex}[v]$. Since $v$ had $\mathit{minIndex}\leq k$, by condition $(6)$ we had that $v$ was in a singleton set of the DSU data structure. Thus, since we performed no DSU operations, $(6)$ and $(5)$ are maintainted. Since $v$ was on the tree-path $T[x,y)$, we have that $p(v)$ is on the tree-path $T[x,y]$. Thus, $(3)$ is also true. Property $(2)$ is obviously maintained, and $(4)$ is still true because we have correctly established that $(x,y)$ is the $\mathit{high}_i$-edge of $v$. Thus, the inductive hypothesis still holds.

Now let us consider case $(ii)$. By property $(5)$ we have $\mathit{minIndex}[v]=k+1$. Thus, the condition in Line~\ref{line:high-if} is satisfied. First, suppose that the condition of the \textbf{while} loop in Line~\ref{line:high-while-2} is not true. Then we simply perform $v\leftarrow p(v)$ in Line~\ref{line:high-go-to-p(v)}, and we go to Line~\ref{line:high-break}. Then, properties $(1)$, $(2)$, $(5)$ and $(6)$ are obviously maintained. By property $(3)$, $v$ was on the tree-path $T[x,y)$ (since $v\neq y$). Thus, $p(v)$ is on the tree-path $T[x,y]$, and so $(3)$ is also true. By property $(5)$ we have that the $\mathit{high}_i$-edge of $v$ was correctly computed, for every $i\in\{1,\dots,k\}$. Thus, property $(4)$ is also maintained. Thus, all the invariants are maintained, and so, when we reach Line~\ref{line:high-break}, we can argue as previously, in order to show that when reach the condition of the \textbf{while} loop in Line~\ref{line:high-while} again, our inductive hypothesis will still be true. 

So let us suppose that the condition of the \textbf{while} loop in Line~\ref{line:high-while-2} is true. Then we will unite the set that contains $v$ with the set that contains its parent, with a call to $\mathtt{unite}(v,p(v))$. By property $(5)$ we have that both those sets are subtrees of $T$. Thus, joining two subtrees with a parent-edge maintaints this invariant. Notice that invariant $(6)$ is maintained. Next, due to the convention we made in the main text concerning the calls to $\mathtt{find}$, we have that $\mathtt{find}(v)$ will return the root of the new subtree that is formed. Now we repeat the \textbf{while} loop in Line~\ref{line:high-while-2}, until we reach a vertex $v$ such that $\mathit{minIndex}[p(v)]\neq k+1$. Since this process does not change the values $\mathit{minIndex}$ of the vertices, by property $(5)$ we have $\mathit{minIndex}[p(v)]\leq k$. Then we go to Line~\ref{line:high-go-to-p(v)}. Notice that the invariants $(1)$ and $(2)$ are maintained, as well as $(5)$ and $(6)$. We claim that, when we reach Line~\ref{line:high-go-to-p(v)}, we have $p(v)\in T[x,y]$. To see this, first observe that the \textbf{while} loop in Line~\ref{line:high-while-2} was ascending the tree-path $T[x,y)$, starting from a vertex $v_0$ on $T[x,y)$ (due to property $(3)$). Then, all vertices from $v_0$ up to $v$ have $\mathit{minIndex}$ $k+1$ (due to $(5)$, $(6)$, and the fact that the condition of the \textbf{while} loop in Line~\ref{line:high-while-2} was repeatedly satisfied). Then, by $(5)$ we have that the $\mathit{high}_i$-edges of $v$ are correctly computed, for every $i\in\{1,\dots,k\}$. Since the \textbf{for} loop in Line~\ref{line:high-for-y} processes the vertices ``$y$" in decreasing order, we have that no back-edge that was processed so far by the \textbf{for} loop in Line~\ref{line:high-for-back-edge} has low enough lower endpoint to be a back-edge in $B(y)$. Thus, no $\mathit{high}_i$-edge of $y$ is computed as yet, for any $i\in\{1,\dots,k\}$, and therefore $\mathit{minIndex}[y]=1$. Since all vertices on the tree-path from $v_0$ to $v$ have $\mathit{minIndex}$ $k+1$, we have that $v$ is still a proper descendant of $y$, and therefore $p(v)\in T[x,y]$. Thus, invariant $(3)$ is maintained. Finally, since all vertices on the tree-path from $v_0$ to $v$ have all their $\mathit{high}_i$-edges computed, for all $i\in\{1,\dots,k\}$, we have that invariant $(4)$ is maintained too. Thus, when we reach Line~\ref{line:high-break}, we have that all the properties of the inductive hypothesis are satisfied. Then, from this point we can argue as previously, in order to show that when we reach the condition of the \textbf{while} loop in Line~\ref{line:high-while} again, our inductive hypothesis will still be true.

This shows the correctness of Algorithm~\ref{algorithm:high}. It remains to analyze its complexity. The sorting of the adjacency lists in Line~\ref{line:high-sort} can be performed in $O(m)$ time with bucket-sort. Whenever we enter the \textbf{while} loop in Line~\ref{line:high-while}, by the inductive hypothesis we have that, when we reach Line~\ref{line:high-break}, we have that either $v=y$ or $v$ is a vertex that has its $\mathit{high}_i$-edge yet to be computed, for some $i\in\{1,\dots,k\}$, but this will be computed correctly in Line~\ref{line:high-assign}. The case $v=y$ can be charged to the back-edge $(x,y)$ that is currently processed by the \textbf{for} loop in Line~\ref{line:high-for-back-edge}. The other case can be charged to the $\mathit{high}_i$-edge of $v$. Thus, the \textbf{while} loop in Line~\ref{line:high-while} will be entered $O(m+kn)$ times in total. Whenever we enter the \textbf{while} loop in Line~\ref{line:high-while}, the first run of the \textbf{while} loop in Line~\ref{line:high-while-2} can be charged to this entry. The remaining entries can be charged to the calls to $\mathtt{unite}$, all of which are non-trivial (i.e., they join sets that were previously not united). This is because the call to $\mathtt{find}(v)$ in Line~\ref{line:high-find} returns the root of the subtree maintained by the DSU data structure and contains $v$. Thus, after the assignment $v\leftarrow\mathtt{find}(v)$ in Line~\ref{line:high-find}, we have that $v$ is not in the same set of the DSU data structure that contains $p(v)$. Therefore, in the next run of the \textbf{while} loop in Line~\ref{line:high-while-2}, the operation $\mathtt{unite}(v,p(v))$ will unite two disjoint sets. Thus, since the number of non-trivial calls to $\mathtt{unite}$ is $O(n)$, we have that the number of times that we enter the \textbf{while} loop in Line~\ref{line:high-while-2} is dominated by the number of times that we enter the \textbf{while} loop in Line~\ref{line:high-while}. The final thing to do is to bound the cost of the DSU operations. Since the tree of the $\mathtt{unite}$ operations is known beforehand (i.e., it coincides with $T$), we can use the data structure of Gabow and Tarjan \cite{DBLP:conf/stoc/GabowT83}. Thus, any sequence of $m'$ $\mathtt{find}$ and $\mathtt{unite}$ operations can be performed in $O(m'+n)$ time in total. In our case, we have $m'=O(m+nk)$. We conclude that Algorithm~\ref{algorithm:high} has an $O(m+kn)$-time implementation.
\end{proof}

\subsection{Computing the leftmost and the rightmost edges}
\label{subsection:leftmost}

Let $v\neq r$ be a vertex, and let $(x_1,y_1),\dots,(x_k,y_k)$ be the list of the back-edges in $B(v)$ sorted in lexicographic order. In other words, we have $x_1\leq\dots\leq x_k$, and if $x_i=x_{i+1}$ then $y_i\leq y_{i+1}$, for any $i\in\{1,\dots,k-1\}$.\footnote{Notice that, since we consider multigraphs, we may have that an entry $(x,y)$ appears several times; then we break ties according to the unique identifiers of the edges of the graph.} Now let $c$ be a descendant of $v$. We are interested in computing some subsets of $B(v)$ that consist of back-edges whose higher endpoint is a descendant of $c$. Notice that, if $i$ is the lowest index in $\{1,\dots,k\}$ such that $x_i\in T(c)$, then, due to the lexicographic order, the set of the back-edges in $B(v)$ whose higher endpoint is a descendant of $c$ is a segment of $B(v)$ starting from $(x_i,y_i)$.

For every $t=1,2,\dots$, we let $L(v,c,t)$ denote the set of the first $t$ back-edges in $B(v)$ whose higher endpoint is a descendant of $c$. More precisely, $L(v,c,t)$ is defined as follows. Let $i$ be the lowest index in $\{1,\dots,k\}$ such that $x_i$ is a descendant of $c$. If such an index does not exist, then we let $L(v,c,t)=\emptyset$. Otherwise, let $j$ be the maximum index in $\{i,\dots,i+t-1\}$ such that $x_j$ is a descendant of $c$. Then $L(v,c,t)=\{(x_i,y_i),\dots,(x_j,y_j)\}$. Similarly, we let $R(v,c,t)$ denote the set of the last $t$ back-edges in $B(v)$ whose higher endpoint is a descendant of $c$. More precisely, $R(v,c,t)$ is defined as follows. Let $i'$ be the greatest index in $\{1,\dots,k\}$ such that $x_{i'}$ is a descendant of $c$. If such an index does not exist, then we let $R(v,c,t)=\emptyset$. Otherwise, let $j'$ be the minimum index in $\{i'-t+1,\dots,i'\}$ such that $x_{j'}$ is a descendant of $c$. Then $R(v,c,t)=\{(x_{j'},y_{j'}),\dots,(x_{i'},y_{i'})\}$.

The challenge in computing the sets $L(v,c,t)$ and $R(v,c,t)$ is that we do not have direct access to the list $B(v)$. Thus, the straightforward way to compute $L(v,c,t)$ is to start processing the vertices in $T(c)$ in increasing order, starting from $c$. For every vertex $x$ that we process, we scan the adjacency list of $x$ for back-edges of the form $(x,y)$. If $y<v$, then we have $(x,y)\in B(v)$, and therefore we collect $(x,y)$. We continue this process until we have collected $t$ back-edges, or we have exhausted the search in the subtree of $c$. Similarly, in order to compute $R(v,c,t)$ we perform the same search, starting from the greatest descendant of $c$ (i.e., $c+\mathit{ND}(c)-1$), and we process the vertices in decreasing order. Obviously, this method may take $O(m')$ time, where $m'$ is the number of edges with one endpoint in $T(c)$. Notice that $m'$ can be as large as $\Omega(m)$. Thus, this method is impractical if the number of $L$ or $R$ sets that we want to compute is $\Omega(n)$. 

Thus, we will assume that the queries for the $L$ and $R$ sets are given in batches, to be performed in an off-line manner. We will focus on computing the $L$ sets. The method and the arguments for the $R$ sets are similar. Thus, let $L(v_1,c_1,t_1),\dots,L(v_N,c_N,t_N)$ be the $L$ sets that we have to compute. (We assume that $c_i$ is a descendant of $v_i$, for every $i\in\{1,\dots,N\}$, and $t_i\geq 1$.) The idea is to apply the straightforward algorithm that we described above, but we process the queries in an order that is convenient for us. Specifically, we processs the vertices $v_i$ in a bottom-up fashion. By doing so, we can avoid vertices that previously were unable to provide a leaping back-edge. More precisely, if we meet a vertex $x$ that is a descendant of $c_i$, during the processing of a query $L(v_i,c_i,t_i)$, then, if we have $l(x)\geq v_i$, we can be certain that there is no back-edge of the form $(x,y)$ in any of $B(v_i),\dots,B(v_N)$. Thus, we mark $x$ as ``inactive", and we attach it to a segment of inactive vertices which we can bypass at once, because they cannot provide a leaping back-edge with low enough lower endpoint anymore. Initially, we assume that all vertices are active.

Thus, for every vertex $x$, we maintain a boolean attribute $\mathit{is\_active}(x)$, which is \emph{true} if and only if $x$ is active. We use a disjoint-set union data structure $\mathit{DSU}$ on the set of inactive vertices, that maintains the partition of the maximal segments (w.r.t. the DFS numbering) of inactive vertices. Thus, if two vertices $x$ and $x+1$ are inactive, then they belong to the same set maintained by the DSU data structure. $\mathit{DSU}$ supports the operations $\mathtt{find}(x)$ and $\mathtt{unite}(x,y)$. We use $\mathtt{find}(x)$, on an inactive vertex $x$, in order to return a representative of the segment of inactive vertices that contains $x$. With the operation $\mathtt{unite}(x,y)$ we unite the segment that contains $x$ with the segment that contains $y$ (we assume that $y=x+1$). Also, for every vertex $x$, we maintain two pointers $\mathit{left}(x)$ and $\mathit{right}(x)$. These are only used when $x$ is an inactive vertex, which is a representative of its segment $S$. In this case, $\mathit{left}(x)$ points to the greatest vertex that is lower than $x$ and not in $S$, and $\mathit{right}(x)$ points to the lowest vertex that is greater than $x$ and not in $S$. Then, by definition, we have that $\mathit{left}(x)$ (resp., $\mathit{right}(x)$) is either an active vertex, or $\bot$. We maintain the invariant that, whenever a DSU operation changes the representative of a segment, it passes the $\mathit{left}$ and the $\mathit{right}$ pointer of the old representative to the new one (so that we can correctly retrieve the endpoints of the corresponding maximal segment).

Now we can describe in more detail the procedure for computing the sets $L(v_1,c_1,t_1),\dots,L(v_N,c_N,t_N)$. Recall that we process these queries in decreasing order w.r.t. the vertices $v_i$. We assume that the edges in the adjacency list of every vertex are sorted in increasing order w.r.t. their lower endpoint. Now, in order to answer a query $L(v,c,t)$, we begin the search from the lowest active vertex $x$ that is a descendant of $c$. This is because all descendants of $c$ that are lower than $x$ are incapable of providing a back-edge that leaps over $v$. If $c$ is active, then we have $x=c$. Otherwise, we have that $c$ is inactive, and we use $z\leftarrow\mathtt{find}(c)$ in order to get the representative $z$ of the maximal segment of inactive vertices that contains $c$. Then we have $x=\mathit{right}(z)$. If $x=\bot$ or $x$ is not a descendant of $c$, then we know that there are no back-edges with higher endpoint in $T(c)$ that leap over $v$ (and so we have $L(v,c,t)=\emptyset$). Otherwise, we check whether $l(x)<v$. If that is the case, then we start traversing the adjacency list of $x$, in order to get back-edges of the form $(x,y)$ that leap over $v$. We keep doing that until we have either collected $t$ back-edges for $L(v,c,t)$, or we have reached a back-edge that does not leap over $v$ (because its lower endpoint is not low enough), or we have reached the end of the adjacency list. If we have gathered less than $t$ back-edges that leap over $v$, then we continue the search in the lowest active vertex that is greater than $x$ and still a descendant of $c$. Otherwise, if we have $l(x)\geq v$, then we mark $x$ as inactive, so as not to process it again. Then we have to properly update the segments. To do so, we have to check whether $x-1$ or $x+1$ is an inactive vertex. If none of those vertices is inactive, then there is nothing we have to do. So let us suppose that $x-1$ is an inactive vertex. Then we expand the segment that contains $x-1$ with a call $\mathtt{unite}(x-1,x)$, and we set the $\mathit{right}$ pointer of the representative of this segment to $x+1$. Then, if $x+1$ is inactive, we have to further expand the segment with a call $\mathtt{unite}(x,x+1)$; otherwise, we are done. On the other hand, if only $x+1$ is inactive, then we expand the segment that contains it with a call $\mathtt{unite}(x,x+1)$, and we set the $\mathit{left}$ pointer of the representative of this segment to $x-1$. Now, after updating the segments, we proceed to the lowest active vertex that is greater than $x$ and a descendant of $c$. Then we repeat the same process.

\begin{algorithm}[H]
\caption{\textsf{Compute the sets $L(v_1,c_1,t_1),\dots,L(v_N,c_N,t_N)$, where $c_i$ is a descendant of $v_i$, for every $i\in\{1,\dots,N\}$}}
\label{algorithm:L-sets}
\LinesNumbered
\DontPrintSemicolon
let $Q$ be the list of all triples $(v_1,c_1,t_1),\dots,(v_N,c_N,t_N)$ sorted in decreasing order w.r.t. the first component\;
\label{line:L-sort-1}
sort the adjacency list of every vertex in increasing order w.r.t. the lower endpoints\;
\label{line:L-sort-2}
initialize a boolean array $\mathit{is\_active}$ with $n$ entries\;
\lForEach{vertex $v$}{set $\mathit{is\_active}[v]\leftarrow\mathit{true}$}
\lForEach{vertex $v$}{initialize two pointers $v.\mathit{left}\leftarrow v-1$ and $v.\mathit{right}\leftarrow v+1$}
\label{line:L-pointers}
\ForEach{$(v,c,t)\in Q$}{
\label{line:L-for-query}
  let $L(v,c,t)\leftarrow\emptyset$\;
  set $\mathit{counter}\leftarrow 0$ \tcp{counter for the number of back-edges we have collected}
  \label{line:L-count-0}
  let $x\leftarrow c$\;
  \label{line:set-x-c}
  \If{$\mathit{is\_active}(x)=\mathit{false}$}{
  \label{line:L-if-c-active}
    $x\leftarrow\mathtt{find}(x).\mathit{right}$\;
    \label{line:L-x-next-c}
  }
  \While{$x\neq\bot$ \textbf{and} $x$ is a descendant of $c$}{
  \label{line:L-while}
    \If{$l(x)<v$}{
    \label{line:L-x-provides}
      let $e=(x,y)$ be the first back-edge in the adjacency list of $x$\;
      \label{line:L-x-provides-1}
      \While{$y<v$ \textbf{and} $\mathit{counter}<t$}{
      \label{line:L-x-while-2}  
        insert $(x,y)$ into $L(v,c,t)$\;
        $\mathit{counter}\leftarrow\mathit{counter}+1$\;
        let $e=(x,y)$ be the next back-edge in the adjacency list of $x$\;
        \lIf{$e=\bot$}{\textbf{break}}
      }
      \lIf{$\mathit{counter}=t$}{\textbf{break}}
      \label{line:L-x-provides-2}
    }
    \Else{
      $\mathit{is\_active}(x)\leftarrow\mathit{false}$\;
      \label{line:L-x-deactivate}
      \If{$\mathit{is\_active}(x-1)=\mathit{false}$ \textbf{and} $\mathit{is\_active}(x+1)=\mathit{true}$}{
      \label{line:L-x-deactivate-1}
        $\mathtt{unite}(x-1,x)$\;
        \label{line:L-x-unite-1}
        $\mathtt{find}(x).\mathit{right}\leftarrow x+1$\;  
        \label{line:L-x-set-right}
      }
      \If{$\mathit{is\_active}(x-1)=\mathit{true}$ \textbf{and} $\mathit{is\_active}(x+1)=\mathit{false}$}{
        $\mathtt{unite}(x,x+1)$\;
        $\mathtt{find}(x).\mathit{left}\leftarrow x-1$\; 
      }    
      \If{$\mathit{is\_active}(x-1)=\mathit{false}$ \textbf{and} $\mathit{is\_active}(x+1)=\mathit{false}$}{
        $\mathtt{unite}(x-1,x)$\;
        \label{line:L-x-unite-x-1}
        $\mathtt{unite}(x,x+1)$\;
        \label{line:L-x-unite-x+1}
      }          
      \label{line:L-x-deactivate-2}
    }
    $x\leftarrow x+1$\;
    \label{line:L-x-next-1}
    \lIf{$x=\bot$}{\textbf{break}}
    \If{$\mathit{is\_active}(x)=\mathit{false}$}{
       $x\leftarrow\mathtt{find}(x).\mathit{right}$\;
       \label{line:L-x-next-3}
    }
    \label{line:L-x-next-2}
  }
}
\end{algorithm}

This procedure for computing the sets $L(v_1,c_1,t_1),\dots,L(v_N,c_N,t_N)$ is shown in Algorithm~\ref{algorithm:L-sets}. The proof of correctness and linear complexity is given in Proposition~\ref{proposition:L-sets}. After that, we describe the minor changes that we have to make to Algorithm~\ref{algorithm:L-sets} in order to get an algorithm that computes sets of the form $R(v_1,c_1,t_1),\dots,R(v_N,c_N,t_N)$, with the same time-bound guarantees.

\begin{proposition}
\label{proposition:L-sets}
Let $L(v_1,c_1,t_1),\dots,L(v_N,c_N,t_N)$ be a collection of queries, where $c_i$ is a descendant of $v_i$, for every $i\in\{1,\dots,N\}$, and $t_i\geq 1$. Then Algorithm~\ref{algorithm:L-sets} correctly computes the sets $L(v_1,c_1,t_1),\dots,L(v_N,c_N,t_N)$. Furthermore, it runs in $O(t_1+\dots+t_N+m)$ time.
\end{proposition}
\begin{proof}
We will prove correctness inductively, by establishing the following: Whenever the \textbf{for} loop in Line~\ref{line:L-for-query} processes a triple $(v,c,t)\in Q$, we have that $(1)$ every vertex $x$ with $l(x)<v$ is active, and every inactive vertex $x$ has $l(x)\geq v$, and $(2)$ for every inactive vertex $x$, $\mathtt{find}(x).\mathit{left}$ is the greatest active vertex $x'$ such that $x'<x$, and $\mathtt{find}(x).\mathit{right}$ is the lowest active vertex $x'$ such that $x'>x$. 

Initially, all vertices are active. Thus, the inductive hypothesis is trivially true before entering the \textbf{for} loop in Line~\ref{line:L-for-query}. Now suppose that the inductive hypothesis holds by the time the \textbf{for} loop in Line~\ref{line:L-for-query} processes a triple $(v,c,t)\in Q$. Then we will show that $L(v,c,t)$ will be correctly computed, and the inductive hypothesis will still hold for the next triple $(v',c',t')$ that will be processed by the \textbf{for} loop in Line~\ref{line:L-for-query}.

Now, given the query $L(v,c,t)$, the first thing to do it to find the lowest descendant $x$ of $c$ that can provide a back-edge of the form $(x,y)\in B(v)$. Notice that $x$ satisfies the property $l(x)\leq y<v$, and therefore it is active, according to $(1)$. Thus, we first check whether $c$ is active, in Line~\ref{line:L-if-c-active}. If $c$ is not active, then by $(1)$ it has $l(c)\geq v$, and therefore it is proper to set $x\leftarrow\mathtt{find}(x).\mathit{right}$ in Line~\ref{line:L-x-next-c}. According to $(2)$, now $x$ is the lowest active vertex that is greater than $c$. Otherwise, if $c$ is active, then we have $x=c$, due to the assignment in Line~\ref{line:set-x-c}. Now, if $x=\bot$ or $x$ is great enough to not be a descendant of $c$, then we will not enter the \textbf{while} loop in Line~\ref{line:L-while}, and the computation of $L(v,c,t)$ is over (i.e., we have $L(v,c,t)=\emptyset$). This is correct, because there is no descendant $x$ of $c$ that can provide a back-edge of the form $(x,y)\in B(v)$. Notice that the inductive hypothesis will still hold for the next query, because we have made no changes in the underlying data structures. 

Otherwise, if we enter the \textbf{while} loop in Line~\ref{line:L-while}, then we have that $x$ is the lowest active descendant of $c$, and therefore in Line~\ref{line:L-x-provides} we check whether it can provide a back-edge of the form $(x,y)\in B(v)$. Notice that this is equivalent to $l(x)<v$. The necessity was already shown. To prove sufficiency, we note that $l(x)<v$ implies that there is a back-edge of the form $(x,y)$ such that $y<v$. We have that $x$ is a descendant of $c$, and therefore a descendant of $v$ (due to the assumption concerning the queries). Then, since $(x,y)$ is a back-edge, we have that $x$ is a descendant of $y$. Thus, $x$ is a common descendant of $v$ and $y$, and therefore $v$ and $y$ are related as ancestor and descendant. Thus, $y<v$ implies that $y$ is a proper ancestor of $v$. This shows that $(x,y)\in B(v)$. Now, if the condition in Line~\ref{line:L-x-provides} is satisfied, then we have to gather as many back-edges of the form $(x,y)\in B(v)$ as $x$ can provide, provided that we do not exceed $t$. This is precisely the purpose of Lines~\ref{line:L-x-provides-1} to \ref{line:L-x-provides-2}. Correctness follows from the fact that the adjacency list of $x$ is sorted in increasing order w.r.t. the lower endpoints. The variable $\mathit{counter}$ counts precisely the number of back-edges $(x',y)$ that we have gathered, where $x'$ is a descendant of $c$ and $(x',y)\in B(v)$. ($\mathit{counter}$ has been initialized to $0$ in Line~\ref{line:L-count-0}.) Thus, if the condition in Line~\ref{line:L-x-provides-2} is satisfied, then it is proper to stop the computation of $L(v,c,t)$, because it has been correctly computed. Notice that the inductive hypothesis holds true, because we have made no changes in the underlying data structure.
Otherwise, if $\mathit{counter}<t$, then we have to proceed to the lowest active descendant of $c$ that is greater than $x$. This is precisely the purpose of Lines~\ref{line:L-x-next-1} to \ref{line:L-x-next-2}. Thus, first we move to $x+1$. Now, if $x\neq\bot$ and $x$ is active, then we go to the condition of the \textbf{while} loop in Line~\ref{line:L-while}. Otherwise, if $x$ is inactive, then it is correct to set $x\leftarrow\mathtt{find}(x).\mathit{right}$ in Line~\ref{line:L-x-next-3}, because now $x$ has moved to its lowest active successor, due to $(2)$ of the inductive hypothesis.

Now suppose that the condition in Line~\ref{line:L-x-provides} is false. Then we have $l(x)\geq v$, and therefore we set the mode of $x$ to inactive, in Line~\ref{line:L-x-deactivate}. Notice that point of $(1)$ of the inductive hypothesis is maintained, because we process the triples $(v',c',t')$ in decreasing order w.r.t. their first component. Thus, the next such triple has $v'\leq v$, and therefore we have $l(x)\geq v\geq v'$. Now, since $x$ is made inactive, we have to maintain invariant $(2)$. This is the purpose of Lines~\ref{line:L-x-deactivate-1} to \ref{line:L-x-deactivate-2}. The idea in those lines is the following. First, if both $x-1$ and $x+1$ are active, then there is nothing to do, because $\mathtt{find}(x)=x$, and the pointers $x.\mathit{left}$ and $x.\mathit{right}$ have been initialized to $x-1$ and $x+1$, respectively, in Line~\ref{line:L-pointers}. Otherwise, if $x-1$ is inactive and $x+1$ is active, then we join $x$ to the segment of inactive vertices that precede it with a call $\mathtt{unite}(x-1,x)$ in Line~\ref{line:L-x-unite-1}. Then, we have to set $\mathtt{find}(x).\mathit{right}\leftarrow x+1$, which is done in Line~\ref{line:L-x-set-right}. We work similarly, if $x-1$ is active and $x+1$ is inactive. Finally, if both $x-1$ and $x+1$ are inactive, then it is sufficient to join $x$ to the segments of inactive vertices that precede it and succeed it, which is done in Lines~\ref{line:L-x-unite-x-1} and \ref{line:L-x-unite-x+1}. We only have to make sure that the $\mathit{left}$ pointer of the new representative is the same as the $\mathit{left}$ pointer of what was previously the representative of the segment that contained $x-1$, and the $\mathit{right}$ pointer of the new representative is the same as the $\mathit{right}$ pointer of what was previously the representative of the segment that contained $x+1$. Thus, the point $(2)$ of the inductive hypothesis is maintained.

This establishes the correctness of Algorithm~\ref{algorithm:L-sets} (taking also into account our presentation of the general idea in the main text). It remains to argue about the complexity of Algorithm~\ref{algorithm:L-sets}. First, the sorting of the triples in Line~\ref{line:L-sort-1} can be performed in $O(n+N)$ time in total with bucket-sort. Also, the sorting of the adjacency lists in Line~\ref{line:L-sort-2} can be performed in $O(m)$ time in total with bucket-sort. Whenever the \textbf{for} loop in Line~\ref{line:L-for-query} processes a triple $(v,c,t)$, we have that the \textbf{while} loop in Line~\ref{line:L-while} will only process those $x$ that either have $l(x)<v$, and so they will provide at least one back-edge for $L(v,c,t)$, or they have $l(x)\geq v$, in which case they will be made inactive, and will not be accessed again in this \textbf{while} loop for any further triple. Thus, the total number of runs of the \textbf{while} loop in Line~\ref{line:L-while} is $O(n+t_1+\dots+t_N)$. Every $x$ that we encounter that has $l(x)<v$, will initiate the \textbf{while} loop in Line~\ref{line:L-x-while-2}. But the number of runs of this \textbf{while} loop will not exceed $t$ (for $(v,c,t)$). Thus, the total number of runs of the \textbf{while} loop in Line~\ref{line:L-x-while-2} is $O(t_1+\dots+t_N)$. Finally, the $\mathtt{unite}$ operations that we perform with the DSU data structure have the form $\mathtt{unite}(x,x+1)$. Since the tree-structure of the calls to $\mathtt{unite}$ is predetermined (it is essentially a path), we can use the data structure of Gabow and Tarjan~\cite{DBLP:conf/stoc/GabowT83} that performs a sequence of $m'$ DSU operations in $O(m'+n)$ time, when applied on a set of $n$ elements. Notice that the number of calls to the DSU data structure is $O(n+t_1+\dots+t_N)$. Thus, the most expensive time expressions that we have gathered are $O(m)$, $O(n+N)$ and $O(n+t_1+\dots+t_N)$. Since $t_i\geq 1$, for every $i\in\{1,\dots,N\}$, we have $N\leq t_1+\dots+t_N$. We conclude that Algorithm~\ref{algorithm:L-sets} runs in $O(m+t_1+\dots+t_N)$ time.
\end{proof}

With only minor changes to Algorithm~\ref{algorithm:L-sets}  we can also answer queries of the form $R(v_1,c_1,t_1),\dots,R(v_N,c_N,t_N)$, where $c_i$ is a descendant of $v_i$, for every $i\in\{1,\dots,N\}$, and $t_i\geq 1$, in $O(t_1+\dots+t_N+m)$ time in total. These changes are as follows. First, in order to compute a query of the form $R(v,c,t)$, we start the search from the greatest active descendant of $c$. Thus, we replace Line~\ref{line:set-x-c} with ``$x\leftarrow c+\mathit{ND}(c)-1$", and Line~\ref{line:L-x-next-c} with ``$x\leftarrow\mathtt{find}(x).\mathit{left}$". Then, since we want to process the active descendants of $c$ in decreasing order, we replace Line~\ref{line:L-x-next-1} with ``$x\leftarrow x-1$", and Line~\ref{line:L-x-next-3} with $x\leftarrow\mathtt{find}(x).\mathit{left}$". Now the proof of correctness is similar as in Proposition~\ref{proposition:L-sets}. In particular, we can argue using the same inductive hypothesis.

\subsection{Computing the $M$ points}
\label{subsection:M}

Given a vertex $v\neq r$, we will need an efficient method to compute the values $M(v)$ and $\widetilde{M}(v)$, as well as values of the form $M(v,c)$ and $M(B(v)\setminus S)$, where $c$ is a descendant of $v$, and $S$ is a subset of $B(v)$. In \cite{DBLP:conf/esa/GeorgiadisIK21} it was shown how to compute the values $M(v)$, for all vertices $v\neq r$, in linear time in total. Also, in \cite{DBLP:conf/esa/GeorgiadisIK21} it was shown how to compute the values $\widetilde{M}(v)$, for all vertices $v\neq r$, in linear time in total. Alternatively, we have $M(v)=M(v,v)$. Also, let $c_1$ and $c_2$ be the $\mathit{low1}$ and the $\mathit{low2}$ child of $M(v)$, respectively. Then it is easy to see that $\widetilde{M}(v)=M(v)$ if ($c_2\neq\bot$ and) $\mathit{low}(c_2)<v$, and $\widetilde{M}(v)=M(v,c_1)$ otherwise. Thus, the computation of both $M(v)$ and $\widetilde{M}(v)$ can be reduced to the computation of values of the form $M(v,c)$, where $c$ is a descendant of $v$. 

Now let $v\neq r$ be a vertex, and let $c$ be a descendant of $v$. Let $(x_1,y_1),\dots,(x_k,y_k)$ be the list of the back-edges in $B(v)\cap B(c)$, sorted in lexicographic order (and thus in increasing order w.r.t. their higher endpoint). Then, following the notation in Section~\ref{subsection:leftmost}, we let $L(v,c,t)$ denote the set $\{(x_1,y_1),\dots,(x_t,y_t)\}$. Similarly, we let $R(v,c,t)$ denote the set $\{(x_k,y_k),\dots,(x_{k-t+1},y_{k-t+1})\}$. Then we have the following.

\begin{proposition}
\label{proposition:computing-M(v,c)}
Let $(v_1,c_1),\dots,(v_N,c_N)$ be a sequence of pairs of vertices such that $v_i\neq r$ and $c_i$ is a descendant of $v_i$, for every $i\in\{1,\dots,N\}$. Then the values $M(v_1,c_1),\dots,M(v_N,c_N)$ can be computed in $O(m+N)$ time in total.
\end{proposition}
\begin{proof}
First, we compute the sets $L(v_1,c_1,1),\dots,L(v_N,c_N,1)$ and $R(v_1,c_1,1),\dots,R(v_N,c_N,1)$. According to Proposition~\ref{proposition:L-sets} (and the comments after Algorithm~\ref{algorithm:L-sets}), this takes $O(m+N)$ time in total. Then, for every $i\in\{1,\dots,N\}$, we gather the higher endpoint $L(v_i,c_i)$ of the back-edge in $L(v_i,c_i,1)$, and the higher endpoint $R(v_i,c_i)$ of the back-edge in $R(v_i,c_i,1)$. We claim that $M(v_i,c_i)=\mathit{nca}\{L(v_i,c_i),R(v_i,c_i)\}$. To see this, let $(x_1,y_1),\dots,(x_k,y_k)$ be the list of the back-edges in $B(v_i)\cap B(c_i)$, sorted in lexicographic order, so that $L(v_i,c_i)=x_1$ and $R(v_i,c_i)=x_k$. By definition, we have $M(v_i,c_i)=\mathit{nca}\{x_1,\dots,x_k\}$. Thus, $z=\mathit{nca}\{x_1,x_k\}$ is a descendant of $M(v_i,c_i)$. Since $z=\mathit{nca}\{x_1,x_k\}$, we have that $z$ is an ancestor of both $x_1$ and $x_k$. This implies that $z\leq x_1\leq x_k\leq z+\mathit{ND}(z)-1$. Thus, for every $j\in\{1,\dots,k\}$ we have $z\leq x_j\leq z+\mathit{ND}(z)-1$. This implies that $z$ is an ancestor of all vertices in $\{x_1,\dots,x_k\}$, and thus $z$ is an ancestor of $M(v_i,c_i)$. This shows that $z=M(v_i,c_i)$.

Thus, we can compute the values $M(v_1,c_1),\dots,M(v_N,c_N)$, by answering the $\mathit{nca}$ queries $\mathit{nca}\{L(v_1,c_1),R(v_1,c_1)\},\dots\mathit{nca}\{L(v_N,c_N),R(v_N,c_N)\}$. By \cite{DBLP:journals/siamcomp/HarelT84} or \cite{DBLP:journals/siamcomp/BuchsbaumGKRTW08}, we know that there is a linear-time preprocessing of $T$, so that we can answer a collection of $N$ $\mathit{nca}$ queries in $O(N)$ time in total. We conclude that the values $M(v_1,c_1),\dots,M(v_N,c_N)$ can be computed in $O(m+N)$ time in total. 
\end{proof}

Similarly, the computation of the values of the form $M(B(v)\setminus S)$ utilizes the leftmost and the rightmost points, as shown by the following.

\begin{lemma}
\label{lemma:M(B(v)-S)}
Let $v\neq r$ be a vertex, let $S$ be a subset of $B(v)$ with $|S|=k$, and let $D$ be the multiset of the higher endpoints of the back-edges in $S$. Then $M(B(v)\setminus S)=\mathit{nca}(\{L_1(v),\dots,L_{k+1}(v),R_1(v),\dots,R_{k+1}(v)\}\setminus D)$.
\end{lemma}
\begin{proof}
Let $x$ and $y$ be the the minimum and the maximum, respectively, among the higher endpoints of the back-edges in $B(v)\setminus S$. We claim that $\mathit{nca}\{x,y\}=M(B(v)\setminus S)$. First, it is clear that $M(B(v)\setminus S)$ is an ancestor of $\mathit{nca}\{x,y\}$. Conversely, let $z$ be the higher endpoint of an edge in $B(v)\setminus S$. Then we have $x\leq z\leq y$. Thus, since $\mathit{nca}\{x,y\}$ is an ancestor of both $x$ and $y$, we have that $\mathit{nca}\{x,y\}$ is an ancestor of $z$. Due to the generality of $z$, this implies that $\mathit{nca}\{x,y\}$ is an ancestor of $M(B(v)\setminus S)$. This shows that $M(B(v)\setminus S)=\mathit{nca}\{x,y\}$.

Now let $D$ be the \emph{multiset} of the higher endpoints of the back-edges in $S$. (That is, if there are $t$ distinct back-edges of the form $(x,y_1),\dots,(x,y_t)$ in $S$, for some $t\geq 1$, then $D$ contains at least $t$ multiple entries for $x$.) We also consider $\{L_1(v),\dots,L_{k+1}(v),R_1(v),\dots,R_{k+1}(v)\}$ as a multiset.
  
It is clear that $M(B(v)\setminus S)$ is an ancestor of $\mathit{nca}(\{L_1(v),\dots,L_{k+1}(v),R_1(v),\dots,R_{k+1}(v)\}\setminus D)$. To see the converse, notice that $x\in\{L_1(v),\dots,L_{k+1}(v)\}\setminus D$ and $y\in\{R_1(v),\dots,R_{k+1}(v)\}\setminus D$, since $|D|=k$. Thus, we have that $\mathit{nca}(\{L_1(v),\dots,L_{k+1}(v),R_1(v),\dots,R_{k+1}(v)\}\setminus D)$ is an ancestor of $\mathit{nca}\{x,y\}$, and therefore an ancestor of $M(B(v)\setminus S)$. This shows that $M(B(v)\setminus S)=\mathit{nca}(\{L_1(v),\dots,L_{k+1}(v),R_1(v),\dots,R_{k+1}(v)\}\setminus D)$.
\end{proof}

\begin{proposition}
\label{proposition:computing-M(B(v)-S)}
Let $(v_1,S_1),\dots,(v_N,S_N)$ be a collection of pairs of vertices and sets of back-edges, such that $v_i\neq r$ and $\emptyset\neq S_i\subseteq B(v_i)$ for every $i\in\{1,\dots,N\}$. Then the values $M(B(v_1)\setminus{S_1}),\dots,M(B(v_N)\setminus{S_N})$ can be computed in $O(m+|S_1|+\dots+|S_N|)$ time in total.
\end{proposition}
\begin{proof}
For every $i\in\{1,\dots,N\}$, let $k_i=|S_i|$. Then we compute the sets $L(v_i,v_i,k_{i+1})$ and $R(v_i,v_i,k_{i+1})$, for every $i\in\{1,\dots,N\}$. According to Proposition~\ref{proposition:L-sets} (and the comments after Algorithm~\ref{algorithm:L-sets}), this takes $O(m+|S_1|+\dots+|S_N|)$ time in total. Then, for every $i\in\{1,\dots,N\}$, we can compute the $L_1(v),\dots,L_{k+1}(v)$ and the $R_1(v),\dots,R_{k+1}(v)$ values, by gathering the higher endpoints of the back-edges in $L(v_i,v_i,k_{i+1})\cup R(v_i,v_i,k_{i+1})$. Let $D_i$ be the set of the higher endpoints of the back-edges in $S_i$. Then, by Lemma~\ref{lemma:M(B(v)-S)} we have $M(B(v_i)\setminus S_i)=\mathit{nca}(\{L_1(v_i),\dots,L_{k+1}(v_i),R_1(v_i),\dots,R_{k+1}(v_i)\}\setminus D_i)$. (We note that the sets $\{L_1(v_i),\dots,L_{k+1}(v_i),R_1(v_i),\dots,R_{k+1}(v_i)\}\setminus D_i$ can be computed in $O(n+|S_1|+\dots+|S_N|)$ time in total with bucket-sort.) Notice that $\mathit{nca}(\{L_1(v_i),\dots,L_{k+1}(v_i),R_1(v_i),\dots,R_{k+1}(v_i)\}\setminus D_i)$ can be broken up into $O(k_i)$ $\mathit{nca}$ queries. Thus, we can compute all values $M(B(v_i)\setminus S_i)$ with the use of $O(k_1+\dots+k_N)=O(|S_1|+\dots+|S_N|)$ $\mathit{nca}$ queries.  By \cite{DBLP:journals/siamcomp/HarelT84} or \cite{DBLP:journals/siamcomp/BuchsbaumGKRTW08}, we know that there is a linear-time preprocessing of $T$, so that we can answer a collection of $N'$ $\mathit{nca}$ queries in $O(N')$ time in total. We conclude that the values $M(B(v_1)\setminus{S_1}),\dots,M(B(v_N)\setminus{S_N})$ can be computed in $O(m+|S_1|+\dots+|S_N|)$ time in total.
\end{proof}

\subsection{Two lemmata concerning paths}
\label{subsection:paths}

\begin{lemma}
\label{lemma:path-to-non-related-nca}
Let $u$ and $v$ be two vertices, and let $P$ be a path in $G$ from $u$ to $v$. Then, $P$ passes from an ancestor of $\mathit{nca}\{u,v\}$.
\end{lemma}
\begin{proof}
Let $w$ be the lowest vertex that is used by $P$. Let us suppose, for the sake of contradiction, that $w$ is not an ancestor of $\mathit{nca}\{u,v\}$. Then, either $w$ is not an ancestor of $u$, or $w$ is not an ancestor of $v$. Let us assume w.l.o.g. that $w$ is not an ancestor of $u$. Since $u$ is not a descendant of $w$, we may consider the first predecessor $z$ of $w$ in $P$ that is not a descendant of $w$. Let $z'$ be the successor of $z$ in $P$. Then, we have that $P$ uses the edge $(z,z')$, and $z'$ is a descendant of $w$. Let us suppose, first, that $(z,z')$ is a tree-edge. Then, since $z'$ is a descendant of $w$, but $z$ is not, we have that $z$ cannot be a child of $z'$. Thus, $z$ is the parent of $z'$. But since $z'$ is a descendant of $w$ and its parent is not, we have that $z'=w$, and therefore $z$ is the parent of $w$. But this contradicts the minimality of $w$. Thus, we have that $(z,z')$ is a back-edge. Then, since $z'$ is a descendant of $w$, but $z$ is not, we have that $z$ cannot be a descendant of $z'$, and therefore it is an ancestor of $z'$. Then, $z'$ is a common descendant of $w$ and $z$, and therefore $w$ and $z$ are related as ancestor and descendant. But since $z$ is not a descendant of $w$, it must be a proper ancestor of $w$, and therefore $z<w$. This again contradicts the minimality of $w$. Thus, our initial supposition cannot be true, and therefore $w$ is an ancestor of $\mathit{nca}\{u,v\}$.
\end{proof}

\begin{lemma}
\label{lemma:back-edge-or-tree-edge}
Let $u$ be a vertex and let $v$ be a proper ancestor of $u$. Let $P$ be a path that starts from a descendant of $u$ and ends in $v$. Then, the first occurrence of an edge that is used by $P$ and leads outside of the subtree of $u$ is either a back-edge that leaps over $u$ or the tree-edge $(u,p(u))$.
\end{lemma}
\begin{proof}
Let $(x,y)$ be the first occurrence of an edge that is used by $P$ and leads outside of the subtree of $u$. We may assume w.l.o.g. that $x$ is a descendant of $u$ and $y$ is not a descendant of $u$. Then, we have that $y$ is not a descendant of $x$, because otherwise it would be a descendant of $u$. Suppose first that $(x,y)$ is a tree-edge. Then, since $y$ is not a descendant of $x$, it must be the parent of $x$. Thus, since $x$ is a descendant of $u$ but its parent is not, we have that $x=u$ and therefore $y=p(u)$. Now let us suppose that $(x,y)$ is a back-edge. Then, since $y$ is not a descendant of $x$, it must be a proper ancestor of $x$. Thus, $x$ is a common descendant of $u$ and $y$, and therefore $u$ and $y$ are related as ancestor and descendant. Then, since $y$ is not a descendant of $u$, we have that $y$ is a proper ancestor of $u$. This shows that $(x,y)$ is a back-edge that leaps over $u$.
\end{proof}

\subsection{An oracle for back-edge queries}
\label{subsection:back-edge-oracle}

Our goal in this section is to prove the following.

\begin{lemma}
\label{lemma:back-edge-oracle}
Let $T$ be a DFS-tree of a connected graph $G$. We can construct in linear time a data structure of size $O(n)$ that we can use in order to answer in constant time queries of the form: given three vertices $u,v,w$, such that $u$ is a proper descendant of $v$, and $v$ is a proper descendant of $w$, is there a back-edge $(x,y)\in B(v)\setminus B(u)$ such that $y\leq w$? 
\end{lemma}
\begin{proof}
First we compute the $\mathit{low}$ points of all vertices. This takes linear time. Then we compute, for every vertex $v$ that is not a leaf, a child $c(v)$ of $v$ that has the lowest $\mathit{low}$ point among all the children of $v$ (breaking ties arbitrarily). We call this the $\mathit{low}$ child of $v$. Then, starting from any vertex $v$ that is either $r$ or a vertex that is not the $\mathit{low}$ child of its parent, we consider the path that starts from $v$ and ends in a leaf by following the $\mathit{low}$ children. In other words, this is the path $v,c(v),c(c(v)),\dots$. Notice that every vertex $v$ of $G$ belongs to precisely one such path. We call this the $\mathit{low}$ path that contains $v$, and we maintain a pointer from $v$ to the $\mathit{low}$ path that contains it. Furthermore, we consider those paths indexed, starting from their lowest vertex. In other words, if $v,c(v),c(c(v)),\dots$ is a $\mathit{low}$ path, then $v$ has index $1$, $c(v)$ has index $2$, and so on, on this path. We also maintain, for every vertex $v$, a pointer to the index of $v$ on the $\mathit{low}$ path that contains it. 

Now, for every vertex $v$, we compute a value $m(v)$ that is defined as follows. If $v$ has less than two children, then $m(v):=l(v)$. Otherwise, let $c_1,\dots,c_k$ be the list of the children of $v$, excluding $c(v)$. Then, $m(v):=\mathit{min}\{l(v),\mathit{low}(c_1),\dots,\mathit{low}(c_k)\}$. Notice that the $m$ values of all vertices can be easily computed in total linear time. Now, for every $\mathit{low}$ path, we initialize a data structure for answering range-minimum queries w.r.t. the $m$ values. (We consider a $\mathit{low}$ path as an array, corresponding to the indexes of its vertices.) More precisely, for every $\mathit{low}$ path $P$ we initialize a  range-minimum query data structure $\mathit{RMQ_P}$. We can use $\mathit{RMQ_P}$ in order to answer queries of the form: given two vertices $u$ and $v$ on $P$ with indices $i$ and $j$, respectively, such that $i\leq j$, return the minimum of $m(u_1),\dots,m(u_{j-i+1})$, where $u_1,\dots,u_{j-i+1}$ is the set of the vertices on $P$ with indices $i,i+1,\dots,j$. Using the RMQ data structure described, e.g., in \cite{DBLP:conf/latin/BenderF00}, the initialization of all those data structures takes $O(n)$ time in total (because the total size of the $\mathit{low}$ paths is $O(n)$), and every range-minimum query on every such data structure can be answered in $O(1)$ worst-case time. This completes the description of the data structure and its construction.

Now let $u,v,w$ be three vertices such that $u$ is a proper descendant of $v$, and $v$ is a proper descendant of $w$. We will show how we can determine in constant time whether there is a back-edge $(x,y)\in B(v)\setminus B(u)$ such that $y\leq w$. First, suppose that $\mathit{low}(v)>w$. Then we know that there is no back-edge $(x,y)\in B(v)$ such that $y\leq w$, and we are done. So let us assume that $\mathit{low}(v)\leq w$. If $l(v)\leq w$, then $(v,l(v))$ is a back-edge in $B(v)\setminus B(u)$ such that $l(v)\leq w$, and we are done. So let us assume that $l(v)>w$. 


First, suppose that $u$ does not belong to the same $\mathit{low}$ path as $v$. Then we claim that there is a proper descendant $c'$ of $v$ on the $\mathit{low}$ path $P$ that contains $v$ such that $l(c')\leq w$. To see this, let us assume the contrary. Now, since $l(v)>w$ and $\mathit{low}(v)\leq w$, we have that $\mathit{low}(c(v))\leq w$ (because $c(v)$ has the lowest $\mathit{low}$ point among all the children of $v$). Then, by assumption, we have $l(c(v))>w$. Thus, since $\mathit{low}(c(v))\leq w$, we have that $c(c(v))$ exists, and $\mathit{low}(c(c(v)))\leq w$. Then, again by assumption, we have $l(c(c(v)))>w$. Therefore, since $\mathit{low}(c(c(v)))\leq w$, we have that $c(c(c(v)))$ exists, and $\mathit{low}(c(c(c(v))))\leq w$. We can see that this process must continue endlessly, in contradiction to the fact that the graph contains a finite number of vertices. Thus, there is indeed a proper descendant $c'$ of $v$ on $P$ such that $l(c')\leq w$. Then we can see that $(c',l(c'))$ is a back-edge in $B(v)\setminus B(u)$ such that $l(c')\leq w$, and we are done.

So let us assume that $u$ and $v$ belong to the same $\mathit{low}$ path $P$. Let $i$ be the index of $v$ on $P$, and let $j$ be the index of $p(u)$ on $P$. (We note that $i\leq j$.) Let $m'$ be the answer to the range-minimum query on $\mathit{RMQ_P}$ on the range $[i,j]$. Then we claim that there is a back-edge $(x,y)\in B(v)\setminus B(u)$ such that $y\leq w$ if and only if $m'\leq w$. So let us suppose, first, that there is a back-edge $(x,y)\in B(v)\setminus B(u)$ such that $y\leq w$. Then $x$ is a descendant of $v$, but not a descendant of $u$. Now let $z$ be the maximum vertex on $T[v,p(u)]$ (i.e, the one closest to $p(u)$) such that $x$ is a descendant of $z$. Then, we have that either $x=z$, or $x$ is a descendant of a child $c'$ of $z$ such that $c'\neq c(z)$ (since all the vertices on $T[v,p(u)]$ are part of the $\mathit{low}$ path that contains $u$ and $v$). If $x=z$, then we have that $l(z)\leq y\leq w$, and therefore $m(z)\leq w$. Therefore, since $m'\leq m(z)$, we have $m'\leq w$, as desired. Otherwise, suppose that $x$ is a descendant of a child $c'$ of $z$ such that $c'\neq c(z)$. Then, we have that $(x,y)\in B(c')$, and therefore $\mathit{low}(c')\leq y\leq w$. Then, since $m(z)\leq\mathit{low}(c')$, we have that $m(z)\leq w$. And since $m'\leq m(z)$, this shows that $m'\leq w$. 

Conversely, let us suppose that $m'\leq w$. There is a vertex $z$ on the tree-path $T[v,p(u)]$ such that $m'=m(z)$. Then, we have that either $m'=l(z)$, or there is a child $c'$ of $z$, with $c'\neq c(z)$, such that $\mathit{low}(c')=m'$. If $m'=l(z)$, then, since $m'\leq w$ and $w$ is a proper ancestor of $v$, we have that there is a back-edge $(z,l(z))$. Then we have that $(z,l(z))\in B(v)$, but $(z,l(z))\notin B(u)$ (since $z$ is a proper ancestor of $u$). Thus, $(z,l(z))$ is the desired back-edge. Otherwise, suppose that there is a child $c'$ of $z$, with $c'\neq c(z)$, such that $\mathit{low}(c')=m'$. Then, there is a back-edge $(x,y)\in B(c')$ such that $y=\mathit{low}(c')$. Since $\mathit{low}(c')=m'\leq w$, we can see that $(x,y)\in B(v)$. But since $c'$ and $c(z)$ are two different children of $z$, we have that $x$ is cannot be a descendant of $u$ (because $u$ is a descendant of $c(z)$ and $x$ is a descendant of $c'$), and therefore we have $(x,y)\notin B(u)$. Thus, $(x,y)$ is a back-edge in $B(v)\setminus B(u)$ such that $y\leq w$. 

This concludes the method by which we can determine in constant time whether there exists a back-edge $(x,y)\in B(v)\setminus B(u)$ such that $y\leq w$. This process is shown in Algorithm~\ref{algorithm:back-edge-query}.
\end{proof}

\begin{algorithm}[H]
\caption{\textsf{Determine whether there is a back-edge $(x,y)\in B(v)\setminus B(u)$ such that $y\leq w$, where $u$ is a proper descendant of $v$, and $v$ is a proper descendant of $w$}}
\label{algorithm:back-edge-query}
\LinesNumbered
\DontPrintSemicolon
\lIf{$\mathit{low}(v)>w$}{
\textbf{return false}
}
\lIf{$l(v)\leq w$}{
\textbf{return true}
}
\lIf{$u$ and $v$ do not belong to the same $\mathit{low}$ path}{
\textbf{return true}
}
let $P$ be the $\mathit{low}$ path that contains $u$ and $v$\;
let $i$ be the index of $v$ on $P$, and let $j$ be the index of $p(u)$ on $P$\;
let $m'$ be the answer to the range-minimum query on $\mathit{RMQ_R}$ on the range $[i,j]$\;
\lIf{$m'\leq w$}{
\textbf{return true}
}
\textbf{return false}\;
\end{algorithm}

\subsection{Segments of vertices that have the same $\mathit{high}$ point}
\label{subsection:segments}

Throughout this section, we assume that $G$ is a $3$-edge-connected graph. According to Proposition~\ref{proposition:3-edge-conn}, this implies that $|B(v)|>2$ for every vertex $v\neq r$, and therefore the $\mathit{high}_1$ and $\mathit{high}_2$ points of $v$ are defined.

Let $x$ be a vertex of $G$, and let $H(x)$ be the list of all vertices $v\neq r$ such that $\mathit{high}(v)=x$, sorted in decreasing order. For a vertex $v\in H(x)$, we let $S(v)$ denote the segment of $H(x)$ that contains $v$ and is maximal w.r.t. the property that its elements are related as ancestor and descendant. The collection of those segments constitutes a partition of $H(x)$, as shown in the following.

\begin{lemma}
\label{lemma:segments:partition}
Let $x$ be a vertex. Then, the collection $\mathcal{S}$ of all segments of $H(x)$ that are maximal w.r.t. the property that their elements are related as ancestor and descendant is a partition of $H(x)$.
\end{lemma}
\begin{proof}
Every vertex $v\in H(x)$ is contained in $S(v)\in\mathcal{S}$, and therefore the collection of all segments in $\mathcal{S}$ covers $H(x)$. Now let $S$ and $S'$ be two distinct segments in $\mathcal{S}$. Let us suppose, for the sake of contradiction, that $S\cap S'\neq\emptyset$. Then there is a vertex $z\in S\cap S'$. 

We will show that all vertices in $S\cup S'$ are related as ancestor and descendant. So let $u$ and $u'$ be two vertices in $S\cup S'$. If both $u$ and $u'$ are either in $S$ or in $S'$, then we have that $u$ and $u'$ are related as ancestor and descendant. So let assume w.l.o.g. that $u\in S$ and $u'\in S'$. Since $u,z\in S$, we have that $u$ and $z$ are related as ancestor and descendant. Also, since $u',z\in S'$, we have that $u'$ and $z$ are related as ancestor and descendant. Thus, we have the following cases to consider: either $(1)$ both $u$ and $u'$ are ancestors of $z$, or $(2)$ one of $u$ and $u'$ is an ancestor of $z$, and the other is a descendant of $z$, or $(3)$ both $u$ and $u'$ are descendants of $z$. 

In case $(1)$, we have that $z$ is a common descendant of $u$ and $u'$, and therefore $u$ and $u'$ are related as ancestor and descendant. In case $(2)$, we may assume w.l.o.g. that $u$ is an ancestor of $z$, and $u'$ is a descendant of $z$. Then, we obviously have that $u$ is an ancestor of $u'$ (due to the transitivity of the ancestry relation). So let us consider case $(3)$. This implies that $u\geq z$ and $u'\geq z$. If $u\geq u'$, then we have $u\geq u'\geq z$. Then, since $H(x)$ is sorted in decreasing order and $S$ is a segment of $H(x)$ and $u,z\in S$, we have $u'\in S$. Then, $u\geq u'$ implies that $u'$ is an ancestor of $u$. Otherwise, supppose that $u< u'$. Then we have $u'> u\geq z$. Then, since $H(x)$ is sorted in decreasing order and $S'$ is a segment of $H(x)$ and $u',z\in S'$, we have $u\in S'$. Then, $u< u'$ implies that $u$ is a proper ancestor of $u'$. Thus, in any case we have shown that $u$ and $u'$ are related as ancestor and descendant.

Now we will show that $S\cup S'$ is a segment of $H(x)$. So let us suppose, for the sake of contradiction, that $S\cup S'$ is not a segment of $H(x)$. Since $H(x)$ is sorted in decreasing order, this means that there are two vertices $u$ and $u'$ in $S\cup S'$, with $u>u'$, and a vertex $w\in H(x)$, such that $u>w>u'$ and $w\notin S\cup S'$. Notice that we cannot have that both $u$ and $u'$ are either in $S$ or in $S'$, because $S$ and $S'$ are segments of $H(x)$, and therefore $u>w>u'$ implies that $w\in S$ or $w\in S'$, respectively. Thus, we may assume w.l.o.g. that $u\in S$ and $u'\in S'$. Since $u$ and $z$ are in $S$, we have that $u$ and $z$ are related as ancestor and descendant. First, let us suppose that $z$ is a descendant of $u$. This implies that $z\geq u$. Thus, we have $z\geq u>w>u'$. Since $S'$ is a segment of $H(x)$ that contains both $z$ and $u'$, this implies that $w\in S'$, a contradiction. Thus, we have that $z$ is a proper ancestor of $u$, and therefore $u> z$. Since $u'$ and $z$ are in $S'$, we have that $u'$ and $z$ are related as ancestor and descendant. Let us suppose that $u'$ is a descendant of $z$. This implies that $u'\geq z$. Then, we have $u>w>u'\geq z$. Since $S$ is a segment of $H(x)$ that contains both $u$ and $z$, this implies that $w\in S$, a contradiction. Thus, we have that $u'$ is a proper ancestor of $z$, and therefore $z> u'$. Thus, since $u>w>u'$ and $u> z> u'$, we have that either $u>w\geq z$ or $z\geq w>u'$. Any of those cases implies that either $w\in S$ or $w\in S'$, since $S$ and $S'$ are segments of $H(x)$. A contradiction. This shows that $S\cup S'$ is a segment of $H(x)$.

Thus, we have shown that $S\cup S'$ is a segment of $H(x)$ with the property that its elements are related as ancestor and descendant. But since $S\neq S'$, this contradicts the maximality of both $S$ and $S'$ with this property. We conclude that the segments in $\mathcal{S}$ partition $H(x)$. 
\end{proof}

For every vertex $x$, we will need to compute the collection of the segments of $H(x)$ that are maximal w.r.t. the property that their elements are related as ancestor and descendant. This can be done with a straightforward method that is shown in Algorithm~\ref{algorithm:segments}. The idea is to traverse the list $H(x)$, and greedily collect all consecutive vertices that are related as ancestor and descendant in order to get a segment. The proof of correctness in given in Lemma~\ref{lemma:segments}.

\begin{algorithm}[H]
\caption{\textsf{Compute the collection $\mathcal{S}$ of the segments of $H(x)$ that are maximal w.r.t. the property that their elements are related as ancestor and descendant}}
\label{algorithm:segments}
\LinesNumbered
\DontPrintSemicolon
let $\mathcal{S}\leftarrow\emptyset$\;
let $z$ be the first element of $H(x)$\;
\While{$z\neq\bot$}{
\label{line:segments-while}
  let $S\leftarrow\{z\}$\;
  \label{line:segments-while-1}
  let $z'\leftarrow\mathit{next}_\mathit{H(x)}(z)$\;
  \While{$z'\neq\bot$ \textbf{and} $z'$ is an ancestor of $z$}{
  \label{line:segments-while-2}
    insert $z'$ into $S$\;
    \label{line:segments-ins}
    $z'\leftarrow\mathit{next}_\mathit{H(x)}(z')$\;
  }
  insert $S$ into $\mathcal{S}$\;
  \label{line:segments-S}
  $z\leftarrow z'$\;
}
\end{algorithm}

\begin{lemma}
\label{lemma:segments}
Let $x$ be a vertex. Then, Algorithm~\ref{algorithm:segments} correctly computes the collection $\mathcal{S}$ of the segments of $H(x)$ that are maximal w.r.t. the property that their elements are related as ancestor and descendant. The running time of Algorithm~\ref{algorithm:segments} is $O(|H(x)|)$.
\end{lemma}
\begin{proof}
The \textbf{while} loop in Line~\ref{line:segments-while} begins the processing of $H(x)$ from its first vertex $z$. Then, the \textbf{while} loop in Line~\ref{line:segments-while-2} collects all the consecutive successors $z'$ of $z$ that are ancestors of $z$, and stops until it reaches a vertex $z'$ that is not an ancestor of $z$. Let $S$ be the resulting set (in Line~\ref{line:segments-S}). Then, we have that all vertices in $S$ have $z$ as a common descendant, and therefore all of them are related as ancestor and descendant. Furthermore, by construction, $S$ is a segment of $H(x)$. Then, notice that $S$ is a maximal segment of $H(x)$ with the property that its elements are related as ancestor and descendant, because the successor of the lowest element in $S$ is not an ancestor of $z$, and therefore it is not related as ancestor and descendant with $z$ (because, if it was, it would be an ancestor of $z$, due to the ordering of $H(x)$). By Lemma~\ref{lemma:segments:partition}, we have that no other segment of $\mathcal{S}$ intersects with $S$. Thus, it is proper to move on to the processing of $z'$, and always move forward in processing the vertices of $H(x)$. Thus, we can see that the \textbf{while} loop in Line~\ref{line:segments-while} correctly computes the segment $S(z)$, for every $z$ that it processes, and every vertex in $H(x)$ will be inserted in such a segment eventually (either at the beginning of the \textbf{while} loop of Line~\ref{line:segments-while} in Line~\ref{line:segments-while-1}, or by the \textbf{while} loop of Line~\ref{line:segments-while-2} in Line~\ref{line:segments-ins}). It is easy to see that the number of steps performed by Algorithm~\ref{algorithm:segments} is $O(|H(x)|)$.
\end{proof}

For every vertex $x$, we also define the list $\widetilde{H}(x)$ that consists of all vertices $v\neq r$ such that either $\mathit{high}_1(v)=x$ or $\mathit{high}_2(v)=x$, sorted in decreasing order. Notice that, for every vertex $v\neq r$, there are at most two distinct $x$ and $x'$ such that $v\in\widetilde{H}(x)$ and $v\in\widetilde{H}(x')$. More precisely, if a vertex $v\neq r$ satisfies that $\mathit{high}_1(v)\neq\mathit{high}_2(v)$, then $v$ belongs to both $\widetilde{H}(\mathit{high}_1(v))$ and $\widetilde{H}(\mathit{high}_2(v))$. But there is no other set of the form $\widetilde{H}(x)$ that contains $v$. Thus, the collection $\{\widetilde{H}(x)\mid x\mbox{ is a vertex}\}$ has total size $O(n)$. 
For every vertex $v\neq r$, we let $\widetilde{S}_1(v)$ denote the segment of $\widetilde{H}(\mathit{high}_1(v))$ that contains $v$ and is maximal w.r.t. the property that its elements are related as ancestor and descendant. Similarly, we let $\widetilde{S}_2(v)$ denote the segment of $\widetilde{H}(\mathit{high}_2(v))$ that contains $v$ and is maximal w.r.t. the property that its elements are related as ancestor and descendant. 
We can see that the collection of the segments of $\widetilde{H}(x)$ that are maximal w.r.t. the property that their elements are related as ancestor and descendant constitutes a partition of $\widetilde{H}(x)$. The proof of this property is precisely the same as in Lemma~\ref{lemma:segments:partition}, because it only relies on the fact that $\widetilde{H}(x)$ is sorted in decreasing order. Furthermore, we can apply a procedure as that shown in Algorithm~\ref{algorithm:segments}, in order to compute the collection of all those maximal segments in $O(|\widetilde{H}(x)|)$ time. We state this result in the following lemma, which has the same proof as Lemma~\ref{lemma:segments}.

\begin{lemma}
\label{lemma:segments-2}
Let $x$ be a vertex. Then, in $O(|\widetilde{H}(x)|)$ time, we can compute the collection of the segments of $\widetilde{H}(x)$ that are maximal w.r.t. the property that their elements are related as ancestor and descendant.
\end{lemma}
\begin{proof}
We can use Algorithm~\ref{algorithm:segments}, where we have replaced every occurrence of ``$H(x)$" with ``$\widetilde{H}(x)$". The proof of correctness is the same as in Lemma~\ref{lemma:segments}.
\end{proof}

Since every vertex $v\neq r$ belongs to at most two sets of the form $\widetilde{H}(x)$, for a vertex $x$, the collection $\bigcup{\{\mathcal{S}(x)\mid x \mbox{ is a vertex }\}}$, where $\mathcal{S}(x)$ is the collection of the segments of $\widetilde{H}(x)$ that are maximal w.r.t. the property that their elements are related as ancestor and descendant, has total size $O(n)$. That is, $\sum_{x}{\sum_{S\in\mathcal{S}(x)} |S|}=O(n)$.

We conclude this section with the following lemma, which shows that the vertices in every segment from $\mathcal{S}(x)$ are sorted in decreasing order w.r.t. their $\mathit{low}$ point.

\begin{lemma}
\label{lemma:vertices_on_segment-2}
Let $x$ be a vertex, and let $u$ and $v$ be two vertices in $\widetilde{H}(x)$ such that $u$ is a descendant of $v$. Then $\mathit{low}(u)\geq\mathit{low}(v)$.
\end{lemma}
\begin{proof}
Since $u\in\widetilde{H}(x)$, we have that either $\mathit{high}_1(u)=x$ or $\mathit{high}_2(u)=x$. Since $v\in\widetilde{H}(x)$, we have that either $\mathit{high}_1(v)=x$ or $\mathit{high}_2(v)=x$. In either case then, we have that $x$ is a proper ancestor of $v$. 

Let us suppose first that $\mathit{high}_1(u)=x$.  Let $(z,w)$ be a back-edge in $B(u)$. Then $z$ is a descendant of $u$, and therefore a descendant of $v$. Furthermore, we have that $w$ is an ancestor of $\mathit{high}_1(u)$, and therefore an ancestor of $x$, and therefore a proper ancestor of $v$. This shows that $(z,w)\in B(v)$. Due to the generality of $(z,w)\in B(u)$, this implies that $B(u)\subseteq B(v)$. From this we infer that $\mathit{low}(u)\geq\mathit{low}(v)$.

Now let us suppose that $\mathit{high}_1(u)\neq x$. This implies that $\mathit{high}_2(u)=x$. Let $(z,w)$ be a back-edge in $B(u)\setminus\{e_\mathit{high}(u)\}$ (such a back-edge exists, because the graph is $3$-edge-connected, and therefore $|B(u)|>1$). Then, we have that $z$ is a descendant of $u$, and therefore a descendant of $v$. Furthermore, $w$ is an ancestor of $\mathit{high}_2(u)$, and therefore an ancestor of $x$, and therefore a proper ancestor of $v$. This shows that $(z,w)\in B(v)$. Due to the generality of $(z,w)\in B(u)\setminus\{e_\mathit{high}(u)\}$, this implies that $B(u)\setminus\{e_\mathit{high}(u)\}\subseteq B(v)$. Since $\mathit{high}_1(u)\neq x$ and $\mathit{high}_2(u)=x$, we have $\mathit{high}_1(u)\neq\mathit{high}_2(u)$. This implies that $\mathit{high}_1(u)>\mathit{high}_2(u)$, and therefore $\mathit{low}(u)<\mathit{high}_1(u)$. Thus, the $\mathit{low}$ point of $u$ is given by the lowest lower endpoint of all back-edges in $B(u)\setminus\{e_\mathit{high}(u)\}$. Since this set is a subset of $B(v)$, we conclude that $\mathit{low}(v)\leq\mathit{low}(u)$.
\end{proof}

\section{Connectivity queries under $4$ edge failures}
\label{section:4eoracle}

Our goal in this section is to prove the following.

\begin{proposition}
\label{proposition:4e-fault}
Let $G$ be a connected graph with $n$ vertices and $m$ edges. Then there is a data structure with size $O(n)$, that we can use in order to answer connectivity queries in the presence of at most four edge-failures in constant time. Specifically, given an edge-set $E'$ with $|E'|\leq 4$, and two vertices $x$ and $y$, we can determine whether $x$ and $y$ are connected in $G\setminus E'$ in $O(1)$ time. The data structure can be constructed in $O(m+n)$ time.
\end{proposition}

Let $G$ be a connected graph, and let $T$ be a rooted spanning tree of $G$.
Let $E'$ be a set of edges of $G$, and let $k$ be the number of tree-edges in $E'$. Then $T\setminus E'$ is split into $k+1$ connected components. Every connected component of $T\setminus E'$ is a subtree of $T$, and the connectivity in $G\setminus E'$ can be reduced to the connectivity of those subtrees as follows. Let $x$ and $y$ be two vertices of $G$, let $C_1$ be the connected component of $T\setminus E'$ that contains $x$, and let $C_2$ be the connected component of $T\setminus E'$ that contains $y$. Then we have that $x$ and $y$ are connected in $G\setminus E'$ if and only if $C_1$ and $C_2$ are connected in $G\setminus E'$. In particular, since $C_1$ and $C_2$ are subtrees of the rooted tree $T$, we can consider the roots $r_1$ and $r_2$ of $C_1$ and $C_2$ as representatives of $C_1$ and $C_2$, respectively. Then, we have that $x$ and $y$ are connected in $G\setminus E'$ if and only if $r_1$ and $r_2$ are connected in $G\setminus E'$. Thus, the connectivity relation of $G\setminus E'$ can be captured by the connectivity relation of the roots of the connected components of $T\setminus E'$ in $G\setminus E'$. In order to capture this connectivity relation, we introduce the concept of a \emph{connectivity graph} for $G\setminus E'$.

\begin{definition}
\label{definition:connectivity-graph}
\normalfont
Let $T$ be a fixed rooted spanning tree of $G$, and let $E'$ be a set of edges of $G$. Let $\{C_1,\dots,C_k\}$ be the set of the connected components of $T\setminus E'$. Thus, for every $i\in\{1,\dots,k\}$, $C_i$ is a rooted subtree of $T$, and let $r_i$ be its root. Then, a connectivity graph for $G\setminus E'$ is a graph $\mathcal{R}$ with vertex set $\{\bar{r_i}\mid i\in\{1,\dots,k\}\}$ such that: for every $i,j\in\{1,\dots,k\}$, $r_i$ is connected with $r_j$ in $G\setminus E'$ if and only if $\bar{r_i}$ is connected with $\bar{r_j}$ in $\mathcal{R}$.  
\end{definition}

We allow a connectivity graph $\mathcal{R}$ for $G\setminus E'$ to be a multigraph. (The important thing is that it captures the connectivity relation of the roots of the connected components of $T\setminus E'$.) In particular, if we shrink every connected component of $T\setminus E'$ with root $z$ into a node $\bar{z}$, then the quotient graph of $G\setminus E'$ that we get is a connectivity graph for $G\setminus E'$. The main challenge is to compute a connectivity graph for $G\setminus E'$ without explicitly computing the connected components of $T\setminus E'$. We can achieve this if we assume that $T$ is a DFS-tree of the graph. Then, with a creative use of the DFS-concepts that we defined in Section~\ref{section:DFS}, we can determine enough edges of $G\setminus E'$ between the connected components of $T\setminus E'$, so that we can construct a connectivity graph for $G\setminus E'$ in constant time, if $|E'|\leq 4$.





We distinguish five different cases, depending on the number of tree-edges contained in $E'$. Then, for each case, we consider all the different possibilities for the edges from $E'$ on $T$ (i.e., all the different topologies of their endpoints w.r.t. the ancestry relation). In order to reduce the number of cases considered, we will use the following three facts (which we make precise and prove in the following paragraphs). First, if we have established that the endpoints of a tree-edge $e$ in $E'$ remain connected in $G\setminus E'$, then it is sufficient to set $E'\leftarrow E'\setminus\{e\}$, and then revert to the previous case, where the number of tree-edges is that of $E'$ minus $1$ (see Lemma~\ref{lemma:4conn-edge-connected}). Second, if there are at least two tree-edges in $E'$ that are not descendants of other tree-edges in $E'$, then we can handle the subtrees induced by those tree-edges separately, by reverting to previous cases (see Lemma~\ref{lemma:4conn-non-related-edges}). And third, if for a tree-edge $(u,p(u))$ in $E'$ we have that $G\setminus E'$ contains no back-edges that leap over $u$, then we can handle the subtree induced by this tree-edge separately from the rest of the tree, by reverting to previous cases (see Lemma~\ref{lemma:4conn-no-back-edge}).




In the following, we assume that $T$ is a fixed DFS-tree of $G$ with root $r$. All connectivity graphs refer to this tree. First, we will need the following technical lemma.

\begin{lemma}
\label{lemma:edge-fault-technical}
Let $E'$ be a set of edges, and let $E''$ be a subset of $E'$ that contains all the back-edges from $E'$, and has the property that no tree-edge from $E'\setminus E''$ is related as ancestor and descendant with a tree-edge from $E''$. Let $U$ be the collection of the higher endpoints of the tree-edges in $E''$, and let $z$ and $z'$ be two vertices in $U\cup\{r\}$. Suppose that $z$ and $z'$ are connected in $G\setminus E''$. Then $z$ and $z'$ are connected in $G\setminus E'$. 
\end{lemma}
\begin{proof}
Since $z$ and $z'$ are connected in $G\setminus E''$, there is a path $P$ from $z$ to $z'$ in $G\setminus E''$. If $P$ does not use any tree-edge from $E'\setminus E''$, then we have that $P$ is a path in $G\setminus E'$, and therefore $z$ and $z'$ are connected in $G\setminus E'$. So let us assume that $P$ uses a tree-edge from $E'\setminus E''$. Let $C$ be the connected component of $T\setminus E'$ that contains $r$. 
Then, we claim that $P$ has the form $P_1+Q+P_2$, where $P_1$ is path from $z$ to a vertex $w\in C$ that does not use any tree-edge from $E'\setminus E''$, $Q$ is a path from $w$ to a vertex $w'\in C$ (that uses tree-edges from $E'\setminus E''$), and $P_2$ is a path from $w'$ to $z'$ that does not use any tree-edge from $E'\setminus E''$ $(*)$. This implies that $P_1$ is a path from $z$ to $w$ in $G\setminus E'$, and $P_2$ is a path from $w'$ to $z'$ in $G\setminus E'$. Then, since $w$ and $w'$ lie in the same connected component of $T\setminus E'$, we have that there is a path $Q'$ from $w$ to $w'$ in $G\setminus E'$. Thus, $P_1+Q'+P_2$ is a path from $z$ to $z'$ in $G\setminus E'$, and therefore we have that $z$ and $z'$ are connected in $G\setminus E'$.

Now we will prove $(*)$. Let $U'$ be the collection of the higher endpoints of the tree-edges in $E'\setminus E''$. Then, we have that no vertex from $U$ is  related as ancestor and descendant with a vertex from $U'$.
Now let $(v,p(v))$ be the first occurrence of a tree-edge from $E'\setminus E''$ that is used by $P$, and let $(v',p(v'))$ be the last occurrence of a tree-edge from $E'\setminus E''$ that is used by $P$. Then, since $P$ starts from $z$ and visits $v$, Lemma~\ref{lemma:path-to-non-related-nca} implies that $P$ contains a subpath from an ancestor $w$ of $\mathit{nca}\{z,v\}$ to $v$, and let $w$ be the first vertex visited by $P$ with this property. Let $P_1$ be the initial part of $P$ from $z$ to the first occurrence of $w$. Then, we have that $P_1$ does not use any tree-edge from $E'\setminus E''$. Notice that there is no $u\in U$ such that $u$ is an ancestor of $w$, because otherwise we would have that $u$ is an ancestor of $v$. Also, there is no $u'\in U'$ such that $u'$ is an ancestor of $w$, because otherwise $u'$ would be an ancestor of $z$. Thus, there is no tree-edge from $E'$ on the tree-path $T[r,w]$.
Similarly, since $P$ visits $v'$ and ends in $z'$, Lemma~\ref{lemma:path-to-non-related-nca} implies that $P$ contains a subpath from $v'$ to an ancestor $w'$ of $\mathit{nca}\{v',z'\}$, and let $w'$ be the last vertex visited by $P$ with this property. Let $P_2$ be the final part of $P$ from the last occurrence of $w'$ to $z'$. Then, we have that $P_2$ does not use any tree-edge from $E'\setminus E''$. Again, we can see that there is no tree-edge from $E'$ on the tree-path $T[r,w']$. Now we can consider the part $Q$ of $P$ from the first occurrence of $w$ to the last occurrence of $w'$, and the proof is complete, because $P=P_1+Q+P_2$, and the endpoints of $Q$ lie in the connected component of $T\setminus E'$ that contains $r$.
\end{proof}

\begin{lemma}
\label{lemma:4conn-edge-connected}
Let $E'$ be a set of edges, and let $(u,p(u))$ and $(v,p(v))$ be two distinct tree-edges in $E'$. Suppose that $u$, $p(u)$ and $v$ are connected in $G\setminus E'$. Let $E''=E'\setminus\{(u,p(u))\}$, and let $\mathcal{R}'$ be a connectivity graph for $G\setminus E''$. Then, $\mathcal{R}'\cup\{(\bar{u},\bar{v})\}$ is a connectivity graph for $G\setminus E'$.
\end{lemma}
\begin{proof}
Let $(u_1,p(u_1)),\dots,(u_k,p(u_k))$ be the tree-edges in $E'$, where $u_1=u$. Let $E''=E'\setminus\{(u,p(u))\}$, and let $\mathcal{R}'$ be a connectivity graph for $G\setminus E''$. Then we have that $V(\mathcal{R}')=\{\bar{u}_2,\dots,\bar{u}_k,\bar{r}\}$. Let $\mathcal{R}$ be the graph with $V(\mathcal{R})=V(\mathcal{R}')\cup\{\bar{u}\}$ and $E(\mathcal{R})=E(\mathcal{R}')\cup\{(\bar{u},\bar{v})\}$. We will show that $\mathcal{R}$ is a connectivity graph for $G\setminus E'$.

Let $z$ and $z'$ be two vertices in $\{u_2,\dots,u_k,r\}$. First, suppose that $z$ and $z'$ are connected in $G\setminus E'$. Then, since $E''\subset E'$, we have that $z$ and $z'$ are connected in $G\setminus E''$. Thus, $\bar{z}$ and $\bar{z}'$ are connected in $\mathcal{R}'$, and therefore they are connected in $\mathcal{R}$. Conversely, suppose that $\bar{z}$ and $\bar{z}'$ are connected in $\mathcal{R}$. Then, it is easy to see that $\bar{z}$ and $\bar{z}'$ are connected in $\mathcal{R}'$. This implies that $z$ and $z'$ are connected in $G\setminus E''$. Thus, there is a path $P$ from $z$ to $z'$ in $G\setminus E''$. If $P$ does not use the edge $(u,p(u))$, then $P$ is a path in $G\setminus E'$, and therefore $z$ and $z'$ are connected in $G\setminus E'$. Otherwise, since $u$ and $p(u)$ are connected in $G\setminus E'$, there is a path $Q$ from $u$ to $p(u)$ in $G\setminus E'$ that avoids the edge $(u,p(u))$. Now replace every occurrence of $(u,p(u))$ in $P$ with $Q$, and let $P'$ be the resulting path. Then, we have that $P'$ is a path from $z$ to $z'$ in $G\setminus E'$. This shows that $z$ and $z'$ are connected in $G\setminus E'$.

Now let $z$ be a vertex in $\{u_2,\dots,u_k,r\}$. First, suppose that $u$ and $z$ are connected in $G\setminus E'$. Then, since $E''\subset E'$, we have that $u$ and $z$ are connected in $G\setminus E''$. Furthermore, since $u$ and $v$ are connected in $G\setminus E'$, we have that $u$ and $v$ are connected in $G\setminus E''$. Thus, $v$ and $z$ are connected in $G\setminus E''$, and therefore $\bar{v}$ and $\bar{z}$ are connected in $\mathcal{R}'$. Thus, the existence of the edge $(\bar{u},\bar{v})$ in $\mathcal{R}$ implies that $\bar{u}$ and $\bar{z}$ are connected in $\mathcal{R}$. Conversely, suppose that $\bar{u}$ and $\bar{z}$ are connected in $\mathcal{R}$. Then, since $(\bar{u},\bar{v})$ is the only edge of $\mathcal{R}$ that is incident to $\bar{u}$, we have that $\bar{v}$ is connected with $\bar{z}$ in $\mathcal{R}$ through a path that avoids $\bar{u}$. Therefore, $\bar{v}$ is connected with $\bar{z}$ in $\mathcal{R}'$. Thus, we have that $v$ and $z$ are connected in $G\setminus E''$. So let $P$ be a path from $v$ to $z$ in $G\setminus E''$. Since $u$ and $p(u)$ are connected in $G\setminus E'$, we have that there is a path $Q$ from $u$ to $p(u)$ in $G\setminus E'$. Now, if $P$ uses the edge $(u,p(u))$, then we replace every occurrence of $(u,p(u))$ in $P$ with the path $Q$. Let $P'$ be the resulting path. Then, $P'$ is a path from $v$ to $z$ in $G\setminus E'$, and therefore $v$ is connected with $z$ in $G\setminus E'$. Therefore, since $u$ is connected with $v$ in $G\setminus E'$, we have that $u$ is connected with $z$ in $G\setminus E'$.

Thus, we have shown that, for every pair of vertices $z$ and $z'$ in $\{u_1,\dots,u_k,r\}$, we have that $z$ and $z'$ are connected in $G\setminus E'$ if and only if $\bar{z}$ and $\bar{z}'$ are connected in $\mathcal{R}$. This means that $\mathcal{R}$ is a connectivity graph for $G\setminus E'$.  
\end{proof}

\begin{lemma}
\label{lemma:4conn-non-related-edges}
Let $E'$ be a set of edges, and let $(u,p(u))$ be a tree-edge in $E'$ with the property that no tree-edge in $E'$ is a proper ancestor of $(u,p(u))$. Let $E_1$ be the set of the tree-edges in $E'$ that are descendants of $(u,p(u))$, plus the back-edges in $E'$, and let $E_2$ be the set of the tree-edges in $E'$ that are not descendants of $(u,p(u))$, plus the back-edges in $E'$. Let $\mathcal{R}_1$ be a connectivity graph for $G\setminus E_1$, and let $\mathcal{R}_2$ be a connectivity graph for $G\setminus E_2$. Then, $\mathcal{R}_1\cup\mathcal{R}_2$ is a connectivity graph for $G\setminus E'$.
\end{lemma}
\begin{proof}
Let $(u_1,p(u_1)),\dots,(u_t,p(u_t))$ be the tree-edges in $E_1$, and let $(v_1,p(v_1)),\dots,(v_s,p(v_s))$ be the tree-edges in $E_2$. (Notice that we may have $s=0$, in which case the conclusion of the lemma follows trivially.) Then, we have that all vertices in $\{u_1,\dots,u_t\}$ are descendants of $u$, and none of the vertices in $\{v_1,\dots,v_s\}$ is related as ancestor and descendant with $u$. This implies that no vertex from $\{u_1,\dots,u_t\}$ is related as ancestor and descendant with a vertex from $\{v_1,\dots,v_s\}$. Let $\mathcal{R}_1$ be a connectivity graph for $G\setminus E_1$, and let $\mathcal{R}_2$ be a connectivity graph for $G\setminus E_2$. Then we have that $V(\mathcal{R}_1)=\{\bar{u}_1,\dots,\bar{u}_t,\bar{r}\}$ and $V(\mathcal{R}_2)=\{\bar{v}_1,\dots,\bar{v}_s,\bar{r}\}$. Let $\mathcal{R}=\mathcal{R}_1\cup \mathcal{R}_2$. We will show that $\mathcal{R}$ is a connectivity graph for $G\setminus E'$. 

Let $z$ and $z'$ be two vertices in $\{u_1,\dots,u_t,r\}$. First, suppose that $z$ and $z'$ are connected in $G\setminus E'$. Then, since $E_1\subseteq E'$, we have that $z$ and $z'$ are connected in $G\setminus E_1$. This implies that $\bar{z}$ and $\bar{z}'$ are connected in $\mathcal{R}_1$. Therefore, $\bar{z}$ and $\bar{z}'$ are connected in $\mathcal{R}$. Conversely, suppose that $\bar{z}$ and $\bar{z}'$ are connected in $\mathcal{R}$. 
Since $\bar{z}$ and $\bar{z}'$ are connected in $\mathcal{R}$ and are both in $\{u_1,\dots,u_t,r\}$, it is easy to see that $\bar{z}$ and $\bar{z}'$ are connected in $\mathcal{R}_1$. This implies that $z$ and $z'$ are connected in $G\setminus E_1$. Then, Lemma~\ref{lemma:edge-fault-technical} implies that $z$ and $z'$ are connected in $G\setminus E'$.

With the analogous argument we can show that, if $z$ and $z'$ are two vertices in $\{v_1,\dots,v_s,r\}$, then $z$ and $z'$ are connected in $G\setminus E'$ if and only if they are connected in $\mathcal{R}$.

Now let $z$ be a vertex in $\{u_1,\dots,u_t\}$, and let $z'$ be a vertex in $\{v_1,\dots,v_s\}$. First, suppose that $z$ and $z'$ are connected in $G\setminus E'$. Then there is a path $P$ from $z$ to $z'$ in $G\setminus E'$. By Lemma~\ref{lemma:path-to-non-related-nca}, we have that $P$ passes from an ancestor $w$ of $\mathit{nca}\{z,z'\}$. Notice that there is no $i\in\{1,\dots,t\}$ such that $u_i$ is an ancestor of $w$, because otherwise we would have that $u_i$ is an ancestor of $z'$. Similarly, there is no $i\in\{1,\dots,s\}$ such that $v_i$ is an ancestor of $w$, because otherwise we would have that $v_i$ is an ancestor of $z$. Thus, there is no tree-edge from $E'$ on the tree-path $T[w,r]$, and therefore $w$ is connected with $r$  in $G\setminus E'$. Thus, both $z$ and $z'$ are connected with $r$ in $G\setminus E'$. Since $E_1\subseteq E'$, we have that $z$ is connected with $r$ in $G\setminus E_1$. This implies that $\bar{z}$ is connected with $\bar{r}$ in $\mathcal{R}_1$, and therefore $\bar{z}$ is connected with $\bar{r}$ in $\mathcal{R}$. Similarly, since $E_2\subset E'$, we have that $z'$ is connected with $r$ in $G\setminus E_2$. This implies that $\bar{z}'$ is connected with $\bar{r}$ in $\mathcal{R}_2$, and therefore $\bar{z}'$ is connected with $\bar{r}$ in $\mathcal{R}$. Thus, we infer that $\bar{z}$ is connected with $\bar{z}'$ in $\mathcal{R}$.

Conversely, suppose that $\bar{z}$ and $\bar{z}'$ are connected in $\mathcal{R}$. Since $\bar{z}\in V(\mathcal{R}_1)$ and $\bar{z}'\in V(\mathcal{R}_2)$, this implies that $\bar{z}$ is connected with $\bar{r}$ in $\mathcal{R}_1$, and $\bar{z}'$ is connected with $\bar{r}$ in $\mathcal{R}_2$. Therefore, $z$ is connected with $r$ in $G\setminus E_1$, and $z'$ is connected with $r$ in $G\setminus E_2$. Then, Lemma~\ref{lemma:edge-fault-technical} implies that $z$ is connected with $r$ in $G\setminus E'$, and $z'$ is connected with $r$ in $G\setminus E'$. Therefore, $z$ is connected with $z'$ in $G\setminus E'$. 

Thus, we have shown that, for every pair of vertices $z$ and $z'$ in $\{u_1,\dots,u_t,v_1,\dots,v_s,r\}$, we have that $z$ and $z'$ are connected in $G\setminus E'$ if and only if $\bar{z}$ and $\bar{z}'$ are connected in $\mathcal{R}$. Since the tree-edges in $E'$ are given by $\{(u_1,p(u_1)),\dots,(u_t,p(u_t)),(v_1,p(v_1)),\dots,(v_s,p(v_s))\}$, this means that $\mathcal{R}$ is a connectivity graph for $G\setminus E'$.
\end{proof}

\begin{lemma}
\label{lemma:4conn-no-back-edge}
Let $E'$ be a set of edges, and let $(u,p(u))$ be a tree-edge in $E'$ with the property that $B(u)\setminus E'=\emptyset$. Let $E_1$ be the set of the tree-edges in $E'$ that are proper descendants of $(u,p(u))$, plus the back-edges in $E'$, and let $E_2$ be the set of the tree-edges in $E'$ that are not descendants of $(u,p(u))$, plus the back-edges in $E'$. Let $\mathcal{R}_1$ be a connectivity graph for $G\setminus E_1$, and let $\mathcal{R}_2$ be a connectivity graph for $G\setminus E_2$. Let $\mathcal{R}_1'$ be the graph that is derived from $\mathcal{R}_1$ by replacing every occurrence of $\bar{r}$ with $\bar{u}$. Then, $\mathcal{R}_1'\cup\mathcal{R}_2$ is a connectivity graph for $G\setminus E'$.
\end{lemma}
\begin{proof}
Let $(u_1,p(u_1)),\dots,(u_t,p(u_t))$ be the tree-edges in $E_1$, and let $(v_1,p(v_1)),\dots,(v_s,p(v_s))$ be the tree-edges in $E_2$. (We note that $u\notin\{u_1,\dots,u_t,v_1,\dots,v_s\}$, and both $t$ and $s$ may be $0$.) Then, we have that all vertices in $\{u_1,\dots,u_t\}$ are proper descendants of $u$, and no vertex from $\{v_1,\dots,v_s\}$ is a descendant of $u$. 
Now let $\mathcal{R}_1$, $\mathcal{R}_1'$, $\mathcal{R}_2$, and $\mathcal{R}$ be as in the statement of the lemma. We will show that $\mathcal{R}$ is a connectivity graph for $G\setminus E'$.

Since $V(\mathcal{R}_1)=\{\bar{u}_1,\dots,\bar{u}_t,\bar{r}\}$ and $V(\mathcal{R}_2)=\{\bar{v}_1,\dots,\bar{v}_s,\bar{r}\}$, we have that $V(\mathcal{R}_1)\cap V(\mathcal{R}_2)=\{\bar{r}\}$. This implies that $V(\mathcal{R}_1')\cap V(\mathcal{R}_2)=\emptyset$. Therefore, it is easy to see that no vertex from $\{\bar{u}_1,\dots,\bar{u}_t,\bar{u}\}$ is connected with a vertex from $\{\bar{v}_1,\dots,\bar{v}_s,\bar{r}\}$ in $\mathcal{R}$.

Now let $z$ be a vertex in $\{u_1,\dots,u_t,u\}$ and let $z'$ be a vertex in $\{v_1,\dots,v_s,r\}$. Let us suppose, for the sake of contradiction, that $z$ is connected with $z'$ in $G\setminus E'$. Then there is a path $P$ from $z$ to $z'$ in $G\setminus E'$. Lemma~\ref{lemma:path-to-non-related-nca} implies that $P$ passes from an ancestor $w$ of $\mathit{nca}\{z,z'\}$. We have that $z$ is a common descendant of $w$ and $u$. Thus, $u$ and $w$ are related as ancestor and descendant. But $u$ cannot be an ancestor of $w$, because otherwise it would be an ancestor of $z'$. Thus, we have that $u$ is a proper descendant of $w$. Since $P$ starts from a descendant of $u$ and reaches a proper ancestor of $u$, by Lemma~\ref{lemma:back-edge-or-tree-edge} we have that $P$ must either use a back-edge that leaps over $u$, or it passes from the tree-edge $(u,p(u))$. But both of these cases are impossible, since $B(u)\setminus E'=\emptyset$ and $(u,p(u))\in E'$. This shows that $z$ is not connected with $z'$ in $G\setminus E'$.    

Now let $z$ and $z'$ be two vertices in $\{u_1,\dots,u_t\}$. First, suppose that $z$ and $z'$ are connected in $G\setminus E'$. Then, since $E_1\subset E'$, we have that $z$ and $z'$ are connected in $G\setminus E_1$. This implies that $\bar{z}$ and $\bar{z}'$ are connected in $\mathcal{R}_1$, and therefore they are connected in $\mathcal{R}_1'$, and therefore they are connected in $\mathcal{R}$. Conversely, suppose that $\bar{z}$ and $\bar{z}'$ are connected in $\mathcal{R}$. Then, we have that $\bar{z}$ and $\bar{z}'$ are connected in $\mathcal{R}_1'$, and therefore they are connected in $\mathcal{R}_1$. This implies that there is a path $P$ from $z$ to $z'$ in $G\setminus E_1$. If $P$ does not use a tree-edge from $E_2'=E_2\cup\{(u,p(u))\}$, then we have that $P$ is a path in $G\setminus E'$, and this shows that $z$ is connected with $z'$ in $G\setminus E'$. Otherwise, let us suppose that $P$ uses a tree-edge from $E_2'$. Then we claim that $P$ has the form $P_1+(u,p(u))+Q+(p(u),u)+P_2$, where $P_1$ is a path from $z$ to $u$ that does not use tree-edges from $E_2'$, $Q$ is a path from $p(u)$ to $p(u)$ that may use tree-edges from $E_2'$, and $P_2$ is a path from $u$ to $z'$ that does not use tree-edges from $E_2'$ $(*)$. This implies that $P_1$ is a path from $z$ to $u$ in $G\setminus E'$, and $P_2$ is a path from $u$ to $z'$ in $G\setminus E'$. Thus, $P_1+P_2$ is a path from $z$ to $z'$ in $G\setminus E'$, and therefore $z$ is connected with $z'$ in $G\setminus E'$.  

Now we will prove $(*)$. Let $(v,p(v))$ be the first occurrence of a tree-edge from $E_2'$ in $P$. Then, since $P$ starts from $z$, by Lemma~\ref{lemma:path-to-non-related-nca} we have that $P$ passes from an ancestor $w$ of $\mathit{nca}\{z,v\}$. Then, we have that $z$ is a common descendant of $u$ and $w$, and therefore $u$ and $w$ are related as ancestor and descendant. Notice that we cannot have that $u$ is a proper ancestor of $w$, because this would imply that $u$ is a proper ancestor of $v$ (and we have that either $v\in\{v_1,\dots,v_s\}$, or $v=u$). Thus, $u$ is a descendant of $w$. Then, since $P$ starts from a descendant of $u$ and reaches a proper ancestor of $u$ (which is either $w$ or $p(u)$), by Lemma~\ref{lemma:back-edge-or-tree-edge} we have that $P$ must either use a back-edge from $B(u)$, or the tree-edge $(u,p(u))$, before it leaves the subtree of $u$. Since $P$ is a path in $G\setminus E_1$ and $E_1$ contains all the back-edges from $E'$ and $B(u)\setminus E'=\emptyset$, the only viable option is that $P$ uses the tree-edge $(u,p(u))$ in order to leave the subtree of $u$. Thus, there is a part $P_1$ of $P$ from $z$ to $u$ that lies entirely within the subtree of $u$. In particular, we have that $P_1$ does not use a tree-edge from $E_2'$. Similarly, by considering the last occurrence of a tree-edge from $E_2'$ in $P$, we can show that there is a part $P_2$ of $P$ from $u$ to $z'$ that does not use a tree-edge from $E_2'$. This establishes $(*)$.

Now let $z$ be a vertex in $\{u_1,\dots,u_t\}$. First, suppose that $u$ and $z$ are connected in $G\setminus E'$. Then there is a path $P$ from $u$ to $z$ in $G\setminus E'$. Since $E_1\subset E'$, we have that $P$ is a path from $u$ to $z$ in $G\setminus E_1$. We have that there is no tree-edge from $E_1$ on the tree-path $T[r,u]$. Thus, $T[r,u]+P$ is a path from $r$ to $z$ in $G\setminus E_1$, and therefore $r$ is connected with $z$ in $G\setminus E_1$. This implies that $\bar{r}$ is connected with $\bar{z}$ in $\mathcal{R}_1$, and therefore $\bar{u}$ is connected with $\bar{z}$ in $\mathcal{R}_1'$. Thus, $\bar{u}$ is connected with $\bar{z}$ in $\mathcal{R}$. Conversely, suppose that $\bar{u}$ is connected with $\bar{z}$ in $\mathcal{R}$. This implies that $\bar{u}$ is connected with $\bar{z}$ in $\mathcal{R}_1'$, and therefore $\bar{r}$ is connected with $\bar{z}$ in $\mathcal{R}_1$.
This implies that $r$ is connected with $z$ in $G\setminus E_1$. Thus, there is a path $P$ from $r$ to $z$ in $G\setminus E_1$. Since $r$ is a proper ancestor of $u$ and $z$ is a descendant of $u$, Lemma~\ref{lemma:back-edge-or-tree-edge} implies that $P$ must either use a back-edge from $B(u)$, or the tree-edge $(p(u),u)$, and then it finally lies entirely within the subtree of $u$. Thus, since $E_1$ contains all the back-edges from $E'$ and $B(u)\setminus E'=\emptyset$, we have that $P$ eventually passes from the tree-edge $(p(u),u)$, and stays within the subtree of $u$. Thus, we have that $u$ is connected with $z$ in $G\setminus E_1$, through a path $P'$ that lies entirely within the subtree of $u$. This implies that $P'$ does not use tree-edges from $E_2$. Thus, $P'$ is a path from $u$ to $z$ in $G\setminus E'$, and therefore $u$ is connected with $z$ in $G\setminus E'$.

Now let $z$ and $z'$ be two vertices in $\{v_1,\dots,v_s,r\}$. First, suppose that $z$ and $z'$ are connected in $G\setminus E'$. Then, since $E_2\subset E'$, we have that $z$ and $z'$ are connected in $G\setminus E_2$. This implies that $\bar{z}$ is connected with $\bar{z}'$ in $\mathcal{R}_2$. Thus, we have that $\bar{z}$ is connected with $\bar{z}'$ in $\mathcal{R}$. Conversely, suppose that $\bar{z}$ and $\bar{z}'$ are connected in $\mathcal{R}$. Then we have that $\bar{z}$ and $\bar{z}'$ are connected in $\mathcal{R}_2$. This implies that $z$ and $z'$ are connected in $G\setminus E_2$. Thus, there is a path $P$ from $z$ to $z'$ in $G\setminus E_2$. If $P$ does not use any tree-edge from $E_1'=E_1\cup\{(u,p(u))\}$, then $P$ is a path in $G\setminus E'$, and therefore $z$ and $z'$ are connected in $G\setminus E'$. So let us assume that $P$ uses a tree-edge from $E_1'$. Then we claim that $P$ has the form $P_1+Q+P_2$, where $P_1$ is a path from $z$ to $p(u)$ that does not use tree-edges from $E_1'$, $Q$ is a path from $p(u)$ to $p(u)$ (that uses tree-edges from $E_1$), and $P_2$ is a path from $p(u)$ to $z'$ that does not use tree-edges from $E_1'$ $(**)$. This implies that $P_1+P_2$ is a path from $z$ to $z'$ in $G\setminus E'$, and therefore $z$ and $z'$ are connected in $G\setminus E'$.

Now we will prove $(**)$. Since $P$ uses a tree-edge from $E_1'$, we may consider the first occurrence $(u',p(u'))$ of an edge from $E_1'$ that is used by $P$. Then, we have that $u'$ is a descendant of $u$. Since $P$ starts from $z$, Lemma~\ref{lemma:path-to-non-related-nca} implies that $P$ contains a subpath from $z$ to $u'$ that passes from an ancestor $w$ of $\mathit{nca}\{z,u'\}$, and let $w$ be the first vertex visited by $P$ with this property. Let $P'$ be the initial part of $P$ from $z$ to the first occurrence of $w$. Then we have that $P'$ does not use any tree-edge from $E_1'$. Since $u'$ is a descendant of $u$ but $z$ is not, we have that $\mathit{nca}\{z,u'\}$ is a proper ancestor of $u$, and therefore $w$ is a proper ancestor of $u$. Then, Lemma~\ref{lemma:back-edge-or-tree-edge} implies that the part of $P$ from $w$ to $u'$ must either use the tree-edge $(u,p(u))$, or a back-edge that leaps over $u$. Since $P$ is a path in $G\setminus E_2$ and $E_2$ contains all the back-edges from $E'$ and $B(u)\setminus E'=\emptyset$, we infer that the part of $P$ from $w$ to $u'$ must use the tree-edge $(u,p(u))$ in order to enter the subtree of $u$. Now let $P''$ be the part of $P$ from the first occurrence of $w$ to the first occurrence of $p(u)$ that is followed by $(u,p(u))$. Then, we have that $P_1=P'+P''$ is a path from $z$ to $p(u)$ that does not use any tree-edge from $E_1'$. Similarly, since $P$ ends in $z'$, we can show that the final part of $P$ is a subpath $P_2$ that starts from $p(u)$ and ends in $z'$. We let $Q$ denote the middle part of $P$ (i.e., the one between $P_1$ and $P_2$), and this establishes $(**)$.

Thus, we have shown that, for any two vertices $z$ and $z'$ in $\{u_1,\dots,u_t,u,v_1,\dots,v_s,r\}$, we have that $z$ is connected with $z'$ in $G\setminus E'$ if and only if $\bar{z}$ is connected with $\bar{z}'$ in $\mathcal{R}$. This means that $\mathcal{R}$ is a connectivity graph for $G\setminus E'$. 
\end{proof}

Given a set of edges $E'$ with $|E'|\leq 4$, we will show how to construct a connectivity graph $\mathcal{R}$ for $G\setminus E'$. We distinguish five cases, depending on the number of tree-edges in $E'$. Whenever possible, we use Lemmata~\ref{lemma:4conn-edge-connected}, \ref{lemma:4conn-non-related-edges} and \ref{lemma:4conn-no-back-edge}, in order to revert to previous cases. This simplifies the analysis a lot, because the number of cases that may appear is very large. If $A$ and $B$ are two subtrees of $T$, we use $(A,B)$ to denote the set of the back-edges that connect $A$ and $B$.

\subsection{$E'$ contains zero tree-edges}
In this case, $T\setminus E'$ is connected, and therefore $G\setminus E'$ is also connected. Thus, we know that every connectivity query in $G\setminus E'$ is positive. (The connectivity graph $\mathcal{R}$ of $G\setminus E'$ consists of a single vertex $\bar{r}$.)

\subsection{$E'$ contains one tree-edge}
Let $(u,p(u))$ be the tree-edge contained in $E'$. The connected components of $T\setminus\{(u,p(u))\}$ are $A=T(u)$ and $B=T(r)\setminus T(u)$. Thus, $\mathcal{R}$ consists of the vertices $\{\bar{u},\bar{r}\}$. We can see that $A$ is connected with $B$ in $G\setminus E'$ if and only if there is a back-edge in $B(u)\setminus{E'}$. Since $E'$ contains at most three back-edges, it is sufficient to have collected at most four back-edges of $B(u)$ in a set $B_4(u)$, and then check whether $B_4(u)\setminus E'=\emptyset$. We may pick the four lowest $\mathit{low}$-edges of $u$ in order to build $B_4(u)$, since these are easy to compute. (I.e., $B_4(u)$ consists of the non-null $\mathit{low}_i$-edges of $u$, for all $i\in\{1,2,3,4\}$.) If $B_4(u)\setminus E'\neq\emptyset$, then we add the edge $(\bar{u},\bar{r})$ to $\mathcal{R}$. Otherwise, $\bar{u}$ is disconnected from $\bar{r}$ in $\mathcal{R}$. 


\subsection{$E'$ contains two tree-edges}
Let $(u,p(u))$ and $(v,p(v))$ be the two tree-edges contained in $E'$. Then we have that $\mathcal{R}$ consists of the vertices $\{\bar{u},\bar{v},\bar{r}\}$. Let us assume w.l.o.g. that $u>v$. Suppose that $u$ and $v$ are not related as ancestor and descendant. Then, we set $E_1=E'\setminus\{(u,p(u))\}$ and $E_2=E'\setminus\{(v,p(v))\}$. Each of the sets $E_1$ and $E_2$ contains only one tree-edge, and therefore we can revert to the previous case in order to build a connectivity graph $\mathcal{R}_1$ and $\mathcal{R}_2$ for $G\setminus E_1$ and $G\setminus E_2$, respectively. Then, by Lemma~\ref{lemma:4conn-non-related-edges} we have that $\mathcal{R}_1\cup\mathcal{R}_2$ is a connectivity graph for $G\setminus E'$.
So let us assume that the two tree-edges in $E'$ are related as ancestor and descendant. Since $u>v$, this implies that $u$ is a descendant of $v$.  
Let $A=T(u)$, $B=T(v)\setminus T(u)$, and $C=T(r)\setminus T(v)$. 

First, we will check whether there is a back-edge from $A$ to $B$ in $G\setminus E'$. Since $E'$ contains at most two back-edges, we have that $(A,B)\setminus E'\neq\emptyset$ if and only if at least one of the three highest $\mathit{high}$-edges of $u$ is not in $E'$ and has its lower endpoint in $B$. In other words, $(A,B)\setminus E'\neq\emptyset$ if and only if the $\mathit{high}_i$-edge of $u$ (exists, and) is not in $E'$, and $\mathit{high}_i(u)\in B$, for some $i\in\{1,2,3\}$. If that is the case, then we know that $u$ is connected with $p(u)$ in $G\setminus E'$. Thus, we add the edge $(\bar{u},\bar{v})$ to $\mathcal{R}$. Then we can set $E'\leftarrow E'\setminus\{(u,p(u))\}$, and revert to the previous case (where $E'$ contains one tree-edge), according to Lemma~\ref{lemma:4conn-edge-connected}.

Otherwise, we have that all the back-edges in $B(u)\setminus E'$ (if there are any), have their lower endpoint in $C$. In this case, we have that there is a back-edge from $A$ to $C$ if and only if $B(u)\setminus E'\neq\emptyset$. Since $E'$ contains at most two back-edges, we have that $B(u)\setminus E'\neq\emptyset$ if and only if at least one of the $\mathit{low}_i$-edges of $u$, for $i\in\{1,2,3\}$, (exists and) does not lie in $E'$. If we have that $B(u)\setminus E'=\emptyset$, then we know that $A$ is not connected with the rest of the graph in $G\setminus E'$. Thus, we can set $E'\leftarrow E'\setminus\{(u,p(u))\}$ and revert to the previous case (where $E'$ contains one tree-edge), according to Lemma~\ref{lemma:4conn-no-back-edge}. 

Thus, let us assume that there is a back-edge between $A$ and $C$ in $G\setminus E'$. Then we add the edge $(\bar{u},\bar{r})$ to $\mathcal{R}$. 
Now it remains to check whether there is a back-edge from $B$ to $C$ in $G\setminus E'$. Since we assume that there is no back-edge from $A$ to $B$ in $G\setminus E'$, we can distinguish three possibilities: either $(1)$ no edge from $E'$ is in $(A,B)$, or $(2)$ precisely one edge $e$ from $E'$ is in $(A,B)$, or $(3)$ both the back-edges $e$ and $e'$ from $E'$ are in $(A,B)$. In case $(1)$, we have that $B(u)\subseteq B(v)$. Thus, there is a back-edge from $B$ to $C$ in $G\setminus E'$ if and only if $\mathit{bcount}(v)-\mathit{bcount}(u)>|E'\cap(B(v)\setminus B(u))|$. We note that it is easy to compute $|E'\cap (B(v)\setminus B(u))|$: we simply count how many back-edges from $E'$ leap over $v$ but not over $u$. Now, in case $(2)$ we have $B(u)\setminus\{e\}\subseteq B(v)$. Thus, there is a back-edge from $B$ to $C$ in $G\setminus E'$ if and only if $\mathit{bcount}(v)-\mathit{bcount}(u)+1>|E'\cap(B(v)\setminus B(u))|$. Finally, in case $(3)$ we have $B(u)\setminus\{e,e'\}\subseteq B(v)$, and $E'$ contains no back-edge from $B(v)\setminus B(u)$. Thus, there is a back-edge from $B$ to $C$ in $G\setminus E'$ if and only if $\mathit{bcount}(v)-\mathit{bcount}(u)+2>0$. We note that it is easy to determine which case $(1)-(3)$ applies: we simply count how many back-edges from $E'$ (if $E'$ contains back-edges) leap over $u$, but not over $v$. If we have determined that there is a back-edge from $B$ to $C$ in $G\setminus E'$, then we add the edge $(\bar{v},\bar{r})$ to $\mathcal{R}$. 


\subsection{$E'$ contains three tree-edges}
Let $(u,p(u))$, $(v,p(v))$  and $(w,p(w))$ be the three tree-edges contained in $E'$. Then we have that $\mathcal{R}$ consists of the vertices $\{\bar{u},\bar{v},\bar{w},\bar{r}\}$. Let us assume w.l.o.g. that $u>v>w$. We may also assume that one of the three tree-edges in $E'$ is an ancestor of the other two. Otherwise, we can use Lemma~\ref{lemma:4conn-non-related-edges} in order to revert to the previous case (where $E'$ contains two tree-edges), because then there are at least two tree-edges in $E'$ with the property that no tree-edge in $E'$ is a proper ancestor of them.

Thus, since one of the three tree-edges in $E'$ is an ancestor of the other two and $w<v<u$, we have that $w$ is a common ancestor of $u$ and $v$. Now there are two cases to consider: either $(1)$ $u$ and $v$ are not related as ancestor and descendant, or $(2)$ $u$ and $v$ are related as ancestor and descendant. Since $u>v$, in case $(2)$ we have that $v$ is an ancestor of $u$. 

\subsubsection{$u$ and $v$ are not related as ancestor and descendant}
Let $A=T(u)$, $B=T(v)$, $C=T(w)\setminus(T(u)\cup T(v))$, and $D=T(r)\setminus T(w)$.

If there is no back-edge in $G\setminus E'$ that connects $A$, or $B$, or $A\cup B\cup C$, with the rest of the graph in $G\setminus E'$, then we can use Lemma~\ref{lemma:4conn-no-back-edge} in order to revert to the previous case (where $E'$ contains two tree-edges). Every one of those conditions is equivalent to [$\mathit{bcount}(u)=0$ or $B(u)=\{e\}$, where $e$ is the back-edge in $E'$], or [$\mathit{bcount}(v)=0$ or $B(v)=\{e\}$, where $e$ is the back-edge in $E'$], or [$\mathit{bcount}(w)=0$ or $B(w)=\{e\}$, where $e$ is the back-edge in $E'$], respectively, and so we can check them easily in constant time.

Otherwise, if there is a back-edge that connects $A$ with $C$ in $G\setminus E'$, or a back-edge that connects $B$ with $C$ in $G\setminus E'$, then we add the edge $(\bar{u},\bar{w})$, or $(\bar{v},\bar{w})$, respectively, to $\mathcal{R}$. Then we set $E'\leftarrow E'\setminus\{(u,p(u))\}$, or $E'\leftarrow E'\setminus\{(v,p(v))\}$, respectively, and we revert to the previous case (where $E'$ contains two tree-edges) according to Lemma~\ref{lemma:4conn-edge-connected}. Notice that there is a back-edge that connects $A$ with $C$ in $G\setminus E'$ if and only if: either $(i)$ $\mathit{high}_1(u)\in C$ and the $\mathit{high}_1$-edge of $u$ is not in $E'$, or $(ii)$ $\mathit{high}_2(u)\in C$ and the $\mathit{high}_2$-edge of $u$ is not in $E'$. Similarly, we can easily check whether there is a back-edge that connects $B$ with $C$ in $G\setminus E'$. 

So let us assume that none of the above is true. This means that, in $G\setminus E'$, there is a back-edge from $A$ to $D$, a back-edge from $B$ to $D$, no back-edge from $A$ to $C$, and no back-edge from $B$ to $C$. Then, we add the edges $(\bar{u},\bar{r})$ and $(\bar{v},\bar{r})$ to $\mathcal{R}$. Now it remains to determine whether there is a back-edge from $C$ to $D$ in $G\setminus E'$. Suppose first that $E'$ contains a back-edge $e$, that connects either $A$ and $C$, or $B$ and $C$. Then, there is a back-edge from $C$ to $D$ in $G\setminus E'$ if and only if $\mathit{bcount}(w)>\mathit{bcount}(u)+\mathit{bcount}(v)-1$. (This is because $B(u)\sqcup B(v)\subseteq B(w)\sqcup\{e\}$ in this case.) Thus, if this condition is satisfied, then we add the edge $(\bar{w},\bar{r})$ to $\mathcal{R}$. Now, let us assume that $E'$ contains a back-edge from $C$ to $D$. Then, there is a back-edge from $C$ to $D$ in $G\setminus E'$ if and only if $\mathit{bcount}(w)-1>\mathit{bcount}(u)+\mathit{bcount}(v)$. Thus, if this condition is satisfied, then we add the edge $(\bar{w},\bar{r})$ to $\mathcal{R}$. Finally, let us assume that the back-edge in $E'$ (if it contains a back-edge) does not connect $A$ and $C$, or $B$ and $C$, or $C$ and $D$. Then, there is a back-edge from $C$ to $D$ in $G\setminus E'$ if and only if $\mathit{bcount}(w)>\mathit{bcount}(u)+\mathit{bcount}(v)$. Thus, if this condition is satisfied, then we add the edge $(\bar{w},\bar{r})$ to $\mathcal{R}$.  

\subsubsection{$v$ is an ancestor of $u$}
Let $A=T(u)$, $B=T(v)\setminus T(u)$, $C=T(w)\setminus T(v)$, and $D=T(r)\setminus T(w)$.

If there is no back-edge in $G\setminus E'$ that connects $A$, or $A\cup B$, or $A\cup B\cup C$, with the rest of the graph in $G\setminus E'$, then we can use Lemma~\ref{lemma:4conn-no-back-edge} in order to revert to the previous case (where $E'$ contains either one or two tree-edges). Every one of those conditions is equivalent to [$\mathit{bcount}(u)=0$ or $B(u)=\{e\}$, where $e$ is the back-edge in $E'$], or [$\mathit{bcount}(v)=0$ or $B(v)=\{e\}$, where $e$ is the back-edge in $E'$], or [$\mathit{bcount}(w)=0$ or $B(w)=\{e\}$, where $e$ is the back-edge in $E'$], respectively, and so we can check them easily in constant time.

If in $G\setminus E'$ there is a back-edge that connects $A$ and $B$, then we add the edge $(\bar{u},\bar{v})$ to $\mathcal{R}$. Then we set $E'\leftarrow E'\setminus\{(u,p(u))\}$, and we revert to the previous case (where $E'$ contains two tree-edges) according to Lemma~\ref{lemma:4conn-edge-connected}. Notice that there is a back-edge in $G\setminus E'$ that connects $A$ and $B$ if and only if: either $(i)$ $\mathit{high}_1(u)\in B$ and the $\mathit{high}_1$-edge of $u$ is not in $E'$, or $(ii)$ $\mathit{high}_2(u)\in B$ and the $\mathit{high}_2$-edge of $u$ is not in $E'$. Thus, we can easily check this condition in constant time.

So let us assume that there is no back-edge that connects $A$ and $B$ in $G\setminus E'$. If in $G\setminus E'$ there is a back-edge that connects $A$ and $C$, and a back-edge that connects $A$ and $D$, then the parts $C$ and $D$ are connected in $G\setminus E'$ through the mediation of $A$. Thus, we add the edge $(\bar{w},\bar{r})$ to $\mathcal{R}$, we set $E'\leftarrow E'\setminus\{(w,p(w))\}$, and we revert to the previous case (where $E'$ contains two tree-edges) according to Lemma~\ref{lemma:4conn-edge-connected}. Since there is no back-edge that connects $A$ and $B$ in $G\setminus E'$, notice that there is a back-edge in $G\setminus E'$ that connects $A$ and $C$ if and only if: either $(i)$ $\mathit{high}_1(u)\in C$ and the $\mathit{high}_1$-edge of $u$ is not in $E'$, or $(ii)$ $\mathit{high}_2(u)\in C$ and the $\mathit{high}_2$-edge of $u$ is not in $E'$. Also, there is a back-edge in $G\setminus E'$ that connects $A$ and $D$ if and only if: either $(i)$ $\mathit{low}_1(u)\in D$ and the $\mathit{low}_1$-edge of $u$ is not in $E'$, or $(ii)$ $\mathit{low}_2(u)\in D$. Thus, we can easily check those conditions in constant time.

So let us assume that none of the above is true. Thus, there are two cases to consider in $G\setminus E'$: either $(1)$ there is a back-edge from $A$ to $C$, but no back-edge from $A$ to $D$, or $(2)$ there is a back-edge from $A$ to $D$, but no back-edge from $A$ to $C$. 

Let us consider case $(1)$ first. Then, we add the edge $(\bar{u},\bar{w})$ to $\mathcal{R}$. First, we will determine whether there is a back-edge from $B$ to $D$ in $G\setminus E'$. Since there is no back-edge from $A$ to $D$ in $G\setminus E'$, notice that there is a back-edge from $B$ to $D$ in $G\setminus E'$ if and only if: either $\mathit{low}_1(v)\in D$ and the $\mathit{low}_1$-edge of $v$ is not in $E'$, or $\mathit{low}_2(v)\in D$ and the $\mathit{low}_2$-edge of $v$ is not in $E'$. Thus, we can check in constant time whether there is a back-edge from $B$ to $D$ in $G\setminus E'$. If we have determined that there is no back-edge from $B$ to $D$ in $G\setminus E'$, then, since we have supposed that there is a back-edge in $B(w)\setminus E'$, we have that there is a back-edge from $C$ to $D$ in $G\setminus E'$ (since $(A,D)\setminus E'=(B,D)\setminus E'=\emptyset$). This implies that $w$ remains connected with $p(w)$ in $G\setminus E'$. Thus, we add the edge $(\bar{w},\bar{r})$ to $\mathcal{R}$, we set $E'\leftarrow E'\setminus\{(w,p(w))\}$, and we revert to the previous case (where $E'$ contains two tree-edges) according to Lemma~\ref{lemma:4conn-edge-connected}. 
So let us suppose that there is a back-edge from $B$ to $D$ in $G\setminus E'$.  Then, we add the edge $(\bar{v},\bar{r})$ to $\mathcal{R}$. Now it remains to determine if $(B,C)\setminus E'\neq\emptyset$ or $(C,D)\setminus E'\neq\emptyset$. Notice that either of those cases implies that $G\setminus E'$ is connected.

We have $B(u)=(A,B)\cup (A,C)\cup (A,D)$, $B(v)=(A,C)\cup (A,D)\cup (B,C)\cup (B,D)$ and $B(w)=(A,D)\cup (B,D)\cup (C,D)$. Thus, we have $N=\mathit{bcount}(w)-\mathit{bcount}(v)+\mathit{bcount}(u)=|(C,D)|-|(B,C)|+|(A,B)|+|(A,D)|$ and $s=\mathit{SumAnc}(w)-\mathit{SumAnc}(v)+\mathit{SumAnc}(u)=\mathit{SumAnc}((C,D))-\mathit{SumAnc}((B,C))+\mathit{SumAnc}((A,B))+\mathit{SumAnc}((A,D))$, where we let $\mathit{SumAnc}(S)$, for a set $S$ of back-edges, be the sum of the lower endpoints of the back-edges in $S$. 

Here we distinguish two cases: either $(1.1)$ the back-edge in $E'$ (if it exists) does not lie in $(A,B)\cup (A,D)$, or $(1.2)$ the back-edge $e$ in $E'$ (exists and) lies in $(A,B)\cup (A,D)$. (It is easy to determine in constant time which case applies.)

First, let us consider case $(1.1)$. Then we have $N=|(C,D)|-|(B,C)|$ and $s=\mathit{SumAnc}((C,D))-\mathit{SumAnc}((B,C))$. Here we distinguish two cases: either $(1.1.1)$ the back-edge in $E'$ (if it exists) does not lie in $(C,D)\cup (B,C)$, or $(1.1.2)$ $E'$ contains a back-edge $e$ that lies in $(C,D)\cup (B,C)$. (Again, it is easy to determine in constant time which case applies.) So let us consider case $(1.1.1)$ first. Thus, if we have $N\neq 0$, then at least one of $(C,D)$ and $(B,C)$ is not empty, and therefore $G\setminus E'$ is connected. Thus, it is sufficient to add one more edge to $\mathcal{R}$ in order to make it connected (e.g., we may add $(\bar{w},\bar{r})$). Otherwise, suppose that $N=0$. Then we have $|(C,D)|=|(B,C)|$. Since the lower endpoint of every back-edge in $(C,D)$ is lower than the lower endpoint of every back-edge in $(B,C)$, this implies that $s=\mathit{SumAnc}((C,D))-\mathit{SumAnc}((B,C))<0$ if and only if $(C,D)\neq\emptyset$ (and $(B,C)\neq\emptyset$). Thus, it is sufficient to add e.g. the edge $(\bar{w},\bar{r})$ to $\mathcal{R}$ if and only if $s<0$.

Now let us consider case $(1.1.2)$. Let $z$ be the lower endpoint of $e$. First, suppose that $e\in (C,D)$. Thus, if $N>1$, then $(C,D)\setminus E'$ is not empty, and therefore $G\setminus E'$ is connected. Similarly, if $N<1$, then $(B,C)$ is not empty, and therefore $G\setminus E'$ is connected. Thus, in those cases, it is sufficient to add one more edge to $\mathcal{R}$ in order to make it connected (e.g., we may add $(\bar{w},\bar{r})$). Otherwise, suppose that $N=1$. Then we have $|(C,D)|=|(B,C)|+1$. Thus, if $|(B,C)|=0$, then we have $(C,D)=\{e\}$, and therefore $s$ coincides with $z$. Otherwise, if $|(B,C)|>0$, then, since the lower endpoint of every back-edge in $(C,D)$ is lower than the lower endpoint of every back-edge in $(B,C)$, we have that $s=\mathit{SumAnc}((C,D))-\mathit{SumAnc}((B,C))<z$. Thus, it is sufficient to add e.g. the edge $(\bar{w},\bar{r})$ to $\mathcal{R}$ if and only if $s<z$. Now let us suppose that $e\in (B,C)$. Thus, if $N>-1$, then $(C,D)$ is not empty, and therefore $G\setminus E'$ is connected. Similarly, if $N<-1$, then $(B,C)\setminus E'$ is not empty, and therefore $G\setminus E'$ is connected. Thus, in those cases, it is sufficient to add one more edge to $\mathcal{R}$ in order to make it connected (e.g., we may add $(\bar{w},\bar{r})$). Otherwise, suppose that $N=-1$. Then we have $|(C,D)|+1=|(B,C)|$. Thus, if $|(C,D)|=0$, then we have $(B,C)=\{e\}$, and therefore $s$ coincides with $-z$. Otherwise, if $|(C,D)|>0$, then, since the lower endpoint of every back-edge in $(C,D)$ is lower than the lower endpoint of every back-edge in $(B,C)$, we have that $s=\mathit{SumAnc}((C,D))-\mathit{SumAnc}((B,C))<-z$. Thus, it is sufficient to add e.g. the edge $(\bar{w},\bar{r})$ to $\mathcal{R}$ if and only if $s<-z$. 

Now let us consider case $(1.2)$. Then, since there is no back-edge from $A$ to $B$ in $G\setminus E'$, and no back-edge from $A$ to $D$ in $G\setminus E'$, we have that one of $(A,B)$ and $(A,D)$ coincides with $\{e\}$, and the other is empty. Then, we have $N=|(C,D)|-|(B,C)|+1$ and $s=\mathit{SumAnc}((C,D))-\mathit{SumAnc}((B,C))+\mathit{SumAnc}(\{e\}$). Thus, if we have $N\neq 1$, then at least one of $(C,D)$ and $(B,C)$ is not empty, and therefore $G\setminus E'$ is connected. Thus, it is sufficient to add one more edge to $\mathcal{R}$ in order to make it connected (e.g., we may add $(\bar{w},\bar{r})$). Otherwise, suppose that $N=1$. Then we have $|(C,D)|-|(B,C)|=0$. Since the lower endpoint of every back-edge in $(C,D)$ is lower than the lower endpoint of every back-edge in $(B,C)$, this implies that $\mathit{SumAnc}((C,D))-\mathit{SumAnc}((B,C))<0$ if and only if $(C,D)\neq\emptyset$ (and $(B,C)\neq\emptyset$). Thus, it is sufficient to add e.g. the edge $(\bar{w},\bar{r})$ to $\mathcal{R}$ if and only if $s<\mathit{SumAnc}(\{e\})$. (Notice that it is easy to compute $\mathit{SumAnc}(\{e\})$: this is just the (DFS number of the) lower endpoint of $e$.)

Now let us consider case $(2)$. Then, we add the edge $(\bar{u},\bar{r})$ to $\mathcal{R}$. Now, if there is a back-edge that connects $B$ and $C$ in $G\setminus E'$, then we add the edge $(\bar{v},\bar{w})$ to $\mathcal{R}$, we set $E'\leftarrow E'\setminus\{(v,p(v)\}$, and we revert to the previous case (where $E'$ contains two tree-edges) according to Lemma~\ref{lemma:4conn-edge-connected}. Notice that, since there is no back-edge from $A$ to $C$ in $G\setminus E'$, this condition is equivalent to: either $\mathit{high}_1(v)\in C$ and the $\mathit{high}_1$-edge of $v$ is not in $E'$, or $\mathit{high}_2(v)\in C$ and the $\mathit{high}_1$-edge of $v$ is not in $E'$. Thus, we can easily check whether there is a back-edge that connects $B$ and $C$ in $G\setminus E'$, in constant time. 

Now let us assume that there is no back-edge that connects $B$ and $C$ in $G\setminus E'$. Then, if there is a back-edge that connects $B$ and $D$ in $G\setminus E'$ $(*)$, then we have that $A$ and $B$ remain connected in $G\setminus E'$ through the mediation of $D$. Thus, we add the edge $(\bar{u},\bar{v})$ to $\mathcal{R}$, we set $E'\leftarrow E'\setminus\{(u,p(u))\}$, and we revert to the previous case (where $E'$ contains two tree-edges) according to Lemma~\ref{lemma:4conn-edge-connected}. In order to check condition $(*)$, i.e., whether there is a back-edge that connects $B$ and $D$ in $G\setminus E'$, we distinguish the following cases. First, suppose that the back-edge in $E'$ (if it exists), is neither in $B(u)$ nor in $B(v)$. Then, since all the back-edges in $B(u)$ connect $A$ and $D$, and since all the back-edges in $B(v)$ connect either $A$ and $D$ or $B$ and $D$, we have that $(*)$ is true if and only if $\mathit{bcount}(v)>\mathit{bcount}(u)$. Now, suppose that the back-edge $e$ in $E'$ is in $B(u)$, but not in $B(v)$. Then, all the back-edges in $B(u)$, except $e$, connect $A$ and $D$, and all back-edges in $B(v)$ connect either $A$ and $D$ or $B$ and $D$. Thus, we have that $(*)$ is true if and only if $\mathit{bcount}(v)\geq\mathit{bcount}(u)$. Now, suppose that the back-edge $e$ in $E'$ is in both $B(u)$ and $B(v)$. Then, we have that all the back-edges in $B(u)$, except possibly $e$, connect $A$ and $D$, and all the back-edges in $B(v)$, except possibly $e$, connect $A$ and $D$ or $B$ and $D$. Thus, we have that $(*)$ is true if and only if $\mathit{bcount}(v)>\mathit{bcount}(u)$. Finally, suppose that the back-edge $e$ in $E'$ is in $B(v)$, but not in $B(u)$. Then, all the back-edges in $B(u)$ connect $A$ and $D$, and all the back-edges in $B(v)$, except possibly $e$, connect $A$ and $D$ or $B$ and $D$. Thus, we have that $(*)$ is true if and only if $\mathit{bcount}(v)>\mathit{bcount}(u)+1$. Thus, we can easily check whether there is a back-edge that connects $B$ and $D$ in $G\setminus E'$, in constant time. 

Finally, suppose that neither of the above two is the case (i.e., we have $(B,C)\setminus E'=\emptyset$ and $(B,D)\setminus E'=\emptyset$). Then we only have to check whether there is a back-edge in $G\setminus E'$ that connects $C$ and $D$. We distinguish the following cases: either $(2.1)$ $E'$ contains a back-edge $e\in (C,D)$, or $(2.2)$ $E'$ contains a back-edge $e\in (A,C)\cup(B,C)$, or $(2.3)$ none of the previous is true. (Notice that it is easy to determine in constant time which case holds.) In case $(2.1)$ we have $B(w)=(A,D)\cup((C,D)\setminus\{e\})\cup\{e\}$. Thus, there is a back-edge from $C$ to $D$ in $G\setminus E'$ if and only if $\mathit{bcount}(w)>\mathit{bcount}(u)+1$. In case $(2.2)$ we have $B(w)=(B(v)\setminus\{e\})\cup (C,D)$, and $e\in B(v)\setminus (C,D)$. Thus, there is a back-edge from $C$ to $D$ in $G\setminus E'$ if and only if $\mathit{bcount}(w)>\mathit{bcount}(v)-1$. In case $(2.3)$ we have $B(w)=B(v)\cup (C,D)$ and $e\notin (C,D)$. Thus, there is a back-edge from $C$ to $D$ in $G\setminus E'$ if and only if $\mathit{bcount}(w)>\mathit{bcount}(v)$.
Thus, in either of those cases, if we determine that there is a back-edge from $C$ to $D$ in $G\setminus E'$, then we add the edge $(\bar{w},\bar{r})$ to $\mathcal{R}$.

\subsection{$E'$ contains four tree-edges}
Let $(u,p(u))$, $(v,p(v))$, $(w,p(w))$ and $(z,p(z))$ be the tree-edges that are contained in $E'$. Then we have that $\mathcal{R}$ consists of the vertices $\{\bar{u},\bar{v},\bar{w},\bar{z},\bar{r}\}$. We may assume w.l.o.g. that $u>v>w>z$. Suppose that there are at least two distinct edges $(u',p(u'))$ and $(v',p(v'))$ in $E'$ with the property that no edge in $E'$ is a proper ancestor of them. Let $E_1$ be the subset of $E'$ that consists of the descendants of $(u',p(u'))$, and let $E_2$ be the subset of $E'$ that consists of the non-descendants of $(u',p(u'))$. Then, we have that $(v',p(v'))\notin E_1$, and $(u',p(u'))\notin E_2$. Thus, we can construct connectivity graphs $\mathcal{R}_1$ and $\mathcal{R}_2$ for $G\setminus E_1$ and $G\setminus E_2$, respectively, by reverting to the previous cases. Then, by Lemma~\ref{lemma:4conn-non-related-edges}, we have that $\mathcal{R}_1\cup\mathcal{R}_2$ is a connectivity graph for $G\setminus E'$.

Thus, we may assume that one of the tree-edges in $E'$ is an ancestor of the other three, because otherwise we can construct the graph $\mathcal{R}$ by reverting to the previous cases. Then, since $u>v>w>z$, we have that $z$ is a common ancestor of all vertices in $\{u,v,w\}$. Now, there are four cases to consider: either $(1)$ no two vertices in $\{u,v,w\}$ are related as ancestor and descendant, or $(2)$ only two among $\{u,v,w\}$ are related as ancestor and descendant, or $(3)$ one of $\{u,v,w\}$ is an ancestor of the other two, but the other two are not related as ancestor and descendant, or $(4)$ every two vertices in $\{u,v,w\}$ are related as ancestor and descendant.

In case $(2)$, we can either have that $w$ is an ancestor of $v$ (and $u$ is not related as ancestor and descendant with $w$ and $v$), or $v$ is an ancestor of $u$ (and $w$ is not related as ancestor and descendant with $u$ and $v$). Both of these cases can be handled with essentially the same argument (it is just that the roles of $w$, $v$ and $u$ in the first case are exchanged with $v$, $u$ and $w$, respectively, in the second case), and so we will assume w.l.o.g. that $v$ is an ancestor of $u$ in this case. In cases $(3)$ and $(4)$, we have that the ancestry relation between the vertices in $\{u,v,w\}$ is fixed by the assumption $u>v>w$. Thus, in case $(3)$ we have that $w$ is an ancestor of both $v$ and $u$ (but $u,v$ are not related as ancestor and descendant), and in case $(4)$ we have that $w$ is an ancestor of $v$, and $v$ is an ancestor of $u$. 

In any case, notice that there are no back-edges in $E'$ (since we have assumed that $|E'|\leq 4$, and therefore $E'$ consists of four tree-edges). 

\subsubsection{No two vertices in $\{u,v,w\}$ are related as ancestor and descendant}
Let $A=T(u)$, $B=T(v)$, $C=T(w)$, $D=T(z)\setminus(T(u)\cup T(v)\cup T(w))$ and $E=T(r)\setminus T(z)$.

If we have that either $\mathit{high}_1(u)=\bot$, or $\mathit{high}_1(v)=\bot$, or $\mathit{high}_1(w)=\bot$, then there is no back-edge in $G\setminus E'$ that connects $A$, or $B$, or $C$, respectively, with the rest of the graph. Therefore, we can use Lemma~\ref{lemma:4conn-no-back-edge} in order to revert to the previous case (where $E'$ contains three tree-edges).

Otherwise, if we have that either $\mathit{high}_1(u)\in D$, or $\mathit{high}_1(v)\in D$, or $\mathit{high}_1(w)\in D$, then we have that $A$ is connected with $D$, or $B$ is connected with $D$, or $C$ is connected with $D$, respectively, through a back-edge in $G\setminus E'$. Therefore, we add the edge $(\bar{u},\bar{z})$, or $(\bar{v},\bar{z})$, or $(\bar{w},\bar{z})$, respectively, to $\mathcal{R}$, we set $E'\leftarrow E'\setminus\{(u,p(u))\}$, or $E'\leftarrow E'\setminus\{(v,p(v))\}$, or $E'\leftarrow E'\setminus\{(w,p(w))\}$, respectively, and we revert to the previous case (where $E'$ contains three tree-edges), according to Lemma~\ref{lemma:4conn-edge-connected}.

Thus, let us assume that $\mathit{high}_1(u)\in E$, and $\mathit{high}_1(v)\in E$, and $\mathit{high}_1(w)\in E$. Then, we add the edges $(\bar{u},\bar{r})$, $(\bar{v},\bar{r})$ and $(\bar{w},\bar{r})$ to $\mathcal{R}$. Now, if $D$ is connected with the rest of the graph in $G\setminus E'$, then this can only be through a back-edge that starts from $D$ and ends in $E$ (since $\mathit{high}_1(u)\in E$ and $\mathit{high}_1(v)\in E$ and $\mathit{high}_1(w)\in E$, implies that $(A,D)=(B,D)=(C,D)=\emptyset$). Then, the existence of a back-edge in $(D,E)$ is equivalent to the condition $\mathit{bcount}(z)>\mathit{bcount}(u)+\mathit{bcount}(v)+\mathit{bcount}(w)$. Thus, if this condition is satisfied, then we add the edge $(\bar{z},\bar{r})$ to $\mathcal{R}$.

\subsubsection{$w$ and $v$ are not related as ancestor and descendant, and $v$ is an ancestor of $u$}
Let $A=T(u)$, $B=T(v)\setminus T(u)$, $C=T(w)$, $D=T(z)\setminus(T(v)\cup T(w))$ and $E=T(r)\setminus T(z)$.

If we have that either $\mathit{high}_1(u)=\bot$, or $\mathit{high}_1(w)=\bot$, then there is no back-edge in $G\setminus E'$ that connects $A$, or $C$, respectively, with the rest of the graph. Therefore, we can use Lemma~\ref{lemma:4conn-no-back-edge} in order to revert to the previous case (where $E'$ contains three tree-edges).

Otherwise, if we have that either $\mathit{high}_1(u)\in B$, or $\mathit{high}_1(w)\in D$, then we have that $A$ is connected with $B$, or $C$ is connected with $D$, respectively, through a back-edge in $G\setminus E'$. Therefore, we add the edge $(\bar{u},\bar{v})$, or $(\bar{w},\bar{z})$, respectively, to $\mathcal{R}$, we set $E'\leftarrow E'\setminus\{(u,p(u))\}$, or $E'\leftarrow E'\setminus\{(w,p(w))\}$, respectively, and we revert to the previous case (where $E'$ contains three tree-edges), according to Lemma~\ref{lemma:4conn-edge-connected}.

Similarly, if we have $\mathit{high}_1(u)\in D$ and $\mathit{low}_1(u)\in E$, then we have that $A$ is connected with both $D$ and $E$, through back-edges in $G\setminus E'$. Therefore, we add the edge $(\bar{z},\bar{r})$ to $\mathcal{R}$, we set $E'\leftarrow E'\setminus\{(z,p(z))\}$, and we revert to the previous case (where $E'$ contains three tree-edges), according to Lemma~\ref{lemma:4conn-edge-connected}.

So let us assume that none of the above holds. This means that $\mathit{high}_1(w)\in E$, and either $(1)$ $\mathit{high}_1(u)\in D$ and $\mathit{low}_1(u)\in D$, or $(2)$ $\mathit{high}_1(u)\in E$. 

Let us consider case $(1)$ first. Then we add the edges $(\bar{w},\bar{r})$ and $(\bar{u},\bar{z})$ to $\mathcal{R}$, and it remains to determine the connectivity between $B$ and $D$ with the rest of the graph.

First, we check whether there is a back-edge that stems from $B$ and ends in either $D$ or $E$. This is equivalent to checking whether $\mathit{bcount}(v)>\mathit{bcount}(u)$ (since $\mathit{high}_1(u)\in D$ implies that $B(u)\subseteq B(v)$). If we have $\mathit{bcount}(v)=\mathit{bcount}(u)$, then we have that $B$ is isolated from the rest of the graph in $G\setminus E'$. Thus, it remains to determine whether $D$ is connected with $E$ through a back-edge. This is equivalent to the condition $\mathit{bcount}(z)>\mathit{bcount}(w)$ (since $\mathit{low}_1(u)\notin E$ implies that $B(u)\cap B(z)=\emptyset$, and then $\mathit{high}_1(w)\in E$ implies that $B(w)\subseteq B(z)$). Thus, if this is satisfied, then we add the edge $(\bar{z},\bar{r})$ to $\mathcal{R}$.

Otherwise, suppose that $\mathit{bcount}(v)>\mathit{bcount}(u)$. Let us assume, first, that $\mathit{low}(v)\in D$. Then we have that $B(v)=B(u)\sqcup (B,D)$. Thus, there is a back-edge from $B$ to $D$, and therefore we add the edge $(\bar{v},\bar{z})$ to $\mathcal{R}$. Then we set $E'\leftarrow E'\setminus\{(v,p(v))\}$, and we revert to the previous case (where $E'$ contains three tree-edges), according to Lemma~\ref{lemma:4conn-edge-connected}.
Now let us assume that $\mathit{low}(v)\in E$. Then, since $\mathit{low}(u)\in D$, we have that there is a back-edge from $B$ to $E$. Thus, we insert the edge $(\bar{v},\bar{r})$ to $\mathcal{R}$. 
Now it remains to determine whether $(B,D)\neq\emptyset$ or $(D,E)\neq\emptyset$. Notice that either of those cases implies that $G\setminus E'$ is connected (because $\mathcal{R}$ already contains the edges $(\bar{u},\bar{z})$, $(\bar{w},\bar{r})$ and $(\bar{v},\bar{r})$, and either of those cases implies that we have to add the edge $(\bar{v},\bar{z})$ or $(\bar{z},\bar{r})$, respectively). Since $\mathit{high}_1(u)\in D$ and $\mathit{low}_1(u)\in D$, we have $B(u)=(A,D)$, $B(v)=(A,D)\cup(B,D)\cup(B,E)$, and $B(z)=(B,E)\cup(C,E)\cup(D,E)$. Since $\mathit{high}_1(w)\in E$, we have $B(w)=(C,E)$. Thus, we have the following:

\begin{itemize}
\item{$\mathit{bcount}(u)=|(A,D)|$.}
\item{$\mathit{bcount}(v)=|(A,D)|+|(B,D)|+|(B,E)|$.}
\item{$\mathit{bcount}(w)=|(C,E)|$.}
\item{$\mathit{bcount}(z)=|(B,E)|+|(C,E)|+|(D,E)|$.}
\end{itemize}
 
This implies that $N=\mathit{bcount}(z)-\mathit{bcount}(w)-\mathit{bcount}(v)+\mathit{bcount}(u)=|(D,E)|-|(B,D)|$. Also, we have $s=\mathit{SumAnc}(z)-\mathit{SumAnc}(w)-\mathit{SumAnc}(v)+\mathit{SumAnc}(u)=\mathit{SumAnc}((D,E))-\mathit{SumAnc}((B,D))$. Now, if $N\neq 0$, then at least one of $(D,E)$ and $(B,D)$ is not empty, and therefore $G\setminus E'$ is connected. Thus, it is sufficient to add one more edge to $\mathcal{R}$ in order to make it connected (e.g., we may add $(\bar{v},\bar{z})$). Otherwise, we have $|(D,E)|=|(B,D)|$. Then, since the lower endpoint of every back-edge in $(D,E)$ is lower than the lower endpoint of every back-edge in $(B,D)$, we have that $s<0$ if and only if $|(D,E)|>0$ (and $|(B,D)|>0$). Thus, we add one more edge to $\mathcal{R}$ in order to make it connected (e.g., $(\bar{v},\bar{z})$), if and only if $s<0$.

Now let us consider case $(2)$. Then we add the edges $(\bar{w},\bar{r})$ and $(\bar{u},\bar{r})$ to $\mathcal{R}$.
Since $\mathit{high}_1(u)\in E$, we have that there is a back-edge from $B$ to $D$ if and only if $\mathit{high}_1(v)\in D$. In this case, we add the edge $(\bar{v},\bar{z})$ to $\mathcal{R}$, we set $E'\leftarrow E'\setminus\{(v,p(v))\}$, and we revert to the previous case (where $E'$ contains three tree-edges), according to Lemma~\ref{lemma:4conn-edge-connected}. So let us suppose that $\mathit{high}_1(v)\in E$. In this case, it may be that there is a back-edge from $B$ to $E$. Since $\mathit{high}_1(u)\in E$ and $\mathit{high}_1(v)\in E$, the existence of a back-edge from $B$ to $E$ is equivalent to $\mathit{bcount}(v)>\mathit{bcount}(u)$. If that is the case, then we have that $A$ is connected with $B$ in $G\setminus E'$ through the mediation of $E$. Thus, we add the edge $(\bar{u},\bar{v})$ to $\mathcal{R}$, we set $E'\leftarrow E'\setminus\{(u,p(u))\}$, and we revert to the previous case (where $E'$ contains three tree-edges), according to Lemma~\ref{lemma:4conn-edge-connected}. So let us assume that there is no back-edge from $B$ to $E$. Since $\mathit{high}_1(u)\in E$ and $\mathit{high}_1(v)\in E$, this implies that $B$ is isolated from the rest of the graph in $G\setminus E'$.
Then, it remains to check whether there is a back-edge from $D$ to $E$. Since $\mathit{high}_1(u)\in E$ and $\mathit{high}_1(w)\in E$ (and $B(v)=B(u)$), we have that $B(z)=B(u)\sqcup B(w)$. Thus, there is a back-edge from $D$ to $E$ if and only if $\mathit{bcount}(z)>\mathit{bcount}(u)+\mathit{bcount}(w)$. In this case, we simply add the edge $(\bar{z},\bar{r})$ to $\mathcal{R}$.
 
\subsubsection{$w$ is an ancestor of both $u$ and $v$, and $\{u,v\}$ are not related as ancestor and descendant}
Let $A=T(u)$, $B=T(v)$, $C=T(w)\setminus(T(u)\cup T(v))$, $D=T(z)\setminus T(w)$, and $E=T(r)\setminus T(z)$.

If we have that either $\mathit{high}_1(u)=\bot$, or $\mathit{high}_1(v)=\bot$, then there is no back-edge in $G\setminus E'$ that connects $A$, or $B$, respectively, with the rest of the graph. Therefore, we can use Lemma~\ref{lemma:4conn-no-back-edge} in order to revert to the previous case (where $E'$ contains three tree-edges).

Otherwise, if we have that either $\mathit{high}_1(u)\in C$, or $\mathit{high}_1(v)\in C$, then we have that $A$ is connected with $C$, or $B$ is connected with $C$, respectively, through a back-edge in $G\setminus E'$. Therefore, we add the edge $(\bar{u},\bar{w})$, or $(\bar{v},\bar{w})$, respectively, to $\mathcal{R}$. Then we set $E'\leftarrow E'\setminus\{(u,p(u))\}$, or $E'\leftarrow E'\setminus\{(v,p(v))\}$, respectively, and we revert to the previous case (where $E'$ contains three tree-edges), according to Lemma~\ref{lemma:4conn-edge-connected}.

Similarly, if we have that either $\mathit{high}_1(u)\in D$ and $\mathit{low}_1(u)\in E$, or $\mathit{high}_1(v)\in D$ and $\mathit{low}_1(v)\in E$, 
then we have that $D$ remains connected with $E$ in $G\setminus E'$, through the mediation of $A$ or $B$, respectively. Therefore, we add the edge $(\bar{z},\bar{r})$ to $\mathcal{R}$, we set $E'\leftarrow E'\setminus\{(z,p(z))\}$, and we revert to the previous case (where $E'$ contains three tree-edges), according to Lemma~\ref{lemma:4conn-edge-connected}.
Also, if we have that $M(z)\in D$, then $D$ is connected with $E$ through a back-edge in $G\setminus E'$. Therefore, we add the edge $(\bar{z},\bar{r})$ to $\mathcal{R}$, we set $E'\leftarrow E'\setminus\{(z,p(z))\}$, and we revert to the previous case (where $E'$ contains three tree-edges), according to Lemma~\ref{lemma:4conn-edge-connected}.

Thus, we may assume that none of the above is true. Therefore, we have that $M(z)\notin D$, and there are four possibilities to consider: either $(1)$ $\mathit{high}_1(u)\in D$, $\mathit{low}_1(u)\in D$, $\mathit{high}_1(v)\in D$ and $\mathit{low}_1(v)\in D$, or $(2)$ $\mathit{high}_1(u)\in D$, $\mathit{low}_1(u)\in D$, $\mathit{high}_1(v)\in E$ and $\mathit{low}_1(v)\in E$, or $(3)$ $\mathit{high}_1(u)\in E$, $\mathit{low}_1(u)\in E$, $\mathit{high}_1(v)\in D$ and $\mathit{low}_1(v)\in D$, or $(4)$ $\mathit{high}_1(u)\in E$, $\mathit{low}_1(u)\in E$, $\mathit{high}_1(v)\in E$ and $\mathit{low}_1(v)\in E$. In either of those cases, the problem will be to determine whether $C$ is connected with either $D$ or $E$ through back-edges.

Let us consider case $(1)$ first. In this case, we add the edges $(\bar{u},\bar{z})$ and $(\bar{v},\bar{z})$ to $\mathcal{R}$. If we have $\mathit{low}_1(w)\in D$, then there are no back-edges from $C$ to $E$. Thus, in order to check whether there is a back-edge from $C$ to $D$, it is sufficient to check whether $\mathit{bcount}(w)>\mathit{bcount}(u)+\mathit{bcount}(v)$ (because $\mathit{high}_1(u)\in D$ and $\mathit{high}_1(v)\in D$ imply that $B(u)\cup B(v)\subseteq B(w)$; this can be strengthened to $B(u)\sqcup B(v)\subseteq B(w)$, since $u$ and $v$ are not related as ancestor and descendant). If that is the case, then we add the edge $(\bar{w},\bar{z})$ to $\mathcal{R}$. (Otherwise, there is nothing to do.) On the other hand, if we have $\mathit{low}_1(w)\in E$, then there is a back-edge from $C$ to $E$ (since $\mathit{low}_1(u)\in D$ and $\mathit{low}_1(v)\in D$). Thus, we add the edge $(\bar{w},\bar{r})$ to $\mathcal{R}$. Now it remains to determine whether there is a back-edge that connects $C$ with $D$. We claim that this is equivalent to checking whether $\mathit{bcount}(w)>\mathit{bcount}(u)+\mathit{bcount}(v)+\mathit{bcount}(z)$. To see this, first notice that $B(u)\sqcup B(v)\subseteq B(w)$ (since $\mathit{high}_1(u)\in D$ and $\mathit{high}_1(v)\in D$). We also have that $(B(u)\cup B(v))\cap B(z)=\emptyset$ (since $\mathit{low}_1(u)\in D$ and $\mathit{low}_1(v)\in D$). We also have $B(z)\subseteq B(w)$ (since $M(z)\notin D$). This shows that $B(u)\sqcup B(v)\sqcup B(z)\subseteq B(w)$. Finally, notice that $B(w)\setminus(B(u)\cup B(v)\cup B(z))$ contains precisely the back-edges that connect $C$ with $D$. Thus, it is sufficient to check whether $\mathit{bcount}(w)>\mathit{bcount}(u)+\mathit{bcount}(v)+\mathit{bcount}(z)$. If that is the case, then we add the edge $(\bar{w},\bar{z})$ to $\mathcal{R}$.

Now let us consider case $(2)$. In this case, we add the edges $(\bar{u},\bar{z})$ and $(\bar{v},\bar{r})$ to $\mathcal{R}$. 
Now we have to determine whether there is a back-edge from $C$ to $D$, or from $C$ to $E$.
We claim that there is a back-edge from $C$ to $E$ if and only if $M(z)\in C$. The necessity is obvious (since $M(z)\notin D$). To see the sufficiency, notice that, since $\mathit{low}_1(u)\in D$, we have that all back-edges in $B(z)$ have their higher endpoint either in $B$, or in $C$, or in $D$. The last case is rejected, since $M(z)\notin D$. If all the back-edges in $B(z)$ have their higher endpoint in $B$, then $M(z)\in B$. Thus, if we have $M(z)\in C$, then at least one back-edge in $B(z)$ must stem from $C$. Thus, if we have $M(z)\in C$, then we add the edge $(\bar{w},\bar{r})$ to $\mathcal{R}$. Now it remains to determine whether there is a back-edge from $C$ to $D$. Notice that $B(w)$ can be partitioned into: the back-edges in $B(u)$ (since $\mathit{high}_1(u)\in D$), the back-edges in $B(v)$ (since $\mathit{high}_1(v)\in E$), the back-edges from $C$ to $D$, and the back-edges from $C$ to $E$. Since $\mathit{low}_1(u)\in D$ and $M(z)\notin D$, we have $(C,E)=B(z)\setminus B(v)$. Thus, in order to determine whether there is a back-edge from $C$ to $D$, it is sufficient to check whether $\mathit{bcount}(w)>\mathit{bcount}(u)+\mathit{bcount}(v)+(\mathit{bcount}(z)-\mathit{bcount}(v))=\mathit{bcount}(u)+\mathit{bcount}(z)$. If that is the case, then we add the edge $(\bar{w},\bar{z})$ to $\mathcal{R}$. (Otherwise, there is nothing to do.) On the other hand, if $M(z)\notin C$, then we know that there is no back-edge from $C$ to $E$. Thus, in order to determine whether there is a back-edge from $C$ to $D$, it is sufficient to check whether $\mathit{bcount}(w)>\mathit{bcount}(u)+\mathit{bcount}(v)$. If that is the case, then we add the edge $(\bar{w},\bar{z})$ to $\mathcal{R}$.
Case $(3)$ is treated with a similar argument.

Finally, let us consider case $(4)$. In this case, we add the edges $(\bar{u},\bar{r})$ and $(\bar{v},\bar{r})$ to $\mathcal{R}$. Then, notice that there is a back-edge from $C$ to $D$ if and only if $\mathit{high}_1(w)\in D$ (since $\mathit{high}_1(u)\in E$ and $\mathit{high}_1(v)\in E$). If that is the case, then we add the edge $(\bar{w},\bar{z})$ to $\mathcal{R}$, we set $E'\leftarrow E'\setminus\{(w,p(w))\}$, and we revert to the previous case (where $E'$ contains three tree-edges), according to Lemma~\ref{lemma:4conn-edge-connected}. Otherwise, all the back-edges in $B(w)$ that stem from $C$ (if there are any), end in $E$. Thus, in order to determine whether such a back-edge exists, we simply check whether $\mathit{bcount}(w)>\mathit{bcount}(u)+\mathit{bcount}(v)$ (because $B(u)\sqcup B(v)\subseteq B(w)$). If that is the case, then we add the edge $(\bar{w},\bar{r})$ to $\mathcal{R}$.

\subsubsection{$w$ is an ancestor of $v$, and $v$ is an ancestor of $u$}
Let $A=T(u)$, $B=T(v)\setminus T(u)$, $C=T(w)\setminus T(v)$, $D=T(z)\setminus T(w)$, and $E=T(r)\setminus T(z)$.

If we have that either $\mathit{high}_1(u)\in B$, or $M(z)\in D$, then we have that $A$ is connected with $B$, or $D$ is connected with $E$, respectively, through a back-edge. Thus, we insert the edge $(\bar{u},\bar{v})$, or $(\bar{z},\bar{r})$, respectively, to $\mathcal{R}$. Then we set $E'\leftarrow E'\setminus\{(u,p(u))\}$, or $E'\leftarrow E'\setminus\{(z,p(z))\}$, respectively, and we revert to the previous case (where $E'$ contains three tree-edges), according to Lemma~\ref{lemma:4conn-edge-connected}.

If we have that either $\mathit{high}_1(u)\in C$ and $\mathit{low}_1(u)\in D$, or $\mathit{high}_1(u)\in D$ and $\mathit{low}_1(u)\in E$, then we have that $C$ is connected with $D$, or $D$ is connected with $E$, respectively, through the mediation of $A$. Thus, we insert the edge $(\bar{w},\bar{z})$, or $(\bar{z},\bar{r})$, respectively, to $\mathcal{R}$. Then we set $E'\leftarrow E'\setminus\{(w,p(w))\}$, or $E'\leftarrow E'\setminus\{(z,p(z))\}$, respectively, and we revert to the previous case (where $E'$ contains three tree-edges), according to Lemma~\ref{lemma:4conn-edge-connected}.

So let us assume that neither of the above is true. (In particular, we have $M(z)\notin D$, and therefore $B(z)=(A,E)\cup (B,E)\cup (C,E)$.) Then there are four cases to consider. Either $(1)$ $\mathit{high}_1(u)\in C$ and $\mathit{low}_1(u)\in C$, or $(2)$ $\mathit{high}_1(u)\in C$ and $\mathit{low}_1(u)\in E$, or $(3)$ $\mathit{high}_1(u)\in D$ and $\mathit{low}_1(u)\in D$, or $(4)$ $\mathit{high}_1(u)\in E$ and $\mathit{low}_1(u)\in E$.

Let us consider case $(1)$ first. Then we add the edge $(\bar{u},\bar{w})$ to $\mathcal{R}$. Notice that $\mathit{high}_1(v)\in C$. Thus, there are three different cases to consider. Either $(1.1)$ $\mathit{low}_1(v)\in C$, or $(1.2)$ $\mathit{low}_1(v)\in D$, or $(1.3)$ $\mathit{low}_1(v)\in E$. Let us consider case $(1.1)$. Then, there is a back-edge from $B$ to $C$ if and only if $\mathit{bcount}(v)>\mathit{bcount}(u)$. If that is the case, then we add the edge $(\bar{v},\bar{w})$ to $\mathcal{R}$, we set $E'\leftarrow E'\setminus\{(v,p(v))\}$, and we revert to the previous case (where $E'$ contains three tree-edges), according to Lemma~\ref{lemma:4conn-edge-connected}. Otherwise, we have that $B$ is isolated from the rest of the graph in $G\setminus E'$. It remains to determine whether $C$ is connected with $D$ or $E$. If $\mathit{bcount}(z)=0$, then there is no back-edge from $C$ to $E$. Then, since $\mathit{low}(v)\in C$, we have that $B(w)=(C,D)$. Thus, there is a back-edge from $C$ to $D$ if and only if $\mathit{bcount}(w)>0$. If that is the case, then we add the edge $(\bar{w},\bar{z})$ to $\mathcal{R}$ (and we are done). So let us assume that $\mathit{bcount}(z)>0$. Then, since $M(z)\notin D$ and $\mathit{low}(v)\in C$, we have that $M(z)\in C$. Thus, there is a back-edge from $C$ to $E$, and so we add the edge $(\bar{w},\bar{r})$ to $\mathcal{R}$. Since $M(z)\in C$ and $\mathit{low}(v)\in C$, we have that $B(z)=(C,E)$. Furthermore, we have $B(w)=(C,D)\cup(C,E)$. Thus, there is a back-edge from $C$ to $D$ if and only if $\mathit{bcount}(w)>\mathit{bcount}(z)$. If that is the case, then we add the edge $(\bar{w},\bar{z})$ to $\mathcal{R}$.

Now let us consider case $(1.2)$. This means that $B$ is connected with $D$ with a back-edge, and so we add the edge $(\bar{v},\bar{z})$ to $\mathcal{R}$. Notice that $\mathit{high}_1(w)\in D$. Thus, there are two cases to consider. Either $(1.2.1)$ $\mathit{low}_1(w)\in D$, or $(1.2.2)$ $\mathit{low}_1(w)\in E$. Let us consider case $(1.2.1)$. Then, since $M(z)\notin D$, we have that there are no back-edges from $D$ to $E$, and therefore $B(z)=\emptyset$ (since $\mathit{low}(w)\in D$). This means that $E$ is isolated from the rest of the graph in $G\setminus E'$. Thus, we can use Lemma~\ref{lemma:4conn-no-back-edge} in order to revert to the previous case (where $E'$ contains three tree-edges).
%
Now let us consider case $(1.2.2)$. Then there is a back-edge from $C$ to $E$, and so we add the edge $(\bar{w},\bar{r})$ to $\mathcal{R}$. 
Thus far, $\mathcal{R}$ contains the edges $(\bar{u},\bar{w})$, $(\bar{v},\bar{z})$ and $(\bar{w},\bar{r})$. It remains to determine whether $(B,C)\neq\emptyset$ or $(C,D)\neq\emptyset$. Observe that either of those cases implies that $G\setminus E'$ is connected, and therefore it is sufficient to add one more edge to $\mathcal{R}$ in order to make it connected (e.g., we may add $(\bar{v},\bar{w})$). Since $\mathit{high}_1(u)\in C$ and $\mathit{low}_1(u)\in C$, we have $B(u)=(A,C)$. Then, since $\mathit{low}_1(v)\in D$, we have $B(v)=(A,C)\cup(B,C)\cup(B,D)$. We also have $B(w)=(B,D)\cup(C,D)\cup(C,E)$, and $B(z)=(C,E)$ (since $M(z)\notin D$). Thus, we have the following:

\begin{itemize}
\item{$\mathit{bcount}(u)=|(A,C)|$.}
\item{$\mathit{bcount}(v)=|(A,C)|+|(B,C)|+|(B,D)|$.}
\item{$\mathit{bcount}(w)=|(B,D)|+|(C,D)|+|(C,E)|$.}
\item{$\mathit{bcount}(z)=|(C,E)|$.}
\end{itemize}

This implies that $N=\mathit{bcount}(w)-\mathit{bcount}(z)-\mathit{bcount}(v)+\mathit{bcount}(u)=|(C,D)|-|(B,C)|$. Furthermore, we have $s=\mathit{SumAnc}(w)-\mathit{SumAnc}(z)-\mathit{SumAnc}(v)+\mathit{SumAnc}(u)=\mathit{SumAnc}((C,D))-\mathit{SumAnc}((B,C))$. Thus, if $N\neq 0$, then at least one of $(C,D)$ and $(B,C)$ is not empty, and therefore it is sufficient to add e.g. $(\bar{v},\bar{w})$ to $\mathcal{R}$ in order to make it connected. Otherwise, if $N=0$, then we have $|(C,D)|=|(B,C)|$. Then, since the lower endpoint of every back-edge in $(C,D)$ is lower than the lower endpoint of every back-edge in $(B,C)$, we have that $s=\mathit{SumAnc}((C,D))-\mathit{SumAnc}((B,C))<0$ if and only if $(C,D)\neq\emptyset$ (and $(B,C)\neq\emptyset$). Thus, in the case $N=0$, we add the edge $(\bar{v},\bar{w})$ to $\mathcal{R}$ if and only if $s<0$.

Now let us consider case $(1.3)$. Then, there is a back-edge from $B$ to $E$, and so we add the edge $(\bar{v},\bar{r})$ to $\mathcal{R}$. Now there are two cases to consider. Either $(1.3.1)$ $\mathit{high}_1(w)\in D$, or $(1.3.2)$ $\mathit{high}_1(w)\in E$. Let us consider case $(1.3.1)$. Then, since $\mathit{low}_1(u)\in C$, we have that the $\mathit{high}_1$-edge of $w$ either stems from $B$ or from $C$. If the $\mathit{high}_1$-edge of $w$ stems from $B$, then we have that $D$ and $E$ remain connected in $G\setminus E'$ through the mediation of $B$. Thus, we add the edge $(\bar{z},\bar{r})$ to $\mathcal{R}$, we set $E'\leftarrow E'\setminus\{(z,p(z))\}$, and we revert to the previous case (where $E'$ contains three tree-edges), according to Lemma~\ref{lemma:4conn-edge-connected}. Otherwise, if the $\mathit{high}_1$-edge of $w$ stems from $C$, then we have that this is a back-edge from $C$ to $D$. Thus, we add the edge $(\bar{w},\bar{z})$ to $\mathcal{R}$, we set $E'\leftarrow E'\setminus\{(w,p(w))\}$, and we revert to the previous case (where $E'$ contains three tree-edges), according to Lemma~\ref{lemma:4conn-edge-connected}. Now let us consider case $(1.3.2)$. In this case, we have that $D$ is isolated from the rest of the graph in $G\setminus E'$, and therefore $\bar{z}$ should be isolated in $\mathcal{R}$. Furthermore, all the back-edges in $B(w)$ that stem from $C$ (if there are any), end in $E$. If $M(z)\in C$, then there is a back-edge from $C$ to $E$. Thus, we add the edge $(\bar{w},\bar{r})$ to $\mathcal{R}$, and we are done, because now $\mathcal{R}$ has enough edges to make the vertices $\bar{u}$, $\bar{v}$, $\bar{w}$ and $\bar{r}$ connected. Otherwise, let us assume that $M(z)\notin C$. Then there is no back-edge from $C$ to $E$. 
Since $\mathit{high}(w)\in E$, there is no back-edge from $B$ to $D$, or from $C$ to $D$. Notice that we have $B(u)=(A,C)$, $B(v)=(A,C)\cup(B,C)\cup(B,E)$ and $B(w)=(B,E)$. Thus, there is a back-edge from $B$ to $C$, if and only if $\mathit{bcount}(v)>\mathit{bcount}(u)+\mathit{bcount}(w)$. If that is the case, then we add the edge $(\bar{v},\bar{w})$ to $\mathcal{R}$. Thus, we have exhausted all possibilities for case $(1.3)$.

Now let us consider case $(2)$. Then, we have that $A$ is connected with both $C$ and $E$ in $G\setminus E'$, and so we add the edges $(\bar{u},\bar{w})$ and $(\bar{u},\bar{r})$ to $\mathcal{R}$. First, we check whether there is back-edge from $B$ to $E$. Such a back-edge exists if and only if there is a back-edge $(x,y)\in B(v)\setminus B(u)$ such that $y\leq p(z)$. Thus, we can determine the existence of such a back-edge in constant time, using the data structure from Lemma~\ref{lemma:back-edge-oracle} (we assume that we have performed the linear-time preprocessing that is required in order to build this data structure). If there is a back-edge from $B$ to $E$, then we have that $A$ and $B$ are connected in $G\setminus E'$ through the mediation of $E$. Thus, we add the edge $(\bar{u},\bar{v})$ to $\mathcal{R}$, we set $E'\leftarrow E'\setminus\{(u,p(u))\}$, and we revert to the previous case (where $E'$ contains three tree-edges), according to Lemma~\ref{lemma:4conn-edge-connected}. Otherwise, suppose that there is no back-edge from $B$ to $E$. Then, we check whether there is a back-edge $(x,y)\in B(v)\setminus B(u)$ such that $y\leq p(w)$. (Again, we can perform this check in constant time using Lemma~\ref{lemma:back-edge-oracle}.) If that is the case, then, since there is no back-edge from $B$ to $E$, we have that $(x,y)$ is a back-edge from $B$ to $D$. Otherwise, we have that there is no back-edge from $B$ to $D$. So here we can distinguish in constant time two cases: either $(2.1)$ there is a back-edge from $B$ to $D$, or $(2.2)$ there is no back-edge from $B$ to $D$.

Let us consider case $(2.1)$. Then, we add the edge $(\bar{v},\bar{z})$ to $\mathcal{R}$. We will determine whether there is a back-edge from $A$ to $D$, or a back-edge from $B$ to $C$, or a back-edge from $C$ to $D$. Notice that, in either of those cases, we have that $G\setminus E'$ is connected, and so it is sufficient to return a connected graph $\mathcal{R}$. Otherwise, we have that $\mathcal{R}$ is a connectivity graph for $G\setminus E'$. (It is irrelevant whether there is a back-edge from $C$ to $E$, because this does not add any new connectivity information, since we know that the parts $C$ and $E$ are connected in $G\setminus E'$ through the mediation of $A$.) Since $\mathit{high}_1(u)\in C$, we have that $B(u)=(A,C)\cup(A,D)\cup(A,E)$. Since there is no back-edge from $B$ to $E$, we have that $B(v)=(A,C)\cup(A,D)\cup(A,E)\cup(B,C)\cup(B,D)$ and $B(w)=(A,D)\cup(A,E)\cup(B,D)\cup(C,D)\cup(C,E)$. And since $M(z)\notin D$ and there is no back-edge from $B$ to $E$, we have that $B(z)=(A,E)\cup(C,E)$. Thus, we have the following:
\begin{itemize}
\item{$\mathit{bcount}(u)=|(A,C)|+|(A,D)|+|(A,E)|$.}
\item{$\mathit{bcount}(v)=|(A,C)|+|(A,D)|+|(A,E)|+|(B,C)|+|(B,D)|$.}
\item{$\mathit{bcount}(w)=|(A,D)|+|(A,E)|+|(B,D)|+|(C,D)|+|(C,E)|$.}
\item{$\mathit{bcount}(z)=|(A,E)|+|(C,E)|$.}
\end{itemize}
This implies that $\mathit{bcount}(z)-\mathit{bcount}(w)+\mathit{bcount}(v)-\mathit{bcount}(u)=|(B,C)|-|(C,D)|-|(A,D)|$. Thus, if we have that $\mathit{bcount}(z)-\mathit{bcount}(w)+\mathit{bcount}(v)-\mathit{bcount}(u)\neq 0$, then at least one of $(B,C)$, $(C,D)$ or $(A,D)$ is not empty. Thus, it suffices to return a connected graph $\mathcal{R}$. Otherwise, suppose that $\mathit{bcount}(z)-\mathit{bcount}(w)+\mathit{bcount}(v)-\mathit{bcount}(u)= 0$. Then we have that $|(B,C)|=|(C,D)|+|(A,D)|$. Consider the value $\mathit{SumAnc}(z)-\mathit{SumAnc}(w)+\mathit{SumAnc}(v)-\mathit{SumAnc}(u)=\mathit{SumAnc}(B,C)-\mathit{SumAnc}(C,D)-\mathit{SumAnc}(A,D)$. If we have that $|(B,C)|=0$, then we also have that $|(C,D)|+|(A,D)|=0$, and therefore $\mathit{SumAnc}(B,C)-\mathit{SumAnc}(C,D)-\mathit{SumAnc}(A,D)=0$. Otherwise, if $|(B,C)|>0$, then we have that $\mathit{SumAnc}(B,C)>\mathit{SumAnc}(C,D)+\mathit{SumAnc}(A,D)$, because $|(B,C)|=|(C,D)|+|(A,D)|$ and the lower endpoint of every back-edge in $(B,C)$ is greater than, or equal to, $w$, whereas the lower endpoint of every back-edge in $(C,D)\cup(A,D)$ is lower than $w$. This implies that $\mathit{SumAnc}(B,C)-\mathit{SumAnc}(C,D)-\mathit{SumAnc}(A,D)>0$. Thus, we have shown that $|(B,C)|>0$ if and only if $\mathit{SumAnc}(B,C)-\mathit{SumAnc}(C,D)-\mathit{SumAnc}(A,D)>0$. Thus, if $\mathit{bcount}(z)-\mathit{bcount}(w)+\mathit{bcount}(v)-\mathit{bcount}(u)= 0$, then it is sufficient to check whether $\mathit{SumAnc}(z)-\mathit{SumAnc}(w)+\mathit{SumAnc}(v)-\mathit{SumAnc}(u)>0$. If that is the case, then we only have to add one more edge to $\mathcal{R}$ in order to make it connected (e.g., $(\bar{u},\bar{z})$). Otherwise, we have that all of the sets $(A,D)$, $(B,C)$ and $(C,D)$ are empty, and $\mathcal{R}$ is already a connectivity graph for $G\setminus E'$.
 
Now let us consider case $(2.2)$. Since there is no back-edge from $B$ to $E$ or from $B$ to $D$, it remains to check whether there is a back-edge from $B$ to $C$. Since $\mathit{high}_1(u)\in C$, we have that $B(u)\subseteq B(v)$. Thus, there is a back-edge in $B(v)\setminus B(u)$ (and therefore a back-edge from $B$ to $C$), if and only if $\mathit{bcount}(v)>\mathit{bcount}(u)$. If that is the case, then we add the edge $(\bar{v},\bar{w})$ to $\mathcal{R}$. Then we set $E'\leftarrow E'\setminus\{(v,p(v))\}$, and we revert to the previous case (where $E'$ contains three tree-edges), according to Lemma~\ref{lemma:4conn-edge-connected}. Otherwise, let us assume that $B(v)=B(u)$. This implies that $B$ is isolated from the rest of the graph in $G\setminus E'$. Now it remains to determine whether there is a back-edge from $A$ to $D$, or from $C$ to $D$. Since $B(v)=B(u)$, this is equivalent to $\mathit{high}_1(w)\in D$. If that is the case, then at least one of $(A,D)$ and $(C,D)$ is not empty. Thus, we have that the parts $A$, $C$, $D$ and $E$, are connected in $G\setminus E'$. Therefore, it is sufficient to add the edge $(\bar{u},\bar{z})$ to $\mathcal{R}$, in order to have a connectivity graph for $G\setminus E'$. Otherwise, if $\mathit{high}_1(w)\notin D$, then $\mathcal{R}$ is already a connectivity graph for $G\setminus E'$.

Now let us consider case $(3)$. Then we add the edge $(\bar{u},\bar{z})$ to $\mathcal{R}$. Since $\mathit{high}_1(u)\in D$, we have that, if $\mathit{high}_1(v)\in C$, then there is a back-edge from $B$ to $C$. Then, we add the edge $(\bar{v},\bar{w})$ to $\mathcal{R}$, we set $E'\leftarrow E'\setminus\{(v,p(v))\}$, and we revert to the previous case (where $E'$ contains three tree-edges), according to Lemma~\ref{lemma:4conn-edge-connected}. So let us assume that $\mathit{high}_1(v)\notin C$. Here we distinguish two cases: either $(3.1)$ $\mathit{low}(v)\in D$, or $(3.2)$ $\mathit{low}(v)\in E$. 

Let us consider case $(3.1)$. Then we have that $B(v)=(A,D)\cup(B,D)$. Thus, we have that there is a back-edge from $B$ to $D$ if and only if $\mathit{bcount}(v)>\mathit{bcount}(u)$. If that is the case, then we have that $A$ remains connected with $B$ in $G\setminus E'$ through the mediation of $D$. Thus, we add the edge $(\bar{u},\bar{v})$ to $\mathcal{R}$, we set $E'\leftarrow E'\setminus\{(u,p(u))\}$, and we revert to the previous case (where $E'$ contains three tree-edges), according to Lemma~\ref{lemma:4conn-edge-connected}. So let us assume that there is no back-edge from $B$ to $D$. Thus, $B$ is isolated from the rest of the graph in $G\setminus E'$. Now it remains to determine whether there is a back-edge from $C$ to $D$, or from $C$ to $E$. We distinguish two cases: $(3.1.1)$ $\mathit{low}(w)\in D$, or $(3.1.2)$ $\mathit{low}(w)\in E$. Let us consider case $(3.1.1)$. Then, we have that $B(w)=(A,D)\cup(C,D)$. Thus, there is a back-edge from $(C,D)$ if and only if $\mathit{bcount}(w)>\mathit{bcount}(u)$. If that is the case, then we simply add the edge $(\bar{w},\bar{z})$ to $\mathcal{R}$. Now let us consider case $(3.1.2)$. Then, since $\mathit{low}(v)\in D$ and $M(z)\notin D$, we have that there is a back-edge from $C$ to $E$. Thus, we add $(\bar{w},\bar{r})$ to $\mathcal{R}$. Notice that we have $B(z)=(C,E)$, and $B(w)=(A,D)\cup(C,D)\cup(C,E)$. Thus, there is a back-edge from $C$ to $D$ if and only if $\mathit{bcount}(w)>\mathit{bcount}(u)+\mathit{bcount}(z)$. If that is the case, then we add the edge $(\bar{w},\bar{z})$ to $\mathcal{R}$. 

Now let us consider case $(3.2)$. Then, since $\mathit{low}(u)\in D$, there is a back-edge from $B$ to $E$. Thus, we add the edge $(\bar{v},\bar{r})$ to $\mathcal{R}$. If $M(z)\in C$, then there is a back-edge from $C$ to $E$, and therefore $B$ and $C$ remain connected in $G\setminus E'$ through the mediation of $E$.
Thus, we add the edge $(\bar{v},\bar{w})$ to $\mathcal{R}$, we set $E'\leftarrow E'\setminus\{(v,p(v))\}$, and we revert to the previous case (where $E'$ contains three tree-edges), according to Lemma~\ref{lemma:4conn-edge-connected}.
So let us suppose that $M(z)\notin C$. Then it remains to determine whether there is a back-edge from $B$ to $D$, or from $C$ to $D$. Since $\mathit{high}_1(u)\in D$ and $\mathit{low}_1(u)\in D$, we have $B(u)=(A,D)$. Since $M(z)\notin C$ and $\mathit{low}(v)\in E$, we have that $B(z)=(B,E)$. Furthermore, we have $B(v)=(A,D)\cup(B,D)\cup(B,E)$. Thus, there is a back-edge from $B$ to $D$ if and only if $\mathit{bcount}(v)>\mathit{bcount}(u)+\mathit{bcount}(z)$. If that is the case, then we add the edge $(\bar{v},\bar{z})$ to $\mathcal{R}$. Since $M(z)\notin C$, we have that $B(w)=(A,D)\cup(B,D)\cup(B,E)\cup(C,D)$. Thus, there is a back-edge from $C$ to $D$ if and only if $\mathit{bcount}(w)>\mathit{bcount}(v)$. If that is the case, then we add the edge $(\bar{w},\bar{z})$ to $\mathcal{R}$.

Now let us consider case $(4)$. Then there is a back-edge from $A$ to $E$, and so we add the edge $(\bar{u},\bar{r})$ to $\mathcal{R}$. If we have that $\mathit{high}_1(v)\in C$, then $B$ is connected with $C$ in $G\setminus E'$, and so we add the edge $(\bar{v},\bar{w})$ to $\mathcal{R}$. Then we set $E'\leftarrow E'\setminus\{(v,p(v))\}$, and we revert to the previous case (where $E'$ contains three tree-edges), according to Lemma~\ref{lemma:4conn-edge-connected}. Thus, we may assume that $\mathit{high}_1(v)\notin C$. Then there are two cases to consider: either $(4.1)$ $\mathit{high}_1(v)\in D$, or $(4.2)$ $\mathit{high}_1(v)\in E$.

Now let us consider case $(4.1)$. Then there is a back-edge from $B$ to $D$, and so we add the edge $(\bar{v},\bar{z})$ to $\mathcal{R}$. Now we distinguish three cases, depending on the location of $M(z)$. Since $M(z)\notin D$, we have that either $(4.1.1)$ $M(z)\in A$, or $(4.1.2)$ $M(z)\in B$, or $(4.1.3)$ $M(z)\in C$. Let us consider case $(4.1.1)$. Then we have that $B(v)=(A,E)\cup (B,D)$. Furthermore, there is no back-edge from $C$ to $E$. Thus, it remains to determine whether there is a back-edge from $C$ to $D$. Notice that we have $B(w)=(C,D)\cup(B,D)\cup(A,E)$. Thus, $(C,D)=B(w)\setminus B(v)$. Thus, there is a back-edge from $C$ to $D$, if and only if $\mathit{bcount}(w)>\mathit{bcount}(v)$. In this case, we simply add the edge $(\bar{w},\bar{z})$ to $\mathcal{R}$. Now let us consider case $(4.1.2)$. 
This implies that there is a back-edge from $B$ to $E$, and therefore $A$ and $B$ are connected in $G\setminus E'$ through the mediation of $E$. Thus, we add the edge $(\bar{u},\bar{v})$ to $\mathcal{R}$, we set $E'\leftarrow E'\setminus\{(u,p(u))\}$, and we revert to the previous case (where $E'$ contains three tree-edges), according to Lemma~\ref{lemma:4conn-edge-connected}.
Now let us consider case $(4.1.3)$. This implies that there is a back-edge from $C$ to $E$, and so we add the edge $(\bar{w},\bar{r})$ to $\mathcal{R}$. It remains to determine whether there is a back-edge from $B$ to $E$, or a back-edge from $C$ to $D$. Notice that either case implies that $G\setminus E'$ is connected. Thus, we only need to check whether $(B,E)\cup (C,D)\neq\emptyset$. Since $\mathit{high}_1(u)\in E$, we have that $B(u)=(A,E)$. Since $\mathit{high}_1(v)\in D$, we have that $B(v)=(A,E)\cup(B,D)\cup(B,E)$. Furthermore, we have $B(w)=(A,E)\cup(B,D)\cup(B,E)\cup(C,D)\cup(C,E)$, and $B(z)=(A,E)\cup(B,E)\cup(C,E)$. Thus, we have the following:
\begin{itemize}
\item{$|B(u)|=|(A,E)|$.}
\item{$|B(v)|=|(A,E)|+|(B,D)|+|(B,E)|$.}
\item{$|B(w)|=|(A,E)|+|(B,D)|+|(B,E)|+|(C,D)|+|(C,E)|$.}
\item{$|B(z)|=|(A,E)|+|(B,E)|+|(C,E)|$.}
\end{itemize}
This implies that $|B(z)|-|B(w)|+|B(v)|-|B(u)|=|(B,E)|-|(C,D)|$. Thus, if we have that $\mathit{bcount}(z)-\mathit{bcount}(w)+\mathit{bcount}(v)-\mathit{bcount}(u)\neq 0$, then we can be certain that one of $(B,E)$ and $(C,D)$ is not empty, and therefore $G\setminus E'$ is connected. Thus, it suffices to add either $(\bar{v},\bar{r})$ or $(\bar{w},\bar{z})$ to $\mathcal{R}$, in order to make it connected. Otherwise, if $|(B,E)|-|(C,D)|=0$, then we must use other means in order to determine whether one of $(B,E)$ and $(C,D)$ is non-empty. For this purpose, we consider the values $\mathit{SumAnc}$ of those sets. Thus, we have that $\mathit{SumAnc}(B,E)-\mathit{SumAnc}(C,D)=\mathit{SumAnc}B(z)-\mathit{SumAnc}B(w)+\mathit{SumAnc}B(v)-\mathit{SumAnc}B(u)$. Then, if both $(B,E)$ and $(C,D)$ are empty, we have that $\mathit{SumAnc}(B,E)-\mathit{SumAnc}(C,D)=0$. Otherwise, since $(B,E)$ and $(C,D)$ have the same number of back-edges, and since the lower endpoints of all back-edges in $(B,E)$ are lower than the lower endpoints of all back-edges in $(C,D)$, we have that $\mathit{SumAnc}(B,E)-\mathit{SumAnc}(C,D)<0$. Thus, we have that both $(B,E)$ and $(C,D)$ are non-empty if and only if $\mathit{SumAnc}(B,E)-\mathit{SumAnc}(C,D)<0$. If that is the case, then it is sufficient to return a connected graph $\mathcal{R}$.

Now let us consider case $(4.2)$. Since $\mathit{high}_1(u)\in E$, we have that $B(u)\subseteq B(v)$. And since $\mathit{high}_1(v)\in E$, we have that $B(v)=B(u)\sqcup(B,E)$. Thus, there is a back-edge from $B$ to $E$ if and only if $\mathit{bcount}(v)>\mathit{bcount}(u)$. If that is the case, then we have that $A$ is connected with $B$ in $G\setminus E'$ through the mediation of $E$. Thus, we add the edge $(\bar{u},\bar{v})$ to $\mathcal{R}$, we set $E'\leftarrow E'\setminus\{(u,p(u))\}$, and we revert to the previous case (where $E'$ contains three tree-edges), according to Lemma~\ref{lemma:4conn-edge-connected}. So let us assume that there is no back-edge from $B$ to $E$. This implies that $B$ is isolated from the remaining parts in $G\setminus E'$. Now, since $M(z)\notin D$, we have that either $(4.2.1)$ $M(z)\in A$, or $(4.2.2)$ $M(z)\in C$. (The case $M(z)\in B$ is rejected, since $(B,E)=\emptyset$.) Let us consider case $(4.2.1)$. Then, there is no back-edge from $C$ to $E$, and it just remains to determine whether there is a back-edge from $C$ to $D$. We have that $B(u)=(A,E)$ and $B(w)=(A,E)\cup(C,D)$. Thus, there is a back-edge from $C$ to $D$ if and only if $\mathit{bcount}(w)>\mathit{bcount}(u)$. If that is the case, then we simply add the edge $(\bar{w},\bar{z})$ to $\mathcal{R}$. Now let us consider case $(4.2.2)$. In this case, there is a back-edge from $C$ to $E$, and so we add the edge $(\bar{w},\bar{r})$ to $\mathcal{R}$. It remains to determine whether there is a back-edge from $C$ to $D$. 
Since $\mathit{high}_1(u)\in E$ and $\mathit{high}_1(v)\in E$, we have that there is a back-edge from $C$ to $D$ if and only if $\mathit{high}_1(w)\in D$. If that is the case, then we simply add the edge $(\bar{w},\bar{z})$ to $\mathcal{R}$.

\subsection{The data structure}
According to the preceding analysis, in order to be able to answer connectivity queries in the presence of at most four edge-failures in constant time, it is sufficient to have computed the following items:
\begin{itemize}
\item{A DFS-tree of the graph rooted at a vertex $r$.}
\item{The values $\mathit{ND}(v)$, $\mathit{bcount}(v)$, $M(v)$ and $\mathit{SumAnc}(v)$, for all vertices $v\neq r$.}
\item{The $\mathit{low}_1$, $\mathit{low}_2$, $\mathit{low}_3$, $\mathit{low}_4$, $\mathit{high}_1$, $\mathit{high}_2$ and $\mathit{high}_3$ edges of $v$, for every vertex $v\neq r$.}
\item{The data structure in Lemma~\ref{lemma:back-edge-oracle} for answering back-edge queries.}
\end{itemize}
Thus, we need $O(n)$ space to store all these items, and the results of Section~\ref{section:DFS} imply that we can compute all of them in linear time in total.

Finally, let us describe how to answer the queries. First, given the set of edges $E'$ that failed (with $|E'|\leq 4$), we build a connectivity graph $\mathcal{R}$ of $G\setminus E'$, by going through the case analysis that is described in the preceding subsections. This takes $O(1)$ time in total. Now let $x$ and $y$ be the two query vertices. Then we determine the root of the connected component of $T\setminus E'$ that contains $x$. This is given either by the largest higher endpoint of the tree-edges in $E'$ that is an ancestor of $x$, or by $r$ if there is no tree-edge in $E'$ whose higher endpoint is an ancestor of $x$. Thus, retrieving this root takes $O(1)$ time. Then we do the same for the other query vertex too, and let $r_1$ and $r_2$ be the two roots that we have gathered. Then we have that $x$ is connected with $y$ in $G\setminus E'$ if and only if $\bar{r}_1$ is connected with $\bar{r}_2$ in $\mathcal{R}$. Since the connected components of $\mathcal{R}$ can be computed in $O(1)$ time, we can have the answer in constant time.

\section{Computing a complete collection of $4$-cuts}
\label{section:computing-4-cuts}

The purpose of this chapter is to provide a summary of the methods that we use in order to establish Theorem~\ref{theorem:main}. Let $G$ be a $3$-edge-connected graph.
The idea is to classify the $4$-cuts of $G$ on a DFS-tree, in order to make it easy for us to compute enough of them efficiently. This results in several algorithms, each of which specializes in computing a specific type of $4$-cuts. In this section we provide our classification of $4$-cuts, down to all the subcases, and we provide an overview of the methods that we use in order to handle each case. We also provide figures with detailed captions, that we consider an organic part of our exposition. The complete analysis, the proofs and the algorithms, are given in the chapters that follow (in Sections~\ref{section:type-2}, \ref{section:type-3a} and \ref{section:type-3b}). We conclude with a technical result (in Section~\ref{subsection:mim-max-queries}), that we will need in the following chapters.
Throughout this chapter we assume familiarity with the DFS-based concepts that we defined in Section~\ref{subsection:basic-dfs}. 

\subsection{A typology of $4$-cuts on a DFS-tree}
\label{subsection:the-general-idea}
Let $r$ be a vertex of $G$, and let $T$ be a DFS-tree of $G$ rooted at $r$. Initially, we classify the $4$-cuts of $G$ according to the number of tree-edges that they contain. Thus, we distinguish Type-$1$, Type-$2$, Type-$3$ and Type-$4$ $4$-cuts, depending on whether they contain one, two, three or four tree-edges, respectively. Notice that there are no $4$-cuts that consist entirely of non-tree edges, because the removal of any set of non-tree edges is insufficient to disconnect the graph, due to the existence of $T$.

We found that it is very difficult to compute enough Type-$4$ $4$-cuts directly. Thus, we use an idea from \cite{DBLP:conf/esa/NadaraRSS21}, in order to reduce the case of those $4$-cuts to the previous cases. Specifically, we first establish the following result.

\begin{proposition}
\label{proposition:implying_type123-4cuts}
Let $G$ be a $3$-edge-connected graph with $n$ vertices, and let $T$ be a DFS-tree of $G$. Then there is a linear-time algorithm that computes a collection $\mathcal{C}$ of $4$-cuts of $G$, that has size $O(n)$ and implies the collection of all $4$-cuts of $G$ that contain at least one back-edge w.r.t. $T$.
\end{proposition}

We establish Proposition~\ref{proposition:implying_type123-4cuts} by essentially following the framework of classification and the techniques of \cite{DBLP:conf/esa/GeorgiadisIK21} for computing all $3$-cuts of a $3$-edge-connected graph (although we have to extend the concepts and techniques significantly). However, for Type-$4$ $4$-cuts, it seems extremely complicated to apply the framework of \cite{DBLP:conf/esa/GeorgiadisIK21}.
Thus, here we rely on the reduction used by \cite{DBLP:conf/esa/NadaraRSS21}. In more detail, \cite{DBLP:conf/esa/NadaraRSS21} also provided a linear-time algorithm for computing all $3$-cuts of a $3$-edge connected graph, by using techniques very similar to \cite{DBLP:conf/esa/GeorgiadisIK21} for the case where there is at least one back-edge in the $3$-cut. Contrary to \cite{DBLP:conf/esa/GeorgiadisIK21}, however, they do not deal directly with the $3$-cuts that consist of three tree-edges. Instead, they show how to reduce the case of $3$-cuts that consist of three tree-edges to the previous cases. They do this through a ``contraction" technique that works as follows. First, we remove all the tree-edges from $G$, and we compute the connected components of the resulting graph. Then we shrink every connected component into a single node, and we re-insert the tree-edges that join different nodes. This results in a graph $Q$ that is $3$-edge-connected; its $k$-cuts, for $k\geq 3$, coincide with those of $G$ that consist only of tree-edges, and its number of edges is at most $2/3$ that of $G$. Thus, we can reduce the computation of $k$-cuts to $Q$. We believe that it is important to formalize and prove this result.

\begin{definition}[Contracted Graph]
\label{definition:contracted-graph}
\normalfont
Let $G$ be a connected graph, and let $T$ be a spanning tree of $G$. Let $C_1,\dots,C_k$ be the connected components of $G\setminus E(T)$. Now we have a function $q:V(G)\rightarrow\{C_1,\dots,C_k\}$ (the quotient map), that maps every vertex of $G$ to the connected component of $G\setminus E(T)$ that contains it. This function induces naturally a map between edges of $G$, and edges between the connected components of $G\setminus E(T)$. Specifically, given an edge $e=(x,y)$ of $G$, we let $q(e)=(q(x),q(y))$.\footnote{To be more precise, the image of $e=(x,y)$ through $q$ should be an edge $q(e)$ (possibly a self-loop) with endpoints $q(x)$ and $q(y)$, associated with a unique identifier that signifies that $q(e)$ is derived from $e$. This is because another edge $e'=(x',y')$ of $G$ may also satisfy that $(q(x),q(y))=(q(x'),q(y'))$, but we want to distinguish between $q(e)$ and $q(e')$.} Then we define the contracted (quotient) graph $Q$ as follows. The vertex set of $Q$ is $\{C_1,\dots,C_k\}$, and the edge set of $Q$ consists of all edges of the form $q(e)$, where $e$ is a tree-edge of $T$ that connects two different connected components of $G\setminus E(T)$. 
\end{definition}

\begin{lemma}[Implicit in \cite{DBLP:conf/esa/NadaraRSS21}]
\label{lemma:contracted-graph}
Let $k\geq 3$ be an integer, let $G$ be a $k$-edge-connected graph with $n$ vertices and $m$ edges, let $T$ be a spanning tree of $G$, let $Q$ be the resulting contracted graph of the connected components of $G\setminus E(T)$, and let $q$ be the corresponding quotient map. Then $Q$ is $k$-edge-connected and it has at most $2m/k$ edges. Furthermore, let $k'\geq k$ be a positive integer. Then, a $k'$-element subset $C$ of $E(T)$ is a $k'$-cut of $G$ if and only if $q(C)$ is a $k'$-cut of $Q$.
\end{lemma}
\begin{proof}
It is easy to bound the number of edges of $Q$. First, since $G$ is $k$-edge-connected, every vertex of $G$ has degree at least $k$. The sum of the degrees of all vertices is $2m$. Thus, we have $kn\leq 2m$, and therefore $n\leq 2m/k$. By construction, $Q$ has at most $|E(T)|=n-1$ edges. Thus, the number of edges of $Q$ is bounded by $2m/k$.

The remaining part of the lemma follows essentially from a correspondence between paths of $G$ and paths of $Q$. Specifically, let $P$ be a path from $x$ to $y$ in $G$. Then there is a contracted path $\widetilde{P}$ from $q(x)$ to $q(y)$ in $Q$ with the property that, for every tree-edge $e$ used by $P$ such that $e$ connects two different connected components of $G\setminus E(T)$, there is an instance of $q(e)$ in $\widetilde{P}$. Furthermore, these are all the edges that appear in $\widetilde{P}$. We note that $\widetilde{P}$ is formed by contracting every part of $P$ that lies entirely within a connected component $z$ of $G\setminus E(T)$ into $z$ (viewed as a vertex of $Q$). Conversely, for every path $P'$ in $Q$, there is a path $P$ in $G$ such that $\widetilde{P}=P'$ (which is formed basically by expanding every vertex $z$ that is used by $P'$ into a path within $z$).  


This explains why $Q$ is $k$-edge-connected. To see this, let $C$ be a set of less than $k$ edges of $Q$, and let $u$ and $v$ be two vertices of $Q$. (Recall that, by definition, we have that $u$ and $v$ are connected components of $G\setminus E(T)$.) Then let $x$ be a vertex of $G$ in $u$, and let $y$ be a vertex of $G$ in $v$. Then, since $G$ is $k$-edge-connected, we have that $G\setminus q^{-1}(C)$ is connected, and therefore there is a path $P$ from $x$ to $y$ in $G\setminus q^{-1}(C)$. Then, $\widetilde{P}$ is a path from $q(x)$ to $q(y)$ in $Q\setminus C$, and therefore $u$ and $v$ are connected in $Q\setminus C$. This shows that no set of less than $k$ edges of $Q$ is sufficient to disconnect $Q$ upon removal. This means that $Q$ is $k$-edge-connected.

Now let $k'\geq k$ be an integer, and let $C$ be a $k'$-cut of $G$ that consists only of tree-edges. We will show that $q(C)$ is $k$-cut of $Q$. First, we have to show that $q(C)$ is well-defined (i.e., it is a set of edges of $Q$). So let $e$ be an edge in $C$. Then, since $C$ is a $k'$-cut of $G$, we have that the endpoints of $e$ lie in different connected components of $G\setminus C$. Therefore, since $C\subseteq E(T)$, we have that the endpoints of $e$ lie in different connected components of $G\setminus E(T)$. This shows that $q(e)$ is an edge of $Q$. Now we will show that $Q\setminus q(C)$ is disconnected. So let us suppose, for the sake of contradiction, that $Q\setminus q(C)$ is connected. Since $C$ is a $k'$-cut of $G$, we have that $G\setminus C$ is disconnected. Thus, there are vertices $x$ and $y$ of $G\setminus C$ that lie in different connected components of $G\setminus C$. Since $Q\setminus q(C)$ is connected, there is a path $P'$ from $q(x)$ to $q(y)$ in $Q\setminus q(C)$. Then there is a path $P$ in $G$ such that $\widetilde{P}=P'$. This implies that $P$ avoids all the edges from $C$, and therefore it is a path in $G\setminus C$. Furthermore, the start of $P$ is in the connected component of $G\setminus E(T)$ that contains $x$, and the end of $P$ is in the connected component of $G\setminus E(T)$ that contains $y$. There is a path $P_x$ in $G\setminus E(T)$ from $x$ to the start of $P$. Similarly, there is a path $P_y$ in $G\setminus E(T)$ from the end of $P$ to $y$. Now, since $C\subseteq E(T)$, the concatenation $P_x+P+P_y$ is a path in $G\setminus C$ from $x$ to $y$. But this contradicts the fact that $x$ and $y$ are disconnected in $G\setminus C$. This shows that $Q\setminus q(C)$ is disconnected. Finally, let $C'$ be a proper subset of $C$. We will show that $Q\setminus q(C')$ is connected. 
So let $u$ and $v$ be two vertices of $Q$. Then there is a vertex $x$ of $G$ in $u$, and there is a vertex $y$ of $G$ in $v$. Since $C$ is a $k$-cut of $G$, we have that $G\setminus C'$ is connected. Thus, there is path $P$ from $x$ to $y$ in $G\setminus C'$. Then, $\widetilde{P}$ is a path from $q(x)$ to $q(y)$ in $Q\setminus q(C')$. This shows that $u$ and $v$ are connected in $Q\setminus C'$. Due to the generality of $u$ and $v$ in $Q$, this shows that $Q\setminus C'$ is connected. Thus, we have that $q(C)$ is a $k$-cut of $Q$.

Conversely, let $C$ be a $k'$-element subset of $E(T)$ such that $q(C)$ is a $k'$-cut of $Q$. We will show that $C$ is a $k'$-cut of $G$. First, we will show that $G\setminus C$ is disconnected. So let us suppose, for the sake of contradiction, that $G\setminus C$ is connected. Since $q(C)$ is a $k'$-cut of $Q$, we have that $Q\setminus q(C)$ is disconnected. Thus, there are vertices $u$ and $v$ of $Q$ such that $u$ and $v$ are disconnected in $Q\setminus q(C)$. Now let $x$ be a vertex in $u$, and let $y$ be a vertex in $v$. Then, since $G\setminus C$ is connected, there is a path $P$ from $x$ to $y$ in $G\setminus C$. But then $\widetilde{P}$ is a path from $q(x)$ to $q(y)$ in $Q\setminus q(C)$, contradicting the fact that $u$ and $v$ are not connected in $Q\setminus q(C)$. This shows that $G\setminus C$ is disconnected. Now let $C'$ be a proper subset of $C$. We will show that $G\setminus C'$ is connected. So let $x$ and $y$ be two vertices of $G$. Since $q(C)$ is a $k'$-cut of $Q$, we have that $Q\setminus q(C')$ is connected. Thus, there is a path $P'$ from $q(x)$ to $q(y)$ in $Q\setminus q(C')$. Then, there is a path $P$ in $G$ such that $\widetilde{P}=P'$. This implies that $P$ is a path in $G\setminus C'$. Furthermore, $P$ starts from a vertex in the connected component of $G\setminus E(T)$ that contains $x$, and ends in a vertex in the connected component of $G\setminus E(T)$ that contains $y$. Then there is a path $P_x$ in $G\setminus E(T)$ from $x$ to the start of $P$. Furthermore, there is a path $P_y$ in $G\setminus E(T)$ from the end of $P$ to $y$. Then, since $C'\subseteq E(T)$, the concatenation $P_x+P+P_y$ is a path from $x$ to $y$ in $G\setminus C'$. Due to the generality of $x$ and $y$ in $G$, this shows that $G\setminus C'$ is connected. We conclude that $C$ is a $k'$-cut of $G$.
\end{proof}

Now, given Proposition~\ref{proposition:implying_type123-4cuts}, we show how to derive Theorem~\ref{theorem:main} by a repeated application of Lemma~\ref{lemma:contracted-graph}.

\begin{theorem}
\label{theorem:main}
Let $G$ be a $3$-edge-connected graph with $n$ vertices. There is a linear-time algorithm that computes a complete collection of $4$-cuts of $G$ with size $O(n)$.
\end{theorem}
\begin{proof}
Let us assume that $G$ has at least two vertices, because otherwise there is nothing to show.
We define a sequence of graphs $G_0,G_1,G_2,\dots$ as follows. First, $G_0=G$. Now suppose that $G_i$ is defined, for some $i\geq 0$, and that it has at least two vertices. Let $T_i$ be an arbitrary DFS-tree of $G_i$. Then we let $G_{i+1}$ be the contracted graph of $G_i$ w.r.t. $T_i$, and we let $q_i$ be the corresponding quotient map (see Definition~\ref{definition:contracted-graph}). Let $N$ be the largest index such that $G_N$ has at least two vertices. Since $G_0$ is $3$-edge-connected, Lemma~\ref{lemma:contracted-graph} implies that $G_0,G_1,\dots,G_N$ is a sequence of $3$-edge-connected graphs. Let $n_i=|V(G_i)|$ and $m_i=|E(G_i)|$, for every $i\in\{0,\dots,N\}$.
By construction, we have $m_1\leq n-1$. Then, Lemma~\ref{lemma:contracted-graph} implies that $m_i\leq n(\frac{2}{3})^{i-1}$, for every $i\in\{1,\dots,N\}$. Since $G_i$ is a $3$-edge-connected graph, for every $i\in\{0,\dots,N\}$, we have $n_i\leq m_i$. This implies that $n_i\leq n(\frac{2}{3})^{i-1}$, for every $i\in\{1,\dots,N\}$.  

Now, for every $i\in\{0,\dots,N\}$, we apply Proposition~\ref{proposition:implying_type123-4cuts} in order to derive, in $O(m_i+n_i)$ time, a collection $\mathcal{C}_i$ of $4$-cuts of $G_i$, that has size $O(n_i)$ and implies all the $4$-cuts of $G_i$ that contain at least one back-edge w.r.t. $T_i$. Notice that this whole process takes time $O(m_0+n_0)+\dots+O(m_N+n_N)=O(m_0+\dots+m_N)=O(m+n\sum_{i=1}^{N}(\frac{2}{3})^{i-1})=O(m+n)$.
Let $i$ be an index in $\{0,\dots,N-1\}$. By Lemma~\ref{lemma:contracted-graph} we have that $q_i^{-1}(\mathcal{C}_{i+1})$ is a collection of $4$-cuts of $G_i$. Furthermore, by repeated application of Lemma~\ref{lemma:contracted-graph} we have that $\mathcal{C}_{i+1}'=q_0^{-1}(q_1^{-1}(\dots q_i^{-1}(\mathcal{C}_{i+1})\dots))$ is a collection of $4$-cuts of $G$.

Now let $\mathcal{C}$ be the collection $\mathcal{C}_0\cup \mathcal{C}_1'\cup\dots\cup\mathcal{C}_N'$. Notice that, for every $i\in\{1,\dots,N\}$, we can construct $\mathcal{C}_i$ in time $O(i|\mathcal{C}_i|)=O(ni(\frac{2}{3})^{i-1})$. Thus, the collection $\mathcal{C}$ can be constructed in time $O(n\sum_{i=1}^{N}{i(\frac{2}{3})^{i-1}})=O(n)$. We have that $\mathcal{C}$ is a collection of $4$-cuts of $G$. We claim that every $4$-cut of $G$ is implied by $\mathcal{C}$. 

So let $C$ be a $4$-cut of $G$. If $C$ contains at least one back-edge w.r.t. $T_0$, then by construction of $\mathcal{C}_0$ (due to Proposition~\ref{proposition:implying_type123-4cuts}) we have that $\mathcal{C}_0$ implies $C$. Therefore, $\mathcal{C}$ also implies $C$ (since $\mathcal{C}_0\subseteq\mathcal{C}$). Otherwise, suppose that $C$ consists only of tree-edges from $T_0$. Then, by Lemma~\ref{lemma:contracted-graph} we have that $q_0(C)$ is a $4$-cut of $G_1$. Now, if $q_0(C)$ contains at least one back-edge w.r.t. $T_1$, then by construction of $\mathcal{C}_1$ (due to Proposition~\ref{proposition:implying_type123-4cuts}) we have that $\mathcal{C}_1$ implies $q_0(C)$. This means that there is a sequence $C_1,\dots,C_k$ of $4$-cuts from $\mathcal{C}_1$, and a sequence $p_1,\dots,p_{k+1}$ of pairs of edges of $G_1$, such that $C_i=p_i\cup p_{i+1}$ for every $i\in\{1,\dots,k\}$, and $p_1\cup p_{k+1}=q_0(C)$ (Definition~\ref{definition:implied_4cut}). Now consider the sequence $q_0^{-1}(C_1),\dots,q_0^{-1}(C_k)$. Then this is a sequence of $4$-cuts from $\mathcal{C}_1'$. Furthermore, we have $q_0^{-1}(C_i)=q_0^{-1}(p_i)\cup q_0^{-1}(p_{i+1})$ for every $i\in\{1,\dots,k\}$, and $C=q_0^{-1}(q_0(C))=q_0^{-1}(p_1\cup p_{k+1})=q_0^{-1}(p_1)\cup q_0^{-1}(p_{k+1})$. This shows that $q_0^{-1}(C_1),\dots,q_0^{-1}(C_k)$ is an implicating sequence of $\mathcal{C}_1'$ that demonstrates that $C$ is implied from $\mathcal{C}_1'$. Thus, $\mathcal{C}$ implies $C$ (since $\mathcal{C}_1'\subseteq\mathcal{C}$). 

Otherwise, suppose that $q_0(C)$ consists only of tree-edges from $T_1$. Then, let $t$ be the maximum index in $\{0,\dots,N\}$ such that $q_{t'}(q_{t'-1}(\dots q_0(C)\dots))$ consists only of tree-edges from $T_{t'+1}$, for every $t'\in\{0,\dots,t\}$. Then, Lemma~\ref{lemma:contracted-graph} implies that $C'=q_t(q_{t-1}(\dots q_0(C)\dots))$ is a $4$-cut of $G_{t+1}$. Furthermore, since $C'$ consists only of tree-edges from $T_{t+1}$, Lemma~\ref{lemma:contracted-graph} implies that $q_{t+1}(C')$ is a $4$-cut of $G_{t+2}$. Due to the maximality of $t$, we have that $q_{t+1}(C')$ must contain at least one back-edge w.r.t. $T_{t+2}$. Then, by construction of $\mathcal{C}_{t+2}$ (due to Proposition~\ref{proposition:implying_type123-4cuts}), we have that $\mathcal{C}_{t+2}$ implies $q_{t+1}(C')$. This means that there is a sequence $C_1,\dots,C_k$ of $4$-cuts from $\mathcal{C}_{t+2}$, and a sequence $p_1,\dots,p_{k+1}$ of pairs of edges of $G_{t+2}$, such that $C_i=p_i\cup p_{i+1}$ for every $i\in\{1,\dots,k\}$, and $p_1\cup p_{k+1}=q_{t+1}(C')$. Now consider the sequence $q_0^{-1}(q_1^{-1}(\dots q_{t+1}^{-1}(C_1)\dots)),\dots,q_0^{-1}(q_1^{-1}(\dots q_{t+1}^{-1}(C_k)\dots))$. Then, it is not difficult to see that this is an implicating sequence of $\mathcal{C}_{t+2}'$, that demonstrates that $q_0^{-1}(q_1^{-1}(\dots q_{t+1}^{-1}(q_{t+1}(C'))\dots)) =C$ is implied from $\mathcal{C}_{t+2}'$. Thus, $\mathcal{C}$ implies $C$ (since $\mathcal{C}_{t+2}'\subseteq\mathcal{C}$). 
\end{proof}

\begin{figure}[h!]\centering
\includegraphics[trim={0cm 23cm 0 0}, clip=true, width=0.8\linewidth]{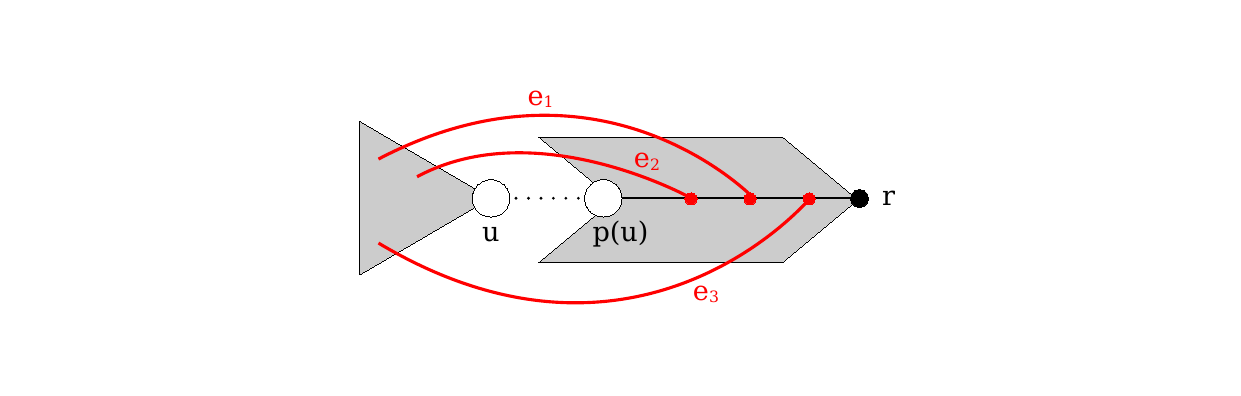}
\caption{\small{A Type-$1$ $4$-cut of the form $\{(u,p(u)),e_1,e_2,e_3\}$, where $e_1,e_2,e_3$ are back-edges. In this case, we have $B(u)=\{e_1,e_2,e_3\}$.}}\label{figure:type1}
\end{figure}

The purpose of everything that follows is to establish Proposition~\ref{proposition:implying_type123-4cuts}. 
The case of Type-$1$ $4$-cuts is the easiest one. So let $C$ be a Type-$1$ $4$-cut of $G$, let $(u,p(u))$ be the tree-edge that is contained in $C$, and let $e_1,e_2,e_3$ be the back-edges that are contained in $C$. Then, by removing $C$ from $G$, we have that each of the subtrees $T(u)$ and $T(r)\setminus T(u)$ of $T$ remains connected (see Figure~\ref{figure:type1}). Thus, these are the two connected components of $G\setminus C$, and therefore the back-edges in $C$ are all the non-tree edges that connect $T(u)$ with $T(r)\setminus T(u)$. Notice that these are precisely the back-edges in $B(u)$. Thus, we have $B(u)=\{e_1,e_2,e_3\}$. Therefore, it is easy to identify all Type-$1$ $4$-cuts: we only have to check, for every vertex $u\neq r$, whether $\mathit{bcount}(u)=3$, and, if yes, we mark $\{(u,p(u)),e_1,e_2,e_3\}$ as a $4$-cut, where $e_1,e_2,e_3$ are the three back-edges that leap over $u$. In order to find $e_1$, $e_2$ and $e_3$, it is sufficient to maintain three distinct back-edges from $B(u)$, for every vertex $u\neq r$. The $\mathit{low}_1$, $\mathit{low}_2$ and $\mathit{low}_3$ edges of $u$ are sufficient for this purpose. By Proposition~\ref{proposition:low}, we can have those edges computed for all vertices $\neq r$, in linear time in total. Thus, all Type-$1$ $4$-cuts can be computed in linear time in total. Notice that every one of them corresponds to a unique vertex $u\neq r$. Thus, the total number of Type-$1$ $4$-cuts is $O(n)$.

\subsection{Type-$2$ $4$-cuts}
\begin{figure}[h!]\centering
\includegraphics[trim={0cm 8cm 0 0}, clip=true, width=0.8\linewidth]{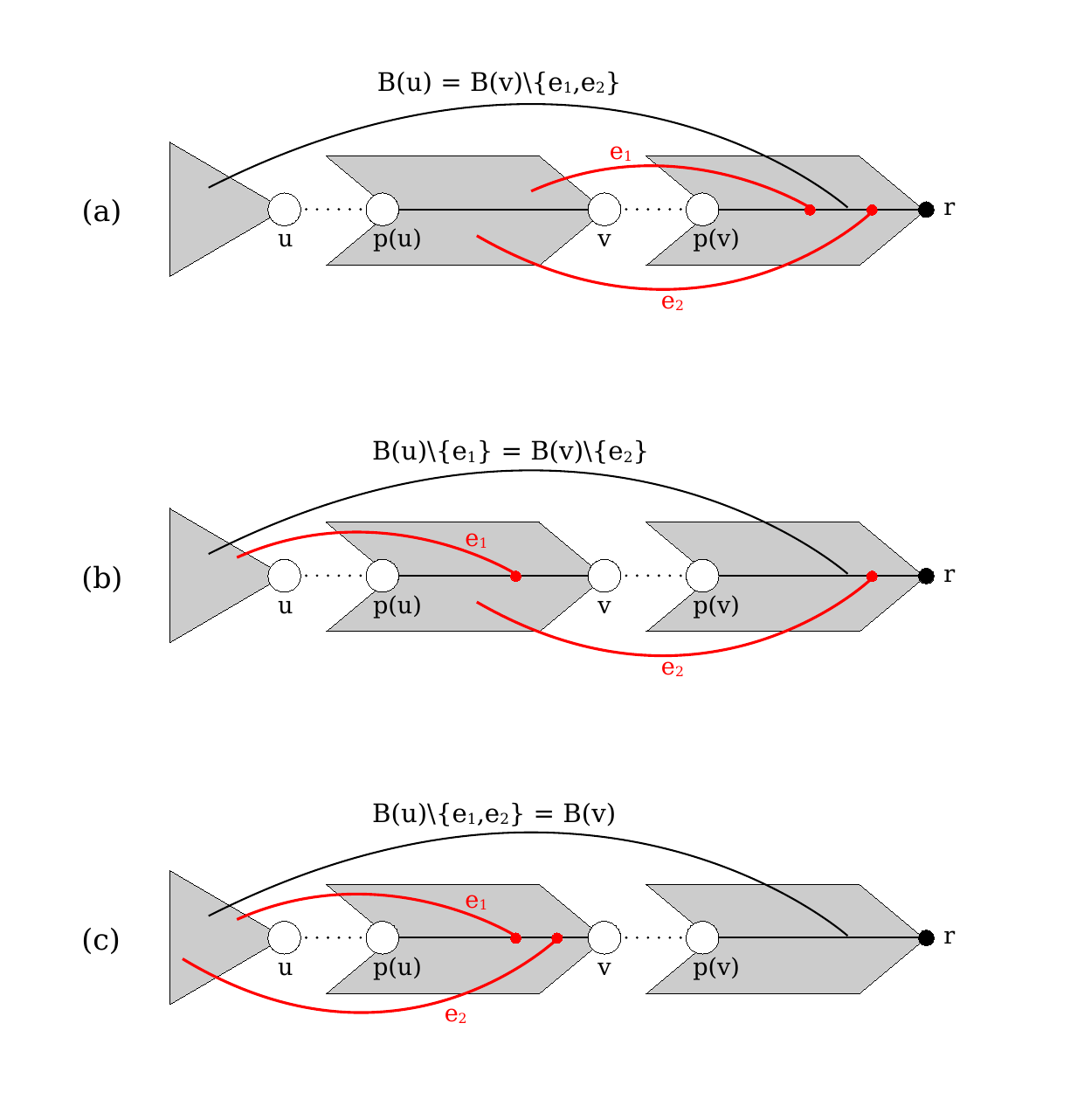}
\caption{\small{All different cases for Type-$2$ $4$-cut of the form $\{(u,p(u)),(v,p(v)),e_1,e_2\}$, where $u$ is a descendant of $v$. In $(a)$ we have $B(v)=B(u)\sqcup\{e_1,e_2\}$. In $(b)$ we have $B(v)\sqcup\{e_1\}=B(u)\sqcup\{e_2\}$. In $(c)$ we have $B(v)\sqcup\{e_1,e_2\}=B(u)$.}}\label{figure:type2}
\end{figure}

Now we consider the case of Type-$2$ $4$-cuts. Let $C$ be a Type-$2$ $4$-cut, and let $(u,p(u))$ and $(v,p(v))$ be the tree-edges that are contained in $C$. Then Lemma~\ref{lemma:type3-4cuts} implies that $u$ and $v$ are related as ancestor and descendant. So let us assume w.l.o.g. that $u$ is a descendant of $v$. Then, Lemma~\ref{lemma:type2cuts} shows that there are three distinct cases to consider (see Figure~\ref{figure:type2}): either $(1)$ $B(v)=B(u)\sqcup\{e_1,e_2\}$, or $(2)$ $B(v)\sqcup\{e_1\}=B(u)\sqcup\{e_2\}$, or $(3)$ $B(u)=B(v)\sqcup\{e_1,e_2\}$, where $e_1$ and $e_2$ are the back-edges in $C$. We call the $4$-cuts in those cases Type-$2i$, Type-$2ii$, and Type-$2iii$, respectively.

All Type-$2i$ and Type-$2iii$ $4$-cuts can be computed explicitly in linear time, and their total number is $O(n)$. On the other hand, the number of Type-$2ii$ $4$-cuts can be as high as $\Omega(n^2)$, and so we cannot compute all of them in linear time. Instead, we compute only a collection of $O(n)$ Type-$2ii$ $4$-cuts, so that the rest of them are implied from this collection. The Type-$2ii$ $4$-cuts are particularly interesting, because their existence is basically the reason that we can have $\Omega(n^2)$ $4$-cuts of Type-$2$ and Type-$3$. More precisely, we show that every Type-$3$ $4$-cut that we have not explicitly computed, is implied by the collection of Type-$3$ $4$-cuts that we have computed, plus that of the Type-$2ii$ $4$-cuts that we have computed.

First, let us consider the Type-$2i$ $4$-cuts. So let $\{(u,p(u)),(v,p(v)),e_1,e_2\}$ be a $4$-cut such that $u$ is a descendant of $v$ and $B(v)=B(u)\sqcup\{e_1,e_2\}$. Then, Lemma~\ref{lemma:case_B(v)=B(u)cup2-cases} shows that there are basically three different cases for the back-edges $e_1$ and $e_2$. That is, $e_1$ and $e_1$ are either $(1)$ the first and second leftmost edges of $v$, or $(2)$ the first leftmost and rightmost edges of $v$, or $(3)$ the first and the second rightmost edges of $v$. In either case, by Lemma~\ref{lemma:case_B(v)=B(u)cup2-inference} we have that $u$ is uniquely determined by $v$ and $e_1,e_2$: that is, $u$ is the lowest proper descendant of $v$ that has $M(u)=M(B(v)\setminus\{e_1,e_2\})$. Thus, the idea is basically to compute all three different values $M(B(v)\setminus\{e_1,e_2\})$, for all different cases $(1)$, $(2)$ and $(3)$. Then, we can precisely determine $u$ in every one of those cases, and then we apply Lemma~\ref{lemma:case_B(v)=B(u)cup2-criterion} in order to verify that we indeed have a $4$-cut. Thus, by Proposition~\ref{proposition:algorithm:type2-1} we have that we can compute all Type-$2i$ $4$-cuts in linear time. Notice that their number is $O(n)$.

Now let us consider the Type-$2iii$ $4$-cuts. So let $\{(u,p(u)),(v,p(v)),e_1,e_2\}$ be a $4$-cut such that $u$ is a descendant of $v$ and $B(u)=B(v)\sqcup\{e_1,e_2\}$. Then Lemma~\ref{lemma:case_B(v)=B(u)cup2} shows that $e_1$ and $e_2$ are completely determined by $u$: i.e., these are the $\mathit{high}_1$ and $\mathit{high}_2$ edges of $u$. Then, by Lemma~\ref{lemma:case_B(v)=B(u)cup2-necessary} we have that $v$ is either the greatest or the second-greatest proper ancestor of $u$ with $M(v)=M(B(u)\setminus\{e_1,e_2\})$. This shows that the number of Type-$2iii$ $4$-cuts is $O(n)$. By Proposition~\ref{proposition:algorithm:type2-3}, we can compute all of them in linear time in total.

\begin{figure}[h!]\centering
\includegraphics[trim={0cm 22cm 0 0}, clip=true, width=0.8\linewidth]{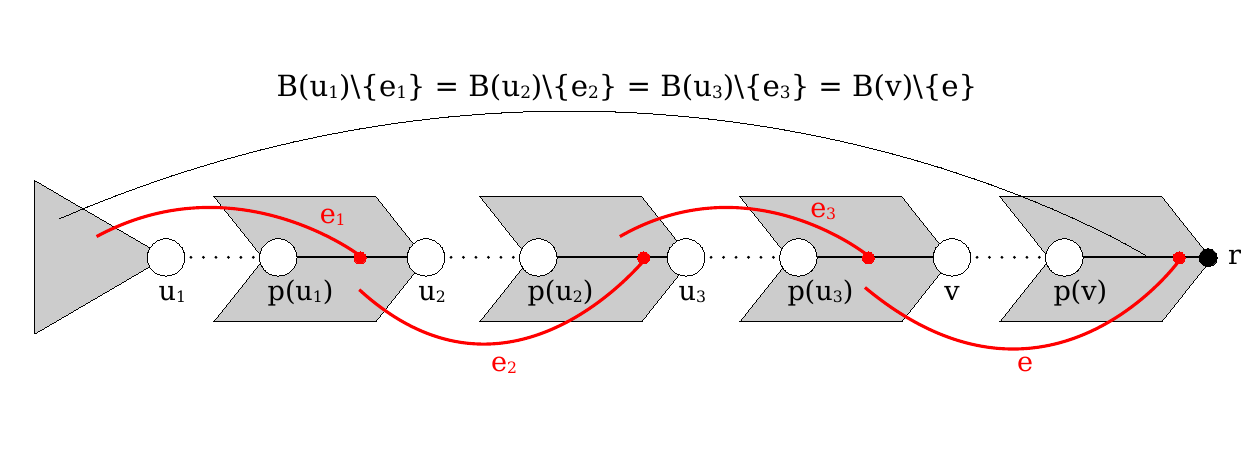}
\caption{\small{With this figure we can see why there can be $\Omega(n^2)$ Type-$2ii$ $4$-cuts in a graph with $n$ vertices. Any of the pairs of edges $\{(v,p(v)),e\}$, $\{(u_1,p(u_1)),e_1\}$, $\{(u_2,p(u_2)),e_2\}$ and $\{(u_3,p(u_3)),e_3\}$ forms a $4$-cut with any of the rest. For example, $\{(u_1,p(u_1)),(v,p(v)),e_1,e\}$ and $\{(u_2,p(u_2)),(u_3,p(u_3)),e_2,e_3\}$ are two $4$-cuts in this figure.}}\label{figure:type2ii}
\end{figure}

Finally, let us consider the Type-$2ii$ $4$-cuts. So let $\{(u,p(u)),(v,p(v)),e_1,e_2\}$ be a $4$-cut such that $u$ is a descendant of $v$ and $B(v)\sqcup\{e_1\}=B(u)\sqcup\{e_2\}$. Then, Lemma~\ref{lemma:case_B(v)cup=B(u)cup-cases} shows that $e_1$ is the $\mathit{high}_1$ edge of $u$, and $e_2$ is either the first leftmost or the first rightmost edge of $v$. Thus, there are two different cases to consider for the back-edge in $B(v)\setminus B(u)$. Let us assume that we have fixed a case for the back-edge $e\in B(v)\setminus B(u)$ (e.g., let $e$ be the first leftmost edge of $v$). 
As we can see in Figure~\ref{figure:type2ii}, the number of proper descendants $u$ of $v$ with the property that $B(v)\sqcup\{e'\}=B(u)\sqcup\{e\}$ can be $\Omega(n)$, and this can be true for $\Omega(n)$ vertices $v$. Thus, the idea is to properly select one such vertex $u$, for every vertex $v$, for every one of the two choices for the back-edge $e$. Thus, we compute $O(n)$ Type-$2ii$ $4$-cuts in total. Furthermore, we can show that one such selection, for every $v$ and $e$, is enough to produce a collection of Type-$2ii$ $4$-cuts that implies all Type-$2ii$ $4$-cuts. We denote the vertex $u$ that we select as $\mathit{lowestU}(v,e)$. As its name suggests, this is the lowest $u$ that has the property that $u$ is a proper descendant of $v$ and there is a back-edge $e'$ such that $B(v)\sqcup\{e'\}=B(u)\sqcup\{e\}$. The reason for selecting this vertex, is that it is convenient to compute. Specifically, for technical reasons we distinguish two cases: either $M(u)=M(B(u)\setminus\{e_\mathit{high}(u)\})$, or $M(u)\neq M(B(u)\setminus\{e_\mathit{high}(u)\})$.
In the first case, by Lemma~\ref{lemma:type-2ii-firstcase-u} we have that $u$ is either the lowest or the second-lowest proper descendant of $v$ such that $M(u)=M(B(v)\setminus\{e\})$. In the second case, by Lemma~\ref{lemma:type-2ii-secondcase-u} we have that $u$ is the lowest proper descendant of $v$ such that $M(u)\neq M(B(u)\setminus\{e_\mathit{high}(u)\})=M(B(v)\setminus\{e\})$. With this information, Proposition~\ref{proposition:type-2-2} establishes that we can compute in linear time a collection $\mathcal{C}$ of $O(n)$ Type-$2ii$ $4$-cuts that implies all Type-$2ii$ $4$-cuts. More precisely, every Type-$2ii$ $4$-cut of the form $\{(u,p(u)),(v,p(v)),e_1,e_2\}$, where $B(v)\sqcup\{e_1\}=B(u)\sqcup\{e_2\}$, is implied by $\mathcal{C}$ through $\{(u,p(u)),e_1\}$ (or equivalently, through $\{(v,p(v)),e_2\}$). This is an important property that allows us also to compute a collection of $O(n)$ Type-$3$ $4$-cuts, that, together with $\mathcal{C}$, implies all Type-$3$ $4$-cuts. 

\subsection{Type-$3$ $4$-cuts}
Let $C$ be a Type-$3$ $4$-cut, and let $(u,p(u))$, $(v,p(v))$ and $(w,p(w))$ be the tree-edges in $C$. We may assume w.l.o.g. that $u>v>w$. Then, Lemma~\ref{lemma:type3-4cuts} implies that $w$ is a common ancestor of $u$ and $v$. Thus, we distinguish two cases: either $(1)$ $u$ and $v$ are not related as ancestor and descendant, or $(2)$ $u$ and $v$ are related as ancestor and descendant. In case $(1)$, we call $C$ a Type-$3\alpha$ $4$-cut. In case $(2)$, we call $C$ a Type-$3\beta$ $4$-cut. Both of these cases are much more involved than the case of Type-$2$ $4$-cuts, but the case of Type-$3\beta$ $4$-cuts is the most challenging.

\subsubsection{Type-$3\alpha$ $4$-cuts}
\begin{figure}[h!]\centering
\includegraphics[trim={0cm 3cm 0 0}, clip=true, width=0.7\linewidth]{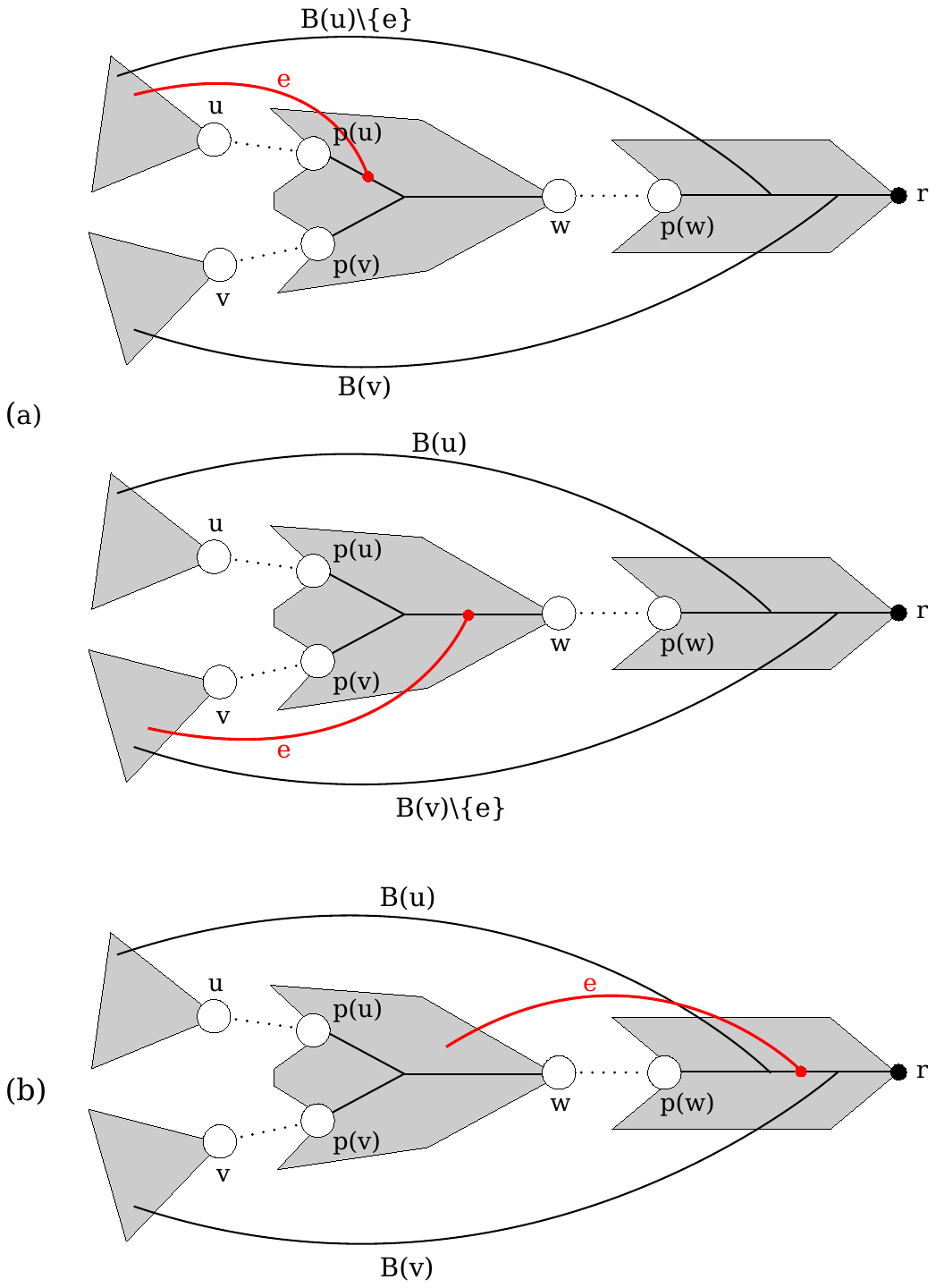}
\caption{\small{The two different cases of a Type-$3\alpha$ $4$-cut $\{(u,p(u)),(v,p(v)),(w,p(w)),e\}$. In $(a)$ we have $e\in B(u)\cup B(v)$ and $B(w)\sqcup\{e\}=B(u)\sqcup B(v)$. Although this implies that $e$ is either in $B(u)$ or in $B(v)$, these cases are symmetric. In $(b)$ we have $B(w)=(B(u)\sqcup B(v))\sqcup\{e\}$. }}\label{figure:type3a}
\end{figure}

Let $C=\{(u,p(u)),(v,p(v)),(w,p(w)),e\}$ be a Type-$3\alpha$ $4$-cut, where $w$ is a common ancestor of $u$ and $v$. Then Lemma~\ref{lemma:type-3a-types} implies that either $(i)$ $e\in B(u)\cup B(v)$ and $B(w)\sqcup\{e\}=B(u)\sqcup B(v)$, or $(ii)$ $B(w)=(B(u)\sqcup B(v))\sqcup\{e\}$ (see Figure~\ref{figure:type3a}). In case $(i)$, $C$ is called a Type-$3\alpha i$ $4$-cut. In case $(ii)$, $C$ is called a Type-$3\alpha ii$ $4$-cut. We treat those cases differently.

\noindent\\
\textbf{Type-$3\alpha i$ $4$-cuts}\\

Let $C=\{(u,p(u)),(v,p(v)),(w,p(w)),e\}$ be a Type-$3\alpha i$ $4$-cut, where $w$ is a common ancestor of $u$ and $v$. Then we have $e\in B(u)\cup B(v)$ and $B(w)\sqcup\{e\}=B(u)\sqcup B(v)$. We may assume w.l.o.g. that $e\in B(u)$. Then Lemma~\ref{lemma:type-3a-i-M} implies that $e=e_\mathit{high}(u)$. By Lemma~\ref{lemma:type-3a-lowchild_desc} we have that one of $u$ and $v$ is a descendant of the $\mathit{low1}$ child of $M(w)$, and the other is a descendant of the $\mathit{low2}$ child of $M(w)$. Either of those cases can be true, regardless of whether $e$ is in $B(u)$ or $B(v)$. So let us assume that $u$ is a descendant of the $\mathit{low1}$ child $c_1$ of $M(w)$, and $v$ is a descendant of the $\mathit{low2}$ child $c_2$ of $M(w)$. Then Lemma~\ref{lemma:type-3a-i-v-lowest} implies that $v$ is the lowest proper descendant of $w$ with $M(v)=M(w,c_2)$.  

Concerning the higher endpoint of $e$, we distinguish two cases, depending on whether $M(u)=M(B(u)\setminus\{e_\mathit{high}(u)\})$ or $M(u)\neq M(B(u)\setminus\{e_\mathit{high}(u)\})$. In the first case, we can compute all such $4$-cuts in linear time, because Lemma~\ref{lemma:type-3a-i-special} implies that $u$ is either the lowest or the second-lowest proper descendant of $w$ such that $M(u)=M(w,c_1)$. Thus, given $w$, we have that $v$ is completely determined, and there are only two options for $u$. This implies that the number of those $4$-cuts is $O(n)$, and by Proposition~\ref{proposition:algorithm:type3-a-i-special} we can compute all of them in linear time in total.

\begin{figure}[h!]\centering
\includegraphics[trim={0cm 31cm 0 0}, clip=true, width=1\linewidth]{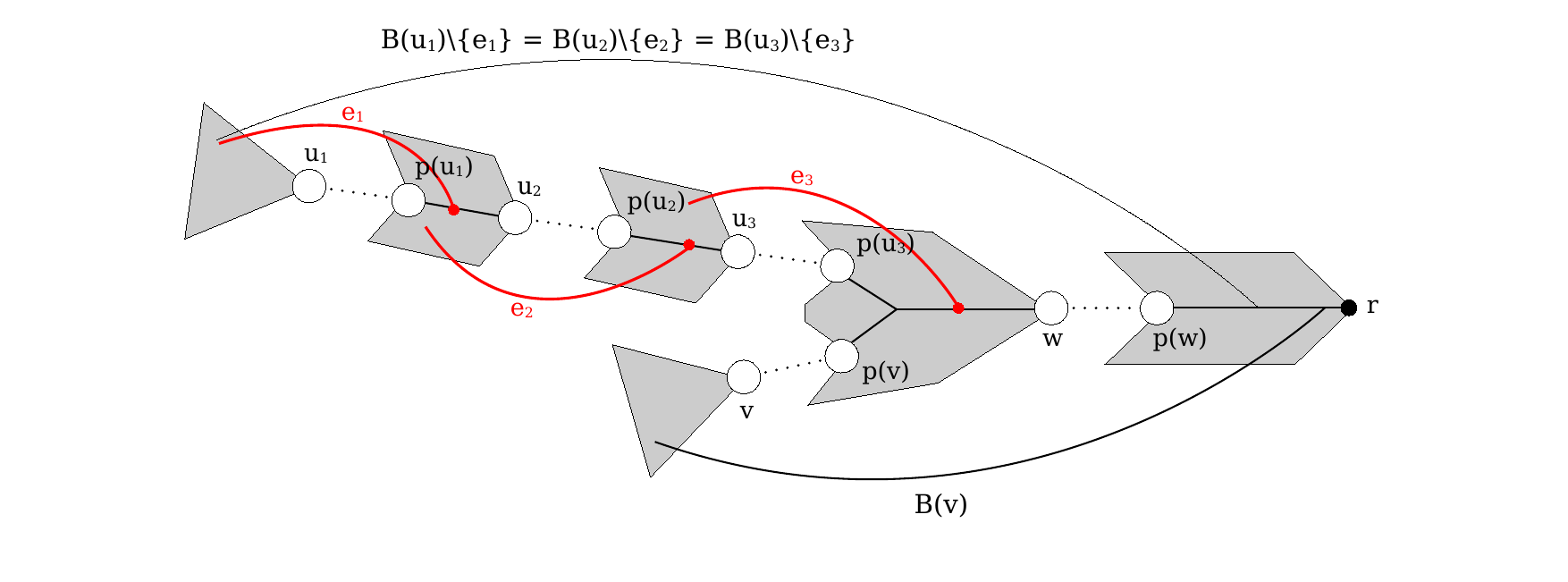}
\caption{\small{With this figure we can see why the number of Type-$3\alpha i$ $4$-cuts can be $\Omega(n^2)$. For a particular $w$, there may be a sequence $(u_1,v),(u_2,v),\dots$ of $\Omega(n)$ pairs of vertices, and a corresponding sequence of back-edges $e_1,e_2,\dots$, such that $e_i\in B(u_i)$ and $B(w)=(B(u_i)\setminus\{e_i\})\sqcup B(v)$, for every $i=1,2,\dots$ -- and this can be true for $\Omega(n)$ vertices $w$. In this example, we have that $C_i=\{(u_i,p(u_i)),(v,p(v)),(w,p(w)),e_i\}$ is a $4$-cut, for every $i\in\{1,2,3\}$. We have $e_i=e_\mathit{high}(u_i)$ for every $i\in\{1,2,3\}$, $M(u_2)\in T(u_2)\setminus T(u_1)$, $M(u_3)\in T(u_3)\setminus T(u_2)$, and $M(B(u_3)\setminus\{e_3\})=M(B(u_2)\setminus\{e_2\})=M(B(u_1)\setminus\{e_1\})\in T(u_1)$. This implies that $M(u_3)\neq M(B(u_3)\setminus\{e_\mathit{high}(u_3)\})$ and $M(u_2)\neq M(B(u_2)\setminus\{e_\mathit{high}(u_2)\})$. Notice that it is enough to have computed the collection $\mathcal{C}=\{\{(u_1,p(u_1)),(u_2,p(u_2)),e_1,e_2\}, \{(u_2,p(u_2)),(u_3,p(u_3)),e_2,e_3\}\}$ of Type-$2ii$ $4$-cuts, and the $4$-cut $C_3$. Then, $C_1$ and $C_2$ are implied from $\mathcal{C}\cup\{C_3\}$.}}\label{figure:type3ai}
\end{figure}

In the case where $M(u)\neq M(B(u)\setminus\{e_\mathit{high}(u)\})$, the number of $4$-cuts can be $\Omega(n^2)$, as shown in Figure~\ref{figure:type3ai}. This is because, although for fixed $w$ we have that $v$ is determined, there may be $\Omega(n)$ options for $u$, and this may be true for $\Omega(n)$ vertices $w$. Then the idea is basically the same as that for computing the Type-$2ii$ $4$-cuts: we only select a proper $u$ for every $w$. Specifically, we select the lowest proper descendant $u$ of $w$ such that $M(u)\neq M(B(u)\setminus\{e_\mathit{high}(u)\})=M(w,c_1)$. It turns out that this is sufficient, in the sense that the $4$-cuts of this type that we compute, are able to imply, together with the collection of Type-$2ii$ $4$-cuts that we have computed, all $4$-cuts of this type. This result is given in Proposition~\ref{proposition:type-3-a-i-non-special}.

After this procedure, we may simply reverse the roles of $u$ and $v$. Thus, we assume that the descendant $u$ of $w$ that has $e\in B(u)$ is a descendant of the $\mathit{low2}$ child of $M(w)$, and $v$ is a descendant of the $\mathit{low1}$ child of $M(w)$. Then we follow a similar procedure to compute all $4$-cuts of this type. The arguments are essentially the same.

\noindent\\
\textbf{Type-$3\alpha ii$ $4$-cuts}\\

\begin{figure}[h!]\centering
\includegraphics[trim={0cm 4cm 0 0}, clip=true, width=0.8\linewidth]{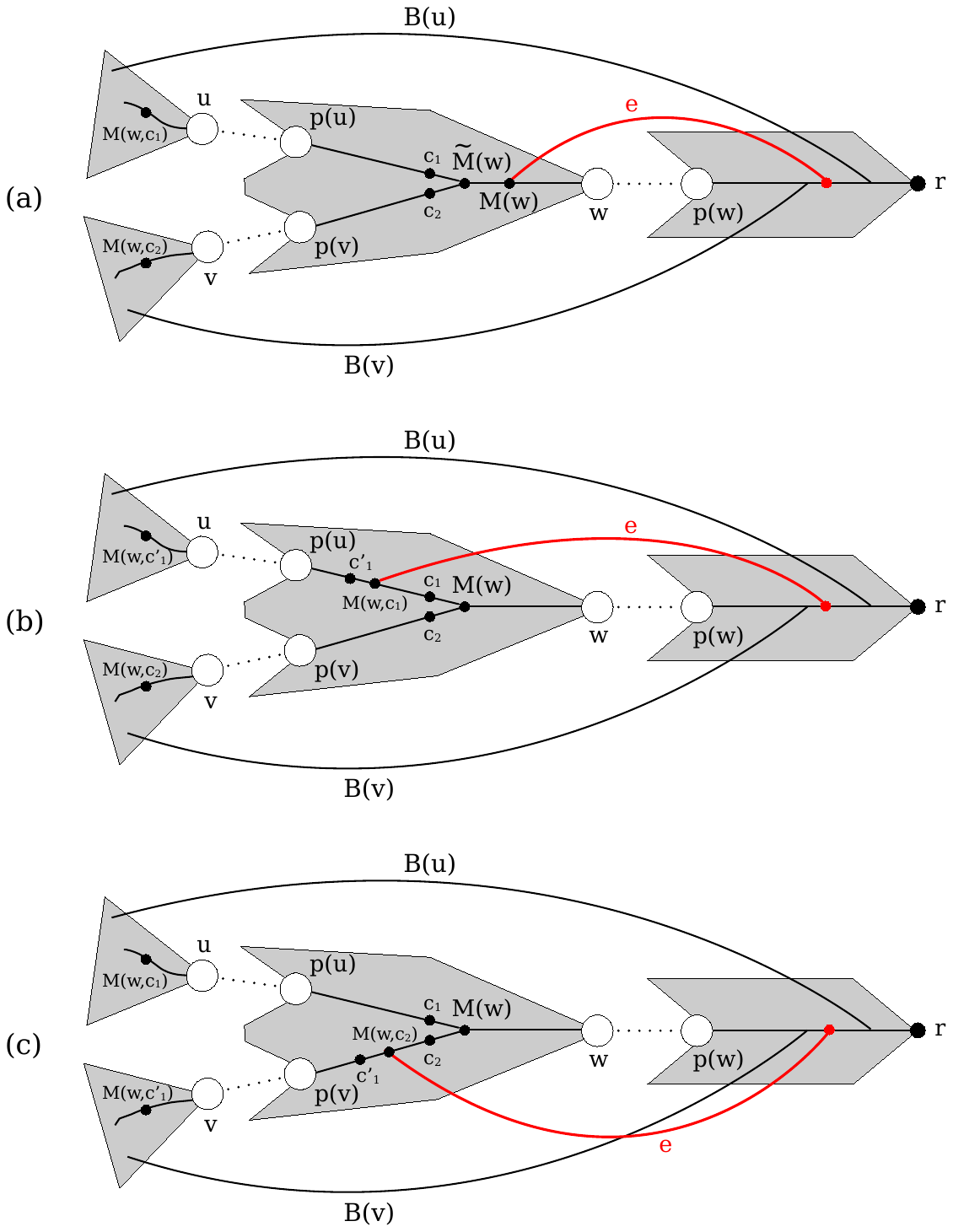}
\caption{\small{$(a)$-$(c)$ correspond to cases $(1)$-$(3)$ of Lemma~\ref{lemma:type-3-a-ii-cases}. These are the cases in which the higher endpoint of $e$ is related as ancestor and descendant with either $u$ or $v$. In $(a)$, the higher endpoint of $e$ is an ancestor of both $u$ and $v$. $c_1$ and $c_2$ are the $\mathit{low1}$ and $\mathit{low2}$ children of $\widetilde{M}(w)$ (not necessarily in that order). We have $M(u)=M(w,c_1)$ and $M(v)=M(w,c_2)$. In $(b)$, the higher endpoint of $e$ is an ancestor of $u$, but not of $v$. $c_1$ and $c_2$ are the $\mathit{low1}$ and $\mathit{low2}$ children of $M(w)$ (not necessarily in that order). We have $M(u)=M(w,c_1')$ and $M(v)=M(w,c_2)$, where $c_1'$ is the $\mathit{low1}$ child of $M(w,c_1)$. In $(c)$, the higher endpoint of $e$ is an ancestor of $v$, but not of $u$. We note that cases $(b)$ and $(c)$ are essentially equivalent; to see this, just switch the labels of $u$ and $v$.}}\label{figure:type3aii}
\end{figure}

\begin{figure}[h!]\centering
\includegraphics[trim={0cm 2.5cm 0 0}, clip=true, width=0.8\linewidth]{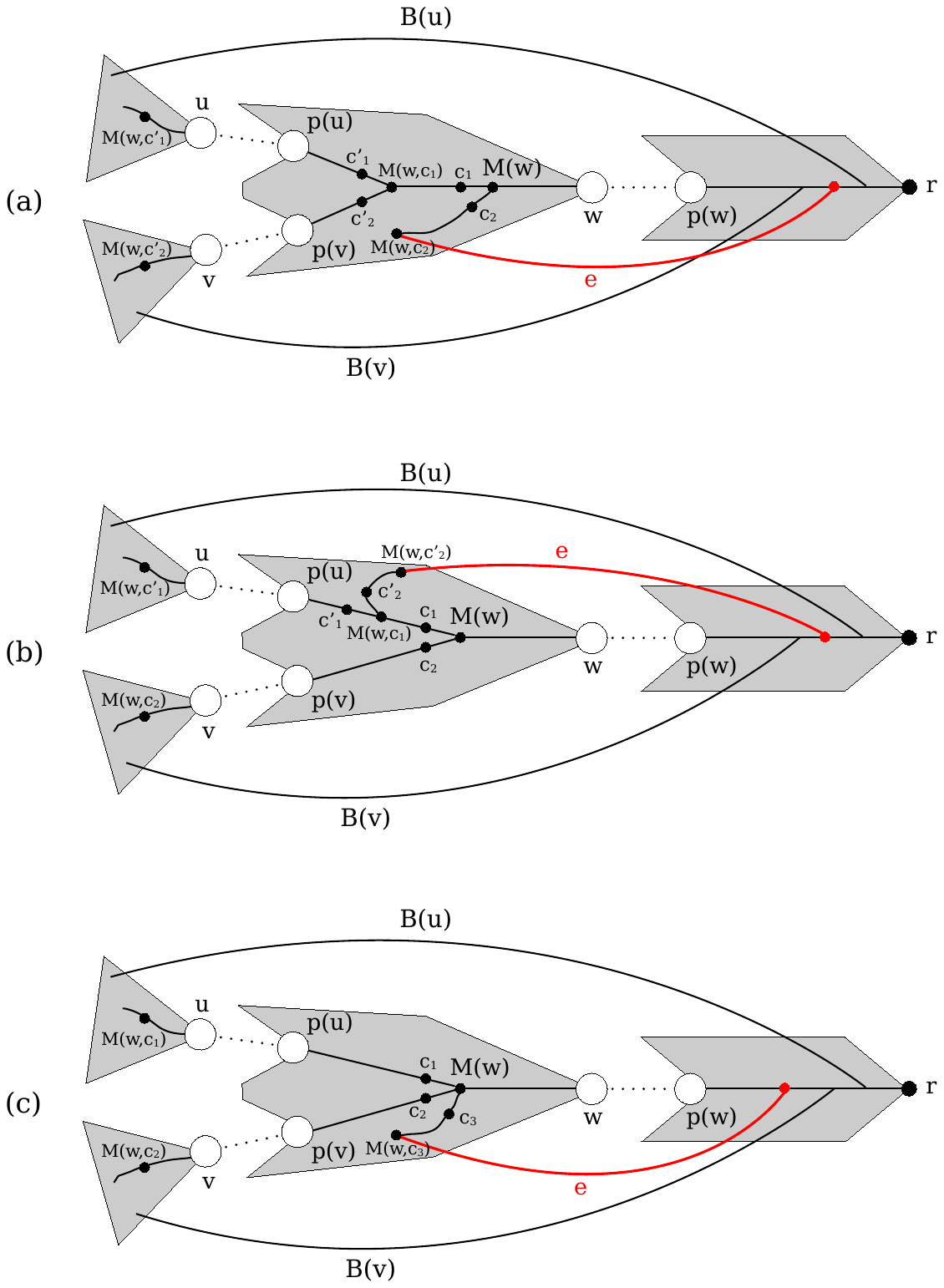}
\caption{\small{$(a)$ and $(b)$ correspond to case $(4.1)$ of Lemma~\ref{lemma:type-3-a-ii-cases}, and $(c)$ corresponds to case $(4.2)$ of Lemma~\ref{lemma:type-3-a-ii-cases}. These are the cases in which the higher endpoint of $e$ is not related as ancestor and descendant with $u$ and $v$. In $(a)$, we get all the different possibilities for case $(4.1.1)$ of Lemma~\ref{lemma:type-3-a-ii-case4.1} by swapping the labels $c_1',c_2'$. In $(b)$, we get all the different possibilities for cases $(4.1.2)$ and $(4.1.3)$ of Lemma~\ref{lemma:type-3-a-ii-case4.1} by swapping the labels $c_1,c_2$ and $c_1',c_2'$. In $(c)$, we get all the different possibilities for case $(4.2)$ of Lemma~\ref{lemma:type-3-a-ii-cases} by permuting the labels $c_1$, $c_2$ and $c_3$.}}\label{figure:type3aii-4-1}
\end{figure}

Let $C=\{(u,p(u)),(v,p(v)),(w,p(w)),e\}$ be a Type-$3\alpha ii$ $4$-cut, where $w$ is a common ancestor of $u$ and $v$. Then we have  $B(w)=(B(u)\sqcup B(v))\sqcup \{e\}$. Let $e=(x,y)$. Then there are various cases to consider, according to the relation of $x$ with $u$ and $v$. Specifically, by Lemma~\ref{lemma:type-3-a-ii-cases} we have the following cases (see Figure~\ref{figure:type3aii} for cases $(1)$-$(3)$, and Figure~\ref{figure:type3aii-4-1} for case $(4)$).

\begin{enumerate}[label*=(\arabic*)]
\item{$x$ is an ancestor of both $u$ and $v$.}
\item{$x$ is an ancestor of $u$, but not an ancestor of $v$.}
\item{$x$ is an ancestor of $v$, but not an ancestor of $u$.}
\item{$x$ is neither an ancestor of $u$ nor an ancestor of $v$.}
\end{enumerate}

In any case, Lemma~\ref{lemma:type-3-a-ii-inference} implies that $u$ and $v$ are uniquely determined by $w$, and therefore the number of all Type-$3\alpha ii$ $4$-cuts is $O(n)$. Furthermore, in any case we have $y=l(x)$.

Now, in case $(1)$, by Lemma~\ref{lemma:type-3-a-ii-case1} we have that $x=M(w)$, one of $u$ and $v$ is a descendant of the $\mathit{low1}$ child $c_1$ of $\widetilde{M}(w)$, and the other is a descendant of the $\mathit{low2}$ child $c_2$ of $\widetilde{M}(w)$. Since these cases are symmetric, we may assume w.l.o.g. that $u$ is a descendant of $c_1$ and $v$ is a descendant of $c_2$. Then Lemma~\ref{lemma:type-3-a-ii-case1} implies that $M(u)=M(w,c_1)$ and $M(v)=M(w,c_2)$. Then, by Lemma~\ref{lemma:type-3-a-ii-inference} we can determine precisely $u$ and $v$. Thus, by Proposition~\ref{proposition:algorithm:type3-a-ii-1} we can compute all those $4$-cuts in linear time in total.

Notice that cases $(2)$ and $(3)$ are essentially the same (to see this, just switch the labels of $u$ and $v$). So let us consider case $(2)$.
Then by Lemma~\ref{lemma:type-3-a-ii-cases} we have that one of $x$ and $v$ is a descendant of the $\mathit{low1}$ child $c_1$ of $M(w)$, and the other is a descendant of the $\mathit{low2}$ child $c_2$ of $M(w)$. Thus, w.l.o.g. we may assume that $x$ is a descendant of $c_1$ and $v$ is a descendant of $c_2$. Then Lemma~\ref{lemma:type-3-a-ii-case2} implies that $x=M(w,c_1)$, $M(u)=M(w,c_1')$ and $M(v)=M(w,c_2)$, where $c_1'$ is the $\mathit{low1}$ child of $M(w,c_1)$. Then, by Lemma~\ref{lemma:type-3-a-ii-inference} we can determine precisely $u$ and $v$. Thus, by Proposition~\ref{proposition:algorithm:type3-a-ii-2} we can compute all those $4$-cuts in linear time in total.

Now let us consider case $(4)$. According to Lemma~\ref{lemma:type-3-a-ii-cases}, this case is further subdivided into the following two cases (see Figure~\ref{figure:type3aii-4-1}).

\begin{enumerate}[label*={}]
\item{(4.1) Two of $\{u,v,x\}$ are descendants of the $\mathit{low1}$ child of $M(w)$ and the other is a descendant of the $\mathit{low2}$ child of $M(w)$, or reversely: two of $\{u,v,x\}$ are descendants of the $\mathit{low2}$ child of $M(w)$ and the other is a descendant of the $\mathit{low1}$ child of $M(w)$.}
\item{(4.2) There is a permutation $\sigma$ of $\{1,2,3\}$ such that $u$ is a descendant of the $\mathit{low\sigma_1}$ child of $M(w)$, $v$ is a descendant of the $\mathit{low\sigma_2}$ child of $M(w)$, and $x$ is a descendant of the $\mathit{low\sigma_3}$ child of $M(w)$.}
\end{enumerate}

Let us consider case $(4.1)$ first. Let $c_1$ be the $\mathit{low1}$ child of $M(w)$, and let $c_2$ be the $\mathit{low2}$ child of $M(w)$. Let us assume that two of $\{u,v,x\}$ are descendants of $c_1$, and the other is a descendant of $c_2$. (The reverse case is treated similarly.) Now let $c_1'$ be the $\mathit{low1}$ child of $M(w,c_1)$, and let $c_2'$ be the $\mathit{low2}$ child of $M(w,c_1)$. Then Lemma~\ref{lemma:type-3-a-ii-case4.1} implies that we have the following three subcases.

\begin{enumerate}[label*={}]
\item{(4.1.1) $u$ and $v$ are descendants of $c_1$, and $x$ is a descendant of $c_2$.}
\item{(4.1.2) $u$ and $x$ are descendants of $c_1$, and $v$ is a descendant of $c_2$.}
\item{(4.1.3) $v$ and $x$ are descendants of $c_1$, and $u$ is a descendant of $c_2$.}
\end{enumerate}

In case $(4.1.1)$ we have $M(u)=M(w,c_1')$ and $M(v)=M(w,c_2')$ (or reversely), and $x=M(w,c_2)$.
In case $(4.1.2)$ we have $M(u)=M(w,c_1')$ and $x=M(w,c_2')$ (or reversely), and $M(v)=M(w,c_2)$.
And in case $(4.1.3)$ we have $M(v)=M(w,c_1')$ and $x=M(w,c_2')$ (or reversely), and $M(u)=M(w,c_2)$.

Thus, we have to consider all the different possibilities (which are $O(1)$ in total), in order to find all $4$-cuts of this type. In either case, by Lemma~\ref{lemma:type-3-a-ii-inference} we can determine precisely $u$ and $v$. Thus, by Proposition~\ref{proposition:algorithm:type3-a-ii-4-1} we can compute all those $4$-cuts in linear time in total. 

Finally, let us consider case $(4.2)$. Let $c_1$, $c_2$ and $c_3$ be the $\mathit{low1}$, $\mathit{low2}$ and $\mathit{low3}$ children of $M(w)$, respectively. Thus, there is a permutation $\sigma$ of $\{1,2,3\}$ such that $u$ is a descendant of $c_{\sigma(1)}$, $v$ is a descendant of $c_\mathit{\sigma(2)}$ and $x$ is a descendant of $c_\mathit{\sigma(3)}$. By Lemma~\ref{lemma:type-3-a-ii-4.2} we have that $M(u)=M(w,c_{\sigma(1)})$, $M(v)=M(w,c_{\sigma(2)})$ and $x=M(w,c_{\sigma(3)})$. Thus, we consider all the different combinations for $\sigma$, and in each case we can determine precisely $u$ and $v$ by Lemma~\ref{lemma:type-3-a-ii-inference}. Proposition~\ref{proposition:algorithm:type3-a-ii-4-2} establishes that we can compute all those $4$-cuts in linear time in total. 

\subsubsection{Type-$3\beta$ $4$-cuts}
Let $C=\{(u,p(u)),(v,p(v)),(w,p(w)),e\}$ be a Type-$3\beta$ $4$-cut of $G$, where $u$ is a descendant of $v$, and $v$ is a descendant of $w$. Then Lemma~\ref{lemma:no-common-three-edges} implies that $e$ is the unique back-edge with the property that $\{(u,p(u)),(v,p(v)),(w,p(w)),e\}$ is a $4$-cut of $G$. Thus, we say that $(u,v,w)$ \emph{induces} the $4$-cut $C$. Whenever we say that a triple of vertices $(u,v,w)$ induces a $4$-cut, we always assume that $u$ is a proper descendant of $v$, and $v$ is a proper descendant of $w$.

\begin{figure}[h!]\centering
\includegraphics[trim={0cm 4cm 0 0}, clip=true, width=0.8\linewidth]{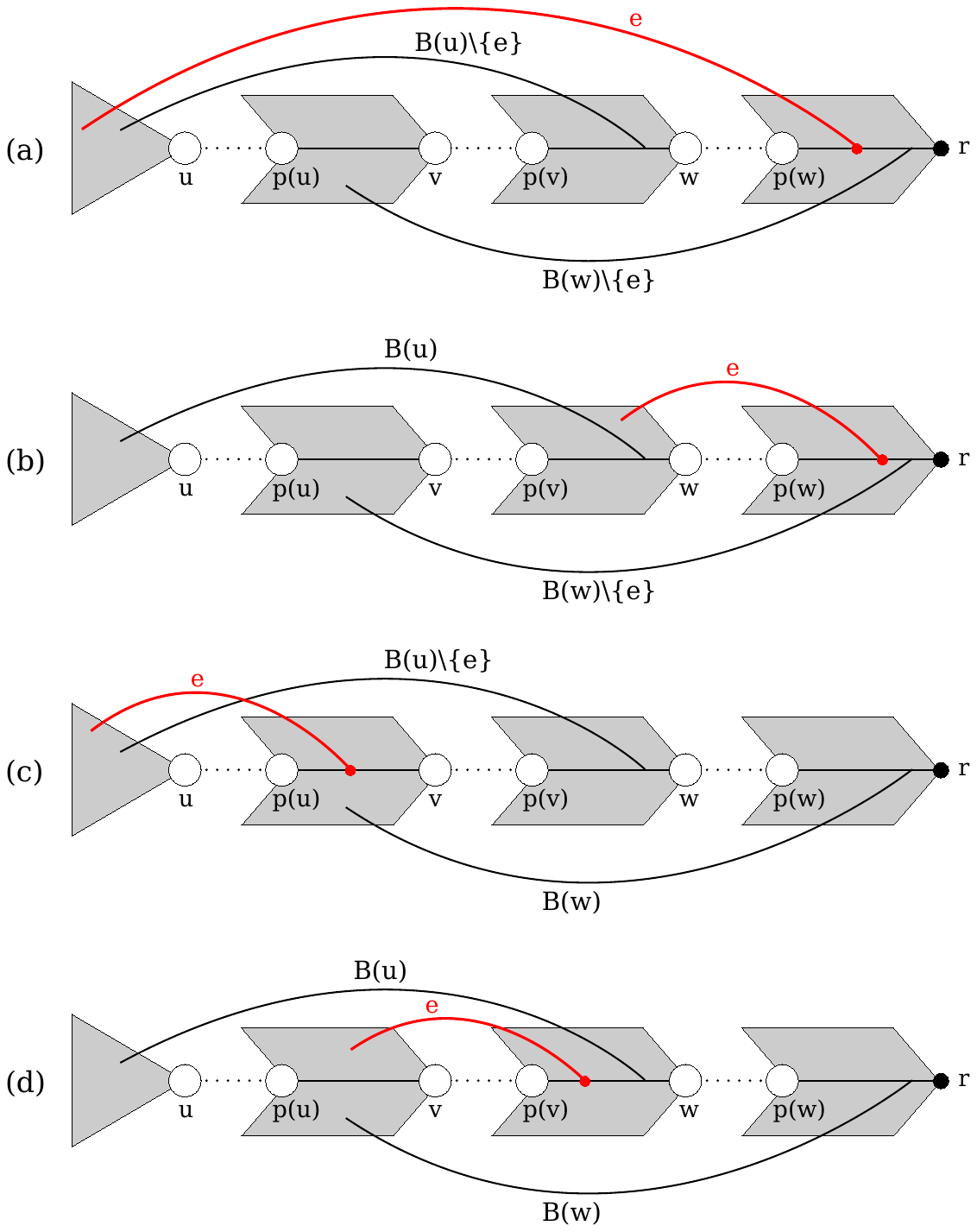}
\caption{\small{All different cases of Type-$3\beta$ $4$-cuts. $(a)$-$(d)$ correspond to cases $(1)$-$(4)$ of Lemma~\ref{lemma:type-3b-cases}. In $(a)$ we have $e\in B(u)\cap B(v)\cap B(w)$ and $B(v)\setminus\{e\}=(B(u)\setminus\{e\})\sqcup(B(w)\setminus\{e\})$. In $(b)$ we have $e\in B(w)$, $e\notin B(v)\cup B(u)$, and $B(v)=B(u)\sqcup(B(w)\setminus\{e\})$. In $(c)$ we have $e\in B(u)$, $e\notin B(v)\cup B(w)$, and $B(v)=(B(u)\setminus\{e\})\sqcup B(w)$. In $(d)$ we have $e\in B(v)$ and $B(v)=(B(u)\sqcup B(w))\sqcup\{e\}$.}}\label{figure:type3b}
\end{figure}

Since $C$ is a Type-$3\beta$ $4$-cut, Lemma~\ref{lemma:type-3b-cases} implies that there are four distinct cases to consider (see Figure~\ref{figure:type3b}):

\begin{enumerate}[label=(\arabic*)]
\item{$e\in B(u)\cap B(v)\cap B(w)$ and $B(v)\setminus\{e\}=(B(u)\setminus\{e\})\sqcup(B(w)\setminus\{e\})$.}
\item{$e\in B(w)$, $e\notin B(v)\cup B(u)$, and $B(v)=B(u)\sqcup(B(w)\setminus\{e\})$.}
\item{$e\in B(u)$, $e\notin B(v)\cup B(w)$, and $B(v)=(B(u)\setminus\{e\})\sqcup B(w)$.}
\item{$e\in B(v)$ and $B(v)=(B(u)\sqcup B(w))\sqcup\{e\}$.}
\end{enumerate}
For technical reasons, we make a distinction into Type-$3\beta i$ and Type-$3\beta ii$ $4$-cuts.

In case $(1)$ we have $B(w)\setminus\{e\}\subset B(v)\setminus\{e\}$, and therefore $M(B(w)\setminus\{e\})$ is a descendant of $M(B(v)\setminus\{e\})$. If $M(B(w)\setminus\{e\})\neq M(B(v)\setminus\{e\})$, then we say that $C$ is a Type-$3\beta i$ $4$-cut. 
Otherwise, we say that $C$ is a Type-$3\beta ii$ $4$-cut. 

In case $(2)$ we have $B(w)\setminus\{e\}\subset B(v)$, and therefore $M(B(w)\setminus\{e\})$ is a descendant of $M(v)$. If $M(B(w)\setminus\{e\})\neq M(v)$, then we say that $C$ is a Type-$3\beta i$ $4$-cut. 
Otherwise, we say that $C$ is a Type-$3\beta ii$ $4$-cut. 

In case $(3)$ we have $B(w)\subset B(v)$, and therefore $M(w)$ is a descendant of $M(v)$. If $M(w)\neq M(v)$, then we say that $C$ is a Type-$3\beta i$ $4$-cut. 
Otherwise, we say that $C$ is a Type-$3\beta ii$ $4$-cut. 

In case $(4)$ we have $B(w)\subset B(v)\setminus\{e\}$, and therefore $M(w)$ is a descendant of $M(B(v)\setminus\{e\})$. If $M(w)\neq M(B(v)\setminus\{e\})$, then we say that $C$ is a Type-$3\beta i$ $4$-cut. 
Otherwise, we say that $C$ is a Type-$3\beta ii$ $4$-cut.


In each of those cases, the Type-$3\beta i$ $4$-cuts are easier to compute than the respective Type-$3\beta ii$ $4$-cuts. This is because we have more information in order to determine (some possible values of) $u$ and $w$ given $v$. More specifically, given $v$, we have that one of $u$ and $w$ is completely determined, and only the other may vary, but only in a very orderly manner. For Type-$3\beta ii$ $4$-cuts, many possible combinations of pairs $u$ and $w$ may exist, given $v$, and this makes things much more complicated. 

When we consider case $(4)$ of Lemma~\ref{lemma:type-3b-cases}, in either Type-$3\beta i$ or Type-$3\beta ii$ $4$-cuts, we perceive a distinct difficulty (which is significantly more involved for Type-$3\beta ii$ $4$-cuts). This is because the back-edge $e$ leaps over $v$, which is ``between" $u$ and $w$ (i.e., $v$ is an ancestor of $u$, but also a descendant of $w$). This forces us to distinguish several subcases, by considering the different possibilities for the higher or the lower endpoint of $e$. (Whereas, in the previous cases, we have some predetermined options for $e$, which then allow us to compute either $u$ or $w$.) In some of those subcases, we cannot even identify beforehand the endpoints of $e$, and we can only retrieve them after having first computed both $u$ and $w$ (see Lemma~\ref{lemma:type-3-b-i-4-edge}). 

\noindent\\
\textbf{Type-$3\beta i$ $4$-cuts}\\

First, let us consider case $(1)$ of Lemma~\ref{lemma:type-3b-cases}. Then by Lemma~\ref{lemma:type-3-b-i-1-info} we have that $u$ is the lowest proper descendant of $v$ with $M(u)=M(v,c)$, where $c$ is either the $\mathit{low1}$ or the $\mathit{low2}$ child of $M(v)$. Furthermore, we have that $e$ is the $\mathit{low}$ edge of $u$, and $M(w)=M(v)$. Also, $w$ satisfies $\mathit{bcount}(w)=\mathit{bcount}(v)-\mathit{bcount}(u)+1$. Thus, since $M(w)=M(v)$, we have that $w$ is completely determined by this property (because the vertices with the same $M$ point have distinct $\mathit{bcount}$). Then, notice that the number of those $4$-cuts is $O(n)$ (because $u$ and $w$ are completely determined by $v$). Proposition~\ref{proposition:algorithm:type-3-b-i-1} shows that we can compute all of them in linear time in total.

\begin{figure}[h!]\centering
\includegraphics[trim={2cm 34cm 0 0}, clip=true, width=1.1\linewidth]{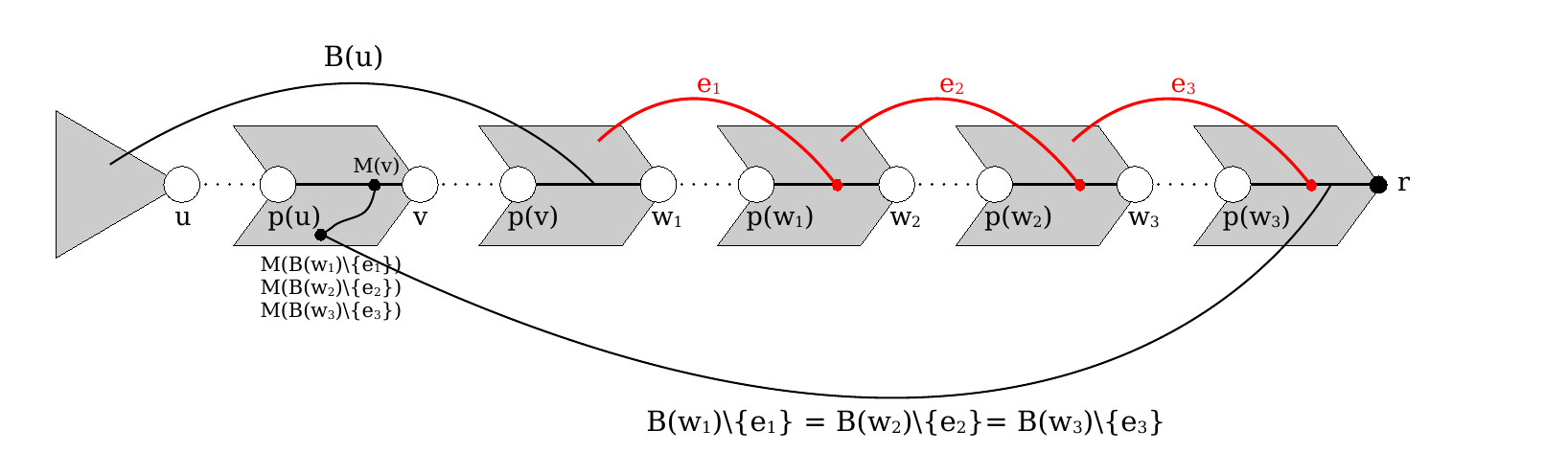}
\caption{\small{Here we have that $\{(u,p(u)),(v,p(v)),(w_i,p(w_i)),e_i\}$ is a $4$-cut, for every $i\in\{1,2,3\}$. This example shows why the number of Type-$3\beta i$ $4$-cuts that satisfy $(2)$ of Lemma~\ref{lemma:type-3b-cases} can be $\Omega(n^2)$. For a particular $v$, we can have $\Omega(n)$ vertices $w$ such that $B(v)=B(u)\sqcup(B(w)\setminus\{e\})$, for a vertex $u$ and a back-edge $e$, and this can be true for $\Omega(n)$ vertices $v$. However, notice that it is sufficient to have computed only $\{(u,p(u)),(v,p(v)),(w_1,p(w_1)),e_1\}$ and the Type-$2ii$ $4$-cuts $\{(w_1,p(w_1)),(w_2,p(w_2)),e_1,e_2\}$ and $\{(w_2,p(w_2)),(w_3,p(w_3)),e_2,e_3\}$, because the remaining $4$-cuts are implied from this selection.}}\label{figure:type3bi-2}
\end{figure}

Now let us consider case $(2)$ of Lemma~\ref{lemma:type-3b-cases}. The number of those $4$-cuts can be $\Omega(n^2)$, as shown in Figure~\ref{figure:type3bi-2}, and so we only compute a subcollection of them, that, together with the collection of Type-$2ii$ $4$-cuts that we have computed, implies all $4$-cuts of this kind.
Specifically, let $c_1$ and $c_2$ be the $\mathit{low1}$ and $\mathit{low2}$ child of $M(v)$, respectively. Then by Lemma~\ref{lemma:type-3-b-i-2-info} we have that $u$ is the lowest proper descendant of $v$ such that $M(u)=M(v,c_2)$. Also, we have that $w$ is an ancestor of $\mathit{low}(u)$ and $M(w)\neq M(B(w)\setminus\{e\})=M(v,c_1)$. Then, Lemma~\ref{lemma:type-3-b-i-2-implied} implies that it is sufficient to have computed the greatest ancestor $w'$ of $\mathit{low}(u)$ for which there is a back-edge $e'\in B(w')$ such that $M(w')\neq M(B(w')\setminus\{e'\})=M(v,c_1)$. Thus, for every vertex $v$, there is only a specific pair of vertices $u$ and $w$ that we have to check, and so the number of $4$-cuts that we will collect is $O(n)$. Proposition~\ref{proposition:algorithm:type-3-b-i-2} establishes that this collection, plus that of the Type-$2ii$ $4$-cuts that we have computed, is enough in order to imply all $4$-cuts of this kind. Furthermore, we can compute all of them in linear time in total.

\begin{figure}[h!]\centering
\includegraphics[trim={2cm 35cm 0 0}, clip=true, width=1.1\linewidth]{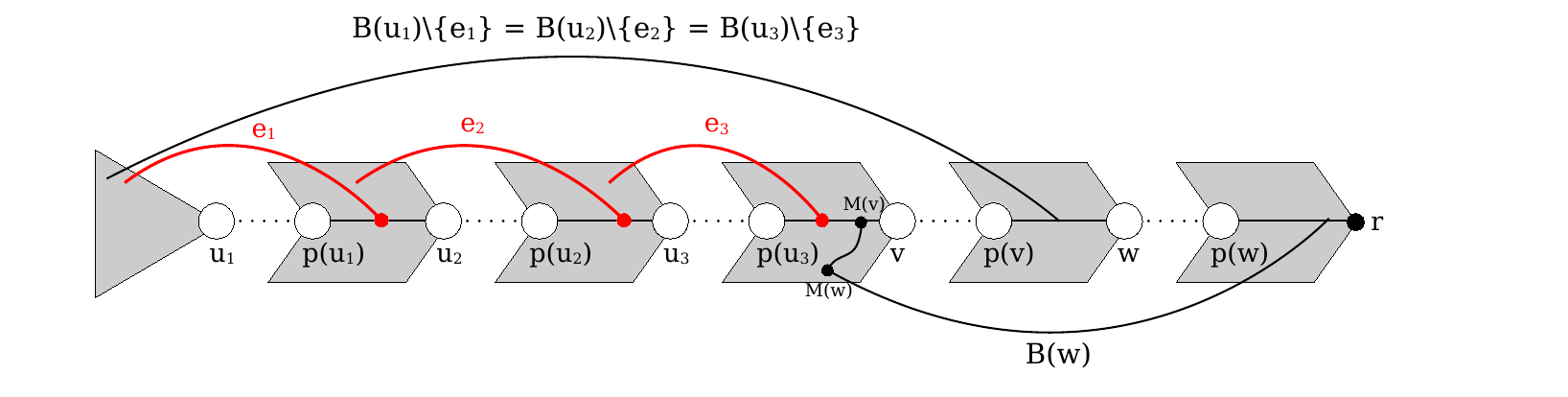}
\caption{\small{Here we have that $\{(u_i,p(u_i)),(v,p(v)),(w,p(w)),e_i\}$ is a $4$-cut, for every $i\in\{1,2,3\}$. This example shows why the number of Type-$3\beta i$ $4$-cuts that satisfy $(3)$ of Lemma~\ref{lemma:type-3b-cases} can be $\Omega(n^2)$. For a particular $v$, we can have $\Omega(n)$ vertices $u$ such that $B(v)=(B(u)\setminus\{e\})\sqcup B(w)$, for a vertex $w$ and a back-edge $e$, and this can be true for $\Omega(n)$ vertices $v$. However, notice that it is sufficient to have computed only $\{(u_3,p(u_3)),(v,p(v)),(w,p(w)),e_3\}$ and the Type-$2ii$ $4$-cuts $\{(u_1,p(u_1)),(u_2,p(u_2)),e_1,e_2\}$ and $\{(u_2,p(u_2)),(u_3,p(u_3)),e_2,e_3\}$, because the remaining $4$-cuts are implied from this selection. Notice that $e_i=e_\mathit{high}(u_i)$, for $i\in\{1,2,3\}$, $M(u_2)\in T(u_2)\setminus T(u_1)$, $M(u_3)\in T(u_3)\setminus T(u_2)$, and $M(B(u_1)\setminus\{e_1\})=M(B(u_2)\setminus\{e_2\})=M(B(u_3)\setminus\{e_3\})$.}}\label{figure:type3bi-3}
\end{figure}

Now let us consider case $(3)$ of Lemma~\ref{lemma:type-3b-cases}. Let $c_1$ and $c_2$ be the $\mathit{low1}$ and the $\mathit{low2}$ child of $M(v)$, respectively. Then by Lemma~\ref{lemma:type-3-b-i-3-info} we have that $w$ is the greatest proper ancestor of $v$ such that $M(w)=M(v,c_1)$. Here we distinguish two cases for $u$, depending on whether $M(u)=M(B(u)\setminus\{e\})$, or $M(u)\neq M(B(u)\setminus\{e\})$. In any case, by Lemma~\ref{lemma:type-3-b-i-3-info} we have that $e$ is the $\mathit{high}$ edge of $u$. Now, in the first case, we can compute all such $4$-cuts explicitly. This is because by Lemma~\ref{lemma:type-3-b-i-3-info} we have that $M(B(u)\setminus\{e_\mathit{high}(u)\})=M(v,c_2)$, and by Lemma~\ref{lemma:type-3-b-i-3-special} we have that $u$ is either the lowest or the second-lowest proper descendant of $v$ with this property. Thus, there are only $O(n)$ $4$-cuts in this case (because, given $v$, there is only one candidate $w$, and at most two candidates $u$).
In the case $M(u)\neq M(B(u)\setminus\{e\})$, the number of $4$-cuts can be $\Omega(n^2)$, as shown in Figure~\ref{figure:type3bi-3}. However, it is sufficient to consider only the lowest proper descendant $u'$ of $v$ that satisfies $M(u')\neq M(B(u')\setminus\{e_\mathit{high}(u')\})=M(v,c_2)$, according to Lemma~\ref{lemma:type-3-b-i-3-non-special}. Thus, in any case, we compute $O(n)$ $4$-cuts in total, and Proposition~\ref{proposition:algorithm:type-3-b-i-3} establishes that these, together with the collection of Type-$2ii$ $4$-cuts that we have computed, are enough in order to imply all $4$-cuts of this kind. Furthermore, this computation can be performed in linear time in total.

Finally, let us consider case $(4)$ of Lemma~\ref{lemma:type-3b-cases}. This branches into several subcases. First, we distinguish two cases, depending on whether $M(v)\neq M(B(v)\setminus\{e\})$ or $M(v)=M(B(v)\setminus\{e\})$. In the first case, by Lemma~\ref{lemma:e_L-e_R} we have that $e$ is either $e_L(v)$ or $e_R(v)$. Then, by Lemma~\ref{lemma:type-3-b-i-4-cases-simple}, we know precisely $u$ and $w$: that is, $u$ is the lowest proper descendant of $v$ such that $M(u)=M(v,c_2)$, and $w$ is the greatest proper ancestor of $v$ such that $M(w)=M(v,c_1)$, where $c_1$ and $c_2$ are the $\mathit{low1}$ and $\mathit{low2}$ children of $M(B(v)\setminus\{e\})$, respectively. Thus, the number of all $4$-cuts of this kind is $O(n)$, and Proposition~\ref{proposition:algorithm:type-3-b-i-4-simple} shows that we can compute all of them in linear time in total.
The case that $M(v)=M(B(v)\setminus\{e\})$ is more involved, because we cannot immediately determine the back-edge $e$. Thus, we first determine $u$ and $w$ according to Lemma~\ref{lemma:type-3-b-i-4-cases} and Lemma~\ref{lemma:type-3-b-i-4-extremities}, by considering all the different cases of Lemma~\ref{lemma:type-3-b-i-4-cases}. 
Notice that the total number of pairs of $u$ and $w$ that we check are $O(1)$ for a given $v$. Then $e$ can be determined by  Lemma~\ref{lemma:type-3-b-i-4-edge}. Thus, the number of all such $4$-cuts is $O(n)$, and Proposition~\ref{proposition:algorithm:type-3-b-i-4} shows that we can compute all of them in linear time in total.

\noindent\\
\textbf{Type-$3\beta ii$ $4$-cuts}\\

In the case of Type-$3\beta ii$ $4$-cuts, given a vertex $v$, there may be many pairs of $u$ and $w$ such that $(u,v,w)$ induces a $4$-cut, for any of the cases $(1)$-$(4)$ of Lemma~\ref{lemma:type-3b-cases}. For all those cases, we follow a common strategy. First, we define a set $U(v)$, for every vertex $v$, that contains some candidates $u$ with the property that there may exist a $w$ such that $(u,v,w)$ induces a $4$-cut of the type we consider. Then, for every $u\in U(v)$, we determine a $w$ (if it exists) such that $(u,v,w)$ induces a $4$-cut. We make sure that the selection of $4$-cuts that we have computed in each case is enough to imply, together with the collection of Type-$2ii$ $4$-cuts that we have computed, all $4$-cuts of the kind that we consider. Since the general strategy is the same, there are a lot of similarities in all those cases on a high level. In particular, the sets $U(v)$ that we define in each particular case have similar definitions, satisfy similar properties, and can be computed with similar methods. However, each particular case presents unique challenges, and demands special care in order to ensure correctness. Thus, we distinguish between Type-$3\beta ii$-$1$, Type-$3\beta ii$-$2$, Type-$3\beta ii$-$3$ and Type-$3\beta ii$-$4$ $4$-cuts, depending on whether they satisfy $(1)$, $(2)$, $(3)$ or $(4)$ of Lemma~\ref{lemma:type-3b-cases}, respectively.

\noindent\\
\textbf{Type-$3\beta ii$-$1$ $4$-cuts}\\

For the case of Type-$3\beta ii$-$1$ $4$-cuts, we define the set $U_1(v)$, for every vertex $v\neq r$, as a segment of $H(\mathit{high}_1(v))$ that consists of the proper descendants of $v$ that have low enough $\mathit{low}$ point in order to be possible to participate in a triple of vertices $(u,v,w)$ that induces a Type-$3\beta ii$-$1$ $4$-cut. 
The sets $U_1(v)$ have total size $O(n)$, and they have the property that, if there is a $w$ such that $(u,v,w)$ induces a Type-$3\beta ii$-$1$ $4$-cut, then $u\in U_1(v)$ (Lemma~\ref{lemma:type3-b-ii-1-U}). By Lemma~\ref{lemma:algorithm:type3-b-ii-1-U}, we can compute all sets $U_1(v)$ in linear time in total. Given $v$ and $u\in U_1(v)$, by Lemma~\ref{lemma:type3-b-ii-1-greatest-w} we have that there is at most one $w$ such that $(u,v,w)$ induces a Type-$3\beta ii$-$1$ $4$-cut: i.e., $w$ is the greatest proper ancestor of $v$ with $M(w)=M(v)$ and $w\leq\mathit{low}_2(u)$. Thus, the number of all Type-$3\beta ii$-$1$ $4$-cuts is $O(n)$. Proposition~\ref{proposition:algorithm:type3-b-ii-1} shows that we can compute all of them in linear time in total.

\noindent\\
\textbf{Type-$3\beta ii$-$2$ $4$-cuts}\\

According to Lemma~\ref{lemma:type3-b-ii-2-info}, if a triple of vertices $(u,v,w)$ induces a Type-3$\beta$ii-$2$ $4$-cut, then the back-edge $e$ of this $4$-cut is either $e_L(w)$ or $e_R(w)$. Here we discuss the case where $e=e_L(w)$. The arguments for the case $e=e_R(w)$ are similar. For convenience, we distinguish two cases, depending on whether $L_1(w)$ is a descendant of $\mathit{high}(v)$.

First, we consider the case that $L_1(w)$ is not a descendant of $\mathit{high}(v)$. The number of $4$-cuts of this kind can be $\Omega(n^2)$, and so we do not compute all of them explicitly. 
Instead, we compute a subcollection of $O(n)$ of them with the property that, together with the collection of Type-$2ii$ $4$-cuts that we have computed, it implies all $4$-cuts of this kind. To do this, we define two sets, $W(v)$ and $U_2(v)$, for every vertex $v\neq r$. The set $W(v)$ contains all candidates $w$ with the property that there may exist a $u$ such that $(u,v,w)$ induces a Type-$3\beta ii$-$2$ $4$-cut, and $U_2(v)$ contains all candidates $u$ with the property that there may exist a $w$ such that $(u,v,w)$ induces a Type-$3\beta ii$-$2$ $4$-cut (see Lemma~\ref{lemma:type3-b-ii-2-u-w}). We do not explicitly compute the sets $W(v)$, but only the greatest and the lowest vertices that are contained in them (denoted as $\mathit{firstW}(v)$ and $\mathit{lastW}(v)$, respectively). The sets $U_2(v)$ have total size $O(n)$, and by Lemma~\ref{lemma:algorithm:type3-b-ii-2-U} we can compute all of them in linear time in total. Then, given $v$ and $u\in U_2(v)$, it is sufficient to compute the greatest $w$ such that $(u,v,w)$ induces a Type-$3\beta ii$-$2$ $4$-cut, according to Lemma~\ref{lemma:type-3-b-ii-2-L-rel}. Thus, we compute a collection of $O(n)$ $4$-cuts of this kind, which, together with the collection of Type-$2ii$ $4$-cuts that we have computed, implies all $4$-cuts of this kind. This result is summarized in Proposition~\ref{proposition:algorithm:type-3-b-ii-2-L}.

Now we consider the case where $L_1(w)$ is a descendant of $\mathit{high}(v)$. By Lemma~\ref{lemma:type-3bii-2-2} we have that $w$ is uniquely determined by $u$ and $v$. Lemma~\ref{lemma:type-3bii-2-2} motivates the definition of the sets $\widetilde{W}(v)$, that contain all possible candidates $w$ with the property that there may exist a $u$ such that $(u,v,w)$ induces a $4$-cut of this kind. 
By Lemma~\ref{lemma:tilde-w-disjoint} we have that the total size of those sets is $O(n)$. By Lemma~\ref{lemma:algorithm:type3-b-ii-w}, we can compute all of them in linear time in total. Then, by Lemma~\ref{lemma:type3-b-ii-high} we have that, if $(u,v,w)$ induces a $4$-cut of this kind, then $u$ belongs to $S(v)$, and $\mathit{bcount}(u)=\mathit{bcount}(v)-\mathit{bcount}(w)+1$. Here we can exploit the fact that all vertices in $S(v)$ have different $\mathit{bcount}$ (see Lemma~\ref{lemma:same_high_dif_bcount}). Thus, for every $w\in\widetilde{W}(v)$, we have that $u$ is uniquely determined by the properties $u\in S(v)$ and $\mathit{bcount}(u)=\mathit{bcount}(v)-\mathit{bcount}(w)+1$. Then, Proposition~\ref{proposition:algorithm:type3-b-ii-high} establishes that we can compute all such $4$-cuts in linear time in total. Notice that their total number is bounded by $O(n)$.

\noindent\\
\textbf{Type-$3\beta ii$-$3$ $4$-cuts}\\

For the case of Type-$3\beta ii$-$3$ $4$-cuts, we define the set $U_3(v)$, for every vertex $v\neq r$, as a segment of $\widetilde{H}(\mathit{high}(v))$ that consists of proper descendants of $v$ that have low enough $\mathit{low}$ point in order to be possible to participate in a triple of vertices $(u,v,w)$ that induces a Type-$3\beta ii$-$3$ $4$-cut. 
The set $U_3(v)$ does not contain all possible candidates $u$ with the property that there may exist a $w$ such that $(u,v,w)$ induces a Type-$3\beta ii$-$3$ $4$-cut. However, by Lemma~\ref{lemma:type-3-b-ii-3-rel} we have that, if $(u,v,w)$ is a triple of vertices that induces a Type-$3\beta ii$-$3$ $4$-cut, then $(\tilde{u},v,w)$ also has this property, where $\tilde{u}$ is the greatest vertex in $U_3(v)$. This is very useful, because the total number of triples $(u,v,w)$ that induce a Type-$3\beta ii$-$3$ $4$-cut can be $\Omega(n)$, and this can be true for $\Omega(n)$ vertices $v$. (Thus, the actual number of Type-$3\beta ii$-$3$ $4$-cuts can be $\Omega(n^2)$.)
The sets $U_3(v)$ have total size $O(n)$, and Lemma~\ref{lemma:algorithm:type3-b-ii-3-U} establishes that we can compute all of them in linear time in total. Given $v$ and $u\in U_3(v)$, by Lemma~\ref{lemma:type-3-b-ii-3-criterion} we have that there is at most one $w$ such that $(u,v,w)$ induces a Type-$3\beta ii$-$3$ $4$-cut: i.e.,  $w$ is the greatest proper ancestor of $v$ such that $M(w)=M(v)$ and $w\leq\mathit{low}(u)$. Thus, the number of all Type-$3\beta ii$-$3$ $4$-cuts that we compute is $O(n)$. Proposition~\ref{proposition:type-3-b-ii-3} shows that we can compute this selection in linear time in total, and this has the property that, together with the collection of Type-$2ii$ $4$-cuts that we have computed, it implies all Type-$3\beta ii$-$3$ $4$-cuts.

\noindent\\
\textbf{Type-$3\beta ii$-$4$ $4$-cuts}\\

In the case of Type-$3\beta ii$-$4$ $4$-cuts we distinguish four different subcases, depending on the location of the endpoints of the back-edge $e$. Specifically, we consider the cases:

\begin{enumerate}
\item{$M(B(v)\setminus\{e\})\neq M(v)$ and $\mathit{high}_1(v)>\mathit{high}(u)$.}
\item{$M(B(v)\setminus\{e\})\neq M(v)$ and $\mathit{high}_1(v)=\mathit{high}(u)$.}
\item{$M(B(v)\setminus\{e\})= M(v)$ and $\mathit{high}_1(v)>\mathit{high}(u)$.}
\item{$M(B(v)\setminus\{e\})= M(v)$ and $\mathit{high}_1(v)=\mathit{high}(u)$.}
\end{enumerate}

In Case $1$ we know precisely the back-edge $e$: i.e., we have $e=e_\mathit{high}(v)$ (due to $\mathit{high}_1(v)>\mathit{high}(u)$, as a consequence of Lemma~\ref{lemma:type3-b-ii-4-info}). Here there may be several vertices $v$ for which there is a specific pair of vertices $u$ and $w$ such that $(u,v,w)$ induces a $4$-cut of this kind. That is, we may have two distinct $v$ and $v'$ such that $(u,v,w)$ and $(u,v',w)$ induce a $4$-cut of this kind. 
Here we define a collection of vertices $V(v)$, for every vertex $v$, that has the property that, if there is a triple of vertices $(u,v,w)$ that induces a $4$-cut of this kind, then, for every $v'\in V(v)$, we have that $(u,v',w)$ also induces a $4$-cut of this kind (see Lemma~\ref{lemma:type-3-b-ii-4-1-in-U}). Furthermore, if two vertices $v$ and $v'$ belong to the same such collection, then $\{(v,p(v)),(v',p(v')),e_\mathit{high}(v),e_\mathit{high}(v')\}$ is a Type-$2ii$ $4$-cut (as a consequence of Lemma~\ref{lemma:V-sets-implied-4cuts}). This is very useful in order to establish that it is sufficient to have computed only a selection of $O(n)$ size of those $4$-cuts, so that the rest of them are implied from this selection, plus the collection of Type-$2ii$ $4$-cuts that we have computed.

Now the idea is to pick one representative vertex from every one of the collections of vertices $V$; thus, we form a collection of vertices $\mathcal{V}$. Then, for every vertex $v\in\mathcal{V}$, we construct a set $U_4^1(v)$, that is a subset of $\widetilde{S}_2(v)$, and consists of proper descendants $u$ of $v$ that have low enough $\mathit{low}$ point in order to be possible to participate in a triple of the form $(u,v,w)$ that induces a $4$-cut of the kind that we consider (see Lemma~\ref{lemma:type-3-b-ii-4-1-in-U}). The sets $U_4^1(v)$ have total size $O(n)$, and by Lemma~\ref{lemma:algorithm:type3-b-ii-4-1-U} we can compute all of them in linear time in total (for the specific selection of representatives $\mathcal{V}$). Then, given $v\in\mathcal{V}$ and $u\in U_4^1(v)$, Lemma~\ref{lemma:type-3-b-ii-4-1-criterion} implies that there is at most one $w$ such that $(u,v,w)$ induces a $4$-cut of the kind that we consider. Thus, we compute a selection of $O(n)$ $4$-cuts of this kind. Proposition~\ref{proposition:algorithm:type3-b-ii-4-1} establishes that we can compute this selection in linear time in total, and this is enough in order to imply, together with the collection of Type-$2ii$ $4$-cuts that we have computed, all $4$-cuts of this kind.

In Case $2$, we have that either $e=e_L(v)$ or $e=e_R(v)$, as a consequence of $M(B(v)\setminus\{e\})\neq M(v)$ (see Lemma~\ref{lemma:e_L-e_R}). Then we only consider the case that $e=e_L(v)$, because the other case is treated with similar arguments and methods. Then we define the set $U_4^2(v)$, for every vertex $v$ that has the potential to participate in a triple of vertices $(u,v,w)$ that induces a $4$-cut of this kind. The sets $U_4^2(v)$, for all vertices $v$ for which they are defined, have total size $O(n)$, and Lemma~\ref{lemma:algorithm:type3-b-ii-4-2-U} shows that we can compute all of them in linear time in total. Then Lemma~\ref{lemma:type3-b-ii-4-2-in-U} shows that if we have a triple of vertices $(u,v,w)$ that induces a $4$-cut of the kind that we consider, then $u\in U_4^2(v)$. By Lemma~\ref{lemma:type3-b-ii-4-2-criterion} we have that $w$ is uniquely determined by $u$ and $v$. Then Proposition~\ref{proposition:algorithm:type3-b-ii-4-2} shows that we can compute all $4$-cuts of this kind, in linear time in total.

In Case $3$ we again know precisely the back-edge $e$, due to $\mathit{high}_1(v)>\mathit{high}(u)$. Then we follow the same idea as in Case $2$ (by properly defining the sets $U_4^3(v)$), and Proposition~\ref{proposition:algorithm:type3-b-ii-4-3} establishes that we can compute all $4$-cuts of this kind in linear time in total.

In Case $4$, the conditions $M(B(v)\setminus\{e\})= M(v)$ and $\mathit{high}_1(v)=\mathit{high}(u)$ are not sufficient in order to determine the endpoints of $e$. However, we can follow the same idea as previously if we assume that the lower endpoint of $e$ is distinct from $\mathit{high}(u)$. In this case, we define the sets $U_4^4(v)$ appropriately, as those segments of $S(v)$ that contain enough vertices $u$ that have the potential to participate in a triple $(u,v,w)$ that induces a $4$-cut of this kind. The total size of all sets $U_4^4(v)$ is $O(n)$, and by Lemma~\ref{lemma:algorithm:type3-b-ii-4-4-U} we can compute all of them in linear time in total. Then, by Lemma~\ref{lemma:type3-b-ii-4-4-in-U} we have that if $(u,v,w)$ is a triple of vertices that induces a $4$-cut of the kind that we consider, then either $u\in U_4^4(v)$, or $u$ is the predecessor of the greatest vertex of $U_4^4(v)$ in $S(v)$. Furthermore, Lemma~\ref{lemma:type3-b-ii-4-4-criterion} shows that $w$ is either the greatest or the second-greatest proper ancestor of $v$ such that $M(w)=M(v)$ and $w\leq\mathit{low}(u)$. Thus, the total number of triples that we have to check is $O(n)$. Proposition~\ref{proposition:algorithm:type3-b-ii-4-4} establishes that we can compute all $4$-cuts of this kind in linear time in total.

Finally, it remains to consider the case where the lower endpoint of $e$ coincides with $\mathit{high}(u)$. Since we are in Case $(4)$ of Lemma~\ref{lemma:type-3b-cases}, we have $B(v)=(B(u)\sqcup B(w))\sqcup\{e\}$. This implies that $e\notin B(u)$, and therefore $e\neq e_\mathit{high}(u)$. This means that $e$ and $e_\mathit{high}(u)$ are two distinct back-edges that have the same lower endpoint. If we could eliminate this possibility, then we could revert to any of the previous cases of $4$-cuts that we have considered. The idea is precisely that: we compute a ``$4$-cut equivalent" graph, in which there is a DFS-tree with the property that no two back-edges that correspond to edges of the original graph can have the same lower endpoint. To do this, we split every vertex $z$ that has at least two incoming back-edges of the form $(x,z)$ and $(y,z)$, into two vertices $z_1$ and $z_2$ that are connected with five multiple edges $(z_1,z_2)$. We make $z_2$ the parent of $z_1$, and $z_1$ inherits the back-edge $(x,z)$ (in the form $(x,z_1)$), whereas $z_2$ inherits the remaining incoming back-edges to $z$. We continue this process until no more such splittings are possible. We show that the $4$-cuts of the original graph are in a bijective correspondence with the $4$-cuts of the resulting graph, and we show how to construct it in linear time. Thus, it is sufficient to perform one more pass on the resulting graph, of all the algorithms that we have developed for computing $4$-cuts. (Or we may perform this computation directly on the resulting graph from the start.) Then we translate the computed $4$-cuts to those of the original graph. We conclude with a post-processing step, that eliminates repetitions of $4$-cuts.

\subsection{Min-max vertex queries}
\label{subsection:mim-max-queries}

As we saw in the preceding section, in order to implement the algorithms in the following chapters we need an oracle for answering min-max vertex queries. Specifically, at various places we have to answer queries of the form ``find the lowest (resp., the greatest) vertex that is greater (resp., lower) than a particular vertex, and belongs to a specific collection of vertices". More precisely, we are given a collection $W_1,\dots,W_k$ of pairwise disjoint sets of vertices, and a set of $N$ queries of the form $q(i,z)\equiv$ ``given an index $i\in\{1,\dots,k\}$ and a vertex $z$, find the greatest (resp., the lowest) vertex $w\in W_i$ such that $w\leq z$ (resp., $w\geq z$)". We can answer all those queries simultaneously, in $O(n+N)$ time in total. The idea is to sort the vertices in the sets $W_1,\dots,W_k$ properly -- in increasing or decreasing order --, depending on whether the queries ask for the greatest or the lowest vertex, respectively. Then we also sort the queries in the same order (w.r.t. the vertices that appear in them). Now we can basically answer all the queries for which the answer lies in a specific collection $W$, independently of the others, by simply traversing the list $W$ and the list of the queries whose answer lies in $W$. The precise procedure is shown in Algorithm~\ref{algorithm:W-queries}, and the explication as well as the proof of correctness is given in Lemma~\ref{lemma:W-queries}. It is straightforward to modify Algorithm~\ref{algorithm:W-queries} appropriately, so that we get an algorithm for answering the reverse type of queries (i.e., those that ask for the lowest vertex that is greater than another vertex), with or without equality. 

\begin{lemma}
\label{lemma:W-queries}
Let $W_1,\dots,W_k$ be a collection of pairwise disjoint sets of vertices. Let $q(i_1,z_1),\dots,q(i_N,z_N)$ be a set of queries of the form $q(i,z)\equiv$`` given an index $i\in\{1,\dots,k\}$ and a vertex $z$, find the greatest vertex $w\in W_i$ such that $w\leq z$". Then, Algorithm~\ref{algorithm:W-queries} answers all those queries in $O(n+N)$ time. 
\end{lemma}
\begin{proof}
The idea is basically to collect all queries of the form $q(i,\cdot)$, for every $i\in\{1,\dots,k\}$, and then process them simultaneously with the list $W_i$. More precicely, we first sort all $W_i$, for $i\in\{1,\dots,k\}$, in decreasing order. Since the sets of vertices $W_1,\dots,W_k$ are pairwise disjoint, we have that $|W_1|+\dots+|W_k|=O(n)$. Thus, all these sortings can be performed in $O(n)$ time in total with bucket-sort. Then, for every $i\in\{1,\dots,k\}$, we collect in a list $Q(i)$ all tuples of the form $(z,t)$, such that $i_t=i$ and $z_t=z$. In other words, if the $t$-th query $q(i_t,z_t)$ has $i_t=i$, then $Q(i)$ contains the entry $(z_t,t)$. Then we sort the lists $Q(i)$ in decreasing order w.r.t. the first component of the tuples that are contained in them. This can be done in $O(n+N)$ time in total with bucket-sort. 

Now, in order to answer a query $q(i,z)$, we have to traverse the list $W_i$, until we reach the first $w\in W_i$ that has $w\leq z$ (since $W_i$ is sorted in decreasing order). If we performed this process for every individual query, we would possibly make an excessive amount of steps in total, because each time we would process the list $W_i$ from the beginning. Thus, the idea in sorting the queries too, is that we can pick up the search from the last entry of $W_i$ that we accessed. Therefore, we process the entries in $Q(i)$ in order, for every index $i\in\{1,\dots,k\}$. Since the first components of the tuples in $Q(i)$ are vertices in decreasing order, it is sufficient to start the search in $W_i$ from the last entry that we accessed. The second component of every tuple $(z,t)$ in $Q(i)$ is a pointer to the corresponding query that is being answered (i.e., this corresponds to the $t$-th query, $q(i_t,z_t)$).

It is easy to see that this is the procedure that is implemented by Algorithm~\ref{algorithm:W-queries}. The \textbf{for} loop in Line~\ref{line:W-queries-for} takes $O(n+N)$ time in total, because it traverses the entire list of the queries, and possibly the entire lists $W_1,\dots,W_k$. Thus, Algorithm~\ref{algorithm:W-queries} runs in $O(n+N)$ time.
\end{proof}

\begin{algorithm}[t!]
\caption{\textsf{Given a collection $W_1,\dots,W_k$ of pairwise disjoint sets of vertices, answer a set of queries $q(i_1,z_1),\dots,q(i_N,z_N)$, where $q(i_t,z_t)$, for every $t\in\{1,\dots,N\}$, asks for the greatest vertex $w\in W_{i_t}$ such that $w\leq z_t$.}}
\label{algorithm:W-queries}
\LinesNumbered
\DontPrintSemicolon
\tcp{$W_1,\dots,W_k$ is a collection of pairwise disjoint sets of vertices}
\ForEach{$i\in\{1,\dots,k\}$}{
  sort $W_i$ in decreasing order\;
}
\ForEach{$i\in\{1,\dots,k\}$}{
  initialize $Q(i)\leftarrow\emptyset$\;
}
\ForEach{$t\in\{1,\dots,N\}$}{
  insert a tuple $(z_t,t)$ into $Q(i_t)$\;
}
\tcp{$Q(i)$ contains a tuple of the form $(z,t)$ if and only if the query $q(i_t,z_t)$ has $i_t=i$ and $z_t=z$. The first component of a tuple in $Q(i)$ stores the vertex of the respective query, and the second component stores a pointer to the query. The information, that we have to search in the set $W_i$ for the answer to this query, is given precisely by the index $i$ of this bucket of tuples}
\ForEach{$i\in\{1,\dots,k\}$}{
  sort $Q(i)$ in decreasing order w.r.t. the first component of its elements\; 
}
\ForEach{$i\in\{1,\dots,k\}$}{
\label{line:W-queries-for}
  let $w$ be the first element of $W_i$\;
  let $p$ be the first element of $Q(i)$\;
  \While{$p\neq\bot$}{
    let $p=(z,t)$\;
    \While{$w\neq\bot$ \textbf{and} $w>z$}{
      $w\leftarrow\mathit{next}_{W_i}(w)$\;
    }
    the answer to $q(i_t,z_t)$ is $w$\;
    $p\leftarrow\mathit{next}_{Q(i)}(p)$\;
  }
}
\end{algorithm}

\section{Computing Type-2 $4$-cuts}
\label{section:type-2}

Throughout this section, we assume that $G$ is a $3$-edge-connected graph with $n$ vertices and $m$ edges. All graph-related elements (e.g., vertices, edges, cuts, etc.) refer to $G$. Furthermore, we assume that we have computed a DFS-tree $T$ of $G$ rooted at a vertex $r$.

\begin{lemma}
\label{lemma:type2cuts}
Let $u,v$ be two vertices such that $v$ is a proper ancestor of $u$ with $v\neq r$. Then there exist two distinct back-edges $e_1,e_2$ such that $\{(u,p(u)),(v,p(v)),e_1,e_2\}$ is a $4$-cut if and only if either $(1)$ $B(v)=B(u)\sqcup\{e_1,e_2\}$, or $(2)$ $B(v)\sqcup\{e_1\}=B(u)\sqcup\{e_2\}$, or $(3)$ $B(u)=B(v)\sqcup\{e_1,e_2\}$. 
\end{lemma}
\begin{proof}
($\Rightarrow$) Let $C=\{(u,p(u)),(v,p(v)),e_1,e_2\}$. First we will show that $e_1$ and $e_2$ are back-edges in $B(u)\cup B(v)$. So let us suppose the contrary. Then we may assume w.l.o.g. that $e_1\notin B(u)\cup B(v)$. Let $e_1=(x,y)$. Then $e_1\notin B(u)\cup B(v)$ means that neither $u$ nor $v$ lies on the tree-path $T[x,y)$. This implies that the tree-path $T[x,y)$ remains intact in $G\setminus C$. But then we have that the endpoints of $e_1$ remain connected in $G\setminus C$, in contradiction to the fact that $C$ is a $4$-cut of $G$. This shows that $e_1\in B(u)\cup B(v)$. Similarly, we can show that $e_2\in B(u)\cup B(v)$.

Since $C$ is a $4$-cut of $G$, we have that $G'=G\setminus\{(u,p(u)),(v,p(v))\}$ is connected. We define the three parts $X=T(u)$, $Y=T(v)\setminus T(u)$, and $Z=T(r)\setminus T(v)$. Notice that these parts remain connected in $G'$.

Suppose first that both $e_1$ and $e_2$ leap over $v$. Let us suppose, for the sake of contradiction, that there is a back-edge $e=(x,y)$ from $X$ to $Y$. Then $e$ leaps over $u$, but not over $v$. Thus, $e\notin\{e_1,e_2\}$. But then we have that $u$ is connected with $p(u)$ in $G'\setminus\{e_1,e_2\}$ through the path $T[u,x],(x,y),T[y,p(u)]$, contradicting the fact that $C$ is a $4$-cut of $G$. This means that there is no back-edge in $B(u)\setminus B(v)$, and therefore $B(u)\subseteq B(v)$. Now let us suppose, for the sake of contradiction, that there is a back-edge $e=(x,y)$ that leaps over $v$, does not leap over $u$, and is distinct from $e_1$ and $e_2$. Since $e$ leaps over $v$, but not over $u$, we have that it is a back-edge from $Y$ to $Z$. But then, since $e\notin\{e_1,e_2\}$, we have that $v$ is connected with $p(v)$ in $G'\setminus\{e_1,e_2\}$ through the path $T[v,x],(x,y),T[y,p(v)]$, contradicting the fact that $C$ is a $4$-cut of $G$. This means that $e_1$ and $e_2$ are the only edges in $B(v)\setminus B(u)$, and so we have $B(v)=B(u)\cup\{e_1,e_2\}$. Notice that it cannot be that both $e_1$ and $e_2$ are in $B(u)$, because otherwise we have $B(v)=B(u)$, in contradiction to the fact that $G$ is $3$-edge-connected. Now let us suppose, for the sake of contradiction, that one of $e_1,e_2$ is in $B(u)$. We may assume w.l.o.g. that $e_1$ is in $B(u)$. Since $C$ is a $4$-cut of $G$, we have that $G'\setminus e_2$ is connected. In particular, $v$ is connected with $p(v)$ in $G'\setminus e_2$. There are two possible ways in which this can be true: either $(i)$ there is a back-edge from $Y$ to $Z$ in $G'\setminus e_2$, or $(ii)$ there is a back-edge from $Y$ to $X$, and a back-edge from $X$ to $Z$, in $G'\setminus e_2$. Case $(i)$ is rejected, because the only back-edge that leaps over $v$ but not over $u$ is $e_2$. Case $(ii)$ is rejected, because $B(u)\subset B(v)$ (and so there is no back-edge from $X$ to $Y$). Since neither of $(i)$ and $(ii)$ can be true, we have arrived at a contradiction. Thus, we have that neither of $e_1,e_2$ can be in $B(u)$, and so it is correct to write $B(v)=B(u)\sqcup\{e_1,e_2\}$.

Now let us suppose that only one of $e_1,e_2$ leaps over $v$. Then we may assume w.l.o.g. that $e_1\notin B(v)$ and $e_2\in B(v)$. Since each one of $e_1,e_2$ leaps over either $u$ or $v$, we have that $e_1\in B(u)$. Since $e_1\in B(u)\setminus B(v)$, we have that $e_1$ is back-edge from $X$ to $Y$. Now let us suppose, for the sake of contradiction, that there is also another back-edge $e=(x,y)$ from $X$ to $Y$ (i.e., $e\neq e_1$). Notice that, since $e_2\in B(v)$, it is impossible that $e=e_2$. But now we have that $u$ remains connected with $p(u)$ in $G'\setminus\{e_1,e_2\}$ through the path $T[u,x],(x,y),T[y,p(u)]$, contradicting the fact that $C$ is a $4$-cut of $G$. Thus we have that $e_1$ is the only back-edge from $X$ to $Y$. Since $e_2\in B(v)$, we have that $e_2$ is either a back-edge from $X$ to $Z$, or a back-edge from $Y$ to $Z$. Let us suppose, for the sake of contradiction, that $e_2$ is a back-edge from $X$ to $Z$. Since $C$ is a $4$-cut of $G$, we have that $G'\setminus e_1$ is connected. In particular, $u$ is connected with $p(u)$ in $G'\setminus e_1$. There are two possible ways for this to be true: either $(i)$ there is a back-edge from $X$ to $Y$ in $G'\setminus e_1$, or $(ii)$ there is a back-edge from $X$ to $Z$, and a back-edge from $Z$ to $Y$ in $G'\setminus e_1$. Case $(i)$ is rejected, since $e_1$ is the only back-edge from $X$ to $Y$. Thus, $(ii)$ implies that there is a back-edge $e=(x,y)$ from $Y$ to $Z$ in $G'\setminus e_1$. Since $e_2$ is a back-edge from $X$ to $Z$, we have that $e\notin\{e_1,e_2\}$. But then $v$ is connected with $p(v)$ in $G'\setminus\{e_1,e_2\}$ through the path $T[v,x],(x,y),T[y,p(v)]$, contradicting the fact that $C$ is a $4$-cut of $G$. Thus we have that $e_2$ is not a back-edge from $X$ to $Z$, and therefore it is a back-edge from $Y$ to $Z$. Similarly, we can show that $e_2$ is the unique back-edge from $Y$ to $Z$. Thus far we have that $e_1\in B(u)\setminus B(v)$ and $e_2\in B(v)\setminus B(u)$, and both $e_1$ and $e_2$ are unique with this property. Now, since $e_1$ is the only back-edge in $B(u)\setminus B(v)$, we have that $B(u)\setminus\{e_1\}\subseteq B(v)$. And since $e_2$ is the only back-edge in $B(v)\setminus B(u)$, we have that $B(v)\setminus\{e_2\}\subseteq B(u)$. Now let $e$ be a back-edge in $B(u)\sqcup\{e_2\}$. Then, either $e=e_1$, or $e\in B(v)$. Thus we get $B(u)\sqcup\{e_2\}\subseteq B(v)\sqcup\{e_1\}$. Similarly, we get the reverse inclusion, and so we have $B(u)\sqcup\{e_2\}=B(v)\sqcup\{e_1\}$.

Finally, let us suppose that neither of $e_1,e_2$ leaps over $v$. Then, since each one of $e_1,e_2$ leaps over either $u$ or $v$, we have that $\{e_1,e_2\}\subseteq B(u)$. Notice that both $e_1$ and $e_2$ are back-edges from $X$ to $Y$. Now we can argue as above, in order to establish that $e_1$ and $e_2$ are the only back-edges from $X$ to $Y$. Thus, the remaining back-edges in $B(u)$ must also be in $B(v)$. Again, arguing as above, we can establish that there is no back-edge from $Y$ to $Z$. Thus, all the back-edges in $B(v)$ are also in $B(u)$. Thus we have $B(u)\setminus\{e_1,e_2\}\subseteq B(v)$, and $B(v)\subseteq B(u)$. Therefore, since $\{e_1,e_2\}\subseteq B(u)\setminus B(v)$, we have that $B(v)\sqcup\{e_1,e_2\}=B(u)$.

($\Leftarrow$) Let $C=\{(u,p(u)),(v,p(v)),e_1,e_2\}$. Since the graph is $3$-edge-connected, we have that $G'=G\setminus\{(u,p(u)),(v,p(v))\}$ is connected. We define the three parts $X=T(u)$, $Y=T(v)\setminus T(u)$, and $Z=T(r)\setminus T(v)$. Notice that these parts are connected in $G'$. Now it is easy to see that either of $(1)$, $(2)$, or $(3)$, implies that $C$ is a $4$-cut of $G$: $(1)$ means that $e_1$ and $e_2$ are the only back-edges from $Y$ to $Z$, and there are no back-edges from $X$ to $Y$; $(2)$ means that $e_1$ is the only back-edge from $X$ to $Y$, and $e_2$ is the only back-edge from $Y$ to $Z$; and $(3)$ means that $e_1$ and $e_2$ are the only back-edges from $X$ to $Y$, and there are no back-edges from $Y$ to $Z$. In either case, we can see that $Y$ becomes disconnected from the rest of the graph in $G\setminus C$, but $G\setminus C'$ remains connected for every proper subset $C'$ of $C$. 
\end{proof}

Based on Lemma~\ref{lemma:type2cuts}, we distinguish three different cases for Type-2 $4$-cuts of the form $\{(u,p(u)),(v,p(v)),e_1,e_2\}$, where $v$ is ancestor of $u$ and $e_1,e_2$ are back-edges: either $(1)$ $B(v)=B(u)\sqcup\{e_1,e_2\}$, or $(2)$ $B(v)\sqcup\{e_1\}=B(u)\sqcup\{e_2\}$, or $(3)$ $B(u)=B(v)\sqcup\{e_1,e_2\}$. We show how to find all $4$-cuts in cases $(1)$ and $(3)$ in linear time, in Sections \ref{subsubsection:case_B(v)=B(u)cup2} and \ref{subsubsection:case_B(u)=B(v)cup2}, respectively. The $4$-cuts in case $(2)$ cannot be computed in linear time, since there can be $\Omega(n^2)$ of them. Instead, we calculate only a specific selection of $O(n)$ of them, so that the rest of them are implied from this selection. We show how we can handle this case in Section~\ref{subsubsection:case_B(v)cup=B(u)cup}.

\subsection{The case $B(v)=B(u)\sqcup\{e_1,e_2\}$}
\label{subsubsection:case_B(v)=B(u)cup2}

\begin{lemma}
\label{lemma:case_B(v)=B(u)cup2-criterion}
Let $u$ and $v$ be two vertices such that $v$ is a proper ancestor of $u$ with $v\neq r$. Then there exist two back-edges $e_1$ and $e_2$ such that $B(v)=B(u)\sqcup\{e_1,e_2\}$ if and only if $\mathit{bcount}(v)=\mathit{bcount}(u)+2$ and $\mathit{high}_1(u)<v$.
\end{lemma}
\begin{proof}
($\Rightarrow$) $\mathit{bcount}(v)=\mathit{bcount}(u)+2$ is an immediate consequence of $B(v)=B(u)\sqcup\{e_1,e_2\}$. Now let $(x,y)$ be a back-edge in $B(u)$. Then $B(v)=B(u)\sqcup\{e_1,e_2\}$ implies that $(x,y)$ is a back-edge in $B(v)$, and therefore $y$ is a proper ancestor of $v$, and therefore $y<v$. Due to the generality of $(x,y)\in B(u)$, this implies that $\mathit{high}_1(u)<v$.

($\Leftarrow$) Since $u$ is a common descendant of $v$ and $\mathit{high}_1(u)$, we have that $v$ and $\mathit{high}_1(u)$ are related as ancestor and descendant. Thus, $\mathit{high}_1(u)<v$ implies that $\mathit{high}_1(u)$ is a proper ancestor of $v$. Now let $(x,y)$ be a back-edge in $B(u)$. Then $x$ is a descendant of $u$, and therefore a descendant of $v$. Furthermore, $y$ is an ancestor of $\mathit{high}_1(u)$, and therefore a proper ancestor of $v$. 
This shows that $(x,y)\in B(v)$, and therefore we have $B(u)\subseteq B(v)$. Now $\mathit{bcount}(v)=\mathit{bcount}(u)+2$ implies that $|B(v)\setminus B(u)|=2$, and so there are two back-edges $e_1,e_2$ such that $B(v)=B(u)\sqcup\{e_1,e_2\}$.
\end{proof}

\begin{lemma}
\label{lemma:case_B(v)=B(u)cup2-cases}
Let $u$ and $v$ be two vertices such that $v$ is a proper ancestor of $u$ with $v\neq r$, and there exist two back-edges $e_1=(x_1,y_1)$ and $e_2=(x_2,y_2)$ such that $B(v)=B(u)\sqcup\{e_1,e_2\}$. Then, neither of $x_1,x_2$ is a descendant of $u$, and one of the following is true:
\begin{enumerate}[label=(\arabic*)]
\item{$L_1(v)=L_2(v)$, $x_1=x_2=L_1(v)$ and $\{y_1,y_2\}=\{l_1(L_1(v)),l_2(L_1(v))\}$}
\item{$L_1(v)\neq L_2(v)$, $\{x_1,x_2\}=\{L_1(v),L_2(v)\}$ and $\{y_1,y_2\}=\{l_1(L_1(v)),l_1(L_2(v))\}$}
\item{$\{x_1,x_2\}=\{L_1(v),R_1(v)\}$ and $\{y_1,y_2\}=\{l_1(L_1(v)),l_1(R_1(v))\}$}
\item{$R_1(v)=R_2(v)$, $x_1=x_2=R_1(v)$ and $\{y_1,y_2\}=\{l_1(R_1(v)),l_2(R_1(v))\}$}
\item{$R_1(v)\neq R_2(v)$, $\{x_1,x_2\}=\{R_1(v),R_2(v)\}$ and $\{y_1,y_2\}=\{l_1(R_1(v)),l_1(R_2(v))\}$}
\end{enumerate}
\end{lemma}
\begin{proof}
Let us suppose, for the sake of contradiction, that at least one of $x_1,x_2$ is a descendant of $u$. We may assume w.l.o.g. that $x_1$ is a descendant of $u$. Then, since $(x_1,y_1)\in B(v)$, we have that $y_1$ is a proper ancestor of $v$, and therefore it is a proper ancestor of $u$. But this implies that $(x_1,y_1)\in B(u)$, a contradiction. Thus we have that neither of $x_1,x_2$ is a descendant of $u$.

Now let $(x'_1,y'_1),\dots,(x'_k,y'_k)$ be the back-edges in $B(v)$ sorted in increasing order w.r.t. their higher endpoint. (Notice that $L_1(v)=x'_1$, $L_2(v)=x'_2$, $R_1(v)=x'_k$, and $R_2(v)=x'_{k-1}$.) Let $i,j\in\{1,\dots,k\}$ be two indices such that $i\leq j$. Suppose that $x'_i$ and $x'_j$ are descendants of $u$. Since we have $x'_i\leq\dots\leq x'_j$, this implies that all $x'_i,\dots,x'_j$ are descendants of $u$. Furthermore, since $(x'_i,y'_i),\dots,(x'_j,y'_j)$ are back-edges in $B(v)$, we have that $y'_i,\dots,y'_j$ are proper ancestors of $v$, and therefore they are proper ancestors of $u$. This shows that all the back-edges $(x'_i,y'_i),\dots,(x'_j,y'_j)$ are in $B(u)$. Now, since $B(v)=B(u)\sqcup\{e_1,e_2\}$, we have that only two back-edges in $B(v)$ are not in $B(u)$. Therefore, $e_1$ and $e_2$ can either be $(i)$ $(x'_1,y'_1)$ and $(x'_2,y'_2)$, or $(ii)$ $(x'_1,y'_1)$ and $(x'_k,y'_k)$, or $(iii)$ $(x'_{k-1},y'_{k-1})$ and $(x'_k,y'_k)$. Thus we get that $(1)$-$(5)$ is an exhaustive list for the different combinations for $x_1$ and $x_2$. 

Suppose first that $L_1(v)=L_2(v)$ and $x_1=x_2=L_1(v)$. Let us suppose, for the sake of contradiction, that $l_3(L_1(v))<v$. Then this means that there are at least three different back-edges $(L_1(v),z_1)$, $(L_1(v),z_2)$, $(L_1(v),z_3)$ in $B(v)$. Since $x_1=L_1(v)$ is not a descendant of $u$, we have that these three back-edges are in $B(v)\setminus B(u)$. But this contradicts $B(v)=B(u)\sqcup\{e_1,e_2\}$. Thus we have that $l_3(L_1(v))\geq v$. Since $e_1,e_2\in B(v)$, we have that $y_1<v$ and $y_2<v$. Thus we get $\{y_1,y_2\}=\{l_1(L_1(v)),l_2(L_1(v))\}$.

Now suppose that $L_1(v)\neq L_2(v)$ and $\{x_1,x_2\}=\{L_1(v),L_2(v)\}$. We may assume w.l.o.g. that $x_1=L_1(v)$ and $x_2=L_2(v)$. Let us suppose, for the sake of contradiction, that $l_2(L_1(v))<v$. This implies that there are at least two different back-edges $(L_1(v),z_1)$ and $(L_1(v),z_2)$ in $B(v)$. Since $x_1=L_1(v)$ is not a descendant of $u$, we have that there are at least three back-edges in $B(v)\setminus B(u)$ (these being $(L_1(v),z_1)$, $(L_1(v),z_2)$, and $(L_2(v),y_2)$), a contradiction. Thus we have $l_2(L_1(v))\geq v$, and so $y_1=l_1(L_1(v))$, since $(x_1,y_1)\in B(v)$. Similarly, we can show that $y_2=l_1(L_2(v))$.

With similar arguments we get the results for $y_1$ and $y_2$ for the cases $\{x_1,x_2\}=\{L_1(v),R_1(v)\}$ and $\{x_1,x_2\}=\{R_1(v),R_2(v)\}$ (whether $R_1(v)=R_2(v)$ or $R_1(v)\neq R_2(v)$).
\end{proof}

\begin{lemma}
\label{lemma:case_B(v)=B(u)cup2-inference}
Let $u$ and $v$ be two vertices such that $v$ is a proper ancestor of $u$ with $v\neq r$, and there exist two back-edges $e_1$ and $e_2$ such that $B(v)=B(u)\sqcup\{e_1,e_2\}$. Then $u$ is the lowest proper descendant of $v$ that has $M(u)=M(B(v)\setminus\{e_1,e_2\})$.
\end{lemma}
\begin{proof}
First we observe that $M(u)=M(B(v)\setminus\{e_1,e_2\})$, as an immediate consequence of $B(u)=B(v)\setminus\{e_1,e_2\}$. Thus we may consider the lowest proper descendant $u'$ of $v$ that has $M(u')=M(B(v)\setminus\{e_1,e_2\})$. Let us suppose, for the sake of contradiction, that $u'\neq u$. Then, since $M(u')=M(u)$ and $u'$ is lower than $u$, we have that $u'$ is a proper ancestor of $u$, and Lemma~\ref{lemma:same_m_subset_B} implies that $B(u')\subseteq B(u)$. Since the graph is $3$-edge-connected, this can be strengthened to $B(u')\subset B(u)$. Thus there is a back-edge $(x,y)\in B(u)\setminus B(u')$. Then, we have that $x$ is a descendant of $u$, and therefore a descendant of $u'$. Furthermore, $B(v)=B(u)\sqcup\{e_1,e_2\}$ implies that $(x,y)\in B(v)$, and therefore $y$ is a proper ancestor of $v$. But then $y$ is also a proper ancestor of $u'$, and so $(x,y)\in B(u')$, a contradiction. We conclude that $u$ is the lowest proper descendant of $v$ with $M(u)=M(B(v)\setminus\{e_1,e_2\})$.
\end{proof}

Now the idea to find all $4$-cuts of the form $\{(u,p(u)),(v,p(v)),e_1,e_2\}$, where $v$ is a proper ancestor of $u$ with $B(v)=B(u)\sqcup\{e_1,e_2\}$, is the following. According to Lemma~\ref{lemma:case_B(v)=B(u)cup2-cases}, there are three different cases for the back-edges $e_1$ and $e_2$: either $\{e_1,e_2\}=\{e_\mathit{L1}(v),e_\mathit{L2}(v)\}$, or $\{e_1,e_2\}=\{e_\mathit{L1}(v),e_\mathit{R1}(v)\}$, or $\{e_1,e_2\}=\{e_\mathit{R1}(v),e_\mathit{R2}(v)\}$. We consider all these cases in turn. For every one of those cases, we seek the lowest proper descendant $u$ of $v$ that satisfies $M(u)=M(B(v)\setminus\{e_1,e_2\})$, according to Lemma~\ref{lemma:case_B(v)=B(u)cup2-inference}.
Then we can check whether we have a $4$-cut using the criterion provided by Lemma~\ref{lemma:case_B(v)=B(u)cup2-criterion}. This procedure is shown in Algorithm~\ref{algorithm:type2-1}. The proof of correctness is given in Proposition~\ref{proposition:algorithm:type2-1}.

\begin{algorithm}[h!]
\caption{\textsf{Compute all $4$-cuts of the form $\{(u,p(u)),(v,p(v)),e_1,e_2\}$, where $v$ is an ancestor of $u$ and $B(v)=B(u)\sqcup\{e_1,e_2\}$}}
\label{algorithm:type2-1}
\LinesNumbered
\DontPrintSemicolon
\ForEach{vertex $v\neq r$}{
\label{line:type-2-1-for}
  compute $M_\mathit{LL}(v)\leftarrow M(B(v)\setminus\{e_\mathit{L1}(v),e_\mathit{L2}(v)\})$\;
  compute $M_\mathit{LR}(v)\leftarrow M(B(v)\setminus\{e_\mathit{L1}(v),e_\mathit{R1}(v)\})$\;
  compute $M_\mathit{RR}(v)\leftarrow M(B(v)\setminus\{e_\mathit{R1}(v),e_\mathit{R2}(v)\})$\;
}
let $u$ be the lowest proper descendant of $v$ such that $M(u)=M_\mathit{LL}(v)$\;
\label{line:type-2-1-u-1}
\If{$\mathit{bcount}(v)=\mathit{bcount}(u)+2$ \textbf{and} $\mathit{high}_1(u)<v$}{
    mark $\{(u,p(u)),(v,p(v)),e_\mathit{L1}(v),e_\mathit{L2}(v)\}$ as a $4$-cut\;
    \label{line:type-2-1-mark-1}
}
let $u$ be the lowest proper descendant of $v$ such that $M(u)=M_\mathit{LR}(v)$\;
\label{line:type-2-1-u-2}
\If{$\mathit{bcount}(v)=\mathit{bcount}(u)+2$ \textbf{and} $\mathit{high}_1(u)<v$}{
    mark $\{(u,p(u)),(v,p(v)),e_\mathit{L1}(v),e_\mathit{R1}(v)\}$ as a $4$-cut\;
    \label{line:type-2-1-mark-2}
}
let $u$ be the lowest proper descendant of $v$ such that $M(u)=M_\mathit{RR}(v)$\;
\label{line:type-2-1-u-3}
\If{$\mathit{bcount}(v)=\mathit{bcount}(u)+2$ \textbf{and} $\mathit{high}_1(u)<v$}{
    mark $\{(u,p(u)),(v,p(v)),e_\mathit{R1}(v),e_\mathit{R2}(v)\}$ as a $4$-cut\;
    \label{line:type-2-1-mark-3}
}
\end{algorithm}

\begin{proposition}
\label{proposition:algorithm:type2-1}
Algorithm~\ref{algorithm:type2-1} correctly computes all $4$-cuts of the form $\{(u,p(u)),(v,p(v)),e_1,e_2\}$, where $v$ is an ancestor of $u$ and $B(v)=B(u)\sqcup\{e_1,e_2\}$. Furthermore, it has a linear-time implementation.
\end{proposition}
\begin{proof}
Let $C=\{(u,p(u)),(v,p(v)),e_1,e_2\}$ be a $4$-cut such that $u$ is a descendant of $v$ and $B(v)=B(u)\sqcup\{e_1,e_2\}$. Then, Lemma~\ref{lemma:case_B(v)=B(u)cup2-cases} implies that either $\{e_1,e_2\}=\{e_\mathit{L1}(v),e_\mathit{L2}(v)\}$, or $\{e_1,e_2\}=\{e_\mathit{L1}(v),e_\mathit{R1}(v)\}$, or $\{e_1,e_2\}=\{e_\mathit{R1}(v),e_\mathit{R2}(v)\}$. By Lemma~\ref{lemma:case_B(v)=B(u)cup2-inference}, we have that $u$ is the lowest proper descendant of $v$ such that $M(u)=M(B(v)\setminus\{e_1,e_2\})$. Lemma~\ref{lemma:case_B(v)=B(u)cup2-criterion} implies that $\mathit{bcount}(v)=\mathit{bcount}(u)+2$ and $\mathit{high}_1(u)<v$. Thus, we can see that $C$ will be marked in Line~\ref{line:type-2-1-mark-1}, or \ref{line:type-2-1-mark-2}, or \ref{line:type-2-1-mark-3}. 

Conversely, suppose that a $4$-element set $C=\{(u,p(u)),(v,p(v)),e_1',e_2'\}$ is marked by Algorithm~\ref{algorithm:type2-1} in Line~\ref{line:type-2-1-mark-1}, or \ref{line:type-2-1-mark-2}, or \ref{line:type-2-1-mark-3}. In either case, we have that $u$ is a proper descendant of $v$ such that $\mathit{bcount}(v)=\mathit{bcount}(u)+2$ and $\mathit{high}_1(u)<v$. Thus, Lemma~\ref{lemma:case_B(v)=B(u)cup2-criterion} implies that there are two back-edges $e_1$ and $e_2$ such that $B(v)=B(u)\sqcup\{e_1,e_2\}$. Lemma~\ref{lemma:case_B(v)=B(u)cup2-cases} implies that the higher endpoints of $e_1$ and $e_2$ are not descendants of $u$. Thus, if $S$ is a subset of $B(v)$ that contains either $e_1$ or $e_2$, then we have that $M(S)\neq M(u)$. To see this, suppose the contrary. Then, w.l.o.g. we may assume that $e_1=(x,y)\in S$, and $M(S)=M(u)$. Then we have that $x$ is a descendant of $M(S)$, and therefore a descendant of $M(u)$, and therefore a descendant of $u$. Furthermore, since $e_1\in B(v)$, we have that $y$ is a proper ancestor of $v$, and therefore a proper ancestor of $u$. This shows that $(x,y)\in B(u)$, a contradiction. Thus, we have that $M(S)\neq M(u)$. 
Now, since $B(v)=B(u)\sqcup\{e_1,e_2\}$, Lemma~\ref{lemma:case_B(v)=B(u)cup2-cases} implies that either $\{e_1,e_2\}=\{e_\mathit{L1}(v),e_\mathit{L2}(v)\}$, or $\{e_1,e_2\}=\{e_\mathit{L1}(v),e_\mathit{R1}(v)\}$, or $\{e_1,e_2\}=\{e_\mathit{R1}(v),e_\mathit{R2}(v)\}$. Let us assume that $\{e_1,e_2\}=\{e_\mathit{L1}(v),e_\mathit{L2}(v)\}$ (the other cases are similar). Then, if $C$ is marked in Line~\ref{line:type-2-1-mark-1}, we have that $\{e_1',e_2'\}=\{e_1,e_2\}$, and so it is correct to mark $C$ as a $4$-cut. Now, if $\{e_\mathit{L1}(v),e_\mathit{R1}(v)\}\neq\{e_1,e_2\}$, then we have that $M(u)\neq M_\mathit{LR}(v)$ (because $B(v)\setminus\{e_\mathit{L1}(v),e_\mathit{R1}(v)\}$ contains either $e_1$ or $e_2$), and therefore $C$ is not marked in Line~\ref{line:type-2-1-mark-2}. Similarly, if $\{e_\mathit{R1}(v),e_\mathit{R2}(v)\}\neq\{e_1,e_2\}$, then we have that $M(u)\neq M_\mathit{RR}(v)$, and therefore $C$ is not marked in Line~\ref{line:type-2-1-mark-3}. This shows that $C$ is indeed a $4$-cut of $G$.

Now we will show that Algorithm~\ref{algorithm:type2-1} runs in linear time. By Proposition~\ref{proposition:computing-M(B(v)-S)}, we have that the values $M(B(v)\setminus\{e_\mathit{L1}(v),e_\mathit{L2}(v)\})$, $M(B(v)\setminus\{e_\mathit{L1}(v),e_\mathit{R1}(v)\})$ and $M(B(v)\setminus\{e_\mathit{R1}(v),e_\mathit{R2}(v)\})$ can be computed in linear time in total, for all vertices $v\neq r$. Thus, the \textbf{for} loop in Line~\ref{line:type-2-1-for} can be performed in linear time. In order to compute the $u$ in Lines~\ref{line:type-2-1-u-1}, \ref{line:type-2-1-u-2} and \ref{line:type-2-1-u-3}, we use Algorithm~\ref{algorithm:W-queries}. Specifically, let us discuss the implementation of Line~\ref{line:type-2-1-u-1} (the argument for Lines~\ref{line:type-2-1-u-2} and \ref{line:type-2-1-u-3} is similar). Then, for every vertex $v\neq r$, we generate a query $q(M^{-1}(M_\mathit{LL}(v)),v)$. This query will return the lowest vertex $u$ with $M(u)=M_\mathit{LL}(v)$ that is greater than $v$. Notice that $M(u)=M_\mathit{LL}(v)$ implies that $M(u)$ is a common descendant of $u$ and $v$, and therefore $u$ and $v$ are related as ancestor and descendant. Thus, $u$ is the lowest proper descendant of $v$ that satisfies $M(u)=M_\mathit{LL}(v)$. Then, since the number of all those queries is $O(n)$, Algorithm~\ref{algorithm:W-queries} can answer them in $O(n)$ time in total, according to Lemma~\ref{lemma:W-queries}. This shows that Algorithm~\ref{algorithm:type2-1} runs in linear time. 
\end{proof}

\subsection{The case $B(v)\sqcup\{e_1\}=B(u)\sqcup\{e_2\}$}
\label{subsubsection:case_B(v)cup=B(u)cup}
Let $\{(u,p(u)),(v,p(v)),e_1,e_2\}$ be a Type-$2$ $4$-cut such that $e_1\in B(u)$ and $e_2\in B(v)$. Then, by Lemma~\ref{lemma:type2cuts} we have that $B(u)\sqcup\{e_2\}=B(v)\sqcup\{e_1\}$, and we call this a Type-$2ii$ $4$-cut. Our goal in this section is to prove that we can compute enough such $4$-cuts in linear time, so that the rest of them are implied from the collection we have computed. For a precise statement of our result, see Proposition~\ref{proposition:type-2-2}. The Type-$2ii$ $4$-cuts are the most significant Type-$2$ $4$-cuts, because their existence is the reason why we may have a quadratic number of $4$-cuts overall. This will become clear in the following sections, where we will see that there are some subtypes of $4$-cuts whose number can be quadratic, but they are involved in an implicating sequence with Type-$2ii$ $4$-cuts, and so we can compute in linear time a subcollection of them that implies them all.

The following two lemmata establish conditions under which we have a Type-$2ii$ $4$-cut.

\begin{lemma}
\label{lemma:type-2-2-criterionn}
Let $u$ and $v$ be two vertices $\neq r$ such that $u$ is a proper descendant of $v$ with $\mathit{bcount}(u)=\mathit{bcount}(v)$. Suppose that there is a back-edge $e\in B(v)$ such that $M(B(u)\setminus\{e_\mathit{high}(u)\})=M(B(v)\setminus\{e\})$. Then $B(v)\sqcup\{e_\mathit{high}(u)\}=B(u)\sqcup\{e\}$.
\end{lemma}
\begin{proof}
Let $(x,y)$ be a back-edge in $B(v)\setminus\{e\}$. Then $x$ is a descendant of $M(B(v)\setminus\{e\})$, and therefore a descendant of $M(B(u)\setminus\{e_\mathit{high}(u)\})$, and therefore a descendant of $u$. Furthermore, $y$ is a proper ancestor of $v$, and therefore a proper ancestor of $u$. This shows that $(x,y)\in B(u)$. Due to the generality of $(x,y)\in B(v)\setminus\{e\}$, this implies that $B(v)\setminus\{e\}\subseteq B(u)$. Since $e\in B(v)$ and $\mathit{bcount}(u)=\mathit{bcount}(v)$, this implies that there is a back-edge $e'\in B(u)$ such that $B(v)\setminus\{e\}=B(u)\setminus\{e'\}$. If we assume that $e\in B(u)$, then we have $e'=e$, and therefore $B(v)=B(u)$, contradicting the fact that the graph is $3$-edge-connected. Similarly, we have $e'\notin B(v)$. Thus, $B(v)\setminus\{e\}=B(u)\setminus\{e'\}$ implies that $B(v)\sqcup\{e'\}=B(u)\sqcup\{e\}$, and $e'$ is the unique back-edge in $B(u)\setminus B(v)$.

Let us suppose, for the sake of contradiction, that $e'\neq e_\mathit{high}(u)$. Since $e'$ is the unique back-edge in $B(u)\setminus B(v)$, this implies that $e_\mathit{high}(u)\in B(v)$. Therefore, $\mathit{high}(u)$ is a proper ancestor of $v$. Now let $(x,y)$ be a back-edge in $B(u)$. Then we have that $x$ is a descendant of $u$, and therefore a descendant of $v$. Furthermore, $y$ is an ancestor of $\mathit{high}(u)$, and therefore a proper ancestor of $v$. This shows that $(x,y)\in B(v)$. Due to the generality of $(x,y)\in B(u)$, this implies that $B(u)\subseteq B(v)$. But this contradicts the fact that there is a back-edge in $B(u)\setminus B(v)$. This shows that $e'=e_\mathit{high}(u)$. Therefore, we have $B(v)\sqcup\{e_\mathit{high}(u)\}=B(u)\sqcup\{e\}$.
\end{proof}

\begin{lemma}
\label{lemma:case_B(v)cup=B(u)cup-criterion}
Let $u$ and $v$ be two vertices such that $v$ is a proper ancestor of $u$ with $v\neq r$. Then there exist two distinct back-edges $e_1$ and $e_2$ such that $B(v)\sqcup\{e_1\}=B(u)\sqcup\{e_2\}$ if and only if $\mathit{bcount}(v)=\mathit{bcount}(u)$ and $\mathit{high}_2(u)<v$.
\end{lemma}
\begin{proof}
($\Rightarrow$) $\mathit{bcount}(v)=\mathit{bcount}(u)$ is an immediate consequence of $B(v)\sqcup\{e_1\}=B(u)\sqcup\{e_2\}$. Now let $(x_1,y_1),\dots,(x_k,y_k)$ be the back-edges in $B(u)$ sorted in decreasing order w.r.t. their lower endpoint. (Thus, we have $\mathit{high}_i(u)=y_i$, for every $i\in\{1,\dots,k\}$.) Let $i\in\{1,\dots,k\}$. If $(x_i,y_i)\in B(v)$, then $y_i$ is a proper ancestor of $v$. This implies that every $y_j$ with $j\in\{i,\dots,k\}$ is also a proper ancestor of $v$, which implies that $(x_j,y_j)\in B(v)$ (since all of $x_1,\dots,x_k$ are descendants of $v$). Thus, we cannot have $(x_1,y_1)\in B(v)$, for otherwise we would have $B(u)\subseteq B(v)$, in contradiction to $B(v)\sqcup\{e_1\}=B(u)\sqcup\{e_2\}$ and $e_1\neq e_2$. Since $B(v)\sqcup\{e_1\}=B(u)\sqcup\{e_2\}$ implies that only one back-edge from $B(u)$ is not in $B(v)$, we thus have that $(x_2,y_2)\in B(v)$, and so $\mathit{high}_2(u)<v$.

($\Leftarrow$) Let $(x_1,y_1),\dots,(x_k,y_k)$ be the back-edges in $B(u)$ sorted in decreasing order w.r.t. their lower endpoint. Then $\mathit{high}_2(u)<v$ implies that $(x_2,y_2)\in B(v)$. Therefore, with the same argument that we used above we can infer that $\{(x_2,y_2),\dots,(x_k,y_k)\}\subseteq B(v)$. If we had that $(x_1,y_1)\in B(v)$, then $\mathit{bcount}(v)=\mathit{bcount}(u)$ would imply that $B(v)=B(u)$, in contradiction to the fact that the graph is $3$-edge-connected. Thus we have that $e_1=(x_1,y_1)$ is the only back-edge in $B(u)$ that is not in $B(v)$. Then, $\mathit{bcount}(v)=\mathit{bcount}(u)$ implies that there must be exactly one back-edge $e_2$ in $B(v)$ that is not in $B(u)$. Thus we get $B(v)\sqcup\{e_1\}=B(u)\sqcup\{e_2\}$.
\end{proof}

The following lemma characterizes the back-edges that participate in a Type-$2ii$ $4$-cut.

\begin{lemma}
\label{lemma:case_B(v)cup=B(u)cup-cases}
Let $u$ and $v$ be two vertices such that $v$ is a proper ancestor of $u$ with $v\neq r$, and there exist two distinct back-edges $e_1$ and $e_2$ such that $B(v)\sqcup\{e_1\}=B(u)\sqcup\{e_2\}$. Then we have $e_1=(\mathit{highD}_1(u),\mathit{high}_1(u))$ and either $(1)$ $e_2=(L_1(v),l_1(L_1(v)))$, or $(2)$ $e_2=(R_1(v),l_1(R_1(v)))$.
\end{lemma}
\begin{proof}
The fact that $e_1=(\mathit{highD}_1(u),\mathit{high}_1(u))$ has essentially been proved in the proof of Lemma~\ref{lemma:case_B(v)cup=B(u)cup-criterion}. Now let $(x_1,y_1),\dots,(x_k,y_k)$ be the back-edges in $B(v)$ sorted in increasing order w.r.t. their higher endpoint. (Then we have $x_1=L_1(v)$ and $x_k=R_1(v)$.) Let $i,j\in\{1,\dots,k\}$ be two indices such that $i\leq j$. If both $(x_i,y_i)$ and $(x_j,y_j)$ are in $B(u)$, then we have that both $x_i$ and $x_j$ are descendants of $u$. This implies that all of $x_i,\dots,x_j$ are descendants of $u$, and therefore all the back-edges $(x_i,y_i),\dots,(x_j,y_j)$ are in $B(u)$ (since all of $y_1,\dots,y_k$ are proper ancestors of $u$). Since $B(v)\sqcup\{e_1\}=B(u)\sqcup\{e_2\}$ implies that exactly one back-edge in $B(v)$ is not in $B(u)$ (and that is $e_2$), we thus have that the higher endpoint of $e_2$ is either $L_1(v)$ or $R_1(v)$.

Let us consider the case that $e_2=(L_1(v),y)$, for some vertex $y$. Let us suppose, for the sake of contradiction, that $l_2(L_1(v))<v$. Then there exist at least two back-edges of the form $(L_1(v),z_1)$ and $(L_1(v),z_2)$ that are in $B(v)$. Since $e_2\notin B(u)$, we have that $L_1(v)$ cannot be a descendant of $u$ (for otherwise, since $y$ is a proper ancestor of $v$, we would have that $e_2\in B(u)$). Thus we have that none of $(L_1(v),z_1)$ and $(L_1(v),z_2)$ is in $B(u)$, in contradiction to the fact that there is only one back-edge in $B(v)\setminus B(u)$. This shows that $l_2(L_1(v))\geq v$. Since $e_2\in B(v)$, we have $y<v$, and so $y$ must be $l_1(L_1(v))$. The case that $e_2=(R_1(v),z)$, for some vertez $z$, is treated with a similar argument.
\end{proof}

Let $u$ and $v$ be two vertices such that $u$ is a proper descendant of $v$ with $B(v)\sqcup\{e\}=B(u)\sqcup\{e'\}$, where $e$ and $e'$ are two distinct back-edges. This implies that $B(v)\setminus\{e'\}=B(u)\setminus\{e\}$, and therefore $M(B(v)\setminus\{e'\})=M(B(u)\setminus\{e\})$. By Lemma~\ref{lemma:case_B(v)cup=B(u)cup-cases}, we have $e=e_\mathit{high}(u)$. Furthermore, $B(v)\sqcup\{e\}=B(u)\sqcup\{e'\}$ implies that $\mathit{bcount}(v)=\mathit{bcount}(u)$. Then, we let $\mathit{lowestU}(v,e')$ denote the lowest proper descendant $u'$ of $v$ such that $e_\mathit{high}(u')\notin B(v)$, $e'\notin B(u')$, $M(B(u')\setminus\{e_\mathit{high}(u')\})=M(B(v)\setminus\{e'\})$, and $\mathit{bcount}(u')=\mathit{bcount}(v)$. 

\begin{lemma}
\label{lemma:type-2-2-lowestU}
Let $u$ and $v$ be two vertices such that $u=\mathit{lowestU}(v,e)$, where $e$ is a back-edge in $B(v)$. Then $B(v)\sqcup\{e_\mathit{high}(u)\}=B(u)\sqcup\{e\}$.
\end{lemma}
\begin{proof}
By definition, we have that $u$ is a proper descendant of $v$ such that $e_\mathit{high}(u)\notin B(v)$, $e\notin B(u)$, $M(B(u)\setminus\{e_\mathit{high}(u)\})=M(B(v)\setminus\{e\})$, and $\mathit{bcount}(u)=\mathit{bcount}(v)$. Let $(x,y)$ be a back-edge in $B(v)\setminus\{e\}$. Then $x$ is a descendant of $M(B(v)\setminus\{e\})=M(B(u)\setminus\{e_\mathit{high}(u)\})$, and therefore a descendant of $u$. Furthermore, $y$ is a proper ancestor of $v$, and therefore a proper ancestor of $u$. This shows that $(x,y)\in B(u)$. Due to the generality of $(x,y)\in B(v)\setminus\{e\}$, this implies that $B(v)\setminus\{e\}\subseteq B(u)$. Since $e_\mathit{high}(u)\notin B(v)$, this can be strengthened to $B(v)\setminus\{e\}\subseteq B(u)\setminus\{e_\mathit{high}(u)\}$. Since $\mathit{bcount}(v)=\mathit{bcount}(u)$, this implies that $B(v)\setminus\{e\}=B(u)\setminus\{e_\mathit{high}(u)\}$. Since $e_\mathit{high}(u)\notin B(v)$ and $e\notin B(u)$, this implies that $B(v)\sqcup\{e_\mathit{high}(u)\}=B(u)\sqcup\{e\}$.
\end{proof}

\begin{lemma}
\label{lemma:type-2-2-lowest-for-implying}
Let $u$ and $v$ be two vertices such that $u$ is a proper descendant of $v$ with $B(v)\sqcup\{e\}=B(u)\sqcup\{e'\}$, where $e$ and $e'$ are two distinct back-edges. Let $u'=\mathit{lowestU}(v,e')$. If $u'\neq u$, then $B(u')\sqcup\{e_\mathit{high}(u)\}=B(u)\sqcup\{e_\mathit{high}(u')\}$.
\end{lemma}
\begin{proof}
Since $B(v)\sqcup\{e\}=B(u)\sqcup\{e'\}$, we have $B(v)\setminus\{e'\}=B(u)\setminus\{e\}$, and therefore $M(B(v)\setminus\{e'\})=M(B(u)\setminus\{e\})$. By Lemma~\ref{lemma:case_B(v)cup=B(u)cup-cases}, we have $e=e_\mathit{high}(u)$. Furthermore, $B(v)\sqcup\{e\}=B(u)\sqcup\{e'\}$ implies that $\mathit{bcount}(v)=\mathit{bcount}(u)$. Thus, it makes sense to consider the lowest proper descendant $u'$ of $v$ such that $e_\mathit{high}(u')\notin B(v)$, $e'\notin B(u')$, $M(B(u')\setminus\{e_\mathit{high}(u')\})=M(B(v)\setminus\{e'\})$, and $\mathit{bcount}(u')=\mathit{bcount}(v)$. By definition, we have $u'=\mathit{lowestU}(v,e')$. 

Let us assume that $u'\neq u$.
Lemma~\ref{lemma:type-2-2-lowestU} implies that $B(v)\sqcup\{e_\mathit{high}(u')\}=B(u')\sqcup\{e'\}$. From this we infer that $B(v)\setminus\{e'\}=B(u')\setminus\{e_\mathit{high}(u')\}$. Since $B(v)\setminus\{e'\}=B(u)\setminus\{e_\mathit{high}(u)\}$, this implies that $B(u')\setminus\{e_\mathit{high}(u')\}=B(u)\setminus\{e_\mathit{high}(u)\}$. If we assume that $e_\mathit{high}(u')\in B(u)$, then we get $e_\mathit{high}(u)=e_\mathit{high}(u')$ and $B(u')=B(u)$, in contradiction to the fact that the graph is $3$-edge-connected. Similarly, we cannot have $e_\mathit{high}(u)\in B(u')$. Thus, $B(u')\setminus\{e_\mathit{high}(u')\}=B(u)\setminus\{e_\mathit{high}(u)\}$ implies that $B(u')\sqcup\{e_\mathit{high}(u)\}=B(u)\sqcup\{e_\mathit{high}(u')\}$.
\end{proof}

The following two lemmata show how we can determine the vertex $u=\mathit{lowestU}(v,e)$. They correspond to two distinct cases, depending on whether $M(u)=M(B(u)\setminus\{e_\mathit{high}(u)\})$, or $M(u)\neq M(B(u)\setminus\{e_\mathit{high}(u)\})$.

\begin{lemma}
\label{lemma:type-2ii-firstcase-u}
Let $u$ and $v$ be two vertices such that $u=\mathit{lowestU}(v,e)$, where $e$ is a back-edge in $B(v)$. Suppose that $M(u)=M(B(u)\setminus\{e_\mathit{high}(u)\})$. Then $u$ is either the lowest or the second-lowest proper descendant of $v$ such that $M(u)=M(B(v)\setminus\{e\})$.
\end{lemma}
\begin{proof}
Let $z=M(B(v)\setminus\{e\})$. Since $u=\mathit{lowestU}(v,e)$, we have that $M(B(u)\setminus\{e_\mathit{high}(u)\})=z$. Since $M(u)=M(B(u)\setminus\{e_\mathit{high}(u)\})$, this implies that $M(u)=z$. Now let us suppose, for the sake of contradiction, that $u$ is neither the lowest nor the second-lowest proper descendant of $v$ such that $M(u)=z$. This means that there are two proper descendants $u'$ and $u''$ of $v$, such that $u>u'>u''$ and $M(u')=M(u'')=z$. Then we have that $z$ is a common descendant of $u$, $u'$ and $u''$, and therefore $u$, $u'$ and $u''$ are related as ancestor and descendant. Thus, $u>u'>u''$ implies that $u$ is a proper descendant of $u'$, and $u'$ is a proper descendant of $u''$. Then, since $M(u)=M(u')=M(u'')$, Lemma~\ref{lemma:same_m_subset_B} implies that $B(u'')\subseteq B(u')\subseteq B(u)$. Since the graph is $3$-edge-connected, this can be strengthened to $B(u'')\subset B(u')\subset B(u)$. Thus, there is a back-edge $(x,y)\in B(u')\setminus B(u'')$, and a back-edge $(x',y')\in B(u)\setminus B(u')$. Since $(x,y)\in B(u')$ and $B(u')\subset B(u)$, we have that $(x,y)\in B(u)$. And since $(x',y')\notin B(u')$ and $(x,y)\in B(u')$, we have $(x,y)\neq (x',y')$. Thus, $(x,y)$ and $(x',y')$ are two distinct back-edges in $B(u)$. We have that $x$ is a descendant of $u$, and therefore a descendant of $u''$. Thus, $y$ cannot be a proper ancestor of $v$, because otherwise it is a proper ancestor of $u''$, and therefore $(x,y)\in B(u'')$. This shows that $(x,y)\notin B(v)$. Similarly, $x'$ is a descendant of $u$, and therefore a descendant of $u'$. Thus, $y'$ cannot be a proper ancestor of $v$, because otherwise it is a proper ancestor of $u'$, and therefore $(x',y')\in B(u')$. This shows that $(x',y')\notin B(v)$. Since $u=\mathit{lowestU}(v,e)$, Lemma~\ref{lemma:type-2-2-lowestU} implies that $B(v)\sqcup\{e_\mathit{high}(u)\}=B(u)\sqcup\{e\}$. This implies that there is only one back-edge in $B(u)\setminus B(v)$. But this contradicts the fact that $(x,y)$ and $(x',y')$ are two distinct back-edges in $B(u)\setminus B(v)$. Thus, we conclude that $u$ is either the lowest or the second-lowest proper descendant of $v$ such that $M(u)=M(B(v)\setminus\{e\})$.
\end{proof}

\begin{lemma}
\label{lemma:type-2ii-secondcase-u}
Let $u$ and $v$ be two vertices such that $u=\mathit{lowestU}(v,e)$, where $e$ is a back-edge in $B(v)$. Suppose that $M(u)\neq M(B(u)\setminus\{e_\mathit{high}(u)\})$. Then $u$ is the lowest proper descendant of $v$ such that $M(u)\neq M(B(u)\setminus\{e_\mathit{high}(u)\})=M(B(v)\setminus\{e\})$.
\end{lemma}
\begin{proof}
Let $z=M(B(v)\setminus\{e\})$. Since $u=\mathit{lowestU}(v,e)$, we have that $M(B(u)\setminus\{e_\mathit{high}(u)\})=z$. Now let us suppose, for the sake of contradiction, that $u$ is not the lowest proper descendant of $v$ such that $M(u)\neq M(B(u)\setminus\{e_\mathit{high}(u)\})=z$. This means that there is a proper descendant $u'$ of $v$, such that $u>u'$ and $M(u')\neq M(B(u')\setminus\{e_\mathit{high}(u')\})=z$. Then we have that $z$ is a common descendant of $u$ and $u'$, and therefore $u$ and $u'$ are related as ancestor and descendant. Thus, $u>u'$ implies that $u$ is a proper descendant of $u'$.

Let us suppose, for the sake of contradiction, that $e_\mathit{high}(u)\in B(u')$. This implies that $\mathit{high}(u)$ is a proper ancestor of $u'$. Let $(x,y)$ be a back-edge in $B(u)$. Then $x$ is a descendant of $u$, and therefore a descendant of $u'$. Furthermore, $y$ is an ancestor of $\mathit{high}(u)$, and therefore a proper ancestor of $u'$. This shows that $(x,y)\in B(u')$. Due to the generality of $(x,y)\in B(u)$, this implies that $B(u)\subseteq B(u')$. Since $M(u)\neq M(B(u)\setminus\{e_\mathit{high}(u)\})=z$, we have that the higher endpoint of $e_\mathit{high}(u)$ is not a descendant of $z$. Similarly, since $M(u')\neq M(B(u')\setminus\{e_\mathit{high}(u')\})=z$, we have that the higher endpoint of $e_\mathit{high}(u')$ is not a descendant of $z$. Furthermore, $e_\mathit{high}(u')$ is the only back-edge in $B(u')$ with this property. Since $e_\mathit{high}(u)\in B(u')$, this implies that $e_\mathit{high}(u')=e_\mathit{high}(u)$. Now let $(x,y)$ be a back-edge in $B(u')$. Then, if $(x,y)=e_\mathit{high}(u')$, we have that $(x,y)=e_\mathit{high}(u)\in B(u)$. Otherwise, we have that $x$ is a descendant of $z$, and therefore a descendant of $u$. Furthermore, $y$ is a proper ancestor of $u'$, and therefore a proper ancestor of $u$. This shows that $(x,y)\in B(u)$. Due to the generality of $(x,y)\in B(u')$, this implies that $B(u')\subseteq B(u)$. Thus, we have $B(u')=B(u)$, in contradiction to the fact that the graph is $3$-edge-connected. This shows that $e_\mathit{high}(u)\notin B(u')$.

Let us suppose, for the sake of contradiction, that $e_\mathit{high}(u')\in B(u)$. Since $M(u')\neq M(B(u')\setminus\{e_\mathit{high}(u')\})=z$, we have that the higher endpoint of $e_\mathit{high}(u')$ is not a descendant of $z$. Since $M(u)\neq M(B(u)\setminus\{e_\mathit{high}(u)\})=z$, we have that $e_\mathit{high}(u)$ is the only back-edge in $B(u)$ with the property that its higher endpoint is not a descendant of $z$. Thus, since $e_\mathit{high}(u')\in B(u)$, we have that $e_\mathit{high}(u)=e_\mathit{high}(u')$. But this implies that $e_\mathit{high}(u)\in B(u')$, a contradiction. Thus, we have $e_\mathit{high}(u')\notin B(u)$.

Let us suppose, for the sake of contradiction, that $e_\mathit{high}(u')\in B(v)$. This implies that $\mathit{high}(u')$ is a proper ancestor of $v$. Since $u=\mathit{lowestU}(v,e)$, by Lemma~\ref{lemma:type-2-2-lowestU} we have that $B(v)\sqcup\{e_\mathit{high}(u)\}=B(u)\sqcup\{e\}$. This implies that $e$ is the only back-edge in $B(v)$ that is not in $B(u)$. Since $e_\mathit{high}(u')\notin B(u)$ and $e_\mathit{high}(u')\in B(v)$, this implies that $e=e_\mathit{high}(u')$. Now let $(x,y)$ be a back-edge in $B(u')$. Then, $x$ is a descendant of $u'$, and therefore a descendant of $v$. Furthermore, $y$ is an ancestor of $\mathit{high}(u')$, and therefore a proper ancestor of $v$. This shows that $(x,y)\in B(v)$. Due to the generality of $(x,y)\in B(u')$, this implies that $B(u')\subseteq B(v)$. Conversely, let $(x,y)$ be a back-edge in $B(v)$. If $(x,y)=e$, then $e=e_\mathit{high}(u')\in B(u')$. Otherwise, $x$ is a descendant of $M(B(v)\setminus\{e\})=z$, and therefore a descendant of $u'$. Furthermore, $y$ is a proper ancestor of $v$, and therefore a proper ancestor of $u'$. This shows that $(x,y)\in B(u')$. Due to the generality of $(x,y)\in B(v)$, this implies that $B(v)\subseteq B(u')$. Thus, we have $B(u')=B(v)$, in contradiction to the fact that the graph is $3$-edge-connected. This shows that $e_\mathit{high}(u')\notin B(v)$.

Let us suppose, for the sake of contradiction, that $e\in B(u')$. Since $u=\mathit{lowestU}(v,e)$, by Lemma~\ref{lemma:type-2-2-lowestU} we have that $B(v)\sqcup\{e_\mathit{high}(u)\}=B(u)\sqcup\{e\}$. This implies that $e\notin B(u)$. The lower endpoint of $e$ is a proper ancestor of $v$, and therefore a proper ancestor of $u$. Thus, since $e\notin B(u)$, we have that the higher endpoint of $e$ is not a descendant of $u$. This implies that the higher endpoint of $e$ is not a descendant of $z$ either. Since $M(u')\neq M(B(u')\setminus\{e_\mathit{high}(u')\})=z$, we have that $e_\mathit{high}(u')$ is the only back-edge in $B(u')$ whose higher endpoint is not a descendant of $z$. Since $e\in B(u')$, this shows that $e_\mathit{high}(u')=e$. But this implies that $e_\mathit{high}(u')\in B(v)$, a contradiction. Thus, we have $e\notin B(u')$.

Now let $(x,y)$ be a back-edge in $B(u')\setminus\{e_\mathit{high}(u')\}$. Then $x$ is a descendant of $z$, and therefore a descendant of $u$. Furthermore, $y$ is a proper ancestor of $u'$, and therefore a proper ancestor of $u$. This shows that $(x,y)\in B(u)$.  Due to the generality of $(x,y)\in B(u')\setminus\{e_\mathit{high}(u')\}$, this implies that $B(u')\setminus\{e_\mathit{high}(u')\}\subseteq B(u)$. And since $e_\mathit{high}(u)\notin B(u')$, this can be strengthened to $B(u')\setminus\{e_\mathit{high}(u')\}\subseteq B(u)\setminus\{e_\mathit{high}(u)\}$. Since $u=\mathit{lowestU}(v,e)$, by Lemma~\ref{lemma:type-2-2-lowestU} we have that $B(v)\sqcup\{e_\mathit{high}(u)\}=B(u)\sqcup\{e\}$. This implies that $B(u)\setminus\{e_\mathit{high}(u)\}=B(v)\setminus\{e\}$. Thus, we have $B(u')\setminus\{e_\mathit{high}(u')\}\subseteq B(v)\setminus\{e\}$. Conversely, let $(x,y)$ be a back-edge in $B(v)\setminus\{e\}$. Then $x$ is a descendant of $z$, and therefore a descendant of $u'$. Furthermore, $y$ is a proper ancestor of $v$, and therefore a proper ancestor of $u'$. This shows that $(x,y)\in B(u')$. Due to the generality of $(x,y)\in B(v)\setminus\{e\}$, this implies that $B(v)\setminus\{e\}\subseteq B(u')$. And since $e_\mathit{high}(u')\notin B(v)$, this can be strengthened to $B(v)\setminus\{e\}\subseteq B(u')\setminus\{e_\mathit{high}(u')\}$. Thus, we have $B(v)\setminus\{e\}=B(u')\setminus\{e_\mathit{high}(u')\}$. This implies that $\mathit{bcount}(v)=\mathit{bcount}(u')$. Thus, we have that $u'$ is a proper descendant of $v$ such that $e_\mathit{high}(u')\notin B(v)$, $e\notin B(u')$, $M(B(u')\setminus\{e_\mathit{high}(u')\})=M(B(v)\setminus\{e\})$, and $\mathit{bcount}(v)=\mathit{bcount}(u')$. But since $u'$ is lower than $u$, we have a contradiction to the minimality of $u=\mathit{lowestU}(v,e)$. 

Thus, we conclude that $u$ is the lowest proper descendant of $v$ such that $M(u)\neq M(B(u)\setminus\{e_\mathit{high}(u)\})=M(B(v)\setminus\{e\})$. 
\end{proof}

\begin{algorithm}[h!]
\caption{\textsf{Compute a collection of Type-$2ii$ $4$-cuts, which implies all Type-$2ii$ $4$-cuts}}
\label{algorithm:type2-2}
\LinesNumbered
\DontPrintSemicolon
\ForEach{vertex $v\neq r$}{
\label{line:type-2-2-compute1}
  compute $M_L(v)=M(B(v)\setminus\{e_L(v)\})$ and $M_R(v)=M(B(v)\setminus\{e_R(v)\})$\;
}
\ForEach{vertex $u\neq r$}{
\label{line:type-2-2-compute2}
  compute $M_\mathit{high}(u)=M(B(u)\setminus\{e_\mathit{high}(u)\})$\;  
}
\ForEach{vertex $v\neq r$}{
  compute the lowest proper descendant $u$ of $v$ with $M(u)\neq M_\mathit{high}(u)=M_L(v)$; denote this vertex as $u_\mathit{highL}(v)$\;
  \label{line:type-2-2-lowest-u}
  compute the lowest proper descendant $u$ of $v$ with $M(u)\neq M_\mathit{high}(u)=M_R(v)$; denote this vertex as $u_\mathit{highR}(v)$\;  
  \label{line:type-2-2-lowest-u-2}
  compute the lowest proper descendant $u$ of $v$ with $M(u)=M_L(v)$; denote this vertex as $u_L(v)$\;
  \label{line:type-2-2-lowest-u-3}
  compute the lowest proper descendant $u$ of $v$ with $M(u)=M_R(v)$; denote this vertex as $u_R(v)$\;
  \label{line:type-2-2-lowest-u-4}
}
\ForEach{vertex $v\neq r$}{
  let $u\leftarrow u_\mathit{highL}(v)$\;
  \If{$\mathit{bcount}(u)=\mathit{bcount}(v)$}{
    mark $\{(u,p(u)),(v,p(v)),e_\mathit{high}(u),e_L(v)\}$ as a Type-$2ii$ $4$-cut\;
    \label{line:type-2-2-mark1}
  }
  let $u\leftarrow u_L(v)$\;
  \If{$M_\mathit{high}(u)=M_L(v)$ \textbf{and} $\mathit{bcount}(u)=\mathit{bcount}(v)$}{
  \label{line:type-2-2-cond1}
    mark $\{(u,p(u)),(v,p(v)),e_\mathit{high}(u),e_L(v)\}$ as a Type-$2ii$ $4$-cut\;
    \label{line:type-2-2-mark2}
  }
  let $u\leftarrow\mathit{prevM}(u)$\;
  \label{line:type-2-2-previous}
  \If{$M_\mathit{high}(u)=M_L(v)$ \textbf{and} $\mathit{bcount}(u)=\mathit{bcount}(v)$}{
  \label{line:type-2-2-cond2}
    mark $\{(u,p(u)),(v,p(v)),e_\mathit{high}(u),e_L(v)\}$ as a Type-$2ii$ $4$-cut\;
    \label{line:type-2-2-mark3}
  }
  let $u\leftarrow u_\mathit{highR}(v)$\;
  \If{$\mathit{bcount}(u)=\mathit{bcount}(v)$}{
    mark $\{(u,p(u)),(v,p(v)),e_\mathit{high}(u),e_R(v)\}$ as a Type-$2ii$ $4$-cut\;
    \label{line:type-2-2-mark1'}
  }
  let $u\leftarrow u_R(v)$\;
  \If{$M_\mathit{high}(u)=M_R(v)$ \textbf{and} $\mathit{bcount}(u)=\mathit{bcount}(v)$}{
  \label{line:type-2-2-cond3}
    mark $\{(u,p(u)),(v,p(v)),e_\mathit{high}(u),e_R(v)\}$ as a Type-$2ii$ $4$-cut\;
    \label{line:type-2-2-mark2'}   
  }
  let $u\leftarrow\mathit{prevM}(u)$\;
  \If{$M_\mathit{high}(u)=M_R(v)$ \textbf{and} $\mathit{bcount}(u)=\mathit{bcount}(v)$}{
  \label{line:type-2-2-cond4}
    mark $\{(u,p(u)),(v,p(v)),e_\mathit{high}(u),e_R(v)\}$ as a Type-$2ii$ $4$-cut\;
    \label{line:type-2-2-mark3'}   
  }
}
\end{algorithm}

\begin{proposition}
\label{proposition:type-2-2}
Algorithm~\ref{algorithm:type2-2} computes a collection $\mathcal{C}$ of Type-$2ii$ $4$-cuts, such that every Type-$2ii$ $4$-cut of the form $\{(u,p(u)),(v,p(v)),e_1,e_2\}$, where $B(v)\sqcup\{e_1\}=B(u)\sqcup\{e_2\}$, is implied by $\mathcal{C}$ through the pair of edges $\{(u,p(u)),e_1\}$. Furthermore, Algorithm~\ref{algorithm:type2-2} has a linear-time implementation.
\end{proposition}
\begin{proof}
First, we observe that all $4$-element sets marked by Algorithm~\ref{algorithm:type2-2} are Type-$2ii$ $4$-cuts. To see this, notice that the markings take place in Lines~\ref{line:type-2-2-mark1}, \ref{line:type-2-2-mark2}, \ref{line:type-2-2-mark3}, \ref{line:type-2-2-mark1'}, \ref{line:type-2-2-mark2'}, or \ref{line:type-2-2-mark3'}. In Lines~\ref{line:type-2-2-mark1} and \ref{line:type-2-2-mark1'}, we have that $u=u_\mathit{highL}(v)$ or $u=u_\mathit{highR}(v)$, respectively, and so $u$ is a proper descendant of $v$ such that $M(B(u)\setminus\{e_\mathit{high}(u)\})=M(B(v)\setminus\{e\})$, where $e=e_L(v)$ or $e=e_R(v)$, respectively, and $\mathit{bcount}(u)=\mathit{bcount}(v)$. Thus, Lemma~\ref{lemma:type-2-2-criterionn} implies that $B(v)\sqcup\{e_\mathit{high}(u)\}=B(u)\sqcup\{e\}$, and therefore Lemma~\ref{lemma:type2cuts} implies that $\{(u,p(u)),(v,p(v)),e_\mathit{high}(u),e\}$ is a Type-$2ii$ $4$-cut. In Lines~\ref{line:type-2-2-mark2}, \ref{line:type-2-2-mark3}, \ref{line:type-2-2-mark2'}, and \ref{line:type-2-2-mark3'}, we have that $u=u_L(v)$, $u=\mathit{prevM}(u_L(v))$, $u=u_R(v)$, or $u=\mathit{prevM}(u_R(v))$, respectively, and so $u$ is also a proper descendant of $v$. Furthermore, since the conditions in Lines~\ref{line:type-2-2-cond1}, \ref{line:type-2-2-cond2}, \ref{line:type-2-2-cond3}, or \ref{line:type-2-2-cond4}, respectively, are satisfied, we have that $M(B(u)\setminus\{e_\mathit{high}(u)\})=M(B(v)\setminus\{e\})$, where $e=e_L(v)$ or $e=e_R(v)$, and $\mathit{bcount}(u)=\mathit{bcount}(v)$. Thus, Lemma~\ref{lemma:type-2-2-criterionn} implies that $B(v)\sqcup\{e_\mathit{high}(u)\}=B(u)\sqcup\{e\}$, and therefore Lemma~\ref{lemma:type2cuts} implies that $\{(u,p(u)),(v,p(v)),e_\mathit{high}(u),e\}$ is a Type-$2ii$ $4$-cut. Thus, the collection $\mathcal{C}$ of $4$-element sets marked by Algorithm~\ref{algorithm:type2-2} is a collection of Type-$2ii$ $4$-cuts.

Now let $\{(u,p(u)),(v,p(v)),e_1,e_2\}$ be a Type-$2ii$ $4$-cut. Then we may assume w.l.o.g. that $u$ is a proper descendant of $v$, and $B(v)\sqcup\{e_1\}=B(u)\sqcup\{e_2\}$. This implies that $B(u)\setminus\{e_1\}=B(v)\setminus\{e_2\}$, and therefore $M(B(u)\setminus\{e_1\})=M(B(v)\setminus\{e_2\})$. Furthermore, we have $e_1\neq e_2$, and $B(u)\sqcup\{e_2\}=B(v)\sqcup\{e_1\}$ implies that $\mathit{bcount}(v)=\mathit{bcount}(u)$. By Lemma~\ref{lemma:case_B(v)cup=B(u)cup-cases} we have that $e_1=e_\mathit{high}(u)$. Thus, it makes sense to consider the lowest proper descendant $u'$ of $v$ such that $e_\mathit{high}(u')\notin B(v)$, $e_2\notin B(u')$, $M(B(u')\setminus\{e_\mathit{high}(u')\})=M(B(v)\setminus\{e_2\})$ and $\mathit{bcount}(u')=\mathit{bcount}(v)$. In other words, we have that $u'=\mathit{lowestU(v,e_2)}$ is defined. Then, Lemma~\ref{lemma:type-2-2-lowestU} implies that $\{(u',p(u')),(v,p(v)),e_\mathit{high}(u'),e_2\}$ is also a Type-$2ii$ $4$-cut. We will show that $\{(u',p(u')),(v,p(v)),e_\mathit{high}(u'),e_2\}$ is marked by Algorithm~\ref{algorithm:type2-2}.

By Lemma~\ref{lemma:case_B(v)cup=B(u)cup-cases} we have that either $e_2=e_L(v)$, or $e_2=e_R(v)$. Let us assume that $e_2=e_L(v)$ (the argument for $e_2=e_R(v)$ is similar). If $M(u')\neq M(B(u')\setminus\{e_\mathit{high}(u')\})$, then Lemma~\ref{lemma:type-2ii-secondcase-u} implies that $u'$ is the lowest proper descendant of $v$ such that $M(u')\neq M(B(u')\setminus\{e_\mathit{high}(u')\})=M(B(v)\setminus \{e_L(v)\})$. Thus, we have $u'=u_\mathit{highL}(v)$ (see Line~\ref{line:type-2-2-lowest-u}), and therefore $\{(u',p(u')),(v,p(v)),e_\mathit{high}(u'),e_L(v)\}$ will be marked in Line~\ref{line:type-2-2-mark1}. On the other hand, if $M(u')=M(B(u')\setminus\{e_\mathit{high}(u')\})$, then Lemma~\ref{lemma:type-2ii-firstcase-u} implies that $u'$ is either the lowest or the second-lowest proper descendant of $v$ such that $M(u')=M(B(v)\setminus\{e_L(v)\})$. Thus, we have that either $u'=u_L(v)$ or $u'=\mathit{prevM}(u_L(v))$ (see Line~\ref{line:type-2-2-lowest-u-3}), and therefore $\{(u',p(u')),(v,p(v)),e_\mathit{high}(u'),e_L(v)\}$ will be marked in either Line~\ref{line:type-2-2-mark2} or Line~\ref{line:type-2-2-mark3}.

Let us rephrase our result so far in a more succinct notation. let $v\neq r$ be a vertex and let $\tilde{e}$ be a back-edge in $B(v)$, such that there is a Type-$2ii$ $4$-cut of the form $\{(u,p(u)),(v,p(v)),e,\tilde{e}\}$, where $u$ is a proper descendant of $v$. Then, we may consider the lowest proper descendant $u'$ of $v$ such that there is a Type-$2ii$ $4$-cut of the form $\{(u',p(u')),(v,p(v)),e',\tilde{e}\}$. Let $C(v,\tilde{e})$ denote $\{(u',p(u')),(v,p(v)),e',\tilde{e}\}$. Then, we have basically shown that $C(v,\tilde{e})\in\mathcal{C}$. We denote $u'$ as $U(v,\tilde{e})$. Notice that, by Lemma~\ref{lemma:case_B(v)cup=B(u)cup-cases}, we have $e'=e_\mathit{high}(u')$.

Now, let $C=\{(u,p(u)),(v,p(v)),e,\tilde{e}\}$ be a Type-2$ii$ $4$-cut such that $u$ is a proper descendant of $v$ and $B(u)\sqcup\{\tilde{e}\}=B(v)\sqcup\{e\}$. If $C\in\mathcal{C}$, then we have that $\mathcal{C}$ implies $C$ through any partition of $C$ into pairs of edges. So let us suppose that $C\notin\mathcal{C}$. By Lemma~\ref{lemma:case_B(v)cup=B(u)cup-cases}, we have $e=e_\mathit{high}(u)$. Now consider the $4$-cut $C_1=C(v,\tilde{e})\in\mathcal{C}$, and let $u_1=U(v,\tilde{e})$. Then we have $C_1=\{(u_1,p(u_1)),(v,p(v)),e_\mathit{high}(u_1),\tilde{e}\}$ and $B(u_1)\sqcup\{\tilde{e}\}=B(v)\sqcup\{e_\mathit{high}(u_1)\}$. Since $C\notin\mathcal{C}$, we have $u\neq u_1$. Thus, Lemma~\ref{lemma:type-2-2-lowest-for-implying} implies that $B(u)\sqcup\{e_\mathit{high}(u_1)\}=B(u_1)\sqcup\{e_\mathit{high}(u)\}$. Notice that $u$ and $u_1$ are related as ancestor and descendant. (To see this, consider any back-edge $(x,y)\in B(u)\setminus\{e_\mathit{high}(u)\}=B(u_1)\setminus\{e_\mathit{high}(u_1)\}$. Then we have that $x$ is a common descendant of $u$ and $u_1$, and therefore $u$ and $u_1$ are related as ancestor and descendant.) Therefore, Lemma~\ref{lemma:type2cuts} implies that $C_1'=\{(u,p(u)),(u_1,p(u_1)),e,e_\mathit{high}(u_1)\}$ is a Type-$2ii$ $4$-cut. If $C_1'\in\mathcal{C}$, then we have that $C$ is implied by $\mathcal{C}$ through the pair of edges $\{(u,p(u)),e\}$, since $\{C_1,C_1'\}\subseteq\mathcal{C}$. So let us suppose that $C_1'\notin\mathcal{C}$. 

Due to the mimimality of $u_1$, we have $u_1<u$. Thus, $u$ is a proper descendant of $u_1$. Therefore, we can consider the $4$-cut $C_2=C(u_1,e_\mathit{high}(u_1))$, and let $u_2=U(u_1,e_\mathit{high}(u_1))$. Then we have $C_2=\{(u_2,p(u_2)),(u_1,p(u_1)),e_\mathit{high}(u_2),e_\mathit{high}(u_1)\}$ and $B(u_2)\sqcup\{e_\mathit{high}(u_1)\}=B(u_1)\sqcup\{e_\mathit{high}(u_2)\}$. Since $C_1\notin\mathcal{C}$, we have $u\neq u_2$. Thus, Lemma~\ref{lemma:type-2-2-lowest-for-implying} implies that $B(u)\sqcup\{e_\mathit{high}(u_2)\}=B(u_2)\sqcup\{e_\mathit{high}(u)\}$. Therefore, Lemma~\ref{lemma:type2cuts} implies that $C_2'=\{(u,p(u)),(u_2,p(u_2)),e,e_\mathit{high}(u_2)\}$ is a Type-$2ii$ $4$-cut. If $C_2'\in\mathcal{C}$, then $C$ is implied by $\mathcal{C}$ through the pair of edges $\{(u,p(u)),e\}$, since $\{C_1,C_2,C_2'\}\subseteq\mathcal{C}$. Otherwise, we can proceed in the same manner, and eventually this process must terminate, because we consider a proper descendant $u_1$ of $v$, then a proper descendant $u_2$ of $u_1$, and so on. Termination implies that we will have arrived at a sequence of $4$-cuts $C_1,C_2,\dots,C_k,C_k'$ in $\mathcal{C}$, such that $C_1=\{(u_1,p(u_1)),(v,p(v)),e_\mathit{high}(u_1),\tilde{e}\}$, $C_i=\{(u_i,p(u_i)),(u_{i-1},p(u_{i-1})),e_\mathit{high}(u_i),e_\mathit{high}(u_{i-1})\}$ for every $i\in\{2,\dots,k\}$, and $C_k'=\{(u,p(u)),(u_k,p(u_k)),e,e_\mathit{high}(u_k)\}$. Thus, $C$ is implied by $\mathcal{C}$ through the pair of edges $\{(u,p(u)),e\}$.

It remains to establish the linear complexity of Algorithm~\ref{algorithm:type2-2}. First, Proposition~\ref{proposition:L-sets} implies that we can compute the edges $e_L(v)$ and $e_R(v)$, for all vertices $v\neq r$, in linear time in total. Also, Proposition~\ref{proposition:high} implies that we can compute the edges $e_\mathit{high}(u)$, for all vertices $u\neq r$, in linear time in total. Then, by Proposition~\ref{proposition:computing-M(B(v)-S)} we can compute the values $M(B(v)\setminus\{e_L(v)\})$ and  $M(B(v)\setminus\{e_R(v)\})$, for all vertices $v\neq r$, in linear time in total. Thus, the \textbf{for} loop in Line~\ref{line:type-2-2-compute1} can be performed in linear time. Similarly, the \textbf{for} loop in Line~\ref{line:type-2-2-compute2} can also be performed in linear time.
In order to compute the vertices $u_\mathit{highL}(v)$ and $u_\mathit{highR}(v)$ in Lines~\ref{line:type-2-2-lowest-u} and \ref{line:type-2-2-lowest-u-2}, respectively, we use Algorithm~\ref{algorithm:W-queries}. Specifically, for every vertex $x$, let $M_\mathit{high}^{-1}(x)$ denote the set of all vertices $u\neq r$ such that $M(u)\neq M(B(u)\setminus\{e_\mathit{high}(u)\})=x$. Then, if $x$ and $y$ are two distinct vertices, we have that the sets $M_\mathit{high}^{-1}(x)$ and $M_\mathit{high}^{-1}(y)$ are disjoint. Now let $v\neq r$ be a vertex, and let $x=M(B(v)\setminus\{e_L(v)\})$. Then we generate a query $q(M_\mathit{high}^{-1}(x),v)$. This is to return the lowest vertex $u$ such that $u\in M_\mathit{high}^{-1}(x)$ and $u>v$. This implies that  $M(B(u)\setminus\{e_\mathit{high}(u)\})=x=M(B(v)\setminus\{e_L(v)\})$, and therefore we have that $x$ is a common descendant of $u$ and $v$, and therefore $u$ and $v$ are related as ancestor and descendant. Then, $u>v$ implies that $u$ is a proper descendant of $v$. Thus, we have that $u$ is the lowest proper descendant of $v$ such that $u\in M_\mathit{high}^{-1}(x)$, and therefore $u=u_\mathit{highL}(v)$. Since the number of all those queries is $O(n)$, Algorithm~\ref{algorithm:W-queries} can answer all of them in linear time in total, according to Lemma~\ref{lemma:W-queries}. Similarly, we can compute all vertices $u_\mathit{highR}(v)$, for $v\neq r$, in linear time in total. Also, we can similarly compute the vertices $u_L(v)$ and $u_R(v)$ in Lines~\ref{line:type-2-2-lowest-u-3} and \ref{line:type-2-2-lowest-u-4}, respectively, in linear time in total. We conclude that Algorithm~\ref{algorithm:type2-3} runs in linear time.
\end{proof}

\subsection{The case $B(u)=B(v)\sqcup\{e_1,e_2\}$}
\label{subsubsection:case_B(u)=B(v)cup2}

\begin{lemma}
\label{lemma:case_B(v)=B(u)cup2}
Let $u$ and $v$ be two vertices such that $v$ is a proper ancestor of $u$ with $v\neq r$. Then there exist two distinct back-edges $e_1,e_2$ such that $B(u)=B(v)\sqcup\{e_1,e_2\}$ if and only if: $\mathit{bcount}(u)=\mathit{bcount}(v)+2$ and $M(B(u)\setminus\{e_1,e_2\})=M(v)$, where $e_1=(\mathit{highD}_1(u),\mathit{high}_1(u))$ and $e_2=(\mathit{highD}_2(u),\mathit{high}_2(u))$ (or reversely). 
\end{lemma}
\begin{proof} 
($\Rightarrow$) $\mathit{bcount}(u)=\mathit{bcount}(v)+2$ and $M(B(u)\setminus\{e_1,e_2\})=M(v)$ are immediate consequences of $B(u)=B(v)\sqcup\{e_1,e_2\}$. Now let $(x_1,y_1),\dots,(x_k,y_k)$ be all the back-edges in $B(u)$ sorted in decreasing order w.r.t. their lower endpoint. (We note that $(x_i,y_i)=(\mathit{highD}_i(u),\mathit{high}_i(u))$, for every $i\in\{1,\dots,k\}$.) Let $i\in\{1,\dots,k\}$ be an index such that $(x_i,y_i)\in B(v)$. Then $y_i$ is a proper ancestor of $v$, and therefore $y_i<v$. This implies that every $y_j$, for $j\in\{i,\dots,k\}$, has $y_j<v$. Since $y_j$ is a proper ancestor of $u$ and $u$ is a descendant of $v$, this implies that $y_j$ is a proper ancestor of $v$. Thus we have that all back-edges $(x_i,y_i),\dots,(x_k,y_k)$ are in $B(v)$, since all $x_1,\dots,x_k$ are descendants of $v$. Since $B(v)=B(u)\setminus\{e_1,e_2\}$, we have that exactly two back-edges in $B(u)$ are not in $B(v)$ (i.e., $e_1$ and $e_2$). Thus we have $\{e_1,e_2\}=\{(x_1,y_1),(x_2,y_2)\}$.\\
($\Leftarrow$) Let $e_1=(\mathit{highD}_1(u),\mathit{high}_1(u))$ and $e_2=(\mathit{highD}_2(u),\mathit{high}_2(u))$. We have that $M(B(u)\setminus\{e_1,e_2\})$ is a descendant of $M(u)$, and therefore $M(v)$ is a descendant of $M(u)$. Thus, since $v$ is an ancestor of $u$, by Lemma~\ref{lemma:same_m_subset_B} we have that $B(v)\subseteq B(u)$. Since $\mathit{bcount}(u)=\mathit{bcount}(v)+2$, this implies that there exist two back-edges $e_1',e_2'\in B(u)\setminus B(v)$ such that $B(u)=B(v)\sqcup\{e_1',e_2'\}$. By the $\Rightarrow$ direction we have $\{e_1',e_2'\}=\{e_1,e_2\}$.
\end{proof}

\begin{lemma}
\label{lemma:case_B(v)=B(u)cup2-necessary}
Let $u$ and $v$ be two vertices such that $v$ is a proper ancestor of $u$ and $B(u)=B(v)\sqcup\{e_1,e_2\}$, for two back-edges $e_1,e_2$. Then $v$ is either the greatest or the second-greatest proper ancestor of $u$ with $M(v)=M(B(u)\setminus\{e_1,e_2\})$.
\end{lemma}
\begin{proof}
First, by Lemma~\ref{lemma:case_B(v)=B(u)cup2} we have that $M(v)=M(B(u)\setminus\{e_1,e_2\})$. Thus, we may consider the greatest proper ancestor $v'$ of $u$ with $M(v')=M(B(u)\setminus\{e_1,e_2\})$. If $v'=v$, then we are done. Otherwise, let us suppose, for the sake of contradiction, that $v'\neq v$ and $v$ is not the second-greatest proper ancestor of $u$ with $M(v)=M(B(u)\setminus\{e_1,e_2\})$. Then there is a proper descendant $v''$ of $v$ that is a proper ancestor of $v'$, such that $M(v'')=M(B(u)\setminus\{e_1,e_2\})$. Since $M(v')=M(v'')=M(v)$, by Lemma~\ref{lemma:same_m_subset_B} we have that $B(v)\subseteq B(v'')\subseteq B(v')$. This can be strengthened to $B(v)\subset B(v'')\subset B(v')$, since the graph is $3$-edge-connected. This implies that there is a back-edge $e\in B(v'')\setminus B(v)$ and a back-edge $e'\in B(v')\setminus B(v'')$. Then neither of $e$ and $e'$ is in $B(v)$, but both of them are in $B(v')$. 

Now let $(x,y)$ be a back-edge in $B(v')$. Then we have that $y$ is a proper ancestor of $v'$, and therefore a proper ancestor of $u$. Furthermore, $x$ is a descendant of $M(v')$, and therefore it is a descendant of $M(B(u)\setminus\{e_1,e_2\})$ (since $M(v)=M(B(u)\setminus\{e_1,e_2\})$), and therefore a descendant of $u$. This shows that $(x,y)$ is in $B(u)$, and thus we have $B(v')\subseteq B(u)$. In particular, we have that both $e$ and $e'$ are in $B(u)$. But since none of them is in $B(v)$, by $B(u)=B(v)\sqcup\{e_1,e_2\}$ we have that $\{e,e'\}=\{e_1,e_2\}$. Now let $(x,y)$ be a back-edge in $B(u)$. If $(x,y)=e_1$ or $(x,y)=e_2$, then $(x,y)\in B(v')$. Otherwise, $B(u)=B(v)\sqcup\{e_1,e_2\}$ implies that $(x,y)\in B(v)$, and therefore  $B(v)\subseteq B(v'')\subseteq B(v')$ implies that $(x,y)\in B(v')$. This shows that $B(u)\subseteq B(v')$. Thus we have $B(v')=B(u)$, in contradiction to the fact that the graph is $3$-edge-connected. Thus, we have that $v$ is the second-greatest proper ancestor of $u$ with $M(v)=M(B(u)\setminus\{e_1,e_2\})$. 
\end{proof}

Now we can describe the method to compute all $4$-cuts of the form $\{(u,p(u)),(v,p(v)),e_1,e_2\}$, where $v$ is an ancestor of $u$ and $B(u)=B(v)\sqcup\{e_1,e_2\}$. The idea is to find, for every vertex $u$, a good candidate proper ancestor $v$ of $u$ that may provide such a $4$-cut. According to Lemma~\ref{lemma:case_B(v)=B(u)cup2-necessary}, $v$ must be either the greatest or the second-greatest proper ancestor of $u$ that satisfies $M(v)=M(B(u)\setminus\{e_1,e_2\})$, where $e_i$ is the $\mathit{high}_i$-edge of $u$, for $i\in\{1,2\}$. Then, if such a $v$ exists, we can simply apply Lemma~\ref{lemma:case_B(v)=B(u)cup2} in order to check whether $u$ and $v$ satisfy $B(u)=B(v)\sqcup\{e_1,e_2\}$.
This procedure is implemented in Algorithm~\ref{algorithm:type2-3}. The proof of correctness and linear complexity is given in Proposition~\ref{proposition:algorithm:type2-3}.

\noindent\\
\begin{algorithm}[H]
\caption{\textsf{Compute all $4$-cuts of the form $\{(u,p(u)),(v,p(v)),e_1,e_2\}$, where $v$ is an ancestor of $u$ and $B(u)=B(v)\sqcup\{e_1,e_2\}$}}
\label{algorithm:type2-3}
\LinesNumbered
\DontPrintSemicolon
\ForEach{vertex $u\neq r$}{
\label{line:type-2-3-for}
  let $e_i(u)\leftarrow (\mathit{highD}_i(u),\mathit{high}_i(u))$, for $i\in\{1,2\}$\;
  compute $M(B(u)\setminus\{e_1(u),e_2(u)\})$\;
}
\ForEach{vertex $u\neq r$}{
  let $v$ be the greatest proper ancestor of $u$ such that $M(v)=M(B(u)\setminus\{e_1(u),e_2(u)\})$\;
  \label{line:type-2-3-v}
  \If{$\mathit{bcount}(u)=\mathit{bcount}(v)+2$}{
    mark $\{(u,p(u)),(v,p(v)),e_1(u),e_2(u)\}$ as a $4$-cut\;
    \label{line:type-2-3-mark-1}
  }
  let $v\leftarrow\mathit{prevM}(v)$\;
  \If{$v$ is an ancestor of $u$ \textbf{and} $\mathit{bcount}(u)=\mathit{bcount}(v)+2$}{
    mark $\{(u,p(u)),(v,p(v)),e_1(u),e_2(u)\}$ as a $4$-cut\;
    \label{line:type-2-3-mark-2}
  }  
}
\end{algorithm}

\begin{proposition}
\label{proposition:algorithm:type2-3}
Algorithm~\ref{algorithm:type2-3} computes all $4$-cuts of the form $\{(u,p(u)),(v,p(v)),e_1,e_2\}$, where $v$ is an ancestor of $u$ and $B(u)=B(v)\sqcup\{e_1,e_2\}$. Furthermore, it has a linear-time implementation.
\end{proposition}
\begin{proof}
Let $C=\{(u,p(u)),(v,p(v)),e_1,e_2\}$ be a $4$-cut such that $u$ is a descendant of $v$ and $B(u)=B(v)\sqcup\{e_1,e_2\}$. Then Lemma~\ref{lemma:case_B(v)=B(u)cup2} implies that $\{e_1,e_2\}=\{e_\mathit{high1}(u),e_\mathit{high2}(u)\}$ and $\mathit{bcount}(u)=\mathit{bcount}(v)+2$. Furthermore, Lemma~\ref{lemma:case_B(v)=B(u)cup2-necessary} implies that $v$ is either the greatest or the second-greatest proper ancestor of $u$ such that $M(v)=M(B(u)\setminus\{e_1,e_2\})$. Thus, it is clear that, if $v$ is the greatest proper ancestor of $u$ with $M(v)=M(B(u)\setminus\{e_1,e_2\})$, then $C$ will be marked in Line~\ref{line:type-2-3-mark-1}. Otherwise, we have that $v$ is the predecessor of $v'$ in $M^{-1}(M(v))$, where $v'$ is the greatest proper ancestor of $u$ such that $M(v')=M(B(u)\setminus\{e_1(u),e_2(u)\})$. Thus, $C$ will be marked in Line~\ref{line:type-2-3-mark-2}.

Conversely, let $C=\{(u,p(u)),(v,p(v)),e_1,e_2\}$ be a $4$-element set that is marked in Line~\ref{line:type-2-3-mark-1} or \ref{line:type-2-3-mark-2}. In either case, we have that $v$ is a proper ancestor of $u$ such that $M(v)=M(B(u)\setminus\{e_\mathit{high1}(u),e_\mathit{high2}(u)\})$ and $\mathit{bcount}(u)=\mathit{bcount}(v)+2$. Thus, Lemma~\ref{lemma:case_B(v)=B(u)cup2} implies that $B(u)=B(v)\sqcup\{e_\mathit{high1}(u),e_\mathit{high2}(u)\}$, and therefore $C$ is correctly marked as a $4$-cut. 

Now we will show that Algorithm~\ref{algorithm:type2-3} runs in linear time. By Proposition~\ref{proposition:computing-M(B(v)-S)} we have that the values $M(B(u)\setminus\{e_\mathit{high1}(u),e_\mathit{high2}(u)\})$ can be computed in linear time in total, for all vertices $u\neq r$. Thus, the \textbf{for} loop in Line~\ref{line:type-2-3-for} is performed in linear time. In order to compute the vertex $v$ in Line~\ref{line:type-2-3-v}, we use Algorithm~\ref{algorithm:W-queries}. Specifically, let $u\neq r$ be a vertex, and let $x=M(B(u)\setminus\{e_\mathit{high1}(u),e_\mathit{high2}(u)\})$. Then we generate a query $q(M^{-1}(x),u)$. This returns the greatest vertex $v$ such that $M(v)=x$ and $v<u$. Notice that, since $M(v)=x$, we have that $M(v)$ is a common descendant of $u$ and $v$, and therefore $u$ and $v$ are related as ancestor and descendant. Then, $v<u$ implies that $v$ is a proper ancestor of $u$. Thus, we have that $v$ is the greatest proper ancestor of $u$ such that $M(v)=M(B(u)\setminus\{e_\mathit{high1}(u),e_\mathit{high2}(u)\})$. Since the number of all those queries is $O(n)$, Algorithm~\ref{algorithm:W-queries} can answer all of them in linear time in total, according to Lemma~\ref{lemma:W-queries}. We conclude that Algorithm~\ref{algorithm:type2-3} runs in linear time.
\end{proof}

\section{Computing Type-3$\alpha$ $4$-cuts}
\label{section:type-3a}

Throughout this section, we assume that $G$ is a $3$-edge-connected graph with $n$ vertices and $m$ edges. All graph-related elements (e.g., vertices, edges, cuts, etc.) refer to $G$. Furthermore, we assume that we have computed a DFS-tree $T$ of $G$ rooted at a vertex $r$.

\begin{lemma}
\label{lemma:type-3a-types}
Let $u,v,w$ be three vertices $\neq r$ such that $w$ is a common ancestor of $\{u,v\}$, and $u,v$ are not related as ancestor and descendant. Then there is a back-edge $e$ such that $\{(u,p(u)),(v,p(v)),(w,p(w)),e\}$ is a $4$-cut if and only if either $(i)$ $e\in B(u)\cup B(v)$ and $B(w)\sqcup\{e\}=B(u)\sqcup B(v)$, or $(ii)$ $B(w)=(B(u)\sqcup B(v))\sqcup \{e\}$.
\end{lemma}
\begin{proof}
($\Rightarrow$) 
Let $C=\{(u,p(u)),(v,p(v)),(w,p(w)),e\}$. Since $w$ is a common ancestor of $\{u,v\}$, and $u,v$ are not related as ancestor and descendant, we have that the subtrees $T(u)$ and $T(v)$, and the tree-paths $T[p(u),w]$ and $T[p(v),w]$, remain intact in $G\setminus C$. Let $e=(x,y)$. Since $C$ is a $4$-cut and $e$ is a back-edge in $C$, by Lemma~\ref{lemma:type3-4cuts} we have that $e$ is either in $B(u)$, or in $B(v)$, or in $B(w)$. 

Let us assume first that $e\in B(u)$. Then $x$ is a descendant of $u$, and therefore it cannot be a descendant of $v$ (since $u$ and $v$ are not related as ancestor and descendant). Thus we have $e\notin B(v)$. Now let us assume, for the sake of contradiction, that $e\in B(w)$. Since $C$ is a $4$-cut, we have that $G'=G\setminus\{(u,p(u)),(v,p(v)),(w,p(w))\}$ is connected. In particular, $u$ is connected with $p(u)$ in $G'$. Suppose first that there is a back-edge $(x',y')\in B(u)$ such that $y'\in T[p(u),w]$. Then $u$ is still connected with $p(u)$ in $G'\setminus e$, contradicting the fact that $C$ is a $4$-cut of $G$. Thus, every back-edge $(x',y')\in B(u)$ must satisfy that $y'$ is a proper ancestor of $w$. Similarly, we have that every back-edge $(x',y')\in B(v)$ must satisfy that $y'$ is a proper ancestor of $w$. Thus, since $u$ is connected with $p(u)$ in $G'$, there must exist a back-edge $(x',y')\in B(w)$ such that $x'$ is not a descendant of $u$ or $v$. In particular, $(x',y')\neq e$. But now, by removing $e$ from $G'$, we can see that $w$ remains connected with $p(w)$ through the path $T[w,x'],(x',y'),T[y',p(w)]$, in contradiction to the fact that $C$ is a $4$-cut of $G$. This shows that $e\notin B(w)$. Now it is not difficult to see that every back-edge in $(B(u)\setminus\{e\})\cup B(v)$ must be in $B(w)$, for otherwise $C$ is not a $4$-cut of $G$. And conversely, every back-edge in $B(w)$ must be in $(B(u)\setminus\{e\})\cup B(v)$, for otherwise $C$ is not a $4$-cut of $G$.
Thus we have shown that $B(w)\sqcup\{e\}=B(u)\sqcup B(v)$. Similarly, if we assume that $e\in B(v)$, we can use the analogous argument to show that $B(w)\sqcup\{e\}=B(u)\sqcup B(v)$. In any case, we observe that we cannot have that $e$ is both in $B(u)\cup B(v)$ and in $B(w)$ $(*)$.

Now let us assume that $e\in B(w)$. Then, by $(*)$ we have that $e\notin B(u)\cup B(v)$. Now let $(x',y')$ be a back-edge in $B(u)$. Let us assume, for the sake of contradiction, that $y'\in T[p(u),w]$. Then notice that $u$ is connected with $p(u)$ in the graph $G\setminus C$ through the path $T[u,x'],(x',y'),T[y',p(u)]$, in contradiction to the fact that $C$ is a $4$-cut of $G$. Thus we have that $y'$ is a proper ancestor of $w$, and therefore $(x',y')\in B(w)$. This shows that $B(u)\subseteq B(w)$. Similarly, we have $B(v)\subseteq B(w)$ using the analogous argument. This shows that $B(u)\cup B(v)\subseteq B(w)$. Since $e$ cannot be in $B(u)\cup B(v)$, this is strengthened to $B(u)\cup B(v)\subseteq B(w)\setminus\{e\}$. Conversely, let $(x',y')$ be a back-edge in $B(w)\setminus\{e\}$. Let us assume that $(x',y')\notin B(u)$. This implies that $x'$ is not a descendant of $u$ (for otherwise we would have $(x',y')\in B(u)$, because $y'$ is a proper ancestor of $w$, and therefore a proper ancestor of $u$). Let us suppose, for the sake of contradiction, that $x'$ is not a descendant of $v$. But then we have that $w$ is connected with $p(w)$ in $G\setminus C$ through the path $T[w,x'],(x',y'),T[y',p(w)]$, contradicting the fact that $C$ is a $4$-cut of $G$. Thus we have that $x'$ is a descendant of $v$, and therefore $(x',y')\in B(v)$ (since $y'$ is a proper ancestor of $w$, and therefore a proper ancestor of $v$). This means that $B(w)\setminus\{e\}\subseteq B(u)\cup B(v)$, and therefore we have $B(w)\setminus\{e\}=B(u)\cup B(v)$. Since $e\in B(w)\setminus(B(u)\cup B(v))$, this implies that $B(w)=(B(u)\cup B(v))\sqcup\{e\}$.

Finally, notice that the expression ``$B(u)\cup B(v)$" can be strengthened to ``$B(u)\sqcup B(v)$" everywhere, because $u$ and $v$ are not related as ancestor and descendant.

($\Leftarrow$) In the following, we let $G'$ denote the graph $G\setminus\{(u,p(u)),(v,p(v)),(w,p(w))\}$. In every case, we will first show that all $3$-set subsets of $\{(u,p(u)),(v,p(v)),(w,p(w)),e\}$ are not $3$-cuts of $G$. Then we will show that $G'\setminus e$ is disconnected.

Let us assume first that there is a back-edge $e\in B(u)\cup B(v)$ such that $B(w)\sqcup\{e\}=B(u)\sqcup B(v)$. We may assume w.l.o.g. that $e\in B(u)$. Then we have that $e\notin B(v)$ and $e\notin B(w)$. Let $e=(x,y)$. Then this means that $x$ is a descendant of $u$ and $y$ is a proper ancestor of $u$. Since $u$ and $v$ are not related as ancestor and descendant, we have that $x$ is not a descendant of $v$. Therefore, the tree-path $T[x,u]$ remains intact in $G'$. Since $e\notin B(w)$, we have that $y$ cannot be a proper ancestor of $w$ (because otherwise we would have $(x,y)\in B(w)$, since $x$ is a descendant of $u$, and therefore a descendant of $w$). Thus we have that $u$ remains connected with $p(u)$ in $G'$ through the path $T[u,x],(x,y),T[y,p(u)]$, and so $\{(u,p(u)),(v,p(v)),(w,p(w))\}$ is not a $3$-cut of $G$. Since $e\notin B(v)\cup B(w)$, we have that the tree-path $T[x,y]$ remains intact in $G\setminus\{(v,p(v)),(w,p(w)),e\}$. Thus, the endpoints of $e$ remain connected in $G\setminus\{(v,p(v)),(w,p(w)),e\}$, and so $\{(v,p(v)),(w,p(w)),e\}$ is not a $3$-cut of $G$. Since the graph is $3$-edge-connected, we have that there is a back-edge $(x',y')\in B(v)$. Since $u$ and $v$ are not related as ancestor and descendant, we have that the tree-paths $T[x',v]$ and $T[p(v),y']$ remain intact in $G\setminus\{(u,p(u)),(v,p(v)),e\}$. Thus, $v$ remains connected with $p(v)$ in $G\setminus\{(u,p(u)),(v,p(v)),e\}$ through the path $T[v,x'],(x',y'),T[y',p(v)]$, and therefore $\{(u,p(u)),(v,p(v)),e\}$ is not a $3$-cut of $G$. Since $w$ is a proper ancestor of $u$, we have that $w$ does not lie on the tree-path $T[x,u]$. And since $y$ is not a proper ancestor of $w$, we have that $y$ lies on the tree-path $T[p(u),w]$. Thus, $u$ remains connected with $p(u)$ in $G\setminus\{(u,p(u)),(w,p(w)),e\}$ through the path $T[u,x],(x,y),T[y,p(u)]$. This shows that $\{(u,p(u)),(w,p(w)),e\}$ is not a $3$-cut of $G$.

Now suppose that we remove $e$ from $G'$. Consider the four parts $A=T(u)$, $B=T(v)$, $C=T(w)\setminus(T(u)\cup T(v))$ an $D=T(r)\setminus T(w)$. Observe that these parts are connected in $G'\setminus e$. Now, the only ways in which $u$ can remain connected with $p(u)$ in $G'\setminus e$ are: $(1)$ there is a back-edge from $A$ to $C$, or $(2)$ there is back-edge from $A$ to $D$ and a back-edge from $D$ to $C$, or $(3)$ there is a back-edge from $A$ to $D$, a back-edge from $D$ to $B$, and a back-edge from $B$ to $C$. Possibility $(1)$ is precluded from the fact that $B(u)\setminus \{e\}\subseteq B(w)$. Possibility $(2)$ is precluded from the fact that $B(w)\subset B(u)\sqcup B(v)$ (i.e., there is no back-edge from $C$ to $D$). And possibility $(3)$ is precluded from $B(v)\subset B(w)$ (i.e., there is no back-edge from $B$ to $C$). Thus, we conclude that $u$ is not connected with $p(u)$ in $G'\setminus e$, and so $\{(u,p(u)),(v,p(v)),(w,p(w)),e\}$ is a $4$-cut of $G$.

Now let us assume that there is a back-edge $e=(x,y)$ such that $B(w)=(B(u)\sqcup B(v))\sqcup \{e\}$. Then we have that $e\notin B(u)\cup B(v)$. This implies that $x$ is not a descendant of $u$ (because otherwise we would have $(x,y)\in B(u)$, since $y$ is a proper ancestor of $w$, and therefore a proper ancestor of $u$). Similarly, we have that $x$ is not a descendant of $v$. Thus, the tree-path $T[x,w]$ remains intact in $G'$. Furthermore, the tree-path $T[y,p(w)]$ remains intact in $G'$. This implies that $w$ remains connected with $p(w)$ in $G'$ through the path $T[w,x],(x,y),T[y,p(w)]$. Therefore, $\{(u,p(u)),(v,p(v)),(w,p(w))\}$ is not a $3$-cut of $G$. Since $e\notin B(u)\cup B(v)$, we have that neither $u$ nor $v$ lies on the tree-path $T[x,y]$. Thus, the endpoints of $e$ remain connected in $G\setminus\{(u,p(u)),(v,p(v)),e\}$, and so $\{(u,p(u)),(v,p(v)),e\}$ is not a $3$-cut of $G$. Now consider the graph $G\setminus\{(u,p(u)),(w,p(w)),e\}$. Since $G$ is $3$-edge-connected, we have that there is a back-edge $(x',y')\in B(v)$. Then, $B(w)=(B(u)\sqcup B(v))\sqcup \{e\}$ implies that $(x',y')\in B(w)$ and $(x',y')\neq e$.   
Since $u$ and $v$ are not related as ancestor and descendant, we have that $x'$ is not a descendant of $u$ (because otherwise, $x'$ would be a common descendant of $u$ and $v$). Thus, $u$ does not lie on the tree-path $T[x',w]$. Furthermore, we have that $y'$ is a proper ancestor of $w$. Thus, $w$ remains connected with $p(w)$ in $G\setminus\{(u,p(u)),(w,p(w)),e\}$, through the path $T[w,x'],(x',y'),T[y',p(w)]$. This shows that $\{(u,p(u)),(w,p(w)),e\}$ is not a $3$-cut of $G$. Similarly, we can show that $\{(v,p(v)),(w,p(w)),e\}$ is not a $3$-cut of $G$.

Now suppose that we remove $e$ from $G'$. Consider the four parts $A=T(u)$, $B=T(v)$, $C=T(w)\setminus(T(u)\cup T(v))$ an $D=T(r)\setminus T(w)$. Observe that these parts are connected in $G'\setminus e$. Now, the only ways in which $w$ can remain connected with $p(w)$ in $G'\setminus e$ are: $(1)$ there is a back-edge from $C$ to $D$, or $(2)$ there is back-edge from $C$ to $A$ and a back-edge from $A$ to $D$, or $(3)$ there is a back-edge from $C$ to $B$ and a back-edge from $B$ to $D$. Possibility $(1)$ is precluded from the fact that $B(w)\setminus\{e\}=B(u)\sqcup B(v)$. Possibility $(2)$ is precluded from the fact that $B(u)\subset B(w)$ (i.e., there is no back-edge from $A$ to $C$). And possibility $(3)$ is precluded from $B(v)\subset B(w)$ (i.e., there is no back-edge from $B$ to $C$). Thus, we conclude that $w$ is not connected with $p(w)$ in $G'\setminus e$, and so $\{(u,p(u)),(v,p(v)),(w,p(w)),e\}$ is a $4$-cut of $G$.
\end{proof}

According to Lemma~\ref{lemma:type-3a-types}, we distinguish two types of Type-3$\alpha$ $4$-cuts: Type-3$\alpha i$ and Type-3$\alpha ii$. In both cases, the $4$-cuts that we consider have the form $\{(u,p(u)),(v,p(v)),(w,p(w)),e\}$, where $w$ is a common ancestor of $\{u,v\}$, and $u,v$ are not related as ancestor and descendant. In the case of Type-3$\alpha i$ $4$-cuts, we have that $e\in B(u)\sqcup B(v)$ and $B(w)\sqcup\{e\}=B(u)\sqcup B(v)$. In Type-3$\alpha ii$ $4$-cuts, we have $e\in B(w)\setminus(B(u)\sqcup B(v))$ and $B(w)=(B(u)\sqcup B(v))\sqcup\{e\}$.

\subsection{Type-3$\alpha i$ $4$-cuts}

\begin{lemma}
\label{lemma:type-3a-lowchild_desc}
Let $u,v,w$ be three vertices such that $w$ is a common ancestor of $\{u,v\}$, and $u,v$ are not related as ancestor and descendant. Suppose that there is a back-edge $e\in B(u)\sqcup B(v)$ such that $\{(u,p(u)),(v,p(v)),(w,p(w)),e\}$ is a $4$-cut. Then, either $u$ is a descendant of the $\mathit{low1}$ child of $M(w)$ and $v$ is a descendant of the $\mathit{low2}$ child of $M(w)$, or reversely.
\end{lemma}
\begin{proof}
By the conditions of the lemma, we have that $\{(u,p(u)),(v,p(v)),(w,p(w)),e\}$ is a Type-3$\alpha i$ $4$-cut, and by Lemma~\ref{lemma:type-3a-types} we have that $B(w)\sqcup\{e\}=B(u)\sqcup B(v)$.

First we will show that both $u$ and $v$ are descendants of $M(w)$. Since $u$ and $v$ are not related as ancestor and descendant, we have that either none of them is an ancestor of $M(w)$, or one of them is an ancestor of $M(w)$, but the other is not related with $M(w)$ as ancestor or descendant. Suppose for the sake of contradiction that the second case is true, and assume w.l.o.g. that $u$ is an ancestor of $M(w)$. This implies that $v$ is not related as ancestor and descendant with $M(w)$. Since the graph is $3$-edge-connected, we have that $|B(v)|\geq 2$. And since all back-edges in $B(v)$ are also in $B(w)$, except possibly $e$, we have that there is a back-edge $(x,y)\in B(v)\cap B(w)$. But then we have that $x$ is a descendant of both $v$ and $M(w)$, which implies that $v$ and $M(w)$ are related as ancestor and descendant, a contradiction. Thus, we have that none of $u$ and $v$ is an ancestor of $M(w)$. The same argument also shows that it cannot be the case that one of $u$ and $v$ is not related as ancestor and descendant with $M(w)$. Thus we have that both $u$ and $v$ are proper descendants of $M(w)$. 

Now let us assume, for the sake of contradiction, that $M(w)$ has less than two children. This implies that there is at least one back-edge of the form $(M(w),y)$ in $B(w)$, for a vertex $y$ with $y<w$. Then $B(w)\sqcup\{e\}=B(u)\sqcup B(v)$ implies that $e\notin B(w)$, and therefore $e\neq (M(w),y)$. But then, since both $u$ and $v$ are descendants of $M(w)$, we have that $\{(u,p(u)),(v,p(v)),(w,p(w)),e\}$ cannot be a $4$-cut, since $w$ is connected with $p(w)$ through the path $T[w,M(w)],(M(w),y),T[y,p(w)]$, a contradiction. This shows that $M(w)$ has at least two children. Furthermore, the same argument shows that there is no back-edge of the form $(M(w),y)$ in $B(w)$. Let $c_1$ and $c_2$ be the $\mathit{low1}$ and $\mathit{low2}$ children of $M(w)$, respectively.

Now let us assume, for the sake of contradiction, that $u$ is neither a descendant of $c_1$ nor a descendant of $c_2$. We may assume w.l.o.g. that $v$ is not a descendant of $c_1$ (because otherwise we have that $v$ is not a descendant of $c_2$, and we can reverse the roles of $c_1$ and $c_2$ in the following). Since there is no back-edge of the form $(M(w),y)$ in $B(w)$, we have that there are at least two back-edges $(x_1,y_1),(x_2,y_2)\in B(w)$ such that $x_1$ is a descendant of $c_1$ and $x_2$ is a descendant of $c_2$. Since neither $u$ nor $v$ is a descendant of $c_1$, we have that $\{(u,p(u)),(v,p(v)),(w,p(w)),e\}$ cannot be a $4$-cut, since $w$ is connected with $p(w)$ through the path $T[w,x_1],(x_1,y_1),T[y_1,p(w)]$, a contradiction. Thus we have shown that $u$ is either a descendant of $c_1$ or a descendant of $c_2$. A similar argument shows that $v$ is either a descendant of $c_1$ or a descendant of $c_2$. Furthermore, the same argument shows that it cannot be the case that both $u$ and $v$ are descendants of $c_1$, or that both of them are descendants of $c_2$. Thus the lemma follows. 
\end{proof}

In the following, we will assume w.l.o.g. that the back-edge of the Type-3$\alpha i$ $4$-cuts that we consider leaps over $u$. Furthermore, we will consider the case that $u$ is a descendant of the $\mathit{low1}$ child of $M(w)$. The other case is treated similarly. Also, throughout this section we let $e(u)$ denote the back-edge $e_\mathit{high}(u)$.

\begin{lemma}
\label{lemma:type-3a-i-M}
Let $u,v,w$ be three vertices such that $w$ is a common ancestor of $\{u,v\}$, and $u,v$ are not related as ancestor and descendant. Suppose that there is a back-edge $e\in B(u)$ such that $\{(u,p(u)),(v,p(v)),(w,p(w)),e\}$ is a $4$-cut. Then $e=(\mathit{highD}_1(u),\mathit{high}_1(u))$. Furthermore, suppose that $u$ is a descendant of the $\mathit{low1}$ child $c$ of $M(w)$. Then $M(B(u)\setminus\{e\})=M(w,c)$.
\end{lemma}
\begin{proof}
By Lemma~\ref{lemma:type-3a-types} we have that $B(w)=(B(u)\setminus\{e\})\sqcup B(v)$. Let $(x_1,y_1),\dots,(x_k,y_k)$ be the list of the back-edges that leap over $u$ sorted in decreasing order w.r.t. their lower endpoint, so that $(x_1,y_1)=(\mathit{highD}_1(u),\mathit{high}_1(u))$. Let us assume, for the sake of contradiction, that $e=(x_i,y_i)$, for some $i\in\{2,\dots,k\}$. Since $u$ is a descendant of $w$, we have that $x_i$ is also a descendant of $w$. But since $e\notin B(w)$, we cannot have that $y_i$ is a proper ancestor of $w$. But then we have that $(x_j,y_j)\notin B(w)$, for any $j\in\{1,\dots,i\}$, since $y_j$ is a descendant of $y_i$. Since $i>1$, this means that there are at least two back-edges in $B(u)\setminus B(w)$, in contradiction to $B(w)=(B(u)\setminus\{e\})\sqcup B(v)$. Thus we have that $e=(x_1,y_1)$. 

Now let $(x,y)\in B(w)$ be a back-edge such that $x$ is a descendant of $c$. By $B(w)=(B(u)\setminus\{e\})\sqcup B(v)$ we have that either $(x,y)\in B(u)\setminus\{e\}$, or $(x,y)\in B(v)$. By Lemma~\ref{lemma:type-3a-lowchild_desc} we have that $v$ is a descendant of the $\mathit{low2}$ child of $M(w)$, and so the second case is impossible (because otherwise we would have that $x$ is a descendant of the $\mathit{low2}$ child of $M(w)$). Thus we have that $(x,y)\in B(u)\setminus\{e\}$. Conversely, let $(x,y)$ be a back-edge in $B(u)\setminus\{e\}$. Then by $B(w)=(B(u)\setminus\{e\})\sqcup B(v)$ we have that $(x,y)\in B(w)$. And since $u$ is a descendant of $c$, we have that $x$ is also a descendant of $c$.
Thus we have shown that $B(u)\setminus\{e\}=\{(x,y)\in B(w)\mid x \mbox{ is a descendant of } c\}$, and so we get $M(B(u)\setminus\{e\})=M(w,c)$.
\end{proof}

\begin{remark}
\normalfont
Notice that the argument in the proof of Lemma~\ref{lemma:type-3a-i-M} works independently of the fact that $c$ is the $\mathit{low1}$ child of $M(w)$. In other words, we could have assumed that $u$ is a descendant of the $\mathit{low2}$ child $c$ of $M(w)$. In this case, Lemma~\ref{lemma:type-3a-lowchild_desc} would imply that $v$ is a descendant of the $\mathit{low1}$ child of $M(w)$, and we would still get $M(B(u)\setminus\{e\})=M(w,c)$ with the same argument.
\end{remark}

\begin{lemma}
\label{lemma:type-3a-i-v-lowest}
Let $u,v,w$ be three vertices $\neq r$, such that $w$ is a proper ancestor of $\{u,v\}$, and $u,v$ are not related as ancestor and descendant. Let $c_1$ and $c_2$ be the $\mathit{low1}$ and $\mathit{low2}$ children of $M(w)$, respectively. Suppose that $\{(u,p(u)),(v,p(v)),(w,p(w)),e(u)\}$ is a $4$-cut, and let us assume that $u$ is a descendant of $c_1$. Then $v$ is the lowest vertex with $M(v)=M(w,c_2)$ that is a proper descendant of $w$.
\end{lemma}
\begin{proof}
Since $\{(u,p(u)),(v,p(v)),(w,p(w)),e(u)\}$ is a $4$-cut, by Lemma~\ref{lemma:type-3a-types} we have that $B(w)=(B(u)\setminus\{e(u)\})\sqcup B(v)$. By Lemma~\ref{lemma:type-3a-lowchild_desc} we have that $v$ is a descendant of $c_2$. Let $S=\{(x,y)\in B(w)\mid x\mbox{ is a descendant of } c_2\}$. Then we have $M(w,c_2)=M(S)$. Now let $(x,y)$ be a back-edge in $B(v)$. By $B(w)=(B(u)\setminus\{e(u)\})\sqcup B(v)$ we have that $(x,y)\in B(w)$. And since $v$ is a descendant of $c_2$, we have $(x,y)\in S$. Conversely, let $(x,y)$ be a back-edge in $S$. Then by $B(w)=(B(u)\setminus\{e(u)\})\sqcup B(v)$ we have that either $(x,y)\in B(u)\setminus\{(e(u))\}$ or $(x,y)\in B(v)$. But since $x$ is a descendant of $c_2$, it cannot be the case that $x$ is a descendant of $u$, because $u$ is a descendant of $c_1$. Thus we have $(x,y)\in B(v)$. This shows that $S=B(v)$, and so we get $M(w,c_2)=M(v)$.

Now let us assume, for the sake of contradiction, that there is a proper ancestor $v'$ of $v$ with $M(v')=M(w,c_2)$, which is also a proper descendant of $w$. Then, since $M(v')=M(v)$, Lemma~\ref{lemma:same_m_subset_B} implies that $B(v')\subseteq B(v)$. Now let $(x,y)$ be a back-edge in $B(v)$. Then, as previously, we have $(x,y)\in B(w)$. Thus, $y$ is proper ancestor of $w$, and therefore a proper ancestor of $v'$. Furthermore, $x$ is a descendant of $v$, and therefore a descendant of $v'$. This shows that $(x,y)\in B(v')$. Due to the generality of $(x,y)\in B(v)$, this implies that $B(v)\subseteq B(v')$. But then we get $B(v')=B(v)$, in contradiction to the fact that the graph is $3$-edge-connected. Thus, $v$ is the lowest proper descendant of $w$ such that $M(v)=M(w,c_2)$.
\end{proof}

Here we distinguish two cases of Type-$3\alpha i$ $4$-cuts, depending on whether $M(B(u)\setminus\{e(u)\})=M(u)$ or $M(B(u)\setminus\{e(u)\})\neq M(u)$. In the first case, we show how to compute all such $4$-cuts in linear time. In the second case, the number of $4$-cuts can be $\Omega(n^2)$. However, we show how to compute a collection of such $4$-cuts in linear time, so that the rest of them are implied by this collection, plus that computed by Algorithm~\ref{algorithm:type2-2}. A vertex $u$ such that $M(B(u)\setminus\{e(u)\})=M(u)$ is called a ``special" vertex.

\subsubsection{The case where $M(B(u)\setminus\{e_\mathit{high}(u)\})= M(u)$}

\begin{lemma}
\label{lemma:type-3a-i-special}
Let $w,v$ be two vertices $\neq r$ such that $w$ is a proper ancestor of $v$. Then there is at most one vertex $u$ such that: $u$ is a special vertex, it is a proper descendant of $w$, and $\{(u,p(u)),(v,p(v)),(w,p(w)),e(u)\}$ is a Type-3$\alpha$ $4$-cut. If that is the case, assume w.l.o.g. that $u$ is a descendant of the $\mathit{low1}$ child $c$ of $M(w)$. Then $u$ is either the lowest or the second-lowest proper descendant of $w$ such that $M(u)=M(w,c)$.
\end{lemma}
\begin{proof}
Let $u$ be proper descendant of $w$ such that $u$ is a special vertex and $\{(u,p(u)),(v,p(v)),(w,p(w)),e(u)\}$ is a Type-3$\alpha$ $4$-cut $(*)$. By Lemma~\ref{lemma:type-3a-types} we have that $B(w)=(B(u)\setminus\{e(u)\})\sqcup B(v)$. By Lemma~\ref{lemma:type-3a-lowchild_desc} we may assume w.l.o.g. that $u$ is a descendant of the $\mathit{low1}$ child $c$ of $M(w)$, and $v$ is a descendant of the $\mathit{low2}$ child of $M(w)$. By Lemma~\ref{lemma:type-3a-i-M} we have that $M(B(u)\setminus\{e(u)\})=M(w,c)$. Since $u$ is a special vertex, we have $M(u)=M(B(u)\setminus\{e(u)\})$. Thus, $M(B(u)\setminus\{e(u)\})=M(w,c)$ implies that $M(u)=M(w,c)$. 

Now let us assume, for the sake of contradiction, that there is another vertex $u'$ such that: $u'$ is a special vertex, it is a proper descendant of $w$, and $\{(u',p(u')),(v,p(v)),(w,p(w)),e(u')\}$ is a Type-3$\alpha$ $4$-cut. Then, since $v$ is a descendant of the $\mathit{low2}$ child of $M(w)$, by Lemma~\ref{lemma:type-3a-lowchild_desc} we have that $u'$ is a descendant of $c$. Thus, $u'$ satisfies the same properties as $u$. This implies that $M(u')=M(w,c)$. Since $u$ is an ancestor of $M(u)=M(w,c)$ and $u'$ is an ancestor of $M(u')=M(w,c)$, we have that $u$ and $u'$ are related as ancestor and descendant (because they have a common descendant). Let us assume w.l.o.g. that $u'$ is a proper ancestor of $u$. Then, since $M(u)=M(u')$, Lemma~\ref{lemma:same_m_subset_B} implies that $B(u')\subseteq B(u)$. Since the graph is $3$-edge-connected, this can be strengthened to $B(u')\subset B(u)$. Thus, there is a back-edge $(x,y)$ in $B(u)\setminus B(u')$. This implies that $x$ is a descendant of $u$, and therefore a descendant of $u'$. Thus, $y$ cannot be a proper ancestor of $u'$. Therefore, since $y$ and $u'$ are related as ancestor and descendant (since they have $x$ as a common descendant), we have that $y$ is a descendant of $u'$. Thus, since $u'$ is a proper descendant of $w$, we have that $y$ cannot be a proper ancestor of $w$, and so $(x,y)\notin B(w)$. Thus, since $(x,y)\in B(u)$, $B(w)=(B(u)\setminus\{e(u)\})\sqcup B(v)$ implies that $(x,y)=e(u)$. Now, since $B(u')\subset B(u)$, we have $e(u')\in B(u)$. Since $(x,y)\notin B(u')$, we have $e(u')\neq (x,y)$. Also, we have $e(u')\notin B(w)$ (since $B(w)=(B(u')\setminus\{e(u')\})\sqcup B(v)$, and $B(u')\cap B(v)=\emptyset$, because $u'$ and $v$ are descendants of different children of $M(w)$, and therefore they are not related as ancestor and descendant). Thus $B(u)\setminus B(w)$ contains at least two back-edges, in contradiction to $B(w)=(B(u)\setminus\{e(u)\})\sqcup B(v)$. This shows that $u$ is unique in satisfying property $(*)$. 

Now let us suppose, for the sake of contradiction, that there are two distinct vertices $u'$ and $u''$ that are lower than $u$, they are proper descendants of $w$, and satisfy $M(u')=M(u'')=M(u)$. Then we have that all $u,u',u''$ are related as ancestor and descendant. We may assume w.l.o.g. that $u''<u'$. Thus, we have that $u''$ is a proper ancestor of $u'$, and $u'$ is a proper ancestor of $u$. Then, by Lemma~\ref{lemma:same_m_subset_B} we have $B(u'')\subseteq B(u')\subseteq B(u)$. Since the graph is $3$-edge-connected, this can be strengthened to $B(u'')\subset B(u')\subset B(u)$. Thus, there is a back-edge $(x,y)\in B(u)\setminus B(u')$, and a back-edge $(x',y')\in B(u')\setminus B(u'')$. Since $(x,y)\in B(u)$, we have that $x$ is a descendant of $u$, and therefore a descendant of $u'$. Thus, since $(x,y)\notin B(u')$, it cannot be that $y$ is a proper ancestor of $u'$. Similarly, since $(x',y')\in B(u')\setminus B(u'')$, it cannot be that $y'$ is a proper ancestor of $u''$. Since both $u'$ and $u''$ are proper descendants of $w$, this implies that neither $y$ nor $y'$ is a proper ancestor of $w$. Thus, $(x,y)\notin B(w)$ and $(x',y')\notin B(w)$. Since $(x,y)\notin B(u')$ and $(x',y')\in B(u')$, we have $(x,y)\neq (x',y')$. And since $(x',y')\in B(u')$ and $B(u')\subset B(u)$, we have $(x',y')\in B(u)$. But then $(x,y)$ and $(x',y')$ are two distinct back-edges in $B(u)\setminus B(w)$, in contradiction to $B(w)=((B(u)\setminus\{e(u)\})\sqcup B(v)$ (which implies that $B(u)\setminus B(w)$ consists of $e(u)$). This shows that $u$ is either the lowest or the second-lowest proper descendant of $w$ such that $M(u)=M(w,c)$. 
\end{proof}

Lemma~\ref{lemma:type-3a-i-special} gives enough information to be able to compute efficiently all $4$-cuts of the form $\{(u,p(u)),(v,p(v)),(w,p(w)),e\}$, where $w$ is common ancestor of $\{u,v\}$, $u,v$ are not related as ancestor and descendant, $e\in B(u)$, and $u$ is a special vertex. 

This method is shown in Algorithm~\ref{algorithm:type3-a-i-special}, for the case where $u$ is a descendant of the $\mathit{low1}$ child of $M(w)$. The case where $u$ is a descendant of the $\mathit{low2}$ child of $M(w)$ is treated similarly, by simply changing the roles of $c_1$ and $c_2$ in Lines~\ref{line:type3-a-i-special-c1} and \ref{line:type3-a-i-special-c2}, respectively. (I.e., we set ``$c_1\leftarrow\mathit{low2}$ child of $M(w)$" and ``$c_2\leftarrow\mathit{low1}$ child of $M(w)$".) The proof of correctness and linear complexity is given in Proposition~\ref{proposition:algorithm:type3-a-i-special}.

\begin{lemma}
\label{lemma:type-3a-i-criterion}
Let $u,v,w$ be three vertices such that $w$ is a common ancestor of $\{u,v\}$, and $u,v$ are not related as ancestor and descendant. Then, there is a back-edge $e\in B(u)$ such that $\{(u,p(u)),(v,p(v)),(w,p(w)),e\}$ is a Type-3$\alpha$i $4$-cut if and only if: $\mathit{high}_2(u)<w$, $\mathit{high}_1(v)<w$, and $\mathit{bcount}(w)=\mathit{bcount}(u)+\mathit{bcount}(v)-1$.
\end{lemma}
\begin{proof}
($\Rightarrow$)
Since $\{(u,p(u)),(v,p(v)),(w,p(w)),e\}$ is a Type-3$\alpha$i $4$-cut where $e\in B(u)$, we have that $B(w)\sqcup\{e\}=B(u)\sqcup B(v)$. This implies that $\mathit{bcount}(w)+1=\mathit{bcount}(u)+\mathit{bcount}(v)$, from which we infer that $\mathit{bcount}(w)=\mathit{bcount}(u)+\mathit{bcount}(v)-1$. Let $(x,y)$ be a back-edge in $B(v)$. Then $B(w)\sqcup\{e\}=B(u)\sqcup B(v)$ implies that $(x,y)\in B(w)\sqcup\{e\}$. Since $B(u)\cap B(v)=\emptyset$, we have that $(x,y)\neq e$. Thus, we have $(x,y)\in B(w)$. This implies that $y$ is a proper ancestor of $w$, and therefore $y<w$. Due to the generality of $(x,y)\in B(v)$, this implies that $\mathit{high}_1(v)<w$. Now let us suppose, for the sake of contradiction, that $\mathit{high}_2(u)\geq w$. This implies that the $\mathit{high1}$ and the $\mathit{high2}$ edges of $u$ are not in $B(w)$. But $e\in B(u)$ and $B(w)\sqcup\{e\}=B(u)\sqcup B(v)$ imply that precisely one back-edge from $B(u)$ is not in $B(w)$, a contradiction. Thus, we have $\mathit{high}_2(u)<w$.\\
($\Leftarrow$)
Since $v$ is a common descendant of $\mathit{high}_1(v)$ and $w$, we have that $\mathit{high}_1(v)$ and $w$ are related as ancestor and descendant. Thus, $\mathit{high}_1(v)<w$ implies that $\mathit{high}_1(v)$ is a proper ancestor of $w$. Similarly, we have that $\mathit{high}_2(u)$ is a proper ancestor of $w$. 
Now, let $(x,y)$ be a back-edge in $B(v)$. Then, $x$ is a descendant of $v$, and therefore a descendant of $w$. Furthermore, $y$ is an ancestor of $\mathit{high}_1(v)$, and therefore a proper ancestor of $w$. This shows that $(x,y)\in B(w)$. Due to the generality of $(x,y)\in B(v)$, this implies that $B(v)\subseteq B(w)$.

Now let $(x_1,y_1),\dots,(x_k,y_k)$ be the list of the back-edges in $B(u)$ sorted in decreasing order w.r.t. their lower endpoint, so that we have $(x_1,y_1)=e(u)$. Let $i$ be an index in $\{2,\dots,k\}$. Then, we have that $x_i$ is a descendant of $u$, and therefore a descendant of $w$. Furthermore, we have that $y_i\leq\mathit{high}_2(u)$, and therefore $y_i$ is an ancestor of $\mathit{high}_2(u)$, and therefore $y_i$ is a proper ancestor of $w$. This shows that $(x_i,y_i)\in B(w)$. Thus, we have shown that $B(u)\setminus\{e(u)\}\subseteq B(w)$.

Since $u$ and $v$ are not related as ancestor and descendant, we have that $B(u)\cap B(v)=\emptyset$ (because otherwise, if there existed a back-edge in $B(u)\cap B(v)$, we would have that its higher endpoint would be a common descendant of both $u$ and $v$). Thus, we have $(B(u)\setminus\{e(u)\})\sqcup B(v)\subseteq B(w)$. Therefore, $\mathit{bcount}(w)=(\mathit{bcount}(u)-1)+\mathit{bcount}(v)$ implies that $(B(u)\setminus\{e(u)\})\sqcup B(v)= B(w)$. Since $B(u)\cap B(v)=\emptyset$, we have that $e(u)\notin B(v)$, and therefore $(B(u)\setminus\{e(u)\})\sqcup B(v)= B(w)$ implies that $e(u)\notin B(w)$. Thus, $(B(u)\setminus\{e(u)\})\sqcup B(v)= B(w)$ and $e(u)\notin B(v)$ imply that $B(u)\sqcup B(v)=B(w)\sqcup\{e(u)\}$. Thus, by Lemma~\ref{lemma:type-3a-types} we have that $\{(u,p(u)),(v,p(v)),(w,p(w)),e(u)\}$ is a Type-3$\alpha$i $4$-cut.
\end{proof}

\noindent\\
\begin{algorithm}[H]
\caption{\textsf{Compute all Type-3$\alpha$i $4$-cuts of the form $\{(u,p(u)),(v,p(v)),(w,p(w)),e\}$, where $w$ is a common ancestor of $\{u,v\}$, $e\in B(u)$, and $u$ is a special vertex.}}
\label{algorithm:type3-a-i-special}
\LinesNumbered
\DontPrintSemicolon
\ForEach{vertex $w\neq r$ such that $M(w)$ has at least two children}{
\label{line:type3-a-i-special-for}
compute $M(w,c_1)$ and $M(w,c_2)$, where $c_1$ and $c_2$ are the $\mathit{low1}$ and $\mathit{low2}$ children of $M(w)$, respectively\;
}
\tcp{the case where $u$ is a descendant of the $\mathit{low1}$ child of $M(w)$; for the other case, simply reverse the roles of $c_1$ and $c_2$}
\ForEach{vertex $w\neq r$}{
   \lIf{$M(w)$ has less than two children}{\textbf{continue}}
   let $c_1\leftarrow\mathit{low1}$ child of $M(w)$\;
   \label{line:type3-a-i-special-c1}
   let $c_2\leftarrow\mathit{low2}$ child of $M(w)$\;
   \label{line:type3-a-i-special-c2}
   \lIf{$M(w,c_1)=\bot$ \textbf{or} $M(w,c_2)=\bot$}{\textbf{continue}}
   let $u$ be the lowest proper descendant of $w$ such that $M(u)=M(w,c_1)$\;
   \label{line:type3-a-i-special-u}
   let $v$ be the lowest proper descendant of $w$ such that $M(v)=M(w,c_2)$\;
   \label{line:type3-a-i-special-v}
   \If{$u$ is a special vertex \textbf{and} $\mathit{bcount}(w)=(\mathit{bcount}(u)-1)+\mathit{bcount}(v)$ \textbf{and} $\mathit{high}_2(u)<w$ \textbf{and} $\mathit{high}_1(v)<w$}{
     mark $\{(u,p(u)),(v,p(v)),(w,p(w)),e(u)\}$ as a $4$-cut\;
     \label{line:type3-a-i-special-mark1}
   }
   $u\leftarrow\mathit{prevM}(u)$\;
   \If{$u$ is a special vertex \textbf{and} $\mathit{bcount}(w)=(\mathit{bcount}(u)-1)+\mathit{bcount}(v)$ \textbf{and} $\mathit{high}_2(u)<w$ \textbf{and} $\mathit{high}_1(v)<w$}{
     mark $\{(u,p(u)),(v,p(v)),(w,p(w)),e(u)\}$ as a $4$-cut\;
     \label{line:type3-a-i-special-mark2}
   }   
}
\end{algorithm}

\begin{proposition}
\label{proposition:algorithm:type3-a-i-special}
Algorithm~\ref{algorithm:type3-a-i-special} correctly computes all $4$-cuts of the form $\{(u,p(u)),(v,p(v)),(w,p(w)),e\}$, where $w$ is a common ancestor of $\{u,v\}$, $e\in B(u)$, and $u$ is a special vertex. Furthermore, it has a linear-time implementation.
\end{proposition}
\begin{proof}
Let $C=\{(u,p(u)),(v,p(v)),(w,p(w)),e\}$ be a Type-$3\alpha i$ $4$-cut where $w$ is a common ancestor of $u$ and $v$, $u$ is a special vertex, and $e\in B(u)$. Let $c_1$ and $c_2$ be the $\mathit{low1}$ and the $\mathit{low2}$ child of $M(w)$, respectively. Lemma~\ref{lemma:type-3a-lowchild_desc} implies that either $u$ is a descendant of $c_1$ and $v$ is a descendant of $c_2$, or reversely. So let us assume w.l.o.g. that $u$ is a descendant of $c_1$. Then Lemma~\ref{lemma:type-3a-i-v-lowest} implies that $v$ is the lowest proper descendant of $w$ such that $M(v)=M(w,c_2)$. Lemma~\ref{lemma:type-3a-i-M} implies that $e=e_\mathit{high}(u)$. Lemma~\ref{lemma:type-3a-i-criterion} implies that $\mathit{bcount}(w)=\mathit{bcount}(u)+\mathit{bcount}(v)-1$, $\mathit{high}_2(u)<w$ and $\mathit{high}_1(v)<w$. If we have that $u$ is the lowest proper descendant of $w$ such that $M(u)=M(w,c_1)$, then we can see that $C$ will be marked in Line~\ref{line:type3-a-i-special-mark1}. Otherwise, by Lemma~\ref{lemma:type-3a-i-special} we have that $u$ is the second-lowest proper descendant of $w$ such that $M(u)=M(w,c_1)$. This implies that $u=\mathit{prevM}(u')$, where $u'$ is the lowest proper descendant of $w$ such that $M(u')=M(w,c_1)$. Thus, $C$ will be marked in Line~\ref{line:type3-a-i-special-mark2}.

Conversely, let $C=\{(u,p(u)),(v,p(v)),(w,p(w)),e(u)\}$ be a $4$-element set that is marked in Line~\ref{line:type3-a-i-special-mark1} or \ref{line:type3-a-i-special-mark2}. Then, in either case we have $\mathit{bcount}(w)=\mathit{bcount}(u)+\mathit{bcount}(v)-1$, $\mathit{high}_2(u)<w$ and $\mathit{high}_1(v)<w$. Furthermore, in either case we have $M(u)=M(w,c_1)$ and $M(v)=M(w,c_2)$. Therefore, Lemma~\ref{lemma:w-v-u-not-related} implies that $u$ and $v$ are not related as ancestor and descendant. Thus, Lemma~\ref{lemma:type-3a-i-criterion} implies that $C$ is indeed a Type-$3\alpha i$ $4$-cut.

Now we will argue about the complexity of Algorithm~\ref{algorithm:type3-a-i-special}. By Proposition~\ref{proposition:computing-M(v,c)} we have that the values $M(w,c_1)$ and $M(w,c_2)$ can be computed in linear time in total, for all vertices $w\neq r$ such that $M(w)$ has at least two children, where $c_1$ and $c_2$ are the $\mathit{low1}$ and $\mathit{low2}$ children of $M(w)$ respectively. Thus, Line~\ref{line:type3-a-i-special-for} can be performed in linear time. The vertices $u$ and $v$ in Lines~\ref{line:type3-a-i-special-u} and \ref{line:type3-a-i-special-v} can be computed with Algorithm~\ref{algorithm:W-queries}. Specifically, whenever we reach Line~\ref{line:type3-a-i-special-u}, we generate a query $q(M^{-1}(M(w,c_1)),w)$. This will return the lowest vertex $u$ with $M(u)=M(w,c_1)$ and $u>w$. Since $M(u)=M(w,c_1)$ implies that $M(u)$ is a common descendant of $u$ and $w$, we have that $u$ and $w$ are related as ancestor and descendant. Thus, $u>w$ implies that $u$ is a proper descendant of $w$. Thus, $u$ is the lowest proper descendant of $w$ such that $M(u)=M(w,c_1)$. We generate the analogous query to get $v$. Since the number of all those queries is $O(n)$, Algorithm~\ref{algorithm:W-queries} can answer all of them in $O(n)$ time, according to Lemma~\ref{lemma:W-queries}. We conclude that Algorithm~\ref{algorithm:type3-a-i-special} runs in linear time.
\end{proof}

\subsubsection{The case where $M(B(u)\setminus\{e_\mathit{high}(u)\})\neq M(u)$}

\begin{lemma}
\label{lemma:non-special-inducing-4cuts}
Let $u$ and $u'$ be two distinct vertices $\neq r$ such that $M(u)\neq M(B(u)\setminus\{e(u)\})=M(B(u')\setminus\{e(u')\})\neq M(u')$. Then, $e(u)\notin B(u')$ and $e(u')\notin B(u)$. Furthermore, if $\mathit{high}_2(u)=\mathit{high}_2(u')$, then $B(u)\sqcup\{e(u')\}=B(u')\sqcup\{e(u)\}$.
\end{lemma}
\begin{proof}
Let us assume w.l.o.g. that $u'<u$. Since the graph is $3$-edge-connected, we have that $|B(u')|>1$. Thus, there is a back-edge $(x,y)\in B(u')\setminus\{e(u')\}$. Then, we have that $x$ is a descendant of $M(B(u')\setminus\{e(u')\})=M(B(u)\setminus\{e(u)\})$, and therefore a descendant of $M(u)$, and therefore a descendant of $u$. Thus, $x$ is a common descendant of $u'$ and $u$, and therefore $u$ and $u'$ are related as ancestor and descendant. Thus, $u'<u$ implies that $u'$ is a proper ancestor of $u$.

Let us suppose, for the sake of contradiction, that $e(u')\in B(u)$. Then, since $M(u')\neq M(B(u')\setminus\{e(u')\})$, we have that the higher endpoint of $e(u')$ is not a descendant of $M(B(u')\setminus\{e(u')\})$, and therefore it is not a descendant of $M(B(u)\setminus\{e(u)\})$. Furthermore, since $M(u)\neq M(B(u)\setminus\{e(u)\})$, we have that the higher endpoint of $e(u)$ is not a descendant of $M(B(u)\setminus\{e(u)\})$, and that this is the only back-edge in $B(u)$ with this property. Thus, since $e(u')\in B(u)$, we have that $e(u)=e(u')$. This implies that $\mathit{high}_1(u)=\mathit{high}_1(u')$. Thus, since $u'$ is a proper ancestor of $u$, by Lemma~\ref{lemma:same_high} we have that $B(u)\subseteq B(u')$. Since the graph is $3$-edge-connected, this can be strengthened to $B(u)\subset B(u')$. Thus, there is a back-edge $(x,y)\in B(u')\setminus B(u)$. Since $e(u')\in B(u)$ and $(x,y)\notin B(u)$, we have that $(x,y)\neq e(u')$. Thus, we have $(x,y)\in B(u')\setminus\{e(u')\}$, and therefore $x$ is a descendant of $M(B(u')\setminus\{e(u')\})=M(B(u)\setminus\{e(u)\})$, and therefore a descendant of $M(u)$. Furthermore, $y$ is a proper ancestor of $u'$, and therefore a proper ancestor of $u$. This shows that $(x,y)\in B(u)$, a contradiction. Thus, we have shown that $e(u')\notin B(u)$. This implies that $e(u')\neq e(u)$.

Let us suppose, for the sake of contradiction, that $e(u)\in B(u')$. Then, since $M(u')\neq M(B(u')\setminus\{e(u')\})$, we have that the higher endpoint of $e(u')$ is not a descendant of $M(B(u')\setminus\{e(u')\})$, and therefore it is not a descendant of $M(B(u)\setminus\{e(u)\})$. Furthermore, we have that $e(u')$ is the only back-edge in $B(u')$ with this property. Now, since $M(u)\neq M(B(u)\setminus\{e(u)\})$, we have that the higher endpoint of $e(u)$ is not a descendant of $M(B(u)\setminus\{e(u)\})$. Since $e(u)\in B(u')$, this implies that $e(u)=e(u')$, a contradiction. Thus, we have shown that $e(u)\notin B(u')$. 

Now let $(x,y)$ be a back-edge in $B(u)\setminus\{e(u)\}$. Then, $x$ is a descendant of $M(B(u)\setminus\{e(u)\})=M(B(u')\setminus\{e(u')\})$, and therefore a descendant of $M(u')$. Furthermore, $y$ is an ancestor of $\mathit{high}_2(u)=\mathit{high}_2(u')$, and therefore a proper ancestor of $u'$. This shows that $(x,y)\in B(u')$. Due to the generality of $(x,y)\in B(u)\setminus\{e(u)\}$, this implies that $B(u)\setminus\{e(u)\}\subseteq B(u')$. And since $e(u')\notin B(u)$, this can be strengthened to $B(u)\setminus\{e(u)\}\subseteq B(u')\setminus\{e(u')\}$. Conversely, let $(x,y)$ be a back-edge in $B(u')\setminus\{e(u')\}$. Then $x$ is a descendant of $M(B(u')\setminus\{e(u')\})=M(B(u)\setminus\{e(u)\})$, and therefore a descendant of $M(u)$. Furthermore, $y$ is a proper ancestor of $u'$, and therefore a proper ancestor of $u$. This shows that $(x,y)\in B(u)$. Due to the generality of $(x,y)\in B(u')\setminus\{e(u')\}$, this implies that $B(u')\setminus\{e(u')\}\subseteq B(u)$. And since $e(u)\notin B(u')$, this can be strengthened to $B(u')\setminus\{e(u')\}\subseteq B(u)\setminus\{e(u)\}$. Thus, we have $B(u)\setminus\{e(u)\}= B(u')\setminus\{e(u')\}$. Since $e(u')\notin B(u)$ and $e(u)\notin B(u')$, this implies that $B(u)\sqcup\{e(u')\}=B(u')\sqcup\{e(u)\}$.
\end{proof}

\begin{lemma}
\label{lemma:type-3-a-i-lowest}
Let $u,v,w$ be three vertices $\neq r$ such that $w$ is a common ancestor of $\{u,v\}$, and $u,v$ are not related as ancestor and descendant. Suppose that $u$ is a non-special vertex such that $\{(u,p(u)),(v,p(v)),(w,p(w)),e(u)\}$ is a Type-3$\alpha i$ $4$-cut. Let $c$ be the child of $M(w)$ that is an ancestor of $u$, and let $u'$ be the lowest non-special vertex such that $M(B(u')\setminus\{e(u')\})=M(w,c)$ and $u'$ is a proper descendant of $w$. Then, $\{(u',p(u')),(v,p(v)),(w,p(w)),e(u')\}$ is a Type-3$\alpha i$ $4$-cut.
\end{lemma}
\begin{proof}
By Lemma~\ref{lemma:type-3a-lowchild_desc} we have that $u$ is a descendant of a child $c$ of $M(w)$. Then, by Lemma~\ref{lemma:type-3a-i-M} we have that $M(B(u)\setminus\{e(u)\})=M(w,c)$. Thus, since $u$ is a non-special vertex that is a proper descendant of $w$, it makes sense to consider the lowest non-special vertex $u'$ such that $M(B(u')\setminus\{e(u')\})=M(w,c)$ and $u'$ is a proper descendant of $w$. If $u'=u$, then by assumption we have that $\{(u',p(u')),(v,p(v)),(w,p(w)),e(u')\}$ is a Type-3$\alpha i$ $4$-cut. So let us assume that $u'<u$. Notice that $M(w,c)$ is a common descendant of $u'$ and $u$, and therefore $u'$ and $u$ are related as ancestor and descendant. Thus, $u'<u$ implies that $u'$ is a proper ancestor of $u$. 

Since $\{(u,p(u)),(v,p(v)),(w,p(w)),e(u)\}$ is a Type-3$\alpha i$ $4$-cut, we have that $B(w)=(B(u)\setminus\{e(u)\})\sqcup B(v)$. Let $(x,y)$ be a back-edge in $B(u)\setminus\{e(u)\}$. Then, $x$ is a descendant of $M(B(u)\setminus\{e(u)\})$, and therefore a descendant of $M(w,c)$, and therefore a descendant of $M(B(u')\setminus\{e(u')\})$, and therefore a descendant of $M(u')$. Furthermore, $B(w)=(B(u)\setminus\{e(u)\})\sqcup B(v)$ implies that $(x,y)\in B(w)$, and therefore $y$ is a proper ancestor of $w$, and therefore a proper ancestor of $u'$. This shows that $B(u)\setminus\{e(u)\}\subseteq B(u')$. Since Lemma~\ref{lemma:non-special-inducing-4cuts} implies that $e(u')\notin B(u)$, this can be strengthened to $B(u)\setminus\{e(u)\}\subseteq B(u')\setminus\{e(u')\}$. Conversely, let $(x,y)$ be a back-edge in $B(u')\setminus\{e(u')\}$. Then $x$ is a descendant of $M(B(u')\setminus\{e(u')\})$, and therefore a descendant of $M(w,c)$, and therefore a descendant of $M(B(u)\setminus\{e(u)\})$, and therefore a descendant of $M(u)$. Furthermore, $y$ is a proper ancestor of $u'$, and therefore a proper ancestor of $u$. This shows that $B(u')\setminus\{e(u')\}\subseteq B(u)$. Since Lemma~\ref{lemma:non-special-inducing-4cuts} implies that $e(u)\notin B(u')$, this can be strengthened to $B(u')\setminus\{e(u')\}\subseteq B(u)\setminus\{e(u)\}$. Thus, we have $B(u)\setminus\{e(u)\}= B(u')\setminus\{e(u')\}$. Therefore, $B(w)=(B(u)\setminus\{e(u)\})\sqcup B(v)$ implies that $B(w)=(B(u')\setminus\{e(u')\})\sqcup B(v)$. 

Let us suppose, for the sake of contradiction, that $e(u')\in B(v)$. Then, the higher endpoint of $e(u')$ is a common descendant of $u'$ and $v$, and therefore $u'$ and $v$ are related as ancestor and descendant. Therefore, since $u'$ is an ancestor of $u$, but $u$ and $v$ are not related as ancestor and descendant, we have that $u'$ is an ancestor of both $u$ and $v$. Since the graph is $3$-edge-connected, we have that $|B(v)|>1$. Thus, there is a back-edge $(x,y)\in B(v)\setminus\{e(u')\}$. Then, $x$ is a descendant of $v$, and therefore a descendant of $u'$. Furthermore, $B(w)=(B(u)\setminus\{e(u)\})\sqcup B(v)$ implies that $(x,y)\in B(w)$, and therefore $y$ is a proper ancestor of $w$, and therefore a proper ancestor of $u'$. This shows that $(x,y)\in B(u')$, in contradiction to (the disjointness of the sets in) $B(w)=(B(u')\setminus\{e(u')\})\sqcup B(v)$. Thus, we have $e(u')\notin B(v)$. Therefore, $B(w)=(B(u')\setminus\{e(u')\})\sqcup B(v)$ implies that $B(w)\sqcup\{e(u')\}=B(u)\sqcup B(v)$.
Thus, by Lemma~\ref{lemma:type-3a-types} we have that $\{(u',p(u')),(v,p(v)),(w,p(w)),e(u')\}$ is a Type-3$\alpha i$ $4$-cut.
\end{proof}

\begin{lemma}
\label{lemma:type-3-a-i-implied}
Let $u,v,w$ be three vertices $\neq r$ such that $w$ is a common ancestor of $\{u,v\}$, and $u,v$ are not related as ancestor and descendant. Suppose that $u$ is a non-special vertex and $C=\{(u,p(u)),(v,p(v)),(w,p(w)),e(u)\}$ is a Type-3$\alpha i$ $4$-cut. Then, every other Type-3$\alpha i$ $4$-cut $C'$ of the form $\{(u',p(u')),(v,p(v)),(w,p(w)),e(u')\}$, where $u'$ is a non-special vertex that is a proper descendant of $w$, is implied by $C$ and some Type-2$ii$ $4$-cuts that are computed by Algorithm~\ref{algorithm:type2-2}.
\end{lemma}
\begin{proof}
Since $C$ is a Type-3$\alpha i$ $4$-cut where $w$ is a common ancestor of $\{u,v\}$, we have $B(w)\sqcup\{e(u)\}=B(u)\sqcup B(v)$. This implies that $B(u)\setminus\{e(u)\}=B(w)\setminus B(v)$. Similarly, for the $4$-cut $C'$ we have $B(u')\setminus\{e(u')\}=B(w)\setminus B(v)$. Thus, we have $B(u)\setminus\{e(u)\}=B(u')\setminus\{e(u')\}$. Notice that we cannot have $e(u)\in B(u')$, because otherwise we would have $B(u)=B(u')$, in contradiction to the fact that the graph is $3$-edge-connected. Similarly, we cannot have $e(u')\in B(u)$. Thus, $B(u)\setminus\{e(u)\}=B(u')\setminus\{e(u')\}$ implies that $B(u)\sqcup\{e(u')\}=B(u')\sqcup\{e(u)\}$. Then, Lemma~\ref{lemma:type2cuts} implies that $C''=\{(u,p(u)),(u',p(u')),e(u),e(u')\}$ is a Type-$2ii$ $4$-cut. Notice that $C'$ is implied by $C$ and $C''$ through the pair of edges $\{(u',p(u')),e(u')\}$. Let $\mathcal{C}$ be the collection of $4$-cuts computed by Algorithm~\ref{algorithm:type2-2}. Then, by Proposition~\ref{proposition:type-2-2} we have that $C''$ is implied by $\mathcal{C}$ through the pair of edges $\{(u',p(u')),e(u')\}$. Thus, by Lemma~\ref{lemma:implied_from_union} we have that $C'$ is implied by $\mathcal{C}\cup\{C\}$ through the pair of edges $\{(u',p(u')),e(u')\}$. 
\end{proof}

\noindent\\
\begin{algorithm}[H]
\caption{\textsf{Compute a collection of Type-3$\alpha$i $4$-cuts of the form $\{(u,p(u)),(v,p(v)),(w,p(w)),e\}$, where $w$ is a common ancestor of $\{u,v\}$, $u$ is a non-special vertex, and $e\in B(u)$, so that all Type-3$\alpha$i $4$-cuts of this form are implied from this collection, plus that of the Type-$2ii$ $4$-cuts computed by Algorithm~\ref{algorithm:type2-2}}}
\label{algorithm:type3-a-i-non-special}
\LinesNumbered
\DontPrintSemicolon
\ForEach{vertex $u\neq r$}{
\label{line:type3-a-i-non-special-for}
  compute $M(B(u)\setminus\{e(u)\})$\;
}
\ForEach{vertex $x$}{
  initialize a collection $\widetilde{U}(x)\leftarrow\emptyset$\;
}
\ForEach{vertex $u\neq r$}{
  let $x\leftarrow M(B(u)\setminus\{e(u)\})$\;
  \If{$M(u)\neq x$}{    
    insert $u$ into $\widetilde{U}(x)$\;
  }
}
\tcp{$\widetilde{U}(x)$ contains all non-special vertices $u$ with $M(B(u)\setminus\{e(u)\})=x$}
\ForEach{vertex $w\neq r$ such that $M(w)$ has at least two children}{
\label{line:type3-a-i-non-special-for-2}
compute $M(w,c_1)$ and $M(w,c_2)$, where $c_1$ and $c_2$ are the $\mathit{low1}$ and $\mathit{low2}$ children of $M(w)$, respectively\;
}
\tcp{the case where $u$ is a descendant of the $\mathit{low1}$ child of $M(w)$; for the other case, simply reverse the roles of $c_1$ and $c_2$}
\ForEach{vertex $w\neq r$}{
   \lIf{$M(w)$ has less than two children}{\textbf{continue}}
   let $c_1\leftarrow\mathit{low1}$ child of $M(w)$\;
   let $c_2\leftarrow\mathit{low2}$ child of $M(w)$\;
   \lIf{$M(w,c_1)=\bot$ \textbf{or} $M(w,c_2)=\bot$}{\textbf{continue}}
   let $u$ be the lowest proper descendant of $w$ in $\widetilde{U}(M(w,c_1))$\;
   \label{line:type3-a-i-non-special-u}
   let $v$ be the lowest proper descendant of $w$ such that $M(v)=M(w,c_2)$\;
   \label{line:type3-a-i-non-special-v}
   \lIf{$u$ and $v$ are related as ancestor and descendant}{\textbf{continue}}
   \If{$\mathit{bcount}(w)=(\mathit{bcount}(u)-1)+\mathit{bcount}(v)$ \textbf{and} $\mathit{high}_2(u)<w$ \textbf{and} $\mathit{high}_1(v)<w$}{
     mark $\{(u,p(u)),(v,p(v)),(w,p(w)),e(u)\}$ as a $4$-cut\;
     \label{line:type3-a-i-non-special-mark}
   }
}
\end{algorithm}

\begin{proposition}
\label{proposition:type-3-a-i-non-special}
Algorithm~\ref{algorithm:type3-a-i-non-special} computes a collection $\mathcal{C}$ of Type-3$\alpha$i $4$-cuts of the form $\{(u,p(u)),(v,p(v)),(w,p(w)),e\}$, where $w$ is a common ancestor of $\{u,v\}$, $u$ is a non-special vertex, and $e\in B(u)$, and it runs in linear time. Furthermore, let $\mathcal{C}'$ be the collection of Type-$2ii$ $4$-cuts computed by Algorithm~\ref{algorithm:type2-2}. Then, every Type-3$\alpha$i $4$-cut of the form $\{(u,p(u)),(v,p(v)),(w,p(w)),e\}$, where $w$ is a common ancestor of $\{u,v\}$, $u$ is a non-special vertex, and $e\in B(u)$ is implied by $\mathcal{C}\cup\mathcal{C}'$.
\end{proposition}
\begin{proof}
Let $C=\{(u,p(u)),(v,p(v)),(w,p(w)),e(u)\}$ be a $4$-element set that is marked in Line~\ref{line:type3-a-i-non-special-mark}. Then we have that $u$ and $v$ are proper descendants of $w$ such that $\mathit{bcount}(w)=(\mathit{bcount}(u)-1)+\mathit{bcount}(v)$, $\mathit{high}_2(u)<w$ and $\mathit{high}_1(v)<w$. Furthermore, we have that $u$ and $v$ are not related as ancestor and descendant. Thus, Lemma~\ref{lemma:type-3a-i-criterion} implies that there is a back-edge $e\in B(u)$ such that $\{(u,p(u)),(v,p(v)),(w,p(w)),e\}$ is a Type-$3\alpha i$ $4$-cut. By Lemma~\ref{lemma:type-3a-i-M}, this implies that $e=e(u)$. Thus, $C$ is indeed a $4$-cut. Let $\mathcal{C}$ be the collection of all $4$-cuts marked in Line~\ref{line:type3-a-i-non-special-mark}.

Let $C=\{(u,p(u)),(v,p(v)),(w,p(w)),e\}$ be a Type-3$\alpha$i $4$-cut such that $w$ is a common ancestor of $\{u,v\}$, $u$ is a non-special vertex, and $e\in B(u)$. Let $c_1$ and $c_2$ be the $\mathit{low1}$ and $\mathit{low2}$ children of $M(w)$, respectively. Lemma~\ref{lemma:type-3a-lowchild_desc} implies that either $u$ is a descendant of $c_1$ and $v$ is a descendant of $c_2$, or $u$ is a descendant of $c_2$ and $v$ is a descendant of $c_1$. Let us assume that $u$ is a descendant of $c_1$. Then, Lemma~\ref{lemma:type-3a-i-v-lowest} implies that $v$ is the lowest proper descendant of $v$ such that $M(v)=M(w,c_2)$. Lemma~\ref{lemma:type-3a-i-M} implies that $M(B(u)\setminus\{e(u)\})=M(w,c_1)$. Thus, we may consider the lowest proper descendant $u'$ of $w$ that is a non-special vertex such that $M(B(u')\setminus\{e(u')\})=M(w,c_1)$. Then, Lemma~\ref{lemma:type-3-a-i-lowest} implies that $C'=\{(u',p(u')),(v,p(v)),(w,p(w)),e(u')\}$ is a Type-$3\alpha i$ $4$-cut. Then, Lemma~\ref{lemma:type-3a-i-criterion} implies that $\mathit{bcount}(w)=(\mathit{bcount}(u')-1)+\mathit{bcount}(v)$, $\mathit{high}_2(u')<w$ and $\mathit{high}_1(v)<w$. Thus, notice that $C'$ will be marked in Line~\ref{line:type3-a-i-non-special-mark}, and therefore $C'\in\mathcal{C}$. Now, if $C'=C$, then it is trivially true that $C$ is implied by $\mathcal{C}$. Otherwise, by Lemma~\ref{lemma:type-3-a-i-implied} we have that $C$ is implied by $\mathcal{C}'\cup\{C'\}$. Thus, we have that $C$ is implied by $\mathcal{C}\cup\mathcal{C}'$.

Now we will argue about the complexity of Algorithm~\ref{algorithm:type3-a-i-non-special}. By Proposition~\ref{proposition:computing-M(B(v)-S)} we have that the values $M(B(u)\setminus\{e(u)\})$ can be computed in linear time in total, for all vertices $u\neq r$. Thus, the \textbf{for} loop in Line~\ref{line:type3-a-i-non-special-for} can be performed in linear time. By Proposition~\ref{proposition:computing-M(v,c)}, we have that the values $M(w,c_1)$ and $M(w,c_2)$ can be computed in linear time in total, for all vertices $w\neq r$ such that $M(w)$ has at least two children, where $c_1$ and $c_2$ are the $\mathit{low1}$ and $\mathit{low2}$ children of $M(w)$. Thus, the \textbf{for} loop in Line~\ref{line:type3-a-i-non-special-for-2} can be performed in linear time. The vertices $u$ and $v$ in Lines~\ref{line:type3-a-i-non-special-u} and \ref{line:type3-a-i-non-special-v} can be computed with Algorithm~\ref{algorithm:W-queries}. Specifically, whenever we reach Line~\ref{line:type3-a-i-non-special-u}, we generate a query $q(\widetilde{U}(M(w,c_1)),w)$, which returns the lowest vertex $u$ in $\widetilde{U}(M(w,c_1))$ such that $u>w$. $u\in\widetilde{U}(M(w,c_1))$ implies that $M(B(u)\setminus\{e(u)\})=M(w,c_1)$, and therefore we have that $M(w,c_1)$ is a common descendant of $u$ and $w$, and therefore $u$ and $w$ are related as ancestor and descendant. Thus, $u>w$ implies that $u$ is a proper descendant of $w$. Thus, $u$ is the lowest proper descendant of $w$ that lies in $\widetilde{U}(M(w,c_1))$. Since the sets $\widetilde{U}$ are disjoint, and the total number of those queries is $O(n)$, Lemma~\ref{lemma:W-queries} implies that Algorithm~\ref{algorithm:W-queries} can answer all those queries in $O(n)$ time in total. Similarly, the vertices $v$ in Line~\ref{line:type3-a-i-non-special-v} can be computed in $O(n)$ time in total. We conclude that Algorithm~\ref{algorithm:type3-a-i-non-special} runs in linear time. 
\end{proof}

\subsection{Type-3$\alpha ii$ $4$-cuts}

We will distinguish the Type-3$\alpha ii$ $4$-cuts according to the following. 

\begin{lemma}
\label{lemma:type-3-a-ii-cases}
Let $u,v,w$ be three vertices $\neq r$ such that $w$ is a common ancestor of $\{u,v\}$, and $u,v$ are not related as ancestor and descendant. Suppose that there is a back-edge $e=(x,y)$ such that $B(w)=(B(u)\sqcup B(v))\sqcup \{e\}$. Then we have the following cases (see also Figure~\ref{figure:type3aii}).
\begin{enumerate}[label*=\arabic*.]
\item{\textbf{$x$ is an ancestor of both $u$ and $v$.} In this case, $x=M(w)$. Furthermore, $u$ is a descendant of the $\mathit{low1}$ child of $\widetilde{M}(w)$ and $v$ is a descendant of the $\mathit{low2}$ child of $\widetilde{M}(w)$, or reversely.}
\item{\textbf{$x$ is an ancestor of $u$, but not an ancestor of $v$.} In this case, $x$ is a descendant of the $\mathit{low1}$ child of $M(w)$ and $v$ is a descendant of the $\mathit{low2}$ child of $M(w)$, or reversely.}
\item{\textbf{$x$ is an ancestor of $v$, but not an ancestor of $u$.} In this case, $x$ is a descendant of the $\mathit{low1}$ child of $M(w)$ and $u$ is a descendant of the $\mathit{low2}$ child of $M(w)$, or reversely.}
\item{\textbf{$x$ is neither an ancestor of $u$ nor an ancestor of $v$.} In this case, we have the following two subcases.
\begin{enumerate}[label*=\arabic*]
\item{Two of $\{u,v,x\}$ are descendants of the $\mathit{low1}$ child of $M(w)$ and the other is a descendant of the $\mathit{low2}$ child of $M(w)$, or reversely: two of $\{u,v,x\}$ are descendants of the $\mathit{low2}$ child of $M(w)$ and the other is a descendant of the $\mathit{low1}$ child of $M(w)$.}
\item{There is a permutation $\sigma$ of $\{1,2,3\}$ such that $u$ is a descendant of the $\mathit{low\sigma_1}$ child of $M(w)$, $v$ is a descendant of the $\mathit{low\sigma_2}$ child of $M(w)$, and $x$ is a descendant of the $\mathit{low\sigma_3}$ child of $M(w)$.}
\end{enumerate}
In cases $4.1, 4.2$ we have $l_2(x)\geq w$ and $\mathit{low}(c_1(x))\geq w$ (if $c_1(x)\neq\bot$).
} 

In all cases $1-4$, we have $y=l_1(x)$.
\end{enumerate}
\end{lemma}
\begin{proof}
Before we consider the four cases in turn, we will show that $M(w)$ is the nearest common ancestor of $\{u,v,x\}$. The fact that $M(w)$ is an ancestor of $x$ is an obvious consequence of $(x,y)\in B(w)$. Now, since the graph is $3$-edge-connected, neither $B(u)$ nor $B(v)$ is empty. Thus there are back-edges $(x',y')\in B(u)$ and $(x'',y'')\in B(v)$. Then $B(w)=(B(u)\sqcup B(v))\sqcup \{e\}$ implies that $(x',y')\in B(w)$ and $(x'',y'')\in B(w)$, and therefore $M(w)$ is a common ancestor of $\{x',x''\}$. Since $x'$ is a descendant of $u$ and $x''$ is a descendant of $v$, we have that $M(w)$ is related as ancestor and descendant with both $u$ and $v$. But since $u,v$ are not related as ancestor and descendant, $M(w)$ must be an ancestor of both $u$ and $v$. Thus far we have that $M(w)$ is a common ancestor of $\{u,v,x\}$. Now let us suppose, for the sake of contradiction, that $M(w)$ is not the nearest common ancestor of $\{u,v,x\}$. This means that there is a proper descendant $c$ of $M(w)$ that is a common ancestor of $\{u,v,x\}$. Now let $(x',y')$ be a back-edge in $B(w)$. Then $B(w)=(B(u)\sqcup B(v))\sqcup \{e\}$ implies that either $x'=x$, or $x'$ is a descendant of $u$, or $x'$ is a descendant of $v$. In any case, $x'$ is a descendant of $c$. But due to the generality of $(x',y')\in B(w)$, this shows that $M(w)$ is a descendant of $c$, a contradiction. Thus we have that $M(w)$ is the nearest common ancestor of $\{u,v,x\}$.

Furthermore, we will show that $u$ and $v$ are proper descendants of $M(w)$. Otherwise, let us assume w.l.o.g. that $u$ is not a proper descendant of $M(w)$. Since $M(w)$ is an ancestor of $u$, this means that $u=M(w)$. Now, since $x$ is a descendant of $M(w)$, it is also a descendant of $u$. And since $y$ is a proper ancestor of $w$, it is also a proper ancestor of $u$. But then we have $(x,y)\in B(u)$, contradicting (the disjointness of the union in) $B(w)=(B(u)\sqcup B(v))\sqcup \{e\}$.

$(1)$ Suppose that $x$ is an ancestor of both $u$ and $v$. Then, since $M(w)$ is the nearest common ancestor of $\{u,v,x\}$, we have that $x=M(w)$. Furthermore, since $u$ and $v$ are proper descendants of $M(w)$, we have $u\neq x$ and $v\neq x$. Let $S=\{(x',y')\in B(w)\mid x'\neq M(w)\}$. Then we have $\widetilde{M}(w)=M(S)$. We will show that $\widetilde{M}(w)$ is the nearest common ancestor of $\{u,v\}$. Let $(x',y')$ be a back-edge in $S$. Then $(x',y')\in B(w)\setminus\{e\}$, and so $B(w)=(B(u)\sqcup B(v))\sqcup \{e\}$ implies that $x'$ is either a descendant of $u$ or a descendant of $v$. This implies that $x'$ is a descendant of $\mathit{nca}(u,v)$. Due to the generality of $(x',y')\in S$, we have that $M(S)$ is a descendant of $\mathit{nca}(u,v)$. Conversely, let $(x',y')$ be a back-edge in $B(u)$ and let $(x'',y'')$ be a back-edge in $B(v)$ (such back-edges exist, because the graph is $3$-edge-connected). Then $B(w)=(B(u)\sqcup B(v))\sqcup \{e\}$ implies that $x'$ and $x''$ are in $S$, and so $M(S)$ is a common ancestor of $x'$ and $x''$. Then, since $x'$ is a descendant of $u$ and $x''$ is a descendant of $v$, we have that $M(S)$ is related to both $u$ and $v$ as ancestor and descendant. But since $u$ and $v$ are not related as ancestor and descendant, we have that $M(S)$ is an ancestor of both $u$ and $v$. Thus, since $M(S)$ is a descendant of $\mathit{nca}(u,v)$, we have that $M(S)$ is the nearest common ancestor of $\{u,v\}$. 

Since $u$ and $v$ are not related as ancestor and descendant, we have that there are children $c_1$ and $c_2$ of $M(S)$ such that $u$ is a descendant of $c_1$ and $v$ is a descendant of $c_2$. Then, since $B(u)$ and $B(v)$ are non-empty and $B(u)\cup B(v)\subset B(w)$, we have that $\mathit{low}(c_1)<w$ and $\mathit{low}(c_2)<w$. Now let us suppose, for the sake of contradiction, that there is also another child $c$ of $M(w)$ that has $\mathit{low}(c)<w$ (i.e., $c\notin\{c_1,c_2\}$). Then neither $u$ nor $v$ is a descendant of $c$. Then $\mathit{low}(c)<w$ implies that there is a back-edge $(x',y')$ such that $x'$ is a descendant of $c$ and $y$ is a proper ancestor of $w$. Since $c$ is a descendant of $M(S)$, which is a descendant of $M(w)$, we thus have that $(x',y')\in B(w)$. Then  $B(w)=(B(u)\sqcup B(v))\sqcup \{e\}$ implies that either $(x',y')\in B(u)\cup B(v)$, or $(x',y')=e$. The case $(x',y')=e$ is rejected, because $x'$ is a descendant of $c$, but $c$ is not an ancestor of either $u$ or $v$ (whereas $x$ is an ancestor of both $u$ and $v$). Thus, we have $(x',y')\in B(u)\cup B(v)$, which implies that either $x'$ is a descendant of $u$, or $x'$ is a descendant of $v$. But this contradicts the fact that $x'$ is a descendant of $c$ (which is not related as ancestor and descendant with either $u$ or $v$). Thus we have that $c_1$ and $c_2$ are the only children of $M(S)$ that have $\mathit{low}(c_i)<w$, for $i\in\{1,2\}$, and so these must coincide with the $\mathit{low1}$ and the $\mathit{low2}$ children of $M(S)$ (not necessarily in that order).

$(2)$ Suppose that $x$ is an ancestor of $u$, but not an ancestor of $v$. Then, since $M(w)$ is a common ancestor of $\{u,v,x\}$, we have that $x$ is a proper descendant of $M(w)$ (otherwise $x$ would be an ancestor of $v$). Since $v$ is also a proper descendant of $M(w)$, we have that both $x$ and $v$ are descendants of children of $M(w)$. Furthermore, since $u$ is a descendant of $x$, we have that $u$ is a descendant of the same child of $M(w)$ as $x$.

Let us suppose, for the sake of contradiction, that $x$ and $v$ are descendants of the same child $c$ of $M(w)$. Let $(x',y')$ be a back-edge in $B(w)$. Then $B(w)=(B(u)\sqcup B(v))\sqcup \{e\}$ implies that either $x'=x$, or $x'$ is a descendant of $u$, or $x'$ is a descendant of $v$. In either case, we have that $x'$ is a descendant of $c$. Due to the generality of $(x',y')$, this means that $M(w)$ is a descendant of $c$, a contradiction. Thus, $x$ and $v$ are descendants of different children of $M(w)$. Let $c_1$ be the child of $M(w)$ that is an ancestor of $x$, and let $c_2$ be the child of $M(w)$ that is an ancestor of $v$. Then the existence of the back-edge $(x,y)$ implies that $\mathit{low}(c_1)<w$. And the fact that the graph is $3$-edge-connected implies that $B(v)\neq\emptyset$, which further implies that $\mathit{low}(c_2)<w$, due to $B(v)\subset B(w)$. 

Now let us suppose, for the sake of contradiction, that there is another child $c$ of $M(w)$ (i.e., with $c\notin\{c_1,c_2\}$) that has $\mathit{low}(c)<w$. Then neither $v$ nor $x$ (and therefore neither $u$) is a descendant of $c$. Then $\mathit{low}(c)<w$ implies that there is a back-edge $(x',y')$ such that $x'$ is a descendant of $c$ and $y$ is a proper ancestor of $w$. Since $c$ is a descendant of $M(w)$, we thus have that $(x',y')\in B(w)$. Then  $B(w)=(B(u)\sqcup B(v))\sqcup \{e\}$ implies that either $x'=x$, or $x'$ is a descendant of $u$, or $x'$ is a descendant of $v$. But this contradicts the fact that $x'$ is a descendant of $c$ (which is not related as ancestor and descendant with either $x$, or $u$, or $v$). Thus we have that $c_1$ and $c_2$ are the only children of $M(w)$ that have $\mathit{low}(c_i)<w$, for $i\in\{1,2\}$, and so these must coincide with the $\mathit{low1}$ and the $\mathit{low2}$ children of $M(w)$ (not necessarily in that order).

$(3)$ The argument for this case is analogous to that for case $(2)$. 

$(4)$ Suppose that $x$ is neither an ancestor of $u$ nor an ancestor of $v$. Then, since $M(w)$ is a common ancestor of $\{u,v,x\}$, we have that $x$ is a proper descendant of $M(w)$ (otherwise $x$ would be an ancestor of both $u$ and $v$). Let us assume first that $u$ and $v$ are descendants of the same child $c_1$ of $M(w)$. Then we cannot have that $x$ is also a descendant of $c_1$, because $M(w)$ is the nearest common ancestor of $\{u,v,x\}$ (and therefore we would have that $M(w)$ is a descendant of $c_1$). So let $c_2$ be the child of $M(w)$ that is an ancestor of $x$. Now we can argue as in $(2)$, in order to demonstrate that $c_1$ and $c_2$ are the only children of $M(w)$ with $\mathit{low}(c_i)<w$, for $i\in\{1,2\}$, and so these must coincide with the $\mathit{low1}$ and the $\mathit{low2}$ children of $M(w)$ (not necessarily in that order).

Now let us assume that $u$ and $v$ are not descendants of the same child of $M(w)$. Let $c_1$ be the child of $M(w)$ that is an ancestor of $u$, and let $c_2$ be the child of $M(w)$ that is an ancestor of $v$. If we assume that $x$ is a descendant of either $c_1$ or $c_2$, then we can argue as in $(2)$, in order to demonstrate that $c_1$ and $c_2$ are the only children of $M(w)$ with $\mathit{low}(c_i)<w$, for $i\in\{1,2\}$, and so these must coincide with the $\mathit{low1}$ and the $\mathit{low2}$ children of $M(w)$ (not necessarily in that order). So let us assume that $x$ is neither a descendant of $c_1$, nor a descendant of $c_2$, and let $c_3$ be the child of $M(w)$ that is an ancestor of $x$. Then we can argue as in $(2)$ in order to demonstrate that $c_1$, $c_2$ and $c_3$ are the only children of $M(w)$ with $\mathit{low}(c_i)<w$, for $i\in\{1,2,3\}$, and so these must coincide with the $\mathit{low1}$, the $\mathit{low2}$, and the $\mathit{low3}$ children of $M(w)$ (not necessarily in that order).

Since $(x,y)\in B(w)$, we have that $e$ is a back-edge in $B(x)$ whose lower endpoint is lower than $w$. Now let us suppose, for the sake of contradiction, that there is one more back-edge $e'=(x',y')\in B(x)$ such that $y'<w$. Then $B(w)=(B(u)\sqcup B(v))\sqcup\{e\}$ implies that either $e'\in B(u)$, or $e'\in B(v)$, or $e'=e$. The last case is rejected by assumption. If $e'\in B(u)$, then $x'$ is a descendant of $u$. Therefore, $x$ and $u$ are related as ancestor and descendant, since they have $x'$ as a common descendant. Then, since $x$ is not an ancestor of $u$, we have that $x$ is a descendant of $u$. But since $y$ is a proper ancestor of $w$, it is also a proper ancestor of $u$, and therefore $(x,y)\in B(u)$, which is impossible. Thus, the case $e'\in B(u)$ is rejected. Similarly, the case $e'\in B(v)$ is also rejected. But then there are no viable options left, and so we are led to a contradiction. This shows that $e=(x,y)$ is the unique back-edge in $B(x)\cap B(w)$. Thus, we have $e=(x,l_1(x))$, and we have $l_2(x)\geq w$ and $\mathit{low}(c_1(x))\geq w$ (if $c_1(x)\neq\bot$).

Finally, let us suppose, for the sake of contradiction, that there is a back-edge of the form $(x,y')\in B(w)$ such that $(x,y')\neq (x,y)$. Then, $B(w)=(B(u)\sqcup B(v))\sqcup\{e\}$ implies that either $(x,y')\in B(u)$ or $(x,y')\in B(v)$. This implies that $x$ is a descendant of $u$ or $v$, respectively. Thus, since $y$ is a proper ancestor of $w$, we have that $(x,y)\in B(u)$ or $(x,y)\in B(v)$, respectively, in contradiction to (the disjointness in) $B(w)=(B(u)\sqcup B(v))\sqcup\{e\}$. This shows that $(x,y)$ is the only back-edge with higher endpoint $x$ such that $y$ is a proper ancestor of $w$. Let us suppose, for the sake of contradiction, that $y\neq l_1(x)$. Then there is a back-edge $(x,y')\neq (x,y)$ such that $y'\leq y$. Since $x$ is a common descendant of $y$ and $y'$, we have that $y$ and $y'$ are related as ancestor and descendant. Thus, $y'\leq y$ implies that $y'$ is an ancestor of $y$, and therefore $y'$ is a proper ancestor of $w$, a contradiction. Thus, we have $y=l_1(x)$.
\end{proof}

\begin{lemma}
\label{lemma:type-3-a-ii-inference}
Let $u,v,w$ be three vertices $\neq r$ such that $w$ is a common ancestor of $\{u,v\}$ and $u,v$ are not related as ancestor and descendant. If there is a back-edge $e$ such that $B(w)=(B(u)\sqcup B(v))\sqcup\{e\}$, then $u$ is the lowest proper descendant of $w$ in $M^{-1}(M(u))$. Similarly, $v$ is the lowest proper descendant of $w$ in $M^{-1}(M(v))$.
\end{lemma}
\begin{proof}
We will provide the argument for $u$, since that for $v$ is similar. Let us suppose, for the sake of contradiction, that there is a proper descendant $u'$ of $w$ with $M(u')=M(u)$ such that $u'$ is lower than $u$. Then we have that $u'$ is a proper ancestor of $u$, and Lemma~\ref{lemma:same_m_subset_B} implies that $B(u')\subseteq B(u)$. This can be strengthened to $B(u')\subset B(u)$, since the graph is $3$-edge-connected. Thus, there is a back-edge $(x,y)\in B(u)\setminus B(u')$. Then $x$ is a descendant of $M(u)$, and therefore a descendant of $M(u')$. Thus, it cannot be that $y$ is a proper ancestor of $u'$, for otherwise we would have $(x,y)\in B(u')$. This implies that $y$ cannot be a proper ancestor of $w$, for otherwise it would be a proper ancestor of $u'$. Thus we have that $(x,y)\notin B(w)$. But this contradicts $B(w)=(B(u)\sqcup B(v))\sqcup\{e\}$, which implies that $B(u)\subset B(w)$. Thus we have that $u$ is the lowest proper descendant of $w$ in $M^{-1}(M(u))$.
\end{proof}

\begin{lemma}
\label{lemma:type-3-a-ii-criterion}
Let $u,v,w$ be three vertices $\neq r$ such that $w$ is a common ancestor of $\{u,v\}$ and $u,v$ are not related as ancestor and descendant. Then there is a back-edge $e$ such that $B(w)=(B(u)\sqcup B(v))\sqcup\{e\}$ if and only if $\mathit{bcount}(w)=\mathit{bcount}(u)+\mathit{bcount}(v)+1$ and $\mathit{high}_1(u)<w$ and $\mathit{high}_1(v)<w$.
\end{lemma}
\begin{proof}
($\Rightarrow$) $\mathit{bcount}(w)=\mathit{bcount}(u)+\mathit{bcount}(v)+1$ is an immediate consequence of $B(w)=(B(u)\sqcup B(v))\sqcup\{e\}$. Now let $(x,y)$ be a back-edge in $B(u)$. Then $B(w)=(B(u)\sqcup B(v))\sqcup\{e\}$ implies that $(x,y)$ is in $B(w)$, and therefore $y$ is a proper ancestor of $w$, and therefore $y<w$. Due to the generality of $(x,y)\in B(u)$, this shows that $\mathit{high}_1(u)<w$. Similarly, we get $\mathit{high}_1(v)<w$.

($\Leftarrow$) Let $(x,y)$ be a back-edge in $B(u)$. Then, $x$ is a descendant of $u$, and therefore a descendant of $w$. Furthermore, $y$ is a proper ancestor of $u$. Thus, since $y$ and $w$ have $u$ as a common descendant, we have that $y$ and $w$ are related as ancestor and descendant. Since $(x,y)\in B(u)$, we have that $y$ is an ancestor of $\mathit{high}_1(u)$, and therefore $y\leq\mathit{high}_1(u)$. Thus, $\mathit{high}_1(u)<w$ implies that $y$ is a proper ancestor of $w$. This shows that $(x,y)\in B(w)$. Due to the generality of $(x,y)\in B(u)$, this implies that $B(u)\subseteq B(w)$. Similarly, we have $B(v)\subseteq B(w)$. Since $u$ and $v$ are not related as ancestor and descendant, we have $B(u)\cap B(v)=\emptyset$. Thus, $B(u)\sqcup B(v)\subseteq B(w)$. Now $\mathit{bcount}(w)=\mathit{bcount}(u)+\mathit{bcount}(v)+1$ implies that $|B(w)\setminus(B(u)\sqcup B(v))|=1$, and so there is a back-edge $e$ such that $B(w)\setminus(B(u)\sqcup B(v))=\{e\}$. This means that $B(w)=(B(u)\sqcup B(v))\sqcup\{e\}$.
\end{proof}

First, we consider case $(1)$ of Lemma~\ref{lemma:type-3-a-ii-cases}.

\begin{lemma}
\label{lemma:type-3-a-ii-case1}
Let case $(1)$ of Lemma~\ref{lemma:type-3-a-ii-cases} be true. Then $l_1(M(w))<w$, $l_2(M(w))\geq w$, and $e=(M(w),l_1(M(w)))$. Let $c_1$ and $c_2$ be the $\mathit{low1}$ and $\mathit{low2}$ children of $\widetilde{M}(w)$, respectively. Assume w.l.o.g. that $u$ is a descendant of $c_1$ and $v$ is a descendant of $c_2$. Then $M(u)=M(w,c_1)$ and $M(v)=M(w,c_2)$.
\end{lemma}
\begin{proof}
Since $x=M(w)$ and $(x,y)$ is a back-edge in $B(w)$, we have that $l_1(M(w))<w$. Let us suppose, for the sake of contradiction, that $l_2(M(w))<w$. Then there is a back-edge $(x,y')\neq (x,y)$ such that $(x,y')\in B(w)$. Since $u$ is a descendant of $c_1$ and $v$ is a descendant of $c_2$, we have that $x$ is not a descendant of either $u$ or $v$. Thus, $(x,y)$ and $(x,y')$ are two distinct back-edges that leap over $w$ and none of them is in $B(u)$ or $B(v)$. This contradicts the fact that $B(w)=(B(u)\sqcup B(v))\sqcup\{e\}$, which implies that exactly one back-edge in $B(w)$ is not in $B(u)\cup B(v)$. Thus, we have that $l_2(M(w))\geq w$. Since $e=(M(w),y)$ satisfies $y<w$, we have that $y=l_1(M(w))$.

Now we will provide the arguments for $u$, since those for $v$ are similar. Let $S=\{(x',y')\in B(w)\mid x' \mbox{ is a descendant of } c_1\}$. Then $M(S)=M(w,c_1)$. Let $(x',y')$ be a back-edge in $B(u)$. Then $B(w)=(B(u)\sqcup B(v))\sqcup\{e\}$ implies that $(x',y')\in B(w)$. Since $x'$ is a descendant of $u$ and $u$ is a descendant of $c_1$, we have that $x'$ is a descendant of $c_1$. Thus we have $(x',y')\in S$. Due to the generality of $(x',y')\in B(u)$, this shows that $M(u)$ is a descendant of $M(S)$. Conversely, let $(x',y')$ be a back-edge in $S$. Then we have that $x'$ is a descendant of $c_1$ and $y'$ is a proper ancestor of $w$. Furthermore, $B(w)=(B(u)\sqcup B(v))\sqcup\{e\}$ implies that either $x'=x$, or $x'$ is a descendant of $u$, or $x'$ is a descendant of $v$. The case $x'=x$ is rejected, because $x=M(w)$. Also, $x'$ cannot be a descendant of $v$, for otherwise $x'$ would be a descendant of $c_2$. Thus $x'$ is a descendant of $u$. Then, since $y'$ is a proper ancestor of $w$, we infer that $(x',y')\in B(u)$. Due to the generality of $(x',y')\in S$, this shows that $M(S)$ is a descendant of $M(u)$. This concludes the proof that $M(u)=M(w,c_1)$. 
\end{proof}

According to Lemma~\ref{lemma:type-3-a-ii-case1}, we can compute all $4$-cuts in case $(1)$ of Lemma~\ref{lemma:type-3-a-ii-cases} as follows. First, we only have to consider those $w\neq r$ such that $l_1(M(w))<w$ and $\widetilde{M}(w)$ has at least two children. In this case, let $c_1$ and $c_2$ be the $\mathit{low1}$ and $\mathit{low2}$ children of $\widetilde{M}(w)$. Then we compute $M(w,c_1)$ and $M(w,c_2)$. If none of $M(w,c_1)$ and $M(w,c_2)$ is $\bot$, then, according to Lemma~\ref{lemma:type-3-a-ii-inference}, we have that $u$ is the lowest proper descendant of $w$ that has $M(u)=M(w,c_1)$, and $v$ is the lowest proper descendant of $w$ that has $M(v)=M(w,c_2)$. Then, according to Lemma~\ref{lemma:type-3-a-ii-criterion}, we have that $\{(u,p(u)),(v,p(v)),(w,p(w)),(M(w),l_1(M(w)))\}$ is a $4$-cut if and only if $\mathit{high}(u)<w$, $\mathit{high}(v)<w$, and $\mathit{bcount}(w)=\mathit{bcount}(u)+\mathit{bcount}(v)+1$. The procedure for computing those $4$-cuts is given in Algorithm~\ref{algorithm:type3-a-ii-1}. The proof of correctness and linear complexity is given in Proposition~\ref{proposition:algorithm:type3-a-ii-1}.

\noindent\\
\begin{algorithm}[H]
\caption{\textsf{Compute all Type-3$\alpha$ii $4$-cuts of the form $\{(u,p(u)),(v,p(v)),(w,p(w)),e\}$, where $w$ is a common ancestor of $\{u,v\}$, and $e$ satisfies $(1)$ of Lemma~\ref{lemma:type-3-a-ii-cases}.}}
\label{algorithm:type3-a-ii-1}
\LinesNumbered
\DontPrintSemicolon
\ForEach{vertex $w\neq r$}{
\label{algorithm:type3-a-ii-1-for-1}
  \If{$l_1(M(w))<w$}{
    let $e(w)=(M(w),l_1(M(w)))$\;
    compute $\widetilde{M}(w)$\;
  }
}
\ForEach{vertex $w\neq r$}{
\label{algorithm:type3-a-ii-1-for-2}
  \If{$\widetilde{M}(w)$ has at least two children}{
    let $c_1$ and $c_2$ be the $\mathit{low1}$ and $\mathit{low2}$ children of $\widetilde{M}(w)$\;
    compute $M(w,c_1)$ and $M(w,c_2)$\;
  }
}
\ForEach{vertex $w\neq r$}{
   \lIf{$l_1(M(w))\geq w$}{\textbf{continue}}
   \lIf{$\widetilde{M}(w)$ has less than two children}{\textbf{continue}}
   let $c_1\leftarrow\mathit{low1}$ child of $\widetilde{M}(w)$\;
   let $c_2\leftarrow\mathit{low2}$ child of $\widetilde{M}(w)$\;   
   \lIf{$\mathit{low}(c_1)\geq w$ \textbf{or} $\mathit{low}(c_2)\geq w$}{\textbf{continue}}
   let $u$ be the lowest proper descendant of $w$ such that $M(u)=M(w,c_1)$\;
   \label{algorithm:type3-a-ii-1-u}
   let $v$ be the lowest proper descendant of $w$ such that $M(v)=M(w,c_2)$\;
   \label{algorithm:type3-a-ii-1-v}
   \tcp{$u$ and $v$ are not related as ancestor and descendant}
   \If{$\mathit{bcount}(w)=\mathit{bcount}(u)+\mathit{bcount}(v)+1$ \textbf{and} $\mathit{high}_1(u)<w$ \textbf{and} $\mathit{high}_1(v)<w$}{
     mark $\{(u,p(u)),(v,p(v)),(w,p(w)),e(w)\}$ as a $4$-cut\;
     \label{algorithm:type3-a-ii-1-mark}
   }
}
\end{algorithm}

\begin{proposition}
\label{proposition:algorithm:type3-a-ii-1}
Algorithm~\ref{algorithm:type3-a-ii-1} correctly computes all Type-3$\alpha$ii $4$-cuts of the form $\{(u,p(u)),(v,p(v)),(w,p(w)),e\}$, where $w$ is a common ancestor of $\{u,v\}$, and $e$ satisfies $(1)$ of Lemma~\ref{lemma:type-3-a-ii-cases}. Furthermore, it has a linear-time implementation.
\end{proposition}
\begin{proof}
Let $C=\{(u,p(u)),(v,p(v)),(w,p(w)),e\}$ be a Type-3$\alpha$ii $4$-cut, where $w$ is a common ancestor of $\{u,v\}$, and $e$ satisfies $(1)$ of Lemma~\ref{lemma:type-3-a-ii-cases}. Then, Lemma~\ref{lemma:type-3-a-ii-case1} implies that $e=(M(w),l_1(M(w)))$. Furthermore, let $c_1$ and $c_2$ be the $\mathit{low1}$ and $\mathit{low2}$ children of $\widetilde{M}(w)$, respectively. Then w.l.o.g. we have that $M(u)=M(w,c_1)$ and $M(v)=M(w,c_2)$. Then, Lemma~\ref{lemma:type-3-a-ii-inference} implies that $u$ is the lowest proper descendant of $w$ with $M(u)=M(w,c_1)$, and $v$ is the lowest proper descendant of $w$ with $M(v)=M(w,c_2)$. Lemma~\ref{lemma:type-3-a-ii-criterion} implies that $\mathit{bcount}(w)=\mathit{bcount}(u)+\mathit{bcount}(v)+1$, $\mathit{high}_1(u)<w$ and $\mathit{high}_1(v)<w$. Thus, all conditions are satisfied for $C$ to be marked in Line~\ref{algorithm:type3-a-ii-1-mark}.

Conversely, suppose that a $4$-element set $C=\{(u,p(u)),(v,p(v)),(w,p(w)),e(w)\}$ is marked in Line~\ref{algorithm:type3-a-ii-1-mark}. Then we have that $u$ and $v$ are descendants of $w$ such that $\mathit{bcount}(w)=\mathit{bcount}(u)+\mathit{bcount}(v)+1$, $\mathit{high}_1(u)<w$ and $\mathit{high}_1(v)<w$. Since $M(u)=M(w,c_1)$ and $M(v)=M(w,c_2)$, where $c_1$ and $c_2$ are different children of $M(w)$, by Lemma~\ref{lemma:w-v-u-not-related} we have that $u$ and $v$ are not related as ancestor and descendant.
Therefore, Lemma~\ref{lemma:type-3-a-ii-criterion} implies that there is a back-edge $e$ such that $B(w)=(B(u)\sqcup B(v))\sqcup\{e\}$. Since $l_1(M(w))<w$, we have that $(M(w),l_1(M(w)))$ is a back-edge in $B(w)$. Since $u$ and $v$ are proper descendants of $M(w)$, we have that this back-edge does not belong to $B(u)\cup B(v)$. Thus, $e=(M(w),l_1(M(w)))$, and therefore $C$ is indeed a Type-3$\alpha$ii $4$-cut.

Now we will argue about the complexity of Algorithm~\ref{algorithm:type3-a-ii-1}. By Proposition~\ref{proposition:computing-M(v,c)}, we have that the values $\widetilde{M}(w)$ can be computed in linear time in total, for all vertices $w\neq r$ (see the first paragraph in Section~\ref{subsection:M}). Thus, the \textbf{for} loop in Line~\ref{algorithm:type3-a-ii-1-for-1} can be performed in linear time. By Proposition~\ref{proposition:computing-M(v,c)} we have that the values $M(w,c_1)$ and $M(w,c_2)$ can be computed in linear time in total, for all vertices $w\neq r$ such that $\widetilde{M}(w)$ has at least two children, where $c_1$ and $c_2$ are the $\mathit{low1}$ and $\mathit{low2}$ children of $\widetilde{M}(w)$, respectively. Thus, the \textbf{for} loop in Line~\ref{algorithm:type3-a-ii-1-for-2} can be performed in linear time. In order to compute the vertices $u$ and $v$ in Lines~\ref{algorithm:type3-a-ii-1-u} and \ref{algorithm:type3-a-ii-1-v}, respectively, we use Algorithm~\ref{algorithm:W-queries}. Specifically, whenever we reach Line~\ref{algorithm:type3-a-ii-1-u}, we generate a query $q(M^{-1}(M(w,c_1)),w)$. This will return the lowest vertex $u$ with $M(u)=M(w,c_1)$ such that $u>w$. Since $M(u)=M(w,c_1)$ is a common descendant of $u$ and $w$, we have that $u$ and $w$ are related as ancestor and descendant. Thus, $u>w$ implies that $u$ is a proper descendant of $w$. Thus, $u$ is the lowest vertex with $M(u)=M(w,c_1)$ such that $u$ is a proper descendant of $w$. Since the number of all those queries is $O(n)$, Algorithm~\ref{algorithm:W-queries} can compute them in linear time in total, according to Lemma~\ref{lemma:W-queries}. The same is true for the queries for $v$ in Line~\ref{algorithm:type3-a-ii-1-v}. We conclude that Algorithm~\ref{algorithm:type3-a-ii-1} has a linear-time implementation.
\end{proof}

Now we consider case $(2)$ of Lemma~\ref{lemma:type-3-a-ii-cases}. Notice that due to the symmetry between cases $(2)$ and $(3)$ of Lemma~\ref{lemma:type-3-a-ii-cases}, case $(3)$ essentially coincides with case $(2)$ (after switching the labels of $u$ and $v$), and thus we do not have to provide a different algorithm for case $(3)$. 

\begin{lemma}
\label{lemma:type-3-a-ii-case2}
Let case $(2)$ of Lemma~\ref{lemma:type-3-a-ii-cases} be true. Let $c_1$ and $c_2$ be the $\mathit{low1}$ and $\mathit{low2}$ children of $M(w)$, respectively. Assume w.l.o.g. that $x$ is a descendant of $c_1$ and $v$ is a descendant of $c_2$. Then $e=(M(w,c_1),l_1(M(w,c_1)))$. Let $S=\{(x',y')\in B(w)\mid x' \mbox{ is a descendant of the }\mathit{low1} \mbox{ child of } x\}$. Then $M(u)=M(S)$ and $M(v)=M(w,c_2)$.
\end{lemma}
\begin{proof}
First we will show that $e=(M(w,c_1),l_1(M(w,c_1)))$. Let $(x',y')$ be a back-edge in $B(w)$ such that $x'$ is a descendant of $c_1$. Then $B(w)=(B(u)\sqcup B(v))\sqcup \{e\}$ implies that either $(x',y')=e$, or $(x',y')\in B(u)$, or $(x',y')\in B(v)$. Only the last case is rejected, since $v$ is a descendant of $c_2$, and thus it cannot be an ancestor of $x'$. Thus we have that the nearest common ancestor of $x$ and $M(u)$ is an ancestor of $x'$. Due to the generality of $(x',y')\in B(w)$ with $x'$ a descendant of $c_1$, this shows that the nearest common ancestor of $x$ and $M(u)$ is an ancestor of $M(w,c_1)$. Conversely, if $(x',y')$ is a back-edge such that either $(x',y')=e$ or $(x',y')\in B(u)$, then $x'$ is a descendant of $c_1$, and $B(w)=(B(u)\sqcup B(v))\sqcup \{e\}$ implies that $(x',y')\in B(w)$. Thus, $x'$ is a descendant of $M(w,c_1)$, and so the nearest common ancestor of $x$ and $M(u)$ is a descendant of $M(w,c_1)$. This shows that $M(w,c_1)=\mathit{nca}\{x,M(u)\}$. Since $x$ is an ancestor of $u$, it is also an ancestor of $M(u)$, and so $\mathit{nca}\{x,M(u)\}=x$. This shows that $x=M(w,c_1)$. Since $(x,y)\in B(w)$, we have that $l_1(M(w,c_1))<w$. Now let us suppose, for the sake of contradiction, that $l_2(M(w,c_1))<w$. Then there is a back-edge $(x,y')\neq (x,y)$ such that $y'<w$, and thus we have $(x,y')\in B(w)$. Notice that since $(x,y)\notin B(u)$, it cannot be the case that $x$ is a descendant of $u$. Furthermore, $x$ cannot be a descendant of $v$, because $v$ is a descendant of $c_2$ whereas $x$ is a descendant of $c_1$. Thus, none of $(x,y)$ and $(x,y')$ is in $B(u)$ or $B(v)$. But this contradicts $B(w)=(B(u)\sqcup B(v))\sqcup\{e\}$, which implies that there is only one back-edge in $B(w)$ that is not in $B(u)$ or $B(v)$. This shows that $l_2(M(w,c_1))\geq w$. Thus, since $y<w$, we have that $y=l_1(M(w,c_1))$.

Now we will provide the arguments for $u$, since those for $v$ are basically given in the proof of Lemma~\ref{lemma:type-3-a-ii-case1}. Since $(x,y)\in B(w)$, it cannot be the case that $x=u$, for otherwise we would have that $(x,y)\in B(u)$, contradicting (the disjointness of the union in) $B(w)=(B(u)\sqcup B(v))\sqcup \{e\}$. Thus, $u$ is a proper descendant of $x$. Let $c$ be the child of $x$ that is an ancestor of $u$. Then, since $B(u)$ is non-empty and $B(u)\subset B(w)$, we have that $\mathit{low}(c)<w$. Now let us suppose, for the sake of contradiction, that there is also another child $c'$ of $x$ that has $\mathit{low}(c')<w$ (i.e., $c'\neq c$). This means that there is a back-edge $(x',y')$ such that $x'$ is a descendant of $c'$ and $y'$ is a proper ancestor of $w$. Then $B(w)=(B(u)\sqcup B(v))\sqcup \{e\}$ implies that either $x'=x$, or $x'$ is a descendant of $u$, or $x'$ is a descendant of $v$. $x'=x$ is rejected, since $x'$ is a descendant of $c'$. Furthermore, $x'$ cannot be a descendant of $v$, because this would imply that $x$ and $v$ are related as ancestor and descendant, contradicting the fact that $x$ and $v$ are descendants of different children of $M(w)$. Thus we have that $x'$ is a descendant of $u$, and therefore a descendant of $c$, a contradiction. Thus $c$ is the unique child of $x$ that satisfies $\mathit{low}(c)<w$, and so it must be the $\mathit{low1}$ child of $x$.

Now let $(x',y')$ be a back-edge in $B(u)$. Then we have that $x'$ is a descendant of $u$, and therefore a descendant of the $\mathit{low1}$ child of $x$. Furthermore, $B(w)=(B(u)\sqcup B(v))\sqcup \{e\}$ implies that $(x',y')\in B(w)$. This shows that $(x',y')\in S$. Due to the generality of $(x',y')\in B(u)$, this implies that $B(u)\subseteq S$. Conversely, let $(x',y')$ be a back-edge in $S$. Then $(x',y')$ is in $B(w)$, and so $B(w)=(B(u)\sqcup B(v))\sqcup \{e\}$ implies that either $x'=x$, or $x'$ is a descendant of $u$, or $x'$ is a descendant of $v$. Since $x'$ is a descendant of the $\mathit{low1}$ child of $x$, the only viable option is that $x'$ is a descendant of $u$. Since $y'$ is a proper ancestor of $w$, it is also a proper ancestor of $u$. This shows that $(x',y')\in B(u)$. Due to the generality of $(x',y')\in S$, this implies that $S\subseteq B(v)$. Thus we have shown that $B(u)=S$, and so $M(u)=M(S)$ is derived.
\end{proof}

According to Lemma~\ref{lemma:type-3-a-ii-case2}, we can compute all $4$-cuts in case $(2)$ of Lemma~\ref{lemma:type-3-a-ii-cases} as follows. First, we only have to consider those $w\neq r$ such that $M(w)$ has at least two children. In this case, let $c_1$ and $c_2$ be the $\mathit{low1}$ and $\mathit{low2}$ children of $M(w)$, respectively. Then we compute $M(w,c_1)$ and $M(w,c_2)$. If none of $M(w,c_1)$ and $M(w,c_2)$ is $\bot$, then we keep considering $w$ only if $l_1(M(w,c_1))<w$ and $M(w,c_1)$ has at least one child. In this case, let $c_1'$ be the $\mathit{low1}$ child of $M(w,c_1)$. Then, we have that $e=(M(w,c_1),l_1(M(w,c_1)))$, and, according to Lemma~\ref{lemma:type-3-a-ii-inference}, we have that $u$ is the lowest proper descendant of $w$ that has $M(u)=M(w,c_1')$, and $v$ is the lowest proper descendant of $w$ that has $M(v)=M(w,c_2)$. Then, according to Lemma~\ref{lemma:type-3-a-ii-criterion}, we have that $\{(u,p(u)),(v,p(v)),(w,p(w)),e\}$ is a $4$-cut if and only if $\mathit{high}(u)<w$, $\mathit{high}(v)<w$, and $\mathit{bcount}(w)=\mathit{bcount}(u)+\mathit{bcount}(v)+1$. The procedure for computing those $4$-cuts is given in Algorithm~\ref{algorithm:type3-a-ii-2}. The proof of correctness and linear complexity is given in Proposition~\ref{proposition:algorithm:type3-a-ii-2}.

\noindent\\
\begin{algorithm}[H]
\caption{\textsf{Compute all Type-3$\alpha$ii $4$-cuts of the form $\{(u,p(u)),(v,p(v)),(w,p(w)),e\}$, where $w$ is a common ancestor of $\{u,v\}$, and $e$ satisfies $(2)$ of Lemma~\ref{lemma:type-3-a-ii-cases}.}}
\label{algorithm:type3-a-ii-2}
\LinesNumbered
\DontPrintSemicolon
\tcp{We assume that the higher endpoint of $e$ is a descendant of the $\mathit{low1}$ child of $M(w)$; the other case is treated similarly, by reversing the roles of $c_1$ and $c_2$}
\ForEach{vertex $w\neq r$}{
\label{line:type3-a-ii-2-for}
   \lIf{$M(w)$ has less than two children}{\textbf{continue}}
   let $c_1\leftarrow\mathit{low1}$ child of $M(w)$\;
   let $c_2\leftarrow\mathit{low2}$ child of $M(w)$\;
   compute $M(w,c_1)$ and $M(w,c_2)$\;
   \lIf{$M(w,c_1)=\bot$ \textbf{or} $l_1(M(w,c_1))\geq w$ \textbf{or} $M(w,c_1)$ has no children}{\textbf{continue}}
   let $e(w)\leftarrow(M(w,c_1),l_1(M(w,c_1)))$\;
   let $c'_1\leftarrow\mathit{low1}$ child of $M(w,c_1)$\;
   compute $M(w,c'_1)$\;   
}
\ForEach{vertex $w\neq r$}{
   \lIf{$M(w)$ has less than two children}{\textbf{continue}}
   let $c_1\leftarrow\mathit{low1}$ child of $M(w)$\;
   let $c_2\leftarrow\mathit{low2}$ child of $M(w)$\;
   \lIf{$\mathit{low}(c_1)\geq w$ \textbf{or} $\mathit{low}(c_2)\geq w$}{\textbf{continue}}
   \lIf{$l_1(M(w,c_1))\geq w$ \textbf{or} $M(w,c_1)$ has no children}{\textbf{continue}}
   let $c_1'\leftarrow\mathit{low1}$ child of $M(w,c_1)$\;
   \lIf{$\mathit{low}(c_1')\geq w$}{\textbf{continue}}
   let $u$ be the lowest proper descendant of $w$ such that $M(u)=M(w,c_1')$\;
   \label{line:type3-a-ii-2-u}
   let $v$ be the lowest proper descendant of $w$ such that $M(v)=M(w,c_2)$\;
   \label{line:type3-a-ii-2-v}
   \tcp{$u$ and $v$ are not related as ancestor and descendant}
   \label{line:type3-a-ii-2-not-rel}
   \If{$\mathit{bcount}(w)=\mathit{bcount}(u)+\mathit{bcount}(v)+1$ \textbf{and} $\mathit{high}_1(u)<w$ \textbf{and} $\mathit{high}_1(v)<w$}{
     mark $\{(u,p(u)),(v,p(v)),(w,p(w)),e(w)\}$ as a $4$-cut\;
     \label{line:type3-a-ii-2-mark}
   }
}
\end{algorithm}

\begin{proposition}
\label{proposition:algorithm:type3-a-ii-2}
Algorithm~\ref{algorithm:type3-a-ii-2} correctly computes all Type-3$\alpha$ii $4$-cuts of the form $\{(u,p(u)),(v,p(v)),(w,p(w)),e\}$, where $w$ is a common ancestor of $\{u,v\}$, and $e$ satisfies $(2)$ of Lemma~\ref{lemma:type-3-a-ii-cases}. Furthermore, it has a linear-time implementation.
\end{proposition}
\begin{proof}
Let $C=\{(u,p(u)),(v,p(v)),(w,p(w)),e\}$ be a Type-3$\alpha$ii $4$-cut where $w$ is a common ancestor of $u$ and $v$, and $e$ satisfies $(2)$ of Lemma~\ref{lemma:type-3-a-ii-cases}. Let us also assume that the higher endpoint $x$ of $e$ is a descendant of the $\mathit{low1}$ child of $M(w)$ (the other case, where $x$ is a descendant of the $\mathit{low2}$ child of $M(w)$, is treated similarly). Let $c_1$ and $c_2$ be the $\mathit{low1}$ and $\mathit{low2}$ children of $M(w)$, respectively. Furthermore, let $c_1'$ be the $\mathit{low1}$ child of $M(w,c_1)$. Then Lemma~\ref{lemma:type-3-a-ii-case2} implies that $e=(M(w,c_1),l_1(M(w,c_1)))$, $M(u)=M(w,c_1')$ and $M(v)=M(w,c_2)$. Then, Lemma~\ref{lemma:type-3-a-ii-inference} implies that $u$ is the lowest proper descendant of $w$ with $M(u)=M(w,c_1')$, and $v$ is the lowest proper descendant of $w$ with $M(v)=M(w,c_2)$. Finally, Lemma~\ref{lemma:type-3-a-ii-criterion} implies that $\mathit{bcount}(w)=\mathit{bcount}(u)+\mathit{bcount}(v)+1$, $\mathit{high}_1(u)<w$ and $\mathit{high}_1(v)<w$. Thus, we have that $C$ will be marked in Line~\ref{line:type3-a-ii-2-mark}.

Conversely, let $C=\{(u,p(u)),(v,p(v)),(w,p(w)),e\}$ be a $4$-element set that is marked in Line~\ref{line:type3-a-ii-2-mark}. Then we have that $u$ and $v$ are descendants of $w$ such that $\mathit{bcount}(w)=\mathit{bcount}(u)+\mathit{bcount}(v)+1$, $\mathit{high}_1(u)<w$ and $\mathit{high}_1(v)<w$. We will show that the comment in Line~\ref{line:type3-a-ii-2-not-rel} is true: i.e., $u$ and $v$ are not related as ancestor and descendant. This is a consequence of the fact that $M(u)=M(w,c_1')$, $M(v)=M(w,c_2)$, and $c_1'$ is a descendant of the $\mathit{low1}$ child $c_1$ of $M(w)$, whereas $c_2$ is the $\mathit{low2}$ child of $M(w)$. Thus, Lemma~\ref{lemma:w-v-u-not-related} implies that $u$ and $v$ are not related as ancestor and descendant. Therefore, Lemma~\ref{lemma:type-3-a-ii-criterion} implies that there is a back-edge $e$ such that $B(w)=(B(u)\sqcup B(v))\sqcup\{e\}$. Since we have that $l_1(M(w,c_1))<w$, we have that the back-edge $e(w)=(M(w,c_1),l_1(M(w,c_1)))$ is in $B(w)$. Since $M(u)$ is a proper descendant of $M(w,c_1)$, we have that $e(w)\notin B(u)$. And since $M(v)=M(w,c_2)$, we have that $e(w)\notin B(v)$. Thus, $e=e(w)$, and therefore $C$ is indeed a Type-3$\alpha$ii $4$-cut.

Now we will argue about the complexity of Algorithm~\ref{algorithm:type3-a-ii-2}. For every $w\neq r$ such that $M(w)$ has at least two children, we have to compute $M(w,c_1)$ and $M(w,c_2)$, where $c_1$ and $c_2$ are the $\mathit{low1}$ and $\mathit{low1}$ children of $M(w)$. By Proposition~\ref{proposition:computing-M(v,c)} this can be done in linear time in total, for all such vertices $w$. Then, for every such $w$, if $M(w,c_1)\neq\bot$ and $M(w,c_1)$ has at least one child, we have to compute $M(w,c_1')$, where $c_1'$ is the $\mathit{low1}$ child of $M(w,c_1)$. Again, by Proposition~\ref{proposition:computing-M(v,c)}, all these calculations  take linear time in total. Thus, the \textbf{for} loop in Line~\ref{line:type3-a-ii-2-for} can be performed in linear time. In order to compute the vertices $u$ and $v$ in Lines~\ref{line:type3-a-ii-2-u} and \ref{line:type3-a-ii-2-v}, we can use Algorithm~\ref{algorithm:W-queries}, as explained e.g. in the proof of Proposition~\ref{proposition:algorithm:type3-a-ii-1}. According to Lemma~\ref{lemma:W-queries}, all these computations take $O(n)$ time in total. We conclude that Algorithm~\ref{algorithm:type3-a-ii-2} has a linear-time implementation.
\end{proof}

Now we consider case $(4.1)$ of Lemma~\ref{lemma:type-3-a-ii-cases}.

\begin{lemma}
\label{lemma:type-3-a-ii-case4.1}
Let case $(4.1)$ of Lemma~\ref{lemma:type-3-a-ii-cases} be true. Let $c_1$ be the $\mathit{low1}$ child of $M(w)$, and let $c_2$ be the $\mathit{low2}$ child of $M(w)$. Let us assume that two of $\{u,v,x\}$ are descendants of $c_1$. Let $c_1'$ and $c_2'$ be the $\mathit{low1}$ and the $\mathit{low2}$ child of $M(w,c_1)$, respectively. Then we have the following three cases.
\begin{enumerate}[label=(\arabic*)]
\item{\textbf{$u$ and $v$ are descendants of $c_1$, and $x$ is a descendant of $c_2$.} Then we have $M(u)=M(w,c_1')$ and $M(v)=M(w,c_2')$ (or reversely). Furthermore, we have $(x,y)=(M(w,c_2),l_1(M(w,c_2)))$.}
\item{\textbf{$u$ and $x$ are descendants of $c_1$, and $v$ is a descendant of $c_2$.} Then we have $M(u)=M(w,c_1')$ and $(x,y)=(M(w,c_2'),l_1(M(w,c_2')))$ (or $M(u)=M(w,c_2')$ and $(x,y)=(M(w,c_1'),l_1(M(w,c_1')))$). Furthermore, we have $M(v)=M(w,c_2)$.}
\item{\textbf{$v$ and $x$ are descendants of $c_1$, and $u$ is a descendant of $c_2$.} Then we have $M(v)=M(w,c_1')$ and $(x,y)=(M(w,c_2'),l_1(M(w,c_2')))$ (or $M(v)=M(w,c_2')$ and $(x,y)=(M(w,c_1'),l_1(M(w,c_1')))$). Furthermore, we have $M(u)=M(w,c_2)$.}
\end{enumerate}
\end{lemma}
\begin{proof}
Let us consider case $(1)$ first. Let $S=\{(x',y')\in B(w)\mid x' \mbox{ is a descendant of } c_1\}$. Then we have $M(S)=M(w,c_1)$. Let $(x',y')$ be a back-edge in $B(w)$. Then, $B(w)=(B(u)\sqcup B(v))\sqcup\{e\}$ implies that either $(x',y')\in B(u)\sqcup B(v)$ or $(x',y')=(x,y)$. The case $(x',y')=(x,y)$ is rejected, because $x'$ is a descendant of $c_1$, whereas $x$ is a descendant of $c_2$ (and $c_1,c_2$ are not related as ancestor and descendant). Thus, we have $(x',y')\in B(u)\sqcup B(v)$. Due to the generality of $(x',y')\in S$, this implies that $S\subseteq B(u)\sqcup B(v)$. Conversely, let $(x',y')\in B(u)$. Then $x'$ is a descendant of $u$, and therefore a descendant of $c_1$. Furthermore, $B(w)=(B(u)\sqcup B(v))\sqcup\{e\}$ implies that $(x',y')\in B(w)$. Thus, we have $(x',y')\in S$. Due to the generality of $(x',y')\in B(u)$, this implies that $B(u)\subseteq S$. Similarly, we can show that $B(v)\subseteq S$. Thus, we have $S\subseteq B(u)\sqcup B(v)$. Since $B(u)\sqcup B(v)\subseteq S$, this can be strengthened to $S=B(u)\sqcup B(v)$. Therefore, $M(S)$ is an ancestor of both $M(u)$ and $M(v)$. Let us suppose, for the sake of contradiction, that $M(u)=M(S)$. Then, $M(u)$ is an ancestor of $M(v)$, and therefore $u$ is an ancestor of $M(v)$. Thus, $M(v)$ is a common descendant of $v$ and $u$, in contradiction to the fact that $u$ and $v$ are not related as ancestor and descendant. Thus, we have that $M(u)$ is a proper descendant of $M(S)$. Similarly, we can show that $M(v)$ is a proper descendant of $M(S)$.

Let us suppose, for the sake of contradiction, that there is a back-edge of the form $(M(S),z)$ in $S$. Then, since $S=B(u)\sqcup B(v)$, we have that either $(M(S),z)\in B(u)$, or $(M(S),z)\in B(v)$. The first case implies that $M(S)$ is a descendant of $M(u)$, and therefore $M(S)=M(u)$ (since $M(S)$ is an ancestor of $M(u)$), which is impossible. Thus, the case $(M(S),z)\in B(u)$ is rejected. Similarly, we can reject $(M(S),z)\in B(v)$. Therefore, there are no viable options left, and so we have arrived at a contradiction. This shows that there is no back-edge of the form $(M(S),z)$ in $S$. Thus, there are at least two back-edges $(x_1,y_1)$ and $(x_2,y_2)$ in $S$ such that $x_1$ is a descendant of the $\mathit{low1}$ child $c_1'$ of $M(S)$, and $x_2$ is a descendant of the $\mathit{low2}$ child $c_2'$ of $M(S)$. Since $S=B(u)\sqcup B(v)$, we have that $(x_1,y_1)\in B(u)\sqcup B(v)$ and $(x_2,y_2)\in B(u)\sqcup B(v)$. Notice that we cannot have that both $(x_1,y_1)\in B(u)$ and $(x_2,y_2)\in B(u)$, because this would imply that $M(u)$ is an ancestor of $\mathit{nca}\{x_1,x_2\}=M(S)$, and so $M(S)=M(u)$ (since $M(S)$ is an ancestor of $M(u)$), which is impossible. Similarly, it cannot be that both $(x_1,y_1)$ and $(x_2,y_2)$ are in $B(v)$. Thus, we have that one of $(x_1,y_1)$ and $(x_2,y_2)$ is in $B(u)$, and the other is in $B(v)$. Let us assume w.l.o.g. that $(x_1,y_1)\in B(u)$ and $(x_2,y_2)\in B(v)$. This implies that $M(u)$ is an ancestor of $x_1$, and $M(v)$ is an ancestor of $x_2$. Since $x_1$ is a common descendant of $c_1'$ and $M(u)$, we have that $c_1'$ and $M(u)$ are related as ancestor and descendant. Similarly, since $x_2$ is a common descendant of $c_2'$ and $M(v)$, we have that $c_2'$ and $M(v)$ are related as ancestor and descendant.

Now let $S_1=\{(x',y')\in B(w)\mid x' \mbox{ is a descendant of } c_1'\}$. Then we have $M(S_1)=M(w,c_1')$. Let $(x',y')$ be a back-edge in $S_1$. Then, $x'$ is a descendant of $c_1'$, and therefore a descendant of $M(S)$. Thus, since $(x',y')\in B(w)$, we have that $(x',y')\in S$. Since $S=B(u)\sqcup B(v)$, this implies that either $(x',y')\in B(u)$ or $(x',y')\in B(v)$. Let us suppose, for the sake of contradiction, that $(x',y')\in B(v)$. Then, $x'$ is a descendant of $M(v)$. Thus, we have that $x'$ is a common descendant of $c_1'$ and $M(v)$, and therefore $c_1'$ and $M(v)$ are related as ancestor and descendant. Since $M(v)$ is related as ancestor and descendant with $c_2'$, but $c_1'$ and $c_2'$ are not related as ancestor and descendant (because they have the same parent), we have that $M(v)$ is an ancestor of both $c_1'$ and $c_2'$. This implies that $M(v)$ is an ancestor of $\mathit{nca}\{c_1',c_2'\}=M(S)$, and therefore $M(S)=M(v)$ (since $M(S)$ is an ancestor of $M(v)$), which is impossible. Thus, the case $(x',y')\in B(v)$ is rejected, and so we have $(x',y')\in B(u)$. Due to the generality of $(x',y')\in S_1$, this implies that $S_1\subseteq B(u)$. Conversely, let $(x',y')$ be a back-edge in $B(u)$. Then $S=B(u)\sqcup B(v)$ implies that $(x',y')\in S$. Thus, since there is no back-edge of the form $(M(S),z)$ in $S$, we have that $x'$ is a descendant of a child $c$ of $M(S)$. Let us suppose, for the sake of contradiction, that $c\neq c_1'$. Since $(x',y')\in B(u)$, we have that $x'$ is a descendant of $M(u)$. Thus, since $x'$ is a common descendant of $c$ and $M(u)$, we have that $c$ and $M(u)$ are related as ancestor and descendant. Since $M(u)$ is related as ancestor and descendant with $c_1'$, but $c$ and $c_1'$ are not related as ancestor and descendant (since they have the same parent), we have that $M(u)$ is a common ancestor of $c_1'$ and $c$, and thus $M(u)$ is an ancestor of $\mathit{nca}\{c,c_1'\}=M(S)$. Since $M(S)$ is an ancestor of $M(u)$, this implies that $M(S)=M(u)$, which is impossible. Thus, we have that $c=c_1'$, and therefore $x'$ is a descendant of $c_1'$. Furthermore, since $(x',y')\in B(w)$, we have that $(x',y')\in S_1$. Due to the generality of $(x',y')\in B(u)$, this implies that $B(u)\subseteq S_1$. Thus, since $S_1\subseteq B(u)$, we have that $S_1=B(u)$. This implies that $M(S_1)=M(u)$, and therefore $M(w,c_1')=M(u)$. Similarly, we can show that $M(w,c_2')=M(v)$.

Now let $S'=\{(x',y')\in B(w)\mid x' \mbox{ is a descendant of } c_2\}$. Notice that $M(w,c_2)=M(S')$, and $(x,y)\in S'$. Let us suppose, for the sake of contradiction, that there is a back-edge $(x',y')$ in $S'$ such that $(x',y')\neq (x,y)$. Then, $B(w)=(B(u)\sqcup B(v))\sqcup\{e\}$ implies that $(x',y')\in B(u)\sqcup B(v)$, and therefore $(x',y')\in S$. This implies that $x'$ is a descendant of $c_1$, which is impossible, since $x'$ is a descendant of $c_2$ (and $c_1,c_2$ are not related as ancestor and descendant). Thus, we have that $S'=\{(x,y)\}$, and therefore $M(S')=x$, and therefore $M(w,c_2)=x$. Lemma~\ref{lemma:type-3-a-ii-cases} implies that $(x,y)=(x,l_1(x))$, and so $e=(M(w,c_2),l_1(M(w,c_2)))$.

The arguments for cases $(2)$ and $(3)$ are similar to those we have used for case $(1)$.
\end{proof}

Based on Lemma~\ref{lemma:type-3-a-ii-case4.1}, we can compute all Type-3$\alpha$ii $4$-cuts that satisfy $(4.1)$ of Lemma~\ref{lemma:type-3-a-ii-cases} as follows. First, it is sufficient to consider only those $w\neq r$ such that $M(w)$ has at least two children. Let $c_1$ and $c_2$ be the $\mathit{low1}$ and $\mathit{low2}$ children of $M(w)$. Let $\{(u,p(u)),(v,p(v)),(w,p(w)),(x,y)\}$ be a $4$-cut that satisfies $(4.1)$ of Lemma~\ref{lemma:type-3-a-ii-cases}, where $u$ and $v$ are both descendants of $w$, and $(x,y)$ is the back-edge in $B(w)\setminus(B(u)\sqcup B(v))$. Then there are six different possibilities:

\begin{enumerate}[label=(\arabic*)]
\item{$u$ and $v$ are descendants of $c_1$, and $x$ is a descendant of $c_2$.}
\item{$u$ and $x$ are descendants of $c_1$, and $v$ is a descendant of $c_2$.}
\item{$v$ and $x$ are descendants of $c_1$, and $u$ is a descendant of $c_2$.}
\item{$u$ and $v$ are descendants of $c_2$, and $x$ is a descendant of $c_1$.}
\item{$u$ and $x$ are descendants of $c_2$, and $v$ is a descendant of $c_1$.}
\item{$v$ and $x$ are descendants of $c_2$, and $u$ is a descendant of $c_1$.}
\end{enumerate}

Notice that cases $(2)$-$(3)$ and $(5)$-$(6)$ are equivalent from an algorithmic perspective, because the names of the variables do not matter (i.e., the names of $u$ and $v$ can be exchanged). Thus, the possible cases that we have to consider are reduced to four.

First, we may have that two of the vertices from $\{u,v,x\}$ are descendants of $c_1$, and the other is a descendant of $c_2$. In this case, we need to have computed $M(w,c_1')$ and $M(w,c_2')$, where $c_1'$ and $c_2'$ are the $\mathit{low1}$ and $\mathit{low2}$ children of $M(w,c_1)$. Furthermore, we need to have computed $M(w,c_2)$. Now, we may have that both $u$ and $v$ are descendants of $c_1$, and $x$ is a descendant of $c_2$. In this case, we have that $M(u)=M(w,c_1')$ and $M(v)=M(w,c_2')$ (or reversely), and $(x,y)=(M(w,c_2),l_1(M(w,c_2)))$. Otherwise, we have that both $u$ and $x$ are descendants of $c_1$, and $v$ is a descendant of $c_2$. Then, we have that either $M(u)=M(w,c_1')$ and $(x,y)=(M(w,c_2'),l_1(M(w,c_2')))$, or $M(u)=M(w,c_2')$ and $(x,y)=(M(w,c_1'),l_1(M(w,c_1')))$. In either case, we have $M(v)=M(w,c_2)$. On the other hand, we may have that two of the vertices from $\{u,v,x\}$ are descendants of $c_2$, and the other is a descendant of $c_1$. This case is treated similarly (we just reverse the roles of $c_1$ and $c_2$).
The process that we follow in order to compute all those $4$-cuts is shown in Algorithm~\ref{algorithm:type3-a-ii-4-1}. The proof of correctness and linear complexity is given in Proposition~\ref{proposition:algorithm:type3-a-ii-4-1}.

\noindent\\
\begin{algorithm}[H]
\caption{\textsf{Compute all Type-3$\alpha$ii $4$-cuts of the form $\{(u,p(u)),(v,p(v)),(w,p(w)),e\}$, where $w$ is a common ancestor of $\{u,v\}$, and $e$ satisfies $(4.1)$ of Lemma~\ref{lemma:type-3-a-ii-cases}.}}
\label{algorithm:type3-a-ii-4-1}
\LinesNumbered
\DontPrintSemicolon
\tcp{We consider the case in which two vertices from $\{u,v,x\}$ are descendants of the $\mathit{low1}$ child of $M(w)$ and the other is a descendant of the $\mathit{low2}$ child of $M(w)$, where $x$ is the higher endpoint of $e$; the other case is treated similarly, by reversing the roles of $c_1$ and $c_2$ below}
\ForEach{vertex $w\neq r$}{
  \lIf{$M(w)$ has less than two children}{\textbf{continue}}
  let $c_1$ and $c_2$ be the $\mathit{low1}$ and $\mathit{low2}$ children of $M(w)$\;
  \lIf{$M(w,c_1)=\bot$ \textbf{or} $M(w,c_2)=\bot$}{\textbf{continue}}
  \lIf{$M(w,c_1)$ has less than two children}{\textbf{continue}}
  let $c_1'$ and $c_2'$ be the $\mathit{low1}$ and $\mathit{low2}$ children of $M(w,c_1)$\;
  \lIf{$M(w,c_1')=\bot$ \textbf{or} $M(w,c_2')=\bot$}{\textbf{continue}}
  let $u$ be the lowest proper descendant of $w$ that has $M(u)=M(w,c_1')$\;
  \label{algorithm:type3-a-ii-4-1-u-1}
  let $v$ be the lowest proper descendant of $w$ that has $M(v)=M(w,c_2')$\;
  \label{algorithm:type3-a-ii-4-1-v-1}
  \If{$\mathit{high}(u)<w$ \textbf{and} $\mathit{high}(v)<w$ \textbf{and} $\mathit{bcount}(w)=\mathit{bcount}(u)+\mathit{bcount}(v)+1$}{
    mark $\{(u,p(u)),(v,p(v)),(w,p(w)),(M(w,c_2),l_1(M(w,c_2)))\}$ as a Type-3$\alpha$ii $4$-cut\;
    \label{algorithm:type3-a-ii-4-1-mark1}
  }
  let $u$ be the lowest proper descendant of $w$ that has $M(u)=M(w,c_1')$\;
  \label{algorithm:type3-a-ii-4-1-u-2}
  let $v$ be the lowest proper descendant of $w$ that has $M(v)=M(w,c_2)$\;
  \label{algorithm:type3-a-ii-4-1-v-2}
  \If{$\mathit{high}(u)<w$ \textbf{and} $\mathit{high}(v)<w$ \textbf{and} $\mathit{bcount}(w)=\mathit{bcount}(u)+\mathit{bcount}(v)+1$}{
    mark $\{(u,p(u)),(v,p(v)),(w,p(w)),(M(w,c_2'),l_1(M(w,c_2')))\}$ as a Type-3$\alpha$ii $4$-cut\;
    \label{algorithm:type3-a-ii-4-1-mark2}
  }
  let $u$ be the lowest proper descendant of $w$ that has $M(u)=M(w,c_2')$\;
  \label{algorithm:type3-a-ii-4-1-u-3}
  let $v$ be the lowest proper descendant of $w$ that has $M(v)=M(w,c_2)$\;
  \label{algorithm:type3-a-ii-4-1-v-3}
  \If{$\mathit{high}(u)<w$ \textbf{and} $\mathit{high}(v)<w$ \textbf{and} $\mathit{bcount}(w)=\mathit{bcount}(u)+\mathit{bcount}(v)+1$}{
    mark $\{(u,p(u)),(v,p(v)),(w,p(w)),(M(w,c_1'),l_1(M(w,c_1')))\}$ as a Type-3$\alpha$ii $4$-cut\;
    \label{algorithm:type3-a-ii-4-1-mark3}
  }
}
\end{algorithm}

\begin{proposition}
\label{proposition:algorithm:type3-a-ii-4-1}
Algorithm~\ref{algorithm:type3-a-ii-4-1} correctly computes all Type-3$\alpha$ii $4$-cuts of the form $\{(u,p(u)),(v,p(v)),(w,p(w)),e\}$, where $w$ is a common ancestor of $\{u,v\}$, and $e$ satisfies $(4.1)$ of Lemma~\ref{lemma:type-3-a-ii-cases}. Furthermore, it has a linear-time implementation. 
\end{proposition}
\begin{proof}
Let $C=\{(u,p(u)),(v,p(v)),(w,p(w)),e\}$ be a Type-3$\alpha$ii $4$-cut, where $w$ is a common ancestor of $\{u,v\}$, and $e$ satisfies $(4.1)$ of Lemma~\ref{lemma:type-3-a-ii-cases}. Let the higher endpoint of $e$ be $x$, and let us assume that two of $\{u,v,x\}$ are descendants of the $\mathit{low1}$ child $c_1$ of $M(w)$, and the other is a descendant of the $\mathit{low2}$ child $c_2$ of $M(w)$. (The other case is treated similarly, by reversing the roles of $c_1$ and $c_2$.) Let $c_1'$ be the $\mathit{low1}$ child of $M(w,c_1)$, and let $c_2'$ be the $\mathit{low2}$ child of $M(w,c_1)$. The possible cases here are: $(1)$ $u$ and $v$ are descendants of $c_1$, or $(2)$ $u$ and $x$ are descendants of $c_1$, or $(3)$ $v$ and $x$ are descendants of $c_1$. We note that case $(3)$ can be subsumed by case $(2)$ (by exchanging the names of $u$ and $v$), and thus we may ignore it. In case $(1)$, Lemma~\ref{lemma:type-3-a-ii-case4.1} implies (w.l.o.g.) that $M(u)=M(w,c_1')$, $M(v)=M(w,c_2')$ and $e=(M(w,c_2),l_1(M(w,c_2)))$. Then, by Lemma~\ref{lemma:type-3-a-ii-inference} we have that $u$ is the lowest proper descendant of $w$ that has $M(u)=M(w,c_1')$, and $v$ is the lowest proper descendant of $w$ that has $M(v)=M(w,c_2')$. Furthermore, according to Lemma~\ref{lemma:type-3-a-ii-criterion}, we have that $\mathit{bcount}(w)=\mathit{bcount}(u)+\mathit{bcount}(v)+1$, $\mathit{high}(u)<w$ and $\mathit{high}(v)<w$. Thus, $C$ will be marked in Line~\ref{algorithm:type3-a-ii-4-1-mark1}. In case $(2)$, Lemma~\ref{lemma:type-3-a-ii-case4.1} implies that either $M(u)=M(w,c_1')$, $e=(M(w,c_2'),l_1(M(w,c_2')))$ and $M(v)=M(w,c_2)$, or $M(u)=M(w,c_2')$, $e=(M(w,c_1'),l_1(M(w,c_1')))$ and $M(v)=M(w,c_2)$. Thus, the same argument as before implies that $C$ will be marked in Line~\ref{algorithm:type3-a-ii-4-1-mark2} or \ref{algorithm:type3-a-ii-4-1-mark3}, respectively.

Conversely, let $C=\{(u,p(u)),(v,p(v)),(w,p(w)),(x,y)\}$ be a $4$-element set that is marked in Line~\ref{algorithm:type3-a-ii-4-1-mark1}, or \ref{algorithm:type3-a-ii-4-1-mark2}, or \ref{algorithm:type3-a-ii-4-1-mark3}. Suppose first that $C$ is marked in Line~\ref{algorithm:type3-a-ii-4-1-mark1}. Then we have $(x,y)=(M(w,c_2),l_1(M(w,c_2)))$, $\mathit{bcount}(w)=\mathit{bcount}(u)+\mathit{bcount}(v)+1$, $\mathit{high}(u)<w$ and $\mathit{high}(v)<w$. Furthermore, we have $M(u)=M(w,c_1')$ and $M(v)=M(w,c_2')$, and therefore we can use a similar argument as in the proof of Lemma~\ref{lemma:w-v-u-not-related} in order to show that $u$ and $v$ are not related as ancestor and descendant. (The argument hinges on the fact that $c_1'$ and $c_2'$ are different children of the same vertex, and $w$ is an ancestor of both $u$ and $v$.) Thus, Lemma~\ref{lemma:type-3-a-ii-criterion} implies that there is a back-edge $e$ such that $B(w)=(B(u)\sqcup B(v))\sqcup\{e\}$. Since $M(w,c_2)\neq\bot$, we have that there is a back-edge $(x',y')\in B(w)$ such that $x'$ is a descendant of $c_2$. Thus, we have $(x',y')\notin B(u)\cup B(v)$ (because otherwise, we would have that $x'$ is a descendant of either $M(u)$ or $M(v)$, and therefore a descendant of either $M(w,c_1')$ or $M(w,c_2')$, and therefore a descendant of $c_1$). 
Thus, we have that $(x',y')=e$, and that this is the only back-edge in $B(w)$ that stems from $T(c_2)$. Thus, we have $e=(M(w,c_2),l_1(M(w,c_2)))$, and therefore $e=(x,y)$. This shows that $C$ is indeed a Type-3$\alpha$ii $4$-cut. With similar arguments we can show that, if $C$ is marked in Lines~\ref{algorithm:type3-a-ii-4-1-mark2} or \ref{algorithm:type3-a-ii-4-1-mark3}, then $C$ is a Type-3$\alpha$ii $4$-cut.

Now we will argue about the complexity of Algorithm~\ref{algorithm:type3-a-ii-4-1}. Notice that for every $w\neq r$ such that $M(w)$ has at least two children, we have to compute the values $M(w,c_1)$ and $M(w,c_2)$, where $c_1$ and $c_2$ are the $\mathit{low1}$ and $\mathit{low2}$ children of $M(w)$. According to Proposition~\ref{proposition:computing-M(v,c)}, these computations take linear time in total, for all such vertices $w$. Then, if $M(w,c_1)\neq\bot$ and $M(w,c_1)$ has at least two children, we have to compute the values $M(w,c_1')$ and $M(w,c_2')$, where $c_1'$ and $c_2'$ are the $\mathit{low1}$ and $\mathit{low2}$ children of $M(w,c_1)$. According to Proposition~\ref{proposition:computing-M(v,c)}, these computation take linear time in total, for all such vertices $w$. Finally, the vertices $u$ and $v$ in Lines~\ref{algorithm:type3-a-ii-4-1-u-1}, \ref{algorithm:type3-a-ii-4-1-v-1}, \ref{algorithm:type3-a-ii-4-1-u-2}, \ref{algorithm:type3-a-ii-4-1-v-2}, \ref{algorithm:type3-a-ii-4-1-u-3} and \ref{algorithm:type3-a-ii-4-1-v-3}, can be computed with Algorithm~\ref{algorithm:W-queries}, as explained e.g. in the proof of Proposition~\ref{proposition:algorithm:type3-a-ii-1}. According to Lemma~\ref{lemma:W-queries}, all these computations take $O(n)$ time in total. We conclude that Algorithm~\ref{algorithm:type3-a-ii-4-1} has a linear-time implementation. 
\end{proof}

Now we consider case $(4.2)$ of Lemma~\ref{lemma:type-3-a-ii-cases}.

\begin{lemma}
\label{lemma:type-3-a-ii-4.2}
Let case $(4.2)$ of Lemma~\ref{lemma:type-3-a-ii-cases} be true. Let $c_1$, $c_2$, and $c_3$, be the $\mathit{low1}$, $\mathit{low2}$, and $\mathit{low3}$ children of $M(w)$ (not necessarily in that order). Let us assume that $u$ is a descendant of $c_1$, $v$ is a descendant of $c_2$, and $x$ is a descendant of $c_3$. Then $M(u)=M(w,c_1)$, $M(v)=M(w,c_2)$, and $e=(M(w,c_3),l_1(M(w,c_3)))$.
\end{lemma}
\begin{proof}
Let $S_1=\{(x',y')\in B(w)\mid x' \mbox{ is a descendant of } c_1\}$. Then, $M(w,c_1)=M(S_1)$. Let $(x',y')$ be a back-edge in $B(u)$. Then $x'$ is a descendant of $u$, and therefore a descendant of $c_1$. Furthermore, $B(w)=(B(u)\sqcup B(v))\sqcup\{e\}$ implies that $(x',y')\in B(w)$. Thus, we have $(x',y')\in S_1$. This implies that $x'$ is a descendant of $M(S_1)$. Due to the generality of $(x',y')\in B(u)$, this implies that $M(u)$ is a descendant of $M(S_1)$. Conversely, let $(x',y')$ be a back-edge in $S_1$. Then $(x',y')\in B(w)$, and so $B(w)=(B(u)\sqcup B(v))\sqcup\{e\}$ implies that either $(x',y')\in B(u)$, or $(x',y')\in B(v)$, or $(x',y')=(x,y)$. Since $x'$ is a descendant of $c_1$, we have that $x'$ cannot be a descendant of either $c_2$ or $c_3$. Thus, the cases $(x',y')\in B(v)$ and $(x',y')=(x,y)$ are rejected. (Because $(x',y')\in B(v)$ would imply that $x'$ is a descendant of $v$, and therefore a descendant of $c_2$; and $x'=x$ would imply that $x'$ is a descendant of $c_3$.) Thus, we are left with the case $(x',y')\in B(u)$. This implies that $x'$ is a descendant of $M(u)$. Due to the generality of $(x',y')\in S_1$, this implies that $M(S_1)$ is a descendant of $M(u)$. Since $M(u)$ is a descendant of $M(S_1)$, this shows that $M(u)=M(S_1)$, and therefore $M(u)=M(w,c_1)$. Similarly, we can show that $M(v)=M(w,c_2)$.

Now let $S_3=\{(x',y')\in B(w)\mid x' \mbox{ is a descendant of } c_3\}$. Then, $M(w,c_3)=M(S_3)$. We obviously have $(x,y)\in S_3$, and so $x$ is a descendant of $M(S_3)$. Let us suppose, for the sake of contradiction, that there is a back-edge $(x',y')\in S_3$ such that $(x',y')\neq (x,y)$. Then we have that $(x',y')\in B(w)$, and so $B(w)=(B(u)\sqcup B(v))\sqcup\{e\}$ implies that either $(x',y')\in B(u)$, or $(x',y')\in B(v)$, or $(x',y')=(x,y)$. The last case is rejected by assumption. If $(x',y')\in B(u)$, then we have that $x'$ is a descendant of $u$, and so $u$ and $c_3$ are related as ancestor and descendant (since they have $x'$ as a common descendant). Since $u$ is not a descendant of $c_3$, we have that $u$ is a proper ancestor of $c_3$, and therefore an ancestor of $M(w)$. But this is impossible, since $u$ is a descendant of $c_1$. Thus, the case $(x',y')\in B(u)$ is rejected. Similarly, the case $(x',y')\in B(v)$ is also rejected. But then there are no viable options left, and so we have a contradiction. Thus, we have that $(x,y)$ is the only back-edge in $S_3$, and so $M(S_3)=x$. By Lemma~\ref{lemma:type-3-a-ii-cases}, we have that $e=(x,l_1(x))$.
\end{proof}

In order to compute all Type-3$\alpha$ii $4$-cuts that satisfy $(4.2)$ of Lemma~\ref{lemma:type-3-a-ii-cases}, we can apply the information provided by Lemma~\ref{lemma:type-3-a-ii-4.2} as follows.
First, we notice that we need to process only those vertices $w\neq r$ such that $M(w)$ has at least three children. Let $c_1$, $c_2$ and $c_3$ be the $\mathit{low1}$, the $\mathit{low2}$ and the $\mathit{low3}$ child of $M(w)$. We assume that we have $M(w,c_1)\neq\bot$, $M(w,c_2)\neq\bot$ and $M(w,c_3)\neq\bot$. First, we find the lowest proper descendant $u$ of $w$ that has $M(u)=M(w,c_1)$, and the lowest proper descendant $v$ of $w$ that has $M(v)=M(w,c_2)$ (this is according to Lemma~\ref{lemma:type-3-a-ii-inference}). Then we check whether $\mathit{high}(u)<w$, $\mathit{high}(v)<w$, and $\mathit{bcount}(w)=\mathit{bcount}(u)+\mathit{bcount}(v)+1$, in order to establish with the use of Lemma~\ref{lemma:type-3-a-ii-criterion} that we indeed have a Type-3$\alpha$ii $4$-cut. If that is the case, then we know that the back-edge of this $4$-cut is $(M(w,c_3),l_1(M(w,c_3)))$. Otherwise, we find the lowest proper descendant $u$ of $w$ that has $M(u)=M(w,c_1)$, and the lowest proper descendant $v$ of $w$ that has $M(v)=M(w,c_3)$, and we perform the same checks. The back-edge in this case is $(M(w,c_2),l_1(M(w,c_2)))$. Finally, we find the lowest proper descendant $u$ of $w$ that has $M(u)=M(w,c_2)$, and the lowest proper descendant $v$ of $w$ that has $M(v)=M(w,c_3)$. Again, we perform the same checks; the back-edge in this case is $(M(w,c_1),l_1(M(w,c_1)))$. The procedure for finding all those $4$-cuts is shown in Algorithm~\ref{algorithm:type3-a-ii-4-2}. The proof of correctness and linear complexity is given in Proposition~\ref{proposition:algorithm:type3-a-ii-4-2}.

\noindent\\
\begin{algorithm}[H]
\caption{\textsf{Compute all Type-3$\alpha$ii $4$-cuts of the form $\{(u,p(u)),(v,p(v)),(w,p(w)),e\}$, where $w$ is a common ancestor of $\{u,v\}$, and $e$ satisfies $(4.2)$ of Lemma~\ref{lemma:type-3-a-ii-cases}.}}
\label{algorithm:type3-a-ii-4-2}
\LinesNumbered
\DontPrintSemicolon
\ForEach{vertex $w\neq r$}{
  \lIf{$M(w)$ has less than three children}{\textbf{continue}}
  let $c_1$, $c_2$ and $c_3$ be the $\mathit{low1}$, $\mathit{low2}$ and $\mathit{low3}$ children of $M(w)$\;
  \lIf{either of $M(w,c_1)$, $M(w,c_2)$ or $M(w,c_3)$ is $\bot$}{\textbf{continue}}
  let $u$ be the lowest proper descendant of $w$ such that $M(u)=M(w,c_1)$\;
  \label{algorithm:type3-a-ii-4-2-u-1}
  let $v$ be the lowest proper descendant of $w$ such that $M(v)=M(w,c_2)$\;
  \label{algorithm:type3-a-ii-4-2-v-1}
  \If{$\mathit{high}(u)<w$ \textbf{and} $\mathit{high}(v)<w$ \textbf{and} $\mathit{bcount}(w)=\mathit{bcount}(u)+\mathit{bcount}(v)+1$}{
    mark $\{(u,p(u)),(v,p(v)),(w,p(w)),(M(w,c_3),l_1(M(w,c_3)))\}$ as a Type-3$\alpha$ii $4$-cut\;
    \label{algorithm:type3-a-ii-4-2-mark1}
  }
  let $u$ be the lowest proper descendant of $w$ such that $M(u)=M(w,c_1)$\;
  \label{algorithm:type3-a-ii-4-2-u-2}
  let $v$ be the lowest proper descendant of $w$ such that $M(v)=M(w,c_3)$\;
  \label{algorithm:type3-a-ii-4-2-v-2}
  \If{$\mathit{high}(u)<w$ \textbf{and} $\mathit{high}(v)<w$ \textbf{and} $\mathit{bcount}(w)=\mathit{bcount}(u)+\mathit{bcount}(v)+1$}{
    mark $\{(u,p(u)),(v,p(v)),(w,p(w)),(M(w,c_2),l_1(M(w,c_2)))\}$ as a Type-3$\alpha$ii $4$-cut\;
    \label{algorithm:type3-a-ii-4-2-mark2}
  }
  let $u$ be the lowest proper descendant of $w$ such that $M(u)=M(w,c_2)$\;
  \label{algorithm:type3-a-ii-4-2-u-3}
  let $v$ be the lowest proper descendant of $w$ such that $M(v)=M(w,c_3)$\;
  \label{algorithm:type3-a-ii-4-2-v-3}
  \If{$\mathit{high}(u)<w$ \textbf{and} $\mathit{high}(v)<w$ \textbf{and} $\mathit{bcount}(w)=\mathit{bcount}(u)+\mathit{bcount}(v)+1$}{
    mark $\{(u,p(u)),(v,p(v)),(w,p(w)),(M(w,c_1),l_1(M(w,c_1)))\}$ as a Type-3$\alpha$ii $4$-cut\;
    \label{algorithm:type3-a-ii-4-2-mark3}
  }
}
\end{algorithm}

\begin{proposition}
\label{proposition:algorithm:type3-a-ii-4-2}
Algorithm~\ref{algorithm:type3-a-ii-4-2} correctly computes all Type-3$\alpha$ii $4$-cuts of the form $\{(u,p(u)),(v,p(v)),(w,p(w)),e\}$, where $w$ is a common ancestor of $\{u,v\}$, and $e$ satisfies $(4.2)$ of Lemma~\ref{lemma:type-3-a-ii-cases}. Furthermore, it has a linear-time implementation. 
\end{proposition}
\begin{proof}
Let $C=\{(u,p(u)),(v,p(v)),(w,p(w)),e\}$ be a Type-3$\alpha$ii $4$-cut, where $w$ is a common ancestor of $\{u,v\}$, and $e$ satisfies $(4.2)$ of Lemma~\ref{lemma:type-3-a-ii-cases}. Let $x$ be the higher endpoint of $e$, and let $c_1$, $c_2$ and $c_3$ be the $\mathit{low1}$, $\mathit{low2}$ and $\mathit{low3}$ children of $M(w)$, respectively. Let us suppose first that $u$ is a descendant of $c_1$, $v$ is a descendant of $c_2$, and $x$ is a descendant of $c_3$. Then Lemma~\ref{lemma:type-3-a-ii-4.2} implies that $M(u)=M(w,c_1)$, $M(v)=M(w,c_2)$ and $e=(M(w,c_3),l_1(M(w,c_3)))$. Then, Lemma~\ref{lemma:type-3-a-ii-inference} implies that $u$ is the lowest proper descendant of $w$ such that $M(u)=M(w,c_1)$, and $v$ is the lowest proper descendant of $w$ such that $M(v)=M(w,c_2)$. Lemma~\ref{lemma:type-3-a-ii-criterion} implies that $\mathit{bcount}(w)=\mathit{bcount}(u)+\mathit{bcount}(v)+1$, $\mathit{high}(u)<w$ and $\mathit{high}(v)<w$. Thus, $C$ will be marked in Line~\ref{algorithm:type3-a-ii-4-2-mark1}. Similarly, if we assume that $u$ is a descendant of $c_1$, $v$ is a descendant of $c_3$, and $x$ is a descendant of $c_2$, or that $u$ is a descendant of $c_2$, $v$ is a descendant of $c_3$, and $x$ is a descendant of $c_1$, then we have that $C$ will be marked in Line~\ref{algorithm:type3-a-ii-4-2-mark2}, or \ref{algorithm:type3-a-ii-4-2-mark3}, respectively. (The other cases that we have tacitly ignored, e.g., the case where $u$ is a descendant of $c_2$ and $v$ is a descendant of $c_1$, are basically subsumed in the cases that we considered; to see this, just exchange the names of the variables $u$ and $v$.)

Conversely, let $C=\{(u,p(u)),(v,p(v)),(w,p(w)),(x,y)\}$ be a $4$-element set that is marked in Line~\ref{algorithm:type3-a-ii-4-2-mark1}, or \ref{algorithm:type3-a-ii-4-2-mark2}, or \ref{algorithm:type3-a-ii-4-2-mark3}. Let us suppose first that $C$ is marked in Line~\ref{algorithm:type3-a-ii-4-2-mark1}. Then we have $(x,y)=(M(w,c_3),l_1(M(w,c_3)))$, $\mathit{bcount}(w)=\mathit{bcount}(u)+\mathit{bcount}(v)+1$, $\mathit{high}(u)<w$ and $\mathit{high}(v)<w$. Furthermore, since $M(u)=M(w,c_1)$ and $M(v)=M(w,c_2)$, and $c_1,c_2$ are different children of $M(w)$, Lemma~\ref{lemma:w-v-u-not-related} implies that $u$ and $v$ are not related as ancestor and descendant. Thus, Lemma~\ref{lemma:type-3-a-ii-criterion} implies that there is a back-edge $e$ such that $B(w)=(B(u)\sqcup B(v))\sqcup\{e\}$. Then, since $M(w,c_3)\neq\bot$, we have that there is a back-edge $(x',y')\in B(w)$ such that $x'$ is a descendant of $c_3$. Then, we have $(x',y')\notin B(u)\cup B(v)$ (because otherwise, we would have that $x'$ is a descendant of either $M(u)$ or $M(v)$, and therefore a descendant of either $M(w,c_1)$ or $M(w,c_2)$, and therefore a descendant of either $c_1$ or $c_2$, which is impossible). This implies that $(x',y')=e$, and that $(x',y')$ is the only back-edge in $B(w)$ that stems from $T(c_3)$. Thus, we have that $e=(M(w,c_3),l_1(M(w,c_3)))$, and therefore $e=(x,y)$. This shows that $C$ is indeed a Type-3$\alpha$ii $4$-cut. Similarly, if we have that $C$ is marked in Line~\ref{algorithm:type3-a-ii-4-2-mark2} or \ref{algorithm:type3-a-ii-4-2-mark3}, then we can use a similar argument in order to show that $C$ is a Type-3$\alpha$ii $4$-cut.

Now we will argue about the complexity of Algorithm~\ref{algorithm:type3-a-ii-4-2}. According to Proposition~\ref{proposition:computing-M(v,c)}, we can compute the values $M(w,c_1)$, $M(w,c_2)$ and $M(w,c_3)$ in linear time in total, for every vertex $w\neq r$ such that $M(w)$ has at least three children, where $c_1$, $c_2$ and $c_3$ are the $\mathit{low1}$, $\mathit{low2}$ and $\mathit{low3}$ children of $M(w)$, respectively. The values $u$ and $v$ in Lines~\ref{algorithm:type3-a-ii-4-2-u-1}, \ref{algorithm:type3-a-ii-4-2-v-1}, \ref{algorithm:type3-a-ii-4-2-u-2}, \ref{algorithm:type3-a-ii-4-2-v-2}, \ref{algorithm:type3-a-ii-4-2-u-3} and \ref{algorithm:type3-a-ii-4-2-v-3}, can be computed with Algorithm~\ref{algorithm:W-queries}, as explained e.g. in the proof of Proposition~\ref{proposition:algorithm:type3-a-ii-1}. According to Lemma~\ref{lemma:W-queries}, all these computations take $O(n)$ time in total.
We conclude that Algorithm~\ref{algorithm:type3-a-ii-4-2} runs in linear time.
\end{proof}

\section{Computing Type-3$\beta$ $4$-cuts}
\label{section:type-3b}

Throughout this section, we assume that $G$ is a $3$-edge-connected graph with $n$ vertices and $m$ edges. All graph-related elements (e.g., vertices, edges, cuts, etc.) refer to $G$. Furthermore, we assume that we have computed a DFS-tree $T$ of $G$ rooted at a vertex $r$.

\begin{lemma}
\label{lemma:type-3b-cases}
Let $u,v,w$ be three vertices $\neq r$ such that $w$ is proper ancestor of $v$ and $v$ is a proper ancestor of $u$, and let $e$ be a back-edge. Then $C'=\{(u,p(u)),(v,p(v)),(w,p(w)),e\}$ is a $4$-cut if and only if one of the following is true. (See Figure~\ref{figure:type3b}.)
\begin{enumerate}[label=(\arabic*)]
\item{$e\in B(u)\cap B(v)\cap B(w)$ and $B(v)\setminus\{e\}=(B(u)\setminus\{e\})\sqcup(B(w)\setminus\{e\})$}
\item{$e\in B(w)$, $e\notin B(v)\cup B(u)$, and $B(v)=B(u)\sqcup(B(w)\setminus\{e\})$}
\item{$e\in B(u)$, $e\notin B(v)\cup B(w)$, and $B(v)=(B(u)\setminus\{e\})\sqcup B(w)$}
\item{$e\in B(v)$ and $B(v)=(B(u)\sqcup B(w))\sqcup\{e\}$}
\end{enumerate}
\end{lemma}
\begin{proof}
($\Rightarrow$) Consider the parts $A=T(u)$, $B=T(v)\setminus T(u)$, $C=T(w)\setminus T(v)$, and $D=T(r)\setminus T(w)$. Observe that every one of those parts remains connected in $G\setminus \{(u,p(u)),(v,p(v)),(w,p(w))\}$. Since $C'$ is a $4$-cut, by Lemma~\ref{lemma:type3-4cuts} we have that $e\in B(u)\cup B(v)\cup B(w)$.

Let us suppose, first, that $e\in B(u)\cap B(v)\cap B(w)$. This means that $e$ connects $A$ and $D$. Thus, since $C'$ is a $4$-cut, we have that $A$ and $D$ must be disconnected in $G\setminus C'$, and so $e$ is the unique back-edge that connects $A$ and $D$. Furthermore, since $e$ connects $A$ and $D$, notice that there is no back-edge from $A$ to $B$, or from $B$ to $C$, or from $C$ to $D$, for otherwise $u$ would be connected with $p(u)$, or $v$ would be connected with $p(v)$, or $w$ would be connected with $p(w)$, respectively, in $G\setminus C'$, in contradiction to the fact that $C'$ is a $4$-cut of $G$. 
Let $e'$ be a back-edge in $B(v)\setminus\{e\}$. Then this can be a back-edge from $A$ to $C$, or from $B$ to $D$. The first case implies that $e'$ is in $B(u)$, and the second case implies that $e'$ is in $B(w)$. This shows that $B(v)\setminus\{e\}\subseteq B(u)\cup B(w)$, which implies that $B(v)\setminus\{e\}\subseteq (B(u)\setminus\{e\})\cup(B(w)\setminus\{e\})$. Conversely, let $e'$ be a back-edge in $(B(u)\cup B(w))\setminus\{e\}$. Then this is either a back-edge from $A$ to $C$ (if $e'\in B(u)\setminus\{e\}$), or from $B$ to $D$ (if $e'\in B(w)\setminus\{e\}$). Thus we have that $(B(u)\cup B(w))\setminus\{e\}\subseteq B(v)$, which implies that $(B(u)\setminus\{e\})\cup(B(w)\setminus\{e\})\subseteq B(v)\setminus\{e\}$. We infer that $B(v)\setminus\{e\}=(B(u)\setminus\{e\})\cup(B(w)\setminus\{e\})$.
Now let $e'$ be a back-edge in $B(u)\setminus \{e\}$. Then this can only be a back-edge from $A$ to $C$. Thus, $e'$ cannot be in $B(w)\setminus \{e\}$, because all the back-edges in $B(w)$ have their lower endpoint in $D$. Thus we have $(B(u)\setminus\{e\})\cap(B(w)\setminus\{e\})=\emptyset$, and so it is proper to write $B(v)\setminus\{e\}=(B(u)\setminus\{e\})\sqcup(B(w)\setminus\{e\})$ (case $(1)$). 

From now on, let us suppose that $e\notin B(u)\cap B(v)\cap B(w)$. We will show that $e$ belongs to exactly one of $B(u)$, $B(v)$, or $B(w)$. So let us suppose, for the sake of contradiction, that this is not true. Then there are three possible cases to consider: either $(I)$ $e\in B(u)\cap B(v)$ and $e\notin B(w)$, or $(II)$ $e\in B(u)\cap B(w)$ and $e\notin B(v)$, or $(III)$ $e\in B(v)\cap B(w)$ and $e\notin B(u)$. 
Suppose that $(I)$ is true. Then $e$ is either from $A$ to $C$, or from $A$ to $D$. The second case is rejected since $e\notin B(w)$. Thus, $e$ connects $A$ and $C$. Since $C'$ is a $4$-cut of $G$, we have that $e$ is the only back-edge from $A$ to $C$. Furthermore, we have that there are no back-edges from $B$ to $C$, or from $C$ to $D$. Thus, $C$ becomes disconnected from the rest of the graph in $G\setminus\{(v,p(v)),(w,p(w)),e\}$, in contradiction to the fact that $C'$ is a $4$-cut of $G$. This shows that $(I)$ cannot be true.      
Now suppose that $(II)$ is true. Since $e\in B(u)\cap B(w)$, we have that $e$ must be a back-edge from $A$ to $D$. But this contradicts $e\notin B(v)$. Thus, $(II)$ cannot be true.
Finally, suppose that $(III)$ is true. Then $e\in B(v)\cap B(w)$ implies that $e$ either connects $A$ and $D$, or $B$ and $D$. The first case is rejected, since $e\notin B(u)$. Thus, $e$ connects $B$ and $D$. Since $C'$ is a $4$-cut of $G$, we have that $e$ is the only back-edge from $B$ to $D$. Furthermore, we have that there are no back-edges from $A$ to $B$, or from $B$ to $C$. Thus, $B$ becomes disconnected from the rest of the graph in $G\setminus\{(u,p(u)),(v,p(v)),e\}$, in contradiction to the fact that $C'$ is a $4$-cut of $G$. This shows that $(III)$ cannot be true. Since all cases $(I)$-$(III)$ lead to a contradiction, we have that $e$ belongs to exactly one of $B(u)$, $B(v)$, or $B(w)$.

Now suppose that $e\in B(w)$. Since $e\notin B(u)\cup B(v)$, we have that $e$ connects $C$ and $D$. Since $C'$ is a $4$-cut, we have that $e$ is the only back-edge that connects $C$ and $D$. Furthermore, there are no back-edges from $A$ to $B$, or from $B$ to $C$. Now let $e'$ be a back-edge in $B(v)$. Then $e'$ is either a back-edge in $B(u)$, or it connects $B$ and $D$. Thus we have that $e'\in B(u)\cup (B(w)\setminus\{e\})$. This shows that $B(v)\subseteq B(u)\cup (B(w)\setminus\{e\})$. Conversely, let $e'$ be a back-edge in $B(u)\cup (B(w)\setminus\{e\})$. If $e'\in B(u)$, then $e'$ is also in $B(v)$, because there is no back-edge from $A$ to $B$. And if $e'\in B(w)\setminus\{e\}$, we have that $e'\in B(v)$ because $e$ is the only back-edge from $C$ to $D$. This shows that $B(u)\cup (B(w)\setminus\{e\})\subseteq B(v)$, and so we have $B(v)=B(u)\cup (B(w)\setminus\{e\})$.
Let us suppose, for the sake of contradiction, that there is a back-edge in $B(u)\cap B(w)$. Then this is a back-edge from $A$ to $D$. Since there is no back-edge from $A$ to $B$, we have that $B(u)\subseteq B(v)$. Since the graph is $3$-edge-connected, we have that $B(u)\neq B(v)$. Thus, there is a back-edge in $B(v)\setminus B(u)$. Since there are no back-edges from $B$ to $C$, this must be a back-edge from $B$ to $D$. Since $e$ is the unique back-edge from $C$ to $D$, we have that there must be at least one back-edge from $A$ to $C$, because otherwise $C$ becomes disconnected from the rest of the graph in $G\setminus\{(v,p(v)),(w,p(w)),e\}$. But the existence of back-edges from $A$ to $D$, from $B$ to $D$, and from $A$ to $C$, (and the fact that $e$ is a back-edge from $C$ to $D$), implies that all parts $A$-$D$ remain connected in $G\setminus C'$, a contradiction. 
This shows that $B(u)\cap B(w)=\emptyset$, and therefore it is correct to write $B(v)=B(u)\sqcup (B(w)\setminus\{e\})$ (case $(2)$).

Now suppose that $e\in B(u)$. Since $e\notin B(v)\cup B(w)$, we have that $e$ connects $A$ and $B$. Since $C'$ is a $4$-cut, we have that $e$ is the only back-edge that connects $A$ and $B$. Furthermore, there are no back-edges from $B$ to $C$, or from $C$ to $D$. Now let $e'$ be a back-edge in $B(v)$. Then $e'$ is either a back-edge in $B(u)$, or a back-edge from $B$ to $C$, or a back-edge from $B$ to $D$. The case that $e'$ connects $B$ and $C$ is forbidden, and so $e'$ is either in $B(u)$ or in $B(w)$. This shows that $B(v)\subseteq B(u)\cup B(w)$, which can be strengthened to $B(v)\subseteq (B(u)\setminus\{e\})\cup B(w)$, since $e\notin B(v)$. Conversely, let $e'$ be a back-edge in $(B(u)\setminus\{e\})\cup B(w)$. If $e'\in B(u)\setminus\{e\}$, then $e'$ is in $B(v)$, because $e$ is the only back-edge from $A$ to $B$. And if $e'\in B(w)$, then we have $e'\in B(v)$, because there are no back-edges from $C$ to $D$. This shows that $(B(u)\setminus\{e\})\cup B(w) \subseteq B(v)$, and thus we have $B(v)=(B(u)\setminus\{e\})\cup B(w)$. 
Now let us suppose, for the sake of contradiction, that there is a back-edge in $B(u)\cap B(w)$. Then this is a back-edge from $A$ to $D$. Since $e$ the only back-edge from $A$ to $B$ and there is no back-edge from $B$ to $C$, we have that there must exist a back-edge from $B$ to $D$, because otherwise $B$ becomes disconnected from the rest of the graph in $G\setminus\{(u,p(u)),(v,p(v)),e\}$, contradicting the fact that $C'$ is a $4$-cut of $G$. Since the graph is $3$-edge-connected, we have that $B(w)\neq B(v)$. Thus, since there are no back-edges from $C$ to $D$ or from $B$ to $C$, we have that there must exist a back-edge from $A$ to $C$. But now, since there is a back-edge from $A$ to $D$, a back-edge from $B$ to $D$, and a back-edge from $A$ to $C$, (and $e$ is a back-edge from $A$ to $B$), we have that all parts $A$-$D$ remain connected in $G\setminus C'$, a contradiction. 
Thus we have that $B(u)\cap B(w)=\emptyset$, and so it correct to write $B(v)=(B(u)\setminus\{e\})\sqcup B(w)$ (case $(3)$).

Finally, let us suppose that $e\in B(v)$. Since $e\notin B(u)\cup B(w)$, we have that $e$ connects $B$ and $C$. Since $C'$ is a $4$-cut, we have that $e$ is the only back-edge that connects $B$ and $C$. Furthermore, there are no back-edges from $A$ to $B$, or from $C$ to $D$. Now let $e'$ be a back-edge in $B(v)\setminus\{e\}$. Then $e'$ is either a back-edge in $B(u)$, or a back-edge from $B$ to $C$, or a back-edge from $B$ to $D$. The case that $e'$ is a back-edge from $B$ to $C$ is rejected, since $e$ is the only back-edge with this property. Thus we have that $e'$ is either in $B(u)$ or in $B(w)$. This shows that $e'\in B(u)\cup B(w)$, and so we have $B(v)\setminus\{e\}\subseteq B(u)\cup B(w)$. This is equivalent to $B(v)\subseteq (B(u)\cup B(w))\sqcup\{e\}$, since $e\notin B(u)\cup B(w)$. Conversely, let $e'$ be a back-edge in $B(u)\cup B(w)$. If $e'\in B(u)$, then, since there is no back-edge from $A$ to $B$, we have that $e'\in B(v)$. And if $e'\in B(w)$, then, since there is no back-edge from $C$ to $D$, we have that $e'\in B(v)$. This shows that $B(u)\cup B(w)\subseteq B(v)$, which can be strengthened to $(B(u)\cup B(w))\sqcup\{e\}\subseteq B(v)$, since $e\in B(v)$ and $e\notin B(u)\cup B(w)$. Thus we have that $B(v)=(B(u)\cup B(w))\sqcup\{e\}$.
Now let us suppose, for the sake of contradiction, that $B(u)\cap B(w)\neq\emptyset$. Then there exists at least one back-edge from $A$ to $D$. Since there is no back-edge from $A$ to $B$, and $e$ is the only back-edge from $B$ to $C$, we have that there must exist a back-edge from $B$ to $D$, because otherwise $B$ would become disconnected from the rest of the graph in $G\setminus\{(u,p(u)),(v,p(v)),e\}$, in contradiction to the fact that $C'$ is a $4$-cut of $G$. Furthermore, since there is no back-edge from $C$ to $D$, and $e$ is the only back-edge from $B$ to $C$, we have that there must exist a back-edge from $A$ to $C$, because otherwise $C$ would become disconnected from the rest of the graph in $G\setminus\{(v,p(v)),(w,p(w)),e\}$, in contradiction to the fact that $C'$ is a $4$-cut of $G$. But now, since there is a back-edge from $A$ to $D$, and a back-edge from $B$ to $D$, and a back-edge from $A$ to $C$, (and $e$ is a back-edge from $B$ to $C$), we have that all parts $A$-$D$ remain connected in $G\setminus C'$, a contradiction. Thus we have that $B(u)\cap B(w)=\emptyset$, and so it is correct to write $B(v)=(B(u)\sqcup B(w))\sqcup\{e\}$ (case $(4)$).

($\Leftarrow$) We have to show that $G\setminus C'$ is disconnected in every one of cases $(1)$-$(4)$, but $G\setminus C''$ is connected for every proper subset $C''$ of $C'$. Since the graph is $3$-edge-connected, it is sufficient to prove that no $3$-element subset of $C'$ is a $3$-cut of $G$. Furthermore, since the graph is $3$-edge-connected, we have that $|B(u)|>1$, $|B(v)|>1$ and $|B(w)|>1$. Consider the parts $A=T(u)$, $B=T(v)\setminus T(u)$, $C=T(w)\setminus T(v)$, and $D=T(r)\setminus T(w)$. Observe that every one of those parts remains connected in $G\setminus \{(u,p(u)),(v,p(v)),(w,p(w))\}$. Notice that there are six different types of back-edges that connect the parts $A$-$D$: back-edges from $A$ to $B$, from $A$ to $C$, from $A$ to $D$, from $B$ to $C$, from $B$ to $D$, and from $C$ to $D$. Notice that such a back-edge is contained either in $B(u)$, or in $B(v)$, or in $B(w)$ (or in intersections or unions between those sets).

Suppose first that $(1)$ is true. Then, since $e\in B(u)\cap B(w)$, we have that $e$ connects $A$ and $D$. The disjointness of the union in $B(v)\setminus\{e\}=(B(u)\setminus\{e\})\sqcup(B(w)\setminus\{e\})$ implies that $e$ is the only back-edge in $B(u)\cap B(w)$, and so it is the only back-edge from $A$ to $D$. Let $e'$ be a back-edge in $B(u)\setminus\{e\}$. Then, since $e'\in B(u)$, we have that $e'$ is either from $A$ to $B$, or from $A$ to $C$, or from $A$ to $D$. The latter case is rejected, since we have that $e$ is the only back-edge with this property. The case that $e'$ is from $A$ to $B$ is rejected, because $B(u)\setminus\{e\}\subseteq B(v)$. Thus, $e'$ is a back-edge from $A$ to $C$. Let $e''$ be a back-edge in $B(w)\setminus\{e\}$. Then, since $e''\in B(w)$, we have that $e''$ is either a back-edge from $A$ to $D$, or from $B$ to $D$, or from $C$ to $D$. The case that $e''$ is from $A$ to $D$ is rejected, because $e$ is the only back-edge with this property. And the case that $e''$ is from $C$ to $D$ is rejected, since we have $B(w)\setminus\{e\}\subseteq B(v)$. Thus, $e''$ is a back-edge from $B$ to $D$. Since $B(v)\setminus\{e\}=(B(u)\setminus\{e\})\sqcup(B(w)\setminus\{e\})$ and $e\in B(u)\cap B(v)\cap B(w)$, we have exhausted all possibilities for the back-edges that connect the parts $A$-$D$.

Now we have collected enough information to see why $C'$ is a $4$-cut of $G$. The only different types of back-edges that connect the parts $A$-$D$ are: at least one back-edge $e'$ from $A$ to $C$, at least one back-edge $e''$ from $B$ to $D$, and $e$ is the unique back-edge from $A$ to $D$. In this situation, observe that, if we remove $C'$ from $G$, then $G$ becomes disconnected into the connected components $A\cup C$ and $B\cup D$. If we remove $\{(u,p(u)),(v,p(v)),e\}$ from $G$, then there remains a path $A\xrightarrow{e'} C\xrightarrow{(w, p(w))} D$, and so this is not a $3$-cut of $G$ (because the endpoints of $e$ remain connected). If we remove $\{(u,p(u)),(w,p(w)),e\}$ from $G$, then there remains a path $A\xrightarrow{e'} C\xrightarrow{(p(v),v)} B\xrightarrow{e''} D$, and so this is not a $3$-cut of $G$ (because the endpoints of $e$ remain connected). If we remove $\{(v,p(v)),(w,p(w)),e\}$ from $G$, then there remains a path $A\xrightarrow{(u,p(u))} B\xrightarrow{e''} D$, and so this is not a $3$-cut of $G$ (because the endpoints of $e$ remain connected). Finally, if we remove $\{(u,p(u)),(v,p(v)),(w,p(w))\}$ from $G$, then there remains a path $C\xrightarrow{e'} A\xrightarrow{e} D\xrightarrow{e''} B$, and so this is not a $3$-cut of $G$ (because all parts $A$-$D$ remain connected). Thus, we have that $C'$ is a $4$-cut of $G$.

Now suppose that $(2)$ is true. Then, since $e\in B(w)$ and $e\notin B(u)\cup B(v)$, we have that $e$ connects $C$ and $D$. $B(v)=B(u)\sqcup(B(w)\setminus\{e\})$ implies that $B(w)\setminus\{e\}\subseteq B(v)$, and therefore $e$ is the unique back-edge from $C$ to $D$ (because all other back-edges in $B(w)$ lie in $B(v)$). Notice that $B(u)\cap (B(w)\setminus\{e\})=\emptyset$ and $e\notin B(u)$, implies that there is no back-edge in $B(u)\cap B(w)$. Let $e'$ be a back-edge in $B(u)$. Then $e'$ is either from $A$ to $B$, or from $A$ to $C$, or from $A$ to $D$. The latter case is rejected, because there is no back-edge in $B(u)\cap B(w)$. The case that $e'$ is from $A$ to $B$ is also rejected, because $e'\in B(v)$. Thus, $e'$ is a back-edge from $A$ to $C$. Let $e''$ be a back-edge in $B(w)\setminus\{e\}$. Then $e''$ is either a back-edge from $A$ to $D$, or from $B$ to $D$, or from $C$ to $D$. The latter case is rejected, because $e$ is the only back-edge with this property. The case that $e''$ is from $A$ to $D$ is rejected, since there is no back-edge in $B(u)\cap B(w)$. Thus, $e''$ is a back-edge from $B$ to $D$. Since $B(v)=B(u)\sqcup(B(w)\setminus\{e\}$, we have exhausted all different combinations for back-edges that connect the parts $A$-$D$.

Now we have collected enough information to see why $C'$ is a $4$-cut of $G$. The only different types of back-edges that connect the parts $A$-$D$ are: at least one back-edge $e'$ from $A$ to $C$, at least one back-edge $e''$ from $B$ to $D$, and $e$ is the unique back-edge from $C$ to $D$. In this situation, observe that, if we remove $C'$ from $G$, then $G$ becomes disconnected into the connected components $A\cup C$ and $B\cup D$. Then we can argue as before, in order to show that no $3$-element subset of $C'$ is a $3$-cut of $G$. Thus, we have that $C'$ is a $4$-cut of $G$.

Now suppose that $(3)$ is true. Then, since $e\in B(u)$ and $e\notin B(v)\cup B(w)$, we have that $e$ connects $A$ and $B$. $B(v)=(B(u)\setminus\{e\})\sqcup B(w)$ implies that $B(u)\setminus\{e\}\subseteq B(v)$, and therefore $e$ is the unique back-edge from $A$ to $B$ (because all other back-edges in $B(u)$ lie in $B(v)$). Notice that $(B(u)\setminus\{e\})\cap B(w)=\emptyset$ and $e\notin B(w)$, implies that there is no back-edge in $B(u)\cap B(w)$. Let $e'$ be a back-edge in $B(u)\setminus\{e\}$. Then $e'$ is either from $A$ to $B$, or from $A$ to $C$, or from $A$ to $D$. The latter case is rejected, because there is no back-edge in $B(u)\cap B(w)$. The case that $e'$ is from $A$ to $B$ is also rejected, because $e$ is the unique back-edge with this property. Thus, $e'$ is a back-edge from $A$ to $C$. Let $e''$ be a back-edge in $B(w)$. Then $e''$ is either a back-edge from $A$ to $D$, or from $B$ to $D$, or from $C$ to $D$. The latter case is rejected, because we have $B(w)\subseteq B(v)$, and therefore all back-edges in $B(w)$ lie in $B(v)$. The case that $e''$ is from $A$ to $D$ is rejected, since there is no back-edge in $B(u)\cap B(w)$. Thus, $e''$ is a back-edge from $B$ to $D$. Since $B(v)=(B(u)\setminus\{e\})\sqcup B(w)$, we have exhausted all different combinations for back-edges that connect the parts $A$-$D$.

Now we have collected enough information to see why $C'$ is a $4$-cut of $G$. The only different types of back-edges that connect the parts $A$-$D$ are: at least one back-edge $e'$ from $A$ to $C$, at least one back-edge $e''$ from $B$ to $D$, and $e$ is the unique back-edge from $A$ to $B$. In this situation, observe that, if we remove $C'$ from $G$, then $G$ becomes disconnected into the connected components $A\cup C$ and $B\cup D$. Then we can argue as before, in order to show that no $3$-element subset of $C'$ is a $3$-cut of $G$. Thus, we have that $C'$ is a $4$-cut of $G$.

Finally, suppose that $(4)$ is true. Then, since $e\in B(v)$ and $e\notin B(u)\cup B(w)$, we have that $e$ connects $B$ and $C$. Since $B(v)=(B(u)\sqcup B(w))\sqcup\{e\}$, we have that all back-edges in $B(v)$, except $e$, are either in $B(u)$ or in $B(w)$. Therefore, $e$ is the unique back-edge from $B$ to $C$. Since $B(u)\cap B(w)=\emptyset$, we have that there is no back-edge from $A$ to $D$. Let $e'$ be a back-edge in $B(u)$. Then $e'$ is either from $A$ to $B$, or from $A$ to $C$, or from $A$ to $D$. The latter case is rejected, because there is no back-edge in $B(u)\cap B(w)$. The case that $e'$ is from $A$ to $B$ is also rejected, because $e\in B(v)$. Thus, $e'$ is a back-edge from $A$ to $C$. Let $e''$ be a back-edge in $B(w)$. Then $e''$ is either a back-edge from $A$ to $D$, or from $B$ to $D$, or from $C$ to $D$. The latter case is rejected, because we have $B(w)\subseteq B(v)$. The case that $e''$ is from $A$ to $D$ is rejected, since there is no back-edge in $B(u)\cap B(w)$. Thus, $e''$ is a back-edge from $B$ to $D$. Since $B(v)=(B(v)\sqcup B(w))\sqcup\{e\}$, we have exhausted all different combinations for back-edges that connect the parts $A$-$D$.

Now we have collected enough information to see why $C'$ is a $4$-cut of $G$. The only different types of back-edges that connect the parts $A$-$D$ are: at least one back-edge $e'$ from $A$ to $C$, at least one back-edge $e''$ from $B$ to $D$, and $e$ is the unique back-edge from $B$ to $C$. In this situation, observe that, if we remove $C'$ from $G$, then $G$ becomes disconnected into the connected components $A\cup C$ and $B\cup D$. Then we can argue as before, in order to show that no $3$-element subset of $C'$ is a $3$-cut of $G$. Thus, we have that $C'$ is a $4$-cut of $G$.
\end{proof}

We distinguish two types of Type-3$\beta$ $4$-cuts -- Type-3$\beta i$ and Type-3$\beta ii$ -- depending on whether the $M$ points of $v$ and $w$ (after the removal of $e$) coincide. The Type-3$\beta i$ $4$-cuts have the following classification, corresponding to the cases in Lemma~\ref{lemma:type-3b-cases}.
\begin{enumerate}
\small
\item{$e\in B(u)\cap B(v)\cap B(w)$, $B(v)\setminus\{e\}=(B(u)\setminus\{e\})\sqcup(B(w)\setminus\{e\})$ and $M(B(v)\setminus\{e\})\neq M(B(w)\setminus\{e\})$}
\item{$e\in B(w)$, $e\notin B(u)\cup B(v)$, $B(v)=B(u)\sqcup(B(w)\setminus\{e\})$ and $M(v)\neq M(B(w)\setminus\{e\})$}
\item{$e\in B(u)$, $e\notin B(v)\cup B(w)$, $B(v)=(B(u)\setminus\{e\})\sqcup B(w)$ and $M(v)\neq M(w)$}
\item{$e\in B(v)$, $B(v)=(B(u)\sqcup B(w))\sqcup\{e\}$ and $M(B(v)\setminus\{e\})\neq M(w)$}
\end{enumerate}

The Type-3$\beta ii$ $4$-cuts are classified as the Type-3$\beta i$ $4$-cuts above, the only difference being that the inequalities are replaced with equalities in each case.
 
The Type-3$\beta i$ $4$-cuts are easier to handle. This is because, given $v$, we have enough information to find $u$ and $w$ relatively easily. On the other hand, for Type-3$\beta ii$ $4$-cuts we have to apply more sophisticated methods, because it is not straightforward what are the possible $u$ and $w$ that may induce with $v$ a $4$-cut of this type. The general idea to compute those $4$-cuts is basically to calculate a set of candidates $u$, for each $v$, that may induce with $v$ and a $w$ a $4$-cut of the desired kind. The search space for $w$, given $v$, is known, but not computed (in most cases). Then, given $v$ and $u$, we can relatively easily determine a $w$ that may induce a desired $4$-cut with $u$ and $v$. Then we apply a criterion in order to check that we indeed get a $4$-cut. The remaining $4$-cuts (if it is impossible to compute all of them in linear time), are implied from the collection we have computed, plus that returned by Algorithm~\ref{algorithm:type2-2}.

\subsection{Type-3$\beta i$ $4$-cuts}

\subsubsection{Case $(1)$ of Lemma~\ref{lemma:type-3b-cases}}

\begin{lemma}
\label{lemma:type-3-b-i-1-info}
Let $u,v,w$ be three vertices $\neq r$ such that $w$ is proper ancestor of $v$, $v$ is a proper ancestor of $u$, and there is a back-edge $e$ such that $e\in B(u)\cap B(v)\cap B(w)$, $B(v)\setminus\{e\}=(B(u)\setminus\{e\})\sqcup(B(w)\setminus\{e\})$, and $M(B(v)\setminus\{e\})\neq M(B(w)\setminus\{e\})$. Then $e=(\mathit{lowD}(u),\mathit{low}(u))$, $\mathit{high}(u)<v$, $\mathit{low}_2(u)\geq w$, and $M(w)=M(v)$. Furthermore, $M(u)=M(v,c)$, where $c$ is either the $\mathit{low1}$ or the $\mathit{low2}$ child of $M(w)$. Finally, $u$ is the lowest proper descendant of $v$ that has $M(u)=M(v,c)$.
\end{lemma}
\begin{proof}
Since $B(v)\setminus\{e\}=(B(u)\setminus\{e\})\sqcup(B(w)\setminus\{e\})$ and $e\in B(v)\cap B(u)\cap B(w)$, we have that $B(u)\subseteq B(v)$ and $B(w)\subseteq B(v)$.
Let $(x,y)$ be a back-edge in $B(u)$. Then $B(u)\subseteq B(v)$ implies that $(x,y)\in B(v)$, and therefore $y$ is a proper ancestor of $v$, and therefore $y<v$. Due to the generality of $(x,y)\in B(u)$, this implies that $\mathit{high}(u)<v$.

The disjoint union in $B(v)\setminus\{e\}=(B(u)\setminus\{e\})\sqcup(B(w)\setminus\{e\})$, and the fact that $e\in B(u)\cap B(w)$, implies that $B(u)$ and $B(w)$ intersect only at $e$. Let us suppose, for the sake of contradiction, that $\mathit{low}_2(u)<w$. Let $(x,y)$ be the $\mathit{low2}$ back-edge of $u$. Then $x$ is a descendant of $u$, and therefore a descendant of $v$, and therefore a descendant of $w$. Since $(x,y)$ is a back-edge, we have that $x$ is a descendant of $y$. Thus, $x$ is a common descendant of $y$ and $w$, and therefore $y$ and $w$ are related as ancestor and descendant. Since $y=\mathit{low_2}(u)$ and $\mathit{low_2}(u)<w$, we have that $y<w$, and so $y$ is a proper ancestor of $w$. This shows that $(x,y)\in B(w)$. Now let $(x',y')$ be the $\mathit{low1}$ back-edge of $u$. Then we have $\mathit{low}_1(u)\leq\mathit{low}_2(u)$, and therefore $y'\leq y$. Since $y<w$, this implies that $y'<w$. But then we can show as previously that $(x',y')\in B(w)$, and so $B(u)\cap B(w)$ contains at least two back-edges (i.e., $(x,y)$ and $(x',y')$), a contradiction. Thus, we have that $\mathit{low}_2(u)\geq w$. This implies that only the $\mathit{low1}$ back-edge of $u$ may be in $B(w)$, and so we have that $e$ is the $\mathit{low1}$ back-edge of $u$.

Now we will show that $M(w)=M(v)$. Since $e\in B(u)\cap B(w)$, we have that the higher endpoint of $e$ is a descendant of both $u$ and $M(w)$. Thus, $u$ and $M(w)$ are related as ancestor and descendant. Let us suppose, for the sake of contradiction, that $M(w)$ is a descendant of $u$. Since the graph is $3$-edge-connected, there is a back-edge $(x,y)\in B(w)\setminus\{e\}$. Then, we have that $x$ is a descendant of $M(w)$, and therefore a descendant of $u$. Furthermore, $y$ is a proper ancestor of $w$, and therefore a proper ancestor of $v$, and therefore a proper ancestor of $u$. This shows that $(x,y)\in B(u)$. But this contradicts the disjointness of the union in $B(v)\setminus\{e\}=(B(u)\setminus\{e\})\sqcup(B(w)\setminus\{e\})$, which implies that there is at most one back-edge in $B(u)\cap B(w)$. This shows that $M(w)$ is a proper ancestor of $u$. Now, since $B(w)\subseteq B(v)$, we have that $M(w)$ is a descendant of $M(v)$. Conversely, let $(x,y)$ be a back-edge in $B(v)$. If $(x,y)=e$, then we have that $(x,y)\in B(w)$, and therefore $x$ is a descendant of $M(w)$. Otherwise, $B(v)\setminus\{e\}=(B(u)\setminus\{e\})\sqcup(B(w)\setminus\{e\})$ implies that either $(x,y)\in B(u)$ or $(x,y)\in B(w)$. If $(x,y)\in B(u)$, then $x$ is a descendant of $u$, and therefore a descendant of $M(w)$. And if $(x,y)\in B(w)$, then $x$ is a descendant of $M(w)$. Thus, in either case we have that $x$ is a descendant of $M(w)$. Due to the generality of $(x,y)\in B(v)$, this shows that $M(v)$ is a descendant of $M(w)$. Thus, we have $M(w)=M(v)$.

Since $B(u)\subseteq B(v)$, we have that $M(u)$ is a descendant of $M(v)$. Let us suppose, for the sake of contradiction, that $M(u)=M(v)$. Then, since $u$ is a proper descendant of $v$, Lemma~\ref{lemma:same_m_subset_B} implies that $B(v)\subseteq B(u)$. Therefore, $B(u)\subseteq B(v)$ implies that $B(u)=B(v)$, in contradiction to the fact that the graph is $3$-edge-connected. Thus, we have that $M(u)\neq M(v)$, and therefore $M(u)$ is a proper descendant of $M(v)$. Let $c$ be the child of $M(v)$ that is an ancestor of $M(u)$.

Let us suppose, for the sake of contradiction, that $c$ is neither the $\mathit{low1}$ nor the $\mathit{low2}$ child of $M(v)$. Let $e=(x,y)$. Since $e\in B(u)$, we have that $x$ is a descendant of $M(u)$, and therefore a descendant of $c$. Furthermore, we have $e\in B(v)$. Thus, $y$ is a proper ancestor of $v$, and therefore a proper ancestor of $M(v)$, and therefore a proper ancestor of $c$. This shows that $e\in B(c)$. Since $e\in B(w)$, we have that $y$ is a proper ancestor of $w$, and therefore $y<w$. Thus, since $e\in B(c)$, we have that $\mathit{low}(c)\leq y<w$. Let $c_1$ and $c_2$ be the $\mathit{low1}$ and the $\mathit{low2}$ child of $M(v)$, respectively. Since $c$ is neither $c_1$ nor $c_2$, we have that $\mathit{low}(c_1)\leq\mathit{low}(c_2)\leq\mathit{low}(c)<w$. Now let $(x',y')$ be a back-edge in $B(c_1)$ such that $y'=\mathit{low}(c_1)$. Then, $x'$ is a descendant of $c_1$, and therefore a descendant of $M(v)$, and therefore a descendant of $M(w)$, and therefore a descendant of $w$. Furthermore, since $(x',y')$ is a back-edge, $y'$ is an ancestor of $x'$. Thus, since $x'$ is a common descendant of $y'$ and $w$, we have that $y'$ and $w$ are related as ancestor and descendant. Then, $y'=\mathit{low}(c_1)<w$ implies that $y'$ is a proper ancestor of $w$. This shows that $(x',y')\in B(w)$. Since $x$ is a descendant of $c$, and $x'$ is a descendant of $c_1$, and $c\neq c_1$, we have that $x\neq x'$ (because otherwise $x$ would be a common descendant of $c$ and $c_1$, and so $c$ and $c_1$ would be related as ancestor and descendant, which is absurd). Thus, $(x',y')\in B(w)$ can be strengthend to $(x',y')\in B(w)\setminus\{e\}$. This implies that $x'$ is a descendant of $M(B(w)\setminus\{e\})$. Thus, $M(B(w)\setminus\{e\})$ is an ancestor of a descendant of $c_1$. Similarly, we can show that $M(B(w)\setminus\{e\})$ is an ancestor of a descendant of $c_2$. Since $c_1$ and $c_2$ are not related as ancestor and descendant, this implies that $M(B(w)\setminus\{e\})$ is an ancestor of $\mathit{nca}\{c_1,c_2\}=M(v)=M(w)$. Since $M(B(w)\setminus\{e\})$ is a descendant of $M(w)$, this implies that $M(B(w)\setminus\{e\})=M(w)=M(v)$. Since $B(v)\setminus\{e\}=(B(u)\setminus\{e\})\sqcup(B(w)\setminus\{e\})$, we have $B(w)\setminus\{e\}\subseteq B(v)\setminus\{e\}$, and therefore $M(B(w)\setminus\{e\})$ is a descendant of $M(B(v)\setminus\{e\})$, which is a descendant of $M(v)$. Thus, since $M(B(w)\setminus\{e\})=M(v)$, we have $M(B(w)\setminus\{e\})=M(B(v)\setminus\{e\})$, in contradiction to the assumption of the lemma. Thus, we have that $c$ is either the $\mathit{low1}$ or the $\mathit{low2}$ child of $M(v)$.

Let us suppose, for the sake of contradiction, that there is a proper descendant $u'$ of $v$, that is lower than $u$ and has $M(u')=M(v,c)$. Then, since $M(u')=M(u)$ and $u'<u$, we have that $u'$ is a proper ancestor of $u$, and Lemma~\ref{lemma:same_m_subset_B} implies that $B(u')\subseteq B(u)$. Since the graph is $3$-edge-connected, this can be strengthened to $B(u')\subset B(u)$. Thus, there is a back-edge $(x,y)\in B(u)\setminus B(u')$. Then, $x$ is a descendant of $M(u)=M(u')$. Furthermore, $B(u)\subseteq B(v)$ implies that $(x,y)\in B(v)$, and therefore $y$ is a proper ancestor of $v$, and therefore a proper ancestor of $u'$. This shows that $(x,y)\in B(u')$, a contradiction. Thus, we have that $u$ is the lowest proper descendant of $v$ that has $M(u)=M(v,c)$.
\end{proof}

Lemma~\ref{lemma:type-3-b-i-1-info} provides enough information to guide us into the search for all Type-3$\beta i$ $4$-cuts that satisfy $(1)$ of Lemma~\ref{lemma:type-3b-cases}. According to Lemma~\ref{lemma:type-3-b-i-1-info}, we have to find, for every vertex $v\neq r$, the lowest proper descendant $u$ of $v$ that has $M(u)=M(v,c)$, where $c$ is either the $\mathit{low1}$ or the $\mathit{low2}$ child of $M(v)$. Then, $w$ should satisfy that $M(w)=M(v)$ and $\mathit{bcount}(w)=\mathit{bcount}(v)-\mathit{bcount}(u)+1$. We note that this $w$, if it exists, is unique, because it has the same $M$ point as $v$. Then, once we collect the triple $u,v,w$ we apply the criterion provided by Lemma~\ref{lemma:type-3-b-i-1-criterion}, in order to check that we indeed get a $4$-cut. This procedure is shown in Algorithm~\ref{algorithm:type-3-b-i-1}. The proof of correctness and linear complexity is given in Proposition~\ref{proposition:algorithm:type-3-b-i-1}.

\begin{lemma}
\label{lemma:type-3-b-i-1-criterion}
Let $v\neq r$ be a vertex such that $M(v)$ has a child $c$. Let $u$ be a proper descendant of $u$ such that $M(u)$ is a descendant of $c$, and let $w$ be a proper ancestor of $v$ such that $M(w)=M(v)$. Then there is a back-edge $e\in B(u)\cap B(v)\cap B(w)$ such that $B(v)\setminus\{e\}=(B(u)\setminus\{e\})\sqcup(B(w)\setminus\{e\})$ if and only if: $\mathit{high}(u)<v$, $w\leq\mathit{low}_2(u)$, and $\mathit{bcount}(v)=\mathit{bount}(u)+\mathit{bcount}(w)-1$.
\end{lemma}
\begin{proof}
($\Rightarrow$)
$\mathit{bcount}(v)=\mathit{bount}(u)+\mathit{bcount}(w)-1$ is an immediate consequence of $B(v)\setminus\{e\}=(B(u)\setminus\{e\})\sqcup(B(w)\setminus\{e\})$ and $e\in B(u)\cap B(v)\cap B(w)$. Since $B(u)\setminus\{e\}\subseteq B(v)\setminus\{e\}$ and $e\in B(u)\cap B(v)$, we have that $B(u)\subseteq B(v)$. This implies that $\mathit{high}(u)<v$ (because every back-edge $(x,y)\in B(u)$ must have $y<v$). Since $B(u)\cap B(w)\neq\emptyset$ and $w$ is a proper ancestor of $u$, we have that the $\mathit{low}$-edge of $u$ is in $B(w)$. And since $(B(u)\setminus\{e\})\cap(B(w)\setminus\{e\})=\emptyset$ and $e\in B(u)\cap B(w)$, we have that the $\mathit{low}$-edge of $u$ is precisely $e$. Then, since $(B(u)\setminus\{e\})\cap(B(w)\setminus\{e\})=\emptyset$ and $u$ is a proper descendant of $w$, we have that no back-edge $e'$ in $B(u)\setminus\{e\}$ has lower point that is lower than $w$ (because this would imply that $e'\in B(w)$). This implies that $\mathit{low}_2(u)\geq w$.\\
($\Leftarrow$)
Let $(x,y)$ be a back-edge in $B(u)$. Then, $x$ is a descendant of $M(u)$, and therefore a descendant of $c$, and therefore a descendant of $M(v)$, and therefore a descendant of $v$. Furthermore, since $(x,y)$ is a back-edge, $y$ is an ancestor of $x$. Thus, $x$ is a common descendant of $v$ and $y$, and therefore $v$ and $y$ are related as ancestor and descendant. Since $(x,y)\in B(u)$, we have that $y$ is an ancestor of $\mathit{high}(u)$, and therefore $y\leq\mathit{high}(u)$. Thus, $\mathit{high}(u)<v$ implies that $y<v$, and therefore $y$ is a proper ancestor of $v$. This shows that $(x,y)\in B(v)$. Due to the generality of $(x,y)\in B(u)$, this implies that $B(u)\subseteq B(v)$. Since $w$ is a proper ancestor of $v$ with $M(w)=M(v)$, by Lemma~\ref{lemma:same_m_subset_B} we have that $B(w)\subseteq B(v)$.

Let us suppose, for the sake of contradiction, that $B(w)\cap B(v)=\emptyset$. Then, $B(u)\subseteq B(v)$ and $B(w)\subseteq B(v)$ imply that $B(u)\sqcup B(w)\subseteq B(v)$, and therefore $\mathit{bcount}(u)+\mathit{bcount}(w)\leq\mathit{bcount}(v)$, in contradiction to $\mathit{bcount}(v)=\mathit{bcount}(u)+\mathit{bcount}(w)-1$. Thus, we have that $B(w)\cap B(v)\neq\emptyset$.

Let $e=(x,y)$ be a back-edge in $B(u)$ such that $y=\mathit{low}_1(u)$. Let us suppose, for the sake of contradiction, that $e\notin B(w)$. Consider a back-edge $(x',y')\in B(u)\setminus\{(x,y)\}$. Then, we have that $y'\geq\mathit{low}_2(u)$. Thus, $\mathit{low}_2(u)\geq w$ implies that $y'\geq w$. This implies that $y'$ cannot be a proper ancestor of $w$, and therefore $(x',y')\notin B(w)$. Due to the generality of $(x',y')\in B(u)\setminus\{e\}$, this implies that $B(w)\cap(B(u)\setminus\{e\})=\emptyset$. Therefore, since $e\notin B(w)$, we have that $B(w)\cap B(u)=\emptyset$, a contradiction. Thus, we have that $e\in B(w)$. Furthermore, since $\mathit{low}_2(u)\geq w$, this is the only back-edge in $B(u)$ that is also in $B(w)$.

Thus, since $B(u)\subseteq B(v)$, $B(w)\subseteq B(v)$, $B(u)\cap B(w)=\{e\}$ and $\mathit{bcount}(v)=\mathit{bcount}(u)+\mathit{bcount}(w)-1$, we have that $B(v)=((B(u)\setminus\{e\})\sqcup(B(w)\setminus\{e\}))\sqcup\{e\}$, and therefore $B(v)\setminus\{e\}=(B(u)\setminus\{e\})\sqcup(B(w)\setminus\{e\})$.
\end{proof}

\noindent\\
\begin{algorithm}[H]
\caption{\textsf{Compute all Type-3$\beta i$ $4$-cuts that satisfy $(1)$ of Lemma~\ref{lemma:type-3b-cases}}}
\label{algorithm:type-3-b-i-1}
\LinesNumbered
\DontPrintSemicolon
\ForEach{vertex $v\neq r$}{
\label{line:type-3-b-i-1-for}
    let $c_1$ be the $\mathit{low1}$ child of $M(v)$\;
    let $c_2$ be the $\mathit{low2}$ child of $M(v)$\;
    compute $M(v,c_1)$ and $M(v,c_2)$\;  
}
initialize an array $A$ of size $m$\;
\ForEach{vertex $x\neq r$}{
\label{line:type-3-b-i-1-x}
  let $c_1$ be the $\mathit{low1}$ child of $x$\;
  let $c_2$ be the $\mathit{low2}$ child of $x$\;
  \ForEach{$z\in M^{-1}(x)$}{
  \label{line:type-3-b-i-1-A}
    set $A[\mathit{bcount}(z)]\leftarrow z$\;   
  }
  \ForEach{$v\in M^{-1}(x)$}{
  \label{line:type-3-b-i-1-v}
  \tcp{consider the case where $u$ is a descendant of the $\mathit{low1}$ child of $x$; the other case is treated similarly, by substituting $c_1$ with $c_2$}
    let $u$ be the lowest proper descendant of $v$ with $M(u)=M(v,c_1)$\;
    \label{line:type-3-b-i-1-u}
    let $w\leftarrow A[\mathit{bcount}(v)-\mathit{bcount}(u)+1]$\;
    \label{line:type-3-b-i-1-w}
    \If{$w\neq\bot$ \textbf{and} $w<v$ \textbf{and} $\mathit{high}(u)<v$ \textbf{and} $w\leq\mathit{low}_2(u)$}{ 
    \label{line:type-3-b-i-1-cond} 
      mark $\{(u,p(u)),(v,p(v)),(w,p(w)),(\mathit{lowD}(u),\mathit{low}(u))\}$ as a $4$-cut\;
      \label{line:type-3-b-i-1-mark} 
    }
  }
  \ForEach{$z\in M^{-1}(x)$}{
    set $A[\mathit{bcount}(z)]\leftarrow\bot$\;
  }  
}
\end{algorithm}

\begin{proposition}
\label{proposition:algorithm:type-3-b-i-1}
Algorithm~\ref{algorithm:type-3-b-i-1} correctly computes all Type-3$\beta i$ $4$-cuts that satisfy $(1)$ of Lemma~\ref{lemma:type-3b-cases}. Furthermore, it has a linear-time implementation.
\end{proposition}
\begin{proof}
Let $C=\{(u,p(u)),(v,p(v)),(w,p(w)),e\}$ be a Type-3$\beta i$ $4$-cut such that $w$ is a proper ancestor of $v$, and $v$ is a proper ancestor of $u$, where the back-edge $e$ satisfies $(1)$ of Lemma~\ref{lemma:type-3b-cases}. Then, by Lemma~\ref{lemma:type-3-b-i-1-info} we have that $M(u)=M(v,c)$, where $c$ is either the $\mathit{low1}$ or the $\mathit{low2}$ child of $M(v)$. Let us assume that $c$ is the $\mathit{low1}$ child of $M(v)$ (the other case is treated similarly). Then, Lemma~\ref{lemma:type-3-b-i-1-info} implies that $u$ is the lowest proper descendant of $v$ such that $M(u)=M(v,c)$, $e=(\mathit{lowD}(u),\mathit{low}(u))$, and $M(w)=M(v)$. Lemma~\ref{lemma:type-3-b-i-1-criterion} implies that $\mathit{bcount}(w)=\mathit{bcount}(v)-\mathit{bcount}(u)+1$, $\mathit{high}(u)<v$ and $w\leq\mathit{low}_2(u)$.

Now, when the \textbf{for} loop in Line~\ref{line:type-3-b-i-1-x} processes $x=M(v)$, we will eventually reach the \textbf{for} loop in Line~\ref{line:type-3-b-i-1-v}, because $v\in M^{-1}(x)$. Notice that when we reach Line~\ref{line:type-3-b-i-1-u} when the \textbf{for} loop in Line~\ref{line:type-3-b-i-1-v} processes $v$, we have that the variable ``$u$" is assigned precisely $u$. Now consider the variable ``$w$" in Line~\ref{line:type-3-b-i-1-w}. We claim that the $\mathit{bcount}(v)-\mathit{bcount}(u)+1$ entry of the array $A$ is precisely $w$. To see this, observe that the \textbf{for} loop in Line~\ref{line:type-3-b-i-1-A} has processed $w$ (because $M(w)=M(v)$), and at some point inserted into the $\mathit{bcount}(v)-\mathit{bcount}(u)+1$ entry of $A$ the value $w$. But then, this entry was not altered afterwards, because Lemma~\ref{lemma:same_M_same_high} implies that all vertices with the same $M$ point have different $\mathit{bcount}$ values (since our graph is $3$-edge-connected). Thus, the variable ``$w$" in Line~\ref{line:type-3-b-i-1-w} holds the value $w$, and therefore $C$ will be marked in Line~\ref{line:type-3-b-i-1-mark}, since all the conditions to reach this line are satisfied.

Conversely, suppose that a $4$-element set $C=\{(u,p(u)),(v,p(v)),(w,p(w)),(x,y)\}$ is marked in Line~\ref{line:type-3-b-i-1-mark}. Then, we have that $(x,y)=(\mathit{lowD}(u),\mathit{low}(u))$, $u$ is a proper descendant of $v$, $M(u)$ is a descendant of the $\mathit{low1}$ child of $M(v)$, $w$ is a proper ancestor of $v$ with $M(w)=M(v)$ (due to $w<v$ in Line~\ref{line:type-3-b-i-1-cond}),  $\mathit{bcount}(w)=\mathit{bcount}(v)-\mathit{bcount}(u)+1$, $\mathit{high}(u)<v$ and $w<\mathit{low}_2(u)$. Thus, Lemma~\ref{lemma:type-3-b-i-1-criterion} implies that there is a back-edge $e\in B(u)\cap B(v)\cap B(w)$ such that $B(v)\setminus\{e\}=(B(u)\setminus\{e\})\sqcup(B(w)\setminus\{e\})$. Since $B(u)\cap B(w)\neq\emptyset$ and $w$ is a proper ancestor of $u$, we have that the $\mathit{low}$-edge of $u$ is in $B(w)$. And since $(B(u)\setminus\{e\})\cap(B(w)\setminus\{e\})=\emptyset$, we have that the $\mathit{low}$-edge of $u$ is precisely $e$. Thus, $C$ is indeed a $4$-cut (that satisfies $(1)$ of Lemma~\ref{lemma:type-3b-cases}).

Now we will argue about the complexity of Algorithm~\ref{algorithm:type-3-b-i-1}. By Proposition~\ref{proposition:computing-M(v,c)}, we have that the values $M(v,c_1)$ and $M(v,c_2)$ can be computed in linear time in total, for all vertices $v\neq r$, where $c_1$ and $c_2$ are the $\mathit{low1}$ and $\mathit{low2}$ children of $M(v)$. Thus, the \textbf{for} loop in Line~\ref{line:type-3-b-i-1-for} can be performed in linear time. In order to compute $u$ in Line~\ref{line:type-3-b-i-1-u}, we use Algorithm~\ref{algorithm:W-queries}. Specifically, whenever we reach this line, we generate a query $q(M^{-1}(M(v,c_1)),v)$. This returns the lowest vertex $u$ with $M(u)=M(v,c_1)$ such that $u>v$. Since $M(u)=M(v,c_1)$ implies that $M(u)$ is a common descendant of $v$ and $u$, we have that $v$ and $u$ are related as ancestor and descendant. Thus, $u>v$ implies that $u$ is a proper descendant of $v$. Therefore, we have that $u$ is the lowest proper descendant of $v$ such that $M(u)=M(v,c_1)$. Since the number of all those queries is $O(n)$, by Lemma~\ref{lemma:W-queries} we have that Algorithm~\ref{algorithm:W-queries} answers all of them in linear time in total. It is easy to see that the remaining operations of Algorithm~\ref{algorithm:type-3-b-i-1} take $O(n)$ time in total. We conclude that Algorithm~\ref{algorithm:type-3-b-i-1} runs in linear time.
\end{proof}

\subsubsection{Case $(2)$ of Lemma~\ref{lemma:type-3b-cases}}

\begin{lemma}
\label{lemma:type-3-b-i-2-info}
Let $u,v,w$ be three vertices $\neq r$ such that $w$ is proper ancestor of $v$, $v$ is a proper ancestor of $u$, and there is a back-edge $e\in B(w)$ such that $e\notin B(v)\cup B(u)$, $B(v)=B(u)\sqcup(B(w)\setminus\{e\})$, and $M(v)\neq M(B(w)\setminus\{e\})$. Then $w$ is an ancestor of $\mathit{low}(u)$, $M(w)\neq M(v)$, $M(B(w)\setminus\{e\})\neq M(w)$, and $e$ is either $(L_1(w),l(L_1(w)))$ or $(R_1(w),l(R_1(w)))$. Furthermore, let $c_1$ be the $\mathit{low1}$ child of $M(v)$, and let $c_2$ be the $\mathit{low2}$ child of $M(v)$. Then $M(B(w)\setminus\{e\})=M(v,c_1)$, and $u$ is the lowest proper descendant of $v$ such that $M(u)=M(v,c_2)$.
\end{lemma}
\begin{proof}
Let us suppose, for the sake of contradiction, that $w$ is not an ancestor of $\mathit{low}(u)$. $B(v)=B(u)\sqcup(B(w)\setminus\{e\})$ implies that $B(u)\subseteq B(v)$. Thus, the $\mathit{low}$-edge of $u$ is in $B(v)$. This implies that $\mathit{low}(u)$ is a proper ancestor of $v$. Then, since $v$ is a common descendant of $\mathit{low}(u)$ and $w$, we have that $\mathit{low}(u)$ and $w$ are related as ancestor and descendant. Since $w$ is not an ancestor of $\mathit{low}(u)$, this implies that $w$ is a proper descendant of $\mathit{low}(u)$. Now let $(x',y')$ be the $\mathit{low}$-edge of $u$. Then we have that $x'$ is a descendant of $u$, and therefore a descendant of $w$. Furthermore, $y'=\mathit{low}(u)$ is a proper ancestor of $w$. This shows that $(x',y')\in B(w)$. But since $B(u)\cap (B(w)\setminus\{e\})=\emptyset$, this implies that $(x',y')=e$, contradicting the fact that $e\notin B(u)$. This shows that $w$ is an ancestor of $\mathit{low}(u)$.

Let $e=(x,y)$. Since $e\in B(w)$, we have that $y$ is a proper ancestor of $w$, and therefore a proper ancestor of $v$. Thus, since $e\notin B(v)$, we have that $x$ cannot be a descendant of $M(v)$. Since $M(w)$ is an ancestor of $x$, this implies that $M(w)$ cannot be a descendant of $M(v)$, and therefore $M(w)\neq M(v)$. Since $B(w)\setminus\{e\}\subseteq B(v)$, we have that $M(B(w)\setminus\{e\})$ is a descendant of $M(v)$. Thus, since $M(w)$ cannot be a descendant of $M(v)$, we have $M(w)\neq M(B(w)\setminus\{e\})$. By Lemma~\ref{lemma:e_L-e_R}, this implies that either $e=e_L(w)$ or $e=e_R(w)$.

$B(v)=B(u)\sqcup(B(w)\setminus\{e\})$ implies that $B(u)\subseteq B(v)$, and therefore $M(u)$ is a descendant of $M(v)$. Thus, since $M(u)$ is a common descendant of $u$ and $M(v)$, we have that $u$ and $M(v)$ are related as ancestor and descendant. Let us suppose, for the sake of contradiction, that $u$ is not proper descendant of $M(v)$. Then, $u$ is an ancestor of $M(v)$. Let $(x',y')$ be a back-edge in $B(v)$. Then $x'$ is a descendant of $M(v)$, and therefore a descendant of $u$. Furthermore, $y'$ is a proper ancestor of $v$, and therefore a proper ancestor of $u$. This shows that $(x',y')\in B(u)$. Due to the generality of $(x',y')\in B(v)$, this implies that $B(v)\subseteq B(u)$. But $B(v)=B(u)\sqcup(B(w)\setminus\{e\})$ implies that $B(u)\subseteq B(v)$, and therefore we have $B(u)=B(v)$, in contradiction to the fact that the graph is $3$-edge-connected. This shows that $u$ is a proper descendant of $M(v)$.

Let us suppose, for the sake of contradiction, that there is a back-edge of the form $(M(v),z)$ in $B(v)$. Then, $B(v)=B(u)\sqcup(B(w)\setminus\{e\})$ implies that either $(M(v),z)\in B(u)$, or $(M(v),z)\in B(w)\setminus\{e\}$. The first case is rejected, because it implies that $M(v)$ is a descendant of $u$. Thus, we have $(M(v),z)\in B(w)\setminus\{e\}$. This implies that $M(v)$ is a descendant of $M(B(w)\setminus\{e\})$. But $B(v)=B(u)\sqcup(B(w)\setminus\{e\})$ implies that $B(w)\setminus\{e\}\subseteq B(v)$, and therefore $M(B(w)\setminus\{e\})$ is a descendant of $M(v)$, and therefore we have $M(v)=M(B(w)\setminus\{e\})$, contradicting one of the assumptions in the statement of the lemma. Thus, we have that there is no back-edge of the form $(M(v),z)$ in $B(v)$. 
This implies that $\mathit{low}(c_1)<v$ and $\mathit{low}(c_2)<v$, where $c_1$ and $c_2$ is the $\mathit{low1}$ and the $\mathit{low2}$ child of $M(v)$, respectively.

Let us suppose, for the sake of contradiction, that there is a back-edge $(x',y')\in B(w)\setminus\{e\}$ such that $x'$ is not a descendant of $c_1$. $B(v)=B(u)\sqcup(B(w)\setminus\{e\})$ implies that $(x',y')\in B(v)$. Thus, since there is no back-edge of the form $(M(v),z)$ in $B(v)$, we have that $x'$ is a descendant of a child $c$ of $M(v)$. We have that $y'$ is a proper ancestor of $v$, and therefore a proper ancestor of $M(v)$, and therefore a proper ancestor of $c$. This shows that $(x',y')\in B(c)$. Thus, we have $\mathit{low}(c)<w$. Since $c\neq c_1$, we have that $\mathit{low}(c_1)\leq\mathit{low}(c)$. Thus, $\mathit{low}(c)<w$ implies that $\mathit{low}(c_1)<w$. This means that there is a back-edge $(x'',y'')\in B(c_1)$ such that $y''<w$. Then $x''$ is a descendant of $M(v)$, and therefore a descendant of $v$, and therefore a descendant of $w$. Furthermore, since $x''$ is a common descendant of $y''$ and $w$, we have that $y''$ and $w$ are related as ancestor and descendant. Thus, $y''<w$ implies that $y''$ is a proper ancestor of $w$. This shows that $(x'',y'')\in B(w)$. Since $x''$ is a descendant of $M(v)$ and $y''$ is a proper ancestor of $v$, we also have that $(x'',y'')\in B(v)$. Thus, $e\notin B(v)$ implies that $(x'',y'')\neq e$. Therefore, $(x'',y'')\in B(w)$ can be strengthened to $(x'',y'')\in B(w)\setminus\{e\}$. This implies that $x''$ is a descendant of $M(B(w)\setminus\{e\})$. Since $(x',y')\in B(w)\setminus\{e\}$, we also have that $x'$ is a descendant of $M(B(w)\setminus\{e\})$. This implies that $M(B(w)\setminus\{e\})$ is an ancestor of $\mathit{nca}\{x',x''\}$. Since $x'$ and $x''$ are descendants of different children of $M(v)$, we have that $\mathit{nca}\{x',x''\}=M(v)$. But then we have that $M(B(w)\setminus\{e\})$ is an ancestor of $M(v)$, and therefore we have $M(B(w)\setminus\{e\})=M(v)$, since $B(v)=B(u)\sqcup(B(w)\setminus\{e\})$ implies that $M(B(w)\setminus\{e\})$ is a descendant of $M(v)$. This contradicts the assumption $M(B(w)\setminus\{e\})\neq M(v)$ in the statement of the lemma. Thus, we have that all back-edges in $B(w)\setminus\{e\}$ have their higher endpoint in $T(c_1)$.

Since $\mathit{low}(c_2)<v$, we have that there is a back-edge $(x',y')\in B(c_2)$ such that $y'<v$. Then, $x'$ is a descendant of $c_2$, and therefore a descendant of $M(v)$, and therefore a descendant of $v$. Thus, since $x'$ is a common descendant of $y'$ and $v$, we have that $y'$ and $v$ are related as ancestor and descendant, and therefore $y'<v$ implies that $y'$ is a proper ancestor of $v$. This shows that $(x',y')\in B(v)$. Thus, $B(v)=B(u)\sqcup(B(w)\setminus\{e\})$ implies that either $(x',y')\in B(u)$, or $(x',y')\in B(w)\setminus\{e\}$. The second case is rejected, since all back-edges in $B(w)\setminus\{e\}$ have their higher endpoint in $T(c_1)$. Thus, $(x',y')\in B(u)$. This implies that $x'$ is a descendant of $u$. Since $u$ is a proper descendant of $M(v)$, we have that $u$ is a descendant of a child $c$ of $M(v)$. Then, we have that $x'$ is a descendant of $c_2$ (by assumption) and also a descendant of $c$ (since it is a descendant of $u$). Thus, we have that $c$ and $c_2$ are related as ancestor and descendant. Therefore, since they have the same parent, they must coincide. In other words, we have $c=c_2$. 

Let $S_1=\{(x',y')\in B(v)\mid x' \mbox{ is a descendant of } c_1\}$. Then, $M(S_1)=M(v,c_1)$. Let $(x',y')$ be a back-edge in $S_1$. Then $(x',y')\in B(v)$, and therefore $B(v)=B(u)\sqcup(B(w)\setminus\{e\})$ implies that either $(x',y')\in B(u)$, or $(x',y')\in B(w)\setminus\{e\}$. Let us suppose, for the sake of contradiction, that $(x',y')\in B(u)$. Then, $x'$ is a descendant of $u$, and therefore a descendant of $c_2$. Thus, $x'$ is a common descendant of $c_1$ and $c_2$, and therefore $c_1$ and $c_2$ are related as ancestor and descendant, which is impossible. Thus, the case $(x',y')\in B(u)$ is rejected, and so we have $(x',y')\in B(w)\setminus\{e\}$. Due to the generality of $(x',y')\in S_1$, this implies that $S_1\subseteq B(w)\setminus\{e\}$. Conversely, let $(x',y')$ be a back-edge in $B(w)\setminus\{e\}$. Then, we have shown that $x'$ is a descendant of $c_1$. Furthermore, $B(v)=B(u)\sqcup(B(w)\setminus\{e\})$ implies that $(x',y')\in B(v)$. This shows that $(x',y')\in S_1$. Due to the generality of $(x',y')\in B(w)\setminus\{e\}$, this implies that $B(w)\setminus\{e\}\subseteq S_1$. Thus, since $S_1\subseteq B(w)\setminus\{e\}$, we have $S_1= B(w)\setminus\{e\}$. This implies that $M(S_1)=M(B(w)\setminus\{e\})$, and therefore $M(v,c_1)=M(B(w)\setminus\{e\})$.

Let $S_2=\{(x',y')\in B(v)\mid x' \mbox{ is a descendant of } c_2\}$. Then, $M(S_2)=M(v,c_2)$. Let $(x',y')$ be a back-edge in $S_2$. Then $(x',y')\in B(v)$, and therefore $B(v)=B(u)\sqcup(B(w)\setminus\{e\})$ implies that either $(x',y')\in B(u)$, or $(x',y')\in B(w)\setminus\{e\}$. The case $(x',y')\in B(w)\setminus\{e\}$ is rejected, because it implies that $(x',y')\in S_1$ (and obviously we have $S_1\cap S_2=\emptyset$). Thus, we have $(x',y')\in B(u)$. Due to the generality of $(x',y')\in S_2$, this implies that $S_2\subseteq B(u)$. Conversely, let $(x',y')$ be a back-edge in $B(u)$. Then, $x'$ is a descendant of $u$, and therefore a descendant of $c_2$. Furthermore, $B(v)=B(u)\sqcup(B(w)\setminus\{e\})$ implies that $(x',y')\in B(v)$. This shows that $(x',y')\in S_2$. Due to the generality of $(x',y')\in B(u)$, this implies that $B(u)\subseteq S_2$. Thus, since $S_2\subseteq B(u)$, we have $S_2=B(u)$. This implies that $M(S_2)=M(u)$, and therefore $M(v,c_2)=M(u)$.

Let us suppose, for the sake of contradiction, that there is a proper descendant $u'$ of $v$ that is lower than $u$ and such that $M(u')=M(v,c_2)$. Then, since $M(u')=M(u)$ and $u'<u$, we have that $u'$ is a proper ancestor of $u$, and Lemma~\ref{lemma:same_m_subset_B} implies that $B(u')\subseteq B(u)$. Since the graph is $3$-edge-connected, this can be strengthened to $B(u')\subset B(u)$. Thus, there is a back-edge $(x',y')\in B(u)\setminus B(u')$. Then, $x'$ is a descendant of $u$, and therefore a descendant of $u'$. Furthermore, $B(v)=B(u)\sqcup(B(w)\setminus\{e\})$ implies that $(x',y')\in B(v)$, and therefore $y'$ is a proper ancestor of $v$, and therefore a proper ancestor of $u'$. This shows that $(x',y')\in B(u')$, a contradiction. Thus, we have that $u$ is the lowest proper descendant of $v$ such that $M(u)=M(v,c_2)$. 
\end{proof}

\begin{lemma}
\label{lemma:type-3-b-i-2-implied}
Let $u,v,w$ be three vertices $\neq r$ such that $w$ is proper ancestor of $v$, $v$ is a proper ancestor of $u$, and there is a back-edge $e\in B(w)$ such that $e\notin B(v)\cup B(u)$, $B(v)=B(u)\sqcup(B(w)\setminus\{e\})$, and $M(v)\neq M(B(w)\setminus\{e\})$. Let $c_1$ be the $\mathit{low1}$ child of $M(v)$, and let $w'$ be the greatest ancestor of $\mathit{low}(u)$ with the property that there is a back-edge $e'\in B(w')$ such that $M(w')\neq M(B(w')\setminus\{e'\})=M(v,c_1)$. Then, we have that $e'\notin B(v)\cup B(u)$, $B(v)=B(u)\sqcup(B(w')\setminus\{e'\})$ and $M(v)\neq M(B(w')\setminus\{e'\})$. Furthermore, if $w'\neq w$, then $B(w)\sqcup\{e'\}=B(w')\sqcup\{e\}$.
\end{lemma}
\begin{proof}
By Lemma~\ref{lemma:type-3-b-i-2-info} we have that $w$ is an ancestor of $\mathit{low}(u)$, $M(B(w)\setminus\{e\})\neq M(w)$, and $M(B(w)\setminus\{e\})=M(v,c_1)$. Thus, we may consider the greatest ancestor $w'$ of $\mathit{low}(u)$ with the property that there is a back-edge $e'\in B(w')$ such that $M(w')\neq M(B(w')\setminus\{e'\})=M(v,c_1)$. We may assume that $w'\neq w$, because otherwise there is nothing to show. Thus we have $w'>w$. Notice that $B(v)=B(u)\sqcup(B(w)\setminus\{e\})$ implies that $B(u)\subseteq B(v)$, and therefore the $\mathit{low}$-edge of $u$ is in $B(v)$. This implies that $\mathit{low}(u)$ is a proper ancestor of $v$, and therefore $w'$ is also a proper ancestor of $v$.

Let us suppose, for the sake of contradiction, that the higher endpoint of $e'$ is a descendant of $M(B(w')\setminus\{e'\})$. Let $(x,y)$ be a back-edge in $B(w')$. If $(x,y)=e'$, then $x$ is a descendant of $M(B(w')\setminus\{e'\})$. Otherwise, if $(x,y)\neq e'$, then we have that $(x,y)\in B(w')\setminus\{e'\}$, and therefore $x$ is a descendant of $M(B(w')\setminus\{e'\})$. Thus, in any case we have that $x$ is a descendant of $M(B(w')\setminus\{e'\})$. Due to the generality of $(x,y)\in B(w')$, this implies that $M(w')$ is a descendant of $M(B(w')\setminus\{e'\})$. But we have $B(w')\setminus\{e'\}\subseteq B(w')$, and therefore $M(B(w')\setminus\{e'\})$ is a descendant of $M(w')$, and therefore $M(B(w')\setminus\{e'\})=M(w')$, a contradiction. This shows that the higher endpoint of $e'$ is not a descendant of $M(B(w')\setminus\{e'\})$. Similarly, we can see that the higher endpoint of $e$ is not a descendant of $M(B(w)\setminus\{e\})$.

Since $v$ is a common descendant of $w$ and $w'$, we have that $w$ and $w'$ are related as ancestor and descendant. Thus, $w'>w$ implies that $w'$ is a proper descendant of $w$. Let $(x,y)$ be a back-edge in $B(w)\setminus\{e\}$. Then $x$ is a descendant of $M(B(w)\setminus\{e\})$, and therefore a descendant of $M(v,c_1)$, and therefore a descendant of $v$, and therefore a descendant of $w'$. Furthermore, $y$ is a proper ancestor of $w$, and therefore a proper ancestor of $w'$. This shows that $(x,y)\in B(w')$. Since $x$ is a descendant of $M(B(w)\setminus\{e\})=M(B(w')\setminus\{e'\})$, we have that $x$ is not the higher endpoint of $e'$, and therefore $(x,y)\neq e'$. Thus, $(x,y)\in B(w')$ can be strengthened to $(x,y)\in B(w')\setminus\{e'\}$. Due to the generality of $(x,y)\in B(w)\setminus\{e\}$, this implies that $B(w)\setminus\{e\}\subseteq B(w')\setminus\{e'\}$. Conversely, let $(x,y)$ be a back-edge in $B(w')\setminus\{e'\}$. Then $x$ is a descendant of $M(B(w')\setminus\{e'\})$, and therefore a descendant of $M(v,c_1)$, and therefore a descendant of $v$. Furthermore, $y$ is a proper ancestor of $w'$, and therefore a proper ancestor of $v$. This shows that $(x,y)\in B(v)$. Then, $B(v)=B(u)\sqcup(B(w)\setminus\{e\})$ implies that either $(x,y)\in B(u)$, or $(x,y)\in B(w)\setminus\{e\}$. The case $(x,y)\in B(u)$ is rejected, because $w'$ is an ancestor of $\mathit{low}(u)$ (and therefore there is no back-edge in $B(u)$ whose lower endpoint is low enough to be a proper ancestor of $w$). Thus, we have $(x,y)\in B(w)\setminus\{e\}$. Due to the generality of $(x,y)\in B(w')\setminus\{e'\}$, this implies that $B(w')\setminus\{e'\}\subseteq B(w)\setminus\{e\}$. Thus, since we have showed the reverse inclusion too, we have that $B(w')\setminus\{e'\}=B(w)\setminus\{e\}$. Then, $B(v)=B(u)\sqcup(B(w)\setminus\{e\})$ implies that $B(v)=B(u)\sqcup(B(w')\setminus\{e'\})$. 

Let us suppose, for the sake of contradiction, that $e\in B(w')$. Then, since $e\in B(w)$ and $B(w')\setminus\{e'\}=B(w)\setminus\{e\}$, we have that $e=e'$ and $B(w')=B(w)$, which contradicts the fact that the graph is $3$-edge-connected. Thus, $e\notin B(w')$. Similarly, we have $e'\notin B(w)$. Thus, $B(w')\setminus\{e'\}=B(w)\setminus\{e\}$ implies that $B(w')\sqcup\{e\}=B(w)\sqcup\{e'\}$. Since $w'$ is an ancestor of $\mathit{low}(u)$, we have $B(u)\cap B(w')=\emptyset$. In particular, this implies that $e'\notin B(u)$. Therefore, $B(v)=B(u)\sqcup(B(w')\setminus\{e'\})$ implies that $e'\notin B(u)\cup B(v)$. $M(B(w')\setminus\{e'\})=M(v,c_1)$ implies that $M(B(w')\setminus\{e'\})$ is a descendant of $c_1$, and therefore a proper descendant of $M(v)$. Thus, $M(v)\neq M(B(w')\setminus\{e'\})$.
\end{proof}

Now we have collected enough information in order to show how to compute a collection of Type-3$\beta i$ $4$-cuts that satisfy $(2)$ of Lemma~\ref{lemma:type-3b-cases}, so that all $4$-cuts of this type are implied by this collection, plus that returned by Algorithm~\ref{algorithm:type2-2}. Based on Lemma~\ref{lemma:type-3-b-i-2-implied}, it is sufficient to compute all Type-3$\beta i$ $4$-cuts of the form $\{(u,p(u)),(v,p(v)),(w,p(w)),e\}$, such that $e$ satisfies $(2)$ of Lemma~\ref{lemma:type-3b-cases}, where $w$ is the greatest ancestor of $\mathit{low}(u)$ such that $M(w)\neq M(B(w)\setminus\{e\})=M(v,c_1)$, where $c_1$ is the $\mathit{low1}$ child of $M(v)$. By Lemma~\ref{lemma:type-3-b-i-2-info} we have that either $e=e_L(w)$, or $e=e_R(w)$. Thus, we distinguish two cases, depending on whether $e=e_L(w)$ or $e=e_R(w)$. By Lemma~\ref{lemma:type-3-b-i-2-info}, we have that $u$ is the lowest proper descendant of $v$ such that $M(u)=M(v,c_2)$, where $c_2$ is the $\mathit{low2}$ child of $M(v)$. Thus, given $v\neq r$, we know exactly what are the $u$ and $w$ thay may possibly induce a $4$-cut with $v$. We use the criterion provided by Lemma~\ref{lemma:type-3-b-i-2-criterion} in order to check whether we get a $4$-cut from $v$, $u$ and $w$. The whole procedure is shown in Algorithm~\ref{algorithm:type-3-b-i-2}. The proof of correctness and linear complexity is given in Proposition~\ref{proposition:algorithm:type-3-b-i-2}.

\begin{lemma}
\label{lemma:type-3-b-i-2-criterion}
Let $v\neq r$ be a vertex such that $M(v)$ has at least two distinct children $c_1$ and $c_2$. Let $w$ be a proper ancestor of $v$ with the property that $M(w)\neq M(v)$ and there is a back-edge $e$ such that $M(B(w)\setminus\{e\})$ is a descendant of $c_1$, and let $u$ be a proper descendant of $v$ such that $M(u)$ is a descendant of $c_2$. Suppose that $\mathit{high}(u)<v$, and $\mathit{bcount}(v)=\mathit{bcount}(u)+\mathit{bcount}(w)-1$. Then, $e\notin B(u)\cup B(v)$ and $B(v)=B(u)\sqcup(B(w)\setminus\{e\})$. 
\end{lemma}
\begin{proof}
Let $(x,y)$ be a back-edge in $B(u)$. Then $x$ is a descendant of $u$, and therefore a descendant of $v$. Since $(x,y)$ is a back-edge, we have that $x$ is a descendant of $y$. Thus, $x$ is a common descendant of $v$ and $y$, and therefore $v$ and $y$ are related as ancestor and descendant. Since $(x,y)\in B(u)$, we have that $y$ is an ancestor of $\mathit{high}(u)$, and therefore $y\leq\mathit{high}(u)$. Thus, $\mathit{high}(u)<v$ implies that $y<v$, and therefore $y$ is a proper ancestor of $v$. This shows that $(x,y)\in B(v)$. Due to the generality of $(x,y)\in B(u)$, this implies that $B(u)\subseteq B(v)$.

Let $(x,y)$ be a back-edge in $B(w)\setminus\{e\}$. Then $x$ is a descendant of $M(B(w)\setminus\{e\})$, and therefore a descendant of $c_1$, and therefore a descendant of $M(v)$. Furthermore, $y$ is a proper ancestor of $w$, and therefore a proper ancestor of $v$. This shows that $(x,y)\in B(v)$. Due to the generality of $(x,y)\in B(w)\setminus\{e\}$, this implies that $B(w)\setminus\{e\}\subseteq B(v)$.

Let us suppose, for the sake of contradiction, that there is a back-edge $(x,y)\in B(u)\cap(B(w)\setminus\{e\})$. Then $x$ is a descendant of both $M(u)$ and $M(B(w)\setminus\{e\})$, and therefore a descendant of both $c_2$ and $c_1$. This implies that $c_1$ and $c_2$ are related as ancestor and descendant, which is absurd. Thus, we have that $B(u)\cap (B(w)\setminus\{e\})=\emptyset$.

Then, since $B(u)\subseteq B(v)$ and $B(w)\setminus\{e\}\subseteq B(v)$ and $B(u)\cap (B(w)\setminus\{e\})=\emptyset$ and $\mathit{bcount}(v)=\mathit{bcount}(u)+\mathit{bcount}(w)-1$, we have that $B(v)=B(u)\sqcup(B(w)\setminus\{e\})$.

Let us suppose, for the sake of contradiction, that $e\in B(u)$. Let $x$ be the higher endpoint of $e$. Notice that $M(w)=\mathit{nca}\{x,M(B(w)\setminus\{e\})\}$. Since $e\in B(u)$, we have that $x$ is a descendant of $M(u)$, and therefore a descendant of $c_2$. Since $M(B(w)\setminus\{e\})=M(v,c_1)$, we have that $M(B(w)\setminus\{e\})$ is a descendant of $c_1$. Thus, we have that $\mathit{nca}\{x,M(B(w)\setminus\{e\})\}=M(v)$, in contradiction to the assumption $M(w)\neq M(v)$. This shows that $e\notin B(u)$. Thus, $B(v)=B(u)\sqcup(B(w)\setminus\{e\})$ implies that $e\notin B(u)\cup B(v)$.
\end{proof}

\noindent\\
\begin{algorithm}[H]
\caption{\textsf{Compute a collection of Type-3$\beta i$ $4$-cuts that satisfy $(2)$ of Lemma~\ref{lemma:type-3b-cases}, so that all of them are implied by this collection, plus that returned by Algorithm~\ref{algorithm:type2-2}}}
\label{algorithm:type-3-b-i-2}
\LinesNumbered
\DontPrintSemicolon
\tcp{We deal with the case that the back-edge of the $4$-cut is $e_L(w)$; the other case is treated similarly}
\ForEach{vertex $w\neq r$}{
\label{line:type-3-b-i-2-for-1}
  compute $M(B(w)\setminus\{e_L(w)\})$\;
}
\ForEach{vertex $x$}{
  let $W_L(x)$ be the collection of all $w\neq r$ such that $M(w)\neq M(B(w)\setminus\{e_L(w)\})=x$\;
}
\ForEach{vertex $v\neq r$}{
\label{line:type-3-b-i-2-for-2}
  \If{$M(v)$ has at least two children}{
    let $c_1$ be the $\mathit{low1}$ child of $M(v)$\;
    let $c_2$ be the $\mathit{low2}$ child of $M(v)$\;
    compute $M(v,c_1)$ and $M(v,c_2)$\;
  }
}
\ForEach{vertex $v\neq r$}{
  \lIf{$M(v)$ has less than two children}{\textbf{continue}}
  let $c_1$ be the $\mathit{low1}$ child of $M(v)$\;
  let $c_2$ be the $\mathit{low2}$ child of $M(v)$\;
  \lIf{$\mathit{low}(c_1)\geq v$ \textbf{or} $\mathit{low}(c_2)\geq v$}{\textbf{continue}}
  \label{line:type-3-b-i-2-check}
  let $u$ be the lowest proper descendant of $v$ such that $M(u)=M(v,c_2)$\;
  \label{line:type-3-b-i-2-u}
  let $w$ be the greatest ancestor of $\mathit{low}(u)$ such that $w\in W_L(M(v,c_1))$\;
  \label{line:type-3-b-i-2-v}
  \If{$M(w)\neq M(v)$ \textbf{and} $\mathit{high}(u)<v$ \textbf{and} $\mathit{bcount}(v)=\mathit{bcount}(u)+\mathit{bcount}(w)-1$}{
    mark $\{(u,p(u)),(v,p(v)),(w,p(w)),e_L(w)\}$ as a $4$-cut\;
    \label{line:type-3-b-i-2-mark}
  }
}
\end{algorithm}

\begin{proposition}
\label{proposition:algorithm:type-3-b-i-2}
Algorithm~\ref{algorithm:type-3-b-i-2} computes a collection $\mathcal{C}$ of Type-3$\beta i$ $4$-cuts that satisfy $(2)$ of Lemma~\ref{lemma:type-3b-cases}, and it runs in linear time. Furthermore, every Type-3$\beta i$ $4$-cut that satisfies $(2)$ of Lemma~\ref{lemma:type-3b-cases} is implied by $\mathcal{C}\cup\mathcal{C}'$, where $\mathcal{C}'$ is the collection of Type-$2ii$ $4$-cuts returned by Algorithm~\ref{algorithm:type2-2}.
\end{proposition}
\begin{proof}
Let $C=\{(u,p(u)),(v,p(v)),(w,p(w)),e_L(w)\}$ be a $4$-element set that is marked in Line~\ref{line:type-3-b-i-2-mark}. Then we have that $\mathit{bcount}(v)=\mathit{bcount}(u)+\mathit{bcount}(w)-1$, $M(w)\neq M(v)$ and $\mathit{high}(u)<v$. Since $M(u)=M(v,c_2)$ we have that $M(u)$ is a descendant of $c_2$. Since $w\in W_L(M(v,c_1))$ we have that $M(B(w)\setminus\{e_L(w)\})=M(v,c_1)$, and therefore $M(B(w)\setminus\{e_L(w)\})$ is a descendant of $c_1$. We have that $u$ is a proper descendant of $v$, and therefore $\mathit{high}(u)<v$ implies that $\mathit{high}(u)$ is a proper ancestor of $v$. This implies that $\mathit{low}(u)$ is also a proper ancestor of $v$, and therefore $w$ being an ancestor of $\mathit{low}(u)$ implies that $w$ is a proper ancestor of $v$. Thus, all the conditions of Lemma~\ref{lemma:type-3-b-i-2-criterion} are satisfied, and therefore we have that $e_L(w)\notin B(u)\cup B(v)$ and $B(v)=B(u)\sqcup(B(w)\setminus\{e_L(w)\})$. Since $M(B(w)\setminus\{e_L(w)\})$ is a descendant of $c_1$, we have that $M(B(w)\setminus\{e_L(w)\})\neq M(v)$. Thus, we have that $C$ is a Type-3$\beta i$ $4$-cut, that satisfies $(2)$ of Lemma~\ref{lemma:type-3b-cases}. So let $\mathcal{C}$ be the collection of all $4$-cuts marked in Line~\ref{line:type-3-b-i-2-mark}.

Now let $C=\{(u,p(u)),(v,p(v)),(w,p(w)),e\}$ be a Type-3$\beta i$ $4$-cut such that $w$ is an ancestor of $v$, and $v$ is an ancestor of $u$, and $e$ satisfies $(2)$ of Lemma~\ref{lemma:type-3b-cases}. Let $c_1$ and $c_2$ be the $\mathit{low1}$ and $\mathit{low2}$ children of $M(v)$, respectively. Then, Lemma~\ref{lemma:type-3-b-i-2-info} implies that $w$ is an ancestor of $\mathit{low}(u)$, $M(w)\neq M(B(w)\setminus\{e\})=M(v,c_1)$, and $u$ is the lowest proper descendant of $v$ such that $M(u)=M(v,c_2)$. Thus, we may consider the greatest ancestor $w'$ of $\mathit{low}(u)$ with the property that there is a back-edge $e'\in B(w')$ such that $M(w')\neq M(B(w')\setminus\{e'\})=M(v,c_1)$. Since $M(w')\neq M(B(w')\setminus\{e'\})$, Lemma~\ref{lemma:e_L-e_R} implies that either $e'=e_L(w')$ or $e'=e_R(w')$. Let us assume that $e'=e_L(w')$. (The other case is treated similarly.) Then, Lemma~\ref{lemma:type-3-b-i-2-implied} implies that $e_L(w')\notin B(u)\cup B(v)$ and $B(v)=B(u)\sqcup(B(w')\setminus\{e_L(w')\})$. Thus, $C'=\{(u,p(u)),(v,p(v)),(w',p(w')),e_L(w')\}$ is a $4$-cut that satisfies $(2)$ of Lemma~\ref{lemma:type-3b-cases}. Furthermore, since $M(B(w')\setminus\{e'\})=M(v,c_1)$, we have that $M(B(w')\setminus\{e'\})$ is a descendant of $c_1$, and therefore $M(B(w')\setminus\{e'\})\neq M(v)$. Thus, $C'$ is a Type-3$\beta i$ $4$-cut. Then, Lemma~\ref{lemma:type-3-b-i-2-info} implies that $M(w')\neq M(v)$. Since, $e_L(w')\notin B(u)\cup B(v)$ and $B(v)=B(u)\sqcup(B(w')\setminus\{e_L(w')\})$, we have $\mathit{bcount}(v)=\mathit{bcount}(u)+\mathit{bcount}(w)-1$. Furthermore, we have $B(u)\subseteq B(v)$, and therefore $\mathit{high}(u)<v$ (because the lower endpoints of all back-edges in $B(u)$ are proper ancestors of $v$). Thus, all the conditions are satisfied for $C'$ to be marked in Line~\ref{line:type-3-b-i-2-mark}, and therefore we have $C\in\mathcal{C}$. Now, if $C'=C$, then there is nothing to show. Otherwise, we have $w'\neq w$, and therefore Lemma~\ref{lemma:type-3-b-i-2-implied} implies that $B(w)\sqcup\{e_L(w')\}=B(w')\sqcup\{e\}$. Thus, Lemma~\ref{lemma:type2cuts} implies that $C''=\{(w,p(w)),(w',p(w')),e,e_L(w')\}$ is a Type-$2ii$ $4$-cut. Notice that $C$ is implied by $C'$ and $C''$ through the pair of edges $\{(w,p(w)),e\}$. By Proposition~\ref{proposition:type-2-2} we have that $C''$ is implied by $\mathcal{C}'$ through the pair of edges $\{(w,p(w)),e\}$, where $\mathcal{C}'$ is the collection of Type-$2ii$ $4$-cuts computed by Algorithm~\ref{algorithm:type2-2}. Therefore, by Lemma~\ref{lemma:implied_from_union} we have that $C$ is implied by $\mathcal{C}'\cup\{C'\}$. This shows that $\mathcal{C}\cup\mathcal{C}'$ implies all Type-3$\beta i$ $4$-cuts that satisfy $(2)$ of Lemma~\ref{lemma:type-3b-cases}.

Now we will argue about the complexity of Algorithm~\ref{algorithm:type-3-b-i-2}. By Proposition~\ref{proposition:computing-M(B(v)-S)} we have that the values $M(B(w)\setminus\{e_L(w)\})$ can be computed in linear time in total, for all vertices $w\neq r$. Thus, the \textbf{for} loop in Line~\ref{line:type-3-b-i-2-for-1} can be performed in linear time. By Proposition~\ref{proposition:computing-M(v,c)} we have that the values $M(v,c_1)$ and $M(v,c_2)$ can be computed in linear time in total, for all vertices $v\neq r$ such that $M(v)$ has at last two children, where $c_1$ and $c_2$ are the $\mathit{low1}$ and $\mathit{low2}$ children of $M(v)$. Thus, the \textbf{for} loop in Line~\ref{line:type-3-b-i-2-for-2} can be performed in linear time. In order to compute the vertex $u$ in Line~\ref{line:type-3-b-i-2-u} we use Algorithm~\ref{algorithm:W-queries}. Specifically, whenever we reach Line~\ref{line:type-3-b-i-2-u}, we generate a query $q(M^{-1}(M(v,c_2)),v)$. This will return the lowest vertex $u$ with $M(u)=M(v,c_2)$ such that $u>v$. Since $M(u)=M(v,c_2)$, we have that $M(u)$ is a common descendant of $v$ and $u$, and therefore $v$ and $u$ are related as ancestor and descendant. Thus, $u>v$ implies that $u$ is a proper descendant of $v$, and therefore we have that $u$ is the greatest proper descendant of $v$ such that $M(u)=M(v,c_2)$. Similarly, in order to compute the vertex $v$ in Line~\ref{line:type-3-b-i-2-v} we use Algorithm~\ref{algorithm:W-queries}. Specifically, whenever we reach Line~\ref{line:type-3-b-i-2-v}, we generate a query $q'(W_L(M(v,c_1)),\mathit{low}(u))$. This is to return the greatest $w$ in $W_L(M(v,c_1))$ such that $w\leq\mathit{low}(u)$. Since $w\in W_L(M(v,c_1))$, we have that $M(B(w)\setminus\{e_L(w)\})=M(v,c_1)$. This implies that $M(v,c_1)$ is a common descendant of $v$ and $w$, and therefore $v$ and $w$ are related as ancestor and descendant. Notice that, since we have reached Line~\ref{line:type-3-b-i-2-v}, we have that $u$ in Line~\ref{line:type-3-b-i-2-u} satisfies $M(u)=M(v,c_2)$. This implies that $\mathit{low}(u)<v$. To see this, let $(x,y)$ be a back-edge in $B(v)$ such that $x$ is a descendant of $c_2$. (Such a back-edge exists, because $\mathit{low}(c_2)<v$, since we have passed the check in Line~\ref{line:type-3-b-i-2-check}.) Then $x$ is a descendant of $M(v,c_2)$, and therefore a descendant of $M(u)$. Furthermore, $y$ is a proper ancestor of $v$, and therefore a proper ancestor of $u$. This shows that $(x,y)\in B(u)$. Therefore, since $\mathit{low}(u)\leq y$ and $y$ is a proper ancestor of $v$, we have $\mathit{low}(u)<v$. Furthermore, we have that $\mathit{low}(u)$ is an ancestor of $y$, and therefore an ancestor of $v$. Then, since $v$ is a common descendant of $\mathit{low}(u)$ and $w$, we have that $\mathit{low}(u)$ and $w$ are related as ancestor and descendant, and then $w\leq\mathit{low}(u)$ implies that $w$ is an ancestor of $\mathit{low}(u)$. Thus, $w$ is the greatest ancestor of $\mathit{low}(u)$ such that $w\in W_L(M(v,c_1))$. Now, since the total number of all $q'$ queries is $O(n)$, and since the sets $W_L$ are pairwise disjoint, by Lemma~\ref{lemma:W-queries} we have that Algorithm~\ref{algorithm:W-queries} can answer all those queries in linear time in total.
We conclude that Algorithm~\ref{algorithm:type-3-b-i-2} runs in linear time.
\end{proof}

\subsubsection{Case $(3)$ of Lemma~\ref{lemma:type-3b-cases}}

\begin{lemma}
\label{lemma:type-3-b-i-3-info}
Let $u,v,w$ be three vertices $\neq r$ such that $w$ is proper ancestor of $v$, $v$ is a proper ancestor of $u$, and there is a back-edge $e\in B(u)$ such that $e\notin B(v)\cup B(w)$, $B(v)=(B(u)\setminus\{e\})\sqcup B(w)$, and $M(v)\neq M(w)$. Then $e=(\mathit{highD}(u),\mathit{high}(u))$ and $\mathit{high}_2(u)<v$. Furthermore, let $c_1$ and $c_2$ be the $\mathit{low1}$ and $\mathit{low2}$ child of $M(v)$, respectively. Then, $M(B(u)\setminus\{e_\mathit{high}(u)\})=M(v,c_2)$, and $w$ is the greatest proper ancestor of $v$ such that $M(w)=M(v,c_1)$.
\end{lemma}
\begin{proof}
$B(v)=(B(u)\setminus\{e\})\sqcup B(w)$ and $e\notin B(v)$ imply that all back-edges in $B(u)$, except $e$, are in $B(v)$. Let us suppose, for the sake of contradiction, that $e_\mathit{high}(u)\in B(v)$. This implies that $\mathit{high}(u)$ is a proper ancestor of $v$. Let $(x,y)$ be a back-edge in $B(u)$. Then $x$ is a descendant of $u$, and therefore a descendant of $v$. Furthermore, $y$ is an ancestor of $\mathit{high}(u)$, and therefore a proper ancestor of $v$. This shows that $(x,y)\in B(v)$. Due to the generality of $(x,y)\in B(u)$, this implies that $B(u)\subseteq B(v)$, in contradiction to the fact that $e\in B(u)\setminus B(v)$. This shows that $e_\mathit{high}(u)\notin B(v)$, and therefore $e=e_\mathit{high}(u)$. Since $B(u)\setminus\{e\}\subseteq B(v)$, we have that the $\mathit{high2}$ back-edge of $B(u)$ is in $B(v)$, and therefore $\mathit{high}_2(u)$ is a proper ancestor of $v$, and therefore $\mathit{high}_2(u)<v$.

$B(v)=(B(u)\setminus\{e\})\sqcup B(w)$ implies that $B(w)\subseteq B(v)$, and therefore $M(w)$ is a descendant of $M(v)$. Thus, since $M(v)\neq M(w)$, we have that $M(w)$ is a proper descendant of $M(v)$. Let $c$ be the child of $M(v)$ that is an ancestor of $M(w)$. Let $(x,y)$ be a back-edge in $B(w)$. Then $x$ is a descendant of $M(w)$, and therefore a descendant of $c$, and therefore a descendant of $M(v)$. Furthermore, $y$ is a proper ancestor of $v$, and therefore a proper ancestor of $M(v)$, and therefore a proper ancestor of $c$. This shows that $(x,y)\in B(c)$. Since $y$ is a proper ancestor of $w$, we have $y<w$. Thus, since $(x,y)\in B(c)$, we have $\mathit{low}(c)\leq y<w$. 

Let us suppose, for the sake of contradiction, that $c\neq c_1$. Since $\mathit{low}(c_1)\leq\mathit{low}(c)$, $\mathit{low}(c)<w$ implies that $\mathit{low}(c_1)<w$. Thus, there is a back-edge $(x,y)\in B(c_1)$ such that $y<w$. Then, $x$ is a descendant of $c_1$, and therefore a descendant of $M(v)$, and therefore a descendant of $v$, and therefore a descendant of $w$. Since $(x,y)$ is a back-edge, we have that $x$ is a descendant of $y$. Thus, $x$ is a common descendant of $w$ and $y$, and therefore $w$ and $y$ are related as ancestor and descendant. Then, since $y<w$, we have that $y$ is a proper ancestor of $w$. This shows that $(x,y)\in B(w)$. This implies that $x$ is a descendant of $M(w)$, and therefore $x$ is a descendant of $c$. Thus, $x$ is a common descendant of $c$ and $c_1$, and therefore $c$ and $c_1$ are related as ancestor and descendant, which is absurd. Thus, we have that $c=c_1$. 

$B(v)=(B(u)\setminus\{e\})\sqcup B(w)$ implies that $B(u)\setminus\{e\}\subseteq B(v)$, and therefore $M(B(u)\setminus\{e\})$ is a descendant of $M(v)$. Let us suppose, for the sake of contradiction, that $M(B(u)\setminus\{e\})=M(v)$. Let $(x,y)$ be a back-edge in $B(w)$. Then $B(v)=(B(u)\setminus\{e\})\sqcup B(w)$ implies that $(x,y)\in B(v)$. Therefore, $x$ is a descendant of $M(v)$, and therefore a descendant of $M(B(u)\setminus\{e\})$, and therefore a descendant of $u$. Furthermore, $y$ is a proper ancestor of $v$, and therefore a proper ancestor of $u$. This shows that $(x,y)\in B(u)$. Since $(x,y)\in B(w)$ and $e\notin B(w)$, we have that $(x,y)\neq e$. Therefore, $(x,y)\in B(u)$ can be strengthened to $(x,y)\in B(u)\setminus\{e\}$. But then we have a contradiction to (the disjointness of the union in) $B(v)=(B(u)\setminus\{e\})\sqcup B(w)$. This shows that $M(B(u)\setminus\{e\})\neq M(v)$, and therefore $M(B(u)\setminus\{e\})$ is a proper descendant of $M(v)$. So let $c'$ be the child of $M(v)$ that is an ancestor of $M(B(u)\setminus\{e\})$.

Let $(x,y)$ be a back-edge in $B(u)\setminus\{e\}$. Then $x$ is a descendant of $M(B(u)\setminus\{e\})$, and therefore a descendant of $c'$. Furthermore, $B(v)=(B(u)\setminus\{e\})\sqcup B(w)$ implies that $(x,y)\in B(v)$, and therefore $y$ is a proper ancestor of $v$, and therefore a proper ancestor of $M(v)$, and therefore a proper ancestor of $c'$. This shows that $(x,y)\in B(c')$. Since $y$ is a proper ancestor of $v$, we have $y<v$. Thus, since $(x,y)\in B(c')$, we have $\mathit{low}(c')\leq y< v$.

Let us suppose, for the sake of contradiction, that there is a back-edge of the form $(M(v),z)$ in $B(v)$. Then $B(v)=(B(u)\setminus\{e\})\sqcup B(w)$ implies that either $(M(v),z)\in B(u)\setminus\{e\}$, or $(M(v),z)\in B(w)$. The first case implies that $M(v)$ is a descendant of $M(B(u)\setminus\{e\})$, and therefore a descendant of $c'$, which is a absurd. The second case implies that $M(v)$ is a descendant of $M(w)$, and therefore a descendant of $c_1$, which is also absurd. This shows that there is no back-edge of the form $(M(v),z)$ in $B(v)$. Thus, Lemma~\ref{lemma:no_M(v)} implies that $\mathit{low}(c_2)<v$.

Let us suppose, for the sake of contradiction, that $c'\neq c_2$. Since $\mathit{low}(c_2)<v$, there is a back-edge $(x,y)\in B(c_2)$ such that $y<v$. Then, $x$ is a descendant of $c_2$, and therefore a descendant of $M(v)$, and therefore a descendant of $v$. Since $(x,y)$ is a back-edge, we have that $x$ is a descendant of $y$. Thus, since $x$ is a common descendant of both $v$ and $y$, we have that $v$ and $y$ are related as ancestor and descendant. Thus, $y<v$ implies that $y$ is a proper ancestor of $v$. This shows that $(x,y)\in B(v)$. Then, $B(v)=(B(u)\setminus\{e\})\sqcup B(w)$ implies that either $(x,y)\in B(u)\setminus\{e\}$, or $(x,y)\in B(w)$. The first case implies that $x$ is a descendant of $M(B(u)\setminus\{e\})$, and therefore a descendant of $c'$, which is absurd, since $x$ is a descendant of $c_2$ (and $c',c_2$ are not related as ancestor and descendant, since they have the same parent and $c'\neq c_2$). The second case implies that $x$ is a descendant of $M(w)$, and therefore a descendant of $c_1$, which is absurd, since $x$ is a descendant of $c_2$ (and $c_1,c_2$ are not related as ancestor and descendant, since they have the same parent). There are no viable options left, and so we have arrived at a contradiction. This shows that $c'=c_2$.

Let $S_1=\{(x,y)\in B(v)\mid x \mbox{ is a descendant of } c_1\}$. Then, $M(S_1)=M(v,c_1)$. Let $(x,y)$ be a back-edge in $S_1$. Then, $(x,y)\in B(v)$, and therefore $B(v)=(B(u)\setminus\{e\})\sqcup B(w)$ implies that either $(x,y)\in B(u)\setminus\{e\}$, or $(x,y)\in B(w)$. Let us suppose, for the sake of contradiction, that $(x,y)\in B(u)\setminus\{e\}$. Then $x$ is a descendant of $M(B(u)\setminus\{e\})$, and therefore a descendant of $c_2$. Thus, $x$ is a common descendant of $c_1$ and $c_2$, and so $c_1$ and $c_2$ are related as ancestor and descendant, which is absurd. Thus, the case $(x,y)\in B(u)\setminus\{e\}$ is rejected, and so we have $(x,y)\in B(w)$. Due to the generality of $(x,y)\in S_1$, this implies that $S_1\subseteq B(w)$.
Conversely, let $(x,y)$ be a back-edge in $B(w)$. Then, $x$ is a descendant of $M(w)$, and therefore a descendant of $c_1$, and therefore a descendant of $M(v)$. Furthermore, $y$ is a proper ancestor of $w$, and therefore a proper ancestor of $v$. This shows that $(x,y)\in B(v)$, and so we have $(x,y)\in S_1$. Due to the generality of $(x,y)\in B(w)$, this implies that $B(w)\subseteq S_1$. Thus, since $S_1\subseteq B(w)$, we have $S_1=B(w)$, and therefore $M(S_1)=M(w)$, and therefore $M(v,c_1)=M(w)$.

Let $S_2=\{(x,y)\in B(v)\mid x \mbox{ is a descendant of } c_2\}$. Then, $M(S_2)=M(v,c_2)$. Let $(x,y)$ be a back-edge in $S_2$. Then, $(x,y)\in B(v)$, and therefore $B(v)=(B(u)\setminus\{e\})\sqcup B(w)$ implies that either $(x,y)\in B(u)\setminus\{e\}$, or $(x,y)\in B(w)$. Let us suppose, for the sake of contradiction, that $(x,y)\in B(w)$. Then $x$ is a descendant of $M(w)$, and therefore a descendant of $c_1$. Thus, $x$ is a common descendant of $c_1$ and $c_2$, and so $c_1$ and $c_2$ are related as ancestor and descendant, which is absurd. Thus, the case $(x,y)\in B(w)$ is rejected, and so we have $(x,y)\in B(u)\setminus\{e\}$. Due to the generality of $(x,y)\in S_2$, this implies that $S_2\subseteq B(u)\setminus\{e\}$.
Conversely, let $(x,y)$ be a back-edge in $B(u)\setminus\{e\}$. Then, $x$ is a descendant of $M(B(u)\setminus\{e\})$, and therefore a descendant of $c_2$. Furthermore, since $B(v)=(B(u)\setminus\{e\})\sqcup B(w)$, we have that $(x,y)\in B(v)$. This shows that $(x,y)\in S_2$. Due to the generality of $(x,y)\in B(u)\setminus\{e\}$, this implies that $B(u)\setminus\{e\}\subseteq S_2$. Thus, since $S_2\subseteq B(u)\setminus\{e\}$, we have $S_2=B(u)\setminus\{e\}$, and therefore $M(S_2)=M(B(u)\setminus\{e\})$, and therefore $M(v,c_2)=M(B(u)\setminus\{e\})$.

Let us suppose, for the sake of contradiction, that there is a proper ancestor $w'$ of $v$ that is greater than $w$ and has $M(w')=M(v,c_1)$. Since $M(w')=M(w)$ and $w'>w$, we have that $w'$ is a proper descendant of $w$, and Lemma~\ref{lemma:same_m_subset_B} implies that $B(w)\subseteq B(w')$. Since the graph is $3$-edge-connected, this can be strengthend to $B(w)\subset B(w')$. This implies that there is a back-edge $(x,y)\in B(w')\setminus B(w)$. Then, $x$ is a descendant of $M(w')$, and therefore a descendant of $M(v,c_1)$, and therefore a descendant of $v$. Furthermore, $y$ is a proper ancestor of $w'$, and therefore a proper ancestor of $v$. This shows that $(x,y)\in B(v)$. Then, $B(v)=(B(u)\setminus\{e\})\sqcup B(w)$ implies that either $(x,y)\in B(u)\setminus\{e\}$, or $(x,y)\in B(w)$. The case $(x,y)\in B(w)$ is rejected (since $(x,y)\in B(w')\setminus B(w)$), and therefore $(x,y)\in B(u)\setminus\{e\}$. This implies that $x$ is a descendant of $M(B(u)\setminus\{e\})=M(v,c_2)$, and therefore a descendant of $c_2$. But then we have that $x$ is a descendant of both $c_2$ and $c_1$, and therefore $c_1$ and $c_2$ are related as ancestor and descendant, which is absurd. This shows that $w$ is the greatest proper ancestor of $v$ that has $M(w)=M(v,c_1)$.
\end{proof}

We distinguish two cases for the Type-3$\beta i$ $4$-cuts of the form $\{(u,p(u)),(v,p(v)),(w,p(w)),e\}$, where $u$ is a descendant of $v$, $v$ is a descendant of $w$, and $e$ satisfies $(3)$ of Lemma~\ref{lemma:type-3b-cases}: either $M(u)=M(B(u)\setminus\{e\})$, or $M(u)\neq M(B(u)\setminus\{e\})$. In the first case, we can compute all such $4$-cuts explicitly. In the second case, we compute only a subcollection $\mathcal{C}$ of them, so that the rest are implied by $\mathcal{C}\cup\mathcal{C}'$, where $\mathcal{C}'$ is the collection of Type-$2ii$ $4$-cuts that are computed by Algorithm~\ref{algorithm:type2-2}. Our result is summarized and proved in Proposition~\ref{proposition:algorithm:type-3-b-i-3}.

The reason that the $4$-cuts in the first case can be computed explicitly is the following.

\begin{lemma}
\label{lemma:type-3-b-i-3-special}
Let $u,v,w$ be three vertices $\neq r$ such that $u$ is a proper descendant of $v$, $v$ is a proper descendant of $w$, and $B(v)=(B(u)\setminus\{e\})\sqcup B(w)$, where $e$ is a back-edge in $B(u)$ such that $e\notin B(v)$. Suppose that $M(u)=M(B(u)\setminus\{e\})$. Then $u$ is either the lowest or the second-lowest proper descendant of $v$ in $M^{-1}(M(u))$.
\end{lemma}
\begin{proof}
Let us suppose, for the sake of contradiction, that $u$ is neither the lowest nor the second-lowest proper descendant of $v$ in $M^{-1}(M(u))$. This means that there are two proper descendants $u'$ and $u''$ of $v$, such that $u>u'>u''$ and $M(u')=M(u'')=M(u)$. Since $M(u)=M(u')=M(u'')$, we have that $u$, $u'$ and $u''$ are related as ancestor and descendant, and therefore $u>u'>u''$ implies that $u$ is a proper descendant of $u'$, and $u'$ is a proper descendant of $u''$. Then, Lemma~\ref{lemma:same_m_subset_B} implies that $B(u'')\subseteq B(u')\subseteq B(u)$. Since the graph is $3$-edge-connected, this can be strengthened to $B(u'')\subset B(u')\subset B(u)$. Thus, there is a back-edge $(x,y)\in B(u)\setminus B(u')$, and a back-edge $(x',y')\in B(u')\setminus B(u'')$. Since $(x,y)\in B(u)$, we have that $x$ is a descendant of $M(u)$, and therefore a descendant of $u'$. Thus, since $(x,y)\notin B(u')$, we have that $y$ is not a proper ancestor of $u'$. This implies that $y$ is not a proper ancestor of $v$ either, and therefore $(x,y)\notin B(v)$. Similarly, since $(x',y')\in B(u')\setminus B(u'')$, we have that $y'$ is not a proper ancestor of $u''$, and therefore not a proper ancestor of $v$, and therefore $(x',y')\notin B(v)$. Notice that both $(x,y)$ and $(x',y')$ are in $B(u)$ (since $B(u'')\subset B(u')\subset B(u)$). Furthermore, since $(x',y')\in B(u')$ and $(x,y)\notin B(u')$, we have that $(x,y)\neq (x',y')$. Thus, $(x,y)$ and $(x',y')$ are two distinct back-edges in $B(u)$ that are not in $B(v)$. But this contradicts the fact $B(v)=(B(u)\setminus\{e\})\sqcup B(w)$, which implies that $e$ is the only back-edge from $B(u)$ that is not in $B(v)$. This shows that $u$ is  either the lowest or the second-lowest proper descendant of $v$ in $M^{-1}(M(u))$.
\end{proof}

In the case where $M(u)\neq M(B(u)\setminus\{e\})$, we consider the lowest proper descendant $u'$ of $v$ that satisfies $M(u')\neq M(B(u')\setminus\{e_\mathit{high}(u')\})=M(B(u)\setminus\{e\})$. Then, the following shows that we still get a Type-$3\beta i$ $4$-cut with $v$ and $w$.

\begin{lemma}
\label{lemma:type-3-b-i-3-non-special}
Let $u,v,w$ be three vertices $\neq r$ such that $w$ is a proper ancestor of $v$, $v$ is a proper ancestor of $u$, and there is a back-edge $e\in B(u)$ such that $e\notin B(v)\cup B(w)$ and $B(v)=(B(u)\setminus\{e\})\sqcup B(w)$. Suppose that $M(u)\neq M(B(u)\setminus\{e\})$. Let $u'$ be the lowest proper descendant of $v$ such that $M(u')\neq M(B(u')\setminus\{e_\mathit{high}(u')\})=M(B(u)\setminus\{e\})$. Then we have $e_\mathit{high}(u')\notin B(v)\cup B(w)$ and $B(v)=(B(u')\setminus\{e_\mathit{high}(u')\})\sqcup B(w)$. Furthermore, if $u'\neq u$, then $B(u)\sqcup\{e_\mathit{high}(u')\}=B(u')\sqcup\{e\}$.
\end{lemma}
\begin{proof}
By the proof of Lemma~\ref{lemma:type-3-b-i-3-info}, we have $e=e_\mathit{high}(u)$. (This result does not rely on the supposition $M(w)\neq M(v)$, which is included in the statement of Lemma~\ref{lemma:type-3-b-i-3-info}.) Thus, it makes sense to consider the lowest proper descendant $u'$ of $v$ that satisfies $M(u')\neq M(B(u')\setminus\{e_\mathit{high}(u')\})=M(B(u)\setminus\{e\})$. If $u'=u$, then there is nothing to show. So let us assume that $u'\neq u$. Then, due to the minimality of $u'$, we have $u'<u$. Since $M(B(u')\setminus\{e_\mathit{high}(u')\})=M(B(u)\setminus\{e\})$, we have that $M(B(u)\setminus\{e\})$ is a common descendant of $u$ and $u'$. Therefore, $u$ and $u'$ are related as ancestor and descendant, and then $u'<u$ implies that $u'$ is a proper ancestor of $u$.

Since $M(u')\neq M(B(u')\setminus\{e_\mathit{high}(u')\})$, we have that $e_\mathit{high}(u')$ is the only back-edge in $B(u')$ whose higher endpoint is not a descendant of $M(B(u')\setminus\{e_\mathit{high}(u')\})$. Similarly, since $M(u)\neq M(B(u)\setminus\{e\})$, we have that $e$ is the only back-edge in $B(u)$ whose higher endpoint is not a descendant of $M(B(u)\setminus\{e\})$.

Let $(x,y)$ be a back-edge in $B(u')\setminus\{e_\mathit{high}(u')\}$. Then $x$ is a descendant of $M(B(u')\setminus\{e_\mathit{high}(u')\})$, and therefore a descendant of $M(B(u)\setminus\{e\})$, and therefore a descendant of $M(u)$. Furthermore, $y$ is a proper ancestor of $u'$, and therefore a proper ancestor of $u$. This shows that $(x,y)\in B(u)$. Due to the generality of $(x,y)\in B(u')\setminus\{e_\mathit{high}(u')\}$, this implies that $B(u')\setminus\{e_\mathit{high}(u')\}\subseteq B(u)$. Since the higher endpoint of $e$ is not a descendant of $M(B(u)\setminus\{e\})=M(B(u')\setminus\{e_\mathit{high}(u')\})$, we have that $e\notin B(u')\setminus\{e_\mathit{high}(u')\}$. Thus, $B(u')\setminus\{e_\mathit{high}(u')\}\subseteq B(u)$ can be strengthened to $B(u')\setminus\{e_\mathit{high}(u')\}\subseteq B(u)\setminus\{e\}$. Conversely, let $(x,y)$ be a back-edge in $B(u)\setminus\{e\}$. Then $x$ is a descendant of $M(B(u)\setminus\{e\})$, and therefore a descendant of $M(B(u')\setminus\{e_\mathit{high}(u')\})$, and therefore a descendant of $M(u')$. Furthermore, $B(v)=(B(u)\setminus\{e\})\sqcup B(w)$ implies that $(x,y)\in B(v)$, and therefore $y$ is a proper ancestor of $v$, and therefore a proper ancestor of $u'$. This shows that $(x,y)\in B(u')$. Due to the generality of $(x,y)\in B(u)\setminus\{e\}$, this implies that $B(u)\setminus\{e\}\subseteq B(u')$. Since the higher endpoint of $e_\mathit{high}(u')$ is not a descendant of $M(B(u')\setminus\{e_\mathit{high}(u')\})=M(B(u)\setminus\{e\})$, we have that $e_\mathit{high}(u')\notin B(u)\setminus\{e\}$. Thus, $B(u)\setminus\{e\}\subseteq B(u')$ can be strengthened to $B(u)\setminus\{e\}\subseteq B(u')\setminus\{e_\mathit{high}(u')\}$. 

Thus, we have shown that $B(u)\setminus\{e\}= B(u')\setminus\{e_\mathit{high}(u')\}$. Then, notice that we cannot have $e\in B(u')$ or $e_\mathit{high}(u')\in B(u)$, because otherwise we get $B(u)=B(u')$, in contradiction to the fact that the graph is $3$-edge-connected. Thus, $B(u)\setminus\{e\}= B(u')\setminus\{e_\mathit{high}(u')\}$ implies that $B(u)\sqcup\{e_\mathit{high}(u')\}=B(u')\sqcup\{e\}$. Furthermore, since $B(v)=(B(u)\setminus\{e\})\sqcup B(w)$ and $B(u)\setminus\{e\}= B(u')\setminus\{e_\mathit{high}(u')\}$, we infer that $B(v)=(B(u')\setminus\{e_\mathit{high}(u')\})\sqcup B(w)$. 

Since the graph is $3$-edge-connected, we have $|(B(u')|>1$. Thus, there is a back-edge $(x,y)\in B(u')\setminus\{e_\mathit{high}(u')\}$. Then we have that $x$ is a descendant of $u'$, and therefore a descendant of $v$, and therefore a descendant of $w$. Thus, since $(B(u')\setminus\{e_\mathit{high}(u')\})\cap B(w)=\emptyset$, we have that $y$ is not a proper ancestor of $w$. Since $(x,y)$ is a back-edge, $x$ is a descendant of $y$. Thus, $x$ is a common descendant of $y$ and $w$, and therefore $y$ and $w$ are related as ancestor and descendant. Thus, since $y$ is not a proper ancestor of $w$, we have that $y$ is a descendant of $w$, and therefore $y\geq w$. Since $(x,y)\in B(u')$, we have that $\mathit{high}_1(u')\geq y$, and therefore $\mathit{high}_1(u')\geq w$. This implies that $e_\mathit{high}(u')\notin B(w)$ (because the lower endpoint of $e_\mathit{high}(u')$ is not low enough to be a proper ancestor of $w$). Then,  $B(v)=(B(u')\setminus\{e_\mathit{high}(u')\})\sqcup B(w)$ implies that $e_\mathit{high}(u')\notin B(v)\cup B(w)$.
\end{proof}

\begin{lemma}
\label{lemma:type-3-b-i-3-criterion}
Let $v\neq r$ be a vertex such that $M(v)$ has at least two children. Let $u$ be a proper descendant of $v$ such that $M(B(u)\setminus\{e_\mathit{high}(u)\})$ is a proper descendant of $M(v)$, and let $w$ be a proper ancestor of $v$ such that $M(w)$ is a proper descendant of $M(v)$. Suppose that $M(u)$ and $M(w)$ are not related as ancestor and descendant, $\mathit{high}_2(u)<v$, and $\mathit{bcount}(u)=\mathit{bcount}(v)-\mathit{bcount}(w)+1$. Then, $e_\mathit{high}(u)\notin B(v)\cup B(w)$ and $B(v)=(B(u)\setminus\{e_\mathit{high}(u)\})\sqcup B(w)$.
\end{lemma}
\begin{proof} 
Let $(x,y)$ be a back-edge in $B(u)\setminus\{e_\mathit{high}(u)\}$. Then, $x$ is a descendant of $M(B(u)\setminus\{e_\mathit{high}(u)\})$, and therefore a descendant of $M(v)$, and therefore a descendant of $v$. Since $(x,y)$ is a back-edge, $x$ is a descendant of $y$. Thus, $x$ is a common descendant of $y$ and $v$, and therefore $y$ and $v$ are related as ancestor and descendant. Since $(x,y)\in B(u)\setminus\{e_\mathit{high}(u)\}$, we have that $y$ is an ancestor of $\mathit{high}_2(u)$, and therefore $y\leq\mathit{high}_2(u)$. Thus, since $\mathit{high}_2(u)<v$, we have that $y$ is a proper ancestor of $v$. This shows that $(x,y)\in B(v)$. Due to the generality of $(x,y)\in B(u)\setminus\{e_\mathit{high}(u)\}$, this implies that $B(u)\setminus\{e_\mathit{high}(u)\}\subseteq B(v)$.

Let $(x,y)$ be a back-edge in $B(w)$. Then, $x$ is a descendant of $M(w)$, and therefore a descendant of $M(v)$. Furthermore, $y$ is a proper ancestor of $w$, and therefore a proper ancestor of $v$. This shows that $(x,y)\in B(v)$. Due to the generality of $(x,y)\in B(w)$, this implies that $B(w)\subseteq B(v)$.

Let us suppose, for the sake of contradiction, that $B(u)\cap B(w)\neq\emptyset$. Then there is a back-edge $(x,y)\in B(u)\cap B(w)$. This implies that $x$ is a descendant of both $M(u)$ and $M(w)$, and therefore $M(u)$ and $M(w)$ are related as ancestor and descendant. A contradiction. Thus, we have that $B(u)\cap B(w)=\emptyset$.

Now, since $B(u)\setminus\{e_\mathit{high}(u)\}\subseteq B(v)$, $B(w)\subseteq B(v)$ and $B(u)\cap B(w)=\emptyset$, we have that $(B(u)\setminus\{e_\mathit{high}(u)\})\sqcup B(w)\subseteq B(v)$. Then, $\mathit{bcount}(u)=\mathit{bcount}(v)-\mathit{bcount}(w)+1$ implies that $(B(u)\setminus\{e_\mathit{high}(u)\})\sqcup B(w)= B(v)$. Since $B(u)\cap B(w)=\emptyset$, we have that $e_\mathit{high}(u)\notin B(w)$. Thus,  $(B(u)\setminus\{e_\mathit{high}(u)\})\sqcup B(w)= B(v)$ implies that $e_\mathit{high}(u)\notin B(v)$, and so we have $e_\mathit{high}(u)\notin B(v)\cup B(w)$.
\end{proof}

\noindent\\
\begin{algorithm}[H]
\caption{\textsf{Compute a collection of Type-3$\beta i$ $4$-cuts that satisfy $(3)$ of Lemma~\ref{lemma:type-3b-cases}, so that all of them are implied by this collection, plus that returned by Algorithm~\ref{algorithm:type2-2}}}
\label{algorithm:type-3-b-i-3}
\LinesNumbered
\DontPrintSemicolon
\ForEach{vertex $u\neq r$}{
\label{line:type-3-b-i-3-for1}
  compute $M(B(u)\setminus\{e_\mathit{high}(u)\})$\;
}
\ForEach{vertex $v\neq r$}{
\label{line:type-3-b-i-3-for2}
  \If{$M(v)$ has at least two children}{
    let $c_1$ be the $\mathit{low1}$ child of $M(v)$\;
    let $c_2$ be the $\mathit{low2}$ child of $M(v)$\;
    compute $M(v,c_1)$ and $M(v,c_2)$\;
  }
}
\ForEach{vertex $v\neq r$}{
  \lIf{$M(v)$ has less than two children}{\textbf{continue}}
  let $c_1$ be the $\mathit{low1}$ child of $M(v)$\;
  let $c_2$ be the $\mathit{low2}$ child of $M(v)$\;
  let $u$ be the lowest proper descendant of $v$ that has $M(u)\neq M(B(u)\setminus\{e_\mathit{high}(u)\})=M(v,c_2)$\;
  \label{line:type-3-b-i-3-u}
  let $w$ be the greatest proper ancestor of $v$ that has $M(w)=M(v,c_1)$\;
  \label{line:type-3-b-i-3-v}
  \If{$\mathit{high}_2(u)<v$ \textbf{and} $\mathit{bcount}(v)=\mathit{bcount}(u)+\mathit{bcount}(w)-1$}{
    mark $\{(u,p(u)),(v,p(v)),(w,p(w)),e_\mathit{high}(u)\}$ as a $4$-cut\;
    \label{line:type-3-b-i-3-mark1}
  }
  let $u$ be the lowest proper descendant of $v$ that has $M(u)= M(B(u)\setminus\{e_\mathit{high}(u)\})=M(v,c_2)$\;
  \label{line:type-3-b-i-3-u-1}
  \If{$\mathit{high}_2(u)<v$ \textbf{and} $\mathit{bcount}(v)=\mathit{bcount}(u)+\mathit{bcount}(w)-1$}{
    mark $\{(u,p(u)),(v,p(v)),(w,p(w)),e_\mathit{high}(u)\}$ as a $4$-cut\;
    \label{line:type-3-b-i-3-mark2}
  }
  $u\leftarrow\mathit{prevM}(u)$\;
  \label{line:type-3-b-i-3-u-2}
  \If{$\mathit{high}_2(u)<v$ \textbf{and} $\mathit{bcount}(v)=\mathit{bcount}(u)+\mathit{bcount}(w)-1$}{
    mark $\{(u,p(u)),(v,p(v)),(w,p(w)),e_\mathit{high}(u)\}$ as a $4$-cut\;
    \label{line:type-3-b-i-3-mark3}
  }
}
\end{algorithm}

\begin{proposition}
\label{proposition:algorithm:type-3-b-i-3}
Algorithm~\ref{algorithm:type-3-b-i-3} computes a collection $\mathcal{C}$ of Type-3$\beta i$ $4$-cuts that satisfy $(3)$ of Lemma~\ref{lemma:type-3b-cases}, and it runs in linear time. Furthermore, every Type-3$\beta i$ $4$-cut that satisfies $(3)$ of Lemma~\ref{lemma:type-3b-cases} is implied by $\mathcal{C}\cup\mathcal{C}'$, where $\mathcal{C}'$ is the collection of Type-$2ii$ $4$-cuts returned by Algorithm~\ref{algorithm:type2-2}.
\end{proposition}
\begin{proof}
Let $C=\{(u,p(u)),(v,p(v)),(w,p(w)),e_\mathit{high}(u)\}$ be a $4$-element set that is marked in Lines~\ref{line:type-3-b-i-3-mark1}, or \ref{line:type-3-b-i-3-mark2}, or \ref{line:type-3-b-i-3-mark3}. Then we have that $\mathit{bcount}(v)=\mathit{bcount}(u)+\mathit{bcount}(w)-1$ and $\mathit{high}_2(u)<v$. Furthermore, $w$ is a proper ancestor of $v$ such that $M(w)$ is a proper descendant of $c_1$ (since $M(w)=M(v,c_1)$). In Lines~\ref{line:type-3-b-i-3-u} and \ref{line:type-3-b-i-3-u-1}, it is clear that $u$ is a proper descendant of $v$ such that $M(B(u)\setminus\{e_\mathit{high}(u)\})$ is a proper descendant of $c_2$. Then, in Line~\ref{line:type-3-b-i-3-u-2}, we still have that $M(B(u)\setminus\{e_\mathit{high}(u)\})$ is a proper descendant of $M(v)$, and now $u$ is even greater than previously. Furthermore, since $M(u)=M(B(u)\setminus\{e_\mathit{high}(u)\})$ and $M(B(u)\setminus\{e_\mathit{high}(u)\})=M(v,c_2)$, we have that $u$ is an ancestor of $M(v,c_2)$, and therefore $u$ is still a proper descendant of $v$. Thus, all the conditions of Lemma~\ref{lemma:type-3-b-i-3-criterion} are satisfied, regardless of whether $C$ is marked in Line~\ref{line:type-3-b-i-3-mark1}, or \ref{line:type-3-b-i-3-mark2}, or \ref{line:type-3-b-i-3-mark3}, and therefore we have that $e_\mathit{high}(u)\notin B(v)\cup B(w)$ an $B(v)=(B(u)\setminus\{e_\mathit{high}(u)\})\sqcup B(w)$. Thus, $C$ is indeed a $4$-cut that satisfies $(3)$ of Lemma~\ref{lemma:type-3b-cases}. Furthermore, since $M(w)=M(v,c_1)$, we have that $M(w)$ is a proper descendant of $M(v)$, and therefore $M(w)\neq M(v)$. This implies that $C$ is a Type-3$\alpha i$ $4$-cut. Thus, the collection $\mathcal{C}$ of all $4$-element sets that are marked in Lines~\ref{line:type-3-b-i-3-mark1} or \ref{line:type-3-b-i-3-mark2} is a collection of Type-3$\beta i$ $4$-cuts that satisfy $(3)$ of Lemma~\ref{lemma:type-3b-cases}.

Now let $C=\{(u,p(u)),(v,p(v)),(w,p(w)),e\}$ be a Type-3$\beta i$ $4$-cut, where $w$ is a proper ancestor of $v$, and $v$ is a proper ancestor of $u$, and $e$ satisfies $(3)$ of Lemma~\ref{lemma:type-3b-cases}. Let $c_1$ and $c_2$ be the $\mathit{low1}$ and $\mathit{low2}$ children of $M(v)$, respectively. Then Lemma~\ref{lemma:type-3-b-i-3-info} implies that $e=e_\mathit{high}(u)$, $\mathit{high}_2(u)<v$, $M(B(u)\setminus\{e\})=M(v,c_2)$, and $w$ is the greatest proper ancestor of $v$ that has $M(w)=M(v,c_1)$. Furthermore, since $C$ satisfies $(3)$ of Lemma~\ref{lemma:type-3b-cases}, we have that $B(v)=(B(u)\setminus\{e\})\sqcup B(w)$, which implies that $\mathit{bcount}(v)=\mathit{bcount}(u)+\mathit{bcount}(w)-1$. 

First, suppose that $M(u)=M(B(u)\setminus\{e_\mathit{high}(u)\})$. Then, by Lemma~\ref{lemma:type-3-b-i-3-special} we have that $u$ is either the lowest or the second-lowest proper descendant of $v$ such that $M(u)=M(B(u)\setminus\{e_\mathit{high}(u)\})=M(v,c_2)$. Thus, notice that $C$ will be marked in Line~\ref{line:type-3-b-i-3-mark2} or \ref{line:type-3-b-i-3-mark3}, respectively, and therefore $C\in\mathcal{C}$. So let us assume that $M(u)\neq M(B(u)\setminus\{e_\mathit{high}(u)\})$.

Now, it makes sense to consider the lowest proper descendant $u'$ of $v$ that has $M(u')\neq M(B(u')\setminus\{e_\mathit{high}(u')\})=M(v,c_2)$. If $u'=u$, then notice that $C$ satisfies enough conditions to be marked in Line~\ref{line:type-3-b-i-3-mark1}, and therefore $C\in\mathcal{C}$. Otherwise, Lemma~\ref{lemma:type-3-b-i-3-non-special} implies that $e_\mathit{high}(u')\notin B(v)\cup B(w)$ and $B(v)=(B(u')\setminus\{e_\mathit{high}(u')\})\sqcup B(w)$. Therefore, $C'=\{(u',p(u')),(v,p(v)),(w,p(w)),e_\mathit{high}(u')\}$ is a $4$-cut that satisfies $(3)$ of Lemma~\ref{lemma:type-3b-cases}. Furthermore, since $M(w)=M(v,c_1)$, we have that $M(w)\neq M(v)$, and therefore $C'$ is a Type-3$\beta i$ $4$-cut. Then, since $u'$ is the lowest proper descendant of $v$ that has $M(u')\neq M(B(u')\setminus\{e_\mathit{high}(u')\})=M(v,c_2)$, we have that $C'$ will be marked in Line~\ref{line:type-3-b-i-3-mark1}, and therefore $C'\in\mathcal{C}$. Furthermore, Lemma~\ref{lemma:type-3-b-i-3-non-special} implies that $B(u)\sqcup\{e_\mathit{high}(u')\}=B(u')\sqcup\{e_\mathit{high}(u)\}$, and therefore Lemma~\ref{lemma:type2cuts} implies that $C''=\{(u,p(u)),(u',p(u')),e_\mathit{high}(u),e_\mathit{high}(u')\}$ is a Type-$2ii$ $4$-cut. Observe that $C$ is implied by $C'$ and $C''$ through the pair of edges $\{(u,p(u)),e_\mathit{high}(u)\}$. According to Proposition~\ref{proposition:type-2-2}, we have that $C''$ is implied by $\mathcal{C}'$ through the pair of edges $\{(u,p(u)),e_\mathit{high}(u)\}$, where $\mathcal{C}'$ is the collection of Type-$2ii$ $4$-cuts computed by Algorithm~\ref{algorithm:type2-2}. Then, by Lemma~\ref{lemma:implied_from_union} we have that $C$ is implied by $\mathcal{C}'\cup\{C'\}$. Thus, we have shown that all Type-3$\beta i$ $4$-cuts that satisfy $(3)$ of Lemma~\ref{lemma:type-3b-cases} are implied by $\mathcal{C}\cup\mathcal{C}'$.

Now we will argue about the complexity of Algorithm~\ref{algorithm:type-3-b-i-3}. By Proposition~\ref{proposition:computing-M(B(v)-S)} we have that the values $M(B(u)\setminus\{e_\mathit{high}(u)\})$ can be computed in linear time in total, for all vertices $u\neq r$. Thus, the \textbf{for} loop in Line~\ref{line:type-3-b-i-3-for1} can be performed in linear time. By Proposition~\ref{proposition:computing-M(v,c)} we have that the values $M(v,c_1)$ and $M(v,c_2)$ can be computed in linear time in total, for all vertices $v\neq r$ such that $M(v)$ has at least two children, where $c_1$ and $c_2$ are the $\mathit{low1}$ and $\mathit{low2}$ children of $M(v)$. Thus, the \textbf{for} loop in Line~\ref{line:type-3-b-i-3-for2} can be performed in linear time. The vertices $u$ and $w$ in Lines~\ref{line:type-3-b-i-3-u}, \ref{line:type-3-b-i-3-u-1} and \ref{line:type-3-b-i-3-v} can be computed with the use of Algorithm~\ref{algorithm:W-queries}. Specifically, for every vertex $x$, we compute the collection $U(x)$ of all vertices $u\neq r$ such that $M(u)\neq M(B(u)\setminus\{e_\mathit{high}(u)\})=x$. Then, the $U$ sets are pairwise disjoint, and we can compute them easily in $O(n)$ time, once we have computed all $M(B(u)\setminus\{e_\mathit{high}(u)\})$ values. Now, when we reach Line~\ref{line:type-3-b-i-3-u}, we generate a query $q(U(M(v,c_2)),v)$. This is to return the lowest vertex $u$ in $U(M(v,c_2))$ that has $u>v$. Since $u\in U(M(v,c_2))$, we have that $M(B(u)\setminus\{e_\mathit{high}(u)\})=M(v,c_2)$. This implies that $M(v,c_2)$ is a common descendant of $u$ and $v$, and therefore $u$ and $v$ are related as ancestor and descendant. Thus, $u>v$ implies that $u$ is a proper descendant of $v$, and therefore we have that $u$ is the lowest proper descendant of $v$ such that $M(u)\neq M(B(u)\setminus\{e_\mathit{high}(u)\})=M(v,c_2)$. Since the number of all those queries is $O(n)$, Lemma~\ref{lemma:W-queries} implies that Algorithm~\ref{algorithm:W-queries} can compute all of them in linear time in total. Similarly, we can have the answers for the $w$ vertices in Line~\ref{line:type-3-b-i-3-v} and the $u$ vertices in Line~\ref{line:type-3-b-i-3-u-1} in $O(n)$ time in total. We conclude that Algorithm~\ref{algorithm:type-3-b-i-3} has a linear-time implementation.
\end{proof}

\subsubsection{Case $(4)$ of Lemma~\ref{lemma:type-3b-cases}}

For the Type-3$\beta i$ $4$-cuts that satisfy $(4)$ of Lemma~\ref{lemma:type-3b-cases}, we distinguish two cases, depending on whether $M(v)\neq M(B(v)\setminus\{e\})$ or $M(v)=M(B(v)\setminus\{e\})$. In the first case, by Lemma~\ref{lemma:e_L-e_R} we have that $e$ is either $e_L(v)$ or $e_R(v)$. By Lemma~\ref{lemma:type-3-b-i-4-cases-simple} below, we know precisely how to locate $u$ and $w$: $u$ is the lowest proper descendant of $v$ such that $M(u)=M(v,c_2)$, and $w$ is the greatest proper ancestor of $v$ such that $M(w)=M(w,c_1)$, where $c_1$ and $c_2$ are the $\mathit{low1}$ and $\mathit{low2}$ children of $M(B(v)\setminus\{e\})$, respectively. Thus, the procedure for computing all Type-3$\beta i$ $4$-cuts that satisfy $(4)$ of Lemma~\ref{lemma:type-3b-cases} and $M(v)\neq M(B(v)\setminus\{e\})$ is shown in Algorithm~\ref{algorithm:type-3-b-i-4-simple}. The proof of correctness and linear complexity is given in Proposition~\ref{proposition:algorithm:type-3-b-i-4-simple}.

In the second case, we first determine $u$ and $w$ according to Lemma~\ref{lemma:type-3-b-i-4-cases} and Lemma~\ref{lemma:type-3-b-i-4-extremities} (by considering all the different cases), and then $e$ is uniquely determined by the relation $B(v)=(B(u)\sqcup B(w))\sqcup\{e\}$. We use the criterion provided by Lemma~\ref{lemma:type-3-b-i-4-criterion} in order to check whether we indeed have a $4$-cut. Thus, the procedure for computing all Type-3$\beta i$ $4$-cuts that satisfy $(4)$ of Lemma~\ref{lemma:type-3b-cases} and $M(v)=M(B(v)\setminus\{e\})$ is shown in Algorithm~\ref{algorithm:type-3-b-i-4}. The proof of correctness is given in Proposition~\ref{proposition:algorithm:type-3-b-i-4}.

Let $u,v,w$ be three vertices $\neq r$. Then we let $e(u,v,w)$ denote the pair $(\mathit{XorDesc}(u)\oplus\mathit{XorDesc}(v)\oplus\mathit{XorDesc}(w),\mathit{XorAnc}(u)\oplus\mathit{XorAnc}(v)\oplus\mathit{XorAnc}(w))$. (We note that $e(u,v,w)$ is not necessarily an edge of the graph.)

\begin{lemma}
\label{lemma:type-3-b-i-4-edge}
Let $u,v,w$ be three vertices $\neq r$ such that there is a back-edge $e\in B(v)$ with $B(v)=(B(u)\sqcup B(w))\sqcup\{e\}$. Then $e=e(u,v,w)$.
\end{lemma}
\begin{proof}
Let $e=(x,y)$. Then $B(v)=(B(u)\sqcup B(v))\sqcup\{e\}$ implies that $\mathit{XorDesc}(v)=\mathit{XorDesc}(u)\oplus\mathit{XorDesc}(w)\oplus{x}$ and $\mathit{XorAnc}(v)=\mathit{XorAnc}(u)\oplus\mathit{XorAnc}(w)\oplus{y}$. This implies that $x=\mathit{XorDesc}(u)\oplus\mathit{XorDesc}(v)\oplus\mathit{XorDesc}(w)$ and $y=\mathit{XorAnc}(u)\oplus\mathit{XorAnc}(v)\oplus\mathit{XorAnc}(w)$.
\end{proof}

\begin{lemma}
\label{lemma:type-3-b-i-4-cases-simple}
Let $u,v,w$ be three vertices $\neq r$ such that $w$ is proper ancestor of $v$, $v$ is a proper ancestor of $u$, and there is a back-edge $e\in B(v)$ such that $B(v)=(B(u)\sqcup B(w))\sqcup\{e\}$ and $M(B(v)\setminus\{e\})\neq M(w)$. Suppose that $M(v)\neq M(B(v)\setminus\{e\})$. Let $c_1$ and $c_2$ be the $\mathit{low1}$ and $\mathit{low2}$ children of $M(B(v)\setminus\{e\})$, respectively. Then, $u$ is the lowest proper descendant of $v$ such that $M(u)=M(v,c_2)$, and $w$ is the greatest proper ancestor of $v$ such that $M(w)=M(v,c_1)$.
\end{lemma}
\begin{proof} 
Let $z=M(B(v)\setminus\{e\})$. Notice that, since $M(v)\neq M(B(v)\setminus\{e\})$, we have that the higher endpoint of $e$ is not a descendant of $z$. 

Since $B(v)=(B(u)\sqcup B(w))\sqcup\{e\}$, we have $B(w)\subseteq B(v)\setminus\{e\}$, and therefore $M(w)$ is a descendant of $M(B(v)\setminus\{e\})=z$. Since by assumption we have that $M(w)\neq z$, this implies that $M(w)$ is a proper descendant of $z$. 
Let $c$ be the child of $z$ that is an ancestor of $M(w)$. Let us suppose, for the sake of contradiction, that $c\neq c_1$. Let $(x,y)$ be a back-edge in $B(w)$. Then $x$ is a descendant of $M(w)$, and therefore a descendant of $c$. Furthermore, $y$ is a proper ancestor of $w$, and therefore a proper ancestor of $v$, and therefore a proper ancestor of $M(v)$, and therefore a proper ancestor of $z$, and therefore a proper ancestor of $c$. This shows that $(x,y)\in B(c)$, and therefore we have $\mathit{low}(c)\leq y$. Since $y$ is a proper ancestor of $w$, we have $y<w$. Thus, $\mathit{low}(c)\leq y$ implies that $\mathit{low}(c)<w$. Now, since $c_1$ is the $\mathit{low1}$ child of $z$, we have $\mathit{low}(c_1)\leq\mathit{low}(c)$, and therefore $\mathit{low}(c_1)<w$. This implies that there is a back-edge $(x',y')\in B(c_1)$ such that $y'<w$. Then we have that $x'$ is a descendant of $c_1$, and therefore a descendant of $z$, and therefore a descendant of $v$, and therefore a descendant of $w$. Since $(x',y')$ is a back-edge, we have that $x'$ is a descendant of $y'$. Thus, $x'$ is a common descendant of $y'$ and $w$, and therefore $y'$ and $w$ are related as ancestor and descendant. Thus, $y'<w$ implies that $y'$ is a proper ancestor of $w$. This shows that $(x',y')\in B(w)$, and therefore we have that $x'$ is a descendant of $M(w)$, and therefore a descendant of $c$. Thus, $x'$ is a common descendant of $c$ and $c_1$, and therefore $c$ and $c_1$ are related as ancestor and descendant. But this is impossible, since $c$ and $c_1$ are supposed to be distinct children of $z$. Thus, we have $c=c_1$. 

Since $B(v)=(B(u)\sqcup B(w))\sqcup\{e\}$, we have $B(u)\subseteq B(v)\setminus\{e\}$, and therefore $M(u)$ is a descendant of $M(B(v)\setminus\{e\})=z$. Let us suppose, for the sake of contradiction, that $M(u)=z$. Let $(x,y)$ be a back-edge in $B(w)$. Then $B(v)=(B(u)\sqcup B(w))\sqcup\{e\}$ implies that $(x,y)\in B(v)\setminus\{e\}$, and therefore $x$ is a descendant of $M(B(v)\setminus\{e\})=z$, and therefore a descendant of $M(u)$. Furthermore, $y$ is a proper ancestor of $w$, and therefore a proper ancestor of $v$, and therefore a proper ancestor of $u$. This shows that $(x,y)\in B(u)$, in contradiction to the fact that $B(u)\cap B(w)=\emptyset$. Thus, we have that $M(u)$ is a proper descendant of $z$. Let $c'$ be the child of $z$ that is an ancestor of $M(u)$. 

Let us suppose, for the sake of contradiction, that $c'$ is neither $c_1$ nor $c_2$. Let $(x,y)$ be a back-edge in $B(u)$. This implies that $x$ is a descendant of $M(u)$, and therefore a descendant of $c'$. Then $B(v)=(B(u)\sqcup B(w))\sqcup\{e\}$ implies that $(x,y)\in B(v)$. Then, $y$ is a proper ancestor of $v$, and therefore a proper ancestor of $M(v)$, and therefore a proper ancestor of $z$, and therefore a proper ancestor of $c'$. This shows that $(x,y)\in B(c')$, and therefore $\mathit{low}(c')\leq y$. Since $y$ is a proper ancestor of $v$, we have $y<v$. Thus, $\mathit{low}(c')\leq y$ implies that $\mathit{low}(c')<v$. Since $c'$ is neither $c_1$ nor $c_2$, we have that $\mathit{low}(c_2)\leq\mathit{low}(c')$, and therefore $\mathit{low}(c_2)<v$. This implies that there is a back-edge $(x,y)\in B(c_2)$ such that $y<v$. Then $x$ is a descendant of $c_2$, and therefore a descendant of $z$, and therefore a descendant of $v$. Since $(x,y)$ is a back-edge, we have that $x$ is a descendant of $y$. Thus, $x$ is a common descendant of $v$ and $y$, and therefore $v$ and $y$ are related as ancestor and descendant. Thus, $y<v$ implies that $y$ is a proper ancestor of $v$. This shows that $(x,y)\in B(v)$. Then, $B(v)=(B(u)\sqcup B(w))\sqcup\{e\}$ implies that either $(x,y)\in B(u)$, or $(x,y)\in B(w)$, or $(x,y)=e$. Let us suppose first that $(x,y)\in B(u)$. Then $x$ is a descendant of $M(u)$, and therefore a descendant of $c'$. But then $x$ is a common descendant of $c_2$ and $c'$, which is impossible (since $c_2$ and $c'$ are two distinct children of $z$). Now let us suppose that $(x,y)\in B(w)$. Then $x$ is a descendant of $M(w)$, and therefore a descendant of $c_1$. But then $x$ is a common descendant of $c_1$ and $c_2$, which is impossible (since $c_1$ and $c_2$ are two distinct children of $z$). The case $(x,y)=e$ is also rejected, because the higher endpoint of $e$ is not a descendant of $z$. Thus, there are no viable options left, and so we have arrived at a contradiction. This shows that $c'$ is either $c_1$ or $c_2$. 

Let us suppose, for the sake of contradiction, that $c'=c_1$. Now let $(x,y)$ be a back-edge in $B(v)\setminus\{e\}$. Then $B(v)=(B(u)\sqcup B(w))\sqcup\{e\}$ implies that either $(x,y)\in B(u)$ or $(x,y)\in B(w)$. If $(x,y)\in B(u)$, then $x$ is a descendant of $M(u)$, and therefore a descendant of $c_1$. If $(x,y)\in B(w)$, then $x$ is a descendant of $M(w)$, and therefore a descendant of $c_1$. In either case, then, we have that $x$ is a descendant of $c_1$. Due to the generality of $(x,y)\in B(v)\setminus\{e\}$, this implies that $M(B(v)\setminus\{e\})$ is a descendant of $c_1$. But this is impossible, because $c_1$ is a child of $z=M(B(v)\setminus\{e\})$. Thus, we have $c'\neq c_1$, and therefore we infer that $c'=c_2$.

Let $S_1=\{(x,y)\in B(v)\mid x \mbox{ is a descendant of } c_1 \}$. Then $M(S_1)=M(v,c_1)$. Let $(x,y)$ be a back-edge in $S_1$. Then $x$ is a descendant of $c_1$ and $(x,y)\in B(v)$. $B(v)=(B(u)\sqcup B(w))\sqcup\{e\}$ implies that either $(x,y)\in B(u)$, or $(x,y)\in B(w)$, or $(x,y)=e$. Since the higher endpoint of $e$ is not a descendant of $z$, the case $(x,y)=e$ is rejected. Let us suppose, for the sake of contradiction, that $(x,y)\in B(u)$. Then $x$ is a descendant of $M(u)$, and therefore a descendant of $c_2$. Thus, $x$ is a common descendant of $c_1$ and $c_2$, and therefore $c_1$ and $c_2$ are related as ancestor and descendant. But this is absurd, and therefore the case $(x,y)\in B(u)$ is rejected. Thus, we are left with the case $(x,y)\in B(w)$. Due to the generality of $(x,y)\in S_1$, this implies that $S_1\subseteq B(w)$. Conversely, let $(x,y)$ be a back-edge in $B(w)$. Then $x$ is a descendant of $M(w)$, and therefore a descendant of $c_1$. Furthermore, $B(v)=(B(u)\sqcup B(w))\sqcup\{e\}$ implies that $(x,y)\in B(v)$. This shows that $(x,y)\in S_1$. Due to the generality of $(x,y)\in B(w)$, this implies that $B(w)\subseteq S_1$. Since we have shown the reverse inclusion too, we infer that $B(w)=S_1$. This implies that $M(w)=M(S_1)$, and therefore $M(w)=M(v,c_1)$.

Let $S_2=\{(x,y)\in B(v)\mid x \mbox{ is a descendant of } c_2\}$. Then $M(S_2)=M(v,c_2)$. Let $(x,y)$ be a back-edge in $S_2$. Then $x$ is a descendant of $c_2$ and $(x,y)\in B(v)$. $B(v)=(B(u)\sqcup B(w))\sqcup\{e\}$ implies that either $(x,y)\in B(u)$, or $(x,y)\in B(w)$, or $(x,y)=e$. Since the higher endpoint of $e$ is not a descendant of $z$, the case $(x,y)=e$ is rejected. The case $(x,y)\in B(w)$ is also rejected, because this implies that $(x,y)\in S_1$ (whereas the sets $S_1$ and $S_2$ are disjoint, since $c_1$ and $c_2$ are not related as ancestor and descendant). Thus, we are left with the case $(x,y)\in B(u)$. Due to the generality of $(x,y)\in S_2$, this implies that $S_2\subseteq B(u)$. Conversely, let $(x,y)$ be a back-edge in $B(u)$. Then $x$ is a descendant of $M(u)$, and therefore a descendant of $c_2$. Furthermore, $B(v)=(B(u)\sqcup B(w))\sqcup\{e\}$ implies that $(x,y)\in B(v)$. This shows that $(x,y)\in S_2$. Due to the generality of $(x,y)\in B(u)$, this implies that $B(u)\subseteq S_2$. Since we have shown the reverse inclusion too, we infer that $B(u)=S_2$. This implies that $M(u)=M(S_2)$, and therefore $M(u)=M(v,c_2)$. 

Let us suppose, for the sake of contradiction, that $u$ is not the lowest proper descendant of $v$ such that $M(u)=M(v,c_2)$. This means that there is a proper descendant $u'$ of $v$ such that $u'<u$ and $M(u')=M(u)$. This implies that $u'$ is a proper ancestor of $u$, and therefore Lemma~\ref{lemma:same_m_subset_B} implies that $B(u')\subseteq B(u)$. Since the graph is $3$-edge-connected, this can be strengthened to $B(u')\subset B(u)$. Thus, there is a back-edge $(x,y)\in B(u)\setminus B(u')$. Then $B(v)=(B(u)\sqcup B(w))\sqcup\{e\}$ implies that $(x,y)\in B(v)$. Since $(x,y)\in B(u)$, we have that $x$ is a descendant of $M(u)=M(u')$. Since $(x,y)\in B(v)$, we have that $y$ is a proper ancestor of $v$, and therefore a proper ancestor of $u'$. This shows that $(x,y)\in B(u')$, a contradiction. Thus, we have shown that $u$ is the lowest proper descendant of $v$ in $M^{-1}(M(u))$.

Let us suppose, for the sake of contradiction, that $w$ is not the greatest proper ancestor of $v$ such that $M(w)=M(v,c_1)$. This means that there is a proper ancestor $w'$ of $v$ such that $w'>w$ and $M(w')=M(w)$. This implies that $w'$ is a proper descendant of $w$, and therefore Lemma~\ref{lemma:same_m_subset_B} implies that $B(w)\subseteq B(w')$. Since the graph is $3$-edge-connected, this can be strengthened to $B(w)\subset B(w')$. Thus, there is a back-edge $(x,y)\in B(w')\setminus B(w)$. Since $(x,y)\in B(w')$, we have that $x$ is a descendant of $M(w')=M(w)$, and therefore a descendant of $c_1$, and therefore a descendant of $z$, and therefore a descendant of $M(v)$. Furthermore, we have that $y$ is a proper ancestor of $w'$, and therefore a proper ancestor of $v$. This shows that $(x,y)\in B(v)$. Then $B(v)=(B(u)\sqcup B(w))\sqcup\{e\}$ implies that either $(x,y)\in B(u)$, or $(x,y)\in B(w)$, or $(x,y)=e$. The case $(x,y)\in B(w)$ is rejected, since $(x,y)\in B(w')\setminus B(w)$. The case $(x,y)\in B(u)$ implies that $x$ is a descendant of $M(u)$, and therefore a descendant of $c_2$. But this is absurd, since $x$ is a descendant of $c_1$ (and $c_1,c_2$ cannot have a common descendant). The case $(x,y)=e$ is also rejected, because the higher endpoint of $e$ is not a descendant of $z$. Thus, there are no viable options left, and so we have arrived at a contradiction. This shows that $w$ is the greatest proper ancestor of $v$ such that $M(w)=M(v,c_1)$. 
\end{proof}

\begin{lemma}
\label{lemma:type-3-b-i-4-cases}
Let $u,v,w$ be three vertices $\neq r$ such that $w$ is proper ancestor of $v$, $v$ is a proper ancestor of $u$, and there is a back-edge $e\in B(v)$ such that $B(v)=(B(u)\sqcup B(w))\sqcup\{e\}$ and $M(B(v)\setminus\{e\})\neq M(w)$. Suppose that $M(v)= M(B(v)\setminus\{e\})$. Let $c_1$, $c_2$ and $c_3$ be the $\mathit{low1}$, the $\mathit{low2}$ and the $\mathit{low3}$ child of $M(v)$, respectively. Then we have that either \textbf{(w.1)} $M(w)=M(v,c_1)$, or \textbf{(w.2)} $M(w)=M(v,c_1')$, where $c_1'$ is the $\mathit{low1}$ child of $M(v,c_1)$. Furthermore, we have that either \textbf{(u.1)} $M(u)=M(v,c_2)$, or \textbf{(u.2)} $M(u)=M(v,c_1'')$, where $c_1''$ is the $\mathit{low1}$ child of $M(v,c_2)$, or \textbf{(u.3)} $M(u)=M(v,c_2'')$, where $c_2''$ is the $\mathit{low2}$ child of $M(v,c_2)$, or \textbf{(u.4)} $M(u)=M(v,c_3)$.
\end{lemma}
\begin{proof} 
Since $B(v)=(B(u)\sqcup B(w))\sqcup\{e\}$, we have $B(w)\subseteq B(v)$, and therefore $M(w)$ is a descendant of $M(v)$. Since  $M(v)= M(B(v)\setminus\{e\})$ and $M(B(v)\setminus\{e\})\neq M(w)$, we have $M(v)\neq M(w)$. Therefore, $M(w)$ is a proper descendant of $M(v)$.
Let $c$ be the child of $M(v)$ that is an ancestor of $M(w)$. Let us suppose, for the sake of contradiction, that $c\neq c_1$. Let $(x,y)$ be a back-edge in $B(w)$. Then $x$ is a descendant of $M(w)$, and therefore a descendant of $c$. Furthermore, $y$ is a proper ancestor of $w$, and therefore a proper ancestor of $v$, and therefore a proper ancestor of $M(v)$, and therefore a proper ancestor of $c$. This shows that $(x,y)\in B(c)$, and therefore we have $\mathit{low}(c)\leq y$. Since $y$ is a proper ancestor of $w$, we have $y<w$. Thus, $\mathit{low}(c)\leq y$ implies that $\mathit{low}(c)<w$. Now, since $c_1$ is the $\mathit{low1}$ child of $M(v)$, we have $\mathit{low}(c_1)\leq\mathit{low}(c)$, and therefore $\mathit{low}(c_1)<w$. This implies that there is a back-edge $(x',y')\in B(c_1)$ such that $y'<w$. Then we have that $x'$ is a descendant of $c_1$, and therefore a descendant of $M(v)$, and therefore a descendant of $v$, and therefore a descendant of $w$. Since $(x',y')$ is a back-edge, we have that $x'$ is a descendant of $y'$. Thus, $x'$ is a common descendant of $y'$ and $w$, and therefore $y'$ and $w$ are related as ancestor and descendant. Thus, $y'<w$ implies that $y'$ is a proper ancestor of $w$. This shows that $(x',y')\in B(w)$, and therefore we have that $x'$ is a descendant of $M(w)$, and therefore a descendant of $c$. Thus, $x'$ is a common descendant of $c$ and $c_1$, and therefore $c$ and $c_1$ are related as ancestor and descendant. But this is impossible, since $c$ and $c_1$ are supposed to be distinct children of $M(v)$. Thus, we have $c=c_1$. 

Since $B(v)=(B(u)\sqcup B(w))\sqcup\{e\}$, we have $B(u)\subseteq B(v)$, and therefore $M(u)$ is a descendant of $M(v)$. Let us suppose, for the sake of contradiction, that $M(u)=M(v)$. Let $(x,y)$ be a back-edge in $B(w)$. Then $B(v)=(B(u)\sqcup B(w))\sqcup\{e\}$ implies that $(x,y)\in B(v)$, and therefore $x$ is a descendant of $M(v)$, and therefore a descendant of $M(u)$. Furthermore, $y$ is a proper ancestor of $w$, and therefore a proper ancestor of $v$, and therefore a proper ancestor of $u$. This shows that $(x,y)\in B(u)$, in contradiction to the fact that $B(u)\cap B(w)=\emptyset$. Thus, we have that $M(u)$ is a proper descendant of $M(v)$. Let $c'$ be the child of $M(v)$ that is an ancestor of $M(u)$. 

Let us suppose, for the sake of contradiction, that $c'$ is neither $c_1$, nor $c_2$, nor $c_3$. Let $(x,y)$ be a back-edge in $B(u)$. Then $x$ is a descencant of $M(u)$, and therefore a descendant of $c'$. Furthermore, $B(v)=(B(u)\sqcup B(w))\sqcup\{e\}$ implies that $(x,y)\in B(v)$, and therefore $y$ is a proper ancestor of $M(v)$, and therefore a proper ancestor of $c'$. This shows that $(x,y)\in B(c')$, and therefore $\mathit{low}(c')\leq y$. Since $(x,y)\in B(v)$, we have that $y$ is a proper ancestor of $v$, and therefore $y<v$. Thus, $\mathit{low}(c')\leq y$ implies that $\mathit{low}(c')<v$. Now, since $c'$ is neither $c_1$, nor $c_2$, nor $c_3$, we have $\mathit{low}(c_1)\leq\mathit{low}(c_2)\leq\mathit{low}(c_3)\leq\mathit{low}(c')<v$. Since $\mathit{low}(c_2)<v$, there is a back-edge $(x,y)\in B(c_2)$ such that $y<v$. Then $x$ is a descendant of $c_2$, and therefore a descendant of $M(v)$, and therefore a descendant of $v$. Since $(x,y)$ is a back-edge, we have that $x$ is a descendant of $y$. Thus, $x$ is a common descendant of $v$ and $y$, and therefore $v$ and $y$ are related as ancestor and descendant. Then, $y<v$ implies that $y$ is a proper ancestor of $y$. This shows that $(x,y)\in B(v)$. Then $B(v)=(B(u)\sqcup B(w))\sqcup\{e\}$ implies that either $(x,y)\in B(u)$, or $(x,y)\in B(w)$, or $(x,y)=e$. The case $(x,y)\in B(u)$ is rejected, because it implies that $x$ is a descendant of $M(u)$, and therefore a descendant of $c'$, and therefore $c'$ and $c_2$ have $x$ as a common descendant, which is absurd. The case $(x,y)\in B(w)$ is also rejected, because it implies that $x$ is a descendant of $M(w)$, and therefore a descendant of $c_1$, and therefore $c_1$ and $c_2$ have $x$ as a common descendant, which is absurd. Thus, the only viable option is $(x,y)=e$. Now, since $\mathit{low}(c_3)<v$, there is a back-edge $(x',y')\in B(c_3)$ such that $y'<v$. We can follow the same reasoning as for $(x,y)$, in order to infer that $(x',y')=e$. But then we have $x'=x$, and therefore $x$ is a common descendant of $c_2$ and $c_3$, which is impossible. This shows that $c'$ is either $c_1$, or $c_2$, or $c_3$.

Let us suppose, for the sake of contradiction, that $c'=c_1$. Let $(x,y)$ be a back-edge in $B(v)\setminus\{e\}$. Then $B(v)=(B(u)\sqcup B(w))\sqcup\{e\}$ implies that either $(x,y)\in B(u)$, or $(x,y)\in B(w)$. If $(x,y)\in B(u)$, then $x$ is a descendant of $M(u)$, and therefore a descendant of $c_1$. If $(x,y)\in B(w)$, then $x$ is a descendant of $M(w)$, and therefore a descendant of $c_1$. In any case, then, we have that $x$ is a descendant of $c_1$. Due to the generality of $(x,y)\in B(v)\setminus\{e\}$, this implies that $M(B(v)\setminus\{e\})$ is a descendant of $c_1$. But this is impossible, because by assumption we have $M(B(v)\setminus\{e\})=M(v)$, and $c_1$ is a child of $M(v)$. This shows that $c'\neq c_1$. Thus, we have that either $c'=c_2$, or $c'=c_3$.

Let $(x,y)$ be a back-edge in $B(w)$. Then $x$ is a descendant of $M(w)$, and therefore a descendant of $c_1$. Furthermore, $B(v)=(B(u)\sqcup B(w))\sqcup\{e\}$ implies that $(x,y)\in B(v)$. This shows that $x$ is a descendant of $M(v,c_1)$. Due to the generality of $(x,y)\in B(w)$, this implies that $M(w)$ is a descendant of $M(v,c_1)$. If $M(w)=M(v,c_1)$, then we get \textbf{(w.1)}. So let us assume that $M(w)$ is a proper descendant of $ M(v,c_1)$.
Then there is a back-edge $(x_1,y_1)\in B(v)$, such that $x_1$ is a descendant of $c_1$, but not a descendant of $M(w)$. Since $x_1$ is a descendant of $c_1$, it cannot be a descendant of $c_2$ or $c_3$, and therefore it cannot be a descendant of $M(u)$. Thus, we have $(x_1,y_1)\notin B(u)$. And since $x_1$ is not a descendant of $M(w)$, we have $(x_1,y_1)\notin B(w)$. Then $B(v)=(B(u)\sqcup B(w))\sqcup\{e\}$ implies that $(x_1,y_1)=e$.

Now, since $M(w)$ is a proper descendant of $M(v,c_1)$, we have that $M(w)$ is a descendant of a child $\tilde{c}$ of $M(v,c_1)$. Let $c_1'$ be the $\mathit{low1}$ child of $M(v,c_1)$. Let us suppose, for the sake of contradiction, that $\tilde{c}\neq c_1'$. Let $(x,y)$ be a back-edge in $B(w)$. Then $x$ is a descendant of $M(w)$, and therefore a descendant of $\tilde{c}$. Furthermore, $B(v)=(B(u)\sqcup B(w))\sqcup\{e\}$ implies that $(x,y)\in B(v)$, and therefore $y$ is a proper ancestor of $v$, and therefore a proper ancestor of $M(v)$, and therefore a proper ancestor of $c_1$, and therefore a proper ancestor of $M(v,c_1)$, and therefore a proper ancestor of $\tilde{c}$. This shows that $(x,y)\in B(\tilde{c})$, and therefore $\mathit{low}(\tilde{c})\leq y$. Since $y$ is a proper ancestor of $w$, we have $y<w$. Therefore, $\mathit{low}(\tilde{c})\leq y$ implies that $\mathit{low}(\tilde{c})<w$. Since $\tilde{c}\neq c_1'$ and $c_1'$ is the $\mathit{low1}$ child of $M(v,c_1)$, we have $\mathit{low}(c_1')\leq\mathit{low}(\tilde{c})<w$. This implies that there is a back-edge $(x,y)\in B(c_1')$ such that $y<w$. Then $x$ is a descendant of $c_1'$, and therefore a descendant of $M(v,c_1)$, and therefore a descendant of $v$. Since $(x,y)$ is a back-edge, we have that $x$ is a descendant of $y$. Thus, $x$ is a common descendant of $v$ and $y$, and therefore $v$ and $y$ are related as ancestor and descendant. Since $w$ is a proper ancestor of $v$, we have $w<v$. Then, $y<w$ implies that $y<v$, and therefore we have that $y$ is a proper ancestor of $v$. This shows that $(x,y)\in B(v)$. Then $B(v)=(B(u)\sqcup B(w))\sqcup\{e\}$  implies that either $(x,y)\in B(u)$, or $(x,y)\in B(w)$, or $(x,y)=e$. The case $(x,y)\in B(u)$ is rejected, because it implies that $x$ is a descendant of $M(u)$, and therefore a descendant of either $c_2$ or $c_3$, whereas $x$ is a descendant of $c_1$. The case $(x,y)\in B(w)$ is also rejected, because it implies that $x$ is a descendant of $M(w)$, and therefore a descendant of $\tilde{c}$, whereas $x$ is a descendant of $c_1'$. Thus, we are left with the case $(x,y)=e$. Since $e\in B(v)$, we have that $x$ is a descendant of $v$, and therefore a descendant of $w$. Since $(x,y)$ is a back-edge, we have that $x$ is a descendant of $y$. Thus, $x$ is a common descendant of $w$ and $y$, and therefore $w$ and $y$ are related as ancestor and descendant. Then, $y<w$ implies that $y$ is a proper ancestor of $w$. But this shows that $(x,y)\in B(w)$, in contradiction to the fact that $e\notin B(w)$. This shows that $\tilde{c}=c_1'$.

Let us suppose, for the sake of contradiction, that the higher endpoint of $e$ is a descendant of $c_1'$. Let $(x,y)$ be a back-edge in $B(v)$ such that $x$ is a descendant of $c_1$. Then $B(v)=(B(u)\sqcup B(w))\sqcup\{e\}$ implies that either $(x,y)\in B(u)$, or $(x,y)\in B(w)$, or $(x,y)=e$. The case $(x,y)\in B(u)$ is rejected, because it implies that $x$ is a descendant of $M(u)$, and therefore a descendant of either $c_2$ or $c_3$, whereas $x$ is a descendant of $c_1$. Now, if $(x,y)\in B(w)$, then $x$ is a descendant of $M(w)$, and therefore a descendant of $c_1'$. And if $(x,y)=e$, then by supposition we have that $x$ is a descendant of $c_1'$. In either case, then, we have that $x$ is a descendant of $c_1'$. Due to the generality of $(x,y)\in B(v)$ such that $x$ is a descendant of $c_1$, this implies that $M(v,c_1)$ is a descendant of $c_1'$. But this is impossible, since $c_1'$ is a child of $M(v,c_1)$. This shows that the higher endpoint of $e$ is not a descendant of $c_1'$.

Now let $S=\{(x,y)\in B(v)\mid x \mbox{ is a descendant of } c_1'\}$. Then we have $M(S)=M(v,c_1')$. Let $(x,y)$ be a back-edge in $S$. Then we have that $x$ is a descendant of $c_1'$ and $(x,y)\in B(v)$. Then $B(v)=(B(u)\sqcup B(w))\sqcup\{e\}$ implies that either $(x,y)\in B(u)$, or $(x,y)\in B(w)$, or $(x,y)=e$. The case $(x,y)\in B(u)$ is rejected, because it implies that $x$ is a descendant of $M(u)$, and therefore a descendant of either $c_2$ or $c_3$, whereas $x$ is a descendant of $c_1'$, and therefore a descendant of $M(v,c_1)$, and therefore a descendant of $c_1$. The case $(x,y)=e$ is also rejected, because we have shown that the higher endpoint of $e$ is not a descendant of $c_1'$. Thus, we are left with $(x,y)\in B(w)$. Due to the generality of $(x,y)\in S$, this shows that $S\subseteq B(w)$. Conversely, let $(x,y)$ be a back-edge in $B(w)$. Then $x$ is a descendant of $M(w)$, and therefore a descendant of $c_1'$. Furthermore, $B(v)=(B(u)\sqcup B(w))\sqcup\{e\}$ implies that $(x,y)\in B(v)$. This shows that $(x,y)\in S$. Due to the generality of $(x,y)\in B(w)$, this implies that $B(w)\subseteq S$. Since we have showed the reverse inclusion too, we infer that $B(w)=S$. This implies that $M(w)=M(S)$, and therefore $M(w)=M(v,c_1')$. Thus, we have established \textbf{(w.2)}.

Now, let us first consider the case that $c'=c_3$ (this is the shortest one). Let $(x,y)$ be a back-edge in $B(u)$. Then $x$ is a descendant of $M(u)$, and therefore a descendant of $c_3$. Furthermore, $B(v)=(B(u)\sqcup B(w))\sqcup\{e\}$ implies that $(x,y)\in B(v)$, and therefore $y$ is a proper ancestor of $v$, and therefore a proper ancestor of $M(v)$, and therefore a proper ancestor of $c_3$. This shows that $(x,y)\in B(c_3)$, and therefore $\mathit{low}(c_3)\leq y$. Since $y$ is a proper ancestor of $v$, we have $y<v$. Therefore, $\mathit{low}(c_3)\leq y$ implies that $\mathit{low}(c_3)<v$. Then, we have $\mathit{low}(c_2)\leq\mathit{low}(c_3)<v$. This implies that there is a back-edge $(x,y)\in B(c_2)$ such that $y<v$. Then $x$ is a descendant of $c_2$, and therefore a descendant of $M(v)$, and therefore a descendant of $v$. Since $(x,y)$ is a back-edge, we have that $x$ is a descendant of $y$. Thus, $x$ is a common descendant of $v$ and $y$, and therefore $v$ and $y$ are related as ancestor and descendant. Then, $y<v$ implies that $y$ is a proper ancestor of $v$. This shows that $(x,y)\in B(v)$. Then $B(v)=(B(u)\sqcup B(w))\sqcup\{e\}$ implies that either $(x,y)\in B(u)$, or $(x,y)\in B(w)$, or $(x,y)=e$. The case $(x,y)\in B(u)$ is rejected, because it implies that $x$ is a descendant of $M(u)$, and therefore a descendant of $c_3$, whereas $x$ is a descendant of $c_2$. The case $(x,y)\in B(w)$ is also rejected, because it implies that $x$ is a descendant of $M(w)$, and therefore a descendant of $c_1$, whereas $x$ is a descendant of $c_2$. Thus, $(x,y)=e$ is the only viable option. This implies that the higher endpoint of $e$ is a descendant of $c_2$.

Now let $S_3=\{(x,y)\in B(v)\mid x \mbox{ is a descendant of } c_3\}$. Then we have $M(S_3)=M(v,c_3)$. Let $(x,y)$ be a back-edge in $S_3$. Then $x$ is a descendant of $c_3$ and $(x,y)\in B(v)$. Then $B(v)=(B(u)\sqcup B(w))\sqcup\{e\}$ implies that either $(x,y)\in B(u)$, or $(x,y)\in B(w)$, or $(x,y)=e$. The case $(x,y)\in B(w)$ is rejected, because it implies that $x$ is a descendant of $M(w)$, and therefore a descendant of $c_1$, whereas $x$ is a descendant of $c_3$. The case $(x,y)=e$ is also rejected, because we have shown that the higher endpoint of $e$ is a descendant of $c_2$. Thus, $(x,y)\in B(u)$ is the only acceptable option. Due to the generality of $(x,y)\in S_3$, this implies that $S_3\subseteq B(u)$. Conversely, let $(x,y)$ be a back-edge in $B(u)$. Then $x$ is a descendant of $M(u)$, and therefore a descendant of $c_3$. Furthermore, $B(v)=(B(u)\sqcup B(w))\sqcup\{e\}$ implies that $(x,y)\in B(v)$. This shows that $(x,y)\in S_3$. Due to the generality of $(x,y)\in B(u)$, this implies that $B(u)\subseteq S_3$. Since we have shown the reverse inclusion too, we infer that $B(u)=S_3$. This implies that $M(u)=M(S_3)$, and therefore $M(u)=M(v,c_3)$. Thus, we get \textbf{(u.4)}.

So let us assume that $c'=c_2$ (which is the only case left). Let $(x,y)$ be a back-edge in $B(u)$. Then $x$ is a descendant of $M(u)$, and therefore a descendant of $c_2$. Furthermore, $B(v)=(B(u)\sqcup B(w))\sqcup\{e\}$ implies that $(x,y)\in B(v)$. This shows that $x$ is a descendant of $M(v,c_2)$. Due to the generality of $(x,y)\in B(u)$, this implies that $M(u)$ is a descendant of $M(v,c_2)$. Now, if $M(u)=M(v,c_2)$, then we get \textbf{(u.1)}. So let us assume that $M(u)$ is a proper descendant of $M(v,c_2)$. Then there is a child $c''$ of $M(v,c_2)$ that is an ancestor of $M(u)$. 

Let $c_1''$ be the $\mathit{low1}$ child of $M(v,c_2)$, and let $c_2''$ be the $\mathit{low2}$ child of $M(v,c_2)$. Let us suppose, for the sake of contradiction, that $c''$ is neither the $\mathit{low1}$ nor the $\mathit{low2}$ child of $M(v,c_2)$. Let $(x,y)$ be a back-edge in $B(u)$. Then $x$ is a descendant of $M(u)$, and therefore a descendant of $c''$. Furthermore, $B(v)=(B(u)\sqcup B(w))\sqcup\{e\}$ implies that $(x,y)\in B(v)$, and therefore $y$ is a proper ancestor of $v$, and therefore a proper ancestor of $M(v)$, and therefore a proper ancestor of $M(v,c_2)$, and therefore a proper ancestor of $c''$. This shows that $(x,y)\in B(c'')$, and therefore $\mathit{low}(c'')\leq y$. Since $y$ is a proper ancestor of $v$, we have $y<v$. Therefore, $\mathit{low}(c'')\leq y$ implies that $\mathit{low}(c'')<v$. Therefore, we have $\mathit{low}(c_1'')\leq\mathit{low}(c'')<v$. This implies that there is a back-edge $(x,y)\in B(c_1'')$ such that $y<v$. Then $x$ is a descendant of $c_1''$, and therefore a descendant of $M(v,c_2)$, and therefore a descendant of $v$. Since $(x,y)$ is a back-edge, we have that $x$ is a descendant of $y$. Thus, $x$ is a common descendant of $v$ and $y$, and therefore $v$ and $y$ are related as ancestor and descendant. Then, $y<v$ implies that $y$ is a proper ancestor of $v$. This shows that $(x,y)\in B(v)$. Then $B(v)=(B(u)\sqcup B(w))\sqcup\{e\}$ implies that either $(x,y)\in B(u)$, or $(x,y)\in B(w)$, or $(x,y)=e$. The case $(x,y)\in B(u)$ is rejected, because it implies that $x$ is a descendant of $M(u)$, and therefore a descendant of $c''$, whereas $x$ is a descendant of $c_1''$. The case $(x,y)\in B(w)$ is also rejected, because it implies that $x$ is a descendant of $M(w)$, and therefore a descendant of $c_1$, whereas $x$ is a descendant of $c_1''$, and therefore a descendant of $M(v,c_2)$, and therefore a descendant of $c_2$. Thus, $(x,y)=e$ is the only acceptable option, and thus the higher endpoint of $e$ is a descendant of $c_1''$. Similarly, since $c''$ is neither $c_1''$ nor $c_2''$, we have that $\mathit{low}(c_2'')\leq\mathit{low}(c'')<v$, and therefore we can show with the same argument that the higher endpoint of $e$ is a descendant of $c_2''$, which is absurd. Thus, our initial supposition cannot be true, and therefore we have that $c''$ is either $c_1''$ or $c_2''$. 

Now let us suppose, for the sake of contradiction, that the higher endpoint of $e$ is a descendant of $c''$. Let $(x,y)$ be a back-edge in $B(v)$ such that $x$ is a descendant of $c_2$. Then $B(v)=(B(u)\sqcup B(w))\sqcup\{e\}$ implies that either $(x,y)\in B(u)$, or $(x,y)\in B(w)$, or $(x,y)=e$. The case $(x,y)\in B(w)$ is rejected, because it implies that $x$ is a descendant of $M(w)$, and therefore a descendant of $c_1$, whereas $x$ is a descendant of $c_2$. Thus, we have that either $(x,y)\in B(u)$, or $(x,y)=e$. If $(x,y)\in B(u)$, then $x$ is a descendant of $M(u)$, and therefore a descendant of $c''$. If $(x,y)=e$, then by supposition we have that $x$ is a descendant of $c''$. In either case, then, we have that $x$ is a descendant of $c''$. Due to the generality of $(x,y)\in B(v)$ such that $x$ is a descendant of $c_2$, this implies that $M(v,c_2)$ is a descendant of $c''$. But this is impossible, because $c''$ is a child of $M(v,c_2)$. Thus, we have that the higher endpoint of $e$ is not a descendant of $c''$.

Now let $S'=\{(x,y)\in B(v)\mid x \mbox{ is a descendant of } c''\}$. Then we have $M(S')=M(v,c'')$. Let $(x,y)$ be a back-edge in $S'$. Then $x$ is a descendant of $c''$ and $(x,y)\in B(v)$. Then $B(v)=(B(u)\sqcup B(w))\sqcup\{e\}$ implies that either $(x,y)\in B(u)$, or $(x,y)\in B(w)$, or $(x,y)=e$. The case $(x,y)\in B(w)$ is rejected, because it implies that $x$ is a descendant of $M(w)$, and therefore a descendant of $c_1$, whereas $x$ is a descendant of $c''$, and therefore a descendant of $M(v,c_2)$, and therefore a descendant of $c_2$. The case $(x,y)=e$ is also rejected, because we have shown that the higher endpoint of $e$ is not a descendant of $c''$. Thus, $(x,y)\in B(u)$ is the only acceptable option. Due to the generality of $(x,y)\in S'$, this implies that $S'\subseteq B(u)$. Conversely, let $(x,y)$ be a back-edge in $B(u)$. Then $x$ is a descendant of $M(u)$, and therefore a descendant of $c''$. Furthermore, $B(v)=(B(u)\sqcup B(w))\sqcup\{e\}$ implies that $(x,y)\in B(v)$. This shows that $(x,y)\in S'$. Due to the generality of $(x,y)\in B(u)$, this implies that $B(u)\subseteq S'$. Since we have shown the reverse inclusion too, we infer that $B(u)=S'$. This implies that $M(u)=M(S')$, and therefore $M(u)=M(v,c'')$. Thus, if $c''=c_1''$, then we get \textbf{(u.2)}. And if $c''=c_2''$, then we get \textbf{(u.3)}.
\end{proof}

\begin{lemma}
\label{lemma:type-3-b-i-4-extremities}
Let $u,v,w$ be three vertices $\neq r$ such that $w$ is proper ancestor of $v$, $v$ is a proper ancestor of $u$, and there is a back-edge $e\in B(v)$ such that $B(v)=(B(u)\sqcup B(w))\sqcup\{e\}$ and $M(B(v)\setminus\{e\})\neq M(w)$. Suppose that $M(v)= M(B(v)\setminus\{e\})$. Then $u$ is the lowest proper descendant of $v$ in $M^{-1}(M(u))$. In case \textbf{(w.1)} of Lemma~\ref{lemma:type-3-b-i-4-cases} we have that $w$ is either the greatest or the second-greatest proper ancestor of $v$ in $M^{-1}(M(w))$. In case \textbf{(w.2)} of Lemma~\ref{lemma:type-3-b-i-4-cases} we have that $w$ is the greatest proper ancestor of $v$ in $M^{-1}(M(w))$.
\end{lemma}
\begin{proof}
Let us suppose, for the sake of contradiction, that $u$ is not the lowest proper descendant of $v$ in $M^{-1}(M(u))$. This means that there is a proper descendant $u'$ of $v$ such that $u'<u$ and $M(u')=M(u)$. This implies that $u'$ is a proper ancestor of $u$, and then Lemma~\ref{lemma:same_m_subset_B} implies that $B(u')\subseteq B(u)$. Since the graph is $3$-edge-connected, this can be strengthened to $B(u')\subset B(u)$. Thus, there is a back-edge $(x,y)\in B(u)\setminus B(u')$. Then $x$ is a descendant of $M(u)=M(u')$. Furthermore, $B(v)=(B(u)\sqcup B(w))\sqcup\{e\}$ implies that $(x,y)\in B(v)$, and therefore $y$ is a proper ancestor of $v$, and therefore a proper ancestor of $u'$. But this shows that $(x,y)\in B(u')$, a contradiction. Thus, $u$ is the lowest proper descendant of $v$ in $M^{-1}(M(u))$.

Let $c_1$, $c_2$ and $c_3$ be the $\mathit{low1}$, $\mathit{low2}$ and $\mathit{low3}$ children of $M(v)$, respectively. Then Lemma~\ref{lemma:type-3-b-i-4-cases} implies that $M(u)$ is either a descendant of $c_2$ or a descendant of $c_3$. In either case, then, we have that no descendant of $M(u)$ can be a descendant of $c_1$. Furthermore, Lemma~\ref{lemma:type-3-b-i-4-cases} implies that either \textbf{(w.1)} $M(w)=M(v,c_1)$, or \textbf{(w.2)} $M(w)=M(v,c_1')$, where $c_1'$ is the $\mathit{low1}$ child of $M(v,c_1)$. 

Let us assume first that \textbf{(w.1)} is true. That is, we have $M(w)=M(v,c_1)$. Let us suppose, for the sake of contradiction, that $w$ is neither the greatest nor the second-greatest proper ancestor of $v$ in $M^{-1}(M(w))$. This means that there are proper ancestors $w'$ and $w''$ of $v$ such that $w''>w'>w$ and $M(w'')=M(w')=M(w)$. This implies that $w''$ is a proper descendant of $w'$, and $w'$ is a proper descendant of $w$. Then, Lemma~\ref{lemma:same_m_subset_B} implies that $B(w)\subseteq B(w')\subseteq B(w'')$. Since the graph is $3$-edge-connected, this can be strengthened to $B(w)\subset B(w')\subset B(w'')$. Thus, there are back-edges $(x,y)\in B(w')\setminus B(w)$ and $(x',y')\in B(w'')\setminus B(w')$. Then $x$ is a descendant of $M(w')=M(w)$, and therefore a descendant of $M(v,c_1)$, and therefore a descendant of $v$. Furthermore, $y$ is a proper ancestor of $w'$, and therefore a proper ancestor of $v$. This shows that $(x,y)\in B(v)$. Then $B(v)=(B(u)\sqcup B(w))\sqcup\{e\}$ implies that either $(x,y)\in B(u)$, or $(x,y)\in B(w)$, or $(x,y)=e$. The case $(x,y)\in B(u)$ is rejected, because it implies that $x$ is a descendant of $M(u)$, and therefore it is not a descendant of $c_1$, whereas $x$ is a descendant of $M(v,c_1)$, and therefore a descendant of $c_1$. The case $(x,y)\in B(w)$ is rejected because $(x,y)\in B(w')\setminus B(w)$. Thus, the only viable option is $(x,y)=e$. Then, with the same argument we can show that $(x',y')=e$, and therefore we have $(x,y)= (x',y')$. But this contradicts the fact that $(x,y)\in B(w')$ and $(x',y')\notin B(w')$. Thus, we have shown that $w$ is either the greatest or the second-greatest proper ancestor of $v$ in $M^{-1}(M(w))$.  

Now let us assume that \textbf{(w.2)} is true. That is, we have $M(w)=M(v,c_1')$, where $c_1'$ is the $\mathit{low1}$ child of $M(v,c_1)$. Let us suppose, for the sake of contradiction, that the higher endpoint of $e$ is a descendant of $c_1'$. Let $(x,y)$ be a back-edge in $B(v)$ such that $x$ is a descendant of $c_1$. Then $B(v)=(B(u)\sqcup B(w))\sqcup\{e\}$ implies that either $(x,y)\in B(u)$, or $(x,y)\in B(w)$, or $(x,y)=e$. The case $(x,y)\in B(u)$ is rejected, because it implies that $x$ is a descendant of $M(u)$, and therefore it is not a descendant of $c_1$, whereas we have that $x$ is a descendant of $c_1$. Thus, we have that either $(x,y)\in B(w)$ or $(x,y)=e$. If $(x,y)\in B(w)$, then $x$ is a descendant of $M(w)$, and therefore a descendant of $M(v,c_1')$, and therefore a descendant of $c_1'$. If $(x,y)=e$, then by supposition we have that $x$ is a descendant of $c_1'$. In either case, then, we have that $x$ is a descendant of $c_1'$. Due to the generality of $(x,y)\in B(v)$ such that $x$ is a descendant of $c_1$, this implies that $M(v,c_1)$ is a descendant of $c_1'$. But this is impossible, because $c_1'$ is a child of $M(v,c_1)$. Thus, we have that the higher endpoint of $e$ is not a descendant of $c_1'$.

Now let us suppose, for the sake of contradiction, that $w$ is not the greatest proper ancestor of $v$ in $M^{-1}(M(w))$. This means that there is a proper ancestor $w'$ of $v$ such that $w'>w$ and $M(w')=M(w)=M(v,c_1')$. This implies that $w'$ is a proper descendant of $w$, and therefore Lemma~\ref{lemma:same_m_subset_B} implies that $B(w)\subseteq B(w')$. Since the graph is $3$-edge-connected, this can be strengthened to $B(w)\subset B(w')$. Thus, there is a back-edge $(x,y)\in B(w')\setminus B(w)$. Then $x$ is a descendant of $M(w')=M(w)=M(v,c_1')$, and therefore a descendant of $v$. Furthermore, $y$ is a proper ancestor of $w'$, and therefore a proper ancestor of $v$. This shows that $(x,y)\in B(v)$. Then $B(v)=(B(u)\sqcup B(w))\sqcup\{e\}$ implies that either $(x,y)\in B(u)$, or $(x,y)\in B(w)$, or $(x,y)=e$. The case $(x,y)\in B(u)$ is rejected, because it implies that $x$ is a descendant of $M(u)$, and therefore it is not a descendant of $c_1$, whereas $x$ is a descendant of $M(v,c_1')$, and therefore a descendant of $c_1'$, and therefore a descendant of $M(v,c_1)$, and therefore a descendant of $c_1$. The case $(x,y)=e$ is also rejected, because we have shown that the higher endpoint of $e$ is not a descendant of $c_1'$, whereas $x$ is a descendant of $M(v,c_1')$, and therefore a descendant of $c_1'$. Thus, $(x,y)\in B(w)$ is the only viable option left. But this contradicts the fact that $(x,y)\in B(w')\setminus B(w)$. Thus, we conclude that $w$ is the greatest proper ancestor of $v$ in $M^{-1}(M(w))$. 
\end{proof}

\begin{lemma}
\label{lemma:type-3-b-i-4-criterion}
Let $u,v,w$ be three vertices $\neq r$ such that $u$ is a proper descendant of $v$, and $v$ is a proper descendant of $w$. Suppose that $(1)$ $M(u)$ and $M(w)$ are descendants of $M(v)$ that are not related as ancestor and descendant, $(2)$ $\mathit{high}(u)<v$, and $(3)$ $\mathit{bcount}(v)=\mathit{bcount}(u)+\mathit{bcount}(w)+1$. Then there is a back-edge $e$ such that $B(v)=(B(u)\sqcup B(w))\sqcup\{e\}$. 
\end{lemma}
\begin{proof}
Let $(x,y)$ be a back-edge in $B(u)$. Then $x$ is a descendant of $M(u)$, and therefore a descendant of $M(v)$, and therefore a descendant of $v$. Furthermore, $y$ is an ancestor of $\mathit{high}(u)$, and therefore $y\leq\mathit{high}(u)$. Then, $\mathit{high}(u)<v$ implies that $y<v$. Since $(x,y)$ is a back-edge, we have that $x$ is a descendant of $y$. Thus, $x$ is a common descendant of $v$ and $y$, and therefore $v$ and $y$ are related as ancestor and descendant. Then, $y<v$ implies that $y$ is a proper ancestor of $v$. This shows that $(x,y)\in B(v)$. Due to the generality of $(x,y)\in B(u)$, this implies that $B(u)\subseteq B(v)$.
Let $(x,y)$ be a back-edge in $B(w)$. Then $x$ is a descendant of $M(w)$, and therefore a descendant of $M(v)$. Furthermore, $y$ is a proper ancestor of $w$, and therefore a proper ancestor of $v$. This shows that $(x,y)\in B(v)$. Due to the generality of $(x,y)\in B(w)$, this implies that $B(w)\subseteq B(v)$. 
Let us suppose, for the sake of contradiction, that $B(u)\cap B(w)\neq\emptyset$. Then there is a back-edge $(x,y)\in B(u)\cap B(w)$. This implies that $x$ is a descendant of both $M(u)$ and $M(w)$, contradicting the fact that $M(u)$ and $M(w)$ are not related as ancestor and descendant. Thus, we have $B(u)\cap B(w)=\emptyset$.
Now, since $B(u)\subseteq B(v)$, $B(w)\subseteq B(v)$, $B(u)\cap B(w)=\emptyset$, and $\mathit{bcount}(v)=\mathit{bcount}(u)+\mathit{bcount}(w)+1$, we infer that there is a back-edge $e$ such that $B(v)=(B(u)\sqcup B(w))\sqcup\{e\}$.
\end{proof}

\noindent\\
\begin{algorithm}[H]
\caption{\textsf{Compute all Type-3$\beta i$ $4$-cuts that satisfy $(4)$ of Lemma~\ref{lemma:type-3b-cases} and $M(v)\neq M(B(v)\setminus\{e\})$}}
\label{algorithm:type-3-b-i-4-simple}
\LinesNumbered
\DontPrintSemicolon
\ForEach{vertex $v\neq r$}{
\label{line:type-3-b-i-4-simple-for1}
  compute $M(B(v)\setminus\{e_L(v)\})$ and $M(B(v)\setminus\{e_R(v)\})$\;
}
\ForEach{vertex $v\neq r$}{
\label{line:type-3-b-i-4-simple-for2}
  \If{$M(B(v)\setminus\{e_L(v)\})$ has at least two children}{
    let $c_1$ be the $\mathit{low1}$ child of $M(B(v)\setminus\{e_L(v)\})$\;
    let $c_2$ be the $\mathit{low2}$ child of $M(B(v)\setminus\{e_L(v)\})$\;
    compute $M(v,c_1)$ and $M(v,c_2)$\;
  }
  \If{$M(B(v)\setminus\{e_R(v)\})$ has at least two children}{
    let $c_1$ be the $\mathit{low1}$ child of $M(B(v)\setminus\{e_R(v)\})$\;
    let $c_2$ be the $\mathit{low2}$ child of $M(B(v)\setminus\{e_R(v)\})$\;
    compute $M(v,c_1)$ and $M(v,c_2)$\;
  }
}
\ForEach{vertex $v\neq r$}{
  \If{$M(B(v)\setminus\{e_L(v)\})$ has at least two children}{
    let $c_1$ be the $\mathit{low1}$ child of $M(B(v)\setminus\{e_L(v)\})$\;
    let $c_2$ be the $\mathit{low2}$ child of $M(B(v)\setminus\{e_L(v)\})$\;
    let $u$ be the lowest proper descendant of $v$ such that $M(u)=M(v,c_2)$\;
    \label{line:type-3-b-i-4-simple-u1}
    let $w$ be the greatest proper ancestor of $v$ such that $M(w)=M(v,c_1)$\;
    \label{line:type-3-b-i-4-simple-w1}
    \If{$\mathit{high}(u)<v$ \textbf{and} $\mathit{bcount}(v)=\mathit{bcount}(u)+\mathit{bcount}(w)+1$}{
      mark $\{(u,p(u)),(v,p(v)),(w,p(w)),e(u,v,w)\}$ as a $4$-cut\;
      \label{line:type-3-b-i-4-simple-mark1}
    }
  }
  \If{$M(B(v)\setminus\{e_R(v)\})$ has at least two children}{
    let $c_1$ be the $\mathit{low1}$ child of $M(B(v)\setminus\{e_R(v)\})$\;
    let $c_2$ be the $\mathit{low2}$ child of $M(B(v)\setminus\{e_R(v)\})$\;
    let $u$ be the lowest proper descendant of $v$ such that $M(u)=M(v,c_2)$\;
    \label{line:type-3-b-i-4-simple-u2}
    let $w$ be the greatest proper ancestor of $v$ such that $M(w)=M(v,c_1)$\;
    \label{line:type-3-b-i-4-simple-w2}
    \If{$\mathit{high}(u)<v$ \textbf{and} $\mathit{bcount}(v)=\mathit{bcount}(u)+\mathit{bcount}(w)+1$}{
      mark $\{(u,p(u)),(v,p(v)),(w,p(w)),e(u,v,w)\}$ as a $4$-cut\;
      \label{line:type-3-b-i-4-simple-mark2}
    }
  }
}
\end{algorithm}

\begin{proposition}
\label{proposition:algorithm:type-3-b-i-4-simple}
Algorithm~\ref{algorithm:type-3-b-i-4-simple} correctly computes all Type-3$\beta i$ $4$-cuts that satisfy $(4)$ of Lemma~\ref{lemma:type-3b-cases} and $M(v)\neq M(B(v)\setminus\{e\})$. Furthermore, it has a linear-time implementation. 
\end{proposition}
\begin{proof}
Let $C=\{(u,p(u)),(v,p(v)),(w,p(w)),e\}$ be a Type-3$\beta i$ $4$-cut, where $w$ is a proper ancestor of $v$, $v$ is a proper ancestor of $u$, $e$ satisfies $(4)$ of Lemma~\ref{lemma:type-3b-cases} and $M(v)\neq M(B(v)\setminus\{e\})$. Since $M(v)\neq M(B(v)\setminus\{e\})$, Lemma~\ref{lemma:e_L-e_R} implies that either $e=e_L(v)$ or $e=e_R(v)$. Let us assume that $e=e_L(v)$ (the other case is treated similarly). Let $c_1$ be the $\mathit{low1}$ child of $M(B(v)\setminus\{e_L(v)\})$, and let $c_2$ be the $\mathit{low2}$ child of $M(B(v)\setminus\{e_L(v)\})$. Then Lemma~\ref{lemma:type-3-b-i-4-cases-simple} implies that $u$ is the lowest proper descendant of $v$ such that $M(u)=M(v,c_2)$, and $w$ is the greatest proper ancestor of $v$ such that $M(w)=M(v,c_1)$. Since $(4)$ of Lemma~\ref{lemma:type-3b-cases} is satisfied, we have $B(v)=(B(u)\sqcup B(w))\sqcup\{e\}$. This implies that $\mathit{bcount}(v)=\mathit{bcount}(u)+\mathit{bcount}(w)+1$. Furthermore, this implies that $B(u)\subseteq B(v)$, and therefore $\mathit{high}(u)<v$ (because the lower endpoint of every back-edge in $B(u)$ is a proper ancestor of $v$). Lemma~\ref{lemma:type-3-b-i-4-edge} implies that $e=e(u,v,w)$. Thus, $C$ will be marked in Line~\ref{line:type-3-b-i-4-simple-mark1}.

Conversely, let $C=\{(u,p(u)),(v,p(v)),(w,p(w)),e(u,v,w)\}$ be a $4$-element set that is marked in Line~\ref{line:type-3-b-i-4-simple-mark1} or \ref{line:type-3-b-i-4-simple-mark2}. Let us assume that $C$ is marked in Line~\ref{line:type-3-b-i-4-simple-mark1} (the other case is treated similarly). Then we have that $u$ is a proper descendant of $v$ such that $M(u)=M(v,c_2)$ and $w$ is a proper ancestor of $v$ such that $M(w)=M(v,c_1)$, where $c_1$ and $c_2$ are the $\mathit{low1}$ and $\mathit{low2}$ children of $M(B(v)\setminus\{e_L(v)\})$, respectively. Then we have that both $M(u)$ and $M(w)$ are descendants of $M(v)$. Furthermore, $M(u)$ is a descendant of $c_1$ and $M(w)$ is a descendant of $c_2$. Thus, $M(u)$ and $M(w)$ are not related as ancestor and descendant. Then, we also have that $\mathit{high}(u)<v$ and $\mathit{bcount}(v)=\mathit{bcount}(u)+\mathit{bcount}(w)+1$. Thus, all the conditions of Lemma~\ref{lemma:type-3-b-i-4-criterion} are satisfied, and therefore we have $B(v)=(B(u)\sqcup B(w))\sqcup\{e\}$. Then, Lemma~\ref{lemma:type-3-b-i-4-edge} implies that $e=e(u,v,w)$. Thus, $C$ is a $4$-cut that satisfies $(4)$ of Lemma~\ref{lemma:type-3b-cases}. Therefore, $C$ is correctly marked in Line~\ref{line:type-3-b-i-4-simple-mark1} as a $4$-cut.

Now will argue about the complexity of Algorithm~\ref{algorithm:type-3-b-i-4-simple}. By Proposition~\ref{proposition:computing-M(B(v)-S)}, we have that the values $M(B(v)\setminus\{e_L(v)\})$ and $M(B(v)\setminus\{e_R(v)\})$ can be computed in linear time in total, for every vertex $v\neq r$. Thus, the \textbf{for} loop in Line~\ref{line:type-3-b-i-4-simple-for1} can be performed in linear time. Proposition~\ref{proposition:computing-M(v,c)} implies that all values $M(v,c_1)$ and $M(v,c_2)$ can be computed in linear time in total, for every vertex $v\neq r$ such that $M(B(v)\setminus\{e_L(v)\})$ (resp., $M(B(v)\setminus\{e_R(v)\})$) has at least two children, where $c_1$ and $c_2$ are the $\mathit{low1}$ and $\mathit{low2}$ children of $M(B(v)\setminus\{e_L(v)\})$ (resp., $M(B(v)\setminus\{e_R(v)\})$). Thus, the \textbf{for} loop in Line~\ref{line:type-3-b-i-4-simple-for2} can be performed in linear time. The vertices $u$ and $w$ in Lines~\ref{line:type-3-b-i-4-simple-u1}, \ref{line:type-3-b-i-4-simple-w1}, \ref{line:type-3-b-i-4-simple-u2} and \ref{line:type-3-b-i-4-simple-w2} can be computed in linear time in total with Algorithm~\ref{algorithm:W-queries} (see e.g. the proof of Proposition~\ref{proposition:algorithm:type-3-b-i-2}, on how we generate the queries that provide the vertices $u$ and $w$). We conclude that Algorithm~\ref{algorithm:type-3-b-i-4-simple} runs in linear time. 
\end{proof}

\noindent\\
\begin{algorithm}[H]
\caption{\textsf{Compute all Type-3$\beta i$ $4$-cuts that satisfy $(4)$ of Lemma~\ref{lemma:type-3b-cases} and $M(v)= M(B(v)\setminus\{e\})$}}
\label{algorithm:type-3-b-i-4}
\LinesNumbered
\DontPrintSemicolon
\ForEach{vertex $v\neq r$}{
  let $c_1(v)$, $c_2(v)$ and $c_3(v)$ denote the $\mathit{low1}$, $\mathit{low2}$ and $\mathit{low3}$ children of $M(v)$, respectively\;
  \label{line:type-3-b-i-4-def1}
  let $c_1'(v)$ denote the $\mathit{low1}$ child of $M(v,c_1)$\;
  \label{line:type-3-b-i-4-def2}
  let $c_1''(v)$ and $c_2''(v)$ denote the $\mathit{low1}$ and $\mathit{low2}$ children of $M(v,c_2)$, respectively\;
  \label{line:type-3-b-i-4-def3}
}
\ForEach{vertex $v\neq r$}{
  \tcp{case \textbf{(w.1)} of Lemma~\ref{lemma:type-3-b-i-4-cases}}
  let $w$ be the greatest proper ancestor of $v$ such that $M(w)=M(v,c_1(v))$\;
  \label{line:type-3-b-i-4-w}
  \If{$w\neq\bot$}{
  \label{line:type-3-b-i-4-w-case1}
    \tcp{case \textbf{(u.1)} of Lemma~\ref{lemma:type-3-b-i-4-cases}}
    let $u$ be the lowest proper descendant of $v$ such that $M(u)=M(v,c_2(v))$\;
    \label{line:type-3-b-i-4-u}
    \If{$\mathit{high}(u)<v$ \textbf{and} $\mathit{bcount}(v)=\mathit{bcount}(u)+\mathit{bcount}(w)+1$}{
      mark $\{(u,p(u)),(v,p(v)),(w,p(w)),e(u,v,w)\}$ as a $4$-cut\;
    }   
    \tcp{case \textbf{(u.2)} of Lemma~\ref{lemma:type-3-b-i-4-cases}}
    let $u$ be the lowest proper descendant of $v$ such that $M(u)=M(v,c_1''(v))$\;
    \If{$\mathit{high}(u)<v$ \textbf{and} $\mathit{bcount}(v)=\mathit{bcount}(u)+\mathit{bcount}(w)+1$}{
      mark $\{(u,p(u)),(v,p(v)),(w,p(w)),e(u,v,w)\}$ as a $4$-cut\;
    }
    \tcp{case \textbf{(u.3)} of Lemma~\ref{lemma:type-3-b-i-4-cases}}
    let $u$ be the lowest proper descendant of $v$ such that $M(u)=M(v,c_2''(v))$\;
    \If{$\mathit{high}(u)<v$ \textbf{and} $\mathit{bcount}(v)=\mathit{bcount}(u)+\mathit{bcount}(w)+1$}{
      mark $\{(u,p(u)),(v,p(v)),(w,p(w)),e(u,v,w)\}$ as a $4$-cut\;
    }
    \tcp{case \textbf{(u.4)} of Lemma~\ref{lemma:type-3-b-i-4-cases}}
    let $u$ be the lowest proper descendant of $v$ such that $M(u)=M(v,c_3(v))$\;
    \If{$\mathit{high}(u)<v$ \textbf{and} $\mathit{bcount}(v)=\mathit{bcount}(u)+\mathit{bcount}(w)+1$}{
      mark $\{(u,p(u)),(v,p(v)),(w,p(w)),e(u,v,w)\}$ as a $4$-cut\;
    }
    \label{line:type-3-b-i-4-fin}
  }
  $w\leftarrow\mathit{prevM}(w)$\; 
  \If{$w\neq\bot$ \textbf{and} $w$ is a proper ancestor of $v$}{
  \label{line:type-3-b-i-4-w-case2}
    perform the same steps as in Lines~\ref{line:type-3-b-i-4-u} to \ref{line:type-3-b-i-4-fin}\;
  }
  \tcp{case \textbf{(w.2)} of Lemma~\ref{lemma:type-3-b-i-4-cases}}
  let $w$ be the greatest proper ancestor of $v$ such that $M(w)=M(v,c_1'(v))$\;
  \If{$w\neq\bot$}{
  \label{line:type-3-b-i-4-w-case3}
    perform the same steps as in Lines~\ref{line:type-3-b-i-4-u} to \ref{line:type-3-b-i-4-fin}\;
  }
}
\end{algorithm}

\begin{proposition}
\label{proposition:algorithm:type-3-b-i-4}
Algorithm~\ref{algorithm:type-3-b-i-4} correctly computes all Type-3$\beta i$ $4$-cuts that satisfy $(4)$ of Lemma~\ref{lemma:type-3b-cases} and $M(v)= M(B(v)\setminus\{e\})$. Furthermore, it has a linear-time implementation. 
\end{proposition}
\begin{proof}
Let $C=\{(u,p(u)),(v,p(v)),(w,p(w)),e\}$ be a Type-3$\beta i$ $4$-cut, where $w$ is a proper ancestor of $v$, $v$ is a proper ancestor of $u$, $e$ satisfies $(4)$ of Lemma~\ref{lemma:type-3b-cases} and $M(v)= M(B(v)\setminus\{e\})$. We will use the notation that is introduced in Lines~\ref{line:type-3-b-i-4-def1}, \ref{line:type-3-b-i-4-def2} and \ref{line:type-3-b-i-4-def3}: i.e., $c_1(v)$ is the $\mathit{low1}$ child of $M(v)$, $c_2(v)$ is the $\mathit{low2}$ child of $M(v)$, $c_3(v)$ is the $\mathit{low3}$ child of $M(v)$, $c_1'(v)$ is the $\mathit{low1}$ child of $M(v,c_1)$, $c_1''(v)$ is the $\mathit{low1}$ child of $M(v,c_2)$, and $c_2''(v)$ is the $\mathit{low2}$ child of $M(v,c_2)$. (We note that some of those values may be $\mathit{null}$.) 

Then, according to Lemma~\ref{lemma:type-3-b-i-4-cases}, we have two cases for $w$: either \textbf{(w.1)} $M(w)=M(v,c_1(v))$, or \textbf{(w.2)} $M(w)=M(v,c_1'(v))$. Furthermore, we have four cases for $u$: either \textbf{(u.1)} $M(u)=M(v,c_2(v))$, or \textbf{(u.2)} $M(u)=M(v,c_1''(v))$, or \textbf{(u.3)} $M(u)=M(v,c_2''(v))$, or \textbf{(u.4)} $M(u)=M(v,c_3(v))$. By Lemma~\ref{lemma:type-3-b-i-4-extremities}, we have that $u$, in any case, is the lowest proper descendant of $v$ with the respective property. Furthermore, $w$ in case \textbf{(w.2)} is the greatest proper ancestor of $v$ with this property, whereas in case \textbf{(w.1)} it is either the greatest or the second-greatest proper ancestor of $v$. Thus, there are twelve distinct cases in total. 

Since $C$ satisfies $(4)$ of Lemma~\ref{lemma:type-3b-cases}, we have that $B(v)=(B(u)\sqcup B(w))\sqcup\{e\}$. This implies that $\mathit{bcount}(v)=\mathit{bcount}(u)+\mathit{bcount}(w)+1$. Furthermore, we have $B(u)\subseteq B(v)$, and therefore $\mathit{high}(u)<v$ (because the lower endpoint of every back-edge in $B(u)$ is a proper ancestor of $v$). By Lemma~\ref{lemma:type-3-b-i-4-edge} we have $e=e(u,v,w)$. Thus, it is clear that, if $w$ satisfies \textbf{(w.1)} and is the greatest proper ancestor of $v$ with this property, or it satisfies \textbf{(w.2)}, then the condition in Line~\ref{line:type-3-b-i-4-w-case1} or Line~\ref{line:type-3-b-i-4-w-case3} will be satisfied, and therefore $C$ will be marked at some point between Lines~\ref{line:type-3-b-i-4-u} to \ref{line:type-3-b-i-4-fin}. Otherwise, we have that $w$ satisfies \textbf{(w.1)} and is the second-greatest proper ancestor of $v$ with this property. So let $w'$ be the greatest proper ancestor of $v$ that satisfies $M(w')=M(v,c_1(v))$. Then we have $w=\mathit{prevM}(w')$, because $\mathit{prevM}(w')$ is the lowest vertex in $M^{-1}(M(w'))$ that is greater than $w'$. Thus, Line~\ref{line:type-3-b-i-4-w-case2} will be satisfied, and therefore $C$ will be marked at some point between Lines~\ref{line:type-3-b-i-4-u} to \ref{line:type-3-b-i-4-fin}. 

Conversely, let $C=\{(u,p(u)),(v,p(v)),(w,p(w)),e(u,v,w)\}$ be a $4$-element set that is marked at some point between Lines~\ref{line:type-3-b-i-4-u} to \ref{line:type-3-b-i-4-fin}, where we have entered the condition in Line~\ref{line:type-3-b-i-4-w-case1}, or \ref{line:type-3-b-i-4-w-case2}, or \ref{line:type-3-b-i-4-w-case3}. Then, in either case, we have the following facts. First, $u$ is a proper descendant of $v$, and $w$ is a proper ancestor of $v$. Second, $M(w)$ is a descendant of $c_1(v)$, and $M(u)$ is a descendant of either $c_2(v)$ or $c_3(v)$. Thus, both $M(w)$ and $M(u)$ are descendants of $M(v)$, but they are not related as ancestor and descendant. And third, we have $\mathit{high}(u)<v$ and $\mathit{bcount}(v)=\mathit{bcount}(u)+\mathit{bcount}(w)+1$. Thus, all the conditions of Lemma~\ref{lemma:type-3-b-i-4-criterion} are satisfied, and therefore we have that there is a back-edge $e$ such that $B(v)=(B(u)\sqcup B(w))\sqcup\{e\}$. Then Lemma~\ref{lemma:type-3-b-i-4-edge} implies that $e=e(u,v,w)$. Thus, $C$ is a $4$-cut that satisfies $(4)$ of Lemma~\ref{lemma:type-3b-cases}, and therefore it is correctly marked as a $4$-cut.

Now we will argue about the complexity of Algorithm~\ref{algorithm:type-3-b-i-4}. For every vertex $v\neq r$, we generate queries for computing $M(v,c_1(v))$, $M(v,c_2(v))$ and $M(v,c_3(v))$ (for those of $c_1(v)$, $c_2(v)$ and $c_3(v)$ that exist). By Proposition~\ref{proposition:computing-M(v,c)}, we can have the answer to all those queries in linear time in total. Then, for every $v\neq r$ such that $M(v,c_1(v))\neq\bot$ and $c_1'(v)\neq\bot$, we generate a query for computing $M(v,c_1'(v))$. Similarly, if $M(v,c_2(v))\neq\bot$, then we generate queries for computing $M(v,c_1''(v))$ and $M(v,c_2''(v))$ (for those of $c_1''(v)$ and $c_2''(v)$ that exist). According to Proposition~\ref{proposition:computing-M(v,c)}, all those queries can be answered in linear time in total. We keep pointers from every vertex $v$ to all its respective values that were computed (i.e., a pointer to $M(v,c_1(v))$, a pointer to $M(v,c_1''(v))$, etc.). In order to get the vertices $u$ and $w$ throughout (e.g., in Line~\ref{line:type-3-b-i-4-w} and Line~\ref{line:type-3-b-i-4-u}), we first collect all the queries for those vertices, and we make appropriate use of Algorithm~\ref{algorithm:W-queries}. The way to do this (and the guarantee of correctness) has been already explained in previous algorithms (e.g., in the proof of Proposition~\ref{proposition:algorithm:type-3-b-i-2}). By Lemma~\ref{lemma:W-queries}, we can have the answer to all those queries in linear time in total. It is easy to see that all other operations in Algorithm~\ref{algorithm:type-3-b-i-4} take $O(n)$ time in total. We conclude that Algorithm~\ref{algorithm:type-3-b-i-4} runs in linear time. 
\end{proof}

\subsection{Type-3$\beta ii$ $4$-cuts}
\label{subsection:type-3b-ii}

\subsubsection{Type-3$\beta ii$-$1$ $4$-cuts}

Now we consider case $(1)$ of Lemma~\ref{lemma:type-3b-cases}. 

Let $u,v,w$ be three vertices $\neq r$ such that $w$ is proper ancestor of $v$, $v$ is a proper ancestor of $u$, and there is a back-edge $e\in B(u)\cap B(v)\cap B(w)$ such that $B(v)\setminus\{e\}=(B(u)\setminus\{e\})\sqcup(B(w)\setminus\{e\})$ and $M(B(v)\setminus\{e\})=M(B(w)\setminus\{e\})$. By Lemma~\ref{lemma:type-3b-cases}, we have that $C=\{(u,p(u)),(v,p(v)),(w,p(w)),e\}$ is a $4$-cut; we call this a Type-3$\beta$ii-$1$ $4$-cut.

The following lemma provides some useful information concerning this type of $4$-cuts.

\begin{lemma}
\label{lemma:type3-b-ii-1-info}
Let $u,v,w$ be three vertices such that $(u,v,w)$ induces a Type-3$\beta$ii-$1$ $4$-cut, and let $e$ be the back-edge of the $4$-cut induced by $(u,v,w)$. Then $e=(\mathit{lowD}(u),\mathit{low}(u))$. Furthermore, $\mathit{low}(u)<w$, $\mathit{low}_2(u)\geq w$, $\mathit{high}(u)=\mathit{high}(v)$, $w$ is an ancestor of $\mathit{high}(v)$ and $M(w)=M(v)$. Finally, if $u'$ is a vertex such that $u\geq u'\geq v$ and $\mathit{high}(u')=\mathit{high}(v)$, then $u'$ is an ancestor of $u$.
\end{lemma}
\begin{proof}
Since $(u,v,w)$ induces a Type-3$\beta$ii-$1$ $4$-cut, we have that $e\in B(u)\cap B(v)\cap B(w)$. Since $e\in B(w)$, we have that the lower endpoint of $e$ is strictly lower than $w$. And since $e\in B(u)$, we have that $\mathit{low}(u)$ is at least as low as the lower endpoint of $e$. This shows that $\mathit{low}(u)<w$. 

Let us suppose, for the sake of contradiction, that $\mathit{low}_2(u)<w$. 
Let $(x,y)$ be a back-edge in $B(u)$ such that $y=\mathit{low}_2(u)$. Then $x$ is a descendant of $u$, and therefore a descendant of $v$, and therefore a descendant of $w$. Since $(x,y)$ is a back-edge, we have that $x$ is a descendant of $y$. Thus, $x$ is a common descendant of $w$ and $y$, and therefore $w$ and $y$ are related as ancestor and descendant. Since $y=\mathit{low}_2(u)$ and $\mathit{low}_2(u)<w$, we have $y<w$. This implies that $y$ is a proper ancestor of $w$. This shows that $(x,y)\in B(w)$. With the same argument, we have that the $\mathit{low}$-edge of $u$ is also in $B(w)$. Since $(u,v,w)$ induces a Type-3$\beta$ii-$1$ $4$-cut, we have $B(v)\setminus\{e\}=(B(u)\setminus\{e\})\sqcup(B(w)\setminus\{e\})$. Since $e\in B(u)\cap B(w)$, this implies that there is only one back-edge in $B(u)\cap B(w)$, a contradiction. Thus, we have $w\leq\mathit{low}_2(u)$. Furthermore, since $e\in B(u)\cap B(w)$, we have that $e$ is the only back-edge in $B(u)$ whose lower endpoint is low enough to be lower than $w$, and therefore $e$ is the $\mathit{low}$-edge of $u$.

Since $B(v)\setminus\{e\}=(B(u)\setminus\{e\})\sqcup(B(w)\setminus\{e\})$ and $e\in B(u)\cap B(v)$, we have $B(u)\subseteq B(v)$. This implies that $\mathit{high}(v)\geq\mathit{high}(u)$. Let us suppose, for the sake of contradiction, that $\mathit{high}(v)>\mathit{high}(u)$. Let $(x,y)$ be a back-edge in $B(v)$ such that $y=\mathit{high}(v)$. Then $y>\mathit{high}(u)$, and therefore $(x,y)\notin B(u)$. Thus, $B(v)\setminus\{e\}=(B(u)\setminus\{e\})\sqcup(B(w)\setminus\{e\})$ implies that $(x,y)\in B(w)$. Then we have that $y$ is a proper ancestor of $w$, and therefore $y<w$. But we have $w\leq\mathit{low}(u)\leq\mathit{high}(u)<\mathit{high}(v)=y$, a contradiction. This shows that $\mathit{high}(u)=\mathit{high}(v)$.

Since both $w$ and $\mathit{high}(v)$ are ancestors of $v$, we have that $w$ and $\mathit{high}(v)$ are related as ancestor and descendant. Now let us suppose, for the sake of contradiction, that $w$ is a proper descendant of $\mathit{high}(v)$. Let $(x,y)$ be a back-edge in $B(u)$ such that $y=\mathit{high}(u)$. Then, since $\mathit{high}(u)=\mathit{high}(v)$, we have that $y=\mathit{high}(v)$, and therefore $y$ is a proper ancestor of $w$. But since $y=\mathit{high}(u)$, this implies that $\mathit{low}_2(u)<w$, a contradiction. This shows that $w$ is an ancestor of $\mathit{high}(v)$.

Since $B(w)\setminus\{e\}\subseteq B(v)\setminus\{e\}$ and $e\in B(v)\cap B(w)$, we have that $B(w)\subseteq B(v)$. This implies that $M(w)$ is a descendant of $M(v)$. Since $e\in B(u)$, we have that $M(u)$ is an ancestor of the higher endpoint of $e$. Furthermore, since $e\in B(w)$, we have that $M(w)$ is an ancestor of the higher endpoint of $e$. Thus, since $M(u)$ and $M(w)$ have a common descendant, they are related as ancestor and descendant. Since $u$ is an ancestor of $M(u)$, this implies that $M(w)$ and $u$ are also related as ancestor and descendant. Let us suppose, for the sake of contradtion, that $M(w)$ is not an ancestor of $u$. Then we have that $M(w)$ is a proper descendant of $u$. Since the graph is $3$-edge-connected, we have that there is a back-edge $(x,y)\in B(w)\setminus\{e\}$. Then $x$ is a descendant of $M(w)$, and therefore a descendant of $u$. Furthermore, $y$ is a proper ancestor of $w$, and therefore a proper ancestor of $u$. This shows that $(x,y)\in B(u)$. But then we have $(x,y)\in (B(u)\setminus\{e\})\cap(B(w)\setminus\{e\})$, a contradiction.
Thus, we have that $M(w)$ is an ancestor of $u$, and therefore an ancestor of $M(u)$. Now let $(x,y)$ be a back-edge in $B(v)$. If $(x,y)=e$, then $e\in B(w)$, and therefore $x$ is a descendant of $M(w)$. Otherwise, since $B(v)\setminus\{e\}=(B(u)\setminus\{e\})\sqcup(B(w)\setminus\{e\})$, we have that either $(x,y)\in B(u)\setminus\{e\}$, or $(x,y)\in B(w)\setminus\{e\}$. If $(x,y)\in B(u)\setminus\{e\}$, then $x$ is a descendant of $u$, and therefore $x$ is a descendant of $M(w)$. If $(x,y)\in B(w)\setminus\{e\}$, then $M(w)$ is an ancestor of $x$. Thus, in every case we have that $M(w)$ is an ancestor of $x$. Due to the generality of $(x,y)\in B(v)$, this shows that $M(w)$ is an ancestor of $M(v)$. Thus, since $M(w)$ is a descendant of $M(v)$, we infer that $M(w)=M(v)$.

Now let $u'$ be a vertex such that $u\geq u'\geq v$ and $\mathit{high}(u')=\mathit{high}(v)$. Since $u$ is a descendant of $v$ and $u\geq u'\geq v$, we have that $u'$ is also a descendant of $v$. Furthermore, since $u\geq u'$, we have that either $u'$ is an ancestor of $u$, or it is not related as ancestor and descendant with $u$. Now let us suppose, for the sake of contradiction, that $u'$ is not an ancestor of $u$. Then $u'$ is not related as ancestor and descendant with $u$. Let $(x,y)$ be a back-edge in $B(u')$ with $y=\mathit{high}(u')$. Then $x$ is a descendant of $u'$, and therefore a descendant of $v$. Thus, $y=\mathit{high}(u')=\mathit{high}(v)$ implies that $(x,y)$ is in $B(v)$. Since $u'$ is not related as ancestor and descendant with $u$, we have that $x$ is not a descendant of $u$. Thus, $(x,y)\notin B(u)$. This implies that $e\neq (x,y)$, and thus $B(v)\setminus\{e\}=(B(u)\setminus\{e\})\sqcup(B(w)\setminus\{e\})$ implies that $(x,y)\in B(w)$, and therefore $y$ is a proper ancestor of $w$. But $y=\mathit{high}(v)$ and $\mathit{high}(v)$ is a descendant of $w$, a contradiction. This shows that $u'$ is an ancestor of $u$.
\end{proof}

We will provide a method to compute all Type-3$\beta$ii-$1$ $4$-cuts in linear time. The idea is to compute, for every vertex $v$, a set $U_1(v)$ of proper descendants $u$ of $v$ that have the potential to participate in a triple $(u,v,w)$ that induces a  Type-3$\beta$ii-$1$ $4$-cut. (These sets have the property that their total size is $O(n)$.) Then, for every $u\in U_1(v)$, we search for all $w$ with $M(w)=M(v)$ that may participate in a triple $(u,v,w)$ that induces a  Type-3$\beta$ii-$1$ $4$-cut. (In fact, we can show that such a $w$, if it exists, is unique.) It is sufficient to restrict our search to $w$ with $M(w)=M(v)$, according to Lemma~\ref{lemma:type3-b-ii-1-info}. 

Now let $v\neq r$ be a vertex with $\mathit{nextM}(v)\neq\bot$. Let $S$ be the segment of $H(\mathit{high}(v))$ that contains $v$ and is maximal w.r.t. the property that its elements are related as ancestor and descendant (i.e., we have $S=S(v)$). Then we let $U_1(v)$ denote the collection of all vertices $u\in S$ such that either $(1)$ $u$ is a proper descendant of $v$ with $\mathit{nextM}(v)>\mathit{low}_2(u)\geq\mathit{lastM}(v)$, or $(2)$ $u$ is the lowest proper descendant of $v$ in $S$ such that $\mathit{low}_2(u)\geq\mathit{nextM}(v)$.

\begin{lemma}
\label{lemma:type3-b-ii-1-U}
Let $(u,v,w)$ be a triple of vertices that induces a Type-3$\beta ii$-$1$ $4$-cut. Then $u\in U_1(v)$.
\end{lemma}
\begin{proof}
Since $(u,v,w)$ induces a Type-3$\beta ii$-$1$ $4$-cut, we have that $u$ is a proper descendant of $v$, and by Lemma~\ref{lemma:type3-b-ii-1-info} we have $\mathit{high}(u)=\mathit{high}(v)$. Let $u'$ be a vertex such that $u\geq u'\geq v$ and $\mathit{high}(u')=\mathit{high}(v)$. Then Lemma~\ref{lemma:type3-b-ii-1-info} implies that $u'$ is an ancestor of $u$. This shows that $u$ and $v$ belong to a segment of $H(\mathit{high}(v))$ with the property that its elements are related as ancestors and descendants. Thus, we have $u\in S(v)$. 

By Lemma~\ref{lemma:type3-b-ii-1-info} we have that $w$ is a proper ancestor of $v$ with $M(w)=M(v)$. Thus, $\mathit{nextM}(v)\neq\bot$ and $w\leq\mathit{nextM}(v)$.
Now, if $u$ satisfies $\mathit{nextM}(v)>\mathit{low}_2(u)\geq\mathit{lastM}(v)$, then $u$ satisfies enough conditions to be in $U_1(v)$. Otherwise, if $\mathit{nextM}(v)>\mathit{low}_2(u)\geq\mathit{lastM}(v)$ is not true, then either $\mathit{low}_2(u)\geq\mathit{nextM}(v)$ or $\mathit{low}_2(u)<\mathit{lastM}(v)$. Since $\mathit{lastM}(v)\leq w$, the case $\mathit{low}_2(u)<\mathit{lastM}(v)$ is rejected by Lemma~\ref{lemma:type3-b-ii-1-info} (because this ensures that $\mathit{low}_2(u)\geq w$, and we have that $w\geq\mathit{lastM}(v)$). Thus we have $\mathit{low}_2(u)\geq\mathit{nextM}(v)$. 

Now let us suppose, for the sake of contradiction, that there is a vertex $u'\in S(v)$ that is a proper descendant of $v$, it is lower than $u$, and satisfies $\mathit{low}_2(u')\geq\mathit{nextM}(v)$. 
This implies that $u'$ is a proper ancestor of $u$ (because all vertices in $S(v)$ are related as ancestor and descendant). Now let $(x,y)$ be a back-edge in $B(u)$. Then $x$ is a descendant of $u$, and therefore a descendant of $u'$. Furthermore, $y$ is an ancestor of $\mathit{high}(u)=\mathit{high}(u')$, and therefore it is a proper ancestor of $u'$. This shows that $(x,y)\in B(u')$, and therefore we have $B(u)\subseteq B(u')$. This can be strengthened to $B(u)\subset B(u')$, since the graph is $3$-edge-connected. Thus, there is a back-edge $(x,y)\in B(u')\setminus B(u)$. Then, $x$ is a descendant of $u'$, and therefore a descendant of $v$. Furthermore, $y$ is an ancestor of $\mathit{high}(u')$, and therefore it is a proper ancestor of $v$ (since $\mathit{high}(u')=\mathit{high}(v)$). This shows that $(x,y)\in B(v)$. 
Since $(u,v,w)$ induces a Type-3$\beta$ii-$1$ $4$-cut, we have $B(v)\setminus\{e\}\subseteq B(u)\cup B(w)$ and $e\in B(u)\cap B(v)\cap B(w)$, where $e$ is the back-edge of the $4$-cut induced by $(u,v,w)$. Since $(x,y)\in B(v)$ and $(x,y)\notin B(u)$, this implies that $(x,y)\neq e$ and $(x,y)\in B(w)$. Thus, since $\mathit{low}_2(u')\geq\mathit{nextM}(v)$ and $w\leq\mathit{nextM}(v)$, we have that $(x,y)=(\mathit{lowD}_1(u'),\mathit{low}_1(u'))$, and $(x,y)$ is the only back-edge in $B(u')$ whose lower endpoint is lower than $w$. Now, since $e\in B(u)$, we have that the higher endpoint of $e$ is a descendant of $u$, and therefore a descendant of $u'$. 
Then, since $e\in B(w)$, we have that the lower endpoint of $e$ is a proper ancestor of $w$, and therefore a proper ancestor of $v$, and therefore a proper ancestor of $u'$. This shows that $e\in B(u')$. But we have that $(x,y)$ is the only back-edge in $B(u')$ whose lower endpoint is lower than $w$, and therefore $(x,y)=e$, a contradiction. We conclude that $u$ is the lowest vertex in $S(v)$ that is a proper descendant of $v$ with $\mathit{low}_2(u)\geq\mathit{nextM}(v)$. Thus, $u$ satisfies enough conditions to be in $U_1(v)$.
\end{proof}

\begin{lemma}
\label{lemma:type3-b-ii-1-W}
Let $v$ and $v'$ be two vertices $\neq r$ such that $\mathit{nextM}(v)\neq\bot$ and $\mathit{nextM}(v')\neq\bot$. Suppose that $v'$ is a proper descendant of $v$ with $\mathit{high}(v')=\mathit{high}(v)$. Then $\mathit{nextM}(v')<\mathit{lastM}(v)$.
\end{lemma}
\begin{proof}
Since $\mathit{high}(v')=\mathit{high}(v)$ and $v'$ is a proper descendant of $v$, by Lemma~\ref{lemma:same_high} we have that $B(v')\subseteq B(v)$. Since the graph is $3$-edge-connected, this can be strengthened to $B(v')\subset B(v)$. This implies that $M(v')$ is a descendant of $M(v)$. Since $\mathit{high}(v')=\mathit{high}(v)$, Lemma~\ref{lemma:same_M_same_high} implies that $M(v')\neq M(v)$ (for otherwise we would have $B(v')=B(v)$). Thus, $M(v')$ is a proper descendant of $M(v)$. 

Now let $w$ be a proper ancestor of $v$ with $M(w)=M(v)$, and let $w'$ be a proper ancestor of $v'$ with $M(w')=M(v')$. Then $\mathit{nextM}(v)\geq w\geq\mathit{lastM}(v)$ and $\mathit{nextM}(v')\geq w'\geq\mathit{lastM}(v')$. Let us suppose, for the sake of contradiction, that $w\leq w'$. Since $w$ is an ancestor of $v$, it is also an ancestor of $v'$. Thus, $w$ and $w'$ have $v'$ as a common descendant, and therefore they are related as ancestor and descendant. Since $w\leq w'$, this implies that $w$ is an ancestor of $w'$. Let us suppose, for the sake of contradiction, that $w'$ is a descendant of $v$. This implies that $w'$ is a proper descendant of $\mathit{high}(v)=\mathit{high}(v')$. Now let $(x,y)$ be a back-edge in $B(v')$. Then $x$ is a descendant of $M(v')=M(w')$. Furthermore, $y$ is an ancestor of $\mathit{high}(v')$, and therefore a proper ancestor of $w'$. This shows that $(x,y)\in B(w')$, and thus we have $B(v')\subseteq B(w')$. Conversely, since $M(v')=M(w')$ and $w'$ is a proper ancestor of $v'$, Lemma~\ref{lemma:same_m_subset_B} implies that $B(w')\subseteq B(v')$. Thus we have $B(v')=B(w')$, a contradiction. This shows that $w'$ is not a descendant of $v$. Since $w'$ and $v$ have $v'$ as a common descendant, we have that $w'$ and $v$ are related as ancestor and descendant. Thus, $w'$ is a proper ancestor of $v$. Therefore, $M(v)=M(w)$ is a descendant of $w'$. Thus, since $w$ is an ancestor of $w'$ and $M(w)$ is a descendant of $w'$, by Lemma~\ref{lemma:M} we have that $M(w)$ is a descendant of $M(w')$. But $M(w)=M(v)$ and $M(w')=M(v')$, and so we have a contradiction to the fact that $M(v')$ is a proper descendant of $M(v)$. This shows that $w>w'$. Due to the generality of $w$ and $w'$, we conclude that $\mathit{lastM}(v)>\mathit{nextM}(v')$.
\end{proof}

\begin{lemma}
\label{lemma:type3-b-ii-1-relation-between-u1}
Let $v$ and $v'$ be two vertices with $\mathit{nextM}(v)\neq\bot$ and $\mathit{nextM}(v')\neq\bot$, such that $S(v')=S(v)$ and $v'$ is a proper descendant of $v$. If $U_1(v')=\emptyset$, then $U_1(v)=\emptyset$. If $U_1(v')\neq\emptyset$, then the lowest vertex in $U_1(v)$ (if it exists) is greater than, or equal to, the greatest vertex in $U_1(v')$.
\end{lemma}
\begin{proof}
First, let us suppose that there is a vertex $u$ in $U_1(v)$. We will show that $u$ is a proper descendant of $v'$. So let us suppose, for the sake of contradiction, that $u$ is not a proper descendant of $v'$. Since $u\in U_1(v)$, we have $u\in S(v)$. Thus, since $v'\in S(v')=S(v)$, we have that $u$ and $v'$ are related as ancestor and descendant. Then, since $u$ is not a proper descendant of $v'$, we have that $u$ is an ancestor of $v'$. Since $u$ and $v'$ are in $S(v)$, we have $\mathit{high}(u)=\mathit{high}(v)=\mathit{high}(v')$. Thus, Lemma~\ref{lemma:same_high} implies that $B(v')\subseteq B(u)$. Since $\mathit{nextM}(v)\neq\bot$ and $\mathit{nextM}(v')\neq\bot$ and $v'$ is a proper descendant of $v$, Lemma~\ref{lemma:type3-b-ii-1-W} implies that $\mathit{nextM}(v')<\mathit{lastM}(v)$. Let $w=\mathit{nextM}(v')$. Since $M(w)=M(v')$ and $w$ is a proper ancestor of $v'$, Lemma~\ref{lemma:same_m_subset_B} implies that $B(w)\subseteq B(v')$. Thus, we have $B(w)\subseteq B(u)$. Since the graph is $3$-edge-connected, we have $|B(w)|\geq 2$. Notice that the lower endpoint of every back-edge in $B(w)$ is lower than $w$, and therefore lower than $\mathit{lastM}(v)$. Thus, $B(w)\subseteq B(u)$ implies that $\mathit{low}_2(u)<\mathit{lastM}(v)$, contradicting the fact that $u\in U_1(v)$. This shows that $u$ is a proper descendant of $v'$.

Now let us suppose, for the sake of contradiction, that $U_1(v')=\emptyset$ and $U_1(v)\neq\emptyset$. Let $u$ be a vertex in $U_1(v)$. Then we have shown that $u$ is a proper descendant of $v'$. Since $u\in U_1(v)$, we have $u\in S(v)$, and therefore $u\in S(v')$. Furthermore, we have $\mathit{low}_2(u)\geq\mathit{lastM}(v)$. By Lemma~\ref{lemma:type3-b-ii-1-W} we have $\mathit{nextM}(v')<\mathit{lastM}(v)$. This implies that $\mathit{low}_2(u)\geq\mathit{nextM}(v')$. Thus, we can consider the lowest proper descendant $u'$ of $v'$ in $S(v')$ that satisfies $\mathit{low}_2(u')\geq\mathit{nextM}(v')$. But then we have $u'\in U_1(v')$, a contradiction. This shows that, if $U_1(v')=\emptyset$, then $U_1(v)=\emptyset$.

Now let us assume that $U_1(v)\neq\emptyset$. This implies that $U_1(v')\neq\emptyset$. Let $u$ be a vertex in $U_1(v)$, and let $u'$ be a vertex in $U_1(v')$. We have shown that $u$ is a proper descendant of $v'$. Since $u\in U_1(v)$, we have $u\in S(v)=S(v')$. Furthermore, we have $\mathit{low}_2(u)\geq\mathit{lastM}(v)$. By Lemma~\ref{lemma:type3-b-ii-1-W} we have $\mathit{nextM}(v')<\mathit{lastM}(v)$. This implies that $\mathit{low}_2(u)\geq\mathit{nextM}(v')$. Since $u'\in U_1(v')$, we have that either $(1)$ $\mathit{low}_2(u')<\mathit{nextM}(v')$, or $(2)$ $u'$ is the lowest vertex in $S(v')$ that is a proper descendant of $v'$ such that $\mathit{low}_2(u')\geq\mathit{nextM}(v')$. Case $(2)$ implies that $u'\leq u$ (due to the minimality of $u'$). So let us assume that case $(1)$ is true. Let us suppose, for the sake of contradiction, that $u\leq u'$. Since $u\in S(v)$ and $u'\in S(v')$ and $S(v)=S(v')$, we have that $u$ and $u'$ are related as ancestor and descendant. Thus, $u\leq u'$ implies that $u$ is an ancestor of $u'$. Furthermore, we have that $u$ and $u'$ have the same $\mathit{high}$ point. Thus, Lemma~\ref{lemma:same_high} implies that $B(u')\subseteq B(u)$. This implies that $\mathit{low}_2(u)\leq\mathit{low}_2(u')$. But we have $\mathit{low}_2(u)\geq\mathit{nextM}(v')$ and $\mathit{low}_2(u')<\mathit{nextM}(v')$, a contradiction. This shows that case $(1)$ implies too that $u'\leq u$. Due to the generality of $u\in U_1(v)$ and $u'\in U_1(v')$, this implies that the lowest vertex in $U_1(v)$ (if it exists) is greater than, or equal to, the greatest vertex in $U_1(v')$. 
\end{proof}

Based on Lemma~\ref{lemma:type3-b-ii-1-relation-between-u1}, we can provide an efficient algorithm for computing the sets $U_1(v)$, for all vertices $v\neq r$ such that $\mathit{nextM}(v)\neq\bot$. The computation takes place on segments of $H(x)$ that are maximal w.r.t. the property that their elements are related as ancestor and descendant. Specifically, let $v\neq r$ be a vertex such that $\mathit{nextM}(v)\neq\bot$. Then we have that $U_1(v)\subset S(v)$. In other words, $U_1(v)$ is a subset of the segment of $H(\mathit{high}(v))$ that contains $v$ and is maximal w.r.t. the property that its elements are related as ancestor and descendant. So let $z_1,\dots,z_k$ be the vertices of $S(v)$, sorted in decreasing order. Then, we have that $v=z_i$, for an $i\in\{1,\dots,k\}$. By definition, $U_1(v)$ contains every vertex $u$ in $\{z_1,\dots,z_{i-1}\}$ such that either $\mathit{nextM}(v)>\mathit{low}_2(u)\geq\mathit{lastM}(v)$, or $u$ is the lowest vertex in this set such that $\mathit{low}_2(u)\geq\mathit{nextM}(v)$.

As an implication of Lemma~\ref{lemma:same_high}, we have that the vertices in $\{z_1,\dots,z_{i-1}\}$ are sorted in increasing order w.r.t. their $B$ set, and therefore they are sorted in decreasing order w.r.t. their $\mathit{low}_2$ point. In other words, we have $B(z_1)\subseteq\dots\subseteq B(z_{i-1})$, and therefore $\mathit{low}_2(z_1)\geq\dots\geq\mathit{low}_2(z_{i-1})$. 
Thus, it is sufficient to process the vertices from $\{z_1,\dots,z_{i-1}\}$ in reverse order, in order to find the first vertex $u$ that has $\mathit{low}_2(u)\geq\mathit{lastM}(v)$. Then, we keep traversing this set in reverse order, and, as long as the $\mathit{low}_2$ point of every vertex $u$ that we meet is lower than $\mathit{nextM}(v)$, we insert $u$ into $U_1(v)$. Then, once we reach a vertex with $\mathit{low}_2$ point no lower than $\mathit{nextM}(v)$, we also insert it into $U_1(v)$, and we are done.

Now, if there is a proper ancestor $v'$ of $v$ in $S(v)$ such that $\mathit{high}(v')=\mathit{high}(v)$, then we have that $S(v)=S(v')$. If $\mathit{nextM}(v')\neq\bot$, then we have that $U_1(v')$ is defined. Then we can follow the same process as above in order to compute $U_1(v')$. Furthermore, according to Lemma~\ref{lemma:type3-b-ii-1-relation-between-u1}, it is sufficient to start from the greatest element of $U_1(v)$ (i.e., the one that was inserted last into $U_1(v)$). In particular, if $U_1(v)=\emptyset$, then it is certain that $U_1(v')=\emptyset$, and therefore we are done. Otherwise, we just pick up the computation from the greatest vertex in $U_1(v)$. In order to perform efficiently those computations, we first compute, for every vertex $x$, the collection $\mathcal{S}(x)$ of the segments of $H(x)$ that are maximal w.r.t. the property that their elements are related as ancestor and descendant. For every vertex $x$, this computation takes $O(|H(x)|)$ time using Algorithm~\ref{algorithm:segments}, according to Lemma~\ref{lemma:segments}. Since every vertex $\neq r$ participates in exactly one set of the form $H(x)$, we have that the total size of all $\mathcal{S}(x)$, for all vertices $x$, is $O(n)$. Then it is sufficient to process separately all segments of $\mathcal{S}(x)$, for every vertex $x$, as described above, by starting the computation each time from the first vertex $v$ of the segment that satisfies $\mathit{nextM}(v)\neq\bot$. The whole procedure is shown in Algorithm~\ref{algorithm:type3-b-ii-1-U}. The result is formally stated in Lemma~\ref{lemma:algorithm:type3-b-ii-1-U}. 

\begin{algorithm}[H]
\caption{\textsf{Compute the sets $U_1(v)$, for all vertices $v$ such that $\mathit{nextM}(v)\neq\bot$}}
\label{algorithm:type3-b-ii-1-U}
\LinesNumbered
\DontPrintSemicolon
\ForEach{vertex $x$}{
  compute the collection $\mathcal{S}(x)$ of the segments of $H(x)$ that are maximal w.r.t. the property
  that their elements are related as ancestor and descendant\;
}
\ForEach{$v\neq r$ such that $\mathit{nextM}(v)\neq\bot$}{
  set $U_1(v)\leftarrow\emptyset$\;
}
\ForEach{vertex $x$}{
  \ForEach{segment $S\in\mathcal{S}(x)$}{
    let $v$ be the first vertex in $S$\;
    \While{$v\neq\bot$ \textbf{and} $\mathit{nextM}(v)=\bot$}{
      $v\leftarrow\mathit{next}_S(v)$\;
    }
    \lIf{$v=\bot$}{\textbf{continue}}
    let $u\leftarrow\mathit{prev}_S(v)$\;
    \While{$v\neq\bot$}{
      \While{$u\neq\bot$ \textbf{and} $\mathit{low}_2(u)<\mathit{lastM}(v)$}{
        $u\leftarrow\mathit{prev}_S(u)$\;
      }
      \While{$u\neq\bot$ \textbf{and} $\mathit{low}_2(u)<\mathit{nextM}(v)$}{
        insert $u$ into $U_1(v)$\;
        $u\leftarrow\mathit{prev}_S(u)$\;
      }
      \If{$u\neq\bot$}{
        insert $u$ into $U_1(v)$\;
      }
      $v\leftarrow\mathit{next}_S(v)$\;
      \While{$v\neq\bot$ \textbf{and} $\mathit{nextM}(v)=\bot$}{
        $v\leftarrow\mathit{next}_S(v)$\;
      }      
    }
  }
}
\end{algorithm}

\begin{lemma}
\label{lemma:algorithm:type3-b-ii-1-U}
Algorithm~\ref{algorithm:type3-b-ii-1-U} correctly computes the collections of vertices $U_1(v)$, for all $v\neq r$ with $\mathit{nextM}(v)\neq\bot$. Furthermore, it has a linear-time implementation.
\end{lemma}
\begin{proof}
This was discussed in the main text, in the three paragraphs above Algorithm~\ref{algorithm:type3-b-ii-1-U}. It is easy to see that Algorithm~\ref{algorithm:type3-b-ii-1-U} implements precisely the idea that we described in those paragraphs.
\end{proof}

Now we will show how to use the sets $U_1$ in order to compute all Type-3$\beta$ii-$1$ $4$-cuts. 

\begin{corollary}
\label{corollary:type3-b-ii-1-uniqueness}
Let $(u,v,w)$ and $(u,v,w')$ be two triples of vertices that induce a Type-3$\beta$ii-$1$ $4$-cut. Then $w=w'$.
\end{corollary}
\begin{proof}
By Lemma~\ref{lemma:type3-b-ii-1-info} we have that the $4$-cuts induced by $(u,v,w)$ and $(u,v,w')$ have the same back-edge (that is, $(\mathit{lowD}(u),\mathit{low}(u))$). Thus, Lemma~\ref{lemma:no-common-three-edges} implies that $w=w'$.
\end{proof}

According to Corollary~\ref{corollary:type3-b-ii-1-uniqueness}, for every $u\in U_1(v)$, where $v\neq r$ is a vertex with $\mathit{nextM}(v)\neq\bot$, there is at most one $w$ such that $(u,v,w)$ induces a Type-3$\beta$ii-$1$ $4$-cut. Thus, the idea is to process all $u\in U_1(v)$, in order to find the $w$ in $M^{-1}(M(v))$ (if it exists) such that $(u,v,w)$ induces a $4$-cut. 

Given a $w$ such that $M(w)=M(v)$ and $w<v$, the following lemma provides a criterion in order to check whether $(u,v,w)$ induces a $4$-cut.

\begin{lemma}
\label{lemma:type3-b-ii-1-criterion}
Let $u,v,w$ be three vertices $\neq r$ such that $u$ is a proper descendant of $v$, $v$ is a proper descendant of $w$, $\mathit{high}(u)=\mathit{high}(v)$, $M(w)=M(v)$, $w\leq\mathit{low}_2(u)$, $\mathit{low}(u)<w$, and $\mathit{bcount}(v)=\mathit{bcount}(u)+\mathit{bcount}(w)-1$. Then there is a back-edge $e\in B(u)\cap B(v)\cap B(w)$ such that $B(v)\setminus\{e\}=(B(u)\setminus\{e\})\sqcup(B(w)\setminus\{e\})$.
\end{lemma}
\begin{proof}
Since $u$ is a descendant of $v$ such that $\mathit{high}(u)=\mathit{high}(v)$, Lemma~\ref{lemma:same_high} implies that $B(u)\subseteq B(v)$. Since $w$ is an ancestor of $v$ such that $M(w)=M(v)$, Lemma~\ref{lemma:same_m_subset_B} implies that $B(w)\subseteq B(v)$. Since $w\leq\mathit{low}_2(u)$, we have that there is at most one back-edge in $B(u)$ that may also be in $B(w)$ (i.e., the $\mathit{low}$-edge of $u$). Let $(x,y)$ be a back-edge in $B(u)$ such that $y=\mathit{low}(u)$. Then $x$ is a descendant of $u$, and therefore a descendant of $v$, and therefore a descendant of $w$. Since $(x,y)$ is a back-edge, we have that $x$ is a descendant of $y$. Thus, $x$ is a common descendant of $w$ and $y$, and therefore $w$ and $y$ are related as ancestor and descendant. Then, $y=\mathit{low}(u)<w$ implies that $y$ is a proper ancestor of $w$. This shows that $(x,y)\in B(w)$. Therefore, we have that the $\mathit{low}$-edge $e$ of $u$ is in $B(w)$. Furthermore, since $\mathit{low}(u)<w<v$, the same argument shows that $e\in B(v)$. Now, since $B(u)\subseteq B(v)$, $B(w)\subseteq B(v)$, $B(u)\cap B(w)=\{e\}$, $e\in B(v)$ and $\mathit{bcount}(v)=\mathit{bcount}(u)+\mathit{bcount}(w)-1$, we have that $B(v)\setminus\{e\}=(B(u)\setminus\{e\})\sqcup(B(w)\setminus\{e\})$.
\end{proof}

\begin{lemma}
\label{lemma:type3-b-ii-1-greatest-w}
Let $(u,v,w)$ be a triple of vertices that induces a Type-3$\beta$ii-$1$ $4$-cut. Then $w$ is the greatest proper ancestor of $v$ with $M(w)=M(v)$ and $w\leq\mathit{low}_2(u)$.
\end{lemma}
\begin{proof}
Let us suppose, for the sake of contradiction, that there is a proper ancestor $w'$ of $v$ such that $M(w')=M(v)$, $w'\leq\mathit{low}_2(u)$, and $w'>w$. Then we have that $M(w')=M(w)$, and so $w'$ is related as ancestor and descendant with $w$. Since $w'>w$, we have that $w'$ is a proper descendant of $w$. Thus, Lemma~\ref{lemma:same_m_subset_B} implies that $B(w)\subseteq B(w')$. Since the graph is $3$-edge-connected, this can be strengthened to $B(w)\subset B(w')$. Similarly, since $w'$ is a proper ancestor of $v$ with $M(w')=M(v)$, we get $B(w')\subset B(v)$. Now, since $B(w)\subset B(w')$, there is a back-edge $(x,y)\in B(w')\setminus B(w)$. Since $B(w')\subset B(v)$, we have that $(x,y)\in B(v)$. Since $(u,v,w)$ induces a Type-3$\beta$ii-$1$ $4$-cut, we have that $B(v)\setminus\{e\}=(B(u)\setminus\{e\})\sqcup(B(w)\setminus\{e\})$ and $e\in B(u)\cap B(v)\cap B(w)$, where $e$ is the back-edge of the $4$-cut induced by $(u,v,w)$. Thus, $(x,y)\in B(v)$ implies that either $(1)$ $(x,y)=e$, or $(2)$ $(x,y)\in B(u)$, or $(3)$ $(x,y)\in B(w)$. Case $(3)$ is immediately rejected, because $(x,y)\in B(w')\setminus B(w)$. Thus, $(1)$ is also rejected (since $e\in B(w)$). Therefore, only $(2)$ can be true. Then, since $\mathit{low}_2(u)\geq w'$ and $(x,y)\in B(w')$, we have that $(x,y)=(\mathit{lowD}(u),\mathit{low}(u))$. But then Lemma~\ref{lemma:type3-b-ii-1-info} implies that $e=(\mathit{lowD}(u),\mathit{low}(u))$, a contradiction. This shows that $w$ is the greatest proper ancestor of $v$ with $M(w)=M(v)$ and $w\leq\mathit{low}_2(u)$. 
\end{proof}

Lemma~\ref{lemma:type3-b-ii-1-greatest-w} suggests the following algorithm in order to find all Type-3$\beta$ii-$1$ $4$-cuts: for every vertex $v\neq r$ such that $\mathit{nextM}(v)\neq\bot$, and every $u\in U_1(v)$, find the greatest proper ancestor $w$ of $v$ such that $M(w)=M(v)$ and $\mathit{low}_2(u)\geq w$, and then check whether $(u,v,w)$ induces a Type-3$\beta$ii-$1$ $4$-cut using Lemma~\ref{lemma:type3-b-ii-1-criterion}. 
This procedure is shown in Algorithm~\ref{algorithm:type3-b-ii-1}. The proof of correctness and linear complexity is given in Proposition~\ref{proposition:algorithm:type3-b-ii-1}.

\begin{algorithm}[H]
\caption{\textsf{Compute all Type-3$\beta ii$-$1$ $4$-cuts}}
\label{algorithm:type3-b-ii-1}
\LinesNumbered
\DontPrintSemicolon
\ForEach{vertex $v\neq r$ such that $\mathit{nextM}(v)\neq\bot$}{
\label{line:type-3-b-ii-1-for}
  compute the set $U_1(v)$\;
}
\ForEach{vertex $v\neq r$ such that $\mathit{nextM}(v)\neq\bot$}{
  \ForEach{$u\in U_1(v)$}{
    let $w$ be the greatest proper ancestor of $v$ such that $w\leq\mathit{low}_2(u)$ and $M(w)=M(v)$\;
    \label{line:type-3-b-ii-1-w}
    \If{$\mathit{low}(u)<w$ \textbf{and} $\mathit{bcount}(v)=\mathit{bcount}(u)+\mathit{bcount}(w)-1$}{
    \label{line:type-3-b-ii-1-cond}
      mark $\{(u,p(u)),(v,p(v)),(w,p(w)),(\mathit{lowD}(u),\mathit{low}(u))\}$ as a $4$-cut\;
      \label{line:type-3-b-ii-1-mark}
    }
  }
}
\end{algorithm}

\begin{proposition}
\label{proposition:algorithm:type3-b-ii-1}
Algorithm~\ref{algorithm:type3-b-ii-1} computes all Type-3$\beta ii$-$1$ $4$-cuts. Furthermore, it has a linear-time implementation.
\end{proposition}
\begin{proof}
Let $C=\{(u,p(u)),(v,p(v)),(w,p(w)),e\}$ be a Type-3$\beta ii$-$1$ $4$-cut, where $w$ is a proper ancestor of $v$, and $v$ is a proper ancestor of $u$. Then we have $B(v)\setminus\{e\}=(B(u)\setminus\{e\})\sqcup(B(w)\setminus\{e\})$, and therefore $\mathit{bcount}(v)=\mathit{bcount}(u)+\mathit{bcount}(w)-1$. Lemma~\ref{lemma:type3-b-ii-1-info} implies that $e=(\mathit{lowD}(u),\mathit{low}(u))$ and $\mathit{low}(u)<w$. Lemma~\ref{lemma:type3-b-ii-1-U} implies that $u\in U_1(v)$. Lemma~\ref{lemma:type3-b-ii-1-greatest-w} implies that $w$ be the greatest proper ancestor of $v$ such that $w\leq\mathit{low}_2(u)$ and $M(w)=M(v)$. Thus, $C$ satisfies all the conditions to be marked in Line~\ref{line:type-3-b-ii-1-mark}.

Conversely, let $C=\{(u,p(u)),(v,p(v)),(w,p(w)),(\mathit{lowD}(u),\mathit{low}(u))\}$ be a $4$-element set that is marked in Line~\ref{line:type-3-b-ii-1-mark}. Since $u\in U_1(v)$, we have that $u$ is a proper descendant of $v$ with $\mathit{high}(u)=\mathit{high}(v)$. Since $w$ is derived in Line~\ref{line:type-3-b-ii-1-w}, we have that $w$ is a proper ancestor of $v$ with $M(w)=M(v)$ and $w\leq\mathit{low}_2(u)$. Since the condition in Line~\ref{line:type-3-b-ii-1-cond} is satisfied, we have $\mathit{low}(u)<w$ and $\mathit{bcount}(v)=\mathit{bcount}(u)+\mathit{bcount}(w)-1$. Thus, all the conditions of Lemma~\ref{lemma:type3-b-ii-1-criterion} are satisfied, and therefore we have that there is a back-edge $e$ such that $e\in B(u)\cap B(v)\cap B(w)$ and $B(v)\setminus\{e\}=(B(u)\setminus\{e\})\sqcup(B(w)\setminus\{e\})$. By the proof of Lemma~\ref{lemma:type3-b-ii-1-info}, we have $e=(\mathit{lowD}(u),\mathit{low}(u))$ (this result is independent of the condition $M(B(w)\setminus\{e\})=M(B(v)\setminus\{e\})$ that is implicit in the statement of this lemma). Thus, we have that $C$ is a $4$-cut that satisfies $(1)$ of Lemma~\ref{lemma:type-3b-cases}.

Now we will argue about the complexity of Algorithm~\ref{algorithm:type3-b-ii-1}. By Lemma~\ref{lemma:algorithm:type3-b-ii-1-U} we have that the sets $U_1(v)$ can be computed in linear time in total, for all vertices $v\neq r$ such that $\mathit{nextM}(v)\neq\bot$. Thus, the \textbf{for} loop in Line~\ref{line:type-3-b-ii-1-for} can be performed in linear time. In order to compute the vertex $w$ in Line~\ref{line:type-3-b-ii-1-w} we use Algorithm~\ref{algorithm:W-queries}. Specifically, whenever we reach Line~\ref{line:type-3-b-ii-1-w}, we generate a query of the form $q(M^{-1}(M(v)),\mathit{min}\{p(v),\mathit{low}_2(u)\})$. This is to return the greatest $w$ in $M^{-1}(M(v))$ such that $w\leq p(v)$ and $w\leq\mathit{low}_2(u)$. Since $M(w)=M(v)$, $w\leq p(v)$ implies that $w$ is a proper ancestor of $v$. Thus, $w$ is the greatest proper ancestor of $v$ with $M(w)=M(v)$ such that $w\leq\mathit{low}_2(u)$. Since the number of all those queries is $O(n)$, Lemma~\ref{lemma:W-queries} implies that they can be answered in linear time in total, using Algorithm~\ref{algorithm:W-queries}. We conclude that Algorithm~\ref{algorithm:type3-b-ii-1} runs in linear time.
\end{proof}

\subsubsection{Type-3$\beta ii$-$2$ $4$-cuts}

Now we consider case $(2)$ of Lemma~\ref{lemma:type-3b-cases}. 

Let $u,v,w$ be three vertices $\neq r$ such that $w$ is proper ancestor of $v$, $v$ is a proper ancestor of $u$, and there is a back-edge $e\in B(w)$ such that $e\notin B(v)\cup B(u)$, $B(v)=B(u)\sqcup(B(w)\setminus\{e\})$ and $M(v)=M(B(w)\setminus\{e\})$. By Lemma~\ref{lemma:type-3b-cases}, we have that $\{(u,p(u)),(v,p(v)),(w,p(w)),e\}$ is a $4$-cut, and we call this a Type-3$\beta$ii-$2$ $4$-cut.

The following lemma provides some useful information concerning this type of $4$-cuts.

\begin{lemma}
\label{lemma:type3-b-ii-2-info}
Let $u,v,w$ be three vertices such that $(u,v,w)$ induces a Type-3$\beta$ii-$2$ $4$-cut, and let $e$ be the back-edge of the $4$-cut induced by $(u,v,w)$. Then $e$ is either $(L_1(w),l(L_1(w)))$ or $(R_1(w),l(R_1(w)))$. Furthermore, $\mathit{high}(u)=\mathit{high}(v)$, $M(w)\neq M(B(w)\setminus\{e\})$, $w$ is an ancestor of $\mathit{high}(v)$, $\mathit{low}(v)<w$ and $w\leq\mathit{low}(u)$. Finally, if $u'$ is a vertex such that $u\geq u'\geq v$ and $\mathit{high}(u')=\mathit{high}(v)$, then $u'$ is an ancestor of $u$.
\end{lemma}
\begin{proof}
Since $(u,v,w)$ induces a Type-3$\beta$ii-$2$ $4$-cut, we have that $B(v)=B(u)\sqcup(B(w)\setminus\{e\})$, $M(v)=M(B(w)\setminus\{e\})$, $e\in B(w)$ and $e\notin B(v)\cup B(u)$. Let us suppose, for the sake of contradiction, that both $(L_1(w),l(L_1(w)))$ and $(R_1(w),l(R_1(w)))$ are back-edges in $B(v)$. Then $\mathit{nca}(L_1(w),R_1(w))$ is a descendant of $M(v)$, which means that $M(w)$ is a descendant of $M(v)$. Since $M(w)$ is an ancestor of $M(B(w)\setminus\{e\})$ and $M(B(w)\setminus\{e\})=M(v)$, we have that $M(w)$ is an ancestor of $M(v)$. Thus we have that $M(w)=M(v)$. Since $w$ is an ancestor of $v$, this implies that $B(w)\subseteq B(v)$, which implies that $e\in B(v)$, a contradiction. Thus we have that at least one of $(L_1(w),l(L_1(w)))$ and $(R_1(w),l(R_1(w)))$  is not in $B(v)$. Due to $B(v)=B(u)\sqcup(B(w)\setminus\{e\})$, we have that $e$ is the only back-edge in $B(w)$ that cannot be in $B(v)$, and therefore this coincides with either $(L_1(w),l(L_1(w)))$ or $(R_1(w),l(R_1(w)))$. Observe that in the argument that we used we arrived at a contradiction from $M(w)=M(v)$. Thus, $M(w)\neq M(v)$, and therefore $M(w)\neq M(B(w)\setminus\{e\})$.

Now let $(x,y)$ be a back-edge in $B(u)$. Then $B(v)=B(u)\sqcup(B(w)\setminus\{e\})$ implies that $(x,y)\in B(v)$, and therefore $y\leq\mathit{high}(v)$. Due to the generality of $(x,y)\in B(u)$, this shows that $\mathit{high}(u)\leq\mathit{high}(v)$. Now let us suppose, for the sake of contradiction, that there is a back-edge $(x,y)$ in $B(v)$ such that $y>\mathit{high}(u)$. This implies that $(x,y)\notin B(u)$. Thus, $B(v)=B(u)\sqcup(B(w)\setminus\{e\})$ implies that $(x,y)\in B(w)$, which implies that $y<w$. Then $y>\mathit{high}(u)$ implies that $\mathit{high}(u)<w$. Now let $(x',y')$ be a back-edge in $B(u)$. Then $x'$ is a descendant of $u$, and therefore a descendant of $v$, and therefore a descendant of $w$. Furthermore, we have $y'\leq\mathit{high}(u)<w$. Since $(x',y')$ is a back-edge, we have that $x'$ is a descendant of $y'$. Thus, $x'$ is a common descendant of $y'$ and $w$, and therefore $y'$ and $w$ are related as ancestor and descendant, and therefore $y'<w$ implies that $y'$ is a proper ancestor of $w$. This shows that $(x',y')\in B(w)$, which is impossible, since $B(u)\cap B(w)=\emptyset$. Thus we have that every back-edge $(x,y)\in B(v)$ has $y\leq\mathit{high}(u)$, and therefore $\mathit{high}(v)\leq\mathit{high}(u)$. This shows that $\mathit{high}(u)=\mathit{high}(v)$.

Since both $w$ and $\mathit{high}(v)$ are ancestors of $v$, we have that $w$ and $\mathit{high}(v)$ are related as ancestor and descendant. Now let us suppose, for the sake of contradiction, that $w$ is a proper descendant of $\mathit{high}(v)$. Let $(x,y)$ be a back-edge in $B(u)$ such that $y=\mathit{high}(u)$. Then, since $\mathit{high}(u)=\mathit{high}(v)$, we have that $y=\mathit{high}(v)$, and therefore $y$ is a proper ancestor of $w$. Furthermore, $x$ is a descendant of $u$, and therefore a descendant of $w$. Thus we have $(x,y)\in B(w)$, contradicting $B(u)\cap B(w)=\emptyset$. This shows that $w$ is an ancestor of $\mathit{high}(v)$.

Since $u$ is a descendant of $w$ such that $B(u)\cap B(w)=\emptyset$, we have that $\mathit{low}(u)\geq w$. Since $B(w)\setminus\{e\}$ is not empty, $B(v)=B(u)\sqcup(B(w)\setminus\{e\})$ implies that there is a back-edge $(x,y)\in B(v)\cap B(w)$, and therefore $y<w$, and therefore $\mathit{low}(v)<w$.

Now let $u'$ be a vertex such that $u\geq u'\geq v$ and $\mathit{high}(u')=\mathit{high}(v)$. Since $u$ is a descendant of $v$ and $u\geq u'\geq v$, we have that $u'$ is also a descendant of $v$. Furthermore, since $u\geq u'$, we have that either $u'$ is an ancestor of $u$, or it is not related as ancestor and descendant with $u$. Now let us suppose, for the sake of contradiction, that $u'$ is not an ancestor of $u$. Then $u'$ is not related as ancestor and descendant with $u$. Let $(x,y)$ be a back-edge in $B(u')$ with $y=\mathit{high}(u')$. Then $x$ is a descendant of $u'$, and therefore a descendant of $v$. Thus, $y=\mathit{high}(u')=\mathit{high}(v)$ implies that $(x,y)$ is in $B(v)$. Since $u'$ is not related as ancestor and descendant with $u$, we have that $x$ is not a descendant of $u$. Thus, $(x,y)\notin B(u)$. Now $B(v)=B(u)\sqcup(B(w)\setminus\{e\})$ implies that $(x,y)\in B(w)\setminus\{e\}$, and therefore $y$ is a proper ancestor of $w$. But $y=\mathit{high}(v)$ and $\mathit{high}(v)$ is a descendant of $w$, a contradiction. This shows that $u'$ is an ancestor of $u$.
\end{proof}

According to Lemma~\ref{lemma:type3-b-ii-2-info}, if a triple of vertices $(u,v,w)$ induces a Type-3$\beta$ii-$2$ $4$-cut, then the back-edge $e$ of this $4$-cut is either $(L_1(w),l(L_1(w)))$ or $(R_1(w),l(R_1(w)))$. In the following we will show how to handle the case where $e=(L_1(w),l(L_1(w)))$. To be specific, we will provide an algorithm that computes a collection of Type-3$\beta$ii-$2$ $4$-cuts of the form $\{(u,p(u)),(v,p(v)),(w,p(w)),e\}$, where $e=(L_1(w),l(L_1(w)))$, so that all $4$-cuts of this form are implied from this collection, plus that returned by Algorithm~\ref{algorithm:type2-2}. The algorithms, the propositions and the arguments for the case $e=(R_1(w),l(R_1(w)))$ are similar. Thus, in this section, for every triple $(u,v,w)$ that we consider that induces a Type-3$\beta$ii-$2$ $4$-cut, we assume that $e=e_L(w)$.

For convenience, we distinguish two cases of Type-3$\beta$ii-$2$ $4$-cuts. First, we have the case where $L_1(w)$ is not a descendant of $\mathit{high}(v)$. In this case, we compute only a subcollection of the $4$-cuts, which, together with the collection of Type-$2ii$ $4$-cuts returned by Algorithm~\ref{algorithm:type2-2}, implies all the Type-3$\beta$ii-$2$ $4$-cuts of this type (see Proposition~\ref{proposition:algorithm:type-3-b-ii-2-L}). Then we have the case where $L_1(w)$ is a descendant of $\mathit{high}(v)$. In this case we can compute all those Type-3$\beta$ii-$2$ $4$-cuts in linear time explicitly (see Proposition~\ref{proposition:algorithm:type3-b-ii-high}). 

\noindent\\
\textbf{The case where $L_1(w)$ is not a descendant of $\mathit{high}(v)$}\\

Let $(u,v,w)$ be a triple of vertices that induces a Type-3$\beta$ii-$2$ $4$-cut, and let $e$ be the back-edge of this $4$-cut. Then we have $e\in B(w)$ and $M(B(w)\setminus\{e\})=M(v)$. Furthermore, by Lemma~\ref{lemma:type3-b-ii-2-info} we have that $M(w)\neq M(B(w)\setminus\{e\})$. Now, for every vertex $v\neq r$, let $W(v)$ be the collection of all vertices $w\neq r$ such that: $(1)$ $M(B(w)\setminus\{e_L(w)\})\neq M(w)$, $(2)$ $M(B(w)\setminus\{e_L(w)\})=M(v)$, and $(3)$ $L_1(w)$ is not a descendant of $\mathit{high}(v)$. In particular, if $W(v)\neq\emptyset$, then we define $\mathit{firstW}(v)=\mathit{max}(W(v))$ and $\mathit{lastW}(v)=\mathit{min}(W(v))$.

\begin{lemma}
\label{lemma:w(v)_ancestor_of_high}
Let $v$ and $w$ be two vertices such that $w\in W(v)$. Then $w$ is a proper ancestor of $\mathit{high}(v)$.
\end{lemma}
\begin{proof}
Since $w\in W(v)$, we have that $M(B(w)\setminus\{e_L(w)\})=M(v)$. Since the graph is $3$-edge-connected, there is a back-edge $(x,y)\in B(w)\setminus\{e_L(w)\}$. Thus, $x$ is a descendant of $M(B(w)\setminus\{e_L(w)\})=M(v)$, and therefore a descendant of $v$, and therefore a descendant of $\mathit{high}(v)$. Since $(x,y)\in B(w)$, we have that $x$ is a descendant of $w$. Thus, $x$ is a common descendant of $\mathit{high}(v)$ and $w$, and therefore $\mathit{high}(v)$ and $w$ are related as ancestor and descendant. 

Now let us suppose, for the sake of contradiction, that $w$ is not a proper ancestor of $\mathit{high}(v)$. Thus, we have that $w$ is a descendant of $\mathit{high}(v)$. Then, since $L_1(w)$ is a descendant of $w$, we have that $L_1(w)$ is a descendant of $\mathit{high}(v)$, in contradiction to the fact that $w\in W(v)$. This shows that $w$ is a proper ancestor of $\mathit{high}(v)$.
\end{proof}

We will show how to compute the values $\mathit{firstW}(v)$ and $\mathit{lastW}(v)$, for all vertices $v$. With those values we can determine in constant time whether $W(v)\neq\emptyset$, for a vertex $v$, by simply checking whether $\mathit{firstW}(v)\neq\bot$. First, for every vertex $x$, we let $W_0(x)$ denote the list of all vertices $w\neq r$ with $M(w)\neq M(B(w)\setminus\{e_L(w)\})=x$, sorted in decreasing order. Notice that, for every vertex $v\neq r$, we have $W(v)\subseteq W_0(M(v))$. Now we have the following.

\begin{lemma}
\label{lemma:W_0}
Let $v$ be a vertex such that $W(v)\neq\emptyset$. Then $\mathit{lastW}(v)$ is the last entry in $W_0(M(v))$.
\end{lemma}
\begin{proof}
Let $w=\mathit{lastW}(v)$, and let $w'$ be the last entry in $W_0(M(v))$. Thus, we have $w'\leq w$. Let us suppose, for the sake of contradiction, that $w'\neq w$. Since $w=\mathit{lastW}(v)$, we have that $w$ is the lowest vertex in $W_0(M(v))$ such that $L_1(w)$ is not a descendant of $\mathit{high}(v)$. Thus, since $w'\neq w$, we have that $L_1(w')$ is a descendant of $\mathit{high}(v)$. Since $w\in W_0(M(v))$, we have that $M(v)=M(B(w)\setminus\{e_L(w)\})\neq M(w)$. Thus, $e_L(w)$ is the only back-edge in $B(w)$ whose higher endpoint (i.e., $L_1(w)$) is not a descendant of $M(v)$. Similarly, since $w'\in W_0(M(v))$, we have that $e_L(w')$ is the only back-edge in $B(w')$ whose higher endpoint (i.e., $L_1(w')$) is not a descendant of $M(v)$. Notice that $M(B(w)\setminus\{e_L(w)\})=M(B(w')\setminus\{e_L(w')\})=M(v)$. This implies that $M(v)$ is a common descendant of $w$ and $w'$, and therefore $w$ and $w'$ are related as ancestor and descendant. Thus, $w'\leq w$ implies that $w'$ is an ancestor of $w$.

Since $w=\mathit{lastW}(v)$, by Lemma~\ref{lemma:w(v)_ancestor_of_high} we have that $w$ is a proper ancestor of $\mathit{high}(v)$. Now let $(x,y)=e_L(w')$. This implies that $x=L_1(w')$. Then we have that $x$ is a descendant of $\mathit{high}(v)$, and therefore a descendant of $w$. Furthermore, $y$ is a proper ancestor of $w'$, and therefore a proper ancestor of $w$. This shows that $(x,y)\in B(w)$. Since $(x,y)=e_L(w')$, we have that $x$ is not a descendant of $M(v)$. Thus, since $(x,y)$ is a back-edge in $B(w)$ such that $x$ is not a descendant of $M(v)$, we have that $(x,y)=e_L(w)$. But this contradicts the fact that $L_1(w)$ is not a descendant of $\mathit{high}(v)$. We conclude that $w'=w$.
\end{proof}

\begin{lemma}
\label{lemma:W_0-1}
Let $v$ and $v'$ be two vertices with $M(v')=M(v)$ such that $v'$ is a proper ancestor of $v$. Suppose that $\mathit{firstW}(v')\neq\bot$. Then $\mathit{firstW}(v)\neq\bot$, and $\mathit{firstW}(v')\leq\mathit{firstW}(v)$.
\end{lemma}
\begin{proof}
Let $w=\mathit{firstW}(v')$. Then we have that $M(w)\neq M(B(w)\setminus\{e_L(w)\})=M(v')$ and $L_1(w)$ is not a descendant of $\mathit{high}(v')$. Since $v'\neq v$ and the graph is $3$-edge-connected, we have that $B(v)\neq B(v')$. Thus, since $v$ is a proper descendant of $v'$ with $M(v)=M(v')$, Lemma~\ref{lemma:same_M_dif_B_lower} implies that $v'$ is an ancestor of $\mathit{high}(v)$. Since $\mathit{high}(v')$ is a proper ancestor of $v'$, this implies that $\mathit{high}(v')$ is a proper ancestor of $\mathit{high}(v)$. Thus, since $L_1(w)$ is not a descendant of $\mathit{high}(v')$, we have that $L_1(w)$ is not a descendant of $\mathit{high}(v)$. This shows that $w\in W(v)$, and therefore $\mathit{firstW}(v)\neq\bot$ and $\mathit{firstW}(v)\geq w$.
\end{proof}

Using the information provided by Lemmata~\ref{lemma:W_0} and \ref{lemma:W_0-1}, we can provide an efficient algorithm for computing the values $\mathit{firstW}(v)$ and $\mathit{lastW}(v)$, for all vertices $v$. First, for every vertex $x$, we collect all vertices $w\neq r$ such that $M(w)\neq M(B(w)\setminus\{e_L(w)\})=x$ into a list $W_0(x)$, and we have $W_0(x)$ sorted in decreasing order. The computation of the values $M(B(w)\setminus\{e_L(w)\})$, for all $w\neq r$, takes linear time in total, according to Proposition~\ref{proposition:computing-M(B(v)-S)}. Then, the construction of the lists $W_0(x)$ takes $O(n)$ time in total, using bucket-sort. Now, for every vertex $v\neq r$, it is sufficient to check the last entry $w$ in $W_0(M(v))$ in order to see if $w=\mathit{lastW}(v)$, according to Lemma~\ref{lemma:W_0}. Since $M(w)\neq M(B(w)\setminus\{e_L(w)\})=M(v)$, we have that $w\in W(v)$ if and only if $L_1(w)$ is not a descendant of $\mathit{high}(v)$. So this can be easily checked in constant time. 

In order to compute the values $\mathit{firstW}$, we traverse the lists $M^{-1}(x)$ and $W_0(x)$ simultaneously, for every vertex $x$. To be specific, let $v\neq r$ be a vertex. Then, if $w=\mathit{firstW}(v)$ exists, we have that $w\in W_0(M(v))$. (Notice that $v$ itself lies in $M^{-1}(M(v))$.) Then, by Lemma~\ref{lemma:w(v)_ancestor_of_high} we have that $w$ is a proper ancestor of $\mathit{high}(v)$, and therefore a proper ancestor of $v$. Thus, it is sufficient to reach the greatest $w'$ in $W_0(M(v))$ that is a proper ancestor of $v$. Then, as long as $w'$ does not satisfy the property that $L_1(w')$ is not a descendant of $\mathit{high}(v)$, we keep traversing the list $W_0(M(v))$. Eventually we will reach $w$. Now, if there is a proper ancestor $v'$ of $v$ with $M(v')=M(v)$ such that $\mathit{firstW}(v')\neq\bot$, then by Lemma~\ref{lemma:W_0-1} we have that $\mathit{firstW}(v')\leq\mathit{firstW}(v)$. Thus, it is sufficient to pick up the search for $\mathit{firstW}(v')$ in $W_0(M(v'))$ from the last entry in $W_0(M(v'))$ that we accessed. This procedure for computing the values $\mathit{firstW}(v)$ and $\mathit{lastW}(v)$, for all vertices $v$, is shown in Algorithm~\ref{algorithm:compute-sets-w}. Our result is summarized in Lemma~\ref{lemma:compute-sets-w}.

\begin{algorithm}[H]
\caption{\textsf{Compute the values $\mathit{firstW}(v)$ and $\mathit{lastW}(v)$, for all vertices $v\neq r$}}
\label{algorithm:compute-sets-w}
\LinesNumbered
\DontPrintSemicolon
\ForEach{vertex $w\neq r$}{
  compute the value $M(B(w)\setminus\{e_L(w)\})$\;
}
\ForEach{vertex $x$}{
  let $W_0(x)$ be the list of all vertices $w\neq r$ with $M(w)\neq M(B(w)\setminus\{e_L(w)\})=x$, sorted in decreasing order\;
  let $M^{-1}(x)$ be the list of all vertices $v\neq r$ with $M(v)=x$, sorted in decreasing order\;
}
\ForEach{vertex $v$}{
  let $\mathit{firstW}(v)\leftarrow\bot$ and $\mathit{lastW}(v)\leftarrow\bot$\;
}
\ForEach{vertex $v\neq r$}{
  let $w$ be the last entry in $W_0(M(v))$\;
  \If{$L_1(w)$ is not a descendant of $\mathit{high}(v)$}{
    set $\mathit{lastW}(v)\leftarrow w$\;
  }
}
\ForEach{vertex $x$}{
  let $v$ be the first entry in $M^{-1}(x)$\;
  let $w$ be the first entry in $W_0(x)$\;
  \While{$v\neq\bot$}{
    \While{$w\neq\bot$ \textbf{and} $w\geq v$}{
      $w\leftarrow\mathit{next}_{W_0(x)}(w)$\;
    }
    \While{$w\neq\bot$ \textbf{and} $L_1(w)$ is a descendant of $\mathit{high}(v)$}{
      $w\leftarrow\mathit{next}_{W_0(x)}(w)$\;
    }
    \If{$w\neq\bot$}{
      set $\mathit{firstW}(v)\leftarrow w$\;
    }
    $v\leftarrow\mathit{nextM}(v)$\;
  }
}
\end{algorithm}

\begin{lemma}
\label{lemma:compute-sets-w}
Algorithm~\ref{algorithm:compute-sets-w} correctly computes the values $\mathit{firstW}(v)$ and $\mathit{lastW}(v)$, for all vertices $v$, in total linear time. If for a vertex $v$ we have $W(v)=\emptyset$, then $\mathit{firstW}(v)=\bot$.
\end{lemma}

Recall that, for every vertex $x$, we let $H(x)$ denote the list of all vertices $v\neq r$ such that $\mathit{high}(v)=x$, sorted in decreasing order. Then, for every vertex $v\neq r$, we let $S(v)$ denote the segment of $H(\mathit{high}(v))$ that contains $v$ and is maximal w.r.t. the property that its elements are related as ancestor and descendant. Furthermore, we let $U(v)$ denote the subsegment of $S(v)$ that contains all the proper descendants of $v$.
Now, for every vertex $v$ such that $W(v)\neq\emptyset$, we let $U_2(v)$ be the collection of all $u\in U(v)$ such that: either $(1)$ $\mathit{firstW}(v)>\mathit{low}(u)\geq\mathit{lastW}(v)$, or $(2)$ $u$ is the lowest vertex in $U(v)$ such that $\mathit{low}(u)\geq\mathit{firstW}(v)$.

\begin{lemma}
\label{lemma:type3-b-ii-2-u-w}
Let $(u,v,w)$ be a triple of vertices that induces a Type-3$\beta$ii-$2$ $4$-cut, where $L_1(w)$ is not a descendant of $\mathit{high}(v)$. Then $u\in U_2(v)$ and $w\in W(v)$. 
\end{lemma}
\begin{proof}
Let $e$ be the back-edge in the $4$-cut induced by $(u,v,w)$. Due to our assumption in this subsection, we have that $e=e_L(w)$. Furthermore, we have $M(B(w)\setminus\{e\})=M(v)$, and Lemma~\ref{lemma:type3-b-ii-2-info} implies that $M(w)\neq M(B(w)\setminus\{e\})$. Thus, since $L_1(w)$ is not a descendant of $\mathit{high}(v)$, $w$ satisfies all the conditions to be in $W(v)$. 

Since $(u,v,w)$ induces a Type-3$\beta$ii-$2$ $4$-cut, Lemma~\ref{lemma:type3-b-ii-2-info} implies that $\mathit{high}(u)=\mathit{high}(v)$. In other words, we have $u\in H(\mathit{high}(v))$. Now let $u'$ be a vertex such that $u\geq u'\geq v$ and $u'\in H(\mathit{high}(v))$. This means that we have $\mathit{high}(u')=\mathit{high}(v)$, and therefore Lemma~\ref{lemma:type3-b-ii-2-info} implies that $u'$ is an ancestor of $u$. This shows that $u\in S(v)$. Since $u$ is a proper descendant of $v$, this implies that $u\in U(v)$.
  
Since $w\in W(v)$, we have that $w\geq\mathit{lastW}(v)$. By Lemma~\ref{lemma:type3-b-ii-2-info} we have $w\leq\mathit{low}(u)$. Thus, $w\geq\mathit{lastW}(v)$ implies that $\mathit{low}(u)\geq\mathit{lastW}(v)$. If $\mathit{low}(u)<\mathit{firstW}(v)$, then by definition we have $u\in U_2(v)$, and the proof is complete. Otherwise, we have $\mathit{low}(u)\geq\mathit{firstW}(v)$. Let us suppose, for the sake of contradiction, that there is a vertex $u'\in U(v)$ that is lower than $u$ and satisfies $\mathit{low}(u')\geq\mathit{firstW}(v)$. Since $u'\in U(v)$, we have that $u'$ is a proper descendant of $v$, and so we have $u'>v$. Furthermore, since $u'\in U(v)$, we have that $\mathit{high}(u')=\mathit{high}(v)$. Then, since $\mathit{high}(u')=\mathit{high}(v)$ and $u>u'>v$, by Lemma~\ref{lemma:type3-b-ii-2-info} we have that $u'$ is an ancestor of $u$. Thus, since $\mathit{high}(u')=\mathit{high}(v)=\mathit{high}(u)$, by Lemma~\ref{lemma:same_high} we have that $B(u)\subseteq B(u')$. Since the graph is $3$-edge-connected, this can be strengthened to $B(u)\subset B(u')$. Thus, there is a back-edge $(x,y)\in B(u')\setminus B(u)$. Then $x$ is a descendant of $u'$, and therefore a descendant of $v$. Furthermore, $y$ is an ancestor of $\mathit{high}(u')=\mathit{high}(v)$, and therefore it is a proper ancestor of $v$. This shows that $(x,y)\in B(v)$. Then, since $(u,v,w)$ induces a Type-3$\beta$ii-$2$ $4$-cut, we have that $B(v)=B(u)\sqcup(B(w)\setminus\{e\})$. Since $(x,y)\notin B(u)$, this implies that $(x,y)\in B(w)\setminus\{e\}$. Since $w\in W(v)$, we have that $w\leq\mathit{firstW}(v)$. But we have supposed that $\mathit{low}(u')\geq\mathit{firstW}(v)$. This implies that $\mathit{low}(u')\geq w$, which implies that $y\geq w$ (since $(x,y)\in B(u')$, and therefore $\mathit{low}(u')\leq y$). This means that $y$ cannot be a proper ancestor of $w$, in contradiction to the fact that $(x,y)\in B(w)$.
This shows that $u$ is the lowest vertex in $U(v)$ that has $\mathit{low}(u)\geq\mathit{firstW}(v)$. Thus, by definition, we have $u\in U_2(v)$. 
\end{proof}

\begin{lemma}
\label{lemma:type-3-b-ii-2-L-rel}
Let $(u,v,w)$ be a triple of vertices that induces a Type-3$\beta$ii-$2$ $4$-cut, where $L_1(w)$ is not a descendant of $\mathit{high}(v)$. Let $w'$ be the greatest vertex in $W(v)$ that has $w'\leq\mathit{low}(u)$. Then $(u,v,w')$ also induces a Type-3$\beta$ii-$2$ $4$-cut. Furthermore, if $w'\neq w$, then $B(w)\sqcup\{e_L(w')\}=B(w')\sqcup\{e_L(w)\}$.
\end{lemma}
\begin{proof}
By the assumption throughout this subsection, we have that the back-edge in the $4$-cut induced by $(u,v,w)$ is $e=e_L(w)$. By Lemma~\ref{lemma:type3-b-ii-2-u-w} we have that $w\in W(v)$, and by Lemma~\ref{lemma:type3-b-ii-2-info} we have that $w\leq\mathit{low}(u)$. Thus, we may consider the greatest vertex $w'$ in $W(v)$ that has $w'\leq\mathit{low}(u)$. We will assume that $w'\neq w$, because otherwise there is nothing to show.

Let $(x,y)$ be a back-edge in $B(w')\setminus\{e_L(w')\}$. Since $w'\in W(v)$, we have $M(B(w')\setminus\{e_L(w')\})=M(v)$. This implies that $x$ is a descendant of $M(v)$. Furthermore, since $w'\in W(v)$, by Lemma~\ref{lemma:w(v)_ancestor_of_high} we have that $w'$ is a proper ancestor of $\mathit{high}(v)$, and therefore a proper ancestor of $v$. This implies that $y$ is a proper ancestor of $v$. Thus we have that $(x,y)\in B(v)$. Since $w'\leq\mathit{low}(u)$, we have that $B(u)\cap B(w')=\emptyset$. Thus, we have that $(x,y)\notin B(u)$, and therefore $B(v)=B(u)\sqcup(B(w)\setminus\{e\})$ implies that $(x,y)\in B(w)\setminus\{e\}$. Due to the generality of $(x,y)\in B(w')\setminus\{e_L(w')\}$, this shows that $B(w')\setminus\{e_L(w')\}\subseteq B(w)\setminus\{e_L(w)\}$. Conversely, let $(x,y)$ be a back-edge in $B(w)\setminus\{e\}$. Since $w\in W(v)$, we have that $M(B(w)\setminus\{e\})=M(v)$. This implies that $x$ is a descendant of $M(v)$, and therefore a descendant of $v$, and therefore a descendant of $\mathit{high}(v)$. Since $w'\in W(v)$, by Lemma~\ref{lemma:w(v)_ancestor_of_high} we have that $w'$ is a proper ancestor of $\mathit{high}(v)$. Thus, we have that $x$ is a descendant of $w'$. Since $w$ and $w'$ are both in $W(v)$, Lemma~\ref{lemma:w(v)_ancestor_of_high} implies that they are both proper ancestors of $\mathit{high}(v)$. Thus, $w$ and $w'$ are related as ancestor and descendant. Due to the maximality of $w'$, we have that $w'>w$, and therefore $w'$ is a proper descendant of $w$. Then, since $(x,y)\in B(w)$, we have that $y$ is a proper ancestor of $w$, and therefore a proper ancestor of $w'$. Since $x$ is a descendant of $w'$, this shows that $(x,y)\in B(w')$. Since $w'\in W(v)$, we have that the higher endpoint of $e_L(w')$ is not a descendant of $\mathit{high}(v)$. Thus, since $x$ is a descendant of $\mathit{high}(v)$, we have that $(x,y)\neq e_L(w')$, and therefore $(x,y)\in B(w')\setminus\{e_L(w')\}$. Thus we have shown that $B(w')\setminus\{e_L(w')\}=B(w)\setminus\{e\}$. This implies that $B(v)=B(u)\sqcup(B(w)\setminus\{e\})$ is equivalent to $B(v)=B(u)\sqcup(B(w')\setminus\{e_L(w')\})$, and therefore, by Lemma~\ref{lemma:type-3b-cases}, $(u,v,w')$ induces a $4$-cut. By definition, this is a Type-3$\beta$ii-$2$ $4$-cut.

Let us suppose, for the sake of contradiction, that $e_L(w')=e_L(w)$. Then, since $B(w')\setminus\{e_L(w')\}=B(w)\setminus\{e\}$, we have that $B(w')=B(w)$, in contradiction to the fact that the graph is $3$-edge-connected. Thus, we have $e_L(w')\neq e_L(w)$. Therefore,  $B(w')\setminus\{e_L(w')\}=B(w)\setminus\{e\}$ implies that $B(w)\sqcup\{e_L(w')\}=B(w')\sqcup\{e_L(w)\}$.
\end{proof}

\begin{lemma}
\label{lemma:lastW-firstW}
Let $v$ and $v'$ be two vertices with $W(v)\neq\emptyset$ and $W(v')\neq\emptyset$ such that $v$ is a proper ancestor of $v'$ and $\mathit{high}(v)=\mathit{high}(v')$. Then $\mathit{lastW}(v)>\mathit{firstW}(v')$.
\end{lemma}
\begin{proof}
Let $w$ and $w'$ be two vertices such that $w\in W(v)$ and $w'\in W(v')$. Then it is sufficient to show that $w>w'$ (because $\mathit{lastW}(v)=\mathit{min}(W(v))$ and $\mathit{firstW}(v')=\mathit{max}(W(v'))$). So let us suppose, for the sake of contradiction, that $w\leq w'$. Since $v$ is a proper ancestor of $v'$ with $\mathit{high}(v)=\mathit{high}(v')$, Lemma~\ref{lemma:same_high} implies that $B(v')\subseteq B(v)$. This implies that $M(v')$ is a descendant of $M(v)$. But we cannot have that $M(v')=M(v)$, because the graph is $3$-edge-connected (and otherwise, $\mathit{high}(v)=\mathit{high}(v')$ would imply $B(v)=B(v')$, by Lemma~\ref{lemma:same_M_same_high}). Thus, $M(v')$ is a proper descendant of $M(v)$. Notice that we cannot have $w=w'$, because $w\in W(v)$ and $w'\in W(v')$ imply that $M(B(w)\setminus\{e_L(w)\})=M(v)$ and $M(B(w')\setminus\{e_L(w')\})=M(v')$ (and we showed that we cannot have $M(v)= M(v')$). Thus, we have $w<w'$. Since $w\in W(v)$, by Lemma~\ref{lemma:w(v)_ancestor_of_high} we have that $w$ is an ancestor of $\mathit{high}(v)$, and therefore a proper ancestor of $v$, and therefore a proper ancestor of $v'$. And since $w'\in W(v')$, by Lemma~\ref{lemma:w(v)_ancestor_of_high} we have that $w'$ is an ancestor of $\mathit{high}(v')$, and therefore a proper ancestor of $v'$. Thus, $v'$ is a common descendant of $w$ and $w'$, and therefore $w$ and $w'$ are related as ancestor and descendant. Thus, $w<w'$ implies that $w$ is a proper ancestor of $w'$.

Since $w\in W(v)$, we have that $M(B(w)\setminus\{e_L(w)\})=M(v)$. Let us suppose, for the sake of contradiction, that all back-edges in $B(w)\setminus\{e_L(w)\}$ have their higher endpoint in $T(M(v'))$. (We note that $B(w)\setminus\{e_L(w)\}$ is not empty, since the graph is $3$-edge-connected.) Then we have that $M(v')$ is an ancestor of $M(B(w)\setminus\{e_L(w)\})=M(v)$, contradicting the fact that $M(v')$ is a proper descendant of $M(v)$. This shows that there is at least one back-edge $(x,y)\in B(w)\setminus\{e_L(w)\}$ such that $x$ is not a descendant of $M(v')$. Since $M(B(w)\setminus\{e_L(w)\})=M(v)$, we have that $x$ is a descendant of $M(v)$. Therefore, $x$ is a descendant of $v$, and therefore a descendant of $\mathit{high}(v)=\mathit{high}(v')$. Thus, it cannot be the case that $(x,y)=e_L(w')$ (because $w'\in W(v')$ implies that $L_1(w')$ is not a descendant of $\mathit{high}(v')$). Now, since $x$ is a descendant of $\mathit{high}(v')$, and $\mathit{high}(v')$ is a descendant of $w'$ (by Lemma~\ref{lemma:w(v)_ancestor_of_high}), we have that $x$ is a descendant of $w'$. Furthermore, $y$ is a proper ancestor of $w$, and therefore a proper ancestor of $w'$. This shows that $(x,y)\in B(w')$. But since $(x,y)\neq e_L(w')$, we have that $(x,y)\in B(w')\setminus\{e_L(w')\}$. Thus, since $w'\in W(v')$, we have that $M(B(w')\setminus\{e_L(w')\})=M(v')$, and therefore $x$ is a descendant of $M(v')$, contradicting the fact that $x$ is not a descendant of $M(v')$. We conclude that $w>w'$. Due to the generality of $w\in W(v)$ and $w'\in W(v')$, this implies that $\mathit{lastW}(v)>\mathit{firstW}(v')$. 
\end{proof}

\begin{lemma}
\label{lemma:type3-b-ii-2-relation-between-u2}
Let $v$ and $v'$ be two vertices with $W(v)\neq\emptyset$ and $W(v')\neq\emptyset$ such that $v'$ is a proper descendant of $v$. Suppose that $v$ and $v'$ belong to the same segment $S$ of $H(\mathit{high}(v))$ that is maximal w.r.t. the property that its elements are related as ancestor and descendant. If $U_2(v')=\emptyset$, then $U_2(v)=\emptyset$. If $U_2(v)\neq\emptyset$, then the lowest vertex in $U_2(v)$ is at least as great as the greatest vertex in $U_2(v')$.
\end{lemma}
\begin{proof}
Let us suppose, for the sake of contradiction, that there is a vertex $u\in U_2(v)$, but $U_2(v')$ is empty. Since $u\in U_2(v)$ we have that $u\in S$, and therefore $u$ is related as ancestor and descendant with $v'$. Let us suppose, for the sake of contradiction, that $u$ is an ancestor of $v'$. Since $W(v')\neq\emptyset$, there is a vertex $w\in W(v')$. This implies that $M(B(w)\setminus\{e_L(w)\})=M(v')$. Since the graph is $3$-edge-connected, we have that $|B(w)|>1$. Thus, there is a back-edge $(x,y)\in B(w)\setminus\{e_L(w)\}$. Then we have that $x$ is a descendant of $M(B(w)\setminus\{e_L(w)\})$, and therefore a descendant of $M(v')$, and therefore a descendant of $v'$, and therefore a descendant of $u$. Furthermore, we have that $y$ is a proper ancestor of $w$. Since $w\in W(v')$, by Lemma~\ref{lemma:w(v)_ancestor_of_high} we have that $w$ is an ancestor of $\mathit{high}(v')=\mathit{high}(v)$. Since $u\in U_2(v)$, we have that $\mathit{high}(u)=\mathit{high}(v)$. Thus, we have that $y$ is a proper ancestor of $w$, which is an ancestor of $\mathit{high}(u)$, which is a proper ancestor of $u$. Since $x$ is a descendant of $u$, this shows that $(x,y)\in B(u)$. Therefore, $\mathit{low}(u)\leq y$. Since $(x,y)\in B(w)$, we have that $y$ is a proper ancestor of $w$, and therefore $y<w$. Thus, $\mathit{low}(u)\leq y$ implies that $\mathit{low}(u)<w$. Now, Lemma~\ref{lemma:lastW-firstW} implies that $\mathit{firstW}(v')<\mathit{lastW}(v)$. Thus, since $w\leq\mathit{firstW}(v')$, we have that $w<\mathit{lastW}(v)$. But then $\mathit{low}(u)<w$ implies that $\mathit{low}(u)<\mathit{lastW}(v)$, in contradiction to $u\in U_2(v)$. Thus, our last supposition cannot be true, and therefore we have that $u$ is a proper descendant of $v'$.

Now let us gather the information we have concerning $u$. We know that $u$ is a proper descendant of $v'$, and it belongs to $S$. Furthermore, since $u\in U_2(v)$ we have that $\mathit{low}(u)\geq\mathit{lastW}(v)$, and by Lemma~\ref{lemma:lastW-firstW} we have that $\mathit{lastW}(v)>\mathit{firstW}(v')$. This implies that $\mathit{low}(u)\geq \mathit{firstW}(v')$. Thus, we can consider the lowest proper descendant $u'$ of $v'$ in $S$ that has $\mathit{low}(u')\geq\mathit{firstW}(v')$, and so $U_2(v')$ cannot be empty. A contradiction. Thus, we have shown that $U_2(v')=\emptyset$ implies that $U_2(v)=\emptyset$.

Now let us assume that $U_2(v)\neq\emptyset$. This implies that $U_2(v')\neq\emptyset$. Let $u$ be any vertex in $U_2(v)$, and let $u'$ be any vertex in $U_2(v')$. Let us suppose, for the sake of contradiction, that $u<u'$. Since $u\in U_2(v)$, we have that $u\in S$. And since $u'\in U_2(v')$, we have that $u'\in S$. Thus, $u$ and $u'$ are related as ancestor and descendant. Therefore, $u<u'$ implies that $u$ is a proper ancestor of $u'$. Since $u'\in U_2(v')$, we have that either $\mathit{low}(u')<\mathit{firstW}(v')$, or $u'$ is the lowest proper descendant of $v'$ in $S$ such that $\mathit{low}(u')\geq\mathit{firstW}(v')$.

Let us suppose, first, that $\mathit{low}(u')<\mathit{firstW}(v')$. Since $u$ is an ancestor of $u'$ with $\mathit{high}(u)=\mathit{high}(u')$ (since $u$ and $u'$ are in $S$), by Lemma~\ref{lemma:same_high} we have that $B(u')\subseteq B(u)$. This implies that $\mathit{low}(u)\leq\mathit{low}(u')$. Therefore, $\mathit{low}(u')<\mathit{firstW}(v')$ implies that $\mathit{low}(u)<\mathit{firstW}(v')$. Lemma~\ref{lemma:lastW-firstW} implies that $\mathit{firstW}(v')<\mathit{lastW}(v)$. Therefore, we have $\mathit{low}(u)<\mathit{lastW}(v)$, in contradiction to the fact that $u\in U_2(v)$. Thus, our last supposition cannot be true, and therefore we have that $u'$ is the lowest proper descendant of $v'$ in $S$ such that $\mathit{low}(u')\geq\mathit{firstW}(v')$. Now, since $u\in U_2(v)$, we can argue as above in order to establish that $u$ is a proper descendant of $v'$ (the argument above did not make use of the assumption $U_2(v')=\emptyset$, and so it can be applied here too). But then, since $u$ is a proper descendant of $v'$ in $S$ such that $u<u'$, the minimality of $u'$ implies that $\mathit{low}(u)<\mathit{firstW}(v')$, and so we can arrive again at $\mathit{low}(u)<\mathit{lastW}(v)$, in contradiction to $u\in U_2(v)$. This shows that $u\geq u'$. Due to the generality of $u\in U_2(v)$, this implies that the lowest vertex in $U_2(v)$ is at least as great as $u'$. And due to the generality of $u'\in U_2(v')$, this implies that the lowest vertex in $U_2(v)$ is at least as great as the greatest vertex in $U_2(v')$. 
\end{proof}

Based on Lemma~\ref{lemma:type3-b-ii-2-relation-between-u2}, we can provide an efficient algorithm for computing the sets $U_2(v)$, for all vertices $v\neq r$ such that $W(v)\neq\emptyset$. The computation takes place on segments of $H(x)$ that are maximal w.r.t. the property that their elements are related as ancestor and descendant. Specifically, let $v\neq r$ be a vertex such that $W(v)\neq\emptyset$. Then we have that $U_2(v)\subset S(v)$. In other words, $U_2(v)$ is a subset of the segment of $H(\mathit{high}(v))$ that contains $v$ and is maximal w.r.t. the property that its elements are related as ancestor and descendant. So let $z_1,\dots,z_k$ be the vertices of $S(v)$, sorted in decreasing order. Then, we have that $v=z_i$, for an $i\in\{1,\dots,k\}$. By definition, $U_2(v)$ contains every vertex $u$ in $\{z_1,\dots,z_{i-1}\}$ such that either $\mathit{firstW}(v)>\mathit{low}(u)\geq\mathit{lastW}(v)$, or $u$ is the lowest vertex in this set such that $\mathit{low}(u)\geq\mathit{firstW}(v)$. As an implication of Lemma~\ref{lemma:high_and_low}, we have that the vertices in $\{z_1,\dots,z_{i-1}\}$ are sorted in decreasing order w.r.t. their $\mathit{low}$ point. Thus, it is sufficient to process the vertices from $\{z_1,\dots,z_{i-1}\}$ in reverse order, in order to find the first vertex $u$ that has $\mathit{low}(u)\geq\mathit{lastW}(v)$. Then, we keep traversing this set in reverse order, and, as long as the $\mathit{low}$ point of every vertex $u$ that we meet is lower than $\mathit{firstW}(v)$, we insert $u$ into $U_2(v)$. Then, once we reach a vertex with $\mathit{low}$ point no lower than $\mathit{firstW}(v)$, we also insert it into $U_2(v)$, and we are done.

Now, if there is a proper ancestor $v'$ of $v$ in $S(v)$ such that $\mathit{high}(v')=\mathit{high}(v)$, then we have that $S(v)=S(v')$. If $W(v')\neq\emptyset$, then we have that $U_2(v')$ is defined. Then we can follow the same process as above in order to compute $U_2(v')$. Furthermore, according to Lemma~\ref{lemma:type3-b-ii-2-relation-between-u2}, it is sufficient to start from the greatest element of $U_2(v)$ (i.e., the one that was inserted last into $U_2(v)$). In particular, if $U_2(v)=\emptyset$, then it is certain that $U_2(v')=\emptyset$, and therefore we are done. Otherwise, we just pick up the computation from the greatest vertex in $U_2(v)$. In order to perform efficiently those computations, first we compute, for every vertex $x$, the collection $\mathcal{S}(x)$ of the segments of $H(x)$ that are maximal w.r.t. the property that their elements are related as ancestor and descendant. For every vertex $x$, this computation takes $O(|H(x)|)$ time using Algorithm~\ref{algorithm:segments}, according to Lemma~\ref{lemma:segments}. Since every vertex $\neq r$ participates in exactly one set of the form $H(x)$, we have that the total size of all $\mathcal{S}(x)$, for all vertices $x$, is $O(n)$. Then it is sufficient to process separately all segments of $\mathcal{S}(x)$, for every vertex $x$, as described above, by starting the computation each time from the first vertex $v$ of the segment that satisfies $W(v)\neq\emptyset$. The whole procedure is shown in Algorithm~\ref{algorithm:type3-b-ii-2-U}. The result is formally stated in Lemma~\ref{lemma:algorithm:type3-b-ii-2-U}. 

\begin{algorithm}[H]
\caption{\textsf{Compute the sets $U_2(v)$, for all vertices $v$ such that $W(v)\neq\emptyset$}}
\label{algorithm:type3-b-ii-2-U}
\LinesNumbered
\DontPrintSemicolon
let $\mathcal{V}$ be the collection of all vertices $v$ such that $W(v)\neq\emptyset$\; 
\ForEach{vertex $x$}{
  compute the collection $\mathcal{S}(x)$ of the segments of $H(x)$ that are maximal w.r.t. the property
  that their elements are related as ancestor and descendant\;
}
\ForEach{$v\in\mathcal{V}$}{
  set $U_2(v)\leftarrow\emptyset$\;
}
\ForEach{vertex $x$}{
  \ForEach{segment $S\in\mathcal{S}(x)$}{
    let $v$ be the first vertex in $S$\;
    \While{$v\neq\bot$ \textbf{and} $v\notin\mathcal{V}$}{
      $v\leftarrow\mathit{next}_S(v)$\;
    }
    \lIf{$v=\bot$}{\textbf{continue}}
    let $u\leftarrow\mathit{prev}_S(v)$\;
    \While{$v\neq\bot$}{
      \While{$u\neq\bot$ \textbf{and} $\mathit{low}(u)<\mathit{lastW}(v)$}{
        $u\leftarrow\mathit{prev}_S(u)$\;
      }
      \While{$u\neq\bot$ \textbf{and} $\mathit{low}(u)<\mathit{firstW}(v)$}{
        insert $u$ into $U_2(v)$\;
        $u\leftarrow\mathit{prev}_S(u)$\;
      }
      \If{$u\neq\bot$}{
        insert $u$ into $U_2(v)$\;
      }
      $v\leftarrow\mathit{next}_S(v)$\;
      \While{$v\neq\bot$ \textbf{and} $v\notin\mathcal{V}$}{
        $v\leftarrow\mathit{next}_S(v)$\;
      }      
    }
  }
}
\end{algorithm}

\begin{lemma}
\label{lemma:algorithm:type3-b-ii-2-U}
Algorithm~\ref{algorithm:type3-b-ii-2-U} correctly computes the sets $U_2(v)$, for all vertices $v\neq r$ such that $W(v)\neq\emptyset$. Furthermore, it runs in linear time.
\end{lemma}
\begin{proof}
The idea behind Algorithm~\ref{algorithm:type3-b-ii-2-U} and its correctness has been discussed in the main text above.
It is easy to see that Algorithm~\ref{algorithm:type3-b-ii-2-U} runs in $O(n)$ time, provided that we have computed the vertices $\mathit{firstW}(v)$ and $\mathit{lastW}(v)$ for every vertex $v$ (if they exist). This can be achieved in linear time according to Lemma~\ref{lemma:compute-sets-w}. Thus, Algorithm~\ref{algorithm:type3-b-ii-2-U} has a linear-time implementation.
\end{proof}

\begin{lemma}
\label{lemma:type-3-b-ii-2-L-criterion}
Let $u,v,w$ be three vertices such that $u\in U_2(v)$, $w\in W(v)$, $w\leq\mathit{low}(u)$, and $\mathit{bcount}(v)=\mathit{bcount}(u)+\mathit{bcount}(w)-1$. Then, $e_L(w)\notin B(u)\cup B(v)$, and $B(v)=B(u)\sqcup(B(w)\setminus\{e_L(w)\})$. 
\end{lemma}
\begin{proof}
Since $u\in U_2(v)$ we have that $u$ is a proper descendant of $v$ with $\mathit{high}(u)=\mathit{high}(v)$. Thus, Lemma~\ref{lemma:same_high} implies that $B(u)\subseteq B(v)$. Since $w\in W(v)$, we have that $M(B(w)\setminus\{e_L(w)\})=M(v)$. Now let $(x,y)$ be a back-edge in $B(w)\setminus\{e_L(w)\}$. Then, $x$ is a descendant of $M(B(w)\setminus\{e_L(w)\})=M(v)$. Furthermore, $y$ is a proper ancestor of $w$. Since $w\in W(v)$, by Lemma~\ref{lemma:w(v)_ancestor_of_high} we have that $w$ is an ancestor of $\mathit{high}(v)$. This implies that $y$ is a proper ancestor of $\mathit{high}(v)$, and therefore a proper ancestor of $v$. This shows that $(x,y)\in B(v)$. Due to the generality of $(x,y)\in B(w)\setminus\{e_L(w)\}$, this implies that $B(w)\setminus\{e_L(w)\}\subseteq B(v)$. Let $(x,y)$ be a back-edge in $B(u)$. Then $\mathit{low}(u)\leq y$, and therefore $w\leq\mathit{low}(u)$ implies that $w\leq y$. Thus, $y$ cannot be a proper ancestor of $w$, and therefore $(x,y)\notin B(w)$. Due to the generality of $(x,y)\in B(u)$, this shows that $B(u)\cap B(w)=\emptyset$. Now, since $B(u)\subseteq B(v)$ and $B(w)\setminus\{e_L(w)\}\subseteq B(v)$ and $B(u)\cap B(w)=\emptyset$ and $\mathit{bcount}(v)=\mathit{bcount}(u)+\mathit{bcount}(w)-1$, we have that $B(v)=B(u)\sqcup(B(w)\setminus\{e_L(w)\})$. Furthermore, since $B(u)\cap B(w)=\emptyset$, we have that $e_L(w)\notin B(u)$, and therefore $B(v)=B(u)\sqcup(B(w)\setminus\{e_L(w)\})$ implies that $e_L(w)\notin B(v)$. Thus, we have $e_L(w)\notin B(u)\cup B(v)$.
\end{proof}

Let $\mathcal{C}_\mathit{3\beta ii2}$ denote the collection of all Type-3$\beta ii$-$2$ $4$-cuts of the form $\{(u,p(u)),(v,p(v)),(w,p(w)),e_L(w)\}$, such that $u$ is a descendant of $v$, $v$ is a descendant of $w$, and $L_1(w)$ is not a descendant of $\mathit{high}(v)$.
Now we are ready to describe the algorithm for computing a collection $\mathcal{C}$ of enough $4$-cuts in $\mathcal{C}_\mathit{3\beta ii2}$, so that all $4$-cuts in $\mathcal{C}_\mathit{3\beta ii2}$ are implied from this collection, plus that computed by Algorithm~\ref{algorithm:type2-2}. So let $(u,v,w)$ be a triple of vertices that induces a $4$-cut $C\in\mathcal{C}_\mathit{3\beta ii2}$. Then, by Lemma~\ref{lemma:type3-b-ii-2-u-w} we have that $u\in U_2(v)$ and $w\in W(v)$, and by Lemma~\ref{lemma:type3-b-ii-2-info} we have that $w\leq\mathit{low}(u)$. Now let $w'$ be the greatest vertex in $W(v)$ that satisfies $w'\leq\mathit{low}(u)$. Then, by Lemma~\ref{lemma:type-3-b-ii-2-L-rel} we have that $(u,v,w')$ also induces a $4$-cut $C'\in \mathcal{C}_\mathit{3\beta ii2}$. Furthermore, if $w'\neq w$, then Lemma~\ref{lemma:type-3-b-ii-2-L-rel} implies that $B(w)\sqcup\{e_L(w')\}=B(w')\sqcup\{e_L(w)\}$. Thus, we have that $C$ is implied by $C'$, plus some Type-$2ii$ $4$-cuts that are computed by Algorithm~\ref{algorithm:type2-2} (see Proposition~\ref{proposition:algorithm:type-3-b-ii-2-L}). Thus, it is sufficient to have computed, for every vertex $v$ such that $W(v)\neq\emptyset$, and every $u\in U_2(v)$, the greatest proper ancestor $w$ of $v$ that satisfies $w\leq\mathit{low}(u)$, and then check if the triple $(u,v,w)$ induces a Type-3$\beta ii$-$2$ $4$-cut. This procedure is shown in Algorithm~\ref{algorithm:type-3-b-ii-2-L}. Our result is summarized in Proposition~\ref{proposition:algorithm:type-3-b-ii-2-L}.

\begin{algorithm}[H]
\caption{\textsf{Compute a collection of Type-3$\beta ii$-$2$ $4$-cuts of the form $\{(u,p(u)),(v,p(v)),(w,p(w)),e_L(w)\}$, where $u$ is a descendant of $v$, $v$ is a descendant of $w$, and $L_1(w)$ is not a descendant of $\mathit{high}(v)$, so that all Type-3$\beta ii$-$2$ $4$-cuts of this form are implied from this collection, plus that of the Type-$2$ii $4$-cuts returned by Algorithm~\ref{algorithm:type2-2}}}
\label{algorithm:type-3-b-ii-2-L}
\LinesNumbered
\DontPrintSemicolon
\ForEach{vertex $w\neq r$}{
\label{line:type-3-b-ii-2-L-for}
  compute $M(B(w)\setminus\{e_L(w)\})$\;
}
\ForEach{vertex $x$}{
\label{line:type-3-b-ii-2-L-for-2}
  let $W_0(x)$ be the list of all vertices $w\neq r$ such that $M(w)\neq M(B(w)\setminus\{e_L(w)\})=x$, sorted in decreasing order\;
}
\ForEach{vertex $v$}{
\label{line:type-3-b-ii-2-L-for-3}
  compute the set $U_2(v)$\;
}
\ForEach{vertex $v$ such that $U_2(v)\neq\emptyset$}{
  \ForEach{$u\in U_2(v)$}{
    let $w$ be the greatest proper ancestor of $v$ in $W_0(M(v))$ such that $w\leq\mathit{low}(u)$ and $w\leq\mathit{firstW}(v)$\;
    \label{line:type-3-b-ii-2-L-w}
    \If{$\mathit{bcount}(v)=\mathit{bcount}(u)+\mathit{bcount}(w)-1$ \textbf{and} $L_1(w)$ is not a descendant of $\mathit{high}(v)$}{
    \label{line:type-3-b-ii-2-L-if}
      mark $\{(u,p(u)),(v,p(v)),(w,p(w)),e_L(w)\}$ as a Type-3$\beta ii$-$2$ $4$-cut\;
      \label{line:type-3-b-ii-2-L-mark}
    } 
  }
}
\end{algorithm}

\begin{proposition}
\label{proposition:algorithm:type-3-b-ii-2-L}
Algorithm~\ref{algorithm:type-3-b-ii-2-L} computes a collection of $4$-cuts $\mathcal{C}\subseteq\mathcal{C}_\mathit{3\beta ii2}$, and it runs in linear time. Furthermore, let $\mathcal{C}'$ be the collection of Type-$2ii$ $4$-cuts computed by Algorithm~\ref{algorithm:type2-2}. Then, every $4$-cut in $\mathcal{C}_\mathit{3\beta ii2}$ is implied by $\mathcal{C}\cup\mathcal{C}'$.
\end{proposition}
\begin{proof}
Let $\{(u,p(u)),(v,p(v)),(w,p(w)),e_L(w)\}$ be a $4$-set that is marked in Line~\ref{line:type-3-b-ii-2-L-mark}. This implies that $w\in W_0(M(v))$. Thus, we have that $M(w)\neq M(B(w)\setminus\{e_L(w)\})=M(v)$. Furthermore, since the condition in Line~\ref{line:type-3-b-ii-2-L-if} is satisfied, we have that $L_1(w)$ is not a descendant of $\mathit{high}(v)$. This shows that $w\in W(v)$. Then, we also have that $u\in U_2(v)$, and $w$ is a proper ancestor of $v$, $w\leq\mathit{low}(u)$ and $\mathit{bcount}(v)=\mathit{bcount}(u)+\mathit{bcount}(w)-1$. Thus, Lemma~\ref{lemma:type-3-b-ii-2-L-criterion} implies that $e_L(w)\notin B(u)\cup B(v)$ and $B(v)=B(u)\sqcup(B(w)\setminus\{e_L(w)\})$. In other words, we have that $\{(u,p(u)),(v,p(v)),(w,p(w)),e_L(w)\}$ is a Type-3$\beta ii$-$2$ $4$-cut. Thus, since $L_1(w)$ is not a descendant of $\mathit{high}(v)$, we have that this $4$-cut is in $\mathcal{C}_\mathit{3\beta ii2}$. This shows that the collection $\mathcal{C}$ of $4$-sets marked by Algorithm~\ref{algorithm:type-3-b-ii-2-L} is a collection of $4$-cuts such that $\mathcal{C}\subseteq\mathcal{C}_\mathit{3\beta ii2}$.

According Proposition~\ref{proposition:computing-M(B(v)-S)}, we can compute the values $M(B(w)\setminus\{e_L(w)\})$, for all vertices $w\neq r$, in total linear time. Thus, the \textbf{for} loop in Line~\ref{line:type-3-b-ii-2-L-for} can be performed in linear time. Then, the construction of the lists $W_0(x)$ in Line~\ref{line:type-3-b-ii-2-L-for-2} can be performed in $O(n)$ in total, for all vertices $x$, with bucket-sort. (Notice that all these lists are pairwise disjoint.) The sets $U_2(v)$, for all vertices $v$ such that $W(v)\neq\emptyset$, can be computed in linear time in total, according to Lemma~\ref{lemma:algorithm:type3-b-ii-2-U}. For the remaining vertices $v$, we let $U_2(v)\leftarrow\emptyset$. Thus, the \textbf{for} loop in Line~\ref{line:type-3-b-ii-2-L-for-3} can be performed in linear time.
It remains to show how we can compute all $w$ in Line~\ref{line:type-3-b-ii-2-L-w}. For this, we can use Algorithm~\ref{algorithm:W-queries}. More specifically, for every vertex $v$ such that $U_2(v)\neq\emptyset$, and for every $u\in U_2(v)$, we generate a query $q(W_0(M(v)),\mathit{min}\{p(v),\mathit{low}(u),\mathit{firstW}(v)\})$. This is to return the greatest $w\in W_0(M(v))$ such that $w\leq p(v)$, $w\leq\mathit{low}(u)$ and $w\leq\mathit{firstW}(v)$. Since $w\in W_0(M(v))$, we have that $M(B(w)\setminus\{e_L(w)\})=M(v)$, and therefore $w$ is an ancestor of $M(v)$. Thus, $M(v)$ is a common descendant of $w$ and $v$, and therefore $w$ and $v$ are related as ancestor and descendant. Thus, $w\leq p(v)$ implies that $w$ is a proper ancestor of $v$. Therefore, $w$ is the greatest proper ancestor of $v$ in $W_0(M(v))$ such that $w\leq\mathit{low}(u)$ and $w\leq\mathit{firstW}(v)$. Since the total size of the $U_2$ sets is $O(n)$, we can answer all those queries in $O(n)$ time using Algorithm~\ref{algorithm:W-queries} according to Lemma~\ref{lemma:W-queries}. Thus, we can see that Algorithm~\ref{algorithm:type-3-b-ii-2-L} runs in linear time.
 
Now let $(u,v,w)$ be a triple of vertices that induces a $4$-cut $C\in\mathcal{C}_\mathit{3\beta ii2}$. This means that $C$ is a Type-3$\beta ii$-$2$ $4$-cut such that $L_1(w)$ is not a descendant of $\mathit{high}(v)$. Therefore, Lemma~\ref{lemma:type3-b-ii-2-u-w} implies that $u\in U_2(v)$ and $w\in W(v)$. Furthermore, Lemma~\ref{lemma:type3-b-ii-2-info} implies that $w\leq\mathit{low}(u)$. So let $w'$ be the greatest vertex in $W(v)$ such that $w'\leq\mathit{low}(u)$. Then, Lemma~\ref{lemma:type-3-b-ii-2-L-rel} implies that $(u,v,w')$ induces a $4$-cut $C'\in\mathcal{C}_\mathit{3\beta ii2}$. This implies that $B(v)=B(u)\sqcup(B(w')\setminus\{e_L(w')\})$, and therefore $\mathit{bcount}(v)=\mathit{bcount}(u)+\mathit{bcount}(w')-1$. Since $w'\in W(v)$ we have $w'\in W_0(M(v))$ and $w'\leq\mathit{firstW}(v)$. 

In what follows, let $\tilde{w}=\mathit{firstW}(v)$. Now let $w''$ be the greatest proper ancestor of $v$ in $W_0(M(v))$ such that $w''\leq\mathit{low}(u)$ and $w''\leq\tilde{w}$. We will show that $w'=w''$. By Lemma~\ref{lemma:w(v)_ancestor_of_high} we have that $\tilde{w}$ is a proper ancestor of $\mathit{high}(v)$. Since $\tilde{w}\in W_0(M(v))$ and $w''\in W_0(M(v))$, we have that $M(v)$ is a common descendant of $\tilde{w}$ and $w''$, and therefore $\tilde{w}$ and $w''$ are related as ancestor and descendant. Thus, $w''\leq\tilde{w}$ implies that $w''$ is an ancestor of $\tilde{w}$. Now let us suppose, for the sake of contradiction, that $w''\notin W(v)$. Since $w''\in W_0(M(v))$, this implies that $L_1(w'')$ is a descendant of $\mathit{high}(v)$ (because otherwise $w''$ would satisfy all the conditions to be in $W(v)$). Since $w''\in W_0(M(v))$, we have $M(v)=M(B(w'')\setminus\{e_L(w'')\})\neq M(w'')$. Thus, $e_L(w'')$ is the only back-edge in $B(w'')$ whose higher endpoint is not a descendant of $M(v)$. Similarly, since $\tilde{w}\in W_0(M(v))$, we have that $e_L(\tilde{w})$ is the only back-edge in $B(\tilde{w})$ whose higher endpoint is not a descendant of $M(v)$. Now let $(x,y)=e_L(w'')$. Then $x=L_1(w'')$ is a descendant of $\mathit{high}(v)$, and therefore a descendant of $\tilde{w}$. Furthermore, $y$ is a proper ancestor of $w''$, and therefore a proper ancestor of $\tilde{w}$. This shows that $(x,y)\in B(\tilde{w})$. But since the higher endpoint of $(x,y)$ is not a descendant of $M(v)$, we have that $(x,y)=e_L(\tilde{w})$, contradicting the fact that $L_1(\tilde{w})$ is not a descendant of $\mathit{high}(v)$ (which is implied by $\tilde{w}\in W(v)$). Thus, we have $w''\in W(v)$, which is a strengthening of the condition $w''\in W_0(M(v))$. Thus, $w''$ is the greatest vertex in $W(v)$ such that $w''\leq\mathit{low}(u)$, and so we have $w''=w'$.

Now, since $w'\in W(v)$, we have that $L_1(w')$ is not a descendant of $\mathit{high}(v)$. And since $w'=w''$, we have that $w'$ will be the value of the variable ``$w$'' when we reach Line~\ref{line:type-3-b-ii-2-L-w} during the processing of $v$ and $u$. Thus, the $4$-cut induced by $(u,v,w')$ satisfies all the conditions to be marked in Line~\ref{line:type-3-b-ii-2-L-mark}, and therefore we have that $C'\in\mathcal{C}$. If $w'=w$, then we have that $C=C'$, and thus it is trivially true that $C$ is implied by $\mathcal{C}$. So let us assume that $w'\neq w$. Then, Lemma~\ref{lemma:type-3-b-ii-2-L-rel} implies that $B(w)\sqcup\{e_L(w')\}=B(w')\sqcup\{e_L(w)\}$. Since both $w$ and $w'$ are in $W(v)$, we have that $M(B(w)\setminus\{e_L(w)\})=M(B(w')\setminus\{e_L(w')\})=M(v)$. Thus, $M(v)$ is a common descendant of $w$ and $w'$. Therefore, since the maximality of $w'$ (w.r.t. to $w'\leq\mathit{low}(u)$ and $w'\in W(v)$) implies that $w'>w$, we have that $w'$ is a proper descendant of $w$. Thus, since $B(w)\sqcup\{e_L(w')\}=B(w')\sqcup\{e_L(w)\}$, Lemma~\ref{lemma:type2cuts} implies that $C''=\{(w,p(w)),(w',p(w')),e_L(w),e_L(w')\}$ is a Type-$2ii$ $4$-cut. Since $C=\{(u,p(u)),(v,p(v)),(w,p(w)),e_L(w)\}$ and $C'=\{(u,p(u)),(v,p(v)),(w',p(w')),e_L(w')\}$, notice that $C$ is implied by $C'$ and $C''$ through the pair of edges $\{(w,p(w)),e_L(w)\}$. Let $\mathcal{C}'$ be the collection of Type-$2ii$ $4$-cuts computed by Algorithm~\ref{algorithm:type2-2}. Then, by Proposition~\ref{proposition:type-2-2} we have that $C''$ is implied by $\mathcal{C}'$ through the pair of edges $\{(w,p(w)),e_L(w)\}$. Thus, by Lemma~\ref{lemma:implied_from_union} we have that $C$ is implied by $\mathcal{C}\cup\mathcal{C}'$.
\end{proof}

\noindent\\
\textbf{The case where $L_1(w)$ is a descendant of $\mathit{high}(v)$}\\

\begin{lemma}
\label{lemma:type-3bii-2-2}
Let $(u,v,w)$ be a triple of vertices that induces a Type-3$\beta$ii-$2$ $4$-cut, where $L_1(w)$ is a descendant of $\mathit{high}(v)$. Then $w$ is the greatest ancestor of $\mathit{high}(v)$ such that $M(w)\neq M(B(w)\setminus\{e_L(w)\})=M(v)$ and $w\leq\mathit{low}(u)$. 
\end{lemma}
\begin{proof}
Let $e$ be the back-edge in the $4$-cut induced by $(u,v,w)$. By the assumption we have made for the $4$-cuts in this subsection, we have that $e=e_L(w)$. Then, by Lemma~\ref{lemma:type3-b-ii-2-info} we have that $w$ is an ancestor of $\mathit{high}(v)$, $M(w)\neq M(B(w)\setminus\{e_L(w)\})=M(v)$ and $w\leq\mathit{low}(u)$.  Now let us suppose, for the sake of contradiction, that there is an ancestor $w'$ of $\mathit{high}(v)$ with $M(w')\neq M(B(w')\setminus\{e_L(w')\})=M(v)$ and $w'\leq\mathit{low}(u)$, such that $w'>w$. Since $w'$ and $w$ have $\mathit{high}(v)$ as a common descendant, they are related as ancestor and descendant. Thus, $w'>w$ implies that $w'$ is a proper descendant of $w$. 

Since $L_1(w)$ is a descendant of $\mathit{high}(v)$, we have that the higher endpoint of $e_L(w)$ is a descendant of $\mathit{high}(v)$, and therefore a descendant of $w'$. Furthermore, the lower endpoint of $e_L(w)$ is a proper ancestor of $w$, and therefore a proper ancestor of $w'$. This shows that $e_L(w)\in B(w')$. Since $M(w)\neq M(B(w)\setminus\{e_L(w)\})=M(v)$, we have that $e_L(w)$ is the only back-edge in $B(w)$ whose higher endpoint is not a descendant of $M(v)$. Similarly, since $M(w')\neq M(B(w')\setminus\{e_L(w')\})=M(v)$, we have that $e_L(w')$ is the only back-edge in $B(w')$ whose higher endpoint is not a descendant of $M(v)$. Thus, since $e_L(w)\in B(w')$, we have $e_L(w)=e_L(w')$.

Since $(u,v,w)$ induces a Type-3$\beta$ii-$2$ $4$-cut, we have that $B(v)=B(u)\sqcup(B(w)\setminus\{e_L(w)\})$. This implies that $B(w)\setminus\{e_L(w)\}=B(v)\setminus B(u)$. Now let $(x,y)$ be a back-edge in $B(w)$. If $(x,y)=e_L(w)$, then $(x,y)\in B(w')$. Otherwise, $B(w)\setminus\{e_L(w)\}=B(v)\setminus B(u)$ implies that $(x,y)\in B(v)$. Therefore, $x$ is a descendant of $v$, and therefore a descendant of $\mathit{high}(v)$, and therefore a descendant of $w'$. Furthermore, $y$ is a proper ancestor of $w$, and therefore a proper ancestor of $w'$. This shows that $(x,y)\in B(w')$. Conversely, let $(x,y)$ be a back-edge in $B(w')$. If $(x,y)=e_L(w')$, then $(x,y)\in B(w)$. Otherwise, $M(B(w')\setminus\{e_L(w')\})=M(v)$ implies that $x$ is a descendant of $M(v)$, and therefore a descendant of $v$. Furthermore, $y$ is a proper ancestor of $w'$, and therefore a proper ancestor of $\mathit{high}(v)$, and therefore a proper ancestor of $v$. This shows that $(x,y)\in B(v)$. Since $w'\leq\mathit{low}(u)$, we have that $B(u)\cap B(w')=\emptyset$ (because no back-edge in $B(u)$ has low enough lower endpoint in order to leap over $w'$). Thus, we have $(x,y)\notin B(u)$, and therefore $(x,y)\in B(v)\setminus B(u)$. Since $B(w)\setminus\{e_L(w)\}=B(v)\setminus B(u)$, this implies that $(x,y)\in B(w)$. Thus, we have that $B(w')=B(w)$, in contradiction to the fact that the graph is $3$-edge-connected. We conclude that $w$ is the greatest ancestor of $\mathit{high}(v)$ such that $M(w)\neq M(B(w)\setminus\{e_L(w)\})=M(v)$ and $w\leq\mathit{low}(u)$. 
\end{proof}

Lemma~\ref{lemma:type-3bii-2-2} motivates the following definition. Let $v$ be a vertex $\neq r$. Then we let $\widetilde{W}(v)$ denote the collection of all vertices $w$ such that $w$ is an ancestor of $\mathit{high}(v)$, $L_1(w)$ is a descendant of $\mathit{high}(v)$, and $M(w)\neq M(B(w)\setminus\{e_L(w)\})=M(v)$. Then, Lemma~\ref{lemma:type-3bii-2-2} implies that, if there is a triple of vertices $(u,v,w)$ that induces a Type-3$\beta$ii-$2$ $4$-cut such that $L_1(w)$ is a descendant of $\mathit{high}(v)$, then $w\in\widetilde{W}(v)$. Thus, the sets $\widetilde{W}(v)$ can guide us into the search for such $4$-cuts.

\begin{lemma}
\label{lemma:tilde-w-disjoint}
Let $v$ and $v'$ be two distinct vertices $\neq r$. Then $\widetilde{W}(v)\cap\widetilde{W}(v')=\emptyset$.
\end{lemma}
\begin{proof}
If $M(v)\neq M(v')$, then obviously $\widetilde{W}(v)\cap\widetilde{W}(v')=\emptyset$. (Because every vertex $w\in\widetilde{W}(v)$ has $M(B(w)\setminus\{e_L(w)\})=M(v)$, and every vertex $w'\in\widetilde{W}(v')$ has $M(B(w')\setminus\{e_L(w')\})=M(v')$.) Thus, we may assume that $M(v)=M(v')$. 

Now let us suppose, for the sake of contradiction, that $\widetilde{W}(v)\cap\widetilde{W}(v')\neq\emptyset$. Since $v\neq v'$, we may assume w.l.o.g. that $v>v'$. Therefore, since $v$ and $v'$ have $M(v)=M(v')$ as a common descendant, we have that $v$ is a proper descendant of $v'$. Let $w$ be a vertex in $\widetilde{W}(v)\cap\widetilde{W}(v')$. Then, since $w\in\widetilde{W}(v)$, we have that $L_1(w)$ is a descendant of $\mathit{high}(v)$. Furthermore, since $w\in\widetilde{W}(v')$, we have that $w$ is an ancestor of $\mathit{high}(v')$. Notice that, since $M(w)\neq M(B(w)\setminus\{e_L(w)\})=M(v)$, we have that $L_1(w)$ (i.e., the higher endpoint of $e_L(w)$) is not a descendant of $M(v)$. Since $v'$ is a proper ancestor of $v$ with $M(v')=M(v)$, by Lemma~\ref{lemma:same_m_subset_B} we have that $B(v')\subseteq B(v)$. 

Let us suppose, for the sake of contradiction, that $\mathit{high}(v)$ is a proper ancestor of $v'$. Let $(x,y)$ be a back-edge in $B(v)$. Then $x$ is a descendant of $v$, and therefore a descendant of $v'$. Furthermore, $y$ is an ancestor of $\mathit{high}(v)$, and therefore a proper ancestor of $v'$. This shows that $(x,y)\in B(v')$. Due to the generality of $(x,y)\in B(v)$, this implies that $B(v)\subseteq B(v')$. Thus, $B(v')\subseteq B(v)$ implies that $B(v)=B(v')$, in contradiction to the fact that the graph is $3$-edge-connected. Thus, we have that $\mathit{high}(v)$ is not a proper ancestor of $v'$. Since $\mathit{high}(v)$ is a proper ancestor of $v$, and $v'$ is also an ancestor of $v$, we have that $\mathit{high}(v)$ and $v'$ are related as ancestor and descendant. Thus, since $\mathit{high}(v)$ is not a proper ancestor of $v'$, we have that $\mathit{high}(v)$ is a descendant of $v'$. 

Now, since $L_1(w)$ is a descendant of $\mathit{high}(v)$, it is a descendant of $v'$. Furthermore, since $w$ is an ancestor of $\mathit{high}(v')$, we have that $w$ is a proper ancestor of $v'$. Therefore, the lower endpoint of $e_L(w)$ is a proper ancestor of $v'$ (since it is a proper ancestor of $w$). This shows that $e_L(w)\in B(v')$, and therefore $L_1(w)$ is a descendant of $M(v')$. But $L_1(w)$ is not a descendant of $M(v)$, in contradiction to the fact that $M(v')=M(v)$. Thus, we conclude that $\widetilde{W}(v)\cap\widetilde{W}(v')=\emptyset$.
\end{proof}   

\begin{lemma}
\label{lemma:sets-tilde-w}
For every vertex $x$, let $\widetilde{L}(x)$ be the list of all $w$ such that $M(w)\neq M(B(w)\setminus\{e_L(w)\})=x$, sorted in decreasing order. Let $v$ be a vertex such that $M(v)=x$ and $\widetilde{W}(v)\neq\emptyset$. Let $w=\mathit{max}(\widetilde{W}(v))$. Then, $w$ is the greatest vertex in $\widetilde{L}(x)$ such that $w\leq\mathit{high}(v)$. Furthermore, $\widetilde{W}(v)$ is a segment of $\widetilde{L}(x)$. 
\end{lemma}
\begin{proof}
By definition of $\widetilde{W}(v)$, we have that $w=\mathit{max}(\widetilde{W}(v))$ is an ancestor of $\mathit{high}(v)$, and therefore $w\leq\mathit{high}(v)$. Now let us suppose, for the sake of contradiction, that there is a vertex $w'\in \widetilde{L}(x)$ with $w'>w$, such that $w'\leq\mathit{high}(v)$. Since $M(B(w')\setminus\{e_L(w')\})=M(B(w)\setminus\{e_L(w)\})=x$, we have that $w'$ and $w$ have $x$ as a common descendant. Thus, $w'$ and $w$ are related as ancestor and descendant. Therefore, $w'>w$ implies that $w'$ is a proper descendant of $w$. 

Since $w'\in\widetilde{L}(x)$, we have that $M(B(w')\setminus\{e_L(w')\})=x=M(v)$. Thus, since $M(B(w')\setminus\{e_L(w')\})$ is a descendant of $w'$ and $M(v)$ is a descendant of $v$, we have that $w'$ and $v$ have $x$ as a common descendant, and therefore they are related as ancestor and descendant. Thus, since $w'\leq\mathit{high}(v)$ and $\mathit{high}(v)$ is an ancestor of $v$, we have that $w'$ is an ancestor of $v$. Thus, since $w'$ and $\mathit{high}(v)$ have $v$ as a common descendant, we have that $w'$ is related as ancestor and descendant with $\mathit{high}(v)$. Therefore, $w'\leq\mathit{high}(v)$ implies that $w'$ is an ancestor of $\mathit{high}(v)$. Since $w=\mathit{max}(\widetilde{W}(v))$ and $w'>w$, we have that $w'\notin\widetilde{W}(v)$. Thus, since $w'$ is an ancestor of $\mathit{high}(v)$, we have that $L_1(w')$ is not a descendant of $\mathit{high}(v)$ (because this is the only condition that prevents $w'$ to be in $\widetilde{W}(v)$). Notice that, since $M(w')\neq M(B(w')\setminus\{e_L(w')\})=x$, we have that $L_1(w')$ (i.e., the higher endpoint of $e_L(w')$) is not a descendant of $x$.

Since $w\in\widetilde{W}(v)$, we have that $L_1(w)$ is a descendant of $\mathit{high}(v)$. Thus, $e_L(w)\neq e_L(w')$. Since $L_1(w)$ is a descendant of $\mathit{high}(v)$, we have that $L_1(w)$ is a descendant of $w'$. And since the lower endpoint of $e_L(w)$ is a proper ancestor of $w$, we have that the lower endpoint of $e_L(w)$ is a proper ancestor of $w'$. This shows that $e_L(w)\in B(w')$. Notice that, since $M(w)\neq M(B(w)\setminus\{e_L(w)\})=x$, we have that $L_1(w)$ (i.e., the higher endpoint of $e_L(w)$) is not a descendant of $x$. Since $e_L(w)\neq e_L(w')$ and $e_L(w)\in B(w')$, we have that $e_L(w)\in B(w')\setminus\{e_L(w')\}$. Thus, $B(w')\setminus\{e_L(w')\}$ contains a back-edge whose higher endpoint is not a descendant of $x$, and therefore $M(B(w')\setminus\{e_L(w')\})\neq x$, a contradiction. Thus, we have shown that there is no vertex $w'\in \widetilde{L}(x)$ with $w'>w$, such that $w'\leq\mathit{high}(v)$. This shows that $w$ is the greatest vertex in $\widetilde{L}(x)$ such that $w\leq\mathit{high}(v)$.

By definition, we have that $\widetilde{W}(v)\subseteq \widetilde{L}(x)$. (Since $\widetilde{L}(x)$ contains every vertex $w$ such that $M(w)\neq M(B(w)\setminus\{e_L(w)\})=M(v)$.) Now let us suppose, for the sake of contradiction, that $\widetilde{W}(v)$ is not a segment of $\widetilde{L}(x)$. Since $w=\mathit{max}(\widetilde{W}(v))$, this means that there are $w'$ and $w''$ in $\widetilde{L}(x)$, with $w>w'>w''$, such that $w'\notin\widetilde{W}(v)$ and $w''\in\widetilde{W}(v)$. Since $w,w',w''\in \widetilde{L}(x)$, we have that $M(B(w)\setminus\{e_L(w)\})=M(B(w')\setminus\{e_L(w')\})=M(B(w'')\setminus\{e_L(w'')\})=x$. Thus, $\{w,w',w''\}$ have $x$ as a common descendant, and therefore all three of them are related as ancestor and descendant. Thus, $w>w'>w''$ implies that $w$ is a proper descendant of $w'$, and $w'$ is a proper descendant of $w''$. 

Since $w\in\widetilde{W}(v)$, we have that $w$ is an ancestor of $\mathit{high}(v)$. This implies that $w'$ is also an ancestor of $\mathit{high}(v)$. Thus, since $w'\in \widetilde{L}(x)$ and $w'\notin\widetilde{W}(v)$, we have that $L_1(w')$ is not a descendant of $\mathit{high}(v)$. Now, since $w''\in\widetilde{W}(v)$, we have that $L_1(w'')$ is a descendant of $\mathit{high}(v)$. Thus, $e_L(w')\neq e_L(w'')$. Since $L_1(w'')$ is a descendant of $\mathit{high}(v)$ and $w'$ is an ancestor of $\mathit{high}(v)$, we have that $L_1(w'')$ is a descendant of $w'$. Furthermore, the lower endpoint of $e_L(w'')$ is a proper ancestor of $w''$, and therefore a proper ancestor of $w'$. This shows that $e_L(w'')\in B(w')$. As previously, notice that $e_L(w'')$ is not a descendant of $x$ (since $M(w'')\neq M(B(w'')\setminus\{e_L(w'')\})=x$). Since $e_L(w')\neq e_L(w'')$ and $e_L(w'')\in B(w')$, we have that $e_L(w'')\in B(w')\setminus\{e_L(w')\}$. Thus, $B(w')\setminus\{e_L(w')\}$ contains a back-edge whose higher endpoint is not a descendant of $x$, and therefore $M(B(w')\setminus\{e_L(w')\})\neq x$ -- in contradiction to $w'\in\widetilde{L}(x)$. Thus, we conclude that $\widetilde{W}(v)$ is a segment of $\widetilde{L}(x)$.
\end{proof}

Algorithm~\ref{algorithm:type3-b-ii-w} shows how we can compute all sets $\widetilde{W}(v)$, for all $v\neq r$, in total linear time. The idea is to find, for every vertex $v$, the greatest $w$ that has $M(w)\neq M(B(w)\setminus\{e_L(w)\})=M(v)$ and $w\leq\mathit{high}(v)$. According to Lemma~\ref{lemma:sets-tilde-w}, if $\widetilde{W}(v)\neq\emptyset$, then this $w$ must satisfy $w\in\widetilde{W}(v)$ (i.e., it also has that $L_1(w)$ is a descendant of $\mathit{high}(v)$). Then, still according to Lemma~\ref{lemma:sets-tilde-w}, we have that $\widetilde{W}(v)$ is a segment of the decreasingly sorted list that consists of all vertices $w'$ with $M(w')\neq M(B(w')\setminus\{e_L(w')\})=M(v)$. Thus, we keep traversing this list, and we greedily insert as many vertices as we can into $\widetilde{W}(v)$, until we reach a $w'$ that no longer satisfies that $L_1(w')$ is a descendant of $\mathit{high}(v)$.
The full proof of correctness and linear complexity of Algorithm~\ref{algorithm:type3-b-ii-w} is given in Lemma~\ref{lemma:algorithm:type3-b-ii-w}. The proof of linear complexity relies on Lemma~\ref{lemma:tilde-w-disjoint}: i.e., the sets in $\{\widetilde{W}(v) \mid v \mbox{ is a vertex }\neq r\}$ are pairwise disjoint.

\begin{algorithm}[H]
\caption{\textsf{Compute the sets $\widetilde{W}(v)$, for all $v\neq r$}}
\label{algorithm:type3-b-ii-w}
\LinesNumbered
\DontPrintSemicolon
compute $M(B(w)\setminus\{e_L(w)\})$, for every $w\neq r$\;
\label{line:type3-b-ii-w-1}
initialize an empty list $\widetilde{L}(x)\leftarrow\emptyset$, for every vertex $x$\;
\For{$w\leftarrow n$ to $w=2$}{
\label{line:type3-b-ii-w-for-1}
  \If{$M(w)\neq M(B(w)\setminus\{e_L(w)\})$}{
    insert $w$ into $\widetilde{L}(M(B(w)\setminus\{e_L(w)\}))$\;
  }
}
\tcp{$\widetilde{L}(x)$ contains all vertices $w$ with $M(w)\neq M(B(w)\setminus\{e_L(w)\})=x$, sorted in decreasing order, for every vertex $x$}
initialize an array $\mathit{currentVertex}[x]$, for every vertex $x$\;
\label{line:type3-b-ii-w-L}
\ForEach{vertex $x$}{
  let $\mathit{currentVertex}[x]\leftarrow$ first element of $\widetilde{L}(x)$\;
  \label{line:type3-b-ii-w-first}
}
initialize $\widetilde{W}(v)\leftarrow\emptyset$, for every vertex $v\neq r$\;
\For{$v\leftarrow n$ to $v=2$}{
\label{line:type3-b-ii-w-for}
  let $x\leftarrow M(v)$\;
  let $w\leftarrow\mathit{currentVertex}[x]$\;
  \label{line:type3-b-ii-w-current}
  \While{$w>\mathit{high}(v)$}{
  \label{line:type3-b-ii-w-while}
    $w\leftarrow\mathit{next}_{\widetilde{L}(x)}(w)$\;
  }
  \While{$w\neq\bot$ \textbf{and} $L_1(w)$ is a descendant of $\mathit{high}(v)$}{
  \label{line:type3-b-ii-w-while-2}
    insert $w$ into $\widetilde{W}(v)$\;
    \label{line:type3-b-ii-w-current-insert}
    $w\leftarrow\mathit{next}_{\widetilde{L}(x)}(w)$\;
  }
  $\mathit{currentVertex}[x]\leftarrow w$\; 
  \label{line:type3-b-ii-w-current-last}
}
\end{algorithm}

\begin{lemma}
\label{lemma:algorithm:type3-b-ii-w}
Algorithm~\ref{algorithm:type3-b-ii-w} correctly computes the sets $\widetilde{W}(v)$, for all vertices $v\neq r$. Furthermore, it has a linear-time implementation.
\end{lemma}
\begin{proof}
It should be clear that, when we reach Line~\ref{line:type3-b-ii-w-L}, we have that $\widetilde{L}(x)$, for every vertex $x$, is the decreasingly sorted list of all vertices $w$ that have $M(w)\neq M(B(w)\setminus\{e_L(w)\})=x$. The idea in the \textbf{for} loop in Line~\ref{line:type3-b-ii-w-for} is to process all vertices $v\neq r$ in a bottom-up fashion. Then, for every vertex $v\neq r$, we start searching in $\widetilde{L}(M(v))$ for the greatest vertex $w$ that has $w\leq\mathit{high}(v)$ (see the \textbf{while} loop in Line~\ref{line:type3-b-ii-w-while}). This search for this $w$ starts from the last vertex that we accessed in $\widetilde{L}(M(v))$. This is ensured by the use of variable $\mathit{currentVertex}[x]$ (where $x=M(v)$), which we use in order to initialize $w$ in Line~\ref{line:type3-b-ii-w-current}, and then we update $\mathit{currentVertex}[x]$ in Line~\ref{line:type3-b-ii-w-current-last}. We only have to explain why this is sufficient.

First, we have that, if $v'$ and $v$ are two vertices with $M(v')=M(v)$ and $v'<v$, then $\mathit{high}(v')$ is an ancestor of $\mathit{high}(v)$. To see this, suppose the contrary. Since $M(v')=M(v)$, we have that $v'$ and $v$ have $M(v)$ as a common descendant. Therefore, $v'<v$ implies that $v'$ is a proper ancestor of $v$. Then, since $\mathit{high}(v')$ is a proper ancestor of $v'$, we have that $\mathit{high}(v')$ is a proper ancestor of $v$. Thus, since $\mathit{high}(v)$ is also a proper ancestor of $v$, we have that $\mathit{high}(v')$ and $\mathit{high}(v)$ are related as ancestor and descendant. Then, since $\mathit{high}(v')$ is not an ancestor of $\mathit{high}(v)$, it must be a proper descendant of $\mathit{high}(v)$. This implies that $\mathit{high}(v')>\mathit{high}(v)$. Since $v'$ is a proper ancestor of $v$ with $M(v')=M(v)$, Lemma~\ref{lemma:same_m_subset_B} implies that $B(v')\subseteq B(v)$. This implies that $\mathit{high}(v)\geq \mathit{high}(v')$, a contradiction. Therefore, we have indeed that $\mathit{high}(v')$ is an ancestor of $\mathit{high}(v)$. This implies that $\mathit{high}(v')\leq \mathit{high}(v)$.

Now we can see inductively that $\widetilde{W}(v)$ is computed correctly by the \textbf{for} loop in Line~\ref{line:type3-b-ii-w-for}, for every vertex $v\neq r$. We use induction on the number of times that a vertex $v$ with $M(v)=x$ is processed, for every fixed $x$. The first time that the \textbf{for} loop in Line~\ref{line:type3-b-ii-w-for} processes a vertex $v$ with $M(v)=x$, we begin the search for vertices $w\in\widetilde{W}(v)$ from the beginning of $\widetilde{L}(x)$ (due to the initialization of $\mathit{currentVertex}[x]$ in Line~\ref{line:type3-b-ii-w-first}). The \textbf{while} loop in Line~\ref{line:type3-b-ii-w-while} traverses the list $\widetilde{L}(x)$, until it reaches a vertex $w$ such that $w\leq\mathit{high}(v)$. Thus, this is the greatest vertex in $\widetilde{L}(x)$ such that $w\leq\mathit{high}(v)$. According to Lemma~\ref{lemma:sets-tilde-w}, if $\widetilde{W}(v)\neq\emptyset$, then $w=\mathit{max}(\widetilde{W}(v))$. Since $w\in\widetilde{L}(x)$ and $w\leq\mathit{high}(v)$, it is sufficient to check whether $L_1(w)$ is a descendant of $\mathit{high}(v)$. If that is the case, then we correctly insert $w$ into $\widetilde{W}(v)$, in Line~\ref{line:type3-b-ii-w-current-insert}. Furthermore, by Lemma~\ref{lemma:sets-tilde-w} we have that $\widetilde{W}(v)$ is a segment of $\widetilde{L}(x)$. Thus, since $w$ is the leftmost element of this segment, it is sufficient to keep traversing $\widetilde{L}(x)$, and keep inserting all the vertices that we meet into $\widetilde{W}(v)$, until we meet a vertex that is provably no longer in $\widetilde{W}(v)$. Since $\widetilde{L}(x)$ is sorted in decreasing order, all the vertices $w'$ that we meet have $w'\leq w\leq\mathit{high}(v)$. Thus, the only reason that may prevent such a $w'$ to be in $\widetilde{W}(v)$, is that $L_1(w')$ is not a descendant of $\mathit{high}(v)$. This shows that the \textbf{while} loop in Line~\ref{line:type3-b-ii-w-while-2} will correctly compute $\widetilde{W}(v)$. Then, in Line~\ref{line:type3-b-ii-w-current-last} we set ``$\mathit{currentVertex}[x]\leftarrow w$". This implies that the next time that we meet a vertex $v'$ that has $M(v')=x$, we will start the search for $\widetilde{W}(v')$ from the last entry $w$ of $\widetilde{L}(x)$ that we accessed while processing $v$. This is sufficient for the following reasons. First, since the \textbf{for} loop in Line~\ref{line:type3-b-ii-w-for} processes the vertices in a bottom-up fashion, we have that the next $v'$ with $M(v')=x$ that it will process is a proper ancestor of $v$. Thus, as shown above, we have that $\mathit{high}(v')\leq \mathit{high}(v)$. Then, the search from $\widetilde{L}(x)$ will be picked up from the vertex $w$ that is either the greatest that satisfies $w\leq\mathit{high}(v)$, or it is the lowest in $\widetilde{W}(v)$. In the first case, it is sufficient to start the search in $\widetilde{L}(x)$ for $\widetilde{W}(v')$ from $w$. In the second case, we note that by Lemma~\ref{lemma:tilde-w-disjoint} we have that $\widetilde{W}(v)\cap\widetilde{W}(v')=\emptyset$. Thus, no vertex that is greater than $w$ can be in $\widetilde{W}(v')$: because no vertex in $\widetilde{W}(v)$ is in $\widetilde{W}(v')$, and by Lemma~\ref{lemma:sets-tilde-w} we have that the greatest vertex in $\widetilde{W}(v)$ is the greatest vertex $w'$ in $\widetilde{L}(x)$ that has $w'\leq\mathit{high}(v)$. Thus, with the same argument that we used for $v$ we can see that $\widetilde{W}(v')$ will be correctly computed, and the same is true for any future vertex $v''$ with $M(v'')=x$ that we will meet. This demonstrates the correctness of Algorithm~\ref{algorithm:type3-b-ii-w}.

By Proposition~\ref{proposition:computing-M(B(v)-S)}, we have that the values $M(B(w)\setminus\{e_L(w)\})$, for all vertices $w\neq r$, can be computed in linear time in total. Thus, Line~\ref{line:type3-b-ii-w-1} can be performed in linear time. All other steps take $O(n)$ time in total. In particular, observe that, in the worst case, the \textbf{for} loop in Line~\ref{line:type3-b-ii-w-for} may have to traverse the entire lists $M^{-1}(x)$ and $\widetilde{L}(x)$, for all vertices $x$. Still, since all lists $M^{-1}(x)$ are pairwise disjoint, and all lists $\widetilde{L}(x)$ are pairwise disjoint, we have that the \textbf{for} loop in Line~\ref{line:type3-b-ii-w-for} takes $O(n)$ time. Thus, Algorithm~\ref{algorithm:type3-b-ii-w} runs in linear time.
\end{proof}

\begin{lemma}
\label{lemma:type3-b-ii-high}
A triple of vertices $(u,v,w)$ induces a Type-3$\beta$ii-$2$ $4$-cut, where $L_1(w)$ is a descendant of $\mathit{high}(v)$, if and only if: $(1)$ $u$ is a proper descendant of $v$, $(2)$ $u$ and $v$ belong to a segment of $H(\mathit{high}(v))$ that is maximal w.r.t. the property that all its elements are related as ancestor and descendant, $(3)$ $w\in\widetilde{W}(v)$, $(4)$ $w\leq\mathit{low}(u)$, and $(5)$ $\mathit{bcount}(v)=\mathit{bcount}(u)+\mathit{bcount}(w)-1$.
\end{lemma}
\begin{proof}
$(\Rightarrow)$ Due to the convention we have made in this subsection, we have that the back-edge in the $4$-cut induced by $(u,v,w)$ is $e_L(w)$. Since $(u,v,w)$ induces a Type-3$\beta$ii-$2$ $4$-cut, by definition we have that $u$ is a proper descendant of $v$. Let $u'$ be a vertex such that $u\geq u'\geq v$ and $\mathit{high}(u')=\mathit{high}(v)$. Then, Lemma~\ref{lemma:type3-b-ii-2-info} implies that $u'$ is an ancestor of $u$. Thus, the segment from $u$ to $v$ in $H(\mathit{high}(v))$ consists of vertices that are related as ancestor and descendant (since all of them are ancestors of $u$). This implies $(2)$. Lemma~\ref{lemma:type-3bii-2-2} implies that $w$ is an ancestor of $\mathit{high}(v)$ and $M(w)\neq M(B(w)\setminus\{e_L(w)\})=M(v)$. By assumption we have that $L_1(w)$ is a descendant of $\mathit{high}(v)$. Thus, $w$ satisfies all the conditions to be in $\widetilde{W}(v)$. $(4)$ is an implication of Lemma~\ref{lemma:type3-b-ii-2-info}. $(5)$ is an immediate implication of the fact that $B(v)=B(u)\sqcup(B(w)\setminus\{e_L(w)\})$.

$(\Leftarrow)$ $(2)$ implies that $\mathit{high}(u)=\mathit{high}(v)$. Since by $(1)$ we have that $u$ is a proper descendant of $v$, Lemma~\ref{lemma:same_high} implies that $B(u)\subseteq B(v)$. Let $(x,y)$ be a back-edge in $B(w)\setminus\{e_L(w)\}$. Since $w\in\widetilde{W}(v)$, we have that $M(B(w)\setminus\{e_L(w)\})=M(v)$. Thus, since $(x,y)\in B(w)\setminus\{e_L(w)\}$, we have that $x$ is a descendant of $M(B(w)\setminus\{e_L(w)\})$, and therefore a descendant of $M(v)$. Furthermore, $y$ is a proper ancestor of $w$, and therefore a proper ancestor of $\mathit{high}(v)$ (since $w\in\widetilde{W}(v)$ implies that $w$ is an ancestor of $\mathit{high}(v)$). This shows that $(x,y)\in B(v)$. Due to the generality of $(x,y)\in B(w)\setminus\{e_L(w)\}$, this implies that $B(w)\setminus\{e_L(w)\}\subseteq B(v)$. Let $(x,y)$ be a back-edge in $B(u)$. Then we have $\mathit{low}(u)\leq y$. Therefore, $w\leq\mathit{low}(u)$ implies that $w\leq y$. Thus, $y$ cannot be a proper ancestor of $w$, and therefore $(x,y)\notin B(w)$. Due to the generality of $(x,y)\in B(u)$, this implies that $B(u)\cap B(w)=\emptyset$. Thus, since $B(u)\subseteq B(v)$ and $B(w)\setminus\{e_L(w)\}\subseteq B(v)$ and $B(u)\cap B(w)=\emptyset$ and  $\mathit{bcount}(v)=\mathit{bcount}(u)+\mathit{bcount}(w)-1$, we have that $B(v)=B(u)\sqcup(B(w)\setminus\{e_L(w)\})$. Thus, $(u,v,w)$ induces a Type-3$\beta$ii-$2$ $4$-cut.
\end{proof}

\begin{lemma}
\label{lemma:same_high_dif_bcount}
Let $u$ and $v$ be two distinct vertices with $\mathit{high}(u)=\mathit{high}(v)$ that are related as ancestor and descendant. Then, $\mathit{bcount}(u)\neq\mathit{bcount}(v)$.
\end{lemma}
\begin{proof}
We may assume w.l.o.g. that $v$ is a proper ancestor of $u$. Then, since $\mathit{high}(u)=\mathit{high}(v)$, Lemma~\ref{lemma:same_high} implies that $B(u)\subseteq B(v)$. Thus, if we suppose that $\mathit{bcount}(u)=\mathit{bcount}(v)$, then we have that $B(u)=B(v)$, in contradiction to the fact that the graph is $3$-edge-connected. This shows that $\mathit{bcount}(u)\neq\mathit{bcount}(v)$.
\end{proof}

Now we will show how to compute all Type-3$\beta$ii-$2$ $4$-cuts of the form $\{(u,p(u)),(v,p(v)),(w,p(w)),e_L(w)\}$, where $B(v)=B(u)\sqcup(B(w)\setminus\{e_L(w)\})$ and $L_1(w)$ is a descendant of $\mathit{high}(v)$. So let $(u,v,w)$ be a triple of vertices that induces a $4$-cut of this form. Then, by Lemma~\ref{lemma:type3-b-ii-high} we have that $w\in\widetilde{W}(v)$, and $u\in S(v)$ (i.e., $u$ belongs to the segment of $H(\mathit{high}(v))$ that contains $v$ and is maximal w.r.t. the property that all its elements are related as ancestor and descendant). Thus, given $v$ and $w\in\widetilde{W}(v)$, it is sufficient to find all $u\in S(v)$ that provide a triple $(u,v,w)$ that induces a $4$-cut of this form. By Lemma~\ref{lemma:type3-b-ii-high} we have that $\mathit{bcount}(v)=\mathit{bcount}(u)+\mathit{bcount}(w)-1$. Then,  Lemma~\ref{lemma:same_high_dif_bcount} implies that $u$ is the unique vertex in $S(v)$ that has $\mathit{bcount}(u)=\mathit{bcount}(v)-\mathit{bcount}(w)+1$. Thus, the idea is to process separately all segments of $H(x)$ that are maximal w.r.t. the property that their elements are related as ancestor and descendant, for every vertex $x$. Let $S$ be such a segment. Then, we store the $\mathit{bcount}$ values for all vertices in $S$. (By Lemma~\ref{lemma:same_high_dif_bcount}, all these are distinct.) Then, for every $v\in S$ such that $\widetilde{W}(v)\neq\emptyset$, and every $w\in\widetilde{W}(v)$, we seek the unique $u$ in $S$ that has  $\mathit{bcount}(u)=\mathit{bcount}(v)-\mathit{bcount}(w)+1$, and, if it exists, then we check whether all conditions in Lemma~\ref{lemma:type3-b-ii-high}  are satisfied, in order to have that $(u,v,w)$ induces a $4$-cut of the desired form. This procedure is shown in Algorithm~\ref{algorithm:type3-b-ii-high}. The proof of correctness and linear complexity is given in Proposition~\ref{proposition:algorithm:type3-b-ii-high}.

\begin{algorithm}[H]
\caption{\textsf{Compute all Type-3$\beta$ii-$2$ $4$-cuts of the form $\{(u,p(u)),(v,p(v)),(w,p(w)),e_L(w)\}$, where $B(v)=B(u)\sqcup(B(w)\setminus\{e_L(w)\})$ and $L_1(w)$ is a descendant of $\mathit{high}(v)$}}
\label{algorithm:type3-b-ii-high}
\LinesNumbered
\DontPrintSemicolon
compute the set $\widetilde{W}(v)$, for every vertex $v\neq r$\;
\label{line:type3-b-ii-high-w-tild}
for every vertex $x$, let $H(x)$ be the list of all vertices $z$ with $\mathit{high}(z)=x$, sorted in decreasing order\;
\ForEach{vertex $x$}{
\label{line:type3-b-ii-high-w-maximal}
  compute the collection $\mathcal{S}(x)$ of the segments of $H(x)$ that are maximal w.r.t. the property
  that their elements are related as ancestor and descendant\;
}
initialize an array $A$ of size $m$\;
\ForEach{vertex $x$}{
\label{line:type3-b-ii-high-main-for}
  \ForEach{$S\in\mathcal{S}(x)$}{
  \label{line:type3-b-ii-high-S}
    \ForEach{$z\in S$}{
    \label{line:type3-b-ii-high-for-A}
      set $A[\mathit{bcount}(z)]\leftarrow z$\;
      \label{line:type3-b-ii-high-A}
    }
    \ForEach{$v\in S$}{
      \ForEach{$w\in\widetilde{W}(v)$}{
      \label{line:type3-b-ii-high-w}
        let $u\leftarrow A[\mathit{bcount}(v)-\mathit{bcount}(w)+1]$\;
        \label{line:type3-b-ii-high-u}
        \If{$u\neq\bot$ \textbf{and} $u$ is a proper descendant of $v$ \textbf{and} $w\leq\mathit{low}(u)$}{
        \label{line:type3-b-ii-high-if}
          mark $\{(u,p(u)),(v,p(v)),(w,p(w)),e_L(w)\}$ as a Type-3$\beta$ii-$2$ $4$-cut\;
          \label{line:type3-b-ii-high-mark}
        }
      }
    }
    \ForEach{$z\in S$}{
      set $A[\mathit{bcount}(z)]\leftarrow\bot$\;
      \label{line:type3-b-ii-high-A-flush}
    }
  }
}
\end{algorithm}

\begin{proposition}
\label{proposition:algorithm:type3-b-ii-high}
Algorithm~\ref{algorithm:type3-b-ii-high} correctly computes all Type-3$\beta$ii-$2$ $4$-cuts of the form $\{(u,p(u)),(v,p(v)),(w,p(w)),e_L(w)\}$, where $B(v)=B(u)\sqcup(B(w)\setminus\{e_L(w)\})$ and $L_1(w)$ is a descendant of $\mathit{high}(v)$. Furthermore, it has a linear-time implementation.
\end{proposition}
\begin{proof}
For every vertex $x$, let $\mathcal{S}(x)$ be the collection of the segments of $H(x)$ that are maximal w.r.t. the property that their elements are related as ancestor and descendant.
Let $\{(u,p(u)),(v,p(v)),(w,p(w)),e_L(w)\}$ be a Type-3$\beta$ii-$2$ $4$-cut such that $B(v)=B(u)\sqcup(B(w)\setminus\{e_L(w)\})$ and $L_1(w)$ is a descendant of $\mathit{high}(v)$. Lemma~\ref{lemma:type3-b-ii-high} implies that $u$ and $v$ belong to the same segment $S$ of $\mathcal{S}(\mathit{high}(v))$, $w\in\widetilde{W}(v)$, $w\leq\mathit{low}(u)$, and $\mathit{bcount}(v)=\mathit{bcount}(u)+\mathit{bcount}(w)-1$. Let $x=\mathit{high}(v)$. Then, during the processing of $S\in\mathcal{S}(x)$ (in the \textbf{for} loop in Line~\ref{line:type3-b-ii-high-S}), we will eventually reach Line~\ref{line:type3-b-ii-high-u} for this particular $w$ (during the \textbf{for} loop in Line~\ref{line:type3-b-ii-high-w}). We will show that the $\mathit{bcount}(v)-\mathit{bcount}(w)+1$ entry of the $A$ array is precisely $u$. Notice that the entries of $A$ are filled with vertices in Line~\ref{line:type3-b-ii-high-A}, for every segment $S\in\mathcal{S}(x)$, for every vertex $x$. And then, after the processing of $S$, the entries of $A$ that were filled with vertices are set again to $\mathit{null}$ in Line~\ref{line:type3-b-ii-high-A-flush}. Thus, when we reach Line~\ref{line:type3-b-ii-high-w}, we have that the non-$\mathit{null}$ entries of $A$ contain vertices from $S$. More precisely, for every $z\in S$, we have that $A[\mathit{bcount}(z)]=z$. Thus, since all vertices in $S$ are related as ancestor and descendant and have the same $\mathit{high}$ point, Lemma~\ref{lemma:same_high_dif_bcount} implies that $A[\mathit{bcount}(z)]=z$, for every $z\in S$, when the \textbf{for} loop in Line~\ref{line:type3-b-ii-high-for-A} is completed (during the processing of $S$); and also, we have $A[c]=\bot$, if there is no vertex $z\in S$ with $\mathit{bcount}(z)=c$. Therefore, since $\mathit{bcount}(v)=\mathit{bcount}(u)+\mathit{bcount}(w)-1$, we have that $\mathit{bcount}(u)=\mathit{bcount}(v)-\mathit{bcount}(w)+1$, and therefore $A[\mathit{bcount}(v)-\mathit{bcount}(w)+1]=u$. Thus, when we reach Line~\ref{line:type3-b-ii-high-u}, we have that the variable ``u" contains the value $u$, and so the $4$-cut $\{(u,p(u)),(v,p(v)),(w,p(w)),e_L(w)\}$ will be correctly marked in Line~\ref{line:type3-b-ii-high-mark} (since the condition in Line~\ref{line:type3-b-ii-high-if} is satisfied).

Conversely, whenever the condition in Line~\ref{line:type3-b-ii-high-if} is satisfied, we have that: $(1)$ $u$ is a proper descendant of $v$, $(2)$ $u$ and $v$ belong to a segment of $H(\mathit{high}(v))$ that is maximal w.r.t. the property that all its elements are related as ancestor and descendant, $(3)$ $w\in\widetilde{W}(v)$, $(4)$ $w\leq\mathit{low}(u)$, and $(5)$ $\mathit{bcount}(v)=\mathit{bcount}(u)+\mathit{bcount}(w)-1$. Thus, Lemma~\ref{lemma:type3-b-ii-high} implies that $\{(u,p(u)),(v,p(v)),(w,p(w)),e_L(w)\}$ is a Type-3$\beta$ii-$2$ $4$-cut such that $B(v)=B(u)\sqcup(B(w)\setminus\{e_L(w)\})$ and $L_1(w)$ is a descendant of $\mathit{high}(v)$. Therefore, it is correct to mark it in Line~\ref{line:type3-b-ii-high-mark}.

Now let us establish the linear-time complexity of Algorithm~\ref{algorithm:type3-b-ii-high}. First, by Lemma~\ref{lemma:algorithm:type3-b-ii-w} we have that Line~\ref{line:type3-b-ii-high-w-tild} has a linear-time implementation. In particular, Lemma~\ref{lemma:tilde-w-disjoint} implies that the total size of all sets $\widetilde{W}(v)$, for $v\neq r$, is $O(n)$. Then, by Lemma~\ref{lemma:segments} we have that the collections $\mathcal{S}(x)$, for every vertex $x$, can be computed in linear time in total. Thus, the \textbf{for} loop in Line~\ref{line:type3-b-ii-high-w-maximal} has a linear-time implementation. Finally, the \textbf{for} loop in Line~\ref{line:type3-b-ii-high-main-for} is completed in linear time, precisely because all segments in the collection $\{S\in\mathcal{S}(x)\mid x \mbox{ is a vertex }\}$ are pairwise disjoint, and so they have total size $O(n)$, and the sets $\widetilde{W}(v)$, for all $v\neq r$, have total size $O(n)$.
\end{proof}

\subsubsection{Type-3$\beta ii$-$3$ $4$-cuts}

Now we consider case $(3)$ of Lemma~\ref{lemma:type-3b-cases}. 

Let $u,v,w$ be three vertices $\neq r$ such that $w$ is proper ancestor of $v$, $v$ is a proper ancestor of $u$, and there is a back-edge $e\in B(u)$ such that $e\notin B(v)\cup B(w)$, $B(v)=(B(u)\setminus\{e\})\sqcup B(w)$ and $M(w)=M(v)$. By Lemma~\ref{lemma:type-3b-cases}, we have that $C=\{(u,p(u)),(v,p(v)),(w,p(w)),e\}$ is a $4$-cut, and we call this a Type-3$\beta$ii-$3$ $4$-cut.

The following lemma provides some useful information concerning this type of $4$-cuts.

\begin{lemma}
\label{lemma:type3-b-ii-3-info}
Let $(u,v,w)$ be a triple of vertices that induces a Type-3$\beta$ii-$3$ $4$-cut, and let $e$ be the back-edge of this $4$-cut. Then $e=(\mathit{highD}(u),\mathit{high}(u))$. Furthermore, $\mathit{low}(u)\geq w$ and $\mathit{high}(u)\neq\mathit{high}_2(u)=\mathit{high}(v)$. Finally, if $u'$ is a vertex such that $u\geq u'\geq v$ and either $\mathit{high}(u')=\mathit{high}(v)$ or $\mathit{high}_2(u')=\mathit{high}(v)$, then $u'$ is an ancestor of $u$. 
\end{lemma}
\begin{proof}
Since $(u,v,w)$ induces a Type-3$\beta$ii-$3$ $4$-cut, we have that $e\in B(u)$, $e\notin B(v)\cup B(w)$, and $B(v)=(B(u)\setminus\{e\})\sqcup B(w)$ $(*)$. This implies that $e$ is the only back-edge in $B(u)$ that is not in $B(v)$. Now let $(x_1,y_1),\dots,(x_k,y_k)$ be all the back-edges in $B(u)$ sorted in decreasing order w.r.t. their lower endpoint, so that we have $(x_i,y_i)=(\mathit{highD}_i(u),\mathit{high}_i(u))$, for every $i\in\{1,\dots,k\}$. Let $i\in\{1,\dots,k\}$ be an index such that $(x_i,y_i)\in B(v)$. Then we have that $y_i$ is a proper ancestor of $v$. This implies that all $y_j$, with $j\in\{i,\dots,k\}$ are proper ancestors of $v$. Furthermore, we have that all $x_j$, for $j\in\{1,\dots,k\}$, are descendants of $u$, and therefore all of them are descendants of $v$. This shows that all the back-edges $(x_i,y_i),\dots,(x_k,y_k)$ are in $B(v)$. Thus, we cannot have that $(x_1,y_1)\in B(v)$, because otherwise we would have $B(u)\subseteq B(v)$. Since $e$ is the unique back-edge in $B(u)\setminus B(v)$, this shows that $e=(x_1,y_1)$, and therefore $e=(\mathit{highD}(u),\mathit{high}(u))$. Furthermore, this argument also shows that $y_2\neq y_1$, and therefore $\mathit{high}_2(u)\neq\mathit{high}(u)$.

Since $e=(\mathit{highD}(u),\mathit{high}(u))$ and $B(u)\setminus\{e\}\subseteq B(v)$ (and $B(u)\setminus\{e\}$ contains at least one back-edge, since the graph is $3$-edge-connected), we have that $(x_2,y_2)\in B(v)$, and therefore $\mathit{high}(v)\geq\mathit{high}_2(u)$. Conversely, let $(x,y)$ be a back-edge in $B(v)$ such that $y=\mathit{high}(v)$. We can use a similar argument as above, in order to conclude that, if $\mathit{high}(v)<w$, then $B(v)\subseteq B(w)$. But this is impossible to be the case, since there are back-edges from $B(u)$ in $B(v)$, and we have $B(u)\cap B(w)=\emptyset$. Thus, we have that $(x,y)\notin B(w)$. Then $B(v)=(B(u)\setminus\{e\})\sqcup B(w)$ implies that $(x,y)\in B(u)\setminus\{e\}$. Since $e=(\mathit{highD}(u),\mathit{high}(u))$ and $\mathit{high}_2(u)\neq\mathit{high}(u)$, this implies that $y_2\geq y$. Thus we have $\mathit{high}_2(u)\geq\mathit{high}(v)$. This shows that $\mathit{high}(v)=\mathit{high}_2(u)$.

Let us suppose, for the sake of contradiction, that $\mathit{low}(u)<w$. Consider a back-edge $(x,y)\in B(u)$ such that $y=\mathit{low}(u)$. Then $x$ is a descendant of $u$, and therefore a descendant of $v$, and therefore a descendant of $w$. Furthermore, since $\mathit{low}(u)<w$, we have that $y$ is a proper ancestor of $w$ (because $y$ and $w$ are related as ancestor and descendant, since both of them are ancestors of $u$). This shows that $(x,y)\in B(w)$. But this contradicts $B(u)\cap B(w)=\emptyset$, which is a consequence of $(*)$. Thus we have shown that $\mathit{low}(u)\geq w$.

Now let $u'$ be a vertex with $u\geq u'\geq v$ such that there is a back-edge $(x,y)\in B(u')$ with $y=\mathit{high}(v)$ (notice that this includes the cases $\mathit{high}(u')=\mathit{high}(v)$ and $\mathit{high}_2(u')=\mathit{high}(v)$). Since $u$ is a descendant of $v$, $u\geq u'\geq v$ implies that $u'$ is a descendant of $v$. Then, $x$ is a descendant of $v$. Thus, since $y=\mathit{high}(v)$, we have that $(x,y)\in B(v)$. Since $(u,v,w)$ induces a Type-3$\beta$ii-$3$ $4$-cut, by definition we have that $w$ is a proper ancestor of $v$ with $M(w)=M(v)$. Then, Lemma~\ref{lemma:same_M_dif_B_lower} implies that $\mathit{high}(v)$ is a descendant of $w$. Thus, $y=\mathit{high}(v)$ is not a proper ancestor of $w$, and therefore $(x,y)\notin B(w)$. Thus, $B(v)=(B(u)\setminus\{e\})\sqcup B(w)$ implies that $(x,y)\in B(u)$. Then we have that $x$ is a common descendant of $u$ and $u'$, and therefore $u$ and $u'$ are related as ancestor and descendant. Thus, $u\geq u'$ implies that $u'$ is an ancestor of $u$. 
\end{proof}

Recall that, for every vertex $x$, we let $\widetilde{H}(x)$ denote the list of all vertices $z$ such that either $\mathit{high}_1(z)=x$ or $\mathit{high}_2(z)=x$, sorted in decreasing order. For every vertex $v\neq r$, let $\widetilde{S}_1(v)$ denote the segment of $\widetilde{H}(\mathit{high}(v))$ that contains $v$ and is maximal w.r.t. the property that its elements are related as ancestor and descendant. Then, for every vertex $v\neq r$ with $\mathit{nextM}(v)\neq\bot$, we let $U_3(v)$ denote the collection of all $u\in\widetilde{S}_1(v)$ such that: $(1)$ $u$ is a proper descendant of $v$, $(2)$ $\mathit{high}_1(u)\neq\mathit{high}_2(u)=\mathit{high}_1(v)$, $(3)$ $\mathit{low}(u)\geq\mathit{lastM}(v)$, and $(4)$ either $\mathit{low}(u)<\mathit{nextM}(v)$, or $u$ is the lowest vertex in $\widetilde{S}_1(v)$ that satisfies $(1)$, $(2)$ and $\mathit{low}(u)\geq\mathit{nextM}(v)$. 

\begin{lemma}
\label{lemma:type-3-b-ii-3-rel}
Let $(u,v,w)$ be a triple of vertices that induces a Type-3$\beta$ii-$3$ $4$-cut. Then $U_3(v)\neq\emptyset$, and let $\tilde{u}$ be the greatest vertex in $U_3(v)$. Then, if $u\notin U_3(v)$, we have that $(\tilde{u},v,w)$ induces a Type-3$\beta$ii-$3$ $4$-cut, and $B(\tilde{u})\sqcup\{e_\mathit{high}(u)\}=B(u)\sqcup\{e_\mathit{high}(\tilde{u})\}$.
\end{lemma}
\begin{proof}
Let $u'$ be a vertex in $\widetilde{H}(\mathit{high}(v))$ such that $u\geq u'\geq v$. Since $u'\in\widetilde{H}(\mathit{high}(v))$, we have that either $\mathit{high}_1(u')=\mathit{high}(v)$ or $\mathit{high}_2(u')=\mathit{high}(v)$. Thus, since $(u,v,w)$ induces a Type-3$\beta$ii-$3$ $4$-cut, by Lemma~\ref{lemma:type3-b-ii-3-info} we have that $u'$ is an ancestor of $u$. This implies that all vertices from $u$ to $v$ in $\widetilde{H}(\mathit{high}(v))$ are related as ancestor and descendant (since all of them are ancestors of $u$). This shows that $u\in\widetilde{S}_1(v)$. Since $(u,v,w)$ induces a Type-3$\beta$ii-$3$ $4$-cut, we have that $u$ is a proper descendant of $v$. Furthermore, by Lemma~\ref{lemma:type3-b-ii-3-info} we have that $\mathit{high}_1(u)\neq\mathit{high}_2(u)=\mathit{high}_1(v)$, and $\mathit{low}(u)\geq w\geq \mathit{lastM}(v)$. Thus, if $\mathit{low}(u)<\mathit{nextM}(v)$, then we have $u\in U_3(v)$. Otherwise, we have $\mathit{low}(u)\geq\mathit{nextM}(v)$, and therefore we have $U_3(v)\neq\emptyset$, because we can consider the lowest vertex $\tilde{u}\in\widetilde{S}_1(v)$ that is a proper descendant of $v$ and satisfies $\mathit{high}_1(\tilde{u})\neq\mathit{high}_2(\tilde{u})=\mathit{high}_1(v)$ and $\mathit{low}(\tilde{u})\geq\mathit{nextM}(v)$. 

This shows that $U_3(v)\neq\emptyset$. Furthermore, this shows that, if $u\notin U_3(v)$, then we can define $\tilde{u}$ as previously, and we have $\tilde{u}<u$ (due to the minimality of $\tilde{u}$), and therefore $\tilde{u}$ is a proper ancestor of $u$ (since both $\tilde{u}$ and $u$ are in $\widetilde{S}_1(v)$, and therefore they are related as ancestor and descendant). 
Now we will show that $\tilde{u}$ is the greatest vertex in $U_3(v)$. Notice that $\mathit{low}(\tilde{u})\geq\mathit{nextM}(v)$, and every other vertex $u'\in U_3(v)$ (except $\tilde{u}$, that is) satisfies $\mathit{low}(u')<\mathit{nextM}(v)$. So let us suppose, for the sake of contradiction, that $\tilde{u}$ is not the greatest vertex in $U_3(v)$. Thus, there is a vertex $u'\in U_3(v)$ such that $u'>\tilde{u}$. Since both $u'$ and $\tilde{u}$ are in $\widetilde{S}_1(v)$, this implies that $u'$ is a proper descendant of $\tilde{u}$. Now let $(x,y)$ be a back-edge in $B(u')$ such that $y=\mathit{low}(u')$. Then $x$ is a descendant of $u'$, and therefore a descendant of $\tilde{u}$. We also have $y=\mathit{low}(u')<\mathit{nextM}(v)\leq\mathit{low}(\tilde{u})$. Since $(x,y)$ is a back-edge, we have that $y$ is an ancestor of $x$. Furthermore, we have that $\mathit{low}(\tilde{u})$ is an ancestor of $\tilde{u}$, and therefore an ancestor of $u'$, and therefore an ancestor of $x$. Thus, $x$ is a common descendant of $y$ and $\mathit{low}(\tilde{u})$, and therefore $y$ and $\mathit{low}(\tilde{u})$ are related as ancestor and descendant, and therefore $y$ is a proper ancestor of $\mathit{low}(\tilde{u})$ (since $y<\mathit{low}(\tilde{u})$). Since $x$ is a descendant of $\tilde{u}$, this implies that $(x,y)\in B(\tilde{u})$. But $y$ is lower than $\mathit{low}(\tilde{u})$, a contradiction. This shows that $\tilde{u}$ is the greatest vertex in $U_3(v)$.

Now let us assume that $u\notin U_3(v)$. Let us suppose, for the sake of contradiction, that $e_\mathit{high}(\tilde{u})= e_\mathit{high}(u)$. Then, since $\tilde{u}$ is an ancestor of $u$, by Lemma~\ref{lemma:same_high} we have $B(u)\subseteq B(\tilde{u})$. This can be strengthened to $B(u)\subset B(\tilde{u})$, since the graph is $3$-edge-connected. Thus, there is a back-edge $(x,y)\in B(\tilde{u})\setminus B(u)$. In particular, we have $(x,y)\neq e_\mathit{high}(u)=e_\mathit{high}(\tilde{u})$. Since $\tilde{u}\in U_3(v)$, we have $\mathit{high}_2(\tilde{u})=\mathit{high}(v)$. Thus, since $(x,y)\neq e_\mathit{high}(\tilde{u})$, we have that $y$ is an ancestor of $\mathit{high}_2(\tilde{u})=\mathit{high}(v)$, and therefore it is a proper ancestor of $v$. Furthermore, $x$ is a descendant of $\tilde{u}$, and therefore a descendant of $v$. This shows that $(x,y)\in B(v)$. Then, since $(u,v,w)$ induces a Type-3$\beta$ii-$3$ $4$-cut, we have $B(v)=(B(u)\setminus\{e_\mathit{high}(u)\})\sqcup B(w)$. Thus, since $(x,y)\in B(v)$ and $(x,y)\notin B(u)$, we have $(x,y)\in B(w)$. But we have $\mathit{low}(\tilde{u})\geq\mathit{nextM}(v)$, and therefore $\mathit{low}(\tilde{u})\geq w$, and therefore $y\geq w$ (since $(x,y)\in B(\tilde{u})$ implies that $y\geq \mathit{low}(\tilde{u})$). Thus, $y$ cannot be a proper ancestor of $w$, a contradiction. This shows that $e_\mathit{high}(\tilde{u})\neq e_\mathit{high}(u)$.

Now let us suppose, for the sake of contradiction, that $e_\mathit{high}(\tilde{u})\in B(u)$. Since $\tilde{u}\in U_3(v)$, we have $\mathit{high}_1(\tilde{u})\neq\mathit{high}_2(\tilde{u})=\mathit{high}(v)$. This implies that the lower endpoint of $e_\mathit{high}(\tilde{u})$ is greater than $\mathit{high}(v)$. Since  $\mathit{high}_1(u)\neq\mathit{high}_2(u)=\mathit{high}(v)$, we have that $e_\mathit{high}(u)$ is the only back-edge in $B(u)$ whose lower endpoint is greater than $\mathit{high}(v)$. Thus, since $e_\mathit{high}(\tilde{u})\in B(u)$, we have $e_\mathit{high}(\tilde{u})=e_\mathit{high}(u)$, a contradiction. This shows that $e_\mathit{high}(\tilde{u})\notin B(u)$. Similarly, we can show that $e_\mathit{high}(u)\notin B(\tilde{u})$.

Now let $(x,y)$ be a back-edge in $B(\tilde{u})\setminus\{e_\mathit{high}(\tilde{u})\}$. Then we have that $x$ is a descendant of $\tilde{u}$, and therefore a descendant of $v$. Furthermore, since $\tilde{u}\in U_3(v)$, we have $\mathit{high}_2(\tilde{u})=\mathit{high}_1(v)$, and therefore $y$ is an ancestor of $\mathit{high}_1(v)$. This shows that $(x,y)\in B(v)$. Then, $B(v)=(B(u)\setminus\{e_\mathit{high}(u)\})\sqcup B(w)$ implies that either $(x,y)\in B(u)\setminus\{e_\mathit{high}(u)\}$, or $(x,y)\in B(w)$. The case $(x,y)\in B(w)$ is rejected, since $\mathit{low}(\tilde{u})\geq\mathit{nextM}(v)\geq w$. Thus, we have $(x,y)\in B(u)\setminus\{e_\mathit{high}(u)\}$. Due to the generality of $(x,y)\in B(\tilde{u})\setminus\{e_\mathit{high}(\tilde{u})\}$, this shows that $B(\tilde{u})\setminus\{e_\mathit{high}(\tilde{u})\}\subseteq B(u)\setminus\{e_\mathit{high}(u)\}$. Conversely, let $(x,y)$ be a back-edge in $B(u)\setminus\{e_\mathit{high}(u)\}$. Then we have that $x$ is a descendant of $u$, and therefore a descendant of $\tilde{u}$. Furthermore, we have $(x,y)\in B(v)$, and therefore $y$ is a proper ancestor of $v$, and therefore a proper ancestor of $\tilde{u}$. This shows that $(x,y)\in B(\tilde{u})$. Since $(x,y)\in B(u)$ and $e_\mathit{high}(\tilde{u})\notin B(u)$, we have $(x,y)\neq e_\mathit{high}(\tilde{u})$. Thus, $(x,y)\in B(\tilde{u})\setminus\{e_\mathit{high}(\tilde{u})\}$. Due to the generality of $(x,y)\in B(u)\setminus\{e_\mathit{high}(u)\}$, this shows that $B(u)\setminus\{e_\mathit{high}(u)\}\subseteq B(\tilde{u})\setminus\{e_\mathit{high}(\tilde{u})\}$. Thus, we have shown that $B(\tilde{u})\setminus\{e_\mathit{high}(\tilde{u})\}= B(u)\setminus\{e_\mathit{high}(u)\}$. 

Then, since $e_\mathit{high}(\tilde{u})\notin B(u)$ and $e_\mathit{high}(u)\notin B(\tilde{u})$, we have $B(\tilde{u})\sqcup\{e_\mathit{high}(u)\}=B(u)\sqcup\{e_\mathit{high}(\tilde{u})\}$. Furthermore, since $B(v)=(B(u)\setminus\{e_\mathit{high}(u)\})\sqcup B(w)$, we have $B(v)=(B(\tilde{u})\setminus\{e_\mathit{high}(\tilde{u})\})\sqcup B(w)$. Finally, since $\mathit{high}_1(\tilde{u})\neq\mathit{high}_2(\tilde{u})=\mathit{high}(v)$, we have $\mathit{high}_1(\tilde{u})>\mathit{high}(v)$, and therefore $e_\mathit{high}(\tilde{u})\notin B(v)$. Then, by Lemma~\ref{lemma:type-3b-cases}, we conclude that $(\tilde{u},v,w)$ induces a Type-3$\beta$ii-$3$ $4$-cut.
\end{proof}

\begin{lemma}
\label{lemma:type3-b-ii-3-relation-between-u3}
Let $v$ and $v'$ be two vertices with $\mathit{nextM}(v)\neq\bot$ and $\mathit{nextM}(v')\neq\bot$, such that $v'$ is a proper descendant of $v$ with $\mathit{high}(v)=\mathit{high}(v')$ and $\widetilde{S}_1(v)=\widetilde{S}_1(v')$. If $U_3(v')=\emptyset$, then $U_3(v)=\emptyset$. If $U_3(v')\neq\emptyset$, then the lowest vertex in $U_3(v)$ (if it exists) is greater than, or equal to, the greatest vertex in $U_3(v')$.
\end{lemma}
\begin{proof}
Since $\mathit{high}(v)=\mathit{high}(v')$ and $v'$ is a proper descendant of $v$, by Lemma~\ref{lemma:same_high} we have that $B(v')\subseteq B(v)$. Since the graph is $3$-edge-connected, this can be strengthened to $B(v')\subset B(v)$. This implies that $M(v')$ is a descendant of $M(v)$. Since $\mathit{high}(v)=\mathit{high}(v')$, we cannot have $M(v')=M(v)$, because otherwise Lemma~\ref{lemma:same_M_same_high} implies that $B(v')=B(v)$. Thus, we have that $M(v')$ is a proper descendant of $M(v)$. Since $\mathit{lastM}(v)$ is an ancestor of $M(v)$ and $\mathit{nextM}(v')$ is an ancestor of $M(v')$, we have that $\mathit{lastM}(v)$ and $\mathit{nextM}(v')$ are related as ancestor and descendant. Let us suppose, for the sake of contradiction, that $\mathit{nextM}(v')$ is a descendant of $\mathit{lastM}(v)$. Since $M(v')$ is a proper descendant of $M(v)$ and $M(\mathit{lastM}(v))=M(v)$, there is a back-edge $(x,y)\in B(\mathit{lastM}(v))$ such that $x$ is not a descendant of $M(v')$. Then, we have that $x$ is a descendant of $M(v)$, and therefore a descendant of $v$, and therefore a descendant of $\mathit{high}(v)=\mathit{high}(v')$. By Lemma~\ref{lemma:lower_than_high}, we have that $\mathit{high}(v')$ is a descendant of $\mathit{nextM}(v')$. Therefore, since $x$ is a descendant of $\mathit{high}(v')$, we have that $x$ is a descendant of $\mathit{nextM}(v')$. Furthermore, $y$ is a proper ancestor of $\mathit{lastM}(v)$, and therefore a proper ancestor of $\mathit{nextM}(v')$. This shows that $(x,y)\in B(\mathit{nextM}(v'))$. But $x$ is not a descendant of $M(v')=M(\mathit{nextM}(v'))$, a contradiction. Thus, we have that $\mathit{nextM}(v')$ is not a descendant of $\mathit{lastM}(v)$. Therefore, since $\mathit{lastM}(v)$ and $\mathit{nextM}(v')$ are related as ancestor and descendant, we have that $\mathit{nextM}(v')$ is a proper ancestor of $\mathit{lastM}(v)$.

Let us suppose, for the sake of contradiction, that $U_3(v')=\emptyset$ and $U_3(v)\neq\emptyset$. Let $u$ be a vertex in $U_3(v)$. Let us suppose, for the sake of contradiction, that $u$ is not a proper descendant of $v'$. Since $u\in U_3(v)$, we have $u\in\widetilde{S}_1(v)$. Since $v'$ is also in $\widetilde{S}_1(v)$, this implies that $u$ and $v'$ are related as ancestor and descendant. Thus, since $u$ is not a proper descendant of $v'$, we have that $u$ is an ancestor of $v'$. Let $(x,y)$ be a back-edge in $B(v')$ such that $y=\mathit{low}(v')$. Lemma~\ref{lemma:same_m_subset_B} implies that $B(\mathit{nextM}(v'))\subseteq B(v')$. Thus, we have that $\mathit{low}(v')$ is an ancestor of $\mathit{low}(\mathit{nextM}(v'))$, and therefore a proper ancestor of $\mathit{nextM}(v')$. Since $\mathit{nextM}(v')$ is a proper ancestor of $\mathit{lastM}(v)$, this implies that $\mathit{low}(v')$ is a proper ancestor of $\mathit{lastM}(v)$. Now, since $(x,y)\in B(v')$, we have that $x$ is a descendant of $v'$, and therefore a descendant of $u$. Furthermore, $y=\mathit{low}(v')$ is a proper ancestor of $\mathit{lastM}(v)$, and therefore a proper ancestor of $v$, and therefore a proper ancestor of $u$. This shows that $(x,y)\in B(u)$. But then we have that $\mathit{low}(u)\leq\mathit{low}(v')<\mathit{lastM}(v)$, in contradiction to the fact that $u\in U_3(v)$. Thus, we have that $u$ is a proper descendant of $v'$. Then, since $u\in U_3(v)$, we have that $\mathit{high}_1(u)\neq\mathit{high}_2(u)=\mathit{high}(v)=\mathit{high}(v')$. Furthermore, we have that $\mathit{low}(u)\geq\mathit{lastM}(v)$, and therefore $\mathit{low}(u)\geq\mathit{nextM}(v')$. This implies that $U_3(v')$ is not empty (because we can consider the lowest proper descendant $u'$ of $v'$ in $\widetilde{S}_1(v')=\widetilde{S}_1(v)$ such that $\mathit{high}_1(u')\neq\mathit{high}_2(u')=\mathit{high}_1(v')$ and $\mathit{low}(u')\geq\mathit{nextM}(v')$). This contradicts our supposition that $U_3(v')\neq\emptyset$. Thus, we have shown that $U_3(v')=\emptyset$ implies that $U_3(v)=\emptyset$.

Now let us assume that $U_3(v)\neq\emptyset$. This implies that $U_3(v')$ is not empty. Let us suppose, for the sake of contradiction, that there is a vertex $u\in U_3(v)$ that is lower than the greatest vertex $u'$ in $U_3(v')$.
Since $u\in U_3(v)$ and $u'\in U_3(v')$, we have $u\in\widetilde{S}_1(v)$ and $u'\in\widetilde{S}_1(v')$, respectively. Thus, since $\widetilde{S}_1(v)=\widetilde{S}_1(v')$, this implies that $u$ and $u'$ are related as ancestor and descendant. Thus, since $u$ is lower than $u'$, we have that $u$ is a proper ancestor of $u'$. Let us suppose, for the sake of contradiction, that $\mathit{low}(u')$ is a proper ancestor of $\mathit{nextM}(v')$. Then, since $\mathit{nextM}(v')$ is a proper ancestor of $\mathit{lastM}(v)$, we have that $\mathit{low}(u')$ is a proper ancestor of $\mathit{lastM}(v)$. Now let $(x,y)$ be a back-edge in $B(u')$ such that $y=\mathit{low}(u')$. Then $x$ is a descendant of $u'$, and therefore a descendant of $u$. Furthermore, $y$ is a proper ancestor of $\mathit{lastM}(v)$, and therefore a proper ancestor of $v$, and therefore a proper ancestor of $u$. This shows that $(x,y)\in B(u)$. Thus, we have $\mathit{low}(u)\leq y<\mathit{lastM}(v)$, in contradiction to the fact that $u\in U_3(v)$. Thus, our last supposition is not true, and therefore we have that $\mathit{low}(u')$ is not a proper ancestor of $\mathit{nextM}(v')$. Thus, since $u'\in U_3(v')$, we have that $\mathit{low}(u')\geq\mathit{nextM}(v')$, and $u'$ is the lowest vertex in $\widetilde{S}_1(v')$ with $\mathit{high}_1(u')\neq\mathit{high}_2(u')=\mathit{high}_1(v')$ that has this property $(*)$. 

Now we will trace the implications of $u\in U_3(v)$. First, we have that $u\in\widetilde{S}_1(v)=\widetilde{S}_1(v')$. Then, we have $\mathit{high}_1(u)\neq\mathit{high}_2(u)=\mathit{high}_1(v)=\mathit{high}_1(v')$. Furthermore, we have that $\mathit{low}(u)\geq\mathit{lastM}(v)$, and therefore $\mathit{low}(u)>\mathit{nextM}(v')$ (since $\mathit{nextM}(v')$ is a proper ancestor of $\mathit{lastM}(v)$). Finally, we can show as above that $u$ is a proper descendant of $v'$ (the proof of this fact above did not rely on $U_3(v')=\emptyset$). But then, since $u$ is lower than $u'$, we have a contradiction to $(*)$.
Thus, we have shown that every vertex in $U_3(v)$ is at least as great as the greatest vertex in $U_3(v')$. In particular, this implies that the lowest vertex in $U_3(v)$ is greater than, or equal to, the greatest vertex in $U_3(v')$.
\end{proof}

Based on Lemma~\ref{lemma:type3-b-ii-3-relation-between-u3}, we can provide an efficient algorithm for computing the sets $U_3(v)$, for all vertices $v\neq r$ such that $\mathit{nextM}(v)\neq\bot$. The computation takes place on segments of $\widetilde{H}(x)$ that are maximal w.r.t. the property that their elements are related as ancestor and descendant. Specifically, let $v\neq r$ be a vertex such that $\mathit{nextM}(v)\neq\bot$. Then we have $U_3(v)\subset\widetilde{S}_1(v)$. In other words, $U_3(v)$ is a subset of the segment of $\widetilde{H}(\mathit{high}_1(v))$ that contains $v$ and is maximal w.r.t. the property that its elements are related as ancestor and descendant. So let $z_1,\dots,z_k$ the vertices of $\widetilde{S}_1(v)$, sorted in decreasing order. Then, we have that $v=z_i$, for an $i\in\{1,\dots,k\}$. By definition, $U_3(v)$ contains every vertex $u$ in $\{z_1,\dots,z_{i-1}\}$ such that either $\mathit{high}_1(u)\neq\mathit{high}_2(u)=\mathit{high}_1(v)$ and $\mathit{nextM}(v)>\mathit{low}(u)\geq\mathit{lastM}(v)$, or $u$ is the lowest vertex in this set with $\mathit{high}_1(u)\neq\mathit{high}_2(u)=\mathit{high}_1(v)$ and $\mathit{low}(u)\geq\mathit{nextM}(v)$. As an implication of Lemma~\ref{lemma:vertices_on_segment-2}, we have that the vertices in $\{z_1,\dots,z_{i-1}\}$ are sorted in decreasing order w.r.t. their $\mathit{low}$ point. Thus, it is sufficient to process the vertices from $\{z_1,\dots,z_{i-1}\}$ in reverse order, in order to find the first vertex $u$ that has $\mathit{low}(u)\geq\mathit{lastM}(v)$. Then, we keep traversing this set in reverse order, and, as long as the $\mathit{low}$ point of every vertex $u$ that we meet is lower than $\mathit{nextM}(v)$, we insert $u$ into $U_3(v)$, provided that it satisfies $\mathit{high}_1(u)\neq\mathit{high}_2(u)=\mathit{high}_1(v)$. Then, once we reach a vertex with $\mathit{low}$ point no lower than $\mathit{nextM}(v)$, we keep traversing this set in reverse order, until we meet one more $u$ that satisfies $\mathit{high}_1(u)\neq\mathit{high}_2(u)=\mathit{high}_1(v)$, which we also insert into $U_3(v)$, and we are done.

Now, if there is a proper ancestor $v'$ of $v$ in $\widetilde{S}_1(v)$ such that $\mathit{high}_1(v')=\mathit{high}_1(v)$, then we have that $\widetilde{S}_1(v')=\widetilde{S}_1(v)$. If $\mathit{nextM}(v')\neq\bot$, then we have that $U_3(v')$ is defined. Then we can follow the same process as above in order to compute $U_3(v')$. Furthermore, according to Lemma~\ref{lemma:type3-b-ii-3-relation-between-u3}, it is sufficient to start from the greatest element of $U_3(v)$ (i.e., the one that was inserted last into $U_3(v)$). In particular, if $U_3(v)=\emptyset$, then it is certain that $U_3(v')=\emptyset$, and therefore we are done. Otherwise, we just pick up the computation from the greatest vertex in $U_3(v)$. In order to perform efficiently those computations, first we compute, for every vertex $x$, the collection $\mathcal{S}(x)$ of the segments of $\widetilde{H}(x)$ that are maximal w.r.t. the property that their elements are related as ancestor and descendant. For every vertex $x$, this computation takes $O(|\widetilde{H}(x)|)$ time, according to Lemma~\ref{lemma:segments-2}. Since every vertex participates in at most two sets of the form $\widetilde{H}(x)$, we have that the total size of all $\mathcal{S}(x)$, for all vertices $x$, is $O(n)$. Then it is sufficient to process separately all segments of $\mathcal{S}(x)$, for every vertex $x$, as described above, by starting the computation each time from the first vertex $v$ of the segment that satisfies $\mathit{nextM}(v)\neq\bot$ and $\mathit{high}_1(v)=x$. The whole procedure is shown in Algorithm~\ref{algorithm:type3-b-ii-3-U}. The result is formally stated in Lemma~\ref{lemma:algorithm:type3-b-ii-3-U}.

\begin{algorithm}[H]
\caption{\textsf{Compute the sets $U_3(v)$, for all vertices $v\neq r$ such that $\mathit{nextM}(v)\neq\bot$}}
\label{algorithm:type3-b-ii-3-U}
\LinesNumbered
\DontPrintSemicolon
\ForEach{vertex $x$}{
  compute the collection $\mathcal{S}(x)$ of the segments of $\widetilde{H}(x)$ that are maximal w.r.t. the property
  that their elements are related as ancestor and descendant\;
}
\ForEach{vertex $v\neq r$ such that $\mathit{nextM}(v)\neq\bot$}{
  set $U_3(v)\leftarrow\emptyset$\;
}
\ForEach{vertex $x$}{
  \ForEach{segment $S\in\mathcal{S}(x)$}{
    let $v$ be the first vertex in $S$\;
    \While{$v\neq\bot$ \textbf{and} ($\mathit{high}_1(v)\neq x$ \textbf{or} $\mathit{nextM}(v)=\bot$)}{
      $v\leftarrow\mathit{next}_S(v)$\;
    }
    \lIf{$v=\bot$}{\textbf{continue}}
    let $u\leftarrow\mathit{prev}_S(v)$\;
    \While{$v\neq\bot$}{
      \While{$u\neq\bot$ \textbf{and} $\mathit{low}(u)<\mathit{lastM}(v)$}{
        $u\leftarrow\mathit{prev}_S(u)$\;
      }
      \While{$u\neq\bot$ \textbf{and} $\mathit{low}(u)<\mathit{nextM}(v)$}{
        \If{$\mathit{high}_1(u)\neq\mathit{high}_2(u)$ \textbf{and} $\mathit{high}_2(u)=x$}{
          insert $u$ into $U_3(v)$\;
        }
        $u\leftarrow\mathit{prev}_S(u)$\;
      }
      \While{$u\neq\bot$ \textbf{and} ($\mathit{high}_1(u)=\mathit{high}_2(u)$ \textbf{or} $\mathit{high}_2(u)\neq x$)}{
        $u\leftarrow\mathit{prev}_S(u)$\;
      }
      \If{$u\neq\bot$}{
        insert $u$ into $U_3(v)$\;
      }
      $v\leftarrow\mathit{next}_S(v)$\;
      \While{$v\neq\bot$ \textbf{and} ($\mathit{high}_1(v)\neq x$ \textbf{or} $\mathit{nextM}(v)=\bot$)}{
        $v\leftarrow\mathit{next}_S(v)$\;
      }      
    }
  }
}
\end{algorithm}

\begin{lemma}
\label{lemma:algorithm:type3-b-ii-3-U}
Algorithm~\ref{algorithm:type3-b-ii-3-U} correctly computes the sets $U_3(v)$, for all vertices $v\neq r$ such that $\mathit{nextM}(v)\neq\bot$. Furthermore, it runs in $O(n)$ time.
\end{lemma}
\begin{proof}
This was basically given in the main text, in the two paragraphs above Algorithm~\ref{algorithm:type3-b-ii-3-U}.
\end{proof}

\begin{lemma}
\label{lemma:type-3-b-ii-3-criterion}
Let $(u,v,w)$ be a triple of vertices such that $u\in U_3(v)$. Then, $(u,v,w)$ induces a Type-3$\beta ii$-$3$ $4$-cut if and only if: $(1)$ $w$ is the greatest proper ancestor of $v$ such that $M(w)=M(v)$ and $w\leq\mathit{low}(u)$, and $(2)$ $\mathit{bcount}(v)=\mathit{bcount}(u)+\mathit{bcount}(w)-1$.
\end{lemma}
\begin{proof}
$(\Rightarrow)$ By definition of Type-3$\beta ii$-$3$ $4$-cuts, we have that $w$ is a proper ancestor of $v$ with $M(w)=M(v)$. Furthermore, we have $B(v)=(B(u)\setminus\{e\})\sqcup B(w)$, where $e$ is the back-edge of the $4$-cut induced by $(u,v,w)$. This implies that $\mathit{bcount}(v)=\mathit{bcount}(u)+\mathit{bcount}(w)-1$. Lemma~\ref{lemma:type3-b-ii-3-info} implies that $w\leq\mathit{low}(u)$. 

Now let us suppose, for the sake of contradiction, that there is a proper ancestor $w'$ of $v$ with $w'>w$ such that $M(w')=M(v)$ and $w'\leq\mathit{low}(u)$. Since $M(w')=M(w)$ and $w'>w$, we have that $w'$ is a proper descendant of $w$, and Lemma~\ref{lemma:same_m_subset_B} implies that $B(w)\subseteq B(w')$. Since the graph is $3$-edge-connected, this can be strengthened to $B(w)\subset B(w')$. Thus, there is a back-edge $(x,y)\in B(w')\setminus B(w)$. Then we have that $x$ is a descendant of $M(w')=M(v)$. Furthermore, we have that $y$ is a proper ancestor of $w'$, and therefore a proper ancestor of $v$. This shows that $(x,y)\in B(v)$. Then, since $(x,y)\notin B(w)$, $B(v)=(B(u)\setminus\{e\})\sqcup B(w)$ implies that $(x,y)\in B(u)\setminus\{e\}$. Since $y$ is a proper ancestor of $w'$, we have that $y<w'$. Then, $w'\leq\mathit{low}(u)$ implies that $y<\mathit{low}(u)$, in contradiction to $(x,y)\in B(u)$. This shows that $w$ is the greatest proper ancestor of $v$ such that $M(w)=M(v)$ and $w\leq\mathit{low}(u)$.

$(\Leftarrow)$ Since $u\in U_3(v)$, we have that $u$ is a proper descendant of $v$ with $\mathit{high}_1(u)\neq\mathit{high}_2(u)=\mathit{high}(v)$. Let $(x,y)$ be a back-edge in $B(u)\setminus\{e_\mathit{high}(u)\}$. Then, we have that $x$ is a descendant of $u$, and therefore a descendant of $v$. Furthermore, we have that $y\leq\mathit{high}_2(u)=\mathit{high}(v)$, and therefore $y<v$. Since $(x,y)$ is a back-edge, we have that $x$ is a descendant of $y$. Thus, $x$ is a common descendant of $v$ and $y$, and therefore $v$ and $y$ are related as ancestor and descendant. Thus, $y<v$ implies that $y$ is a proper ancestor of $v$. Therefore, since $x$ is a descendant of $v$, we have $(x,y)\in B(v)$. Due to the generality of $(x,y)\in B(u)\setminus\{e_\mathit{high}(u)\}$, this implies that $B(u)\setminus\{e_\mathit{high}(u)\}\subseteq B(v)$. Since $w$ is a proper ancestor of $v$ with $M(w)=M(v)$, Lemma~\ref{lemma:same_m_subset_B} implies that $B(w)\subseteq B(v)$. Let $(x,y)$ be a back-edge in $B(w)$. Then we have that $y$ is a proper ancestor of $w$, and therefore $y<w$. Thus, $w\leq\mathit{low}(u)$ implies that $y<\mathit{low}(u)$. Therefore, we cannot have $(x,y)\in B(u)$. This shows that $B(u)\cap B(w)=\emptyset$. 

Thus, since $B(u)\setminus\{e_\mathit{high}(u)\}\subseteq B(v)$ and $B(w)\subseteq B(v)$ and $B(u)\cap B(w)=\emptyset$ and $\mathit{bcount}(v)=\mathit{bcount}(u)+\mathit{bcount}(w)-1$, we have that $B(v)=(B(u)\setminus\{e_\mathit{high}(u)\})\sqcup B(w)$. Furthermore, since $B(u)\cap B(w)=\emptyset$, we have that $e_\mathit{high}(u)\notin B(w)$, and therefore $B(v)=(B(u)\setminus\{e_\mathit{high}(u)\})\sqcup B(w)$ implies that $e_\mathit{high}(u)\notin B(v)\cup B(w)$. Thus, since $M(w)=M(v)$, we have that $(u,v,w)$ induces a Type-3$\beta ii$-$3$ $4$-cut.
\end{proof}

Now we are ready to describe the algorithm for computing a collection of enough Type-3$\beta ii$-$3$ $4$-cuts, so that the rest of them are implied from this collection, plus that computed by Algorithm~\ref{algorithm:type2-2}. So let $(u,v,w)$ be a triple of vertices that induces a Type-3$\beta ii$-$3$ $4$-cut. Then, Lemma~\ref{lemma:type-3-b-ii-3-rel} implies that either $u\in U_3(v)$, or $(\tilde{u},v,w)$ induces a Type-3$\beta ii$-$3$ $4$-cut, where $\tilde{u}$ is the greatest vertex in $U_3(v)$. Furthermore, if $u\notin U_3(v)$, then $B(u)\sqcup\{e_\mathit{high}(\tilde{u})\}=B(\tilde{u})\sqcup\{e_\mathit{high}(u)\}$. Thus, if $u\notin U_3(v)$, then it is sufficient to have computed the $4$-cut $C$ induced by $(\tilde{u},v,w)$, because then the one induced by $(u,v,w)$ is implied by $C$, plus some Type-$2ii$ $4$-cuts that are computed by Algorithm~\ref{algorithm:type2-2} (see Proposition~\ref{proposition:type-3-b-ii-3}). Now, if $u\in U_3(v)$, then by Lemma~\ref{lemma:type-3-b-ii-3-criterion} we have that $w$ is the greatest proper ancestor of $v$ such that $w\leq\mathit{low}(u)$ and $M(w)=M(v)$. Thus, we can use Algorithm~\ref{algorithm:W-queries} in order to get $w$ from $v$ and $u$.

The full procedure for computing enough Type-3$\beta ii$-$3$ $4$-cuts is shown in Algorithm~\ref{algorithm:type-3-b-ii-3}. The proof of correctness and linear complexity is given in Proposition~\ref{proposition:type-3-b-ii-3}.

\begin{algorithm}[H]
\caption{\textsf{Compute a collection of Type-3$\beta ii$-$3$ $4$-cuts, so that all Type-3$\beta ii$-$3$ $4$-cuts are implied from this collection, plus that of the Type-$2$ii $4$-cuts returned by Algorithm~\ref{algorithm:type2-2}}}
\label{algorithm:type-3-b-ii-3}
\LinesNumbered
\DontPrintSemicolon
\ForEach{vertex $v\neq r$ such that $\mathit{nextM}(v)\neq\bot$}{
\label{line:type-3-b-ii-3-U}
  compute $U_3(v)$\;  
}
\ForEach{vertex $v\neq r$ such that $\mathit{nextM}(v)\neq\bot$}{
  \ForEach{$u\in U_3(v)$}{
    let $w$ be the greatest proper ancestor of $v$ such that $w\leq\mathit{low}(u)$ and $M(w)=M(v)$\;
    \label{line:type-3-b-ii-3-w}
    \If{$\mathit{bcount}(v)=\mathit{bcount}(u)+\mathit{bcount}(w)-1$}{
      mark $\{(u,p(u)),(v,p(v)),(w,p(w)),e_\mathit{high}(u)\}$ as a Type-3$\beta ii$-$3$ $4$-cut\;
      \label{line:type-3-b-ii-3-mark}
    }
  }
}
\end{algorithm}

\begin{proposition}
\label{proposition:type-3-b-ii-3}
Algorithm~\ref{algorithm:type-3-b-ii-3} computes a collection $\mathcal{C}$ of Type-3$\beta ii$-$3$ $4$-cuts, and it runs in $O(n)$ time. Furthermore, let $\mathcal{C}'$ be the collection of Type-$2ii$ $4$-cuts computed by Algorithm~\ref{algorithm:type2-2}. Then, every Type-3$\beta ii$-$3$ $4$-cut is implied by $\mathcal{C}\cup\mathcal{C}'$.
\end{proposition}
\begin{proof}
When we reach Line~\ref{line:type-3-b-ii-3-mark}, notice that the following conditions are true. $(1)$ $u\in U_3(v)$, $(2)$ $w$ is the greatest proper ancestor of $v$ such that $w\leq\mathit{low}(u)$ and $M(w)=M(v)$, and $(3)$ $\mathit{bcount}(v)=\mathit{bcount}(u)+\mathit{bcount}(w)-1$. Thus, Lemma~\ref{lemma:type-3-b-ii-3-criterion} implies that $(u,v,w)$ induces a Type-3$\beta ii$-$3$ $4$-cut. By Lemma~\ref{lemma:type3-b-ii-3-info}, we have that the back-edge in this $4$-cut is $e_\mathit{high}(u)$. Thus, it is correct to mark $\{(u,p(u)),(v,p(v)),(w,p(w)),e_\mathit{high}(u)\}$ as a Type-3$\beta ii$-$3$ $4$-cut. This shows that the collection $\mathcal{C}$ of the $4$-element sets marked by Algorithm~\ref{algorithm:type-3-b-ii-3} is a collection of Type-3$\beta ii$-$3$ $4$-cuts.

According to Lemma~\ref{lemma:algorithm:type3-b-ii-3-U}, the computation of all sets $U_3(v)$, for all vertices $v\neq r$ such that $\mathit{nextM}(v)\neq\bot$, can be performed in $O(n)$ time in total, using Algorithm~\ref{algorithm:type3-b-ii-3-U}. Thus, the \textbf{for} loop in Line~\ref{line:type-3-b-ii-3-U} can be performed in $O(n)$ time. In particular, this implies that the total size of all $U_3$ sets is $O(n)$. It remains to explain how to compute the $w$ in Line~\ref{line:type-3-b-ii-3-w}, for every vertex $u\in U_3(v)$, for every vertex $v\neq r$ such that $\mathit{nextM}(v)\neq\bot$. For this purpose, we can simply use Algorithm~\ref{algorithm:W-queries}. First, we let $M^{-1}(x)$, for every vertex $x$, be the collection of all vertices $w$ such that $M(w)=x$. Thus, if $x\neq x'$, then $M^{-1}(x)\cap M^{-1}(x')=\emptyset$. Then, for every vertex $v\neq r$ such that $\mathit{nextM}(v)\neq\bot$, and every vertex $u\in U_3(v)$, we generate a query $q(M^{-1}(M(v)),\mathit{min}\{\mathit{low}(u),p(v)\})$. This is to return the greatest $w$ that has $M(w)=M(v)$, $w\leq\mathit{low}(u)$ and $w\leq p(v)$. In particular, since $M(w)=M(v)$ and $w\leq p(v)$, we have that $w$ is a proper ancestor of $v$. Thus, the $w$ returned is the greatest proper ancestor of $v$ with $M(w)=M(v)$ such that $w\leq\mathit{low}(u)$. The number of all those queries is bounded by the total size of the $U_3$ sets, which is bounded by $O(n)$. Thus, Lemma~\ref{lemma:W-queries} implies that all these queries can be answered in $O(n)$ time in total. 

Now let $(u,v,w)$ be a triple of vertices that induces a Type-3$\beta ii$-$3$ $4$-cut $C$. If $u\in U_3(v)$, then by Lemma~\ref{lemma:type-3-b-ii-3-criterion}  we have that $w$ is uniquely determined by $u$ and $v$, and therefore $C$ has been marked at some point in Line~\ref{line:type-3-b-ii-3-mark}. So let us assume that $u\notin U_3(v)$. Then, Lemma~\ref{lemma:type-3-b-ii-3-rel} implies that $(\tilde{u},v,w)$ induces a Type-3$\beta ii$-$3$ $4$-cut $C'$, where $\tilde{u}$ is the greatest vertex in $U_3(v)$. Thus, $C'\in\mathcal{C}$. Furthermore, Lemma~\ref{lemma:type-3-b-ii-3-rel} implies $B(u)\sqcup\{e_\mathit{high}(\tilde{u})\}=B(\tilde{u})\sqcup\{e_\mathit{high}(u)\}$. Thus, by Lemma~\ref{lemma:type2cuts} we have that $C''=\{(u,p(u)),(\tilde{u},p(\tilde{u})),e_\mathit{high}(u),e_\mathit{high}(\tilde{u})\}$ is a Type-$2ii$ $4$-cut. By Lemma~\ref{lemma:type3-b-ii-3-info}, we have that the $4$-cuts induced by $(u,v,w)$ and $(\tilde{u},v,w)$ are $\{(u,p(u)),(v,p(v)),(w,p(w)),e_\mathit{high}(u)\}$ and $\{(\tilde{u},p(\tilde{u})),(v,p(v)),(w,p(w)),e_\mathit{high}(\tilde{u})\}$, respectively. Notice that $C$ is implied by $C'$ and $C''$ through the pair of edges $\{(u,p(u)),e_\mathit{high}(u)\}$. According to Proposition~\ref{proposition:type-2-2}, we have that $C''$ is implied by $\mathcal{C}'$ through the pair of edges $\{(u,p(u)),e_\mathit{high}(u)\}$. Thus, by Lemma~\ref{lemma:implied_from_union} we have that $C$ is implied by $\mathcal{C}\cup\mathcal{C}'$. 
\end{proof} 

\subsubsection{Type-3$\beta ii$-$4$ $4$-cuts}

Now we consider case $(4)$ of Lemma~\ref{lemma:type-3b-cases}. 

Let $u,v,w$ be three vertices $\neq r$ such that $w$ is proper ancestor of $v$, $v$ is a proper ancestor of $u$, and there is a back-edge $e\in B(v)$ such  $B(v)=(B(u)\sqcup B(w))\sqcup\{e\}$ and $M(B(v)\setminus\{e\})=M(w)$. By Lemma~\ref{lemma:type-3b-cases}, we have that $C=\{(u,p(u)),(v,p(v)),(w,p(w)),e\}$ is a $4$-cut, and we call this a Type-3$\beta$ii-$4$ $4$-cut.

The following lemma provides some useful information concerning this type of $4$-cuts.

\begin{lemma}
\label{lemma:type3-b-ii-4-info}
Let $(u,v,w)$ be a triple of vertices that induces a Type-3$\beta$ii-$4$ $4$-cut, and let $e$ be the back-edge in the $4$-cut induced by $(u,v,w)$. Then $w\leq\mathit{low}(u)$, and either $\mathit{high}_1(v)=\mathit{high}(u)$, or $\mathit{high}_1(v)>\mathit{high}(u)$ and $\mathit{high}_2(v)=\mathit{high}(u)$. If $\mathit{high}_1(v)\neq\mathit{high}(u)$, then $e=e_\mathit{high}(v)$.
\end{lemma}
\begin{proof}
Since $(u,v,w)$ induces a Type-3$\beta$ii-$4$ $4$-cut, we have that $B(v)=(B(u)\sqcup B(w))\sqcup\{e\}$, where $e$ is the back-edge in the $4$-cut induced by $(u,v,w)$. This implies that $B(u)\cap B(w)=\emptyset$.
Since $\mathit{low}(u)$ and $w$ have $u$ as a common descendant, we have that $\mathit{low}(u)$ and $w$ are related as ancestor and descendant. Let us suppose, for the sake of contradiction, that $\mathit{low}(u)<w$. Then, we have that $\mathit{low}(u)$ is a proper ancestor of $w$. Let $(x,y)$ be a back-edge in $B(u)$ such that $y=\mathit{low}(u)$. Then $x$ is a descendant of $u$, and therefore a descendant of $w$. Furthermore, $y$ is a proper ancestor of $w$. This shows that $(x,y)\in B(w)$, in contradiction to the fact that $B(u)\cap B(w)=\emptyset$. This shows that $\mathit{low}(u)\geq w$.

Now let us assume that $\mathit{high}_1(v)\neq\mathit{high}(u)$. (If $\mathit{high}_1(v)=\mathit{high}(u)$, then there is nothing further to show.) Since $B(u)\subseteq B(v)$, we have that $\mathit{high}_1(v)\geq\mathit{high}(u)$. Thus, $\mathit{high}_1(v)\neq\mathit{high}(u)$ implies that $\mathit{high}_1(v)>\mathit{high}(u)$. Let $(x,y)$ be a back-edge in $B(v)$ such that $y>\mathit{high}(u)$. Then, $B(v)=(B(u)\sqcup B(w))\sqcup\{e\}$ implies that either $(x,y)\in B(u)$, or $(x,y)\in B(w)$, or $(x,y)=e$. Since $y>\mathit{high}(u)$, the case $(x,y)\in B(u)$ is rejected. Furthermore, since $w\leq\mathit{low}(u)\leq\mathit{high}(u)<y$, the case $(x,y)\in B(w)$ is rejected too. Thus, we have $(x,y)=e$. Therefore, since $\mathit{high}_1(v)>\mathit{high}(u)$, we have that $e=\mathit{high}(v)$. 

Now let $(x',y')$ be a back-edge in $B(v)$ such that $y'=\mathit{high}_2(v)$ and $(x',y')\neq e$. Then we cannot have $y'>\mathit{high}(u)$, because otherwise we would conclude as previously that $(x',y')=e$, which is impossible. Thus, we have $y'\leq\mathit{high}(u)$. Let us suppose, for the sake of contradiction, that $y'<\mathit{high}(u)$. Since $B(v)=(B(u)\sqcup B(w))\sqcup\{e\}$, we have that the $\mathit{high}$-edge of $B(u)$ is in $B(v)$. Thus, there is a back-edge $(x'',y'')$ in $B(v)$ such that $y''=\mathit{high}(u)$. Then we have $\mathit{high}_1(v)>y''>\mathit{high}_2(v)$, which contradicts the definition of the $\mathit{high}_2$ point. Thus, we conclude that $\mathit{high}_2(v)=\mathit{high}(u)$.
\end{proof}

\begin{lemma}
\label{lemma:type3-b-ii-4-seg-1}
Let $(u,v,w)$ be a triple of vertices that induces a Type-3$\beta$ii-$4$ $4$-cut, such that $\mathit{high}_1(v)\neq\mathit{high}(u)$. Let $u'$ be a vertex in $\widetilde{H}(\mathit{high}_2(v))$ such that $u\geq u'\geq v$. Then $u'$ is an ancestor of $u$. 
\end{lemma}
\begin{proof}
Since $(u,v,w)$ induces a Type-3$\beta$ii-$4$ $4$-cut, we have that $u$ is a descendant of $v$. Thus, $u\geq u'\geq v$ implies that $u'$ is also a descendant of $v$. Since $\mathit{high}_1(v)\neq\mathit{high}(u)$, by Lemma~\ref{lemma:type3-b-ii-4-info} we have that $\mathit{high}_2(v)=\mathit{high}(u)$ and $e=e_\mathit{high}(v)$, where $e$ is the back-edge in the $4$-cut induced by $(u,v,w)$. 

Since $u'\in\widetilde{H}(\mathit{high}_2(v))$, we have that either $\mathit{high}_1(u')=\mathit{high}_2(v)$, or $\mathit{high}_2(u')=\mathit{high}_2(v)$. In either case then, there is a back-edge $(x,y)\in B(u)$ such that $y=\mathit{high}_2(v)$. Then, we have that $x$ is a descendant of $u'$, and therefore a descendant of $v$. Furhermore, $y=\mathit{high}_2(v)$ is a proper ancestor of $v$. This shows that $(x,y)\in B(v)$. Since $(u,v,w)$ induces a Type-3$\beta$ii-$4$ $4$-cut, by Lemma~\ref{lemma:type3-b-ii-4-info} we have that $w\leq\mathit{low}(u)$. Then, we have that $w\leq\mathit{low}(u)\leq\mathit{high}(u)=\mathit{high}_2(v)=y$. This implies that $y$ cannot be a proper ancestor of $w$, and therefore we have that $(x,y)\notin B(w)$. Since $(u,v,w)$ induces a Type-3$\beta$ii-$4$ $4$-cut, we have that $B(v)=(B(u)\sqcup B(w))\sqcup\{e\}$. Thus, since $(x,y)\in B(v)$ and $(x,y)\notin B(w)$, we have that either $(x,y)\in B(u)$, or $(x,y)=e$.

Let us suppose, for the sake of contradiction, that $(x,y)=e$. Since $e=e_\mathit{high}(v)$, we have that $y=\mathit{high}_1(v)$. Since $\mathit{high}_1(v)\neq\mathit{high}(u)$ and $\mathit{high}(u)=\mathit{high}_2(v)$, we have that $\mathit{high}_1(v)\neq\mathit{high}_2(v)$. But then $y=\mathit{high}_1(v)$ contradicts the fact that $y=\mathit{high}_2(v)$. This shows that $(x,y)\neq e$. Thus, we have that $(x,y)\in B(u)$. This implies that $x$ is a common descendant of $u$ and $u'$. Thus, we have that $u$ and $u'$ are related as ancestor and descendant. Then, $u\geq u'$ implies that $u'$ is an ancestor of $u$.
\end{proof}

\begin{lemma}
\label{lemma:type3-b-ii-4-seg-2}
Let $(u,v,w)$ be a triple of vertices that induces a Type-3$\beta$ii-$4$ $4$-cut, such that $M(B(v)\setminus\{e\})\neq M(v)$, where $e$ is the back-edge in the $4$-cut induced by $(u,v,w)$. Let $u'$ be a vertex in $H(\mathit{high}_1(v))$ such that $u\geq u'\geq v$. Then $u'$ is an ancestor of $u$. 
\end{lemma}
\begin{proof}
Since $(u,v,w)$ induces a Type-3$\beta$ii-$4$ $4$-cut, we have that $u$ is a descendant of $v$. Thus, $u\geq u'\geq v$ implies that $u'$ is also a descendant of $v$. Lemma~\ref{lemma:type3-b-ii-4-info} implies that either $\mathit{high}_1(v)=\mathit{high}(u)$, or $\mathit{high}_2(v)=\mathit{high}(u)$. In any case, then, since $\mathit{high}_2(v)\leq\mathit{high}_1(v)$, we have that $\mathit{high}(u)\leq\mathit{high}_1(v)$.

Since $u'\in H(\mathit{high}_1(v))$, we have that $\mathit{high}(u')=\mathit{high}_1(v)$. Thus, there is a back-edge $(x,y)\in B(u)$ such that $y=\mathit{high}_1(v)$. Then, we have that $x$ is a descendant of $u'$, and therefore a descendant of $v$. Furhermore, $y=\mathit{high}_1(v)$ is a proper ancestor of $v$. This shows that $(x,y)\in B(v)$. Since $(u,v,w)$ induces a Type-3$\beta$ii-$4$ $4$-cut, by Lemma~\ref{lemma:type3-b-ii-4-info} we have that $w\leq\mathit{low}(u)$. Then, we have that $w\leq\mathit{low}(u)\leq\mathit{high}(u)\leq\mathit{high}_1(v)=y$. This implies that $y$ cannot be a proper ancestor of $w$, and therefore we have that $(x,y)\notin B(w)$. Since $(u,v,w)$ induces a Type-3$\beta$ii-$4$ $4$-cut, we have that $B(v)=(B(u)\sqcup B(w))\sqcup\{e\}$. Thus, since $(x,y)\in B(v)$ and $(x,y)\notin B(w)$, we have that either $(x,y)\in B(u)$, or $(x,y)=e$. If $(x,y)\in B(u)$, then we have that $x$ is a common descendant of $u$ and $u'$. Thus, $u$ and $u'$ are related as ancestor and descendant. Then, $u\geq u'$ implies that $u'$ is an ancestor of $u$. So let us assume that $(x,y)\notin B(u)$. This implies that $(x,y)=e$.

Let us suppose, for the sake of contradiction, that $u'$ is not an ancestor of $M(v)$ $(*)$. Since $e=(x,y)\in B(u')\cap B(v)$, we have that $x$ is a descendant of both $u'$ and $M(v)$. Thus, $u'$ and $M(v)$ are related as ancestor and descendant. Then, since $u'$ is not an ancestor of $M(v)$, we have that $u'$ is a proper descendant of $M(v)$. Let $c$ be the child of $M(v)$ that is an ancestor of $u'$. Since $M(B(v)\setminus\{e\})\neq M(v)$, we have that $e$ is the only back-edge in $B(v)$ whose higher endpoint is not a descendant of $M(B(v)\setminus\{e\})$. Furthermore, since $M(B(v)\setminus\{e\})\neq M(v)$, we have that $M(B(v)\setminus\{e\})$ is a proper descendant of $M(v)$. Let $c'$ be the child of $M(v)$ that is an ancestor of $M(B(v)\setminus\{e\})$. 

Let $(x',y')$ be a back-edge in $B(v)$. Then, if $(x',y')=e$, we have that $(x',y')=(x,y)$, and therefore $x'$ is a descendant of $u'$, and therefore a descendant of $c$. Otherwise, if $(x',y')\neq e$, then we have that $(x',y')\in B(v)\setminus\{e\}$, and therefore $x'$ is a descendant of $M(B(v)\setminus\{e\})$, and therefore a descendant of $c'$. This shows that there is no back-edge of the form $(M(v),z)\in B(v)$. Let us suppose, for the sake of contradiction, that $c=c'$. Then, the previous argument shows that all back-edges in $B(v)$ stem from $T(c')$, and therefore $M(v)$ is a descendant of $c'$, which is absurd. Thus, we have that $c\neq c'$.

Since the graph is $3$-edge-connected, we have that $|B(u')|>1$. Thus, there is a back-edge $(x',y')\in B(u')\setminus\{e\}$. Since $u'$ is a descendant of $v$ with $\mathit{high}(u')=\mathit{high}_1(v)$, by Lemma~\ref{lemma:same_high} we have that $B(u')\subseteq B(v)$. Thus, $(x',y')\in B(u')$ implies that $(x',y')\in B(v)$. Then, since $(x',y')\neq e$, we have that $(x',y')\in B(v)\setminus\{e\}$, and therefore $x'$ is a descendant of $M(B(v)\setminus\{e\})$, and therefore a descendant of $c'$. But since $(x',y')\in B(u')$, we have that $x'$ is a descendant of $u'$, and therefore a descendant of $c$. Thus, $x'$ is a common descendant of $c$ and $c'$, and therefore $c$ and $c'$ are related as ancestor and descendant, which is absurd. Thus, starting from $(*)$, we have arrived at a contradiction. This shows that $u'$ is an ancestor of $M(v)$.

Since $B(v)=(B(u)\sqcup B(w))\sqcup\{e\}$, we have that $B(u)\subseteq B(v)$, and therefore $M(u)$ is a descendant of $M(v)$. Since $u'$ is an ancestor of $M(v)$, this implies that $M(u)$ is a descendant of $u'$. Thus, $M(u)$ is a common descendant of $u$ and $u'$, and therefore $u$ and $u'$ are related as ancestor and descendant. Thus, $u\geq u'$ implies that $u'$ is an ancestor of $u$.
\end{proof}

\begin{lemma}
\label{lemma:type3-b-ii-4-seg-3}
Let $(u,v,w)$ be a triple of vertices that induces a Type-3$\beta$ii-$4$ $4$-cut, such that $\mathit{high}_1(v)=\mathit{high}(u)$ and $y\neq\mathit{high}_1(v)$, where $y$ is the lower endpoint of the back-edge in the $4$-cut induced by $(u,v,w)$. Let $u'$ be a vertex in $H(\mathit{high}_1(v))$ such that $u\geq u'\geq v$. Then $u'$ is an ancestor of $u$. 
\end{lemma}
\begin{proof}
Let $e=(x,y)$ be the back-edge in the $4$-cut induced by $(u,v,w)$.
Since $(u,v,w)$ induces a Type-3$\beta$ii-$4$ $4$-cut, we have that $u$ is a descendant of $v$. Thus, $u\geq u'\geq v$ implies that $u'$ is also a descendant of $v$. Since $u'\in H(\mathit{high}_1(v))$, we have $\mathit{high}(u')=\mathit{high}_1(v)$. This implies that there is a back-edge $(x',y')\in B(u')$ such that $y'=\mathit{high}_1(v)$. Then, $x'$ is a descendant of $u'$, and therefore a descendant of $v$. Furthermore, $y=\mathit{high}_1(v)$ is a proper ancestor of $v$. This shows that $(x',y')\in B(v)$. Since $(u,v,w)$ induces a Type-3$\beta$ii-$4$ $4$-cut, by Lemma~\ref{lemma:type3-b-ii-4-info} we have $w\leq\mathit{low}(u)$. Thus, we have $w\leq\mathit{low}(u)\leq\mathit{high}(u)=\mathit{high}_1(v)=y'$. This implies that $y'$ cannot be a proper ancestor of $w$, and therefore we have $(x',y')\notin B(w)$. Since $(u,v,w)$ induces a Type-3$\beta$ii-$4$ $4$-cut, we have $B(v)=(B(u)\sqcup B(w))\sqcup \{e\}$. Thus, since $(x',y')\in B(v)$ and $(x',y')\notin B(w)$, we have that either $(x',y')\in B(u)$, or $(x',y')=e$. The case $(x',y')=e$ is rejected, because it implies that $y'=y$, contradicting the fact that $y'=\mathit{high}_1(v)$ and $y\neq\mathit{high}_1(v)$. Thus, we have $(x',y')\in B(u)$. This implies that $x'$ is a descendant of $u$. Thus, $x'$ is a common descendant of $u$ and $u'$, and therefore $u$ and $u'$ are related as ancestor and descendant. Since $u\geq u'$, this implies that $u'$ is an ancestor of $u$. 
\end{proof}

Here, we distinguish the following four different cases, depending on the location of the endpoints of the back-edge $e$:

\begin{enumerate}
\item{$M(B(v)\setminus\{e\})\neq M(v)$ and $\mathit{high}_1(v)>\mathit{high}(u)$.}
\item{$M(B(v)\setminus\{e\})\neq M(v)$ and $\mathit{high}_1(v)=\mathit{high}(u)$.}
\item{$M(B(v)\setminus\{e\})= M(v)$ and $\mathit{high}_1(v)>\mathit{high}(u)$.}
\item{$M(B(v)\setminus\{e\})= M(v)$ and $\mathit{high}_1(v)=\mathit{high}(u)$.}
\end{enumerate}

\noindent\\
\textbf{The case where $M(B(v)\setminus\{e\})\neq M(v)$ and $\mathit{high}_1(v)>\mathit{high}(u)$}\\

Let $(u,v,w)$ be a triple of vertices that induces a Type-3$\beta$ii-$4$ $4$-cut, such that $M(B(v)\setminus\{e\})\neq M(v)$ and $\mathit{high}_1(v)>\mathit{high}(u)$, where $e$ is the back-edge in the $4$-cut induced by $(u,v,w)$. Then, by Lemma~\ref{lemma:type3-b-ii-4-info} we have $\mathit{high}_1(v)\neq\mathit{high}_2(v)$ and $e=e_\mathit{high}(v)$.

Let $v\neq r$ be a vertex such that $\mathit{high}_1(v)\neq\mathit{high}_2(v)$ and $M(B(v)\setminus\{e_\mathit{high}(v)\})\neq M(v)$. Then, we let $V(v)$ denote the collection of all vertices $v'\neq r$ such that $\mathit{high}_1(v')\neq\mathit{high}_2(v')=\mathit{high}_2(v)$ and $M(B(v)\setminus\{e_\mathit{high}(v)\})=M(B(v')\setminus\{e_\mathit{high}(v')\})\neq M(v')$. (We note that $v\in V(v)$.) If a vertex $v\neq r$ does not satisfy  $\mathit{high}_1(v)\neq\mathit{high}_2(v)$ and $M(B(v)\setminus\{e_\mathit{high}(v)\})\neq M(v)$, then we let $V(v)=\emptyset$.

\begin{lemma}
\label{lemma:V-sets-implied-4cuts}
Let $v$ and $v'$ be two distinct vertices such that $v'\in V(v)$. Then, $B(v)\sqcup\{e_\mathit{high}(v')\}=B(v')\sqcup\{e_\mathit{high}(v)\}$. 
\end{lemma}
\begin{proof}
Since $v'\in V(v)$, we have $\mathit{high}_1(v)\neq\mathit{high}_2(v)=\mathit{high}_2(v')\neq\mathit{high}_1(v')$ and $M(v)\neq M(B(v)\setminus\{e_\mathit{high}(v)\})=M(B(v')\setminus\{e_\mathit{high}(v')\})\neq M(v')$. Since $\mathit{high}_1(v)\neq\mathit{high}_2(v)$, we have $\mathit{high}_1(v)>\mathit{high}_2(v)$. Similarly, we have $\mathit{high}_1(v')>\mathit{high}_2(v')$. Furthermore, $e_\mathit{high}(v)$ is the unique back-edge in $B(v)$ whose lower endpoint is $\mathit{high}_1(v)$, and $e_\mathit{high}(v')$ is the unique back-edge in $B(v')$ whose lower endpoint is $\mathit{high}_1(v')$. 

Let $(x,y)$ be a back-edge in $B(v)$ such that $y=\mathit{high}_2(v)$. Then we have that $x$ is a descendant of $v$. Furthermore, since $(x,y)\in B(v)\setminus\{e_\mathit{high}(v)\}$ and $M(B(v)\setminus\{e_\mathit{high}(v)\})=M(B(v')\setminus\{e_\mathit{high}(v')\})$, we have that $x$ is a descendant of $M(B(v')\setminus\{e_\mathit{high}(v')\})$, and therefore a descendant of $M(v')$, and therefore a descendant of $v'$. Thus, $x$ is a common descendant of $v$ and $v'$, and therefore we have that $v$ and $v'$ are related as ancestor and descendant. Thus, we may assume w.l.o.g. that $v'$ is a proper ancestor of $v$. 

Let us suppose, for the sake of contradiction, that $e_\mathit{high}(v')\in B(v)$. Then, since $\mathit{high}_1(v')>\mathit{high}_2(v')=\mathit{high}_2(v)$, we have that $e_\mathit{high}(v)=e_\mathit{high}(v')$. This implies that $\mathit{high}_1(v')=\mathit{high}_1(v)$. Thus, since $v'$ is a proper ancestor of $v$, Lemma~\ref{lemma:same_high} implies that $B(v)\subseteq B(v')$. Since the graph is $3$-edge-connected, this can be strengthened to $B(v)\subset B(v')$. Thus, there is a back-edge $(x,y)\in B(v')\setminus B(v)$. Then $y$ is a proper ancestor of $v'$, and therefore a proper ancestor of $v$. Thus, $x$ cannot be a descendant of $v$. Since $M(v')$ is an ancestor of $x$, this implies that $M(v')$ cannot be a descendant of $v$. Let $(x',y')$ be a back-edge in $B(v)$. Then, $x'$ is a descendant of $v$. Furthermore, $B(v)\subseteq B(v')$ implies that $(x',y')\in B(v')$, and therefore $x'$ is a descendant of $M(v')$. Thus, $x'$ is a common descendant of $v$ and $M(v')$, and therefore $v$ and $M(v')$ are related as ancestor and descendant. Since $M(v')$ is not a descendant of $v$, this implies that $M(v')$ is a proper ancestor of $v$. Let $c$ be the child of $M(v')$ that is an ancestor of $v$. Since $e_\mathit{high}(v')\in B(v)$, we have that the higher endpoint of $e_\mathit{high}(v')$ is a descendant of $v$, and therefore a descendant of $c$. 
Now let $(x',y')$ be a back-edge in $B(v')$. If $(x',y')=e_\mathit{high}(v')$, then we have that $x'$ is a descendant of $c$. If $(x',y')\neq e_\mathit{high}(v')$, then we have $(x',y')\in B(v')\setminus\{e_\mathit{high}(v')\}$, and therefore $x$ is a descendant of $M(B(v')\setminus\{e_\mathit{high}(v')\})=M(B(v)\setminus\{e_\mathit{high}(v)\})$, and therefore a descendant of $v$, and therefore a descendant of $c$. Thus, in either case we have that $x'$ is a descendant of $c$. Due to the generality of $(x',y')\in B(v')$, this implies that $M(v')$ is a descendant of $c$. But this is absurd, since $c$ is a child of $M(v')$. This shows that $e_\mathit{high}(v')\notin B(v)$. 

Notice that we cannot have $e_\mathit{high}(v)\in B(v')$. (Because otherwise, since $\mathit{high}_1(v)>\mathit{high}_2(v)=\mathit{high}_2(v')$, we would have $e_\mathit{high}(v)=e_\mathit{high}(v')$, contradicting the fact that $e_\mathit{high}(v')\notin B(v)$.) Now let $(x,y)$ be a back-edge in $B(v)\setminus\{e_\mathit{high}(v)\}$. Then $x$ is a descendant of $v$, and therefore a descendant of $v'$. Furthermore, $y$ is an ancestor of $\mathit{high}_2(v)=\mathit{high}_2(v')$, and therefore a proper ancestor of $v'$. This shows that $(x,y)\in B(v')$. Due to the generality of $(x,y)\in B(v)\setminus\{e_\mathit{high}(v)\}$, this implies that $B(v)\setminus\{e_\mathit{high}(v)\}\subseteq B(v')$. And since $e_\mathit{high}(v')\notin B(v)$, this can be strengthened to $B(v)\setminus\{e_\mathit{high}(v)\}\subseteq B(v')\setminus\{e_\mathit{high}(v')\}$. Conversely, let $(x,y)$ be a back-edge in $B(v')\setminus\{e_\mathit{high}(v')\}$. Then, $x$ is a descendant of $M(B(v')\setminus\{e_\mathit{high}(v')\})$, and therefore a descendant of $M(B(v)\setminus\{e_\mathit{high}(v)\})$, and therefore a descendant of $M(v)$. Furthermore, $y$ is an ancestor of $\mathit{high}_2(v')=\mathit{high}_2(v)$, and therefore a proper ancestor of $v$. This shows that $(x,y)\in B(v)$. Due to the generality of $(x,y)\in B(v')\setminus\{e_\mathit{high}(v')\}$, this implies that $B(v')\setminus\{e_\mathit{high}(v')\}\subseteq B(v)$. And since $e_\mathit{high}(v)\notin B(v')$, this can be strengthened to $B(v')\setminus\{e_\mathit{high}(v')\}\subseteq B(v)\setminus\{e_\mathit{high}(v)\}$. Thus, we have shown that $B(v)\setminus\{e_\mathit{high}(v)\}=B(v')\setminus\{e_\mathit{high}(v')\}$. Since $e_\mathit{high}(v)\notin B(v')$ and $e_\mathit{high}(v')\notin B(v)$, this implies that $B(v)\sqcup\{e_\mathit{high}(v')\}=B(v')\sqcup\{e_\mathit{high}(v)\}$.
\end{proof}

\begin{lemma}
Let $v$ and $v'$ be two distinct vertices such that $v'\in V(v)$. Then, $v$ and $v'$ belong to the same segment of $\widetilde{H}(\mathit{high}_2(v))$ that is maximal w.r.t. the property that its elements are related as ancestor and descendant.
\end{lemma}
\begin{proof}
Since $v'\in V(v)$, we have that $M(v)\neq M(B(v)\setminus\{e_\mathit{high}(v)\})=M(B(v')\setminus\{e_\mathit{high}(v')\})\neq M(v')$ and $\mathit{high}_1(v)\neq \mathit{high}_2(v)=\mathit{high}_2(v')\neq \mathit{high}_1(v')$. Thus, both $v$ and $v'$ belong to $\widetilde{H}(\mathit{high}_2(v))$. Furthermore, we have that $M(B(v)\setminus\{e_\mathit{high}(v)\})=M(B(v')\setminus\{e_\mathit{high}(v')\})$ is a common descendant of $M(v)$ and $M(v')$, and therefore a common descendant of $v$ and $v'$. Thus, $v$ and $v'$ are related as ancestor and descendant. We may assume w.l.o.g. that $v$ is a proper descendant of $v'$. This implies that $v>v'$.

Let us suppose, for the sake of contradiction, that $v$ and $v'$ do not belong to a segment of $\widetilde{H}(\mathit{high}_2(v))$ with the property that its elements are related as ancestor and descendant. Since $\widetilde{H}$ is sorted in decreasing order, this means that there is a vertex $v''\in\widetilde{H}(\mathit{high}_2(v))$ such that $v>v''>v'$, and $v''$ is not an ancestor of $v$. Notice that, since $v>v''>v'$ and $v$ is a descendant of $v'$, we have that $v''$ is also a descendant of $v'$.

Since $v''\in\widetilde{H}(\mathit{high}_2(v))$ we have that either $\mathit{high}_1(v'')=\mathit{high}_2(v)$, or $\mathit{high}_2(v'')=\mathit{high}_2(v)$. Thus, in either case, there is a back-edge $(x,y)$ in $B(v'')$ such that $y=\mathit{high}_2(v)$. Then, we have that $x$ is a descendant of $v''$, and therefore a descendant of $v'$. Furthermore, we have $y=\mathit{high}_2(v)=\mathit{high}_2(v')$, and therefore $y$ is a proper ancestor of $v'$. This shows that $(x,y)\in B(v')$. Notice that we cannot have that $x$ is a descendant of $v$ (because otherwise, $x$ is a common descendant of $v$ and $v''$, and therefore $v$ and $v''$ are related as ancestor and descendant, which is impossible, since $v''<v$ and $v''$ is not an ancestor of $v$). This implies that $(x,y)\notin B(v)$. Thus, we have $(x,y)\in B(v')\setminus B(v)$. Lemma~\ref{lemma:V-sets-implied-4cuts} implies that $B(v')\setminus B(v)=\{e_\mathit{high}(v')\}$. Thus, $(x,y)=e_\mathit{high}(v')$. But $y=\mathit{high}_2(v)=\mathit{high}_2(v')$ and $\mathit{high}_2(v')\neq\mathit{high}_1(v')$, a contradiction. This shows that $v$ and $v'$ belong to a segment of $\widetilde{H}(\mathit{high}_2(v))$ with the property that its elements are related as ancestor and descendant, and so they belong to a maximal such segment.
\end{proof}

Let $v\neq r$ be a vertex such that $\mathit{high}_1(v)\neq\mathit{high}_2(v)$ and $M(B(v)\setminus\{e_\mathit{high}(v)\})\neq M(v)$.
We let $\widetilde{W}(v)$ denote the collection of all ancestors $w$ of $\mathit{high}_2(v)$ such that $M(w)=M(B(v)\setminus\{e_\mathit{high}(v)\})$. We also let $\widetilde{\mathit{firstW}}(v):=\mathit{max}(\widetilde{W}(v))$ and $\widetilde{\mathit{lastW}}(v):=\mathit{min}(\widetilde{W}(v))$. (If $\widetilde{W}(v)=\emptyset$, then we let $\widetilde{\mathit{firstW}}(v):=\bot$ and $\widetilde{\mathit{lastW}}(v):=\bot$.) 

\begin{lemma}
\label{lemma:computing-first-last-W}
The values $\widetilde{\mathit{firstW}}(v)$ and $\widetilde{\mathit{lastW}}(v)$ can be computed in linear time in total, for all vertices $v\neq r$.
\end{lemma}
\begin{proof}
Let $v\neq r$ be a vertex. If $M(B(v)\setminus\{e_\mathit{high}(v)\})=M(v)$ or $\mathit{high}_1(v)=\mathit{high}_2(v)$, then by definition we have $\widetilde{W}(v)=\emptyset$, and therefore $\widetilde{\mathit{firstW}}(v)=\bot$ and $\widetilde{\mathit{lastW}}(v)=\bot$. So let us assume that  $M(B(v)\setminus\{e_\mathit{high}(v)\})\neq M(v)$ and $\mathit{high}_1(v)\neq\mathit{high}_2(v)$.

Let $x=M(B(v)\setminus\{e_\mathit{high}(v)\})$. Then, notice that $\widetilde{W}(v)$ consists of all vertices $w$ with $M(w)=x$ that are ancestors of $\mathit{high}_2(v)$. Let $w=\mathit{lastM}(x)$. If $w>\mathit{high}_2(v)$, then we have $\widetilde{W}(v)=\emptyset$, because this implies that every vertex $w'$ with $M(w')=x$ has $w'>\mathit{high}_2(v)$, and therefore it cannot be an ancestor of $\mathit{high}_2(v)$. Now suppose that $w\leq\mathit{high}_2(v)$. We have that $x$ is a common descendant of $w$ and $v$, and therefore $w$ and $v$ are related as ancestor and descendant. Then $w\leq\mathit{high}_2(v)$ implies that $w<v$, and therefore $w$ is an ancestor of $v$. Then, since $v$ is a common descendant of $w$ and $\mathit{high}_2(v)$, we have that $w$ and $\mathit{high}_2(v)$ are related as ancestor and descendant. Thus, $w\leq\mathit{high}_2(v)$ implies that $w$ is an ancestor of $\mathit{high}_2(v)$. Thus, we have $w\in\widetilde{W}(v)$, and since $\widetilde{\mathit{lastW}}(v)=\mathit{min}(\widetilde{W}(v))$, we have $\widetilde{\mathit{lastW}}(v)=w$. Thus, $\widetilde{\mathit{lastW}}(v)$ can be easily computed in constant time, for every vertex $v$.

If we have established that $\widetilde{\mathit{lastW}}(v)\neq\bot$, then, in order to compute $\widetilde{\mathit{firstW}}(v)$, we use Algorithm~\ref{algorithm:W-queries}. Specifically, we generate a query of the form $q(M^{-1}(x),\mathit{high}_2(v))$. This is to return the greatest $w$ with $M(w)=M(B(v)\setminus\{e_\mathit{high}(v)\})$ such that $w\leq\mathit{high}_2(v)$. Then we can show as previously that $w$ is an ancestor of $\mathit{high}_2(v)$, and therefore we have that $w$ is the greatest ancestor of $\mathit{high}_2(v)$ such that $M(w)=M(B(v)\setminus\{e_\mathit{high}(v)\})$. Thus, we have $w=\widetilde{\mathit{firstW}}(v)$. Since the number of all those queries is $O(n)$, Lemma~\ref{lemma:W-queries} implies that Algorithm~\ref{algorithm:W-queries} can compute all of them in $O(n)$ time in total.
\end{proof}

Let $v\neq r$ be a vertex such that $\mathit{high}_1(v)\neq\mathit{high}_2(v)$ and $M(B(v)\setminus\{e_\mathit{high}(v)\})\neq M(v)$.
Let $\widetilde{S}$ be the segment of $\widetilde{H}(\mathit{high}_2(v))$ that contains $v$ and is maximal w.r.t. the property that all its elements are related as ancestor and descendant (i.e., $\widetilde{S}=\widetilde{S}_2(v)$). Then, $U_4^1(v)$ is the collection of all vertices $u\in \widetilde{S}$ such that: $(1)$ $u$ is a proper descendant of $v$ with $\mathit{high}(u)=\mathit{high}_2(v)$, $(2)$ $\mathit{low}(u)\geq\widetilde{\mathit{lastW}}(v)$, and $(3)$ either $\mathit{low}(u)<\widetilde{\mathit{firstW}(v)}$, or $u$ is the lowest vertex in $\widetilde{S}$ that satisfies $(1)$, $(2)$ and $\mathit{low}(u)\geq\widetilde{\mathit{firstW}}(v)$.

\begin{lemma}
\label{lemma:type-3-b-ii-4-1-U-sets}
Let $v$ and $v'$ be two vertices $\neq r$ such that $v'$ is a proper descendant of $v$ with $\mathit{high}_1(v)\neq\mathit{high}_2(v)=\mathit{high}_2(v')\neq\mathit{high}_1(v')$, $M(B(v)\setminus\{e_\mathit{high}(v)\})\neq M(v)$ and $M(B(v')\setminus\{e_\mathit{high}(v')\})\neq M(v')$. Let us assume that $v$ and $v'$ belong to the same segment $\widetilde{S}$ of $\widetilde{H}(\mathit{high}_2(v))$ that is maximal w.r.t. the property that its elements are related as ancestor and descendant. Let us further assume that $v'\notin V(v)$, $\widetilde{W}(v)\neq\emptyset$ and $\widetilde{W}(v')\neq\emptyset$. If $U_4^1(v')=\emptyset$, then $U_4^1(v)=\emptyset$. If $U_4^1(v)\neq\emptyset$, then the lowest vertex in $U_4^1(v)$ is greater than, or equal to, the greatest vertex in $U_4^1(v')$.
\end{lemma}
\begin{proof}
First we will show that $M(B(v)\setminus\{e_\mathit{high}(v)\})$ is a proper ancestor of $v'$, and $\widetilde{\mathit{firstW}}(v')$ is a proper ancestor of $\widetilde{\mathit{lastW}}(v)$.

Since $\mathit{high}_1(v)\neq\mathit{high}_2(v)=\mathit{high}_2(v')\neq\mathit{high}_1(v')$, $M(B(v)\setminus\{e_\mathit{high}(v)\})\neq M(v)$, $M(B(v')\setminus\{e_\mathit{high}(v')\})\neq M(v')$, and $v'\notin V(v)$, we have $M(B(v)\setminus\{e_\mathit{high}(v)\})\neq M(B(v')\setminus\{e_\mathit{high}(v')\})$ (because this is the only condition that prevents $v'$ from being in $V(v)$).

Since $\mathit{high}_1(v)\neq\mathit{high}_2(v)$, we have that $\mathit{high}_1(v)>\mathit{high}_2(v)$, and $e_\mathit{high}(v)$ is the unique back-edge in $B(v)$ whose lower endpoint is $\mathit{high}_1(v)$. Similarly, we have that $\mathit{high}_1(v')>\mathit{high}_2(v')$, and $e_\mathit{high}(v')$ is the unique back-edge in $B(v')$ whose lower endpoint is $\mathit{high}_1(v')$.

Now let $(x,y)$ be a back-edge in $B(v')$ such that $y=\mathit{high}_2(v')$. Then, $x$ is a descendant of $v'$, and therefore a descendant of $v$. Furthermore, $y=\mathit{high}_2(v')=\mathit{high}_2(v)$, and therefore $y$ is a proper ancestor of $v$. This shows that $(x,y)\in B(v)$. Since $\mathit{high}_2(v)\neq\mathit{high}_1(v)$, we have that $(x,y)\neq e_\mathit{high}(v)$. This implies that $x$ is a descendant of $M(B(v)\setminus\{e_\mathit{high}(v)\})$. Thus, we have that $x$ is a common descendant of $v'$ and $M(B(v)\setminus\{e_\mathit{high}(v)\})$. This shows that $v'$ and $M(B(v)\setminus\{e_\mathit{high}(v)\})$ are related as ancestor and descendant. 

Let us suppose, for the sake of contradiction, that $M(B(v)\setminus\{e_\mathit{high}(v)\})$ is not a proper ancestor of $v'$. Then we have that $M(B(v)\setminus\{e_\mathit{high}(v)\})$ is a descendant of $v'$. 
Let $(x,y)$ be a back-edge in $B(v)\setminus\{e_\mathit{high}(v)\}$. Then $x$ is a descendant of $M(B(v)\setminus\{e_\mathit{high}(v)\})$, and therefore a descendant of $v'$. Furthermore, $y$ is a proper ancestor of $v$, and therefore a proper ancestor of $v'$. This shows that $(x,y)\in B(v')$. Due to the generality of $(x,y)\in B(v)\setminus\{e_\mathit{high}(v)\}$, this implies that $B(v)\setminus\{e_\mathit{high}(v)\}\subseteq B(v')$. Notice that, if $e_\mathit{high}(v')\in B(v)$, then, since $\mathit{high}_1(v')>\mathit{high}_2(v')=\mathit{high}_2(v)$, we have that $e_\mathit{high}(v')=e_\mathit{high}(v)$. Thus, whether $e_\mathit{high}(v')\in B(v)$ or $e_\mathit{high}(v')\notin B(v)$, we infer that $B(v)\setminus\{e_\mathit{high}(v)\}\subseteq B(v')$ can be strengthened to $B(v)\setminus\{e_\mathit{high}(v)\}\subseteq B(v')\setminus\{e_\mathit{high}(v')\}$. Conversely, let $(x,y)$ be a back-edge in $B(v')\setminus\{e_\mathit{high}(v')\}$. Then $x$ is a descendant of $v'$, and therefore a descendant of $v$. Furthermore, $y$ is an ancestor of $\mathit{high}_2(v')=\mathit{high}_2(v)$, and therefore a proper ancestor of $v$. This shows that $(x,y)\in B(v)$. Due to the generality of $(x,y)\in B(v')\setminus\{e_\mathit{high}(v')\}$, this implies that $B(v')\setminus\{e_\mathit{high}(v')\}\subseteq B(v)$. As previously, we can argue that $B(v')\setminus\{e_\mathit{high}(v')\}\subseteq B(v)$ can be strengthened to $B(v')\setminus\{e_\mathit{high}(v')\}\subseteq B(v)\setminus\{e_\mathit{high}(v)\}$. Thus, we have shown that $B(v')\setminus\{e_\mathit{high}(v')\}= B(v)\setminus\{e_\mathit{high}(v)\}$, and therefore we have $M(B(v')\setminus\{e_\mathit{high}(v')\})=M(B(v)\setminus\{e_\mathit{high}(v)\})$, a contradiction. This shows that $M(B(v)\setminus\{e_\mathit{high}(v)\})$ is a proper ancestor of $v'$.

Now let $w$ be a vertex in $\widetilde{W}(v)$, and let $w'$ be a vertex in $\widetilde{W}(v')$. Then we have that $w$ is an ancestor of $\mathit{high}_2(v)$, and $w'$ is an ancestor of $\mathit{high}_2(v')$. Thus, since $\mathit{high}_2(v)=\mathit{high}_2(v')$, we have that $w$ and $w'$ have a common descendant, and therefore they are related as ancestor and descendant. Let us suppose, for the sake of contradiction, that $w'$ is not a proper ancestor of $w$. Then we have that $w'$ is a descendant of $w$. Since $w\in\widetilde{W}(v)$, we have that $M(w)=M(B(v)\setminus\{e_\mathit{high}(v)\})$. And since $w'\in\widetilde{W}(v')$, we have that $M(w')=M(B(v')\setminus\{e_\mathit{high}(v')\})$. Since $M(B(v)\setminus\{e_\mathit{high}(v)\})$ is a proper ancestor of $v'$, and $v'$ is an ancestor of $M(B(v')\setminus\{e_\mathit{high}(v')\})$, we have that $M(w)$ is a proper ancestor of $M(w')$. Thus, there is a back-edge $(x,y)\in B(w)$ such that $x$ is not a descendant of $M(w')$. Then, $x$ is a descendant of $M(w)$, and therefore a descendant of $M(B(v)\setminus\{e_\mathit{high}(v)\})$, and therefore a descendant of $v$, and therefore a descendant of $\mathit{high}_2(v)=\mathit{high}_2(v')$, and therefore a descendant of $w'$ (since $w'\in\widetilde{W}(v')$ implies that $w'$ is an ancestor of $\mathit{high}_2(v')$). Furthermore, we have that $y$ is a proper ancestor of $w$, and therefore a proper ancestor of $w'$. This shows that $(x,y)\in B(w')$. But this implies that $x$ is a descendant of $M(w')$, a contradiction. This shows that $w'$ is a proper ancestor of $w$. Due to the generality of $w'\in\widetilde{W}(v')$, this implies that $\widetilde{\mathit{firstW}}(w')$ is a proper ancestor of $w$. And due to the generality of $w\in\widetilde{W}(v)$, this implies that $\widetilde{\mathit{firstW}}(v')$ is a proper ancestor of $\widetilde{\mathit{lastW}}(v)$. 

Now let us suppose, for the sake of contradiction, that there is a vertex $u\in U_4^1(v)$, and $U_4^1(v')=\emptyset$. Since $u\in U_4^1(v)$, we have that $u\in\widetilde{S}$. Thus, since $v'\in\widetilde{S}$, we have that $u$ and $v'$ are related as ancestor and descendant. Let us suppose, for the sake of contradiction, that $u$ is not a descendant of $v'$. Then, we have that $u$ is a proper ancestor of $v'$. Let $(x,y)$ be a back-edge in $B(v')$ such that $y=\mathit{low}(v')$. Then, $x$ is a descendant of $v'$, and therefore a descendant of $u$. Furthermore, $y$ is an ancestor of $\mathit{high}_2(v')=\mathit{high}_2(v)$, and therefore a proper ancestor of $v$, and therefore a proper ancestor of $u$ (since $u\in U_4^1(v)$ implies that $u$ is a proper descendant of $v$). This shows that $(x,y)\in B(u)$. 


We have that $M(\widetilde{\mathit{firstW}}(v'))=M(B(v')\setminus\{e_\mathit{high}(v')\})$ and $M(B(v')\setminus\{e_\mathit{high}(v')\})$ is a descendant of $M(v')$. Thus, $M(\widetilde{\mathit{firstW}}(v'))$ is a descendant of $M(v')$. Furthermore, since $\widetilde{\mathit{firstW}}(v')$ is an ancestor of $\mathit{high}_2(v')$, we have that $\widetilde{\mathit{firstW}}(v')$ is a proper ancestor of $v'$. Thus, Lemma~\ref{lemma:same_m_subset_B} implies that $B(\widetilde{\mathit{firstW}}(v'))\subseteq B(v')$, and therefore $\mathit{low}(v')<\widetilde{\mathit{firstW}}(v')$. Thus, since $y=\mathit{low}(v')$, we have $y<\widetilde{\mathit{firstW}}(v')$, and therefore $y<\widetilde{\mathit{lastW}}(v)$ (since $\widetilde{\mathit{firstW}}(v')<\widetilde{\mathit{lastW}}(v)$). Since $(x,y)\in B(u)$, we have that $\mathit{low}(u)\leq y$, and therefore $\mathit{low}(u)<\widetilde{\mathit{lastW}}(v)$. But this contradicts the fact that $\mathit{low}(u)\geq\widetilde{\mathit{lastW}}(v)$ (which is an implication of $u\in U_4^1(v)$). This shows that $u$ is a descendant of $v'$. Since $\mathit{high}(u)=\mathit{high}_2(v)=\mathit{high}_2(v')\neq\mathit{high}_1(v')$, we have that $u\neq v'$. Thus, $u$ is a proper descendant of $v'$. Thus, we have the following facts: $u\in\widetilde{S}$, $u$ is a proper descendant of $v'$, $\mathit{high}(u)=\mathit{high}_2(v)=\mathit{high}_2(v')$, and $\mathit{low}(u)\geq\widetilde{\mathit{lastW}}(v)>\widetilde{\mathit{firstW}}(v')$. But this implies that $U_4^1(v')\neq\emptyset$ (because we can consider the lowest $u$ that has those properties). A contradiction. This shows that if $U_4^1(v')=\emptyset$, then $U_4^1(v)=\emptyset$.

Now let us assume that $U_4^1(v)\neq\emptyset$. This implies that $U_4^1(v')\neq\emptyset$. Let us suppose, for the sake of contradiction, that there is a vertex $u\in U_4^1(v)$ that is lower than the greatest vertex $u'$ in $U_4^1(v')$.
Since $u\in U_4^1(v)$, we have that $u\in\widetilde{S}$. Since $u'\in U_4^1(v')$ we have that $u'\in\widetilde{S}$. This implies that $u$ and $u'$ are related as ancestor and descendant. Thus, since $u$ is lower than $u'$, we have that $u$ is a proper ancestor of $u'$. 

Let us suppose, for the sake of contradiction, that $\mathit{low}(u')$ is a proper ancestor of $\widetilde{\mathit{firstW}}(v')$. Then, since $\widetilde{\mathit{firstW}}(v')$ is a proper ancestor of $\widetilde{\mathit{lastW}}(v)$, we have that $\mathit{low}(u')$ is a proper ancestor of $\widetilde{\mathit{lastW}}(v)$. Now let $(x,y)$ be a back-edge in $B(u')$ such that $y=\mathit{low}(u')$. Then $x$ is a descendant of $u'$, and therefore a descendant of $u$. Furthermore, $y$ is a proper ancestor of $\widetilde{\mathit{lastW}}(v)$, and therefore a proper ancestor of $\mathit{high}_2(v)$, and therefore a proper ancestor of $v$, and therefore a proper ancestor of $u$. This shows that $(x,y)\in B(u)$. Thus, we have $\mathit{low}(u)\leq y<\widetilde{\mathit{lastW}}(v)$, in contradiction to the fact that $u\in U_4^1(v)$. Thus, our last supposition is not true, and therefore we have that $\mathit{low}(u')$ is not a proper ancestor of $\widetilde{\mathit{firstW}}(v')$. 

Since $u'\in U_4^1(v')$, we have that $u'$ is a descendant of $v'$, and therefore a descendant of $\mathit{high}_2(v')$, and therefore a descendant of $\widetilde{\mathit{firstW}}(v')$. Thus, $u'$ is a common descendant of $\mathit{low}(u')$ and  $\widetilde{\mathit{firstW}}(v')$, and therefore $\mathit{low}(u')$ and  $\widetilde{\mathit{firstW}}(v')$ are related as ancestor and descendant. Thus, since $\mathit{low}(u')$ is not a proper ancestor of $\widetilde{\mathit{firstW}}(v')$, we have that $\mathit{low}(u')$ is a descendant of $\widetilde{\mathit{firstW}}(v')$, and therefore $\mathit{low}(u')\geq \widetilde{\mathit{firstW}}(v')$. Therefore, since $u'\in U_4^1(v')$, we have that $u'$ is the lowest proper descendant of $v'$ in $\widetilde{S}$ such that $\mathit{high}(u')=\mathit{high}_2(v')$ and $\mathit{low}(u')\geq \widetilde{\mathit{firstW}}(v')$ $(*)$. 

Now we will trace the implications of $u\in U_4^1(v)$. First, we have that $u\in\widetilde{S}$. Then, we have $\mathit{high}(u)=\mathit{high}_2(v)=\mathit{high}_2(v')$. Furthermore, we have that $\mathit{low}(u)\geq\widetilde{\mathit{lastW}}(v)$, and therefore $\mathit{low}(u)>\widetilde{\mathit{firstW}}(v')$ (since $\widetilde{\mathit{firstW}}(v')$ is a proper ancestor of $\widetilde{\mathit{lastW}}(v)$). Finally, we can show as above that $u$ is a proper descendant of $v'$ (the proof of this fact above did not rely on $U_4^1(v')=\emptyset$). But then, since $u$ is lower than $u'$, we have a contradiction to $(*)$.
Thus, we have shown that every vertex in $U_4^1(v)$ is at least as great as the greatest vertex in $U_4^1(v')$. In particular, this implies that the lowest vertex in $U_4^1(v)$ is greater than, or equal to, the greatest vertex in $U_4^1(v')$.
\end{proof}

Due to the similarity of the definitions of the $U_2$ and the $U_4^1$ sets, and the similarity between Lemmata~\ref{lemma:type3-b-ii-2-relation-between-u2} and \ref{lemma:type-3-b-ii-4-1-U-sets}, we can use a similar procedure as Algorithm~\ref{algorithm:type3-b-ii-2-U} in order to compute all $U_4^1$ sets in linear time. This is shown in Algorithm~\ref{algorithm:type3-b-ii-4-1-U}. Our result is summarized in Lemma~\ref{lemma:algorithm:type3-b-ii-4-1-U}.

\begin{algorithm}[H]
\caption{\textsf{Compute all sets $U_4^1(v)$, for a collection of vertices $\mathcal{V}$ that satisfies the properties described in Lemma~\ref{lemma:algorithm:type3-b-ii-4-1-U}}}
\label{algorithm:type3-b-ii-4-1-U}
\LinesNumbered
\DontPrintSemicolon
\ForEach{vertex $x$}{
  compute the collection $\mathcal{S}(x)$ of the segments of $\widetilde{H}(x)$ that are maximal w.r.t. the property
  that their elements are related as ancestor and descendant\;
}
\ForEach{vertex $v\in\mathcal{V}$}{
  set $U_4^1(v)\leftarrow\emptyset$\;
}
\ForEach{vertex $x$}{
  \ForEach{segment $S\in\mathcal{S}(x)$}{
    let $v$ be the first vertex in $S$\;
    \While{$v\neq\bot$ \textbf{and} ($v\notin\mathcal{V}$ \textbf{or} $\mathit{high}_2(v)\neq x$)}{
      $v\leftarrow\mathit{next}_S(v)$\;
    }
    \lIf{$v=\bot$}{\textbf{continue}}
    let $u\leftarrow\mathit{prev}_S(v)$\;
    \While{$v\neq\bot$}{
      \While{$u\neq\bot$ \textbf{and} $\mathit{low}(u)<\widetilde{\mathit{lastW}}(v)$}{
        $u\leftarrow\mathit{prev}_S(u)$\;
      }
      \While{$u\neq\bot$ \textbf{and} $\mathit{low}(u)<\widetilde{\mathit{firstW}}(v)$}{
        \If{$\mathit{high}(u)=x$}{
          insert $u$ into $U_4^1(v)$\;
        }
        $u\leftarrow\mathit{prev}_S(u)$\;
      }
      \While{$u\neq\bot$ \textbf{and} $\mathit{high}(u)\neq x$}{
        $u\leftarrow\mathit{prev}_S(u)$\;
      }
      \If{$u\neq\bot$}{
        insert $u$ into $U_4^1(v)$\;
      }
      $v\leftarrow\mathit{next}_S(v)$\;
      \While{$v\neq\bot$ \textbf{and} ($v\notin\mathcal{V}$ \textbf{or} $\mathit{high}_2(v)\neq x$)}{
        $v\leftarrow\mathit{next}_S(v)$\;
      }      
    }
  }
}
\end{algorithm}

\begin{lemma}
\label{lemma:algorithm:type3-b-ii-4-1-U}
Let $\mathcal{V}$ be a collection of vertices such that: 
\begin{enumerate}[label=(\arabic*)]
\item{For every $v\in\mathcal{V}$, we have $v\neq r$ and $M(B(v)\setminus\{e_\mathit{high}(v)\})\neq M(v)$ and $\mathit{high}_1(v)\neq\mathit{high}_2(v)$.}
\item{For every $v,v'\in\mathcal{V}$ with $v\neq v'$, we have $v'\notin V(v)$.}
\item{For every $v\in\mathcal{V}$, we have $\widetilde{W}(v)\neq\emptyset$, and the vertices $\widetilde{\mathit{firstW}}(v)$ and $\widetilde{\mathit{lastW}}(v)$ are computed.}
\end{enumerate}
Then, Algorithm~\ref{algorithm:type3-b-ii-4-1-U} correctly computes the sets $U_4^1(v)$, for all vertices $v\in\mathcal{V}$. Furthermore, on input $\mathcal{V}$, Algorithm~\ref{algorithm:type3-b-ii-4-1-U} runs in linear time. 
\end{lemma}
\begin{proof}
The proof is the same as that of Lemma~\ref{lemma:algorithm:type3-b-ii-2-U} (which is given in the main text, in the two paragraphs right above Algorithm~\ref{algorithm:type3-b-ii-2-U}). Notice that the difference between Algorithm~\ref{algorithm:type3-b-ii-2-U} and Algorithm~\ref{algorithm:type3-b-ii-4-1-U} is that the second algorithm has a different set $\mathcal{V}$ of vertices for which the $U_4^1$ sets are defined, and the occurrences of ``$\mathit{firstW}$" and ``$\mathit{lastW}$" are replaced with ``$\widetilde{\mathit{firstW}}$" and ``$\widetilde{\mathit{lastW}}$", respectively. Now we can use the argument of Lemma~\ref{lemma:algorithm:type3-b-ii-2-U}, by just replacing the references to Lemma~\ref{lemma:type3-b-ii-2-relation-between-u2} with references to Lemma~\ref{lemma:type-3-b-ii-4-1-U-sets}.
\end{proof}

\begin{lemma}
\label{lemma:type-3-b-ii-4-1-in-U}
Let $(u,v,w)$ be a triple of vertices that induces a Type-3$\beta$ii-$4$ $4$-cut, such that $M(B(v)\setminus\{e\})\neq M(v)$ and $\mathit{high}_1(v)>\mathit{high}(u)$, where $e$ is the back-edge in the $4$-cut induced by $(u,v,w)$. Then $u\in U_4^1(v)$. Furthermore, for every $v'\in V(v)$ we have that $(u,v',w)$ is a triple of vertices that induces a Type-3$\beta$ii-$4$ $4$-cut.
\end{lemma}
\begin{proof}
Since $\mathit{high}_1(v)>\mathit{high}(u)$, by Lemma~\ref{lemma:type3-b-ii-4-info} we have that $\mathit{high}(u)=\mathit{high}_2(v)$. Now let $u'$ be a vertex with $u\geq u'\geq v$ such that $u'\in\widetilde{H}(\mathit{high}_2(v))$. Since $\mathit{high}_1(v)\neq\mathit{high}(u)$, Lemma~\ref{lemma:type3-b-ii-4-seg-1} implies that $u'$ is an ancestor of $u$. Thus, we have that $u$ and $v$ belong to the same segment $\widetilde{S}$ of $\widetilde{H}(\mathit{high}_2(v))$ that is maximal w.r.t. the property that its elements are related as ancestor and descendant.

Since $(u,v,w)$ induces a Type-3$\beta$ii-$4$ $4$-cut, we have that $w$ is a proper ancestor of $v$ with $M(w)=M(B(v)\setminus\{e\})$. By Lemma~\ref{lemma:type3-b-ii-4-info} we have that $e=e_\mathit{high}(v)$ and $w\leq\mathit{low}(u)$. Since $\mathit{high}_1(v)>\mathit{high}(u)$ and $\mathit{high}(u)=\mathit{high}_2(v)$, we have that $\mathit{high}_1(v)\neq\mathit{high}_2(v)$. Since $w$ is a proper ancestor of $v$ with $w\leq\mathit{low}(u)\leq\mathit{high}(u)=\mathit{high}_2(v)$, we have that $w$ is an ancestor of $\mathit{high}_2(v)$. This shows that $w\in\widetilde{W}(v)$, and therefore $w\leq\widetilde{\mathit{firstW}}(v)$. Since $w\leq\mathit{low}(u)$ and $\widetilde{\mathit{lastW}}(v)\leq w$, we have $\widetilde{\mathit{lastW}}(v)\leq\mathit{low}(u)$. Now, if $\mathit{low}(u)<\widetilde{\mathit{firstW}}(v)$, then $u$ satisfies enough conditions to be in $U_4^1(v)$. Otherwise, let us assume that $\mathit{low}(u)\geq\widetilde{\mathit{firstW}}(v)$. 

Let us suppose, for the sake of contradiction, that $u$ is not the lowest vertex in $\widetilde{S}$ that is a proper descendant of $v$ such that $\mathit{high}(u)=\mathit{high}_2(v)$ and $\mathit{low}(u)\geq\widetilde{\mathit{firstW}}(v)$. Then, there is a vertex $u'\in\widetilde{S}$ that is a proper descendant of $v$ such that $u'<u$, $\mathit{high}(u')=\mathit{high}_2(v)$ and $\mathit{low}(u')\geq\widetilde{\mathit{firstW}}(v)$. Since both $u$ and $u'$ are in $\widetilde{S}$, we have that $u$ and $u'$ are related as ancestor and descendant. Thus, $u'<u$ implies that $u'$ is a proper ancestor of $u$. Now let $(x,y)$ be a back-edge in $B(u)$. Then, $x$ is a descendant of $u$, and therefore a descendant of $u'$. Furthermore, since $\mathit{high}(u)=\mathit{high}_2(v)$, we have that $y$ is a proper ancestor of $v$, and therefore a proper ancestor of $u'$. This shows that $(x,y)\in B(u')$. Due to the generality of $(x,y)\in B(u)$, this implies that $B(u)\subseteq B(u')$. Conversely, let $(x,y)$ be a back-edge in $B(u')$. Then $x$ is a descendant of $u'$, and therefore a descendant of $v$. Furthermore, $y$ is an ancestor of $\mathit{high}(u')=\mathit{high}_2(v)$, and therefore a proper ancestor of $v$. This shows that $(x,y)\in B(v)$. Since $(u,v,w)$ induces a Type-3$\beta$ii-$4$ $4$-cut, we have that $B(v)=(B(u)\sqcup B(w))\sqcup\{e\}$. This implies that either $(x,y)\in B(u)$, or $(x,y)\in B(w)$, or $(x,y)=e$. Since $y\geq\mathit{low}(u')\geq\widetilde{\mathit{firstW}}(v)\geq w$, we have that $(x,y)$ cannot be in $B(w)$. Since $\mathit{high}_1(v)\neq\mathit{high}_2(v)$, we have that $\mathit{high}_1(v)>\mathit{high}_2(v)$. Thus, since $e=e_\mathit{high}(v)$ and $\mathit{high}_1(v)>\mathit{high}_2(v)=\mathit{high}(u')$, we have that $e\notin B(u')$. Thus, $(x,y)\in B(u)$ is the only viable option. Due to the generality of $(x,y)\in B(u')$, this implies that $B(u')\subseteq B(u)$. Thus, we have $B(u')=B(u)$, in contradiction to the fact that the graph is $3$-edge-connected. This shows that $u$ is the lowest vertex in $\widetilde{S}$ that is a proper descendant of $v$ such that $\mathit{high}(u)=\mathit{high}_2(v)$ and $\mathit{low}(u)\geq\widetilde{\mathit{firstW}}(v)$. Thus, $u$ satisfies enough conditions to be in $U_4^1(v)$.

Since $\mathit{high}_1(v)\neq\mathit{high}_2(v)$ and $M(B(v)\setminus\{e_\mathit{high}(v)\})\neq M(v)$, we have that $v\in V(v)$. Now let $v'$ be a vertex in $V(v)$ such that $v'\neq v$. Then, by Lemma~\ref{lemma:V-sets-implied-4cuts} we have $B(v)\sqcup\{e_\mathit{high}(v')\}=B(v')\sqcup\{e_\mathit{high}(v)\}$. This implies that $B(v)\setminus\{e_\mathit{high}(v)\}=B(v')\setminus\{e_\mathit{high}(v')\}$. Also, $B(v)=(B(u)\sqcup B(w))\sqcup\{e_\mathit{high}(v)\}$ implies that $B(v)\setminus\{e_\mathit{high}(v)\}=B(u)\sqcup B(w)$. Thus, we infer that $B(v')\setminus\{e_\mathit{high}(v')\}=B(u)\sqcup B(w)$, and therefore $B(v')=(B(u)\sqcup B(w))\cup\{e_\mathit{high}(v')\}$. 

Since $v'\in V(v)$, we have $\mathit{high}_1(v')\neq\mathit{high}_2(v')=\mathit{high}_2(v)$. This implies that $\mathit{high}_1(v')>\mathit{high}_2(v')$, and therefore $\mathit{high}_1(v')>\mathit{high}(u)$ (since $\mathit{high}_2(v')=\mathit{high}(u)$). Thus, we cannot have $e_\mathit{high}(v')\in B(u)$. Furthermore, since $w\leq\mathit{low}(u)\leq\mathit{high}(u)=\mathit{high}_2(v)=\mathit{high}_2(v')<\mathit{high}_1(v')$, we have that $e_\mathit{high}(v')\notin B(w)$. Thus, $B(v')=(B(u)\sqcup B(w))\cup\{e_\mathit{high}(v')\}$ can be strengthened to $B(v')=(B(u)\sqcup B(w))\sqcup\{e_\mathit{high}(v')\}$. Finally, since $v'\in V(v)$, we have that $M(B(v')\setminus\{e_\mathit{high}(v')\})=M(B(v)\setminus\{e_\mathit{high}(v)\})=M(w)$. Thus, we conclude that $(u,v',w)$ induces a Type-3$\beta$ii-$4$ $4$-cut.  
\end{proof}

\begin{lemma}
\label{lemma:type-3-b-ii-4-1-criterion}
Let $(u,v,w)$ be a triple of vertices such that $\mathit{high}_1(v)\neq\mathit{high}_2(v)$, $M(B(v)\setminus\{e_\mathit{high}(v)\})\neq M(v)$ and $u\in U_4^1(v)$. Then, $(u,v,w)$ induces a Type-3$\beta$ii-$4$ $4$-cut if and only if: $\mathit{bcount}(v)=\mathit{bcount}(u)+\mathit{bcount}(w)+1$, and $w$ is the greatest proper ancestor of $v$ with $M(w)=M(B(v)\setminus\{e_\mathit{high}(v)\})$ such that $w\leq\mathit{low}(u)$.
\end{lemma}
\begin{proof}
($\Rightarrow$) Since $(u,v,w)$ induces a Type-3$\beta$ii-$4$ $4$-cut, we have $B(v)=(B(u)\sqcup B(w))\sqcup\{e\}$, where $e$ is the back-edge in the $4$-cut induced by $(u,v,w)$. Thus, we get $\mathit{bcount}(v)=\mathit{bcount}(u)+\mathit{bcount}(w)+1$. Since $\mathit{high}_1(v)\neq\mathit{high}_2(v)$ we have $\mathit{high}_1(v)>\mathit{high}_2(v)$. Since $u\in U_4^1(v)$ we have $\mathit{high}(u)=\mathit{high}_2(v)$. This implies that $\mathit{high}_1(v)>\mathit{high}(u)$. Thus, by Lemma~\ref{lemma:type3-b-ii-4-info} we have $e=e_\mathit{high}(v)$. Since $(u,v,w)$ induces a Type-3$\beta$ii-$4$ $4$-cut, we have $M(w)=M(B(v)\setminus\{e\})$. Thus, $M(w)=M(B(v)\setminus\{e_\mathit{high}(v)\})$. Furthermore, Lemma~\ref{lemma:type3-b-ii-4-info} implies that $w\leq\mathit{low}(u)$.

Now let us suppose, for the sake of contradiction, that there is an ancestor $w'$ of $v$ with $w'>w$, such that $M(w')=M(B(v)\setminus\{e_\mathit{high}(v)\})$ and $w'\leq\mathit{low}(u)$. Then, we have that $v$ is a common descendant of $w$ and $w'$, and therefore $w$ and $w'$ are related as ancestor and descendant. Thus, $w'>w$ implies that $w'$ is a proper descendant of $w$. Then, since $M(w')=M(w)$, Lemma~\ref{lemma:same_m_subset_B} implies that $B(w)\subseteq B(w')$. Since the graph is $3$-edge-connected, this can be strengthened to $B(w)\subset B(w')$. Thus, there is a back-edge $(x,y)\in B(w')\setminus B(w)$. Then, we have that $x$ is a descendant of $M(w')$, and therefore a descendant of $M(B(v)\setminus\{e_\mathit{high}(v)\})$, and therefore a descendant of $M(v)$. Furthermore, we have that $y$ is a proper ancestor of $w'$, and therefore a proper ancestor of $v$. This shows that $(x,y)\in B(v)$. Then, $B(v)=(B(u)\sqcup B(w))\sqcup\{e\}$ implies that either $(x,y)\in B(u)$, or $(x,y)\in B(w)$, or $(x,y)=e$. The case $(x,y)\in B(u)$ is rejected, since $y<w'\leq\mathit{low}(u)$. Furthermore, the case $(x,y)=e$ is rejected, since $e=e_\mathit{high}(v)$, and $\mathit{high}_1(v)>\mathit{high}(u)\geq\mathit{low}(u)$ (but $y<w'\leq\mathit{low}(u)$). Thus, we have $(x,y)\in B(w)$, which is impossible, since $(x,y)\in B(w')\setminus B(w)$. 
This shows that $w$ is the greatest proper ancestor of $v$ with $M(w)=M(B(v)\setminus\{e_\mathit{high}(v)\})$ such that $w\leq\mathit{low}(u)$.

($\Leftarrow$) We have to show that there is a back-edge $e$ such that $B(v)=(B(u)\sqcup B(w))\sqcup\{e\}$, and $M(w)=M(B(v)\setminus\{e\})$.

Since $w\leq\mathit{low}(u)$, we have $B(u)\cap B(w)=\emptyset$ (because, if there existed a back-edge in $B(u)\cap B(w)$, its lower endpoint would be lower than $w$, and therefore lower than $\mathit{low}(u)$, which is impossible). 
Let $(x,y)$ be a back-edge in $B(u)$. Since $u\in U_4^1(v)$, we have that $u$ is a proper descendant of $v$ with $\mathit{high}(u)=\mathit{high}_2(v)$. Thus, $(x,y)\in B(u)$ implies that $x$ is a descendant of $v$. Furthermore, since $y\in B(u)$, we have that $y$ is an ancestor of $\mathit{high}(u)=\mathit{high}_2(v)$, and therefore a proper ancestor of $v$. This shows that $(x,y)\in B(v)$. Due to the generality of $(x,y)\in B(u)$, this implies that $B(u)\subseteq B(v)$.

Let $(x,y)$ be a back-edge in $B(w)$. Then, $x$ is a descendant of $M(w)=M(B(v)\setminus\{e_\mathit{high}(v)\})$, and therefore a descendant of $v$. Furthermore, $y$ is a proper ancestor of $w$, and therefore a proper ancestor of $v$. This shows that $(x,y)\in B(v)$. Due to the generality of $(x,y)\in B(w)$, this implies that $B(w)\subseteq B(v)$.

Since $\mathit{high}_1(v)\neq\mathit{high}_2(v)$, we have $\mathit{high}_1(v)>\mathit{high}_2(v)$. Then, since $u\in U_4^1(v)$, we have $\mathit{high}(u)=\mathit{high}_2(v)$, and therefore $\mathit{high}_1(v)>\mathit{high}(u)$, and therefore we cannot have $e_\mathit{high}(v)\in B(u)$ (because the lower endpoint of $e_\mathit{high}(v)$ is greater than $\mathit{high}(u)$). Furthermore, since $w\leq\mathit{low}(u)\leq\mathit{high}(u)=\mathit{high}_2(v)<\mathit{high}_1(v)$, we have $e_\mathit{high}(v)\notin B(w)$ (because the lower endpoint of $e_\mathit{high}(v)$ is greater than $w$).

Thus, we have $B(u)\subseteq B(v)$, $B(w)\subseteq B(v)$, and the sets $B(u)$, $B(w)$, and $\{e_\mathit{high}(v)\}$ are mutually disjoint. Thus, $\mathit{bcount}(v)=\mathit{bcount}(u)+\mathit{bcount}(w)+1$ implies that $B(v)=(B(u)\sqcup B(w))\sqcup\{e_\mathit{high}(v)\}$. By assumption, we have $M(w)=M(B(v)\setminus\{e_\mathit{high}(v)\})$. This shows that $(u,v,w)$ induces a Type-3$\beta$ii-$4$ $4$-cut. 
\end{proof}

\begin{algorithm}[h!]
\caption{\textsf{Compute a collection of Type-3$\beta$ii-$4$ $4$-cuts of the form $\{(u,p(u)),(v,p(v)),(w,p(w)),e\}$, where $w$ is an ancestor of $v$, $v$ is an ancestor of $u$, $M(B(v)\setminus\{e\})\neq M(v)$ and $\mathit{high}_1(v)\neq\mathit{high}(u)$, such that all $4$-cuts of this form are implied from this collection plus that returned by Algorithm~\ref{algorithm:type2-2}}}
\label{algorithm:type3-b-ii-4-1}
\LinesNumbered
\DontPrintSemicolon
select a representative vertex for every non-empty set in $\{V(v)\mid v \mbox{ is a vertex } \neq r\}$; call this vertex a ``marked" vertex\;
\label{line:type3-b-ii-4-1-repr}
\tcp{If $V(v)\neq\emptyset$, for a vertex $v\neq r$, then the representative vertex of $V(v)$ is a vertex $v'\in V(v)$, and so it has $M(B(v')\setminus\{e_\mathit{high}(v')\})\neq M(v')$ and $\mathit{high}_2(v')\neq\mathit{high}_1(v')$}
\ForEach{marked vertex $v$}{
\label{line:type3-b-ii-4-1-for}
  \If{$\widetilde{W}(v)\neq\emptyset$}{
  \label{line:type3-b-ii-4-1-cond-W}
    compute the set $U_4^1(v)$\;
    \label{line:type3-b-ii-4-1-comp-U}
  }
}
\ForEach{marked vertex $v$}{
  \ForEach{$u\in U_4^1(v)$}{
    let $w$ be the greatest proper ancestor of $v$ such that $M(w)=M(B(v)\setminus\{e_\mathit{high}(v)\})$ and $w\leq\mathit{low}(u)$\;
    \label{line:type3-b-ii-4-1-w}
    \If{$\mathit{bcount}(v)=\mathit{bcount}(u)+\mathit{bcount}(w)+1$}{
    \label{line:type3-b-ii-4-1-cond}
      mark $\{(u,p(u)),(v,p(v)),(w,p(w)),e_\mathit{high}(v)\}$ as a Type-3$\beta$ii-$4$ $4$-cut\;
      \label{line:type3-b-ii-4-1-mark}
    }
  }
}
\end{algorithm}

\begin{proposition}
\label{proposition:algorithm:type3-b-ii-4-1}
Algorithm~\ref{algorithm:type3-b-ii-4-1} computes a collection $\mathcal{C}$ of Type-3$\beta$ii-$4$ $4$-cuts of the form $\{(u,p(u)),(v,p(v)),(w,p(w)),e\}$, where $M(B(v)\setminus\{e\})\neq M(v)$ and $\mathit{high}_1(v)\neq\mathit{high}(u)$. Let $\mathcal{C}'$ be the collection of Type-$2ii$ $4$-cuts returned by Algorithm~\ref{algorithm:type2-2}. Then, every Type-3$\beta$ii-$4$ $4$-cut of the form $\{(u,p(u)),(v,p(v)),(w,p(w)),e\}$, where $M(B(v)\setminus\{e\})\neq M(v)$ and $\mathit{high}_1(v)\neq\mathit{high}(u)$ is implied by $\mathcal{C}\cup\mathcal{C}'$. Finally, Algorithm~\ref{algorithm:type3-b-ii-4-1} has a linear-time implementation.
\end{proposition}
\begin{proof}
Let $C=\{(u,p(u)),(v,p(v)),(w,p(w)),e_\mathit{high}(v)\}$ be a $4$-element set that is marked in Line~\ref{line:type3-b-ii-4-1-mark}. Then we have that $v$ is a marked vertex, and therefore it has $M(B(v)\setminus\{e_\mathit{high}(v)\})\neq M(v)$ and $\mathit{high}_2(v)\neq\mathit{high}_1(v)$. We also have that $u\in U_4^1(v)$, $w$ is the greatest proper ancestor of $v$ such that $M(w)=M(B(v)\setminus\{e_\mathit{high}(v)\})$ and $w\leq\mathit{low}(u)$, and $\mathit{bcount}(v)=\mathit{bcount}(u)+\mathit{bcount}(w)+1$. Thus, all the conditions of Lemma~\ref{lemma:type-3-b-ii-4-1-criterion} are satisfied, and therefore we have that $(u,v,w)$ induces a Type-3$\beta$ii-$4$ $4$-cut. Let $e$ be the back-edge in the $4$-cut induced by $(u,v,w)$. Since $u\in U_4^1(v)$ we have $\mathit{high}(u)=\mathit{high}_2(v)$. Therefore, $\mathit{high}_2(v)\neq\mathit{high}_1(v)$ implies that $\mathit{high}(u)\neq\mathit{high}_1(v)$. Thus, Lemma~\ref{lemma:type3-b-ii-4-info} implies that $e=e_\mathit{high}(v)$. Therefore, we have that $C$ is the Type-3$\beta$ii-$4$ $4$-cut induced by $(u,v,w)$. So let $\mathcal{C}$ be the collection of all Type-3$\beta$ii-$4$ $4$-cuts marked in Line~\ref{line:type3-b-ii-4-1-mark}.

Let $C=\{(u,p(u)),(v,p(v)),(w,p(w)),e\}$ be a Type-3$\beta$ii-$4$ $4$-cut, where $w$ is a proper ancestor of $v$, $v$ is a proper ancestor of $u$, $M(B(v)\setminus\{e\})\neq M(v)$ and $\mathit{high}_1(v)\neq\mathit{high}(u)$. Since $\mathit{high}_1(v)\neq\mathit{high}(u)$, Lemma~\ref{lemma:type3-b-ii-4-info} implies that $\mathit{high}_1(v)>\mathit{high}(u)$ and $e=e_\mathit{high}(v)$. Thus, Lemma~\ref{lemma:type-3-b-ii-4-1-in-U} implies that $u\in U_4^1(v)$. Since $u\in U_4^1(v)$ we have $\mathit{high}(u)=\mathit{high}_2(v)$. Therefore, $\mathit{high}_1(v)>\mathit{high}(u)$ implies that $\mathit{high}_1(v)\neq\mathit{high}_2(v)$. Thus, Lemma~\ref{lemma:type-3-b-ii-4-1-criterion} implies that $\mathit{bcount}(v)=\mathit{bcount}(u)+\mathit{bcount}(w)+1$, and $w$ is the greatest proper ancestor of $v$ such that $M(w)=M(B(v)\setminus\{e_\mathit{high}(v)\})$ and $w\leq\mathit{low}(u)$. Thus, if $v$ is one of the marked vertices, then $C$ satisfies enough conditions to be marked in Line~\ref{line:type3-b-ii-4-1-mark}, and therefore $C\in\mathcal{C}$. So let us assume that $v$ is not one of the marked vertices. 

Since $\mathit{high}_1(v)\neq\mathit{high}_2(v)$ and $M(B(v)\setminus\{e_\mathit{high}(v)\})\neq M(v)$, we have that $V(v)\neq\emptyset$. Let $v'$ be the marked vertex that was picked as a representative of $V(v)$ in Line~\ref{line:type3-b-ii-4-1-repr}. Then, Lemma~\ref{lemma:type-3-b-ii-4-1-in-U} implies that $(u,v',w)$ induces a Type-3$\beta$ii-$4$ $4$-cut $C'$. Then, since $v'\in V(v)$ we have $\mathit{high}_1(v')\neq\mathit{high}_2(v')=\mathit{high}_2(v)=\mathit{high}(u)$ and $M(B(v')\setminus\{e_\mathit{high}(v')\})\neq M(v')$. Then, since $\mathit{high}_1(v')\neq\mathit{high}(u)$, Lemma~\ref{lemma:type3-b-ii-4-info} implies that the back-edge in the $4$-cut induced by $(u,v',w)$ is $e_\mathit{high}(v')$, and $\mathit{high}_1(v')>\mathit{high}(u)$. Then, Lemma~\ref{lemma:type-3-b-ii-4-1-in-U} implies that $u\in U_4^1(v')$, and Lemma~\ref{lemma:type-3-b-ii-4-1-criterion} implies that $\mathit{bcount}(v')=\mathit{bcount}(u)+\mathit{bcount}(w)+1$, and $w$ is the greatest proper ancestor of $v'$ such that $M(w)=M(B(v')\setminus\{e_\mathit{high}(v')\})$ and $w\leq\mathit{low}(u)$. Thus, $C'$ satisfies enough conditions to be marked in Line~\ref{line:type3-b-ii-4-1-mark}, and therefore $C'\in\mathcal{C}$. 

Since $v'\in V(v)$ and $v'\neq v$, Lemma~\ref{lemma:V-sets-implied-4cuts} implies that $B(v)\sqcup\{e_\mathit{high}(v')\}=B(v')\sqcup\{e_\mathit{high}(v)\}$. Then, Lemma~\ref{lemma:type2cuts} implies that $C''=\{(v,p(v)),(v',p(v')),e_\mathit{high}(v),e_\mathit{high}(v')\}$ is a Type-$2ii$ $4$-cut. Since $C=\{(u,p(u)),(v,p(v)),(w,p(w)),e_\mathit{high}(v)\}$ and $C'=\{(u,p(u)),(v',p(v')),(w,p(w)),e_\mathit{high}(v')\}$, notice that $C$ is implied by $C'$ and $C''$ through the pair of edges $\{(v,p(v)),e_\mathit{high}(v)\}$. Let $\mathcal{C}'$ be the collection of Type-$2ii$ $4$-cuts computed by Algorithm~\ref{algorithm:type2-2}. By Proposition~\ref{proposition:type-2-2} we have that $C''$ is implied by $\mathcal{C}'$ through the pair of edges $\{(v,p(v)),e_\mathit{high}(v)\}$. Thus, by Lemma~\ref{lemma:implied_from_union} we have that $C$ is implied by $\mathcal{C}'\cup\{C'\}$ through the pair of edges $\{(v,p(v)),e_\mathit{high}(v)\}$. Therefore, $C$ is implied by $\mathcal{C}'\cup\mathcal{C}$.

Now we will argue about the complexity of Algorithm~\ref{algorithm:type3-b-ii-4-1}. By Proposition~\ref{proposition:computing-M(B(v)-S)}, the values $M(B(v)\setminus\{e_\mathit{high}(v)\})$ can be computed in linear time in total, for all vertices $v\neq r$. Then, for every vertex $v\neq r$ such that $\mathit{high}_1(v)\neq\mathit{high}_2(v)$ and $M(B(v)\setminus\{e_\mathit{high}(v)\})\neq M(v)$, we generate a triple $(v,\mathit{high}_2(v),M(B(v)\setminus\{e_\mathit{high}(v)\}))$. Let $L$ be the collection of all those triples. Then we sort $L$ lexicographically w.r.t. the second and the third component of its elements. We note that this sorting can be performed in $O(n)$ time with bucket-sort. Then, every $V$ set corresponds to a segment of $L$ that is maximal w.r.t. the property that its elements coincide in their second and their third components. Then, we just pick a triple from every such segment, we extract its first component $v$, and we mark it, in order to get a marked representative of the corresponding $V$ set. Thus, the collection of the marked vertices in Line~\ref{line:type3-b-ii-4-1-repr} can be constructed in linear time.

By Lemma~\ref{lemma:computing-first-last-W} we have that the vertices $\widetilde{\mathit{firstW}}(v)$ and $\widetilde{\mathit{lastW}}(v)$ can be computed in linear time in total, for all marked vertices $v$. Then, by Lemma~\ref{lemma:algorithm:type3-b-ii-4-1-U} we have that the sets $U_4^1(v)$ can be computed in linear time in total, for all marked vertices $v$. Thus, the \textbf{for} loop in Line~\ref{line:type3-b-ii-4-1-for} can be performed in linear time.

In order to compute the vertex $w$ in Line~\ref{line:type3-b-ii-4-1-w} we use Algorithm~\ref{algorithm:W-queries}. Specifically, whenever we reach Line~\ref{line:type3-b-ii-4-1-w}, we generate a query of the form $q(M^{-1}(M(B(v)\setminus\{e_\mathit{high}(v)\})),\mathit{min}\{p(v),\mathit{low}(u)\})$. This is to return the greatest vertex $w$ with $M(w)=M(B(v)\setminus\{e_\mathit{high}(v)\})$ such that $w\leq\mathit{min}\{p(v),\mathit{low}(u)\}$. Since $M(B(v)\setminus\{e_\mathit{high}(v)\})=M(w)$ is a common descendant of $w$ and $v$, we have that $w$ and $v$ are related as ancestor and descendant. Then, $w\leq\mathit{min}\{p(v),\mathit{low}(u)\}$ implies that $w\leq p(v)$, and therefore $w$ is a proper ancestor of $v$. Thus, $w$ is the greatest proper ancestor of $v$ with $M(w)=M(B(v)\setminus\{e_\mathit{high}(v)\})$ such that $w\leq\mathit{low}(u)$. Now, since the number of all those queries is $O(n)$, Lemma~\ref{lemma:W-queries} implies that Algorithm~\ref{algorithm:W-queries} can answer all of them in $O(n)$ time in total.
We conclude that Algorithm~\ref{algorithm:type3-b-ii-4-1} runs in linear time.
\end{proof}

\noindent\\
\textbf{The case where $M(B(v)\setminus\{e\})\neq M(v)$ and $\mathit{high}_1(v)=\mathit{high}(u)$}\\

In this case, by Lemma~\ref{lemma:e_L-e_R} we have that either $e=e_L(v)$ or $e=e_R(v)$. In this subsection, we will focus on the case $e=e_L(v)$. Thus, whenever we consider a triple of vertices $(u,v,w)$ that induces a Type-3$\beta$ii-$4$ $4$-cut, such that $M(B(v)\setminus\{e\})\neq M(v)$ and $\mathit{high}_1(v)=\mathit{high}(u)$, we assume that $e=e_L(v)$, where $e$ is the back-edge in the $4$-cut induced by $(u,v,w)$. We will show how to compute all $4$-cuts of this type in linear time. The algorithms and the arguments for the case $e=e_R(v)$ are similar.

For every vertex $v\neq r$, we let $W_L(v)$ denote the collection of all proper ancestors $w$ of $v$ such that $M(w)=M(B(v)\setminus\{e_L(v)\})$. We also let $\mathit{firstW_L}(v):=\mathit{max}(W_L(v))$ and $\mathit{lastW_L}(v):=\mathit{min}(W_L(v))$. (If $W_L(v)=\emptyset$, then we let $\mathit{firstW_L}(v):=\bot$ and $\mathit{lastW_L}(v):=\bot$.) 

\begin{lemma}
\label{lemma:compute_wl}
For all vertices $v\neq r$, the values $\mathit{firstW_L}(v)$ and $\mathit{lastW_L}(v)$ can be computed in total linear time.
\end{lemma}
\begin{proof}
First, we need to have the values $M(B(v)\setminus\{e_L(v)\})$ computed, for all vertices $v\neq r$. According to Proposition~\ref{proposition:computing-M(B(v)-S)}, this can be achieved in linear time in total. Now, in order to compute $\mathit{lastW_L}(v)$, we just need to know whether $w=\mathit{lastM}(M(B(v)\setminus\{e_L(v)\}))$ is a proper ancestor of $v$. If that is the case, then we set $\mathit{lastW_L}(v)\leftarrow w$. Otherwise, $\mathit{lastW_L}(v)$ is left to be $\bot$. Then, for every vertex $v\neq r$, we have that $\mathit{firstW_L}(v)$ is the greatest proper ancestor $w$ of $v$ that has $M(w)=M(B(v)\setminus\{e_L(v)\})$. Thus, we can compute all $\mathit{firstW_L}$ values using Algorithm~\ref{algorithm:W-queries}. Specifically, let $v\neq r$ be a vertex, and let $x=M(B(v)\setminus\{e_L(v)\})$. Then we generate a query of the form $q(M^{-1}(x),p(v))$. This query returns the greatest vertex $w$ in $M^{-1}(x)$ that has $w\leq p(v)$. Since $w\in M^{-1}(x)$, we have that $M(w)=M(B(v)\setminus\{e_L(v)\})$. Thus, $M(w)$ is an ancestor of $M(v)$, and therefore $w$ is an ancestor of $M(v)$. Therefore, $M(v)$ is a common descendant of $w$ and $v$, and so $w$ and $v$ are related as ancestor and descendant. Then, $w\leq p(v)$ implies that $w$ is a proper ancestor of $v$. This shows that $w$ is the greatest proper ancestor of $v$ such that $w\in M^{-1}(x)$. In other words, $w=\mathit{firstW_L}(v)$. By Lemma~\ref{lemma:W-queries} we have that all these queries can be answered in $O(n)$ time in total.
\end{proof}

Now let $v\neq r$ be a vertex such that $M(B(v)\setminus\{e_L(v)\})\neq M(v)$, and let $S$ be the segment of $H(\mathit{high}_1(v))$ that contains $v$ and is maximal w.r.t. the property that all its elements are related as ancestor and descendant (i.e., $S=S(v)$). Then, we let $U_4^2(v)$ denote the collection of all vertices $u\in S$ such that: $(1)$ $u$ is a proper descendant of $v$, $(2)$ $\mathit{low}(u)\geq\mathit{lastW_L}(v)$, and $(3)$ either $\mathit{low}(u)<\mathit{firstW_L}(v)$, or $u$ is the lowest vertex in $S$ that satisfies $(1)$, $(2)$ and $\mathit{low}(u)\geq\mathit{firstW_L}(v)$.

\begin{lemma}
\label{lemma:type-3-b-ii-4-2-U-sets}
Let $v$ and $v'$ be two vertices such that $v'$ is a proper descendant of $v$ with $\mathit{high}_1(v)=\mathit{high}_1(v')$, $M(B(v)\setminus\{e_L(v)\})\neq M(v)$ and $M(B(v')\setminus\{e_L(v')\})\neq M(v')$. Suppose that $W_L(v)\neq\emptyset$, $W_L(v')\neq\emptyset$, and both $v$ and $v'$ belong to the same segment $S$ of $H(\mathit{high}_1(v))$ that is maximal w.r.t. the property that its elements are related as ancestor and descendant. If $U_4^2(v')=\emptyset$, then $U_4^2(v)=\emptyset$. If $U_4^2(v)\neq\emptyset$, then the lowest vertex in $U_4^2(v)$ is greater than, or equal to, the greatest vertex in $U_4^2(v')$.
\end{lemma}
\begin{proof}
First we will show that $M(B(v)\setminus\{e_L(v)\})$ is a proper ancestor of $M(B(v')\setminus\{e_L(v')\})$, and $\mathit{firstW_L}(v')$ is a proper ancestor of $\mathit{lastW_L}(v)$.

Since $v'$ is a proper descendant of $v$ such that $\mathit{high}_1(v')=\mathit{high}_1(v)$, Lemma~\ref{lemma:same_high} implies that $B(v')\subseteq B(v)$.
Since the graph is $3$-edge-connected, we have that $|B(v')|>1$. Thus, there is a back-edge $(x,y)\in B(v')\setminus\{e_L(v)\}$. Then, since $B(v')\subseteq B(v)$, we have that $(x,y)\in B(v)$. Since $(x,y)\neq e_L(v)$, this can be strengthened to $(x,y)\in B(v)\setminus\{e_L(v)\}$. This implies that $x$ is a descendant of $M(B(v)\setminus\{e_L(v)\})$. Thus, we have that $x$ is a common descendant of $v'$ and $M(B(v)\setminus\{e_L(v)\})$, and therefore $v'$ and $M(B(v)\setminus\{e_L(v)\})$ are related as ancestor and descendant.

If $M(B(v)\setminus\{e_L(v)\})$ is a proper ancestor of $v'$, then we have that $M(B(v)\setminus\{e_L(v)\})$ is a proper ancestor of $M(B(v')\setminus\{e_L(v')\})$, since $M(B(v')\setminus\{e_L(v')\})$ is a descendant of $M(v')$, and therefore a descendant of $v'$. So let us assume that $M(B(v)\setminus\{e_L(v)\})$ is a descendant of $v'$. Let $(x,y)$ be a back-edge in $B(v)\setminus\{e_L(v)\}$. Then, we have that $x$ is a descendant of $M(B(v)\setminus\{e_L(v)\})$, and therefore a descendant of $v'$. Furthermore, we have that $y$ is a proper ancestor of $v$, and therefore a proper ancestor of $v'$. This shows that $(x,y)\in B(v')$. Due to the generality of $(x,y)\in B(v)\setminus\{e_L(v)\}$, this shows that $B(v)\setminus\{e_L(v)\})\subseteq B(v')$. Then, since $B(v)\setminus\{e_L(v)\}\subseteq B(v')\subseteq B(v)$ and $B(v)\neq B(v')$ (since the graph is $3$-edge-connected), we have that $B(v)\setminus\{e_L(v)\}= B(v')$. This implies that $M(B(v)\setminus\{e_L(v)\})=M(v')$. Since $M(B(v')\setminus\{e_L(v')\})\neq M(v')$, we have that $M(B(v')\setminus\{e_L(v')\})$ is a proper descendant of $M(v')$. Therefore, since  $M(B(v)\setminus\{e_L(v)\})=M(v')$, we have that $M(B(v')\setminus\{e_L(v')\})$ is a proper descendant of $M(B(v)\setminus\{e_L(v)\})$.

Now let $w$ be a vertex in $W_L(v)$, and let $w'$ be a vertex in $W_L(v')$. Then we have that $M(w)=M(B(v)\setminus\{e_L(v)\})$ and $M(w')=M(B(v')\setminus\{e_L(v')\})$. Thus, we have that $M(w')$ is a proper descendant of $M(w)$, and therefore a proper descendant of $w$. Thus, since $M(w')$ is a common descendant of $w'$ and $w$, we have that $w'$ and $w$ are related as ancestor and descendant. Let us suppose, for the sake of contradiction, that $w'$ is not a proper ancestor of $w$. Then, $w'$ is a descendant of $w$. Since $M(w')$ is a proper descendant of $M(w)$, there is a back-edge $(x,y)$ in $B(w)$ such that $x$ is not a descendant of $M(w')$. Then, we have that $x$ is a descendant of $M(w)=M(B(v)\setminus\{e_L(v)\})$, and therefore a descendant of $v$, and therefore a descendant of $\mathit{high}_1(v)=\mathit{high}_1(v')$, and therefore a descendant of $v'$, and therefore a descendant of $w'$. Furthermore, we have that $y$ is a proper ancestor of $w$, and therefore a proper ancestor of $w'$. This shows that $(x,y)\in B(w')$, and therefore $x$ is a descendant of $M(w')$, a contradiction. Thus, we have that $w'$ is a proper ancestor of $w$. Due to the generality of $w'\in W_L(v')$, this implies that $\mathit{firstW_L}(v')$ is a proper ancestor of $w$. And due to the generality of $w\in W_L(v)$, this implies that $\mathit{firstW_L}(v')$ is a proper ancestor of $\mathit{lastW_L}(v)$.

Now let us suppose, for the sake of contradiction, that there is a vertex $u\in U_4^2(v)$, and $U_4^2(v')=\emptyset$. Since $u\in U_4^2(v)$, we have that $u\in S$. Thus, since $v'\in S$, we have that $u$ and $v'$ are related as ancestor and descendant. Let us suppose, for the sake of contradiction, that $u$ is not a proper descendant of $v'$. Then, we have that $u$ is an ancestor of $v'$. Since $W_L(v')\neq\emptyset$, we let $w'=\mathit{firstW_L}(v')$. Let $(x,y)$ be a back-edge in $B(w')$ such that $y=\mathit{low}(w')$. Then, $x$ is a descendant $M(w')=M(B(v')\setminus\{e_L(v')\})$, and therefore a descendant of $M(v')$, and therefore a descendant of $v'$, and therefore a descendant of $u$. Furthermore, $y$ is a proper ancestor of $w'$, and therefore a proper ancestor of $v'$. Thus, since $x$ is a descendant of $v'$, this shows that $(x,y)\in B(v')$. Therefore, $y$ is an ancestor of $\mathit{high}_1(v')=\mathit{high}_1(v)$, and therefore a proper ancestor of $v$, and therefore a proper ancestor of $u$ (since $u\in U_4^2(v)$ implies that $u$ is a proper descendant of $v$). Thus, since $x$ is a descendant of $u$, this shows that $(x,y)\in B(u)$. Since $y=\mathit{low}(w')$ and $w'=\mathit{firstW_L}(v')$, we have that $y<\mathit{firstW_L}(v')$, and therefore $y<\mathit{lastW_L}(v)$ (since $\mathit{firstW_L}(v')$ is a proper ancestor of $\mathit{lastW_L}(v)$). Since $(x,y)\in B(u)$, we have that $\mathit{low}(u)\leq y$, and therefore $\mathit{low}(u)<\mathit{lastW_L}(v)$. But this contradicts the fact that $\mathit{low}(u)\geq\mathit{lastW_L}(v)$ (which is an implication of $u\in U_4^2(v)$). Thus, our last supposition is not true, and therefore $u$ is a proper descendant of $v'$. Thus, we have the following facts: $u\in S$, $u$ is a proper descendant of $v'$, and $\mathit{low}(u)\geq\mathit{lastW_L}(v)>\mathit{firstW_L}(v')$. But this implies that $U_4^2(v')\neq\emptyset$ (because we can consider the lowest $u$ that has those properties). A contradiction. This shows that if $U_4^2(v')=\emptyset$, then $U_4^2(v)=\emptyset$.

Now let us assume that $U_4^2(v)\neq\emptyset$. This implies that $U_4^2(v')\neq\emptyset$. Let us suppose, for the sake of contradiction, that there is a vertex $u\in U_4^2(v)$ that is lower than the greatest vertex $u'$ in $U_4^2(v')$.
Since $u\in U_4^2(v)$, we have that $u\in S$. Since $u'\in U_4^2(v')$ we have that $u'\in S$. This implies that $u$ and $u'$ are related as ancestor and descendant. Thus, since $u$ is lower than $u'$, we have that $u$ is a proper ancestor of $u'$. Let us suppose, for the sake of contradiction, that $\mathit{low}(u')$ is a proper ancestor of $\mathit{firstW_L}(v')$. Then, since $\mathit{firstW_L}(v')$ is a proper ancestor of $\mathit{lastW_L}(v)$, we have that $\mathit{low}(u')$ is a proper ancestor of $\mathit{lastW_L}(v)$. Now let $(x,y)$ be a back-edge in $B(u')$ such that $y=\mathit{low}(u')$. Then $x$ is a descendant of $u'$, and therefore a descendant of $u$. Furthermore, $y$ is a proper ancestor of $\mathit{lastW_L}(v)$, and therefore a proper ancestor of $v$, and therefore a proper ancestor of $u$. This shows that $(x,y)\in B(u)$. Thus, we have $\mathit{low}(u)\leq y<\mathit{lastW_L}(v)$, in contradiction to the fact that $u\in U_4^2(v)$. Thus, our last supposition is not true, and therefore we have that $\mathit{low}(u')$ is not a proper ancestor of $\mathit{firstW_L}(v')$. 

Since $u'\in U_4^2(v')$, we have that $u'$ is a proper descendant of $v'$, and therefore a proper descendant of $\mathit{firstW_L}(v')$. Thus, $u'$ is a common descendant of $\mathit{low}(u')$ and $\mathit{firstW_L}(v')$, and therefore $\mathit{low}(u')$ and  $\mathit{firstW_L}(v')$ are related as ancestor and descendant. Thus, since $\mathit{low}(u')$ is not a proper ancestor of $\mathit{firstW_L}(v')$, we have that $\mathit{low}(u')$ is a descendant of $\mathit{firstW_L}(v')$, and therefore $\mathit{low}(u')\geq\mathit{firstW_L}(v')$. Thus, since $u'\in U_4^2(v')$, we have that $u'$ is the lowest vertex in $S$ that is a proper descendant of $v'$ such that $\mathit{low}(u')\geq\mathit{firstW_L}(v')$ $(*)$. 

Now we will trace the implications of $u\in U_4^2(v)$. First, we have that $u\in S$. Furthermore, we have that $\mathit{low}(u)\geq\mathit{lastW_L}(v)$, and therefore $\mathit{low}(u)>\mathit{firstW_L}(v')$ (since $\mathit{firstW_L}(v')$ is a proper ancestor of $\mathit{lastW_L}(v)$). Finally, we can show as above that $u$ is a proper descendant of $v'$ (the proof of this fact above did not rely on $U_4^2(v')=\emptyset$). But then, since $u$ is lower than $u'$, we have a contradiction to $(*)$.
Thus, we have shown that every vertex in $U_4^2(v)$ is at least as great as the greatest vertex in $U_4^2(v')$. In particular, this implies that the lowest vertex in $U_4^2(v)$ is greater than, or equal to, the greatest vertex in $U_4^2(v')$.
\end{proof}

Due to the similarity of the definitions of the $U_2$ and the $U_4^2$ sets, and the similarity between Lemmata~\ref{lemma:type3-b-ii-2-relation-between-u2} and \ref{lemma:type-3-b-ii-4-2-U-sets}, we can use a similar procedure as Algorithm~\ref{algorithm:type3-b-ii-2-U} in order to compute all $U_4^2$ sets in linear time. This is shown in Algorithm~\ref{algorithm:type3-b-ii-4-2-U}. Our result is summarized in Lemma~\ref{lemma:algorithm:type3-b-ii-4-2-U}.

\begin{algorithm}[H]
\caption{\textsf{Compute the sets $U_4^2(v)$, for all vertices $v\neq r$ such that $M(B(v)\setminus\{e_L(v)\})\neq M(v)$ and $W_L(v)\neq\emptyset$}}
\label{algorithm:type3-b-ii-4-2-U}
\LinesNumbered
\DontPrintSemicolon
let $\mathcal{V}$ be the collection of all vertices $v\neq r$ such that $M(B(v)\setminus\{e_L(v)\})\neq M(v)$ and $W_L(v)\neq\emptyset$\;
\ForEach{vertex $x$}{
  compute the collection $\mathcal{S}(x)$ of the segments of $H(x)$ that are maximal w.r.t. the property that their elements
  are related as ancestor and descendant\;
}
\ForEach{$v\in\mathcal{V}$}{
  set $U_4^2(v)\leftarrow\emptyset$\;
}
\ForEach{vertex $x$}{
  \ForEach{segment $S\in\mathcal{S}(x)$}{
    let $v$ be the first vertex in $S$\;
    \While{$v\neq\bot$ \textbf{and} $v\notin\mathcal{V}$}{
      $v\leftarrow\mathit{next}_S(v)$\;
    }
    \lIf{$v=\bot$}{\textbf{continue}}
    let $u=\mathit{prev}_S(v)$\;
    \While{$v\neq\bot$}{
      \While{$u\neq\bot$ \textbf{and} $\mathit{low}(u)<\mathit{lastW_L}(v)$}{
        $u\leftarrow\mathit{prev_S}(u)$\;
      }
      \While{$u\neq\bot$ \textbf{and} $\mathit{low}(u)<\mathit{firstW_L}(v)$}{
        insert $u$ into $U_4^2(v)$\;
        $u\leftarrow\mathit{prev_S}(u)$\;
      }
      \If{$u\neq\bot$}{
        insert $u$ into $U_4^2(v)$\;
      }
      $v\leftarrow\mathit{next}_S(v)$\;
      \While{$v\neq\bot$ \textbf{and} $v\notin\mathcal{V}$}{
        $v\leftarrow\mathit{next}_S(v)$\;
      }
    }
  }
}
\end{algorithm}

\begin{lemma}
\label{lemma:algorithm:type3-b-ii-4-2-U}
Let $\mathcal{V}$ be the collection of all vertices $v\neq r$ such that $M(B(v)\setminus\{e_L(v)\})\neq M(v)$ and $W_L(v)\neq\emptyset$, and suppose that the vertices $\mathit{firstW_L}(v)$ and $\mathit{lastW_L}(v)$ are computed for every $v\in\mathcal{V}$. Then, Algorithm~\ref{algorithm:type3-b-ii-4-2-U} correctly computes the sets $U_4^2(v)$, for all vertices $v\in\mathcal{V}$, in total linear time.
\end{lemma}
\begin{proof}
The proof is the same as that of Lemma~\ref{lemma:algorithm:type3-b-ii-2-U} (which is given in the main text, in the two paragraphs right above Algorithm~\ref{algorithm:type3-b-ii-2-U}). Notice that the difference between Algorithm~\ref{algorithm:type3-b-ii-2-U} and Algorithm~\ref{algorithm:type3-b-ii-4-2-U} is that the second algorithm has a different set $\mathcal{V}$ of vertices for which the $U_4^2$ sets are defined, and the occurrences of ``$\mathit{firstW}$" and ``$\mathit{lastW}$" are replaced with ``$\mathit{firstW_L}$" and ``$\mathit{lastW_L}$", respectively. Now we can use the argument of Lemma~\ref{lemma:algorithm:type3-b-ii-2-U}, by just replacing the references to Lemma~\ref{lemma:type3-b-ii-2-relation-between-u2} with references to Lemma~\ref{lemma:type-3-b-ii-4-2-U-sets}.
\end{proof}

\begin{lemma}
\label{lemma:u42_elu}
Let $v\neq r$ be a vertex such that $M(B(v)\setminus\{e_L(v)\})\neq M(v)$, and let $u$ be a proper descendant of $v$ with $\mathit{high}(u)=\mathit{high}(v)$. Then, $e_L(v)\notin B(u)$.
\end{lemma}
\begin{proof}
Since $u$ is a proper descendant of $v$ with $\mathit{high}(u)=\mathit{high}(v)$, Lemma~\ref{lemma:same_high} implies that $B(u)\subseteq B(v)$. This implies that $M(u)$ is a descendant of $M(v)$. Thus, since $M(u)$ is a common descendant of $u$ and $M(v)$, we have that $u$ and $M(v)$ are related as ancestor and descendant.

Let us suppose, for the sake of contradiction, that $u$ is not a proper descendant of $M(v)$. Thus, we have that $u$ is an ancestor of $M(v)$. Let $(x,y)$ be a back-edge in $B(v)$. Then, $x$ is a descendant of $M(v)$, and therefore a descendant of $u$. Furthermore, $y$ is a proper ancestor of $v$, and therefore a proper ancestor of $u$. This shows that $(x,y)\in B(u)$. Due to the generality of $(x,y)\in B(v)$, this implies that $B(v)\subseteq B(u)$. Thus, $B(u)\subseteq B(v)$ implies that $B(u)=B(v)$, in contradiction to the fact that the graph is $3$-edge-connected. This shows that $u$ is a proper descendant of $M(v)$.

Now let us suppose, for the sake of contradiction, that $u$ is an ancestor of $L_1(v)$ (i.e., the higher endpoint of $e_L(v)$). Then, since $u$ is a proper descendant of $M(v)$, we cannot have $M(v)=L_1(v)$. Thus, $L_1(v)$ is a proper descendant of $M(v)$. So let $c$ be the child of $M(v)$ that is an ancestor of $L_1(v)$. Then, since $M(B(v)\setminus\{e_L(v)\})\neq M(v)$, we have that $e_L(v)$ is the unique back-edge in $B(v)$ whose higher endpoint is a descendant of $c$. Now, since $u$ is an ancestor of $L_1(v)$ and a proper descendant of $M(v)$, we have that $u$ is also a descendant of $c$. Since the graph is $3$-edge-connected, we have that $|B(u)|>1$. Thus, there is a back-edge $(x,y)\in B(u)\setminus\{e_L(v)\}$. Then, $x$ is a descendant of $u$, and therefore a descendant of $v$. Furthermore, $y$ is an ancestor of $\mathit{high}(u)=\mathit{high}(v)$, and therefore a proper ancestor of $v$. This shows that $(x,y)\in B(v)$. Since $(x,y)\neq e_L(v)$, this can be strengthened to $(x,y)\in B(v)\setminus\{e_L(v)\}$. Thus, we have that in $B(v)\setminus\{e_L(v)\}$ there is still a back-edge whose higher endpoint is a descendant of $c$ (i.e., $(x,y)$), which is impossible. We conclude that $u$ is not an ancestor of $L_1(v)$, and therefore $e_L(v)\notin B(u)$. 
\end{proof}

\begin{lemma}
\label{lemma:type3-b-ii-4-2-in-U}
Let $(u,v,w)$ be a triple of vertices that induces a Type-3$\beta$ii-$4$ $4$-cut, with back-edge $e_L(v)$, such that $M(B(v)\setminus\{e_L(v)\})\neq M(v)$ and $\mathit{high}_1(v)=\mathit{high}(u)$. Then $u\in U_4^2(v)$. 
\end{lemma}
\begin{proof}
Since $(u,v,w)$ induces a $4$-cut, we have that $u$ is a proper descendant of $v$. Let $u'$ be a vertex such that $u\geq u'\geq v$ and $\mathit{high}(u')=\mathit{high}(u)$. Since $\mathit{high}_1(v)=\mathit{high}(u)$, this implies that $\mathit{high}(u')=\mathit{high}_1(v)$, and therefore $u'\in H(\mathit{high}_1(v))$. Then, since $(u,v,w)$ induces a Type-3$\beta$ii-$4$ $4$-cut with back-edge $e_L(v)$ such that $M(B(v)\setminus\{e_L(v)\})\neq M(v)$, by Lemma~\ref{lemma:type3-b-ii-4-seg-2} we have that $u'$ is an ancestor of $u$. This shows that $u$ and $v$ belong to a segment $S$ of $H(\mathit{high}_1(v))$ that is maximal w.r.t. the property that its elements are related as ancestor and descendant. 

Since $(u,v,w)$ induces a Type-3$\beta$ii-$4$ $4$-cut with back-edge $e_L(v)$, we have that $w$ is a proper ancestor of $v$ with $M(w)=M(B(v)\setminus\{e_L(v)\})$. Thus, since $M(B(v)\setminus\{e_L(v)\})\neq M(v)$, we have that $w\in W_L(v)$, and therefore $w\geq\mathit{lastW_L}(v)$. By Lemma~\ref{lemma:type3-b-ii-4-info}, we have that $w\leq\mathit{low}(u)$, and therefore $\mathit{lastW_L}(v)\leq\mathit{low}(u)$. Thus, if $\mathit{low}(u)<\mathit{firstW_L}(v)$, then $u$ satisfies enough conditions to be in $U_4^2(v)$. So let us assume that $\mathit{low}(u)\geq\mathit{firstW_L}(v)$. 

Let us suppose, for the sake of contradiction, that $u$ is not the lowest vertex in $S$ that is a proper descendant of $v$ such that $\mathit{low}(u)\geq\mathit{firstW_L}(v)$. Then, there is a vertex $u'$ in $S$ that is lower than $u$, it is a proper descendant of $v$, and has $\mathit{low}(u')\geq\mathit{firstW_L}(v)$. Since both $u$ and $u'$ are in $S$, they are related as ancestor and descendant. Thus, since $u'$ is lower than $u$, we have that $u'$ is a proper ancestor of $u$. Since both $u$ and $u'$ are in $S$, we have that $\mathit{high}(u)=\mathit{high}(u')$. Thus, Lemma~\ref{lemma:same_high} implies that $B(u)\subseteq B(u')$. Now let $(x,y)$ be a back-edge in $B(u')$. Then $x$ is a descendant of $u'$, and therefore a descendant of $v$. Furthermore, $y$ is an ancestor of $\mathit{high}(u')$, and therefore an ancestor of $\mathit{high}(v)$ (because $u'\in S$, and therefore $\mathit{high}(u')=\mathit{high}(v)$), and therefore a proper ancestor of $v$. This shows that $(x,y)\in B(v)$. Since $(u,v,w)$ induces a Type-3$\beta$ii-$4$ $4$-cut with back-edge $e_L(v)$, we have that $B(v)=(B(u)\sqcup B(w))\sqcup\{e_L(v)\}$. Thus, $(x,y)\in B(v)$ implies that either $(x,y)\in B(u)$, or $(x,y)\in B(w)$, or $(x,y)=e_L(v)$. Since $y\geq\mathit{low}(u')\geq\mathit{firstW_L}(v)\geq w$, we have that $y$ cannot be a proper ancestor of $w$, and so the case $(x,y)\in B(w)$ is rejected. Then, since $u'$ is a proper descendant of $v$ with $\mathit{high}(u')=\mathit{high}(v)$, Lemma~\ref{lemma:u42_elu} implies that $e_L(v)\notin B(u')$. Therefore $(x,y)\neq e_L(v)$. Thus, since $(x,y)\in B(w)$ is rejected, the only viable option is $(x,y)\in B(u)$. Due to the generality of $(x,y)\in B(u')$, this implies that $B(u')\subseteq B(u)$. Thus, since $B(u)\subseteq B(u')$, we have that $B(u')=B(u)$, in contradiction to the fact that the graph is $3$-edge-connected. Thus, we have that $u$ is the lowest vertex in $S$ that is a proper descendant of $v$ such that $\mathit{low}(u)\geq\mathit{firstW_L}(v)$. This means that $u$ satisfies enough conditions to be in $U_4^2(v)$.
\end{proof}

\begin{lemma}
\label{lemma:type3-b-ii-4-2-criterion}
Let $(u,v,w)$ be a triple of vertices such that $e_L(v)\notin B(u)$, $M(B(v)\setminus\{e_L(v)\})\neq M(v)$ and $u\in U_4^2(v)$. Then, $(u,v,w)$ induces a Type-3$\beta$ii-$4$ $4$-cut with back-edge $e_L(v)$ if and only if: $\mathit{bcount}(v)=\mathit{bcount}(u)+\mathit{bcount}(w)+1$, and $w$ is the greatest proper ancestor of $v$ with $M(w)=M(B(v)\setminus\{e_L(v)\})$ such that $w\leq\mathit{low}(u)$.
\end{lemma}
\begin{proof}
($\Rightarrow$) Since $(u,v,w)$ induces a Type-3$\beta$ii-$4$ $4$-cut with back-edge $e_L(v)$, we have that $B(v)=(B(u)\sqcup B(w))\sqcup\{e_L(v)\}$. Thus, we get $\mathit{bcount}(v)=\mathit{bcount}(u)+\mathit{bcount}(w)+1$. Furthermore, we have $M(w)=M(B(v)\setminus\{e_L(v)\})$ (since $e_L(v)$ is the back-edge of the $4$-cut induced by $(u,v,w)$). By Lemma~\ref{lemma:type3-b-ii-4-info}, we have $w\leq\mathit{low}(u)$.

Now let us suppose, for the sake of contradiction, that there is a proper ancestor $w'$ of $v$ with $M(w')=M(B(v)\setminus\{e_L(v)\})$ and $w'>w$, such that $w'\leq\mathit{low}(u)$. Then, since $M(w)=M(B(v)\setminus\{e_L(v)\})$ and $M(w')=M(B(v)\setminus\{e_L(v)\})$, we have that $M(B(v)\setminus\{e_L(v)\})$ is a common descendant of $w$ and $w'$, and therefore $w$ and $w'$ are related as ancestor and descendant. Thus, $w'>w$ implies that $w'$ is a proper descendant of $w$.
Since $w'$ is a proper descendant of $w$ with $M(w')=M(w)$, Lemma~\ref{lemma:same_m_subset_B} implies that $B(w)\subseteq B(w')$. This can be strengthened to $B(w)\subset B(w')$, since the graph is $3$-edge-connected. Thus, there is a back-edge $(x,y)\in B(w')\setminus B(w)$. Then, $x$ is a descendant of $M(w')$, and therefore a descendant of $M(B(v)\setminus\{e_L(v)\})$, and therefore a descendant of $v$. Furthermore, $y$ is a proper ancestor of $w'$, and therefore a proper ancestor of $v$. This shows that $(x,y)\in B(v)$. Since $B(v)=(B(u)\sqcup B(w))\sqcup\{e_L(v)\}$, this implies that either $(x,y)\in B(u)$, or $(x,y)\in B(w)$, or $(x,y)=e_L(v)$. The case $(x,y)\in B(u)$ is rejected, since $y<w'$ and $w'\leq\mathit{low}(u)$, and therefore $y<\mathit{low}(u)$. The case $(x,y)=e_L(v)$ is also rejected, because $M(B(v)\setminus\{e_L(v)\})\neq M(v)$, and therefore the higher endpoint of $e_L(v)$ is not a descendant of $M(B(v)\setminus\{e_L(v)\})$ ($=M(w')$). Thus, we have that $(x,y)\in B(w)$, a contradiction. This shows that $w$ is the greatest proper ancestor of $v$ with $M(w)=M(B(v)\setminus\{e_L(v)\})$ such that $w\leq\mathit{low}(u)$.

($\Leftarrow$) We have to show that $B(v)=(B(u)\sqcup B(w))\sqcup\{e_L(v)\}$, and $M(w)=M(B(v)\setminus\{e_L(v)\})$.
Since $w\leq\mathit{low}(u)$, we have that $B(u)\cap B(w)=\emptyset$ (because, if there existed a back-edge in $B(u)\cap B(w)$, its lower endpoint would be lower than $\mathit{low}(u)$, which is impossible). 

Let $(x,y)$ be a back-edge in $B(u)$. Since $u\in U_4^2(v)$, we have that $u$ is a proper descendant of $v$ with $\mathit{high}(u)=\mathit{high}_1(v)$. Thus, $(x,y)\in B(u)$ implies that $x$ is a descendant of $v$. Furthermore, we have that $y$ is an ancestor of $\mathit{high}(u)=\mathit{high}_1(v)$, and therefore a proper ancestor of $v$. This shows that $(x,y)\in B(v)$. Due to the generality of $(x,y)\in B(u)$, this implies that $B(u)\subseteq B(v)$.

Let $(x,y)$ be a back-edge in $B(w)$. Then, $x$ is a descendant of $M(w)=M(B(v)\setminus\{e_L(v)\})$, and therefore a descendant of $v$. Furthermore, $y$ is a proper ancestor of $w$, and therefore a proper ancestor of $v$. This shows that $(x,y)\in B(v)$. Due to the generality of $(x,y)\in B(w)$, this implies that $B(w)\subseteq B(v)$.

Since $M(B(v)\setminus\{e_L(v)\})\neq M(v)$, we have that the higher endpoint of $e_L(v)$ is not a descendant of $M(B(v)\setminus\{e_L(v)\})$. Since  $M(B(v)\setminus\{e_L(v)\})= M(w)$, this implies that $e_L(v)\notin B(w)$. By assumption, we have $e_L(v)\notin B(u)$. Therefore, we have $e_L(v)\notin B(u)\cup B(w)$.

Thus, we have $B(u)\subseteq B(v)$, $B(w)\subseteq B(v)$, and $B(u)\cap B(w)=\emptyset$. This implies that $B(u)\sqcup B(v)\subseteq B(v)$. Therefore, since $\mathit{bcount}(v)=\mathit{bcount}(u)+\mathit{bcount}(w)+1$, we have that there is a back-edge $e$, such that $B(v)=(B(u)\sqcup B(w))\sqcup\{e\}$. Since $e_L(v)\notin B(u)\cup B(w)$, this implies that $e=e_L(v)$. By assumption, we have $M(w)=M(B(v)\setminus\{e_L(v)\})$.

\end{proof}

\begin{algorithm}[H]
\caption{\textsf{Compute all Type-3$\beta$ii-$4$ $4$-cuts of the form $\{(u,p(u)),(v,p(v)),(w,p(w)),e_L(v)\}$, where $u$ is a proper descendant of $v$, $v$ is a proper descendant of $w$, $M(B(v)\setminus\{e_L(v)\})\neq M(v)$ and $\mathit{high}_1(v)=\mathit{high}(u)$}}
\label{algorithm:type3-b-ii-4-2}
\LinesNumbered
\DontPrintSemicolon
let $\mathcal{V}$ be the collection of all vertices $v\neq r$ such that $M(B(v)\setminus\{e_L(v)\})\neq M(v)$
and $W_L(v)\neq\emptyset$\;
\label{line:type4-b-ii-4-2-V}
\ForEach{$v\in\mathcal{V}$}{
  compute $U_4^2(v)$\;
  \label{line:type4-b-ii-4-2-U}
}
\ForEach{$v\in\mathcal{V}$}{
\label{line:type4-b-ii-4-2-main-for}
  \ForEach{$u\in U_4^2(v)$}{
    let $w$ be the greatest proper ancestor of $v$ with $M(w)=M(B(v)\setminus\{e_L(v)\})$ such that $w\leq\mathit{low}(u)$\;
    \label{line:type4-b-ii-4-2-w}
    \If{$\mathit{bcount}(v)=\mathit{bcount}(u)+\mathit{bcount}(w)+1$ \textbf{and} $e_L(v)\notin B(u)$}{
      mark $\{(u,p(u)),(v,p(v)),(w,p(w)),e_L(v)\}$ as a Type-3$\beta$ii-$4$ $4$-cut\; 
      \label{line:type4-b-ii-4-2-mark}
    }
  }
}
\end{algorithm}

\begin{proposition}
\label{proposition:algorithm:type3-b-ii-4-2}
Algorithm~\ref{algorithm:type3-b-ii-4-2} correctly computes all Type-3$\beta$ii-$4$ $4$-cuts of the form $\{(u,p(u)),(v,p(v)),(w,p(w)),e_L(v)\}$, where $u$ is a descendant of $v$, $v$ is a descendant of $w$, $M(B(v)\setminus\{e_L(v)\})\neq M(v)$ and $\mathit{high}_1(v)=\mathit{high}(u)$. Furthermore, it has a linear-time implementation.
\end{proposition}
\begin{proof}
Let $(u,v,w)$ be a triple of vertices that induces a Type-3$\beta$ii-$4$ $4$-cut with back-edge $e_L(v)$ such that $M(B(v)\setminus\{e_L(v)\})\neq M(v)$ and $\mathit{high}_1(v)=\mathit{high}(u)$. By the definition of Type-3$\beta$ii-$4$ $4$-cuts, this implies that $e_L(v)\notin B(u)$. Then, by Lemma~\ref{lemma:type3-b-ii-4-2-in-U} we have that $u\in U_4^2(v)$. Then, by Lemma~\ref{lemma:type3-b-ii-4-2-criterion} we have that $w$ is the greatest proper ancestor of $v$ with $M(w)=M(B(v)\setminus\{e_L(v)\})$ such that $w\leq\mathit{low}(u)$. This implies that $w\in W_L(v)$. Thus, since $M(B(v)\setminus\{e_L(v)\})\neq M(v)$ and $W_L(v)\neq\emptyset$, we have that $v$ is in the collection $\mathcal{V}$, computed in Line~\ref{line:type4-b-ii-4-2-V}. Then, notice that $\{(u,p(u)),(v,p(v)),(w,p(w)),e_L(v)\}$ (i.e., the $4$-cut induced by $(u,v,w)$) will be correctly marked by Algorithm~\ref{algorithm:type3-b-ii-4-2} in Line~\ref{line:type4-b-ii-4-2-mark}.

Conversely, suppose that a $4$-element set $\{(u,p(u)),(v,p(v)),(w,p(w)),e_L(v)\}$ is marked by Algorithm~\ref{algorithm:type3-b-ii-4-2} in Line~\ref{line:type4-b-ii-4-2-mark}. Then we have that: $(1)$ $v$ is a vertex such that $M(B(v)\setminus\{e_L(v)\})\neq M(v)$ (due to $v\in\mathcal{V}$), $(2)$ $u\in U_4^2(v)$, $(3)$ $w$ is the greatest proper ancestor of $v$ such that $M(w)=M(B(v)\setminus\{e_L(v)\})$ and $w\leq\mathit{low}(u)$, and $(4)$ $\mathit{bcount}(v)=\mathit{bcount}(u)+\mathit{bcount}(w)+1$ and $e_L(v)\notin B(u)$. Thus, Lemma~\ref{lemma:type3-b-ii-4-2-criterion} implies that $\{(u,p(u)),(v,p(v)),(w,p(w)),e_L(v)\}$ is a Type-3$\beta$ii-$4$ $4$-cut. Since $u\in U_4^2(v)$, we have that $\mathit{high}_1(v)=\mathit{high}(u)$. 

Now we will show that Algorithm~\ref{algorithm:type3-b-ii-4-2} has a linear-time implementation. First, by Proposition~\ref{proposition:computing-M(B(v)-S)}, we can compute $M(B(v)\setminus\{e_L(v)\})$ for all vertices $v\neq r$, in total linear time. By Lemma~\ref{lemma:compute_wl}, we have that the values $\mathit{firstW_L}(v)$ and $\mathit{lastW_L}(v)$ can be computed in linear time in total, for all vertices $v\neq r$. Then, we can check in constant time whether $W_L(v)\neq\emptyset$, for a vertex $v\neq r$, by simply checking whether e.g. $\mathit{firstW_L}(v)\neq\bot$. Thus, the set $\mathcal{V}$ in Line~\ref{line:type4-b-ii-4-2-V} can be computed in linear time. By Lemma~\ref{lemma:algorithm:type3-b-ii-4-2-U}, we have that the sets $U_4^2(v)$ can be computed in total linear time, for all vertices $v\in\mathcal{V}$, using Algorithm~\ref{algorithm:type3-b-ii-4-2-U}. Thus, the \textbf{for} loop in Line~\ref{line:type4-b-ii-4-2-U} can be performed in linear time. In particular, we have that the total size of all $U_4^2$ sets is $O(n)$.

It remains to show how to find the vertex $w$ in Line~\ref{line:type4-b-ii-4-2-w}. To do this, we use Algorithm~\ref{algorithm:W-queries}. Specifically, let $v$ be a vertex in $\mathcal{V}$, let $u$ be a vertex in $U_4^2(v)$, and let $x=M(B(v)\setminus\{e_L(v)\})$. Then we generate a query $q(M^{-1}(x),\mathit{min}\{\mathit{low}(u),p(v)\})$. This is to return the greatest vertex $w$ such that $M(w)=M(B(v)\setminus\{e_L(v)\})$ and $w\leq\mathit{low}(u)$ and $w\leq p(v)$. Since $M(w)=M(B(v)\setminus\{e_L(v)\})$, we have that $M(w)$ is a descendant of $M(v)$, and therefore a descendant of $v$. Thus, $M(w)$ is a common descendant of $v$ and $w$, and therefore $v$ and $w$ are related as ancestor and descendant. Thus, $w\leq p(v)$ implies that $w$ is a proper ancestor of $v$, and so $w$ is the greatest proper ancestor of $v$ such that $M(w)=M(B(v)\setminus\{e_L(v)\})$ and $w\leq\mathit{low}(u)$. Now, using Algorithm~\ref{algorithm:W-queries} we can answer all those queries in total linear time, according to Lemma~\ref{lemma:W-queries}. Thus, the \textbf{for} loop in Line~\ref{line:type4-b-ii-4-2-main-for} can be performed in linear time. We conclude that Algorithm~\ref{algorithm:type3-b-ii-4-2} runs in linear time.
\end{proof}

\noindent\\
\textbf{The case where $M(B(v)\setminus\{e\})= M(v)$ and $\mathit{high}_1(v)>\mathit{high}(u)$}\\

Let $(u,v,w)$ be a triple of vertices that induces a Type-3$\beta$ii-$4$ $4$-cut, such that $M(w)=M(B(v)\setminus\{e\})= M(v)$ and $\mathit{high}_1(v)>\mathit{high}(u)$, where $e$ is the back-edge in the $4$-cut induced by $(u,v,w)$. Then, by Lemma~\ref{lemma:type3-b-ii-4-info} we have that $\mathit{high}_1(v)\neq\mathit{high}_2(v)=\mathit{high}(u)$, and $e=e_\mathit{high}(v)$. Furthermore, by Lemma~\ref{lemma:type3-b-ii-4-info} we have that $w\leq\mathit{low}(u)$. Since $\mathit{high}_1(v)\neq\mathit{high}_2(v)$, we have that $\mathit{high}_2(v)<\mathit{high}_1(v)$. Thus, since $\mathit{low}(u)\leq\mathit{high}(u)=\mathit{high}_2(v)<\mathit{high}_1(v)$, we have $w\leq\mathit{high}_2(v)<\mathit{high}_1(v)$. Since $M(w)=M(v)$, this implies that $w$ is an ancestor of $\mathit{high}_2(v)$, and a proper ancestor of $\mathit{high}_1(v)$.

Now let $v\neq r$ be a vertex such that $\mathit{high}_1(v)\neq\mathit{high}_2(v)$ and $\mathit{lastM}(v)\leq\mathit{high}_2(v)$. We let $\widetilde{\mathit{nextM}}(v)$ denote the greatest proper ancestor $w$ of $\mathit{high}_1(v)$ such that $M(w)=M(v)$. 

\begin{lemma}
\label{lemma:tilde-nextM}
Let $v\neq r$ be a vertex such that $\mathit{high}_1(v)\neq\mathit{high}_2(v)$ and $\mathit{lastM}(v)\leq\mathit{high}_2(v)$. Then $\widetilde{\mathit{nextM}}(v)$ is either $\mathit{nextM}(v)$ or $\mathit{nextM}(\mathit{nextM}(v))$.
\end{lemma}
\begin{proof}
Since $\mathit{high}_1(v)\neq\mathit{high}_2(v)$, we have that $\mathit{high}_2(v)<\mathit{high}_1(v)$. Thus, since $\mathit{high}_1(v)$ and $\mathit{high}_2(v)$ are both ancestors of $v$, we have that $\mathit{high}_2(v)$ is a proper ancestor of $\mathit{high}_1(v)$. Since $\mathit{lastM}(v)$ and $v$ have the same $M$ point, we have that $\mathit{lastM}(v)$ is related as ancestor and descendant with $v$. Due to the minimality of $\mathit{lastM}(v)$ in $M^{-1}(M(v))$, this implies that $\mathit{lastM}(v)$ is an ancestor of $v$. Thus, since $v$ is a common descendant of $\mathit{high}_2(v)$ and $\mathit{lastM}(v)$, we have that $\mathit{high}_2(v)$ and $\mathit{lastM}(v)$ are related as ancestor and descendant. Thus, $\mathit{lastM}(v)\leq\mathit{high}_2(v)$ implies that $\mathit{lastM}(v)$ is an ancestor of $\mathit{high}_2(v)$. Therefore, $\mathit{lastM}(v)$ is a proper ancestor of $\mathit{high}_1(v)$. This shows that $\widetilde{\mathit{nextM}}(v)\neq\bot$.

Now suppose that $\widetilde{\mathit{nextM}}(v)\neq\mathit{nextM}(v)$. By Lemma~\ref{lemma:same_M_dif_B_lower} we have that $\mathit{nextM}(v)$ is an ancestor of $\mathit{high}_1(v)$. Thus, $\widetilde{\mathit{nextM}}(v)\neq\mathit{nextM}(v)$ implies that $\mathit{nextM}(v)=\mathit{high}_1(v)$. By definition, $\mathit{nextM}(\mathit{nextM}(v))$ is the greatest proper ancestor of $\mathit{nextM}(v)$ that has the same $M$ point with $v$. Thus, since $\mathit{nextM}(v)=\mathit{high}_1(v)$, we have that $\mathit{nextM}(\mathit{nextM}(v))$ is the greatest proper ancestor of $\mathit{high}_1(v)$ that has the same $M$ point as $v$. In other words, $\mathit{nextM}(\mathit{nextM}(v))=\widetilde{\mathit{nextM}}(v)$.
\end{proof}

Now let $S$ be the segment of $\widetilde{H}(\mathit{high}_2(v))$ that contains $v$ and is maximal w.r.t. the property that all its elements are related as ancestor and descendant (i.e., we have $S=\widetilde{S}_2(v)$). Then, we let $U_4^3(v)$ denote the collection of all vertices $u\in S$ such that: $(1)$ $u$ is a proper descendant of $v$ with $\mathit{high}(u)=\mathit{high}_2(v)$, $(2)$ $\mathit{low}(u)\geq\mathit{lastM}(v)$, and $(3)$ either $\mathit{low}(u)<\widetilde{\mathit{nextM}}(v)$, or $u$ is the lowest vertex in $S$ that satisfies $(1)$, $(2)$ and $\mathit{low}(u)\geq\widetilde{\mathit{nextM}}(v)$.

\begin{lemma}
\label{lemma:type-3bii-4-3-aux}
Let $v$ and $v'$ be two vertices $\neq r$ such that $v'$ is a proper descendant of $v$ with $\mathit{high}_1(v)\neq \mathit{high}_2(v)=\mathit{high}_2(v')\neq\mathit{high}_1(v')$. Let $w$ and $w'$ be two vertices such that $M(w)=M(v)$, $M(w')=M(v')$, and both $w$ and $w'$ are ancestors of $\mathit{high}_2(v)=\mathit{high}_2(v')$. Then $w'$ is a proper ancestor of $w$.
\end{lemma}
\begin{proof}
Let $(x,y)$ be a back-edge in $B(v')$ such that $y=\mathit{high}_2(v')$. Then $x$ is a descendant of $v'$, and therefore a descendant of $v$. Furthermore, $y=\mathit{high}_2(v')=\mathit{high}_2(v)$ is a proper ancestor of $v$. This shows that $(x,y)\in B(v)$, and therefore $x$ is a descendant of $M(v)$. Thus, since $x$ is a common descendant of $M(v)$ and $v'$, we have that $M(v)$ and $v'$ are related as ancestor and descendant.

Let us suppose, for the sake of contradiction, that $M(v)$ is not a proper ancestor of $v'$. Then we have that $M(v)$ is a descendant of $v'$. We have that $\mathit{high}_1(v')$ and $v$ are related as ancestor and descendant, since both have $v'$ as a common descendant. Let us suppose, for the sake of contradiction, that $\mathit{high}_1(v')$ is not a proper ancestor of $v$. Then we have that $\mathit{high}_1(v')$ is a descendant of $v$. Now consider a back-edge $(x,y)\in B(v)$ such that $y=\mathit{high}_1(v)$. Then we have that $x$ is a descendant of $M(v)$, and therefore a descendant of $v'$. Furthermore, $y$ is a proper ancestor of $v$, and therefore a proper ancestor of $v'$. This shows that $(x,y)\in B(v')$. Since $\mathit{high}_1(v)$ is a proper ancestor of $v$ and $\mathit{high}_1(v')$ is a descendant of $v$, we have that $\mathit{high}_1(v')$ is a proper descendant of $\mathit{high}_1(v)$, and therefore $\mathit{high}_1(v')>\mathit{high}_1(v)$. Since $(x,y)\in B(v')$, this implies that $\mathit{high}_2(v')\geq \mathit{high}_1(v)$. Since $\mathit{high}_2(v')=\mathit{high}_2(v)$, this implies that $\mathit{high}_2(v)\geq\mathit{high}_1(v)$, and therefore $\mathit{high}_2(v)=\mathit{high}_1(v)$. But this contradicts $\mathit{high}_2(v)\neq\mathit{high}_1(v)$. This shows that our last supposition cannot be true, and therefore we have that $\mathit{high}_1(v')$ is a proper ancestor of $v$. 

Now let $(x,y)$ be a back-edge in $B(v')$. Then $x$ is a descendant of $v'$, and therefore a descendant of $v$. Furthermore, $y$ is an ancestor of $\mathit{high}_1(v')$, and therefore a proper ancestor of $v$. This shows that $(x,y)\in B(v)$. Due to the generality of $(x,y)\in B(v')$, this implies that $B(v')\subseteq B(v)$. Conversely, let $(x,y)$ be a back-edge in $B(v)$. Then $x$ is a descendant of $M(v)$, and therefore a descendant of $v'$. Furthermore, $y$ is a proper ancestor of $v$, and therefore a proper ancestor of $v'$. This shows that $(x,y)\in B(v')$. Due to the generality of $(x,y)\in B(v)$, this implies that $B(v)\subseteq B(v')$. Thus we have $B(v')=B(v)$, in contradiction to the fact that the graph is $3$-edge-connected. This shows that $M(v)$ is a proper ancestor of $v'$.

Now let $w$ and $w'$ be two vertices such that $M(w)=M(v)$, $M(w')=M(v')$, and both $w$ and $w'$ are ancestors of $\mathit{high}_2(v)=\mathit{high}_2(v')$. Then $w$ and $w'$ are related as ancestor and descendant. Let us suppose, for the sake of contradiction, that $w'$ is not a proper ancestor of $w$. Then we have that $w'$ is a descendant of $w$. Since $M(v)=M(w)$ is a proper ancestor of $v'$, there is a back-edge $(x,y)\in B(w)$ such that $x$ is not a descendant of $v'$. Then, $x$ is a descendant of $M(w)=M(v)$, and therefore a descendant of $v$, and therefore a descendnant of $\mathit{high}_2(v)$, and therefore a descendant of $w'$. Furthermore, $y$ is a proper ancestor of $w$, and therefore a proper ancestor of $w'$. This shows that $(x,y)\in B(w')$. But we have that $M(w')=M(v')$, and therefore $x$ must be a descendant of $M(v')$, and therefore a descendant of $v'$. This contradicts the fact that $x$ is not a descendant of $v'$. Thus, we have that $w'$ is a proper ancestor of $w$.
\end{proof}

\begin{lemma}
\label{lemma:type-3-b-ii-4-3-U-sets}
Let $v$ and $v'$ be two vertices $\neq r$ such that $\mathit{high}_1(v)\neq \mathit{high}_2(v)=\mathit{high}_2(v')\neq\mathit{high}_1(v')$, $\mathit{lastM}(v)\leq\mathit{high}_2(v)$, $\mathit{lastM}(v')\leq\mathit{high}_2(v')$, $v'$ is a proper descendant of $v$, and both $v$ and $v'$ belong to the same segment $\widetilde{S}$ of $\widetilde{H}(\mathit{high}_2(v))$ that is maximal w.r.t. the property that all its elements are related as ancestor and descendant (i.e., we have $\widetilde{S}=\widetilde{S}(v)=\widetilde{S}(v')$). If $U_4^3(v')=\emptyset$, then $U_4^3(v)=\emptyset$. If $U_4^3(v)\neq\emptyset$, then the lowest vertex in $U_4^3(v)$ is at least as great as the greatest vertex in $U_4^3(v')$.
\end{lemma}
\begin{proof}
Notice that, since $\mathit{lastM}(v)\leq\mathit{high}_2(v)$ and $\mathit{lastM}(v')\leq\mathit{high}_2(v')$, we have that both $\widetilde{\mathit{nextM}}(v)$ and $\widetilde{\mathit{nextM}}(v')$ are well-defined. Furthermore, Lemma~\ref{lemma:type-3bii-4-3-aux} implies that $\widetilde{\mathit{nextM}}(v')$ is a proper ancestor of $\mathit{lastM}(v)$.

Let us suppose, for the sake of contradiction, that $U_4^3(v')=\emptyset$ and $U_4^3(v)\neq\emptyset$. Let $u$ be a vertex in $U_4^3(v)$. Let us suppose, for the sake of contradiction, that $u$ is not a proper descendant of $v'$. Since $u\in U_4^3(v)$, we have that $u\in\widetilde{S}$. Thus, since $v'\in\widetilde{S}$, we have that $u$ and $v'$ are related as ancestor and descendant. Since $u$ is not a proper descendant of $v'$, we have that $u$ is an ancestor of $v'$. Let $(x,y)$ be a back-edge in $B(v')$ such that $y=\mathit{low}(v')$. Lemma~\ref{lemma:same_m_same_low} implies that $\mathit{low}(v')$ is a proper ancestor of $\widetilde{\mathit{nextM}}(v')$. Thus, since $\widetilde{\mathit{nextM}}(v')$ is a proper ancestor of $\mathit{lastM}(v)$, we have that $\mathit{low}(v')$ is a proper ancestor of $\mathit{lastM}(v)$. Now, since $(x,y)\in B(v')$, we have that $x$ is a descendant of $v'$, and therefore a descendant of $u$. Furthermore, $y=\mathit{low}(v')$ is a proper ancestor of $\mathit{lastM}(v)$, and therefore a proper ancestor of $v$, and therefore a proper ancestor of $u$. This shows that $(x,y)\in B(u)$. But then we have that $\mathit{low}(u)\leq y=\mathit{low}(v')<\mathit{lastM}(v)$, in contradiction to the fact that $u\in U_4^3(v)$. Thus, our last supposition is not true, and therefore $u$ is a proper descendant of $v'$. Then, since $u\in U_4^3(v)$, we have that $\mathit{high}(u)=\mathit{high}_2(v)=\mathit{high}_2(v')$. Furthermore, we have that $\mathit{low}(u)\geq\mathit{lastM}(v)$, and therefore $\mathit{low}(u)\geq\widetilde{\mathit{nextM}}(v')$. This implies that $U_4^3(v')$ is not empty (because we can consider the lowest proper descendant $u'$ of $v'$ in $\widetilde{S}(v')=\widetilde{S}(v)$ such that $\mathit{high}(u')=\mathit{high}_2(v')$ and $\mathit{low}(u')\geq\widetilde{\mathit{nextM}}(v')$). This contradicts our supposition that $U_4^3(v')\neq\emptyset$. Thus, we have shown that $U_4^3(v')=\emptyset$ implies that $U_4^3(v)=\emptyset$.

Now let us assume that $U_4^3(v)\neq\emptyset$. This implies that $U_4^3(v')$ is not empty. Let us suppose, for the sake of contradiction, that there is a vertex $u\in U_4^3(v)$ that is lower than the greatest vertex $u'$ in $U_4^3(v')$.
Since $u\in U_4^3(v)$, we have that $u\in\widetilde{S}$. Since $u'\in U_4^3(v')$ we have that $u'\in\widetilde{S}$. This implies that $u$ and $u'$ are related as ancestor and descendant. Thus, since $u$ is lower than $u'$, we have that $u$ is a proper ancestor of $u'$. Let us suppose, for the sake of contradiction, that $\mathit{low}(u')$ is a proper ancestor of $\widetilde{\mathit{nextM}}(v')$. Then, since $\widetilde{\mathit{nextM}}(v')$ is a proper ancestor of $\mathit{lastM}(v)$, we have that $\mathit{low}(u')$ is a proper ancestor of $\mathit{lastM}(v)$. Now let $(x,y)$ be a back-edge in $B(u')$ such that $y=\mathit{low}(u')$. Then $x$ is a descendant of $u'$, and therefore a descendant of $u$. Furthermore, $y$ is a proper ancestor of $\mathit{lastM}(v)$, and therefore a proper ancestor of $v$, and therefore a proper ancestor of $u$. This shows that $(x,y)\in B(u)$. Thus, we have $\mathit{low}(u)\leq y=\mathit{low}(u')<\mathit{lastM}(v)$, in contradiction to the fact that $u\in U_4^3(v)$. Thus, our last supposition is not true, and therefore we have that $\mathit{low}(u')$ is not a proper ancestor of $\widetilde{\mathit{nextM}}(v')$. 

Since $u'\in U_4^3(v')$, we have that $u'$ is a proper descendant of $v'$, and therefore a proper descendant of $\widetilde{\mathit{nextM}}(v')$. Thus, since $u'$ is a common descendant of $\mathit{low}(u')$ and $\widetilde{\mathit{nextM}}(v')$, we have that $\mathit{low}(u')$ and $\widetilde{\mathit{nextM}}(v')$ are related as ancestor and descendant. Therefore, since $\mathit{low}(u')$ is not a proper ancestor of $\widetilde{\mathit{nextM}}(v')$, we have that $\mathit{low}(u')$ is a descendant of $\widetilde{\mathit{nextM}}(v')$, and therefore $\mathit{low}(u')\geq\widetilde{\mathit{nextM}}(v')$. Thus, since $u'\in U_4^3(v')$, we have that $u'$ is the lowest proper descendant of $v'$ in $\widetilde{S}$ with $\mathit{high}(u')=\mathit{high}_2(v')$ and $\mathit{low}(u')\geq\widetilde{\mathit{nextM}}(v')$ $(*)$. 

Now we will trace the implications of $u\in U_4^3(v)$. First, we have that $u\in\widetilde{S}$. Then, we have $\mathit{high}(u)=\mathit{high}_2(v)=\mathit{high}_2(v')$. Furthermore, we have that $\mathit{low}(u)\geq\mathit{lastM}(v)$, and therefore $\mathit{low}(u)>\widetilde{\mathit{nextM}}(v')$ (since $\widetilde{\mathit{nextM}}(v')$ is a proper ancestor of $\mathit{lastM}(v)$). Finally, we can show as above that $u$ is a proper descendant of $v'$ (the proof of this fact above did not rely on $U_4^3(v')=\emptyset$). But then, since $u$ is lower than $u'$, we have a contradiction to $(*)$.
Thus, we have shown that every vertex in $U_4^3(v)$ is at least as great as the greatest vertex in $U_4^3(v')$. In particular, this implies that the lowest vertex in $U_4^3(v)$ is greater than, or equal to, the greatest vertex in $U_4^3(v')$.
\end{proof}

Due to the similarity of the definitions of the $U_3$ and the $U_4^3$ sets, and the similarity between Lemmata~\ref{lemma:type3-b-ii-3-relation-between-u3} and \ref{lemma:type-3-b-ii-4-3-U-sets}, we can use a similar procedure as Algorithm~\ref{algorithm:type3-b-ii-3-U} in order to compute all $U_4^3$ sets in linear time. This is shown in Algorithm~\ref{algorithm:type3-b-ii-4-3-U}. Our result is summarized in Lemma~\ref{lemma:algorithm:type3-b-ii-4-3-U}.

\begin{algorithm}[H]
\caption{\textsf{Compute the sets $U_4^3(v)$, for all vertices $v\neq r$ such that $\mathit{high}_1(v)\neq\mathit{high}_2(v)$ and $\widetilde{\mathit{nextM}}(v)\neq\bot$}}
\label{algorithm:type3-b-ii-4-3-U}
\LinesNumbered
\DontPrintSemicolon
let $\mathcal{V}$ be the collection of all vertices $v\neq r$ such that $\mathit{high}_1(v)\neq\mathit{high}_2(v)$ and $\widetilde{\mathit{nextM}}(v)\neq\bot$\;
\ForEach{vertex $x$}{
  compute the collection $\mathcal{S}(x)$ of the segments of $\widetilde{H}(x)$ that are maximal w.r.t. the property
  that their elements are related as ancestor and descendant\;
}
\ForEach{$v\in\mathcal{V}$}{
  set $U_4^3(v)\leftarrow\emptyset$\;
}
\ForEach{vertex $x$}{
  \ForEach{segment $S\in\mathcal{S}(x)$}{
    let $v$ be the first vertex in $S$\;
    \While{$v\neq\bot$ \textbf{and} ($\mathit{high}_2(v)\neq x$ \textbf{or} $v\notin\mathcal{V}$)}{
    \label{line:type-3-b-ii-4-3-v-1}
      $v\leftarrow\mathit{next}_S(v)$\;
    }
    \lIf{$v=\bot$}{\textbf{continue}}
    let $u\leftarrow\mathit{prev}_S(v)$\;
    \While{$v\neq\bot$}{
      \While{$u\neq\bot$ \textbf{and} $\mathit{low}(u)<\mathit{lastM}(v)$}{
        $u\leftarrow\mathit{prev}_S(u)$\;
      }
      \While{$u\neq\bot$ \textbf{and} $\mathit{low}(u)<\widetilde{\mathit{nextM}}(v)$}{
        \If{$\mathit{high}_1(u)=x$}{
        \label{line:type-3-b-ii-4-3-u-1}
          insert $u$ into $U_4^3(v)$\;
        }
        $u\leftarrow\mathit{prev}_S(u)$\;
      }
      \While{$u\neq\bot$ \textbf{and} $\mathit{high}_1(u)\neq x$}{
      \label{line:type-3-b-ii-4-3-u-2}
        $u\leftarrow\mathit{prev}_S(u)$\;
      }
      \If{$u\neq\bot$}{
        insert $u$ into $U_4^3(v)$\;
      }
      $v\leftarrow\mathit{next}_S(v)$\;
      \While{$v\neq\bot$ \textbf{and} ($\mathit{high}_2(v)\neq x$ \textbf{or}  $v\notin\mathcal{V}$)}{
      \label{line:type-3-b-ii-4-3-v-2}
        $v\leftarrow\mathit{next}_S(v)$\;
      }      
    }
  }
}
\end{algorithm}

\begin{lemma}
\label{lemma:algorithm:type3-b-ii-4-3-U}
Algorithm~\ref{algorithm:type3-b-ii-4-3-U} correctly computes the sets $U_4^3(v)$, for all vertices $v\neq r$ such that $\widetilde{\mathit{nextM}}(v)\neq\emptyset$. Furthermore, it runs in linear time.
\end{lemma}
\begin{proof}
The proof here is similar as that of Lemma~\ref{lemma:algorithm:type3-b-ii-3-U} (which was given in the main text, in the two paragraphs above Algorithm~\ref{algorithm:type3-b-ii-3-U}). The differences are the following. First, the set $\mathcal{V}$ of the vertices for which the sets $U_4^3$ are defined is given by all vertices $v\neq r$ such that $\mathit{high}_1(v)\neq\mathit{high}_2(v)$ and $\widetilde{\mathit{nextM}}(v)\neq\bot$. Due to Lemma~\ref{lemma:tilde-nextM}, it is easy to collect all those vertices in $O(n)$ time. I.e., we have to check, for every vertex $v\neq r$ with $\mathit{high}_1(v)\neq\mathit{high}_2(v)$, whether $\mathit{nextM}(v)$ is a proper ancestor of $\mathit{high}_1(v)$. If that is the case, then $\widetilde{\mathit{nextM}}(v)=\mathit{nextM}(v)$. Otherwise, $\widetilde{\mathit{nextM}}(v)=\mathit{nextM}(\mathit{nextM}(v))$ (which may be $\mathit{null}$). Second, here we process the vertices $v$ in their $\widetilde{S}_2(v)$ segment (instead of the $\widetilde{S}_1(v)$). Thus, in Lines~\ref{line:type-3-b-ii-4-3-v-1} and \ref{line:type-3-b-ii-4-3-v-2}, we check whether $v$ satisfies $\mathit{high}_2(v)=x$ (where $x$ is the vertex for which we process the $\widetilde{H}(x)$ list). And third, we have that every $u\in U_4^3(v)$ satisfies $\mathit{high}_1(u)=\mathit{high}_2(v)$. Thus, in Lines~\ref{line:type-3-b-ii-4-3-u-1} and \ref{line:type-3-b-ii-4-3-u-2} we  have the appropriate condition (where is it checked whether $\mathit{high}_1(u)=x$). Then the proof follows the same reasoning as in Lemma~\ref{lemma:algorithm:type3-b-ii-3-U}. The main difference in the argument here is that every reference to Lemma~\ref{lemma:type3-b-ii-3-relation-between-u3} is replaced with a reference to Lemma~\ref{lemma:type-3-b-ii-4-3-U-sets}.
\end{proof}

\begin{lemma}
\label{lemma:type3-b-ii-4-3-in-U}
Let $(u,v,w)$ be a triple of vertices that induces a Type-3$\beta$ii-$4$ $4$-cut, such that $\mathit{high}_1(v)>\mathit{high}(u)$ and $M(B(v)\setminus\{e_\mathit{high}(v)\})= M(v)$. Then $u\in U_4^3(v)$. 
\end{lemma}
\begin{proof}
Since $\mathit{high}_1(v)>\mathit{high}(u)$, Lemma~\ref{lemma:type3-b-ii-4-info} implies that $\mathit{high}_2(v)=\mathit{high}(u)$. Thus, we may consider the segment $\widetilde{S}$ of $\widetilde{H}(\mathit{high}_2(v))$ from $u$ to $v$. Let $u'$ be a vertex in $\widetilde{S}$. Then, we have that $u\geq u'\geq v$, and  therefore Lemma~\ref{lemma:type3-b-ii-4-seg-1} implies that $u'$ is an ancestor of $u$. Thus, we have that all elements of $\widetilde{S}$ are related as ancestor and descendant (since all of them are ancestors of $u$). 
Since $\mathit{high}_1(v)>\mathit{high}(u)$, Lemma~\ref{lemma:type3-b-ii-4-info} implies that $e=e_\mathit{high}(v)$, where $e$ is the back-edge in the $4$-cut induced by $(u,v,w)$. Furthermore, Lemma~\ref{lemma:type3-b-ii-4-info} implies that $w\leq\mathit{low}(u)$. Thus, since $\mathit{low}(u)\leq\mathit{high}(u)=\mathit{high}_2(v)$, we have that $w\leq\mathit{high}_2(v)$. Then, since $(u,v,w)$ induces a Type-3$\beta$ii-$4$ $4$-cut, we have $M(w)=M(B(v)\setminus\{e\})=M(v)$, and therefore $\widetilde{\mathit{nextM}}(v)$ is defined (and it is greater than, or equal to, $w$). Also, we have $w\geq\mathit{lastM}(v)$, and therefore $\mathit{low}(u)\geq\mathit{lastM}(v)$. Thus, if $\mathit{low}(u)<\widetilde{\mathit{nextM}}(v)$, then $u$ satisfies enough conditions to be in $U_4^3(v)$. So let us assume that $\mathit{low}(u)\geq\widetilde{\mathit{nextM}}(v)$. 

Let us suppose, for the sake of contradiction, that $u$ is not the lowest vertex in $\widetilde{S}$ that is a proper descendant of $v$ such that $\mathit{high}(u)=\mathit{high}_2(v)$, $\mathit{low}(u)\geq\mathit{lastM}(v)$ and $\mathit{low}(u)\geq\widetilde{\mathit{nextM}}(v)$. Then, there is a vertex $u'$ in $\widetilde{S}$, that is a proper ancestor of $u$ and a proper descendant of $v$, such that $\mathit{high}(u')=\mathit{high}_2(v)$, $\mathit{low}(u')\geq\mathit{lastM}(v)$ and $\mathit{low}(u')\geq\widetilde{\mathit{nextM}}(v)$. Since $(u,v,w)$ induces a Type-3$\beta$ii-$4$ $4$-cut, we have that $B(v)=(B(u)\sqcup B(w))\sqcup\{e\}$.
Now let $(x,y)$ be a back-edge in $B(u)$. Then $x$ is a descendant of $u$, and therefore a descendant of $u'$. Furthermore, $B(v)=(B(u)\sqcup B(w))\sqcup\{e\}$ implies that $(x,y)\in B(v)$, and therefore $y$ is a proper ancestor of $v$, and therefore a proper ancestor of $u'$. This shows that $(x,y)\in B(u')$. Due to the generality of $(x,y)\in B(u)$, this shows that $B(u)\subseteq B(u')$. Conversely, let $(x,y)$ be a back-edge in $B(u')$. Then we have that $x$ is a descendant of $u'$, and therefore a descendant of $v$. Furthermore, $y$ is an ancestor of $\mathit{high}(u')=\mathit{high}_2(v)$, and therefore it is a proper ancestor of $v$. This shows that $(x,y)\in B(v)$. Then, $B(v)=(B(u)\sqcup B(w))\sqcup\{e\}$ implies that either $(x,y)\in B(u)$, or $(x,y)\in B(w)$, or $(x,y)=e$. The case $(x,y)\in B(w)$ is rejected, since $y=\mathit{low}(u')\geq\widetilde{\mathit{nextM}}(v)\geq w$. Furthermore, since $e=e_\mathit{high}(v)$ and $\mathit{high}_1(v)>\mathit{high}(u)$ and $\mathit{high}(u)=\mathit{high}_2(v)=\mathit{high}(u')$, we cannot have $e_\mathit{high}(v)\in B(u')$, and therefore the case $(x,y)=e$ is also rejected.
Thus, we have that $(x,y)\in B(u)$. Due to the generality of $(x,y)\in B(u')$, this shows that $B(u')\subseteq B(u)$. Thus, we have $B(u)=B(u')$, in contradiction to the fact that the graph is $3$-edge-connected. Thus, we have shown that $u$ is the lowest vertex in $\widetilde{S}$ that is a proper descendant of $v$ such that $\mathit{high}(u)=\mathit{high}_2(v)$, $\mathit{low}(u)\geq\mathit{lastM}(v)$ and $\mathit{low}(u)\geq\widetilde{\mathit{nextM}}(v)$. We conclude that $u$ satisfies enough conditions to be in $U_4^3(v)$. 
\end{proof}

\begin{lemma}
\label{lemma:type3-b-ii-4-3-criterion}
Let $(u,v,w)$ be a triple of vertices such that $\mathit{high}_1(v)>\mathit{high}(u)$, $M(B(v)\setminus\{e_\mathit{high}(v)\})= M(v)$ and $u\in U_4^3(v)$. Then, $(u,v,w)$ induces a Type-3$\beta$ii-$4$ $4$-cut if and only if: $\mathit{bcount}(v)=\mathit{bcount}(u)+\mathit{bcount}(w)+1$, and $w$ is the greatest proper ancestor of $v$ with $M(w)=M(v)$ such that $w\leq\mathit{low}(u)$.
\end{lemma}
\begin{proof}
($\Rightarrow$) Since $(u,v,w)$ induces a Type-3$\beta$ii-$4$ $4$-cut, we have $B(v)=(B(u)\sqcup B(w))\sqcup\{e\}$, where $e$ is the back-edge in the $4$-cut induced by $(u,v,w)$. Thus, we get $\mathit{bcount}(v)=\mathit{bcount}(u)+\mathit{bcount}(w)+1$. Since $\mathit{high}_1(v)>\mathit{high}(u)$, by Lemma~\ref{lemma:type3-b-ii-4-info} we have $e=e_\mathit{high}(v)$. Furthermore, Lemma~\ref{lemma:type3-b-ii-4-info} implies that $w\leq\mathit{low}(u)$. Since $(u,v,w)$ induces a Type-3$\beta$ii-$4$ $4$-cut, we have $M(w)=M(B(v)\setminus\{e\})$. Thus, since $M(B(v)\setminus\{e_\mathit{high}(v)\})= M(v)$, we have $M(w)=M(v)$.

Now let us suppose, for the sake of contradiction, that there is a proper ancestor $w'$ of $v$ with $M(w')=M(v)$ and $w'\leq\mathit{low}(u)$, such that $w'>w$. Then, since $M(v)=M(w)$ and $M(v)=M(w')$, we have that $M(v)$ is a common descendant of $w$ and $w'$, and therefore $w$ and $w'$ are related as ancestor and descendant. Thus, $w'>w$ implies that $w'$ is a proper descendant of $w$.
Since $w'$ is a proper descendant of $w$ with $M(w')=M(w)$, Lemma~\ref{lemma:same_m_subset_B} implies that $B(w)\subseteq B(w')$. This can be strengthened to $B(w)\subset B(w')$, since the graph is $3$-edge-connected. Thus, there is a back-edge $(x,y)\in B(w')\setminus B(w)$. Then, $x$ is a descendant of $M(w')=M(v)$. Furthermore, $y$ is a proper ancestor of $w'$, and therefore a proper ancestor of $v$. This shows that $(x,y)\in B(v)$. Since $B(v)=(B(u)\sqcup B(w))\sqcup\{e\}$, this implies that either $(x,y)\in B(u)$, or $(x,y)\in B(w)$, or $(x,y)=e$. The case $(x,y)\in B(u)$ is rejected, since $y<w'$ and $w'\leq\mathit{low}(u)$, and therefore $y<\mathit{low}(u)$. The case $(x,y)=e$ is also rejected, because $e=e_\mathit{high}(v)$, and $\mathit{high}_1(v)>\mathit{high}(u)\geq\mathit{low}(u)\geq w'$. Thus, we have that $(x,y)\in B(w)$, a contradiction. This shows that $w$ is the greatest proper ancestor of $v$ with $M(w)=M(v)$ such that $w\leq\mathit{low}(u)$.

($\Leftarrow$) We have to show that there is a back-edge $e$ such that $B(v)=(B(u)\sqcup B(w))\sqcup\{e\}$, and $M(w)=M(B(v)\setminus\{e\})$.

Since $M(B(v)\setminus\{e_\mathit{high}(v)\})= M(v)$ and $M(w)=M(v)$, we have that $M(w)=M(B(v)\setminus\{e_\mathit{high}(v)\})$. Since $\mathit{high}_1(v)>\mathit{high}(u)$, we have that $e_\mathit{high}(v)\notin B(u)$. And since $w\leq\mathit{low}(u)\leq\mathit{high}(u)<\mathit{high}_1(v)$, we have that $e_\mathit{high}(v)\notin B(w)$. Furthermore, since $w\leq\mathit{low}(u)$, we have that $B(u)\cap B(w)=\emptyset$ (because, if there existed a back-edge in $B(u)\cap B(w)$, its lower endpoint would be lower than $\mathit{low}(u)$, which is impossible). This shows that the sets $B(u)$, $B(w)$ and $\{e_\mathit{high}(v)\}$ are pairwise disjoint.

Let $(x,y)$ be a back-edge in $B(u)$. Since $u\in U_4^3(v)$, we have that $u$ is a proper descendant of $v$ with $\mathit{high}(u)=\mathit{high}_2(v)$. Thus, $(x,y)\in B(u)$ implies that $x$ is a descendant of $v$. Furthermore, we have that $y$ is an ancestor of $\mathit{high}(u)=\mathit{high}_2(v)$, and therefore a proper ancestor of $v$. This shows that $(x,y)\in B(v)$. Due to the generality of $(x,y)\in B(u)$, this implies that $B(u)\subseteq B(v)$.

Let $(x,y)$ be a back-edge in $B(w)$. Then, $x$ is a descendant of $M(w)=M(v)$, and therefore a descendant of $v$. Furthermore, $y$ is a proper ancestor of $w$, and therefore a proper ancestor of $v$. This shows that $(x,y)\in B(v)$. Due to the generality of $(x,y)\in B(w)$, this implies that $B(w)\subseteq B(v)$.

Thus, we have $B(u)\subseteq B(v)$, $B(w)\subseteq B(v)$, $e_\mathit{high}(v)\in B(v)$, and the sets $B(u)$, $B(w)$ and $\{e_\mathit{high}(v)\}$ are pairwise disjoint. Thus, $(B(u)\sqcup B(v))\sqcup\{e_\mathit{high}(v)\}\subseteq B(v)$. Since $\mathit{bcount}(v)=\mathit{bcount}(u)+\mathit{bcount}(w)+1$, this implies that $B(v)=(B(u)\sqcup B(w))\sqcup\{e_\mathit{high}(v)\}$. By assumption we have $M(B(v)\setminus\{e_\mathit{high}(v)\})= M(v)$. Therefore, since $M(w)=M(v)$, we have $M(w)=M(B(v)\setminus\{e_\mathit{high}(v)\})$.
\end{proof}

\begin{algorithm}[H]
\caption{\textsf{Compute all Type-3$\beta$ii-$4$ $4$-cuts of the form $\{(u,p(u)),(v,p(v)),(w,p(w)),e\}$, where $M(B(v)\setminus\{e\})= M(v)$ and $\mathit{high}_1(v)>\mathit{high}(u)$}}
\label{algorithm:type3-b-ii-4-3}
\LinesNumbered
\DontPrintSemicolon
let $\mathcal{V}$ be the collection of all vertices $v\neq r$ such that $M(B(v)\setminus\{e_\mathit{high}(v)\})=M(v)$, $\mathit{high}_1(v)\neq\mathit{high}_2(v)$ and $\mathit{lastM}(v)\leq\mathit{high}_2(v)$\;
\label{line:type4-b-ii-4-3-V}
\ForEach{$v\in\mathcal{V}$}{
  compute $U_4^3(v)$\;
  \label{line:type4-b-ii-4-3-U}
}
\ForEach{$v\in\mathcal{V}$}{
  \ForEach{$u\in U_4^3(v)$}{
    let $w$ be the greatest proper ancestor of $v$ with $M(w)=M(v)$ such that $w\leq\mathit{low}(u)$\;
    \label{line:type4-b-ii-4-3-w}    
    \If{$\mathit{bcount}(v)=\mathit{bcount}(u)+\mathit{bcount}(w)+1$ \textbf{and} $\mathit{high}_1(v)>\mathit{high}(u)$}{
      mark $\{(u,p(u)),(v,p(v)),(w,p(w)),e_\mathit{high}(v)\}$ as a Type-3$\beta$ii-$4$ $4$-cut\; 
      \label{line:type4-b-ii-4-3-mark}
    }
  }
}
\end{algorithm}

\begin{proposition}
\label{proposition:algorithm:type3-b-ii-4-3}
Algorithm~\ref{algorithm:type3-b-ii-4-3} correctly computes all Type-3$\beta$ii-$4$ $4$-cuts of the form $\{(u,p(u)),(v,p(v)),(w,p(w)),e\}$, where $u$ is a descendant of $v$, $v$ is a descendant of $w$, $M(B(v)\setminus\{e\})= M(v)$ and $\mathit{high}_1(v)>\mathit{high}(u)$. Furthermore, it has a linear-time implementation.
\end{proposition}
\begin{proof}
Let $(u,v,w)$ be a triple of vertices that induces a Type-3$\beta$ii-$4$ $4$-cut with back-edge $e$ such that $M(B(v)\setminus\{e\})= M(v)$ and $\mathit{high}_1(v)>\mathit{high}(u)$. Since $\mathit{high}_1(v)>\mathit{high}(u)$, by Lemma~\ref{lemma:type3-b-ii-4-info} we have that  $\mathit{high}_1(v)\neq\mathit{high}_2(v)$ and $e=e_\mathit{high}(v)$. Then, by Lemma~\ref{lemma:type3-b-ii-4-3-in-U} we have that $u\in U_4^3(v)$. This implies that $\mathit{lastM}(v)\leq\mathit{high}_2(v)$, and therefore $v$ belongs to the collection $\mathcal{V}$ computed in Line~\ref{line:type4-b-ii-4-3-V}. Then, by Lemma~\ref{lemma:type3-b-ii-4-3-criterion} we have that $\mathit{bcount}(v)=\mathit{bcount}(u)+\mathit{bcount}(w)+1$, and $w$ is the greatest proper ancestor of $v$ with $M(w)=M(v)$ such that $w\leq\mathit{low}(u)$. Then, notice that $\{(u,p(u)),(v,p(v)),(w,p(w)),e_\mathit{high}(v))\}$ (i.e., the $4$-cut induced by $(u,v,w)$) will be correctly marked by Algorithm~\ref{algorithm:type3-b-ii-4-3} in Line~\ref{line:type4-b-ii-4-3-mark}. 

Conversely, suppose that a $4$-element set $\{(u,p(u)),(v,p(v)),(w,p(w)),e_\mathit{high}(v)\}$ is marked by Algorithm~\ref{algorithm:type3-b-ii-4-3} in Line~\ref{line:type4-b-ii-4-3-mark}. Then we have that: $(1)$ $M(B(v)\setminus\{e_\mathit{high}(v)\})= M(v)$, $(2)$ $u\in U_4^3(v)$, $(3)$ $w$ is the greatest proper ancestor of $v$ such that $M(w)=M(v)$ and $w\leq\mathit{low}(u)$, and $(4)$ $\mathit{bcount}(v)=\mathit{bcount}(u)+\mathit{bcount}(w)+1$ and $\mathit{high}_1(v)>\mathit{high}(u)$. Thus, Lemma~\ref{lemma:type3-b-ii-4-3-criterion} implies that $\{(u,p(u)),(v,p(v)),(w,p(w)),e_\mathit{high}(v)\}$ is a Type-3$\beta$ii-$4$ $4$-cut. 

Now we will show that Algorithm~\ref{algorithm:type3-b-ii-4-3} has a linear-time implementation. Computing the values $M(B(v)\setminus\{e_\mathit{high}(v)\})$, for all vertices $v\neq r$, takes linear time in total, according to Proposition~\ref{proposition:computing-M(B(v)-S)}. Thus, the computation of the collection of vertices $\mathcal{V}$ in Line~\ref{line:type4-b-ii-4-3-V} can be performed in linear time.
By Lemma~\ref{lemma:algorithm:type3-b-ii-4-3-U}, we have that the sets $U_4^3(v)$ can be computed in linear time, for all vertices $v\in\mathcal{V}$, using Algorithm~\ref{algorithm:type3-b-ii-4-3-U}. Thus, the \textbf{for} loop in Line~\ref{line:type4-b-ii-4-3-U} can be performed in linear time. In particular, we have that the total size of all sets $U_4^3$ is $O(n)$.
In order to find the vertex $w$ in Line~\ref{line:type4-b-ii-4-3-w}, we can use Algorithm~\ref{algorithm:W-queries}. Specifically, let $v$ and $u$ be two vertices such that $v\in\mathcal{V}$ and $u\in U_4^3(v)$. Then we generate a query $q(M^{-1}(M(v)),\mathit{min}\{\mathit{low}(u),p(v)\})$. This returns the greatest $w$ such that $M(w)=M(v)$ and $w\leq\mathit{low}(u)$ and $w\leq p(v)$. Thus, we have that $w$ is the greatest proper ancestor of $v$ such that $M(w)=M(v)$ and $w\leq\mathit{low}(u)$. Since the number of all those queries is $O(n)$, Lemma~\ref{lemma:W-queries} implies that all of them can be answered in linear time in total, using Algorithm~\ref{algorithm:W-queries}. We conclude that Algorithm~\ref{algorithm:type3-b-ii-4-3} has a linear-time implementation.
\end{proof}

\noindent\\
\textbf{The case where $M(B(v)\setminus\{e\})= M(v)$ and $\mathit{high}_1(v)=\mathit{high}(u)$}\\

Let $(u,v,w)$ be a triple of vertices that induces a Type-3$\beta$ii-$4$ $4$-cut such that $M(B(v)\setminus\{e\})= M(v)$. Then, since $M(w)=M(B(v)\setminus\{e\})$, we have $M(w)=M(v)$.

Now let $v\neq r$ be a vertex such that $\mathit{nextM}(v)\neq\bot$. Then we let $U_4^4(v)$ denote the collection of all vertices $u\in S(v)$ such that: $(1)$ $u$ is a proper descendant of $v$, $(2)$ $\mathit{low}(u)\geq\mathit{lastM}(v)$, and $(3)$ either $\mathit{low}(u)<\mathit{nextM}(v)$, or $u$ is the lowest vertex in $S(v)$ that satisfies $(1)$ and $\mathit{low}(u)\geq\mathit{nextM}(v)$.

\begin{lemma}
\label{lemma:type-3bii-4-4-aux}
Let $v$ and $v'$ be two vertices $\neq r$ with $\mathit{nextM}(v)\neq\bot$ and $\mathit{nextM}(v')\neq\bot$, such that $v'$ is a proper descendant of $v$ with $\mathit{high}_1(v)=\mathit{high}_1(v')$. Then, $\mathit{nextM}(v')$ is a proper ancestor of $\mathit{lastM}(v)$.
\end{lemma}
\begin{proof}
Let $(x,y)$ be a back-edge in $B(v')$ such that and $y=\mathit{high}_1(v')$. Then, we have that $x$ is a descendant of $v'$, and therefore a descendant of $v$. Furthermore, since $\mathit{high}_1(v')=\mathit{high}_1(v)$, we have that $y$ is a proper ancestor of $v$. This shows that $(x,y)\in B(v)$. Thus, we have that $M(v)$ is an ancestor of $x$. Therefore, since $x$ is a common descendant of $v'$ and $M(v)$, we have that $v'$ and $M(v)$ are related as ancestor and descendant.

Let us suppose, for the sake of contradiction, that $M(v)$ is not a proper ancestor of $v'$. Then, we have that $M(v)$ is a descendant of $v'$. Let $(x,y)$ be a back-edge in $B(v')$. Then $x$ is a descendant of $v'$, and therefore a descendant of $v$. Furthermore, $y$ is an ancestor of $\mathit{high}_1(v')=\mathit{high}_1(v)$, and therefore a proper ancestor of $v$. This shows that $(x,y)\in B(v)$. Due to the generality of $(x,y)\in B(v')$, this implies that $B(v')\subseteq B(v)$. Conversely, let $(x,y)$ be a back-edge in $B(v)$. Then, $x$ is a descendant of $M(v)$, and therefore a descendant of $v'$. Furthermore, $y$ is a proper ancestor of $v$, and therefore a proper ancestor of $v'$. This shows that $(x,y)\in B(v')$. Due to the generality of $(x,y)\in B(v)$, this implies that $B(v)\subseteq B(v')$. Thus we have $B(v')=B(v)$, in contradiction to the fact that the graph is $3$-edge-connected. This shows that $M(v)$ is a proper ancestor of $v'$.

Let $w$ and $w'$ be two vertices such that $M(w)=M(v)$, $M(w')=M(v')$, $w\leq\mathit{nextM}(v)$ and $w'\leq\mathit{nextM}(v')$. Then,  Lemma~\ref{lemma:same_M_dif_B_lower} implies that $w$ is an ancestor of $\mathit{high}_1(v)$ and $w'$ is an ancestor of $\mathit{high}_1(v')$. Thus, since $\mathit{high}_1(v)=\mathit{high}_1(v')$, we have that $w$ and $w'$ have a common descendant, and therefore they are related as ancestor and descendant.

Let us suppose, for the sake of contradiction, that $w'$ is not a proper ancestor of $w$. Then, we have that $w'$ is a descendant of $w$. Since $M(w)=M(v)$ is a proper ancestor of $v'$, there is a back-edge $(x,y)$ in $B(w)$ such that $x$ is not a descendant of $v'$. Therefore, $x$ is not a descendant of $M(v')=M(w')$. Since $x$ is a descendant of $M(v)$, we have that $x$ is a descendant of $v$, and therefore a descendant of $\mathit{high}_1(v)=\mathit{high}_1(v')$, and therefore a descendant of $w'$. Furthermore, $y$ is a proper ancestor of $w$, and therefore a proper ancestor of $w'$. This shows that $(x,y)\in B(w')$. But this implies that $x$ is a descendant of $M(w')=M(v')$, a contradiction.
This shows that $w'$ is a proper ancestor of $w$. Due to the generality of $w'$ and $w$, this implies that $\mathit{nextM}(v')$ is a proper ancestor of $\mathit{lastM}(v)$.
\end{proof}

\begin{lemma}
\label{lemma:type-3-b-ii-4-4-u-sets-rel}
Let $v$ and $v'$ be two vertices $\neq r$ with $\mathit{nextM}(v)\neq\bot$ and $\mathit{nextM}(v')\neq\bot$, such that $v'$ is a proper descendant of $v$, $\mathit{high}_1(v)=\mathit{high}_1(v')$, and both $v$ and $v'$ belong to the same segment $S$ of $H(\mathit{high}_1(v))$ that is maximal w.r.t. the property that all its elements are related as ancestor and descendant (i.e., we have $S=S(v)=S(v')$). If $U_4^4(v')=\emptyset$, then $U_4^4(v)=\emptyset$. If $U_4^4(v)\neq\emptyset$, then the lowest vertex in $U_4^4(v)$ is at least as great as the greatest vertex in $U_4^4(v')$.
\end{lemma}
\begin{proof}
By Lemma~\ref{lemma:type-3bii-4-4-aux}, we have that $\mathit{nextM}(v')$ is a proper ancestor of $\mathit{lastM}(v)$.

Let us suppose, for the sake of contradiction, that $U_4^4(v')=\emptyset$ and $U_4^4(v)\neq\emptyset$. Let $u$ be a vertex in $U_4^4(v)$. Let us suppose, for the sake of contradiction, that $u$ is not a proper descendant of $v'$. Since $u\in U_4^4(v)$, we have that $u\in S$. Thus, since $v'\in S$, we have that $u$ and $v'$ are related as ancestor and descendant. Since $u$ is not a proper descendant of $v'$, we have that $u$ is an ancestor of $v'$. Let $(x,y)$ be a back-edge in $B(v')$ such that $y=\mathit{low}(v')$. Lemma~\ref{lemma:same_m_same_low} implies that $\mathit{low}(v')$ is a proper ancestor of $\mathit{nextM}(v')$. Thus, since $\mathit{nextM}(v')$ is a proper ancestor of $\mathit{lastM}(v)$, we have that $\mathit{low}(v')$ is a proper ancestor of $\mathit{lastM}(v)$. Now, since $(x,y)\in B(v')$, we have that $x$ is a descendant of $v'$, and therefore a descendant of $u$. Furthermore, $y=\mathit{low}(v')$ is a proper ancestor of $\mathit{lastM}(v)$, and therefore a proper ancestor of $v$, and therefore a proper ancestor of $u$. This shows that $(x,y)\in B(u)$. But then we have $\mathit{low}(u)\leq y=\mathit{low}(v')<\mathit{lastM}(v)$, in contradiction to the fact that $u\in U_4^4(v)$. Thus, our last supposition is not true, and therefore $u$ is a proper descendant of $v'$. Then, since $u\in U_4^4(v)$, we have  $\mathit{high}(u)=\mathit{high}_1(v)=\mathit{high}_1(v')$. Furthermore, we have $\mathit{low}(u)\geq\mathit{lastM}(v)$, and therefore $\mathit{low}(u)\geq\mathit{nextM}(v')$. This implies that $U_4^4(v')$ is not empty (because we can consider the lowest proper descendant $u'$ of $v'$ in $S(v')=S(v)$ such that $\mathit{high}(u')=\mathit{high}_1(v')$ and $\mathit{low}(u')\geq\mathit{nextM}(v')$). This contradicts our supposition that $U_4^4(v')\neq\emptyset$. Thus, we have shown that $U_4^4(v')=\emptyset$ implies that $U_4^4(v)=\emptyset$.

Now let us assume that $U_4^4(v)\neq\emptyset$. This implies that $U_4^4(v')$ is not empty. Let us suppose, for the sake of contradiction, that there is a vertex $u\in U_4^4(v)$ that is lower than the greatest vertex $u'$ in $U_4^4(v')$.
Since $u\in U_4^4(v)$, we have $u\in S$. Since $u'\in U_4^4(v')$ we have $u'\in S$. This implies that $u$ and $u'$ are related as ancestor and descendant. Thus, since $u$ is lower than $u'$, we have that $u$ is a proper ancestor of $u'$. Let us suppose, for the sake of contradiction, that $\mathit{low}(u')$ is a proper ancestor of $\mathit{nextM}(v')$. Then, since $\mathit{nextM}(v')$ is a proper ancestor of $\mathit{lastM}(v)$, we have that $\mathit{low}(u')$ is a proper ancestor of $\mathit{lastM}(v)$. Now let $(x,y)$ be a back-edge in $B(u')$ such that $y=\mathit{low}(u')$. Then $x$ is a descendant of $u'$, and therefore a descendant of $u$. Furthermore, $y$ is a proper ancestor of $\mathit{lastM}(v)$, and therefore a proper ancestor of $v$, and therefore a proper ancestor of $u$. This shows that $(x,y)\in B(u)$. Thus, we have $\mathit{low}(u)\leq y<\mathit{lastM}(v)$, in contradiction to the fact that $u\in U_4^4(v)$. Thus, our last supposition is not true, and therefore we have that $\mathit{low}(u')$ is not a proper ancestor of $\mathit{nextM}(v')$. 

Since $u'\in U_4^4(v')$, we have that $u'$ is a proper descendant of $v'$, and therefore a proper descendant of $\mathit{nextM}(v')$. Thus, $u'$ is a common descendant of $\mathit{low}(u')$ and $\mathit{nextM}(v')$, and therefore $\mathit{low}(u')$ and $\mathit{nextM}(v')$ are related as ancestor and descendant. Thus, since $\mathit{low}(u')$ is not a proper ancestor of $\mathit{nextM}(v')$, we have that $\mathit{low}(u')$ is a descendant of $\mathit{nextM}(v')$, and therefore $\mathit{low}(u')\geq\mathit{nextM}(v')$. Thus, since $u'\in U_4^4(v')$, we have that $u'$ is the lowest proper descendant of $v'$ in $S$ with $\mathit{high}(u')=\mathit{high}_1(v')$ such that $\mathit{low}(u')\geq\mathit{nextM}(v')$ $(*)$. 

Now we will trace the implications of $u\in U_4^4(v)$. First, we have $u\in S$. Then, we have $\mathit{high}(u)=\mathit{high}_1(v)=\mathit{high}_1(v')$. Furthermore, we have $\mathit{low}(u)\geq\mathit{lastM}(v)$, and therefore $\mathit{low}(u)>\mathit{nextM}(v')$ (since $\mathit{nextM}(v')$ is a proper ancestor of $\mathit{lastM}(v)$). Finally, we can show as above that $u$ is a proper descendant of $v'$ (the proof of this fact above did not rely on $U_4^4(v')=\emptyset$). But then, since $u$ is lower than $u'$, we have a contradiction to $(*)$.
Thus, we have shown that every vertex in $U_4^4(v)$ is at least as great as the greatest vertex in $U_4^4(v')$. In particular, this implies that the lowest vertex in $U_4^4(v)$ is greater than, or equal to, the greatest vertex in $U_4^4(v')$.
\end{proof}

Due to the similarity of the definitions of the $U_1$ and the $U_4^4$ sets, and their properties described in Lemmata~\ref{lemma:type3-b-ii-1-relation-between-u1} and \ref{lemma:type-3-b-ii-4-4-u-sets-rel}, respectively, we can compute all $U_4^4$ sets with a procedure similar to Algorithm~\ref{algorithm:type3-b-ii-1-U}. This is shown in Algorithm~\ref{algorithm:type3-b-ii-4-4-U}. Our result is summarized in Lemma~\ref{lemma:algorithm:type3-b-ii-4-4-U}.

\noindent\\
\begin{algorithm}[H]
\caption{\textsf{Compute the sets $U_4^4(v)$, for all vertices $v\neq r$ such that $\mathit{nextM}(v)\neq\bot$}}
\label{algorithm:type3-b-ii-4-4-U}
\LinesNumbered
\DontPrintSemicolon
\ForEach{vertex $x$}{
  compute the collection $\mathcal{S}(x)$ of the segments of $H(x)$ that are maximal w.r.t. the property that their elements
  are related as ancestor and descendant\;
}
\ForEach{$v\neq r$ such that $\mathit{nextM}(v)\neq\bot$}{
  set $U_4^4(v)\leftarrow\emptyset$\;
}
\ForEach{vertex $x$}{
  \ForEach{segment $S\in\mathcal{S}(x)$}{
    let $v$ be the first vertex in $S$\;
    \While{$v\neq\bot$ \textbf{and} $\mathit{nextM}(v)=\bot$}{
      $v\leftarrow\mathit{next}_S(v)$\;
    }
    \lIf{$v=\bot$}{\textbf{continue}}
    let $u=\mathit{prev}_S(v)$\;
    \While{$v\neq\bot$}{
      \While{$u\neq\bot$ \textbf{and} $\mathit{low}(u)<\mathit{lastM}(v)$}{
        $u\leftarrow\mathit{prev_S}(u)$\;
      }
      \While{$u\neq\bot$ \textbf{and} $\mathit{low}(u)<\mathit{nextM}(v)$}{
        insert $u$ into $U_4^4(v)$\;
        $u\leftarrow\mathit{prev_S}(u)$\;
      }
      \If{$u\neq\bot$}{
        insert $u$ into $U_4^4(v)$\;
      }
      $v\leftarrow\mathit{next}_S(v)$\;
      \While{$v\neq\bot$ \textbf{and} $\mathit{nextM}(v)=\bot$}{
        $v\leftarrow\mathit{next}_S(v)$\;
      }
    }
  }
}
\end{algorithm}

\begin{lemma}
\label{lemma:algorithm:type3-b-ii-4-4-U}
Algorithm~\ref{algorithm:type3-b-ii-4-4-U} correctly computes the sets $U_4^4(v)$, for all vertices $v\neq r$ such that $\mathit{nextM}(v)\neq\bot$. Furthermore, it has a linear-time implementation.
\end{lemma}
\begin{proof}
The argument is almost identical to that provided for Lemma~\ref{lemma:algorithm:type3-b-ii-1-U} in the main text (in the three paragraphs above Algorithm~\ref{algorithm:type3-b-ii-1-U}). The only difference is that here we care about the $\mathit{low}$ point of the vertices $u$ (and not for their $\mathit{low}_2$ point). This does not affect the analysis of correctness, because the computation is performed in segments of $H(x)$ that are maximal w.r.t. the property that their elements are related as ancestor and descendant, and therefore all vertices in those segments are sorted in decreasing order w.r.t. their $\mathit{low}$ point.
\end{proof}

\begin{lemma}
\label{lemma:type3-b-ii-4-4-in-U}
Let $(u,v,w)$ be a triple of vertices that induces a Type-3$\beta$ii-$4$ $4$-cut, such that $\mathit{high}_1(v)=\mathit{high}(u)$ and $M(B(v)\setminus\{e\})= M(v)$, where $e$ is the back-edge in the $4$-cut induced by $(u,v,w)$. Suppose that the lower endpoint of $e$ is distinct from $\mathit{high}_1(v)$. Then $U_4^4(v)\neq\emptyset$, and let $\tilde{u}$ be the greatest vertex in $U_4^4(v)$. If $u\notin U_4^4(v)$, then $u$ is the predecessor of $\tilde{u}$ in $H(\mathit{high}_1(v))$. 
\end{lemma}
\begin{proof}
Since $\mathit{high}_1(v)=\mathit{high}(u)$ and $u$ is a proper descendant of $v$, we may consider the segment $S$ of $H(\mathit{high}_1(v))$ from $u$ to $v$. Let $u'$ be a vertex in $S$. Due to the sorting of $H(\mathit{high}_1(v))$, we have $u\geq u'\geq v$. Thus, since the lower endpoint of $e$ is distinct from $\mathit{high}_1(v)$, Lemma~\ref{lemma:type3-b-ii-4-seg-3} implies that $u'$ is an ancestor of $u$. Thus, we have that all elements of $S$ are related as ancestor and descendant (since all of them are ancestors of $u$), and therefore $S\subseteq S(v)$.
Since $(u,v,w)$ induces a Type-3$\beta$ii-$4$ $4$-cut, we have $M(w)=M(B(v)\setminus\{e\})$. Therefore, since $M(B(v)\setminus\{e\})= M(v)$, we have $M(w)=M(v)$. Since $w$ is a proper ancestor of $v$, this implies that $\mathit{nextM}(v)\neq\bot$, and $w\leq\mathit{nextM}(v)$.

By Lemma~\ref{lemma:type3-b-ii-4-info} we have $w\leq\mathit{low}(u)$, and therefore $\mathit{lastM}(v)\leq\mathit{low}(u)$. Thus, if $\mathit{low}(u)<\mathit{nextM}(v)$, then $u$ satisfies enough conditions to be in $U_4^4(v)$. Otherwise, we have $\mathit{low}(u)\geq\mathit{nextM}(v)$, and therefore $U_4^4(v)$ is not empty, because we can consider the lowest proper descendant $u'$ of $v$ in $S(v)$ such that $\mathit{low}(u')\geq\mathit{nextM}(v)$. So let $\tilde{u}$ be the greatest vertex in $U_4^4(v)$. Let us suppose, for the sake of contradiction, that $u\notin U_4^4(v)$, and $u$ is not the predecessor of $\tilde{u}$ in $H(\mathit{high}_1(v))$.

Since $u$ is a proper descendant of $v$ in $S(v)$ such that $\mathit{low}(u)\geq\mathit{nextM}(v)$, we have that there is a proper descendant $u'$ of $v$ in $S(v)$ that is lower than $u$ and has $\mathit{low}(u')\geq\mathit{nextM}(v)$ (because this is the only condition that prevents $u$ from being in $U_4^4(v)$). Thus, we may consider the lowest vertex $u'$ that has this property. Then, we have that $u'\in U_4^4(v)$, and every other vertex $u''\in U_4^4(v)$ satisfies $\mathit{low}(u'')<\mathit{nextM}(v)$. Let us suppose, for the sake of contradiction, that $u'$ is not the greatest vertex in $U_4^4(v)$. Then there is a vertex $u''\in U_4^4(v)$ such that $u''>u'$. Since $u'\in U_4^4(v)$ and $u''\in U_4^4(v)$, we have $u'\in S(v)$ and $u''\in S(v)$. Therefore, $u''>u'$ implies that $u''$ is a proper descendant of $u'$. Furthermore, we have $\mathit{high}(u')=\mathit{high}_1(v)=\mathit{high}(u'')$. Thus, Lemma~\ref{lemma:same_high} implies that $B(u'')\subseteq B(u')$. This implies that $\mathit{low}(u')\leq\mathit{low}(u'')$. But we have $\mathit{low}(u')\geq\mathit{nextM}(v)$ and $\mathit{low}(u'')<\mathit{nextM}(v)$, a contradiction. This shows that $u'$ is indeed the greatest vertex in $U_4^4(v)$, and therefore we have $u'=\tilde{u}$. 

Let $\tilde{u}'$ be the predecessor of $\tilde{u}$ in $H(\mathit{high}_1(v))$. Then, since $\tilde{u}$ is lower than $u$, and $u$ is neither $\tilde{u}$ nor $\tilde{u}'$, we have $u>\tilde{u}'>\tilde{u}$. Thus, since $u$ and $\tilde{u}$ are in $S(v)$, we have that $\tilde{u}'$ is also in $S(v)$ (because $S(v)$ is a segment of $H(\mathit{high}_1(v))$). Thus, $\tilde{u}'$ is related as ancestor and descendant with both $u$ and $\tilde{u}$. Therefore, we have that $u$ is a proper descendant of $\tilde{u}'$, and $\tilde{u}'$ is a proper descendant of $\tilde{u}$. Then, since $\mathit{high}(u)=\mathit{high}(\tilde{u}')=\mathit{high}(\tilde{u})$, Lemma~\ref{lemma:same_high} implies that $B(u)\subseteq B(\tilde{u}')\subseteq B(\tilde{u})$. Since the graph is $3$-edge-connected, this can be strengthened to $B(u)\subset B(\tilde{u}')\subset B(\tilde{u})$.

Let $(x,y)$ be a back-edge in $B(\tilde{u})$. Then we have that $x$ is a descendant of $\tilde{u}$, and therefore a descendant of $v$. Since $\tilde{u}\in S(v)$, we have $\mathit{high}(\tilde{u})=\mathit{high}_1(v)$. Thus, since $(x,y)\in B(\tilde{u})$, we have that $y$ is an ancestor of $\mathit{high}(\tilde{u})=\mathit{high}_1(v)$, and therefore $y$ is a proper ancestor of $v$. This shows that $(x,y)\in B(v)$. Since $(u,v,w)$ induces a Type-3$\beta$ii-$4$ $4$-cut, we have $B(v)=(B(u)\sqcup B(w))\sqcup\{e\}$. This implies that either $(x,y)\in B(u)$, or $(x,y)\in B(w)$, or $(x,y)=e$. The case $(x,y)\in B(w)$ is rejected, since $y\geq\mathit{low}(\tilde{u})\geq\mathit{nextM}(v)\geq w$. Thus, we have that either $(x,y)\in B(u)$ or $(x,y)=e$. Due to the generality of $(x,y)\in B(\tilde{u})$, this implies that $B(\tilde{u})\subseteq B(u)\sqcup\{e\}$. Thus, we have $B(u)\subset B(\tilde{u}')\subset B(\tilde{u})\subseteq B(u)\sqcup\{e\}$. But this implies that $\mathit{bcount}(u)<\mathit{bcount}(\tilde{u}')<\mathit{bcount}(\tilde{u})\leq\mathit{bcount}(u)+1$, which is impossible (because those numbers are integers).

Thus, we conclude that either $u\in U_4^4(v)$, or $u$ is the predecessor of $\tilde{u}$ in $H(\mathit{high}_1(v))$.
\end{proof}

\begin{lemma}
\label{lemma:type3-b-ii-4-4-criterion}
Let $(u,v,w)$ be a triple of vertices such that $u$ is a proper descendant of $v$, $v$ is a proper descendant of $w$, and $M(w)=M(v)$. Then there is a back-edge $e$ such that $B(v)=(B(u)\sqcup B(w))\sqcup\{e\}$ if and only if: $(1)$ $\mathit{high}(u)<v$, $(2)$ $\mathit{bcount}(v)=\mathit{bcount}(u)+\mathit{bcount}(w)+1$, and $(3)$ $w$ is either the greatest or the second-greatest proper ancestor of $v$ such that $M(w)=M(v)$ and $w\leq\mathit{low}(u)$.
\end{lemma}
\begin{proof}
($\Rightarrow$) $B(v)=(B(u)\sqcup B(w))\sqcup\{e\}$ implies that $B(u)\subseteq B(v)$. Let $(x,y)$ be a back-edge in $B(u)$. Then $B(u)\subseteq B(v)$ implies that $(x,y)\in B(v)$, and therefore $y$ is a proper ancestor of $v$, and therefore $y<v$. Due to the generality of $(x,y)\in B(u)$, this implies that $\mathit{high}(u)<v$. $(2)$ is an immediate consequence of $B(v)=(B(u)\sqcup B(w))\sqcup\{e\}$.

Let us suppose, for the sake of contradiction, that $\mathit{low}(u)<w$. Let $(x,y)$ be a back-edge in $B(u)$ such that $y=\mathit{low}(u)$. Then $x$ is a descendant of $u$, and therefore a descendant of $v$, and therefore a descendant of $w$. Since $(x,y)$ is a back-edge, we have that $x$ is a descendant of $y$. Thus, $x$ is a common descendant of $w$ and $y$, and therefore $w$ and $y$ are related as ancestor and descendant. Then, $y=\mathit{low}(u)<w$ implies that $y$ is a proper ancestor of $w$. But this shows that $(x,y)\in B(w)$, in contradiction to $B(u)\cap B(w)=\emptyset$. This shows that $\mathit{low}(u)\geq w$. 

Thus, it makes sense to consider the greatest proper ancestor $w'$ of $v$ such that $M(w')=M(v)$ and $w'\leq\mathit{low}(u)$. If $w=w'$, then we are done. Otherwise, we can also consider the second-greatest proper ancestor $w''$ of $v$ such that $M(w'')=M(v)$ and $w''\leq\mathit{low}(u)$. 

Let us suppose, for the sake of contradiction, that $w$ is neither $w'$ nor $w''$. Thus, we have $w<w''<w'$. Then, since $M(w)=M(w'')=M(w')$, we have that $w$ is a proper ancestor of $w''$, $w''$ is a proper ancestor of $w'$, and Lemma~\ref{lemma:same_m_subset_B} implies that $B(w)\subseteq B(w'')\subseteq B(w')$. Since the graph is $3$-edge-connected, this can be strengthened to $B(w)\subset B(w'')\subset B(w')$. Notice that, since $w'\leq\mathit{low}(u)$, we have $B(u)\cap B(w')=\emptyset$ (because the lower endpoint of every back-edge in $B(u)$ is not low enough to be a proper ancestor of $w'$).

Let $(x,y)$ be a back-edge in $B(w')$. Then $x$ is a descendant of $M(w')$, and therefore a descendant of $M(v)$, and therefore a descendant of $v$. Furthermore, $y$ is a proper ancestor of $w'$, 
and therefore $y$ is a proper ancestor of $v$. This shows that $(x,y)\in B(v)$. Then $B(v)=(B(u)\sqcup B(w))\sqcup\{e\}$ implies that either $(x,y)\in B(u)$, or $(x,y)\in B(w)$, or $(x,y)=e$. The case $(x,y)\in B(u)$ is rejected, because $B(u)\cap B(w')=\emptyset$. Thus, we have that either $(x,y)\in B(w)$, or $(x,y)=e$. Due to the generality of $(x,y)\in B(w')$, this implies that $B(w')\subseteq B(w)\sqcup\{e\}$. Thus, we have $B(w)\subset B(w'')\subset B(w')\subseteq B(w)\sqcup\{e\}$. But this implies that $\mathit{bcount}(w)<\mathit{bcount}(w'')<\mathit{bcount}(w')\leq\mathit{bcount}(w)+1$, which is impossible (because those number are integers). 

Thus, we have that $w$ is either the greatest or the second-greatest proper ancestor of $v$ such that $M(w)=M(v)$ and $w\leq\mathit{low}(u)$.

($\Leftarrow$) Let $(x,y)$ be a back-edge in $B(u)$. Then $x$ is a descendant of $u$, and therefore a descendant of $v$. Furthermore, $y$ is an ancestor of $\mathit{high}(u)$, and therefore $y\leq\mathit{high}(u)$, and therefore $y<v$ (due to $(1)$). Since $(x,y)$ is a back-edge, we have that $x$ is a descendant of $y$. Thus, $x$ is a common descendant of $v$ and $y$, and therefore $v$ and $y$ are related as ancestor and descendant. Then, $y<v$ implies that $y$ is a proper ancestor of $v$. This shows that $(x,y)\in B(v)$. Due to the generality of $(x,y)\in B(u)$, this implies that $B(u)\subseteq B(v)$.

Let $(x,y)$ be a back-edge in $B(w)$. Then $x$ is a descendant of $M(w)=M(v)$. Furthermore, $y$ is a proper ancestor of $w$, and therefore a proper ancestor of $v$. This shows that $(x,y)\in B(v)$. Due to the generality of $(x,y)\in B(w)$, this implies that $B(w)\subseteq B(v)$.
Since $w\leq\mathit{low}(u)$, we infer that $B(u)\cap B(w)=\emptyset$ (because the lower endpoint of every back-edge in $B(u)$ is not low enough to be a proper ancestor of $w$).

Now, since $B(u)\subseteq B(v)$, $B(w)\subseteq B(v)$, $B(u)\cap B(w)=\emptyset$, and $\mathit{bcount}(v)=\mathit{bcount}(u)+\mathit{bcount}(w)+1$, we infer that there is a back-edge $e$ such that $B(v)=(B(u)\sqcup B(w))\sqcup\{e\}$.
\end{proof}

\begin{algorithm}[H]
\caption{\textsf{Compute all Type-3$\beta$ii-$4$ $4$-cuts of the form $\{(u,p(u)),(v,p(v)),(w,p(w)),e\}$, where $w$ is an ancestor of $v$, $v$ is an ancestor of $u$, $M(B(v)\setminus\{e\})= M(v)$, $\mathit{high}_1(v)=\mathit{high}(u)$, and the lower endpoint of $e$ is distinct from $\mathit{high}_1(v)$}}
\label{algorithm:type3-b-ii-4-4}
\LinesNumbered
\DontPrintSemicolon
\ForEach{$v\neq r$ such that $\mathit{nextM}(v)\neq\bot$}{
\label{line:type4-b-ii-4-4-for-1}
  compute $U_4^4(v)$\;  
}
\ForEach{$v\neq r$ such that $\mathit{nextM}(v)\neq\bot$}{
  let $\widetilde{U}_4^4(v)\leftarrow U_4^4(v)$\;
}
\ForEach{$v\neq r$ such that $\mathit{nextM}(v)\neq\bot$}{
\label{line:type4-b-ii-4-4-for-2}
  \If{$U_4^4(v)\neq\emptyset$}{
    let $\tilde{u}$ be the greatest vertex in $U_4^4(v)$\;
    let $u$ be the predecessor of $\tilde{u}$ in $H(\mathit{high}_1(v))$\;
    \If{$u\neq\bot$ \textbf{and} $u$ is a proper descendant of $v$}{
    \label{line:type4-b-ii-4-4-cond-u}
      insert $u$ into $\widetilde{U}_4^4(v)$\;
    }
  }
}
\ForEach{$v\neq r$ such that $\mathit{nextM}(v)\neq\bot$}{
  \ForEach{$u\in \widetilde{U}_4^4(v)$}{
    let $w$ be the greatest proper ancestor of $v$ with $M(w)=M(v)$ such that $w\leq\mathit{low}(u)$\;
    \label{line:type4-b-ii-4-4-w}    
    \If{$\mathit{bcount}(v)=\mathit{bcount}(u)+\mathit{bcount}(w)+1$}{
      mark $\{(u,p(u)),(v,p(v)),(w,p(w)),e(u,v,w)\}$ as a $4$-cut\; 
      \label{line:type4-b-ii-4-4-mark1}
    }
    $w\leftarrow{nextM}(w)$\;
    \label{line:type4-b-ii-4-4-w-2}    
    \If{$\mathit{bcount}(v)=\mathit{bcount}(u)+\mathit{bcount}(w)+1$}{
    \label{line:type4-b-ii-4-4-cond-2}
      mark $\{(u,p(u)),(v,p(v)),(w,p(w)),e(u,v,w)\}$ as a $4$-cut\; 
      \label{line:type4-b-ii-4-4-mark2}
    }    
  }
}
\end{algorithm}

\begin{proposition}
\label{proposition:algorithm:type3-b-ii-4-4}
Algorithm~\ref{algorithm:type3-b-ii-4-4} computes a collection of Type-3$\beta$ii-$4$ $4$-cuts, which includes all Type-3$\beta$ii-$4$ $4$-cuts of the form $\{(u,p(u)),(v,p(v)),(w,p(w)),e\}$, where $w$ is a proper ancestor of $v$, $v$ is a proper ancestor of $u$, $M(B(v)\setminus\{e\})= M(v)$, $\mathit{high}_1(v)=\mathit{high}(u)$, and the lower endpoint of $e$ is distinct from $\mathit{high}_1(v)$. Furthermore, it has a linear-time implementation.
\end{proposition}
\begin{proof}
Let $C=\{(u,p(u)),(v,p(v)),(w,p(w)),e\}$ be a Type-3$\beta$ii-$4$ $4$-cut such that $w$ is a proper ancestor of $v$, $v$ is a proper ancestor of $u$, $M(B(v)\setminus\{e\})= M(v)$, $\mathit{high}_1(v)=\mathit{high}(u)$, and the lower endpoint of $e$ is distinct from $\mathit{high}_1(v)$. Lemma~\ref{lemma:type-3-b-i-4-edge} implies that $e=e(u,v,w)$. Since $C$ is a Type-3$\beta$ii-$4$ $4$-cut, we have $B(v)=(B(u)\sqcup B(w))\sqcup\{e\}$ and $M(B(v)\setminus\{e\})=M(w)$. Since $M(B(v)\setminus\{e\})=M(v)$, this implies that $M(w)=M(v)$. Since the lower endpoint of $e$ is distinct from $\mathit{high}_1(v)$, Lemma~\ref{lemma:type3-b-ii-4-4-in-U} implies that $U_4^4(v)$ is not empty, and either $u\in U_4^4(v)$, or $u$ is the predecessor of $\tilde{u}$ in $H(\mathit{high}_1(v))$, where $\tilde{u}$ is the greatest vertex in $U_4^4(v)$. Thus, we have $u\in\widetilde{U}_4^4(v)$. By Lemma~\ref{lemma:type3-b-ii-4-4-criterion} we have that $\mathit{bcount}(v)=\mathit{bcount}(u)+\mathit{bcount}(w)+1$, and $w$ is either the greatest or the second-greatest proper ancestor of $v$ such that $M(w)=M(v)$ and $w\leq\mathit{low}(u)$. Thus, if $w$ is the greatest proper ancestor of $v$ such that $M(w)=M(v)$ and $w\leq\mathit{low}(u)$, then $C$ satisfies enough conditions to be marked in Line~\ref{line:type4-b-ii-4-4-mark1}. Otherwise, let $w'$ be the greatest proper ancestor of $v$ such that $M(w')=M(v)$ and $w'\leq\mathit{low}(u)$. Consider the vertex $w''=\mathit{nextM}(w')$. Then we have $w''<w'$ and $M(w'')=M(w')$. Thus, $w''$ is a proper ancestor of $w'$, and therefore $w''$ is a proper ancestor of $v$ with $w''<w'\leq\mathit{low}(u)$. Since $w''$ is the greatest vertex with $M(w'')=M(w')$ that is lower than $w'$, this means that $w''$ is the second-greatest proper ancestor of $v$ with $M(w'')=M(v)$ and $w''\leq\mathit{low}(u)$. Thus, we have $w=w''$, and therefore $C$ satisfies enough condition to be marked in Line~\ref{line:type4-b-ii-4-4-mark2}. 

Conversely, let $C=\{(u,p(u)),(v,p(v)),(w,p(w)),e(u,v,w)\}$ be a $4$-element set that is marked in Line~\ref{line:type4-b-ii-4-4-mark1} or \ref{line:type4-b-ii-4-4-mark2}. In either case, we have that $u$ is in $\widetilde{U}_4^4(v)$. This means that either $u\in U_4^4(v)$, or $u$ is the predecessor of $\tilde{u}$ in $H(\mathit{high}_1(v))$, where $\tilde{u}$ is the greatest vertex in $U_4^4(v)$. Then, since $u\in H(\mathit{high}_1(v))$, we have $\mathit{high}(u)=\mathit{high}_1(v)$, and therefore $\mathit{high}(u)<v$. If $u\in U_4^4(v)$, then by definition we have that $u$ is a proper descendant of $v$. Otherwise, since the condition in Line~\ref{line:type4-b-ii-4-4-cond-u} is satisfied (during the processing of $v$), we have that $u$ is a proper descendant of $v$. 

Let us suppose first that $C$ is marked in Line~\ref{line:type4-b-ii-4-4-mark1}. Then we have $\mathit{bcount}(v)=\mathit{bcount}(u)+\mathit{bcount}(w)+1$, and $w$ is the greatest proper ancestor of $v$ with $M(w)=M(v)$ and $w\leq\mathit{low}(u)$. Thus, all the conditions of Lemma~\ref{lemma:type3-b-ii-4-4-criterion} are satisfied, and so we have that there is a back-edge $e$ such that $B(v)=(B(u)\sqcup B(w))\sqcup\{e\}$. Then Lemma~\ref{lemma:type-3-b-i-4-edge} implies that $e=e(u,v,w)$. Thus, Lemma~\ref{lemma:type-3b-cases} implies that $C$ is a Type-3$\beta$ $4$-cut.

Now let us suppose that $C$ is marked in Line~\ref{line:type4-b-ii-4-4-mark2}. Let $w'$ be the greatest proper ancestor of $v$ with $M(w')=M(v)$ and $w'\leq\mathit{low}(u)$. Then we have $w=\mathit{nextM}(w')$. This means that $w$ is the greatest vertex with $M(w)=M(w')$ and $w<w'$. This implies that $w$ is a proper ancestor of $w'$, and therefore a proper ancestor of $v$. Furthermore, we have $w<w'\leq\mathit{low}(u)$. This shows that $w$ is the second-greatest proper ancestor of $v$ such that $M(w)=M(v)$ and $w\leq\mathit{low}(u)$. Since we have met the condition in Line~\ref{line:type4-b-ii-4-4-cond-2}, we have $\mathit{bcount}(v)=\mathit{bcount}(u)+\mathit{bcount}(w)+1$. Thus, all the conditions of Lemma~\ref{lemma:type3-b-ii-4-4-criterion} are satisfied, and so we have that there is a back-edge $e$ such that $B(v)=(B(u)\sqcup B(w))\sqcup\{e\}$. Then Lemma~\ref{lemma:type-3-b-i-4-edge} implies that $e=e(u,v,w)$. Thus, Lemma~\ref{lemma:type-3b-cases} implies that $C$ is a Type-3$\beta$ $4$-cut. 

Now we will argue about the complexity of Algorithm~\ref{algorithm:type3-b-ii-4-4}. By Lemma~\ref{lemma:algorithm:type3-b-ii-4-4-U} we have that the sets $U_4^4(v)$ can be computed in linear time in total, for all vertices $v\neq r$ such that $\mathit{nextM}(v)\neq\bot$. Thus, the \textbf{for} loop in Line~\ref{line:type4-b-ii-4-4-for-1} can be performed in linear time. In particular, we have that the total size of all $U_4^4$ sets is $O(n)$. Thus, the \textbf{for} loop in Line~\ref{line:type4-b-ii-4-4-for-2} takes $O(n)$ time. In order to compute the vertex $w$ in Line~\ref{line:type4-b-ii-4-4-w}, we use Algorithm~\ref{algorithm:W-queries}. We have showed previously how to generate the appropriate queries that provide $w$ (see e.g., the proof of Proposition~\ref{proposition:algorithm:type3-b-ii-4-3}). Since the number of all those queries is $O(n)$ (because it is bounded by the total size of all sets of the form $\widetilde{U}_4^4(v)$), by Lemma~\ref{lemma:W-queries} we have that Algorithm~\ref{algorithm:W-queries} can answer all of them in $O(n)$ time. We conclude that Algorithm~\ref{algorithm:type3-b-ii-4-4} runs in linear time. 
\end{proof}

\begin{figure}[h!]\centering
\includegraphics[trim={0cm 11cm 0cm 1cm}, clip=true, width=0.8\linewidth]{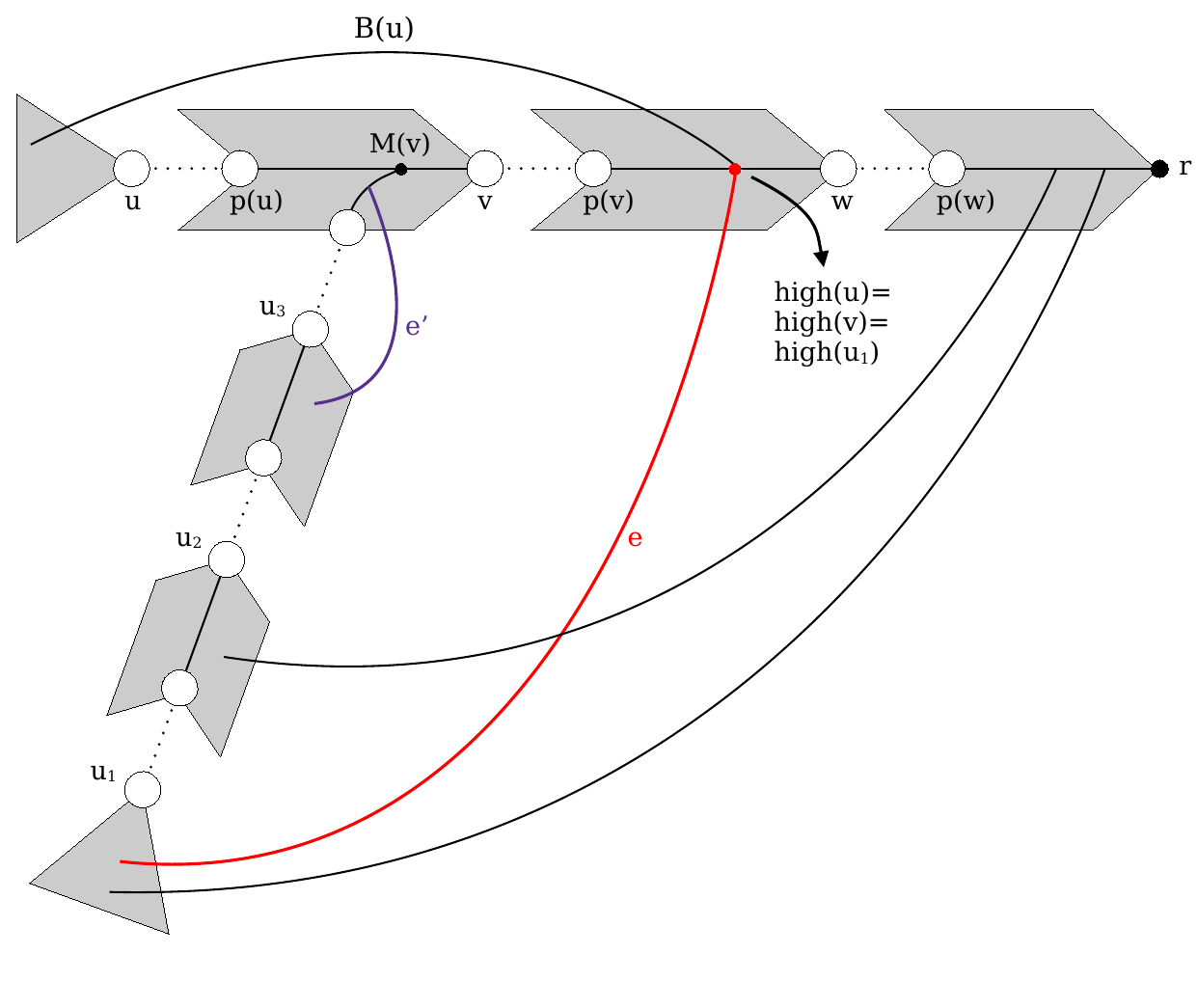}
\caption{\small{In this example we have that $\{(u,p(u)),(v,p(v)),(w,p(w)),e\}$ is a Type-$3\beta$ii-$4$ $4$-cut, such that $M(v)=M(B(v)\setminus\{e\})$, $\mathit{high}_1(v)=\mathit{high}_1(u)$ and the lower endpoint of $e$ is $\mathit{high}_1(v)$. Notice that $u\notin S(v)$, because $u_1$ and $u_2$ also have $\mathit{high}_1(u_1)=\mathit{high}_1(u_2)=\mathit{high}_1(v)$. (I.e., $u$ does not belong to a segment of $H(\mathit{high}_1(v))$ that contains $v$ and has the property that its elements are related as ancestor and descendant.) Also, it is not necessarily true that $\mathit{high}_2(u)=\mathit{high}_2(v)$, because we may have $\mathit{high}_2(u)<\mathit{high}_1(u)$, whereas $\mathit{high}_2(v)=\mathit{high}_1(v)$. However, even if we have $\mathit{high}_2(u)=\mathit{high}_2(v)$, then $u\notin\widetilde{S}(v)$, since $\mathit{high}_2(u_3)=\mathit{high}_2(v)=\mathit{high}_1(v)$. Thus, in this situation we cannot use the same techniques that we used so far in order to compute the Type-$3\beta$ii $4$-cuts.}}\label{figure:type3bii4}
\end{figure}

According to Proposition~\ref{proposition:algorithm:type3-b-ii-4-4}, Algorithm~\ref{algorithm:type3-b-ii-4-4} computes all Type-3$\beta$ii-$4$ $4$-cuts of the form $\{(u,p(u)),(v,p(v)),(w,p(w)),e\}$, where $u$ is a descendant of $v$, $v$ is a descendant of $w$, $M(B(v)\setminus\{e\})= M(v)$, $\mathit{high}_1(v)=\mathit{high}(u)$, and the lower endpoint of $e$ is distinct from $\mathit{high}_1(v)$. It remains to show how to compute all such $4$-cuts in the case where the lower endpoint of $e$ is $\mathit{high}_1(v)$. For this case, we cannot use directly any of our techniques so far, because these rely on the fact that $u$ and $v$ belong to the same segment of $H(\mathit{high}_1(v))$ or $\widetilde{H}(\mathit{high}_2(v))$ that is maximal w.r.t. the property that its elements are related as ancestor and descendant. However, in this particular case, this is not necessarily true. (See Figure~\ref{figure:type3bii4} for an example.)

Notice that, if such a $4$-cut exists, then we have that there are two distinct back-edges with the same lower endpoint: i.e., the back-edge $e$, and one of the back-edges in $B(u)$ whose lower endpoint is $\mathit{high}(u)=\mathit{high}_1(v)$. Thus, if we had the property that no two back-edges have the same lower endpoint, then this case would not arise. We basically rely on this observation. Thus, we will perform the computation on a different -- but ``$4$-cut-equivalent" -- graph, that has a DFS-tree in which no two back-edges that correspond to edges of the original graph can have the same lower endpoint. We construct this graph through repeated application of the following vertex-splitting operation.

\begin{figure}[h!]\centering
\includegraphics[trim={0 21cm 0 0}, clip=true, width=0.8\linewidth]{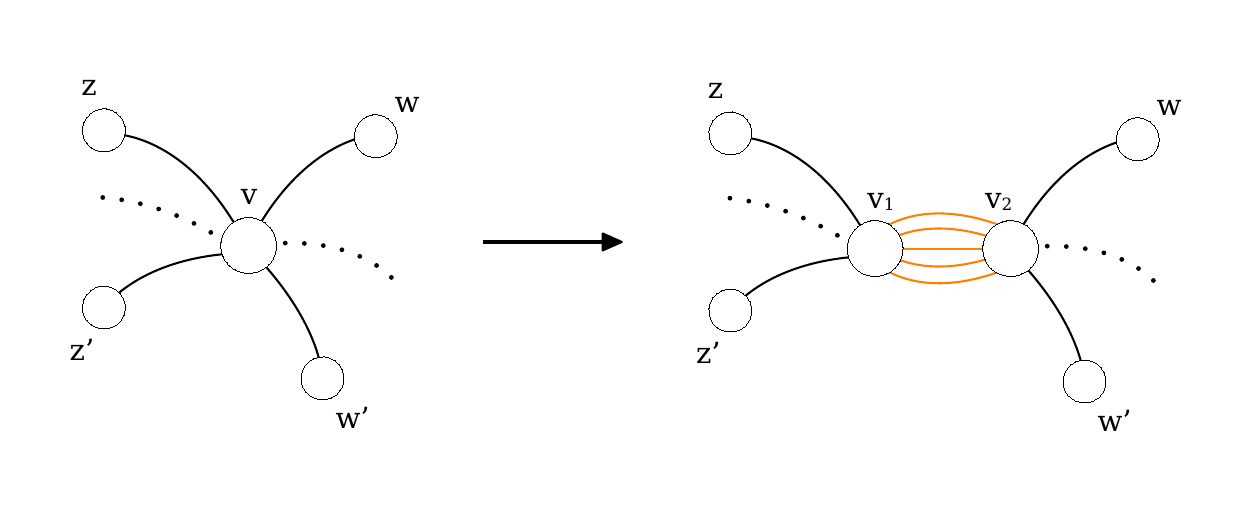}
\caption{\small{Splitting a vertex $v$ at $(E_1,E_2)$ as $v_1$ and $v_2$, where $E_1=\{(v,z),\dots,(v,z')\}$ and $E_2=\{(v,w),\dots,(v,w')\}$.}}\label{figure:split_vertex}
\end{figure}

\begin{definition}[Vertex Splitting]
\normalfont{Let $v$ be a vertex of $G$, and let $(E_1,E_2)$ be an ordered bipartition of $\partial(v)$. Let $G'$ be the graph that is formed from $G$ by replacing $v$ with two vertices $v_1$ and $v_2$, and by inserting five multiple edges of the form $(v_1,v_2)$, one edge $(v_1,z)$ for every $z$ such that there is an edge $(v,z)\in E_1$, and one edge $(v_2,z)$ for every $z$ such that there is an edge $(v,z)\in E_2$. Then, $G'$ is called \emph{the graph that is derived from $G$ by splitting $v$ at $(E_1,E_2)$ as $v_1$ and $v_2$}. (See Figure~\ref{figure:split_vertex}.) 

We also define the corresponding mapping of edges $f:E(G)\rightarrow E(G')$ as follows. If $(x,y)$ is an edge of $G$ such that none of $x,y$ is $v$, then $f((x,y))=(x,y)$. Otherwise, if, say, $x=v$, then $(x,y)$ belongs to one of the sets $E_1$ or $E_2$. If $(x,y)\in E_1$, then $f((x,y))=(v_1,y)$. Otherwise, $f((x,y))=(v_2,y)$.\footnote{A more precise definition of $f$ would require that it maintains the unique identifiers of the multiple edges. For the sake of simplicity, however, we omit this consideration from the definition and from the arguments that follow.}}
\end{definition} 

A graph $G'$ that is derived from $G$ by splitting a vertex maintains all $4$-cuts, as shown in the following.

\begin{lemma}
\label{lemma:splitting}
Let $G'$ be the graph that is derived from $G$ by splitting a vertex $v$ at $(E_1,E_2)$ as $v_1$ and $v_2$, and let $f:E(G)\rightarrow E(G')$ be the corresponding mapping of edges. Then, $G'$ is $3$-edge-connected. Furthermore, if $C$ is a $4$-cut of $G$, then $f(C)$ is a $4$-cut of $G'$. Conversely, if $C'$ is a $4$-cut of $G'$, then none of the edges in $C'$ has the form $(v_1,v_2)$, and $f^{-1}(C')$ is a $4$-cut of $G$.
\end{lemma}
\begin{proof}
In order to prove this lemma, we establish a correspondence between paths in $G$ and paths in $G'$. This basically works by replacing every part of a path in $G$ that passes from $v$, with a part in $G'$ that passes from $v_1$ or $v_2$. More precisely, we define the correspondence as follows.

Let $P=x_1,e_1,x_2,\dots,e_{k-1},x_k$ be a path in $G$. We perform the following substitutions.
\begin{itemize}
\item{If $x_1=v$, then $e_1$ has the form $(v,z)$, and we have that either $e_1\in E_1$ or $e_1\in E_2$. If $e_1\in E_1$, then we replace the part $x_1,e_1$ in $P$ with $v_1,(v_1,z)$. Otherwise, if $e_1\in E_2$, then we replace the part $x_1,e_1$ in $P$ with $v_2,(v_2,z)$.}
\item{If $x_k=v$, then $e_{k-1}$ has the form $(z,v)$, and we have that either $e_{k-1}\in E_1$ or $e_{k-1}\in E_2$. If $e_{k-1}\in E_1$, then we replace the part $e_{k-1},v$ in $P$ with $(z,v_1),v_1$. Otherwise, if $e_{k-1}\in E_2$, then we replace the part $e_{k-1},v$ in $P$ with $(z,v_2),v_2$.}
\item{If there is a part $e_i,v,e_{i+1}$ in $P$, for some $i\in\{1,\dots,k-2\}$, then we have that the edge $e_i$ has the form $(z,v)$, the edge $e_{i+1}$ has the form $(v,z')$, and there are four different cases to consider, depending on whether $e_i$ and $e_{i+1}$ are in $E_1$ or $E_2$. If $e_i\in E_1$ and $e_{i+1}\in E_1$, then we replace the part $(z,v),v,(v,z')$ with $(z,v_1),v_1,(v_1,z')$. If $e_i\in E_1$ and $e_{i+1}\in E_2$, then we replace the part $(z,v),v,(v,z')$ with $(z,v_1),v_1,(v_1,v_2),v_2,(v_2,z')$. If $e_i\in E_2$ and $e_{i+1}\in E_1$, then we replace the part $(z,v),v,(v,z')$ with $(z,v_2),v_2,(v_2,v_1),v_1,(v_1,z')$. And if $e_i\in E_2$ and $e_{i+1}\in E_2$, then we replace the part $(z,v),v,(v,z')$ with $(z,v_2),v_2,(v_2,z')$. }
\end{itemize}
We denote the resulting sequence as $P'$. Observe the following facts.

\begin{enumerate}
\item{$P'$ is a path in $G'$.}
\item{Every occurrence of a vertex $x\neq v$ in $P$ is maintained in $P'$.}
\item{Every occurrence of $v$ in $P$, is substituted with either $v_1$, or $v_2$, or $v_1,(v_1,v_2),v_2$, or $v_2,(v_2,v_1),v_1$.}
\item{Every occurrence of an edge $e$ in $P$, is substituted with $f(e)$.}
\item{Every part $e,v,e'$ in $P$ (where $e$ and $e'$ are edges), is substituted with either $f(e),v_1,f(e')$, or $f(e),v_2,f(e')$, or $f(e),v_1,(v_1,v_2),v_2,f(e')$, or $f(e),v_2,(v_2,v_1),v_1,f(e')$.}
\end{enumerate}

Conversely, let $Q=x_1,e_1,x_2,\dots,e_{k-1},x_k$ be a path in $G'$. We perform the following substitutions.
\begin{itemize}
\item{If $x_i$ is $v_1$ or $v_2$, and $x_{i+1}$ is not $v_1$ or $v_2$, for some $i\in\{1,\dots,k-2\}$, then $x_i,e_i$ is replaced with $v,(v,x_{i+1})$.}
\item{If $x_i$ is $v_1$ or $v_2$, and $x_{i-1}$ is not $v_1$ or $v_2$, for some $i\in\{2,\dots,k-1\}$, then $e_{i-1},x_i$ is replaced with $(x_{i-1},v),v$.}
\item{Every maximal segment of the form $v_i,(v_i,v_j),v_j$, for $i,j\in\{1,2\}$, is replaced with $v$.}
\end{itemize}

We denote the resulting sequence as $\widetilde{Q}$. (Notice that after performing simultaneously the above substitutions, there may appear some segments of the form $v,v,v,\dots$. We replace those maximal segments with $v$, so that we indeed have a path. For example, $Q$ may contain the segment $v_1,(v_1,v_2),v_2,(v_2,z)$, where $z\notin\{v_1,v_2\}$. Then we replace this segment with $v,(v,z)$ in $\widetilde{Q}$.) Observe the following facts.

\begin{enumerate}
\item{$\widetilde{Q}$ is a path in $G$.}
\item{Every occurrence of a vertex $x\notin\{v_1,v_2\}$ in $Q$ is maintained in $\widetilde{Q}$.}
\item{Every occurrence of $v_1$ or $v_2$ in $Q$, is substituted with $v$.}
\item{Every occurrence of an edge $e\neq(v_1,v_2)$ in $Q$, is substituted with $f^{-1}(e)$.}
\end{enumerate}

Now it is easy to see why $G'$ is connected. Let $x$ and $y$ be two distinct vertices in $G'$, none of which is either $v_1$ or $v_2$. Then, since $G$ is connected, there is a path $P$ in $G$ from $x$ to $y$. Then, $P'$ is a path in $G'$ from $x$ to $y$, and so $x$ and $y$ are connected in $G'$. The existence of the edges of the form $(v_1,v_2)$ in $G'$ shows that $v_1$ and $v_2$ are connected in $G'$. Finally, since $\{E_1,E_2\}$ is a bipartition of $\partial(v)$, we have that both $E_1$ and $E_2$ are non-empty. So let $(v,z)$ be an edge in $E_1$. Then, there is an edge of the form $(v_1,z)$ in $G'$, and so $v_1$ is connected with the vertices in $G'\setminus\{v_1,v_2\}$. Furthermore, let $(v,z)$ be an edge in $E_2$. Then, there is an edge of the form $(v_2,z)$ in $G'$, and so $v_2$ is connected with the vertices in $G'\setminus\{v_1,v_2\}$. This shows that $G'$ is connected.

Now let us suppose, for the sake of contradiction, that $G'$ is not $3$-edge-connected. This means that there is a $k$-edge cut $C$ of $G'$, for some $k\leq 2$. This implies that the endpoints of any edge in $C$ are not connected in $G'\setminus C$. Thus, we have that no edge of the form $(v_1,v_2)$ is contained in $C$ (because there are five edges of this form, and so all of them must be removed in order to disconnect $v_1$ from $v_2$). Thus, the set of edges $f^{-1}(C)$ is defined. Then, since $G$ is $3$-edge-connected, we have that $G\setminus f^{-1}(C)$ is connected (since $|f^{-1}(C)|=|C|\leq 2$). Since $C$ is a $k$-edge cut of $G'$, we have that $G'\setminus C$ consists of two connected components $S_1$ and $S_2$. Let $x$ be a vertex in $S_1$ and let $y$ be a vertex in $S_2$. If $x\in\{v_1,v_2\}$ then we let $x'$ denote $v$; otherwise, if $x\notin\{v_1,v_2\}$, then we let $x'$ denote $x$. Similarly, if $y\in\{v_1,v_2\}$, then we let $y'$ denote $v$; otherwise, if $y\notin\{v_1,v_2\}$, then we let $y'$ denote $y$. Then, since $G\setminus f^{-1}(C)$ is connected, we have that there is a path $P$ in $G\setminus f^{-1}(C)$ from $x'$ to $y'$. Then, the path $P'$ in $G'$ avoids the edges in $C$, and demonstrates that $x$ and $y$ are connected in $G'\setminus C$. (To see this, distinguish the following cases. If none of $x$ and $y$ is in $\{v_1,v_2\}$, then $P'$ is a path from $x$ to $y$, and the contradiction is clear. If, say, $x$ is $v_1$, then $P'$ is a path from either $v_1$ or $v_2$; but this distinction has no effect, since the existence of the edges of the form $(v_1,v_2)$ implies that $v_1$ is connected with $v_2$ in $G'\setminus C$. The same holds if $x$ is $v_2$, or if $y$ is either $v_1$ or $v_2$.) Thus, we have arrived at a contradiction. This shows that $G'$ is $3$-edge-connected.

Now let $C$ be a $4$-cut of $G$. This implies that $G\setminus C$ is split into two connected components $S_1$ and $S_2$, but $G\setminus C'$ is connected for every proper subset $C'$ of $C$. Let $x$ be a vertex in $S_1$, and let $y$ be a vertex in $S_2$. Then, there is no path from $x$ to $y$ in $G\setminus C$. Let us suppose, for the sake of contradiction, that $x'$ and $y'$ are connected in $G'\setminus f(C)$, where we let $x'$ denote $v_1$ if $x=v$, or $x$ if $x\neq v$; and similarly, we let $y'$ denote $v_1$ if $y=v$, or $y$ if $y\neq v$. Then there is a path $Q$ in $G'\setminus f(C)$ from $x'$ to $y'$. Consider the path $\widetilde{Q}$ in $G$. First, observe that $\widetilde{Q}$ is a path in $G\setminus C$ (since $Q$ is a path in $G'\setminus f(C)$). Furthermore, notice that $\widetilde{Q}$ is a path from $x$ to $y$. But this is impossible, since $x$ and $y$ are not connected in $G\setminus C$. Thus, we have that $G'\setminus f(C)$ is disconnected. Now let $C'$ be a proper subset of $C$. Let us suppose, for the sake of contradiction, that $G'\setminus f(C')$ is disconnected. Then, let $x$ and $y$ be two vertices that are not connected in $G'\setminus f(C')$. Notice that it cannot be that both $x$ and $y$ are in $\{v_1,v_2\}$, because there are five edges of the form $(v_1,v_2)$, whereas $|f(C')|=|C'|<4$. If none of $x$ and $y$ is either $v_1$ or $v_2$, then there is a path $P$ from $x$ to $y$ in $G\setminus C'$, and therefore there is a path $P'$ from $x$ to $y$ in $G'\setminus f(C')$, which is impossible. Thus, one of $x$ and $y$ is either $v_1$ or $v_2$. Let us assume w.l.o.g. that $x=v_1$. Then we have that $y\notin\{v_1,v_2\}$, and so there is a path $P$ from $v$ to $y$ in $G\setminus C'$, and therefore $P'$ is a path from either $v_1$ or $v_2$ to $y$ in $G'\setminus f(C')$, which is also impossible (since $v_1$ is connected with $v_2$ in $G'\setminus f(C')$, but $v_1$ is not connected with $y$ in $G'\setminus f(C')$). Thus, we have that $G'\setminus f(C')$ is also connected. Since this is true for every proper subset $C'$ of $C$, this shows that $f(C)$ is a $4$-cut of $G'$.

Conversely, let $C$ be a $4$-cut of $G'$. This implies that the endpoints of every edge in $C$ are disconnected in $G'\setminus C$. Thus, since $|C|=4$, we have that $C$ contains no edge of the form $(v_1,v_2)$ (since there are five of them in $G'$), and $v_1$ is connected with $v_2$ in $G'\setminus C$. Thus, $f^{-1}(C)$ is a set of four edges of $G$. Let us suppose, for the sake of contradiction, that $G\setminus f^{-1}(C)$ is connected. Since $C$ is a $4$-cut of $G'$, there are two vertices $x$ and $y$ that are disconnected in $G'\setminus C$. Let $x'$ denote $v$ if $x\in\{v_1,v_2\}$, and $x$ if $x\notin\{v_1,v_2\}$. Similarly, let $y'$ denote $v$ if $y\in\{v_1,v_2\}$, and $y$ if $y\notin\{v_1,v_2\}$. Then, since $G\setminus f^{-1}(C)$ is connected, there is a path $P$ from $x'$ to $y'$ in $G\setminus f^{-1}(C)$. Then, observe that $P'$ is a path in $G'$, that avoids the edges in $C$. Furthermore, if $x'=x$ and $y'=y$, then $P'$ is a path from $x$ to $y$, which is impossible, since $x$ and $y$ are disconnected in $G'\setminus C$. Otherwise, let us assume w.l.o.g. that $x'=v$. Then, $P'$ is a path from either $v_1$ or $v_2$ to $y$, which demonstrates that $x$ is connected with $y$ in $G'\setminus C$ (since $v_1$ is connected with $v_2$ in $G'\setminus C$). This is impossible. Thus, we have that $G\setminus f^{-1}(C)$ is disconnected. Now, since $C$ is a $4$-cut of $G'$, we have that $G'\setminus C'$ is connected, for every proper subset $C'$ of $C$. Let us suppose, for the sake of contradiction, that $G\setminus f^{-1}(C')$ is disconnected, for a proper subset $C'$ of $C$. Then, there are two vertices $x$ and $y$ of $G$ that are disconnected in $G\setminus f^{-1}(C')$. If $x=v$, let $x'$ denote $v_1$; otherwise, let $x'$ denote $x$. Similarly, if $y=v$, let $y'$ denote $v_1$; otherwise, let $y'$ denote $y$. Then, since $G'\setminus C'$ is connected, we have that there is a path $Q$ from $x'$ to $y'$ in $G'\setminus C'$. Then, observe that $\widetilde{Q}$ is a path from $x$ to $y$ in $G\setminus f^{-1}(C')$, in contradiction to our supposition. Thus, we have that $G\setminus f^{-1}(C')$ is connected. Since this is true for every proper subset $C'$ of $C$, this shows that $f^{-1}(C)$ is a $4$-cut of $G$.
\end{proof}

\begin{figure}[h!]\centering
\includegraphics[trim={0cm 25cm 0cm 0cm}, clip=true, width=1\linewidth]{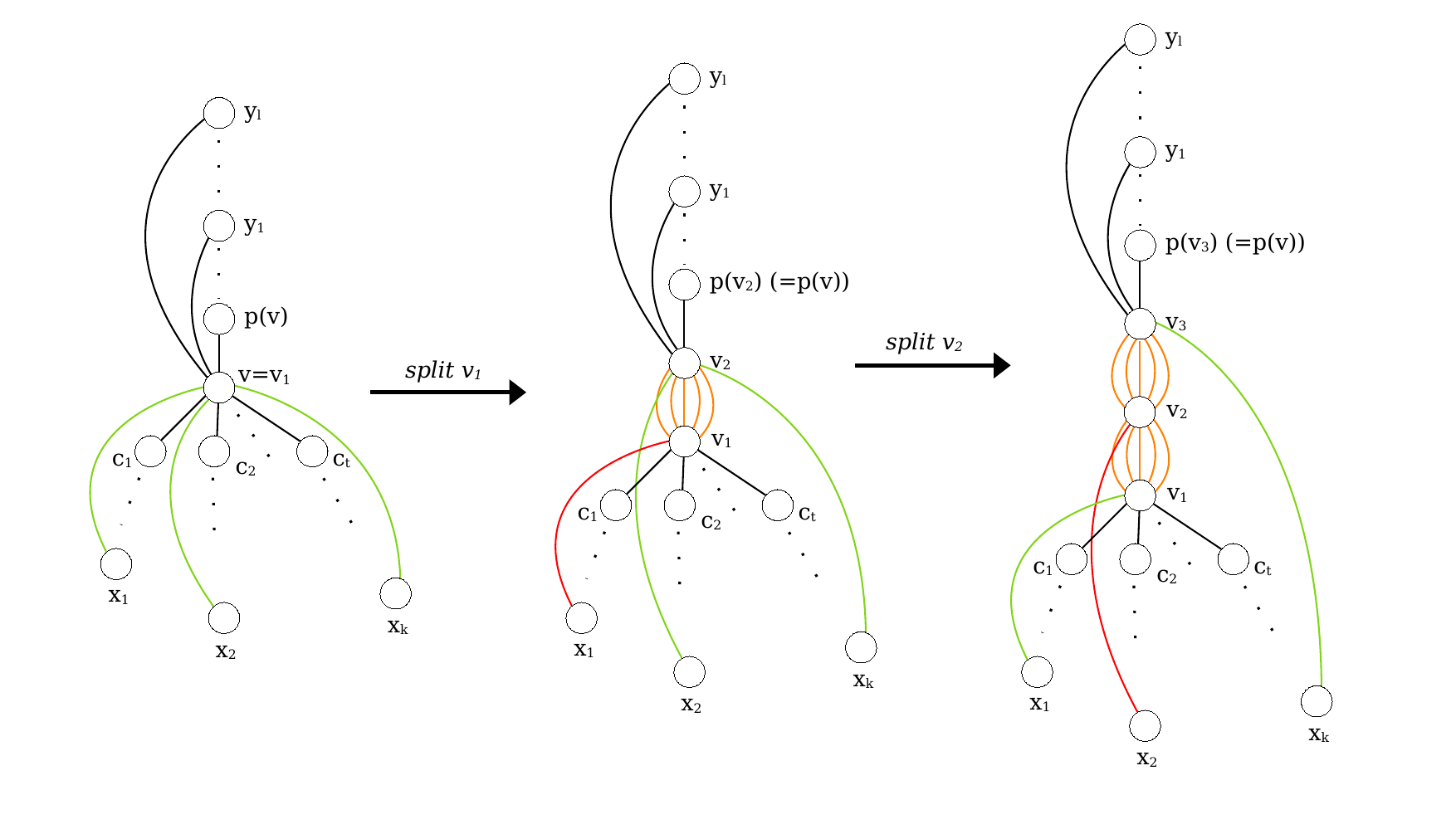}
\caption{\small{Splitting a vertex $v$ on a DFS-tree, so that the number of back-edges with lower endpoint $v$ is reduced by one. With green are shown the target back-edges (incoming to $v$) that we want to separate w.r.t. their lower endpoint. With red is shown the back-edge that was separated from the rest and is now unique with the property of having the lower endpoint that it has. With orange are shown the five multiple edges that join the two vertices into which the vertex was split. We note that one of those multiple edges is a tree-edge, and the rest are back-edges.}}\label{figure:split-dfs}
\end{figure}

Now the idea is to repeatedly split vertices on $T$, so that the resulting DFS-tree has the property that no two back-edges that correspond to back-edges of the original graph can have the same lower endpoint. So let $v$ be a vertex, and let $(x_1,v),\dots,(x_k,v)$ be all the incoming back-edges to $v$. We may assume that $v\neq r$ (because this is sufficient for our purposes). If $k=1$, then there is nothing to do, because there is only one back-edge whose lower endpoint is $v$. Otherwise, let $(c_1,v),\dots,(c_t,v)$ be the parent edges of the children of $v$ (if it has any), and let $(v,y_1),\dots,(v,y_l)$ be the back-edges that stem from $v$. Then, we have $\partial(v)=\{(x_1,v),\dots,(x_k,v)\}\cup\{(c_1,v),\dots,(c_t,v)\}\cup\{(v,y_1),\dots,(v,y_l)\}\cup\{(v,p(v))\}$. Let $E_1=\{(x_1,v)\}\cup\{(c_1,v),\dots,(c_t,v)\}$, and let $E_2=\partial(v)\setminus{E_1}$. Then, we split $v$ at $P=(E_1,E_2)$ as $v_1$ and $v_2$, while maintaining the DFS-tree structure (see Figure~\ref{figure:split-dfs}). To be specific, we detach $v$ from $T$, and we introduce the vertices $v_1$ and $v_2$ that replace $v$, such that $v_2=p(v_1)$. Then, $v_1$ inherits the children of $v$ and the back-edge $(x_1,v)$ (as $(x_1,v_1)$), and $v_2$ inherits the remaining edges from $\partial(v)$. Thus, we introduce the parent edges $(c_1,v_1),\dots,(c_t,v_1)$, four back-edges of the form $(v_1,v_2)$ (because we already have $(v_1,v_2)$ as a parent edge), we set $p(v_2)\leftarrow p(v)$ (where $p(v)$ was the parent of $v$ before its deletion), and we also put back the remaining back-edges from $\partial(v)$ as $(x_2,v_2),\dots,(x_k,v_2)$ and $(v_2,y_1),\dots,(v_2,y_l)$. We refer to Figure~\ref{figure:split-dfs} for a depiction of this process.

In this way, we have achieved the following things. First, we have maintained a DFS-tree for the graph that is derived from $G$ by splitting $v$ at $P$, and second, we have effectively reduced the number of back-edges with lower endpoint $v$ by one (i.e., now all of these, except one, have $v_2$ as their lower endpoint). Now, if $k>2$, then we continue this process, by splitting $v_2$, until eventually we have separated the back-edges with lower endpoint $v$ into back-edges with different lower endpoints. Then, we continue this process for all vertices $\neq r$. Let $G'$ be the resulting graph, and let $T'$ be the corresponding DFS-tree. Notice that, by Lemma~\ref{lemma:splitting}, all $4$-cuts of $G$ that contain at least one back-edge (w.r.t. $T$) whose lower endpoint is not $r$, correspond to $4$-cuts of $G'$ that also have at least one back-edge (w.r.t. $T'$). Thus, we can compute all Type-3$\beta$ii-$4$ $4$-cuts of $G$ of the form $\{(u,p(u)),(v,p(v)),(w,p(w)),e\}$, where $u$ is a descendant of $v$, $v$ is a descendant of $w$, $M(B(v)\setminus\{e\})= M(v)$, $\mathit{high}_1(v)=\mathit{high}(u)$, and the lower endpoint of $e$ is $\mathit{high}_1(v)$, by simply computing all $4$-cuts of $G'$ that contain three tree-edges and one back-edge, using the algorithms that we have developed so far (since our analysis covers all cases, except this one, which may only arise due to the existence of at least two back-edges with the same lower endpoint).

It remains to show how we can efficiently perform all those splittings in linear time in total. This can be done easily, because it is essentially sufficient to determine the final vertices and the edges of $G'$, and the parent relation of the vertices of $G'$. Then we may run a DFS with start vertex $r$, in order to assign a DFS numbering, determine the back-edges, and compute the DFS parameters that we need.

Let us describe in detail how to construct the vertices and the edges of the final graph $G'$. First, for every vertex $v$, let $\mathit{in}(v)$ denote the number of incoming back-edges to $v$. Then, for every vertex $v\neq r$ such that $\mathit{in}(v)>1$, we will perform $\mathit{in}(v)-1$ splittings of $v$. This will substitute $v$ with $\mathit{in}(v)$ copies of it, which we denote as $v_1,v_2,\dots,v_{\mathit{in}(v)}$. Every one of the vertices $v_1,\dots,v_{\mathit{in}(v)}$ will be used in order to inherit one of the incoming back-edges to $v$. For every $i\in\{1,\dots,\mathit{in}(v)-1\}$, we create five multiple edges of the form $(v_i,v_{i+1})$; one of these edges will be the parent edge of $v_i$. Among the $\mathit{in}(v)$ copies of $v$, we let $v_1$ inherit the children edges of $v$, and we let $v_{\mathit{in}(v)}$ inherit the outgoing back-edges from $v$ and the parent edge $(v,p(v))$ (if $v\neq r$). Thus, let $c$ be a child of $v$. Then the parent edge of $c$ is replaced with $(c_{\mathit{in}(c)},v_1)$ if $\mathit{in}(c)>1$, or with $(c,v_1)$ otherwise. Now let $e_1,\dots,e_{\mathit{in}(v)}$ be the list of the incoming back-edges to $v$. Let $i$ be an index in $\{1,\dots,\mathit{in}(v)\}$, and let $e_i=(x,v)$. Then $e_i$ is replaced with $(x_{\mathit{in}(x)},v_i)$ if $\mathit{in}(x)>1$, or with $(x,v_i)$ otherwise. It is not difficult to see that this construction can be completed in linear time.

\bibliography{text}

\end{document}